\documentclass[12pt,a4paper]{article}

\newcommand{\mrm}[1]{\mathrm{#1}}
\newcommand{\mbf}[1]{\mathbf{#1}}
\newcommand{\mtt}[1]{\mathtt{#1}}
\newcommand{\tsc}[1]{\textsc{#1}}
\newcommand{\tbf}[1]{\textbf{#1}}
\newcommand{\ttt}[1]{\texttt{#1}}
 
\newcommand{\Py}{\tsc{Pythia}}
\newcommand{\Je}{\tsc{Jetset}}
 
\newcommand{\alphas}{\alpha_{\mathrm{s}}}
\newcommand{\alphaem}{\alpha_{\mathrm{em}}}
\newcommand{\ssintw}{\sin^2 \! \theta_W}
\newcommand{\scostw}{\cos^2 \! \theta_W}
\newcommand{\kT}{k_{\perp}}
\newcommand{\pT}{p_{\perp}}
\newcommand{\pTmin}{p_{\perp\mathrm{min}}}
\newcommand{\MSbar}{$\overline{\mbox{\textsc{ms}}}$}
\newcommand{\br}[1]{\overline{#1}}

\newcommand{\gtrsim}{\raisebox{-0.8mm}%
{\hspace{1mm}$\stackrel{>}{\sim}$\hspace{1mm}}}
 
\renewcommand{\a}{\mathrm{a}}
\renewcommand{\b}{\mathrm{b}}
\renewcommand{\c}{\mathrm{c}}
\renewcommand{\d}{\mathrm{d}}
\newcommand{\e}{\mathrm{e}}
\newcommand{\f}{\mathrm{f}}
\newcommand{\g}{\mathrm{g}}
\newcommand{\hrm}{\mathrm{h}}

\newcommand{\n}{\mathrm{n}}
\newcommand{\p}{\mathrm{p}}
\newcommand{\q}{\mathrm{q}}
\newcommand{\s}{\mathrm{s}}
\renewcommand{\t}{\mathrm{t}}
\renewcommand{\u}{\mathrm{u}}
\newcommand{\A}{\mathrm{A}}
\newcommand{\B}{\mathrm{B}}
\newcommand{\D}{\mathrm{D}}
\newcommand{\F}{\mathrm{F}}
\newcommand{\G}{\mathrm{G}}
\renewcommand{\H}{\mathrm{H}}
\newcommand{\J}{\mathrm{J}}
\newcommand{\K}{\mathrm{K}}
\renewcommand{\L}{\mathrm{L}}
\newcommand{\Q}{\mathrm{Q}}
\newcommand{\R}{\mathrm{R}}

\newcommand{\W}{\mathrm{W}}
\newcommand{\Z}{\mathrm{Z}}
\newcommand{\bbar}{\overline{\mathrm{b}}}
\newcommand{\cbar}{\overline{\mathrm{c}}}
\newcommand{\dbar}{\overline{\mathrm{d}}}
\newcommand{\fbar}{\overline{\mathrm{f}}}
\newcommand{\pbar}{\overline{\mathrm{p}}}
\newcommand{\qbar}{\overline{\mathrm{q}}}
\newcommand{\sbar}{\overline{\mathrm{s}}}
\newcommand{\tbar}{\overline{\mathrm{t}}}
\newcommand{\ubar}{\overline{\mathrm{u}}}
\newcommand{\Dbar}{\overline{\mathrm{D}}}
\newcommand{\Fbar}{\overline{\mathrm{F}}}
\newcommand{\Qbar}{\overline{\mathrm{Q}}}
\newcommand{\qsea}{\ensuremath{\q_{\mrm{s}}}}
\newcommand{\qval}{\ensuremath{\q_{\mrm{v}}}}
\newcommand{\qcmp}{\ensuremath{\q_{\mrm{c}}}}

\newcommand{\sq}{\tilde{\mathrm q}}
\newcommand{\sqs}{\tilde{\mathrm q}^*}
\newcommand{\sqbar}{\overline{\tilde{\mathrm{q}}}}
\newcommand{\sg}{\tilde{\mathrm{g}}}
\newcommand{\tp}{\tilde{\mathrm t}}
\newcommand{\tm}{\tilde{\mathrm t}^*}
\newcommand{\sd}{\tilde{\mathrm d}}
\newcommand{\su}{\tilde{\mathrm u}}
\newcommand{\sch}{\tilde{\mathrm c}}
\newcommand{\sst}{\tilde{\mathrm s}}
\newcommand{\st}{\tilde{\mathrm t}}
\newcommand{\sbo}{\tilde{\mathrm b}}
\newcommand{\sbs}{\tilde{\mathrm b}^*}
\newcommand{\se}{\tilde{\mathrm e}}
\newcommand{\smu}{\tilde{\mu}}
\newcommand{\stau}{\tilde{\tau}}
\newcommand{\snu}{\tilde{\nu}}
\newcommand{\sell}{\tilde{\ell}}

\newcommand{\glu}{\tilde{\mathrm g}}
\newcommand{\chio}{\tilde{\chi}}
\newcommand{\chip}{\tilde{\chi}^{\pm}}
\newcommand{\chim}{\tilde{\chi}^{\mp}}
\newcommand{\grav}{\tilde{\mathrm G}}
\def\ino{\widetilde \chi}               
\def\winog{\widetilde \chi^\pm}               
\def\zinog{\widetilde \chi^0}               
\def\w3{\widetilde W_3}               
\def\Bino{\widetilde B}               
\def\Wino{\widetilde W}               
\def\higgsino{\widetilde H}               
\def\sbottom{\tilde \b}             
\def\stop{\tilde \t}             
\def\stau{\tilde\tau}             
\def\slepton{\tilde\ell}             
\def\squark{\tilde \q}             
\def\sneutrino{\tilde \nu}          
\def\gluino{\tilde \g}             
\def\gravitino{\widetilde G}          
\def\sgnmu{sign($\mu$)}              

\newcommand{\thw}{\theta_W}
\def\nn{\nonumber}
\def\ts{\thinspace}
\def\tx{\textstyle}
\def\ra{\rightarrow}
\def\be{\begin{equation}} 
\def\bea{\begin{eqnarray}}
\def\eea{\end{eqnarray}}
\def\ba{\begin{array}}
\def\ea{\end{array}}
\def\chipr{\chi^{\ts \prime}}
\def\CC{{\cal C}}

\def\CM{{\cal M}}
\def\CO{{\cal O}}
\def\atro{\alpha_{\tro}}
\def\Ntc{N_{TC}}
\def\suc{{\bf SU(3)}}
\def\uone{{\bf U(1)_1}}
\def\utwo{{\bf U(1)_2}}
\def\suone{{\bf SU(3)_1}}
\def\sutwo{{\bf SU(3)_2}}
\newcommand{\tro}{\rho_{\mrm{tc}}}
 
\newcommand{\troct}{\rho_{\mrm{tc}8}} 
\newcommand{\troctaa}{\rho_{11}} 
\newcommand{\troctbb}{\rho_{22}} 
\newcommand{\troctab}{\rho_{12}} 
\newcommand{\troctabp}{\rho_{12'}} 
\newcommand{\tropm}{\rho_{\mrm{tc}}^\pm}

\newcommand{\troz}{\rho_{\mrm{tc}}^0}
\newcommand{\tom}{\omega_\mrm{tc}}
\newcommand{\tpi}{\pi_\mrm{tc}}
\newcommand{\tpipm}{\pi_\mrm{tc}^\pm}
\newcommand{\tpimp}{\pi_\mrm{tc}^\mp}
\newcommand{\tpip}{\pi_\mrm{tc}^+}
\newcommand{\tpim}{\pi_\mrm{tc}^-}
\newcommand{\tpiz}{\pi_\mrm{tc}^0}
\newcommand{\tpipr}{\pi_\mrm{tc}^{0 \prime}}

\newcommand{\Jpsi}{\mathrm{J}/\psi}
\newcommand{\pomeron}{\mathrm{I}\!\mathrm{P}}
\newcommand{\reggeon}{\mathrm{I}\!\mathrm{R}}
\newcommand{\ee}{\e^+\e^-}
\newcommand{\ep}{\e\p}
\newcommand{\pp}{\p\p}
\newcommand{\ppbar}{\p\pbar}
\newcommand{\gammaZ}{\gamma^* / \Z^0}
\newcommand{\gast}{\gamma^*}
\newcommand{\galep}{\texttt{'gamma/}\textit{lepton}\texttt{'}}
\newcommand{\mmin}{\mathrm{min}}
\newcommand{\mmax}{\mathrm{max}}
\newcommand{\BE}{\mathrm{BE}}

\newcommand{\KF}{\texttt{KF}}
\newcommand{\KC}{\texttt{KC}}
\newcommand{\ISUB}{\texttt{ISUB}}
 
\newenvironment{Itemize}{\begin{list}{$\bullet$}%
{\setlength{\topsep}{0.2mm}\setlength{\partopsep}{0.2mm}%
\setlength{\itemsep}{0.2mm}\setlength{\parsep}{0.2mm}}}%
{\end{list}}
\newcounter{enumct}
\newenvironment{Enumerate}{\begin{list}{\arabic{enumct}.}%
{\usecounter{enumct}\setlength{\topsep}{0.2mm}%
\setlength{\partopsep}{0.2mm}\setlength{\itemsep}{0.2mm}%
\setlength{\parsep}{0.2mm}}}{\end{list}}
 
\newenvironment{entry}%
{\begin{list}{}{\setlength{\topsep}{0mm} \setlength{\itemsep}{0mm}
\setlength{\parskip}{0mm} \setlength{\parsep}{0mm}
\setlength{\leftmargin}{20mm} \setlength{\rightmargin}{0mm}
\setlength{\labelwidth}{18mm} \setlength{\labelsep}{2mm}}}%
{\end{list}}
\newenvironment{subentry}%
{\begin{list}{}{\setlength{\topsep}{0mm} \setlength{\itemsep}{0mm}
\setlength{\parskip}{0mm} \setlength{\parsep}{0mm}
\setlength{\leftmargin}{10mm} \setlength{\rightmargin}{0mm}
\setlength{\labelwidth}{18mm} \setlength{\labelsep}{2mm}}}%
{\end{list}}
\newcommand{\itemc}[1]{\item[\textbf{#1}\hfill]}
\newcommand{\iteme}[1]{\item[\texttt{#1}\hfill]}
\newcommand{\itemn}[1]{\item[{#1}\hfill]}
 
\newcommand{\drawbox}[1]{\vspace{\baselineskip}\noindent%
\fbox{\texttt{#1}}\vspace{0.5\baselineskip}}
\newcommand{\drawboxtwo}[2]{\vspace{\baselineskip}\noindent%
\fbox{\begin{minipage}{150mm}\begin{tabbing}%
{\texttt{#1}}\\{\texttt{#2}}%
\end{tabbing}\end{minipage}}\vspace{0.5\baselineskip}}
\newcommand{\drawboxthree}[3]{\vspace{\baselineskip}\noindent%
\fbox{\begin{minipage}{150mm}\begin{tabbing}%
{\texttt{#1}}\\{\texttt{#2}}\\{\texttt{#3}}%
\end{tabbing}\end{minipage}}\vspace{0.5\baselineskip}}
\newcommand{\drawboxfour}[4]{\vspace{\baselineskip}\noindent%
\fbox{\begin{minipage}{150mm}\begin{tabbing}%
{\texttt{#1}}\\{\texttt{#2}}\\{\texttt{#3}}\\{\texttt{#4}}%
\end{tabbing}\end{minipage}}\vspace{0.5\baselineskip}}
\newcommand{\drawboxseven}[7]{\vspace{\baselineskip}\noindent%
\fbox{\begin{minipage}{150mm}\begin{tabbing}%
{\texttt{#1}}\\{\texttt{#2}}\\{\texttt{#3}}\\{\texttt{#4}}\\%
{\texttt{#5}}\\{\texttt{#6}}\\{\texttt{#7}}%
\end{tabbing}\end{minipage}}\vspace{0.5\baselineskip}}
\newcommand{\boxsep}{\vspace{0.5\baselineskip}} 
 
\newlength{\captivewidth}
\newcommand{\captive}[1]{\rule{5mm}{0mm}%
\begin{minipage}{\captivewidth}%
\caption[small]{#1}\end{minipage}}

\newlength{\tablinsep}
\newlength{\halfpagewid}
\newlength{\abstwidth}
\begin{document}
 
\sloppy
 
\pagestyle{empty}
 
\begin{flushright}
hep-ph/0308153\\
LU TP 03--38\\
August 2003
\end{flushright}
 
\vspace{\fill}
 
\begin{center}

\tbf{{\Huge P}{\LARGE YTHIA}~~{\Huge 6.3}}\\[5mm]
\tbf{{\LARGE Physics and Manual}} \\[10mm]
{\Large Torbj\"orn Sj\"ostrand,$^1$ Leif L\"onnblad,$^1$} \\[2mm]
{\Large Stephen Mrenna,$^2$ Peter Skands$^1$} \\[5mm]
{\large $^1$Department of Theoretical Physics,} \\[1mm]
{\large Lund University, S\"olvegatan 14A,} \\[1mm]
{\large S-223 62 \tsc{Lund}, \tsc{Sweden}}\\[5mm]
{\large $^2$Computing Division, Simulations Group,} \\[1mm]
{\large Fermi National Accelerator Laboratory} \\[1mm]
{\large MS 234, Batavia, IL  60510, USA} 

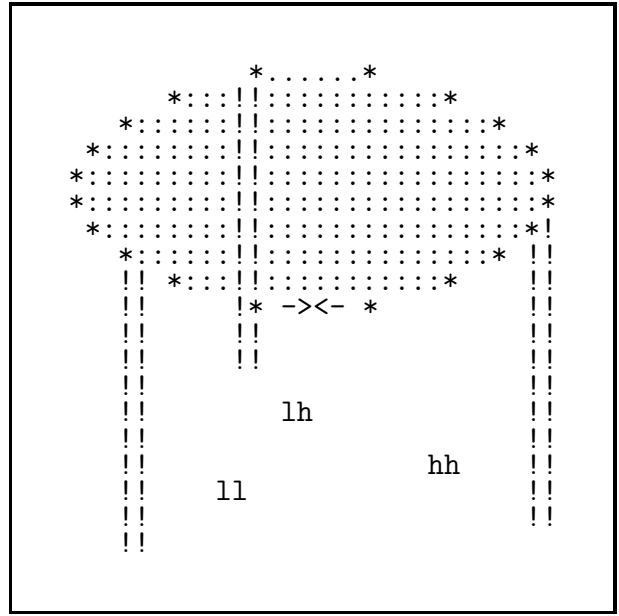
\vspace{\fill}

\fbox{\hspace{1mm}\rule{0mm}{7mm}
\begin{minipage}[t]{150mm}
\begin{tabbing}
{\tt ~~~~~~~~~~~~*......*~~~~~~~~~~~~}\\[-1.1mm]
{\tt ~~~~~~~*:::!!:::::::::::*~~~~~~~}\\[-1.1mm]
{\tt ~~~~*::::::!!::::::::::::::*~~~~}\\[-1.1mm]
{\tt ~~*::::::::!!::::::::::::::::*~~}\\[-1.1mm]
{\tt ~*:::::::::!!:::::::::::::::::*~}\\[-1.1mm]
{\tt ~*:::::::::!!:::::::::::::::::*~}\\[-1.1mm]
{\tt ~~*::::::::!!::::::::::::::::*!~}\\[-1.1mm]
{\tt ~~~~*::::::!!::::::::::::::*~!!~}\\[-1.1mm]
{\tt ~~~~!!~*:::!!:::::::::::*~~~~!!~}\\[-1.1mm]
{\tt ~~~~!!~~~~~!*~-><-~*~~~~~~~~~!!~}\\[-1.1mm]
{\tt ~~~~!!~~~~~!!~~~~~~~~~~~~~~~~!!~}\\[-1.1mm]
{\tt ~~~~!!~~~~~!!~~~~~~~~~~~~~~~~!!~}\\[-1.1mm]
{\tt ~~~~!!~~~~~~~~~~~~~~~~~~~~~~~!!~}\\[-1.1mm]
{\tt ~~~~!!~~~~~~~~lh~~~~~~~~~~~~~!!~}\\[-1.1mm]
{\tt ~~~~!!~~~~~~~~~~~~~~~~~~~~~~~!!~}\\[-1.1mm]
{\tt ~~~~!!~~~~~~~~~~~~~~~~~hh~~~~!!~}\\[-1.1mm]
{\tt ~~~~!!~~~~ll~~~~~~~~~~~~~~~~~!!~}\\[-1.1mm]
{\tt ~~~~!!~~~~~~~~~~~~~~~~~~~~~~~!!~}\\[-1.1mm]
{\tt ~~~~!!~~~~~~~~~~~~~~~~~~~~~~~~~~}\\[-1.1mm]
\end{tabbing}
\end{minipage}
\rule[-6mm]{0mm}{6mm}\hspace{3mm}}
\end{center}

\vspace{\fill}

\phantom{dummy}
 
\clearpage

\begin{center}
{\bf Abstract}\\[2ex]
\begin{minipage}{\abstwidth}
The {\Py} program can be used to generate high-energy-physics `events', 
i.e.\ sets of outgoing particles produced in the interactions between 
two incoming particles. The objective is to provide as accurate as 
possible a representation of event properties in a wide range of 
reactions, with emphasis on those where strong interactions play a 
r\^ole, directly or indirectly, and therefore multihadronic final
states are produced. The physics is then not understood well enough 
to give an exact description; instead the program has to be based on
a combination of analytical results and various QCD-based models. 
This physics input is summarized here, for areas such as hard 
subprocesses, initial- and final-state parton showers, beam remnants 
and underlying events, fragmentation and decays, and much more. 
Furthermore, extensive information is provided on all program elements: 
subroutines and functions, switches and parameters, and particle and
process data. This should allow the user to tailor the generation task 
to the topics of interest.
\\[3mm]
The code and further information may be found on the {\Py} web 
page:\\[1mm]
\ttt{http://www.thep.lu.se/}$\sim$\ttt{torbjorn/Pythia.html}~.
\\[3mm]
The information in this edition of the manual refers to 
{\Py} version 6.301, of 14 August 2003.
\\[3mm]
The official reference to the latest published version is\\
T.~Sj\"ostrand, P.~Ed\'en, C.~Friberg, L.~L\"onnblad, G.~Miu, S.~Mrenna 
and E.~Norrbin, Computer Physics Commun. {\bf 135} (2001) 238.
\end{minipage}
\end{center}

\vspace{\fill}

\phantom{dummy}

\clearpage

\phantom{dummy}
\vspace{\fill}
 
\begin{minipage}{14cm}
\begin{flushright}
\textit{\LARGE Dedicated to}\\[5mm] 
\textit{\LARGE the Memory of}\\[9mm]
\textbf{\textit{\Huge Bo Andersson}}\\[10mm]
\textit{\Large 1937 -- 2002}\\[5mm] 
\textit{\Large originator, inspirator}
\end{flushright}
\end{minipage}

\vspace{\fill}
\vspace{\fill}
\phantom{dummy}

\clearpage

\phantom{dummy}

\clearpage

\section*{Preface}

The {\Py} program is frequently used for event generation
in high-energy physics. The emphasis is on multiparticle production
in collisions between elementary particles. This in particular means
hard interactions in $\ee$, $\pp$ and $\ep$ colliders, although
also other applications are envisaged. The program is intended to 
generate complete events, in as much detail as 
experimentally observable ones, within the bounds of our current
understanding of the underlying physics. Many of the components of
the program represents original research, in the sense that models 
have been developed and implemented for a number of aspects not 
covered by standard theory. 

Historically, the family of event generators from the Lund group was 
begun with {\Je} in 1978. The {\Py} program followed a few years later. 
With time, the two programs so often had to be used together that it 
made sense to merge them. Therefore {\Py}~5.7 and {\Je}~7.4 were the 
last versions to appear individually; as of {\Py}~6.1 all the code is 
collected under the {\Py} heading. At the same time, the \tsc{SPythia} 
sideline of {\Py} was reintegrated. Both programs have a long history, 
and several manuals have come out. The most recent one is \\[1mm]
T. Sj\"ostrand, P. Ed\'en, C. Friberg, L. L\"onnblad, G. Miu, S. Mrenna 
and E. Norrbin,\\
Computer Physics Commun. {\bf 135} (2001) 238, \\[1mm]
so please use this for all official references. Additionally remember 
to cite the original literature on the physics topics of particular 
relevance for your studies. (There is no reason to omit references to 
good physics papers simply because some of their contents have also 
been made available as program code.)

Event generators often have a reputation for being `black boxes'; 
if nothing else, this report should provide you with a glimpse of 
what goes on inside the program. Some such understanding may be of 
special interest for new users, who have no background in the field. 
An attempt has been made to structure the report sufficiently well 
so that many of the sections can be read independently of each other,
so you can pick the sections that interest you. We have tried to
keep together the physics and the manual sections on specific 
topics, where practicable.

A large number of persons should be thanked for their contributions.
Bo Andersson and G\"osta Gustafson are the originators of the Lund 
model, and strongly influenced the early development of the programs.
Hans-Uno Bengtsson is the originator of the {\Py} program. Mats
Bengtsson is the main author of the final-state parton-shower 
algorithm. Patrik Ed\'en has contributed an improved popcorn scenario 
for baryon production. Christer Friberg has helped develop the expanded 
photon physics machinery, Emanuel Norrbin the new matrix-element 
matching of the final-state parton shower algorithm and the handling 
of low-mass strings, and Gabriela Miu the matching of initial-state 
showers.

Further comments on the programs and smaller pieces of code have been 
obtained from users too numerous to be mentioned here, but who are all 
gratefully acknowledged. To write programs of this size and 
complexity would be impossible without a strong support and user 
feedback. So, if you find errors, please let us know. 

The moral responsibility for any remaining errors clearly rests with
the authors. However, kindly note that this is a `University World' 
product, distributed `as is', free of charge, without any binding 
guarantees. And always remember that the program does not represent a 
dead collection of established truths, but rather one of many possible
approaches to the problem of multiparticle production in high-energy
physics, at the frontline of current research. Be critical!

\clearpage

\tableofcontents
 
\clearpage

\pagestyle{plain}
\setcounter{page}{1}
 
\section{Introduction}
 
Multiparticle production is the most characteristic feature
of current high-energy physics. Today, observed particle
multiplicities are typically between ten and a hundred, and
with future machines this range will be extended upward.
The bulk of the multiplicity is found in jets, i.e.\ in 
collimated bunches of hadrons (or decay products of hadrons) produced 
by the hadronization of partons, i.e.\ quarks and gluons. (For some 
applications it will be convenient to extend the parton concept 
also to some non-coloured but showering particles, such as 
electrons and photons.)
 
\subsubsection*{The Complexity of High-Energy Processes}
 
To first approximation, all processes have a simple structure at
the level of interactions between the fundamental objects of nature,
i.e.\ quarks, leptons and gauge bosons. For instance, a lot can
be understood about the structure of hadronic events at LEP just
from the `skeleton' process $\ee \to \Z^0 \to \q \qbar$.
Corrections to this picture can be subdivided, arbitrarily but
conveniently, into three main classes.
 
Firstly, there are bremsstrahlung-type modifications, i.e.\ the
emission of additional final-state particles by branchings such as
$\e \to \e \gamma$ or $\q \to \q \g$. Because of the largeness of
the strong coupling constant $\alphas$, and because of the presence
of the triple gluon vertex, QCD emission off quarks and gluons is
especially prolific. We therefore speak about `parton showers',
wherein a single initial parton may give rise to a whole bunch of
partons in the final state. Also photon emission may give sizable
effects in $\ee$ and $\ep$ processes. The bulk of the bremsstrahlung
corrections are universal, i.e.\ do not depend on the details of
the process studied, but only on one or a few key numbers, such as
the momentum transfer scale of the process. Such universal
corrections may be included to arbitrarily high orders, using a
probabilistic language. Alternatively, exact calculations of
bremsstrahlung corrections may be carried out order by order in
perturbation theory, but rapidly the calculations then become
prohibitively complicated and the answers correspondingly lengthy.
 
Secondly, we have `true' higher-order corrections, which involve a
combination of loop graphs and the soft parts of the
bremsstrahlung graphs above, a combination needed to
cancel some  divergences. In a complete description it is
therefore not possible to consider bremsstrahlung separately,
as assumed here. The
necessary perturbative calculations are usually very difficult;
only rarely have results been presented that include more than one
non-`trivial' order, i.e.\ more than one loop. As above, answers
are usually very lengthy, but some results are sufficiently simple
to be generally known and used, such as the running of $\alphas$, or
the correction factor $1 + \alphas/\pi + \cdots$ in the partial
widths of $\Z^0 \to \q \qbar$ decay channels. For high-precision
studies it is imperative to take into account the results of
loop calculations, but usually effects are minor for the qualitative
aspects of high-energy processes.
 
Thirdly, quarks and gluons are confined. In the two points above,
we have used a perturbative language to describe the short-distance
interactions of quarks, leptons and gauge bosons. For leptons
and colourless bosons this language is sufficient. However, for
quarks and gluons it must be complemented with the structure of 
incoming hadrons, and a picture for the hadronization process, 
wherein the coloured partons are transformed into jets of colourless 
hadrons, photons and leptons. The hadronization can be further 
subdivided into fragmentation and decays, where the former describes 
the way the creation of new quark-antiquark pairs can break up a 
high-mass system into lower-mass ones, ultimately hadrons. (The word 
`fragmentation' is also sometimes used in a broader sense, but we 
will here use it with this specific meaning.) This process is still 
not yet understood from first principles, but has to be based on models. 
In one sense, hadronization effects are overwhelmingly large, since 
this is where the bulk of the multiplicity comes from. In another 
sense, the overall energy flow of a high-energy event is mainly 
determined by the perturbative processes, with only a minor additional 
smearing caused by the hadronization step. One may therefore pick 
different levels of ambition, but in general detailed studies require 
a detailed modelling of the hadronization process.
 
The simple structure that we started out with has now become
considerably more complex --- instead of maybe two final-state
partons we have a hundred final particles. The original physics
is not gone, but the skeleton process has been dressed up and
is no longer directly visible. A direct comparison between theory
and experiment is therefore complicated at best, and
impossible at worst.
 
\subsubsection*{Event Generators}
 
It is here that event generators come to the rescue. In an event
generator, the objective striven for is to use computers to generate 
events as detailed as could be observed by a perfect detector.
This is not done in one step, but rather by `factorizing' the full
problem into a number of components, each of which can be handled
reasonably accurately. Basically, this means that the hard process
is used as input to generate bremsstrahlung corrections, and that
the result of this exercise is thereafter left to hadronize. This
sounds a bit easier than it really is --- else this report would
be a lot thinner. However, the basic idea is there: if the
full problem is too complicated to be solved in one go, try to
subdivide it into smaller tasks of manageable proportions.
In the actual generation procedure, most steps therefore involve
the branching of one object into two, or at least into a very small
number, with the daughters free to branch in their turn. A lot of
book-keeping is involved, but much is of a repetitive nature, and
can therefore be left for the computer to handle.
 
As the name indicates, the output of an event generator should be
in the form of `events', with the same average behaviour and the
same fluctuations as real data. In the data, fluctuations arise from
the quantum mechanics of the underlying theory. In
generators, Monte Carlo techniques are used to select all relevant
variables according to the desired probability distributions,
and thereby ensure randomness in the final events.
Clearly some loss of information is entailed: quantum mechanics is
based on amplitudes, not probabilities. However, only very rarely
do (known) interference phenomena appear that cannot be cast in a
probabilistic language. This is therefore not a more restraining
approximation than many others.
 
Once there, an event generator can be used in many different ways.
The five main applications are probably the following:
\begin{Itemize}
\item To give physicists a feeling for the kind
of events one may expect/hope to find, and at what rates.
\item As a help in the planning of a new detector, so that detector
performance is optimized, within other constraints, for the
study of interesting physics scenarios.
\item As a tool for devising the analysis strategies that should
be used on real data, so that signal-to-background conditions are
optimized.
\item As a method for estimating detector acceptance corrections
that have to be applied to raw data, in order to extract the
`true' physics signal.
\item As a convenient framework within which to interpret the
observed phenomena in terms of a more fundamental
underlying theory (usually the Standard Model).
\end{Itemize}
 
Where does a generator fit into the overall analysis chain of an
experiment? In `real life', the machine produces interactions.
These events are observed by detectors, and the interesting ones
are written to tape by the
data acquisition system. Afterward the events may be reconstructed,
i.e.\ the electronics signals (from wire chambers, calorimeters, and
all the rest) may be
translated into a deduced setup of charged tracks or
neutral energy depositions, in the best of worlds with full knowledge
of momenta and particle species. Based on this cleaned-up
information, one may proceed with the physics analysis.
In the Monte Carlo world, the r\^ole of the machine, namely to produce
events, is taken by the event generators described in this report.
The behaviour of the detectors --- how particles produced by the
event generator traverse the detector, spiral in magnetic
fields, shower in calorimeters, or sneak out through cracks, etc. ---
is simulated in programs such as \tsc{Geant} \cite{Bru89}. 
Traditionally, this latter activity is called event simulation,
which is somewhat unfortunate
since the same words could equally well be applied to what, here, we
call event generation. A more appropriate term is detector
simulation. Ideally, the output of this simulation has exactly the
same format as the real data recorded by the detector, and can
therefore be put through the same event reconstruction and physics
analysis chain, except that here we know what the `right answer'
should be, and so can see how well we are doing.
 
Since the full chain of detector simulation and event
reconstruction is very
time-consuming, one often does `quick and dirty' studies in
which these steps are skipped entirely, or at least replaced by
very simplified procedures which only take into account the geometric
acceptance of the detector and other trivial effects. One may then
use the output of the event generator directly in the physics studies.
 
There are still many holes in our understanding of the full event
structure, despite an impressive amount of work and detailed
calculations. To put together a generator therefore involves making
a choice on what to include, and how to include it. At best, the
spread between generators can be used to give some impression of
the uncertainties involved. A multitude of approximations will
be discussed in the main part of this report, but already here
is should be noted that many major approximations are related to
the almost complete neglect
of the second point above, i.e.\ of the non-`trivial'
higher-order effects. It can therefore only be hoped that
the `trivial' higher order parts give the bulk of the experimental
behaviour. By and large, this seems to be the case; for $\ee$
annihilation it even turns out to be a very good approximation.
 
The necessity to make compromises has one major implication:
to write a good event generator is an art, not an exact science.
It is therefore essential not to blindly trust the
results of any single event generator, but always to make several
cross-checks. In addition, with computer programs of tens of
thousands of lines, the question is not whether bugs exist, but how
many there are, and how critical their positions.
Further, an event generator cannot be thought of as all-powerful,
or able to give intelligent answers to ill-posed questions;
sound judgement and some understanding of a
generator are necessary prerequisites for successful use. In spite
of these limitations, the event generator approach is the most
powerful tool at our disposal if we wish to gain a detailed and
realistic understanding of physics at current or future high-energy
colliders.
 
\subsubsection*{The Origins of the JETSET and PYTHIA Programs}
 
Over the years, many event generators have appeared. Surveys of
generators for $\ee$ physics in general and LEP in particular
may be found in \cite{Kle89,Sjo89,Kno96,Lon96}, for high-energy 
hadron--hadron ($\pp$) physics in \cite{Ans90,Sjo92,Kno93,LHC00}, 
and for $\ep$ physics in \cite{HER92,HER99}. We refer the reader 
to those for additional details and references. In this particular 
report, the two closely connected programs {\Je} and {\Py}, now 
merged under the {\Py} label, will be described.
 
{\Je} has its roots in the efforts of the Lund
group to understand the hadronization process, starting in the late
seventies \cite{And83}. The so-called string fragmentation model
was developed as an explicit and detailed framework, within which
the long-range confinement forces are allowed to distribute the
energies and flavours of a parton configuration among a collection
of primary hadrons, which subsequently may decay further. This model,
known as the Lund string model, or `Lund' for short, contained a
number of specific predictions, which were confirmed by data from
PETRA and PEP, whence the model gained a widespread
acceptance. The Lund string model is still today the most elaborate
and widely used fragmentation model
at our disposal. It remains at the heart of the
{\Py} program.
 
In order to predict the shape of events at PETRA/PEP, and to
study the fragmentation process in detail, it was necessary to start
out from the partonic configurations that were to fragment.
The generation of complete $\ee$ hadronic events was therefore
added, originally based on simple $\gamma$ exchange and
first-order QCD matrix elements, later extended to full $\gammaZ$
exchange with first-order initial-state QED radiation and 
second-order QCD matrix elements. A number of utility routines 
were also provided early on, for everything from event listing 
to jet finding.
 
By the mid-eighties it was clear that the matrix-element approach had
reached the limit of its usefulness, in the sense that it could not
fully describe the multijet topologies of the data. (Later on,
the use of optimized perturbation theory was to lead to a resurgence
of the matrix-element approach, but only for specific applications.)
Therefore a parton-shower description was
developed \cite{Ben87a} as an alternative to the matrix-element one.
The combination of parton showers and string fragmentation has been
very successful, and forms the main approach to the description of
hadronic $\Z^0$ events.
 
In recent years, the {\Je} part of the code has been a fairly stable 
product, covering the four main areas of fragmentation, final-state 
parton showers, $\ee$ event generation and general utilities.
 
The successes of string fragmentation in $\ee$ made it interesting
to try to extend this framework to other processes, and explore
possible physics consequences. Therefore a number of
other programs were written, which combined a process-specific
description of the hard interactions with the general fragmentation
framework of {\Je}. The {\Py} program
evolved out of early studies on fixed-target proton--proton
processes, addressed mainly at issues related to string drawing.
 
With time, the interest shifted towards higher energies, first
to the SPS $\ppbar$ collider, and later to the Tevatron, SSC and LHC, 
in the context of a number of workshops in the USA and Europe. 
Parton showers were added, for final-state radiation by making use 
of the {\Je} routine, for initial-state one by the development
of the concept of `backwards evolution', specifically for
{\Py} \cite{Sjo85}. Also a framework was developed for
minimum-bias and underlying events \cite{Sjo87a}.
 
Another main change was the introduction of an increasing
number of hard processes, within the Standard Model and beyond.
A special emphasis was put on the search for the Standard Model
Higgs, in different mass ranges and in different channels, with due
respect to possible background processes.
 
The bulk of the machinery developed for hard processes actually
depended little on the choice of initial state, as long as the
appropriate parton distributions were there for the incoming
partons and particles. It therefore made sense to extend the
program from being only a $\pp$ generator to working also for
$\ee$ and $\ep$. This process was only completed in 1991,
again spurred on by physics workshop activities. Currently
{\Py} should therefore work equally well for a selection
of different possible incoming beam particles.

An effort independent of the Lund group activities got going to
include supersymmetric event simulation in {\Py}. This resulted
in the \tsc{SPythia} program.
 
While {\Je} was independent of {\Py} until 1996, their ties 
had grown much stronger over the years, and the border-line 
between the two programs had become more and more artificial.
It was therefore decided to merge the two, and also include the
\tsc{SPythia} extensions, starting from {\Py}~6.1. The different 
origins in part still are reflected in this manual,
but the strive is towards a seamless merger.
 
The tasks of including new processes, and of improving the simulation
of parton showers and other aspects of already present processes, are 
never-ending. Work therefore continues apace. 
 
\subsubsection*{About this Report}
 
As we see, {\Je} and {\Py} started out as very
ideologically motivated programs, developed to study specific
physics questions in enough detail that explicit predictions
could be made for experimental quantities. As it was recognized
that experimental imperfections could distort the basic predictions,
the programs were made available for general use by experimentalists.
It thus became feasible to explore the models in more detail
than would otherwise have been possible. As time went by, the
emphasis came to shift somewhat, away from the original strong
coupling to a specific fragmentation model, towards a description of
high-energy multiparticle production processes in general.
Correspondingly, the use expanded from being one of just comparing
data with specific model predictions, to one of extensive use for
the understanding of detector performance, for the derivation of
acceptance correction factors, for the prediction of physics at
future high-energy accelerators, and for the design of related
detectors.
 
While the ideology may be less apparent, it is still there, however.
This is not something unique to the programs discussed here,
but inherent in any event generator, or at least any generator that 
attempts to go beyond the simple parton level skeleton description
of a hard process. Do not accept the myth that everything available
in Monte Carlo form represents ages-old common knowledge, tested
and true. Ideology is present by commissions or omissions
in any number of details. A programs like {\Py} represents
a major amount of original physics research, often on complicated
topics where no simple answers are available.
As a (potential) program user you must be
aware of this, so that you can form your own opinion, not just about
what to trust and what not to trust, but also how much to trust a given
prediction, i.e.\ how uncertain it is likely to be.
{\Py} is particularly well endowed in this respect, since a
number of publications exist where most of the relevant physics is
explained in considerable detail. In fact, the problem may rather be
the opposite, to find the relevant information among all the possible
places. One main objective of the current report is therefore to
collect much of this information in one single place. Not all the
material found in specialized papers is reproduced, by a wide margin,
but at least enough should be found here to understand the general
picture and to know where to go for details.
 
The current report is therefore intended to update and extend the 
previous round of published physics descriptions and program manuals
\cite{Sjo86,Sjo87,Ben87,Sjo94,Mre97,Sjo01}. Make all references to 
the most recent published one in \cite{Sjo01}. Further specification 
could include a statement of 
the type `We use {\Py} version X.xxx'. (If you are a {\LaTeX} fan, 
you may want to know that the program name in this report has been 
generated by the command \verb+\textsc{Pythia}+.)
Kindly do not refer to {\Py} as `unpublished', `private communication' 
or `in preparation': such phrases are incorrect and only create 
unnecessary confusion.

In addition, remember that many of the individual physics components
are documented in separate publications. If some of these contain 
ideas that are useful to you, there is every reason to cite them. 
A reasonable selection would vary as a function of the physics you are
studying. The criterion for which to pick should be simple: imagine 
that a Monte Carlo implementation had not been available. Would you
then have cited a given paper on the grounds of its physics contents 
alone? If so, do not punish the extra effort of turning these ideas 
into publicly available software. (Monte Carlo manuals are good for
nothing in the eyes of many theorists, so often only the acceptance 
of `mainstream' publications counts.) Here follows a list of some
main areas where the $\Py$ programs contain original research:
\begin{Itemize}
\item The string fragmentation model \cite{And83,And98}.
\item The string effect \cite{And80}.
\item Baryon production (diquark/popcorn) \cite{And82,And85,Ede97}.
\item Small-mass string fragmentation \cite{Nor98}.
\item Fragmentation of multiparton systems \cite{Sjo84}.
\item Colour rearrangement \cite{Sjo94a} and Bose-Einstein effects
\cite{Lon95}.
\item Fragmentation effects on $\alphas$ determinations \cite{Sjo84a}.
\item Initial-state parton showers \cite{Sjo85,Miu99}.
\item Final-state parton showers \cite{Ben87a,Nor01}.
\item Photon radiation from quarks \cite{Sjo92c}
\item Deeply Inelastic Scattering \cite{And81a,Ben88}.
\item Photoproduction \cite{Sch93a}, $\gamma\gamma$ \cite{Sch94a}
and $\gast\p/\gast\gamma/\gast\gast$ \cite{Fri00} physics.
\item Parton distributions of the photon \cite{Sch95,Sch96}.
\item Colour flow in hard scatterings \cite{Ben84}.
\item Elastic and diffractive cross sections \cite{Sch94}.
\item Minijets (multiple parton--parton interactions) \cite{Sjo87a}.
\item Rapidity gaps \cite{Dok92}.
\item Jet clustering in $k_{\perp}$ \cite{Sjo83}.
\end{Itemize}
 
In addition to a physics survey, the current report also contains a
complete manual for the program. Such manuals have always been
updated and distributed jointly with the programs, but have grown in 
size with time. A word of warning may therefore be in place. 
The program description is fairly
lengthy, and certainly could not be absorbed in one sitting. This is
not even necessary, since all switches and parameters are provided
with sensible default values, based on our best understanding (of
the physics, and of what you expect to happen if you do not
specify any options). As a new user, you can therefore disregard all
the fancy options, and just run the program with a minimum ado.
Later on, as you gain experience, the options that seem useful can
be tried out. No single user is ever likely to find need for more than
a fraction of the total number of possibilities available,
yet many of them have been added to meet specific user requests.

In some instances, not even this report will provide you with all the 
information you desire. You may wish to find out about recent versions 
of the program, know about related software, pick up a few sample 
main programs to get going, or get hold of related physics papers. 
Some such material can be found on the {\Py} web page:\\[1mm]
 \ttt{http://www.thep.lu.se/}$\sim$\ttt{torbjorn/Pythia.html}~.
 
\subsubsection*{Disclaimer}
 
At all times it should be remembered that this is not a commercial
product, developed and supported by professionals. Instead it is
a `University World' product, developed by a very few physicists
(mainly the current first author)
originally for their own needs, and supplied to other
physicists on an `as-is' basis, free of charge. No guarantees are
therefore given for the proper functioning of the program, nor for
the validity of physics results. In the end, it is always up to you
to decide for yourself whether to trust a given result or not. Usually
this requires comparison either with analytical results or with
results of other programs, or with both. Even this is not necessarily 
foolproof: for instance, if an error
is made in the calculation of a matrix element for a given process,
this error will be propagated both into the analytical results based
on the original calculation and into all the event generators which
subsequently make use of the published formulae. In the end, there
is no substitute for a sound physics judgement.
 
This does not mean that you are all on your own, with a program
nobody feels responsible for. Attempts are made to check processes as
carefully as possible, to write programs that do not invite
unnecessary errors, and to provide a detailed and accurate
documentation. All of this while maintaining the full power and
flexibility, of course, since the physics must always take precedence
in any conflict of interests. If nevertheless any errors or
unclear statements are found, please do communicate them to e-mail 
\ttt{torbjorn@thep.lu.se}, or to another person in charge. For instance, 
all questions on the supersymmetric and technicolor machinery are better 
directed to \ttt{mrenna@fnal.gov}. Every attempt will be made to solve 
problems as soon as is reasonably possible, given that this support is 
by a few persons, who mainly have other responsibilities. 

However, in order to make debugging at all possible, we request 
that any sample code you want to submit as evidence be completely 
self-contained, and peeled off from all irrelevant aspects. Use
simple write statements or the {\Py} histogramming routines to make
your point. Chances are that, if the error cannot be reproduced
by fifty lines of code, in a main program linked only to {\Py}, 
the problem is sitting elsewhere. Numerous errors have been caused
by linking to other (flawed) libraries, e.g.\ collaboration-specific 
frameworks for running {\Py}. Then you should put the blame elsewhere.
 
\subsubsection*{Appendix: The Historical Pythia}
 
The `{\Py}' label may need some explanation.
 
The myth tells how Apollon, the God of Wisdom, killed the powerful
dragon-like monster Python, close to the village of Delphi in Greece.
To commemorate this victory, Apollon founded the Pythic Oracle in
Delphi, on the slopes of Mount Parnassos. Here men could come to
learn the will of the Gods and the course of the future. The oracle
plays an important r\^ole in many of the other Greek myths, such
as those of Heracles and of King Oedipus.
 
Questions were to be put to the Pythia, the `Priestess' or
`Prophetess' of the Oracle. In fact, she was a local woman,
usually a young maiden, of no particular religious schooling.
Seated on a tripod, she inhaled the obnoxious vapours that
seeped up through a crevice in the ground. This brought her
to a trance-like state, in which she would scream seemingly
random words and sounds. It was the task of the professional
priests in Delphi to record those utterings and edit them into
the official Oracle prophecies, which often took the form of
poems in perfect hexameter. In fact, even these edited replies
were often less than easy to interpret. The Pythic oracle
acquired a reputation for ambiguous answers.
 
The Oracle existed already at the beginning of the historical
era in Greece, and was universally recognized as the foremost
religious seat. Individuals and city states came to consult, on
everything from cures for childlessness to matters of war. Lavish
gifts allowed the temple area to be built and decorated. Many
states supplied their own treasury halls, where especially beautiful
gifts were on display. Sideshows included the Omphalos,
a stone reputedly marking the centre of the Earth, and the Pythic
games, second only to the Olympic ones in importance.
 
Strife inside Greece eventually led to a decline in the power of
the Oracle. A serious blow was dealt when the Oracle of Zeus Ammon
(see below) declared Alexander the Great
to be the son of Zeus. The Pythic Oracle lived on, however, and was
only closed by a Roman Imperial decree in 390 \tsc{ad}, at a time 
when Christianity was ruthlessly destroying any religious opposition.
Pythia then had been at the service of man and Gods for a
millennium and a half.
 
The r\^ole of the Pythic Oracle prophecies on the course of history 
is nowhere better described than in `The Histories' by Herodotus
\cite{HerBC}, the classical and captivating description of the
Ancient World at the time of the Great War between Greeks and
Persians. Especially famous is the episode with King Croisus
of Lydia. Contemplating a war against the upstart Persian
Empire, he resolves to ask an oracle what the outcome of a potential
battle would be. However, to have some guarantee for the
veracity of any prophecy, he decides to send embassies to all the
renowned oracles of the known World. The messengers are instructed
to inquire the various divinities, on the hundredth day after
their departure, what King Croisus is doing at that very moment.
{}From the Pythia the messengers bring back the reply
\begin{em}\begin{verse}
I know the number of grains of sand as well as  the expanse of
the sea, \\
And I comprehend the dumb and hear him who does not speak, \\
There came to my mind the smell of the hard-shelled turtle, \\
Boiled in copper together with the lamb, \\
With copper below and copper above.
\end{verse}\end{em}
The veracity of the Pythia is thus established by the crafty ruler,
who had waited until the appointed day, slaughtered a turtle and a
lamb, and boiled them together in a copper cauldron with a
copper lid. Also the Oracle of Zeus Ammon in the Libyan desert
is able to give a correct reply (lost to posterity), while all
others fail. King Croisus now sends a second embassy to Delphi,
inquiring after the outcome of a battle against the Persians.
The Pythia answers
\begin{em}\begin{verse}
If Croisus passes over the Halys he will dissolve a great Empire.
\end{verse}\end{em}
Taking this to mean he would win, the King collects his army and
crosses the border river, only to suffer a crushing defeat and
see his Kingdom conquered. When the victorious King Cyrus allows
Croisus to send an embassy to upbraid the Oracle, the God Apollon
answers through his Prophetess that he has correctly predicted the
destruction of a great empire --- Croisus' own --- and that he
cannot be held responsible if people choose to interpret the
Oracle answers to their own liking.
 
The history of the {\Py} program is neither as long nor as
dignified as that of its eponym. However, some points of contact
exist. You must be very careful when you formulate the questions:
any ambiguities will corrupt the reply you get. And you must be even
more careful not to misinterpret the answers; in particular not to pick
the interpretation that suits you before considering the alternatives.
Finally, even a perfect God has servants that are only human: a priest
might mishear the screams of the Pythia and therefore produce an
erroneous oracle reply; the current author might unwittingly let a
bug free in the program {\Py}.
 
\clearpage
 
\section{Physics Overview}
 
In this section we will try to give an overview of the main physics
features of {\Py}, and also to introduce some
terminology. The details will be discussed in subsequent sections.
 
For the description of a typical high-energy event, an event
generator should contain a simulation of several physics aspects.
If we try to follow the evolution of an event in some semblance of
a time order, one may arrange these aspects as follows:
\begin{Enumerate}
\item Initially two beam particles are coming in towards each other.
      Normally each particle is characterized by a set of parton 
      distributions, which defines the partonic substructure in terms 
      of flavour composition and energy sharing.
\item One shower initiator parton from each beam starts off
      a sequence of branchings, such as $\q \to \q \g$, which build up
      an initial-state shower.
\item One incoming parton from each of the two showers
      enters the hard process, where then a number of
      outgoing partons are produced, usually two.
      It is the nature of this process that determines the main
      characteristics of the event.
\item The hard process may produce a set of short-lived resonances,
      like the $\Z^0/\W^{\pm}$ gauge bosons, whose decay to normal 
      partons has to be considered in close association with the 
      hard process itself.
\item The outgoing partons may branch, just like the incoming did, 
      to build up final-state showers.
\item In addition to the hard process considered above, further
      semihard interactions may occur between the other partons 
      of two incoming hadrons.
\item When a shower initiator is taken out of a beam particle,
      a beam remnant is left behind. This remnant may have
      an internal structure, and a net colour charge that relates
      it to the rest of the final state.
\item The QCD confinement mechanism ensures that the outgoing quarks 
      and gluons are not observable, but instead fragment to colour 
      neutral hadrons.
\item Normally the fragmentation mechanism can be seen as occurring
      in a set of separate colour singlet subsystems, but 
      interconnection effects such as colour rearrangement or 
      Bose--Einstein may complicate the picture.
\item Many of the produced hadrons are unstable and decay further.
\end{Enumerate}
 
Conventionally, only quarks and gluons are counted as partons, while
leptons and photons are not. If pushed {\it ad absurdum} this may 
lead to
some unwieldy terminology. We will therefore, where it does not matter,
speak of an electron or a photon in the `partonic' substructure of an
electron, lump branchings $\e \to \e \gamma$ together with other
`parton shower' branchings such as $\q \to \q \g$, and so on. With
this notation, the division into the above ten points applies equally
well to an interaction between two leptons, between a lepton and a
hadron, and between two hadrons.
 
In the following sections, we will survey the above ten aspects,
not in the same order as given here, but rather in the order in 
which they appear in the program execution, i.e.\ starting with the 
hard process.
 
\subsection{Hard Processes and Parton Distributions}
 
In the original {\Je} code, only two hard processes are available. The 
first and main one is $\ee \to \gammaZ \to \q \qbar$. Here the `$*$' of
$\gamma^*$ is used to denote that the photon must be off the mass shell.
The distinction is of some importance, since a photon on the mass shell
cannot decay. Of course also the $\Z^0$ can be off the mass shell,
but here the  distinction is less relevant (strictly speaking,
a $\Z^0$ is always off the mass shell). In the following we may not
always use `$*$' consistently, but the rule of thumb is to use a `$*$'
only when a process is not kinematically possible for a particle of
nominal mass. The quark $\q$ in the final state of
$\ee \to \gammaZ \to \q \qbar$
may be $\u$, $\d$, $\s$, $\c$, $\b$ or $\t$; the flavour in
each event is picked at random, according to the relative couplings,
evaluated at the hadronic c.m.\ energy. Also the angular distribution of
the final $\q \qbar$ pair is included. No parton-distribution functions 
are needed.
 
The other original {\Je} process is a routine to generate $\g \g \g$ and
$\gamma \g \g$ final states, as expected in onium 1$^{--}$ decays
such as $\Upsilon$. Given the large top mass, toponium decays weakly much 
too fast for these processes to be of any interest, so therefore no new 
applications are expected.

\subsubsection{Hard Processes}
 
The current {\Py} contains a much richer selection, with around 240
different hard processes. These may be classified in
many different ways.
 
One is according to the number of final-state objects: we speak of
`$2 \to 1$' processes, `$2 \to 2$' ones, `$2 \to 3$' ones, etc.
This aspect is very relevant from a programming point of view:
the more particles in the final state, the more complicated the
phase space and therefore the whole generation procedure. In fact,
{\Py} is optimized for $2 \to 1$ and $2 \to 2$ processes.
There is currently no generic treatment of processes with three or
more particles in the final state, but rather a few different
machineries, each tailored to the pole structure of a specific class of
graphs.
 
Another classification is according to the physics scenario. This will
be the main theme of section \ref{s:pytproc}. The following major
groups may be distinguished:
\begin{Itemize}
\item Hard QCD processes, e.g.\ $\q \g \to \q \g$.
\item Soft QCD processes, such as diffractive and elastic scattering,
and minimum-bias events. Hidden in this class is also process 96,
which is used internally for the merging of soft and hard physics,
and for the generation of multiple interactions.
\item Heavy-flavour production, both open and hidden, e.g.\ 
$\g \g \to \t \tbar$ and $\g \g \to \Jpsi \g$.
\item Prompt-photon production, e.g.\ $\q \g \to \q \gamma$.
\item Photon-induced processes, e.g.\ $\gamma \g \to \q \qbar$.
\item Deeply Inelastic Scattering, e.g.\ $\q \ell \to \q \ell$.
\item $\W / \Z$ production, such as the $\ee \to \gammaZ$ or 
$\q \qbar \to \W^+ \W^-$.
\item Standard model Higgs production, where the Higgs is reasonably
light and narrow, and can therefore still be considered as a resonance.
\item Gauge boson scattering processes, such as $\W \W \to \W \W$,  
when the Standard Model Higgs is so heavy and broad that resonant and
non-resonant contributions have to be considered together.
\item Non-standard Higgs particle production, within the framework
of a two-Higgs-doublet scenario with three neutral ($\hrm^0$, $\H^0$
and $\A^0$) and two charged ($\H^{\pm}$) Higgs states. Normally 
associated with SUSY (see below), but does not have to be. 
\item Production of new gauge bosons, such as a $\Z'$, $\W'$ and $\R$
(a horizontal boson, coupling between generations).
\item Technicolor production, as an alternative scenario to the 
standard picture of electroweak symmetry breaking by a fundamental Higgs. 
\item Compositeness is a possibility not only in the Higgs sector, 
but may also apply to fermions, e.g.\ giving $\d^*$ and $\u^*$ production.
At energies below the threshold for new particle production, contact
interactions may still modify the standard behaviour.
\item Left--right symmetric models give rise to doubly charged Higgs
states, in fact one set belonging to the left and one to the right 
{\bf SU(2)} gauge group. Decays involve right-handed $\W$'s and neutrinos. 
\item Leptoquark ($\L_{\Q}$) production is encountered in some 
beyond-the-standard-model scenarios.
\item Supersymmetry (SUSY) is probably the favourite scenario for
physics beyond the standard model. A rich set of processes are allowed,
already if one obeys $R$-parity conservation, and even more so if one 
does not. The supersymmetric machinery and process selection is inherited 
from \textsc{SPythia}~\cite{Mre97}, however with many improvements in the 
event generation chain. Many different SUSY scenarios have been 
proposed, and the program is flexible enough to allow input from several 
of these, in addition to the ones provided internally.
\item The possibility of extra dimensions at low energies has been
a topic of much study in recent years, but has still not settled down to
some standard scenarios. Its inclusion into {\Py} is also only in a very 
first stage.
\end{Itemize}
This is by no means a survey of all interesting physics. Also, within
the scenarios studied, not all contributing graphs have always been
included, but only the more important and/or more interesting ones.
In many cases, various approximations are involved in the matrix
elements coded.

\subsubsection{Resonance Decays}
\label{sss:resdecintro}

As we noted above, the bulk of the processes above are of the 
$2 \to 2$ kind, with very few leading to the production of more 
than two final-state particles.
This may be seen as a major limitation, and indeed is so
at times. However, often one can come quite far with only one
or two particles in the final state, since showers will add the
required extra activity. The classification may also be misleading
at times, since an $s$-channel resonance is considered as a single
particle, even if it is assumed always to decay into two final-state
particles. Thus the process
$\ee \to \W^+ \W^- \to \q_1 \qbar'_1 \, \q_2 \qbar'_2$ is classified
as $2 \to 2$, although the decay treatment of the $\W$ pair includes
the full $2 \to 4$ matrix elements (in the doubly resonant 
approximation, i.e.\ excluding interference with non-$\W\W$ 
four-fermion graphs).

Particles which admit this close connection between the hard process
and the subsequent evolution are collectively called resonances in
this manual. It includes all particles in mass above the
$\b$ quark system, such as $\t$, $\Z^0$, $\W^{\pm}$, $\hrm^0$, 
supersymmetric particles, and many more. Typically their decays are
given by electroweak physics, or physics beyond the Standard Model.
What characterizes a ({\Py}) resonance is that partial widths
and branching ratios can be calculated dynamically, as a function of
the actual mass of a particle. Therefore not only do branching ratios
change between an $\hrm^0$ of nominal mass 100 GeV and one of 200 GeV,
but also for a Higgs of nominal mass 200 GeV, the branching ratios
would change between an actual mass of 190 GeV and 210 GeV, say.
This is particularly relevant for reasonably broad resonances, and
in threshold regions. For an approach like this to work, it is
clearly necessary to have perturbative expressions available for all
partial widths. 

Decay chains can become quite lengthy, e.g.\ for supersymmetric processes,
but follow a straight perturbative pattern. If the simulation is 
restricted to only some set of decays, the corresponding cross section
reduction can easily be calculated. (Except in some rare cases where a
nontrivial threshold behaviour could complicate matters.) It is 
therefore standard in {\Py} to quote cross sections with such reductions 
already included. Note that the branching ratios of a particle is affected
also by restrictions made in the secondary or subsequent decays.
For instance, the branching ratio of $\hrm^0 \to \W^+ \W^-$, relative to
$\hrm^0 \to \Z^0 \Z^0$ and other channels, is changed if the allowed $\W$
decays are restricted. 

The decay products of resonances are typically quarks, leptons, or other 
resonances, e.g.\ $\W \to \q \qbar'$ or $\hrm^0 \to \W^+ \W^-$. Ordinary 
hadrons are not produced in these decays, but only in subsequent 
hadronization steps. In decays to quarks, parton showers are
automatically added to give a more realistic multijet structure, and
one may also allow photon emission off leptons. If the decay products
in turn are resonances, further decays are necessary. Often
spin information is available in resonance decay matrix elements.
This means that the angular orientations in the two decays of a
$\W^+ \W^-$ pair are properly correlated. In other cases, the
information is not available, and then resonances decay isotropically.

Of course, the above `resonance' terminology is arbitrary. A $\rho$,
for instance, could also be called a resonance, but not in the above
sense. The width is not perturbatively calculable, it decays to hadrons
by strong interactions, and so on. From a practical point of view, the
main dividing line is that the values of --- or a change in --- branching 
ratios cannot affect the cross section of a process. For instance, if 
one wanted to consider the decay $\Z^0 \rightarrow \c \cbar$, 
with a $\D$ meson producing a lepton, not only
would there then be the problem of different leptonic branching ratios
for different $\D$'s (which means that fragmentation and decay 
treatments would no longer decouple), but also that of additional
$\c \cbar$ pair production in parton-shower evolution, at a rate
that is unknown beforehand. In practice, it is therefore next to
impossible to force $\D$ decay modes in a consistent manner.
 
\subsubsection{Parton Distributions}
 
The cross section for a process $ij \to k$ is given by
\begin{equation}
\sigma_{ij \to k} = \int \d x_1 \int \d x_2 \, f^1_i(x_1) \,
f^2_j(x_2) \, \hat{\sigma}_{ij \to k} ~.
\end{equation}
Here $\hat{\sigma}$ is the cross section for the hard partonic process,
as codified in the matrix elements for each specific process.
For processes with many particles in the final state 
it would be replaced by an
integral over the allowed final-state phase space. The $f^a_i(x)$ are
the parton-distribution functions, which describe the probability to 
find a parton $i$ inside beam particle $a$, with parton $i$ carrying a
fraction $x$ of the total $a$ momentum. Actually, parton distributions
also depend on some momentum scale $Q^2$ that characterizes the hard
process.
 
Parton distributions are most familiar for hadrons, such as the proton.
Hadrons are inherently composite objects, made up of quarks and
gluons. Since we do not understand QCD, a derivation from first 
principles
of hadron parton distributions does not yet exist, although some
progress is being made in lattice QCD studies. It is therefore
necessary to rely on parameterizations, where experimental data are
used in conjunction with the evolution equations for the $Q^2$
dependence, to pin down the parton distributions. Several different
groups have therefore produced their own fits, based on slightly
different sets of data, and with some variation in the theoretical
assumptions.
 
Also for fundamental particles, such as the electron, is it convenient
to introduce parton distributions. The function $f^{\e}_{\e}(x)$ thus
parameterizes the probability that the electron that takes part in the
hard process retains a fraction $x$ of the original energy, the rest
being radiated (into photons) in the initial state. Of course, such
radiation could equally well be made part of the hard interaction,
but the parton-distribution approach usually is much more convenient.
If need be, a description with fundamental electrons is recovered for
the choice $f_{\e}^{\e}(x, Q^2)=\delta(x-1)$. Note that, contrary to
the proton case, electron parton distributions are calculable from first
principles, and reduce to the $\delta$ function above for $Q^2 \to 0$.
 
The electron may also contain photons, and the photon may in its turn
contain quarks and gluons. The internal structure of the
photon is a bit of a problem, since the photon contains a point-like 
part, which is perturbatively calculable, and a resolved part (with 
further subdivisions), which is not. Normally, the photon
parton distributions are therefore parameterized, just as the hadron
ones. Since the electron ultimately contains quarks and gluons,
hard QCD processes like $\q \g \to \q \g$ therefore not only appear in
$\pp$ collisions, but also in $\ep$ ones (`resolved photoproduction')
and in $\ee$ ones (`doubly resolved 2$\gamma$ events'). The parton 
distribution function approach here makes it much easier to reuse one 
and the same hard process in different contexts.
 
There is also another kind of possible generalization. The two
processes $\q \qbar \to \gammaZ$, studied in hadron colliders,
and $\ee \to \gammaZ$, studied in $\ee$ colliders, are really
special cases of a common process, $\f \fbar \to \gammaZ$,
where $\f$ denotes a fundamental fermion, i.e.\ a quark, lepton or
neutrino. The whole structure is therefore only coded once, and
then slightly different couplings and colour prefactors are used,
depending on the initial state considered. Usually the
interesting cross section is a sum over several different initial
states, e.g.\ $\u \ubar \to \gammaZ$ and
$\d \dbar \to \gammaZ$ in a hadron collider. This kind of
summation is always implicitly done, even when not explicitly
mentioned in the text.
 
\subsection{Initial- and Final-State Radiation}
 
In every process that contains coloured and/or charged objects
in the initial or final state, gluon and/or photon radiation
may give large corrections to the overall topology of events.
Starting from a basic $2 \to 2$ process, this kind of corrections
will generate $2 \to 3$, $2 \to 4$, and so on, final-state
topologies. As the available energies are increased, hard
emission of this kind is increasingly important, relative to
fragmentation, in determining the event structure.
 
Two traditional approaches exist to the modelling of perturbative
corrections. One is the matrix-element method, in
which Feynman diagrams are calculated, order by order. In principle,
this is the correct approach, which takes into account exact
kinematics, and the full interference and helicity structure. The
only problem is that calculations become increasingly difficult in
higher orders, in particular for the loop graphs.
Only in exceptional cases have therefore more than one loop been
calculated in full, and often we do not have any loop corrections
at all at our disposal. On the other hand,
we have indirect but strong evidence that, in fact, the emission
of multiple soft gluons plays a significant r\^ole in building up the
event structure, e.g.\ at LEP, and this sets a limit to the
applicability of matrix elements.
Since the phase space available for gluon emission increases with
the available energy, the matrix-element approach becomes less
relevant for the full structure of events at higher energies.
However, the perturbative expansion is better behaved at higher 
energy scales, owing to the running of $\alphas$. As a consequence, 
inclusive measurements, e.g.\ of the rate of  well-separated jets, 
should yield more reliable results at high energies.
 
The second possible approach is the parton-shower one. Here an
arbitrary number of branchings of one parton into two (or more)
may be combined, to yield a description of multijet events,
with no explicit upper limit on the number of partons involved.
This is possible since the full matrix-element expressions are
not used, but only approximations derived by simplifying the
kinematics, and the interference and helicity structure. Parton showers
are therefore expected to give a good description of the substructure of
jets, but in principle the shower approach has limited predictive power
for the rate of well-separated jets (i.e.\ the 2/3/4/5-jet composition).
In practice, shower programs may be matched to first-order matrix 
elements to describe the hard-gluon emission region reasonably well, 
in particular for the $\ee$ annihilation process.
Nevertheless, the shower description is not optimal for
absolute $\alphas$ determinations.
 
Thus the two approaches are complementary in many respects,
and both have found use. However, because of its simplicity
and flexibility, the parton-shower option is generally the
first choice, while the matrix elements one is mainly used for
$\alphas$ determinations, angular distribution of jets,
triple-gluon vertex studies, and other specialized studies.
Obviously, the ultimate goal would be to have an approach where the
best aspects of the two worlds are harmoniously married. This is
currently a topic of quite some study.
 
\subsubsection{Matrix elements}
 
Matrix elements are especially made use of in the older {\Je}-originated
implementation of the process $\ee \to \gammaZ \to \q \qbar$.
 
For initial-state QED radiation, a first order (un-exponentiated)
description has been adopted. This means that events are subdivided
into two classes, those where a photon is radiated above some
minimum energy, and those without such a photon. In the latter
class, the soft and virtual corrections have been lumped together
to give a total event rate that is correct up to one loop. This 
approach worked fine at PETRA/PEP energies, but does not do so well
for the $\Z^0$ line shape, i.e.\ in regions where the cross section
is rapidly varying and high precision is strived for.
 
For final-state QCD radiation, several options are available. The
default is the parton-shower one (see below), but the matrix-elements
options are also frequently used. In the definition of 3- or
4-jet events, a cut is introduced whereby it is required that
any two partons have an invariant mass bigger than some fraction
of the c.m.\ energy. 3-jet events which do not fulfil this
requirement are lumped with the 2-jet ones. The first-order
matrix-element option, which only contains 3- and 2-jet events
therefore involves no ambiguities. In second order, where also
4-jets have to be considered, a main issue is what to do with
4-jet events that fail the cuts. Depending on the choice of
recombination scheme, whereby the two nearby partons are joined
into one, different 3-jet events are produced. Therefore the
second-order differential 3-jet rate has been the subject of
some controversy, and the program actually contains two
different implementations.
 
By contrast, the normal {\Py} event generation machinery does not 
contain any full higher-order matrix elements, with loop 
contributions included. There are several cases where higher-order 
matrix elements are included at the Born level. Consider the case 
of resonance production at a hadron collider, e.g.\ of a $\W$,
which is contained in the lowest-order process $\q \qbar' \to \W$.
In an inclusive description, additional jets recoiling against the
$\W$ may be generated by parton showers. {\Py} also contains
the two first-order processes $\q \g \to \W \q'$ and
$\q \qbar' \to \W \g$. The cross sections for these processes
are divergent when the $\pT \to 0$. In this region a correct
treatment would therefore have to take into account loop corrections,
which are not available in {\Py}. 

Even without having these accessible, we know approximately what the 
outcome should be. The virtual corrections have to cancel the 
$\pT \to 0$ singularities of the real emission. The total cross 
section of $\W$ production therefore receives finite 
$\mathcal{O}(\alphas)$ corrections to the lowest-order answer. These 
corrections can often be neglected to first approximation, except 
when high precision is required. As for the shape of the $\W$ $\pT$ 
spectrum, the large cross section for low-$\pT$ emission has to be
interpreted as allowing more than one emission to take place. A
resummation procedure is therefore necessary to have matrix element
make sense at small $\pT$. The outcome is a cross section below the
naive one, with a finite behaviour in the $\pT \to 0$ limit.  

Depending on the physics application, one could then use {\Py} in one 
of two ways. In an inclusive description, which is dominated by the 
region of reasonably small $\pT$, the preferred option is 
lowest-order matrix elements combined with parton showers, which
actually is one way of achieving the required resummation. For $\W$
production as background to some other process, say, only the 
large-$\pT$ tail might be of interest. Then the shower approach may 
be inefficient, since only few events will end up in the interesting
region, while the matrix-element alternative allows reasonable cuts to
be inserted from the beginning of the generation procedure. (One would 
probably still want to add showers to describe additional softer 
radiation, at the cost of some smearing of the original cuts.) 
Furthermore, and not less importantly, the matrix elements should give 
a more precise prediction of the high-$\pT$ event rate than the 
approximate shower procedure. 

In the particular case considered here, that of $\W$ production, and a
few similar processes, actually the shower has been improved by a matching 
to first-order matrix elements, thus giving a decent description over the 
whole $\pT$ range. This does not provide the first-order corrections to 
the total $\W$ production  rate, however, nor the possibility to select 
only a high-$\pT$ tail of events.
 
\subsubsection{Parton showers}
 
The separation of radiation into initial- and final-state showers is
arbitrary, but very convenient. There are also situations where it
is appropriate: for instance, the process 
$\ee \to \Z^0 \to \q \qbar$ only
contains final-state QCD radiation (QED radiation, however, is
possible both in the initial and final state), while
$\q \qbar \to \Z^0 \to \ee$ only contains initial-state QCD one.
Similarly, the distinction of emission as coming either from the
$\q$ or from the $\qbar$ is arbitrary. In general, the assignment
of radiation to a given mother parton is a good approximation
for an emission close to the direction of motion of that parton,
but not for the wide-angle emission in between two jets, where
interference terms are expected to be important.
 
In both initial- and final-state showers, the structure is given in
terms of branchings $a \to bc$, specifically $\e \to \e \gamma$,
$\q \to \q \g$, $\q \to \q \gamma$, $\g \to \g \g$, and 
$\g \to \q \qbar$. (Further branchings, like $\gamma \to \e^+ \e^-$ 
and $\gamma \to \q \qbar$, could also have been added, but have not 
yet been of interest.) Each of these processes is characterized by
a splitting kernel $P_{a \to bc}(z)$. The branching rate is
proportional to the integral $\int P_{a \to bc}(z) \, \d z$.
The $z$ value picked for a branching describes the energy sharing,
with daughter $b$ taking a fraction $z$ and daughter $c$ the
remaining $1-z$ of the mother energy. Once formed, the daughters 
$b$ and $c$ may in turn branch, and so on.
 
Each parton is characterized by some virtuality scale $Q^2$,
which gives an approximate sense of time ordering
to the cascade. In the initial-state shower, $Q^2$ values are
gradually increasing as the hard scattering is approached, while
$Q^2$ is decreasing in the final-state showers.
Shower evolution is cut off at some lower
scale $Q_0$, typically around 1 GeV for QCD branchings. From above,
a maximum scale $Q_{\mmax}$ is introduced, where the showers are
matched to the hard interaction itself. The relation between 
$Q_{\mmax}$ and the kinematics of the hard scattering
is uncertain, and the choice made can strongly affect the
amount of well-separated jets.
 
Despite a number of common traits, the initial- and final-state
radiation machineries are in fact quite different, and are described
separately below. 
 
Final-state showers are time-like,
i.e.\ partons have $m^2 = E^2 - \mbf{p}^2 \geq 0$. The evolution
variable $Q^2$ of the cascade is therefore in {\Py}
associated with the $m^2$ of the
branching parton, but this choice is not unique.
Starting from $Q^2_{\mmax}$, an original parton is evolved
downwards in $Q^2$ until a branching occurs. The selected
$Q^2$ value defines the mass of the branching parton, and the $z$
of the splitting kernel the parton energy division between its
daughters. These daughters may now, in turn, evolve
downwards, in this case with maximum virtuality already defined by
kinematics, and so on down to the $Q_0$ cut-off.
 
In QCD showers, corrections to the leading-log picture, so-called 
coherence effects, lead to an ordering of subsequent emissions in terms
of decreasing angles. This does not follow automatically from the
mass-ordering constraint, but is implemented as an additional
requirement on allowed emissions. Photon emission is not affected
by angular ordering. It is also possible to obtain
non-trivial correlations between azimuthal angles in the various
branchings, some of which are implemented as
options. Finally, the theoretical analysis strongly
suggests the scale choice $\alphas = \alphas(\pT^2) =
\alphas(z(1-z)m^2)$, and this is the default in the program.
 
The final-state radiation machinery is normally applied in the c.m.\ 
frame of the hard scattering or a decaying resonance. The total energy 
and momentum of that subsystem is preserved, as is the direction of the 
outgoing partons (in their common rest frame), where applicable.
 
In contrast to final-state showers, initial-state ones are space-like.
This means that, in the sequence of branchings $a \to bc$ that lead
up from the shower initiator to the hard interaction,
particles $a$ and $b$ have $m^2 = E^2 - \mbf{p}^2 <0$.
The `side branch' particle $c$, which does not participate
in the hard scattering, may be on the mass shell, or have a time-like
virtuality. In the latter case a time-like shower will evolve off
it, rather like the final-state radiation described above. To first
approximation, the evolution of the space-like main branch
is characterized by the
evolution variable $Q^2 = -m^2$, which is required to be strictly
increasing along the shower, i.e.\ $Q_b^2 > Q_a^2$. Corrections
to this picture have been calculated,
but are basically absent in {\Py}.
 
Initial-state radiation is handled within the backwards evolution
scheme. In this approach, the
choice of the hard scattering is based on the use of evolved
parton distributions, which means that
the inclusive effects of initial-state radiation are already
included. What remains is therefore to construct the exclusive
showers. This is done starting from the two incoming partons
at the hard interaction, tracing the showers `backwards in time',
back to the two shower initiators. In other words,
given a parton $b$, one tries to find the parton $a$ that branched
into $b$. The evolution in the Monte Carlo is therefore in
terms of a sequence of decreasing space-like virtualities $Q^2$
and increasing momentum fractions $x$. Branchings on
the two sides are interleaved in a common sequence of
decreasing $Q^2$ values.
 
In the above formalism, there is no real distinction between
gluon and photon emission. Some of the details actually do differ,
as will be explained in the full description.
 
The initial- and final-state radiation shifts around the kinematics of
the original hard interaction. In Deeply Inelastic Scattering, this
means that the $x$ and $Q^2$ values that can be derived from the 
momentum of the scattered lepton do not automatically agree with the 
values originally picked. In high-$\pT$ processes, it means that one no 
longer has two jets with opposite and compensating $\pT$, but more 
complicated topologies. Effects of any original kinematics selection 
cuts are therefore smeared out, an unfortunate side-effect of the 
parton-shower approach.
 
\subsection{Beam Remnants and Multiple Interactions}
 
In a hadron--hadron collision, the initial-state radiation algorithm
reconstructs one shower initiator in each beam. This initiator only
takes some fraction of the total beam energy, leaving behind a beam
remnant which takes the rest. For a proton beam, a $\u$ quark
initiator would leave behind a $\u \d$ diquark beam remnant, with an
antitriplet colour charge. The remnant is therefore colour-connected
to the hard interaction, and forms part of the same fragmenting
system. It is further customary to assign a primordial transverse
momentum to the shower initiator, to take into account the motion
of quarks inside the original hadron, at least as required by the
uncertainty principle by the proton size, probably augmented by 
unresolved (i.e.\ not simulated) soft shower activity. This primordial 
$k_{\perp}$ is selected according to some suitable distribution, and 
the recoil is assumed to be taken up by the beam remnant.
 
Often the remnant is more complicated, e.g.\ a gluon initiator
would leave behind a $\u \u \d$ proton remnant system in a colour octet
state, which can conveniently be subdivided into a colour triplet
quark and a colour antitriplet diquark, each of which are 
colour-connected to the hard interaction. The energy sharing between 
these two remnant objects, and their relative transverse momentum,
introduces additional degrees of freedom, which are not understood 
from first principles.
 
Na\"{\i}vely, one would expect an $\ep$ event to have only one beam
remnant, and an $\ee$ event none. This is not always correct, e.g.
a $\gamma \gamma \to \q \qbar$ interaction in an $\ee$ event would
leave behind the $\e^+$ and $\e^-$ as beam remnants, and a
$\q \qbar \to \g \g$ interaction in resolved photoproduction in an
$\ee$ event would leave behind one $\e^{\pm}$ and one $\q$ or $\qbar$
in each remnant. Corresponding complications occur for
photoproduction in $\ep$ events.
 
There is another source of beam remnants. If parton distributions are
used to resolve an electron inside an electron, some of the original
energy is not used in the hard interaction, but is rather associated
with initial-state photon radiation. The initial-state shower is
in principle intended to trace this evolution and reconstruct the
original electron before any radiation at all took place. However,
because of cut-off procedures, some small amount may be left 
unaccounted for.
Alternatively, you may have chosen to switch off initial-state
radiation altogether, but still preserved the resolved electron
parton distributions. In either case the remaining energy is given to
a single photon of vanishing transverse momentum, which is then
considered in the same spirit as `true' beam remnants.
 
So far we have assumed that
each event only contains one hard interaction, i.e.\ that each
incoming particle has only one parton which takes part in hard
processes, and that all other constituents sail through unaffected.
This is appropriate in $\ee$ or $\ep$ events, but not necessarily so in
hadron--hadron collisions. Here each of the beam particles contains a
multitude of partons, and so the probability for several interactions
in one and the same event need not be negligible. In principle these
additional interactions could arise because one single parton from
one beam scatters against several different partons from the other
beam, or because several partons from each beam take place in
separate $2 \to 2$ scatterings. Both are expected, but combinatorics
should favour the latter, which is the mechanism considered in
{\Py}.
 
The dominant $2 \to 2$ QCD cross sections  are
divergent for $\pT \to 0$, and drop rapidly for larger
$\pT$. Probably the lowest-order perturbative cross sections
will be regularized at small $\pT$ by colour coherence effects:
an exchanged gluon of small $\pT$ has a large transverse
wave function and can therefore not resolve the individual colour
charges of the two incoming hadrons; it will only couple to an
average colour charge that vanishes in the limit $\pT \to 0$.
In the program, some effective $\pTmin$ scale is therefore
introduced, below which the perturbative cross section is either
assumed completely vanishing or at least strongly damped.
Phenomenologically, $\pTmin$ comes out to be a number of 
the order of 1.5--2.0~GeV, with some energy dependence.
 
In a typical `minimum-bias' event one therefore expects to find one
or a few scatterings at scales around or a bit above 
$\pTmin$,
while a high-$\pT$ event also may have additional scatterings
at the $\pTmin$ scale. The probability to have several
high-$\pT$ scatterings in the same event is small, since the
cross section drops so rapidly with $\pT$.
 
The understanding of multiple interaction is still very primitive. 
{\Py} therefore contains several different options,
with a fairly advanced one as default. The options differ e.g.\
on the issue of the `pedestal' effect: is there an increased
probability or not for additional interactions in an event which
is known to contain a hard scattering, compared with one that
contains no hard interactions? Other differences concern the level 
of detail in the generation of scatterings after the first one, and 
the model that describes how the scatterings are intercorrelated in 
colour space.
 
\subsection{Hadronization}
 
QCD perturbation theory, formulated in terms of quarks and
gluons, is valid at short distances. At long distances, QCD
becomes strongly interacting and perturbation theory breaks
down. In this confinement regime, the coloured partons are
transformed into colourless hadrons, a process called either
hadronization or fragmentation. In this paper we reserve the
former term for the combination of fragmentation and the subsequent
decay of unstable particles.
 
The fragmentation process has yet to be understood from first
principles, starting from the QCD Lagrangian. This has left the
way clear for the development of a number of different
phenomenological models. Three main schools are usually
distinguished, string fragmentation (SF), independent
fragmentation (IF) and cluster fragmentation (CF),
but many variants and hybrids exist.
Being models, none of them can lay claims
to being `correct', although some may be better founded than
others. The best that can be aimed for is internal consistency, 
a good representation of existing data, and a predictive power for
properties not yet studied or results at higher energies.
 
\subsubsection{String Fragmentation}

The original {\Je} program is intimately connected with string 
fragmentation, in the form of the time-honoured `Lund model'. This 
is the default for all {\Py} applications, but independent
fragmentation options also exist, for applications
where one wishes to study the importance of string effects.
 
All current models are of a probabilistic and iterative nature.
This means that the fragmentation process as a whole is described in
terms of one or a few simple underlying branchings, of the type
jet $\to$ hadron + remainder-jet, string $\to$
hadron + remainder-string, and so on. At each branching,
probabilistic rules are given for the production of new flavours,
and for the sharing of energy and momentum between the products.
 
To understand fragmentation models, it is useful to start with
the simplest possible system, a colour-singlet $\q \qbar$ 2-jet
event, as produced in $\ee$ annihilation. Here lattice QCD studies
lend support to a linear confinement picture (in the absence
of dynamical quarks), i.e.\ the energy stored in the colour
dipole field between a charge and an anticharge increases linearly
with the separation between the charges, if the short-distance
Coulomb term is neglected. This is quite different
from the behaviour in QED, and is related to the presence of a
triple-gluon vertex in QCD. The details are not yet well
understood, however.
 
The assumption of linear confinement provides the starting point for
the string model. As the $\q$ and $\qbar$ partons move apart from
their common production vertex, the physical picture is that of a
colour flux tube (or maybe a colour vortex line) being stretched
between the $\q$ and the $\qbar$. The transverse dimensions
of the tube are of typical hadronic sizes, roughly 1 fm. If
the tube is assumed to be uniform along its length, this
automatically leads to a confinement picture with a linearly
rising potential. In order to obtain a Lorentz covariant and causal
description of the energy flow due to this linear confinement,
the most straightforward way is to use the dynamics of the massless
relativistic string with no transverse degrees of freedom.
The mathematical, one-dimensional string can
be thought of as parameterizing the position of the axis of a
cylindrically symmetric flux tube. From
hadron spectroscopy, the string constant, i.e.\ the amount of
energy per unit length, is deduced to be
$ \kappa \approx 1$ GeV/fm. The
expression `massless' relativistic string is somewhat of a
misnomer: $\kappa$ effectively corresponds to a `mass density' along
the string.
 
Let us now turn to the fragmentation process. As the $\q$ and
$\qbar$ move apart, the potential energy stored in the string
increases, and the string may break by the production of a new
$\q' \qbar'$ pair, so that the system splits into two
colour-singlet systems $\q \qbar'$ and $\q' \qbar$. If the invariant
mass of either of these string pieces is large enough, further
breaks may occur. In the Lund string model, the string break-up
process is assumed to proceed until only on-mass-shell hadrons
remain, each hadron corresponding to a small piece of string
with a quark in one end and an antiquark in the other.
 
In order to generate the quark--antiquark pairs $\q' \qbar'$ which
lead to string break-ups, the Lund model invokes the idea of
quantum mechanical tunnelling. This leads to a flavour-independent
Gaussian spectrum for the $\pT$ of $\q' \qbar'$ pairs.
Since the string is assumed to have no transverse excitations,
this $\pT$ is locally compensated between the quark and the
antiquark of the pair. The total $\pT$ of a hadron is made
up out of the $\pT$ contributions from the quark and
antiquark that together
form the hadron. Some contribution of very soft perturbative gluon
emission may also effectively be included in this description.
 
The tunnelling picture also implies a suppression of heavy-quark
production, $\u : \d : \s : \c \approx 1 : 1 : 0.3 : 10^{-11}$.
Charm and heavier quarks hence are not expected to be produced in
the soft fragmentation, but only in perturbative parton-shower
branchings $\g \to \q \qbar$.
 
When the quark and antiquark from two adjacent string breaks are
combined to form a meson, it is necessary to invoke an algorithm to
choose between the different allowed possibilities, notably
between pseudoscalar and vector mesons.
Here the string model is not particularly predictive. Qualitatively one
expects a $1 : 3$ ratio, from counting the number of spin states,
multiplied by some wave-function normalization factor, which should
disfavour heavier states.
 
A tunnelling mechanism can also be used to explain the production of
baryons. This is still a poorly understood area. In the simplest
possible approach, a diquark in a colour antitriplet state is just
treated like an ordinary antiquark, such that a string can break
either by quark--antiquark or antidiquark--diquark pair production.
A more complex scenario is the `popcorn' one, where
diquarks as such do not exist, but rather quark--antiquark pairs
are produced one after the other. This latter picture gives a less
strong correlation in flavour and momentum space between the
baryon and the antibaryon of a pair.
 
In general, the different string breaks are causally disconnected.
This means that it is possible to describe the breaks in any convenient
order, e.g.\ from the quark end inwards. One therefore is led to write
down an iterative scheme for the fragmentation, as follows.
Assume an initial quark $\q$ moving out along the $+z$ axis, with the
antiquark going out in the opposite direction.
By the production of a $\q_1 \qbar_1$ pair, a meson with flavour content
$\q \qbar_1$ is produced, leaving behind an unpaired quark $\q_1$.
A second pair $\q_2 \qbar_2$ may now be produced, to give a new meson 
with flavours $\q_1 \qbar_2$, etc. At each step the produced
hadron takes some fraction of the available energy and momentum.
This process may be iterated until all energy is used up, with some
modifications close to the $\qbar$ end of the string in order to 
make total energy and momentum come out right.
 
The choice of starting the fragmentation from the quark end is
arbitrary, however. A fragmentation process described in terms of
starting at the $\qbar$ end of the system and fragmenting towards
the $\q$ end should be equivalent.
This `left--right' symmetry constrains the allowed shape of the
fragmentation function $f(z)$, where $z$ is the fraction
of the remaining light-cone momentum $E \pm p_z$ (+ for the $\q$ jet,
$-$ for the $\qbar$ one) taken by each new particle.
The resulting `Lund symmetric fragmentation function' has two free
parameters, which are determined from data.
 
If several partons are moving apart from a common origin, the details
of the string drawing become more complicated. For a $\q \qbar \g$
event, a string is stretched from the
$\q$ end via the $\g$ to the $\qbar$ end, i.e.
the gluon is a kink on the string, carrying energy and momentum.
As a consequence, the gluon has two string pieces attached, and
the ratio of gluon to quark string force is 2, a number which
can be compared with the ratio of colour charge Casimir operators,
$N_C/C_F = 2/(1-1/N_C^2) = 9/4$. In this, as in other
respects, the string model can be viewed as a variant of QCD
where the number of colours $N_C$ is not 3 but infinite.
Note that the factor 2 above does not depend on
the kinematical configuration: a smaller opening angle between
two partons corresponds to a smaller
string length drawn out per unit time, but also to an increased
transverse velocity of the string piece, which gives an exactly
compensating boost factor in the energy density per unit string
length.
 
The $\q \qbar \g$ string will fragment along its length. To first
approximation this means that there is
one fragmenting string piece between
$\q$ and $\g$ and a second one between $\g$ and $\qbar$. One hadron
is straddling both string pieces, i.e.\ sitting around the gluon
corner. The rest of the particles are produced as in two simple
$\q \qbar$ strings, but strings boosted with respect to the overall
c.m.\ frame. When considered in detail, the string motion and
fragmentation is more complicated, with the appearance of
additional string regions during the time evolution of the system.
These corrections are especially important for soft and
collinear gluons, since they provide a smooth transition between
events where such radiation took place and events where it did not.
Therefore the string fragmentation scheme is `infrared safe' with
respect to soft or collinear gluon emission.
 
For events that involve many partons, there may be several possible
topologies for their ordering along the string.
An example would be a $\q \qbar \g_1 \g_2$ (the gluon indices are here
used to label two different gluon-momentum vectors), where the
string can connect the partons in either of the sequences
$\q - \g_1 - \g_2 - \qbar$ and $\q - \g_2 - \g_1 - \qbar$.
The matrix elements that are calculable in perturbation theory
contain interference terms between these two possibilities, which
means that the colour flow is not always well-defined. Fortunately,
the interference terms are down in magnitude by a factor
$1/N_C^2$, where $N_C = 3$ is the number of colours, so
approximate recipes can be found. In the leading log shower
description, on the other hand, the rules for the colour flow are
well-defined. 

A final comment: in the argumentation for the importance of colour flows 
there is a tacit assumption that soft-gluon exchanges between partons 
will not normally mess up the original colour assignment. Colour
rearrangement models provide toy scenarios wherein deviations from this
rule could be studied. Of particular interest has been the process
$\e^+ \e^- \to \W^+ \W^- \to \q_1 \qbar_2 \q_3 \qbar_4$, where the
original singlets $\q_1 \qbar_2$ and $\q_3 \qbar_4$ could be rearranged 
to $\q_1 \qbar_4$ and $\q_3 \qbar_2$. So far, there are no experimental
evidence for dramatic effects of this kind, but the more realistic models
predict effects sufficiently small that these have not been ruled out.
Another example of nontrivial effects is that of Bose--Einstein correlations
between identical final-state particles, which reflect the true quantum 
nature of the hadronization process. 
 
\subsubsection{Decays}
 
A large fraction of the particles produced by fragmentation are
unstable and subsequently decay into the observable stable (or
almost stable) ones. It  is therefore important to include all
particles with their proper mass distributions and decay properties. 
Although involving little deep physics, this is less trivial than 
it may sound: while a lot of experimental information is available, 
there is also very much that is missing. For charm mesons,
it is necessary to put together measured exclusive branching ratios
with some inclusive multiplicity distributions to obtain a consistent
and reasonably complete set of decay channels, a rather delicate
task. For bottom even less is known, and for some $\B$ baryons only 
a rather simple phase-space type of generator has been used for hadronic 
decays.
 
Normally it is assumed that decay products are distributed according
to phase space, i.e.\ that there is no dynamics involved in their
relative distribution. However, in many cases additional assumptions
are necessary, e.g.\ for semileptonic decays of charm and bottom
hadrons one needs to include the proper weak matrix elements.
Particles may also be produced polarized and impart a non-isotropic
distribution to their decay products. Many of these effects are
not at all treated in the program. In fact, spin information is not
at all carried along, but has to be reconstructed explicitly when
needed.
 
This normal decay treatment makes use of a set of tables where 
branching ratios and decay modes are stored. It encompasses all
hadrons made out of $\d$, $\u$, $\s$, $\c$ and $\b$ quarks, and also
the leptons. The decay products are hadrons, leptons and photons.
Some $\b\bbar$ states are sufficiently heavy that they are allowed to 
decay to partonic states, like $\Upsilon \to \g\g\g$, which 
subsequently fragment, but these are exceptions.   

You may at will change the particle properties, decay channels or 
branching ratios of the above particles. There is no censorship what is
allowed or not allowed, beyond energy--momentum and (electrical and
colour) charge conservation. There is also no impact e.g.\ on the cross 
section of processes, since there is no way of knowing e.g.\ if the
restriction to one specific decay of a particle is because that decay
is of particular interest to us, or because recent measurement have 
shown that this indeed is the only channel. Furthermore, the number of
particles produced of each species in the hadronization process is not
known beforehand, and so cannot be used to correctly bias the preceding 
steps of the generation chain. All of this contrasts with the class of
`resonances' described above, in section \ref{sss:resdecintro}.  
 
\clearpage
 
\section{Program Overview}
 
This section contains a diverse collection of information. The first 
part is an overview of previous {\Je} and {\Py} versions.
The second gives instructions for installation of the program and
describes its philosophy: how it is constructed and how it is supposed 
to be used. It also contains some information on how to read this manual.
The third and final part contains several examples of pieces
of code or short programs, to illustrate the general style of
program usage. This last part is mainly intended as an introduction for
completely new users, and can be skipped by more experienced ones.

The combined {\Py} package is completely self-contained.
Interfaces to externally defined subprocesses, parton-distribution 
function libraries, $\tau$ decay libraries, and a time routine 
are provided, however, plus a few other optional interfaces.
 
Many programs written by other persons make use of {\Py}, especially
the string fragmentation machinery. It is not the intention to give a 
complete list here.
A majority of these programs are specific to given collaborations,
and therefore not publicly distributed. Below we give a list of a
few public programs from the `Lund group', which may have a
somewhat wider application. None of them are supported by the
{\Py} author team, so any requests should be directed to the persons
mentioned.
\begin{Itemize}
\item \tsc{Ariadne} is a generator for dipole emission, written
mainly by L. L\"onnblad \cite{Pet88}. 
\item \tsc{Aroma} is a generator for heavy-flavour processes in
leptoproduction, written by G.~Ingelman, J.~Rathsman and G. Schuler 
\cite{Ing88}.
\item \tsc{Fritiof} is a generator for hadron--hadron, hadron--nucleus
and nucleus--nucleus collisions \cite{Nil87}.
\item \tsc{Lepto} is a leptoproduction event generator, written mainly
by G. Ingelman \cite{Ing80}. It can generate parton configurations
in Deeply Inelastic Scattering according to a number of possibilities.
\item \tsc{Pompyt} is a generator for pomeron interactions written by
G. Ingelman and collaborators \cite{Bru96}. 
\end{Itemize}
One should also note that a version of {\Py} has been modified to 
include the effects of longitudinally polarized incoming protons. 
This is the work of St.~G\"ullenstern et al.\ \cite{Gul93}.
 
\subsection{Update History}
  
For the record, in Tables \ref{Jetsetver} and \ref{Pythiaver} we
list the official main versions of {\Je} and {\Py}, respectively, 
with some brief comments.
 
\begin{table}[tp]
\captive{The main versions of {\Je}, with their date of
appearance, published manuals, and main changes from previous
versions. \protect\label{Jetsetver}} \\
\vspace{1ex}
\begin{center}
\begin{tabular}{|c|c|c|l|@{\protect\rule{0mm}{\tablinsep}}}
\hline
No. & Date  & Publ.   &  Main new or improved
features \\[1mm]
\hline
1   & Nov 78 & \cite{Sjo78} &  single-quark jets \\
2   & May 79 & \cite{Sjo79} & heavy-flavour jets \\
3.1 & Aug 79 &   ---   &  2-jets in $\ee$, preliminary 3-jets \\
3.2 & Apr 80 & \cite{Sjo80} &  3-jets in $\ee$ with full matrix
                              elements, \\
    &       &         &  toponium $\to \g \g \g$ decays \\
3.3 & Aug 80 &   ---   &  softer fragmentation spectrum \\
4.1 & Apr 81 &   ---   &  baryon production and diquark
                         fragmentation, \\
    &       &         &  fourth-generation quarks, larger
                         jet systems \\
4.2 & Nov 81 &   ---   &  low-$\pT$ physics \\
4.3 & Mar 82 & \cite{Sjo82} &  4-jets and QFD structure in $\ee$, \\
    & Jul 82 & \cite{Sjo83} &  event-analysis routines \\
5.1 & Apr 83 &   ---   &  improved string fragmentation scheme,
                          symmetric \\
    &       &         &  fragmentation, full 2$^{\mrm{nd}}$ order QCD
                         for $\ee$ \\
5.2 & Nov 83 &   ---   &  momentum-conservation schemes for IF, \\
    &       &         &  initial-state photon radiation in $\ee$ \\
5.3 & May 84 &   ---   &  `popcorn' model for baryon production \\
6.1 & Jan 85 &   ---   &  common blocks restructured, parton showers \\
6.2 & Oct 85 & \cite{Sjo86} &  error detection \\
6.3 & Oct 86 & \cite{Sjo87} &  new parton-shower scheme \\
7.1 & Feb 89 &   ---   &  new particle codes and common block
                         structure, \\
    &       &         &  more mesons, improved decays, vertex
                         information, \\
    &       &         &  Abelian gluon model, Bose--Einstein
                         effects \\
7.2 & Nov 89 &   ---   &  interface to new standard common block, \\
    &         &        &  photon emission in showers \\
7.3 & May 90 & \cite{Sjo92d}  &  expanded support for non-standard
                particles \\
7.4 & Dec 93 & \cite{Sjo94}  &  updated particle data and defaults 
\\[1mm] 
\hline
\end{tabular}
\end{center}
\end{table}
 
\begin{table}[tp]
\captive{The main versions of {\Py}, with their date of
appearance, published manuals, and main changes from previous
versions. \protect\label{Pythiaver}} \\
\vspace{1ex}
\begin{center}
\begin{tabular}{|c|c|c|l|@{\protect\rule{0mm}{\tablinsep}}}
\hline
No. & Date  & Publ.   &  Main new or improved features \\[1mm]
\hline
1   & Dec 82 & \cite{Ben84} &  synthesis of
     predecessors \tsc{Compton}, \tsc{Highpt} and \\
    &       &         &  \tsc{Kassandra} \\
2   &  ---  &         & \\
3.1 &  ---  &         & \\
3.2 &  ---  &         & \\
3.3 & Feb 84 & \cite{Ben84a} & scale-breaking parton distributions \\
3.4 & Sep 84 & \cite{Ben85} & more efficient kinematics selection \\
4.1 & Dec 84 &         &  initial- and final-state parton showers,
      $\W$ and $\Z$ \\
4.2 & Jun 85 &         &  multiple interactions \\
4.3 & Aug 85 &         &  $\W \W$, $\W \Z$, $\Z \Z$ and $\R$
      processes \\
4.4 & Nov 85 &         &  $\gamma \W$, $\gamma \Z$, $\gamma \gamma$
      processes \\
4.5 & Jan 86 &         &  $\H^0$ production, diffractive and
      elastic events \\
4.6 & May 86 &         &  angular correlation in resonance
     pair decays \\
4.7 & May 86 &         &  $\Z'^0$ and $\H^+$ processes \\
4.8 & Jan 87 & \cite{Ben87} &  variable impact parameter
      in multiple interactions \\
4.9 & May 87 &         &  $\g \H^+$ process \\
5.1 & May 87 &         &  massive matrix elements for heavy quarks \\
5.2 & Jun 87 &         &  intermediate boson scattering \\
5.3 & Oct 89 &         &  new particle and subprocess codes,
    new common block \\
    &       &         &  structure, new kinematics selection,
    some \\
    &       &         &  lepton--lepton and lepton--hadron
    interactions, \\
    &       &         &  new subprocesses \\
5.4 & Jun 90 &         &  $s$-dependent widths, resonances not on the
    mass shell, \\
    &       &         &  new processes, new parton distributions \\
5.5 & Jan 91 &         &  improved $\ee$ and $\ep$, several new
    processes \\
5.6 & Sep 91 & \cite{Sjo92d} &  reorganized parton distributions, 
    new processes, \\
    &        &         &  user-defined external processes \\
5.7 & Dec 93 & \cite{Sjo94}  &  new total cross sections, 
    photoproduction, top decay \\
6.1 & Mar 97 & \cite{Sjo01} & 
             merger with {\Je}, double precision, supersymmetry, \\
    &   &  & technicolor, extra dimensions, etc. new processes,  \\
    &   &  & improved initial- and final-state showers, \\
    &   &  & baryon production, virtual photon processes \\
6.2 & Aug 01 & \cite{Sjo01a} & user processes, R-parity 
    violation \\
6.3 & Aug 03 & this & improved multiple interactions \\[1mm]
\hline
\end{tabular}
\end{center}
\end{table}
 
All versions preceding {\Py}~6.1 should now be considered obsolete, 
and are no longer maintained. For stable applications, the earlier 
combination {\Je}~7.4 and {\Py}~5.7 could still be used, however.
(A note on backwards compatibility:  persons who have code that relies 
on the old \ttt{LUJETS} single precision common block could easily write 
a translation routine to copy the \ttt{PYJETS} double precision 
information to \ttt{LUJETS}. In fact, among the old {\Je}~7 routines,
only \ttt{LUGIVE} and \ttt{LULOGO} routines have access to some {\Py}
common blocks, and therefore these are the only ones that need to
be modified if one, for some reason, would like to combine {\Py}~6 with 
the old {\Je}~7 routines.)  

The move from {\Je}~7.4 and {\Py}~5.7 to {\Py}~6.1 was a major one.
For reasons of space, individual points are therefore not listed 
separately below, but only the main ones. The {\Py} web page contains
complete update notes, where all changes are documented by topic and 
subversion. 

The main new features of {\Py}~6.1, either present from the beginning
or added later on, include:
\begin{Itemize}
\item {\Py} and {\Je} have been merged.
\item All real variables are declared in double precision.
\item The internal mapping of particle codes has changed.
\item The supersymmetric process machinery of \tsc{SPythia} has been 
included and further improved, with several new processes.
\item Many new processes of beyond-the-standard-model physics, in
areas such as technicolor and doubly-charged Higgs bosons.
\item An expanded description of QCD processes in virtual-photon
interactions, combined with a new machinery for the flux of virtual 
photons from leptons.
\item Initial-state parton showers are matched to the next-to-leading
order matrix elements for gauge boson production. 
\item Final-state parton showers are matched to a number of different
first-order matrix elements for gluon emission, including full
mass dependence.
\item The hadronization description of low-mass strings has been
improved, with consequences especially for heavy-flavour production.
\item An alternative baryon production model has been introduced.
\item Colour rearrangement is included as a new option, and several
alternative Bose-Einstein descriptions are added.  
\end{Itemize} 

By comparison, the move from {\Py}~6.1 to {\Py}~6.2 was rather less
dramatic. Again update notes tell the full story. Some of the main
new features, present from the beginning or added later on, which 
may affect backwards compatibility, are:
\begin{Itemize}
\item A new machinery to handle user-defined external processes, 
according to the standard in \cite{Boo01}. The old machinery is no 
longer available. Some of the alternatives for the \ttt{FRAME} argument 
in the \ttt{PYINIT} call have also been renamed to make way for a new 
\ttt{'USER'} option.
\item The maximum size of the decay channel table has been increased 
from 4000 to 8000, affecting the \ttt{MDME}, \ttt{BRAT} and \ttt{KFDP} 
arrays in the \ttt{PYDAT3} common block.
\item A number of internally used and passed arrays, such as 
\ttt{WDTP}, \ttt{WDTE}, \ttt{WDTPP}, \ttt{WDTEP}, \ttt{WDTPM}, 
\ttt{WDTEM}, \ttt{XLAM} and \ttt{IDLAM}, have been expanded
from dimension 300 to 400.
\item Lepton- and baryon-number-violating decay channels have been 
included for supersymmetric particles \cite{Ska01, Sjo03}. Thus the 
decay tables have grown considerably longer.
\item The string hadronization scheme has been improved and expanded 
better to handle junction topologies, where three strings meet. This
is relevant for baryon-number-violating processes, and also for the
handling of baryon beam remnants. Thus new routines have been 
introduced, and also e.g.\ new \ttt{K(I,1)} status codes. 
\item A runtime interface to \tsc{Isasusy} has been added, for 
determining the SUSY mass spectrum and mixing parameters more 
accurately than with the internal {\Py} routines.
\item The Technicolor scenario is updated and extended. A new common 
block, \ttt{PYTCSM}, is introduced for the parameters and switches 
in Technicolor and related scenarios, and variables are moved to it 
from a few other common blocks. New processes 381--388 are introduced
for standard QCD $2 \to 2$ interactions with Technicolor (or other 
compositeness) extensions, while the processes 11, 12, 13, 28, 53, 
68, 81 and 82 now revert back to being pure QCD.
\item The \ttt{PYSHOW} timelike showering routine has been expanded
to allow showering inside systems consisting of up to seven particles,
which can be made use of in some resonance decays and in user-defined
processes.
\item The \ttt{PYSSPA} spacelike showering routine has been expanded 
with a $\q \to \q \gamma$ branching.
\item The \ttt{PYSIGH} routine has been split into several, in order to 
make it more manageable. (It had increased to a size of over 7000 
lines, which gave some compilers problems.) All the phase-space and 
parton-density generic weights remain, whereas the process-specific
matrix elements have been grouped into new routines \ttt{PYSGQC}, 
\ttt{PYSGHF}, \ttt{PYSGWZ}, \ttt{PYSGHG}, \ttt{PYSGSU}, \ttt{PYSGTC} 
and \ttt{PYSGEX}. 
\item Some exotic particles and QCD effective states have been moved 
from temporary flavour codes to a PDG-consistent naming, and a few new 
codes have been introduced.
\item The maximum number of documentation lines in the beginning of the 
event record has been expanded from 50 to 100.
\item The default parton distribution set for the proton is now
CTEQ 5L.
\item The default Standard Model Higgs mass has been changed to 115 GeV.
\end{Itemize} 

The move from {\Py}~6.2 to {\Py}~6.3 involves no common block changes or
different call sequences. That is, any program that ran with {\Py}~6.2 
also ought to run with {\Py}~6.3, without any change required. Some news 
should be noted, however.
\begin{Itemize}
\item There is a completely new scenario for multiple interactions as the
new default \cite{Sjo03a}. Thus the physics output is changed whenever 
multiple interactions are allowed, i.e.\ in hadron-hadron collisions 
(where a resolved photon counts as a hadron).
\item A new $\pT$-ordered final-state shower is introduced as a 
non-default option to the existing mass-ordered one \cite{Sjo03b}.
\item The current interface to the \tsc{Isasusy} evolution package is 
based on \tsc{Isajet} version 7.67. The dimensions of the \ttt{SSPAR}, 
\ttt{SUGMG} and \ttt{SUGPAS} common blocks found in the \ttt{PYSUGI}
routine could have to be modified if another \tsc{Isajet} version is
linked. The latest change occured in {\Py}~6.217, so there is a
difference if you upgrade from a version prior to that.
\item The new routine \ttt{PYLHA3} introduces a generic interface to
SUSY spectrum and decay calculators in agreement with the 
SUSY Les Houches Accord \cite{Ska03}, and thus offers a long-term 
solution to the incompatibility problems noted above. Without SUSY switched
on, \ttt{PYLHA3} can still be used stand-alone to read in LHA3 decay tables
for any particle, see section \ref{ss:parapartdat}. 
\end{Itemize} 
The update notes tell exactly with what version a new feature or a 
bug fix is introduced. 

\subsection{Program Installation}
\label{ss:install}
 
The {\Py} `master copy' is the one found on the web page\\[2mm]
\drawbox{\ttt{http://www.thep.lu.se/}$\sim$\ttt{torbjorn/Pythia.html}}\\
There you have, for several subversions \ttt{xx}:
\begin{tabbing}
~~~ \= \ttt{pythia63xx.f}~~~~~~~~~ \= the {\Py}~6.3xx code, \\
\> \ttt{pythia63xx.tex} \> this {\Py} manual, and \\
\> \ttt{pythia63xx.update} \> plain text update notes to the manual.
\end{tabbing}
In addition to these, one may also find older versions of the
program and manuals, sample main programs and other pieces of
related software, and other physics papers.
 
The program is written essentially entirely in standard Fortran 77,
and should run on any platform with such a compiler. To a first 
approximation, program compilation should therefore be straightforward.
 
Unfortunately, experience with many different compilers has been
uniform: the options available for obtaining optimized code may
actually produce erroneous code (e.g.\ operations inside \ttt{DO}
loops are moved out before them, where some of the variables have
not yet been properly set). Therefore the general advice is to
use a low optimization level. Note that this is often not the 
default setting.
 
\ttt{SAVE} statements have been included in accordance with the
Fortran standard. 

All default settings and particle and process data are stored in
\ttt{BLOCK DATA PYDATA}. This subprogram must be linked for a proper
functioning of the other routines. On some platforms this is not done
automatically but must be forced by you, e.g.\ by having a line
\begin{verbatim} 
      EXTERNAL PYDATA
\end{verbatim} 
at the beginning of your main program. This applies in particular if 
{\Py} is maintained as a library from which routines are to be loaded 
only when they are needed. In this connection we note that the library 
approach does not give any significant space advantages over a loading 
of the packages as a whole, since a normal run will call on most of the
routines anyway, directly or indirectly.

With the move towards higher energies, e.g.\ for LHC applications,
single-precision (32 bit) real arithmetic has become inappropriate. 
Therefore a declaration \ttt{IMPLICIT DOUBLE PRECISION(A-H,O-Z)} at 
the beginning of each subprogram is inserted to ensure double-precision
(64 bit) real arithmetic. Remember that this also means that all calls
to {\Py} routines have to be done with real variables declared
correspondingly in the user-written calling program. An
\ttt{IMPLICIT INTEGER(I-N)} is also included to avoid 
problems on some compilers. Integer functions beginning with \ttt{PY}
have to be declared explicitly. In total, therefore all routines
begin with
\begin{verbatim} 
C...Double precision and integer declarations.
      IMPLICIT DOUBLE PRECISION(A-H, O-Z)
      IMPLICIT INTEGER(I-N)
      INTEGER PYK,PYCHGE,PYCOMP
\end{verbatim} 
and you are recommended to do the same in your main programs. 
Note that, in running text and in description of common block default 
values, the more cumbersome double-precision notation is not always 
made explicit, but code examples should be correct.

On a machine where \texttt{DOUBLE PRECISION} would give 128 bits,
it may make sense to use compiler options to revert to 64 bits,
since the program is anyway not constructed to make use of 128
bit precision. 

Fortran~77 makes no provision for double-precision complex numbers.
Therefore complex numbers have been used only sparingly. However,
some matrix element expressions, mainly for supersymmetric and 
technicolor processes, simplify considerably when written in terms
of complex variables. In order to achieve a uniform precision,
such variables have been declared \texttt{COMPLEX*16}, and are 
manipulated with functions such as \texttt{DCMPLX} and 
\texttt{DCONJG}. Affected are \texttt{PYSIGH}, \texttt{PYWIDT} and 
several of the supersymmetry routines. Should the compiler not 
accept this deviation from the standard, or some simple equivalent
thereof (like \texttt{DOUBLE COMPLEX} instead of \texttt{COMPLEX*16})
these code pieces could be rewritten to ordinary \texttt{COMPLEX}, 
also converting the real numbers involved to and from single precision, 
with some drop in accuracy for the affected processes. \texttt{PYRESD} 
already contains some ordinary \texttt{COMPLEX} variables, and should 
not cause any problems.

Several compilers report problems when an odd number of integers
precede a double-precision variable in a common block. Therefore
an extra integer has  been introduced as padding in a few instances,
e.g.\ \texttt{NPAD}, \texttt{MSELPD} and \texttt{NGENPD}. 

Since Fortran~77 provides no date-and-time routine, 
\texttt{PYTIME} allows a system-specific routine to be interfaced,
with some commented-out examples given in the code.
This routine is only used for cosmetic improvements of the output,
however, so can be left at the default with time 0 given.
 
A test program, \ttt{PYTEST}\label{p:PYTEST}, is included in the 
{\Py} package. It is disguised as a subroutine, so you have to run a 
main program
\begin{verbatim}
      CALL PYTEST(1)
      END
\end{verbatim}
This program will generate over a thousand events of different types,
under a variety of conditions. If {\Py} has not been properly
installed, this program is likely to crash, or at least generate a
number of erroneous events. This will then clearly be marked in the
output, which otherwise will just contain a few sample event listings
and a table of the number of different particles produced. To switch
off the output of normal events and final table, use \ttt{PYTEST(0)}
instead of \ttt{PYTEST(1)}. The final tally of errors detected should
read 0.

For a program written to run {\Py}~5 and {\Je}~7, most of the conversion 
required for {\Py}~6 is fairly straightforward, and can be automatized. 
Both a simple Fortran routine and a more sophisticated Perl \cite{Gar98} 
script exist to this end, see the {\Py} web page. Some manual 
checks and interventions may still be required.
 
\subsection{Program Philosophy}
 
The Monte Carlo program is built as a slave system, i.e.\ you,
the user, have to supply the main program. From this the various
subroutines are called on to execute specific tasks, after which
control is returned to the main program. Some of these tasks may
be very trivial, whereas the `high-level' routines by themselves
may make a large number of subroutine calls. Many routines are
not intended to be called directly by you, but only from
higher-level routines such as \ttt{PYEXEC}, \ttt{PYEEVT},
\ttt{PYINIT} or \ttt{PYEVNT}.
 
Basically, this means that there are three ways by which you 
communicate with the
programs. First, by setting common block variables, you specify
the details of how the programs should perform specific tasks,
e.g.\ which subprocesses should be generated, which
particle masses should be assumed, which coupling constants used,
which fragmentation scenarios, and so on with hundreds of options
and parameters. Second, by calling subroutines you tell the programs
to generate events according to the rules established above.
Normally there are few subroutine arguments, and those are usually
related to details of the physical situation, such as what
c.m.\ energy to assume for events. Third, you can either look at the
common block \ttt{PYJETS} to extract information on the generated
event, or you can call on various functions and subroutines
to analyse the event further for you.
 
It should be noted that, while the physics content is obviously at
the centre of attention, the {\Py} package
also contains a very extensive setup of auxiliary service routines.
The hope is that this will provide a comfortable
working environment, where not only events are generated, but where
you also linger on to perform a lot of the subsequent studies.
Of course, for detailed studies, it may be necessary to interface
the output directly to a detector simulation program.
 
The general rule is that all routines have names that are six 
characters long, beginning with \ttt{PY}. There are three exceptions 
the length rules: \ttt{PYK}, \ttt{PYP} and \ttt{PYR}. The
former two functions are strongly coupled to the \ttt{K} and \ttt{P}
matrices in the \ttt{PYJETS} common block, the latter is very
frequently used. Also common block names are six characters long and 
start with \ttt{PY}. There are three integer functions,
\ttt{PYK},\ttt{PYCHGE} and \ttt{PYCOMP}. In all routines where 
they are to be used, they have to be declared \ttt{INTEGER}.
 
On the issue of initialization, the routines of different origin and
functionality behave quite differently. Routines that are intended to 
be called from many different places, such as showers, fragmentation 
and decays, require no specific initialization (except for the one 
implied by the presence of \ttt{BLOCK DATA PYDATA}, see above), i.e.\ 
each event and each task stands on its own. Current common block values 
are used to perform the tasks in specific ways, and those rules can be 
changed from one event to the next (or even within the generation of 
one and the same event) without any penalty. The random-number generator
is initialized at the first call, but usually this is transparent.
 
In the core process generation machinery (e.g.\ selection of the hard 
process kinematics), on the other hand, a sizable 
amount of initialization is performed in the \ttt{PYINIT} call, and 
thereafter the events generated by \ttt{PYEVNT} all obey the rules 
established at that point. This improves the efficiency of the 
generation process, and also ties in with the Monte Carlo integration
of the process cross section over many events. Therefore common block 
variables that specify methods and constraints to be used have to be 
set before the \ttt{PYINIT} call and then not be changed afterwards, 
with few exceptions. Of course, it is possible to perform several 
\ttt{PYINIT} calls in the same run, but there is a significant time 
overhead involved, so this is not something one would do for each new 
event. The two older separate process generation routines \ttt{PYEEVT} 
(and some of the routines called by it) and \ttt{PYONIA} also contain 
some elements of initialization, where there are a few advantages if 
events are generated in a coherent fashion. The cross section is not 
as complicated here, however, so the penalty for reinitialization is 
small, and also does not require any special user calls.
 
Apart from writing a title page, giving a brief initialization 
information, printing error messages if need be,
and responding to explicit requests for listings, all tasks of the
program are performed `silently'. All output is directed to unit
\ttt{MSTU(11)}, by default 6, and it is up to you to set
this unit open for write. The only exceptions are
\ttt{PYRGET}, \ttt{PYRSET} and \ttt{PYUPDA} where, for obvious reasons,
the input/output file number is specified at each call. Here you
again have to see to it that proper read/write access is set.
 
The programs are extremely versatile, but the price to be paid for this 
is having a large number of adjustable parameters and switches for
alternative modes of operation. No single user is ever likely to
need more than a fraction of the available options.
Since all these parameters and switches are assigned sensible default
values, there is no reason to worry about them until the need arises.
 
Unless explicitly stated (or obvious from the context) all switches and
parameters can be changed independently of each other. One should note,
however, that if only a few switches/parameters are changed, this may
result in an artificially bad agreement with data. Many disagreements
can often be cured by a subsequent retuning of some other parameters of
the model, in particular those that were once determined by
a comparison with data in the context of the default scenario.
For example, for $\ee$ annihilation, such a retuning could involve one
QCD parameter ($\alphas$ or $\Lambda$), the longitudinal fragmentation
function, and the average transverse momentum in fragmentation.
 
The program contains a number of checks that requested processes have
been implemented, that flavours specified for jet systems make sense,
that the energy is sufficient to allow hadronization, that the memory 
space in \ttt{PYJETS} is large enough, etc. If anything goes wrong
that the program can catch (obviously this may not always be possible),
an error message will be printed and the treatment of the corresponding
event will be cut short. In serious cases, the program will abort.
As long as no error messages appear on the
output, it may not be worthwhile to look into the rules for error
checking, but if but one message appears, it should be enough cause for
alarm to receive prompt attention. Also warnings are sometimes printed.
These are less serious, and the experienced user might deliberately
do operations which go against the rules, but still can be made to
make sense in their context. Only the first few warnings will be
printed, thereafter the program will be quiet. By default, the program
is set to stop execution after ten errors, after printing
the last erroneous event.
 
It must be emphasized that not all errors will be caught. In particular,
one tricky question is what happens if an integer-valued common block
switch or subroutine/function argument is used with a value that is not
defined. In some subroutine calls, a prompt return will be expedited,
but in most instances the subsequent action is entirely unpredictable,
and often completely haywire. The same goes for real-valued variables
that are assigned values outside the physically sensible range. One 
example will suffice here: if \ttt{PARJ(2)} is defined as the 
$\s / \u$ suppression factor, a value $>1$ will not give more 
profuse production of $\s$ than of $\u$, but actually a spillover 
into $\c$ production. Users, beware!

\subsection{Manual Conventions}
 
In the manual parts of this report, some conventions are used.
All names of subprograms, common blocks and variables are given in
upper-case `typewriter' style, e.g.\ \ttt{MSTP(111)=0}. Also
program examples are given in this style.
 
If a common block variable must have a value set at the beginning of
execution, then a default value is stored in the block data
subprogram \ttt{PYDATA}. Such a default value is
usually indicated by a `(D=\ldots)' immediately after the variable
name, e.g.
\begin{entry}
\iteme{MSTJ(1) :} (D=1) choice of fragmentation scheme.
\end{entry}

All variables in the fragmentation-related common blocks (with very few 
exceptions, clearly marked) can be freely changed from one event to the 
next, or even within the treatment of one single event; see discussion 
on initialization in the previous section. In the process generation
machinery common blocks the situation is more complicated. The values of 
many switches and parameters are used already in the \ttt{PYINIT} call,
and cannot be changed after that. The problem is mentioned in the
preamble to the afflicted common blocks, which in particular means
\ttt{PYPARS} and \ttt{PYSUBS}. For the variables which may still 
be changed from one event to the next, a `(C)' is added after
the `(D=\ldots)' statement.

Normally, variables internal to the program are kept in separate
common blocks and arrays, but in a few cases such internal variables
appear among arrays of switches and parameters, mainly for historical
reasons. These are denoted by `(R)' for variables you may want to 
read, because they contain potentially interesting information, and 
by `(I)' for purely internal variables. In neither case may the
variables be changed by you.

In the description of a switch, the alternatives that this
switch may take are often enumerated, e.g.
\begin{entry}
\iteme{MSTJ(1) :} (D=1) choice of fragmentation scheme.
\begin{subentry}
\iteme{= 0 :} no jet fragmentation at all.
\iteme{= 1 :} string fragmentation according to the Lund model.
\iteme{= 2 :} independent fragmentation, according to specification
in \ttt{MSTJ(2)} and \ttt{MSTJ(3)}.
\end{subentry}
\end{entry}
If you then use any value other than 0, 1 or 2, results are 
undefined. The action could even be different in different parts 
of the program, depending on the order in which the alternatives are
identified. 

It is also up to you to choose physically sensible values for
parameters: there is no check on the allowed ranges of variables.
We gave an example of this at the end of the preceding section.

Subroutines you are expected to use are enclosed in a box at the
point where they are defined:

\drawbox{CALL PYLIST(MLIST)}

\boxsep

This is followed by a description of input or output parameters.
The difference between input and output is not explicitly marked,
but should be obvious from the context. In fact, the event-analysis 
routines of section \ref{ss:evanrout} return values, 
while all the rest only have input variables.  

Routines that are only used internally are not boxed in.
However, we use boxes for all common blocks, so as to 
enhance the readability. 

In running text, often specific switches and parameters will be 
mentioned, without a reference to the place where they are described 
further. The Index at the very end of the document allows you to
find this place. Tables~\ref{t:comblockone} and \ref{t:comblocktwo} 
gives a brief summary of almost all common blocks and the variables 
stored there. Often names for switches begin with \ttt{MST} and 
parameters with \ttt{PAR}. No common block variables begin with 
\ttt{PY}. There is thus no possibility to confuse an array element 
with a function or subroutine call.
 
\begin{table}[tp]
\captive{An almost complete list of common blocks, with brief comments
on their main functions. The listing continues in 
Table~\protect\ref{t:comblocktwo}.
\protect\label{t:comblockone} } 
\begin{verbatim}
C...The event record.
      COMMON/PYJETS/N,NPAD,K(4000,5),P(4000,5),V(4000,5)
C...Parameters.
      COMMON/PYDAT1/MSTU(200),PARU(200),MSTJ(200),PARJ(200)
C...Particle properties + some flavour parameters.
      COMMON/PYDAT2KCHG(500,4),PMAS(500,4),PARF(2000),VCKM(4,4)
C...Decay information.
      COMMON/PYDAT3/MDCY(500,3),MDME(8000,2),BRAT(8000),KFDP(8000,5)
C...Particle names
      COMMON/PYDAT4/CHAF(500,2)
      CHARACTER CHAF*16
C...Random number generator information.
      COMMON/PYDATR/MRPY(6),RRPY(100)
C...Selection of hard scattering subprocesses.
      COMMON/PYSUBS/MSEL,MSELPD,MSUB(500),KFIN(2,-40:40),CKIN(200)
C...Parameters. 
      COMMON/PYPARS/MSTP(200),PARP(200),MSTI(200),PARI(200)
C...Internal variables.
      COMMON/PYINT1/MINT(400),VINT(400)
C...Process information.
      COMMON/PYINT2/ISET(500),KFPR(500,2),COEF(500,20),ICOL(40,4,2)
C...Parton distributions and cross sections.
      COMMON/PYINT3/XSFX(2,-40:40),ISIG(1000,3),SIGH(1000)
C...Resonance width and secondary decay treatment.
      COMMON/PYINT4/MWID(500),WIDS(500,5)
C...Generation and cross section statistics.
      COMMON/PYINT5/NGENPD,NGEN(0:500,3),XSEC(0:500,3)
C...Process names.
      COMMON/PYINT6/PROC(0:500)
      CHARACTER PROC*28
C...Total cross sections.
      COMMON/PYINT7/SIGT(0:6,0:6,0:5)
C...Photon parton distributions: total and valence only.
      COMMON/PYINT8/XPVMD(-6:6),XPANL(-6:6),XPANH(-6:6),XPBEH(-6:6), 
     &XPDIR(-6:6) 
      COMMON/PYINT9/VXPVMD(-6:6),VXPANL(-6:6),VXPANH(-6:6),VXPDGM(-6:6) 
C...Supersymmetry parameters.
      COMMON/PYMSSM/IMSS(0:99),RMSS(0:99)
C...Supersymmetry mixing matrices.
      COMMON/PYSSMT/ZMIX(4,4),UMIX(2,2),VMIX(2,2),SMZ(4),SMW(2),
     &SFMIX(16,4),ZMIXI(4,4),UMIXI(2,2),VMIXI(2,2)
C...R-parity-violating couplings in supersymmetry.
      COMMON/PYMSRV/RVLAM(3,3,3), RVLAMP(3,3,3), RVLAMB(3,3,3)
C...Internal parameters for R-parity-violating processes.
      COMMON/PYRVNV/AB(2,16,2),RMS(0:3),RES(6,5),IDR,IDR2,DCMASS,KFR(3)
      COMMON/PYRVPM/RM(0:3),A(2),B(2),RESM(2),RESW(2),MFLAG
      LOGICAL MFLAG
\end{verbatim}
\end{table}
 
\begin{table}[tp]
\captive{Continuation of Table~\protect\ref{t:comblockone}.
\protect\label{t:comblocktwo} } 
\begin{verbatim}
C...Parameters for Gauss integration of supersymmetric widths.
      COMMON/PYINTS/XXM(20)
      COMMON/PYG2DX/X1
C...Histogram information.
      COMMON/PYBINS/IHIST(4),INDX(1000),BIN(20000)
C...HEPEVT common block.
      PARAMETER (NMXHEP=4000)
      COMMON/HEPEVT/NEVHEP,NHEP,ISTHEP(NMXHEP),IDHEP(NMXHEP),
     &JMOHEP(2,NMXHEP),JDAHEP(2,NMXHEP),PHEP(5,NMXHEP),VHEP(4,NMXHEP)
      DOUBLE PRECISION PHEP,VHEP
C...User process initialization common block.
      INTEGER MAXPUP
      PARAMETER (MAXPUP=100)
      INTEGER IDBMUP,PDFGUP,PDFSUP,IDWTUP,NPRUP,LPRUP
      DOUBLE PRECISION EBMUP,XSECUP,XERRUP,XMAXUP
      COMMON/HEPRUP/IDBMUP(2),EBMUP(2),PDFGUP(2),PDFSUP(2),
     &IDWTUP,NPRUP,XSECUP(MAXPUP),XERRUP(MAXPUP),XMAXUP(MAXPUP),
     &LPRUP(MAXPUP)
C...User process event common block.
      INTEGER MAXNUP
      PARAMETER (MAXNUP=500)
      INTEGER NUP,IDPRUP,IDUP,ISTUP,MOTHUP,ICOLUP
      DOUBLE PRECISION XWGTUP,SCALUP,AQEDUP,AQCDUP,PUP,VTIMUP,SPINUP
      COMMON/HEPEUP/NUP,IDPRUP,XWGTUP,SCALUP,AQEDUP,AQCDUP,IDUP(MAXNUP),
     &ISTUP(MAXNUP),MOTHUP(2,MAXNUP),ICOLUP(2,MAXNUP),PUP(5,MAXNUP),
     &VTIMUP(MAXNUP),SPINUP(MAXNUP)
\end{verbatim}
\end{table}
 
\subsection{Getting Started with the Simple Routines}
\label{ss:JETstarted}

Normally {\Py} is expected to take care of the full event generation 
process. At times, however, one may want to access the more simple 
underlying routines, which allow a large flexibility to `do it
yourself'. We therefore start with a few cases of this kind, 
at the same time introducing some of the more frequently used
utility routines.

As a first example, assume that you want to study the production of
$\u \ubar$ 2-jet systems at 20 GeV energy. To do this, write a main
program
\begin{verbatim}
      IMPLICIT DOUBLE PRECISION(A-H, O-Z)
      CALL PY2ENT(0,2,-2,20D0)
      CALL PYLIST(1)
      END
\end{verbatim}
and run this program, linked together with {\Py}. The routine
\ttt{PY2ENT}
is specifically intended for storing two entries (partons or particles).
The first argument (0) is a command to perform fragmentation and decay
directly after the entries have been stored, the second and third that
the two entries are $\u$ (2) and $\ubar$ ($-2$), and the last that the 
c.m.\ energy of the pair is 20 GeV, in double precision. When this is run, 
the resulting event is stored in the \ttt{PYJETS} common block. This 
information can then be read out by you. No output is produced by 
\ttt{PY2ENT} itself, except for a title page which appears once for 
every {\Py} run.
 
Instead the second command, to \ttt{PYLIST}, provides a simple visible
summary of the information stored in \ttt{PYJETS}. The argument (1)
indicates that the short version should be used, which is suitable
for viewing the listing directly on an 80-column terminal screen.
It might look as shown here.
\begin{verbatim}
                        Event listing (summary)
 
  I  particle/jet KS     KF orig   p_x     p_y     p_z      E       m
 
  1  (u)       A  12      2    0   0.000   0.000  10.000  10.000   0.006
  2  (ubar)    V  11     -2    0   0.000   0.000 -10.000  10.000   0.006
  3  (string)     11     92    1   0.000   0.000   0.000  20.000  20.000
  4  (rho+)       11    213    3   0.098  -0.154   2.710   2.856   0.885
  5  (rho-)       11   -213    3  -0.227   0.145   6.538   6.590   0.781
  6  pi+           1    211    3   0.125  -0.266   0.097   0.339   0.140
  7  (Sigma0)     11   3212    3  -0.254   0.034  -1.397   1.855   1.193
  8  (K*+)        11    323    3  -0.124   0.709  -2.753   2.968   0.846
  9  p~-           1  -2212    3   0.395  -0.614  -3.806   3.988   0.938
 10  pi-           1   -211    3  -0.013   0.146  -1.389   1.403   0.140
 11  pi+           1    211    4   0.109  -0.456   2.164   2.218   0.140
 12  (pi0)        11    111    4  -0.011   0.301   0.546   0.638   0.135
 13  pi-           1   -211    5   0.089   0.343   2.089   2.124   0.140
 14  (pi0)        11    111    5  -0.316  -0.197   4.449   4.467   0.135
 15  (Lambda0)    11   3122    7  -0.208   0.014  -1.403   1.804   1.116
 16  gamma         1     22    7  -0.046   0.020   0.006   0.050   0.000
 17  K+            1    321    8  -0.084   0.299  -2.139   2.217   0.494
 18  (pi0)        11    111    8  -0.040   0.410  -0.614   0.751   0.135
 19  gamma         1     22   12   0.059   0.146   0.224   0.274   0.000
 20  gamma         1     22   12  -0.070   0.155   0.322   0.364   0.000
 21  gamma         1     22   14  -0.322  -0.162   4.027   4.043   0.000
 22  gamma         1     22   14   0.006  -0.035   0.422   0.423   0.000
 23  p+            1   2212   15  -0.178   0.033  -1.343   1.649   0.938
 24  pi-           1   -211   15  -0.030  -0.018  -0.059   0.156   0.140
 25  gamma         1     22   18  -0.006   0.384  -0.585   0.699   0.000
 26  gamma         1     22   18  -0.034   0.026  -0.029   0.052   0.000
                 sum:  0.00        0.000   0.000   0.000  20.000  20.000
\end{verbatim}
(A few blanks have been removed between the columns to make it fit into
the format of this text.) Look in the particle/jet column and note
that the first two lines are the original $\u$ and $\ubar$. 
The parentheses enclosing the names, `\ttt{(u)}' and `\ttt{(ubar)}',
are there as a reminder that these partons actually have been allowed 
to fragment. The partons are still retained so that event histories
can be studied. Also note that the \ttt{KF} (flavour code) column
contains 2 in the first line and $-2$ in the second. These are the
codes actually stored to denote the presence of a $\u$ and a $\ubar$, 
cf. the \ttt{PY2ENT} call, while the names written are just 
conveniences used when producing visible output. The \ttt{A} and 
\ttt{V} near the end of the particle/jet column indicate the beginning 
and end of a string (or cluster, or independent fragmentation) parton 
system; any intermediate entries belonging to the same system would 
have had an \ttt{I} in that column. (This gives a poor man's 
representation of an up-down arrow, $\updownarrow$.)
 
In the \ttt{orig} (origin) column, the zeros indicate that $\u$ and
$\ubar$ are two initial entries. The subsequent line, number 3, denotes
the fragmenting $\u \ubar$ string system as a whole, and has origin 1,
since the first parton of this string system is entry number 1. The
particles in lines 4--10 have origin 3 to denote that they come
directly from the fragmentation of this string. In string
fragmentation it is not meaningful to say that a particle comes from
only the $\u$ quark or only the $\ubar$ one. It is the string system
as a whole that gives a $\rho^+$, a $\rho^-$, a $\pi^+$, a
$\Sigma^0$, a $\K^{*+}$, a $\pbar^-$, and a $\pi^-$. Note that some
of the particle names are again enclosed in parentheses, indicating
that these particles are not present in the final state either, but have 
decayed further. Thus the $\pi^-$ in line 13 and the $\pi^0$ in line
14 have origin 5, as an indication that they come from the decay of the
$\rho^-$ in line 5. Only the names not enclosed in parentheses 
remain at the end of the fragmentation/decay chain, and
are thus experimentally observable. The actual status code used to
distinguish between different classes of entries is given in the
\ttt{KS} column; codes in the range 1--10 correspond to remaining
entries, and those above 10 to those that have fragmented or decayed.
 
The columns with \ttt{p\_x}, \ttt{p\_y}, \ttt{p\_z}, \ttt{E} and
\ttt{m} are quite self-explanatory. All momenta, energies and
masses are given in units of GeV, since the speed of light 
is taken to be $c = 1$.
Note that energy and momentum are conserved at each step of the
fragmentation/decay process (although there exist options where this is
not true). Also note that the $z$ axis plays the r\^ole of preferred
direction, along which the original partons are placed. 
The final line is
intended as a quick check that nothing funny happened. It contains the
summed charge, summed momentum, summed energy and invariant mass of the
final entries at the end of the fragmentation/decay chain, and the
values should agree with the input implied by the \ttt{PY2ENT}
arguments. (In fact, warnings would normally appear on the output if
anything untoward happened, but that is another story.)
 
The above example has illustrated roughly what information is to be had
in the event record, but not so much about how it is stored. This is
better seen by using a 132-column format for listing events. Try e.g.
the following program
\begin{verbatim}
      IMPLICIT DOUBLE PRECISION(A-H, O-Z)
      CALL PY3ENT(0,1,21,-1,30D0,0.9D0,0.7D0)
      CALL PYLIST(2)
      CALL PYEDIT(3)
      CALL PYLIST(2)
      END
\end{verbatim}
where a 3-jet $\d \g \dbar$ event is generated in the first executable
line and listed in the second. This listing will contain the numbers as
directly stored in the common block \ttt{PYJETS}
\begin{verbatim}
      COMMON/PYJETS/N,NPAD,K(4000,5),P(4000,5),V(4000,5)
\end{verbatim}
For particle \ttt{I}, \ttt{K(I,1)} thus gives information on whether
or not a parton or particle has fragmented or decayed, \ttt{K(I,2)}
gives the particle code, \ttt{K(I,3)} its origin, \ttt{K(I,4)} and
\ttt{K(I,5)} the position of fragmentation/decay products, and
\ttt{P(I,1})--\ttt{P(I,5)} momentum, energy and mass. The number
of lines in current use is given by \ttt{N}, i.e.
1 $\leq$ \ttt{I} $\leq$ \ttt{N}. The \ttt{V} matrix contains decay
vertices; to view those \ttt{PYLIST(3)} has to be used. \ttt{NPAD} 
is a dummy, needed to avoid some compiler troubles. It is important 
to learn the rules for how information is stored in \ttt{PYJETS}.
 
The third executable line in the program illustrates another important 
point about {\Py}: a number of routines are available for manipulating 
and analyzing the event record after the event has been generated.
Thus \ttt{PYEDIT(3)} will remove everything except stable charged
particles, as shown by the result of the second \ttt{PYLIST} call.
More advanced possibilities include things like sphericity or
clustering routines. {\Py} also contains some simple routines for
histogramming, used to give self-contained examples of analysis
procedures.

Apart from the input arguments of subroutine calls, control on the
doings of {\Py} may be imposed via many common blocks. Here
sensible default values are always provided. A user might want to
switch off all particle decays by putting \ttt{MSTJ(21)=0} or
increase the $\s / \u$ ratio in fragmentation by putting
\ttt{PARJ(2)=0.40D0}, to give but two examples. It is by exploring
the possibilities offered here that {\Py} can be turned into an
extremely versatile tool, even if all the nice physics is already
present in the default values.
 
As a final, semi-realistic example, assume that the $\pT$
spectrum of $\pi^+$ particles is to be studied in 91.2 GeV
$\ee$ annihilation events, where $\pT$ is to be defined with
respect to the sphericity axis. Using the internal routines for
simple histogramming, a complete program might look like
\begin{verbatim}
C...Double precision and integer declarations.
      IMPLICIT DOUBLE PRECISION(A-H, O-Z)
      IMPLICIT INTEGER(I-N)
      INTEGER PYK,PYCHGE,PYCOMP

C...Common blocks.
      COMMON/PYJETS/N,NPAD,K(4000,5),P(4000,5),V(4000,5)

C...Book histograms.
      CALL PYBOOK(1,'pT spectrum of pi+',100,0D0,5D0)

C...Number of events to generate. Loop over events.
      NEVT=100
      DO 110 IEVT=1,NEVT

C...Generate event. List first one.
        CALL PYEEVT(0,91.2D0)
        IF(IEVT.EQ.1) CALL PYLIST(1)

C...Find sphericity axis.
        CALL PYSPHE(SPH,APL)

C...Rotate event so that sphericity axis is along z axis.
        CALL PYEDIT(31)

C...Loop over all particles, but skip if not pi+.
        DO 100 I=1,N
          IF(K(I,2).NE.211) GOTO 100

C...Calculate pT and fill in histogram.
          PT=SQRT(P(I,1)**2+P(I,2)**2)
          CALL PYFILL(1,PT,1D0)

C...End of particle and event loops.
  100   CONTINUE
  110 CONTINUE

C...Normalize histogram properly and list it.
      CALL PYFACT(1,20D0/NEVT)
      CALL PYHIST

      END
\end{verbatim}
Study this program, try to understand what happens at each step, and
run it to check that it works. You should then be ready to look
at the relevant sections of this report and start writing your own
programs.
 
\subsection{Getting Started with the Event Generation Machinery}
\label{ss:PYTstarted}
 
A run with the full {\Py} event generation machinery has to be more 
strictly organized than the simple examples above, in that it is 
necessary to initialize the
generation before events can be generated, and in that it is
not possible to change switches and parameters freely during
the course of the run. A fairly precise recipe for how a run 
should be structured can therefore be given.
 
Thus, the nowadays normal usage of {\Py} can be subdivided into 
three steps.
\begin{Enumerate}
\item The initialization step. It is here that all the basic
  characteristics of the coming generation are specified.
  The material in this section includes the following.
  \begin{Itemize}
  \item Declarations for double precision and integer variables and
  integer functions:
    \begin{verbatim} 
    IMPLICIT DOUBLE PRECISION(A-H, O-Z)
    IMPLICIT INTEGER(I-N)
    INTEGER PYK,PYCHGE,PYCOMP
    \end{verbatim} 
    \vspace{-\baselineskip}
  \item Common blocks, at least the following, and maybe some more:
    \begin{verbatim}
    COMMON/PYJETS/N,NPAD,K(4000,5),P(4000,5),V(4000,5)
    COMMON/PYDAT1/MSTU(200),PARU(200),MSTJ(200),PARJ(200)
    COMMON/PYSUBS/MSEL,MSELPD,MSUB(500),KFIN(2,-40:40),CKIN(200)
    COMMON/PYPARS/MSTP(200),PARP(200),MSTI(200),PARI(200)
    \end{verbatim}
    \vspace{-\baselineskip}
  \item Selection of required processes. Some fixed `menus'
     of subprocesses can be selected with different \ttt{MSEL} 
     values, but with {\tt MSEL}=0 it is possible to compose 
     `\`a la carte', using the subprocess numbers.
    To generate processes 14, 18 and 29, for instance, one needs
    \begin{verbatim}
    MSEL=0
    MSUB(14)=1
    MSUB(18)=1
    MSUB(29)=1
    \end{verbatim}
    \vspace{-\baselineskip}
  \item Selection of kinematics cuts in the \ttt{CKIN} array.
  To generate hard scatterings with 5~GeV $\leq \pT \leq$
    10~GeV, for instance, use
    \begin{verbatim}
    CKIN(3)=5D0
    CKIN(4)=10D0
    \end{verbatim}
    \vspace{-\baselineskip}
  Unfortunately, initial- and final-state radiation will shift
  around the kinematics of the hard scattering, making the effects
  of cuts less predictable. One therefore always has to be very
  careful that no desired event configurations are cut out.
  \item Definition of underlying physics scenario, e.g.\ Higgs mass.
  \item Selection of parton-distribution sets, $Q^2$ definitions,
    and all other details of the generation.
  \item Switching off of generator parts not needed for toy
    simulations, e.g.\ fragmentation for parton level studies.
  \item Initialization of the event generation procedure. Here
    kinematics is set up, maxima of differential cross sections
    are found for future Monte Carlo generation, and a number of
    other preparatory tasks carried out. Initialization is performed
    by \ttt{PYINIT}, which should be called only after the switches
    and parameters above have been set to their desired values. The
    frame, the beam particles and the energy have to be specified, 
    e.g.
    \begin{verbatim}
    CALL PYINIT('CMS','p','pbar',1800D0)
    \end{verbatim}
    \vspace{-\baselineskip}
  \item Any other initial material required by you, e.g.
    histogram booking.
  \end{Itemize}
\item The generation loop. It is here that events are generated
  and studied. It includes the following tasks:
  \begin{Itemize}
  \item Generation of the next event, with
    \begin{verbatim}
    CALL PYEVNT
    \end{verbatim}
    \vspace{-\baselineskip}
  \item Printing of a few events, to check that everything is
    working as planned, with
    \begin{verbatim}
    CALL PYLIST(1)
    \end{verbatim}
    \vspace{-\baselineskip}
  \item An analysis of the event for properties of interest,
    either directly reading out information from the 
    \ttt{PYJETS} common block
    or making use of the utility routines in {\Py}.
  \item Saving of events on disk or tape, or interfacing to detector
    simulation.
  \end{Itemize}
\item The finishing step. Here the tasks are:
  \begin{Itemize}
  \item Printing a table of deduced cross sections, obtained as a
  by-product of the Monte Carlo generation activity, with the
  command
    \begin{verbatim}
    CALL PYSTAT(1)
    \end{verbatim}
    \vspace{-\baselineskip}
  \item Printing histograms and other user output.
  \end{Itemize}
\end{Enumerate}
 
To illustrate this structure, imagine a toy example, where one wants
to simulate the production of a 300 GeV Higgs particle. In
{\Py}, a program for this might look something like the
following.
\begin{verbatim}
C...Double precision and integer declarations.
      IMPLICIT DOUBLE PRECISION(A-H, O-Z)
      IMPLICIT INTEGER(I-N)
      INTEGER PYK,PYCHGE,PYCOMP

C...Common blocks.
      COMMON/PYJETS/N,NPAD,K(4000,5),P(4000,5),V(4000,5)
      COMMON/PYDAT1/MSTU(200),PARU(200),MSTJ(200),PARJ(200)
      COMMON/PYDAT2/KCHG(500,4),PMAS(500,4),PARF(2000),VCKM(4,4)
      COMMON/PYDAT3/MDCY(500,3),MDME(8000,2),BRAT(8000),KFDP(8000,5)
      COMMON/PYSUBS/MSEL,MSELPD,MSUB(500),KFIN(2,-40:40),CKIN(200)
      COMMON/PYPARS/MSTP(200),PARP(200),MSTI(200),PARI(200)
 
C...Number of events to generate. Switch on proper processes.
      NEV=1000
      MSEL=0
      MSUB(102)=1
      MSUB(123)=1
      MSUB(124)=1
 
C...Select Higgs mass and kinematics cuts in mass.
      PMAS(25,1)=300D0
      CKIN(1)=290D0
      CKIN(2)=310D0
 
C...For simulation of hard process only: cut out unnecessary tasks.
      MSTP(61)=0
      MSTP(71)=0
      MSTP(81)=0
      MSTP(111)=0
 
C...Initialize and list partial widths.
      CALL PYINIT('CMS','p','p',14000D0)
      CALL PYSTAT(2)
 
C...Book histogram.
      CALL PYBOOK(1,'Higgs mass',50,275D0,325D0)
 
C...Generate events. Look at first few.
      DO 200 IEV=1,NEV
        CALL PYEVNT
        IF(IEV.LE.3) CALL PYLIST(1)
 
C...Loop over particles to find Higgs and histogram its mass.
        DO 100 I=1,N
          IF(K(I,1).LT.20.AND.K(I,2).EQ.25) HMASS=P(I,5)
  100   CONTINUE
        CALL PYFILL(1,HMASS,1D0)
  200 CONTINUE
 
C...Print cross sections and histograms.
      CALL PYSTAT(1)
      CALL PYHIST
 
      END
\end{verbatim}
 
Here 102, 123 and 124 are the three main Higgs production
graphs $\g \g \rightarrow \hrm$, $\Z \Z \rightarrow \hrm$, and
$\W \W \rightarrow \hrm$, and \ttt{MSUB(ISUB)=1} is the command to
switch on process {\ISUB}. Full freedom to combine subprocesses
`\`a la carte' is ensured by \ttt{MSEL=0}; ready-made `menus'
can be ordered with other \ttt{MSEL} numbers. The \ttt{PMAS} 
command sets the mass of the Higgs, and the \ttt{CKIN}
variables the desired mass range of the Higgs --- a Higgs with a
300 GeV nominal mass actually has a fairly broad Breit--Wigner type
mass distribution. The \ttt{MSTP} switches that come next are there to
modify the generation procedure, in this case to switch off initial-
and final-state radiation, multiple interactions among beam jets,
and fragmentation, to give only the `parton skeleton' of the hard
process. The \ttt{PYINIT} call initializes {\Py}, by finding
maxima of cross sections, recalculating the Higgs decay properties
(which depend on the Higgs mass), etc. The decay properties can
be listed with \ttt{PYSTAT(2)}.
 
Inside the event loop, \ttt{PYEVNT}
is called to generate an event, and \ttt{PYLIST(1)} to list the event.
The information used by \ttt{PYLIST(1)} is the event record, stored
in the common block \ttt{PYJETS}. Here one finds all produced particles,
both final and intermediate ones, with information on particle
species and event history (\ttt{K} array), particle momenta
(\ttt{P} array) and production vertices (\ttt{V} array).
In the loop over all particles produced, \ttt{1} through \ttt{N},
the Higgs particle is found by its code, \ttt{K(I,2)=25},
and its mass is stored in \ttt{P(I,5)}.
 
After all events have been generated, \ttt{PYSTAT(1)}
gives a summary of the number of events generated in the
various allowed channels, and the inferred cross sections.
 
In the run above, a typical event listing might look like the following.
\begin{verbatim}
                         Event listing (summary)
 
    I  particle/jet     KF    p_x      p_y      p_z       E        m
 
    1  !p+!           2212    0.000    0.000 8000.000 8000.000    0.938
    2  !p+!           2212    0.000    0.000-8000.000 8000.000    0.938
 ======================================================================
    3  !g!              21   -0.505   -0.229   28.553   28.558    0.000
    4  !g!              21    0.224    0.041 -788.073  788.073    0.000
    5  !g!              21   -0.505   -0.229   28.553   28.558    0.000
    6  !g!              21    0.224    0.041 -788.073  788.073    0.000
    7  !H0!             25   -0.281   -0.188 -759.520  816.631  300.027
    8  !W+!             24  120.648   35.239 -397.843  424.829   80.023
    9  !W-!            -24 -120.929  -35.426 -361.677  391.801   82.579
   10  !e+!            -11   12.922   -4.760 -160.940  161.528    0.001
   11  !nu_e!           12  107.726   39.999 -236.903  263.302    0.000
   12  !s!               3  -62.423    7.195 -256.713  264.292    0.199
   13  !cbar!           -4  -58.506  -42.621 -104.963  127.509    1.350
 ======================================================================
   14  (H0)             25   -0.281   -0.188 -759.520  816.631  300.027
   15  (W+)             24  120.648   35.239 -397.843  424.829   80.023
   16  (W-)            -24 -120.929  -35.426 -361.677  391.801   82.579
   17  e+              -11   12.922   -4.760 -160.940  161.528    0.001
   18  nu_e             12  107.726   39.999 -236.903  263.302    0.000
   19  s         A       3  -62.423    7.195 -256.713  264.292    0.199
   20  cbar      V      -4  -58.506  -42.621 -104.963  127.509    1.350
   21  ud_1      A    2103   -0.101    0.176 7971.328 7971.328    0.771
   22  d         V       1   -0.316    0.001  -87.390   87.390    0.010
   23  u         A       2    0.606    0.052   -0.751    0.967    0.006
   24  uu_1      V    2203    0.092   -0.042-7123.668 7123.668    0.771
 ======================================================================
                sum:  2.00     0.00     0.00     0.00 15999.98 15999.98
\end{verbatim}
The above event listing is abnormally short, in part because some
columns of information were removed to make it fit into this text,
in part because all initial- and final-state QCD radiation, all
non-trivial beam jet structure, and all fragmentation was inhibited
in the generation. Therefore only the skeleton of the process is
visible. In lines 1 and 2 one recognizes the two incoming protons.
In lines 3 and 4 are incoming partons before initial-state radiation
and in 5 and 6 after --- since there is no such radiation they coincide
here. Line 7 shows the Higgs produced by $\g \g$ fusion, 8 and 9 its
decay products and 10--13 the second-step decay products. Up to this
point lines give a summary of the event history, indicated by the
exclamation marks that surround particle names (and also reflected in
the \ttt{K(I,1)} code, not shown). From line 14 onwards come the
particles actually produced in the final states, first in lines 14--16
particles that subsequently decayed, which have their names surrounded
by brackets, and finally the particles and partons left in the end,
including beam remnants.
Here this also includes a number of unfragmented partons, since
fragmentation was inhibited. Ordinarily, the listing would have gone
on for a few hundred more lines, with the particles produced in the
fragmentation and their decay products. The final line gives total
charge and momentum, as a convenient check that nothing unexpected
happened. The first column of the listing
is just a counter, the second gives the particle name and information
on status and string drawing (the \ttt{A} and \ttt{V}), the third
the particle-flavour code (which is used to give the name),
and the subsequent columns give the momentum components.
 
One of the main problems is to select kinematics efficiently. Imagine
for instance that one is interested in the production of a single 
$\Z$ with a
transverse momentum in excess of 50 GeV. If one tries to generate
the inclusive sample of $\Z$ events, by the basic production graphs
$\q \qbar \rightarrow \Z$, then most events will have low transverse
momenta and will have to be discarded. That any of the desired events
are produced at all is due to the initial-state generation machinery, 
which can build up transverse momenta for the incoming
$\q$ and $\qbar$. However, the amount
of initial-state radiation cannot be constrained beforehand. To
increase the efficiency, one may therefore turn to the higher-order
processes $\q \g \rightarrow \Z \q$ and $\q \qbar \rightarrow \Z \g$,
where
already the hard subprocess gives a transverse momentum to the $\Z$.
This transverse momentum can be constrained as one wishes, but again
initial- and final-state radiation will smear the picture. If one were
to set a $\pT$ cut at 50 GeV for the hard-process generation,
those events where the $\Z$ was given only 40 GeV in the hard process 
but got the rest from initial-state radiation would be missed. 
Not only therefore would cross sections
come out wrong, but so might the typical event shapes. In the end,
it is therefore necessary to find some reasonable compromise, by
starting the generation at 30 GeV, say, if one knows that only rarely
do events below this value fluctuate up to 50 GeV. Of course, most 
events will therefore not contain a $\Z$ above 50 GeV, and one will 
have to live with some inefficiency. It is not uncommon that only one 
event out of ten can be used, and occasionally it can be even worse.
 
If it is difficult to set kinematics, it is often easier to set the
flavour content of a process. In a Higgs study, one might wish, for
example, to consider the decay $\hrm^0 \rightarrow \Z^0 \Z^0$, with each
$\Z^0 \rightarrow \ee$ or $\mu^+ \mu^-$. It is therefore necessary to
inhibit all other $\hrm^0$ and $\Z^0$ decay channels, and also to adjust
cross sections to take into account this change, all of which is fairly
straightforward. The same cannot be said for decays of ordinary hadrons,
where the number produced in a process is not known beforehand, and 
therefore inconsistencies easily can arise if one tries to force
specific decay channels.

In the examples given above, all run-specific parameters are set in 
the code (in the main program; alternatively it could be in a 
subroutine called by the main program). This approach is allowing 
maximum flexibility to change parameters during the course of the run. 
However, in many experimental collaborations one does not want to allow 
this freedom, but only one set of parameters, to be read in from an 
external file at the beginning of a run and thereafter never changed.
This in particular applies when {\Py} is to be linked with other 
libraries, such as \tsc{Geant} \cite{Bru89} and detector-specific software.
While a linking of a normal-sized main program with {\Py} is 
essentially instantaneous on current platforms (typically less than a 
second), this may not hold for other libraries. For this purpose one 
then needs a parser of {\Py} parameters, the core of which can be 
provided by the \ttt{PYGIVE} routine. 

As an example, consider a main program of the form
\begin{verbatim}
C...Double precision and integer declarations.
      IMPLICIT DOUBLE PRECISION(A-H, O-Z)
      IMPLICIT INTEGER(I-N)
      INTEGER PYK,PYCHGE,PYCOMP
C...Input and output strings.
      CHARACTER FRAME*12,BEAM*12,TARGET*12,PARAM*100

C...Read parameters for PYINIT call.
      READ(*,*) FRAME,BEAM,TARGET,ENERGY

C...Read number of events to generate, and to print.
      READ(*,*) NEV,NPRT

C...Loop over reading and setting parameters/switches.
  100 READ(*,'(A)',END=200) PARAM
      CALL PYGIVE(PARAM)
      GOTO 100

C...Initialize PYTHIA.
  200 CALL PYINIT(FRAME,BEAM,TARGET,ENERGY)

C...Event generation loop
      DO 300 IEV=1,NEV
        CALL PYEVNT
        IF(IEV.LE.NPRT) CALL PYLIST(1)
  300 CONTINUE   

C...Print cross sections.
      CALL PYSTAT(1)

      END
\end{verbatim}
and a file \ttt{indata} with the contents 
\begin{verbatim}
CMS,p,p,14000.
1000,3
! below follows commands sent to PYGIVE
MSEL=0           ! Mix processes freely
MSUB(102)=1      ! g + g -> h0 
MSUB(123)=1      ! Z0 + Z0 -> h0
MSUB(124)=1      ! W+ + W- -> h0 
PMAS(25,1)=300.  ! Higgs mass
CKIN(1)=290.     ! lower cutoff on mass
CKIN(2)=310.     ! upper cutoff on mass
MSTP(61)=0       ! no initial-state showers
MSTP(71)=0       ! no final-state showers    
MSTP(81)=0       ! no multiple interactions   
MSTP(111)=0      ! no hadronization
\end{verbatim}
Here the text following the exclamation marks is interpreted 
as a comment by \ttt{PYGIVE}, and thus purely intended to 
allow better documentation of changes. The main program could 
then be linked to {\Py}, to an executable \ttt{a.out}, 
and run e.g.\ with a Unix command line
\begin{verbatim}
  a.out < indata > output
\end{verbatim}
to produce results on the file \ttt{output}. Here the \ttt{indata}
could be changed without requiring a recompilation. Of course, the
main program would have to be more realistic, e.g.\ with events
saved to disk or tape, but the principle should be clear.  
 
\clearpage
 
\section{Monte Carlo Techniques}
 
Quantum mechanics introduces a concept of randomness in the behaviour
of physical processes. The virtue of event generators is that this
randomness can be simulated by the use of Monte Carlo techniques.
In the process, the program authors have to use some ingenuity to 
find the most efficient way to simulate an assumed probability
distribution. A detailed description of possible techniques would 
carry us too far, but
in this section some of the most frequently used approaches are
presented, since they will appear in discussions in
subsequent sections. Further examples may be found e.g.\ in
\cite{Jam80}.
 
First of all one assumes the existence of a random number generator.
This is a (Fortran) function which, each time it is called, returns 
a number $R$ in the range between 0 and 1, such that the inclusive
distribution of numbers $R$ is flat in the range,
and such that different numbers $R$ are uncorrelated. The random
number generator that comes with {\Py} is described at the
end of this section, and we defer the discussion until then.
 
\subsection{Selection From a Distribution}
\label{ss:MCdistsel}
 
The situation that is probably most common is that we know a 
function $f(x)$ which is non-negative in the allowed $x$ range 
$x_{\mmin} \leq x \leq x_{\mmax}$. We want to select an $x$
`at random' so that the probability in a small interval $\d x$ 
around a given $x$ is proportional to $f(x) \, \d x$. Here $f(x)$ 
might be a fragmentation function, a differential cross section, 
or any of a number of distributions.
 
One does not
have to assume that the integral of $f(x)$ is explicitly normalized
to unity: by the Monte Carlo procedure of picking exactly one
accepted $x$ value, normalization is implicit in the final result.
Sometimes the integral of $f(x)$ does carry a physics content of
its own, as part of an overall weight factor
we want to keep track of. Consider, for instance, the case when $x$
represents one or several phase-space variables and $f(x)$ a
differential cross section; here the integral has a meaning of total
cross section for the process studied. The task of a Monte Carlo is
then, on the one hand, to generate events one at a time, and, on the
other hand, to estimate the total cross section.
The discussion of this important example is
deferred to section \ref{ss:PYTcrosscalc}.
 
If it is possible to find a primitive function $F(x)$ which
has a known inverse $F^{-1}(x)$, an $x$ can be found as
follows (method 1):
\begin{eqnarray}
& \displaystyle{ \int_{x_{\mmin}}^{x} f(x) \, \d x =
R \int_{x_{\mmin}}^{x_{\mmax}} f(x) \, \d x } & \nonumber \\[1mm]
\Longrightarrow & x = F^{-1}(F(x_{\mmin}) + 
R(F(x_{\mmax}) - F(x_{\mmin}))) ~. &
\end{eqnarray}
The statement of the first line is that a fraction $R$ of the
total area under $f(x)$ should be to the left of $x$.
However, seldom are functions of interest so nice that the
method above works. It is therefore necessary to use more complicated
schemes.
 
Special tricks can sometimes be found. Consider e.g.\ the generation
of a Gaussian $f(x) = \exp(-x^2)$. This function is not integrable,
but if we combine it with the same Gaussian distribution of a second
variable $y$, it is possible to transform to polar coordinates
\begin{equation}
f(x) \, \d x \, f(y) \, \d y = \exp(-x^2-y^2) \, \d x \, \d y =
r \exp(-r^2) \, \d r \, \d \varphi ~,
\end{equation}
and now the $r$ and $\varphi$ distributions may be easily generated
and recombined to yield $x$. At the same time we get a second number
$y$, which can also be used. For the generation of transverse momenta
in fragmentation, this is very convenient, since in fact we want to
assign two transverse degrees of freedom.
 
If the maximum of $f(x)$ is known, $f(x) \leq f_{\mmax}$ in the 
$x$ range considered, a hit-or-miss method will always yield the
correct answer (method 2):
\begin{Enumerate}
\item select an $x$ with even probability in the allowed range, i.e.
      $x = x_{\mmin} + R(x_{\mmax} - x_{\mmin})$;
\item compare a (new) $R$ with the ratio $f(x)/f_{\mmax}$;
      if $f(x)/f_{\mmax} \le R$, then reject the $x$ value
      and return to point 1 for a new try;
\item otherwise the most recent $x$ value is retained as final answer.
\end{Enumerate}
The probability that $f(x)/f_{\mmax} > R$ is proportional to 
$f(x)$; hence the correct distribution of retained $x$ values.
The efficiency of this method, i.e.\ the average probability that
an $x$ will be retained, is $(\int \, f(x) \, \d x) 
/ (f_{\mmax}(x_{\mmax} - x_{\mmin}))$.
The method is acceptable if this number is not too low, i.e.\ if
$f(x)$ does not fluctuate too wildly.
 
Very often $f(x)$ does have narrow spikes, and it may not even be
possible to define an $f_{\mmax}$. An example of the former 
phenomenon is a
function with a singularity just outside the allowed region,
an example of the latter an integrable singularity just at the
$x_{\mmin}$ and/or $x_{\mmax}$ borders.
Variable transformations may then be used to make a function
smoother. Thus a function $f(x)$ which blows up as $1/x$ for
$x \rightarrow 0$, with an $x_{\mmin}$ close to 0, would instead
be roughly constant if transformed to the variable $y = \ln x$.
 
The variable transformation strategy may be seen as a combination
of methods 1 and 2, as follows. Assume the existence of a function
$g(x)$, with $f(x) \leq g(x)$ over the $x$ range of interest.
Here $g(x)$ is picked to be a `simple' function, such that the
primitive function $G(x)$ and its inverse $G^{-1}(x)$ are known.
Then (method 3):
\begin{Enumerate}
\item select an $x$ according to the distribution $g(x)$, using
      method 1;
\item compare a (new) $R$ with the ratio $f(x)/g(x)$;
      if $f(x)/g(x) \le R$, then reject the $x$ value
      and return to point 1 for a new try;
\item otherwise the most recent $x$ value is retained as final answer.
\end{Enumerate}
This works, since the first step will select $x$ with a probability
$g(x) \, \d x = \d G(x)$ and the second retain this choice with
probability $f(x)/g(x)$. The total probability to pick a value $x$
is then just the product of the two, i.e.\ $f(x) \, \d x$.
 
If $f(x)$ has several spikes, method 3 may work for each spike
separately, but it may not be possible to find a $g(x)$ that
covers all of them at the same time, and which still has an
invertible primitive function.  However, assume that
we can find a function $g(x) = \sum_i g_i(x)$,
such that $f(x) \leq g(x)$ over
the $x$ range considered, and such that the functions $g_i(x)$
each are non-negative and simple, in the sense that we can find
primitive functions and their inverses. In that case (method 4):
\begin{Enumerate}
\item select an $i$ at random, with relative probability given
      by the integrals
      \begin{equation}
      \int_{x_{\mmin}}^{x_{\mmax}} g_i(x) \, \d x =
      G_i(x_{\mmax}) - G_i(x_{\mmin}) ~; \nonumber
      \end{equation}
\item for the $i$ selected, use method 1 to find an $x$, i.e.
      \begin{equation}
      x = G_i^{-1}(G_i(x_{\mmin}) + 
      R(G_i(x_{\mmax})-G_i(x_{\mmin}))) ~;
      \nonumber
      \end{equation}
\item compare a (new) $R$ with the ratio $f(x)/g(x)$;
      if $f(x)/g(x) \le R$, then reject the $x$ value
      and return to point 1 for a new try;
\item otherwise the most recent $x$ value is retained as final answer.
\end{Enumerate}
This is just a trivial extension of method 3, where steps 1 and 2
ensure that, on the average, each $x$ value picked there is
distributed according to $g(x)$: the first step picks $i$
with relative probability $\int g_i(x) \, \d x$, the second $x$ with
absolute probability $g_i(x) / \int g_i(x) \, \d x$ (this is one place
where one must remember to do normalization correctly); the
product of the two is therefore $g_i(x)$ and the sum over all $i$
gives back $g(x)$.
 
We have now arrived at an approach that is sufficiently powerful for
a large selection of problems. In general,
for a function $f(x)$ which is known to have sharp peaks in a few
different places, the generic behaviour at each peak separately
may be covered by one or a few simple functions $g_i(x)$, to which
one adds a few more $g_i(x)$ to cover the basic behaviour away
from the peaks. By a suitable selection
of the relative strengths of the different $g_i$'s, it is
possible to find a function $g(x)$ that matches well the general
behaviour of $f(x)$, and thus achieve a reasonable Monte Carlo 
efficiency.
 
The major additional complication is when $x$ is a multidimensional
variable. Usually the problem is not so much $f(x)$ itself, but
rather that the phase-space boundaries may be very complicated.
If the boundaries factorize it is possible to pick phase-space
points restricted to the desired region. Otherwise the region may
have to be inscribed in a hyper-rectangle, with points picked within
the whole hyper-rectangle but only retained if they are inside the
allowed region. This may lead to a significant loss in efficiency.
Variable transformations may often make the allowed region easier to
handle.
 
There are two main methods to handle several dimensions, each with its
set of variations. The first method is based on a factorized ansatz,
i.e.\ one attempts to find a function $g(\mbf{x})$ which is
everywhere larger than $f(\mbf{x})$, and which can be factorized
into $g(\mbf{x}) = g^{(1)}(x_1) \, g^{(2)}(x_2) \cdots g^{(n)}(x_n)$,
where $\mbf{x} = (x_1,x_2,\ldots,x_n)$. Here each $g^{(j)}(x_j)$ may
in its turn be a sum of functions $g^{(j)}_i$, as in method 4 above.
First, each $x_j$ is selected independently, and afterwards the ratio
$f(\mbf{x})/g(\mbf{x})$ is used to determine whether to retain the
point.
 
The second method is useful if the boundaries of the allowed region
can be written in a form where the maximum range of $x_1$  is known,
the allowed range of $x_2$ only depends on $x_1$, that of $x_3$ only
on $x_1$ and $x_2$, and so on until $x_n$, whose range may depend on 
all the
preceding variables. In that case it may be possible to find a function
$g(\mbf{x})$ that can be integrated over $x_2$ through $x_n$ to yield
a simple function of $x_1$, according to which $x_1$ is selected.
Having done that, $x_2$ is selected according to a distribution
which now depends on $x_1$, but with $x_3$ through $x_n$ integrated
over. In particular, the allowed range for $x_2$ is known.
The procedure is continued until $x_n$ is reached, where now the
function depends on all the preceding $x_j$ values. In the end, the
ratio $f(\mbf{x})/g(\mbf{x})$ is again used to determine whether to
retain the point.
 
\subsection{The Veto Algorithm}
\label{ss:vetoalg}
 
The `radioactive decay' type of problems is very common,  in particular
in parton showers, but it is also used, e.g.\ in the multiple
interactions description in {\Py}. In this kind of problems
there is one variable $t$, which may be thought of as giving a kind
of time axis along which different events are ordered. The
probability that `something will happen' (a nucleus decay, a
parton branch) at time $t$ is described by a function $f(t)$, which
is non-negative in the range of $t$ values to be studied. However,
this na\"{\i}ve probability is modified by the additional requirement
that something can only happen at time $t$ if it did not happen
at earlier times $t' < t$. (The original nucleus cannot decay once
again if it already did decay; possibly the decay products may decay
in their turn, but that is another question.)
 
The probability that nothing has happened by time $t$ is expressed by
the function ${\cal N}(t)$ and the differential probability that
something happens at time $t$ by ${\cal P}(t)$. The basic equation
then is
\begin{equation}
 {\cal P}(t) = - \frac{\d {\cal N}}{\d t} = f(t) \, {\cal N}(t) ~.
\end{equation}
For simplicity, we shall assume that the process starts at time
$t = 0$, with ${\cal N}(0) = 1$.
 
The above equation can be solved easily if one notes that
$\d {\cal N} / {\cal N} = \d \ln {\cal N}$:
\begin{equation}
 {\cal N}(t) = {\cal N}(0) \exp \left\{ - \int_0^t f(t') \, \d t'
               \right\}
             = \exp \left\{ - \int_0^t f(t') \, \d t' \right\} ~,
\end{equation}
and thus
\begin{equation}
  {\cal P}(t) = f(t) \exp \left\{ - \int_0^t f(t') \, \d t' \right\} ~.
\label{mc:Pveto}
\end{equation}
With $f(t) = c$ this is nothing but the textbook formulae for
radioactive decay. In particular, at small times the correct
decay probability, ${\cal P}(t)$, agrees well with the input
one, $f(t)$, since the exponential factor is close to unity there.
At larger $t$, the exponential gives a dampening which ensures that
the integral of ${\cal P}(t)$ never can exceed unity, even if the
integral of $f(t)$ does. The exponential can be seen as the
probability that nothing happens between the original time 0 and
the final time $t$. In the parton-shower language, this corresponds
to the so-called Sudakov form factor.
 
If $f(t)$ has a primitive function with a known inverse, it is easy
to select $t$ values correctly:
\begin{equation}
  \int_0^t {\cal P}(t') \, \d t' = {\cal N}(0) - {\cal N}(t) =
  1 -  \exp \left\{ - \int_0^t f(t') \, \d t' \right\} = 1 - R ~,
\end{equation}
which has the solution
\begin{equation}
  F(0) - F(t) = \ln R  ~~~ \Longrightarrow ~~~
  t = F^{-1}(F(0) - \ln R) ~.
\end{equation}
 
If $f(t)$ is not sufficiently nice, one may again try to find a
better function $g(t)$, with $f(t) \leq g(t)$ for all $t \geq 0$.
However to use method 3 with this $g(t)$ would not work, since the
method would not correctly take
into account the effects of the exponential term in ${\cal P}(t)$.
Instead one may use the so-called veto algorithm:
\begin{Enumerate}
\item start with $i = 0$ and $t_0 = 0$;
\item add 1 to $i$ and select $t_i = G^{-1}(G(t_{i-1}) - \ln R)$,
      i.e.\ according to $g(t)$, but with the constraint that
      $t_i > t_{i-1}$,
\item compare a (new) $R$ with the ratio $f(t_i)/g(t_i)$;
      if $f(t_i)/g(t_i) \le R$, then return to point 2 for a new try;
\item otherwise $t_{i}$ is retained as final answer.
\end{Enumerate}
 
It may not be apparent why this works. Consider, however, the various
ways in which one can select a specific time $t$. The probability that
the first try works, $t = t_1$, i.e.\ that no intermediate $t$ values
need be rejected, is given by
\begin{equation}
  {\cal P}_0 (t) = \exp \left\{ - \int_0^t g(t') \, \d t' \right\}
  \, g(t) \, \frac{f(t)}{g(t)}
   = f(t) \exp \left\{ - \int_0^t g(t') \, \d t' \right\} ~,
\end{equation}
where the exponential times $g(t)$ comes from eq.~(\ref{mc:Pveto})
applied to $g$, and the ratio $f(t)/g(t)$ is the probability that
$t$ is
accepted. Now consider the case where one intermediate time $t_1$ is
rejected and $t = t_2$ is only accepted in the second step.
This gives
\begin{equation}
  {\cal P}_1 (t) = \int_0^t \d t_1
  \exp \left\{ - \int_0^{t_1} g(t') \, \d t' \right\} g(t_1)
  \left[ 1 - \frac{f(t_1)}{g(t_1)} \right]
  \exp \left\{ - \int_{t_1}^t g(t') \, \d t' \right\} g(t) \,
  \frac{f(t)}{g(t)} ~,
\end{equation}
where the first exponential times $g(t_1)$ gives the probability
that $t_1$ is first selected, the square brackets the probability
that $t_1$
is subsequently rejected, the following piece the probability that
$t = t_2$ is selected when starting from $t_1$, and the final factor
that $t$ is retained. The whole is to be integrated over all
possible intermediate times $t_1$. The exponentials together give an
integral over the range from 0 to $t$, just as in ${\cal P}_0$,
and the factor for the final step being accepted is also the same,
so therefore one finds that
\begin{equation}
  {\cal P}_1 (t) = {\cal P}_0 (t) \int_0^t \d t_1
  \left[ g(t_1) - f(t_1) \right] ~.
\end{equation}
This generalizes.
In ${\cal P}_2$ one has to consider two intermediate times,
$0 \leq t_1 \leq t_2 \leq t_3 = t$, and so
\begin{eqnarray}
{\cal P}_2 (t) & = &  {\cal P}_0 (t)
   \int_0^t \d t_1 \left[ g(t_1) - f(t_1) \right]
   \int_{t_1}^t \d t_2 \left[ g(t_2) - f(t_2) \right]  \nonumber \\
& = & {\cal P}_0 (t) \frac{1}{2} \left(
   \int_0^t \left[ g(t') - f(t') \right] \d t' \right)^2 ~.
\end{eqnarray}
The last equality is most easily seen if one also considers the
alternative region $0 \leq t_2 \leq t_1 \leq t$, where the
r\^oles of $t_1$ and $t_2$ have just been interchanged, and the
integral therefore has the same value as in the region considered.
Adding the two regions, however, the integrals over $t_1$ and
$t_2$ decouple, and become equal. In general, for ${\cal P}_i$,
the $i$ intermediate times can be ordered in $i!$ different ways.
Therefore the total probability to accept $t$, in any step, is
\begin{eqnarray}
{\cal P} (t) & = &
   \displaystyle{ \sum_{i=0}^{\infty} {\cal P}_i (t)
   = {\cal P}_0 (t) \sum_{i=0}^{\infty} \frac{1}{i!}
   \left( \int_0^t \left[ g(t') - f(t') \right] \d t' \right)^i } 
   \nonumber \\
& = & \displaystyle{ f(t) \exp \left\{ - \int_0^t g(t') \, \d t'
   \right\} \exp \left\{ \int_0^t \left[ g(t') - f(t') \right] \d t'
   \right\} }  \nonumber \\
& = & \displaystyle{ f(t) \exp \left\{ - \int_0^t f(t') \, \d t' 
   \right\} ~, }
\end{eqnarray}
which is the desired answer.
 
If the process is to be stopped at some scale $t_{\mmax}$, i.e.
if one would like to remain with a fraction ${\cal N}(t_{\mmax})$
of events where nothing happens at all, this is easy to include
in the veto algorithm: just iterate upwards in $t$ at usual, but
stop the process if no allowed branching is found before 
$t_{\mmax}$.
 
Usually $f(t)$ is a function also of additional variables $x$. The
methods of the preceding section are easy to generalize if one
can find a suitable function $g(t,x)$ with $f(t,x) \leq g(t,x)$.
The $g(t)$ used in the veto algorithm is the integral of $g(t,x)$
over $x$. Each time a $t_i$ has been selected also an $x_i$ is
picked, according to $g(t_i,x) \, dx$, and the $(t,x)$ point is
accepted with probability $f(t_i,x_i)/g(t_i,x_i)$.
 
\subsection{The Random Number Generator}
 
In recent years, progress has been made in constructing portable
generators with large periods and other good properties; see the review
\cite{Jam90}. Therefore the current version contains a random number
generator based on the algorithm proposed by Marsaglia, Zaman and Tsang
\cite{Mar90}. This routine should work on any machine with a mantissa
of at least 48 digits, i.e.\ on computers with a 64-bit (or more) 
representation of double precision real numbers. Given the same 
initial state, the sequence will also be identical on different platforms. 
This need not mean that the same sequence of events will be generated, 
since the different treatments of roundoff errors in numerical 
operations will lead to slightly different real numbers being tested 
against these random numbers in IF statements. Also code optimization 
may lead to a divergence of the event sequence. 

Apart from nomenclature issues, the coding of \ttt{PYR} as 
a function rather than a subroutine, and the extension to double 
precision, the only difference between our code and the code given in
\cite{Jam90} is that slightly different algorithms are used to ensure
that the random number is not equal to 0 or 1 within the machine 
precision. Further developments of the algorithm has been proposed
\cite{Lus94} to remove residual possibilities of small long-range
correlations, at the price of a slower generation procedure. However, 
given that {\Py} is using random numbers for so many different tasks,
without any fixed cycle, this has been deemed unnecessary. 
 
The generator has a period of over $10^{43}$, and the
possibility to obtain almost $10^9$
different and disjoint subsequences, selected
by giving an initial integer number. The price to be paid for the
long period is that the state of the generator at a given moment 
cannot be described by a single integer, but requires about 100 words.
Some of these are real numbers, and are thus not correctly represented
in decimal form. The old-style procedure, which made it possible to 
restart the generation from a seed value written to the run output, 
is therefore
not convenient. The CERN library implementation keeps track of the
number of random numbers generated since the start. With this value
saved, in a subsequent run the random generator can be asked to skip
ahead the corresponding number of random numbers. {\Py} 
is a heavy user of random numbers, however: typically 30\% of the full
run time is spent on random number generation. Of this, half is
overhead coming from the function call administration, but the other
half is truly related to the speed of the algorithm. Therefore a
skipping ahead would take place with 15\% of the time cost of the 
original run, i.e.\ an uncomfortably high figure.
 
Instead a different solution is chosen here. Two special routines are
provided for writing and reading the state of the random number 
generator (plus some initialization information) on a sequential 
file, in a platform-dependent internal representation. The file used 
for this purpose
has to be specified by you, and opened for read and write.
A state is written as a single record, in free format. It is possible
to write an arbitrary number of states on a file, and a record can be
overwritten, if so desired. The event generation loop might then look
something like:
\begin{Enumerate}
\item save the state of the generator on file (using flag set in
point 3 below),
\item generate an event,
\item study the event for errors or other reasons why to regenerate
it later; set flag to overwrite previous generator state if no errors,
otherwise set flag to create new record;
\item loop back to point 1.
\end{Enumerate}
With this procedure, the file will contain the state before each of the
problematical events. These events can therefore be generated in a 
shorter run, where further information can be printed. (Inside {\Py}, 
some initialization may take place in connection with the very first 
event generated in a run, so it may be necessary to generate one
ordinary event before reading in a saved state to generate the
interesting events.) An alternative approach might be to save the state 
every 100 events or so. If the events are subsequently processed through 
a detector simulation, you may have to save also other sets of seeds, 
naturally.

Unfortunately, the procedure is not always going to work. For instance,
if cross section maximum violations have occured before the interesting 
event in the original run, there is a possibility that another event is 
picked in the re-started one, where the maximum weight estimate has not
been updated. Another problem is the multiple interaction machinery,
where some of the options contain an element of learning, which again 
means that the event sequence may be broken.
 
In addition to the service routines, the common block which contains
the state of the generator is available for manipulation,
if you so desire. In particular, the initial seed value is by
default 19780503, i.e.\ different from the Marsaglia/CERN default
54217137. It is possible to change this value before any random numbers
have been generated, or to force re-initialization in mid-run with any
desired new seed. 
 
It should be noted that, of course, the appearance of a random
number generator package inside {\Py} does in no way preclude
the use of other routines. You can easily revert to having
\ttt{PYR} as nothing but an interface to an
arbitrary external random number generator; e.g.\ to call a routine
\ttt{RNDM} all you need to have is
\begin{verbatim}
      FUNCTION PYR(IDUMMY)
      IMPLICIT DOUBLE PRECISION(A-H, O-Z)
  100 PYR=RNDM(IDUMMY)
      IF(PYR.LE.0D0.OR.PYR.GE.1D0) GOTO 100
      RETURN
      END
\end{verbatim}
 
The random generator subpackage consists of the following components.
 
\drawbox{R = PYR(IDUMMY)}\label{p:PYR}
\begin{entry}
\itemc{Purpose:} to generate a (pseudo)random number \ttt{R} uniformly
  in the range 0$<$\ttt{R}$<$1, i.e.\ excluding the endpoints.
\iteme{IDUMMY :} dummy input argument; normally 0.
\end{entry}
 
\drawbox{CALL PYRGET(LFN,MOVE)}\label{p:PYRGET}
\begin{entry}
\itemc{Purpose:} to dump the current state of the random number
  generator on a separate file, using internal representation for
  real and integer numbers. To be precise, the full contents of
  the \ttt{PYDATR} common block are written on the file, with the
  exception of \ttt{MRPY(6)}.
\iteme{LFN :} (logical file number) the file number to which the state
  is dumped. You
  must associate this number with a true file (with a platform-dependent
  name), and see to it that this file is open for write.
\iteme{MOVE :} choice of adding a new record to the file or
  overwriting old record(s). Normally only options 0 or $-$1 should be
  used.
    \begin{subentry}
    \iteme{= 0 (or > 0) :} add a new record to the end of the
      file.
    \iteme{= -1 :} overwrite the last record with a new one (i.e.\ do
      one \ttt{BACKSPACE} before the new write).
    \iteme{= $-n$ :} back up $n$ records before writing the
      new record. The records following after the new one are lost,
      i.e.\ the last $n$ old records are lost and one new added.
    \end{subentry}
\end{entry}
 
\drawbox{CALL PYRSET(LFN,MOVE)}\label{p:PYRSET}
\begin{entry}
\itemc{Purpose:} to read in a state for the random number generator,
  from which the subsequent generation can proceed. The state must
  previously have been saved by a \ttt{PYRGET} call. Again the full
  contents of the \ttt{PYDATR} common block are read, with the
  exception of \ttt{MRPY(6)}.
\iteme{LFN :} (logical file number) the file number from which the
  state is read. You
  must associate this number with a true file previously written with
  a \ttt{PYRGET} call, and see to it that this file is open for read.
\iteme{MOVE :} positioning in file before a record is read. With zero
  value, records are read one after the other for each new call, while
  non-zero values may be used to navigate back and forth, and e.g.
  return to the same initial state several times.
    \begin{subentry}
    \iteme{= 0 :} read the next record.
    \iteme{= $+n$ :} skip ahead $n$ records before reading the record
      that sets the state of the random number generator.
    \iteme{= $-n$ :} back up $n$ records before reading the record that
      sets the state of the random number generator.
    \end{subentry}
\end{entry}
 
\drawbox{COMMON/PYDATR/MRPY(6),RRPY(100)}\label{p:PYDATR}
\begin{entry}
\itemc{Purpose:} to contain the state of the random number generator
  at any moment (for communication between \ttt{PYR}, \ttt{PYRGET}
  and \ttt{PYRSET}), and also to provide you with the
  possibility to initialize different random number sequences, and to
  know how many numbers have been generated.
\iteme{MRPY(1) :}\label{p:MRPY} (D=19780503) the integer number that 
  specifies
  which of the possible subsequences will be initialized in the next
  \ttt{PYR} call for which \ttt{MRPY(2)=0}. Allowed values are
  0$\leq$\ttt{MRPY(1)}$\leq$900\,000\,000, the original Marsaglia
  (and CERN library) seed is 54217137. The \ttt{MRPY(1)} value is not
  changed by any of the {\Py} routines.
\iteme{MRPY(2) :} (D=0) initialization flag, put to 1 in the first
  \ttt{PYR} call of run. A re-initialization of the random number
  generator can be made in mid-run by resetting \ttt{MRPY(2)} to 0
  by hand. In addition, any time the counter \ttt{MRPY(3)} reaches
  1000000000, it is reset to 0 and \ttt{MRPY(2)} is increased by 1.
\iteme{MRPY(3) :} (R) counter for the number of random numbers
  generated from the beginning of the run. To avoid overflow when
  very many numbers are generated, \ttt{MRPY(2)} is used as
  described above.
\iteme{MRPY(4), MRPY(5) :} \ttt{I97} and \ttt{J97} of the CERN
  library implementation; part of the state of the generator.
\iteme{MRPY(6) :} (R) current position, i.e.\ how many records after
  beginning, in the file; used by \ttt{PYRGET} and \ttt{PYRSET}.
\iteme{RRPY(1) - RRPY(97) :}\label{p:RRPY} the \ttt{U} array of the 
  CERN library implementation; part of the state of the generator.
\iteme{RRPY(98) - RRPY(100) :} \ttt{C}, \ttt{CD} and \ttt{CM} of the
  CERN library implementation; the first part of the state of the
  generator, the latter two constants calculated at initialization.
\end{entry}
 
\clearpage
 
\section{The Event Record}
 
The event record is the central repository for information about
the particles produced in the current event: flavours, momenta,
event history, and production vertices. It plays a very central
r\^ole: without a proper understanding of what the record is and
how information is stored, it is meaningless to try to use {\Py}. 
The record is stored in the common block 
\ttt{PYJETS}. Almost all the routines that the user calls
can be viewed as performing some action on the record: fill a
new event, let partons fragment or particles decay, boost it,
list it, find clusters, etc.
 
In this section we will first describe the {\KF} flavour code,
subsequently the \ttt{PYJETS} common block, and then give a few
comments about the r\^ole of the event record in the programs.
 
To ease the interfacing of different event generators, a
\ttt{HEPEVT} standard common block structure for the event record
has been agreed on. For historical reasons the standard common blocks
are not directly used in {\Py}, but a conversion routine comes with
the program, and is described at the end of this section.
 
\subsection{Particle Codes}
\label{ss:codes}
 
The Particle Data Group particle code \cite{PDG88,PDG92,PDG00} is 
used consistently throughout the program. Almost all known 
discrepancies between earlier versions of the PDG standard and the 
{\Py} usage have now been resolved. The one known exception is the 
(very uncertain) classification of $\f_0(980)$, with $\f_0(1370)$
also affected as a consequence. There is also a possible point of 
confusion in the technicolor sector between ${\pi'}^0_{\mrm{tc}}$ 
and $\eta_{\mrm{tc}}$. The latter is retained for historical reasons, 
whereas the former was introduced for consistency in models of
top--color--assisted technicolor.
The PDG standard, with the local {\Py} extensions, is referred to as 
the {\KF} particle code. This code you have to be thoroughly familiar 
with. It is described below.
 
The {\KF} code is not convenient for a direct storing of masses,
decay data, or other particle properties, since the {\KF}
codes are so spread out. Instead a compressed code {\KC} between
1 and 500 is used here. A particle and its antiparticle are
mapped to the same {\KC} code, but else the mapping is unique.
Normally this code is only used at very specific
places in the program, not visible to the user. If need be, the
correspondence can always be obtained by using the function
\ttt{PYCOMP}, i.e.\ \mbox{\ttt{KC = PYCOMP(KF)}}. This mapping is not
hardcoded, but can be changed by user intervention, e.g.
by introducing new particles with the \ttt{PYUPDA} facility.
It is therefore not intended that you should ever want or need to know 
any {\KC} codes at all. It may be useful to know, however, that for codes
smaller than 80, {\KF} and {\KC} agree. Normally a user would never do the 
inverse mapping, but we note that this is stored as 
\ttt{KF = KCHG(KC,4)}, making use of the \ttt{KCHG} array in the
\ttt{PYDAT2} common block. Of course, the sign of a particle could
never be recovered by this inverse operation.
 
The particle names printed in the tables in this section correspond 
to the ones obtained
with the routine \ttt{PYNAME}, which is used extensively, e.g.\ in
\ttt{PYLIST}. Greek characters are spelt out in full, with a capital
first letter to correspond to a capital Greek letter. Generically the
name of a particle is made up of the following pieces:
\begin{Enumerate}
\item The basic root name. This includes a * for most spin 1
($L = 0$) mesons and spin $3/2$ baryons, and a $'$ for some spin
$1/2$ baryons (where there are two states to be distinguished,
cf. $\Lambda$--$\Sigma^0$). The rules for heavy baryon naming are in
accordance with the 1986 Particle Data Group conventions \cite{PDG86}.
For mesons with one unit of orbital angular momentum, K (D, B,
\ldots) is used for quark-spin 0 and K* (D*, B*, \ldots) for
quark-spin 1 mesons; the convention for `*' may here deviate slightly
from the one used by the PDG.
\item Any lower indices, separated from the root by a \_. For
heavy hadrons, this is the additional heavy-flavour content not
inherent in the root itself. For a diquark, it is the spin.
\item The characters `bar' for an antiparticle, wherever the distinction 
between particle and antiparticle is not inherent in the charge 
information.
\item Charge information: $++$, $+$, $0$, $-$, or $--$.
Charge is not given for quarks or diquarks. Some neutral particles
which are customarily given without a 0 also here lack it,
such as neutrinos, $\g$, $\gamma$,
and flavour-diagonal mesons other than $\pi^0$ and $\rho^0$. Note that
charge is included both for the proton and the neutron. While
non-standard, it is helpful in avoiding misunderstandings when
looking at an event listing.
\end{Enumerate}
 
Below follows a list of {\KF} particle codes. The list is not complete;
a more extensive one may be obtained with \ttt{CALL PYLIST(11)}.
Particles are grouped together, and the basic rules are described
for each group. Whenever a distinct antiparticle exists, it is given
the same {\KF} code with a minus sign (whereas {\KC} codes are always
positive).
 
\begin{Enumerate}
 
\item Quarks and leptons, Table \ref{t:codeone}. \\
This group contains the basic building blocks of matter, arranged
according to family, with the lower member of weak isodoublets also
having the smaller code (thus $\d$ precedes $\u$). A fourth generation 
is included as part of the scenarios for exotic physics. The quark codes 
are used as building blocks for the diquark, meson and baryon codes below.
 
\begin{table}[ptb]
\captive{Quark and lepton codes.
\protect\label{t:codeone} } \\
\vspace{1ex}
\begin{center}
\begin{tabular}{|c|c|c||c|c|c|@{\protect\rule{0mm}{\tablinsep}}}
\hline
{\KF} & Name & Printed & {\KF} & Name & Printed \\
\hline
    1 & $\d$ & \ttt{d}   &  11 & $\e^-$ &  \ttt{e-}    \\
    2 & $\u$ & \ttt{u}   &  12 & $\nu_{\e}$ &  \ttt{nu\_e}   \\
    3 & $\s$ & \ttt{s}   &  13 & $\mu^-$ &  \ttt{mu-}   \\
    4 & $\c$ & \ttt{c}   &  14 & $\nu_{\mu}$ & \ttt{nu\_mu}   \\
    5 & $\b$ & \ttt{b}   &  15 & $\tau^-$ & \ttt{tau-}   \\
    6 & $\t$ & \ttt{t}   &  16 & $\nu_{\tau}$ &  \ttt{nu\_tau}   \\
    7 & $\b'$ & \ttt{b'} &  17 & $\tau'$ & \ttt{tau'}  \\
    8 & $\t'$ & \ttt{t'} &  18 & $\nu'_{\tau}$ & \ttt{nu'\_tau}  \\
    9 & & & 19 & & \\
   10 & & & 20 & & \\
\hline
\end{tabular}
\end{center}
\end{table}
 
\item Gauge bosons and other fundamental bosons,
Table \ref{t:codetwo}. \\
This group includes all the gauge and Higgs bosons of the standard
model, as well as some of the bosons appearing in various extensions
of it. They correspond to one extra {\bf U(1)}
and one extra {\bf SU(2)} group, a further Higgs doublet, 
a graviton, a horizontal gauge boson $\R$ (coupling between 
families), and a (scalar) leptoquark $\L_{\Q}$. 
 
\begin{table}[ptb]
\captive{Gauge boson and other fundamental boson codes.
\protect\label{t:codetwo} }  \\
\vspace{1ex}
\begin{center}
\begin{tabular}{|c|c|c||c|c|c|@{\protect\rule{0mm}{\tablinsep}}}
\hline
{\KF} & Name & Printed & {\KF} & Name & Printed \\
\hline
   21 & $\g$ & \ttt{g}             & 31 & &   \\
   22 & $\gamma$ & \ttt{gamma}     & 32 & $\Z'^0$ & \ttt{Z'0} \\
   23 & $\Z^0$ & \ttt{Z0}          & 33 & $\Z''^0$ & \ttt{Z"0}\\
   24 & $\W^+$ & \ttt{W+}          & 34 & $\W'^+$ & \ttt{W'+} \\
   25 & $\hrm^0$ & \ttt{h0}        & 35 & $\H^0$ & \ttt{H0}   \\
   26 & &                          & 36 & $\A^0$ & \ttt{A0}   \\
   27 & &                          & 37 & $\H^+$ & \ttt{H+}   \\
   28 & &                          & 38 & &                   \\
   29 & &                          & 39 & $\mrm{G}$ & \ttt{Graviton} \\
   30 & &                          & 40 & &                   \\
      & &                          & 41 & $\R^0$ & \ttt{R0}   \\
      & &                          & 42 & $\L_{\Q}$ & \ttt{LQ}\\
\hline
\end{tabular}
\end{center}
\end{table}
 
\item Exotic particle codes. \\
The positions 43--80 are used as temporary sites for exotic particles
that eventually may be shifted to a separate code sequence. Currently
this list is empty. The ones not in use are at your disposal (but with 
no guarantees that they will remain so).
 
\item Various special codes, Table \ref{t:codefour}. \\
In a Monte Carlo, it is always necessary to have codes that do not
correspond to any specific particle, but are used to lump together
groups of similar particles for decay treatment (nowadays largely 
obsolete), to specify generic decay products (also obsolete), 
or generic intermediate states in external processes, or 
additional event record information from jet searches. These codes, 
which again are non-standard, are found between numbers 81 and 100.\\
The junction, code 88, is not a physical particle but marks the place
in the event record where three string pieces come together in a point,
e.g. a Y-shaped topology with a quark at each end. No distinction is made 
between a junction and an antijunction, i.e. whether a baryon or an 
antibaryon is going to be produced in the neighbourhood of the junction.
  
\begin{table}[ptb]
\captive{Various special codes.
\protect\label{t:codefour} }   \\
\vspace{1ex}
\begin{center}
\begin{tabular}{|c|c|c|@{\protect\rule{0mm}{\tablinsep}}}
\hline
{\KF} & Printed & Meaning \\
\hline
   81 &  \ttt{specflav} & Spectator flavour; used in decay-product
listings  \\
   82 &  \ttt{rndmflav} & A random $\u$, $\d$, or $\s$ flavour;
possible decay product  \\
   83 &  \ttt{phasespa} & Simple isotropic phase-space decay  \\
   84 &  \ttt{c-hadron} & Information on decay of generic charm
hadron  \\
   85 &  \ttt{b-hadron} & Information on decay of generic bottom
hadron  \\
   86 &   &  \\
   87 &   &  \\
   88 &  \ttt{junction} & A junction of three string pieces \\
   89 &   & (internal use for unspecified resonance data) \\
   90 &  \ttt{system}   & Intermediate pseudoparticle in external process \\
   91 &  \ttt{cluster}  & Parton system in cluster fragmentation  \\
   92 &  \ttt{string}   & Parton system in string fragmentation  \\
   93 &  \ttt{indep.}   & Parton system in independent fragmentation  \\
   94 &  \ttt{CMshower} & Four-momentum of time-like showering system  \\
   95 &  \ttt{SPHEaxis} & Event axis found with \ttt{PYSPHE}  \\
   96 &  \ttt{THRUaxis} & Event axis found with \ttt{PYTHRU}  \\
   97 &  \ttt{CLUSjet}  & Jet (cluster) found with \ttt{PYCLUS}  \\
   98 &  \ttt{CELLjet}  & Jet (cluster) found with \ttt{PYCELL}  \\
   99 &  \ttt{table}    & Tabular output from \ttt{PYTABU}  \\
  100 &           &   \\
\hline
\end{tabular}
\end{center}
\end{table} 

\item Diquark codes, Table \ref{t:codefive}. \\
A diquark made up of a quark with code $i$ and another with code $j$,
where $i \geq j$, and with total spin $s$, is given the code
\begin{equation}
\mtt{KF} = 1000 i + 100 j + 2s + 1  ~,
\end{equation}
i.e.\ the tens position is left empty (cf. the baryon code below).
Some of the most frequently used codes are listed in the table. All 
the lowest-lying spin 0 and 1 diquarks are included in the program.
 
\begin{table}[ptb]
\captive{Diquark codes. For brevity, diquarks containing $\c$ or $\b$
quarks are not listed, but are defined analogously.
\protect\label{t:codefive} }  \\
\vspace{1ex}
\begin{center}
\begin{tabular}{|c|c|c||c|c|c|@{\protect\rule{0mm}{\tablinsep}}}
\hline
{\KF} & Name & Printed & {\KF} & Name & Printed \\
\hline
      &          &             & 1103 & $\d\d_1$ & \ttt{dd\_1}  \\
 2101 & $\u\d_0$ & \ttt{ud\_0} & 2103 & $\u\d_1$ & \ttt{ud\_1}  \\
      &          &             & 2203 & $\u\u_1$ & \ttt{uu\_1}  \\
 3101 & $\s\d_0$ & \ttt{sd\_0} & 3103 & $\s\d_1$ & \ttt{sd\_1}  \\
 3201 & $\s\u_0$ & \ttt{su\_0} & 3203 & $\s\u_1$ & \ttt{su\_1}  \\
      &          &             & 3303 & $\s\s_1$ & \ttt{ss\_1}  \\
\hline
\end{tabular}
\end{center}
\end{table}
 
\item Meson codes, Tables \ref{t:codesixa} and \ref{t:codesixb}. \\
A meson made up of a quark with code $i$ and an antiquark with
code $-j$, $j \neq i$, and with total spin $s$, is given the code
\begin{equation}
\mtt{KF} = \left\{ 100 \max(i,j) + 10 \min(i,j) + 2s + 1 \right\}
\, \mrm{sign}(i-j) \, (-1)^{\max(i,j)} ~,
\label{eq:KFmeson}
\end{equation}
assuming it is not orbitally or radially excited.
Note the presence of an extra $-$ sign if the heaviest quark is a
down-type one. This is in accordance with the particle--antiparticle
distinction adopted in the 1986 Review of Particle Properties
\cite{PDG86}. It means for example that a $\B$ meson contains a 
$\bbar$ antiquark rather than a $\b$ quark.
 
The flavour-diagonal states are arranged in order of ascending 
mass. Thus the obvious generalization of eq.~(\ref{eq:KFmeson})
to $\mtt{KF} = 110 i + 2 s + 1$ is only valid for charm and bottom.
The lighter quark states can appear mixed, e.g. the $\pi^0$ 
(111) is an equal mixture of $\d\dbar$ (naively code 111) and
$\u\ubar$ (naively code 221).

The standard rule of having the last digit of the form
$2s+1$ is broken for the $\K_{\mrm{S}}^0$--$\K_{\mrm{L}}^0$ system, 
where it is 0, and this convention should carry over to mixed states 
in the $\B$ meson system, should one choose to define such. For 
higher multiplets with the same spin, $\pm$10000, $\pm$20000, etc., 
are added to provide the extra distinction needed. Some of the most 
frequently used codes are given below.
 
The full lowest-lying pseudoscalar and vector multiplets are included
in the program, Table \ref{t:codesixa}.
 
\begin{table}[ptb]
\captive{Meson codes, part 1.
\protect\label{t:codesixa} }  \\
\vspace{1ex}
\begin{center}
\begin{tabular}{|c|c|c||c|c|c|@{\protect\rule{0mm}{\tablinsep}}}
\hline
{\KF} & Name & Printed & {\KF} & Name & Printed \\
\hline
  211 & $\pi^+$ & \ttt{pi+}        & 213 & $\rho^+$ & \ttt{rho+}  \\
  311 & $\K^0$ & \ttt{K0}          & 313 & $\K^{*0}$ & \ttt{K*0}  \\
  321 & $\K^+$ & \ttt{K+}          & 323 & $\K^{*+}$ & \ttt{K*+}  \\
  411 & $\D^+$ & \ttt{D+}          & 413 & $\D^{*+}$ & \ttt{D*+}  \\
  421 & $\D^0$ & \ttt{D0}          & 423 & $\D^{*0}$ & \ttt{D*0}  \\
  431 & $\D_{\s}^+$ & \ttt{D\_s+}     &
433 & $\D_{\s}^{*+}$ & \ttt{D*\_s+}  \\
  511 & $\B^0$ & \ttt{B0}          & 513 & $\B^{*0}$ & \ttt{B*0}  \\
  521 & $\B^+$ & \ttt{B+}          & 523 & $\B^{*+}$ & \ttt{B*+}  \\
  531 & $\B_{\s}^0$ & \ttt{B\_s0}     &
533 & $\B_{\s}^{*0}$ & \ttt{B*\_s0}  \\
  541 & $\B_{\c}^+$ & \ttt{B\_c+}     &
543 & $\B_{\c}^{*+}$ & \ttt{B*\_c+}  \\
  111 & $\pi^0$ & \ttt{pi0}        & 113 & $\rho^0$ & \ttt{rho0}  \\
  221 & $\eta$ & \ttt{eta}         & 223 & $\omega$ & \ttt{omega}  \\
  331 & $\eta'$ & \ttt{eta'}       & 333 & $\phi$ & \ttt{phi}  \\
  441 & $\eta_{\c}$ & \ttt{eta\_c} & 443 & $\Jpsi$ & \ttt{J/psi}  \\
  551 & $\eta_{\b}$ & \ttt{eta\_b} &
553 & $\Upsilon$ & \ttt{Upsilon}  \\
\hline
  130 & $\K_{\mrm{L}}^0$ & \ttt{K\_L0} & & &  \\
  310 & $\K_{\mrm{S}}^0$ & \ttt{K\_S0} & & &  \\
\hline
\end{tabular}
\end{center}
\end{table}
 
Also the lowest-lying orbital angular momentum $L = 1$ mesons are
included, Table \ref{t:codesixb}: one pseudovector multiplet
obtained for total quark-spin 0 ($L = 1, S = 0 \Rightarrow J = 1$)
and one scalar, one pseudovector and one tensor multiplet obtained
for total quark-spin 1 ($L = 1, S = 1 \Rightarrow J = 0, 1$ or 2),
where $J$ is what is conventionally called the spin $s$ of the meson.
Any mixing between the two pseudovector multiplets is
not taken into account. Please note that some members of these
multiplets have still not been found, and are included here only based
on guesswork. Even for known ones, the information on particles 
(mass, width, decay modes) is highly incomplete.

Only two radial excitations are included, the $\psi' = \psi(2S)$ and
$\Upsilon' = \Upsilon(2S)$.
 
\begin{table}[ptb]
\captive{Meson codes, part 2. For brevity, states with $\b$ quark
are omitted from this listing, but are defined in the program.
\protect\label{t:codesixb} }  \\
\vspace{1ex}
\begin{center}
\begin{tabular}{|c|c|c||c|c|c|@{\protect\rule{0mm}{\tablinsep}}}
\hline
{\KF} & Name & Printed & {\KF} & Name & Printed \\
\hline
10213 & $\b_1$ & \ttt{b\_1+}              &
10211 & $\a_0^+$ & \ttt{a\_0+}  \\
10313 & $\K_1^0$ & \ttt{K\_10}            &
10311 & $\K_0^{*0}$ & \ttt{K*\_00}  \\
10323 & $\K_1^+$ & \ttt{K\_1+}            &
10321 & $\K_0^{*+}$ & \ttt{K*\_0+}  \\
10413 & $\D_1^+$ & \ttt{D\_1+}            &
10411 & $\D_0^{*+}$ & \ttt{D*\_0+}  \\
10423 & $\D_1^0$ & \ttt{D\_10}            &
10421 & $\D_0^{*0}$ & \ttt{D*\_00}  \\
10433 & $\D_{1 \s}^+$ & \ttt{D\_1s+}      &
10431 & $\D_{0 \s}^{*+}$ & \ttt{D*\_0s+}  \\
10113 & $\b_1^0$ & \ttt{b\_10}            &
10111 & $\a_0^0$ & \ttt{a\_00}  \\
10223 & $\hrm_1^0$ & \ttt{h\_10}            &
10221 & $\f_0^0$ & \ttt{f\_00}  \\
10333 & $\hrm'^0_1$ & \ttt{h'\_10}          &
10331 & $\f'^0_0$ & \ttt{f'\_00}  \\
10443 & $\hrm_{1 \c}^0$ & \ttt{h\_1c0}      &
10441 & $\chi_{0 \c}^0$ & \ttt{chi\_0c0}  \\  \hline
20213 & $\a_1^+$ & \ttt{a\_1+}            &
  215 & $\a_2^+$ & \ttt{a\_2+}  \\
20313 & $\K_1^{*0}$ & \ttt{K*\_10}        &
  315 & $\K_2^{*0}$ & \ttt{K*\_20}  \\
20323 & $\K_1^{*+}$ & \ttt{K*\_1+}        &
  325 & $\K_2^{*+}$ & \ttt{K*\_2+}  \\
20413 & $\D_1^{*+}$ & \ttt{D*\_1+}        &
  415 & $\D_2^{*+}$ & \ttt{D*\_2+}  \\
20423 & $\D_1^{*0}$ & \ttt{D*\_10}        &
  425 & $\D_2^{*0}$ & \ttt{D*\_20}  \\
20433 & $\D_{1 \s}^{*+}$ & \ttt{D*\_1s+}  &
  435 & $\D_{2 \s}^{*+}$ & \ttt{D*\_2s+}  \\
20113 & $\a_1^0$ & \ttt{a\_10}            &
  115 & $\a_2^0$ & \ttt{a\_20}  \\
20223 & $\f_1^0$ & \ttt{f\_10}            &
  225 & $\f_2^0$ & \ttt{f\_20}  \\
20333 & $\f'^0_1$ & \ttt{f'\_10}          &
  335 & $\f'^0_2$ & \ttt{f'\_20}  \\
20443 & $\chi_{1 \c}^0$ & \ttt{chi\_1c0}  &
  445 & $\chi_{2 \c}^0$ & \ttt{chi\_2c0}  \\  \hline
100443 & $\psi'$     & \ttt{psi'}     &   &   &  \\
100553 & $\Upsilon'$ & \ttt{Upsilon'} &   &   &  \\
\hline
\end{tabular}
\end{center}
\end{table}
 
\item Baryon codes, Table \ref{t:codeseven}.  \\
A baryon made up of quarks $i$, $j$ and $k$, with $i \geq j \geq k$,
and total spin $s$, is given the code
\begin{equation}
\mtt{KF} = 1000 i + 100 j + 10 k + 2s + 1  ~.
\end{equation}
An exception is provided by spin $1/2$ baryons made up of three
different types of quarks, where the two lightest quarks form a spin-0
diquark ($\Lambda$-like baryons). Here the order of the $j$ and $k$
quarks is reversed, so as to provide a simple means of distinction
to baryons with the lightest quarks in a spin-1 diquark
($\Sigma$-like baryons).
 
For hadrons with heavy flavours, the root names are Lambda or
Sigma for hadrons with two $\u$ or $\d$ quarks, Xi for those
with one, and Omega for those without $\u$ or $\d$ quarks.
 
Some of the most frequently used codes are given in Table
\ref{t:codeseven}. The full lowest-lying spin $1/2$ and $3/2$
multiplets are included in the program.
 
\begin{table}[ptb]
\captive{Baryon codes. For brevity, some states with $\b$ quarks
or multiple $\c$ ones are omitted from this listing, but are 
defined in the program.
\protect\label{t:codeseven} }  \\
\vspace{1ex}
\begin{center}
\begin{tabular}{|c|c|c||c|c|c|@{\protect\rule{0mm}{\tablinsep}}}
\hline
{\KF} & Name & Printed & {\KF} & Name & Printed \\
\hline
      &  &                            &
 1114 & $\Delta^-$ & \ttt{Delta-} \\
 2112 & $\n$ & \ttt{n0}                     &
 2114 & $\Delta^0$ & \ttt{Delta0} \\
 2212 & $\p$ & \ttt{p+}                     &
 2214 & $\Delta^+$ & \ttt{Delta+} \\
      &  &                            &
 2224 & $\Delta^{++}$ & \ttt{Delta++} \\
 3112 & $\Sigma^-$ & \ttt{Sigma-}           &
 3114 & $\Sigma^{*-}$ & \ttt{Sigma*-} \\
 3122 & $\Lambda^0$ & \ttt{Lambda0}         & & &  \\
 3212 & $\Sigma^0$ & \ttt{Sigma0}           &
 3214 & $\Sigma^{*0}$ & \ttt{Sigma*0} \\
 3222 & $\Sigma^+$ & \ttt{Sigma+}           &
 3224 & $\Sigma^{*+}$ & \ttt{Sigma*+} \\
 3312 & $\Xi^-$ & \ttt{Xi-}                 &
 3314 & $\Xi^{*-}$ & \ttt{Xi*-} \\
 3322 & $\Xi^0$ & \ttt{Xi0}                 &
 3324 & $\Xi^{*0}$ & \ttt{Xi*0} \\
      &  &                            &
 3334 & $\Omega^-$ & \ttt{Omega-} \\
 4112 & $\Sigma_{\c}^0$ & \ttt{Sigma\_c0}   &
 4114 & $\Sigma_{\c}^{*0}$ & \ttt{Sigma*\_c0} \\
 4122 & $\Lambda_{\c}^+$ & \ttt{Lambda\_c+} & & &  \\
 4212 & $\Sigma_{\c}^+$ & \ttt{Sigma\_c+}   &
 4214 & $\Sigma_{\c}^{*+}$ & \ttt{Sigma*\_c+} \\
 4222 & $\Sigma_{\c}^{++}$ & \ttt{Sigma\_c++} &
 4224 & $\Sigma_{\c}^{*++}$ & \ttt{Sigma*\_c++} \\
 4132 & $\Xi_{\c}^0$ & \ttt{Xi\_c0}         & & &  \\
 4312 & $\Xi'^0_{\c}$ & \ttt{Xi'\_c0}       &
 4314 & $\Xi_{\c}^{*0}$ & \ttt{Xi*\_c0}  \\
 4232 & $\Xi_{\c}^+$ & \ttt{Xi\_c+}         & & &  \\
 4322 & $\Xi'^+_{\c}$ & \ttt{Xi'\_c+}        &
 4324 & $\Xi_{\c}^{*+}$ & \ttt{Xi*\_c+}  \\
 4332 & $\Omega_{\c}^0$ & \ttt{Omega\_c0}   &
 4334 & $\Omega_{\c}^{*0}$ & \ttt{Omega*\_c0}  \\
 5112 & $\Sigma_{\b}^-$ & \ttt{Sigma\_b-}     &
 5114 & $\Sigma_{\b}^{*-}$ & \ttt{Sigma*\_b-}  \\
 5122 & $\Lambda_{\b}^0$ & \ttt{Lambda\_b0} & & &  \\
 5212 & $\Sigma_{\b}^0$ &\ttt{Sigma\_b0}   &
 5214 & $\Sigma_{\b}^{*0}$ & \ttt{Sigma*\_b0}  \\
 5222 & $\Sigma_{\b}^+$ & \ttt{Sigma\_b+}   &
 5224 & $\Sigma_{\b}^{*+}$ &  \ttt{Sigma*\_b+}  \\
\hline
\end{tabular}
\end{center}
\end{table}

\item QCD effective states, Table \ref{t:codeeight}.  \\
We here include the pomeron $\pomeron$ and reggeon $\reggeon$ 
`particles', which are important e.g.\ in the description of 
diffractive scattering, but do not have a simple 
correspondence with other particles in the classification scheme.\\
Also included are codes to be used for denoting diffractive states 
in {\Py}, as part of the event history. The first two digits here
are 99 to denote the non-standard character. The second, third and 
fourth last digits give flavour content, while the very last one is 
0, to denote the somewhat unusual character of the code. Only a few 
codes have been introduced with names; depending on circumstances 
these also have to double up for other diffractive states. Other
diffractive codes for strange mesons and baryon beams are also 
accepted by the program, but do not give nice printouts.
 
\begin{table}[ptb]
\captive{QCD effective states.
\protect\label{t:codeeight} }  \\
\vspace{1ex}
\begin{center}
\begin{tabular}{|c|c|c|@{\protect\rule{0mm}{\tablinsep}}}
\hline
{\KF} & Printed & Meaning \\
\hline
  110 & \ttt{reggeon} & reggeon $\reggeon$ \\
  990 & \ttt{pomeron} & pomeron $\pomeron$ \\
9900110 & \ttt{rho\_diff0} & Diffractive 
$\pi^0 / \rho^0 / \gamma$ state \\
9900210 & \ttt{pi\_diffr+} & Diffractive $\pi^+$ state \\
9900220 & \ttt{omega\_di0} & Diffractive $\omega$ state \\
9900330 & \ttt{phi\_diff0} & Diffractive $\phi$ state \\
9900440 & \ttt{J/psi\_di0} & Diffractive $\Jpsi$ state \\
9902110 & \ttt{n\_diffr}   & Diffractive $\n$ state \\
9902210 & \ttt{p\_diffr+}  & Diffractive $\p$ state \\
\hline
\end{tabular}
\end{center}
\end{table}
 
\item Supersymmetric codes, Table \ref{t:codenine}.  \\
SUSY doubles the number of states of the Standard Model (at least).
Fermions have separate superpartners to the left- and right-handed
components. In the third generation these 
are assumed to mix to nontrivial mass eigenstates, while mixing is not 
included in the first two. Note that all sparticle names begin with a tilde.
Default masses are arbitrary and branching ratios not set at all. 
This is taken care of at initialization if \ttt{IMSS(1)} is positive.   

\begin{table}[ptb]
\captive{Supersymmetric codes.
\protect\label{t:codenine} }  \\
\vspace{1ex}
\begin{center}
\begin{tabular}{|c|c|c||c|c|c|@{\protect\rule{0mm}{\tablinsep}}}
\hline
{\KF} & Name & Printed & {\KF} & Name & Printed \\
\hline
 1000001 & $\sd_L$ & $\sim$\ttt{d\_L} &
 2000001 & $\sd_R$ & $\sim$\ttt{d\_R} \\
 1000002 & $\su_L$ & $\sim$\ttt{u\_L} &
 2000002 & $\su_R$ & $\sim$\ttt{u\_R} \\
 1000003 & $\sst_L$ & $\sim$\ttt{s\_L} &
 2000003 & $\sst_R$ & $\sim$\ttt{s\_R} \\
 1000004 & $\sch_L$ & $\sim$\ttt{c\_L} &
 2000004 & $\sch_R$ & $\sim$\ttt{c\_R} \\
 1000005 & $\sbo_1$ & $\sim$\ttt{b\_1} &
 2000005 & $\sbo_2$ & $\sim$\ttt{b\_2} \\
 1000006 & $\st_1$ & $\sim$\ttt{t\_1} &
 2000006 & $\st_2$ & $\sim$\ttt{t\_2} \\
 1000011 & $\se_L$ & $\sim$\ttt{e\_L-} &
 2000011 & $\se_R$ & $\sim$\ttt{e\_R-} \\
 1000012 & $\snu_{{\e}L}$ & $\sim$\ttt{nu\_eL} &
 2000012 & $\snu_{{\e}R}$ & $\sim$\ttt{nu\_eR}  \\
 1000013 & $\smu_L$ & $\sim$\ttt{mu\_L-} &
 2000013 & $\smu_R$ & $\sim$\ttt{mu\_R-} \\
 1000014 & $\snu_{{\mu}L}$ & $\sim$\ttt{nu\_muL} &
 2000014 & $\snu_{{\mu}R}$ & $\sim$\ttt{nu\_muR} \\
 1000015 & $\stau_1$ & $\sim$\ttt{tau\_L-} &
 2000015 & $\stau_2$ & $\sim$\ttt{tau\_R-} \\
 1000016 & $\snu_{{\tau}L}$ & $\sim$\ttt{nu\_tauL} &
 2000016 & $\snu_{{\tau}R}$ & $\sim$\ttt{nu\_tauR} \\
\hline
 1000021 & $\glu$ & $\sim$\ttt{g} &
 1000025 & $\chio^0_3$ & $\sim$\ttt{chi\_30} \\
 1000022 & $\chio^0_1$ & $\sim$\ttt{chi\_10} &
 1000035 & $\chio^0_4$  & $\sim$\ttt{chi\_40}\\
 1000023 & $\chio^0_2$ & $\sim$\ttt{chi\_20} &
 1000037 & $\chio^+_2$  & $\sim$\ttt{chi\_2+} \\
 1000024 & $\chio^+_1$ & $\sim$\ttt{chi\_1+} &
 1000039 & $\grav$ & $\sim$\ttt{Gravitino} \\
\hline
\end{tabular}
\end{center}
\end{table}
 
\item Technicolor codes, Table \ref{t:codeten}.  \\
A set of colourless and coloured technihadrons have been
included, the latter specifically for the case of Topcolor 
assisted Technicolor. Where unclear, indices 1 or 8 denote the
colour multiplet. 
Then there are coloured technirhos and 
technipions that can mix with the Coloron (or $\mathrm{V}_8$)
associated with the breaking of 
\textbf{SU(3)}$_2 \times $\textbf{SU(3)}$_3$ to ordinary 
\textbf{SU(3)}$_\mathrm{C}$ (where the 2 and 3 indices refer to 
the first two and the third generation, respectively).\\
The $\eta_{\mrm{tc}}$ belongs to an older iteration of 
Technicolor modelling than the rest. It was originally given the
3000221 code, and thereby now comes to clash with the
${\pi'}^0_{\mrm{tc}}$ of the current main scenario. Since the
$\eta_{\mrm{tc}}$ is one-of-a-kind, it was deemed better to move 
it to make way for the ${\pi'}^0_{\mrm{tc}}$. This 
leads to a slight inconsistency with the PDG codes.  

\begin{table}[ptb]
\captive{Technicolor codes.
\protect\label{t:codeten} }  \\
\vspace{1ex}
\begin{center}
\begin{tabular}{|c|c|c||c|c|c|@{\protect\rule{0mm}{\tablinsep}}}
\hline
{\KF} & Name & Printed & {\KF} & Name & Printed \\
\hline
3000111 & $\pi^0_{\mrm{tc}}$ & \ttt{pi\_tc0} &
3100021 & $\mathrm{V}_{8,\mrm{tc}}$ & \ttt{V8\_tc}  \\
3000211 & $\pi^+_{\mrm{tc}}$ & \ttt{pi\_tc+} & 
3100111 & $\pi^0_{22,1,\mrm{tc}}$ & \ttt{pi\_22\_1\_tc}   \\
3000221 & ${\pi'}^0_{\mrm{tc}}$ & \ttt{pi'\_tc0} & 
3200111 & $\pi^0_{22,8,\mrm{tc}}$ & \ttt{pi\_22\_8\_tc}   \\
3000113 & $\rho^0_{\mrm{tc}}$ & \ttt{rho\_tc0} & 
3100113 & $\rho^0_{11,\mrm{tc}}$ & \ttt{rho\_11\_tc}   \\
3000213 & $\rho^+_{\mrm{tc}}$ & \ttt{rho\_tc+} & 
3200113 & $\rho^0_{12,\mrm{tc}}$  & \ttt{rho\_12\_tc}   \\
3000223 & $\omega^0_{\mrm{tc}}$ & \ttt{omega\_tc0} & 
3300113 & $\rho^0_{21,\mrm{tc}}$  & \ttt{rho\_21\_tc}   \\
3000331 & $\eta_{\mrm{tc}}$          & \ttt{eta\_tc0} &
3400113 & $\rho^0_{22,\mrm{tc}}$  & \ttt{rho\_22\_tc} \\
\hline
\end{tabular}
\end{center}
\end{table}
 
\item Excited fermion codes, Table \ref{t:codeeleven}.  \\
A first generation of excited fermions are included.

\begin{table}[ptb]
\captive{Excited fermion codes.
\protect\label{t:codeeleven} }  \\
\vspace{1ex}
\begin{center}
\begin{tabular}{|c|c|c||c|c|c|@{\protect\rule{0mm}{\tablinsep}}}
\hline
{\KF} & Name & Printed & {\KF} & Name & Printed \\
\hline
 4000001 & $\u^*$ & \ttt{d*} &
 4000011 & $\e^*$ & \ttt{e*-} \\
 4000002 & $\d^*$ & \ttt{u*} &
 4000012 & $\nu^*_{\e}$ & \ttt{nu*\_e0} \\
\hline
\end{tabular}
\end{center}
\end{table}
 
\item Exotic particle codes, Table \ref{t:codetwelve}.  \\
This section includes the excited graviton, as the first 
(but probably not last) manifestation of the possibility of
large extra dimensions. Although it is not yet in the PDG 
standard, we assume that such states will go in a new series
of numbers.\\
Included is also a set of particles associated with an extra
\tbf{SU(2)} gauge group for righthanded states, as required
in order to obtain a left--right symmetric theory at high 
energies. This includes righthanded (Majorana) neutrinos,
righthanded $\Z_R^0$ and $\W_R^{\pm}$ gauge bosons, and both 
left- and righthanded doubly charged Higgs bosons. 
Such a scenario would also contain other Higgs states, but
these do not bring anything new relative to the ones
already introduced, from an observational point of view.
Here the first two digits are 99 to denote the non-standard 
character.
 
\begin{table}[ptb]
\captive{Exotic particle codes.
\protect\label{t:codetwelve} }  \\
\vspace{1ex}
\begin{center}
\begin{tabular}{|c|c|c||c|c|c|@{\protect\rule{0mm}{\tablinsep}}}
\hline
{\KF} & Name & Printed & {\KF} & Name & Printed \\
\hline
5000039 & $\G^*$ & \ttt{Graviton*} & & & \\
\hline
9900012 & $\nu_{R\mrm{e}}$ & \ttt{nu\_Re}   &
9900023 & $\Z_R^0$         & \ttt{Z\_R0}     \\
9900014 & $\nu_{R\mu}$     & \ttt{nu\_Rmu}  &
9900024 & $\W_R^+$         & \ttt{W\_R+}     \\
9900016 & $\nu_{R\tau}$    & \ttt{nu\_Rtau} &
9900041 & $H_L^{++}$       & \ttt{H\_L++}    \\
        &                  &                &
9900042 & $H_R^{++}$       & \ttt{H\_R++}    \\
\hline
\end{tabular}
\end{center}
\end{table}
  
\end{Enumerate}

A hint on large particle numbers: if you want to avoid mistyping
the number of zeros, it may pay off to define a statement like
\vspace{-0.5\baselineskip}
\begin{verbatim}
      PARAMETER (KSUSY1=1000000,KSUSY2=2000000,KTECHN=3000000,
     &KEXCIT=4000000,KDIMEN=5000000)
\end{verbatim}
\vspace{-0.5\baselineskip}
at the beginning of your program and then refer to particles as
\ttt{KSUSY1+1} = $\sd_L$ and so on. This then also agrees with the 
internal notation (where feasible).
 
\subsection{The Event Record}
\label{ss:evrec}
 
Each new event generated is in its entirety stored in the common block
\ttt{PYJETS}, which thus forms the event record. Here each parton or
particle that appears at some stage of the fragmentation or decay
chain will occupy one line in the matrices. The different components
of this line will tell which parton/particle it is, from where it
originates, its present status (fragmented/decayed or not), its
momentum, energy and mass, and the space--time position of its
production vertex. Note that \ttt{K(I,3)}--\ttt{K(I,5)} and the
\ttt{P} and \ttt{V} vectors may take special meaning for some
specific applications (e.g.\ sphericity or cluster
analysis), as described in those connections.

The event history information stored in \ttt{K(I,3)}--\ttt{K(I,5)}
should not be taken too literally. In the particle decay chains, the
meaning of a mother is well-defined, but the fragmentation description
is more complicated. The primary hadrons produced in string
fragmentation come from the string as a whole, rather than from an
individual parton. Even when the string is not included in the history
(see \ttt{MSTU(16)}), the pointer from hadron to parton is deceptive.
For instance, in a $\q\g\qbar$ event, those hadrons are pointing towards
the $\q$ ($\qbar$) parton that were produced by fragmentation from that
end of the string, according to the random procedure used in the
fragmentation routine. No particles point to the $\g$. This assignment
seldom agrees with the visual impression, and is not intended to. 
 
The common block \ttt{PYJETS} has expanded with time, and can now house
4000 entries. This figure may seem ridiculously large, but actually the
previous limit of 2000 was often reached in studies of high-$\pT$
processes at the LHC (and SSC). This is because the event record
contains not only the final particles, but also all intermediate partons
and hadrons, which subsequently showered, fragmented or decayed. Included
are also a wealth of photons coming from $\pi^0$ decays; the simplest
way of reducing the size of the event record is actually to switch off
$\pi^0$ decays by \ttt{MDCY(PYCOMP(111),1)=0}. Also note that some
routines, such as \ttt{PYCLUS} and \ttt{PYCELL}, use memory after the
event record proper as a working area. Still, to change the size of
the common block, upwards or downwards, is easy: just do a global
substitute in the common block and change the \ttt{MSTU(4)} value to the
new number. If more than 10000 lines are to be used, the packing of
colour information should also be changed, see \ttt{MSTU(5)}.
 
\drawbox{COMMON/PYJETS/N,NPAD,K(4000,5),P(4000,5),V(4000,5)}\label{p:PYJETS}
 
\begin{entry}
 
\itemc{Purpose:} to contain the event record, i.e.\ the complete list
of all partons and particles (initial, intermediate and final) in the 
current event. (By parton we here mean the subclass of particles that
carry colour, for which extra colour flow information is then required.
Normally this means quarks and gluons, which can fragment to hadrons,
but also squarks and other exotic particles fall in this category.)
 
\iteme{N :}\label{p:N}  number of lines in the \ttt{K}, \ttt{P} and 
\ttt{V} matrices occupied by the current event. \ttt{N} is continuously
updated as the definition of the original configuration and the
treatment of fragmentation and decay proceed. In the following,
the individual parton/particle number, running between 1 and
\ttt{N}, is called \ttt{I}.
 
\iteme{NPAD :} dummy to ensure an even number of integers before the
double precision reals, as required by some compilers.
 
\iteme{K(I,1) :}\label{p:K} status code KS, which gives the current 
status of the parton/particle stored in the line. The ground rule is 
that codes 1--10 correspond to currently existing partons/particles, 
while larger codes contain partons/particles which no longer exist, 
or other kinds of event information.
\begin{subentry}
\iteme{= 0 :} empty line.
\iteme{= 1 :} an undecayed particle or an unfragmented parton, the latter
being either a single parton or the last one of a parton system.
\iteme{= 2 :} an unfragmented parton, which is followed by more partons 
in the same colour-singlet parton system.
\iteme{= 3 :} an unfragmented parton with special colour flow information
stored in \ttt{K(I,4)} and \ttt{K(I,5)}, such that adjacent partons
along the string need not follow each other in the event record.
\iteme{= 4 :} a particle which could have decayed, but did not
within the allowed volume around the original vertex.
\iteme{= 5 :} a particle which is to be forced to decay in the next
\ttt{PYEXEC} call, in the vertex position given (this code is only
set by user intervention).
\iteme{= 11 :} a decayed particle or a fragmented parton, the latter 
being either a single parton or the last one of a parton system, cf. \ttt{=1}.
\iteme{= 12 :} a fragmented parton, which is followed by more partons in the
same colour-singlet parton system, cf. \ttt{=2}. Further, a $\B$ meson
which decayed as a $\br{\B}$ one, or vice versa, because of
$\B$--$\br{\B}$ mixing, is marked with this code rather than
\ttt{=11}.
\iteme{= 13 :} a parton which has been removed when special colour flow
information has been used to rearrange a parton system, cf. \ttt{=3}.
\iteme{= 14 :} a parton which has branched into further partons, with
special colour-flow information provided, cf. \ttt{=3}.
\iteme{= 15 :} a particle which has been forced to decay (by user
intervention), cf. \ttt{=5}.
\iteme{= 21 :} documentation lines used to give a compressed story of
the event at the beginning of the event record.
\iteme{= 31 :} lines with information on sphericity, thrust or cluster
search.
\iteme{= 32 :} tabular output, as generated by \ttt{PYTABU}.
\iteme{= 41 :} a junction, with partons arranged in colour, except that
two quark lines may precede or follow a junction. For instance, a 
configuration like $\q_1 \g_1 \, \q_2 \g_2 \,$(junction)$\, \g_3 \q_3$
corresponds to having three strings $\q_1 \g_1$, $\q_2 \g_2$ and
$\q_3 \g_3$ meeting in the junction. The occurence of non-matching colours 
easily reveal the $\q_2$ as not being a continuation of the $\q_1 \g_1$ 
string. Here each $\g$ above is shorthand for an arbitrary number of gluons, 
including none. The most general topology allows two junctions in a system,
i.e. $\q_1 \g_1 \, \q_2 \g_2 \,$(junction)$\, \g_0\, $(junction)$\, \g_3 
\qbar_3 \, \g_4 \qbar_4$. The final $\q/\qbar$ would have status code 1, 
the other partons 2. Thus code \ttt{=41} occurs where \ttt{=2} would 
normally have been used, had the junction been an ordinary parton.
\iteme{= 42 :} a junction, with special colour flow information
stored in \ttt{K(I,4)} and \ttt{K(I,5)}, such that adjacent partons
along the string need not follow each other in the event record.
Thus this code matches the \ttt{=3} of ordinary partons.
\iteme{= 51 :} a junction of strings which have been fragmented,
cf. \ttt{=41}.  Thus this code matches the \ttt{=12} of ordinary partons.      
\iteme{= 52 :} a junction of strings which have been rearranged in
colour, cf. \ttt{=42}.  Thus this code matches the \ttt{=13} of ordinary
partons.  
  
\iteme{< 0 :} these codes are never used by the program, and are
therefore usually not affected by operations on the record, such as
\ttt{PYROBO}, \ttt{PYLIST} and event-analysis routines (the exception
is some \ttt{PYEDIT} calls, where lines are moved but not deleted).
Such codes may therefore be useful in some connections.
\end{subentry}
 
\iteme{K(I,2) :} particle {\KF} code, as described in section
\ref{ss:codes}.
 
\iteme{K(I,3) :} line number of parent particle, where known,
otherwise 0. Note that the assignment of a particle to a given parton in a
parton system is unphysical, and what is given there is only related to
the way the fragmentation was generated.
 
\iteme{K(I,4) :} normally the line number of the first daughter;
it is 0 for an undecayed particle or unfragmented parton.
 
For \ttt{K(I,1) = 3, 13} or \ttt{14}, instead, it  contains special
colour-flow information (for internal use only) of the form \\
\ttt{K(I,4)} = 200000000*MCFR + 100000000*MCTO + 10000*ICFR + ICTO, \\
where ICFR and ICTO give the line numbers of the partons from which
the colour comes and to where it goes, respectively; MCFR and
MCTO originally are 0 and are set to 1 when the corresponding
colour connection has been traced in the \ttt{PYPREP} rearrangement
procedure. (The packing may be changed with \ttt{MSTU(5)}.)
The `from' colour position may indicate a parton which branched
to produce the current parton, or a parton created together with
the current parton but with matched anticolour, while the `to'
normally indicates a parton that the current parton branches
into. Thus, for setting up an initial colour configuration, it
is normally only the `from' part that is used, while the `to' part
is added by the program in a subsequent call to parton-shower
evolution (for final-state radiation; it is the other way around
for initial-state radiation).

For \ttt{K(I,1) = 42} or \ttt{52}, see below. 

{\bf Note:} normally most users never have to worry about the exact
rules for colour-flow storage, since this is used mainly for
internal purposes. However, when it is necessary to define this
flow, it is recommended to use the \ttt{PYJOIN} routine, since it is
likely that this would reduce the chances of making a mistake.
 
\iteme{K(I,5) :} normally the line number of the last daughter;
it is 0 for an undecayed particle or unfragmented parton.
 
For \ttt{K(I,1) = 3, 13} or \ttt{14}, instead, it contains special
colour-flow information (for internal use only) of the form \\
\ttt{K(I,5)} = 200000000*MCFR + 100000000*MCTO + 10000*ICFR + ICTO, \\
where ICFR and ICTO give the line numbers of the partons from which
the anticolour comes and to where it goes, respectively; MCFR
and MCTO originally are 0 and are set to 1 when the corresponding
colour connection has been traced in the \ttt{PYPREP} rearrangement
procedure. For further discussion, see \ttt{K(I,4)}.

For \ttt{K(I,1) = 42} or \ttt{52}, see below. 
 
\iteme{K(I,4), K(I,5) :} For junctions with \ttt{K(I,1) = 42} or \ttt{52}
the colour flow information scheme presented above has to be modified,
since now three colour or anticolour lines meet. Thus the form is\\
\ttt{K(I,4)} = 100000000*MC1 + 10000*\ttt{ITP} + IC1, \\
\ttt{K(I,5)} = 200000000*MC2 + 100000000*MC3 + 10000*IC2 + IC3. \\
The colour flow possibilities are 
\begin{entry}
\iteme{ITP = 1 :} junction of three colours in the final state, with 
positions as stored in IC1, IC2 and IC3. A typical example would be 
neutralino decay to three quarks. Note that the positions need not be 
filled by the line numbers of the final quark themselves, but more likely
by the immediate neutralino decay products that thereafter initiate showers 
and branch further. 
\iteme{ITP = 2 :} junction of three anticolours in the final state,
with positions as stored in IC1, IC2 and IC3. 
\iteme{ITP = 3 :} junction of one incoming anticolour to two outgoing
colours, with the anticolour position stored in IC1 and the two colour 
ones in IC2 and IC3. A typical example would be an antisquark decaying 
to two quarks. 
\iteme{ITP = 4 :} junction of one incoming colour to two outgoing
anticolours, with the colour position stored in IC1 and the two anticolour 
ones in IC2 and IC3. 
\iteme{ITP = 5 :} junction of a colour octet into three colours. The
incoming colour is supposed to pass through unchanged, and so is bookkept 
as usual for the particle itself. IC1 is the position of the incoming 
anticolour, while IC2 and IC3 are the positions of the new colours 
associated with the vanishing of this anticolour. A typical example would 
be gluino decay to three quarks.   
\iteme{ITP = 6 :} junction of a colour octet into three anticolours. The
incoming anticolour is supposed to pass through unchanged, and so is 
bookkept as usual for the particle itself. IC1 is the position of the 
incoming colour, while IC2 and IC3 are the positions of the new anticolours 
associated with the vanishing of this colour. 
\end{entry}
Thus odd (even) \ttt{ITP} code corresponds to a $+1$ ($-1$) change in
baryon number across the junction.\\
The MC1, MC2 and MC3 mark which colour connections have been traced in a  
\ttt{PYPREP} rearrangement procedure, as above.

\iteme{P(I,1) :}\label{p:P} $p_x$, momentum in the $x$ direction, 
in GeV/$c$.
 
\iteme{P(I,2) :} $p_y$, momentum in the $y$ direction, in GeV/$c$.
 
\iteme{P(I,3) :} $p_z$, momentum in the $z$ direction, in GeV/$c$.
 
\iteme{P(I,4) :} $E$, energy, in GeV.
 
\iteme{P(I,5) :} $m$, mass, in GeV/$c^2$. In parton showers, with
space-like virtualities, i.e.\ where $Q^2 = - m^2 > 0$,
one puts \ttt{P(I,5)}$ = -Q$.
 
\iteme{V(I,1) :}\label{p:V} $x$ position of production vertex, in mm.
 
\iteme{V(I,2) :} $y$ position of production vertex, in mm.
 
\iteme{V(I,3) :} $z$ position of production vertex, in mm.
 
\iteme{V(I,4) :} time of production, in mm/$c$
($\approx 3.33 \times 10^{-12}$ s).
 
\iteme{V(I,5) :} proper lifetime of particle, in mm/$c$
($\approx 3.33 \times 10^{-12}$ s). If the particle is not expected to
decay, \ttt{V(I,5)=0}. A line with \ttt{K(I,1)=4}, i.e.\ a
particle that could have decayed, but did not within the
allowed region, has the proper non-zero \ttt{V(I,5)}.
 
In the absence of electric or magnetic fields, or other
disturbances, the decay vertex \ttt{VP} of an unstable particle
may be calculated as \\
\ttt{VP(j) = V(I,j) + V(I,5)*P(I,j)/P(I,5)},
\ttt{j} = 1--4.
 
\end{entry}
 
\subsection{How The Event Record Works}
 
The event record is the main repository for information about an
event. In the generation chain, it is used as a `scoreboard' for
what has already been done and what remains to do.
This information can be studied by you, to access information
not only about the final state, but also about what came before.
 
\subsubsection{A simple example}
 
The first example of section \ref{ss:JETstarted} may help to clarify 
what is going on. When \ttt{PY2ENT} is called to generate a $\q\qbar$
pair, the quarks are stored in lines 1 and 2 of the event record,
respectively. Colour information is set to show that they belong
together as a colour singlet. The counter \ttt{N} is also updated
to the value of 2. At no stage is a previously generated event
removed. Lines 1 and 2 are overwritten, but lines 3 
onwards still contain whatever may have been there before. This does
not matter, since \ttt{N} indicates where the `real' record ends.
 
As \ttt{PYEXEC} is called, explicitly by you or indirectly
by \ttt{PY2ENT}, the first entry is considered and found to be
the first parton of a system. Therefore the second entry is also found, 
and these two together form a colour singlet parton system, which may 
be allowed to fragment. The `string' that fragments is put in line 3
and the fragmentation products in lines 4 through 10 (in this
particular case). At the same time, the $\q$ and $\qbar$ in the
first two lines are marked as having fragmented, and the same for
the string. At this stage, \ttt{N} is 10. Internally in \ttt{PYEXEC}
there is another counter with the value 2, which indicates how far 
down in the record the event has been studied.
 
This second counter is gradually increased by one. If the entry in
the corresponding line can fragment or decay, then fragmentation or
decay is performed.
The fragmentation/decay products are added at the end of the event
record, and \ttt{N} is updated accordingly. The entry is then also
marked as having been treated. For instance, when line 3 is
considered, the `string' entry of this line is seen to have been
fragmented,
and no action is taken. Line 4, a $\rho^+$, is allowed to decay to
$\pi^+ \pi^0$; the decay products are stored in lines 11 and 12,
and line 4 is marked as having decayed. Next, entry 5 is allowed to
decay. The entry in line 6, $\pi^+$, is a stable particle (by
default) and is therefore passed by without any action being taken.
 
In the beginning of the process, entries are usually unstable, and
\ttt{N} grows faster than the second counter of treated entries.
Later on, an increasing fraction of the entries are stable end
products, and the r\^oles are now reversed, with the second counter
growing faster. When the two coincide, the end of the record has 
been reached, and the process can be stopped. All unstable objects 
have now been allowed to fragment or decay. They are still present 
in the record, so as to simplify the tracing of the history.
 
Notice that \ttt{PYEXEC} could well be called a second time.
The second counter would then start all over from the beginning, but
slide through until the end without causing any action, since
all objects that can be treated already have been.
Unless some of the relevant switches were changed meanwhile, that
is. For instance, if $\pi^0$ decays were switched off the first time
around but on the second, all the $\pi^0$'s found in the record
would be allowed to decay in the second call. A particle once
decayed is not `undecayed', however, so if the $\pi^0$ is put back
stable and \ttt{PYEXEC} is called a third time, nothing will happen.
 
\subsubsection{Complete PYTHIA events}
\label{sss:PYrecord}
 
In a full-blown event generated with {\Py}, the usage of \ttt{PYJETS}
is more complicated, although the general principles survive.
\ttt{PYJETS} is used extensively by many of the generation routines; 
indeed it provides the bridge between many of them. The {\Py} event 
listing begins (optionally)
with a few lines of event summary, specific to the hard process
simulated and thus not described in the overview above. These
specific parts are covered in the following.
 
In most instances, only the particles actually produced
are of interest. For \ttt{MSTP(125)=0}, the event record starts
off with the parton configuration existing after hard interaction,
initial- and final-state radiation, multiple interactions and beam
remnants have been considered. The partons are arranged in colour
singlet clusters, ordered as required for string fragmentation.
Also photons and leptons produced as part of the hard interaction
(e.g.\ from $\q\qbar \to \g \gamma$ or $\u\ubar \to \Z^0 \to \ee$)
appear in this part of the event record. These original entries
appear with pointer \ttt{K(I,3)=0}, whereas the products of the
subsequent fragmentation and decay have \ttt{K(I,3)} numbers 
pointing back to the line of the parent.
 
The standard documentation, obtained with \ttt{MSTP(125)=1},
includes a few lines at the beginning of the event record, which
contain a brief summary of the process that has taken place. The
number of lines used depends on the nature of the hard process
and is stored in \ttt{MSTI(4)} for the current event. These lines
all have \ttt{K(I,1)=21}. For all processes, lines 1 and 2 give
the two incoming particles. When listed with \ttt{PYLIST}, these two
lines will be separated from subsequent ones by a sequence of
`\ttt{======}' signs, to improve readability. For diffractive and
elastic events, the two outgoing states in lines 3 and 4 complete the
list. Otherwise, lines 3 and 4 contain the two partons that initiate
the two initial-state parton showers, and 5 and 6 the end products of
these showers, i.e.\ the partons that enter the hard interaction. With
initial-state radiation switched off, lines 3 and 5 and lines 4 and 6
are identical. For a simple $2 \to 2$ hard scattering, lines 7 and 8 give
the two outgoing partons/particles from the hard  interaction, before
any final-state radiation. For $2 \to 2$ processes proceeding via an
intermediate resonance such as $\gammaZ$, $\W^{\pm}$ or $\hrm^0$, the
resonance is found in line 7 and the two outgoing partons/particles in
8 and 9. In some cases one of these may be a resonance in its own 
right, or both of them, so that further pairs of lines are added for
subsequent decays. If the decay of a given resonance has been
switched off, then no decay products are listed either in this
initial summary or in the subsequent ordinary listing. Whenever partons
are listed, they are assumed to be on the mass shell for simplicity.
The fact that effective masses may be generated by initial-
and final-state radiation is taken into account in the actual parton
configuration that is allowed to fragment, however. The listing of the 
event documentation closes with another line made up of `\ttt{======}' 
signs.
 
A few examples may help clarify the picture. For a single diffractive
event $\p \pbar \to \p_{\mrm{diffr}} \pbar$, the event record will start
with \\
\verb& I K(I,1)   K(I,2) K(I,3) &  comment  \\
\verb& 1     21     2212      0 &  incoming $\p$ \\
\verb& 2     21    -2212      0 &  incoming $\pbar$ \\
\verb&========================= &  not part of record; appears in 
listings \\
\verb& 3     21  9902210      1 &  outgoing $\p_{\mrm{diffr}}$ \\
\verb& 4     21    -2212      2 &  outgoing $\pbar$ \\
\verb&========================= &  again not part of record 
 
The typical QCD $2 \to 2$ process would be \\
\verb& I K(I,1)   K(I,2) K(I,3) &  comment \\
\verb& 1     21     2212      0 &  incoming $\p$ \\
\verb& 2     21    -2212      0 &  incoming $\pbar$ \\
\verb&========================= & \\
\verb& 3     21        2      1 &  $\u$ picked from incoming $\p$ \\
\verb& 4     21       -1      2 &  $\dbar$ picked from incoming 
$\pbar$ \\
\verb& 5     21       21      3 &  $\u$ evolved to $\g$ at hard 
scattering \\
\verb& 6     21       -1      4 &  still $\dbar$ at hard scattering \\
\verb& 7     21       21      0 &  outgoing $\g$ from hard 
scattering \\
\verb& 8     21       -1      0 &  outgoing $\dbar$ from hard 
scattering \\
\verb&========================= & 

Note that, where well defined, the \ttt{K(I,3)} code does contain
information as to which side the different partons come from, e.g.
above the gluon in line 5 points back to the $\u$ in line 3,
which points back to the proton in line 1. In the example above, it
would have been possible to associate the scattered g in line 7
with the incoming one in line 5, but this is not possible in the
general case, consider e.g.\ $\g \g \to \g \g$.

A special case is
provided by $\W^+ \W^-$ or $\Z^0 \Z^0$ fusion to an $\hrm^0$. Then the
virtual $\W$'s or $\Z$'s are shown in lines 7 and 8, the $\hrm^0$ in
line 9, and the two recoiling quarks (that emitted the bosons) in 10
and 11, followed by the Higgs decay products. Since the $\W$'s and
$\Z$'s are space-like, what is actually listed as the mass for them
is $-\sqrt{-m^2}$. Thus $\W^+\W^-$ fusion to an $\hrm^0$ in process 8 
(not process 124, which is lengthier) might look like \\
\verb& I K(I,1)   K(I,2) K(I,3) &  comment \\
\verb& 1     21     2212      0 &  first incoming $\p$ \\
\verb& 2     21     2212      0 &  second incoming $\p$ \\
\verb&========================= &  \\
\verb& 3     21        2      1 &  $\u$ picked from first $\p$ \\
\verb& 4     21       21      2 &  $\g$ picked from second $\p$ \\
\verb& 5     21        2      3 &  still $\u$ after initial-state 
radiation \\
\verb& 6     21       -4      4 &  $\g$ evolved to $\cbar$ \\
\verb& 7     21       24      5 &  space-like $\W^+$ emitted by $\u$ 
quark  \\
\verb& 8     21      -24      6 &  space-like $\W^-$ emitted by 
$\cbar$ quark \\
\verb& 9     21       25      0 &  Higgs produced by $\W^+ \W^-$ 
fusion \\
\verb&10     21        1      5 &  $\u$ turned into $\d$ by emission 
of $\W^+$ \\
\verb&11     21       -3      6 &  $\cbar$ turned into $\sbar$ by 
emission of $\W^-$ \\
\verb&12     21       23      9 &  first $\Z^0$ coming from decay 
of $\hrm^0$ \\
\verb&13     21       23      9 &  second $\Z^0$ coming from decay 
of $\hrm^0$ \\
\verb&14     21       12     12 &  $\nu_{\e}$ from first $\Z^0$ 
decay \\
\verb&15     21      -12     12 &  $\br{\nu}_{\e}$ from first 
$\Z^0$ decay  \\
\verb&16     21        5     13 &  $\b$ quark from second $\Z^0$ 
decay  \\
\verb&17     21       -5     13 &  $\bbar$ antiquark from second 
$\Z^0$ decay  \\
\verb&========================= &

Another special case is when a spectrum of virtual photons are generated
inside a lepton beam, i.e.\ when \ttt{PYINIT} is called with one or
two {\galep} arguments. (Where \textit{lepton} could be either of
\ttt{e-}, \ttt{e+}, \ttt{mu-}, \ttt{mu+}, \ttt{tau-} or \ttt{tau+}.) 
Then the documentation section is expanded to reflect the new layer of 
administration. Positions 1 and 2 contain the original beam particles, 
e.g.\ $\e$ and $\p$ (or $\e^+$ and $\e^-$). In position 3 (and 4 for 
$\e^+\e^-$) is (are) the scattered outgoing lepton(s). Thereafter comes 
the normal documentation, but starting from the photon rather
than a lepton. For $\e\p$, this means 4 and 5 are the $\gamma^*$ 
and $\p$, 6 and 7 the shower initiators, 8 and 9 the incoming partons 
to the hard interaction, and 10 and 11 the outgoing ones. Thus the 
documentation is 3 lines longer (4 for $\e^+\e^-$) than normally.

The documentation lines are often helpful to understand in broad
outline what happened in a given event. However, they only provide
the main points of the process, with many intermediate layers of
parton showers omitted. The documentation can therefore appear internally 
inconsistent, if the user does not remember what could have happened 
in between. For instance, the listing above would show the Higgs with the 
momentum it has before radiation off the two recoiling $\u$ and $\cbar$ 
quarks is considered. When these showers are included, the Higgs momentum 
may shift by the changed recoil. However, this update is not visible in the
initial summary, which thus still shows the Higgs before the showering.
When the Higgs decays, on the other hand, it is the real Higgs momentum
further down in the event record that is used, and that thus sets the 
momenta of the decay products that are also copied up to the summary.
Such effects will persist in further decays; e.g. the $\b$ and $\bbar$
shown at the end of the example above are before showers, and may deviate
from the final parton momenta quite significantly. Similar shifts will 
also occur e.g. in a $\t \to \b \W^+ \to \b \q \qbar'$ decays, 
when the gluon radiation off the $\b$ gives a recoil to the $\W$ that is
not visible in the $\W$ itself but well in its decay products. In summary,
the documentation section should never be mistaken for the physically
observable state in the main section of the event record, and never be 
used as part of any realistic event analysis.

(An alternative approach would be in the spirit of the Les Houches 
`parton-level' event record, section \ref{ss:PYnewproc}, where the whole 
chain of decays normally is carried out before starting the parton showers. 
With this approach, one could have an internally consistent summary, but 
then in diverging disagreement with the "real" particles after each layer 
of shower evolution.)

After these lines with the initial information, the event record looks
the same as for \ttt{MSTP(125)=0}, i.e.\ first comes the parton
configuration to be fragmented and, after another separator line
`\ttt{======}' in the output (but not the event record), the products
of subsequent fragmentation and decay chains. This ordinary listing 
begins in position \ttt{MSTI(4)+1}. The \ttt{K(I,3)}
pointers for the partons, as well as leptons and photons produced
in the hard interaction, are now pointing towards the documentation
lines above, however. In particular, beam remnants point to 1 or 2,
depending on which side they belong to, and partons emitted in the
initial-state parton showers point to 3 or 4. In the second example
above, the partons produced by final-state radiation will be pointing
back to 7 and 8; as usual, it should be remembered that a specific
assignment to 7 or 8 need not be unique. For the third example, 
final-state radiation partons will come both from partons 10 and 11 
and from partons 16 and 17, and additionally there will be a
neutrino--antineutrino pair pointing to 14 and 15. 

A hadronic event may contain several (semi)hard interactions, in the
multiple interactions scenario. The hardest interaction of an event 
is shown in the initial section of the event record, while further
ones are not. Therefore these extra partons, documented in the main
section of the event, do not have a documentation copy to point back to, 
and so are assigned \ttt{K(I,3)=0}.
 
There exists a third documentation option, \ttt{MSTP(125)=2}. Here
the history of initial- and final-state parton branchings may be traced,
including all details on colour flow. This information has not been
optimized for user-friendliness, and cannot be recommended for
general usage. With this option, the initial documentation lines
are the same. They are followed by blank lines, \ttt{K(I,1)=0}, up to
line 100 (can be changed in \ttt{MSTP(126)}). From line 101 onwards
each parton with \ttt{K(I,1)=} 3, 13 or 14 appears with special
colour-flow information in the \ttt{K(I,4)} and \ttt{K(I,5)}
positions. For an ordinary $2 \to 2$ scattering, the two incoming
partons at the hard scattering are stored in lines 101 and 102, and the
two outgoing in 103 and 104. The colour flow between these partons has
to be chosen according to the proper relative probabilities in
cases when many alternatives are possible, see section 
\ref{sss:QCDjetclass}.
If there is initial-state radiation, the two partons in lines 101 and 
102 are copied down to lines 105 and 106, from which the initial-state 
showers are reconstructed backwards step by step. The branching
history may be read by noting that, for a branching $a \to b c$,
the \ttt{K(I,3)} codes of $b$ and $c$ point towards the line number
of $a$. Since the showers are reconstructed backwards, this actually
means that parton $b$ would appear in the listing before parton
$a$ and $c$, and hence have a pointer to a position below
itself in the list. Associated time-like partons $c$ may initiate
time-like showers, as may the partons of the hard scattering. Again
a showering parton or pair of partons will be copied down towards
the end of the list and allowed to undergo successive branchings
$c \to d e$, with $d$ and $e$ pointing towards $c$. The mass of
time-like partons is properly stored in \ttt{P(I,5)}; for space-like
partons $-\sqrt{-m^2}$ is stored instead. After this
section, containing all the branchings, comes the final parton
configuration, properly arranged in colour, followed by all
subsequent fragmentation and decay products, as usual.
 
\subsection{The HEPEVT Standard}
\label{ss:HEPEVT}
 
A set of common blocks was developed and agreed on within the
framework of the 1989 LEP physics study, see \cite{Sjo89}.
This standard defines an event record structure which should make
the interfacing of different event generators much simpler.
 
It would be a major work to rewrite {\Py} to agree with this
standard event record structure. More importantly, the standard
only covers quantities which can be defined unambiguously, i.e.
which are independent of the particular program used. There are
thus no provisions for the need for colour-flow information in
models based on string fragmentation, etc., so the standard
common blocks would anyway have to be supplemented with additional
event information. For the moment, the adopted approach is therefore
to retain the \ttt{PYJETS} event record, but supply a routine
\ttt{PYHEPC} which can convert to or from the standard event record.
Owing to a somewhat different content in the two records, some
ambiguities do exist in the translation procedure. \ttt{PYHEPC}
has therefore to be used with some judgement.
 
In this section, the standard event structure is first presented,
i.e.\ the most important points in \cite{Sjo89} are recapitulated.
Thereafter the conversion routine is described, with particular
attention to ambiguities and limitations. 
 
The standard event record is stored in two common blocks. The second
of these is specifically intended for spin information. Since {\Py}
never (explicitly) makes use of spin information, this latter
common block is not addressed here. A third common block for colour
flow information has been discussed, but never formalized. Note that 
a \ttt{CALL PYLIST(5)} can be used to obtain a simple listing of 
the more interesting information in the event record.
 
In order to make the components of the standard more distinguishable
in your programs, the three characters \ttt{HEP} (for High Energy
Physics) have been chosen to be a part of all names.

Originally it was not specified whether real variables should be in 
single or double precision. At the time, this meant that single 
precision became the default choice, but since then the trend has been
towards increasing precision. In connection with the 1995 LEP~2
workshop, it was therefore agreed to adopt \ttt{DOUBLE PRECISION} 
real variables as part of the standard, and also to extend the size 
from 2000 to 4000 entries \cite{Kno96}. If, for some reason, one would
want to revert to single precision, this would only require trivial 
changes to the code of the \ttt{PYHEPC} conversion routine described 
below. 
 
\drawboxfour{~PARAMETER (NMXHEP=4000)}
{~COMMON/HEPEVT/NEVHEP,NHEP,ISTHEP(NMXHEP),IDHEP(NMXHEP),}
{\&JMOHEP(2,NMXHEP),JDAHEP(2,NMXHEP),PHEP(5,NMXHEP),VHEP(4,NMXHEP)}
{~DOUBLE PRECISION PHEP, VHEP}
\label{p:HEPEVT}\begin{entry}
 
\itemc{Purpose:} to contain an event record in a
Monte Carlo-independent format.
 
\iteme{NMXHEP:} maximum numbers of entries (particles) that can
be stored in the common block. The default value of 4000 can be changed
via the parameter construction. In the translation, it is
checked that this value is not exceeded.
 
\iteme{NEVHEP:} is normally the event number, but may have special 
meanings, according to the description below:
\begin{subentry}
\iteme{> 0 :} event number, sequentially increased by 1 for each call 
to the main event generation routine, starting with 1 for the
first event generated.
\iteme{= 0 :} for a program which does not keep track of event numbers,
as some of the {\Py} routines.
\iteme{= -1 :} special initialization record; not used by {\Py}.
\iteme{= -2 :} special final record; not used by {\Py}.
\end{subentry}
 
\iteme{NHEP:} the actual number of entries stored in the current event.
These are found in the first \ttt{NHEP} positions of the respective
arrays below. Index \ttt{IHEP}, 1$\leq$\ttt{IHEP}$\leq$\ttt{NHEP},
is used below to denote a given entry.
 
\iteme{ISTHEP(IHEP):} status code for entry \ttt{IHEP}, with the 
following meanings:
\begin{subentry}
\iteme{= 0 :} null entry.
\iteme{= 1 :} an existing entry, which has not decayed or fragmented.
This is the main class of entries, which represents the
`final state' given by the generator.
\iteme{= 2 :} an entry which has decayed or fragmented and is 
therefore not appearing in the final state, but is retained for
event history information.
\iteme{= 3 :} a documentation line, defined separately from the event
history. This could include the two incoming reacting particles, etc.
\iteme{= 4 - 10 :} undefined, but reserved for future standards.
\iteme{= 11 - 200 :} at the disposal of each model builder for
constructs specific to his program, but equivalent to a null line
in the context of any other program.
\iteme{= 201 - :} at the disposal of users, in particular for event
tracking in the detector.
\end{subentry}
 
\iteme{IDHEP(IHEP) :} particle identity, according to the PDG
standard. The four additional codes 91--94 have been introduced
to make the event history more legible, see section \ref{ss:codes}
and the \ttt{MSTU(16)} description of how daughters can point back 
to them.
 
\iteme{JMOHEP(1,IHEP) :} pointer to the position where the mother
is stored. The value is 0 for initial entries.
 
\iteme{JMOHEP(2,IHEP) :} pointer to position of second mother.
Normally only one mother exists, in which case the value 0 is to be
used. In {\Py}, entries with codes 91--94 are the only ones to have
two mothers. The flavour contents of these objects, as well as
details of momentum sharing, have to be found by looking at the
mother partons, i.e.\ the two partons in positions \ttt{JMOHEP(1,IHEP)}
and \ttt{JMOHEP(2,IHEP)} for a cluster or a shower system, and the
range
\ttt{JMOHEP(1,IHEP)}--\ttt{JMOHEP(2,IHEP)} for a string or an
independent fragmentation parton system.
 
\iteme{JDAHEP(1,IHEP) :} pointer to the position of the first daughter.
If an entry has not decayed, this is 0.
 
\iteme{JDAHEP(2,IHEP) :} pointer to the position of the last daughter.
If an entry has not decayed, this is 0. It is assumed that daughters are
stored sequentially, so that the whole range
\ttt{JDAHEP(1,IHEP)}--\ttt{JDAHEP(2,IHEP)} contains daughters. This
variable should be set also when only one daughter is present, as in
$\K^0 \to \K_{\mrm{S}}^0$ decays, so that looping from the first 
daughter to the last one works transparently.
Normally daughters are stored after mothers, but in backwards
evolution of initial-state radiation the opposite may appear,
i.e.\ that mothers are found below the daughters they branch into.
Also, the two daughters then need not appear one after the other,
but may be separated in the event record.
 
\iteme{PHEP(1,IHEP) :} momentum in the $x$ direction, in GeV/$c$.
 
\iteme{PHEP(2,IHEP) :} momentum in the $y$ direction, in GeV/$c$.
 
\iteme{PHEP(3,IHEP) :} momentum in the $z$ direction, in GeV/$c$.
 
\iteme{PHEP(4,IHEP) :} energy, in GeV.
 
\iteme{PHEP(5,IHEP) :} mass, in GeV/$c^2$. For space-like partons,
it is allowed to use a negative mass, according to
\ttt{PHEP(5,IHEP)}$ = -\sqrt{-m^2}$.
 
\iteme{VHEP(1,IHEP) :} production vertex $x$ position, in mm.
 
\iteme{VHEP(2,IHEP) :} production vertex $y$ position, in mm.
 
\iteme{VHEP(3,IHEP) :} production vertex $z$ position, in mm.
 
\iteme{VHEP(4,IHEP) :} production time, in mm/$c$
($\approx 3.33 \times 10^{-12}$ s).
\end{entry}
\boxsep
 
This completes the brief description of the standard. In {\Py}, the
routine \ttt{PYHEPC} is provided as an interface.
 
\drawbox{CALL PYHEPC(MCONV)}\label{p:PYHEPC}
\begin{entry}
\itemc{Purpose:} to convert between the \ttt{PYJETS} event record and
the \ttt{HEPEVT} event record.
\iteme{MCONV :} direction of conversion.
\begin{subentry}
\iteme{= 1 :} translates the current \ttt{PYJETS} record into the
\ttt{HEPEVT} one, while leaving the original \ttt{PYJETS} one
unaffected.
\iteme{= 2 :} translates the current \ttt{HEPEVT} record into the
\ttt{PYJETS} one, while leaving the original \ttt{HEPEVT} one
unaffected.
\end{subentry}
\end{entry}
\boxsep
 
The conversion of momenta is trivial: it is just a matter of exchanging
the order of the indices. The vertex information is but little more
complicated; the extra fifth component present in \ttt{PYJETS} can be
easily
reconstructed from other information for particles which have decayed.
(Some of the advanced features made possible by this component, such as
the possibility to consider decays within expanding spatial volumes in
subsequent \ttt{PYEXEC} calls, cannot be used if the record is
translated back and forth, however.) Also, the particle codes
\ttt{K(I,2)} and \ttt{IDHEP(I)}
are identical, since they are both based on the PDG codes.
 
The remaining, non-trivial areas deal with the status codes and the 
event history. In moving from \ttt{PYJETS} to \ttt{HEPEVT}, 
information on colour flow is lost. On the other hand, the position 
of a second mother, if any, has to be found; this only affects lines 
with \ttt{K(I,2)=} 91--94. Also, for lines with \ttt{K(I,1)=} 13
or 14, the daughter pointers have to be found. By and large,
however, the translation from \ttt{PYJETS} to \ttt{HEPEVT}
should cause little problem, and there should never be any need for
user intervention. (We assume that {\Py} is run with the default
\ttt{MSTU(16)=1} mother pointer assignments, otherwise some 
discrepancies with respect to the proposed standard event history 
description will be present.)
 
In moving from \ttt{HEPEVT} to \ttt{PYJETS}, information on a second
mother is lost. Any codes \ttt{IDHEP(I)} not equal to 1, 2 or 3 are
translated into \ttt{K(I,1)=0}, and so all entries with
\ttt{K(I,1)}$\geq 30$ are effectively lost in a translation back and
forth. All entries with \ttt{IDHEP(I)=2} are translated
into \ttt{K(I,1)=11}, and so entries of type
\ttt{K(I,1) = 12, 13, 14} or \ttt{15} are never found. There is thus
no colour-flow information available for partons which have
fragmented. For partons with \ttt{IDHEP(I)=1},
i.e.\ which have not fragmented, an attempt is made to subdivide the
partonic system into colour singlets, as required for subsequent
string fragmentation. To this end, it is assumed that partons are
stored sequentially along strings. Normally, a string would then start
at a $\q$ ($\qbar$) or $\qbar\qbar$ ($\q\q$) entry, cover a number
of intermediate gluons, and end at a $\qbar$ ($\q$) or $\q\q$
($\qbar\qbar$) entry. Particles could be interspersed in this list with
no adverse effects, i.e.\ a $\u-\g-\gamma-\ubar$
sequence would be interpreted as a $\u-\g-\ubar$ string plus an
additional photon. A closed gluon loop would be assumed to be made up
of a sequential listing of the gluons, with the string continuing from
the last gluon up back to the first one. Contrary to the previous, open
string case, the appearance of any particle but a gluon would therefore
signal the end of the gluon loop. For example, a $\g-\g-\g-\g$ sequence
would be interpreted as one single four-gluon loop, while a
$\g-\g-\gamma-\g-\g$ sequence would be seen as composed of two
2-gluon systems.
 
If these interpretations, which are not unique, are not to your liking,
it is up to you to correct them, e.g.\ by using
\ttt{PYJOIN} to tell exactly which partons should be joined,
in which sequence, to give a string. Calls to \ttt{PYJOIN}
(or the equivalent) are also necessary if \ttt{PYSHOW} is to be used
to have some partons develop a shower.
 
For practical applications, one should note that $\ee$ events,
which have been allowed to shower but not to fragment, do have partons
arranged in the order assumed above, so that a translation to
\ttt{HEPEVT} and back does not destroy the possibility to perform
fragmentation by a simple \ttt{PYEXEC} call. Also the hard interactions
in hadronic events fulfil this condition, while problems may appear in the
multiple interaction scenario, where several closed $\g\g$ loops may
appear directly following one another, and thus would be
interpreted as a single multigluon loop after translation back and
forth.
 
\clearpage
 
\section{The Old Electron--Positron Annihilation Routines}
\label{s:JETSETproc}
 
{}From the {\Je} package, {\Py} inherits routines for the dedicated
simulation of two hard processes in $\e^+\e^-$ annihilation.
The process of main interest is $\ee \to \gammaZ \to \q \qbar$.
The description provided by the \ttt{PYEEVT} routine has been a main 
staple from PETRA days up to the LEP1 era. Nowadays it is superseded 
by process 1 of the main {\Py} event generation machinery, see section 
\ref{sss:WZclass}. This latter process offers a better description of 
flavour selection, resonance shape and initial-state radiation. It can 
also, optionally, be used with the second-order matrix element machinery 
documented in this section. For backwards compatibility, however, the 
old routines have still been retained here. There are also a few 
features found in the routines in this section, and not in the other 
ones, such as polarized incoming beams.

For the process  $\ee \to \gammaZ \to \q \qbar$,
higher-order QCD corrections can be obtained either with parton
showers or with second-order matrix elements. The details of the
parton-shower evolution are given in section \ref{s:showinfi},
while this section contains the matrix-element description, including
a summary of the older algorithm for initial-state photon radiation 
used here. 
 
The other standalone hard process in this section is $\Upsilon$ decay to
$\g \g \g$ or $\gamma \g \g$, which is briefly commented on.
 
The main sources of information for this chapter are
refs. \cite{Sjo83,Sjo86,Sjo89}.
 
\subsection{Annihilation Events in the Continuum}
\label{ss:eematrix}
 
The description of $\ee$ annihilation into hadronic events involves a
number of components: the $s$ dependence of the total cross section
and flavour composition, multiparton matrix elements, angular
orientation of events,  initial-state photon bremsstrahlung
and effects of initial-state electron polarization.
Many of the published formulae
have been derived for the case of massless outgoing quarks. For each
of the components described in the following, we will begin by
discussing the massless case, and then comment on what is done to
accommodate massive quarks.
 
\subsubsection{Electroweak cross sections}
 
In the standard theory, fermions have the following couplings
(illustrated here for the first generation):
\begin{center}
\begin{tabular}{lll@{\protect\rule{0mm}{\tablinsep}}}
$e_{\nu} = 0$, & $v_{\nu} = 1$, & $a_{\nu} = 1$, \\
$e_{\e} = -1$, & $v_{\e} = -1 + 4\ssintw$, & $a_{\e} = -1$, \\
$e_{\u} = 2/3$, & $v_{\u} = 1 - 8\ssintw /3$, & $a_{\nu} = 1$, \\
$e_{\d} = -1/3$, & $v_{\d} = -1 + 4\ssintw /3$, & $a_{\d} = -1$, \\
\end{tabular}
\end{center}
with $e$ the electric charge, and $v$ and $a$ the vector and axial
couplings to the $\Z^0$. The relative energy dependence of the weak
neutral current to the electromagnetic one is given by
\begin{equation}
   \chi(s) = \frac{1}{16\ssintw\scostw} \;
    \frac{s}{s - m_{\Z}^2 + i m_{\Z}\Gamma_{\Z}} ~,
\label{ee:chis}
\end{equation}
where $s = E_{\mrm{cm}}^2$.
In this section the electroweak mixing parameter $\ssintw$ and the
$\Z^0$ mass $m_{\Z}$ and width $\Gamma_{\Z}$ are considered as
constants to be given by you (while the full {\Py} event generation
machinery itself calculates an $s$-dependent width).
 
Although the incoming $\e^+$ and $\e^-$ beams are normally
unpolarized, we have included the possibility of polarized beams,
following the formalism of \cite{Ols80}. Thus the incoming
$\e^+$ and $\e^-$ are characterized by polarizations
$\mbf{P}^{\pm}$ in the rest frame of the particles:
\begin{equation}
\mbf{P}^{\pm} = P_{\mrm{T}}^{\pm} \hat{\mbf{s}}^{\pm} + 
P_{\mrm{L}}^{\pm} \hat{\mbf{p}}^{\pm} ~,
\end{equation}
where $0 \leq P_{\mrm{T}}^{\pm} \leq 1$ and 
$-1 \leq P_{\mrm{L}}^{\pm} \leq 1$, with the constraint
\begin{equation}
(\mbf{P}^{\pm})^2 = (P_{\mrm{T}}^{\pm})^2 + (P_{\mrm{L}}^{\pm})^2 
\leq 1 ~.
\end{equation}
Here $\hat{\mbf{s}}^{\pm}$ are unit vectors perpendicular to the beam
directions $\hat{\mbf{p}}^{\pm}$. To be specific, we choose a
right-handed coordinate frame with 
$\hat{\mbf{p}}^{\pm} = (0,0, \mp 1)$,
and standard transverse polarization directions (out of the machine
plane for storage rings) $\hat{\mbf{s}}^{\pm} = (0, \pm 1,0)$, the
latter corresponding to azimuthal angles $\varphi^{\pm} = \pm \pi /2$.
As free parameters in the program we choose $P_{\mrm{L}}^+$, 
$P_{\mrm{L}}^-$, $P_{\mrm{T}} = \sqrt{P_{\mrm{T}}^+ P_{\mrm{T}}^-}$ 
and $\Delta \varphi = (\varphi^+ + \varphi^-) /2$.
 
In the massless QED case, the probability to produce a flavour $\f$ is
proportional to $e_{\f}^2$, i.e up-type quarks are four times as likely
as down-type ones. In lowest-order massless QFD (Quantum Flavour Dynamics;
part of the Standard Model) the corresponding
relative probabilities are given by \cite{Ols80}
\begin{eqnarray}
   h_{\f}(s) & = & e_{\e}^2 \, (1 - P_{\mrm{L}}^+ P_{\mrm{L}}^-) 
   \, e_{\f}^2 \, + \, 2 e_{\e} \left\{ v_{\e} 
   (1 - P_{\mrm{L}}^+ P_{\mrm{L}}^-) - a_{\e} 
   (P_{\mrm{L}}^- - P_{\mrm{L}}^+)
   \right\} \, \Re\chi(s) \, e_{\f} v_{\f} \, +     \nonumber \\
   & &  + \, \left\{ (v_{\e}^2 + a_{\e}^2) (1 - P_{\mrm{L}}^+ 
   P_{\mrm{L}}^-) - 2 v_{\e} a_{\e} (P_{\mrm{L}}^- - P_{\mrm{L}}^+) 
   \right\} \,
   \left| \chi(s) \right|^2 \, \left\{ v_{\f}^2 + a_{\f}^2 \right\} ~,
\label{ee:hf}
\end{eqnarray}
where $\Re\chi(s)$ denotes the real part of $\chi(s)$.
The $h_{\f}(s)$ expression depends both on the $s$ value and on the 
longitudinal polarization of the $\e^{\pm}$ beams in a non-trivial way.
 
The cross section for the process $\ee \to \gammaZ \to \f \fbar$
may now be written as
\begin{equation}
  \sigma_{\f}(s) = \frac{4 \pi \alphaem^2}{3 s} R_{\f}(s) ~,
\end{equation}
where $R_{\f}$ gives the ratio to the lowest-order QED cross section for
the process $\ee \to \mu^+ \mu^-$,
\begin{equation}
   R_{\f}(s) = N_C \, R_{\mrm{QCD}} \, h_{\f}(s) ~.
\end{equation}
The factor of $N_C = 3$ counts the number of colour states available
for the $\q\qbar$ pair. The $R_{\mrm{QCD}}$
factor takes into account QCD loop corrections to the cross section.
For $n_f$ effective flavours (normally $n_f =5$)
\begin{equation}
 R_{\mrm{QCD}} \approx 1 + \frac{\alphas}{\pi} + (1.986 - 0.115 n_f)
 \left( \frac{\alphas}{\pi} \right)^2 + \cdots  
\label{ee:RQCD}
\end{equation}
in the $\br{\mrm{MS}}$ renormalization scheme \cite{Din79}.
Note that $R_{\mrm{QCD}}$ does not affect the relative quark-flavour 
composition, and so is of peripheral interest here.
(For leptons the $N_C$ and $R_{\mrm{QCD}}$ factors would be absent, 
i.e.\ $N_C \, R_{\mrm{QCD}} = 1$, but leptonic final states are not 
generated by this routine.)
 
Neglecting higher-order QCD and QFD effects, the corrections for
massive quarks are given in terms of the velocity $\beta_{\f}$ of a
fermion with mass $m_{\f}$, $\beta_{\f} = \sqrt{ 1 - 4 m_{\f}^2 /s}$, 
as follows. The vector quark current terms in $h_{\f}$ (proportional to
$e_{\f}^2$, $e_{\f} v_{\f}$, or $v_{\f}^2$) are multiplied by a
threshold factor $\beta_{\f} (3 - \beta_{\f}^2) /2$, while the axial
vector quark current term (proportional to $a_{\f}^2$) is
multiplied by $\beta_{\f}^3$. While inclusion of quark masses in the
QFD formulae decreases the total cross section, first-order QCD
corrections tend in the opposite direction \cite{Jer81}. Na\"{\i}vely,
one would expect one factor of $\beta_{\f}$ to get cancelled. So far,
the available options are either to include threshold factors
in full or not at all.
 
Given that all five quarks are light at
the scale of the $\Z^0$, the issue of quark masses is not really
of interest at LEP. Here, however, purely weak corrections are
important, in particular since they change the $\b$ quark
partial width differently from that of the other ones \cite{Kuh89}.
No such effects are included in the program.
 
\subsubsection{First-order QCD matrix elements}
 
The Born process $\ee \to \q \qbar$ is modified in first-order 
QCD by the probability for the $\q$ or $\qbar$ to
radiate a gluon, i.e.\ by the process $\ee \to \q \qbar \g$.
The matrix element is conveniently given in terms of scaled energy
variables in the c.m.\ frame of the event, 
$x_1 = 2E_{\q}/E_{\mrm{cm}}$,
$x_2 = 2E_{\qbar}/E_{\mrm{cm}}$, 
and $x_3 = 2E_{\g}/E_{\mrm{cm}}$,
i.e.\ $x_1 + x_2 + x_3 = 2$. For massless
quarks the matrix element reads \cite{Ell76}
\begin{equation}
 \frac{1}{\sigma_0} \, \frac{\d \sigma}{\d x_1 \, \d x_2} =
 \frac{\alphas}{2\pi} \, C_F \,
 \frac{x_1^2 + x_2^2}{(1-x_1)(1-x_2)} ~,
  \label{ee:ME3j}
\end{equation}
where $\sigma_0$ is the lowest-order cross section, $C_F = 4/3$ is the
appropriate colour factor, and
the kinematically allowed region is $0 \leq x_i \leq 1, i = 1, 2, 3$.
By kinematics, the $x_k$ variable for parton $k$ is related to the
invariant mass $m_{ij}$ of the other two partons $i$ and $j$ by
$y_{ij} = m_{ij}^2/E_{\mrm{cm}}^2 = 1 - x_k$.
 
The strong coupling constant $\alphas$ is in first order given by
\begin{equation}
 \alphas(Q^2) = \frac{12\pi}{(33-2n_f) \, \ln(Q^2/\Lambda^2)} ~.
 \label{ee:aS3j}
\end{equation}
Conventionally $Q^2 = s = E_{\mrm{cm}}^2$; we will return to this 
issue below.
The number of flavours $n_f$ is 5 for LEP applications, and so the
$\Lambda$ value determined is $\Lambda_5$  (while e.g.\ most
Deeply Inelastic Scattering studies refer to $\Lambda_4$,
the energies for these experiments being below the bottom threshold).
The $\alphas$ values are matched at flavour thresholds, i.e.
as $n_f$ is changed the $\Lambda$ value is also changed. It is
therefore the derivative of $\alphas$ that changes at a
threshold, not $\alphas$ itself.
 
In order to separate 2-jets from 3-jets, it is useful to
introduce jet-resolution parameters. This can be done in several
different ways. Most famous are the $y$ and $(\epsilon, \delta)$
procedures. We will only refer to the $y$ cut, which is the one
used in the program. Here a 3-parton configuration is called
a 2-jet event if
\begin{equation}
\min_{i,j} (y_{ij}) = \min_{i,j} \left( \frac{m_{ij}^2}{E_{\mrm{cm}}^2}
 \right) < y ~.
\end{equation}
 
The cross section in eq.~(\ref{ee:ME3j}) diverges for
$x_1 \rightarrow 1$ or $x_2 \rightarrow 1$ but, when
first-order propagator and vertex corrections are included,
a corresponding singularity with opposite sign appears in the
$\q \qbar$ cross section, so that the total cross section is finite.
In analytical calculations, the average value of any well-behaved
quantity ${\cal Q}$ can therefore be calculated as
\begin{equation}
   \left\langle {\cal Q} \right\rangle =
   \frac{1}{\sigma_{\mrm{tot}}} \lim_{y \rightarrow 0}
   \left( {\cal Q}(\mrm{2parton}) \, \sigma_{\mrm{2parton}}(y) +
   \int_{y_{ij} > y} {\cal Q}(x_1,x_2) \,
   \frac{\d \sigma_{\mrm{3parton}}}{\d x_1 \, \d x_2} \, 
   \d x_1 \, \d x_2 \right) ~,
  \label{ee:Obs}
\end{equation}
where any explicit $y$ dependence disappears in the limit
$y \rightarrow 0$.
 
In a Monte Carlo program, it is not possible to
work with a negative total 2-jet rate, and thus it is necessary to
introduce a fixed non-vanishing $y$ cut in the 3-jet
phase space. Experimentally, there is evidence for the need of a
low $y$ cut, i.e.\ a large 3-jet rate.
For LEP applications, the recommended value is $y = 0.01$,
which is about as far down as one can go and still retain a positive
2-jet rate. With $\alphas = 0.12$, in full second-order QCD
(see below), the $2:3:4$ jet composition is then approximately
$11 \% : 77 \% : 12 \%$.
 
Note, however, that initial-state QED radiation may
occasionally lower the c.m.\ energy significantly, i.e.\ increase
$\alphas$, and thereby bring the 3-jet fraction above unity
if $y$ is kept fixed at 0.01 also in those events. Therefore,
at PETRA/PEP energies, $y$ values slightly above 0.01 are needed.
In addition to the $y$ cut, the program contains a cut on the
invariant mass $m_{ij}$ between any two partons, which is typically 
required to be larger than 2 GeV. This cut corresponds to the
actual merging of two nearby parton jets, i.e.\ where a treatment with
two separate partons rather than one would be superfluous in view
of the smearing arising from the subsequent fragmentation. Since
the cut-off mass scale $\sqrt{y} E_{\mrm{cm}}$ normally is much larger,
this additional cut only enters for events at low energies.
 
For massive quarks, the amount of QCD radiation is slightly reduced
\cite{Iof78}:
\begin{eqnarray}
  \frac{1}{\sigma_0} \, \frac{\d \sigma}{\d x_1 \, \d x_2} & = &
  \frac{\alphas}{2\pi}
  \, C_F \, \left\{ \frac{x_1^2 + x_2^2}{(1-x_1)(1-x_2)} -
  \frac{4 m_{\q}^2}{s} \left( \frac{1}{1-x_1} + \frac{1}{1-x_2}
  \right) \right.   \nonumber \\[1mm]
  & & - \left. \frac{2 m_{\q}^2}{s} \left( \frac{1}{(1-x_1)^2} +
  \frac{1}{(1-x_2)^2} \right) - \frac{4 m_{\q}^4}{s^2}
  \left( \frac{1}{1-x_1} + \frac{1}{1-x_2} \right)^2 \right\} ~.
\label{ee:threejMEmass}
\end{eqnarray}
Properly, the above expression is only valid for the vector part of the
cross section, with a slightly different expression for the axial part, 
but here the one above is used for it all.
In addition, the phase space for emission is reduced by the
requirement
\begin{equation}
\frac{(1-x_1)(1-x_2)(1-x_3)}{x_3^2} \geq \frac{m_{\q}^2}{s} ~.
\end{equation}
For $\b$ quarks at LEP energies, these corrections are fairly small.
 
\subsubsection{Four-jet matrix elements}
 
Two new event types are added in second-order QCD,
$\ee \to \q \qbar \g \g$ and $\ee \to \q \qbar \q' \qbar'$.
The 4-jet cross section has been calculated by several
groups \cite{Ali80a,Gae80,Ell81,Dan82}, which agree on the result.
The formulae are too lengthy to be quoted here. In one of the
calculations \cite{Ali80a}, quark masses were explicitly included,
but here only the massless expressions are included, as taken
from \cite{Ell81}. Here the angular orientation of the event has been
integrated out, so that five independent internal kinematical
variables remain. These may be related to the six $y_{ij}$ and
the four $y_{ijk}$ variables,
$y_{ij} = m_{ij}^2 / s = (p_i + p_j)^2 / s$ and
$y_{ijk} = m_{ijk}^2 / s = (p_i + p_j + p_k)^2 / s$,
in terms of which the matrix elements are given.
 
The original calculations were for the pure $\gamma$-exchange case;
it has been pointed out \cite{Kni89} that an additional
contribution to the $\ee \to \q \qbar \q' \qbar'$ cross section
arises from the axial part of the $\Z^0$. This term is not included
in the program, but fortunately it is finite and small.
 
Whereas the way the string, i.e.\ the fragmenting colour flux tube,
is stretched is uniquely given in $\q \qbar \g$ event, for
$\q \qbar \g \g$ events there are two possibilities:
\mbox{$\q - \g_1 - \g_2 - \qbar$} or
\mbox{$\q - \g_2 - \g_1 - \qbar$}.
A knowledge of quark and gluon colours, obtained by perturbation
theory, will uniquely specify the stretching of the string, as long
as the two gluons do not have the same colour. The probability for
the latter is down in magnitude by a factor $1 / N_C^2 = 1 / 9$.
One may either choose to neglect these terms entirely, or to keep
them for the choice of kinematical setup, but then drop them at the
choice of string drawing \cite{Gus82}. We have adopted the latter
procedure. Comparing the two possibilities, differences are
typically 10--20\% for a given kinematical configuration,
and less for the total 4-jet cross section, so from a practical
point of view this is not a major problem.
 
In higher orders, results depend on the renormalization scheme;
we will use $\br{\mrm{MS}}$ throughout. In addition to this choice,
several possible forms can be chosen for $\alphas$,
all of which are equivalent to that order but differ in higher
orders. We have picked the recommended standard \cite{PDG88}
\begin{equation}
\label{ee:aS4j}
 \alphas(Q^2) =
 \frac{12\pi}{(33-2n_f) \, \ln (Q^2 / \Lambda^2_{\br{\mrm{MS}}})}
 \left\{ 1 - 6 \, \frac{153-19n_f}{(33-2n_f)^2} \,
 \frac{\ln (\ln ( Q^2 / \Lambda^2_{\br{\mrm{MS}}}))}
 {\ln ( Q^2 / \Lambda^2_{\br{\mrm{MS}}})}
 \right\} ~.
\end{equation}
 
\subsubsection{Second-order three-jet matrix elements}
 
As for first order, a full second-order calculation consists both of
real parton emission terms and of vertex and propagator corrections.
These modify the 3-jet and 2-jet cross sections.
Although there was some initial confusion, everybody soon agreed
on the size of the loop corrections \cite{Ell81,Ver81,Fab82}.
In analytic calculations, the procedure of eq.~(\ref{ee:Obs}),
suitably expanded, can therefore be used unambiguously for a
well-behaved variable.
 
For Monte Carlo event simulation, it is again necessary to impose
some finite jet-resolution criterion. This means that four-parton
events which fail the cuts should be reassigned either to the
3-jet or to the 2-jet event class. It is this area that
caused quite a lot of confusion in the past
\cite{Kun81,Got82,Ali82,Zhu83,Gut84,Gut87,Kra88},
and where full agreement does not exist. Most likely, agreement
will never be reached, since there are indeed ambiguous points
in the procedure, related to uncertainties on the theoretical
side, as follows.
 
For the $y$-cut case, any two partons with an invariant mass
$m_{ij}^2 < y E_{\mrm{cm}}^2$ should be recombined into one. If the
four-momenta are simply added, the sum will correspond to a parton
with a positive mass, namely the original $m_{ij}$.
The loop corrections are given in terms of final
massless partons, however. In order to perform the (partial)
cancellation between the four-parton real and the 3-parton
virtual contributions, it is therefore necessary to get rid of
the bothersome mass in the four-parton states. Several
recombinations are used in practice, which go under names such as
`E', `E0', `p' and `p0' \cite{OPA91}. In the `E'-type schemes,
the energy of a recombined parton is given by  $E_{ij} = E_i + E_j$,
and three-momenta may have to be adjusted accordingly. In the
`p'-type schemes, on the other hand, three-momenta are added,
$\mbf{p}_{ij} = \mbf{p}_i + \mbf{p}_j$, and then energies may have
to be adjusted. These procedures result in different 3-jet
topologies, and therefore in different second-order differential
3-jet cross sections.
 
Within each scheme, a number of lesser points remain to be dealt
with, in particular what to do if a recombination of a nearby parton
pair were to give an event with a non-$\q\qbar\g$ flavour structure.
 
This code contains two alternative second-order 3-jet
implementations, GKS and ERT(Zhu). The latter is the recommended one
and default. Other parameterizations have also been made
available that run together with {\Je}~6 (but not adopted to the
current program), see \cite{Sjo89,Mag89}.
 
The GKS option is based on the GKS \cite{Gut84} calculation, where
some of the original mistakes in FKSS \cite{Fab82} have been
corrected. The GKS formulae have the advantage of giving the
second-order corrections in closed analytic form, as not-too-long
functions of $x_1$, $x_2$, and the $y$ cut. However, it is
today recognized, also by the authors, that important
terms are still missing, and that the matrix elements
should therefore not be taken too seriously. The option is thus
kept mainly for backwards compatibility.
 
The ERT(Zhu) generator \cite{Zhu83} is based on the ERT matrix elements
\cite{Ell81}, with a Monte Carlo recombination procedure suggested
by Kunszt \cite{Kun81} and developed by Ali \cite{Ali82}. It has
the merit of giving corrections in a convenient, parameterized form.
For practical applications, the main limitation is that the
corrections are only given for discrete values of the cut-off
parameter $y$, namely $y$ = 0.01, 0.02, 0.03, 0.04, and 0.05.
At these $y$ values, the full second-order 3-jet cross section is 
written in terms of the `ratio function' $R(X,Y;y)$, defined by
\begin{equation}
\frac{1}{\sigma_0} \frac{\d \sigma_3^{\mrm{tot}}}{\d X \, \d Y} =
\frac{\alphas}{\pi} A_0(X,Y)
\left\{ 1 + \frac{\alphas}{\pi} R(X,Y;y) \right\} ~,
\label{ee:Zhupar}
\end{equation}
where $X = x_1 - x_2 = x_{\q} - x_{\qbar}$, $Y = x_3 = x_g$,
$\sigma_0$ is the lowest-order hadronic cross section,
and $A_0(X,Y)$ the standard first-order 3-jet cross section,
cf. eq.~(\ref{ee:ME3j}).
By Monte Carlo integration, the value of $R(X,Y;y)$ is
evaluated in bins of $(X,Y)$, and the result parameterized
by a simple function $F(X,Y;y)$. Further details are found in 
\cite{Sjo89}.
 
\subsubsection{The matrix-element event generator scheme}
 
The program contains parameterizations, separately, of the total
first-order 3-jet rate, the total second-order 3-jet rate,
and the total 4-jet rate, all as functions of $y$ (with
$\alphas$ as a separate prefactor).
These parameterizations have been obtained as follows:
\begin{Itemize}
\item
The first-order 3-jet matrix element is almost analytically
integrable; some small finite pieces were obtained by a truncated
series expansion of the relevant integrand.
\item
The  GKS second-order 3-jet matrix elements were integrated for
40 different $y$-cut values, evenly distributed in $\ln y$ between
a smallest value $y = 0.001$ and the kinematical limit $y = 1/3$.
For each $y$ value, 250\,000 phase-space points were generated,
evenly in $\d \ln (1-x_i) = \d x_i/(1-x_i)$, $i = 1,2$, and the
second-order 3-jet rate in the point evaluated. The properly
normalized sum of weights in each of the 40 $y$ points were
then fitted to a polynomial in $\ln(y^{-1}-2)$. For the ERT(Zhu)
matrix elements the parameterizations in eq.~(\ref{ee:Zhupar})
were used to perform a corresponding Monte Carlo integration for
the five $y$ values available.
\item
The 4-jet rate was integrated numerically, separately for
$\q \qbar \g \g$ and $\q \qbar \q' \qbar'$ events, by generating large
samples of 4-jet phase-space points
within the boundary $y = 0.001$. Each point was classified according
to the actual minimum $y$ between any two partons. The same
events could then be used to update the summed weights for 40
different counters, corresponding to $y$ values evenly distributed
in $\ln y$ between $y = 0.001$ and the kinematical limit $y = 1/6$.
In fact, since
the weight sums for large $y$ values only received contributions
from few phase-space points, extra (smaller) subsamples of events were
generated with larger $y$ cuts. The summed weights,
properly normalized, were then parameterized in terms of
polynomials in $\ln(y^{-1} - 5)$.
Since it turned out to be difficult to obtain one single good fit
over the whole range of $y$ values, different parameterizations are
used above and below $y=0.018$. As originally given, the
$\q \qbar \q' \qbar'$ parameterization only took into account four
$\q'$ flavours, i.e.\ secondary $\b \bbar$ pairs were not generated,
but this has been corrected for LEP.
\end{Itemize}
 
In the generation stage, each event is treated on its own, which means
that the $\alphas$ and $y$ values may be allowed to vary from event to
event. The main steps are the following.
\begin{Enumerate}
\item
The $y$ value to be used in the current event is determined. If
possible, this is the value given by you, but additional
constraints exist from the validity of the parameterizations
($y \geq 0.001$ for GKS, $0.01 \leq y \leq 0.05$ for ERT(Zhu))
and an extra (user-modifiable) requirement of a minimum absolute
invariant mass between jets (which translates into varying $y$ cuts
due to the effects of initial-state QED radiation).
\item
The $\alphas$ value is calculated.
\item
For the $y$ and $\alphas$ values given, the relative
two/three/four-jet composition is determined. This is achieved by
using the parameterized functions of $y$ for 3- and 4-jet rates,
multiplied by the relevant number of factors of $\alphas$.
In ERT(Zhu), where the second-order 3-jet rate is available
only at a few $y$ values, intermediate results are obtained by linear
interpolation in the ratio of second-order to first-order
3-jet rates. The 3-jet and 4-jet rates are normalized to
the analytically known second-order total event rate, i.e.\ divided
by $R_{\mrm{QCD}}$ of eq.~(\ref{ee:RQCD}). Finally, the 2-jet rate is 
obtained by conservation of total probability.
\item
If the combination of $y$ and $\alphas$ values is such that the total
3- plus 4-jet fraction is larger than unity, i.e.\ the remainder
2-jet fraction negative, the $y$-cut value is raised (for that event),
and the process is started over at point 3.
\item
The choice is made between generating a 2-, 3- or 4-jet event,
according to the relative probabilities.
\item
For the generation of 4-jets, it is first necessary to make a choice
between $\q \qbar \g \g$ and $\q \qbar \q' \qbar'$ events, according to
the relative (parameterized) total cross sections. A phase-space point
is then selected, and the differential cross section at this point is
evaluated and compared with a parameterized maximum weight. If the
phase-space point is rejected, a new one is selected, until an
acceptable 4-jet event is found.
\item
For 3-jets, a phase-space point is first chosen according to the
first-order cross section. For this point, the weight
\begin{equation}
W(x_1,x_2;y) = 1 + \frac{\alphas}{\pi} R(x_1,x_2;y)
\label{ee:WTJS}
\end{equation}
is evaluated. Here $R(x_1,x_2;y)$ is analytically given for GKS
\cite{Gut84}, while it is approximated by the parameterization
$F(X,Y;y)$ of eq.~(\ref{ee:Zhupar}) for ERT(Zhu). Again, linear
interpolation of $F(X,Y;y)$ has to be applied for intermediate $y$
values. The weight $W$ is compared with a maximum weight
\begin{equation}
W_{\mmax}(y) = 1 + \frac{\alphas}{\pi} R_{\mmax}(y) ~,
\end{equation}
which has been numerically determined beforehand and suitably
parameterized. If the phase-space point is rejected, a
new point is generated, etc.
\item
Massive matrix elements are not available for second-order 
QCD (but are in the first-order option). However, if a
3- or 4-jet event determined above falls outside
the phase-space region allowed for massive quarks, the event is
rejected and reassigned to be a 2-jet event. (The way the
$y_{ij}$ and $y_{ijk}$ variables of 4-jet events should be
interpreted for massive quarks is not even unique, so some latitude
has been taken here to provide a reasonable continuity from
3-jet events.) This procedure is known not to give the expected full
mass suppression, but is a reasonable first approximation.
\item
Finally, if the event is classified as a 2-jet event, either
because it was initially so assigned, or because it failed the
massive phase-space cuts for 3- and 4-jets, the
generation of 2-jets is trivial.
\end{Enumerate}
 
\subsubsection{Optimized perturbation theory}
 
Theoretically, it turns out that the second-order corrections to the
3-jet rate are large. It is therefore not unreasonable to expect
large third-order corrections to the 4-jet rate. Indeed, the
experimental 4-jet rate is much larger than second order predicts
(when fragmentation effects have been included),
if $\alphas$ is determined based on the 3-jet rate
\cite{Sjo84a,JAD88}.
 
The only consistent way to resolve this issue is to go ahead and
calculate the full next order. This is a tough task, however, so
people have looked at possible shortcuts.
For example, one can try to minimize the higher-order contributions
by a suitable choice of the renormalization scale \cite{Ste81} ---
`optimized perturbation theory'. This
is equivalent to a different choice for the $Q^2$ scale in
$\alphas$, a scale which is not unambiguous anyway. Indeed
the standard value $Q^2 = s = E_{\mrm{cm}}^2$ is larger than the 
natural physical scale of gluon emission in events, given that most 
gluons are fairly soft. One could therefore pick another scale,
$Q^2 = f s$, with $f < 1$. The
${\cal O}(\alphas)$ 3-jet rate would be increased by
such a scale change, and so would the number of 4-jet
events, including those which collapse into 3-jet ones. The loop
corrections depend on the $Q^2$ scale, however,
and compensate the changes above by giving a larger negative
contribution to the 3-jet rate.
 
The possibility of picking an optimized scale $f$ is implemented
as follows \cite{Sjo89}. Assume that the differential 3-jet
rate at scale $Q^2 = s$ is given by the expression
\begin{equation}
R_3 = r_1 \alphas + r_2 \alphas^2 ~,
\end{equation}
where $R_3$, $r_1$ and $r_2$ are functions of the kinematical
variables $x_1$ and $x_2$ and the $y$ cut, as implied by the 
second-order formulae above, see e.g.\ eq.~(\ref{ee:Zhupar}). 
When the coupling is chosen at a different scale, $Q'^2 = f s$, 
the 3-jet rate has to be changed to
\begin{equation}
R_3' = r_1' \alphas' + r_2 \alphas'^2 ~,
\end{equation}
where $r_1' = r_1$,
\begin{equation}
r_2' = r_2 + r_1 \frac{33-2n_f}{12\pi} \ln f ~,
\label{ee:r2optim}
\end{equation}
and $\alphas' = \alphas(fs)$.
Since we only have the Born term for 4-jets, here the effects of a
scale change come only from the change in the coupling constant.
Finally, the 2-jet cross section can still be calculated from the
difference between the total cross section and the 3- and 4-jet
cross sections.
 
If an optimized scale is used in the program, the default value is
$f=0.002$, which is favoured by the studies in ref. \cite{Bet89}. (In
fact, it is also possible to use a correspondingly optimized
$R_{\mrm{QCD}}$ factor, eq.~(\ref{ee:RQCD}), but then the 
corresponding $f$ is chosen independently and much closer to unity.)  
The success of describing the jet rates should not hide the fact that 
one is dabbling in (educated, hopefully) guesswork, and that any
conclusions based on this method have to be taken with a pinch of
salt.
 
One special problem associated with the use of optimized perturbation
theory is that the differential 3-jet rate may become negative
over large regions of the $(x_1, x_2)$ phase space. This problem
already exists, at least in principle, even for a scale $f = 1$,
since $r_2$ is not guaranteed to be positive definite. Indeed,
depending on the choice of $y$ cut, $\alphas$ value and recombination
scheme, one may observe a small region of negative differential
3-jet rate for the full second-order expression. This region
is centred around $\q \qbar \g$ configurations, where the $\q$ and
$\qbar$ are close together in one hemisphere and the $\g$ is alone in
the other, i.e.\ $x_1 \approx x_2 \approx 1/2$. It is well understood
why second-order corrections should be negative in this region
\cite{Dok89}: the $\q$ and $\qbar$ of a $\q \qbar \g$ state are in a
relative colour octet state, and thus the colour force between them is
repulsive, which translates into a negative second-order term.
 
However, as $f$ is decreased below unity, $r_2'$ receives a negative
contribution from the $\ln f$ term, and the region of negative
differential cross section has a tendency to become larger, also
after taking into account related changes in $\alphas$. In an 
event-generator framework, where all events are supposed to come 
with unit
weight, it is clearly not possible to simulate negative cross sections.
What happens in the program is therefore that no 3-jet events at
all are generated in the regions of negative differential cross section,
and that the 3-jet rate in regions of positive cross sections is
reduced by a constant factor, chosen so that the total number of
3-jet events comes out as it should. This is a consequence of the
way the program works, where it is first decided what kind of event to
generate, based on integrated 3-jet rates in which positive and
negative contributions are added up with sign, and only thereafter
the kinematics is chosen.
 
Based on our physics understanding of the origin of this negative
cross section, the approach adopted is as sensible as any, at least
to that order in perturbation theory (what one might strive for is a
properly exponentiated description of the relevant region). It can
give rise to funny results for low $f$ values, however, as observed
by OPAL \cite{OPA92} for the energy--energy correlation asymmetry.
 
\subsubsection{Angular orientation}
 
While pure $\gamma$ exchange gives a simple $1 + \cos^2\theta$
distribution for the $\q$ (and $\qbar$) direction in $\q \qbar$ events,
$\Z^0$ exchange and $\gammaZ$ interference results in a
forward--backward asymmetry. If one introduces
\begin{eqnarray}
  h'_{\f}(s) & = & 2 e_{\e} \left\{ a_{\e} 
  (1 - P_{\mrm{L}}^+ P_{\mrm{L}}^-) -
  v_{\e} (P_{\mrm{L}}^- - P_{\mrm{L}}^+) \right\} \,  
  \Re\chi(s)  e_{\f} a_{\f}
  \nonumber \\
  & & + \, \left\{ 2 v_{\e} a_{\e} (1 - P_{\mrm{L}}^+ P_{\mrm{L}}^-) -
  (v_{\e}^2 + a_{\e}^2) (P_{\mrm{L}}^- - P_{\mrm{L}}^+) \right\} \,
  |\chi(s)|^2 \, v_{\f} a_{\f} ~,
\end{eqnarray}
then the angular distribution of the quark is given by
\begin{equation}
   \frac{\d \sigma}{\d (\cos\theta_{\f})} \propto
   h_{\f}(s)(1 + \cos^2\theta_{\f}) + 2 h'_{\f}(s) \cos\theta_{\f} ~.
\end{equation}
 
The angular orientation of a 3- or 4-jet event may be described
in terms of three angles $\chi$, $\theta$ and $\varphi$; for 2-jet
events only $\theta$ and $\varphi$ are necessary. From a standard
orientation, with the $\q$ along the $+z$ axis and the $\qbar$ in the
$xz$ plane with $p_x > 0$, an arbitrary orientation may be reached by
the rotations $+\chi$ in azimuthal angle, $+\theta$ in polar angle,
and $+\varphi$ in azimuthal angle, in that order. Differential
cross sections,
including QFD effects and arbitrary beam polarizations have been given
for 2- and 3-jet events in refs. \cite{Ols80,Sch80}. We use the
formalism of ref. \cite{Ols80}, with translation from their 
terminology according to$\chi \to \pi - \chi$ and
$\varphi^- \to - (\varphi + \pi/2)$. The resulting formulae are
tedious, but
straightforward to apply, once the internal jet configuration has been
chosen. 4-jet events are approximated by 3-jet ones, by joining
the two gluons of a $\q \qbar \g \g$ event and the $\q'$ and $\qbar'$
of a $\q \qbar \q' \qbar'$ event into one effective jet. This means
that some angular asymmetries are neglected \cite{Ali80a}, but that weak
effects are automatically included. It is assumed that the second-order
3-jet events have the same angular orientation as the first-order
ones, some studies on this issue may be found in \cite{Kor85}. Further,
the formulae normally refer to the massless case; only for the QED
2- and 3-jet cases are mass corrections available.
 
The main effect of the angular distribution of multijet events
is to smear the lowest-order result, i.e.\ to reduce any anisotropies
present in 2-jet systems. In the parton-shower option of the program,
only the initial $\q \qbar$ axis is determined. The subsequent shower
evolution then {\it de facto} leads to a smearing of the jet axis,
although not necessarily in full agreement with the expectations
from multijet matrix-element treatments.
 
\subsubsection{Initial-state radiation}
 
Initial-state photon radiation has been included using the formalism of
ref. \cite{Ber82}. Here each event contains either no photon or one,
i.e.\ it is a first-order non-exponentiated description.
The main formula for the hard radiative photon
cross section is
\begin{equation}
\frac{\d \sigma}{\d x_{\gamma}} = \frac{\alphaem}{\pi} \,
\left( \ln\frac{s}{m_{\e}^2} -1 \right) \,
\frac{1 + (1-x_{\gamma})^2}{x_{\gamma}} \, \sigma_0 (\hat{s}) ~,
\end{equation}
where $x_{\gamma}$ is the photon energy fraction of the beam energy,
$\hat{s} = (1-x_{\gamma}) s$ is the squared reduced hadronic c.m.\
energy, and $\sigma_0$ is the ordinary annihilation cross section at
the reduced energy. In particular, the selection of jet flavours
should be done according to expectations at the reduced
energy. The cross section is divergent both for $x_{\gamma} \to 1$ and
$x_{\gamma} \to 0$. The former is related to the fact that
$\sigma_0$ has a $1/\hat{s}$ singularity (the real photon pole) for
$\hat{s} \to 0$. An upper cut on $x_{\gamma}$ can here be chosen 
to fit the
experimental setup. The latter is a soft photon singularity, which is
to be compensated in the no-radiation cross section. A requirement
$x_{\gamma} > 0.01$ has therefore been chosen so that the hard-photon
fraction is smaller than unity. In the total cross section, effects
from photons with $x_{\gamma} < 0.01$ are taken into account, together
with vertex and vacuum polarization corrections (hadronic vacuum
polarizations using a simple parameterization of the more complicated
formulae of ref. \cite{Ber82}).
 
The hard photon spectrum can be integrated analytically, for the
full $\gammaZ$ structure including interference terms, provided that
no new flavour thresholds are crossed and that the $R_{\mrm{QCD}}$
term in the cross section can be approximated by a constant over the
range of allowed $\hat{s}$ values. In fact, threshold effects can be
taken into account by standard rejection techniques, at the price of
not obtaining the exact cross section analytically, but only by an
effective Monte Carlo integration taking place in parallel with the
ordinary event generation. In addition to $x_{\gamma}$, the polar
angle $\theta_{\gamma}$ and azimuthal angle $\varphi_{\gamma}$ of
the photons are also to be chosen. Further, for the orientation
of the hadronic system, a choice has to be made whether the photon is
to be considered as having been radiated from the $\e^+$ or from the
$\e^-$.
 
Final-state photon radiation, as well as interference between initial-
and final-state radiation, has been left out of this treatment. The
formulae for $\ee \to \mu^+ \mu^-$ cannot be simply taken over for
the case of outgoing quarks, since the quarks as such only live for
a short while before turning into hadrons. Another simplification in
our treatment is that effects of incoming polarized $\e^{\pm}$ beams
have been completely neglected, i.e.\ neither the effective shift in
azimuthal distribution of photons nor the reduction in polarization
is included. The polarization parameters of the program are to be
thought of as the effective polarization surviving after 
initial-state radiation.
 
\subsubsection{Alternative matrix elements}
 
The program contains two sets of `toy model' matrix elements, one
for an Abelian vector gluon model and one for a scalar gluon model.
Clearly both of these alternatives are already excluded by data,
and are anyway not viable alternatives for a consistent theory of
strong interactions. They are therefore included more as references
to show how well the characteristic features of QCD can be measured
experimentally.
 
Second-order matrix elements are available for the Abelian vector
gluon model. These are easily obtained from the standard QCD matrix
elements by a substitution of the Casimir group factors:
$C_F = 4/3 \to 1$, $N_C = 3 \to 0$, and $T_R = n_{\f}/2 \to 3 n_{\f}$.
First-order matrix elements contain only $C_F$; therefore the
standard first-order QCD results may be recovered by a rescaling
of $\alphas$ by a factor $4/3$. In second order the change of $N_C$
to 0 means that $\g \to \g\g$ couplings are absent from the Abelian
model, while the change of $T_R$ corresponds to an enhancement of the
$\g \to \q'\qbar'$ coupling, i.e.\ to an enhancement of the
$\q\qbar\q'\qbar'$ 4-jet event rate.
 
The second-order corrections to the 3-jet rate
turn out to be strongly negative --- if
$\alphas$ is fitted to get about the right rate of 4-jet events,
the predicted differential 3-jet rate is negative almost
everywhere in the $(x_1, x_2)$ plane. Whether this unphysical
behaviour would be saved by higher orders is unclear. It has been
pointed out that the rate can be made positive by a suitable choice of
scale, since $\alphas$ runs in opposite directions in an Abelian
model and in QCD \cite{Bet89}. This may be seen directly from eq.
(\ref{ee:r2optim}), where the term $33 = 11 N_C$ is absent in the
Abelian model, and therefore the scale-dependent term changes sign.
In the program, optimized scales have not been implemented for this
toy model. Therefore the alternatives provided for you are either
to generate only 4-jet events, or to neglect second-order
corrections to the 3-jet rate, or to have the total 3-jet
rate set vanishing (so that only 2- and 4-jet events are
generated). Normally we would expect the former to be the one of most
interest, since it is in angular (and flavour) distributions of 4-jet
events that the structure of QCD can be tested.
Also note that the `correct' running of $\alphas$ is not included;
you are expected to use the option where $\alphas$ is just given as
a constant number.
 
The scalar gluon model is even more excluded than the Abelian vector
one, since differences appear already in the 3-jet matrix
element \cite{Lae80}:
\begin{equation}
\frac{\d \sigma}{\d x_1 \, \d x_2} \propto \frac{x_3^2}{(1-x_1)(1-x_2)}
\end{equation}
when only $\gamma$ exchange is included. The axial part of the $\Z^0$
gives a slightly different shape; this is included in the program but
does not make much difference. The angular orientation does include
the full $\gammaZ$ interference \cite{Lae80}, but the main interest is
in the 3-jet topology as such \cite{Ell79}. No higher-order
corrections are included. It is recommended to use the option of a
fixed $\alphas$ also here, since the correct running is not available.
 
\subsection{Decays of Onia Resonances}
\label{ss:oniadecays}
 
Many different possibilities are open for the decay of heavy
$J^{PC} = 1^{--}$ onia resonances. Of special interest are
the decays into three gluons or two gluons plus a photon, since these
offer unique possibilities to study a `pure sample' of gluon jets.
A routine for this purpose is included in the program. It was written
at a time where the expectations were to find toponium at PETRA 
energies. Given the large value of the top mass, weak decays
dominate, to the extent that the top quark decays weakly even
before a bound toponium state is formed, and thus the routine will be
of no use for top. The charm system, on the other hand, is far too low
in mass for a jet language to be of any use. The only application is
therefore likely to be for $\Upsilon$, which unfortunately also is on
the low side in mass.
 
The matrix element for $\q \qbar \to \g \g \g$ is (in lowest order)
\cite{Kol78}
\begin{equation}
\frac{1}{\sigma_{\g \g \g}} 
\frac{\d \sigma_{\g \g \g}}{\d x_1 \, \d x_2} =
\frac{1}{\pi^2 - 9} \left\{ \left( \frac{1-x_1}{x_2 x_3} \right)^2 +
\left( \frac{1-x_2}{x_1 x_3} \right)^2 +
\left( \frac{1-x_3}{x_1 x_2} \right)^2 \right\} ~,
\label{ee:Upsilondec}
\end{equation}
where, as before, $x_i = 2 E_i / E_{\mrm{cm}}$ in the c.m.\ frame of 
the event. This is a well-defined expression, without the kind of
singularities encountered in the $\q \qbar \g$ matrix elements.
In principle, no cuts at all would be necessary, but for reasons
of numerical simplicity we implement a $y$ cut as for continuum
jet production, with all events not fulfilling this cut considered
as (effective) $\g \g$ events. For $\g \g \g$ events, each $\g \g$
invariant mass is required to be at least 2 GeV.
 
Another process is $\q \qbar \to \gamma \g \g$, obtained by replacing
a gluon in $\q \qbar \to \g \g \g$ by a photon. This process has the
same normalized cross section as the one above, if e.g.\ $x_1$ is taken
to refer to the photon. The relative rate is \cite{Kol78}
\begin{equation}
\frac{\sigma_{\gamma \g \g}}{\sigma_{\g \g \g}} =
\frac{36}{5} \, \frac{e_{\q}^2 \, \alphaem}
{\alphas(Q^2)} ~.
\label{ee:UpsilonBR}
\end{equation}
Here $e_{\q}$ is the charge of the heavy quark, and the scale in
$\alphas$ has been chosen as the mass of the onium state. If the
mass of the recoiling $\g \g$ system is lower than some cut-off
(by default 2 GeV), the event is rejected.
 
In the present implementation the angular orientation of the
$\g \g \g$ and $\gamma \g \g$ events is given for the
$\ee \to \gamma^* \to$ onium case \cite{Kol78} (optionally with
beam polarization effects included), i.e.\ weak effects have not
been included, since they are negligible at around 10~GeV.
 
It is possible to start a perturbative shower evolution from either
of the two states above. However, for $\Upsilon$ the phase space
for additional evolution is so constrained that not much is to be
gained from that. We therefore do not recommend this possibility.
The shower generation machinery, when starting up from a
$\gamma \g \g$ configuration, is constructed such that the
photon energy is not changed. This means that there is currently no
possibility to use showers to bring the theoretical photon
spectrum in better agreement with the experimental one.
 
In string fragmentation language, a $\g \g \g$ state corresponds to
a closed string triangle with the three gluons at the corners. As
the partons move apart from a common origin, the string triangle
expands. Since the photon does not take part in the fragmentation,
the $\gamma \g \g$ state corresponds to a double string running
between the two gluons.
 
\subsection{Routines and Common Block Variables}
\label{ss:eeroutines}
 
\subsubsection{$\ee$ continuum event generation}
 
The only routine a normal user will call to generate $\ee$ continuum
events is \ttt{PYEEVT}. The other routines listed below, as well as
\ttt{PYSHOW} (see section \ref{ss:showrout}), are called by 
\ttt{PYEEVT}.
 
\drawbox{CALL PYEEVT(KFL,ECM)}\label{p:PYEEVT}
\begin{entry}
\itemc{Purpose:} to generate a complete event
$\ee \to \gammaZ \to \q\qbar \to$ parton shower $\to$ hadrons
according to QFD and QCD cross sections. As an alternative to
parton showers, second-order matrix elements are available for
$\q\qbar + \q\qbar\g + \q\qbar\g\g + \q\qbar\q'\qbar'$ production.
\iteme{KFL :} flavour of events generated.
\begin{subentry}
\iteme{= 0 :} mixture of all allowed flavours according to relevant
probabilities.
\iteme{= 1 - 8 :} primary quarks are only of the specified flavour
\ttt{KFL}.
\end{subentry}
\iteme{ECM :} total c.m.\ energy of system.
\itemc{Remark:} Each call generates one event, which is independent
of preceding ones, with one exception, as follows. If radiative
corrections are included, the shape of the hard photon spectrum is
recalculated only with each \ttt{PYXTEE} call, which normally is done
only if \ttt{KFL}, \ttt{ECM} or \ttt{MSTJ(102)} is changed. A change
of e.g.\ the $\Z^0$ mass in mid-run has to be followed either by a user
call to \ttt{PYXTEE} or by an internal call forced e.g.\ by putting
\ttt{MSTJ(116)=3}.
\end{entry}

\boxsep

\begin{entry} 
\iteme{SUBROUTINE PYXTEE(KFL,ECM,XTOT) :}\label{p:PYXTEE}
to calculate the total hadronic cross section,
including quark thresholds, weak, beam polarization, and QCD effects
and radiative corrections. In the process, variables necessary
for the treatment of hard photon radiation are calculated and
stored.
\begin{subentry}
\iteme{KFL, ECM :} as for \ttt{PYEEVT}.
\iteme{XTOT :} the calculated total cross section in nb.
\end{subentry}
 
\iteme{SUBROUTINE PYRADK(ECM,MK,PAK,THEK,PHIK,ALPK) :}\label{p:PYRADK}
to describe initial-state hard $\gamma$ radiation.
 
\iteme{SUBROUTINE PYXKFL(KFL,ECM,ECMC,KFLC) :}\label{p:PYXKFL}
to generate the primary quark flavour in case this
is not specified by you.
 
\iteme{SUBROUTINE PYXJET(ECM,NJET,CUT) :}\label{p:PYXJET}
to determine the number of jets (2, 3 or 4) to be
generated within the kinematically allowed region (characterized by
\ttt{CUT} $= y_{\mrm{cut}}$) in the matrix-element approach; to be 
chosen such that all probabilities are between 0 and 1.
 
\iteme{SUBROUTINE PYX3JT(NJET,CUT,KFL,ECM,X1,X2) :}\label{p:PYX3JT}
to generate the internal momentum variables of a
3-jet event, $\q\qbar\g$, according to first- or second-order
QCD matrix elements.
 
\iteme{SUBROUTINE PYX4JT(NJET,CUT,KFL,ECM,KFLN,X1,X2,X4,X12,X14) :}%
\label{p:PYX4JT} 
to generate the internal momentum variables for a
4-jet event, $\q\qbar\g\g$ or $\q\qbar\q'\qbar'$, according to
second-order QCD matrix elements.
 
\iteme{SUBROUTINE PYXDIF(NC,NJET,KFL,ECM,CHI,THE,PHI) :}\label{p:PYXDIF}
to describe the angular orientation of the jets.
In first-order QCD the complete QED or QFD formulae are used; in
second  order 3-jets are assumed to have the same orientation
as in first, and 4-jets are approximated by 3-jets.

\end{entry}
 
\subsubsection{A routine for onium decay}
 
In \ttt{PYONIA} we have implemented the decays of heavy onia
resonances into three gluons or two gluons plus a photon, which are
the dominant non-background-like decays of $\Upsilon$.
 
\drawbox{CALL PYONIA(KFL,ECM)}\label{p:PYONIA}
\begin{entry}
\itemc{Purpose:} to simulate the process
$\ee \to \gamma^* \to 1^{--}$ onium resonance $\to (\g\g\g$ or
$\g\g\gamma) \to$ shower $\to$ hadrons.
\iteme{KFL :} the flavour of the quark giving rise to the resonance.
\begin{subentry}
\iteme{= 0 :} generate $\g\g\g$ events alone.
\iteme{= 1 - 8 :} generate $\g\g\g$ and $\g\g\gamma$ events in mixture
determined by the squared charge of flavour \ttt{KFL}, see 
eq.~(\ref{ee:UpsilonBR}). Normally \ttt{KFL=} 5.
\end{subentry}
\iteme{ECM :} total c.m.\ energy of system.
\end{entry}
 
\subsubsection{Common block variables}
 
The status codes and parameters relevant for the $\ee$ routines are
found in the common block \ttt{PYDAT1}. This common block also contains
more general status codes and parameters, described elsewhere.
 
\drawbox{COMMON/PYDAT1/MSTU(200),PARU(200),MSTJ(200),PARJ(200)}
\begin{entry}
 
\itemc{Purpose:} to give access to a number of status codes and
parameters regulating the performance of the $\ee$ event generation
routines.
 
\iteme{MSTJ(101) :}\label{p:MSTJ101} (D=5) gives the type of QCD 
corrections used for continuum events.
\begin{subentry}
\iteme{= 0 :} only $\q\qbar$ events are generated.
\iteme{= 1 :} $\q\qbar + \q\qbar\g$ events are generated according
to first-order QCD.
\iteme{= 2 :} $\q\qbar + \q\qbar\g + \q\qbar\g\g + \q\qbar\q'\qbar'$
events are generated according to second-order QCD.
\iteme{= 3 :} $\q\qbar + \q\qbar\g + \q\qbar\g\g + \q\qbar\q'\qbar'$
events are generated, but without second-order corrections to the
3-jet rate.
\iteme{= 5 :} a parton shower is allowed to develop from an original
$\q\qbar$ pair, see \ttt{MSTJ(38) - MSTJ(50)} for details.
\iteme{= -1 :} only $\q\qbar\g$ events are generated (within same
matrix-element cuts as for \ttt{=1}). Since the change in flavour
composition from mass cuts or radiative corrections is not
taken into account, this option is not intended for
quantitative studies.
\iteme{= -2 :} only $\q\qbar\g\g$ and $\q\qbar\q'\qbar'$ events are
generated (as for \ttt{=2}). The same warning as for \ttt{=-1} applies.
\iteme{= -3 :} only $\q\qbar\g\g$ events are generated (as for
\ttt{=2}). The same warning as for \ttt{=-1} applies.
\iteme{= -4 :} only $\q\qbar\q'\qbar'$ events are generated
(as for \ttt{=2}). The same warning as for \ttt{=-1} applies.
\itemc{Note 1:} \ttt{MSTJ(101)} is also used in \ttt{PYONIA}, with
\iteme{$\leq$ 4 :} $\g\g\g + \gamma\g\g$ events are generated
according to lowest-order matrix elements.
\iteme{$\geq$ 5 :} a parton shower is allowed to develop from the
original $\g\g\g$ or $\g\g\gamma$ configuration, see
\ttt{MSTJ(38) - MSTJ(50)} for details.
\itemc{Note 2:} The default values of fragmentation parameters have
been chosen to work well with the default parton-shower approach
above. If any of the other options are used, or if the parton
shower is used in non-default mode, it is normally necessary to
retune fragmentation parameters. As an example, we note that
the second-order matrix-element approach (\ttt{MSTJ(101)=2}) at
PETRA/PEP energies gives a better description when the $a$ and
$b$ parameters of the symmetric fragmentation function are set to
$a =$\ttt{PARJ(41)=1}, $b =$\ttt{PARJ(42)=0.7}, and the
width of the transverse momentum distribution to
$\sigma =$\ttt{PARJ(21)=0.40}.
In principle, one also ought to change the joining parameter
to \ttt{PARJ(33)=PARJ(35)=1.1} to preserve a flat rapidity
plateau, but if this should be forgotten, it does not make too
much difference. For applications at TRISTAN or LEP, one has to 
change the matrix-element approach
parameters even more, to make up for additional soft gluon
effects not covered in this approach.
\end{subentry}
 
\iteme{MSTJ(102) :} (D=2) inclusion of weak effects ($\Z^0$ exchange)
for flavour production, angular orientation, cross sections and
initial-state photon radiation in continuum events.
\begin{subentry}
\iteme{= 1 :} QED, i.e.\ no weak effects are included.
\iteme{= 2 :} QFD, i.e.\ including weak effects.
\iteme{= 3 :} as \ttt{=2}, but at initialization in \ttt{PYXTEE} the
$\Z^0$ width is calculated from $\ssintw$, $\alphaem$ and 
$\Z^0$ and quark masses (including bottom and top threshold factors for
\ttt{MSTJ(103)} odd), assuming three full generations, and the
result is stored in \ttt{PARJ(124)}.
\end{subentry}
 
\iteme{MSTJ(103) :} (D=7) mass effects in continuum matrix elements,
in the form \ttt{MSTJ(103)} $= M_1 + 2M_2 + 4M_3$, where $M_i = 0$
if no mass effects and $M_i = 1$ if mass effects should be included.
Here;
\begin{subentry}
\iteme{$M_1$ :} threshold factor for new flavour production
according to QFD result;
\iteme{$M_2$ :} gluon emission probability (only applies for
\ttt{|MSTJ(101)|}$\leq 1$, otherwise no mass effects anyhow);
\iteme{$M_3$ :} angular orientation of event (only applies for
\ttt{|MSTJ(101)|}$\leq 1$ and
\ttt{MSTJ(102)=1}, otherwise no mass effects anyhow).
\end{subentry}
 
\iteme{MSTJ(104) :} (D=5) number of allowed flavours, i.e.\ flavours
that can be produced in a continuum event if the energy is enough.
A change to 6 makes top production allowed above the threshold, etc.
Note that in $\q\qbar\q'\qbar'$ events only the first five flavours
are allowed in the secondary pair, produced by a gluon breakup.
 
\iteme{MSTJ(105) :} (D=1) fragmentation and decay in \ttt{PYEEVT} and
\ttt{PYONIA} calls.
\begin{subentry}
\iteme{= 0 :} no \ttt{PYEXEC} calls, i.e.\ only matrix-element
and/or parton-shower treatment, and collapse of small jet
systems into one or two particles (in \ttt{PYPREP}).
\iteme{= 1 :} \ttt{PYEXEC} calls are made to generate fragmentation
and decay chain.
\iteme{= -1 :} no \ttt{PYEXEC} calls and no collapse of small jet
systems into one or two particles (in \ttt{PYPREP}).
\end{subentry}
 
\iteme{MSTJ(106) :} (D=1) angular orientation in \ttt{PYEEVT} and
\ttt{PYONIA}.
\begin{subentry}
\iteme{= 0 :} standard orientation of events, i.e.\ $\q$ along $+z$ axis
and $\qbar$ along $-z$ axis or in $xz$ plane with $p_x > 0$ for
continuum events, and $\g_1\g_2\g_3$ or $\gamma\g_2\g_3$ in $xz$ plane
with $\g_1$ or $\gamma$ along the $+z$ axis for onium events.
\iteme{= 1 :} random orientation according to matrix elements.
\end{subentry}
 
\iteme{MSTJ(107) :} (D=0) radiative corrections to continuum events.
\begin{subentry}
\iteme{= 0 :} no radiative corrections.
\iteme{= 1 :} initial-state radiative corrections (including weak
effects for \ttt{MSTJ(102)=} 2 or 3).
\end{subentry}
 
\iteme{MSTJ(108) :} (D=2) calculation of $\alphas$ for matrix-element
alternatives. The \ttt{MSTU(111)} and \ttt{PARU(112)} values are
automatically overwritten in \ttt{PYEEVT} or \ttt{PYONIA} calls
accordingly.
\begin{subentry}
\iteme{= 0 :} fixed $\alphas$ value as given in \ttt{PARU(111)}.
\iteme{= 1 :} first-order formula is always used, with
$\Lambda_{\mrm{QCD}}$ given by \ttt{PARJ(121)}.
\iteme{= 2 :} first- or second-order formula is used, depending on
value of \ttt{MSTJ(101)}, with $\Lambda_{\mrm{QCD}}$ given by
\ttt{PARJ(121)} or \ttt{PARJ(122)}.
\end{subentry}
 
\iteme{MSTJ(109) :} (D=0) gives a possibility to switch from QCD
matrix elements to some alternative toy models. Is not relevant for
shower evolution, \ttt{MSTJ(101)=5}, where one can use
\ttt{MSTJ(49)} instead.
\begin{subentry}
\iteme{= 0 :} standard QCD scenario.
\iteme{= 1 :} a scalar gluon model. Since no second-order corrections
are available in this scenario, one can only use this with
\ttt{MSTJ(101) = 1} or \ttt{-1}. Also note that the event-as-a-whole
angular distribution is for photon exchange only (i.e.\ no weak
effects), and that no higher-order corrections to the total
cross section are included.
\iteme{= 2 :} an Abelian vector gluon theory, with the colour factors
$C_F = 1$ ($= 4/3$ in QCD), $N_C = 0$ ($= 3$ in QCD) and
$T_R = 3 n_f$ ($= n_f/2$ in QCD). If one selects
$\alpha_{\mrm{Abelian}} = (4/3) \alpha_{\mrm{QCD}}$,
the 3-jet cross section will agree with
the QCD one, and differences are to be found only in 4-jets.
The \ttt{MSTJ(109)=2} option has to be run with
\ttt{MSTJ(110)=1} and \ttt{MSTJ(111)=0}; if need be, the latter
variables will be overwritten by the program. \\
{\bf Warning:} second-order corrections give a large negative
contribution to the 3-jet cross section, so large that
the whole scenario is of doubtful use. In order to make the
second-order options work at all, the 3-jet cross section
is here by hand set exactly equal to zero for \ttt{MSTJ(101)=2}.
It is here probably better to use the option \ttt{MSTJ(101)=3},
although this is not a consistent procedure either.
\end{subentry}
 
\iteme{MSTJ(110) :} (D=2) choice of second-order contributions
to the 3-jet rate.
\begin{subentry}
\iteme{= 1 :} the GKS second-order matrix elements.
\iteme{= 2 :} the Zhu parameterization of the ERT matrix elements,
based on the program of Kunszt and Ali, i.e.\ in historical sequence
ERT/Kunszt/Ali/Zhu. The parameterization is available for
$y =$ 0.01, 0.02, 0.03, 0.04 and 0.05. Values outside this
range are put at the nearest border, while those inside
it are given by a linear interpolation between the
two nearest points. Since this procedure is rather primitive,
one should try to work at one of the values given above.
Note that no Abelian QCD parameterization is available for
this option.
\end{subentry}
 
\iteme{MSTJ(111) :} (D=0) use of optimized perturbation theory for
second-order matrix elements (it can also be used for first-order
matrix elements, but here it only corresponds to a trivial
rescaling of the $\alphas$ argument).
\begin{subentry}
\iteme{= 0 :} no optimization procedure; i.e.\ $Q^2 = E_{\mrm{cm}}^2$.
\iteme{= 1 :} an optimized $Q^2$ scale is chosen as
$Q^2 = f E_{\mrm{cm}}^2$, where $f =$\ttt{PARJ(128)} for the total
cross section $R$ factor, while $f =$\ttt{PARJ(129)} for the
3- and 4-jet rates. This $f$ value enters via the
$\alphas$, and also via a term proportional to $\alphas^2 \ln f$.
Some constraints are imposed; thus the optimized `3-jet'
contribution to $R$ is assumed to be positive (for \ttt{PARJ(128)}), 
the total 3-jet rate is not allowed to be negative
(for \ttt{PARJ(129)}), etc.
However, there is no guarantee that the differential 3-jet
cross section is not negative (and truncated to 0) somewhere
(this can also happen with $f = 1$, but is then less frequent).
The actually obtained $f$ values are stored in \ttt{PARJ(168)} and
\ttt{PARJ(169)}, respectively.
If an optimized $Q^2$ scale is used, then the $\Lambda_{\mrm{QCD}}$
(and $\alphas$) should also be changed. With the value $f = 0.002$,
it has been shown \cite{Bet89} that a $\Lambda_{\mrm{QCD}} = 0.100$
GeV gives a reasonable agreement; the parameter to be changed is
\ttt{PARJ(122)} for a second-order running $\alphas$. Note that,
since the optimized $Q^2$ scale is sometimes below the charm
threshold, the effective number of flavours used in $\alphas$ may
well be 4 only. If one feels that it is still appropriate to use 5
flavours (one choice might be as good as the other), it is
necessary to put \ttt{MSTU(113)=5}.
\end{subentry}
 
\iteme{MSTJ(115) :} (D=1) documentation of continuum or onium
events, in increasing order of completeness.
\begin{subentry}
\iteme{= 0 :} only the parton shower, the fragmenting partons and the
generated hadronic system are stored in the \ttt{PYJETS} common block.
\iteme{= 1 :} also a radiative photon is stored (for continuum events).
\iteme{= 2 :} also the original $\ee$ are stored (with 
\ttt{K(I,1)=21}).
\iteme{= 3 :} also the $\gamma$ or $\gammaZ$ exchanged for continuum
events, the onium state for resonance events is stored (with 
\ttt{K(I,1)=21}).
\end{subentry}
 
\iteme{MSTJ(116) :} (D=1) initialization of total cross section and
radiative photon spectrum in \ttt{PYEEVT} calls.
\begin{subentry}
\iteme{= 0 :} never; cannot be used together with radiative
corrections.
\iteme{= 1 :} calculated at first call and then whenever \ttt{KFL}
or \ttt{MSTJ(102)} is changed or \ttt{ECM} is changed by more than
\ttt{PARJ(139)}.
\iteme{= 2 :} calculated at each call.
\iteme{= 3 :} everything is re-initialized in the next call, but
\ttt{MSTJ(116)} is afterwards automatically put \ttt{=1} for use
in subsequent calls.
\end{subentry}
 
\iteme{MSTJ(119) :} (I) check on need to re-initialize \ttt{PYXTEE}.
 
\iteme{MSTJ(120) :} (R) type of continuum event generated with the
matrix-element option (with the shower one, the result is always
\ttt{=1}).
\begin{subentry}
\iteme{= 1 :} $\q\qbar$.
\iteme{= 2 :} $\q\qbar\g$.
\iteme{= 3 :} $\q\qbar\g\g$ from Abelian (QED-like) graphs in
matrix element.
\iteme{= 4 :} $\q\qbar\g\g$ from non-Abelian (i.e.\ containing
triple-gluon coupling) graphs in matrix element.
\iteme{= 5 :} $\q\qbar\q'\qbar'$.
\end{subentry}
 
\iteme{MSTJ(121) :} (R) flag set if a negative differential
cross section was encountered in the latest \ttt{PYX3JT} call.
Events are still generated, but maybe not quite according to
the distribution one would like (the rate is set to zero in the
regions of negative cross section, and the differential rate
in the regions of positive cross section is rescaled to give
the `correct' total 3-jet rate).

\boxsep 

\iteme{PARJ(121) :}\label{p:PARJ121} (D=1.0 GeV) $\Lambda$ value 
used in first-order
calculation of $\alphas$ in the matrix-element alternative.
 
\iteme{PARJ(122) :} (D=0.25 GeV) $\Lambda$ values used in second-order
calculation of $\alphas$ in the matrix-element alternative.
 
\iteme{PARJ(123) :} (D=91.187 GeV) mass of $\Z^0$ as used in
propagators for the QFD case.
 
\iteme{PARJ(124) :} (D=2.489 GeV) width of $\Z^0$ as used in
propagators for the QFD case. Overwritten at initialization if
\ttt{MSTJ(102)=3}.
 
\iteme{PARJ(125) :} (D=0.01) $y_{\mrm{cut}}$, minimum squared scaled 
invariant mass of any two partons in 3- or 4-jet events; the main
user-controlled matrix-element cut. \ttt{PARJ(126)} provides an
additional constraint. For each new event, it is additionally
checked that the total 3- plus 4-jet fraction does not
exceed unity; if so the effective $y$ cut will be dynamically
increased. The actual $y$-cut value is stored in
\ttt{PARJ(150)}, event by event.
 
\iteme{PARJ(126) :} (D=2. GeV) minimum invariant mass of any two
partons in 3- or 4-jet events; a cut in addition to the one above,
mainly for the case of a radiative photon lowering the hadronic 
c.m.\ energy significantly.
 
\iteme{PARJ(127) :} (D=1. GeV) is used as a safety margin for small
colour-singlet jet systems, cf. \ttt{PARJ(32)}, specifically
$\q\qbar'$ masses in $\q\qbar\q'\qbar'$ 4-jet events and $\g\g$ mass
in onium $\gamma\g\g$ events.
 
\iteme{PARJ(128) :} (D=0.25) optimized $Q^2$ scale for the QCD $R$
(total rate) factor for the \ttt{MSTJ(111)=1} option is given by
$Q^2 = f E_{\mrm{cm}}^2$, where $f =$\ttt{PARJ(128)}. For various 
reasons the actually used $f$ value may be increased compared with 
the nominal one; while \ttt{PARJ(128)} gives the nominal value, 
\ttt{PARJ(168)} gives the actual one for the current event.
 
\iteme{PARJ(129) :} (D=0.002) optimized $Q^2$ scale for the 3-
and 4-jet rate for the \ttt{MSTJ(111)=1} option is given by
$Q^2 = f E_{\mrm{cm}}^2$, where $f =$\ttt{PARJ(129)}. For various 
reasons the actually used $f$ value may be increased compared with 
the nominal one; while \ttt{PARJ(129)} gives the nominal value, 
\ttt{PARJ(169)} gives the actual one for the current event. The 
default value is in agreement with the studies of Bethke \cite{Bet89}.
 
\iteme{PARJ(131), PARJ(132) :} (D=2*0.) longitudinal polarizations
$P_{\mrm{L}}^+$ and $P_{\mrm{L}}^-$ of incoming $\e^+$ and $\e^-$.
 
\iteme{PARJ(133) :} (D=0.) transverse polarization
$P_{\mrm{T}} = \sqrt{P_{\mrm{T}}^+ P_{\mrm{T}}^-}$, with 
$P_{\mrm{T}}^+$ and $P_{\mrm{T}}^-$ transverse
polarizations of incoming $\e^+$ and $\e^-$.
 
\iteme{PARJ(134) :} (D=0.) mean of transverse polarization
directions of incoming $\e^+$ and $\e^-$,
$\Delta \varphi = (\varphi^+ + \varphi^-) /2$, with $\varphi$
the azimuthal angle of polarization, leading to a shift in the
$\varphi$ distribution of jets by $\Delta \varphi$.
 
\iteme{PARJ(135) :} (D=0.01) minimum photon energy fraction
(of beam energy) in initial-state radiation; should normally
never be changed (if lowered too much, the fraction of events
containing a radiative photon will exceed unity, leading to
problems).
 
\iteme{PARJ(136) :} (D=0.99) maximum photon energy fraction
(of beam energy) in initial-state radiation; may be changed
to reflect actual trigger conditions of a detector (but must
always be larger than \ttt{PARJ(135)}).
 
\iteme{PARJ(139) :} (D=0.2 GeV) maximum deviation of $E_{\mrm{cm}}$ 
from the corresponding value at last \ttt{PYXTEE} call, above which 
a new call is made if \ttt{MSTJ(116)=1}.
 
\iteme{PARJ(141) :} (R) value of $R$, the ratio of continuum
cross section to the lowest-order muon pair production cross section,
as given in massless QED (i.e.\ three times the sum of active
quark squared charges, possibly modified for polarization).
 
\iteme{PARJ(142) :} (R) value of $R$ including quark-mass effects
(for \ttt{MSTJ(102)=1}) and/or weak propagator effects
(for \ttt{MSTJ(102)=2}).
 
\iteme{PARJ(143) :} (R) value of $R$ as \ttt{PARJ(142)}, but
including QCD corrections as given by \ttt{MSTJ(101)}.
 
\iteme{PARJ(144) :} (R) value of $R$ as \ttt{PARJ(143)}, but
additionally including corrections from initial-state photon
radiation (if \ttt{MSTJ(107)=1}). Since the effects of heavy
flavour thresholds are not simply integrable, the initial value
of \ttt{PARJ(144)} is updated during the
course of the run to improve accuracy.
 
\iteme{PARJ(145) - PARJ(148) :} (R) absolute cross sections in nb
as for the cases \ttt{PARJ(141) - PARJ(144)} above.
 
\iteme{PARJ(150) :} (R) current effective matrix element cut-off
$y_{\mrm{cut}}$, as given by \ttt{PARJ(125), PARJ(126)} and the
requirements of having non-negative cross sections for 2-,
3- and 4-jet events. Not used in parton showers.
 
\iteme{PARJ(151) :} (R) value of c.m.\ energy \ttt{ECM} at last
\ttt{PYXTEE} call.
 
\iteme{PARJ(152) :} (R) current first-order contribution to the
3-jet fraction; modified by mass effects. Not used in parton
showers.
 
\iteme{PARJ(153) :} (R) current second-order contribution to the
3-jet fraction; modified by mass effects. Not used in parton
showers.
 
\iteme{PARJ(154) :} (R) current second-order contribution to the
4-jet fraction; modified by mass effects. Not used in parton
showers.
 
\iteme{PARJ(155) :} (R) current fraction of 4-jet rate
attributable to $\q\qbar\q'\qbar'$ events rather than $\q\qbar\g\g$
ones; modified by mass effects. Not used in parton showers.
 
\iteme{PARJ(156) :} (R) has two functions when using second-order
QCD. For a 3-jet event, it gives the ratio of the second-order
to the total 3-jet cross section in the given kinematical
point. For a 4-jet event, it gives the ratio of the
modified 4-jet cross section, obtained when neglecting interference 
terms whose colour flow is not well defined, to the full
unmodified one, all evaluated in the given kinematical point.
Not used in parton showers.
 
\iteme{PARJ(157) - PARJ(159) :} (I) used for cross-section
calculations to include mass threshold effects to radiative
photon cross section. What is stored is basic cross section,
number of events generated and number that passed cuts.
 
\iteme{PARJ(160) :} (R) nominal fraction of events that should
contain a radiative photon.
\iteme{PARJ(161) - PARJ(164) :} (I) give shape of radiative photon
spectrum including weak effects.
 
\iteme{PARJ(168) :} (R) actual $f$ value of current event in
optimized perturbation theory for $R$; see \ttt{MSTJ(111)} and
\ttt{PARJ(128)}.
 
\iteme{PARJ(169) :} (R) actual $f$ value of current event in
optimized perturbation theory for 3- and 4-jet rate;
see \ttt{MSTJ(111)} and \ttt{PARJ(129)}.
 
\iteme{PARJ(171) :} (R) fraction of cross section corresponding
to the axial coupling of quark pair to the intermediate $\gammaZ$
state; needed for the Abelian gluon model 3-jet matrix
element.
 
\end{entry}
 
\subsection{Examples}
 
An ordinary $\ee$ annihilation event in the continuum, at a 
c.m.\ energy of 91 GeV, may be generated with
\begin{verbatim}
      CALL PYEEVT(0,91D0)
\end{verbatim}
In this case a $\q\qbar$ event is generated, including weak effects,
followed by parton-shower evolution and fragmentation/decay treatment.
Before a call to \ttt{PYEEVT}, however, a number of default values
may be changed, e.g.\ \ttt{MSTJ(101)=2} to use second-order QCD
matrix elements, giving a mixture of $\q\qbar$, $\q\qbar\g$,
$\q\qbar\g\g$, and $\q\qbar\q'\qbar'$ events, \ttt{MSTJ(102)=1} to
have QED only, \ttt{MSTJ(104)=6} to allow $\t\tbar$ production
as well, \ttt{MSTJ(107)=1} to include initial-state photon radiation
(including a treatment of the $\Z^0$ pole), \ttt{PARJ(123)=92.0} to
change the $\Z^0$ mass, \ttt{PARJ(81)=0.3} to change the 
parton-shower $\Lambda$ value, or \ttt{PARJ(82)=1.5} to change the 
parton-shower cut-off. If initial-state photon radiation is used, some
restrictions apply to how one can alternate the generation of
events at different energies or with different $\Z^0$ mass, etc.
These restrictions are not there for
efficiency reasons (the extra time for recalculating the extra
constants every time is small), but because it ties in with the
cross-section calculations (see \ttt{PARJ(144)}).
 
Most parameters can be changed independently of each other. However,
if just one or a few parameters/switches are changed, one should not
be surprised to find a rather bad agreement with the data, like e.g.
a too low or high average hadron multiplicity. It is therefore usually
necessary to retune one parameter related to the perturbative QCD
description, like $\alphas$ or $\Lambda$, one of the two parameters
$a$ and $b$ of the Lund symmetric fragmentation function (since they
are so strongly correlated, it is often not necessary to retune both
of them), and the average fragmentation transverse momentum --- see
Note~2 of the \ttt{MSTJ(101)} description for an example. For very
detailed studies it may be necessary to retune even more parameters.
 
The three-gluon and gluon--gluon--photon decays of $\Upsilon$ may be
simulated by a call
\begin{verbatim}
      CALL PYONIA(5,9.46D0)
\end{verbatim}
 
A typical program for analysis of $\ee$ annihilation events at
200 GeV might look something like
\begin{verbatim}
      IMPLICIT DOUBLE PRECISION(A-H, O-Z)
      IMPLICIT INTEGER(I-N)
      INTEGER PYK,PYCHGE,PYCOMP
      COMMON/PYJETS/N,NPAD,K(4000,5),P(4000,5),V(4000,5)
      COMMON/PYDAT1/MSTU(200),PARU(200),MSTJ(200),PARJ(200)
      COMMON/PYDAT2/KCHG(500,4),PMAS(500,4),PARF(2000),VCKM(4,4)
      COMMON/PYDAT3/MDCY(500,3),MDME(8000,2),BRAT(8000),KFDP(8000,5)
      MDCY(PYCOMP(111),1)=0           ! put pi0 stable
      MSTJ(107)=1                     ! include initial-state radiation
      PARU(41)=1D0                    ! use linear sphericity
      .....                           ! other desired changes
      CALL PYTABU(10)                 ! initialize analysis statistics
      DO 100 IEV=1,1000               ! loop over events
        CALL PYEEVT(0,200D0)          ! generate new event
        IF(IEV.EQ.1) CALL PYLIST(2)   ! list first event
        CALL PYTABU(11)               ! save particle composition
                                      !   statistics
        CALL PYEDIT(2)                ! remove decayed particles
        CALL PYSPHE(SPH,APL)          ! linear sphericity analysis
        IF(SPH.LT.0D0) GOTO 100       ! too few particles in event for
                                      !   PYSPHE to work on it (unusual)
        CALL PYEDIT(31)               ! orient event along axes above
        IF(IEV.EQ.1) CALL PYLIST(2)   ! list first treated event
        .....                         ! fill analysis statistics
        CALL PYTHRU(THR,OBL)          ! now do thrust analysis
        .....                         ! more analysis statistics
  100 CONTINUE                        !
      CALL PYTABU(12)                 ! print particle composition
                                      !   statistics
      .....                           ! print analysis statistics
      END
\end{verbatim}
 
\clearpage
 
\section{Process Generation}
\label{s:PYTprocgen}
 
Much can be said about the processes in {\Py} and
the way they are generated. Therefore the material has been split
into three sections. In the current one the philo\-sophy underlying
the event generation scheme is presented.  Here we provide a
generic description, where some special cases are swept under the
carpet. In the next section, the existing processes are enumerated,
with some comments about applications and limitations. Finally, in
the third section the generation routines and common block switches
are described.
 
The section starts with a survey of parton distributions, followed
by a detailed description of the simple $2 \to 2$ and $2 \to 1$
hard subprocess generation schemes, including pairs of resonances.
This is followed by a few comments on more complicated configurations,
and on nonperturbative processes.
 
\subsection{Parton Distributions}
\label{ss:structfun}
 
The parton distribution function $f_i^a(x,Q^2)$ parameterizes the 
probability to find a parton $i$ with a fraction $x$ of the beam energy 
when the beam particle $a$ is probed by a hard scattering at virtuality 
scale $Q^2$. Usually the momentum-weighted combination $x f_i^a(x,Q^2)$ 
is used, for which the normalization condition
$\sum_i \int_0^1 dx \, x f_i^a(x,Q^2) \equiv 1$ normally applies.
The $Q^2$ dependence of parton distributions is perturbatively
calculable, see section \ref{sss:initshowstruc}.
 
The parton distributions in {\Py} come in many shapes, as shown in the
following.
 
\subsubsection{Baryons}
 
For protons, many sets exist on the market. These are obtained by fits
to experimental data, constrained so that the $Q^2$ dependence is in
accordance with the standard QCD evolution equations. The current default 
in {\Py} is GRV 94L \cite{Glu95}, a simple leading-order fit. 
Several other sets are found in {\Py}. The complete list is:
\begin{Itemize}
\item EHLQ sets 1 and 2 \cite{Eic84};
\item DO sets 1 and 2 \cite{Duk82};
\item the GRV 92L (updated version) fit \cite{Glu92};
\item the CTEQ 3L, CTEQ 3M and CTEQ 3D fits \cite{Lai95};
\item the GRV 94L, GRV 94M and GRV 94D fits \cite{Glu95}; and
\item the CTEQ 5L and CTEQ 5M1 fits \cite{Lai00}.

\end{Itemize}
Of these, EHLQ, DO, GRV 92L, CTEQ 3L, GRV94L and CTEQ5L are leading-order 
parton distributions, while CTEQ 3D and GRV94D are in the 
next-to-leading-order DIS scheme and the rest
in the next-to-leading order $\br{\mrm{MS}}$ scheme. The EHLQ and DO 
sets are by now rather old, and are kept mainly for backwards
compatibility. Since only Born-level matrix elements
are included in the program, there is no particular reason to use
higher-order parton distributions --- the resulting combination is
anyway only good to leading-order accuracy. (Some higher-order 
corrections are effectively included by the parton-shower
treatment, but there is no exact match.)
 
There is a steady flow of new parton-distribution sets on the market.
To keep track of all of them is a major work on its own. Therefore
{\Py} contains an interface to an external library of parton
distribution functions, \tsc{Pdflib} \cite{Plo93}.
This is a truly encyclopedic collection of almost all proton, pion 
and photon parton distributions proposed since the late 70's.
Three dummy routines come with the {\Py} package, so as to avoid 
problems with unresolved external references if \tsc{Pdflib} is not 
linked. One should also note that {\Py} does not check the results, 
but assumes that sensible answers will be returned, also outside the 
nominal $(x, Q^2)$ range of a set. Only the sets that come with 
{\Py} have been suitably modified to provide reasonable answers 
outside their nominal domain of validity.
 
{}From the proton parton distributions, those of the neutron are obtained
by isospin conjugation, i.e.\ $f_{\u}^{\n} = f_{\d}^{\p}$ and
 $f_{\d}^{\n} = f_{\u}^{\p}$.
 
The program does allow for incoming beams of a number of hyperons:
$\Lambda^0$, $\Sigma^{-,0,+}$, $\Xi^{-,0}$ and $\Omega^-$. Here
one has essentially no experimental information. One could imagine
to construct models in which valence $\s$ quarks are found at larger
average $x$ values than valence $\u$ and $\d$ ones, because of the
larger $\s$-quark mass. However, hyperon beams is a little-used part
of the program, included only for a few specific studies. Therefore
a simple approach has been taken, in which an average valence
quark distribution is constructed as
$f_{\mrm{val}} = (f_{\u,\mrm{val}}^{\p} + f_{\d,\mrm{val}}^{\p})/3$, 
according to
which each valence quark in a hyperon is assumed to be distributed.
Sea-quark and gluon distributions are taken as in the proton.
Any proton parton distribution set may be used with this procedure.
 
\subsubsection{Mesons and photons}
 
Data on meson parton distributions are scarce, so only very few sets
have been constructed, and only for the $\pi^{\pm}$. {\Py} contains
the Owens set 1 and 2 parton distributions \cite{Owe84}, which for a
long time were essentially the only sets on the market, and the
more recent dynamically generated GRV LO (updated version) 
\cite{Glu92a}. The latter one is the default in {\Py}. Further sets 
are found in \tsc{Pdflib} and can therefore be used by {\Py}, just as
described above for protons.

Like the proton was used as a template for simple hyperon sets,
so also the pion is used to derive a crude ansatz for  
$\K^{\pm}/\K_{\mrm{S}}^0/\K_{\mrm{L}}^0$. The procedure is the 
same, except that now $f_{\mrm{val}} = 
(f_{\u,\mrm{val}}^{\pi^+} + f_{\dbar,\mrm{val}}^{\pi^+})/2$.
 
Sets of photon parton distributions have been obtained as for
hadrons; an additional complication comes from the necessity to
handle the matching of the vector meson dominance (VMD) and the
perturbative pieces in a consistent manner. New sets have been 
produced where this division is explicit and therefore especially
well suited for applications to event generation\cite{Sch95}.
The Schuler and Sj\"ostand set 1D is the default. Although the 
vector-meson philosophy is at the base, the details of the fits do 
not rely on pion data, but only on $F_2^{\gamma}$ data. Here 
follows a brief summary of relevant details.

Real photons obey a set of inhomogeneous evolution equations, where the 
inhomogeneous term is induced by $\gamma \to \q\qbar$ branchings.
The solution can be written as the sum of two terms,
\begin{equation}
f_a^{\gamma}(x,Q^2) = f_a^{\gamma,\mrm{NP}}(x,Q^2;Q_0^2) 
+ f_a^{\gamma,\mrm{PT}}(x,Q^2;Q_0^2) ~,
\end{equation}
where the former term is a solution to the homogeneous evolution
with a (nonperturbative) input at $Q=Q_0$ and the latter is a
solution to the full inhomogeneous equation with boundary condition
$f_a^{\gamma,\mrm{PT}}(x,Q_0^2;Q_0^2) \equiv 0$. One possible 
physics interpretation is to let $f_a^{\gamma,\mrm{NP}}$ correspond
to $\gamma \leftrightarrow V$ fluctuations, where 
$V = \rho^0, \omega, \phi, \ldots$ is a set of vector mesons,
and let $f_a^{\gamma,\mrm{PT}}$ correspond to perturbative (`anomalous')
$\gamma \leftrightarrow \q\qbar$ fluctuations. The discrete spectrum 
of vector mesons can be combined with the continuous (in virtuality 
$k^2$) spectrum of $\q\qbar$ fluctuations, to give
\begin{equation}
f_a^{\gamma}(x,Q^2) =
\sum_V \frac{4\pi\alphaem}{f_V^2} f_a^{\gamma,V}(x,Q^2) 
+ \frac{\alphaem}{2\pi} \, \sum_{\q} 2 e_{\q}^2 \, 
\int_{Q_0^2}^{Q^2} \frac{{\d} k^2}{k^2} \, 
f_a^{\gamma,\q\qbar}(x,Q^2;k^2) ~,
\label{eq:decomprealga}
\end{equation}
where each component $f^{\gamma,V}$ and $f^{\gamma,\q\qbar}$ obeys a
unit momentum sum rule.
  
In sets 1 the $Q_0$ scale is picked at a low value, 0.6~GeV, where
an identification of the nonperturbative component with a set of
low-lying mesons appear natural, while sets 2 use a higher value,
2~GeV, where the validity of perturbation theory is better established.
The data are not good enough to allow a precise determination of
$\Lambda_{\mrm{QCD}}$. Therefore we use a fixed value 
$\Lambda^{(4)} = 200$~MeV, in agreement with conventional 
results for proton distributions. In the VMD component the $\rho^0$ 
and $\omega$ have been added coherently, so that
$\u\ubar : \d\dbar = 4 : 1$ at $Q_0$.

Unlike the $\p$, the $\gamma$ has a direct component where the photon 
acts as an unresolved probe. In the definition of $F_2^{\gamma}$ this 
adds a component $C^{\gamma}$, symbolically
\begin{equation}
F_2^{\gamma}(x,Q^2) = \sum_{\q} e_{\q}^2 \left[ f_{\q}^{\gamma} + 
f_{\qbar}^{\gamma} \right] \otimes C_{\q} + 
f_{\g}^{\gamma} \otimes C_{\g} + C^{\gamma} ~.
\end{equation}
Since $C^{\gamma} \equiv 0$ in leading order, and since we stay with 
leading-order fits, it is permissible to neglect this complication.
Numerically, however, it makes a non-negligible difference. We 
therefore make two kinds of fits, one DIS type with $C^{\gamma} = 0$
and one {\MSbar} type including the universal part
of $C^{\gamma}$.

When jet production is studied for real incoming photons, the standard 
evolution approach is reasonable also for heavy flavours, i.e.\ 
predominantly the $\c$, but with a lower cut-off $Q_0 \approx m_{\c}$ 
for $\gamma \to \c\cbar$. Moving to Deeply Inelastic Scattering, 
$\e\gamma \to \e X$, there is an extra kinematical constraint:
$W^2 = Q^2 (1-x)/x > 4 m_{\c}^2$. It is here better to use the
`Bethe-Heitler' cross section for $\gamma^* \gamma \to \c\cbar$.
Therefore each distribution appears in two variants. For applications
to real $\gamma$'s the parton distributions are calculated as the sum 
of a vector-meson part and an anomalous part including all five flavours.
For applications to DIS, the sum runs over the same vector-meson part, 
an anomalous part and possibly a $C^{\gamma}$ part for the three 
light flavours, and a Bethe-Heitler part for $\c$ and $\b$. 

In version 2 of the SaS distributions, which are the ones found here, the 
extension from real to virtual photons was improved, and further options
made available \cite{Sch96}. The resolved components of the photon
are dampened by phenomenologically motivated virtuality-dependent 
dipole factors, while the direct ones are explicitly calculable.
Thus eq.~(\ref{eq:decomprealga}) generalizes to
\begin{eqnarray}
f_a^{\gamma^\star}(x,Q^2,P^2)
& = & \sum_V \frac{4\pi\alphaem}{f_V^2} \left(
\frac{m_V^2}{m_V^2 + P^2} \right)^2 \,
f_a^{\gamma,V}(x,Q^2;\tilde{Q}_0^2)
\nonumber \\
& + & \frac{\alphaem}{2\pi} \, \sum_{\q} 2 e_{\q}^2 \,
\int_{Q_0^2}^{Q^2} \frac{{\d} k^2}{k^2} \, \left(
\frac{k^2}{k^2 + P^2} \right)^2 \, f_a^{\gamma,\q\qbar}(x,Q^2;k^2)
~.
\label{eq:decompvirtga}
\end{eqnarray}
In addition to the introduction of the dipole form factors,
note that the lower input scale for the VMD states is here shifted
from $Q_0^2$ to some $\tilde{Q}_0^2 \geq Q_0^2$. This is based on
a study of the evolution equation \cite{Bor93} that shows that
the evolution effectively starts `later' in $Q^2$ for a virtual
photon. Equation~(\ref{eq:decompvirtga}) is one possible answer. It 
depends on both $Q^2$ and $P^2$ in a non-trivial way, however, so that 
results are only obtained by a time-consuming numerical integration 
rather than as a simple parametrization. Therefore several other
alternatives are offered, that are in some sense equivalent, but can
be given in simpler form. 

In addition to the SaS sets, {\Py} also contains the Drees--Grassie set 
of parton distributions \cite{Dre85} and, as for the proton, there is 
an interface to the \tsc{Pdflib} library \cite{Plo93}. These calls are 
made with photon virtuality $P^2$ below the hard process scale 
$Q^2$. Further author-recommended constrains are implemented in the 
interface to the GRS set \cite{Glu99} which, along with SaS, is among 
the few also to define parton distributions of virtual photons.
However, these sets do not allow a subdivision of the photon parton
distributions into one VMD part and one anomalous part. This 
subdivision is necessary a sophisticated modelling of $\gamma\p$ and
$\gamma\gamma$ events, see above and section \ref{sss:photoprod}. 
As an alternative, for the VMD part alone, the 
$\rho^0$ parton distribution can be found from the assumed equality
\begin{equation}
f^{\rho^0}_i = f^{\pi^0}_i = \frac{1}{2} \, (f^{\pi^+}_i +
f^{\pi^-}_i) ~.
\end{equation}
Thus any $\pi^+$ parton distribution set, from any library, can be 
turned into a VMD $\rho^0$ set. The $\omega$ parton distribution is 
assumed the same, while the $\phi$ and $\Jpsi$ ones are handled 
in the very crude approximation 
$f^{\phi}_{\s,\mrm{val}} = f^{\pi^+}_{\u,\mrm{val}}$ and
$f^{\phi}_{\mrm{sea}} = f^{\pi^+}_{\mrm{sea}}$.
The VMD part needs to be complemented by an 
anomalous part to make up a full photon distribution. The latter
is fully perturbatively calculable, given the lower cut-off scale
$Q_0$. The SaS parameterization of the anomalous part is therefore used 
throughout for this purpose. The $Q_0$ scale can be set freely in the
\ttt{PARP(15)} parameter.

The $f_i^{\gamma,\mrm{anom}}$ distribution can be further decomposed, 
by the flavour and the $\pT$ of the original branching 
$\gamma \to \q\qbar$. The flavour is distributed according to squared
charge (plus flavour thresholds for heavy flavours) and the $\pT$
according to $\d \pT^2 / \pT^2$ in the range $Q_0 < \pT < Q$.
At the branching scale, the photon only consists of a $\q\qbar$ pair,
with $x$ distribution $\propto x^2 + (1-x)^2$. A component
$f_a^{\gamma,\q\qbar}(x,Q^2;k^2)$, characterized by its $k \approx \pT$ 
and flavour, then is evolved homogeneously from $\pT$ to $Q$. For 
theoretical studies it is convenient to be able to access a specific 
component of this kind. Therefore also leading-order parameterizations 
of these decomposed distributions are available \cite{Sch95}.    

\subsubsection{Leptons}
\label{sss:estructfun}
 
Contrary to the hadron case, there is no necessity to introduce the
parton-distribution function concept for leptons. A lepton can be 
considered as a point-like particle, with initial-state radiation 
handled by higher-order matrix elements. However, the parton 
distribution function approach offers a slightly simplified but very 
economical description of initial-state radiation effects for any hard 
process, also those for which higher-order corrections are not yet 
calculated.
 
Parton distributions for electrons have been introduced in {\Py},
and are used also for muons and taus, with a trivial substitution 
of masses. Alternatively, one is free to 
use a simple `unresolved' $\e$, $f_{\e}^{\e}(x, Q^2) = \delta(x-1)$, 
where the $\e$ retains the full original momentum.
 
Electron parton distributions are calculable entirely from first
principles, but different levels of approximation may be used.
The parton-distribution formulae in {\Py} are based on a
next-to-leading-order exponentiated description, see ref.
\cite{Kle89}, p. 34. The approximate behaviour is
\begin{eqnarray}
& & f_{\e}^{\e}(x,Q^2) \approx \frac{\beta}{2}
(1-x)^{\beta/2-1}   ~;              \nonumber \\
& & \beta = \frac{2 \alphaem}{\pi}
\left( \ln \frac{Q^2}{m_{\e}^2} -1 \right) ~.
\end{eqnarray}
The form is divergent but integrable for $x \to 1$, i.e.\ the electron
likes to keep most of the energy. To handle the numerical precision
problems for $x$ very close to unity, the parton distribution is set, by
hand, to zero for $x > 1-10^{-10}$, and is rescaled upwards in the range
$1-10^{-7} < x < 1-10^{-10}$, in such a way that the total area under the
parton distribution is preserved:
\begin{equation}
\left( f_{\e}^{\e}(x,Q^2) \right)_{\mrm{mod}} =
\left\{ \begin{array}{ll}
 f_{\e}^{\e}(x,Q^2) & 0 \leq x \leq 1-10^{-7} \\[2mm]
 \frac{\displaystyle 1000^{\beta/2}}{\displaystyle 1000^{\beta/2}-1}
\, f_{\e}^{\e}(x,Q^2) &  1-10^{-7} < x < 1-10^{-10}  \\[4mm]
0 & x > 1-10^{-10} \, ~.
\end{array} \right.
\end{equation}

A separate issue is that electron beams may not be monochromatic,
more so than for other particles because of the small electron mass.
In storage rings the main mechanism is synchrotron radiation.
For future high-luminosity linear colliders, the beam--beam 
interactions at the collision vertex may induce a quite significant
energy loss --- `beamstrahlung'. Note that neither of these are 
associated with any off-shellness of the electrons, i.e. the resulting
spectrum only depends on $x$ and not $Q^2$. Examples of beamstrahlung
spectra are provided by the \tsc{Circe} program \cite{Ohl97}, with a 
sample interface on the {\Py} webpages.
 
The branchings $\e \to \e \gamma$, which are responsible for the
softening of the $f_{\e}^{\e}$ parton distribution, also gives rise to
a flow of photons. In photon-induced hard processes, the
$f_{\gamma}^{\e}$ parton distribution can be used to describe the
equivalent flow of photons. In the next section, a complete 
differential photon flux machinery is introduced. Here some simpler 
first-order expressions are introduced, for the flux integrated up
to a hard interaction scale $Q^2$. There is some ambiguity in
the choice of $Q^2$ range over which emissions should be included.
The na\"{\i}ve (default) choice is
\begin{equation}
f_{\gamma}^{\e}(x,Q^2) = \frac{\alphaem}{2 \pi} \,
\frac{1+(1-x)^2}{x} \, \ln \left( \frac{Q^2}{m_{\e}^2} \right) ~.
\end{equation}
Here it is assumed that only one scale enters the problem, namely
that of the hard interaction, and that the scale of the branching
$\e \to \e \gamma$ is bounded from above by the hard interaction scale.
For a pure QCD or pure QED shower this is an appropriate procedure,
cf. section \ref{sss:showermatching}, but in other cases it may not be
optimal. In particular, for photoproduction the alternative that is
probably most appropriate is \cite{Ali88}:
\begin{equation}
f_{\gamma}^{\e}(x,Q^2) =
\frac{\alphaem}{2 \pi} \, \frac{1+(1-x)^2}{x}
\, \ln \left( \frac{Q^2_{\mmax} (1-x)}{m_{\e}^2 \, x^2} \right) ~.
\end{equation}
Here $Q^2_{\mmax}$ is a user-defined cut for the range of
scattered electron kinematics that is counted as
photoproduction. Note that we now deal with two different $Q^2$
scales, one related to the hard subprocess itself, which appears
as the argument of the parton distribution, and the other related to
the scattering of the electron, which is reflected in 
$Q^2_{\mmax}$.

Also other sources of photons should be mentioned. One is the 
beamstrahlung photons mentioned above, where again \tsc{Circe} 
provides sample parameterizations. Another, particularly interesting 
one, is laser backscattering, wherein an intense laser pulse is shot
at an incoming high-energy electron bunch. By Compton backscattering
this gives rise to a photon energy spectrum with a peak at a 
significant fraction of the original electron energy \cite{Gin82}.
Both of these sources produce real photons, which can be considered as
photon beams of variable energy (see subsection \ref{ss:PYvaren}), 
decoupled from the production process proper.
 
In resolved photoproduction or resolved $\gamma\gamma$
interactions, one has to include the parton distributions for quarks
and gluons inside the photon inside the electron. This is best done
with the machinery of the next section. However, as an older and 
simpler alternative, $f_{\q,\g}^{\e}$ can be obtained by a numerical
convolution according to
\begin{equation}
f_{\q,\g}^{\e}(x, Q^2) =
\int_x^1 \frac{dx_{\gamma}}{x_{\gamma}} \, 
f_{\gamma}^{\e}(x_{\gamma}, Q^2) \, f^{\gamma}_{\q,\g} \! 
\left( \frac{x}{x_{\gamma}}, Q^2 \right)   ~,
\label{pg:foldqgine}
\end{equation}
with $f^{\e}_{\gamma}$ as discussed above. The necessity for numerical
convolution makes this parton distribution evaluation rather slow
compared with the others; one should therefore only have it
switched on for resolved photoproduction studies.
 
One can obtain the positron distribution inside an electron,
which is also the electron sea parton distribution, by a convolution
of the two branchings $\e \to \e \gamma$ and $\gamma \to \ee$;
the result is \cite{Che75}
\begin{equation}
f_{\e^+}^{\e^-}(x,Q^2) =
\frac{1}{2} \, \left\{ \frac{\alphaem}{2 \pi} \,
\left( \ln \frac{Q^2}{m_{\e}^2} -1 \right) \right\}^2 \,
\frac{1}{x} \, \left( \frac{4}{3} - x^2 - \frac{4}{3} x^3 +
2x(1+x) \ln x \right)   ~.
\end{equation}
 
Finally, the program also contains the distribution of a
transverse $\W^-$ inside an electron
\begin{equation}
f_{\W}^{\e}(x,Q^2) = \frac{\alphaem}{2 \pi} \,
\frac{1}{4 \ssintw} \, \frac{1+(1-x)^2}{x} \,
\ln \left( 1 + \frac{Q^2}{m_{\W}^2} \right)    ~.
\end{equation}
 
\subsubsection{Equivalent photon flux in leptons}
\label{sss:equivgamma}

With the {\galep} option of a \ttt{PYINIT} call,
an $\e\p$ or $\e^+\e^-$ event (or corresponding processes with muons) 
is factorized into the flux of virtual photons and the subsequent
interactions of such photons. While real photons always are transverse, 
the virtual photons also allow a longitudinal component.
This corresponds to cross sections
\begin{equation}
\d\sigma(\e\p\rightarrow\e\mbf{X})=\sum_{\xi=\mathrm{T,L}}
\int \! \int \d y \, \d Q^2 
\;f_{\gamma/\e}^{\xi}(y,Q^2) 
\;\d\sigma(\gast_{\xi}\p\rightarrow\mbf{X})
\end{equation}
and 
\begin{equation}
\d\sigma(\e\e\rightarrow\e\e\mbf{X})=\!\!\!\sum_{\xi_1,\xi_2=\mathrm{T,L}}
\int \! \int \! \int \!\d y_1 \, \d Q_1^2 \, \d y_2 \, \d Q_2^2 
\;f_{\gamma/\e}^{\xi_1}(y_1,Q_1^2) f_{\gamma/\e}^{\xi_2}(y_2,Q_2^2)
\;\d\sigma(\gast_{\xi_1}\gast_{\xi_2}\rightarrow\mbf{X})\;.
\end{equation}
For $\e\p$ events, this factorized ansatz is perfectly general, so long 
as azimuthal distributions in the final state are not studied in detail.
In $\e^+\e^-$ events, it is not a good approximation when the 
virtualities $Q_1^2$ and $Q_2^2$ of both photons become of the order of 
the squared invariant mass $W^2$ of the colliding photons 
\cite{Sch98}. In this region the cross section have terms that depend 
on the relative azimuthal angle of the scattered leptons, and 
the transverse and longitudinal polarizations are non-trivially mixed. 
However, these terms are of order $Q_1^2Q_2^2/W^2$ and can 
be neglected whenever at least one of the photons has low virtuality 
compared to $W^2$. 

When $Q^2/W^2$ is small, one can derive \cite{Bon73,Bud75,Sch98}
\begin{eqnarray}
f_{\gamma/l}^{\mathrm{T}}(y,Q^2) & = & \frac{\alphaem}{2\pi} 
\left( \frac{(1+(1-y)^2}{y} \frac{1}{Q^2}-\frac{2m_{l}^2y}{Q^4} 
\right)\;,\\
f_{\gamma/l}^{\mathrm{L}}(y,Q^2) & = & \frac{\alphaem}{2\pi} 
\frac{2(1-y)}{y} \frac{1}{Q^2}\;, 
\end{eqnarray}
where $l=\e^{\pm},~\mu^{\pm}$ or~$\tau^{\pm}$. 
In $f_{\gamma/l}^{\mathrm{T}}$ the second term, proportional to 
$m_{l}^2/Q^4$, is not leading log and is therefore often 
omitted. Clearly it is irrelevant at large $Q^2$, but around the lower 
cut-off $Q^2_{\mathrm{min}}$ it significantly dampens the small-$y$ rise 
of the first term. (Note that $Q^2_{\mathrm{min}}$ is $y$-dependent,
so properly the dampening is in a region of the $(y,Q^2)$ plane.)
Overall, under realistic conditions, it reduces event rates by 5--10\% 
\cite{Sch98,Fri93}.

The $y$ variable is defined as the light-cone fraction the photon takes 
of the incoming lepton momentum. For instance, for $l^+l^-$ events, 
\begin{equation}
y_i = \frac{q_i k_j}{k_i k_j} ~, \qquad j=2 (1)~\mathrm{for}~i=1 (2) ~,  
\end{equation} 
where the $k_i$ are the incoming lepton four-momenta and the $q_i$
the four-momenta of the virtual photons.

Alternatively, the energy fraction the photon takes in the rest frame 
of the collision can be used,
\begin{equation}
x_i = \frac{q_i (k_1 + k_2)}{k_i (k_1 + k_2)} ~, \qquad i=1,2 ~.  
\end{equation} 
The two are simply related,
\begin{equation}
y_i = x_i + \frac{Q_i^2}{s} ~,
\end{equation}
with $s=(k_1 + k_2)^2$. (Here and in the following formulae we have
omitted the lepton and hadron mass terms when it is not of importance 
for the argumentation.) 
Since the Jacobian $\d(y_i, Q_i^2) / \d(x_i, Q_i^2) = 1$, either variable 
would be an equally valid choice for covering the phase space. 
Small $x_i$ values will be of less interest for us, since they lead 
to small $W^2$, so $y_i/x_i \approx 1$ except in the high-$Q^2$ tail, 
and often the two are used interchangeably. Unless special $Q^2$ cuts 
are imposed, cross sections obtained with 
$f_{\gamma/l}^{\mathrm{T,L}}(x,Q^2) \, \d x$ rather than
$f_{\gamma/l}^{\mathrm{T,L}}(y,Q^2) \, \d y$ differ only at the 
per mil level. For comparisons with experimental cuts, 
it is sometimes relevant to know which of the two is being used in 
an analysis.

In the $\e\p$ kinematics, the $x$ and $y$ definitions give that
\begin{equation}
W^2 = x s = y s - Q^2 ~.
\end{equation}
The $W^2$ expression for $l^+l^-$ is more complicated, especially 
because of the dependence on the relative azimuthal angle of the 
scattered leptons,
$\varphi_{12} = \varphi_1 - \varphi_2$:
\begin{eqnarray}
W^2 & = & x_1 x_2 s + \frac{2 Q_1^2 Q_2^2}{s} - 
2 \sqrt{1 - x_1 - \frac{Q_1^2}{s}} \sqrt{1 - x_2 - \frac{Q_2^2}{s}}
Q_1 Q_2 \cos\varphi_{12} \nonumber \\
& = & y_1 y_2 s - Q_1^2 - Q_2^2 + \frac{Q_1^2 Q_2^2}{s} - 
2 \sqrt{1 - y_1} \sqrt{1 - y_2} Q_1 Q_2 \cos\varphi_{12} ~.
\label{W2gaga}
\end{eqnarray}

The lepton scattering angle $\theta_i$ is related to $Q_i^2$ as
\begin{equation}
Q_i^2 = \frac{x_i^2}{1-x_i} m_i^2 + (1-x_i) \left(
s - \frac{2}{(1-x_i)^2} m_i^2 - 2 m_j^2 \right) \sin^2(\theta_i/2) ~, 
\end{equation}
with $m_i^2 = k_i^2 = {k'}^2_i$ and terms of $O(m^4)$ neglected.
The kinematical limits thus are
\begin{eqnarray}
 (Q_i^2)_{\mathrm{min}} & \approx & \frac{x_i^2}{1 - x_i} m_i^2 ~, \\
 (Q_i^2)_{\mathrm{max}} & \approx & (1 - x_i) s ~,
\end{eqnarray} 
unless experimental conditions reduce the $\theta_i$ ranges.

In summary, we will allow the possibility of experimental cuts in the
$x_i$, $y_i$, $Q_i^2$, $\theta_i$ and $W^2$ variables. Within the 
allowed region, the phase space is Monte Carlo sampled according to
$\prod_i (\d Q_i^2/Q_i^2) \, (\d x_i / x_i) \, \d \varphi_i$, with the 
remaining flux factors combined with the cross section factors to give 
the event weight used for eventual acceptance or rejection. This cross
section in its turn can contain the parton densities of a resolved 
virtual photon, thus offering an effective convolution that gives 
partons inside photons inside electrons.
 
\subsection{Kinematics and Cross Section for a Two-body Process}
\label{ss:kinemtwo}
 
In this section we begin the description of kinematics selection
and cross-section calculation. The example is for the case of a
$2 \to 2$ process, with final-state masses assumed to be vanishing.
Later on we will expand to finite fixed masses, and to resonances.
 
Consider two incoming beam particles in their c.m.\ frame, each with
energy $E_{\mrm{beam}}$. The total squared c.m.\ energy is then
$s = 4 E_{\mrm{beam}}^2$. The two partons that enter the hard 
interaction do not carry the total beam momentum, but only fractions 
$x_1$ and $x_2$, respectively, i.e.\ they have four-momenta
\begin{eqnarray}
   p_1 & = & E_{\mrm{beam}}(x_1; 0, 0, x_1) ~, \nonumber \\
   p_2 & = & E_{\mrm{beam}}(x_2; 0, 0, -x_2) ~.
\end{eqnarray}
There is no reason to put the incoming partons on the mass shell, 
i.e.\ to have time-like incoming four-vectors,
since partons inside a particle are always virtual and thus
space-like. These space-like virtualities are introduced as
part of the initial-state parton-shower description, see section
\ref{sss:initshowtrans}, but do not affect the formalism of this
section, wherefore massless incoming partons is a sensible ansatz. 
The one example where it would be appropriate to put a
parton on the mass shell is for an incoming lepton beam, but even
here the massless kinematics description is adequate as long as
the c.m.\ energy is correctly calculated with masses.
 
The squared invariant mass of the two partons is defined as
\begin{equation}
  \hat{s} = (p_1 + p_2)^2 = x_1 \, x_2 \, s ~.
\end{equation}
Instead of $x_1$ and $x_2$, it is often customary to use $\tau$
and either $y$ or $x_{\mrm{F}}$:
\begin{eqnarray}
  \tau & = & x_1 x_2 = \frac{\hat{s}}{s} ~;  \\
  y & = & \frac{1}{2} \ln \frac{x_1}{x_2} ~; \\
  x_{\mrm{F}} & = & x_1 - x_2 ~.
\end{eqnarray}
 
In addition to $x_1$ and $x_2$, two additional variables are needed
to describe the kinematics of a scattering $1 + 2 \to 3 + 4$.
One corresponds to the azimuthal angle $\varphi$ of the scattering
plane around the beam axis. This angle is always isotropically
distributed for unpolarized incoming beam particles, and so need not
be considered further. The other variable can be picked as
$\hat{\theta}$, the polar angle of parton 3 in the c.m.\ frame of the
hard scattering. The conventional choice is to use the variable
\begin{equation}
  \hat{t} = (p_1 - p_3)^2 = (p_2 - p_4)^2 =
  - \frac{\hat{s}}{2} (1 - \cos \hat{\theta}) ~,
\end{equation}
with $\hat{\theta}$ defined as above. In the following, we will make
use of both $\hat{t}$ and $\hat{\theta}$. It is also customary
to define $\hat{u}$,
\begin{equation}
  \hat{u} = (p_1 - p_4)^2 = (p_2 - p_3)^2 =
  - \frac{\hat{s}}{2} (1 + \cos \hat{\theta}) ~,
\end{equation}
but $\hat{u}$ is not an independent variable since
\begin{equation}
  \hat{s} + \hat{t} + \hat{u} = 0 ~.
\end{equation}
 
If the two outgoing particles have masses $m_3$ and $m_4$,
respectively, then the four-momenta in the c.m.\ frame of the hard
interaction are given by
\begin{equation}
\hat{p}_{3,4} = \left(
\frac{\hat{s} \pm (m_3^2-m_4^2)}{2\sqrt{\hat{s}}} ,
\pm \frac{\sqrt{\hat{s}}}{2} \, \beta_{34} \sin \hat{\theta}, 0,
\pm \frac{\sqrt{\hat{s}}}{2} \, \beta_{34} \cos \hat{\theta}
\right) ~,
\end{equation}
where
\begin{equation}
\beta_{34} = \sqrt{ \left( 1 - \frac{m_3^2}{\hat{s}} -
\frac{m_4^2}{\hat{s}} \right)^2 - 4 \, \frac{m_3^2}{\hat{s}} \,
\frac{m_4^2}{\hat{s}} }    ~.
\label{pg:betafactor}
\end{equation}
Then $\hat{t}$ and $\hat{u}$ are modified to
\begin{equation}
\hat{t}, \hat{u} = - \frac{1}{2} \left\{ ( \hat{s} - m_3^2 - m_4^2 )
\mp \hat{s} \, \beta_{34} \cos \hat{\theta} \right\} ~,
\label{pg:thatuhat}
\end{equation}
with
\begin{equation}
\hat{s} + \hat{t} + \hat{u} = m_3^2 + m_4^2 ~.
\end{equation}
 
The cross section for the process $1 + 2 \to 3 + 4$ may be written as
\begin{eqnarray}
\sigma & = & \int \int \int \d x_1 \, \d x_2 \, \d \hat{t} \,
f_1(x_1, Q^2) \, f_2(x_2, Q^2) \, \frac{\d \hat{\sigma}}{\d \hat{t}}
\nonumber \\
& = & \int \int \int \frac{\d \tau}{\tau} \, \d y \, \d \hat{t} \,
x_1 f_1(x_1, Q^2) \, x_2 f_2(x_2, Q^2) \,
\frac{\d \hat{\sigma}}{\d \hat{t}} ~.
\label{pg:sigma}
\end{eqnarray}
 
The choice of $Q^2$ scale is ambiguous, and several alternatives are
available in the program. For massless outgoing particles the default
is the squared transverse momentum
\begin{equation}
Q^2 = \hat{p}_{\perp}^2 = \frac{\hat{s}}{4} \sin^2\hat{\theta} =
    \frac{\hat{t}\hat{u}}{\hat{s}} ~,
\end{equation}
which is modified to
\begin{equation}
Q^2 = \frac{1}{2} (m_{\perp 3}^2 + m_{\perp 4}^2) =
\frac{1}{2} (m_3^2 + m_4^2) + \hat{p}_{\perp}^2 =
\frac{1}{2} (m_3^2 + m_4^2) +
\frac{\hat{t} \hat{u} - m_3^2 m_4^2}{\hat{s}}
\end{equation}
when masses are introduced in the final state. The mass term is 
selected such that, for $m_3 = m_4 = m$, the expression reduces to 
the squared transverse mass, 
$Q^2 = \hat{m}_{\perp}^2 = m^2 + \hat{p}_{\perp}^2$.
For cases with spacelike virtual incoming photons, of virtuality
$Q_i^2 = - m_i^2 = |p_i^2|$, a further generalization to 
\begin{equation}
Q^2 = \frac{1}{2} (Q_1^2 + Q_2^2 + m_3^2 + m_4^2) + \hat{p}_{\perp}^2
\end{equation}
is offered. 

The $\d \hat{\sigma}/\d \hat{t}$ expresses the differential
cross section for a scattering, as a function of the kinematical
quantities $\hat{s}$, $\hat{t}$ and $\hat{u}$.
It is in this function that the physics of a given process
resides.
 
The performance of a machine is measured in terms of its
luminosity $\cal L$, which is directly proportional to the
number of particles in each bunch and to the bunch crossing
frequency, and inversely proportional to the area of the bunches at
the collision point. For a process with a $\sigma$ as given by
eq.~(\ref{pg:sigma}), the differential event rate is given
by $\sigma {\cal L}$, and the number of events collected
over a given period of time
\begin{equation}
N = \sigma \, \int {\cal L} \, \d t ~.
\end{equation}
The program does not calculate the number of events, but only the
integrated cross sections.
 
\subsection{Resonance Production}
\label{ss:kinemreson}
 
The simplest way to produce a resonance is by a $2 \to 1$ process.
If the decay of the resonance is not considered, the cross-section
formula does not depend on $\hat{t}$, but takes the form
\begin{equation}
\sigma = \int \int \frac{\d \tau}{\tau} \, \d y \,
x_1 f_1(x_1, Q^2) \, x_2 f_2(x_2, Q^2) \,
\hat{\sigma}(\hat{s})  ~.
\label{pg:sigmares}
\end{equation}
Here the physics is contained in the cross section
$\hat{\sigma}(\hat{s})$. The $Q^2$ scale is usually taken to
be $Q^2 = \hat{s}$.
 
In published formulae, cross sections are often given in
the zero-width approximation, i.e.
$\hat{\sigma}(\hat{s}) \propto \delta (\hat{s} - m_R^2)$,
where $m_R$ is the mass of the resonance. Introducing the
scaled mass $\tau_R = m_R^2/s$, this corresponds to a delta
function $\delta (\tau - \tau_R)$, which can be used to
eliminate the integral over $\tau$.
 
However, what we normally want to do is replace the $\delta$
function by the appropriate Breit--Wigner shape. For a resonance
width $\Gamma_R$ this is achieved by the replacement
\begin{equation}
\delta (\tau - \tau_R) \to \frac{s}{\pi} \,
\frac{m_R \Gamma_R}{(s \tau - m_R^2)^2 + m_R^2 \Gamma_R^2}  ~.
\label{pg:resshapeone}
\end{equation}
In this formula the resonance width $\Gamma_R$ is a constant.
 
An improved description of resonance shapes is obtained if
the width is made $\hat{s}$-dependent (occasionally also 
referred to as mass-dependent width, since $\hat{s}$ is not
always the resonance mass), see e.g.\ \cite{Ber89}.
To first approximation, this means that the expression
$m_R \Gamma_R$ is to be replaced by $\hat{s} \Gamma_R / m_R$,
both in the numerator and the denominator. An intermediate
step is to perform this replacement only in the numerator.
This is convenient when not only $s$-channel resonance 
production is simulated but also non-resonance $t$- or
$u$-channel graphs are involved, since mass-dependent widths
in the denominator here may give an imperfect cancellation of
divergences. (More about this below.)  

To be more precise, in the program the quantity $H_R(\hat{s})$
is introduced, and the Breit--Wigner is written as
\begin{equation}
\delta (\tau - \tau_R) \to \frac{s}{\pi} \,
\frac{H_R(s \tau)}{(s \tau - m_R^2)^2 + H_R^2(s \tau)}  ~.
\label{pg:resshapetwo}
\end{equation}
The $H_R$ factor is evaluated as a sum over all possible final-state
channels, $H_R = \sum_f H_R^{(f)}$. Each decay channel may have its
own $\hat{s}$ dependence, as follows.
 
A decay to a fermion pair, $R \to \f \fbar$, gives no contribution
below threshold, i.e.\ for $\hat{s} < 4 m_{\f}^2$. Above threshold,
$H_R^{(f)}$ is proportional to $\hat{s}$, multiplied by a threshold
factor $\beta (3 - \beta^2)/2$ for the vector part of a spin 1
resonance, by $\beta^3$ for the axial vector part, by $\beta^3$ for 
a scalar resonance and by $\beta$ for a pseudoscalar one. Here
$\beta = \sqrt{1 - 4m_{\f}^2/\hat{s}}$.
For the decay into unequal masses, e.g.\ of the $\W^+$, corresponding
but more complicated expressions are used.

For decays into a quark pair, a first-order strong correction factor 
$1 + \alphas(\hat{s}) / \pi$ is included in $H_R^{(f)}$. This is 
the correct choice for all spin 1 colourless resonances, but is here 
used for all resonances where no better knowledge is available.
Currently the major exception is top decay, where the
factor $1 - 2.5 \, \alphas(\hat{s}) / \pi$ is used to approximate 
loop corrections \cite{Jez89}.
The second-order corrections are often known, but then are specific
to each resonance, and are not included. An option exists for the
$\gamma/\Z^0/\Z'^0$ resonances, where threshold effects due to
$\q\qbar$ bound-state formation are taken into account in a
smeared-out, average sense, see eq.~(\ref{pp:threshenh}).
 
For other decay channels, not into fermion pairs, the $\hat{s}$
dependence is typically more complicated. An example would be the 
decay $\hrm^0 \to \W^+ \W^-$, with a nontrivial threshold and a subtle 
energy dependence above that \cite{Sey95a}. Since a Higgs
with $m_{\hrm} < 2 m_{\W}$ could still decay in this channel, it is 
in fact necessary to perform a two-dimensional integral over
the $W^{\pm}$ Breit--Wigner mass distributions to obtain the correct
result (and this has to be done numerically, at least in part).
Fortunately, a Higgs particle lighter than $2 m_{\W}$ is
sufficiently narrow that the integral only needs to be performed
once and for all at initialization (whereas most other partial
widths are recalculated whenever needed). Channels that proceed
via loops, such as $\hrm \to \g \g$, also display complicated
threshold behaviours.
 
The coupling structure within the electroweak sector is usually
(re)expressed in terms of gauge boson masses, $\alphaem$ and
$\ssintw$, i.e.\ factors of $G_{\F}$ are replaced according to
\begin{equation}
\sqrt{2} G_{\F} = \frac{\pi \, \alphaem}{\ssintw \, m_{\W}^2} ~.
\end{equation}
Having done that, $\alphaem$ is allowed to run \cite{Kle89}, 
and is evaluated at the $\hat{s}$ scale. Thereby the relevant
electroweak loop correction factors are recovered at the
$m_{\W}/m_{\Z}$ scale. However, the option exists to
go the other way and eliminate $\alphaem$ in favour of $G_{\F}$.  
Currently $\ssintw$ is not allowed to run.

For Higgs particles and technipions, fermion masses enter not only 
in the kinematics but also as couplings. The latter kind of quark masses
(but not the former, at least not in the program) are running with the 
scale of the process, i.e.\ normally the resonance mass. The expression 
used is \cite{Car96}
\begin{equation}
m(Q^2) = m_0 \left( \frac{\ln(k^2 m_0^2/\Lambda^2)}{\ln(Q^2/\Lambda^2)}
\right)^{12/(33 - 2 n_{\f})} ~.
\label{eq:runqmass}
\end{equation}
Here $m_0$ is the input mass at a reference scale $k m_0$, defined in the
{\MSbar} scheme. Typical choices are either $k=1$ or $k=2$; the latter 
would be relevant if the reference scale is chosen at the $ \Q \Qbar$ 
threshold. Both $\Lambda$ and $n_{\f}$ are as given in $\alphas$.  
 
In summary, we see that an $\hat{s}$ dependence may enter several
different ways into the $H_R^{(f)}$ expressions from which the
total $H_R$ is built up. 
 
When only decays to a specific final state $f$ are considered, the
$H_R$ in the denominator remains the sum over all allowed decay
channels, but the numerator only contains the $H_R^{(f)}$ term
of the final state considered.
 
If the combined production and decay process $i \to R \to f$ is
considered, the same $\hat{s}$ dependence is implicit in the
coupling structure of $i \to R$ as one would have had in
$R \to i$, i.e.\ to first approximation there is a symmetry between
couplings of a resonance to the initial and to the final state.
The cross section $\hat{\sigma}$
is therefore, in the program, written in the form
\begin{equation}
\hat{\sigma}_{i \to R \to f}(\hat{s}) \propto \frac{\pi}{\hat{s}} \,
\frac{H_R^{(i)}(\hat{s}) \, H_R^{(f)}(\hat{s})}
{(\hat{s} - m_R^2)^2 + H_R^2(\hat{s})} ~.
\label{pg:Hinoutsym}
\end{equation}
As a simple example, the cross section for the process
$\e^- \br{\nu}_{\e} \to \W^- \to \mu^- \br{\nu}_{\mu}$
can be written as
\begin{equation}
\hat{\sigma}(\hat{s}) = 12 \, \frac{\pi}{\hat{s}} \,
\frac{H_{\W}^{(i)}(\hat{s}) \, H_{\W}^{(f)}(\hat{s})}
{(\hat{s} - m_{\W}^2)^2 + H_{\W}^2(\hat{s})} ~,
\end{equation}
where
\begin{equation}
H_{\W}^{(i)}(\hat{s}) = H_{\W}^{(f)}(\hat{s}) =
\frac{\alphaem(\hat{s})}{24 \, \ssintw} \, \hat{s} ~.
\end{equation}
If the effects of several initial and/or final states are studied,
it is straightforward to introduce an appropriate summation in the
numerator.
 
The analogy between the $H_R^{(f)}$ and $H_R^{(i)}$ cannot be pushed
too far, however. The two differ in several important aspects.
Firstly, colour factors appear reversed: the decay $R \to \q\qbar$
contains a colour factor $N_C = 3$ enhancement, while
$\q\qbar \to R$ is instead suppressed by a factor $1/N_C = 1/3$.
Secondly, the $1 + \alphas(\hat{s}) / \pi$ first-order correction
factor for the final state has to be replaced by a more complicated
$K$ factor for the initial state. This factor is not known usually, or 
it is known (to first non-trivial order) but too lengthy to be included
in the program. Thirdly, incoming partons as a rule are space-like.
All the threshold suppression factors of the final-state expressions
are therefore irrelevant when production is considered. In sum, the
analogy between $H_R^{(f)}$ and $H_R^{(i)}$ is mainly useful as a
consistency cross-check, while the two usually are
calculated separately. Exceptions include the rather
messy loop structure involved in $\g\g \to \hrm^0$ and $\hrm^0 \to \g\g$,
which is only coded once.
 
It is of some interest to consider the observable resonance shape
when the effects of parton distributions are included. In a
hadron collider, to first approximation, parton distributions tend
to have a behaviour roughly like $f(x) \propto 1/x$ for small $x$
--- this is why $f(x)$ is replaced by $xf(x)$ in eq.
(\ref{pg:sigma}). Instead, the basic parton-distribution behaviour
is shifted into the factor of $1/\tau$ in the integration phase space
$\d \tau/\tau$, cf. eq.~(\ref{pg:sigmares}). When convoluted with the
Breit--Wigner shape, two effects appear. One is that the overall
resonance is tilted: the low-mass tail is enhanced and the high-mass
one suppressed. The other is that an extremely long tail develops
on the low-mass side of the resonance: when $\tau \to 0$, eq.
(\ref{pg:Hinoutsym}) with $H_R(\hat{s}) \propto \hat{s}$ gives
a $\hat{\sigma}(\hat{s}) \propto \hat{s} \propto \tau$, which
exactly cancels the $1/\tau$ factor mentioned above. Na\"{\i}vely, the
integral over $y$, $\int \d y = - \ln \tau$, therefore gives a
net logarithmic divergence of the resonance shape when $\tau \to 0$.
Clearly, it is then necessary to consider the shape of the 
parton distributions in more detail. At not-too-small $Q^2$, 
the evolution
equations in fact lead to parton distributions more strongly
peaked than $1/x$, typically with $xf(x) \propto x^{-0.3}$, and
therefore a divergence like $\tau^{-0.3}$ in the cross-section
expression. Eventually this divergence is regularized by a closing
of the phase space, i.e.\ that $H_R(\hat{s})$ vanishes faster than
$\hat{s}$, and by a less drastic small-$x$ parton-distribution
behaviour when $Q^2 \approx \hat{s} \to 0$.
 
The secondary peak at small $\tau$ may give a rather high
cross section, which can even rival that of the ordinary
peak around the nominal mass. This is the case, for instance,
with $\W$ production. Such a peak has never been observed
experimentally, but this is not surprising, since the background
from other processes is overwhelming at low $\hat{s}$.
Thus a lepton of one or a few GeV of transverse momentum is far
more likely to come from the decay of a charm or bottom hadron than
from an extremely off-shell $\W$ of a mass of a few GeV. When 
resonance production is studied, it is therefore important to set 
limits on the mass of the resonance, so as to agree with the 
experimental definition, at least to first approximation. If not, 
cross-section information given by the program may be very confusing.

Another problem is that often the matrix elements really are valid
only in the resonance region. The reason is that one usually includes 
only the simplest $s$-channel graph in the calculation. It is this 
`signal' graph that has a peak at the position of the resonance, 
where it (usually) gives much larger cross sections than the other
`background' graphs. Away from the resonance position, `signal' and
`background' may be of comparable order, or the `background' may 
even dominate. There is a quantum mechanical interference when some 
of the `signal' and `background' graphs have the same initial and 
final state, and this interference may be destructive or constructive.
When the interference is non-negligible, it is no longer meaningful 
to speak of a `signal' cross section. As an example, consider the 
scattering of longitudinal $\W$'s, 
$\W^+_{\mrm{L}} \W^-_{\mrm{L}} \to \W^+_{\mrm{L}} \W^-_{\mrm{L}}$,
where the `signal' process is $s$-channel exchange of a Higgs.
This graph by itself is ill-behaved away from the resonance region.
Destructive interference with `background' graphs such as $t$-channel 
exchange of a Higgs and $s$- and $t$-channel exchange of a $\gamma/\Z$ 
is required to save unitarity at large energies.  
 
In $\ee$ colliders, the $f_{\e}^{\e}$ parton distribution is peaked
at $x = 1$ rather than at $x = 0$. The situation therefore is the
opposite, if one considers e.g.\ $\Z^0$ production in a machine
running at energies above $m_{\Z}$: the tail towards lower masses
is suppressed and the one towards higher masses enhanced, with a 
sharp secondary peak at around the nominal energy of the machine.
Also in this case, an appropriate definition of cross sections  
therefore is necessary --- with additional complications due to the 
interference between $\gamma^*$ and $\Z^0$. When other processes are
considered, problems of interference with background appears also 
here. Numerically the problems may be less pressing, however,
since the secondary peak is occurring in a high-mass region, rather 
than in a more complicated low-mass one. Further, in $\ee$ there is 
little uncertainty from the shape of the parton distributions. 
 
In $2 \to 2$ processes where a pair of resonances are produced, e.g.
$\ee \to \Z^0 \hrm^0$, cross section are almost always given
in the zero-width approximation for the resonances. Here
two substitutions of the type
\begin{equation}
1 = \int \delta (m^2 - m_R^2) \, dm^2
\to \int \frac{1}{\pi} \,
\frac{m_R \Gamma_R}{(m^2 - m_R^2)^2 + m_R^2 \Gamma_R^2} \, dm^2
\label{pg:resshapethree}
\end{equation}
are used to introduce mass distributions
for the two resonance masses, i.e.\ $m_3^2$ and $m_4^2$.
In the formula, $m_R$ is the nominal mass and $m$ the actually
selected one. The
phase-space integral over $x_1$, $x_1$ and $\hat{t}$ in eq.
(\ref{pg:sigma}) is then extended to involve also $m_3^2$ and
$m_4^2$. The effects of the mass-dependent width is only partly
taken into account, by replacing the nominal masses $m_3^2$ and
$m_4^2$ in the $\d \hat{\sigma} / \d \hat{t}$ expression by the
actually generated ones (also e.g.\ in the relation between
$\hat{t}$ and $\cos\hat{\theta}$), while the widths are evaluated 
at the nominal masses. This is the equivalent of a simple replacement 
of $m_R \Gamma_R$ by $\hat{s} \Gamma_R / m_R$ in the numerator of eq.
(\ref{pg:resshapeone}), but not in the denominator. In addition, the 
full threshold dependence of the widths, i.e.\ the velocity-dependent 
factors, is not reproduced.
 
There is no particular reason why the full mass-dependence could not be 
introduced, except for the extra work and time consumption needed for 
each process. In fact, the matrix elements for several $\gammaZ$ and
$\W^{\pm}$ production processes do contain the full expressions.
On the other hand, the matrix elements given in the literature
are often valid only when the resonances are almost on the mass shell,
since some graphs have been omitted. As an example, the process
$\q\qbar \to \e^- \br{\nu}_{\e} \mu^+ \nu_{\mu}$ is dominated by
$\q\qbar \to \W^- \W^+$ when each of the two lepton pairs is close to
$m_{\W}$ in mass, but in general also receives contributions e.g.\ from
$\q\qbar \to \Z^0 \to \ee$, followed by $\e^+ \to \br{\nu}_{\e} \W^+$
and $\W^+ \to \mu^+ \nu_{\mu}$. The latter contributions are neglected
in cross sections given in the zero-width approximation.

Widths may induce gauge invariance problems, in particular when the
$s$-channel graph interferes with $t$- or $u$-channels. Then there may
be an imperfect cancellation of contributions at high energies, leading
to an incorrect cross section behaviour. The underlying reason is that 
a Breit-Wigner corresponds to a resummation of terms of different
orders in coupling constants, and that therefore effectively the  
$s$-channel contributions are calculated to higher orders than the
$t$- or $u$-channel ones, including interference contributions.
A specific example is $\e^+ \e^- \to \W^+ \W^-$, where $s$-channel 
$\gamma^*/\Z^*$ exchange interferes with $t$-channel $\nu_{\e}$ exchange. 
In such cases, a fixed width is used in the denominator. One could also 
introduce procedures whereby the width is made to vanish completely at high
energies, and theoretically this is the cleanest, but the fixed-width 
approach appears good enough in practice.

Another gauge invariance issue is when two particles of the same kind
are produced in a pair, e.g. $\g \g \to \t \tbar$. Matrix elements are 
then often calculated for one common $m_{\t}$ mass, even though in real
life the masses $m_3 \neq m_4$. The proper gauge invariant procedure 
to handle this would be to study the full six-fermion state obtained 
after the two $\t \to \b \W \to \b \f_i \fbar_j$ decays, but that may
be overkill if indeed the $\t$'s are close to mass shell. Even when only
equal-mass matrix elements are available, Breit-Wigners are therefore
used to select two separate masses $m_3$ and $m_4$. From these two
masses, an average mass $\br{m}$ is constructed so that the 
$\beta_{34}$ velocity factor of eq.~(\ref{pg:betafactor}) is retained, 
\begin{equation}
\beta_{34}(\hat{s},\br{m}^2,\br{m}^2) = 
\beta_{34}(\hat{s},m_3^2, m_4^2) 
~~~\Rightarrow~~~
\br{m}^2 = \frac{m_3^2 + m_4^2}{2} - 
\frac{(m_3^2 - m_4^2)^2}{4 \hat{s}}.
\end{equation} 
This choice certainly is not unique, but normally should provide a sensible 
behaviour, also around threshold. 
Of course, the differential cross section is no longer guaranteed to
be gauge invariant when gauge bosons are involved, or positive definite.  
The program automatically flags the latter situation as unphysical.
The approach may well break down when 
either or both particles are far away from mass shell.
Furthermore, the preliminary choice of scattering 
angle $\hat{\theta}$ is also retained. Instead of the correct $\hat{t}$ 
and $\hat{u}$ of eq.~(\ref{pg:thatuhat}), modified 
\begin{equation}
\br{\hat{t}}, \br{\hat{u}} = 
- \frac{1}{2} \left\{ ( \hat{s} - 2 \br{m}^2 )
\mp \hat{s} \, \beta_{34} \cos \hat{\theta} \right\} =
(\hat{t}, \hat{u}) - \frac{(m_3^2 - m_4^2)^2}{4 \hat{s}}
\end{equation}
can then be obtained. The $\br{m}^2$, $\br{\hat{t}}$ and
$\br{\hat{u}}$ are now used in the matrix elements to decide
whether to retain the event or not. 
 
Processes with one final-state resonance and another ordinary
final-state product, e.g.\ $\q \g \to \W^+ \q'$, are treated in
the same spirit as the $2 \to 2$ processes with two resonances,
except that only one mass need be selected according to a
Breit--Wigner.
 
\subsection{Cross-section Calculations}
\label{ss:PYTcrosscalc}
 
In the program, the variables used in the generation of a
$2 \to 2$ process are $\tau$, $y$ and $z = \cos\hat{\theta}$.
For a $2 \to 1$ process, the $z$ variable can be integrated out,
and need therefore not be generated as part of the hard process,
except when the allowed angular range of decays is restricted.
In unresolved lepton beams, i.e.\ when
$f_{\e}^{\e}(x) = \delta(x-1)$, the variables $\tau$ and/or $y$
may be integrated out. We will cover all these special cases
towards the end of the section, and here concentrate on `standard' 
$2 \to 2$ and $2 \to 1$ processes.
 
\subsubsection{The simple two-body processes}
 
In the spirit of section \ref{ss:MCdistsel}, we want to select simple
functions such that the true $\tau$, $y$ and $z$ dependence of the
cross sections is approximately modelled. In particular, (almost) all
conceivable kinematical peaks should be represented by separate
terms in the approximate formulae. If this can be achieved,
the ratio of the correct to the approximate cross sections will
not fluctuate too much, but allow reasonable Monte Carlo efficiency.
 
Therefore the variables are generated according to the distributions
$h_{\tau}(\tau)$, $h_y(y)$ and $h_z(z)$, where normally
\begin{eqnarray}
h_{\tau}(\tau) & = & \frac{c_1}{{\cal I}_1} \, \frac{1}{\tau} +
\frac{c_2}{{\cal I}_2} \, \frac{1}{\tau^2} +
\frac{c_3}{{\cal I}_3} \, \frac{1}{\tau(\tau + \tau_R)} +
\frac{c_4}{{\cal I}_4} \,
\frac{1}{(s\tau - m_R^2)^2 + m_R^2 \Gamma_R^2} \nonumber \\
& & + \frac{c_5}{{\cal I}_5} \, \frac{1}{\tau(\tau + \tau_{R'})} +
\frac{c_6}{{\cal I}_6} \,
\frac{1}{(s\tau - m_{R'}^2)^2 + m_{R'}^2 \Gamma_{R'}^2} ~,  \\[1mm]
h_y(y) & = & \frac{c_1}{{\cal I}_1} \, (y - y_{\mmin}) +
\frac{c_2}{{\cal I}_2} \, (y_{\mmax} - y) +
\frac{c_3}{{\cal I}_3} \, \frac{1}{\cosh y} ~,  \\[1mm]
h_z(z) & = & \frac{c_1}{{\cal I}_1} +
\frac{c_2}{{\cal I}_2} \, \frac{1}{a-z} +
\frac{c_3}{{\cal I}_3} \, \frac{1}{a+z} +
\frac{c_4}{{\cal I}_4} \, \frac{1}{(a-z)^2} +
\frac{c_5}{{\cal I}_5} \, \frac{1}{(a+z)^2}    ~.
\label{pg:hforms}
\end{eqnarray}
Here each term is separately integrable, with an invertible primitive
function, such that generation of $\tau$, $y$ and $z$ separately
is a standard task, as described in section \ref{ss:MCdistsel}.
In the following we describe the details of the scheme, including
the meaning of the coefficients $c_i$ and ${\cal I}_i$, which are
separate for $\tau$, $y$ and $z$.
 
The first variable to be selected is $\tau$. The range of allowed
values, $\tau_{\mmin} \leq \tau \leq \tau_{\mmax}$, is generally
constrained by a number of user-defined requirements. A cut on the
allowed mass range is directly reflected in $\tau$, a cut on the
$\pT$ range indirectly. The first two terms
of $h_{\tau}$ are intended to represent a smooth $\tau$ dependence,
as generally obtained in processes which do not receive contributions
from $s$-channel resonances. Also $s$-channel exchange of
essentially massless particles ($\gamma$, $\g$, light quarks and
leptons) are accounted for, since these do not produce any separate
peaks at non-vanishing $\tau$. The last four terms of $h_{\tau}$ are
there to catch the peaks in the cross section from resonance
production. These terms are only included when needed. Each resonance
is represented by two pieces, a first to cover the interference with
graphs which peak at $\tau =0$, plus the variation of 
parton distributions, and a second to approximate the
Breit--Wigner shape of the resonance itself. The subscripts $R$ and
$R'$ denote values pertaining to the two resonances, with
$\tau_R = m_R^2/s$. Currently there is only one process where the
full structure with two resonances is used, namely
$\f \fbar \to \gamma^*/\Z^0/\Z'^0$. Otherwise either one or no
resonance peak is taken into account.
 
The kinematically allowed range of $y$ values is
constrained by $\tau$, $|y| \leq - \frac{1}{2} \ln\tau$, and you
may impose additional cuts. Therefore the allowed range
$y_{\mmin} \leq y \leq y_{\mmax}$ is only constructed after 
$\tau$ has been selected. The first two terms of $h_y$ give a fairly 
flat $y$ dependence --- for processes which are symmetric in
$y \leftrightarrow -y$, they will add to give a completely flat $y$
spectrum between the allowed limits. In
principle, the natural subdivision would have been one term flat
in $y$ and one forward--backward asymmetric, i.e.\ proportional to
$y$. The latter is disallowed by the requirement of positivity,
however. The $y - y_{\mmin}$ and $y_{\mmax} - y$ terms 
actually used give the same amount of freedom, but respect positivity.
The third term is peaked at around $y = 0$, and represents the bias
of parton distributions towards this region.
 
The allowed $z = \cos\hat{\theta}$ range is na\"{\i}vely
$-1 \leq z \leq 1$. However, most cross sections are divergent for
$z \to \pm 1$, so some kind of regularization is necessary. Normally
one requires $\pT \geq \pTmin$, which translates into
$z^2 \leq 1 - 4 \pTmin^2/(\tau s)$ for massless outgoing
particles. Since again the limits depend on $\tau$,
the selection of $z$ is done after that of
$\tau$. Additional requirements may constrain the range further.
In particular, a $p_{\perp\mmax}$ constraint may split the allowed
$z$ range into two, i.e.\ $z_{-\mmin} \leq z \leq z_{-\mmax}$ or
$z_{+\mmin} \leq z \leq z_{+\mmax}$. An un-split range is 
represented by $z_{-\mmax} = z_{+\mmin} = 0$. 
For massless outgoing particles
the parameter $a = 1$ in $h_z$, such that the five terms represent
a piece flat in angle and pieces peaked as $1/\hat{t}$, $1/\hat{u}$,
$1/\hat{t}^2$, and $1/\hat{u}^2$, respectively. For non-vanishing
masses one has $a = 1 + 2 m_3^2 m_4^2/\hat{s}^2$.
In this case, the full range $-1 \leq z \leq 1$ is therefore
available --- physically, the standard $\hat{t}$ and $\hat{u}$
singularities are regularized by the masses $m_3$ and $m_4$.
 
For each of the terms, the ${\cal I}_i$ coefficients represent the
integral over the quantity multiplying the coefficient $c_i$; thus,
for instance:
\begin{eqnarray}
h_{\tau}: & & {\cal I}_1 = \int \frac{\d \tau}{\tau} =
     \ln \left( \frac{\tau_{\mmax}}{\tau_{\mmin}} \right) ~, 
     \nonumber \\
     & & {\cal I}_2 = \int \frac{\d \tau}{\tau^2} =
     \frac{1}{\tau_{\mmin}} - \frac{1}{\tau_{\mmax}} ~;  
     \nonumber \\
h_y: & & {\cal I}_1 = \int (y - y_{\mmin}) \, \d y =
     \frac{1}{2} (y_{\mmax} - y_{\mmin})^2 ~; \nonumber \\
h_z: & & {\cal I}_1 = \int \d z = (z_{-\mmax} - z_{-\mmin}) +
     (z_{+\mmax} - z_{+\mmin}) , \nonumber \\
     & & {\cal I}_2 = \int \frac{\d z}{a-z} =
     \ln \left( \frac{(a-z_{-\mmin})(a-z_{+\mmin})}
     {(a-z_{-\mmax})(a-z_{-\mmin})} \right) ~.
\end{eqnarray}
 
The $c_i$ coefficients are normalized to unit sum for $h_{\tau}$, 
$h_y$ and $h_z$ separately. They have a simple interpretation, as
the probability
for each of the terms to be used in the preliminary selection of
$\tau$, $y$ and $z$, respectively. The variation of the cross section
over the allowed phase space is explored in the initialization
procedure of a {\Py} run, and based on this knowledge the $c_i$
are optimized so as to give functions $h_{\tau}$, $h_y$ and $h_z$ that
closely follow the general behaviour of the true cross section.
For instance, the coefficient $c_4$ in $h_{\tau}$ is to be made larger
the more the total cross section is dominated by the region around
the resonance mass.
 
The phase-space points tested at initialization
are put on a grid, with the number of
points in each dimension given by the number of terms in the
respective $h$ expression, and with the position of each point
given by the median value of the distribution of one of the terms.
For instance, the $\d \tau / \tau$ distribution gives a median point at
$\sqrt{\tau_{\mmin}\tau_{\mmax}}$, and $\d \tau / \tau^2$ has the 
median $2 \tau_{\mmin} \tau_{\mmax} 
/ (\tau_{\mmin} + \tau_{\mmax})$.
Since the allowed $y$ and $z$ ranges depend on the $\tau$ value
selected, then so do the median points defined for these two
variables.
 
With only a limited set of phase-space points
studied at the initialization, the `optimal' set of coefficients
is not uniquely defined. To be on the safe side, 40\% of the total
weight is therefore assigned evenly between all allowed $c_i$,
whereas the remaining 60\% are assigned according to the relative
importance surmised, under the constraint that no coefficient is
allowed to receive a negative contribution from this second piece.
 
After a preliminary choice has been made of $\tau$, $y$ and $z$,
it is necessary to find the weight of the event, which is to be used
to determine whether to keep it or generate another one.
Using the relation $\d \hat{t} = \hat{s} \, \beta_{34} \, \d z / 2$, 
eq.~(\ref{pg:sigma}) may be rewritten as
\begin{eqnarray}
\sigma & = & \int \int \int \frac{\d \tau}{\tau} \, \d y \,
\frac{\hat{s} \beta_{34}}{2} \d z \,
x_1 f_1(x_1, Q^2) \, x_2 f_2(x_2, Q^2) \,
\frac{\d \hat{\sigma}}{\d \hat{t}}    \nonumber \\[1mm]
& = & \frac{\pi}{s} \int h_{\tau}(\tau) \, \d \tau
\int h_y(y) \, \d y \int h_z(z) \, \d z \, \beta_{34} \,
\frac{x_1 f_1(x_1, Q^2) \, x_2 f_2(x_2, Q^2)}
{\tau^2 h_{\tau}(\tau) \, h_y(y) \, 2 h_z(z)} \,
\frac{\hat{s}^2}{\pi} \frac{\d \hat{\sigma}}{\d \hat{t}}  
\nonumber \\[1mm]
& = & \left\langle \frac{\pi}{s} \,
\frac{\beta_{34}}{\tau^2 h_{\tau}(\tau) \, h_y(y) \, 2 h_z(z)} \,
x_1 f_1(x_1, Q^2) \, x_2 f_2(x_2, Q^2) \,
\frac{\hat{s}^2}{\pi} \frac{\d \hat{\sigma}}{\d \hat{t}}
\right\rangle ~.
\label{pg:sigmamap}
\end{eqnarray}
In the middle line, a factor of $1 = h_{\tau}/h_{\tau}$
has been introduced to rewrite the $\tau$ integral
in terms of a phase space of unit volume:
$\int h_{\tau}(\tau) \, \d \tau = 1$ according to the relations
above. Correspondingly for the $y$ and $z$ integrals. In addition,
factors of $1 = \hat{s}/ (\tau s)$ and $1 = \pi / \pi$ are used to
isolate the dimensionless cross section
$(\hat{s}^2/\pi) \, \d \hat{\sigma} / \d \hat{t}$.
The content of the last line is that, with $\tau$, $y$ and $z$
selected according to the expressions $h_{\tau}(\tau)$, $h_y(y)$
and $h_z(z)$, respectively, the cross section is obtained as the
average of the final expression over all events. Since the $h$'s
have been picked to give unit volume, there is no need to multiply
by the total phase-space volume.
 
As can be seen, the cross section for a given Monte Carlo event is
given as the product of four factors, as follows:
\begin{Enumerate}
\item The $\pi/s$ factor, which is common to all events, gives the
overall dimensions of the cross section, in GeV$^{-2}$. Since
the final cross section is given in units of mb, the conversion
factor of 1 GeV$^{-2} = 0.3894$ mb is also included here.
\item Next comes the Jacobian, which compensates for the change
from the original to the final phase-space volume.
\item The parton-distribution function weight is obtained by making 
use of the parton distribution libraries in {\Py} or externally. 
The $x_1$ and $x_2$ values are obtained from $\tau$ and $y$ via the 
relations $x_{1,2} = \sqrt{\tau} \exp(\pm y)$.
\item Finally, the dimensionless cross section
$(\hat{s}^2/\pi) \, \d \hat{\sigma} / \d \hat{t}$
is the quantity that has to be coded for each process separately,
and where the physics content is found.
\end{Enumerate}
 
Of course, the expression in the last line is not strictly necessary
to obtain the cross section by Monte Carlo integration. One could
also have used eq.~(\ref{pg:sigma}) directly, selecting phase-space
points evenly in $\tau$, $y$ and $\hat{t}$, and averaging over those
Monte Carlo weights. Clearly this would be much simpler, but the price
to be paid is that the weights of individual events could fluctuate
wildly. For instance, if the cross section contains a narrow
resonance, the few phase-space points that are generated in the
resonance region obtain large weights, while the rest do not.
With our procedure, a resonance would be included in the
$h_{\tau}(\tau)$ factor, so that more events would be generated
at around the appropriate $\tau_R$ value (owing to the $h_{\tau}$
numerator in the phase-space expression), but with these events
assigned a lower, more normal weight (owing to the factor
$1/h_{\tau}$ in the weight expression).
Since the weights fluctuate less, fewer phase-space points
need be selected to get a reasonable cross-section estimate.
 
In the program, the cross section is obtained as the average over all
phase-space points generated. The events actually handed on to
you should have unit weight, however (an option with weighted events
exists, but does not represent the mainstream usage). At
initialization, after the $c_i$ coefficients have been determined,
a search inside the allowed phase-space volume is therefore made
to find the maximum of the weight expression in the last line of
eq.~(\ref{pg:sigmamap}). In the subsequent generation of events,
a selected phase-space point is then retained with a probability
equal to the weight in the point divided by the maximum weight.
Only the retained phase-space points are considered further, and
generated as complete events.
 
The search for the maximum is begun by evaluating the weight in the
same grid of points as used to determine the $c_i$ coefficients.
The point with highest weight is used as starting point for a
search towards the maximum. In unfortunate cases, the convergence
could be towards a local maximum which is not the global one.
To somewhat reduce this risk, also the grid point with 
second-highest weight is used for another search. After 
initialization, when events are generated, a warning message 
will be given by default at any time a phase-space
point is selected where the weight is larger than the maximum,
and thereafter the maximum weight is adjusted to reflect the new
knowledge. This means that events generated before this time have
a somewhat erroneous distribution in phase space, but if the
maximum violation is rather modest the effects should be negligible.
The estimation of the cross section is not affected by any of these
considerations, since the maximum weight does not enter into eq.
(\ref{pg:sigmamap}).

For $2 \to 2$ processes with identical final-state particles,
the symmetrization factor of $1/2$ is explicitly included at the
end of the $\d \hat{\sigma} / \d \hat{t}$ calculation. In the final
cross section, a factor of 2 is retrieved because of integration
over the full phase space (rather than only half of it). That
way, no special provisions are needed in the phase-space 
integration machinery.
 
\subsubsection{Resonance production}
 
We have now covered the simple $2 \to 2$ case. In a $2 \to 1$
process, the $\hat{t}$ integral is absent, and the differential
cross section $\d \hat{\sigma} / \d \hat{t}$ is replaced by
$\hat{\sigma}(\hat{s})$. The cross section may now be written as
\begin{eqnarray}
\sigma & = & \int \int \frac{\d \tau}{\tau} \, \d y \,
x_1 f_1(x_1, Q^2) \, x_2 f_2(x_2, Q^2) \,
\hat{\sigma}(\hat{s})    \nonumber \\
& = & \frac{\pi}{s} \int h_{\tau}(\tau) \, \d \tau  
\int h_y(y) \, \d y \,
\frac{x_1 f_1(x_1, Q^2) \, x_2 f_2(x_2, Q^2)}
{\tau^2 h_{\tau}(\tau) \, h_y(y)} \,
\frac{\hat{s}}{\pi} \hat{\sigma}(\hat{s}) \nonumber \\
& = & \left\langle \frac{\pi}{s} \,
\frac{1}{\tau^2 h_{\tau}(\tau) \, h_y(y)} \,
x_1 f_1(x_1, Q^2) \, x_2 f_2(x_2, Q^2) \,
\frac{\hat{s}}{\pi} \hat{\sigma}(\hat{s})
\right\rangle ~.
\label{pg:sigmamapres}
\end{eqnarray}
The structure is thus exactly the same, but the $z$-related pieces
are absent, and the r\^ole of the dimensionless cross section is
played by $(\hat{s}/\pi) \hat{\sigma}(\hat{s})$.
 
If the range of allowed decay angles of the resonance is restricted,
e.g.\ by requiring the decay products to have a minimum transverse
momentum, effectively this translates into constraints on the
$z = \cos\hat{\theta}$ variable of the $2 \to 2$ process. The
difference is that the angular dependence of a resonance decay is
trivial, and that therefore the $z$-dependent factor can be easily
evaluated. For a spin-0 resonance, which decays isotropically, the
relevant weight is simply
$(z_{-\mmax} - z_{-\mmin})/2 + (z_{+\mmax} - z_{+\mmin})/2$.
For a transversely polarized spin-1 resonance the expression is,
instead,
\begin{equation}
\frac{3}{8}(z_{-\mmax} - z_{-\mmin}) +
\frac{3}{8}(z_{+\mmax} - z_{+\mmin}) +
\frac{1}{8}(z_{-\mmax} - z_{-\mmin})^3 +
\frac{1}{8}(z_{+\mmax} - z_{+\mmin})^3  ~.
\end{equation}
Since the allowed $z$ range could depend on $\tau$ and/or $y$ (it does
for a $\pT$ cut), the factor has to be evaluated for each individual
phase-space point and included in the expression of eq.
(\ref{pg:sigmamapres}).
 
For $2 \to 2$ processes where either of the final-state
particles is a resonance, or both, an additional choice has to 
be made for
each resonance mass, eq.~(\ref{pg:resshapethree}). Since the allowed
$\tau$, $y$ and $z$ ranges depend on $m_3^2$ and $m_4^2$,
the selection of masses have to precede the choice of the other
phase-space variables. Just as for the other variables, masses
are not selected uniformly over the allowed range, but are rather
distributed according to a function $h_m(m^2) \, dm^2$, with a
compensating factor $1/h_m(m^2)$ in the Jacobian. The functional
form picked is normally
\begin{equation}
h_m(m^2) = \frac{c_1}{{\cal I}_1} \, \frac{1}{\pi} \,
\frac{m_R \Gamma_R}{(m^2 - m_R^2)^2 + m_R^2 \Gamma_R^2} +
\frac{c_2}{{\cal I}_2} +
\frac{c_3}{{\cal I}_3} \, \frac{1}{m^2} +
\frac{c_4}{{\cal I}_4} \, \frac{1}{m^4}  ~.
\label{pg:reshdist}
\end{equation}
The definition of the ${\cal I}_i$ integrals is analogous to the one
before. The $c_i$ coefficients are not found by optimization, but
predetermined, normally to $c_1 = 0.8$, $c_2 = c_3 =0.1$, $c_4 = 0$.
Clearly, had the phase space and the cross section been independent
of $m_3^2$ and $m_4^2$, the optimal choice would have been to put
$c_1 = 1$ and have all other $c_i$ vanishing --- then the $1/h_m$
factor of the Jacobian would exactly have cancelled the
Breit--Wigner of eq.~(\ref{pg:resshapethree}) in the cross section.
The second and the third terms are there to cover the possibility
that the cross section does not die away quite as fast as given by
the na\"{\i}ve Breit--Wigner shape. In particular, the third term covers 
the possibility of a secondary peak at small $m^2$, in a spirit 
slightly similar to the one discussed for resonance production in 
$2 \to 1$ processes.
 
The fourth term is only used for processes involving $\gammaZ$ 
production, where the $\gamma$ propagator
guarantees that the cross section does have a significant secondary
peak for $m^2 \to 0$. Therefore here the choice is $c_1 = 0.4$,
$c_2 = 0.05$, $c_3 = 0.3$ and $c_4 = 0.25$.
 
A few special tricks have been included to improve efficiency
when the allowed mass range of resonances is constrained by
kinematics or by user cuts. For instance, if a pair of equal
or charge-conjugate resonances are produced, such as in
$\ee \to \W^+ \W^-$, use is made of the constraint that the lighter
of the two has to have a mass smaller than half the c.m.\ energy.
 
\subsubsection{Lepton beams}
 
Lepton beams have to be handled slightly differently from what has
been described so far. One also has to distinguish between a lepton
for which parton distributions are included and one which is treated
as an unresolved point-like particle. The necessary modifications are
the same for $2 \to 2$ and $2 \to 1$ processes, however, since the
$\hat{t}$ degree of freedom is unaffected.
 
If one incoming beam is an unresolved lepton, the corresponding
parton-distribution piece collapses to a $\delta$ function. This
function can be used to integrate out the $y$ variable:
$\delta(x_{1,2} - 1) = \delta(y \pm (1/2) \ln \tau)$.
It is therefore only necessary to select the $\tau$ and the $z$
variables according to the proper distributions, with compensating
weight factors, and only one set of parton distributions has to be 
evaluated explicitly.
 
If both incoming beams are unresolved leptons, both the $\tau$ and
the $y$ variables are trivially given: $\tau = 1$ and $y = 0$.
Parton-distribution weights disappear completely. For a $2 \to 2$
process, only the $z$ selection remains to be performed, while
a $2 \to 1$ process is completely specified, i.e.\ the cross section
is a simple number that only depends on the c.m.\ energy.
 
For a resolved electron, the $f_{\e}^{\e}$ parton distribution is
strongly peaked towards $x = 1$. This affects both the $\tau$
and the $y$ distributions, which are not well described by
either of the pieces in $h_{\tau}(\tau)$ or $h_y(y)$ in processes with
interacting $\e^{\pm}$. (Processes which involve e.g.\ the $\gamma$
content of the $\e$ are still well simulated, since
$f_{\gamma}^{\e}$ is peaked at small $x$.)
 
If both parton distributions 
are peaked close to 1, the $h_{\tau}(\tau)$ expression in eq.
(\ref{pg:hforms}) is therefore increased with one additional term of
the form $h_{\tau}(\tau) \propto 1 / (1 - \tau)$, with coefficients
$c_7$ and ${\cal I}_7$ determined as before. The divergence when
$\tau \to 1$ is cut off by our regularization procedure for the
$f_{\e}^{\e}$ parton distribution; therefore we only need consider
$\tau < 1 - 2 \times 10^{-10}$.
 
Correspondingly, the $h_y(y)$ expression is expanded with a term
$1/(1 - \exp(y-y_0))$ when incoming beam number 1 consists of a
resolved $\e^{\pm}$, and with a term $1/(1 - \exp(-y-y_0))$
when incoming beam number 2 consists of a resolved $\e^{\pm}$.
Both terms are present for an $\ee$ collider, only one for an
$\ep$ one. The coefficient $y_0 = - (1/2) \ln \tau$ is the na\"{\i}ve
kinematical limit of the $y$ range, $|y| < y_0$. From the
definitions of $y$ and $y_0$ it is easy to see
that the two terms above correspond to $1/(1-x_1)$ and $1/(1-x_2)$,
respectively, and thus are again regularized by our 
parton-distribution function cut-off. Therefore the integration ranges 
are $y < y_0 -10^{-10}$ for the first term and $y > - y_0 + 10^{-10}$
for the second one.
 
\subsubsection{Mixing processes}
\label{sss:mixingproc}
 
In the cross-section formulae given so far, we have deliberately
suppressed a summation over the allowed incoming flavours. For
instance, the process $\f\fbar \to \Z^0$ in a hadron collider
receives contributions from $\u\ubar \to \Z^0$, $\d\dbar \to \Z^0$,
$\s\sbar \to \Z^0$, and so on. These contributions share the
same basic form, but differ in the parton-distribution weights
and (usually) in a few coupling constants in the hard matrix
elements. It is therefore convenient to generate the terms
together, as follows:
\begin{Enumerate}
\item A phase-space point is picked, and all
common factors related to this choice are evaluated, i.e.
the Jacobian and the common pieces of the matrix elements
(e.g.\ for a $\Z^0$ the basic Breit--Wigner shape, excluding
couplings to the initial flavour).
\item The parton-distribution-function library is called to produce 
all the parton distributions, at the relevant $x$ and $Q^2$ values,
for the two incoming beams.
\item A loop is made over the two incoming flavours, one from each
beam particle. For each allowed set of incoming flavours, the full 
matrix-element expression is constructed, using the common pieces and 
the flavour-dependent couplings. This is multiplied by the common 
factors and the parton-distribution weights to obtain a cross-section 
weight.
\item Each allowed flavour combination is stored as a separate entry
in a table, together with its weight. In addition, a summed weight
is calculated.
\item The phase-space point is kept or rejected, according to a
comparison of the summed weight with the maximum weight obtained
at initialization. Also the cross-section Monte Carlo integration
is based on the summed weight.
\item If the point is retained, one of the allowed flavour
combinations is picked according to the relative weights stored
in the full table.
\end{Enumerate}
 
Generally, the flavours of the final state are either completely
specified by those of the initial state, e.g.\ as in $\q\g \to \q\g$,
or completely decoupled from them, e.g.\ as in
$\f\fbar \to \Z^0 \to \f'\fbar'$. In neither case need therefore the
final-state flavours be specified in the cross-section calculation.
It is only necessary, in the latter case, to include an overall
weight factor, which takes into account the summed contribution of
all final states that are to be simulated. For instance, if only
the process $\Z^0 \to \ee$  is studied, the relevant weight factor
is simply $\Gamma_{\e\e} / \Gamma_{\mrm{tot}}$. Once the kinematics 
and the incoming flavours have been selected, the outgoing flavours
can be picked according to the appropriate relative probabilities.
 
In some processes, such as $\g\g \to \g\g$, several different colour
flows are allowed, each with its own kinematical dependence of the
matrix-element weight, see section \ref{sss:QCDjetclass}. Each colour 
flow is then given as a separate entry in the table mentioned above, 
i.e.\ in total an entry is characterized by the two incoming flavours, 
a colour-flow index, and the weight. For an accepted phase-space 
point, the colour flow is selected in the same way as the incoming 
flavours.
 
The program can also allow the mixed generation of two or more
completely different processes, such as $\f\fbar \to \Z^0$ and
$\q\qbar \to \g\g$. In that case, each process is initialized
separately, with its own set of coefficients $c_i$ and so on.
The maxima obtained for the individual cross sections are all
expressed in the same units, even when the dimensionality of the
phase space is different. (This is because we always transform to
a phase space of unit volume, 
$\int h_{\tau}(\tau) \, \d \tau \equiv 1$, etc.) The above 
generation scheme need therefore only be generalized as follows:
\begin{Enumerate}
\item One process is selected among the allowed ones, with a relative
probability given by the maximum weight for this process.
\item A phase-space point is found, using the distributions
$h_{\tau}(\tau)$ and so on, optimized for this particular process.
\item The total weight for the phase-space point is evaluated,
again with Jacobians, matrix elements and allowed incoming flavour
combinations that are specific to the process.
\item The point is retained with a probability given by the ratio of
the actual to the maximum weight of the process. If the point is
rejected, one has to go back to step 1 and pick a new process.
\item Once a phase-space point has been accepted, flavours may be
selected, and the event generated in full.
\end{Enumerate}
It is clear why this works: although phase-space points are selected
among the allowed processes according to relative probabilities given
by the maximum weights, the probability that a point is accepted is
proportional to the ratio of actual to maximum weight. In total,
the probability for a given process to be retained is therefore only
proportional to the average of the actual weights, and any dependence
on the maximum weight is gone.

In $\gamma\p$ and $\gamma\gamma$ physics, the different components
of the photon give different final states, see section 
\ref{sss:photoprod}. Technically, this introduces a further level
of administration, since each event class contains a set of (partly
overlapping) processes. From an ideological point of view, however,
it just represents one more choice to be made, that of event class,
before the selection of process in step 1 above. When a weighting
fails, both class and process have to be picked anew.
 
\subsection{Three- and Four-body Processes}
 
The {\Py} machinery to handle $2 \to 1$ and $2 \to 2$ processes
is fairly sophisticated and generic. The same cannot be said about
the generation of hard scattering processes with more than two 
final-state particles. The number of phase-space variables is 
larger, and
it is therefore more difficult to find and transform away all possible
peaks in the cross section by a suitably biased choice of phase-space
points. In addition, matrix-element expressions for $2 \to 3$
processes are typically fairly lengthy. Therefore {\Py} only contains
a very limited number of $2 \to 3$ and $2 \to 4$ processes, and almost
each process is a special case of its own. It is therefore less
interesting to discuss details, and we only give a very generic
overview.
 
If the Higgs mass is not light, interactions among longitudinal
$\W$ and $\Z$ gauge bosons are of interest. In the program,
$2 \to 1$ processes such as $\W_{\mrm{L}}^+ \W_{\mrm{L}}^- \to \hrm^0$ 
and $2 \to 2$ ones such as 
$\W_{\mrm{L}}^+ \W_{\mrm{L}}^- \to \Z_{\mrm{L}}^0 \Z_{\mrm{L}}^0$ 
are included.
The former are for use when the $\hrm^0$ still is reasonably narrow,
such that a resonance description is applicable, while the latter
are intended for high energies, where different contributions have
to be added up. Since the program does not contain 
$\W_{\mrm{L}}$ or $\Z_{\mrm{L}}$
distributions inside hadrons, the basic hard scattering has
to be convoluted with the $\q \to \q' \W_{\mrm{L}}$ and 
$\q \to \q \Z_{\mrm{L}}$
branchings, to yield effective $2 \to 3$ and $2 \to 4$ processes.
However, it is possible to integrate out the scattering angles of
the quarks analytically, as well as one energy-sharing variable
\cite{Cha85}. Only after an event has been accepted are these other
kinematical variables selected. This involves further choices of
random variables, according to a separate selection loop.
 
In total, it is therefore only necessary to introduce one additional
variable in the basic phase-space selection, which is chosen to be
$\hat{s}'$, the squared invariant mass of the full $2 \to 3$ or
$2 \to 4$ process, while $\hat{s}$ is used for the squared invariant
mass of the inner $2 \to 1$ or $2 \to 2$ process. The $y$ variable
is coupled to the full process, since parton-distribution weights
have to be given for the original quarks at
$x_{1,2} = \sqrt{\tau'} \exp{(\pm y)}$. The $\hat{t}$ variable is
related to the inner process, and thus not needed for the $2 \to 3$
processes. The selection of the $\tau' = \hat{s}'/s$ variable is
done after $\tau$, but before $y$ has been chosen. To improve the
efficiency, the selection is made according to a weighted phase space
of the form $\int h_{\tau'}(\tau') \, \d \tau'$, where
\begin{equation}
h_{\tau'}(\tau') = \frac{c_1}{{\cal I}_1} \frac{1}{\tau'} +
\frac{c_2}{{\cal I}_2} \, \frac{(1- \tau / \tau')^3}{\tau'^2} +
\frac{c_3}{{\cal I}_3} \, \frac{1}{\tau' (1 - \tau')} ~,
\end{equation}
in conventional notation. The $c_i$ coefficients are optimized
at initialization. The $c_3$ term, peaked at $\tau' \approx 1$,
is only used for $\ee$ collisions.
The choice of $h_{\tau'}$ is roughly matched to the
longitudinal gauge-boson flux factor, which is of the form
\begin{equation}
\left( 1 + \frac{\tau}{\tau'} \right) \,
\ln \left( \frac{\tau}{\tau'} \right) -
2 \left( 1 - \frac{\tau}{\tau'} \right)    ~.
\end{equation}
 
For a light $\hrm$ the effective $\W$ approximation above breaks down,
and it is necessary to include the full structure of the
$\q \q' \to \q \q' \hrm^0$ (i.e.\ $\Z\Z$ fusion) and
$\q \q' \to \q'' \q''' \hrm^0$ (i.e.\ $\W\W$ fusion) matrix elements.
The $\tau'$, $\tau$ and $y$ variables are here retained, and selected
according to standard procedures. The Higgs mass is represented by the
$\tau$ choice; normally the $\hrm^0$ is so narrow that the $\tau$
distribution effectively collapses to a $\delta$ function. In addition,
the three-body final-state phase space is rewritten as
\begin{equation}
\left( \prod_{i=3}^5 \frac{1}{(2 \pi)^3} \frac{\d^3 p_i}{2 E_i}
\right) \, (2 \pi)^4 \delta^{(4)} (p_3 + p_4 + p_5 - p_1 - p_2) =
\frac{1}{(2 \pi)^5} \, \frac{\pi^2}{4 \sqrt{\lambda_{\perp 34}}}
\, \d p_{\perp 3}^2 \, \frac{\d \varphi_3}{2 \pi} \,
\d p_{\perp 4}^2 \, \frac{\d \varphi_4}{2 \pi} \, \d y_5    ~,
\end{equation}
where $\lambda_{\perp 34} = (m_{\perp 34}^2 - m_{\perp 3}^2 -
m_{\perp 4}^2)^2 - 4 m_{\perp 3}^2 m_{\perp 4}^2$.
The outgoing quarks are labelled 3 and 4, and the outgoing Higgs 5.
The $\varphi$ angles are selected isotropically, while the two
transverse momenta are picked, with some foreknowledge of the
shape of the $\W / \Z$ propagators in the cross sections, according
to $h_{\perp} (\pT^2) \, \d \pT^2$, where
\begin{equation}
h_{\perp}(\pT^2) = \frac{c_1}{{\cal I}_1} +
\frac{c_2}{{\cal I}_2} \, \frac{1}{m_R^2 + \pT^2} +
\frac{c_3}{{\cal I}_3} \, \frac{1}{(m_R^2 + \pT^2)^2}    ~,
\end{equation}
with $m_R$ the $\W$ or $\Z$ mass, depending on process, and
$c_1 = c_2 = 0.05$, $c_3 = 0.9$. Within the limits given by the
other variable choices, the rapidity $y_5$ is chosen uniformly.
A final choice remains to be made, which comes from a twofold
ambiguity of exchanging the longitudinal momenta of partons 3
and 4 (with minor modifications if they are massive). Here
the relative weight can be obtained exactly from the form of the
matrix element itself.
 
\subsection{Resonance Decays}
\label{ss:resdecay}
 
Resonances (see section \ref{sss:resdecintro})
can be made to decay in two different routines. One is the
standard decay treatment (in \ttt{PYDECY}) that can be used
for any unstable particle, where decay channels
are chosen according to fixed probabilities, and decay angles
usually are picked isotropically in the rest frame of the resonance,
see section \ref{ss:partdecays}.
The more sophisticated treatment (in \ttt{PYRESD}) is the default
one for resonances produced in {\Py}, and is described here.
The ground rule is that everything in mass up to and including $\b$
hadrons is decayed with the simpler \ttt{PYDECY} routine, while
heavier particles are handled with \ttt{PYRESD}. This also includes
the $\gamma^* / \Z^0$, even though here the mass in principle could 
be below the $\b$ threshold. Other resonances include, e.g., $\t$, 
$\W^{\pm}$, $\hrm^0$, $\Z'^0$, $\W'^{\pm}$, $\H^0$, $\A^0$,
$\H^{\pm}$, and technicolor and supersymmetric particles.
 
\subsubsection{The decay scheme}
 
In the beginning of the decay treatment, either one or two
resonances may be present, the former represented by processes
such as $\q \qbar' \to \W^+$ and $\q \g \to \W^+ \q'$, the latter
by $\q \qbar \to \W^+ \W^-$. If the latter is the case, the
decay of the two resonances is considered in parallel
(unlike \ttt{PYDECY}, where one particle at a time is made to decay).
 
First the decay channel of each resonance is selected according to
the relative weights $H_R^{(f)}$, as described above, evaluated at
the actual mass of the resonance, rather than at the nominal one.
Threshold factors are therefore fully taken into account, with
channels automatically switched off below the threshold. Normally
the masses of the decay products are well-defined, but e.g.\ in
decays like $\hrm^0 \to \W^+ \W^-$ it is also necessary to select
the decay product masses. This is done according to two Breit--Wigners
of the type in eq.~(\ref{pg:resshapethree}), multiplied by the
threshold factor, which depends on both masses.
 
Next the decay angles of the resonance are selected isotropically in
its rest frame. Normally the full range of decay angles is available,
but in $2 \to 1$ processes the decay angles of the original resonance
may be restrained by user cuts, e.g.\ on the $\pT$ of the decay
products. Based on the angles, the four-momenta of the decay products
are constructed and boosted to the correct frame. As a rule, matrix
elements are given with quark and lepton masses assumed vanishing.
Therefore the four-momentum vectors constructed at this stage are
actually massless for all quarks and leptons.
 
The matrix elements may now be evaluated. For a process such as
$\q \qbar \to \W^+ \W^- \to \e^+ \nu_{\e} \mu^- \br{\nu}_{\mu}$,
the matrix element is a function of the four-momenta of the two
incoming fermions and of the four outgoing ones. An upper limit for
the event weight can be constructed from the cross section for the
basic process $\q \qbar \to \W^+ \W^-$, as already used to select
the two $\W$ momenta. If the weighting fails, new resonance decay
angles are picked and the procedure is iterated until acceptance.
 
Based on the accepted set of angles, the correct decay product
four-momenta are constructed, including previously neglected 
fermion masses. Quarks and, optionally, leptons are allowed to
radiate, using the standard final-state showering machinery, with
maximum virtuality given by the resonance mass.
 
In some decays new resonances are produced, and these are then
subsequently allowed to decay. Normally only one resonance pair is 
considered at a time, with the possibility of full correlations. 
In a few cases triplets can also appear, but such configurations 
currently are considered without inclusion of correlations. Also 
note that, in a process like
$\q\qbar \to \Z^0 \hrm^0 \to \Z^0 \W^+ \W^- \to 6$ fermions,
the spinless nature of the $\hrm^0$ ensures that the $\W^{\pm}$
decays are decoupled from that of the $\Z^0$ (but not from each other).
 
\subsubsection{Cross-section considerations}
\label{sss:resdecaycross}
 
The cross section for a process which involves the production of one
or several resonances is always reduced to take into account channels
not allowed by user flags. This is trivial for a single $s$-channel
resonance, cf. eq.~(\ref{pg:Hinoutsym}), but can also be included
approximately if several layers of resonance decays are involved.
At initialization, the ratio between the user-allowed width and the
nominally possible one is evaluated and stored, starting from the
lightest resonances and moving upwards. As an example, one first finds
the reduction factors for $\W^+$ and for $\W^-$ decays, which need not
be the same if e.g.\ $\W^+$ is allowed to decay only to quarks and
$\W^-$ only to leptons. These factors enter together as a weight
for the $\hrm^0 \to \W^+ \W^-$ channel, which is thus reduced in
importance compared with other possible Higgs decay channels.
This is also reflected
in the weight factor of the $\hrm^0$ itself, where some channels are open
in full, others completely closed, and finally some (like the one
above) open but with reduced weight. Finally, the weight for the
process $\q\qbar \to \Z^0 \hrm^0$ is evaluated as the product of the
$\Z^0$ weight factor and the $\hrm^0$ one. The standard cross section 
of the process is multiplied with this weight.
 
Since the restriction on allowed decay modes is already included in
the hard process cross section, mixing of different event types is
greatly simplified, and the selection of decay channel chains is
straightforward. There is a price to be paid, however. The reduction
factors evaluated at initialization all refer to resonances at their
nominal masses. For instance, the $\W$ reduction factor is evaluated
at the nominal $\W$ mass, even when that factor is used, later on, in
the description of the decay of a 120 GeV Higgs, where at least one
$\W$ would be produced below this mass. We know of no case where this
approximation has any serious consequences, however.
 
The weighting procedure works because the number of resonances to be
produced, directly or in subsequent decays, can be derived
recursively already from the start. It does not work for particles
which could also be produced at later stages, such as the 
parton-shower evolution and the fragmentation. For instance,
$\D^0$ mesons can be produced fairly late in the event generation
chain, in unknown numbers, and so weights could not be introduced
to compensate, e.g.\ for the forcing of decays only into $\pi^+ \K^-$.
 
One should note that this reduction factor is separate from the
description of the resonance shape itself, where the full
width of the resonance has to be used. This width is based on the
sum of all possible decay modes, not just the simulated ones.
{\Py} does allow the possibility to change also the underlying
physics scenario, e.g.\ to include the decay of a $\Z^0$ into a
fourth-generation neutrino.
 
Normally the evaluation of the reduction factors is straightforward.
However, for decays into a pair of equal or charge-conjugate
resonances, such as $\Z^0 \Z^0$ or $\W^+ \W^-$, it is possible to
pick combinations in such a way that the weight of the pair does
not factorize into a product of the weight of each resonance itself.
To be precise, any decay channel can be given seven different status
codes:
\begin{Itemize}
\item $-1$: a non-existent decay mode, completely switched off and of no
concern to us;
\item 0: an existing decay channel, which is switched off;
\item 1: a channel which is switched on;
\item 2: a channel switched on for particles, but off for antiparticles;
\item 3: a channel switched on for antiparticles, but off for particles;
\item 4: a channel switched on for one of the particles or antiparticles, 
   but not for both;
\item 5: a channel switched on for the other of the particles or
   antiparticles, but not for both.
\end{Itemize}
The meaning of possibilities 4 and 5 is exemplified by the statement
`in a $\W^+\W^-$ pair, one $\W$ decays hadronically and the other
leptonically', which thus covers the cases where either $\W^+$ or
$\W^-$ decays hadronically.
 
Neglecting non-existing channels, each channel belongs to either of the
classes above. If we denote the total branching ratio into channels
of type $i$ by $r_i$, this then translates into the requirement
$r_0 + r_1 + r_2 + r_3 + r_4 + r_5 = 1$. For a single particle the
weight factor is $r_1 + r_2 + r_4$, and for a single antiparticle
$r_1 + r_3 + r_4$. For a pair of identical resonances, the joint weight
is instead
\begin{equation}
(r_1 + r_2)^2 + 2 (r_1 + r_2) (r_4 + r_5) + 2 r_4 r_5 ~,
\end{equation}
and for a resonance--antiresonance pair
\begin{equation}
(r_1 + r_2)(r_1 + r_3) + (2 r_1 + r_2 + r_3) (r_4 + r_5) +
2 r_4 r_5 ~.
\label{eq:WWallchancomb}
\end{equation}
If some channels come with a reduced weight because of restrictions 
on subsequent decay chains, this may be described in terms of
properly reduced $r_i$, so that the sum is less than unity.
For instance, in a $\t\tbar \to \b\W^+ \, \bbar\W^-$ process,
the $\W$ decay modes may be restricted to $\W^+ \to \q\qbar$
and $\W^- \to \e^-\bar{\nu}_{\e}$, in which case
$(\sum r_i)_{\t} \approx 2/3$ and $(\sum r_i)_{\tbar} \approx 1/9$.
With index $\pm$ denoting resonance/antiresonance, 
eq.~(\ref{eq:WWallchancomb}) then generalizes to
\begin{equation}
(r_1 + r_2)^+ (r_1 + r_3)^- + (r_1 + r_2)^+ (r_4 + r_5)^- +
(r_4 + r_5)^+ (r_1 + r_3)^- + r_4^+ r_5^- + r_5^+ r_4^- ~.
\label{eq:WWallchancombgen}
\end{equation}
 
\subsection{Nonperturbative Processes}
\label{ss:nonpertproc}
 
A few processes are not covered by the discussion so far. These are
the ones that depend on the details of hadronic wave functions,
and therefore are not strictly calculable perturbatively
(although perturbation theory may often provide some guidance).
What we have primarily in mind is elastic scattering, diffractive
scattering and low-$\pT$ `minimum-bias' events in hadron--hadron
collisions, but one can also find corresponding processes in
$\gamma \p$ and $\gamma \gamma$ interactions. The description
of these processes is rather differently structured from that of
the other ones, as is explained below. Models for
`minimum-bias' events are discussed in detail in section
\ref{ss:multint}, to which we refer for details on this part
of the program.

\subsubsection{Hadron--hadron interactions}
 
In hadron--hadron interactions, the total hadronic cross section
for $AB \to$ anything, $\sigma^{AB}_{\mrm{tot}}$, is calculated
using the parameterization of Donnachie and Landshoff \cite{Don92}.
In this approach, each cross section appears as the sum of one
pomeron term and one reggeon one
\begin{equation}
\sigma^{AB}_{\mrm{tot}}(s) = X^{AB} \, s^{\epsilon} +  
Y^{AB} \, s^{-\eta} ~,
\label{pg:sigtotpomreg}
\end{equation}
where $s = E_{\mrm{cm}}^2$. The powers $\epsilon = 0.0808$ and
$\eta = 0.4525$ are expected to be universal, whereas the 
coefficients $X^{AB}$ and $Y^{AB}$ are specific to each initial
state. (In fact, the high-energy behaviour given by the pomeron term 
is expected to be the same for particle and antiparticle interactions, 
i.e.\ $X^{\br{A}B} = X^{AB}$.) Parameterizations not provided
in \cite{Don92} have been calculated in the same spirit, making use
of quark counting rules \cite{Sch93a}.  

The total cross section is subdivided according to
\begin{equation}
\sigma^{AB}_{\mrm{tot}}(s) = \sigma^{AB}_{\mrm{el}}(s) + 
\sigma^{AB}_{\mrm{sd}(XB)}(s) + \sigma^{AB}_{\mrm{sd}(AX)}(s) + 
\sigma^{AB}_{\mrm{dd}}(s) + \sigma^{AB}_{\mrm{nd}}(s)   ~.
\label{pg:sigtotsplit}
\end{equation}
Here `el' is the elastic process $AB \to AB$, `sd$(XB)$' the single 
diffractive $AB \to XB$, `sd$(AX)$' the single diffractive
$AB \to AX$, `dd' the double diffractive $AB \to X_1 X_2$,
and `nd' the non-diffractive ones. Higher diffractive topologies,
such as central diffraction, are currently neglected. 
In the following, the elastic and diffractive cross sections 
and event characteristics are described, as given in the model by
Schuler and Sj\"ostrand \cite{Sch94,Sch93a}. The non-diffractive 
component is identified with the `minimum bias' physics already 
mentioned, a practical but not unambiguous choice. Its cross section 
is given by `whatever is left' according to eq.~(\ref{pg:sigtotsplit}), 
and its properties are discussed in section \ref{ss:multint}.

At not too large squared momentum transfers $t$, the elastic cross
section can be approximated by a simple exponential fall-off. If one
neglects the small real part of the cross section, the optical 
theorem then gives
\begin{equation}
\frac{\d\sigma_{\mrm{el}}}{\d t} = 
\frac{\sigma_{\mrm{tot}}^2}{16 \pi} \, \exp(B_{\mrm{el}} t) ~,
\end{equation}
and $\sigma_{\mrm{el}} = \sigma_{\mrm{tot}}^2 / 16 \pi B_{\mrm{el}}$.
The elastic slope parameter is parameterized by
\begin{equation}
B_{\mrm{el}} = B^{AB}_{\mrm{el}}(s) = 2 b_A + 2 b_B + 
4 s^{\epsilon} -4.2 ~, 
\end{equation}
with $s$ given in units of GeV and $B_{\mrm{el}}$ in GeV$^{-2}$.
The constants $b_{A,B}$ are $b_{\p} = 2.3$, 
$b_{\pi,\rho,\omega,\phi} = 1.4$, $b_{\J/\psi} = 0.23$.
The increase of the slope parameter with c.m.\ energy is faster
than the logarithmically one conventionally assumed; that way the 
ratio $\sigma_{\mrm{el}} / \sigma_{\mrm{tot}}$ remains well-behaved
at large energies.

The diffractive cross sections are given by
\begin{eqnarray}
\frac{\d\sigma_{\mrm{sd}(XB)}(s)}{\d t \, \d M^2} & = & 
\frac{g_{3\pomeron}}{16\pi} \, \beta_{A\pomeron} \, 
\beta_{B\pomeron}^2 \, \frac{1}{M^2} \, \exp(B_{\mrm{sd}(XB)}t) 
\, F_{\mrm{sd}} ~, \nonumber \\
\frac{\d\sigma_{\mrm{sd}(AX)}(s)}{\d t \, \d M^2} & = & 
\frac{g_{3\pomeron}}{16\pi} \, \beta_{A\pomeron}^2 \, 
\beta_{B\pomeron} \, \frac{1}{M^2} \, \exp(B_{\mrm{sd}(AX)}t)
\, F_{\mrm{sd}} ~, \nonumber \\
\frac{\d\sigma_{\mrm{dd}}(s)}{\d t \, \d M_1^2 \, \d M_2^2} & = &  
\frac{g_{3\pomeron}^2}{16\pi} \, \beta_{A\pomeron} \, 
\beta_{B\pomeron} \, \frac{1}{M_1^2} \, \frac{1}{M_2^2} \, 
\exp(B_{\mrm{dd}}t) \, F_{\mrm{dd}} ~. 
\end{eqnarray}

The couplings $\beta_{A\pomeron}$ are related to the pomeron term
$X^{AB} s^{\epsilon}$ of the total cross section parameterization,
eq.~(\ref{pg:sigtotpomreg}). Picking a reference scale 
$\sqrt{s_{\mrm{ref}}} = 20$ GeV, the couplings are given by 
$\beta_{A\pomeron}\beta_{B\pomeron} = 
X^{AB} \, s_{\mrm{ref}}^{\epsilon}$. The triple-pomeron coupling is
determined from single-diffractive data to be 
$g_{3\pomeron} \approx 0.318$ mb$^{1/2}$; within the context of the 
formulae in this section.

The spectrum of diffractive masses $M$ is taken to begin
0.28 GeV $\approx 2 m_{\pi}$ above the mass of the respective 
incoming particle and extend to the kinematical limit. The simple
$\d M^2 / M^2$ form is modified by the mass-dependence in the 
diffractive slopes and in the $F_{\mrm{sd}}$ and $F_{\mrm{dd}}$ 
factors (see below). 

The slope parameters are assumed to be
\begin{eqnarray}
B_{\mrm{sd}(XB)}(s) & = & 2b_B + 2\alpha' \ln\left(\frac{s}{M^2}\right)
~, \nonumber \\
B_{\mrm{sd}(AX)}(s) & = & 2b_A + 2\alpha' \ln\left(\frac{s}{M^2}\right)
~, \nonumber \\
B_{\mrm{dd}}(s) & = & 2\alpha' \ln\left(e^4 + \frac{s s_0}{M_1^2 M_2^2}
\right) ~. 
\end{eqnarray}
Here $\alpha' = 0.25$ GeV$^{-2}$ and conventionally $s_0$ is picked as
$s_0 = 1 / \alpha'$. The term $e^4$ in $B_{\mrm{dd}}$ is added by hand
to avoid a breakdown of the standard expression for large values of
$M_1^2 M_2^2$. The $b_{A,B}$ terms protect $B_{\mrm{sd}}$ from breaking
down; however a minimum value of 2~GeV$^{-2}$ is still explicitly required 
for $B_{\mrm{sd}}$, which comes into play e.g.\ for a $\Jpsi$ state
(as part of a VMD photon beam).

The kinematical range in $t$ depends on all the masses of the 
problem. In terms of the scaled variables $\mu_1 = m_A^2/s$,
$\mu_2 = m_B^2/s$, $\mu_3 = M_{(1)}^2/s$ ($=m_A^2/s$ when $A$
scatters elastically), $\mu_4 = M_{(2)}^2/s$ ($=m_B^2/s$ when $B$
scatters elastically), and the combinations
\begin{eqnarray}
C_1 & = & 1 - (\mu_1 + \mu_2 + \mu_3 + \mu_4) +
(\mu_1 - \mu_2) (\mu_3 - \mu_4) ~, \nonumber \\
C_2 & = & \sqrt{(1 - \mu_1 -\mu_2)^2 - 4 \mu_1 \mu_2} \,
\sqrt{(1 - \mu_3 - \mu_4)^2 - 4 \mu_3 \mu_4} ~, \nonumber \\
C_3 & = & (\mu_3 - \mu_1) (\mu_4 - \mu_2) +
(\mu_1 + \mu_4 - \mu_2 - \mu_3) (\mu_1 \mu_4 - \mu_2 \mu_3) ~,
\end{eqnarray}
one has $t_{\mmin} < t < t_{\mmax}$ with
\begin{eqnarray}
t_{\mmin} & = & - \frac{s}{2} (C_1 + C_2) ~, \nonumber \\
t_{\mmax} & = & - \frac{s}{2} (C_1 - C_2)  
= - \frac{s}{2} \, \frac{4C_3}{C_1 + C_2}
= \frac{s^2 C_3}{t_{\mmin}} ~.
\end{eqnarray}

The Regge formulae above for single- and double-diffractive events
are supposed to hold in certain asymptotic regions of the total phase
space.  Of course, there will be diffraction also outside these 
restrictive regions. Lacking a theory which predicts differential cross 
sections at arbitrary $t$ and $M^2$ values, the Regge formulae are used
everywhere, but fudge factors are introduced in order to obtain 
`sensible' behaviour in the full phase space. These factors are:  
\begin{eqnarray}
F_{\mrm{sd}} & = & \left( 1 - \frac{M^2}{s} \right)  
\left( 1 + \frac{c_{\mrm{res}} \, M_{\mrm{res}}^2}
{M_{\mrm{res}}^2 + M^2} \right)  ~, \nonumber \\
F_{\mrm{dd}} & = & 
\left( 1 - \frac{\left( M_1 + M_2 \right)^2}{s} \right) 
\left( \frac{s\, m_{\p}^2}{ s\, m_{\p}^2 + M_1^2\, M_2^2} \right) 
\nonumber  \\
& \times & 
\left( 1 + \frac{c_{\mrm{res}} \, M_{\mrm{res}}^2}
{M_{\mrm{res}}^2 + M_1^2} \right)
\left( 1 + \frac{c_{\mrm{res}} \, M_{\mrm{res}}^2}
{M_{\mrm{res}}^2 + M_2^2} \right) ~.
\end{eqnarray}
The first factor in either expression suppresses production close to 
the kinematical limit. The second factor in $F_{dd}$ suppresses 
configurations where the two diffractive systems overlap in rapidity 
space. The final factors give an enhancement of the low-mass region,
where a resonance structure is observed in the data. Clearly a more
detailed modelling would have to be based on a set of exclusive states
rather than on this smeared-out averaging procedure. A reasonable fit
to $\p\p / \pbar\p$ data is obtained for 
$c_{\mrm{res}} = 2$ and $M_{\mrm{res}} = 2$~GeV, 
for an arbitrary particle $A$ which is diffractively excited 
we use $M_{\mrm{res}}^A = m_A - m_{\p} + 2$~GeV. 

The diffractive cross-section formulae above have been integrated 
for a set of c.m.\ energies, starting at 10 GeV, and the results have 
been parameterized. The form of these parameterizations is given in
ref. \cite{Sch94}, with explicit numbers for the $\p\p/\pbar\p$
case. {\Py} also contains similar parameterizations for 
$\pi\p$ (assumed to be same as $\rho\p$ and $\omega\p$),
$\phi\p$, $\Jpsi\p$, $\rho\rho$ ($\pi\pi$ etc.), $\rho\phi$,
$\rho\Jpsi$, $\phi\phi$, $\phi\Jpsi$ and $\Jpsi\Jpsi$.      
 
The processes above do not obey the ordinary event mixing strategy.
First of all, since their total cross sections are known, it is
possible to pick the appropriate process from the start, and then
remain with that choice. In other words, if the selection of
kinematical variables fails, one would not go back and pick a new
process, the way it was done in section \ref{sss:mixingproc}.
Second, it is not possible to impose any cuts or restrain allowed
incoming or outgoing flavours: if not additional information were to
be provided, it would make the whole scenario ill-defined.
Third, it is not recommended to mix generation of these processes
with that of any of the other ones: normally the other processes
have so small cross sections that they would almost never be
generated anyway. (We here exclude the cases of `underlying events'
and `pile-up events', where mixing is provided for, and even is a
central part of the formalism, see sections \ref{ss:multint} and
\ref{ss:pileup}.)

Once the cross-section parameterizations has been used to pick one
of the processes, the variables $t$ and $M$ are selected according
to the formulae given above.

A $\rho^0$ formed by $\gamma \to \rho^0$ in elastic or diffractive
scattering is polarized, and therefore its decay angular distribution 
in $\rho^0 \to \pi^+ \pi^-$ is taken to be proportional to 
$\sin^2 \theta$, where the reference axis is given by the $\rho^0$ 
direction of motion.

A light diffractive system, with a mass less than 1 GeV above the 
mass of the incoming particle, is allowed to decay isotropically into 
a two-body state. Single-resonance diffractive states, such as a 
$\Delta^+$, are therefore not explicitly generated, but are assumed 
described in an average, smeared-out sense.

A more massive diffractive system is subsequently treated
as a string with the quantum numbers of the original hadron. Since the
exact nature of the pomeron exchanged between the hadrons is unknown,
two alternatives are included. In the first, the pomeron is assumed to
couple to (valence) quarks, so that the string is stretched directly
between the struck quark and the remnant diquark (antiquark) of the
diffractive state. In the second, the interaction is rather with a
gluon, giving rise to a `hairpin' configuration in which the string
is stretched from a quark to a gluon and then back to a diquark
(antiquark). Both of these scenarios could be present in the data;
the default choice is to mix them in equal proportions.
 
There is experimental support for more complicated scenarios
\cite{Ing85}, wherein the pomeron has a partonic substructure,
which e.g.\ can lead to high-$\pT$ jet production in the diffractive
system. The full machinery, wherein a pomeron spectrum is convoluted
with a pomeron-proton hard interaction, is not available in {\Py}.
(But is found in the \tsc{Pompyt} program \cite{Bru96}.)

\subsubsection{Photoproduction and $\gamma\gamma$ physics}
\label{sss:photoprod}
 
The photon physics machinery in {\Py} has been largely expanded in 
recent years. Historically, the model was first developed for
photoproduction, i.e.\ a real photon on a hadron target 
\cite{Sch93,Sch93a}. Thereafter $\gamma\gamma$ physics was added
in the same spirit \cite{Sch94a,Sch97}. Only recently also virtual
photons have been added to the description \cite{Fri00}, including 
the nontrivial transition region between real photons and Deeply 
Inelastic Scattering (DIS). In this section we partly trace
this evolution towards more complex configurations.

The total $\gamma\p$ and $\gamma\gamma$ cross sections can again be 
parameterized in a
form like eq.~(\ref{pg:sigtotpomreg}), which is not so obvious since 
the photon has more complicated structure than an ordinary hadron.
In fact, the structure is still not so well understood. The model we
outline is the one studied by Schuler and Sj\"ostrand 
\cite{Sch93,Sch93a}, and further updated in \cite{Fri00}. In this model 
the physical photon is represented by
\begin{equation}
| \gamma \rangle = \sqrt{Z_3} \, | \gamma_B \rangle +
\sum_{V=\rho^0,\omega,\phi,\Jpsi} \frac{e}{f_V} \, | V \rangle +
\sum_{\q} \frac{e}{f_{\q\qbar}} \, | \q\qbar \rangle + 
\sum_{\ell=\e,\mu,\tau} \frac{e}{f_{\ell\ell}} \, 
| \ell^+ \ell^- \rangle ~.      
\label{pg:gammadecompo}
\end{equation}
 
By virtue of this superposition, one is led to a model of $\gamma\p$ 
interactions, where three different kinds of events may be 
distinguished:
\begin{Itemize}
\item Direct events, wherein the bare photon $| \gamma_B \rangle$
interacts directly with a parton from the proton. The process is 
perturbatively calculable, and no parton distributions of the photon
are involved. The typical event structure is two high-$\pT$ jets and
a proton remnant, while the photon does not leave behind any 
remnant.
\item VMD events, in which the photon fluctuates into a vector meson, 
predominantly a $\rho^0$. All the event classes known from ordinary 
hadron--hadron interactions may thus occur here, such as elastic, 
diffractive, low-$\pT$ and high-$\pT$ events. For the latter, one may 
define (VMD) parton distributions of the photon, and the photon also 
leaves behind a beam remnant. This remnant is smeared in transverse 
momentum by a typical `primordial $k_{\perp}$' of a few hundred MeV.
\item Anomalous or GVMD (Generalized VMD) events, in which the photon 
fluctuates into a $\q\qbar$ pair of larger virtuality than in the VMD 
class. The initial parton distribution is perturbatively calculable, 
as is the subsequent QCD evolution. It gives rise to the so-called 
anomalous part of the parton distributions of the photon, whence one 
name for the class. As long as only real photons were considered,
it made sense to define the cross section of this event class to be
completely perturbatively calculable, given some lower $\pT$ cut-off. 
Thus only high-$\pT$ events could occur. However, alternatively, one 
may view these states as excited higher resonances ($\rho'$ etc.), 
thus the GVMD name. In this case one is lead to a picture which also 
allows a low-$\pT$ cross section, uncalculable in perturbation theory. 
The reality may well interpolate between these two extreme alternatives, 
but the current framework more leans towards the latter point of view.
Either the  $\q$ or the $\qbar$ plays the r\^ole of a beam remnant, 
but this remnant has a larger $\pT$ than in the VMD case, related to 
the virtuality of the $\gamma \leftrightarrow \q\qbar$ fluctuation.
\end{Itemize} 
The $| \ell^+ \ell^- \rangle$ states can only interact strongly with 
partons inside the hadron at higher orders, and can therefore be 
neglected in the study of hadronic final states. 

In order that the above classification is smooth and free of double
counting, one has to introduce scales that separate the three
components. The main one is $k_0$, which separates the low-mass
vector meson region from the high-mass $| \q\qbar \rangle$ one, 
$k_0 \approx m_{\phi}/2 \approx 0.5$ GeV. Given this dividing 
line to VMD states, the anomalous parton distributions are 
perturbatively calculable. The total cross section of a state is not, 
however, since this involves aspects of soft physics and eikonalization 
of jet rates. Therefore an ansatz is chosen where the total cross section 
of a state scales like $k_V^2/\kT^2$, where the adjustable parameter 
$k_V \approx m_{\rho}/2$ for light quarks. The $\kT$ scale is
roughly equated with half the mass of the GVMD state.
The spectrum of GVMD states 
is taken to extend over a range $k_0 < \kT < k_1$, where $k_1$ is 
identified with the $\pTmin(s)$ cut-off of the perturbative jet 
spectrum in hadronic interactions, $\pTmin(s) \approx 1.5$~GeV at 
typical energies, see section \ref{ss:multint} and especially 
eq.~(\ref{eq:ptmin}). Above that range, the states are assumed to be 
sufficiently weakly interacting that no eikonalization procedure is 
required, so that cross sections can be calculated perturbatively 
without any recourse to pomeron phenomenology. There is some 
arbitrariness in that choice, and some simplifications are required 
in order to obtain a manageable description.

The VMD and GVMD/anomalous events are together called resolved ones.
In terms of high-$\pT$ jet production, the VMD and anomalous
contributions can be combined into a total resolved one, and the
same for parton-distribution functions. However, the two classes
differ in the structure of the underlying event and possibly in the
appearance of soft processes.

In terms of cross sections, eq.\ (\ref{pg:gammadecompo}) corresponds 
to
\begin{equation}
\sigma_{\mrm{tot}}^{\gamma\p}(s) = \sigma_{\mrm{dir}}^{\gamma\p}(s) +
\sigma_{\mrm{VMD}}^{\gamma\p}(s) + \sigma_{\mrm{anom}}^{\gamma\p}(s) ~.
\label{pg:gammacrossdecompo}
\end{equation}

The direct cross section is, to lowest order, the perturbative cross 
section for the two processes $\gamma\q \to \q\g$ and 
$\gamma\g \to \q\qbar$, with a lower cut-off $\pT > k_1$, in order to 
avoid double-counting with the interactions of the GVMD states.
Properly speaking, this should be multiplied by the $Z_3$ coefficient, 
\begin{equation}
Z_3 = 1 -  
\sum_{V=\rho^0,\omega,\phi,\Jpsi} \left( \frac{e}{f_V} \right)^2 - 
\sum_{\q} \left( \frac{e}{f_{\q\qbar}} \right)^2 -
\sum_{\ell=\e,\mu,\tau} \left( \frac{e}{f_{\ell\ell}} \right)^2 ~,
\end{equation}
but normally $Z_3$ is so close to unity as to make no difference.

The VMD factor $(e/f_V)^2 = 4\pi\alphaem/f_V^2$ gives the probability 
for the transition $\gamma \to V$. The coefficients $f_V^2/4\pi$ are
determined from data to be (with a non-negligible amount of 
uncertainty) 2.20 for $\rho^0$, 23.6 for $\omega$, 18.4 for $\phi$ 
and 11.5 for $\Jpsi$. Together these numbers imply that the photon
can be found in a VMD state about 0.4\% of the time, dominated by the
$\rho^0$ contribution. All the properties of the VMD interactions
can be obtained by appropriately scaling down $V\p$ physics 
predictions. Thus the whole machinery developed in the previous 
section for hadron--hadron interactions is directly applicable.
Also parton distributions of the VMD component inside the photon 
are obtained by suitable rescaling.

The contribution from the `anomalous' high-mass fluctuations to the
total cross section is obtained by a convolution of the fluctuation
rate 
\begin{equation}
\sum_{\q} \left( \frac{e}{f_{\q\qbar}} \right)^2 \approx 
\frac{\alphaem}{2\pi} \, \left( 2 \sum_{\q} e_{\q}^2 \right) 
\int_{k_0}^{k_1} \frac{\d \kT^2}{\kT^2} ~,
\label{anomintfirst}
\end{equation}
which is to be multiplied by the abovementioned reduction factor
$k_V^2/\kT^2$ for the total cross section, and all scaled by the
assumed real vector meson cross section.

As an illustration of this scenario, the phase space of $\gamma\p$ 
events may be represented by a $(\kT,\pT)$ plane.
Two transverse momentum scales are distinguished: the 
photon resolution scale $\kT$ and the hard interaction scale $\pT$.
Here $\kT$ is a measure of the virtuality of a fluctuation of the 
photon and $\pT$ corresponds to the most virtual rung of the ladder, 
possibly apart from $\kT$. 
As we have discussed above, the low-$\kT$ region corresponds to
VMD and GVMD states that encompasses both perturbative high-$\pT$ and
nonperturbative low-$\pT$ interactions. Above $k_1$, the region is split 
along the line $\kT = \pT$. When $\pT > \kT$ the photon is resolved by
the hard interaction, as described by the anomalous part of the photon 
distribution function. This is as in the GVMD sector, except that we should 
(probably) not worry about multiple parton--parton interactions. In the
complementary region $\kT > \pT$, the $\pT$ scale is just part of the 
traditional evolution of the parton distributions of the proton  up to 
the scale of $\kT$, and 
thus there is no need to introduce an internal structure of the photon. 
One could imagine the direct class of events as extending below $k_1$
and there being the low-$\pT$ part of the GVMD class, only appearing 
when a hard interaction at a larger $\pT$ scale would not preempt it.  
This possibility is implicit in the standard cross section framework.  

In $\gamma\gamma$ physics \cite{Sch94a,Sch97}, the superposition in 
eq.~(\ref{pg:gammadecompo}) applies separately for each of the two 
incoming photons. In total there are therefore $3 \times 3 = 9$ 
combinations. However, trivial symmetry reduces this to six distinct 
classes, written in terms of the total cross section 
(cf. eq.~(\ref{pg:gammacrossdecompo})) as
\begin{eqnarray}
\sigma_{\mrm{tot}}^{\gamma\gamma}(s) & = &
\sigma_{\mrm{dir}\times\mrm{dir}}^{\gamma\gamma}(s) +
\sigma_{\mrm{VMD}\times\mrm{VMD}}^{\gamma\gamma}(s) + 
\sigma_{\mrm{GVMD}\times\mrm{GVMD}}^{\gamma\gamma}(s) \nonumber \\ 
& + & 2 \sigma_{\mrm{dir}\times\mrm{VMD}}^{\gamma\gamma}(s) +
2 \sigma_{\mrm{dir}\times\mrm{GVMD}}^{\gamma\gamma}(s) + 
2 \sigma_{\mrm{VMD}\times\mrm{GVMD}}^{\gamma\gamma}(s) ~.
\label{pg:gagacrossdecompo}
\end{eqnarray}
A parameterization of the total $\gamma\gamma$ cross section is found in
\cite{Sch94a,Sch97}.

The six different kinds of $\gamma\gamma$ events are thus:
\begin{Itemize}
\item The direct$\times$direct events, which correspond to the 
subprocess $\gamma\gamma \to \q\qbar$ (or $\ell^+\ell^-$). 
The typical event structure is two high-$\pT$ jets and no beam
remnants. 
\item The VMD$\times$VMD events, which have the same properties as
the VMD $\gamma\p$ events. There are four by four combinations of
the two incoming vector mesons, with one VMD factor for each meson.
\item The GVMD$\times$GVMD events, wherein each photon
fluctuates into a $\q\qbar$ pair of larger virtuality than in the
VMD class. The `anomalous' classification assumes that
one parton of each pair gives a beam remnant, whereas
the other (or a daughter parton thereof) participates in a high-$\pT$ 
scattering. The GVMD concept implies the presence also of low-$\pT$
events, like for VMD.
\item The direct$\times$VMD events, which have the same properties as
the direct $\gamma\p$ events. 
\item The direct$\times$GVMD events, in which a bare photon 
interacts with a parton from the anomalous photon.
The typical structure is then two high-$\pT$ jets and a beam remnant. 
\item The VMD$\times$GVMD events, which have the same properties 
as the GVMD $\gamma\p$ events. 
\end{Itemize} 

Like for photoproduction events, this can be illustrated in a 
parameter space, but now three-dimensional, with axes
given by the $k_{\perp 1}$, $k_{\perp 2}$ and $p_{\perp}$ scales. Here 
each $k_{\perp i}$ is a measure of the virtuality of a fluctuation of 
a photon, and $p_{\perp}$ corresponds to the most virtual rung on 
the ladder between the two photons, possibly excepting the endpoint 
$k_{\perp i}$ ones. So, to first approximation, the coordinates along the 
$k_{\perp i}$ axes determine the characters of the interacting photons 
while $p_{\perp}$ determines the character of the interaction process. 
Double counting should be avoided by trying to impose a consistent 
classification. Thus, for instance, $p_{\perp} > k_{\perp i}$ 
with $k_{\perp 1} < k_0$ and $k_0 <k_{\perp 2} < k_1$ gives a hard 
interaction between a VMD and a GVMD photon, while 
$k_{\perp 1} > p_{\perp} > k_{\perp 2}$ with $k_{\perp 1} > k_1$ 
and $k_{\perp 2} < k_0$ is a single-resolved process
(direct$\times$VMD; with $p_{\perp}$ now in the parton distribution 
evolution).

In much of the literature, where a coarser classification is used, 
our direct$\times$direct is called direct, our direct$\times$VMD
and direct$\times$GVMD is called single-resolved since they both 
involve one resolved photon which gives a beam remnant, 
and the rest are called double-resolved since both photons are resolved 
and give beam remnants. 
 
If the photon is virtual, it has a reduced probability to fluctuate into 
a vector meson state, and this state has a reduced interaction probability.
This can be modelled by a traditional dipole factor
$(m_V^2/(m_V^2 + Q^2))^2$ for a photon of virtuality $Q^2$, where 
$m_V \to 2 \kT$ for a GVMD state. Putting it all together, the cross
section of the GVMD sector of photoproduction then scales like
\begin{equation}
\int_{k_0^2}^{k_1^2} \frac{\d\kT^2}{\kT^2} \, \frac{k_V^2}{\kT^2} \,
\left( \frac{4\kT^2}{4\kT^2 + Q^2} \right)^2 ~.
\end{equation}

For a virtual photon the DIS process $\gast \q \to \q$ is also possible, 
but by gauge invariance its cross section must vanish in the limit 
$Q^2 \to 0$. At large $Q^2$, the direct processes can be considered 
as the $\mathcal{O}(\alphas)$ correction to the lowest-order 
DIS process, but the direct ones survive for $Q^2 \to 0$. There is no 
unique prescription for a proper combination at all $Q^2$, but we have
attempted an approach that gives the proper limits and minimizes 
double-counting. For large $Q^2$, the DIS $\gast\p$ cross section
is proportional to the structure function $F_2 (x, Q^2)$ with the
Bjorken $x = Q^2/(Q^2 + W^2)$. Since normal parton distribution 
parameterizations are frozen below some $Q_0$ scale and therefore do not
obey the gauge invariance condition, an ad hoc factor 
$(Q^2/(Q^2 + m_{\rho}^2))^2$ is introduced for the conversion from 
the parameterized $F_2(x,Q^2)$ to a $\sigma_{\mrm{DIS}}^{\gast\p}$:
\begin{equation}
\sigma_{\mrm{DIS}}^{\gast\p} \simeq  
\left( \frac{Q^2}{Q^2 + m_{\rho}^2} \right)^2 \,
\frac{4\pi^2\alphaem}{Q^2} F_2(x,Q^2) =
\frac{4\pi^2\alphaem Q^2}{(Q^2+m_{\rho}^2)^2} \,
\sum_{\q} e_{\q}^2 \, \left\{ x  q(x, Q^2) + x \br{q}(x,Q^2) \right\} 
~.
\label{eq:sigDIS}
\end{equation}
Here $m_{\rho}$ is some nonperturbative hadronic mass parameter, 
for simplicity identified with the $\rho$ mass. One of the 
$Q^2/(Q^2+m_{\rho}^2)$ factors is required already to give finite 
$\sigma_{\mrm{tot}}^{\gamma\p}$ for conventional parton distributions, 
and could be viewed as a screening of the individual partons at small 
$Q^2$. The second factor is chosen to give not only a finite but 
actually a vanishing $\sigma_{\mrm{DIS}}^{\gast\p}$ 
for $Q^2 \to 0$ in order to retain the pure photoproduction description 
there. This latter factor thus is more a matter  of convenience, and 
other approaches could have been pursued.

In order to avoid double-counting between DIS and direct events, a 
requirement $\pT > \max(k_1, Q)$ is imposed on direct events. In the 
remaining DIS ones, denoted lowest order (LO) DIS, thus $\pT < Q$. 
This would suggest a subdivision
$\sigma_{\mrm{LO\,DIS}}^{\gast\p} = \sigma_{\mrm{DIS}}^{\gast\p} -
\sigma_{\mrm{direct}}^{\gast\p}$, with $\sigma_{\mrm{DIS}}^{\gast\p}$ 
given by eq.~(\ref{eq:sigDIS}) and $\sigma_{\mrm{direct}}^{\gast\p}$ 
by the perturbative matrix elements. In the limit $Q^2 \to 0$, the 
DIS cross section is now constructed to vanish while the direct is not, 
so this would give $\sigma_{\mrm{LO\,DIS}}^{\gast\p} < 0$. However, 
here we expect the correct answer not to be a negative number but an 
exponentially suppressed one, by a Sudakov form factor. This modifies 
the cross section: 
\begin{equation}
\sigma_{\mrm{LO\,DIS}}^{\gast\p} = \sigma_{\mrm{DIS}}^{\gast\p} -
\sigma_{\mrm{direct}}^{\gast\p}
~~ \longrightarrow ~~ 
\sigma_{\mrm{DIS}}^{\gast\p} \; \exp \left( - \frac{%
\sigma_{\mrm{direct}}^{\gast\p}}{\sigma_{\mrm{DIS}}^{\gast\p}} \right) \;.
\label{eq:LODIS}
\end{equation}
Since we here are in a region where the DIS cross section is no longer the 
dominant one, this change of the total DIS cross section is not essential. 

The overall picture, from a DIS perspective, now requires three scales 
to be kept track of. The traditional DIS region is the strongly ordered 
one, $Q^2 \gg \kT^2 \gg \pT^2$, where DGLAP-style evolution 
\cite{Alt77, Gri72} is responsible for the event 
structure. As always, ideology wants strong ordering, while 
the actual classification is based on ordinary ordering 
$Q^2 > \kT^2 > \pT^2$. The region $\kT^2 > \max(Q^2,\pT^2)$ is also
DIS, but of the $\mathcal{O}(\alphas)$ direct kind. The region 
where $\kT$ is the smallest scale corresponds to 
non-ordered emissions, that then go beyond DGLAP validity,
while the region $\pT^2 > \kT^2 > Q^2$ cover the interactions of a 
resolved virtual photon. Comparing with the plane of real 
photoproduction, we conclude that the whole region
$\pT > \kT$ involves no double-counting, since we have made no
attempt at a non-DGLAP DIS description but can choose to cover this 
region entirely by the VMD/GVMD descriptions. Actually, it is only 
in the corner $\pT < \kT < \min(k_1, Q)$ that an overlap can occur 
between the resolved 
and the DIS descriptions. Some further considerations show that
usually either of the two is strongly suppressed in this region,
except in the range of intermediate $Q^2$ and rather small $W^2$.
Typically, this is the region where $x \approx Q^2/(Q^2 + W^2)$ is not 
close to zero, and where $F_2$ is dominated by the valence-quark 
contribution. The latter behaves roughly $\propto (1-x)^n$, with an 
$n$ of the order of 3 or 4. Therefore we will introduce a corresponding 
damping factor to the VMD/GVMD terms. 

In total, we have now arrived at our ansatz for all $Q^2$:
\begin{equation}
\sigma_{\mrm{tot}}^{\gast\p} = 
\sigma_{\mrm{DIS}}^{\gast\p} \; \exp \left( - 
\frac{\sigma_{\mrm{direct}}^{\gast\p}}{\sigma_{\mrm{DIS}}^{\gast\p}} 
\right) + \sigma_{\mrm{direct}}^{\gast\p} +
\left( \frac{W^2}{Q^2 + W^2} \right)^n \left(
\sigma_{\mrm{VMD}}^{\gast\p} + 
\sigma_{\mrm{GVMD}}^{\gast\p} \right) \;,
\label{eq:gammapallQ}
\end{equation}
with four main components. Most of these in their turn have
a complicated internal structure, as we have seen. 

Turning to $\gast\gast$ processes, finally, the parameter space is now 
five-dimensional: $Q_1$, $Q_2$, $k_{\perp 1}$, $k_{\perp 2}$ and $\pT$. 
As before, an effort is made to avoid double-counting, by having a 
unique classification of each region in the five-dimensional space. 
Remaining double-counting is dealt with as above.
In total, our ansatz for $\gast\gast$ interactions at all $Q^2$ contains 
13 components: 9 when two VMD, GVMD or direct photons interact, as is 
already allowed for real photons, plus a further 4 where a `DIS photon' 
from either side interacts with a VMD or GVMD one. With the label 
resolved used to denote VMD and GVMD, one can write
\begin{eqnarray}
\sigma_{\mrm{tot}}^{\gast\gast} (W^2, Q_1^2, Q_2^2) & = &
\sigma_{\mrm{DIS}\times\mrm{res}}^{\gast\gast} \; 
\exp \left( - \frac{\sigma_{\mrm{dir}\times\mrm{res}}^{\gast\gast}}%
{\sigma_{\mrm{DIS}\times\mrm{res}}^{\gast\gast}} \right) +
\sigma_{\mrm{dir}\times\mrm{res}}^{\gast\gast} 
\nonumber \\
 & + & \sigma_{\mrm{res}\times\mrm{DIS}}^{\gast\gast} \; 
\exp \left( - \frac{\sigma_{\mrm{res}\times\mrm{dir}}^{\gast\gast}}%
{\sigma_{\mrm{res}\times\mrm{DIS}}^{\gast\gast}} \right) +
\sigma_{\mrm{res}\times\mrm{dir}}^{\gast\gast}  
\label{eq:gagaallQ}  \\
 & + & \sigma_{\mrm{dir}\times\mrm{dir}}^{\gast\gast}
+ \left( \frac{W^2}{Q_1^2 + Q_2^2 + W^2} \right)^3 \;
\sigma_{\mrm{res}\times\mrm{res}}^{\gast\gast} \nonumber
\end{eqnarray}
Most of the 13 components in their turn have a complicated internal 
structure, as we have seen. 

An important note is that the $Q^2$ dependence of the DIS and direct 
photon interactions is implemented in the matrix element expressions, 
i.e.\ in processes such as $\gast\gast \to \q\qbar$ or 
$\gast \q \to \q \g$ the photon virtuality explicitly enters. This is 
different from VMD/GVMD, where dipole factors are used to reduce the 
total cross sections and the assumed flux of 
partons inside a virtual photon relative to those of a real one, but 
the matrix elements themselves contain no dependence on the virtuality
either of the partons or of the photon itself. 
Typically results are obtained with the SaS~1D parton distributions 
for the virtual transverse photons
\cite{Sch95,Sch96}, since these are well matched to our framework, e.g.
allowing a separation of the VMD and GVMD/anomalous components. 
Parton distributions of virtual longitudinal photons are by default
given by some $Q^2$-dependent factor times the transverse ones.
The new set by Ch\'yla \cite{Chy00} allows more precise modelling 
here, but first indications are that many studies will not be sensitive 
to the detailed shape.

The photon physics machinery is of considerable complexity, and
so the above is only a brief summary. Further details can be found 
in the literature quoted above. Some topics are also covered in
other places in this manual, e.g. the flux of transverse and
longitudinal photons in subsection \ref{sss:equivgamma}, scale choices 
for parton density evaluation in subsection \ref{ss:kinemtwo}, and 
further aspects of the generation machinery and switches in subsection
\ref{ss:photoanddisclass}.  

\clearpage
 
\section{Physics Processes}
\label{s:pytproc}
 
In this section we enumerate the physics processes that are available 
in {\Py}, introducing the {\ISUB} code that can be used to select
desired processes. A number of comments are made about the
physics scenarios involved, in particular with respect to
underlying assumptions and domain of validity. The section closes
with a survey of interesting processes by machine.
 
\subsection{The Process Classification Scheme}
\label{ss:ISUBcode}
 
A wide selection of fundamental $2 \to 1$ and $2 \to 2$ tree
processes of the Standard Model (electroweak and strong) has been
included in {\Py}, and slots are provided for many more, not yet
implemented. In addition, `minimum-bias'-type processes
(like elastic scattering), loop graphs, box graphs, $2 \to 3$ tree
graphs and many non-Standard Model processes are included. The
classification is not always unique. A process that proceeds only
via an $s$-channel state is classified as a $2 \to 1$ process
(e.g.\ $\q \qbar \to \gammaZ \to \ee$), but a $2 \to 2$
cross section may well have contributions from $s$-channel diagrams
($\g \g \to \g \g$ obtains contributions from
$\g \g \to \g^* \to \g \g$). Also, in the program, $2 \to 1$ and
$2 \to 2$ graphs may sometimes be convoluted with two $1 \to 2$
splittings to form effective $2 \to 3$ or $2 \to 4$ processes
($\W^+ \W^- \to \hrm^0$ is folded with $\q \to \q'' \W^+$ and
$\q' \to \q''' \W^-$ to give $\q \q' \to \q'' \q''' \hrm^0$).

The original classification and numbering scheme feels less relevant
today than when originally conceived. In those days, the calculation
of $2 \to 3$ or $2 \to 4$ matrix elements was sufficiently complicated 
that one would wish to avoid it if at all possible, e.g.\ by having in 
mind to define effective parton densities for all standard model 
particles, such as the $\W^{\pm}$. Today, the improved computational 
techniques and increased computing power implies that people would be 
willing to include a branching $\q \to \q \W$ as part of the hard 
process, i.e.\ not try to factor it off in some approximation. With the 
large top mass and large Higgs mass limits, there is also a natural 
subdivision, such that the $\b$ quark is the heaviest object for which 
the parton distribution concept makes sense. Therefore most of the 
prepared but empty slots are likely to remain empty, or be reclaimed 
for other processes.
  
It is possible to select a combination of subprocesses to simulate,
and also afterwards to know which subprocess was actually selected in
each event. For this purpose, all subprocesses are numbered according
to an {\ISUB} code. The list of possible codes is given in Tables
\ref{t:procone}, \ref{t:proctwo}, \ref{t:procthree}, \ref{t:procfour},
\ref{t:procfive}, \ref{t:procsix}, \ref{t:procseven} and
\ref{t:proceight}. Only processes marked with a `+' sign in the first
column have been implemented in the program to date. Although
{\ISUB} codes were originally designed in a logical fashion,
we must admit that subsequent developments of the program have tended
to obscure the structure. For instance, the process numbers for Higgs
production are spread out, in part as a consequence of the original
classification, in part because further production mechanisms have been
added one at a time, in whatever free slots could be found. At some
future date the subprocess list will therefore be re-organized.
In the thematic descriptions that follow the main tables, the
processes of interest are repeated in a more logical order. If you
want to look for a specific process, it will be easier
to find it there.
 
\begin{table}[pt]
\captive{Subprocess codes, part 1. First column is `+' for processes
implemented and blank for those that are only foreseen. Second is
the subprocess number {\ISUB}, and third the description of the
process. The final column gives references from which the
cross sections have been obtained. See text for further information.
\protect\label{t:procone} }
\begin{center}
\begin{tabular}{|c|r|l|l|@{\protect\rule{0mm}{\tablinsep}}}
\hline
In & No. & Subprocess & Reference \\
\hline
+ &   1 & $\f_i \fbar_i \to \gammaZ$ & \cite{Eic84} \\
+ &   2 & $\f_i \fbar_j \to \W^+$ & \cite{Eic84} \\
+ &   3 & $\f_i \fbar_i \to \hrm^0$ & \cite{Eic84} \\
  &   4 & $\gamma \W^+ \to \W^+$ &  \\
+ &   5 & $\Z^0 \Z^0 \to \hrm^0$ & \cite{Eic84,Cha85} \\
  &   6 & $\Z^0 \W^+ \to \W^+$ &  \\
  &   7 & $\W^+ \W^- \to \Z^0$ &  \\
+ &   8 & $\W^+ \W^- \to \hrm^0$ & \cite{Eic84,Cha85} \\
+ &  10 & $\f_i \f_j \to \f_k \f_l$ (QFD) & \cite{Ing87a}  \\
+ &  11 & $\f_i \f_j \to \f_i \f_j$ (QCD) & 
\cite{Com77,Ben84,Eic84} \\
+ &  12 & $\f_i \fbar_i \to \f_k \fbar_k$ & 
\cite{Com77,Ben84,Eic84} \\
+ &  13 & $\f_i \fbar_i \to \g \g$ & \cite{Com77,Ben84}  \\
+ &  14 & $\f_i \fbar_i \to \g \gamma$ & \cite{Hal78,Ben84}  \\
+ &  15 & $\f_i \fbar_i \to \g \Z^0$ & \cite{Eic84}  \\
+ &  16 & $\f_i \fbar_j \to \g \W^+$ & \cite{Eic84}  \\
  &  17 & $\f_i \fbar_i \to \g \hrm^0$ &   \\
+ &  18 & $\f_i \fbar_i \to \gamma \gamma$ & \cite{Ber84}  \\
+ &  19 & $\f_i \fbar_i \to \gamma \Z^0$ & \cite{Eic84}  \\
+ &  20 & $\f_i \fbar_j \to \gamma \W^+$ & \cite{Eic84,Sam91}  \\
  &  21 & $\f_i \fbar_i \to \gamma \hrm^0$ &   \\
+ &  22 & $\f_i \fbar_i \to \Z^0 \Z^0$ & \cite{Eic84,Gun86}  \\
+ &  23 & $\f_i \fbar_j \to \Z^0 \W^+$ & \cite{Eic84,Gun86}  \\
+ &  24 & $\f_i \fbar_i \to \Z^0 \hrm^0$ & \cite{Ber85}  \\
+ &  25 & $\f_i \fbar_i \to \W^+ \W^-$ & \cite{Bar94,Gun86}  \\
+ &  26 & $\f_i \fbar_j \to \W^+ \hrm^0$ & \cite{Eic84}  \\
  &  27 & $\f_i \fbar_i \to \hrm^0 \hrm^0$ &   \\
+ &  28 & $\f_i \g \to \f_i \g$ & \cite{Com77,Ben84}  \\
+ &  29 & $\f_i \g \to \f_i \gamma$ & \cite{Hal78,Ben84}  \\
+ &  30 & $\f_i \g \to \f_i \Z^0$ & \cite{Eic84}  \\
+ &  31 & $\f_i \g \to \f_k \W^+$ & \cite{Eic84}  \\
+ &  32 & $\f_i \g \to \f_i \hrm^0$ &   \\
+ &  33 & $\f_i \gamma \to \f_i \g$ & \cite{Duk82}  \\
+ &  34 & $\f_i \gamma \to \f_i \gamma$ & \cite{Duk82}  \\
+ &  35 & $\f_i \gamma \to \f_i \Z^0$ & \cite{Gab86}  \\
+ &  36 & $\f_i \gamma \to \f_k \W^+$ & \cite{Gab86}  \\
  &  37 & $\f_i \gamma \to \f_i \hrm^0$ &   \\
  &  38 & $\f_i \Z^0 \to \f_i \g$ &   \\
  &  39 & $\f_i \Z^0 \to \f_i \gamma$ &   \\
\hline
\end{tabular}
\end{center}
\end{table}
 
\begin{table}[pt]
\captive{Subprocess codes, part 2. Comments as before.
\protect\label{t:proctwo} }
\begin{center}
\begin{tabular}{|c|r|l|l|@{\protect\rule{0mm}{\tablinsep}}}
\hline
In & No. & Subprocess & Reference \\
\hline
  &  40 & $\f_i \Z^0 \to \f_i \Z^0$ &   \\
  &  41 & $\f_i \Z^0 \to \f_k \W^+$ &   \\
  &  42 & $\f_i \Z^0 \to \f_i \hrm^0$ &   \\
  &  43 & $\f_i \W^+ \to \f_k \g$ &   \\
  &  44 & $\f_i \W^+ \to \f_k \gamma$ &   \\
  &  45 & $\f_i \W^+ \to \f_k \Z^0$ &   \\
  &  46 & $\f_i \W^+ \to \f_k \W^+$ &   \\
  &  47 & $\f_i \W^+ \to \f_k \hrm^0$ &   \\
  &  48 & $\f_i \hrm^0 \to \f_i \g$ &   \\
  &  49 & $\f_i \hrm^0 \to \f_i \gamma$ &   \\
  &  50 & $\f_i \hrm^0 \to \f_i \Z^0$ &   \\
  &  51 & $\f_i \hrm^0 \to \f_k \W^+$ &   \\
  &  52 & $\f_i \hrm^0 \to \f_i \hrm^0$ &   \\
+ &  53 & $\g \g \to \f_k \fbar_k$ & \cite{Com77,Ben84}  \\
+ &  54 & $\g \gamma \to \f_k \fbar_k$ & \cite{Duk82}  \\
  &  55 & $\g \Z^0 \to \f_k \fbar_k$ &   \\
  &  56 & $\g \W^+ \to \f_k \fbar_l$ &   \\
  &  57 & $\g \hrm^0 \to \f_k \fbar_k$ &   \\
+ &  58 & $\gamma \gamma \to \f_k \fbar_k$ & \cite{Bar90}  \\
  &  59 & $\gamma \Z^0 \to \f_k \fbar_k$ &   \\
  &  60 & $\gamma \W^+ \to \f_k \fbar_l$ &   \\
  &  61 & $\gamma \hrm^0 \to \f_k \fbar_k$ &   \\
  &  62 & $\Z^0 \Z^0 \to \f_k \fbar_k$ &   \\
  &  63 & $\Z^0 \W^+ \to \f_k \fbar_l$ &   \\
  &  64 & $\Z^0 \hrm^0 \to \f_k \fbar_k$ &   \\
  &  65 & $\W^+ \W^- \to \f_k \fbar_k$ &   \\
  &  66 & $\W^+ \hrm^0 \to \f_k \fbar_l$ &   \\
  &  67 & $\hrm^0 \hrm^0 \to \f_k \fbar_k$ &   \\
+ &  68 & $\g \g \to \g \g$ &  \cite{Com77,Ben84} \\
+ &  69 & $\gamma \gamma \to \W^+ \W^-$ & \cite{Kat83}   \\
+ &  70 & $\gamma \W^+ \to \Z^0 \W^+$ & \cite{Kun87}  \\
+ &  71 & $\Z^0 \Z^0 \to \Z^0 \Z^0$ (longitudinal) &  \cite{Abb87} \\
+ &  72 & $\Z^0 \Z^0 \to \W^+ \W^-$ (longitudinal) &  \cite{Abb87} \\
+ &  73 & $\Z^0 \W^+ \to \Z^0 \W^+$ (longitudinal) &  \cite{Dob91} \\
  &  74 & $\Z^0 \hrm^0 \to \Z^0 \hrm^0$ &   \\
  &  75 & $\W^+ \W^- \to \gamma \gamma$ &   \\
+ &  76 & $\W^+ \W^- \to \Z^0 \Z^0$ (longitudinal) & \cite{Ben87b} \\
+ &  77 & $\W^+ \W^{\pm} \to \W^+ \W^{\pm}$ (longitudinal) & 
\cite{Dun86,Bar90a}  \\
  &  78 & $\W^+ \hrm^0 \to \W^+ \hrm^0$ &   \\
  &  79 & $\hrm^0 \hrm^0 \to \hrm^0 \hrm^0$ &   \\
+ &  80 & $\q_i \gamma \to \q_k \pi^{\pm}$ & \cite{Bag82} \\
\hline
\end{tabular}
\end{center}
\end{table}
 
\begin{table}[pt]
\captive{Subprocess codes, part 3. Comments as before
\protect\label{t:procthree} }
\begin{center}
\begin{tabular}{|c|r|l|l|@{\protect\rule{0mm}{\tablinsep}}}
\hline
In & No. & Subprocess & Reference \\
\hline
+ &  81 & $\f_i \fbar_i \to \Q_k \Qbar_k$ & \cite{Com79}  \\
+ &  82 & $\g \g \to \Q_k \Qbar_k$ & \cite{Com79}  \\
+ &  83 & $\q_i \f_j \to \Q_k \f_l$ & \cite{Dic86}  \\
+ &  84 & $\g \gamma \to \Q_k \Qbar_k$ & \cite{Fon81}  \\
+ &  85 & $\gamma \gamma \to \F_k \Fbar_k$ & \cite{Bar90}  \\
+ &  86 & $\g \g \to \Jpsi \g$ & \cite{Bai83}  \\
+ &  87 & $\g \g \to \chi_{0 \c} \g$ & \cite{Gas87}  \\
+ &  88 & $\g \g \to \chi_{1 \c} \g$ & \cite{Gas87}  \\
+ &  89 & $\g \g \to \chi_{2 \c} \g$ & \cite{Gas87}  \\
+ &  91 & elastic scattering               & \cite{Sch94}  \\
+ &  92 & single diffraction ($AB \to XB$) & \cite{Sch94}  \\
+ &  93 & single diffraction ($AB \to AX$) & \cite{Sch94}  \\
+ &  94 & double diffraction               & \cite{Sch94}  \\
+ &  95 & low-$\pT$ production & \cite{Sjo87a}  \\
+ &  96 & semihard QCD $2 \to 2$ & \cite{Sjo87a} \\
+ &  99 & $\gast \q \to \q$ & \\
  & 101 & $\g \g \to \Z^0$ &   \\
+ & 102 & $\g \g \to \hrm^0$ & \cite{Eic84}  \\
+ & 103 & $\gamma \gamma \to \hrm^0$ & \cite{Dre89}  \\
+ & 104 & $\g \g \to \chi_{0 \c}$ & \cite{Bai83}  \\
+ & 105 & $\g \g \to \chi_{2 \c}$ & \cite{Bai83}  \\
+ & 106 & $\g \g \to \Jpsi \gamma$ & \cite{Dre91}  \\
+ & 107 & $\g \gamma \to \Jpsi \g$ & \cite{Ber81}  \\
+ & 108 & $\gamma \gamma \to \Jpsi \gamma$ & \cite{Jun97}  \\
+ & 110 & $\f_i \fbar_i \to \gamma \hrm^0$ & \cite{Ber85a} \\
+ & 111 & $\f_i \fbar_i \to \g \hrm^0$ & \cite{Ell88}  \\
+ & 112 & $\f_i \g \to \f_i \hrm^0$ & \cite{Ell88}  \\
+ & 113 & $\g \g \to \g \hrm^0$ & \cite{Ell88}  \\
+ & 114 & $\g \g \to \gamma \gamma$ & \cite{Con71,Ber84,Dic88}  \\
+ & 115 & $\g \g \to \g \gamma$ & \cite{Con71,Ber84,Dic88}  \\
  & 116 & $\g \g \to \gamma \Z^0$ &   \\
  & 117 & $\g \g \to \Z^0 \Z^0$ &   \\
  & 118 & $\g \g \to \W^+ \W^-$ &   \\
  & 119 & $\gamma \gamma \to \g \g$ &   \\
+ & 121 & $\g \g \to \Q_k \Qbar_k \hrm^0$ & \cite{Kun84}  \\
+ & 122 & $\q_i \qbar_i \to \Q_k \Qbar_k \hrm^0$ & \cite{Kun84}  \\
+ & 123 & $\f_i \f_j \to \f_i \f_j \hrm^0$ ($\Z \Z$ fusion) & 
\cite{Cah84}  \\
+ & 124 & $\f_i \f_j \to \f_k \f_l \hrm^0$ ($\W^+\W^-$ fusion) & 
\cite{Cah84}  \\
+ & 131 & $\f_i \gast_{\mrm{T}} \to \f_i \g$ & \cite{Alt78}  \\
+ & 132 & $\f_i \gast_{\mrm{L}} \to \f_i \g$ & \cite{Alt78} \\
+ & 133 & $\f_i \gast_{\mrm{T}} \to \f_i \gamma$ & \cite{Alt78} \\
\hline
\end{tabular}
\end{center}
\end{table}
 
\begin{table}[pt]
\captive{Subprocess codes, part 4. Comments as before.
\protect\label{t:procfour} }
\begin{center}
\begin{tabular}{|c|r|l|l|@{\protect\rule{0mm}{\tablinsep}}}
\hline
In & No. & Subprocess & Reference \\
\hline
+ & 134 & $\f_i \gast_{\mrm{L}} \to \f_i \gamma$ & \cite{Alt78} \\
+ & 135 & $\g \gast_{\mrm{T}} \to \f_i \fbar_i$ & \cite{Alt78} \\
+ & 136 & $\g \gast_{\mrm{L}} \to \f_i \fbar_i$ & \cite{Alt78} \\
+ & 137 & $\gast_{\mrm{T}} \gast_{\mrm{T}} \to \f_i \fbar_i$ & 
\cite{Bai81} \\
+ & 138 & $\gast_{\mrm{T}} \gast_{\mrm{L}} \to \f_i \fbar_i$ & 
\cite{Bai81} \\
+ & 139 & $\gast_{\mrm{L}} \gast_{\mrm{T}} \to \f_i \fbar_i$ & 
\cite{Bai81} \\
+ & 140 & $\gast_{\mrm{L}} \gast_{\mrm{L}} \to \f_i \fbar_i$ & 
\cite{Bai81} \\
+ & 141 & $\f_i \fbar_i \to \gamma/\Z^0/\Z'^0$ & \cite{Alt89}  \\
+ & 142 & $\f_i \fbar_j \to \W'^+$ & \cite{Alt89}  \\
+ & 143 & $\f_i \fbar_j \to \H^+$ & \cite{Gun87}  \\
+ & 144 & $\f_i \fbar_j \to \R$ & \cite{Ben85a}  \\
+ & 145 & $\q_i \ell_j \to \L_{\Q}$ & \cite{Wud86}  \\
+ & 146 & $\e \gamma \to \e^*$ & \cite{Bau90}  \\
+ & 147 & $\d \g \to \d^*$ & \cite{Bau90}  \\
+ & 148 & $\u \g \to \u^*$ & \cite{Bau90}  \\
+ & 149 & $\g \g \to \eta_{\mrm{tc}}$ & \cite{Eic84,App92}  \\
+ & 151 & $\f_i \fbar_i \to \H^0$ & \cite{Eic84}  \\
+ & 152 & $\g \g \to \H^0$ & \cite{Eic84}  \\
+ & 153 & $\gamma \gamma \to \H^0$ & \cite{Dre89}  \\
+ & 156 & $\f_i \fbar_i \to \A^0$ & \cite{Eic84}  \\
+ & 157 & $\g \g \to \A^0$ & \cite{Eic84}  \\
+ & 158 & $\gamma \gamma \to \A^0$ & \cite{Dre89}  \\
+ & 161 & $\f_i \g \to \f_k \H^+$ & \cite{Bar88}  \\
+ & 162 & $\q_i \g \to \ell_k \L_{\Q}$ & \cite{Hew88}  \\
+ & 163 & $\g \g \to \L_{\Q} \br{\L}_{\Q}$ & 
\cite{Hew88,Eic84} \\
+ & 164 & $\q_i \qbar_i \to \L_{\Q} \br{\L}_{\Q}$ & 
\cite{Hew88}  \\
+ & 165 & $\f_i \fbar_i \to \f_k \fbar_k$ (via $\gammaZ$) & 
\cite{Eic84,Lan91}  \\
+ & 166 & $\f_i \fbar_j \to \f_k \fbar_l$ (via $\W^{\pm}$) & 
\cite{Eic84,Lan91}  \\
+ & 167 & $\q_i \q_j \to \q_k \d^*$ & \cite{Bau90}  \\
+ & 168 & $\q_i \q_j \to \q_k \u^*$ & \cite{Bau90}  \\
+ & 169 & $\q_i \qbar_i \to \e^{\pm} \e^{*\mp}$ & \cite{Bau90}  \\
+ & 171 & $\f_i \fbar_i \to \Z^0 \H^0$ & \cite{Eic84}  \\
+ & 172 & $\f_i \fbar_j \to \W^+ \H^0$ & \cite{Eic84}  \\
+ & 173 & $\f_i \f_j \to \f_i \f_j \H^0$ ($\Z \Z$ fusion) & 
\cite{Cah84}  \\
+ & 174 & $\f_i \f_j \to \f_k \f_l \H^0$ ($\W^+\W^-$ fusion) & 
\cite{Cah84}  \\
+ & 176 & $\f_i \fbar_i \to \Z^0 \A^0$ & \cite{Eic84}  \\
+ & 177 & $\f_i \fbar_j \to \W^+ \A^0$ & \cite{Eic84}  \\
+ & 178 & $\f_i \f_j \to \f_i \f_j \A^0$ ($\Z \Z$ fusion) & 
\cite{Cah84}  \\
+ & 179 & $\f_i \f_j \to \f_k \f_l \A^0$ ($\W^+\W^-$ fusion) & 
\cite{Cah84}  \\
+ & 181 & $\g \g \to \Q_k \Qbar_k \H^0$ & \cite{Kun84}  \\
+ & 182 & $\q_i \qbar_i \to \Q_k \Qbar_k \H^0$ & \cite{Kun84}  \\
\hline
\end{tabular}
\end{center}
\end{table}

\begin{table}[pt]
\caption{Subprocess codes, part 5. Comments as before.
\protect\label{t:procfive} }
\begin{center}
\begin{tabular}{|c|r|l|l|@{\protect\rule{0mm}{\tablinsep}}}
\hline
In & No. & Subprocess & Reference \\
\hline
+ & 183 & $\f_i \fbar_i \to \g \H^0$ & \cite{Ell88}  \\
+ & 184 & $\f_i \g \to \f_i \H^0$ & \cite{Ell88}  \\
+ & 185 & $\g \g \to \g \H^0$ & \cite{Ell88}  \\
+ & 186 & $\g \g \to \Q_k \Qbar_k \A^0$ & \cite{Kun84}  \\
+ & 187 & $\q_i \qbar_i \to \Q_k \Qbar_k \A^0$ & \cite{Kun84}  \\
+ & 188 & $\f_i \fbar_i \to \g \A^0$ & \cite{Ell88}  \\
+ & 189 & $\f_i \g \to \f_i \A^0$ & \cite{Ell88}  \\
+ & 190 & $\g \g \to \g \A^0$ & \cite{Ell88}  \\
+ & 191 & $\f_i \fbar_i \to \rho^0_{\mrm{tc}}$ & \cite{Eic96}  \\
+ & 192 & $\f_i \fbar_j \to \rho^{\pm}_{\mrm{tc}}$ & \cite{Eic96}  \\
+ & 193 & $\f_i \fbar_i \to \omega^0_{\mrm{tc}}$ & \cite{Eic96}  \\
+ & 194 & $\f_i \fbar_i \to \f_k \fbar_k$ & \cite{Eic96,Lan99}  \\
+ & 195 & $\f_i \fbar_j \to \f_k \fbar_l$ & \cite{Eic96,Lan99}  \\
+  & 201 & $\f_i \fbar_i \to \se_L \se_L^*$ & \cite{Bar87,Daw85} \\
+  & 202 & $\f_i \fbar_i \to \se_R \se_R^*$ & \cite{Bar87,Daw85} \\
+   & 203 & $\f_i \fbar_i \to \se_L \se_R^*+\se_L^* \se_R$ &
\cite{Bar87} \\
+  & 204 & $\f_i \fbar_i \to \smu_L \smu_L^*$ & \cite{Bar87,Daw85} \\
+  & 205 & $\f_i \fbar_i \to \smu_R \smu_R^*$ & \cite{Bar87,Daw85} \\
+   & 206 & $\f_i \fbar_i\to\smu_L \smu_R^*+\smu_L^* \smu_R$ &
\cite{Bar87} \\
+  & 207 & $\f_i \fbar_i\to\stau_1 \stau_1^*$ & \cite{Bar87,Daw85} \\
+  & 208 & $\f_i \fbar_i\to\stau_2 \stau_2^*$ & \cite{Bar87,Daw85} \\
+   & 209 & $\f_i \fbar_i\to\stau_1
\stau_2^*+\stau_1^*\stau_2$&\cite{Bar87} \\
+  & 210 & $\f_i \fbar_j\to \sell_L {\snu}_{\ell}^*+
\sell_L^* \snu_{\ell}$&\cite{Daw85} \\
+  & 211 & $\f_i \fbar_j\to \stau_1
\snu_{\tau}^*+\stau_1^*\snu_{\tau}$ & \cite{Daw85} \\
+  & 212 & $\f_i \fbar_j\to \stau_2
\snu_{\tau}{}^*+\stau_2^*\snu_{\tau}$ 
& \cite{Daw85} \\
+  & 213 & $\f_i \fbar_i\to \snu_{\ell} \snu_{\ell}^*$ & 
\cite{Bar87,Daw85} \\
+  & 214 & $\f_i \fbar_i\to \snu_{\tau} \snu_{\tau}^*$ 
& \cite{Bar87,Daw85} \\
+ & 216 & $\f_i \fbar_i \to \chio_1 \chio_1$ & \cite{Bar86a} \\
+ & 217 & $\f_i \fbar_i \to \chio_2 \chio_2$ & \cite{Bar86a} \\
+ & 218 & $\f_i \fbar_i \to \chio_3 \chio_3$ & \cite{Bar86a} \\
+ & 219 & $\f_i \fbar_i \to \chio_4 \chio_4$ & \cite{Bar86a} \\
+ & 220 & $\f_i \fbar_i \to \chio_1 \chio_2$ & \cite{Bar86a} \\
+ & 221 & $\f_i \fbar_i \to \chio_1 \chio_3$ & \cite{Bar86a} \\
+ & 222 & $\f_i \fbar_i \to \chio_1 \chio_4$ & \cite{Bar86a} \\
+ & 223 & $\f_i \fbar_i \to \chio_2 \chio_3$ & \cite{Bar86a} \\
+ & 224 & $\f_i \fbar_i \to \chio_2 \chio_4$ & \cite{Bar86a} \\
+ & 225 & $\f_i \fbar_i \to \chio_3 \chio_4$ & \cite{Bar86a} \\
+ & 226 & $\f_i \fbar_i \to \chip_1 \chim_1$ & \cite{Bar86b} \\
+ & 227 & $\f_i \fbar_i \to \chip_2 \chim_2$ & \cite{Bar86b} \\
+ & 228 & $\f_i \fbar_i \to \chip_1 \chim_2$ & \cite{Bar86b} \\
\hline
\end{tabular}
\end{center}
\end{table}

\begin{table}[pt]
\caption{Subprocess codes, part 6. Comments as before.
\protect\label{t:procsix}}
\begin{center}
\begin{tabular}{|c|r|l|l|@{\protect\rule{0mm}{\tablinsep}}}
\hline
In & No. & Subprocess & Reference \\
\hline
+ & 229 & $\f_i \fbar_j \to \chio_1 \chip_1$ & \cite{Bar86a,Bar86b} \\
+ & 230 & $\f_i \fbar_j \to \chio_2 \chip_1$ & \cite{Bar86a,Bar86b} \\
+ & 231 & $\f_i \fbar_j \to \chio_3 \chip_1$ & \cite{Bar86a,Bar86b} \\
+ & 232 & $\f_i \fbar_j \to \chio_4 \chip_1$ & \cite{Bar86a,Bar86b} \\
+ & 233 & $\f_i \fbar_j \to \chio_1 \chip_2$ & \cite{Bar86a,Bar86b} \\
+ & 234 & $\f_i \fbar_j \to \chio_2 \chip_2$ & \cite{Bar86a,Bar86b} \\
+ & 235 & $\f_i \fbar_j \to \chio_3 \chip_2$ & \cite{Bar86a,Bar86b} \\
+ & 236 & $\f_i \fbar_j \to \chio_4 \chip_2$ & \cite{Bar86a,Bar86b} \\
+ & 237 & $\f_i \fbar_i \to \glu \chio_1$ & \cite{Daw85}     \\
+ & 238 & $\f_i \fbar_i \to \glu \chio_2$ & \cite{Daw85}     \\
+ & 239 & $\f_i \fbar_i \to \glu \chio_3$ & \cite{Daw85}     \\
+ & 240 & $\f_i \fbar_i \to \glu \chio_4$ & \cite{Daw85}     \\
+ & 241 & $\f_i \fbar_j \to \glu \chip_1$ & \cite{Daw85}     \\
+ & 242 & $\f_i \fbar_j \to \glu \chip_2$ & \cite{Daw85}     \\
+ & 243 & $\f_i \fbar_i \to \glu \glu$ & \cite{Daw85} \\
+ & 244 & $\g \g \to \glu \glu$ & \cite{Daw85} \\
+ & 246 & $\f_i \g \to {\sq_i}{}_L \chio_1$ & \cite{Daw85} \\
+ & 247 & $\f_i \g \to {\sq_i}{}_R \chio_1$ & \cite{Daw85} \\
+ & 248 & $\f_i \g \to {\sq_i}{}_L \chio_2$ & \cite{Daw85} \\
+ & 249 & $\f_i \g \to {\sq_i}{}_R \chio_2$ & \cite{Daw85} \\
+ & 250 & $\f_i \g \to {\sq_i}{}_L \chio_3$ & \cite{Daw85} \\
+ & 251 & $\f_i \g \to {\sq_i}{}_R \chio_3$ & \cite{Daw85} \\
+ & 252 & $\f_i \g \to {\sq_i}{}_L \chio_4$ & \cite{Daw85} \\
+ & 253 & $\f_i \g \to {\sq_i}{}_R \chio_4$ & \cite{Daw85} \\
+ & 254 & $\f_i \g \to {\sq_j}{}_L \chip_1$ & \cite{Daw85} \\
+ & 256 & $\f_i \g \to {\sq_j}{}_L \chip_2$ & \cite{Daw85} \\
+ & 258 & $\f_i \g \to {\sq_i}{}_L \glu$ & \cite{Daw85}\\
+ & 259 & $\f_i \g \to {\sq_i}{}_R \glu$ & \cite{Daw85}\\
+ & 261 & $\f_i \fbar_i \to \tp_1 \tm_1$ & \cite{Daw85} \\
+ & 262 & $\f_i \fbar_i \to \tp_2 \tm_2$ & \cite{Daw85} \\
+ & 263 & $\f_i \fbar_i \to \tp_1 \tm_2+\tm_1 \tp_2$ & \cite{Daw85} \\
+ & 264 & $\g \g \to \tp_1 \tm_1$ & \cite{Daw85} \\
+ & 265 & $\g \g \to \tp_2 \tm_2$ & \cite{Daw85} \\
+ & 271 & $\f_i \f_j \to {\sq_i}{}_L {\sq_j}{}_L$ & \cite{Daw85} \\
+ & 272 & $\f_i \f_j \to {\sq_i}{}_R {\sq_j}{}_R$ & \cite{Daw85} \\
+ & 273 & $\f_i \f_j \to {\sq_i}{}_L {\sq_j}{}_R+
{\sq_i}{}_R {\sq_j}{}_L$ & \cite{Daw85} \\
+ & 274 & $\f_i \fbar_j \to {\sq_i}{}_L {\sqs_j}{}_L$ & \cite{Daw85} \\
+ & 275 & $\f_i \fbar_j \to {\sq_i}{}_R {\sqs_j}{}_R$ & \cite{Daw85} \\
+ & 276 & $\f_i \fbar_j \to {\sq_i}{}_L {\sqs_j}{}_R+
{\sq_i}{}_R {\sqs_j}{}_L$ & \cite{Daw85} \\
+ & 277 & $\f_i \fbar_i \to {\sq_j}{}_L {\sqs_j}{}_L$ & \cite{Daw85} \\
\hline
\end{tabular}
\end{center}
\end{table}

\begin{table}[pt]
\caption{Subprocess codes, part 7. Comments as before.
\protect\label{t:procseven} }
\begin{center}
\begin{tabular}{|c|r|l|l|@{\protect\rule{0mm}{\tablinsep}}}
\hline
In & No. & Subprocess & Reference \\
\hline
+ & 278 & $\f_i \fbar_i \to {\sq_j}{}_R {\sqs_j}{}_R$ & \cite{Daw85} \\
+ & 279 & $\g \g \to {\sq_i}{}_L {\sqs_i}{}_L$ & \cite{Daw85} \\
+ & 280 & $\g \g \to {\sq_i}{}_R {\sqs_i}{}_R$ & \cite{Daw85} \\
+ & 281 & $\b \q \to \sbo_1 \sq_L$ ($\q$ not $\b$) &   \\
+ & 282 & $\b \q \to \sbo_2 \sq_R$ &   \\
+ & 283 & $\b \q \to \sbo_1 \sq_R + \sbo2 \sq_L$ &   \\
+ & 284 & $\b \qbar \to \sbo_1 \sqs_L$ &   \\
+ & 285 & $\b \qbar \to \sbo_2 \sqs_R$ &   \\
+ & 286 & $\b \qbar \to \sbo_1 \sqs_R + \sbo_2 \sqs_L$ &   \\
+ & 287 & $\f_i \fbar_i \to \sbo_1 \sbs_1$ &   \\
+ & 288 & $\f_i \fbar_i \to \sbo_2 \sbs_2$ &  \\
+ & 289 & $\g \g \to \sbo_1 \sbs_1$ &  \\
+ & 290 & $\g \g \to \sbo_2 \sbs_2$ &  \\
+ & 291 & $\b \b \to \sbo_1 \sbo_1$ &  \\
+ & 292 & $\b \b \to \sbo_2 \sbo_2$ &  \\
+ & 293 & $\b \b \to \sbo_1 \sbo_2$ &   \\
+ & 294 & $\b \g \to \sbo_1 \glu$ &   \\
+ & 295 & $\b \g \to \sbo_2 \glu$ &   \\
+ & 296 & $\b \bbar \to \sbo_1 \sbs_2 + \sbs_1 \sbo_2$ &   \\
+ & 297 & $\f_i \fbar_j \to \H^{\pm} \hrm^0$ &  \\
+ & 298 & $\f_i \fbar_j \to \H^{\pm} \H^0$ &  \\
+ & 299 & $\f_i \fbar_i \to \A \hrm^0$ &  \\
+ & 300 & $\f_i \fbar_i \to \A \H^0$ &  \\
+ & 301 & $\f_i \fbar_i \to \H^+ \H^-$ &  \\
+ & 341 & $\ell_i \ell_j \to \H_L^{\pm\pm}$ & \cite{Hui97} \\
+ & 342 & $\ell_i \ell_j \to \H_R^{\pm\pm}$ & \cite{Hui97} \\
+ & 343 & $\ell_i \gamma \to \H_L^{\pm\pm} \e^{\mp}$ & \cite{Hui97} \\
+ & 344 & $\ell_i \gamma \to \H_R^{\pm\pm} \e^{\mp}$ & \cite{Hui97} \\
+ & 345 & $\ell_i \gamma \to \H_L^{\pm\pm} \mu^{\mp}$ & \cite{Hui97} \\
+ & 346 & $\ell_i \gamma \to \H_R^{\pm\pm} \mu^{\mp}$ & \cite{Hui97} \\
+ & 347 & $\ell_i \gamma \to \H_L^{\pm\pm} \tau^{\mp}$ & \cite{Hui97} \\
+ & 348 & $\ell_i \gamma \to \H_R^{\pm\pm} \tau^{\mp}$ & \cite{Hui97} \\
+ & 349 & $\f_i \fbar_i \to \H_L^{++} \H_L^{--}$ & \cite{Hui97} \\
+ & 350 & $\f_i \fbar_i \to \H_R^{++} \H_R^{--}$ & \cite{Hui97} \\
+ & 351 & $\f_i \f_j \to \f_k f_l \H_L^{\pm\pm}$ ($\W\W$) fusion) & 
   \cite{Hui97} \\
+ & 352 & $\f_i \f_j \to \f_k f_l \H_R^{\pm\pm}$ ($\W\W$) fusion) & 
   \cite{Hui97} \\
+ & 353 & $\f_i \fbar_i \to \Z_R^0$ & \cite{Eic84}  \\
+ & 354 & $\f_i \fbar_j \to \W_R^+$ & \cite{Eic84}  \\
+ & 361 & $\f_i \fbar_i \to \W^+_{\mrm{L}} \W^-_{\mrm{L}} $ & 
   \cite{Lan99} \\
+ & 362 & $\f_i \fbar_i \to \W^{\pm}_{\mrm{L}} \pi^{\mp}_{\mrm{tc}}$ & 
   \cite{Lan99} \\
\hline
\end{tabular}
\end{center}
\end{table}

\begin{table}[pt]
\caption{Subprocess codes, part 8. Comments as before.
\protect\label{t:proceight} }
\begin{center}
\begin{tabular}{|c|r|l|l|@{\protect\rule{0mm}{\tablinsep}}}
\hline
In & No. & Subprocess & Reference \\
\hline
+ & 363 & $\f_i \fbar_i \to \pi^+_{\mrm{tc}} \pi^-_{\mrm{tc}}$ & 
   \cite{Lan99} \\
+ & 364 & $\f_i \fbar_i \to \gamma \pi^0_{\mrm{tc}} $ & \cite{Lan99} \\
+ & 365 & $\f_i \fbar_i \to \gamma {\pi'}^0_{\mrm{tc}} $ & \cite{Lan99} \\
+ & 366 & $\f_i \fbar_i \to \Z^0 \pi^0_{\mrm{tc}} $ & \cite{Lan99} \\
+ & 367 & $\f_i \fbar_i \to \Z^0 {\pi'}^0_{\mrm{tc}} $ & \cite{Lan99} \\
+ & 368 & $\f_i \fbar_i \to \W^{\pm} \pi^{\mp}_{\mrm{tc}}$ & \cite{Lan99} \\
+ & 370 & $\f_i \fbar_j \to \W^{\pm}_{\mrm{L}} \Z^0_{\mrm{L}}$ & 
   \cite{Lan99} \\
+ & 371 & $\f_i \fbar_j \to \W^{\pm}_{\mrm{L}} \pi^0_{\mrm{tc}}$ & 
   \cite{Lan99} \\
+ & 372 & $\f_i \fbar_j \to \pi^{\pm}_{\mrm{tc}} \Z^0_{\mrm{L}} $ &
   \cite{Lan99} \\
+ & 373 & $\f_i \fbar_j \to \pi^{\pm}_{\mrm{tc}} \pi^0_{\mrm{tc}} $ &
   \cite{Lan99} \\
+ & 374 & $\f_i \fbar_j \to \gamma \pi^{\pm}_{\mrm{tc}} $ & \cite{Lan99} \\
+ & 375 & $\f_i \fbar_j \to \Z^0 \pi^{\pm}_{\mrm{tc}} $ & \cite{Lan99} \\
+ & 376 & $\f_i \fbar_j \to \W^{\pm} \pi^0_{\mrm{tc}} $  & \cite{Lan99} \\
+ & 377 & $\f_i \fbar_j \to \W^{\pm} {\pi'}^0_{\mrm{tc}}$ & \cite{Lan99} \\
+ & 381 & $\q_i \q_j \to \q_i \q_j$ (QCD+TC) & \cite{Chi90,Lan02a} \\
+ & 382 & $\q_i \qbar_i \to \q_k \qbar_k$ (QCD+TC) & \cite{Chi90,Lan02a} \\
+ & 383 & $\q_i \qbar_i \to \g \g$ (QCD+TC) & \cite{Lan02a}  \\
+ & 384 & $\f_i \g \to \f_i \g$ (QCD+TC) & \cite{Lan02a}  \\
+ & 385 & $\g \g \to \q_k \qbar_k$ (QCD+TC) & \cite{Lan02a}  \\
+ & 386 & $\g \g \to \g \g$ (QCD+TC) &  \cite{Lan02a} \\
+ & 387 & $\f_i \fbar_i \to \Q_k \Qbar_k$ (QCD+TC) & \cite{Lan02a}  \\
+ & 388 & $\g \g \to \Q_k \Qbar_k$ (QCD+TC) & \cite{Lan02a}  \\
+ & 391 & $\f \fbar \to \G^*$ & \cite{Ran99} \\
+ & 392 & $\g \g \to \G^*$ & \cite{Ran99} \\
+ & 393 & $\q \qbar \to \g \G^*$ & \cite{Ran99,Bij01} \\
+ & 394 & $\q \g \to \q \G^*$ & \cite{Ran99,Bij01} \\
+ & 395 & $\g \g \to \g \G^*$ & \cite{Ran99,Bij01} \\
\hline
\end{tabular}
\end{center}
\end{table}

In the following, $\f_i$ represents a fundamental fermion of flavour
$i$, i.e.\ $\d$, $\u$, $\s$, $\c$, $\b$, $\t$, $\b'$,
$\t'$, $\e^-$, $\nu_{\e}$, $\mu^-$, $\nu_{\mu}$, $\tau^-$,
$\nu_{\tau}$, ${\tau'}^-$ or ${\nu'}_{\tau}$. A corresponding antifermion
is denoted by $\fbar_i$. In several cases, some classes of fermions 
are explicitly excluded, since they do not couple to the $\g$ or
$\gamma$ (no $\ee \to \g \g$, e.g.). When processes have only been
included for quarks, while leptons might also have been possible,
the notation $\q_i$ is used. A lepton is denoted by $\ell$; in a few
cases neutrinos are also lumped under this heading.
In processes where fermion masses are explicitly included in the 
matrix elements, an $\F$ or $\Q$ is used to denote
an arbitrary fermion or quark. Flavours appearing already in
the initial state are denoted by indices $i$ and $j$, whereas new 
flavours in the final state are denoted by $k$ and~$l$.

In supersymmetric processes, antiparticles of sfermions 
are denoted by $^*$, i.e.\ $\st^*$ rather than the more correct but 
cumbersome $\br{\st}$ or $\tilde{\t}$.
 
Charge-conjugate channels are always assumed included as well (where
separate), and processes involving a $\W^+$ also imply those involving
a $\W^-$. Wherever $\Z^0$ is written, it is understood that $\gamma^*$
and $\gammaZ$ interference should be included as well (with
possibilities to switch off either, if so desired). In some cases
this is not fully implemented, see further below.
Correspondingly, $\Z'^0$ denotes the complete set
$\gamma^*/\Z^0/\Z'^0$ (or some subset of it). Thus the notation 
$\gamma$ is only used for a photon on the mass shell.
 
In the last column of the tables below, references are given to works 
from which formulae have been taken. Sometimes these references are to 
the original works on the subject, sometimes only to the place where 
the formulae are given in the most convenient or accessible form, or
where chance lead us. Apologies to all matrix-element calculators who
are not mentioned. However, remember that this is not a review article
on physics processes, but only a way for readers to know what is
actually found in the program, for better or worse. In several
instances, errata have been obtained from the authors. Often
the formulae given in the literature have been generalized to include
trivial radiative corrections, Breit--Wigner line shapes with
$\hat{s}$-dependent widths (see section \ref{ss:kinemreson}), etc.
 
The following sections contain some useful comments on the processes
included in the program, grouped by physics interest rather than
sequentially by {\ISUB} or \ttt{MSEL} code (see \ref{ss:PYswitchkin} 
for further information on the \ttt{MSEL} code). The different 
{\ISUB} and \ttt{MSEL} codes
that can be used to simulate the different groups are given. {\ISUB}
codes within brackets indicate the kind of processes that indirectly
involve the given physics topic, although only as part of a larger
whole. Some obvious examples, such as the possibility to produce jets 
in just about any process, are not spelled out in detail.
 
The text at times contains information on which special switches or
parameters are of particular interest to a given process. All these
switches are described in detail in sections \ref{ss:PYswitchpar}
\ref{ss:coupcons} and \ref{ss:susycode}, but 
are alluded to here so as to provide a more complete picture of the
possibilities available for the different subprocesses. However, the
list of possibilities is certainly not exhausted by the text below.
 
\subsection{QCD Processes}

Obviously most processes in {\Py} contain QCD physics one way or 
another, so the above title should not be overstressed. One example:
a process like $\ee \to \gammaZ \to \q\qbar$ is also traditionally 
called a QCD event, but is here book-kept as  $\gammaZ$ production.
In this section we discuss scatterings between coloured partons, 
plus a few processes that are close relatives to other processes
of this kind. 

\subsubsection{QCD jets} 
\label{sss:QCDjetclass}
  
\ttt{MSEL} = 1, 2 \\
{\ISUB} = 
\begin{tabular}[t]{rl@{\protect\rule{0mm}{\tablinsep}}}
11 & $\q_i \q_j \to \q_i \q_j$ \\
12 & $\q_i \qbar_i \to \q_k \qbar_k$ \\
13 & $\q_i \qbar_i \to \g \g$ \\
28 & $\q_i \g \to \q_i \g$ \\ 
53 & $\g \g \to \q_k \qbar_k$ \\
68 & $\g \g \to \g \g$ \\
96 & semihard QCD $2 \to 2$ \\
\end{tabular}
 
No higher-order processes are explicitly included,
nor any higher-order loop corrections to the $2 \to 2$ processes.
However, by initial- and final-state QCD radiation, multijet events 
are being generated, starting from the above processes. The shower 
rate of multijet production is clearly uncertain
by some amount, especially for well-separated jets. 

A string-based fragmentation scheme such as the Lund model needs 
cross sections for the different colour flows; these have been 
calculated in \cite{Ben84} and differ from the usual calculations by 
interference terms of the order $1/N_C^2$. By default, the standard 
colour-summed QCD expressions for the differential cross sections 
are used. In this case, the interference terms are distributed among 
the various colour flows according to the pole structure of the terms. 
However, the interference terms can be excluded, by changing 
\ttt{MSTP(34)}
 
As an example, consider subprocess 28, $\q \g \to \q \g$. The total 
cross section for this process, obtained by summing and squaring the 
Feynman $\hat{s}$-, $\hat{t}$-, and $\hat{u}$-channel graphs, is
\cite{Com77}
\begin{equation} 
2 \left( 1 - \frac{\hat{u}\hat{s}}{\hat{t}^2} \right) - 
\frac{4}{9} \left( \frac{\hat{s}}{\hat{u}} + 
\frac{\hat{u}}{\hat{s}} \right) - 1 ~. 
\end{equation}
(An overall factor $\pi \alphas^2/\hat{s}^2$ is ignored.) 
Using the identity of the Mandelstam variables 
for the massless case, $\hat{s} + \hat{t} + \hat{u} = 0$,  
this can be rewritten as 
\begin{equation}
\frac{\hat{s}^2 + \hat{u}^2}{\hat{t}^2} -
\frac{4}{9} \left( \frac{\hat{s}}{\hat{u}} + 
\frac{\hat{u}}{\hat{s}} \right) ~. 
\end{equation}

On the other hand, the cross sections for the two possible colour 
flows of this subprocess are \cite{Ben84}
\begin{eqnarray} 
   A: & & \frac{4}{9} \left( 2 \frac{\hat{u}^2}{\hat{t}^2} -
\frac{\hat{u}}{\hat{s}} \right) ~;   \nonumber \\
   B: & & \frac{4}{9} \left( 2 \frac{\hat{s}^2}{\hat{t}^2} -
\frac{\hat{s}}{\hat{u}} \right) ~.
\end{eqnarray}
Colour configuration $A$ is one in which the original colour
of the $\q$ annihilates with the anticolour of the $\g$,
the $\g$ colour flows through, and a new colour--anticolour is
created between the final $\q$ and $\g$. In colour configuration
$B$, the gluon anticolour flows through, but the $\q$ and $\g$
colours are interchanged. Note that these two colour
configurations have different kinematics dependence.
For \ttt{MSTP(34)=0}, these are the cross sections actually 
used.  

The sum of the $A$ and $B$ contributions is
\begin{equation} 
\frac{8}{9} \frac{\hat{s}^2 + \hat{u}^2}{\hat{t}^2} -
\frac{4}{9} \left( \frac{\hat{s}}{\hat{u}} + 
\frac{\hat{u}}{\hat{s}} \right) ~. 
\end{equation} 
The difference between this expression and that of \cite{Com77}, 
corresponding to the interference between the two colour-flow 
configurations, is then 
\begin{equation} 
\frac{1}{9} \frac{\hat{s}^2 + \hat{u}^2}{\hat{t}^2} ~,
\end{equation}  
which can be naturally divided between colour flows $A$ and $B$: 
\begin{eqnarray}
   A: & & \frac{1}{9} \frac{\hat{u}^2}{\hat{t}^2} ~; \nonumber \\
   B: & & \frac{1}{9} \frac{\hat{s}^2}{\hat{t}^2} ~.
\end{eqnarray} 
For \ttt{MSTP(34)=1}, the standard QCD matrix element is therefore
used, with the same relative importance of the two colour configurations 
as above. Similar procedures are followed also for the other QCD 
subprocesses. 
  
All the matrix elements in this group are for massless quarks 
(although final-state quarks are of course put on the mass shell). 
As a consequence, cross sections are divergent for $\pT \to 0$,
and some kind of regularization is required. Normally you 
are expected to set the desired $\pTmin$ value in 
\ttt{CKIN(3)}. 
  
The new flavour produced in the annihilation processes ({\ISUB} = 12 and
53) is determined by the flavours allowed for gluon splitting into 
quark--antiquark; see switch \ttt{MDME}. 

Subprocess 96 is special among all the ones in the program. In terms of
the basic cross section, it is equivalent to the sum of the other ones, 
i.e. 11, 12, 13, 28, 53 and 68. The phase space is mapped differently,
however, and allows $\pT$ as input variable. This is especially useful
in the context of the multiple interactions machinery, see subsection
\ref{ss:multint}, where potential scatterings are considered in order
of decreasing $\pT$, with a form factor related to the probability of not
having another scattering with a $\pT$ larger than the considered one.
You are not expected to access process 96 yourself. Instead it is 
automatically initialized and used either if process 95 is included or 
if multiple interactions are switched on. The process will then appear 
in the maximization information output, but not in the cross section
table at the end of a run. Instead, the hardest scattering generated within 
the context of process 95 is reclassified as an event of the 11, 12, 13, 
28, 53 or 68  kinds, based on the relative cross section for these in 
the point chosen. Further multiple interactions, subsequent to the hardest 
one, also do not show up in cross section tables. 
     
\subsubsection{Heavy flavours}
\label{sss:heavflavclass} 
  
\ttt{MSEL} = 4, 5, 6, 7, 8 \\
{\ISUB} =
\begin{tabular}[t]{rl@{\protect\rule{0mm}{\tablinsep}}}
81 & $\q_i \qbar_i \to \Q_k \Qbar_k$  \\
82 & $\g \g \to \Q_k \Qbar_k$  \\
(83) & $\q_i \f_j \to Q_k f_l$  \\
(84) & $\g \gamma \to \Q_k \Qbar_k$ \\
(85) & $\gamma \gamma \to \F_k \Fbar_k$ \\
(1) & $\f_i \fbar_i \to \gammaZ \to \F_k \Fbar_k$ \\
(2) & $\f_i \fbar_j \to \W^+\to \F_k \Fbar_l $ \\
(142) & $\f_i \fbar_j \to \W'^+\to \F_k \Fbar_l $ \\
\end{tabular}
  
The matrix elements in this group differ from the corresponding ones 
in the group above in that they correctly take into account the quark 
masses. As a consequence, the cross sections are finite for 
$\pT \to 0$. It is therefore not necessary to introduce any special 
cuts. 
  
The two first processes that appear here are the dominant lowest-order
QCD graphs in hadron colliders --- a few other graphs will be mentioned
later, such as process 83.

The choice of flavour to produce is according to a hierarchy of options:
\begin{Enumerate}
\item if \ttt{MSEL=4-8} then the flavour is set by the \ttt{MSEL} value;
\item else if \ttt{MSTP(7)=1-8} then the flavour is set by the
\ttt{MSTP(7)} value;
\item else the flavour is determined by the heaviest flavour allowed for 
gluon splitting into quark--antiquark; see switch \ttt{MDME}.
\end{Enumerate} 
Note that only one heavy flavour is allowed at a time; if more than one 
is turned on in \ttt{MDME}, only the heaviest will be produced (as 
opposed to the case for {\ISUB} = 12 and 53 above, where more than one 
flavour is allowed simultaneously). 
  
The lowest-order processes listed above just represent one source of 
heavy-flavour production. Heavy quarks can also be present in the 
parton distributions at the $Q^2$ scale of the hard interaction, 
leading to processes like $\Q \g \to \Q \g$, 
so-called flavour excitation, or they can be 
created by gluon splittings $\g \to \Q \Qbar$ in initial- or final-state 
shower evolution. The implementation and importance of these various
production mechanisms is discussed in detail in \cite{Nor98}.
In fact, as the c.m.\ energy is increased, these other processes gain 
in importance relative to the lowest-order production graphs above. 
As as example, only 10\%--20\% of the $\b$ production at LHC 
energies come from the lowest-order graphs. The figure is even smaller 
for charm, while it is well above 50\% for top. At LHC energies, 
the specialized treatment described in this section is therefore 
only of interest for top (and potential fourth-generation quarks) --- 
the higher-order corrections can here be approximated by an effective 
$K$ factor, except possibly in some rare corners of phase space. 

For charm and bottom, on the other hand, it is necessary to simulate the 
full event sample (within the desired kinematics cuts), and then only 
keep those events that contain $\b/\c$, be that either from lowest-order 
production, or flavour excitation, or gluon splitting. Obviously this may 
be a time-consuming enterprise --- although the probability for a high-$\pT$ 
event at collider energies to contain (at least) one charm or bottom pair 
is fairly large, most of these heavy flavours are carrying a small fraction 
of the total $\pT$ flow of the jets, and therefore do not survive normal 
experimental cuts. We note that the lowest-order production of charm or 
bottom with processes 12 and 53, as part of the standard QCD mix, is now
basically equivalent with that offered by processes 81 and 82. For 12 
vs.\ 81 this is rather trivial, since only $s$-channel gluon exchange is 
involved, but for process 53 it requires a separate evaluation of massive
matrix elements for $\c$ and $\b$ in the flavour loop. This is performed 
by retaining the $\hat{s}$ and $\hat{\theta}$ values already preliminarily
selected for the massless kinematics, and recalculating $\hat{t}$ and
$\hat{u}$ with mass effects included. Some of the documentation information
in \ttt{PARI} does not properly reflect this recalculation, but that is
purely a documentation issue. Also process 96, used internally for the
total QCD jet cross section, includes $\c$ and $\b$ masses. Only the hardest 
interaction in a multiple interactions scenario may contain $\c/\b$, 
however, for technical reasons, so that the total rate may be underestimated. 
(Quite apart from other uncertainties, of course.)  

As an aside, it is not only for the lowest-order graphs that events 
may be generated with a guaranteed heavy-flavour content. One may also 
generate the flavour excitation process by itself, in the massless 
approximation, using {\ISUB} = 28 and setting the \ttt{KFIN} array 
appropriately. No trick exists to force the gluon splittings without 
introducing undesirable biases, however. In order to have it all, one 
therefore has to make a full QCD jets run, as already noted.  

Also other processes can generate heavy flavours, all the way up to top,
but then without a proper account of masses. By default, top production
is switched off in those processes where a new flavour pair is produced 
at a gluon or photon vertex, i.e.\ 12, 53, 54, 58, 96 and 135--140, while
charm and bottom is allowed. These defaults can be changed by setting
the \ttt{MDME(IDC,1)} values of the appropriate $\g$ or $\gamma$
`decay channels', see further below.
  
The cross section for a heavy quark pair close to threshold can be 
modified according to the formulae of \cite{Fad90}, see \ttt{MSTP(35)}. 
Here threshold effects due to $\Q\Qbar$ bound-state formation are taken 
into account in a smeared-out, average sense. Then the na\"{\i}ve 
cross section is multiplied by the squared wave function at the origin. 
In a colour-singlet channel this gives a net enhancement of the form
\begin{equation}
|\Psi^{(s)}(0)|^2 = \frac{X_{(s)}}{1 - \exp(- X_{(s)})}  ~ ,
~~~ \mrm{where}~ X_{(s)} = \frac{4}{3} 
\frac{\pi \alphas}{\beta}  ~,
\label{pp:threshenh}
\end{equation}
where $\beta$ is the quark velocity,
while in a colour octet channel there is a net suppression given by
\begin{equation}
|\Psi^{(8)}(0)|^2 = \frac{X_{(8)}}{\exp(- X_{(8)}) -1}  ~ ,
~~~ \mrm{where}~ X_{(8)} = \frac{1}{6} 
\frac{\pi \alphas}{\beta}  ~.
\label{pp:threshsup}
\end{equation}
The $\alphas$ factor in this expression is related to the energy
scale of bound-state formation; it is selected independently from
the one of the standard production cross section.
The presence of a threshold factor affects the total rate and also 
kinematical distributions. 

Heavy flavours can also be produced by secondary decays of gauge 
bosons or new exotic particles. We have listed 1, 2 and 142 above 
as among the most important ones. There is a special point to 
including $\W'$ in this list. Imagine that you want to study 
the electroweak $s$-channel production of a single top, 
$\u \dbar \to \W^+ \to \t \bbar$, and therefore decide to force this 
particular decay mode of the $\W$. But then the same decay channel
is required for the $\W^+$ produced in the decay $\t \to \b \W^+$,
i.e. you have set up an infinite recursion 
$\W \to \t \to \W \to \t \to \ldots$. The way out is to use the 
$\W'$, which has default couplings just like the normal $\W$, 
only a different mass, which then can be changed to agree,
\ttt{PMAS(34,1) = PMAS(24,1)}. 
The $\W'$ is now forced to decay to $\t \bbar$, while the $\W$ 
can decay freely (or also be forced, e.g. to have a leptonic
decay, if desired). (Additionally, it may be necessary to raise 
\ttt{CKIN(1)} to be at least around the top mass, so that the 
program does not get stuck in a region of phase space where
the cross section is vanishing.)  

Heavy flavours, i.e.\ top and fourth generation, are assumed to be so
short-lived that they decay before they have time to hadronize. This
means that the light quark in the decay $\Q \to \W^{\pm} \q$ 
inherits the colour of the heavy one. The current {\Py} description 
represents a change of philosophy compared to older versions, 
formulated at a time when the top was thought to be much lighter 
than we now know it to be.  For event shapes the difference between the 
two time orderings normally has only marginal effects \cite{Sjo92a}.

It should be noted that cross section calculations are different in 
the two cases. The top (or a fourth generation fermion) is 
assumed short-lived,  and is treated like a resonance in the sense 
of section \ref{sss:resdecaycross}, i.e.\ the cross-section is reduced 
so as only to correspond to the channels left open by you.
This also includes the restrictions on secondary decays, i.e.\ on the
decays of a $\W^+$ or a $\H^+$ produced in the top decay.
For $\b$ and $\c$ quarks, which are long-lived enough to form hadrons, 
no such reduction takes place. 
Branching ratios then have to be folded in by hand to get the correct 
cross sections. The logic behind this difference is that when 
hadronization takes place, one would normally decay the 
$\D^0$ and $\D^+$ meson according to different branching ratios.
But which $\D$ mesons are to be formed is not known at the bottom quark
creation, so one could not weight for that. For a $\t$ quark, which
decays rapidly, this ambiguity does not exist, and so a reduction
factor can be introduced directly coupled to the $\t$ quark production
process.

This rule about cross-section calculations applies to all the 
processes explicitly set up to handle heavy flavour creation.
In addition to the ones above, this means all the ones in Tables
\ref{t:procone}--\ref{t:proceight} where the fermion final state is 
given as capital letters (`$\Q$' and `$\F$') and also flavours produced
in resonance decays ($\Z^0$, $\W^{\pm}$, $\hrm^0$, etc., including
processes 165 and 166). However, heavy flavours could also be produced
in a process such as 31, $\q_i \g \to \q_k \W^{\pm}$, where $\q_k$
could be a top quark. In this case, the thrust of the description is
clearly on light flavours --- the kinematics of the process is 
formulated in the massless fermion limit --- so any top production
is purely incidental. Since here the choice of scattered flavour is
only done at a later stage, the top branching ratios are not
correctly folded in to the hard scattering cross section. So, for 
applications like these, it is not recommended to restrict the allowed
top decay modes. Often one might like to get rid of the possibility
of producing top together with light flavours. This can be 
done by switching off (i.e.\ setting \ttt{MDME(I,1)=0}) the 
`channels' $\d \to \W^- \t$, $\s \to \W^- \t$, $\b \to \W^- \t$,
$\g \to \t\tbar$ and $\gamma \to \t\tbar$. Also any heavy flavours 
produced by parton shower evolution would not be correctly weighted
into the cross section. However, currently top production is switched
off both as a beam remnant (see \ttt{MSTP(9)} and in initial 
(see \ttt{KFIN} array) and final (see \ttt{MSTJ(45)}) state radiation. 

In pair production of heavy flavour (top) in processes 81,82, 84 
and 85, matrix elements are only given for one common mass, although
Breit-Wigners are used to select two separate masses. As described in
subsection \ref{ss:kinemreson}, an average mass value is constructed 
for the matrix element evaluation so that the $\beta_{34}$ kinematics 
factor can be retained.

Because of its large mass, it is possible that the top quark can decay 
to some not yet discovered particle. Some such alternatives are included
in the program, such as $\t \to \b \H^+$ or $\t \to \grav \st$. These 
decays are not obtained by default, but can be included as discussed 
for the respective physics scenario.     

\subsubsection{J/$\psi$ and other Hidden Heavy Flavours}
\label{sss:Jpsiclass}

{\ISUB} = 
\begin{tabular}[t]{rl@{\protect\rule{0mm}{\tablinsep}}}
 86 & $\g \g \to \Jpsi \g$ \\
 87 & $\g \g \to \chi_{0 \c} \g$ \\
 88 & $\g \g \to \chi_{1 \c} \g$ \\
 89 & $\g \g \to \chi_{2 \c} \g$ \\
104 & $\g \g \to \chi_{0 \c}$ \\
105 & $\g \g \to \chi_{2 \c}$  \\
106 & $\g \g \to \Jpsi \gamma$ \\
107 & $\g \gamma \to \Jpsi \g$ \\
108 & $\gamma \gamma \to \Jpsi \gamma$ \\
\end{tabular}
  
In {\Py} one may distinguish between three main sources of $\Jpsi$ 
production.
\begin{Enumerate}
\item Decays of $\B$ mesons and baryons.
\item Parton-shower evolution, wherein a $\c$ and a $\cbar$ quark
produced in adjacent branchings 
(e.g.\ $\g \to \g \g \to \c \cbar \c \cbar$)
turn out to have so small an invariant mass that the pair collapses 
to a single particle.
\item Direct production, where a $\c$ quark loop gives a coupling 
between a set of gluons and a $\c\cbar$ bound state. Higher-lying 
states, like the $\chi_c$ ones, may subsequently decay to $\Jpsi$.
\end{Enumerate}

The first two sources are implicit in the production of $\b$ and
$\c$ quarks, although the forcing specifically of $\Jpsi$ production 
is difficult. In this section are given the main processes for the 
third source, intended for applications at hadron colliders. 
Processes 104 and 105 are the equivalents of
87 and 89 in the limit of $\pT \to 0$; note that $\g \g \to \Jpsi$
and $\g \g \to \chi_{1 \c}$ are forbidden and thus absent. As always 
one should beware of double-counting between 87 and 104, and between 
89 and 105, and thus use either the one or the other depending on 
the kinematical domain to be studied. The cross sections depend
on wave function values at the origin, see \ttt{PARP(38)} and
\ttt{PARP(39)}. A review of the physics issues involved may be found 
in \cite{Glo88} (note, however, that the choice of $Q^2$ scale is 
different in {\Py}). 

It is known that the above sources are not enough to explain the 
full $\Jpsi$ rate, and further production mechanisms have been 
proposed, extending on the more conventional treatment here
\cite{Can97}.

While programmed for the charm system, it would be straightforward 
to apply these processes instead to bottom mesons. One needs to change 
the codes of states produced, which is achieved by 
\ttt{KFPR(ISUB,1)=KFPR(ISUB,1)+110} for the processes {\ISUB} above, 
and changing the values of the wave functions at the origin, 
\ttt{PARP(38)} and \ttt{PARP(39)}.

\subsubsection{Minimum bias}
\label{sss:minbiasclass} 
  
\ttt{MSEL} = 1, 2 \\
{\ISUB} = 
\begin{tabular}[t]{rl@{\protect\rule{0mm}{\tablinsep}}}
91 & elastic scattering  \\
92 & single diffraction ($AB \to XB$) \\
93 & single diffraction ($AB \to AX$) \\
94 & double diffraction  \\
95 & low-$\pT$ production  \\
\end{tabular}
  
These processes are briefly discussed in section 
\ref{ss:nonpertproc}. They are mainly intended for interactions 
between hadrons, although one may also consider $\gamma \p$ 
and $\gamma\gamma$ interactions in the options where the incoming 
photon(s) is (are) assumed resolved.

Uncertainties come from a number of sources, e.g.\ from the 
parameterizations of the various cross sections and slope parameters.

In diffractive scattering, the structure of the selected hadronic 
system may be regulated with \ttt{MSTP(101)}. No high-$\pT$ 
jet production in diffractive events is included so far. 
  
The subprocess 95, low-$\pT$ events, is somewhat unique in 
that no meaningful physical border-line to high-$\pT$ events can be 
defined. Even if the QCD $2 \to 2$ high-$\pT$ processes are formally 
switched off, some of the generated events will be classified as 
belonging to this group, with a $\pT$ spectrum of interactions to 
match the `minimum-bias' event sample. The generation of such jets
is performed with the help of the auxiliary subprocess 96, see
subsection \ref{sss:QCDjetclass}. Only with the option 
\ttt{MSTP(82)=0} will subprocess 95 yield strictly low-$\pT$  
events, events which will then probably not be compatible with any 
experimental data. A number of options exist for the detailed 
structure of low-$\pT$ events, see in particular \ttt{MSTP(81)} and 
\ttt{MSTP(82)}. Further details on the model(s) for minimum-bias
events are found in section \ref{ss:multint}.

\subsection{Physics with Incoming Photons}
\label{ss:photoanddisclass}

With recent additions, the machinery for photon physics has become 
rather extensive \cite{Fri00}. The border between the physics of real 
photon interactions and of virtual photon ones is now bridged 
by a description that continuously interpolates between the two 
extremes, as summarized in section \ref{sss:photoprod}. 
Furthermore, the {\galep} option (where \textit{lepton} is 
to be replaced by \ttt{e-}, \ttt{e+}, \ttt{mu-}, \ttt{mu+}, 
\ttt{tau-} or \ttt{tau+} as the case may be) in 
a \ttt{PYINIT} call gives access to an internally generated spectrum 
of photons of varying virtuality. The \ttt{CKIN(61) - CKIN(78)} 
variables can be used to set experimentally motivated $x$ and $Q^2$ 
limits on the photon fluxes. With this option, and the default 
\ttt{MSTP(14)=30}, one automatically obtains a realistic first 
approximation to `all' QCD physics of $\gast\p$ and $\gast\gast$ 
interactions. The word `all' clearly does not mean that a perfect
description is guaranteed, or that all issues are addressed, but 
rather that the intention is to simulate all processes that give
a significant contribution to the total cross section in whatever
$Q^2$ range is being studied: jets, low-$\pT$ events, elastic and
diffractive scattering, etc.

The material to be covered encompasses many options, several of 
which have been superseded by further developments but have been 
retained for backwards compatibility. Therefore it is here split 
into three sections. The first covers the physics of real photons 
and the subsequent one that of (very) virtual ones. Thereafter, in 
the final section, the threads are combined into a machinery
applicable at all $Q^2$.

\subsubsection{Photoproduction and $\gamma\gamma$ physics}
\label{sss:photoprodclass}

\ttt{MSEL} = 1, 2, 4, 5, 6, 7, 8 \\
{\ISUB} = 
\begin{tabular}[t]{rl@{\protect\rule{0mm}{\tablinsep}}}
33 & $\q_i \gamma \to \q_i \g$ \\
34 & $\f_i \gamma \to \f_i \gamma$ \\
54 & $\g \gamma \to \q_k \qbar_k$ \\
58 & $\gamma \gamma \to \f_k \fbar_k$ \\
80 & $\q_i \gamma \to \q_k \pi^{\pm}$ \\
84 & $\g \gamma \to \Q_k \Qbar_k$ \\
85 & $\gamma \gamma \to \F_k \Fbar_k$ \\
\end{tabular}

An (almost) real photon has both a point-like component and a 
hadron-like one. This means that several classes of processes 
may be distinguished, see section \ref{sss:photoprod}.
\begin{Enumerate}
\item The processes listed above are possible when the photon 
interacts as a point-like particle, i.e.\ couples directly to 
quarks and leptons.
\item When the photon acts like a hadron, i.e.\ is resolved in a 
partonic substructure, then high-$\pT$  parton--parton interactions
are possible, as described in sections \ref{sss:QCDjetclass} 
and \ref{sss:promptgammaclass}. These interactions may be further
subdivided into VMD and anomalous (GVMD) ones \cite{Sch93,Sch93a}.
\item A hadron-like photon can also produce the equivalent of the 
minimum bias processes of section \ref{sss:minbiasclass}. Again,
these can be subdivided into VMD and GVMD (anomalous) ones. 
\end{Enumerate}

For $\gamma\p$ events, we believe that the best description can be 
obtained when three separate event classes are combined, one for 
direct, one for VMD and one for GVMD/anomalous events, see the 
detailed description in \cite{Sch93,Sch93a}.
These correspond to \ttt{MSTP(14)} being 0, 2 and 3, respectively.
The direct component is high-$\pT$ only, while VMD and GVMD
contain both high-$\pT$ and low-$\pT$ events. The option
\ttt{MSTP(14)=1} combines the VMD and GVMD/anomalous parts of the 
photon into one single resolved photon concept, which therefore is 
less precise than the full subdivision.

When combining three runs to obtain the totality of $\gamma\p$
interactions, to the best of our knowledge, it is necessary to choose
the $\pT$ cut-offs with some care, so as to represent the expected
total cross section.
\begin{Itemize}
\item The direct processes by themselves only depend on the 
\ttt{CKIN(3)} cut-off of the generation. In older program versions
the preferred value was 0.5 GeV \cite{Sch93,Sch93a}. In the more recent
description in \cite{Fri00}, also eikonalization of direct with
anomalous interactions into the GVMD event class is considered.
That is, given a branching $\gamma \to \q\qbar$, direct interactions 
are viewed as the low-$\pT$ events and anomalous ones as high-$\pT$ 
events that have to merge smoothly. Then the \ttt{CKIN(3)} cut-off is 
increased to the $\pTmin$ of
multiple interactions processes, see \ttt{PARP(81)} (or \ttt{PARP(82)}, 
depending on minijet unitarization scheme). See \ttt{MSTP(18)} for
a possibility to switch back to the older behaviour. However, full
backwards compatibility cannot be assured, so the older scenarios 
are better simulated by using an older {\Py} version.
\item The VMD processes work as ordinary
hadron--hadron ones, i.e.\ one obtains both low- and high-$\pT$ 
events by default, with dividing line set by $\pTmin$ above.
\item Also the GVMD processes work like the VMD ones. Again this is
a change from previous versions, where the anomalous processes only
contained high-$\pT$ physics and the low-$\pT$ part was covered in the
direct event class. See \ttt{MSTP(15)=5} for a possibility to switch 
back to the older behaviour, with comments as above for the direct
class. A GVMD state is book-kept as a diffractive state in the event 
listing, even when it scatters `elastically', since the subsequent 
hadronization descriptions are very similar.
\end{Itemize} 

The processes in points 1 and 2 can be simulated with a photon beam,
i.e.\ when \ttt{'gamma'} appears as argument in the \ttt{PYINIT} call.
It is then necessary to use option \ttt{MSTP(14)} to switch between 
a point-like and a resolved photon --- it is not possible to simulate 
the two sets of processes in a single run. This would be the normal
mode of operation for beamstrahlung photons, which have $Q^2 = 0$ 
but with a nontrivial energy spectrum that would be provided by some 
external routine.

For bremsstrahlung photons, the $x$ and $Q^2$ spectrum can be simulated
internally, with the {\galep} argument in the \ttt{PYINIT} 
call. This is the recommended procedure, wherein direct and resolved
processes can be mixed. An older --- now not recommended --- alternative 
is to use a parton-inside-electron structure function concept, obtainable
with a simple \ttt{'e-'} (or other lepton) argument in \ttt{PYINIT}.
To access these quark and gluon distributions inside the photon (itself 
inside the electron), \ttt{MSTP(12)=1} must then be used. Also the default 
value \ttt{MSTP(11)=1} is required for the preceding step, that
of finding photons inside the electron. Also here the direct and resolved 
processes may be generated together. However, this option only works
for high-$\pT$ physics. It is not possible to have also the low-$\pT$ 
physics (including multiple interactions in high-$\pT$ events) for an 
electron beam. Kindly note that subprocess 34 contains both the scattering 
of an electron off a photon and the scattering of a quark (inside a photon 
inside an electron) off a photon; the former can be switched off with the
help of the \ttt{KFIN} array.

If you are only concerned with standard QCD physics, the option 
\ttt{MSTP(14)=10} or the default \ttt{MSTP(14)=30} gives an automatic 
mixture of the VMD, direct and GVMD/anomalous event classes. The mixture 
is properly given according to
the relative cross sections. Whenever possible, this option is therefore
preferable in terms of user-friendliness. However, it can only work
because of a completely new layer of administration, not found anywhere 
else in {\Py}. For instance, a subprocess like $\q\g \to \q\g$ is 
allowed in several of the classes, but appears with different sets of 
parton distributions and different $\pT$ cut-offs in each of these,
so that it is necessary to switch gears between each event in the 
generation. It is therefore not possible to avoid a number of 
restrictions on what you can do in this case:
\begin{Itemize}
\item The \ttt{MSTP(14)=10} and \ttt{=30} options can only be used for 
incoming photon beams, with or without convolution with the
bremsstrahlung spectrum, i.e.\ when \ttt{'gamma'} or {\galep}
is the argument in the \ttt{PYINIT} call.
\item The machinery has only been set up to generate standard
QCD physics, specifically either `minimum-bias' one or high-$\pT$ jets.
There is thus no automatic mixing of processes only for heavy-flavour 
production, say, or of some exotic particle.
For minimum bias, you are not allowed to use the \ttt{CKIN} variables 
at all. This is not a major limitation, since it is in the spirit of 
minimum-bias physics not to impose any constraints on allowed jet 
production. (If you still do, these cuts will be ineffective for the 
VMD processes but take effect for the other ones, giving 
inconsistencies.) 
The minimum-bias physics option is obtained by default; by switching 
from \ttt{MSEL=1} to \ttt{MSEL=2} also the elastic and diffractive 
components of the VMD and GVMD parts are included. High-$\pT$ jet 
production is obtained by setting the \ttt{CKIN(3)} cut-off larger than 
the $\pTmin(W^2)$ of the multiple interactions scenario. For 
lower input \ttt{CKIN(3)} values the program will automatically switch 
back to minimum-bias physics.
\item Multiple interactions become possible in both the VMD and GVMD
sector, with the average number of interactions given by the 
ratio of the jet to the total cross section. Currently only
the simpler default scenario \ttt{MSTP(82)=1} is implemented, 
however, i.e.\ the more sophisticated variable-impact-parameter ones need 
further physics studies and model development.
\item Some variables are internally recalculated and reset, notably 
\ttt{CKIN(3)}. This is because it must have values that depend on the 
component studied. It can therefore not be modified without changing 
\ttt{PYINPR} and recompiling the program, which obviously is a major 
exercise.  

\item Pileup events are not at all allowed.  
\end{Itemize}

Also, a warning about the usage of \tsc{Pdflib} for photons. So long 
as \ttt{MSTP(14)=1}, i.e.\ the photon is not split up, \tsc{Pdflib} 
is accessed by \ttt{MSTP(56)=2} and \ttt{MSTP(55)} as the parton 
distribution set. However, when the VMD and anomalous pieces are split, 
the VMD part is based on a rescaling of pion distributions by VMD factors
(except for the SaS sets, that already come with a separate VMD piece).
Therefore, to access \tsc{Pdflib} for \ttt{MSTP(14)=10}, it is not 
correct to set \ttt{MSTP(56)=2} and a photon distribution in 
\ttt{MSTP(55)}. Instead, one should put \ttt{MSTP(56)=2}, 
\ttt{MSTP(54)=2} and a pion distribution code in \ttt{MSTP(53)},
while \ttt{MSTP(55)} has no function. The anomalous part is still based on
the SaS parameterization, with \ttt{PARP(15)} as main free parameter.   

Currently, hadrons are not defined with any photonic content. None
of the processes are therefore relevant in hadron--hadron collisions.
In $\e\p$ collisions, the electron can emit an almost real photon,
which may interact directly or be resolved. In $\ee$ collisions, 
one may have direct, singly-resolved or doubly-resolved processes.

The $\gamma\gamma$ equivalent to the $\gamma\p$ description involves 
six different event classes, see section \ref{sss:photoprod}.
These classes can be obtained by setting \ttt{MSTP(14)} to 0, 2, 3, 
5, 6 and 7, respectively. If one combines the VMD and anomalous
parts of the parton distributions of the photon, in a more coarse 
description, it is enough to use the \ttt{MSTP(14)} options 0, 1 and 4. 
The cut-off procedures follows from the ones used for the $\gamma\p$ 
ones above.

As with $\gamma\p$ events, the options \ttt{MSTP(14)=10} or 
\ttt{MSTP(14)=30} give a mixture of the six possible $\gamma\gamma$ 
event classes. The same complications and restrictions exist here as 
already listed above. 

Process 54 generates a mixture of quark flavours; allowed flavours
are set by the gluon \ttt{MDME values}. Process 58 can generate both 
quark and lepton pairs, according to the \ttt{MDME} values of the 
photon. Processes 84 and 85 are variants of these matrix elements, 
with fermion masses included in the matrix elements, but where only 
one flavour can be generated at a time. This flavour is selected as
described for processes 81 and 82 in section \ref{sss:heavflavclass},
with the exception that for process 85 the `heaviest' flavour allowed
for photon splitting takes to place of the heaviest flavour allowed 
for gluon splitting. Since lepton {\KF} codes come after quark ones, 
they are counted as being `heavier', and thus take precedence if 
they have been allowed.

Process 80 is a higher twist one. The theory for such processes 
is rather shaky, so results should not be taken too literally. 
The messy formulae given in \cite{Bag82} have not been programmed
in full, instead the pion form factor has been parameterized as
$Q^2 F_{\pi}(Q^2) \approx 0.55 / \ln Q^2$, with $Q$ in GeV. 

\subsubsection{Deeply Inelastic Scattering and $\gamma^*\gamma^*$ physics}
\label{sss:DISclass}

\ttt{MSEL} = 1, 2, 35, 36, 37, 38 \\
{\ISUB} = 
\begin{tabular}[t]{rl@{\protect\rule{0mm}{\tablinsep}}}
 10 & $\f_i \f_j \to \f_k \f_l$ \\
 83 & $\q_i \f_j \to \Q_k \f_l$ \\
 99 & $\gast \q \to \q$ \\
131 & $\f_i \gast_{\mrm{T}} \to \f_i \g$ \\
132 & $\f_i \gast_{\mrm{L}} \to \f_i \g$ \\
133 & $\f_i \gast_{\mrm{T}} \to \f_i \gamma$ \\
134 & $\f_i \gast_{\mrm{L}} \to \f_i \gamma$ \\
135 & $\g \gast_{\mrm{T}} \to \f_i \fbar_i$ \\
136 & $\g \gast_{\mrm{L}} \to \f_i \fbar_i$ \\
137 & $\gast_{\mrm{T}} \gast_{\mrm{T}} \to \f_i \fbar_i$ \\
138 & $\gast_{\mrm{T}} \gast_{\mrm{L}} \to \f_i \fbar_i$ \\
139 & $\gast_{\mrm{L}} \gast_{\mrm{T}} \to \f_i \fbar_i$ \\
140 & $\gast_{\mrm{L}} \gast_{\mrm{L}} \to \f_i \fbar_i$ \\
\end{tabular}

Among the processes in this section, 10 and 83 are intended to 
stand on their own, while the rest are part of the newer machinery
for $\gast\p$ and $\gast\gast$ physics. We therefore separate 
the description in this section into these two main parts.

The Deeply Inelastic Scattering (DIS) processes, i.e.\ $t$-channel 
electroweak gauge boson exchange, are traditionally associated
with interactions between a lepton or neutrino and a hadron, but 
processes 10  and 83 can equally well be applied for $\q\q$ scattering 
in hadron colliders (with a cross section much smaller than 
corresponding QCD processes, however). If applied to incoming $\ee$
beams, process 10 corresponds to Bhabha scattering.

For process 10 both $\gamma$, $\Z^0$ and $\W^{\pm}$ exchange 
contribute, including interference between $\gamma$ and $\Z^0$.
The switch \ttt{MSTP(21)} may be used to restrict to only some
of these, e.g.\ neutral or charged current only.

The option \ttt{MSTP(14)=10} (see previous section) has now been 
extended so that it also works for DIS of 
an electron off a (real) photon, i.e.\ process 10. What is obtained 
is a mixture of the photon acting as a vector meson and it acting 
as an anomalous state. This should therefore be the sum of what can 
be obtained with \ttt{MSTP(14)=2} and \ttt{=3}. It is distinct from 
\ttt{MSTP(14)=1} in that different sets are used for the parton 
distributions --- in \ttt{MSTP(14)=1} all the contributions to the
photon distributions are lumped together, while they are split in 
VMD and anomalous parts for \ttt{MSTP(14)=10}. Also the beam remnant 
treatment is different, with a simple Gaussian distribution (at least 
by default) for \ttt{MSTP(14)=1} and the VMD part of \ttt{MSTP(14)=10}, 
but a powerlike distribution $\d k_{\perp}^2 / k_{\perp}^2$ between 
\ttt{PARP(15)} and $Q$ for the anomalous part of \ttt{MSTP(14)=10}. 

To access this option for $\e$ and $\gamma$ as incoming beams, it is 
only necessary to set \ttt{MSTP(14)=10} and keep \ttt{MSEL} at its 
default value. Unlike the corresponding option for $\gamma\p$ and 
$\gamma\gamma$, no cuts are overwritten, i.e.\ it is still your 
responsibility to set these appropriately. 

Cuts especially appropriate for DIS usage include either
\ttt{CKIN(21)-CKIN(22)} or 
\ttt{CKIN(23)-CKIN(24)} for the $x$ range (former or latter depending 
on which side is the incoming real photon), \ttt{CKIN(35)-CKIN(36)} for 
the $Q^2$ range, and \ttt{CKIN(39)-CKIN(40)} for the $W^2$ range. 

In principle, the DIS $x$ variable of an event corresponds to the
$x$ value stored in \ttt{PARI(33)} or \ttt{PARI(34)}, depending
on which side the incoming hadron is on, while the DIS
$Q^2 = -\hat{t} = $\ttt{-PARI(15)}. However, just like initial- and 
final-state radiation can shift jet momenta, they can modify
the momentum of the scattered lepton. Therefore the DIS $x$ and 
$Q^2$ variables are not automatically conserved. An option, on by
default, exists in \ttt{MSTP(23)}, where the event can be `modified 
back' so as to conserve $x$ and $Q^2$, but this option is rather 
primitive and should not be taken too literally.

Process 83 is the equivalent of process 10 for $\W^{\pm}$ exchange
only, but with the heavy-quark mass included in the matrix element.
In hadron colliders it is mainly of interest for the production of 
very heavy flavours, where the possibility of producing just one 
heavy quark is kinematically favoured over pair production. The
selection of the heavy flavour is already discussed in section
\ref{sss:heavflavclass}.

Turning to the other processes, part of the $\gast\p$ and $\gast\gast$
process-mixing machineries, 99 has close similarities with the 
above discussed 10 one. Whereas 10 would simulate the full process
$\e \q \to \e \q$, 99 assumes a separate machinery for the flux
of virtual photons, $\e \to \e \gast$ and only covers the second
half of the process, $\gast \q \to \q$. One limitation of this
factorization is that only virtual photons are considered in 
process 99, not contributions from the $\Z^0$ neutral current   
or the $\W^{\pm}$ charged current. 

Note that 99 has no correspondence in the real-photon case, but has 
to vanish in this limit by gauge invariance, or indeed by simple 
kinematics considerations. This, plus the desire to avoid double-counting
with real-photon physics processes, is why the cross section for 
this process is explicitly made to vanish for photon virtuality
$Q^2 \to 0$, eq.~(\ref{eq:sigDIS}), also when parton distributions
have not been constructed to fulfil this, see \ttt{MSTP(19)}. 
(No such safety measures are present in 10, again illustrating how 
the two are intended mainly to be used at large or at small $Q^2$, 
respectively.)

For a virtual photon, processes 131--136 may be viewed as first-order 
corrections to 99.  The three with a transversely polarized photon,
131, 133 and 135, smoothly reduce to the real-photon direct 
(single-resolved for $\gamma\gamma$) processes 33, 34 and 54.
The other three, corresponding to the exchange of a longitudinal
photon, vanish like $Q^2$ for $Q^2 \to 0$. The double-counting issue
with process 99 is solved by requiring the latter process not to 
contain any shower branchings with a $\pT$ above the lower $\pT$
cut-off of processes 131-136. The cross section is then to be reduced
accordingly, see eq.~(\ref{eq:LODIS}) and the discussion there,
and again \ttt{MSTP(19)}.

We thus see that process 99 by default is a low-$\pT$ process in
about the same sense as process 95, giving `what is left' of the total
cross section when jet events have been removed. Therefore, it will be 
switched off in event class mixes such as \ttt{MSTP(14)=30} if 
\ttt{CKIN(3)} is above $\pTmin(W^2)$ and \ttt{MSEL} is not 2. There 
is a difference, however, in that process 99 events still are allowed 
to contain shower evolution (although currently only the final-state 
kind has been implemented), since the border to the other processes 
is at $\pT = Q$ for large $Q$ and thus need not be so small. The $\pT$
scale of the `hard process', stored e.g.\ in \ttt{PARI(17)} always
remains 0, however. (Other \ttt{PARI} variables defined for normal
$2 \to 2$ and $2 \to 1$ processes are not set at all, and may well
contain irrelevant junk left over from previous events.) 

Processes 137--140, finally, are extensions of process 58 from
the real-photon limit to the virtual-photon case, and correspond to 
the direct process of $\gamma^*\gamma^*$ physics. The four cases
correspond to either of the two photons being either transversely or
longitudinally polarized. As above, the cross section of a longitudinal
photon vanishes when its virtuality approaches 0. 

 \subsubsection{Photon physics at all virtualities}
\label{sss:photonallQclass}

{\ISUB} = 
\begin{tabular}[t]{ll@{\protect\rule{0mm}{\tablinsep}}}
direct$\times$direct: & 137, 138, 139, 140 \\
direct$\times$resolved: & 131, 132, 135, 136 \\
DIS$\times$resolved: & 99 \\
resolved$\times$resolved, high-$\pT$: & 11, 12, 13, 28, 53, 68 \\
resolved$\times$resolved, low-$\pT$: & 91, 92, 93, 94, 95 \\
\end{tabular}\\
where `resolved' is a hadron or a VMD or GVMD photon.

At intermediate photon virtualities, processes described in both of 
the sections above are allowed, and have to be mixed appropriately.
The sets are of about equal importance at around 
$Q^2 \sim m_{\rho}^2 \sim 1$~GeV$^2$, but the transition is gradual
over a larger $Q^2$ range. The ansatz for this mixing is given by
eq.~(\ref{eq:gammapallQ}) for $\gast\p$ events and 
eq.~(\ref{eq:gagaallQ}) for $\gast\gast$ ones. In short, for direct
and DIS processes the photon virtuality explicitly enters in the 
matrix element expressions, and thus is easily taken into account.
For resolved photons, perturbation theory does not provide a unique
answer, so instead cross sections are suppressed by dipole factors, 
$(m^2/(m^2 + Q^2))^2$, where $m = m_V$ for a VMD state and $m = 2 \kT$ 
for a GVMD state characterized by a $\kT$ scale of the 
$\gast \to \q\qbar$ branching. These factors appear explicitly for
total, elastic and diffractive cross sections, and are also implicitly
used e.g.\ in deriving the SaS parton distributions for virtual photons.
Finally, some double-counting need to be removed, between direct and
DIS processes as mentioned in the previous section, and between
resolved and DIS at large $x$.

Since the mixing is not trivial, it is recommended to use the default
\ttt{MSTP(14)=30} to obtain it in one go and hopefully consistently,
rather than building it up by combining separate runs. The main issues
still under your control include, among others
\begin{Itemize}
\item The \ttt{CKIN(61) - CKIN(78)} should be used to set the range of
$x$ and $Q^2$ values emitted from the lepton beams. That way one may 
decide between almost real or very virtual photons, say. Also some other
quantities, like $W^2$, can be constrained to desirable ranges.
\item Whether or not minimum bias events are simulated depends on the
\ttt{CKIN(3)} value, just like in hadron physics. The only difference is 
that the initialization energy scale $W_{\mrm{init}}$ is selected in the 
allowed $W$ range rather than to be the full c.m.\ energy.\\
For a high \ttt{CKIN(3)}, \ttt{CKIN(3)}$> \pTmin(W_{\mrm{init}}^2)$, 
only jet production is included. Then further \ttt{CKIN} values can be
set to constrain e.g.\ the rapidity of the jets produced.\\
For a low \ttt{CKIN(3)}, \ttt{CKIN(3)}$< \pTmin(W_{\mrm{init}}^2)$,
like the default value \ttt{CKIN(3) = 0}, low-$\pT$ physics is switched 
on together with jet production, with the latter properly eikonalized to 
be lower than the total one. The ordinary \ttt{CKIN} cuts, not related to
the photon flux, cannot be used here.\\ 
For a low \ttt{CKIN(3)}, when \ttt{MSEL=2} instead of the default 
\ttt{=1}, also elastic and diffractive events are simulated. 
\item The impact of resolved longitudinal photons is not unambiguous,
e.g.\ only recently the first parameterization of parton distributions
appeared \cite{Chy00}. Different simple alternatives can be probed by
changing \ttt{MSTP(17)} and associated parameters.
\item The choice of scales to use in parton distributions for jet rates
is always ambiguous, but depends on even more scales for virtual photons
than in hadronic collisions. \ttt{MSTP(32)} allows a choice between
several alternatives.
\item The matching of $\pT$ generation by shower evolution to that by
primordial $\kT$ is a general problem, for photons with an additional
potential source in the $\gast \to \q\qbar$ vertex. \ttt{MSTP(66)} 
offer some alternatives.
\item \ttt{PARP(15)} is the $k_0$ parameter separating VMD from GVMD.
\item \ttt{PARP(18)} is the $k_{\rho}$ parameter in GVMD total cross
sections.
\item \ttt{MSTP(16)} selects the momentum variable for an 
$\e \to \e \gast$ branching.
\item \ttt{MSTP(18)} regulates the choice of $\pTmin$ for direct 
processes. 
\item \ttt{MSTP(19)} regulates the choice of partonic cross section in
process 99, $\gast\q \to \q$.
\item \ttt{MSTP(20)} regulates the suppression of the resolved cross 
section at large $x$.
\end{Itemize}
The above list is not complete, but gives some impression what can
be done. 

\subsection{Electroweak Gauge Bosons}

This section covers the production and/or exchange of $\gamma$, 
$\Z^0$ and $\W^{\pm}$ gauge bosons, singly and in pairs. The topic 
of longitudinal gauge-boson scattering at high energies is deferred 
to the Higgs section, since the presence or absence of a Higgs here 
makes a big difference.

\subsubsection{Prompt photon production}
\label{sss:promptgammaclass} 
  
\ttt{MSEL} = 10 \\ 
{\ISUB} = 
\begin{tabular}[t]{rl@{\protect\rule{0mm}{\tablinsep}}}
 14 & $\q_i \qbar_i \to \g \gamma$  \\
 18 & $\f_i \fbar_i \to \gamma \gamma$ \\
 29 & $\q_i \g \to \q_i \gamma$ \\
114 & $\g \g \to \gamma \gamma$  \\
115 & $\g \g \to \g \gamma$ \\
\end{tabular}

In hadron colliders, processes {\ISUB} = 14 and 29 give the main source 
of single-$\gamma$ production, with {\ISUB} = 115 giving an additional 
contribution which, in some kinematics regions, may become important. 
For $\gamma$-pair production, the process {\ISUB} = 18 is often 
overshadowed in importance by {\ISUB} = 114. 
  
Another source of photons is bremsstrahlung off incoming or outgoing 
quarks. This has to be treated on an equal footing with QCD parton 
showering. For time-like parton-shower evolution, i.e.\ in the 
final-state showering and in the side branches of the initial-state 
showering, photon emission may be switched on or off with 
\ttt{MSTJ(41)}. Photon radiation off the space-like incoming 
quark legs is not yet included, but should be of lesser importance 
for production at reasonably large $\pT$ values. Radiation off an 
incoming electron is included in a leading-log approximation. 
  
{\bf Warning:} the cross sections for the box graphs 114 and 115 become 
very complicated, numerically unstable and slow when the 
full quark mass dependence is included. For quark masses much 
below the $\hat{s}$ scale, the simplified massless expressions are 
therefore used --- a fairly accurate approximation. However, there 
is another set of subtle numerical cancellations between different 
terms in the massive matrix elements in the region of small-angle 
scattering. The associated problems have not been sorted out yet. 
There are therefore two possible solutions. One is to use the 
massless formulae throughout. The program then becomes faster and 
numerically stable, but does not give, for example, the characteristic 
dip (due to destructive interference) at top threshold. This is the 
current default procedure, with five flavours assumed, but this 
number can be changed in \ttt{MSTP(38)}. The other possibility is 
to impose cuts on the scattering angle of the hard process, see 
\ttt{CKIN(27)} and \ttt{CKIN(28)}, since the numerically unstable 
regions are when $|\cos\hat{\theta}|$ is close to unity. It is then 
also necessary to change \ttt{MSTP(38)} to 0. 
 
\subsubsection{Single $\W/\Z$ production}
\label{sss:WZclass} 
  
\ttt{MSEL} = 11, 12, 13, 14, 15, (21) \\
{\ISUB} = 
\begin{tabular}[t]{rl@{\protect\rule{0mm}{\tablinsep}}}
  1 & $\f_i \fbar_i \to \gammaZ$ \\
  2 & $\f_i \fbar_j \to \W^+$ \\
 15 & $\f_i \fbar_i \to \g (\gammaZ)$ \\
 16 & $\f_i \fbar_j \to \g \W^+$ \\
 19 & $\f_i \fbar_i \to \gamma (\gammaZ)$ \\
 20 & $\f_i \fbar_j \to \gamma \W^+$ \\
 30 & $\f_i \g \to \f_i (\gammaZ)$ \\
 31 & $\f_i \g \to \f_k \W^+$ \\
 35 & $\f_i \gamma \to \f_i (\gammaZ)$ \\
 36 & $\f_i \gamma \to \f_k \W^+$ \\
(141) & $\f_i \fbar_i \to \gamma/\Z^0/\Z'^0$ \\
(142) & $\f_i \fbar_j \to \W'^+$ \\
\end{tabular}
  
This group consists of $2 \to 1$ processes, i.e.\ production of a 
single resonance, and $2 \to 2$ processes, where the resonance 
is recoiling against a jet or a photon. The processes 141 and 142, 
which also are listed here, are described further elsewhere.

With initial-state showers turned on, the $2 \to 1$ processes also 
generate additional jets; in order to avoid double-counting, the 
corresponding $2 \to 2$ processes should therefore not be turned 
on simultaneously. The basic rule is to use the $2 \to 1$ processes 
for inclusive generation of $\W/\Z$, i.e.\ where the bulk of the
events studied have $\pT \ll m_{\W/\Z}$. With the introduction of 
explicit matrix-element-inspired corrections to the parton shower 
\cite{Miu99}, also the high-$\pT$ tail is well described in this
approach, thus offering an overall good description of the full $\pT$
spectrum of gauge bosons \cite{Bal01}. 

If one is interested in the high-$\pT$ tail only, however, the 
generation efficiency will be low. It is here better to start from 
the $2 \to 2$ matrix elements and add showers to these. However, the 
$2 \to 2$ matrix elements are divergent for $\pT \to 0$, and should
not be used down to the low-$\pT$ region, or one may get unphysical
cross sections. As soon as the generated $2 \to 2$ cross section
corresponds to a non-negligible fraction of the total $2 \to 1$ one,
say 10\%--20\%, Sudakov effects are likely to be affecting the 
shape of the $\pT$ spectrum to a corresponding extent, and results
should not be trusted. 

The problems of double-counting and Sudakov effects apply not only 
to $\W/\Z$ production in hadron colliders, but also to a process
like $\ee \to \Z^0 \gamma$, which clearly is part of the 
initial-state radiation corrections to $\ee \to \Z^0$ obtained for 
\ttt{MSTP(11)=1}. As is the case for $\Z$ production in association 
with jets, the $2 \to 2$ process should therefore only be used for the 
high-$\pT$ region. 
  
The $\Z^0$ of subprocess 1 includes the full interference structure 
$\gammaZ$; via \ttt{MSTP(43)} you can select to produce only 
$\gamma^*$, only $\Z^0$, or the full $\gammaZ$. The same holds true 
for the $\Z'^0$ of subprocess 141; via \ttt{MSTP(44)} any combination 
of $\gamma^*$, $\Z^0$ and $\Z'^0$ can be selected. Thus, subprocess 
141 with \ttt{MSTP(44)=4} is essentially equivalent to subprocess 
1 with \ttt{MSTP(43)=3}; however, process 141 also includes the 
possibility of a decay into Higgs bosons. Also processes 
15, 19, 30 and 35 contain the full mixture of $\gammaZ$, with 
\ttt{MSTP(43)} available to change this. 

Note that process 1, with only 
$\q\qbar \to \gamma^* \to \ell^+ \ell^-$ allowed,
and studied in the region well below the $\Z^0$ mass, is what is 
conventionally called Drell--Yan. This latter process therefore does 
not appear under a separate heading, but can be obtained by a
suitable setting of switches and parameters.    
  
A process like $\f_i \fbar_j \to \gamma \W^+$ is only included in
the limit that the $\gamma$ is emitted in the `initial state',
while the possibility of a final-state radiation off the $\W^+$
decay products is not explicitly included (but can be obtained
implicitly by the parton-shower machinery) and various interference
terms are not at all present. Some caution must therefore be
exercised; see also section \ref{sss:WZpairclass} for related 
comments.

For the $2 \to 1$ processes, the Breit--Wigner includes an 
$\hat{s}$-dependent width, which should provide an improved 
description of line shapes. In fact, from a line-shape point of view, 
process 1 should provide a more accurate simulation of $\ee$
annihilation events than the dedicated $\ee$ generation scheme of 
\ttt{PYEEVT} (see section \ref{ss:eematrix}). Another difference is 
that \ttt{PYEEVT} only allows the generation of $\gammaZ \to \q\qbar$, 
while process 1 additionally contains $\gammaZ \to \ell^+ \ell^-$ and 
$\gammaZ \to \nu \br{\nu}$. The parton-shower and fragmentation 
descriptions are the same, but the process 1 implementation only 
contains a partial interface to the first- and second-order 
matrix-element options available in \ttt{PYEEVT}, see \ttt{MSTP(48)}. 
  
All processes in this group have been included with the 
correct angular distribution in the subsequent $\W/\Z \to \f \fbar$
decays. In process 1 also fermion mass effects 
have been included in the angular distributions, while this is not the
case for the other ones. Normally mass effects are not large anyway.

The process $\ee \to \e^+ \e^- \Z^0$ can be simulated in two 
different ways. One is to make use of the $\e$ `sea' distribution 
inside $\e$, i.e.\ have splittings $\e \to \gamma \to \e$. 
This can be obtained, together with ordinary $\Z^0$ production, by 
using subprocess 1, with \ttt{MSTP(11)=1} and \ttt{MSTP(12)=1}. Then 
the contribution of the type above is 5.0 pb for a 500 GeV $\ee$ 
collider, compared with the correct 6.2 pb \cite{Hag91}. Alternatively 
one may use process 35, with \ttt{MSTP(11)=1} and \ttt{MSTP(12)=0}. 
This process has a singularity in the forward direction, regularized by 
the electron mass and also sensitive to the virtuality of the photon.
It is therefore among the few where the incoming masses have been
included in the matrix element expression. Nevertheless, it may be 
advisable to set small lower cut-offs, e.g.\ \ttt{CKIN(3)=CKIN(5)=0.01},
if one should experience problems (e.g.\ at higher energies).
  
Process 36, $\f \gamma \to \f' \W^{\pm}$ may have corresponding 
problems; except that in $\ee$ the forward scattering amplitude for 
$\e \gamma \to \nu \W$ is killed (radiation zero), which means 
that the differential cross section is vanishing for $\pT \to 0$.
It is therefore feasible to use the default \ttt{CKIN(3)} and 
\ttt{CKIN(5)} values in $\ee$, and one also comes closer to the 
correct cross section. 

The process $\g \g \to \Z^0 \b \bbar$, formerly available as process
131, has been removed from the current version, since the implementation
turned out to be slow and unstable. However, process 1 with incoming
flavours set to be $\b \bbar$ (by \ttt{KFIN(1,5)=}
\ttt{KFIN(1,-5)=KFIN(2,5)=KFIN(2,-5)=1} and everything
else \ttt{=0}) provides an alternative description, where the 
additional $\b \bbar$ are generated by $\g \to \b \bbar$ branchings
in the initial-state showers. (Away from the low-$\pT$ region,
process 30 with \ttt{KFIN} values as above except that also incoming 
gluons are allowed, offers yet another description. Here it is in terms
of $\g\b \to \Z^0 \b$, with only one further $\g \to \b \bbar$ branching
constructed by the shower.) At first glance, the shower approach would
seem less reliable than the full $2 \to 3$ matrix element. The relative
lightness of the $\b$ quark will generate large logs of the type
$\ln(m_{\Z}^2/m_{\b}^2)$, however, that ought to be resummed \cite{Car00}. 
This is implicit in the parton-density approach of incoming $\b$ quarks 
but absent from the lowest-order $\g \g \to \Z^0 \b \bbar$ matrix elements.
Therefore actually the shower approach may be the more accurate of the 
two. Within the general range of uncertainty of any leading-order
description, at least it is not any worse.

\subsubsection{$\W/\Z$ pair production}
\label{sss:WZpairclass} 
  
\ttt{MSEL} = 15 \\
{\ISUB} = 
\begin{tabular}[t]{rl@{\protect\rule{0mm}{\tablinsep}}}
 22 & $\f_i \fbar_i \to (\gammaZ) (\gammaZ)$ \\
 23 & $\f_i \fbar_j \to \Z^0 \W^+$ \\
 25 & $\f_i \fbar_i \to \W^+ \W^-$ \\
 69 & $\gamma \gamma \to \W^+ \W^-$ \\
 70 & $\gamma \W^+ \to \Z^0 \W^+$ \\
\end{tabular}
  
In this section we mainly consider the production of $\W/\Z$ pairs
by fermion--antifermion annihilation, but also include two processes
which involve $\gamma/\W$ beams. Scatterings between gauge-boson 
pairs, i.e.\ processes like $\W^+ \W^- \to \Z^0 \Z^0$, depend so 
crucially on the assumed Higgs scenario that they are considered 
separately in section \ref{sss:heavySMHclass}.

The cross sections used for the above processes are those derived
in the narrow-width limit, but have been extended to include 
Breit--Wigner shapes with mass-dependent widths for the final-state 
particles. In process 25, the contribution from $\Z^0$ exchange
to the cross section is now evaluated with the fixed nominal $\Z^0$ 
mass and width in the propagator. If instead the actual mass and the 
running width were to be used, it gives a diverging cross section at 
large energies, by imperfect gauge cancellation. 

However, one should realize that other graphs, not included here, can 
contribute in regions away from the $\W/\Z$ mass. This problem is 
especially important if several flavours coincide in the four-fermion 
final state. Consider, as an example, 
$\ee \to \mu^+ \mu^- \nu_{\mu} \br{\nu}_{\mu}$.
Not only would such a final state receive contributions from 
intermediate $\Z^0\Z^0$ and $\W^+\W^-$ states, but also 
from processes $\ee \to \Z^0 \to \mu^+ \mu^-$, followed
either by $\mu^+ \to \mu^+ \Z^0 \to \mu^+ \nu_{\mu} \br{\nu}_{\mu}$, 
or by 
$\mu^+ \to \br{\nu}_{\mu} \W^+ \to \br{\nu}_{\mu} \mu^+ \nu_{\mu}$.
In addition, all possible interferences should be considered.
Since this is not done, the processes have to be used with some 
sound judgement. Very often, one may wish to constrain a
lepton pair mass to be close to $m_{\Z}$, in which case a number
of the possible `other' processes are negligible.

For the $\W$ pair production graph, one experimental objective is to do 
precision measurements of the cross section near threshold. Then also
other effects enter. One such is Coulomb corrections, induced by 
photon exchange between the two $\W$'s and their decay products.
The gauge invariance issues induced by the finite $\W$ lifetime are not 
yet fully resolved, and therefore somewhat different approximate formulae 
may be derived \cite{Kho96}. The options in \ttt{MSTP(40)} provide a
reasonable range of uncertainty.

Of the above processes, the first contains the full
$\f_i \fbar_i \to (\gammaZ)(\gammaZ)$ structure, obtained by a 
straightforward generalization of the formulae in ref. \cite{Gun86} 
(done by one of the {\Py} authors). Of course, the possibility of
there being significant contributions from graphs that are not 
included is increased, in particular 
if one $\gamma^*$ is very light and therefore could be a 
bremsstrahlung-type photon. It is possible to use \ttt{MSTP(43)} to 
recover the pure $\Z^0$ case, i.e.\ $\f_i \fbar_i \to \Z^0 \Z^0$
exclusively. In processes 23 and 70, only the pure $\Z^0$ contribution
is included.

Full angular correlations are included for the first three processes,
i.e.\ the full $2 \to 2 \to 4$ matrix elements are included in the
resonance decays, including the appropriate $\gammaZ$ interference
in process 22. In the latter two processes no spin 
information is currently preserved, i.e.\ the $\W/\Z$ bosons are 
allowed to decay isotropically.

We remind you that the mass ranges of the two resonances may be set 
with the \ttt{CKIN(41) - CKIN(44)} parameters; this is particularly 
convenient, for instance, to pick one resonance almost on the mass 
shell and the other not. 

\subsection{Higgs Production}
\label{ss:Hclass}

A fair fraction of all the processes in {\Py} deal with Higgs 
production in one form or another. This multiplication is caused by 
the need to consider production by several different processes, 
depending on Higgs mass and machine type. Further, the program 
contains a full two-Higgs-multiplet scenario, as predicted for example
in the Minimal Supersymmetric extension of the Standard Model
(MSSM). Therefore the continued discussion is, somewhat arbitrarily,
subdivided into a few different scenarios. Doubly-charged Higgs
particles appear in left--right symmetric models, and are covered in 
section \ref{sss:LRDCHclass}.

\subsubsection{Light Standard Model Higgs}
\label{sss:lightSMHclass}
  
\ttt{MSEL} = 16, 17, 18 \\
{\ISUB} = 
\begin{tabular}[t]{rl@{\protect\rule{0mm}{\tablinsep}}}
  3 & $\f_i \fbar_i \to \hrm^0$ \\
 24 & $\f_i \fbar_i \to \Z^0 \hrm^0$ \\
 26 & $\f_i \fbar_j \to \W^+ \hrm^0$ \\
 32 & $\f_i \g \to \f_i \hrm^0$   \\
102 & $\g \g \to \hrm^0$ \\
103 & $\gamma \gamma \to \hrm^0$ \\
110 & $\f_i \fbar_i \to \gamma \hrm^0$ \\
111 & $\f_i \fbar_i \to \g \hrm^0$ \\
112 & $\f_i \g \to \f_i \hrm^0$ \\
113 & $\g \g \to \g \hrm^0$ \\
121 & $\g \g \to \Q_k \Qbar_k \hrm^0$ \\
122 & $\q_i \qbar_i \to \Q_k \Qbar_k \hrm^0$ \\
123 & $\f_i \f_j \to \f_i \f_j \hrm^0$ ($\Z^0 \Z^0$ fusion) \\
124 & $\f_i \f_j \to \f_k \f_l \hrm^0$ ($\W^+ \W^-$ fusion) \\
\end{tabular}
  
In this section we discuss the production of a reasonably light 
Standard Model Higgs, below 700 GeV, say, so that the narrow
width approximation can be used with some confidence. Below 400 GeV
there would certainly be no trouble, while above that the narrow
width approximation is gradually starting to break down.

In a hadron collider, the main production processes are 102, 123 
and 124, i.e.\ $\g\g$, $\Z^0 \Z^0$ and $\W^+ \W^-$ fusion. In the
latter two processes, it is also necessary to take into account
the emission of the space-like $\W/\Z$ bosons off quarks, which
in total gives the $2 \to 3$ processes above. 

Further processes of lower cross sections may be of interest
because of easier signals. For instance, processes 24 and 26 give 
associated production of a $\Z$ or a $\W$ together with the $\hrm^0$.
There is also the processes 3 (see below), 121 and 122, which involve 
production of heavy flavours.

Process 3 contains contributions from all flavours, but is 
completely dominated by the subprocess $\t \tbar \to \hrm^0$,
i.e.\ by the contribution from the top sea distributions.
Assuming, of course, that parton densities for top quarks are
available, which is no longer the case in current parameterizations.
This process is by now known to overestimate the cross section 
for Higgs production as compared with a more careful calculation 
based on the subprocess $\g \g \to \t \tbar \hrm^0$, process 121. 
The difference between the two is that in process 3 the
$\t$ and $\tbar$ are added by the initial-state shower, while in 
121 the full matrix element is used. The price to be paid is that
the complicated multibody phase space in process 121 makes the
program run slower than with most other processes. As usual, it 
would be double-counting to include the same flavour both with 
3 and 121. An intermediate step --- in practice probably not so 
useful --- is offered by process 32, $\q \g \to \q \hrm^0$,
where the quark is assumed to be a $\b$ one, with the antiquark 
added by the showering activity.

Process 122 is similar in structure to 121, but is less 
important. In both process 121 and 122 the produced quark is assumed 
to be a $\t$; this can be changed in \ttt{KFPR(121,2)} and 
\ttt{KFPR(122,2)} before initialization, however. For $\b$ quarks it 
could well be that process 3 with $\b \bbar \to \hrm^0$ is more 
reliable than process 121 with $\g \g \to \b \bbar \hrm^0$ 
\cite{Car00}; see the discussion on $\Z^0 \b \bbar$ final states in 
section \ref{sss:WZclass}. Thus it would make sense to run with all 
quarks up to and including $\b$ simulated in process 3 and then 
consider $\t$ quarks separately in process 121. Assuming no $\t$
parton densities, this would actually be the default behaviour,
meaning that the two could be combined in the same run without
double counting. 

The two subprocess 112 and 113, with a Higgs recoiling against a 
quark or gluon jet, are also effectively generated by initial-state 
corrections to subprocess 102. Thus, in order to avoid double-counting, 
just as for the case of $\Z^0/\W^+$ production, section \ref{sss:WZclass}, 
these subprocesses should not be switched on simultaneously. Process 111,
$\q\qbar \to \g\hrm^0$ is different, in the sense that it proceeds 
through an $s$-channel gluon coupling to a heavy-quark loop, and that 
therefore the emitted gluon is necessary in the final state in order 
to conserve colours. It is not to be confused with a gluon-radiation 
correction to the Born-level process 3, like in process 32, since 
processes 3 and 32 vanish for massless quarks while process 111 is mainly 
intended for such. The lack of a matching Born-level process shows up
by process 111 being vanishing in the $\pT \to 0$ limit. Numerically it 
is of negligible importance, except at very large $\pT$ values. 
Process 102, possibly augmented by 111, should thus be used for 
inclusive production of Higgs, and 111--113 for the study 
of the Higgs subsample with high transverse momentum. 

A warning is that the matrix-element expressions for processes 111--113 
are very lengthy and the coding therefore more likely to contain some 
errors and numerical instabilities than for most other processes. 
Therefore the full expressions are only available by setting the 
non-default value \ttt{MSTP(38)=0}. Instead the default is based on 
the simplified expressions obtainable if only the top quark contribution 
is considered, in the $m_{\t} \to \infty$ limit \cite{Ell88}. As a slight
improvement, this expression is rescaled by the ratio of the 
$\g\g \to \hrm^0$ cross sections (or, equivalently, the $\hrm \to \g\g$
partial widths) of the full calculation and that in the 
$m_{\t} \to \infty$ limit. Simple checks show that this approach normally
agrees with the full expressions to within $\sim 20$\%, which is small
compared with other uncertainties. The agreement is worse for process
111 alone, about a factor of 2, but this process is small anyway.
We also note that the matrix element correction factors, used in the
initial-state parton shower for process 102, subsection 
\ref{sss:newinshow}, are based on the same $m_{\t} \to \infty$ limit
expressions, so that the high-$\pT$ tail of process 102 is well matched
to the simple description of process 112 and 113.
 
In $\ee$ annihilation, associated production of an $\hrm^0$ with a 
$\Z^0$, process 24, is usually the dominant one close to threshold,
while the $\Z^0 \Z^0$ and $\W^+ \W^-$ fusion processes 123 and 124
win out at high energies. Process 103, $\gamma\gamma$ fusion, may
also be of interest, in particular when the possibilities of 
beamstrahlung photons and backscattered photons are included
(see subsection \ref{sss:estructfun}). 
Process 110, which gives an $\hrm^0$ in association with a $\gamma$,
is a loop process and is therefore suppressed in rate. It would
have been of interest for a $\hrm^0$ mass above 60 GeV at LEP 1,
since its phase space suppression there is less severe than for the 
associated production with a $\Z^0$. Now it is not likely to be of 
any further interest.
   
The branching ratios of the Higgs are very strongly dependent on 
the mass. In principle, the program is set up to calculate these 
correctly, as a function of the actual Higgs mass, i.e.\ not just
at the nominal mass. However, higher-order corrections may at 
times be important and not fully unambiguous; see for instance 
\ttt{MSTP(37)}. 

Since the Higgs is a spin-0 particle it decays isotropically. In decay
processes such as $\hrm^0 \to \W^+ \W^- / \Z^0 \Z^0 \to 4$ fermions angular 
correlations are included \cite{Lin97}. Also in processes 24 and 26, 
$\Z^0$ and $\W^{\pm}$ decay angular distributions are correctly taken into 
account.

\subsubsection{Heavy Standard Model Higgs}
\label{sss:heavySMHclass}

{\ISUB} = 
\begin{tabular}[t]{rl@{\protect\rule{0mm}{\tablinsep}}}
  5 & $\Z^0 \Z^0 \to \hrm^0$ \\
  8 & $\W^+ \W^- \to \hrm^0$ \\
 71 & $\Z^0 \Z^0 \to \Z^0 \Z^0$ (longitudinal) \\
 72 & $\Z^0 \Z^0 \to \W^+ \W^-$ (longitudinal) \\
 73 & $\Z^0 \W^+ \to \Z^0 \W^+$ (longitudinal) \\
 76 & $\W^+ \W^- \to \Z^0 \Z^0$ (longitudinal) \\
 77 & $\W^+ \W^{\pm} \to \W^+ \W^{\pm}$ (longitudinal) \\
\end{tabular}

Processes 5 and 8 are the simple $2 \to 1$ versions of what is now
available in 123 and 124 with the full $2 \to 3$ kinematics.
For low Higgs masses processes 5 and 8 overestimate the correct
cross sections and should not be used, whereas good agreement between
the $2 \to 1$ and $2 \to 3$ descriptions is observed when heavy 
Higgs production is studied. 

The subprocesses 5 and 8, $V V \to \hrm^0$, which contribute to the 
processes $V V \to V' V'$, show a bad high-energy behaviour. Here 
$V$ denotes a longitudinal intermediate gauge boson, $\Z^0$ or 
$\W^{\pm}$. This  can be cured only by the inclusion of all 
$V V \to V' V'$ graphs, as is done in subprocesses 71, 72, 73, 76 
and 77. In particular, subprocesses 5 and 8 give rise to a fictitious 
high-mass tail of the Higgs. If this tail is thrown away, however, 
the agreement between the $s$-channel graphs only (subprocesses 
5 and 8) and the full set of graphs (subprocesses 71 etc.) is very 
good: for a Higgs of nominal mass 300 (800) GeV, a cut at 600 (1200) 
GeV retains 95\% (84\%) of the total cross section, and differs from 
the exact calculation, cut at the same values, by only 2\% (11\%) 
(numbers for SSC energies). With this prescription there is 
therefore no need to use subprocesses 71 etc. rather than 
subprocesses 5 and 8. 

For subprocess 77, there is an option, see \ttt{MSTP(45)}, to select
the charge combination of the scattering $\W$'s: like-sign, 
opposite-sign (relevant for Higgs), or both.

Process 77 contains a divergence for $\pT \to 0$ due to 
$\gamma$-exchange contributions. This leads to an infinite total 
cross section, which is entirely fictitious, since the simple 
parton-distribution function approach to the longitudinal $\W$ flux 
is not appropriate in this limit. For this process, it is therefore 
necessary to make use of a cut, e.g.\ $\pT > m_{\W}$.
  
For subprocesses 71, 72, 76 and 77, an option is included (see 
\ttt{MSTP(46)}) whereby you can select only the $s$-channel 
Higgs graph; this will then be essentially equivalent to running 
subprocess 5 or 8 with the proper decay channels (i.e.\ $\Z^0\Z^0$ or 
$\W^+\W^-$) set via \ttt{MDME}. The difference is that the 
Breit--Wigners in subprocesses 5 and 8 contain a mass-dependent 
width, whereas the width in subprocesses 71--77 is calculated at 
the nominal Higgs mass; also, higher-order corrections to the widths 
are treated more accurately in subprocesses 5 and 8. Further, 
processes 71--77 assume the incoming $\W/\Z$ to be on the mass shell, 
with associated kinematics factors, while processes 5 and 8 have 
$\W/\Z$ correctly space-like. All this leads to differences in the 
cross sections by up to a factor of 1.5. 
  
In the absence of a Higgs, the sector of longitudinal $\Z$ and $\W$ 
scattering will become strongly interacting at energies above 1 TeV. 
The models proposed by Dobado, Herrero and Terron \cite{Dob91} to 
describe this kind of physics have been included as alternative matrix 
elements for subprocesses 71, 72, 73, 76 and 77, selectable by 
\ttt{MSTP(46)}. From the point of view of the general classification 
scheme for subprocesses, this kind of models should appropriately be 
included as separate subprocesses with numbers above 100, but the 
current solution allows a more efficient reuse of existing code. 
By a proper choice of parameters, it is also here possible to 
simulate the production of a techni-$\rho$ (see subsection
\ref{sss:technicolorclass}). 

Currently, the scattering of transverse gauge bosons has not been 
included, neither that of mixed transverse--longitudinal scatterings.
These are expected to be less important at high energies, and do not
contain an $\hrm^0$ resonance peak, but need not be
entirely negligible in magnitude. As a rule of thumb, processes 
71--77 should not be used for $VV$ invariant masses below 500 GeV.

The decay products of the longitudinal gauge bosons are correctly 
distributed in angle.

\subsubsection{Extended neutral Higgs sector}
\label{sss:extneutHclass} 

\ttt{MSEL} = 19 \\  
{\ISUB} = 
\begin{tabular}[t]{rrrl@{\protect\rule{0mm}{\tablinsep}}}
$\hrm^0$ & $\H^0$ & $\A^0$ &  \\
  3 & 151 & 156 & $\f_i \fbar_i \to X$ \\
102 & 152 & 157 & $\g \g \to X$ \\
103 & 153 & 158 & $\gamma \gamma \to X$ \\
111 & 183 & 188 & $\q \qbar \to \g X$ \\
112 & 184 & 189 & $\q \g \to \q X$ \\
113 & 185 & 190 & $\g \g \to \g X$ \\
 24 & 171 & 176 & $\f_i \fbar_i \to \Z^0 X$ \\
 26 & 172 & 177 & $\f_i \fbar_j \to \W^+ X$ \\
123 & 173 & 178 & $\f_i \f_j \to \f_i \f_j X$ ($\Z \Z$ fusion) \\
124 & 174 & 179 & $\f_i \f_j \to \f_k \f_l X$ ($\W^+\W^-$ fusion) \\
121 & 181 & 186 & $\g \g \to \Q_k \Qbar_k X$ \\
122 & 182 & 187 & $\q_i \qbar_i \to \Q_k \Qbar_k X$ \\
\end{tabular} 
  
In {\Py}, the particle content of a two-Higgs-doublet scenario is 
included: two neutral scalar particles, 25 and 35, one pseudoscalar 
one, 36, and a charged doublet, $\pm 37$. (Of course, these particles 
may also be associated with corresponding Higgs states in larger 
multiplets.) By convention, we choose to call the lighter scalar 
Higgs $\hrm^0$ and the heavier $\H^0$. The pseudoscalar is called 
$\A^0$ and the charged $\H^{\pm}$. Charged-Higgs production is covered 
in section \ref{sss:chHclass}.
   
A number of $\hrm^0$ processes have been duplicated for $\H^0$ and 
$\A^0$. The correspondence between {\ISUB} numbers is shown in the table
above: the first column of {\ISUB} numbers corresponds to
$X = \hrm^0$, the second to $X = \H^0$, and the third to $X = \A^0$. 
Note that several of these processes are not expected to take 
place at all, owing to vanishing Born term couplings. We have still 
included them for flexibility in simulating arbitrary couplings at 
the Born or loop level, or for the case of mixing between the
scalar and pseudoscalar sectors.
  
A few Standard Model Higgs processes have no correspondence in the 
scheme above. These include
\begin{Itemize}
\item 5 and 8, which anyway have been superseded by 123 and 124; 
\item 71, 72, 73, 76 and 77, which deal with what happens if there 
is no light Higgs, and so is a scenario complementary to the one 
above, where several light Higgs bosons are assumed; and
\item 110, which is mainly of interest in Standard Model Higgs 
searches.  
\end{Itemize} 

The processes 111--113, 183--185 and 188--190 have only been worked 
out in full detail for the Standard Model Higgs case, and not when 
e.g. squark loop contributions need be considered. The approximate 
procedure outlined in subsection \ref{sss:lightSMHclass}, based on 
combining the kinematics shape from simple expressions in the 
$m_{\t} \to \infty$ limit with a normalization derived from the 
$\g\g \to X$ cross section, should therefore be viewed as a first 
ansatz only. In particular, it is not recommended to try the 
non-default \ttt{MSTP(38)=0} option, which is incorrect beyond the 
Standard Model.  
  
In processes 121, 122, 181, 182, 186 and 187 the recoiling heavy 
flavour is assumed to be top, which is the only one of interest in 
the Standard Model, and the one where the parton-distribution-function 
approach invoked in processes 3, 151 and 156 is least reliable. 
However, it is possible to change the quark flavour in 121 etc.; 
for each process {\ISUB} this flavour is given by \ttt{KFPR(ISUB,2)}. 
This may become relevant if couplings to $\b\bbar$ states are
enhanced, e.g.\ if $\tan\beta \gg 1$ in the MSSM. The matrix elements 
in this group are based on scalar Higgs couplings; differences for
a pseudoscalar Higgs remains to be worked out, but are proportional
to the heavy quark mass relative to other kinematic quantities.
  
By default, the $\hrm^0$ has the couplings of the Standard Model 
Higgs, while the $\H^0$ and $\A^0$ have couplings set in 
\ttt{PARU(171) - PARU(178)} and \ttt{PARU(181) - PARU(190)}, 
respectively. The default values for the $\H^0$ and $\A^0$ have no 
deep physics motivation, but are set just so that the program will 
not crash due to the absence of any couplings whatsoever. You 
should therefore set the above couplings to your desired values if 
you want to simulate either $\H^0$ or $\A^0$. Also the couplings 
of the $\hrm^0$ particle can be modified, in 
\ttt{PARU(161) - PARU(165)}, provided that \ttt{MSTP(4)} is set to 
1. 
  
For \ttt{MSTP(4)=2}, the mass of the $\hrm^0$ (in \ttt{PMAS(25,1)}) 
and the $\tan\beta$ value (in \ttt{PARU(141)}) are used to derive 
the masses of the other Higgs bosons, as well as all Higgs couplings. 
\ttt{PMAS(35,1) - PMAS(37,1)} and \ttt{PARU(161) - PARU(195)} are 
overwritten accordingly. The relations used are the ones of the 
Born-level MSSM \cite{Gun90}. Loop corrections to those
expressions have been calculated within specific supersymmetric 
scenarios, and are known to have a non-negligible effects on the 
resulting phenomenology. By switching on supersymmetry simulation
and setting parameters appropriately, one will gain access to these
mass formulae, see section~\ref{ss:susycode}.
 
Note that not all combinations of $m_{\hrm}$ and $\tan\beta$ are 
allowed; for \ttt{MSTP(4)=2} the requirement of a finite $\A^0$ mass 
imposes the constraint 
\begin{equation}
m_{\hrm} < m_{\Z} \, \frac{\tan^2\beta - 1}{\tan^2\beta + 1}, 
\end{equation}
or, equivalently, 
\begin{equation}
\tan^2\beta > \frac{m_{\Z} + m_{\hrm}}{m_{\Z} - m_{\hrm}}. 
\end{equation}
If this condition is not fulfilled, the program will crash.

A more realistic approach to the Higgs mass spectrum is to 
include radiative corrections to the Higgs potential. Such a 
machinery has never been implemented as such in {\Py}, but 
appears as part of the Supersymmetry framework described in 
subsection \ref{ss:susyclass}. At tree level, the
minimal set of inputs would be \ttt{IMSS(1)=1} to switch on
SUSY, \ttt{RMSS(5)} to set the $\tan\beta$ value 
(this overwrites the \ttt{PARU(141)} value when SUSY is 
switched on) and \ttt{RMSS(19)} to set $\A^0$ mass. 
However, the significant radiative corrections depend
on the properties of all particles that couple to the
Higgs boson, and the user may want to change the default values
of the relevant \ttt{RMSS} inputs.  In practice, the most
important are those related indirectly to the physical masses of
the third generation supersymmetric quarks and the Higgsino:  
\ttt{RMSS(10)} to set the left--handed doublet SUSY mass 
parameter, \ttt{RMSS(11)} to set the right stop mass parameter, 
\ttt{RMSS(12)} to set the right sbottom mass parameter, 
\ttt{RMSS(4)} to set the Higgsino mass and a portion of the 
squark mixing, and \ttt{RMSS(16)} and \ttt{RMSS(17)} to set the 
stop and bottom trilinear couplings, respectively, which 
specifies the remainder of the squark mixing. From these inputs, 
the Higgs masses and couplings would be derived. Note that 
switching on SUSY also implies that Supersymmetric decays 
of the Higgs particles become possible if kinematically allowed. 
If you do not want this to happen, you may want to increase the
SUSY mass parameters. (Use \ttt{CALL PYSTAT(2)} after 
initialization to see the list of branching ratios.)
 
Pair production of Higgs states may be a relevant source, see
section~\ref{sss:Hpairclass} below.

Finally, heavier Higgs bosons may decay into lighter ones, if 
kinematically allowed, in processes like $\A^0 \to \Z^0 \hrm^0$ or 
$\H^+ \to \W^+ \hrm^0$. Such modes are included as part of the
general mixture of decay channels, but they can be enhanced if 
the uninteresting channels are switched off.

\subsubsection{Charged Higgs sector}
\label{sss:chHclass}

\ttt{MSEL} = 23 \\
{\ISUB} = 
\begin{tabular}[t]{rl@{\protect\rule{0mm}{\tablinsep}}}
143 & $\f_i \fbar_j \to \H^+$ \\
161 & $\f_i \g \to \f_k \H^+$ \\
\end{tabular}

A charged Higgs doublet, $\H^{\pm}$, is included in the program. 
This doublet may be the one predicted in the MSSM scenario,
see section \ref{sss:extneutHclass}, or in any other scenario.
The $\tan\beta$ parameter, which is relevant also 
for charged Higgs couplings, is set via \ttt{PARU(141)} or,
if \ttt{Susy} is switched on, via \ttt{RMSS(5)}. 
  
The basic subprocess for charged Higgs production in hadron
colliders is {\ISUB} = 143. However, this process is dominated 
by $\t \bbar \to \H^+$, and so depends on the choice of $\t$ 
parton distribution, if at all present. A better representation 
is provided by subprocess 161, $\f \g \to \f' \H^+$; i.e.\ actually 
$\bbar \g \to \tbar \H^+$. It is therefore recommended to use 
161 and not 143; to use both would be double-counting. 

Pair production of Higgs states may be a relevant source, see
section~\ref{sss:Hpairclass} below.

A major potential source of charged Higgs production is top decay.
It is possible to switch on the decay channel $\t \to \b \H^+$. Top 
will then decay to $\H^+$ a fraction of the time, whichever way it is 
produced. The branching ratio is automatically calculated, based
on the $\tan\beta$ value and masses.
It is possible to only have the $\H^+$ decay mode switched on,  
in which case the cross section is reduced accordingly. 

\subsubsection{Higgs pairs}
\label{sss:Hpairclass}

{\ISUB} = 
\begin{tabular}[t]{rl@{\protect\rule{0mm}{\tablinsep}}}
(141) & $\f_i \fbar_i \to \gamma/\Z^0/\Z'^0$ \\
297 & $\f_i \fbar_j \to \H^{\pm} \hrm^0$ \\
298 & $\f_i \fbar_j \to \H^{\pm} \H^0$ \\
299 & $\f_i \fbar_i \to \A \hrm^0$ \\
300 & $\f_i \fbar_i \to \A \H^0$ \\
301 & $\f_i \fbar_i \to \H^+ \H^-$ \\
\end{tabular}

The subprocesses 297--301 give the production of a pair of Higgs 
bosons via the $s$-channel exchange of a 
$\gamma^*/\Z^0$ or a $\W^{\pm}$ state.

Note that Higgs pair production is still possible through
subprocess 141, as part of the decay of a generic combination of
 $\gamma^*/\Z^0/\Z'^0$. Thus it can be used to simulate 
$\Z^0 \to \hrm^0 \A^0$ and $\Z^0 \to \H^0 \A^0$ for associated 
neutral Higgs production. The fact that we here make use of the 
$\Z'^0$ can easily be discounted, either by letting the relevant 
couplings vanish, or by the option \ttt{MSTP(44)=4}.

Similarly the decay $\gamma^*/\Z^0/\Z'^0 \to \H^+ \H^-$
allows the production of a pair of charged Higgs particles.
This process is especially important in $\ee$ colliders.
The coupling of the $\gamma^*$ to $\H^+ \H^-$ is determined by
the charge alone (neglecting loop effects), while the $\Z^0$ 
coupling is regulated by \ttt{PARU(142)}, and that of the 
$\Z'^0$ by \ttt{PARU(143)}. The $\Z'^0$ piece can be switched off, 
e.g.\ by \ttt{MSTP(44)=4}. An ordinary $\Z^0$, i.e.\ particle code 23, 
cannot be made to decay into a Higgs pair, however.

The advantage of the explicit pair production processes is the
correct implementation of the pair threshold.

\subsection{Non-Standard Physics}

The number of possible non-Standard Model scenarios is essentially
infinite, but many of the studied scenarios still share a lot of
aspects. For instance, new $\W'$ and $\Z'$ gauge bosons can
arise in a number of different ways. Therefore it still makes sense
to try to cover a few basic classes of particles, with enough
freedom in couplings that many kinds of detailed scenarios can be
accommodated by suitable parameter choices. We have already seen one
example of this, in the extended Higgs sector above. In this section
a few other kinds of non-standard generic physics are discussed.
Supersymmetry is covered separately in the following section,
since it is such a large sector by itself.

\subsubsection{Fourth-generation fermions}
  
\ttt{MSEL} = 7, 8, 37, 38  \\
{\ISUB} = 
\begin{tabular}[t]{rl@{\protect\rule{0mm}{\tablinsep}}}
 1 & $\f_i \fbar_i \to \gammaZ$ \\
 2 & $\f_i \fbar_j \to \W^+$ \\
81 & $\q_i \qbar_i \to \Q_k \Qbar_k$  \\
82 & $\g \g \to \Q_k \Qbar_k$  \\
83 & $\q_i \f_j \to \Q_k \f_l$ \\
84 & $\g \gamma \to \Q_k \Qbar_k$ \\
85 & $\gamma \gamma \to \F_k \Fbar_k$ \\
141 & $\f_i \fbar_i \to \gamma/\Z^0/\Z'^0$ \\
142 & $\f_i \fbar_j \to \W'^+$ \\
\end{tabular}

The prospects of a fourth generation currently seem rather dim, but 
the appropriate flavour content is still found in the program. In 
fact, the fourth generation is included on an equal basis with the
first three, provided \ttt{MSTP(1)=4}. Also processes other than 
the ones above can therefore be used, e.g.\ all other processes with
gauge bosons, including non-standard ones such as the $\Z'^0$. We
therefore do not repeat the descriptions found elsewhere, e.g.\ how
to set only the desired flavour in processes 81--85. Note that it
may be convenient to set \ttt{CKIN(1)} and other cuts such that the
mass of produced gauge bosons is enough for the wanted particle
production --- in principle the program will cope even without that,
but possibly at the expense of very slow execution.
  
\subsubsection{New gauge bosons} 
  
\ttt{MSEL} = 21, 22, 24 \\
{\ISUB} = 
\begin{tabular}[t]{rl@{\protect\rule{0mm}{\tablinsep}}}
141 & $\f_i \fbar_i \to \gamma/\Z^0/\Z'^0$ \\
142 & $\f_i \fbar_j \to \W'^+$ \\
144 & $\f_i \fbar_j \to \R$ \\
\end{tabular}
  
The $\Z'^0$ of subprocess 141 contains the full $\gamma^*/\Z^0/\Z'^0$ 
interference structure for couplings to fermion pairs. With 
\ttt{MSTP(44)} it is possible to pick only a subset, e.g.\ only the
pure $\Z'^0$ piece. The couplings of the $\Z'^0$ to quarks and leptons 
in the first generation can be set via \ttt{PARU(121) - PARU(128)},
in the second via \ttt{PARJ(180) - PARJ(187)} and in the third via
\ttt{PARJ(188) - PARJ(195)}. The eight numbers 
correspond to the vector and axial couplings of down-type quarks,
up-type quarks, leptons and neutrinos, respectively. The default 
corresponds to the same couplings as that of the Standard Model
$\Z^0$, with axial couplings $a_{\f} = \pm 1$ and vector couplings
$v_{\f} = a_{\f} - 4 e_{\f} \ssintw$. This implies a resonance width
that increases linearly with the mass. By a suitable choice of the
parameters, it is possible to simulate just about any imaginable
$\Z'^0$ scenario, with full interference effects in cross sections
and decay angular distributions. Note that also the possibility of
a generation dependence has been included for the $\Z'^0$, which is
normally not the case. 

The coupling to the decay channel $\Z'^0 \to \W^+ \W^-$ is regulated
by \ttt{PARU(129) - PARU(130)}. The former gives the strength of the
coupling, which determines the rate. The default, \ttt{PARU(129)=1.},
corresponds to the `extended gauge model' of \cite{Alt89}, wherein
the $\Z^0 \to \W^+ \W^-$ coupling is used, scaled down by a factor
$m_{\W}^2/m_{\Z'}^2$, to give a $\Z'^0$ partial width into this channel
that again increases linearly. If this factor is cancelled, by having
\ttt{PARU(129)} proportional to $m_{\Z'}^2/m_{\W}^2$, one obtains a
partial width that goes like the fifth power of the $\Z'^0$ mass, the
`reference model' of \cite{Alt89}. In the decay angular distribution
one could imagine a much richer structure than is given by the one
parameter \ttt{PARU(130)}.

Other decay modes include $\Z'^0 \to \Z^0 \hrm^0$, predicted in
left--right symmetric models (see \ttt{PARU(145)} and ref. 
\cite{Coc91}), and a number of other Higgs decay channels, see 
sections \ref{sss:extneutHclass} and \ref{sss:chHclass}.
  
The $\W'^{\pm}$ of subprocess 142 so far does not contain 
interference with the Standard Model $\W^{\pm}$ --- in practice this 
should not be a major limitation. The couplings of the $\W'$ to 
quarks and leptons are set via \ttt{PARU(131) - PARU(134)}.
Again one may set vector and axial couplings freely, separately
for the $\q \qbar'$ and the $\ell \nu_{\ell}$ decay channels.
The defaults correspond to the $V-A$ structure of the Standard Model
$\W$, but can be changed to simulate a wide selection of models.
One possible limitation is that the same Cabibbo--Kobayashi--Maskawa
quark mixing matrix is assumed as for the standard $\W$.

The coupling $\W' \to \Z^0 \W$ can be set via 
\ttt{PARU(135) - PARU(136)}. Further comments on this channel as for 
$\Z'$; in particular, default couplings again agree with 
the `extended gauge model' of \cite{Alt89}. A $\W' \to \W \hrm^0$
channel is also included, in analogy with the $\Z'^0 \to \Z^0 \hrm^0$
one, see \ttt{PARU(146)}.
  
The $\R$ boson (particle code 41) of subprocess 144 represents one 
possible scenario for a horizontal gauge boson, i.e.\ a gauge boson 
that couples between the generations, inducing processes like 
$\s \dbar \to \R^0 \to \mu^- \e^+$. Experimental limits on 
flavour-changing neutral currents forces such a boson to be fairly 
heavy. The model implemented is the one described in \cite{Ben85a}. 

A further example of new gauge groups follows right after this. 
 
\subsubsection{Left--Right Symmetry and Doubly Charged Higgs Bosons}
\label{sss:LRDCHclass} 

{\ISUB} = 
\begin{tabular}[t]{rl@{\protect\rule{0mm}{\tablinsep}}}
341 & $\ell_i \ell_j \to \H_L^{\pm\pm}$ \\
342 & $\ell_i \ell_j \to \H_R^{\pm\pm}$ \\
343 & $\ell_i \gamma \to \H_L^{\pm\pm} \e^{\mp}$ \\
344 & $\ell_i \gamma \to \H_R^{\pm\pm} \e^{\mp}$ \\
345 & $\ell_i \gamma \to \H_L^{\pm\pm} \mu^{\mp}$ \\
346 & $\ell_i \gamma \to \H_R^{\pm\pm} \mu^{\mp}$ \\
347 & $\ell_i \gamma \to \H_L^{\pm\pm} \tau^{\mp}$ \\
348 & $\ell_i \gamma \to \H_R^{\pm\pm} \tau^{\mp}$ \\
349 & $\f_i \fbar_i \to \H_L^{++} \H_L^{--}$ \\
350 & $\f_i \fbar_i \to \H_R^{++} \H_R^{--}$ \\
351 & $\f_i \f_j \to \f_k f_l \H_L^{\pm\pm}$ ($\W\W$ fusion) \\
352 & $\f_i \f_j \to \f_k f_l \H_R^{\pm\pm}$ ($\W\W$ fusion) \\
353 & $\f_i \fbar_i \to \Z_R^0$ \\
354 & $\f_i \fbar_i \to \W_R^{\pm}$ \\
\end{tabular}

At current energies, the world is lefthanded, i.e.\ the Standard
Model contains an {\bf SU(2)}$_L$ group. Left--right symmetry at 
some larger scale implies the need for an {\bf SU(2)}$_R$ group.
Thus the particle content is expanded by righthanded $\Z_R^0$ and
$\W_R^{\pm}$ and righthanded neutrinos. The Higgs fields have to be 
in a triplet representation, leading to doubly-charged Higgs particles, 
one set for each of the two {\bf SU(2)} groups. Also the number of 
neutral and singly-charged Higgs states is increased relative to the 
Standard Model, but a search for the lowest-lying states of this kind 
is no different from e.g.\ the freedom already accorded by the MSSM 
Higgs scenarios. 

{\Py} implements the scenario of \cite{Hui97}. The expanded particle 
content with default masses is:\\
\begin{tabular}{llr}
{\KF} &  name       &  $m$ (GeV)\\
9900012 &  $\nu_{R\e}$  &    500  \\
9900014 &  $\nu_{R\mu}$ &    500  \\
9900016 &  $\nu_{R\tau}$&    500  \\
9900023 &  $\Z_R^0$     &   1200  \\
9900024 &  $\W_R^+$     &    750  \\
9900041 &  $\H_L^{++}$  &    200  \\    
9900042 &  $\H_R^{++}$  &    200  \\         
\end{tabular}\\
The main decay modes implemented are\\
$\H_L^{++} \to \W_L^+ \W_L^+, \ell_i^+ \ell_j^+$ ($i, j$ generation 
             indices); and\\
$\H_R^{++} \to \W_R^+ \W_R^+, \ell_i^+ \ell_j^+$.\\
The physics parameters of the scenario are found in
\ttt{PARP(181) - PARP(192)}. 

The $W_R^{\pm}$ has been implemented as a simple copy of the 
ordinary $\W^{\pm}$, with the exception that it couple to 
righthanded neutrinos instead of the ordinary lefthanded ones. 
Thus the standard CKM matrix is used in the quark sector, and the 
same vector and axial coupling strengths, leaving only the mass as
free parameter. The $\Z_R^0$ implementation (without interference 
with $\gamma$ or the ordinary $\Z^0$) allows decays both to left-
and righthanded neutrinos, as well as other fermions, according to 
one specific model ansatz \cite{Fer00}. Obviously both the $W_R^{\pm}$ 
and the $\Z_R^0$ descriptions are  likely to be simplifications,
but provide a starting point.

The righthanded neutrinos can be allowed to decay further 
\cite{Riz81,Fer00}. Assuming them to have a mass below that of 
$\W_R^+$, they decay to three-body states via a virtual $\W_R^+$,
$\nu_{R\ell} \to \ell^+ \f \fbar'$ and 
$\nu_{R\ell} \to \ell^- \fbar \f'$, where both choices are allowed
owing to the Majorana character of the neutrinos. If there is
a significant mass splitting, also sequential decays 
$\nu_{R\ell} \to \ell^{\pm} {\ell'}^{\mp} {\nu'}_{R\ell}$
are allowed. Currently the decays are isotropic in phase space.
If the neutrino masses are close to or above the $\W_R$ ones, this
description has to be substituted by a sequential decay via
a real $\W_R$ (not implemented, but actually simpler to do than the
one here). 

\subsubsection{Leptoquarks}
\label{sss:LQclass} 
  
\ttt{MSEL} = 25 \\
{\ISUB} = 
\begin{tabular}[t]{rl@{\protect\rule{0mm}{\tablinsep}}}
145 & $\q_i \ell_j \to \L_{\Q}$ \\
162 & $\q \g \to \ell \L_{\Q}$ \\
163 & $\g \g \to \L_{\Q} \br{\L}_{\Q}$ \\
164 & $\q_i \qbar_i \to \L_{\Q} \br{\L}_{\Q}$ \\
\end{tabular}
  
Several processes that can generate a leptoquark have been included. 
Currently only one leptoquark has been implemented, as particle 42,
denoted $\L_{\Q}$. The leptoquark is assumed to carry specific quark 
and lepton quantum numbers, by default $\u$ quark plus electron. 
These flavour numbers are conserved, i.e.\ a process such as 
$\u \e^- \to \L_{\Q} \to \d \nu_{\e}$ is not allowed. This may be a 
bit restrictive, but it represents one of many leptoquark possibilities.
The spin of the leptoquark is assumed to be zero, i.e.\ its decay is
isotropic. 
  
Although only one leptoquark is implemented, its flavours may be 
changed arbitrarily to study the different possibilities. The 
flavours of the leptoquark are defined by the quark and lepton 
flavours in the decay mode list. Since only one decay channel is 
allowed, this means that the quark flavour is stored in 
\ttt{KFDP(MDCY(42,2),1)} and the lepton one in 
\ttt{KFDP(MDCY(42,2),2)}. The former must always be a quark, while 
the latter could be a lepton or an antilepton; a charge-conjugate 
partner is automatically defined by the program. At initialization, 
the charge is recalculated as a function of the flavours defined; 
also the leptoquark name is redefined to be of the type 
\ttt{'LQ\_(q)(l)'}, where actual quark \ttt{(q)} and lepton \ttt{(l)} 
flavours are displayed. 
  
The $\L_{\Q} \to \q \ell$ vertex contains an undetermined Yukawa 
coupling strength, which affects both the width of the leptoquark 
and the cross section for many of the production graphs. This 
strength may be changed in \ttt{PARU(151)}. The definition of 
\ttt{PARU(151)} corresponds to the $k$ factor of \cite{Hew88}, i.e.\ 
to $\lambda^2/(4\pi\alphaem)$, where $\lambda$ is the Yukawa 
coupling strength of \cite{Wud86}. Note that \ttt{PARU(151)} 
is thus quadratic in the coupling. 
  
The leptoquark is likely to be fairly long-lived, in which case it 
has time to fragment into a mesonic- or baryonic-type state, which 
would decay later on. This is a bit tedious to handle; therefore 
the leptoquark is always assumed to decay before fragmentation. 
This may give some imperfections in the event 
generation, but should not be off by much in the final analysis
\cite{Fri97}. 
  
Inside the program, the leptoquark is treated as a resonance. 
Since it carries colour, some extra care is required. 
In particular, it is not allowed to put the leptoquark stable, 
by modifying either \ttt{MDCY(42,1)} or \ttt{MSTP(41)}: then the 
leptoquark would be handed undecayed to {\Py}, which would try 
to fragment it (as it does with any other coloured object), and 
most likely crash. 
  
\subsubsection{Compositeness and anomalous couplings} 
\label{sss:ancoupclass}
  
{\ISUB} = 
\begin{tabular}[t]{rl@{\protect\rule{0mm}{\tablinsep}}}
 20 & $\f_i \fbar_j \to \gamma \W^+$  \\
165 & $\f_i \fbar_i \to \f_k \fbar_k$ (via $\gammaZ$) \\
166 & $\f_i \fbar_j \to \f_k \fbar_l$ (via $\W^{\pm}$) \\
381 & $\f_i \f_j \to \f_i \f_j$ \\
382 & $\f_i \fbar_i \to \f_k \fbar_k$  \\
\end{tabular}
  
Some processes have been set up to allow anomalous coupling to be 
introduced, in addition to the Standard Model ones. These can be 
switched on by \ttt{ITCM(5)}$\geq 1$; the default \ttt{ITCM(5)=0} 
corresponds to the Standard Model behaviour. 
  
In processes 381 and 382, the quark substructure is included in the 
left--left isoscalar model \cite{Eic84,Chi90} for 
\ttt{ITCM(5)=1}, with compositeness 
scale $\Lambda$ given in \ttt{RTCM(41)} (default 1000~GeV) and sign 
$\eta$ of interference term in \ttt{RTCM(42)} (default $+1$; only 
other alternative $-1$). The above model assumes that only $\u$ and 
$\d$ quarks are composite (at least at the scale studied); with 
\ttt{ITCM(5)=2} compositeness terms are included in the 
interactions between all quarks. When \ttt{ITCM(5)=0}, the two 
processes are equivalent with 11 and 12. A consistent set of high-$\pT$ 
jet production processes in compositeness scenarios is thus obtained 
by combining 381 and 382 with 13, 28, 53 and 68.
  
The processes 165 and 166 are basically equivalent to 1 and 2, i.e.\ 
$\gammaZ$ and $\W^{\pm}$ exchange, respectively, but a bit less 
fancy (no mass-dependent width etc.). The reason for this duplication 
is that the resonance treatment formalism of processes 1 and 2 could
not easily be extended to include other than $s$-channel graphs.
In processes 165 and 166, only one final-state flavour 
is generated at the time; this flavour should be set in 
\ttt{KFPR(165,1)} and \ttt{KFPR(166,1)}, respectively. For process 
166 one gives the down-type flavour, and the program will associate 
the up-type flavour of the same generation. Defaults are 11 in both 
cases, i.e.\ $\ee$ and $\e^+ \nu_{\e}$ ($\e^- \br{\nu}_{\e}$) final 
states. While \ttt{ITCM(5)=0} gives the Standard Model results, 
\ttt{ITCM(5)=1} contains the left--left isoscalar model (which 
does not affect process 166), and \ttt{ITCM(5)=3} the 
helicity-non-conserving model (which affects both) \cite{Eic84,Lan91}. 
Both models above assume that only $\u$ and $\d$ quarks are 
composite; with \ttt{ITCM(5)=} 2 or 4, respectively, contact terms 
are included for all quarks in the initial state. Parameters are 
\ttt{RTCM(41)} and \ttt{RTCM(42)}, as above. 
  
Note that processes 165 and 166 are book-kept as $2 \to 2$ processes, 
while 1 and 2 are $2 \to 1$ ones. This means that the default $\Q^2$
scale in parton distributions is $\pT^2$ for the former and $\hat{s}$ 
for the latter. To make contact between the two, it is recommended to 
set \ttt{MSTP(32)=4}, so as to use $\hat{s}$ as scale also for 
processes 165 and 166. 
  
In process 20, for $\W \gamma$ pair production, it is possible to set 
an anomalous magnetic moment for the $\W$ in \ttt{RTCM(46)}
($= \eta = \kappa-1$; where $\kappa = 1$ is the Standard Model 
value). The production process is affected according to the formulae 
of \cite{Sam91}, while $\W$ decay currently remains 
unaffected. It is necessary to set \ttt{ITCM(5)=1} to enable this 
extension. 
  
\subsubsection{Excited fermions}
\label{sss:qlstarclass} 
  
{\ISUB} = 
\begin{tabular}[t]{rl@{\protect\rule{0mm}{\tablinsep}}}
146 & $\e \gamma \to \e^*$ \\
147 & $\d \g \to \d^*$ \\
148 & $\u \g \to \u^*$ \\
167 & $\q_i \q_j \to \q_k \d^*$ \\
168 & $\q_i \q_j \to \q_k \u^*$ \\
169 & $\q_i \qbar_i \to \e^{\pm} \e^{*\mp}$ \\
\end{tabular}
  
Compositeness scenarios may also give rise to sharp resonances of 
excited quarks and leptons. An excited copy of the first generation
is implemented, consisting of spin $1/2$ particles $\d^*$ (code 4000001), 
$\u^*$ (4000002), $\e^*$ (4000011) and $\nu^*_{\e}$ (4000012). 
  
The current implementation contains gauge interaction production
by quark--gluon fusion (processes 147 and 148) or lepton--photon fusion
(process 146) and contact interaction production by quark--quark or 
quark--antiquark scattering (processes 167--169) . The couplings $f$, 
$f'$ and $f_s$ to the {\bf SU(2)}, {\bf U(1)} and {\bf SU(3)} groups 
are stored in \ttt{RTCM(43) - RTCM(45)}, the scale parameter 
$\Lambda$ in \ttt{RTCM(41)}; you are also expected to change the 
$\f^*$ masses in accordance with what is desired --- see \cite{Bau90} 
for details on conventions. Decay processes are of 
the types $\q^* \to \q \g$, $\q^* \to \q \gamma$, 
$\q^* \to \q \Z^0$ or $\q^* \to \q' \W^{\pm}$, with the latter 
three (two) available also for $\e^*$ ($\nu^*_{\e}$).  
A non-trivial angular dependence is included in the $\q^*$
decay for processes 146--148, but has not been included for 
processes 167--169.

\subsubsection{Technicolor}
\label{sss:technicolorclass}
  
\ttt{MSEL} = 50, 51 \\
{\ISUB} = 
\begin{tabular}[t]{rl@{\protect\rule{0mm}{\tablinsep}}}
149 & $\g \g \to \eta_{\mrm{tc}}$ (obsolete)  \\
191 & $\f_i \fbar_i \to \rho_{\mrm{tc}}^0$  (obsolete) \\
192 & $\f_i \fbar_j \to \rho_{\mrm{tc}}^+$ (obsolete)   \\
193 & $\f_i \fbar_i \to \omega_{\mrm{tc}}^0$ (obsolete)  \\
194 & $\f_i \fbar_i \to \f_k \fbar_k$   \\
195 & $\f_i \fbar_j \to \f_k \fbar_l$   \\
361 & $\f_i \fbar_i \to \W^+_{\mrm{L}} \W^-_{\mrm{L}} $  \\
362 & $\f_i \fbar_i \to \W^{\pm}_{\mrm{L}} \pi^{\mp}_{\mrm{tc}}$   \\
363 & $\f_i \fbar_i \to \pi^+_{\mrm{tc}} \pi^-_{\mrm{tc}}$   \\
364 & $\f_i \fbar_i \to \gamma \pi^0_{\mrm{tc}} $   \\
365 & $\f_i \fbar_i \to \gamma {\pi'}^0_{\mrm{tc}} $   \\
366 & $\f_i \fbar_i \to \Z^0 \pi^0_{\mrm{tc}} $   \\
367 & $\f_i \fbar_i \to \Z^0 {\pi'}^0_{\mrm{tc}} $   \\
368 & $\f_i \fbar_i \to \W^{\pm} \pi^{\mp}_{\mrm{tc}}$ \\
370 & $\f_i \fbar_j \to \W^{\pm}_{\mrm{L}} \Z^0_{\mrm{L}} $  \\
371 & $\f_i \fbar_j \to \W^{\pm}_{\mrm{L}} \pi^0_{\mrm{tc}}$   \\
372 & $\f_i \fbar_j \to \pi^{\pm}_{\mrm{tc}} \Z^0_{\mrm{L}} $ \\
373 & $\f_i \fbar_j \to \pi^{\pm}_{\mrm{tc}} \pi^0_{\mrm{tc}} $   \\
374 & $\f_i \fbar_j \to \gamma \pi^{\pm}_{\mrm{tc}} $   \\
375 & $\f_i \fbar_j \to \Z^0 \pi^{\pm}_{\mrm{tc}} $   \\
376 & $\f_i \fbar_j \to \W^{\pm} \pi^0_{\mrm{tc}} $   \\
377 & $\f_i \fbar_j \to \W^{\pm} {\pi'}^0_{\mrm{tc}}$ \\
381 & $\q_i \q_j \to \q_i \q_j$  \\
382 & $\q_i \qbar_i \to \q_k \qbar_k$ \\
383 & $\q_i \qbar_i \to \g \g$  \\
384 & $\f_i \g \to \f_i \g$  \\
385 & $\g \g \to \q_k \qbar_k$   \\
386 & $\g \g \to \g \g$ \\
387 & $\f_i \fbar_i \to \Q_k \Qbar_k$  \\
388 & $\g \g \to \Q_k \Qbar_k$  \\
\end{tabular}

Technicolor (TC) is an alternative way to manifest 
the Higgs mechanism for giving masses to the $\W$ and $\Z$ bosons
using strong 
dynamics instead of weakly--coupled fundamental scalars.
In TC, the breaking of a chiral 
symmetry in a new, strongly interacting gauge theory generates the 
Goldstone bosons necessary for electroweak symmetry breaking. 
Thus three of the technipions assume the r\^ole of the longitudinal 
components of the $\W$ and $\Z$ bosons, but other states can remain 
as separate particles depending on the gauge group: 
technipions ($\pi_{\mrm{tc}}$), technirhos 
($\rho_{\mrm{tc}}$), techniomegas ($\omega_{\mrm{tc}}$), etc. 

No fully--realistic model of strong electroweak--symmetry breaking
has been found so far, and some of the assumptions and simplifications
used in model--building may need to be discarded in the future.
The processes represented here correspond to 
several generations of development. Processes 149, 191, 192 and 193
should be considered obsolete and superseded by the other
processes 194, 195 and 361--377.  
The former processes are kept for cross--checks and
backward--compatibility.
In section \ref{sss:heavySMHclass} it is discussed how processes
71--77 can be used to simulate a scenario 
with techni-$\rho$ resonances in longitudinal gauge boson scattering.

Process 149 describes the production of
a spin--0 techni-$\eta$ meson (particle code 
{\KF} = 3000331), which is an electroweak singlet and a QCD colour octet.
It is one of the possible techni-$\pi$ particles; the name 
``techni-$\eta$'' is not used universally in the literature.
The techni-$\eta$ couples to ordinary fermions proportional to fermion
mass.  The dominant decay mode is therefore $\t \tbar$, if kinematically
allowed.  An effective $\g\g$--coupling arises through an anomaly,
and is roughly comparable in size with
that to $\b \bbar$. 
Techni-$\eta$ production at hadron colliders is therefore 
predominantly through $\g \g$ fusion, as implemented in process 149.
In topcolor--assisted technicolor (discussed below), particles
like the techni-$\eta$ should not have a predominant coupling
to $\t$ quarks.  In this sense, the process is considered obsolete.

(The following discussion borrows liberally from the introduction
to Ref.~\cite{Lan99a} with the author's permission.)
Modern technicolor models of dynamical electroweak symmetry breaking require
walking technicolor~\cite{Hol81} to prevent 
large flavor-changing neutral currents
and the assistance of topcolor (TC2) interactions that are strong near
1~TeV~\cite{Nam88,Hil95,Lan95} to provide the large mass of the
top quark. Both additions to the basic technicolor scenario~\cite{Wei79,Eic80}
tend to require a large number $N_D$ of technifermion doublets
to make the $\beta$--function of walking technicolor
small. They are needed in TC2 to generate
the hard masses of quarks and leptons, to induce the right mixing between
heavy and light quarks, and to break topcolor symmetry down to ordinary
color. A large number of
techni-doublets implies a relatively low technihadron 
mass scale \cite{Lan89,Eic96}, set by the
technipion decay constant $F_T \simeq F_\pi/\sqrt{N_D}$, where 
$F_\pi = 246$ GeV. 

The model adopted in {\sc Pythia} is the ``Technicolor
Straw Man Model'' (TCSM) \cite{Lan99a,Lan02a}. 
The TCSM describes the phenomenology
of color--singlet vector and pseudoscalar technimesons and their
interactions with SM particles.
These technimesons are expected to be the lowest-lying
bound states of the lightest technifermion doublet, $(T_U, T_D)$, 
with components that
transform under technicolor ${\bf SU(\Ntc)}$ as
fundamentals, but are QCD singlets; they have electric charges $Q_U$ and 
$Q_D=Q_U-1$. 
The vector technimesons form a spin-one isotriplet $\tro^{\pm,0}$ and an
isosinglet $\tom$. Since techni-isospin is likely to be a good approximate
symmetry, $\tro$ and $\tom$ should be approximately 
mass--degenerate. The pseudoscalars,
or technipions, also comprise an isotriplet $\Pi_\mrm{tc}^{\pm,0}$ 
and an isosinglet
$\Pi_\mrm{tc}^{0 \prime}$. However, these are not mass eigenstates. In this 
model, they are simple, two-state mixtures of the longitudinal
weak bosons $W_L^\pm$, $Z_L^0$ --- the true Goldstone bosons of dynamical
electroweak symmetry breaking in the limit that the 
{$\bf SU(2) \otimes U(1)$} couplings $g,g'$ vanish --- and mass-eigenstate 
pseudo-Goldstone technipions $\tpi^\pm, \tpiz$:
\be\label{eq:pistates}
 \vert\Pi_\mrm{tc}\rangle = \sin\chi \ts \vert
W_L\rangle + \cos\chi \ts \vert\tpi\rangle\ts;\,
\vert\Pi_\mrm{tc}^{0 \prime} \rangle
= \cos\chipr \ts \vert\tpipr\rangle\ + \cdots,
\end{equation}
where $\sin\chi = F_T/F_\pi \ll 1$, $\chi'$ is another
mixing angle and the ellipsis refer to other technipions needed to 
eliminate the TC anomaly from the $\Pi_\mrm{tc}^{0 \prime}$ chiral 
current. These massive technipions are also expected to be approximately
degenerate.  However, 
there may be appreciable $\tpiz$--$\tpipr$ mixing \cite{Eic96}. If that
happens, the lightest neutral technipions are ideally-mixed $\bar T_U T_U$ 
and $\bar T_D T_D$ bound states.  To simulate this effect, there are
separate factors $C_{\tpiz\to \g\g}$ and $C_{\tpipr\to \g\g}$ to weight
the $\tpi$ and $\tpip$ partial widths for $\g\g$ decays.

Technipion decays are induced mainly by extended technicolor (ETC)
interactions which couple them to quarks and leptons~\cite{Eic80}. 
These couplings are proportional to fermion mass, except for
the top quark, which has most of its mass generation through
TC2 interactions.  Thus,
there is no great
preference for $\tpi$ to decay to top quarks nor for top quarks to decay into
them. Also, because of anomaly cancellation, the constituents of the
isosinglet technipion $\tpipr$ may include colored technifermions as well as
color-singlets, and it decays into a pair of gluons as well as heavy
quarks.  The relevant technipion
decay modes are $\tpip \ra \t \bar \b, \c \bar \b, \u \bar \b$,
$\c \bar \s$, $\c \bar \d$ and $\tau^+ \nu_\tau$; 
$\tpiz \ra \t \bar\t, \b \bar \b$, $\c\bar\c$, and $\tau^+\tau^-$; and 
$\tpipr \ra \g\g$, $\t \bar\t, \b \bar \b$, $\c \bar \c$, and
$\tau^+\tau^-$.  In the numerical evaluation of these widths, the
running mass (see \ttt{PYMRUN}) is used, and all fermion pairs
are considered as final states.  
The decay $\tpip\to \W^+\b\bar\b$ is also included,
with the final state kinematics distributed according to phase
space (i.e. not weighted by the squared matrix element).
The $\tpi$ couplings to fermions can be weighted by parameters
$C_\c$, $C_\b$, $C_\t$ and $C_\tau$ depending on the heaviest
quark involved in the decay.

In the limit of vanishing gauge couplings $g,g' = 0$, 
the $\tro$ and $\tom$ coupling to technipions are:
\bea\label{eq:vt_decays}
\tro &\ra& \Pi_\mrm{tc} \Pi_\mrm{tc} = \cos^2 \chi\ts (\tpi\tpi) + 
2\sin\chi\ts\cos\chi
\ts (\W_L\tpi) + \sin^2 \chi \ts (\W_L \W_L) \ts; \nn \\
\tom &\ra& \Pi_\mrm{tc} \Pi_\mrm{tc} \Pi_\mrm{tc} = 
\cos^3 \chi \ts (\tpi\tpi\tpi) + \cdots \ts.
\eea
The $\tro\to\tpi\tpi$ decay amplitude, then, is given simply by
\be\label{eq:rhopipi}
\CM(\tro(q) \ra \pi_A(p_1) \pi_B(p_2)) = g_{\tro} \ts \CC_{AB}
\ts \epsilon(q)\cdot(p_1 - p_2) \ts,
\end{equation}
where the technirho coupling
$\atro \equiv g_{\tro}^2/4\pi = 2.91(3/\Ntc)$ is scaled naively from
QCD ($\Ntc=4$ by default) and $\CC_{AB} = \cos^2\chi$ for $\tpi\tpi$,
$\sin\chi \cos\chi$ for $\tpi \W_L$, and $\sin^2\chi$ for $\W_L \W_L$.

Walking technicolor enhancements of technipion masses are assumed to close 
off the channel $\tom \ra \tpi\tpi\tpi$ (which is not included) and
to kinematically suppress the channels $\tro \ra \tpi\tpi$  and the
isospin-violating $\tom \ra \tpi\tpi$ (which are allowed with appropriate
choices of mass parameters). 
The rates for the isospin-violating decays $\tom \ra \pi^+_A \pi^-_B =
\W^+_L \W^-_L$, $\W^\pm_L \tpimp$, $\tpip \tpim$ are given by
$\Gamma(\tom \ra \pi^+_A \pi^-_B) = \vert\epsilon_{\rho\omega}\vert^2 \ts
\Gamma(\troz \ra \pi^+_A \pi^-_B)$
where $\epsilon_{\rho\omega}$ is the isospin-violating $\tro$-$\tom$ mixing.
Taking the value $5\%$ in analogy with QCD,
this decay mode is also dynamically suppressed (but is included).
While a light technirho can 
decay to $\W_L \tpi$ or $\W_L \W_L$ through TC dynamics,
a light techniomega decays mainly 
through electroweak dynamics, $\tom \ra \gamma\tpiz$,
$Z^0\tpiz$, $W^\pm \tpimp$, etc., where $\Z$ and $\W$ may are transversely
polarized. Since
$\sin^2\chi \ll 1$, the electroweak decays of $\tro$ to the transverse gauge
bosons $\gamma,\W,\Z$ plus a technipion may be competitive with the
open-channel strong decays. 

Note, the exact meaning of longitudinal or transverse polarizations only
makes sense at high energies, where the Goldstone equivalence theorem
can be applied.
At the moderate energies considered in the TCSM, 
the decay products of the $\W$ and $\Z$ bosons are distributed according 
to phase space, regardless of their designation as longitudinal
$\W_{\mrm{L}}/\Z_{\mrm{L}}$ or ordinary transverse gauge bosons.  

An effective Lagrangian for technivector interactions
can be constructed \cite{Lan99a}, exploiting gauge invariance, 
chiral symmetry, and angular momentum and parity conservation.
For example, the lowest-dimensional operator
mediating the decay $\tom(q) \ra \gamma(p_1) \tpiz(p_2))$
is $(e/M_V)\ts F_{\tro} \cdot \widetilde{F}_\gamma \ts
\tpiz$, where the mass parameter $M_V$ is expected to be
of order several 100~GeV. This leads to the
decay amplitude:
\be\label{eq:omgampi}
\CM(\tom(q) \ra \gamma(p_1) \tpiz(p_2)) = {e \cos\chi\over{M_V}}
\epsilon^{\mu\nu\lambda\rho} \epsilon_\mu(q) \epsilon^*_\nu(p_1) q_\lambda
p_{1\rho} \ts.
\end{equation}
Similar expressions exist for the other amplitudes involving different
technivectors and/or different gauge bosons \cite{Lan99a}, where
the couplings are derived in the valence technifermion 
approximation \cite{Eic96,Lan99}.
In a similar fashion, decays to fermion-antifermion pairs are included.
These partial widths are typically small, but can have important
phenomenological consequences, such as narrow lepton-antilepton
resonances produced with electroweak strength.

Final states containing
Standard Model particles and/or pseudo-Goldstone bosons (technipions)
can be produced at colliders 
through two mechanisms:  technirho and techniomega mixing
with gauge bosons through a vector--dominance mechanisms,
and anomalies \cite{Lan02} involving no techni--resonances.
Processes 191, 192 and 193 are based on $s$-channel production of
the respective resonance \cite{Eic96}
in the narrow width approximation.
All decay modes implemented can 
be simulated separately or in combination, in the standard fashion. 
These include pairs of fermions, of gauge bosons, of technipions, 
and of mixtures of gauge bosons and technipions.
Processes 194,195 and 361--377,
instead, include interference and a correct treatment
of kinematic thresholds, both of which are important effects, but
also are limited to specific final states.  Therefore, several processes
need to be simulated at once to determine the full effect of TC.

Process 194 is intended to more accurately represent the 
mixing between the $\gamma^*$, $\Z^0$, $\rho_{\mrm{tc}}^0$ and
$\omega_{\mrm{tc}}^0$ particles in the Drell--Yan process \cite{Lan99}.  
Process 195 is the analogous charged channel process including 
$\W^{\pm}$ and $\rho_{\mrm{tc}}^{\pm}$ mixing.  By default, the final 
state fermions are $\e^+ \e^-$ and $\e^{\pm} \nu_{\e}$, respectively.  
These can be changed through the parameters \ttt{KFPR(194,1)} and 
\ttt{KFPR(195,1)}, respectively (where the \ttt{KFPR} value should 
represent a charged fermion).

Processes 361--368 describe the pair production of technipions and
gauge bosons through 
$\rho_{\mrm{tc}}^0/\omega_{\mrm{tc}}^0$ resonances and anomaly contributions.
Processes 370--377 describe pair production through the
$\rho_{\mrm{tc}}^{\pm}$ resonance and anomalies.
It is important to note that processes \ttt{361,362,370,371,372}
include final states with 
only longitudinally--polarized $\W$ and $\Z$ bosons,
whereas the others include final states with only transverse $\W$
and $\Z$ bosons.  Thus, {\bf all} processes must be simulated to
get the {\bf full} effect of the TC model under investigation.
All processes 361--377 are obtained by \ttt{MSEL=50}.

Cross sections for neutral charged final states
at virtuality $\sqrt{s}$
are calculated using the full
$\gamma$--$\Z^0$--$\tro$--$\tom$ propagator matrix, $\Delta_0(s)$. With
$\CM^2_V = M^2_V - i \sqrt{s} \ts \Gamma_V(s)$ and $\Gamma_V(s)$ the
energy-dependent width for $V = \Z^0,\tro,\tom$, this
matrix is the inverse of
\begin{equation}
\label{eq:gzprop}
\Delta_0^{-1}(s) =\left(\ba{cccc}
s & 0 & s f_{\gamma\tro} & s f_{\gamma\tom} \\
0 & s  - \CM^2_\Z  & s f_{\Z\tro} & s f_{\Z\tom} \\
s f_{\gamma\tro}  & s f_{\Z\tro}  & s - \CM^2_{\troz} & 0 \\
s f_{\gamma\tom}  & s f_{\Z\tom}  & 0 & s - \CM^2_{\tom} 
\ea\right) \ts,
\end{equation}
with $f_{\gamma\tro} = \xi$, $f_{\gamma\tom} = \xi \ts (Q_U + Q_D)$,
$f_{\Z\tro} = \xi \ts \cot 2\thw$, and $f_{\Z\tom} = - \xi \ts
(Q_U + Q_D) \tan\thw$, and $\xi = \sqrt{\alpha/\atro}$ determining
the strength of the kinetic mixing.
Because of the off-diagonal entries, the propagators resonate
at mass values shifted from the nominal $M_V$ values.
Note that special care is taken in the
limit of very heavy technivectors to reproduce the canonical
$\gamma^*/\Z^*\to\tpip\tpim$ couplings.
Cross sections for charged final states require the $\W^\pm$--$\tropm$ 
matrix $\Delta_{\pm}$:
\be
\label{eq:wprop}
\Delta_{\pm}^{-1}(s) =\left(\ba{cc} s - \CM^2_\W & s f_{\W\tro} \\ s
  f_{\W\tro} & s - \CM^2_{\tropm} \ea\right) \ts,
\end{equation}
where $f_{\W\tro} = \xi/(2\sin\thw)$. 

By default, the TCSM Model has the parameters
$\Ntc$= 4,
$\sin\chi$  = $1\over 3$,
$Q_U$  = $4 \over 3$,
$Q_D = Q_U-1$  = $1\over 3$,
$C_\b=C_\c=C_\tau$= 1,
$C_\t$= $m_\b/m_\t$,
$C_{\tpi}$= $\tx{4\over{3}}$,
$C_{\tpiz\to \g\g}$=0, $C_{\tpipr\to \g\g}$=1,
$\vert\epsilon_{\rho\omega}\vert$ = 0.05,
$F_T = F_\pi \sin\chi$  = $82$ GeV,
$M_{\tropm}= M_{\troz} = M_{\tom}$ = $210$ GeV,
$M_{\tpipm} = M_{\tpiz} = M_{\tpipr}$ = $110$ GeV,
$M_{V} = M_{A}$ = $200$ GeV.
The techniparticle mass parameters are set through
the usual \ttt{PMAS} array.
Parameters regulating production and decay rates are stored in 
the \ttt{RTCM} array in \ttt{PYTCSM}.

In the original TCSM outlined above, the existence of top-color interactions
only affected the coupling of technipions to top quarks, which
is a significant effect only for higher masses.  In general, however,
TC2 requires some new and possibly light colored particles.
In most TC2 models, the existence of a large $\t\bar\t$, but
not $\b\bar\b$, condensate and mass is due to $\suone \otimes \uone$ gauge
interactions which are strong near 1~TeV. The $\suone$ interaction is
$\t$--$\b$ symmetric while $\uone$ couplings are 
$\t$--$\b$ asymmetric. There are
weaker $\sutwo \otimes \utwo$ gauge interactions in which light quarks (and
leptons) may \cite{Hil95}, or may not~\cite{Chi96}, participate. The two 
${\bf U(1)}$'s must be
broken to weak hypercharge ${\bf U(1)_Y}$ at an energy somewhat 
higher than 1~TeV
by electroweak--singlet condensates. 
The full phenomenology of even such a simple model can be quite complicated,
and many (possibly--unrealistic) simplifications are made to reduce
the number of free parameters \cite{Lan02a}.  Nonetheless, it is useful to 
have some benchmark to guide experimental searches.

The two TC2 ${\bf SU(3)}$'s can be broken to their diagonal $\suc$
subgroup by using technicolor and $\uone$ interactions, both strong near
1~TeV. This can be explicitly accomplished
\cite{Lan95} using two
electroweak doublets of technifermions, $T_1= (U_1,D_1)$ and $T_2 =
(U_2,D_2)$, which transform respectively as $(3,1,\Ntc)$ and $(1,3,\Ntc)$
under the two color groups and technicolor. The desired pattern of symmetry
breaking occurs if ${\bf SU(\Ntc)}$ and $\uone$ interactions work together to
induce electroweak and $\suone \otimes \sutwo$ non-invariant condensates
$\langle \ts \bar U_{iL} U_{jR} \rangle$ and $\langle \ts 
\bar D_{iL} D_{jR} \rangle$, $(i,j = 1,2)\ts$.
This minimal TC2 scenario leads to a rich spectrum of color--nonsinglet
states readily accessible in hadron collisions. The lowest--lying ones
include eight ``colorons'', $V_8$, the massive gauge bosons of broken
topcolor ${\bf SU(3)}$; 
four isosinglet $\troct$ formed from $\bar T_i T_j$ and the
isosinglet pseudo-Goldstone technipions formed from $\bar T_2 T_2$. 
In this treatment, the isovector technipions are ignored, because
they must be pair produced in $\troct$ decays, 
and such decays are assumed to be
kinematically suppressed.

The colorons are new fundamental particles with couplings to quarks.
In standard TC2~\cite{Hil95}, top and bottom quarks couple to $\suone$ 
and the four light quarks to $\sutwo$. Because the
$\suone$ interaction is strong and acts exclusively on the third
generation, the residual $V_8$ coupling can be enhanced for $\t$
and $\b$ quarks.  The coupling $g_a=g_c\cot\theta_3$ for $\t$ and $\b$
and $g_a=-g_c\tan\theta_3$ for $\u,\d,\c,\s$,
where $g_c$ is the QCD coupling and $\cot\theta_3$ is related to the
original $g_1$ and $g_2$ couplings.
In flavor--universal TC2~\cite{Chi96} all quarks couple to
$\suone$, not $\sutwo$, so that colorons couple equally and strongly to all
flavors: $g_a = g_c \cot\theta_3$.

Assuming that techni--isospin is not badly broken by ETC interactions, 
the $\troct$ are isosinglets
labeled by the
technifermion content and color index $A$: 
$\rho_{11}^{A}, \rho_{22}^{A}, \rho_{12}^{A}, \rho_{12'}^{A}$.
The first two of these states, $\troctaa$ and $\troctbb$, mix with $V_8$ and
$g$. The topcolor--breaking condensate, $\langle \bar T_{1L} T_{2R} \rangle
\neq 0$, causes them to also mix with $\troctab$ and $\troctabp$. 
Technifermion condensation also leads to a number of (pseudo)Goldstone boson
technipions.  The lightest technipions are expected to be
the isosinglet $\suc$ octet and singlet $\bar T_2
T_2$ states $\pi_{22}^{A}$ and $\pi_{22}^0$.

These technipions can decay into either
fermion--antifermion pairs or two gluons; presently, they are
assumed to decay only into gluons. As noted, walking
technicolor enhancement of technipion 
masses very likely close off the $\troct \ra
\tpi\tpi$ channels. Then the $\troct$ decay into $\q\bar\q$ and $\g\g$.
The rate for the former are proportional to the amount
of kinetic mixing, set by 
$\xi_\g = {g_c / {g_{\tro}}}$.
Additionally, the $\troctbb$ decays to $\g\pi_{22}^{0,A}$.

The $V_8$ colorons are expected to be considerably heavier than the $\troct$,
with mass in the range 0.5--1~TeV. In both the standard and
flavor--universal models, colorons couple strongly to $\bar T_1 T_1$, but with
only strength $g_c$ to $\bar T_2 T_2$. Since relatively light technipions are
$\bar T_2 T_2$ states, it is estimated that
$\Gamma(V_8 \ra \tpi\tpi) = \CO(\alpha_c)$
and $\Gamma(V_8 \ra g\tpi) = \CO(\alpha_c^2)$. Therefore, these decay
modes are ignored, so that
the $V_8$ decay rate is the
sum over open channels of
\be\label{eq:vdecay}
\Gamma(V_8 \ra \q_a \bar \q_a) = {\alpha_a \over{6}} \ts
\left(1 + {2m_a^2 \over{s}} \right) \ts 
\left(s - 4m_a^2\right)^{1\over{2}} \ts,
\end{equation}
where $\alpha_a = g^2_a/4\pi$.

The phenomenological effect of this techniparticle structure is
to modify the gluon propagator in ordinary QCD processes, because
of mixing
between the gluon, $V_8$ and the $\troct$'s. 
The $\g$--$V_8$--$\troctaa$--$\troctbb$--$\troctab$--$\troctabp$ propagator
is the inverse of the symmetric matrix
\be\label{eq:invprop}
D^{-1}(s) = \left(\ba{cccccc}
s & 0 & s \ts \xi_\g & s \ts \xi_\g & 0 & 0 \\ \\
0 & s - \CM^2_{V_8} & s \ts \xi_{\rho_{11}} & s \ts \xi_{\rho_{22}} & s
\ts \xi_{\rho_{12}} & s \ts \xi_{\rho_{12'}}\\ \\
s \ts \xi_\g & s \ts \xi_{\rho_{11}} & s - \CM^2_{11}  & - M^2_{11,22} &
-M^2_{11,12} & -M^2_{11,12'}\\ \\
s \ts \xi_\g & s \ts \xi_{\rho_{22}} & -M^2_{11,22} & s - \CM^2_{22}  &
-M^2_{22,12} & -M^2_{22,12'}\\ \\
0 & s \ts \xi_{\rho_{12}} & -M^2_{11,12} & -M^2_{22,12} &
s - \CM^2_{12} & -M^2_{12,12'}\\ \\
0 & s \ts \xi_{\rho_{12'}} & -M^2_{11,12'} & -M^2_{22,12'} &
 -M^2_{12,12'} & s - \CM^2_{12'} \\
\ea\right) \ts.
\end{equation}
Here, $\CM^2_V = M^2_V - i \sqrt{s} \ts \Gamma_V(s)$ uses the
energy--dependent widths of the octet vector bosons, and
the $\xi_{\rho_{ij}}$ are proportional to $\xi_\g$ and
elements of matrices that describe the pattern of technifermion
condensation.  The mixing terms $M^2_{ij,kl}$, induced by
$\bar T_1 T_2$ condensation are assumed to be real.

This extension of the TCSM is still under development, and
any results should be carefully scrutinized.
The main effects are indirect, in that they modify
the underlying two-parton QCD processes much like compositeness terms,
except that a resonant structure are visible. 
Similar to compositeness, the effects of these colored technihadrons
are simulated by setting \ttt{ITCM(5)=5} for processes 381--388.  
By default, these processes are equivalent to the 11, 12, 13, 28, 53, 
68, 81 and 82 ones, respectively. The last two are specific for 
heavy-flavour production, while the first six could be used to describe
standard or non-standard high-$\pT$ jet production. These six are 
simulated by \ttt{MSEL=51}. 
The parameter dependence of the `model' is encoded
in $\tan\theta_3$ (\ttt{RTCM(21)}) and a mass parameter 
$M_8$ (\ttt{RTCM(27)}), which determines the decay width
$\rho_{22}\to \g\pi_{22}$ analogously to $M_V$ for $\tom\to\gamma\tpi$.
For \ttt{ITCM(2)} equal to 0 (1), the 
standard (flavor universal) TC2 couplings are used.
The mass parameters are set by the \ttt{PMAS} array using the
codes: $V_8$ (3100021), $\pi_{22}^{1}$ (3100111), 
$\pi_{22}^{8}$ (3200111), $\rho_{11}$ (3100113),
$\rho_{12}$ (3200113), $\rho_{21}$ (3300113), and $\rho_{22}$ (3400113).
The mixing parameters $M_{ij,kl}$ take on the (arbitrary) values
$M_{11,22}=100$ GeV, $M_{11,12}=M_{11,21}=M_{22,12}=150$ GeV,
$M_{22,21}=75$ GeV and $M_{12,21}=50$ GeV, while the
kinetic mixing terms $\xi_{\rho_{ij}}$ are calculated assuming
the technicolor condensates are fully mixed, i.e. 
$\langle T_i\bar T_j \rangle \propto {1 / \sqrt{2}}$.  

\subsubsection{Extra Dimensions}
\label{sss:extradimclass}

{\ISUB} = 
\begin{tabular}[t]{rl@{\protect\rule{0mm}{\tablinsep}}}
391 & $\f \fbar \to \G^*$ \\
392 & $\g \g \to \G^*$ \\
393 & $\q \qbar \to \g \G^*$ \\
394 & $\q \g \to \q \G^*$ \\
395 & $\g \g \to \g \G^*$ \\
\end{tabular}

In recent years, the area of observable consequences of extra dimensions
has attracted a strong interest. The field is still in rapid development, 
so there still does not exist a `standard phenomenology'. The topic is
also fairly new in {\Py}, and currently only a first scenario is
available.

The $\G^*$, introduced as new particle code 5000039, is intended 
to represent the lowest excited graviton state in a Randall-Sundrum 
scenario \cite{Ran99} of extra dimensions. The lowest-order production
processes by fermion or gluon fusion are found in 391 and 392. The
further processes 393--395 are intended for the high-$\pT$ tail in hadron
colliders. As usual, it would be double-counting to have both sets of
processes switched on at the same time. Processes 391 and 392, with 
initial-state showers switched on, are appropriate for the full 
cross section at all $\pT$ values, and gives a reasonable description
also of the high-$\pT$ tail. Processes 393--395 could be useful e.g.
for the study of invisible decays of the $\G^*$, where a large $\pT$
imbalance would be required. It also serves to test/confirm the shower
expectations of different $\pT$ spectra for different production
processes \cite{Bij01}.

Decay channels of the $\G^*$ to $\f \fbar$, $\g\g$, $\gamma \gamma$, 
$\Z^0 \Z^0$ and $\W^+ \W^-$ contribute to the total width. The correct 
angular distributions are included for decays to a fermion pair in the
lowest-order processes, whereas other decays currently are taken to be 
isotropic. 

The $\G^*$ mass is to be considered a free parameter. The other degree 
of freedom in this scenario is a dimensionless coupling, see 
\ttt{PARP(50)}.

\subsection{Supersymmetry}
\label{ss:susyclass}

\ttt{MSEL} = 39--45 \\  
{\ISUB} = 201--296 (see tables at the beginning of this chapter)

{\Py} includes the possibility of simulating a large variety of production
and decay processes in the Minimal Supersymmetric extension
of the Standard Model (MSSM). The simulation is based 
on an effective Lagrangian of softly-broken SUSY with parameters 
defined at the weak scale, typically between $m_{\Z}$ and 1
TeV. The relevant parameters should either be supplied directly by the user or
they should be read in from a SUSY Les Houches Accord spectrum file (see
below). Some other possibilities for obtaining the SUSY parameters 
also exist in the code, as described below, but these  
are only intended for backwards compatibility and debugging purposes.

\subsubsection{General Introduction}
In any ($N=1$) 
supersymmetric version of the SM there exists a partner to each SM
state with the same gauge quantum numbers but whose spin differs by one half
unit. Additionally, the dual requirements of generating masses for 
up-- and down--type fermions while preserving SUSY and gauge
invariance, require that the SM Higgs sector be enlarged to two scalar
doublets, with corresponding spin-partners. 

After Electroweak symmetry breaking (EWSB), the bosonic Higgs sector contains
a quintet of physical states: two CP--even scalars, $h^0$ and $H^0$, one
CP--odd pseudoscalar, $A^0$, and a pair of charged scalar Higgs bosons,
$H^\pm$ (naturally, this classification is only correct when CP violation
is absent in the Higgs sector. Non--trivial phases between certain
soft--breaking parameters will induce mixing between the CP eigenstates). 
The fermionic Higgs (called ``Higgsino'') sector is constituted by 
the superpartners of these fields, but these are not normally exact mass
eigenstates, so we temporarily postpone the discussion of them.
  
In the gauge sector, the spin--1/2 partners of the $\bf U(1)_Y$ and $\bf
SU(2)_L$ gauge bosons (called ``gauginos'') are the Bino, $\Bino$, the
neutral Wino, $\w3$, and the charged Winos, $\Wino_1$ and $\Wino_2$, while
the partner of the gluon is the gluino, $\gluino$.  After EWSB, the $\Bino$
and $\w3$ mix with the neutral Higgsinos, $\higgsino_1, \higgsino_2$, to form
four neutral Majorana fermion mass-eigenstates, the neutralinos,
$\zinog_{1-4}$. In addition, the charged Higgsinos, $\higgsino^\pm$, mix with
the charged Winos, $\Wino_1$ and $\Wino_2$, resulting in two charged Dirac
fermion mass eigenstates, the charginos, $\winog_{1,2}$. Note that the
$\tilde{\gamma}$ and $\tilde{\Z}$, which sometimes occur in the literature,
are linear combinations of the $\Bino$ and $\w3$, by exact analogy with the
mixing giving the $\gamma$ and $\Z^0$, but these are not normally mass
eigenstates after EWSB, due to the enlarged mixing caused by the presence of
the Higgsinos.

The spin--0 partners of the SM fermions (so-called ``scalar fermions'', or
``sfermions'') are the squarks $\squark$, sleptons $\slepton$, and sneutrinos
$\sneutrino$.  Each fermion (except the neutrinos) has two scalar partners,
one associated with each of its chirality states. These are named
left--handed and right--handed sfermions, respectively.  Due to their scalar
nature, it is of course impossible for these particles to possess any
intrinsic ``handedness'' themselves, but they inherit their couplings to the
gauge sector from their SM partners, so that e.g.\ a $\tilde \d_R$ does not
couple to {$\bf SU(2)_L$} while a $\tilde \d_L$ does.

Generically, the \ttt{KF} code numbering scheme used in {\Py} reflects the
relationship between particle and sparticle, so that e.g.~for sfermions, the
left--handed (right--handed) superpartners have codes 1000000 (2000000) plus
the code of the corresponding SM fermion.  A complete list of the particle
partners and their {\KF} codes is given in Table~\ref{t:codenine}. Note that,
at times, antiparticles of scalar particles are denoted by $^*$, i.e.\
$\st^*$ rather than the more correct but cumbersome $\br{\st}$ or
$\tilde{\tbar}$.

The MSSM Lagrangian contains interactions between particles
and sparticles, with couplings fixed by SUSY. There are also a number of
soft SUSY--breaking mass parameters.  ``Soft'' here means
that they break the
mass degeneracy between SM particles and their SUSY partners
without reintroducing quadratic divergences in the theory or destroying
its gauge invariance.
In the MSSM, the soft SUSY--breaking parameters are extra mass terms for
gauginos and sfermions and trilinear scalar couplings. Further soft terms may
arise, for instance in models with broken $R$-parity, but we here restrict
our attention to the minimal case. 

The exact number of independent parameters
depends on the detailed mechanism of SUSY breaking.
The general MSSM model in {\Py} assumes only a few relations
between these parameters which seem theoretically difficult to
avoid. Thus, the first two generations of sfermions with
otherwise similar quantum numbers, e.g.\ $\tilde \d_L$ and $\tilde \s_L$, 
have the same masses. 
Despite such simplifications, there are a fairly large number of 
parameters that appear in the SUSY Lagrangian and determine
the physical masses and interactions with Standard Model particles,
though far less than the $>100$  which are allowed in all generality.
The Lagrangian (and, hence, Feynman rules) follows the conventions
set down by Kane and Haber in their Physics Report article  
\cite{Hab85} and the papers of Gunion and Haber \cite{Gun86a}.
Once the parameters of the softly-broken SUSY Lagrangian are
specified, the interactions are fixed, and the sparticle masses can
be calculated. 

\subsubsection{Extended Higgs Sector}

{\Py} already simulates a Two Higgs Doublet Model (2HDM) obeying tree-level
relations fixed by two parameters, which are conveniently taken as the ratio
of doublet vacuum expectation values $\tan\beta$, and the pseudoscalar mass
$M_A$.  The Higgs particles are considered Standard Model fields, since a
2HDM is a straightforward extension of the Standard Model.  The MSSM Higgs
sector is more complicated than that described above in
subsection~\ref{ss:Hclass}, and includes important radiative corrections to
the tree--level relations.  The CP--even Higgs mixing angle $\alpha$ is
shifted as well as the full Higgs mass spectrum.  The properties of the
radiatively--corrected Higgs sector in {\Py} are derived in the effective
potential approach \cite{Car95}.  The effective potential contains an
all--orders resummation of the most important radiative corrections, but
makes approximations to the virtuality of internal propagators.  This is to
be contrasted with the diagrammatic technique, which performs a fixed--order
calculation without approximating propagators.  In practice, both techniques
can be systematically corrected for their respective approximations, so that
there is good agreement between their predictions, though sometimes the
agreement occurs for slightly different values of SUSY--breaking
parameters. The description of Higgs properties in {\Py} is based on the same
\ttt{FORTRAN} code as in \ttt{HDecay} \cite{Djo97}, except that certain
corrections that are particularly important at large values of $\tan\beta$
are included in {\Py}.

There are several notable properties of the MSSM Higgs sector.  As long as
the soft SUSY--breaking parameters are less than about 1.5 TeV, a number
which represents a fair, albeit subjective, limit for where the required
degree of fine-tuning of MSSM parameters becomes unacceptably large, there is
an upper bound of about 135 GeV on the mass of the CP--even Higgs boson most
like the Standard Model one, i.e. the one with the largest couplings to the
$\W$ and $\Z$ bosons, be it the $h$ or $H$.  If it is $h$ that is the
SM--like Higgs boson, then $H$ can be significantly heavier.  On the other
hand, if $H$ is the SM--like Higgs boson, then $h$ must be even lighter.  If
all SUSY particles are heavy, but $M_A$ is small, then the low--energy theory
would look like a two--Higgs--doublet model.  For sufficiently large $M_A$,
the heavy Higgs doublet decouples, and the effective low--energy theory has
only one light Higgs doublet with SM--like couplings to gauge bosons and
fermions.

The Standard Model fermion masses are not fixed by SUSY,
but their Yukawa couplings become 
a function of $\tan \beta$.
For the up-- and down--quark and leptons, 
$m_u = h_u v \sin \beta$, 
$m_d = h_d v \cos \beta$, and 
$m_\ell = h_\ell v \cos \beta$,
where $h_{f=u,d,\ell}$ 
is the corresponding Yukawa coupling and  $v \approx 246$ GeV is the 
order parameter of Electroweak symmetry breaking.  
At large $\tan\beta$, significant corrections can occur
to these relations.  These are included for the $\b$ quark,
which appears to have the most sensitivity to them, and
the $\t$ quark.
The array values \ttt{RMSS(40)} and \ttt{RMSS(41)} are used for 
temporary storage of the corrections $\Delta m_{\t}$ and $\Delta m_{\b}$.  
\ttt{PYPOLE}, based on the updated version of \ttt{SubHpole}, written 
by Carena et al.\ \cite{Car95}, also includes some bug fixes, so that 
it is generally better behaved.

The input parameters that determine the MSSM Higgs sector
in {\Py} are \ttt{RMSS(5)} ($\tan\beta$), \ttt{RMSS(19)} ($M_A$),
\ttt{RMSS(10-12)} (the third generation squark mass parameters),
\ttt{RMSS(15-16)} (the third generation squark trilinear
couplings), and \ttt{RMSS(4)} (the Higgsino mass $\mu$).
Additionally, the large $\tan\beta$ corrections related
to the $\b$ Yukawa coupling depend on \ttt{RMSS(3)}
(the gluino mass).  
Of course, these calculations also 
depend on SM parameters ($m_{\t}, m_{\Z}, \alphas,$ etc.).  Any 
modifications to these quantities from virtual MSSM effects are not 
taken into account. In principle, the sparticle masses also acquire 
loop corrections that depend on all MSSM masses. 

See section \ref{sss:models} for a description how to use the 
loop-improved RGE's of \tsc{Isasusy} to determine the SUSY
mass and mixing spectrum (including also loop
corrections to the Higgs mass spectrum and couplings) with \Py. 

If \ttt{IMSS(4)=0}, an approximate version of the effective potential
calculation can be used.  It is not as accurate as that available for
\ttt{IMSS(4)=1}, but it useful for demonstrating the effects of higher
orders.  Alternatively, for \ttt{IMSS(4)=2}, the physical Higgs masses
are set by their \ttt{PMAS} values while the CP--even Higgs boson mixing
angle $\alpha$ is set by \ttt{RMSS(18)}.  These values and $\tan\beta$
(\ttt{RMSS(5)}) are enough to determine the couplings, provided that
the same tree--level relations are used.

\subsubsection{Superpartners of Gauge and Higgs Bosons}

The chargino and neutralino masses and 
their mixing angles (that is, their gaugino and Higgsino composition)
are determined by the SM gauge boson masses ($M_\W$ and $M_\Z$),
$\tan \beta$,
two soft SUSY--breaking
parameters (the ${\bf SU(2)_L}$ gaugino mass $M_2$ and the ${\bf U(1)_Y}$ 
gaugino mass $M_1$), together with the Higgsino mass parameter $\mu$, 
all evaluated at the electroweak scale $\sim M_\Z$. {\Py} assumes
that the input parameters are evaluated at the ``correct'' scale.
Obviously, more care is needed to set precise experimental limits or
to make a connection to higher--order calculations. 

Explicit solutions of the chargino and neutralino masses and
their mixing angles (which appear in Feynman rules)
are found by diagonalizing the $2\times 2$ chargino 
$\mathbf{M_C}$ and
$4\times 4$ neutralino $\mathbf{M_N}$ mass matrices: 
\begin{eqnarray}
\label{eqn:inomatrices}
&\mathbf{M_{C}} =  \left( \begin{array}{cc}
 M_2 &  \sqrt{2}M_\W s\beta \\
\sqrt{2}M_\W c\beta & \mu \\ \end{array} \right); 
\mathbf{M_{N}} =  \left( \begin{array}{cc}
 \mathbf{M}_i &  \mathbf{Z} \\
 \mathbf{Z^T} & \mathbf{M}_\mu \\ \end{array} \right) &  \\
 & \nonumber \\
&\mathbf{M}_i =  \left( \begin{array}{cc}
 M_1 &  0 \\
 0   &  M_2 \\ \end{array} \right);
\mathbf{M}_\mu =  \left( \begin{array}{cc}
 0 &  -\mu \\
 -\mu   &  0 \\ \end{array} \right);
\mathbf{Z} =  \left( \begin{array}{cc}
 -M_\Z c\beta s_W & M_\Z s\beta s_W   \\
 M_\Z c\beta c_W & -M_\Z s\beta c_W  \\ \end{array} \right) & \nonumber
\end{eqnarray}
$\mathbf{M_C}$ is written in the $(\Wino^+,\higgsino^+)$ basis,
$\mathbf{M_N}$ in the $(\Bino,\Wino^3,\higgsino_1,\higgsino_2)$ basis,
with the notation $s\beta=\sin\beta,c\beta=\cos\beta,s_W=\sin\theta_W$
and $c_W=\cos\theta_W$.
Different sign conventions and bases sometimes appear in
the literature.  In particular, {\Py} agrees with the
{\sc Isasusy} \cite{Bae93} convention for $\mu$, but uses a different 
basis of fields and has different-looking mixing matrices.

In general, the soft SUSY--breaking parameters can be
complex valued, resulting in CP violation in some sector of the theory, but
more directly expanding the possible masses and mixings of sparticles.  
Presently, the consequences of arbitrary phases are
only considered in the chargino and neutralino sector,
though it is well known that they can have a significant
impact on the Higgs sector.  A generalization of the 
Higgs sector is among the plans for the future development
of the program. The chargino and neutralino
input parameters are \ttt{RMSS(5)} ($\tan\beta$),
\ttt{RMSS(1)} (the modulus of
$M_1$) and \ttt{RMSS(30)} (the phase of $M_1$),
\ttt{RMSS(2)} and \ttt{RMSS(31)} (the modulus and
phase of $M_2$), and \ttt{RMSS(4)} and \ttt{RMSS(33)}
(the modulus and phase of $\mu$).  To simulate the
case of real parameters (which is CP--conserving),
the phases are zeroed by default.  In addition,
the moduli parameters can be signed, to make a simpler
connection to the CP--conserving case.  (For example,
\ttt{RMSS(5)=-100.0} and \ttt{RMSS(30)=0.0} represents
$\mu=-100$ GeV.)  

The expressions for the production cross sections
and decay widths
of neutralino and chargino pairs contain the phase dependence,
but ignore possible effects of the phases in the sfermion
masses appearing in propagators.
The production cross sections have been updated to include
the dependence on beam polarization through the
parameters \ttt{PARJ(131,132)} (see Sect.~\ref{ss:polarization}).
There are several approximations made for three--body decays.
The numerical expressions for three--body decay widths 
ignore the effects of finite
fermion masses in the matrix element, but include them
in the phase space.  No
three--body decays $\chi_i^0\to \t\bar \t\chi^0_j$ are simulated,
nor $\chi^+_i (\chi^0_i) \to \t \bar \b \chi^0_j (\chi^-_j)$. 
Finally, the effects of mixing between the third generation interaction
and mass eigenstates for sfermions is ignored, except that the
physical sfermion masses are used. 
The kinematic distributions of the decay products are spin--averaged,
but include the correct matrix--element weighting. 
Note that for the $R$--parity violating decays (see
below), both 
sfermion mixing effects and masses of $\b$, $\t$, and $\tau$ are fully
included. 

Since the $SU(3)_C$ symmetry of the SM is not broken, the
gluinos have masses determined by the $SU(3)_C$ gaugino mass parameter
$M_3$, input through the parameter \ttt{RMSS(3)}.  
The physical gluino mass is shifted from the value
of the gluino mass parameter $M_3$ because of radiative corrections.
As a result, there is an indirect dependence on the squark masses.
Nonetheless, it is sometimes convenient to input the physical
gluino mass, assuming that there is some choice of $M_3$ which
would be shifted to this value.  This can be accomplished through
the input parameter \ttt{IMSS(3)}.  
A phase for the gluino mass can be set using \ttt{RMSS(32)},
and this can influence the gluino decay width (but no effect
is included in the $\gluino+\ino$ production).
Three--body decays of
the gluino to $\t \bar \t$ and $\b \bar \b$ and $\t \bar \b$ 
plus the appropriate neutralino or chargino are allowed and
include the full effects of sfermion mixing.  However, they
do not include the effects of phases arising from complex
neutralino or chargino parameters.  

There is one exception to the above discussion about the input parameters
to the neutralino and chargino mass matrices.  In the case when
$M_2$ is much smaller than other mass parameters (as occurs in models
of anomaly--mediated SUSY breaking), radiative corrections are
very important in keeping the lightest neutralino lighter than the
lightest chargino.  If ever the opposite occurs in solving the
eigenvalue problem numerically, the chargino mass is set to the
neutralino mass plus 2 times the charged pion mass, thus allowing the 
decay $\winog_1\to\pi^\pm\zinog_1$.

\subsubsection{Superpartners of Standard Model Fermions}

The mass eigenstates of squarks and sleptons are, in principle,
mixtures of their left-- and right--handed components, given by:
\begin{eqnarray}
M_{\tilde{f}_L}^2 
= m_{\bf 2}^2 + m^2_f + D_{\tilde{f}_L} & &
M_{\tilde{f}_R}^2 
= m_{\bf 1}^2 + m^2_f + D_{\tilde{f}_R}
\label{eq:LRmasseigen}  
\end{eqnarray}
where $m_{\bf 2}$ are soft SUSY--breaking parameters
for superpartners of ${\bf SU(2)_L}$ doublets, and $m_{\bf 1}$ 
are parameters for singlets.  
The $D$--terms
associated with Electroweak symmetry breaking are
$D_{\tilde{f}_L}= M_Z^2 \cos (2 \beta) (T_{3_{f}} -
Q_f \sin^2\theta_W)$ and 
$D_{\tilde{f}_R}= M_Z^2 \cos( 2 \beta) Q_f\sin^2\theta_W$,
where $T_{3_f}$ is the weak isospin eigenvalue ($=\pm 1/2$) of the fermion and
$Q_f$ is the electric charge. Taking the $D$--terms into account, one easily
sees that the masses of sfermions in ${\bf SU(2)_L}$ doublets are related by
a sum rule: $M_{\tilde{f}_L,T_3=1/2}^2 - M_{\tilde{f}_L,T_3=-1/2}^2 =
M_Z^2\cos( 2 \beta)$.

In many high--energy models, 
the soft SUSY--breaking sfermion mass parameters are
taken to be equal at the high--energy scale, but, in principle, 
they can be different for
each generation or even within a generation.
However, the sfermion flavor dependence can have important effects on
low--energy observables, and it is often strongly constrained.
The suppression of flavor changing neutral currents (FCNC's), such as
$K_L\to\pi^\circ\nu\bar\nu$, requires that either $(i)$ the squark soft
SUSY--breaking mass matrix is diagonal and degenerate, 
or $(ii)$ the masses of the
first-- and second--generation sfermions are very large. Thus we make the
data--motivated simplification of setting $M_{\su_{L}}=M_{\sch_{L}}$, 
$M_{\sd_{L}}=M_{\sst_{L}}$, $M_{\su_{R}}=M_{\sch_{R}}$, 
$M_{\sd_{R}}=M_{\sst_{R}}$. \vspace{1mm}

The left--right 
sfermion mixing is determined by the product of
soft SUSY--breaking parameters and the mass of
the corresponding fermion.
Unless the soft SUSY--breaking parameters for the first two
generations are orders of magnitude greater than for
the third generation, the mixing in 
the first two generations can be neglected. This simplifying assumption is
also made in {\Py}: the sfermions $\squark_{L,R}$, with 
$\squark = \tilde \u,\tilde \d,\tilde \c,\tilde \s$,
and  $\slepton_{L,R},\sneutrino_\ell$,
with $\ell = \e, \mu$, are the real mass eigenstates with
masses $m_{\squark_{L,R}}$ and 
$m_{\slepton_{L,R}}, m_{\sneutrino_{\ell}},$ respectively.
For the third generation sfermions, due to weaker experimental constraints, 
the left--right mixing can be  nontrivial.
The tree-level 
mass matrix for the top squarks (stops) in the ($\stop_L,  \stop_R$) 
basis is given by
\begin{equation}
M^2_{\stop} =  \left( \begin{array}{cc}
 m_{Q_3}^2 + m_\t^2 + D_{\stop_L} &  m_\t ( A_\t - \mu/ \tan \beta) \\
 m_\t ( A_\t - \mu/ \tan \beta) &  m_{U_3}^2 + m_\t^2 + D_{\stop_R} \\
 \end{array} \right), 
\label{stop_matrix}
\end{equation}
where $A_\t$ is a trilinear coupling.
Different sign conventions for $A_\t$
occur in the literature; {\Py} and \tsc{Isasusy} use
opposite signs. Unless there is
a cancellation between $A_\t$ and $\mu/\tan\beta$, left--right mixing
occurs for the stop squarks because of the large top quark mass.
The stop  mass eigenstates are then given by  
\begin{eqnarray}
\stop_1 & = & 
\cos \theta_{\stop} \;\;\stop_L + \sin \theta_{\stop} \;\;
 \stop_R
\nonumber \\
\stop_2 &= & 
 - \sin \theta_{\stop} \;\;\stop_L + \cos \theta_{\stop} \;\;
\stop_R,
\end{eqnarray}
where the masses and mixing angle $\theta_{\stop}$ are fixed by
diagonalizing the squared--mass matrix Eq.~(\ref{stop_matrix}).
Note that different conventions exist also for the mixing angle
$\theta_{\stop}$, and that {\Py} here agrees with \tsc{Isasusy}.
When translating Feynman rules from the (L,R) to (1,2) basis,
we use:
\begin{eqnarray}
\stop_L & = & 
\cos \theta_{\stop} \;\;\stop_1 - \sin \theta_{\stop} \;\;
 \stop_2
\nonumber \\
\stop_R &= & 
 \sin \theta_{\stop} \;\;\stop_1 + \cos \theta_{\stop} \;\;
\stop_2.
\end{eqnarray}

Because of the large mixing, the lightest stop $\stop_1$ can be
one of the lightest sparticles.
For the sbottom, an
analogous formula for the mass matrix holds with $m_{U_3}\to m_{D_3}$, 
$A_\t\to A_\b$, $D_{\stop_{L,R}}\to D_{\sbottom_{L,R}}$, $m_\t\to m_\b$, 
and $\tan\beta\to 1/\tan\beta$.  For the stau, the substitutions
$m_{Q_3}\to m_{L_3}$, 
$m_{U_3}\to m_{E_3}$, $A_\t\to A_\tau$, $D_{\stop_{L,R}}\to
D_{\stau_{L,R}}$,
$m_t\to m_\tau$ and $\tan\beta\to$ 1/$\tan\beta$ are appropriate.
The parameters
$A_\t$, $A_\b$, and $A_\tau$ can be independent,
or they might be related by some underlying principle.
When $m_\b\tan\beta$ or $m_\tau\tan\beta$ is large (${\cal O}(m_\t))$, 
left--right
mixing can also become relevant for the sbottom and stau.

Most of the SUSY input parameters are needed to specify the
properties of the sfermions.  As mentioned earlier, the effects of
mixing between the interaction and mass eigenstates are assumed
negligible for the first two generations.  Furthermore, sleptons
and squarks are treated slightly differently.  The physical
slepton masses $\tilde\ell_L$ and $\tilde\ell_R$ are set by
\ttt{RMSS(6)} and \ttt{RMSS(7)}.  By default, the $\stau$
mixing is set by the parameters \ttt{RMSS(13)}, \ttt{RMSS(14)} and 
\ttt{RMSS(17)}, which represent $M_{L_3}$, $M_{E_3}$ and $A_\tau$, 
respectively, i.e. neither $D$--terms nor $m_\tau$ is included.
However, for \ttt{IMSS(8)=1}, the $\stau$ masses will follow the same 
pattern as for the first two generations. 
Previously, it was assumed that the soft SUSY-breaking parameters 
associated with the stau included $D$--terms.  This is no longer the case, 
and is more consistent with the treatment of the stop and sbottom. 
For the first two generations of squarks, 
the parameters \ttt{RMSS(8)} and \ttt{RMSS(9)} are the mass parameters
$m_{\bf 2}$ and $m_{\bf 1}$,
i.e. without $D$--terms included.  For more generality, the
choice \ttt{IMSS(9)=1} means that $m_{\bf 1}$ for $\tilde \u_R$
is set instead
by \ttt{RMSS(22)}, while $m_{\bf 1}$ for $\tilde \d_R$ is
\ttt{RMSS(9)}.  Note that the left--handed squark mass parameters
must have the same value since they reside in the same ${\bf SU(2)_L}$
doublet.  For the third generation, the parameters \ttt{RMSS(10)},
\ttt{RMSS(11)}, \ttt{RMSS(12)}, \ttt{RMSS(15)} and \ttt{RMSS(16)}
represent $M_{Q_3}$, $M_{D_3}$,
$M_{U_3}$, $A_{\b}$ and $A_{\t}$, respectively.

There is added 
flexibility in the treatment of stops, sbottoms and staus.
With the flag \ttt{IMSS(5)=1}, the properties of the third generation 
sparticles can be specified by their mixing angle and mass eigenvalues 
(instead of being derived from the soft SUSY-breaking parameters).  
The parameters \ttt{RMSS(26) - RMSS(28)} specify the mixing angle 
(in radians) for the sbottom, stop, and stau. The parameters 
\ttt{RMSS(10) - RMSS(14)} specify the two stop masses, the one 
sbottom mass (the other being fixed by the other parameters) and the 
two stau masses.  Note that the masses \ttt{RMSS(10)} and \ttt{RMSS(13)} 
correspond to the left-left entries of the diagonalized matrices, while 
\ttt{RMSS(11), RMSS(12)} and \ttt{RMSS(14)} correspond to the right-right 
entries. These entries need not be ordered in mass. 

\subsubsection{Models \label{sss:models}}

At present, the exact mechanism of SUSY breaking is unknown. 
It is generally assumed that the breaking occurs spontaneously
in a set of fields that are almost entirely disconnected from
the fields of the MSSM; if SUSY is broken explicitly
in the MSSM, then some superpartners must be {\it lighter} than
the corresponding Standard Model particle, a phenomenological
disaster.  The breaking of SUSY in this ``hidden sector''
is then communicated to the MSSM fields through one or
several mechanisms:  gravitational interactions, gauge
interactions, anomalies, etc.  While any one of these
may dominate, it is also possible that all contribute at once.

We may parametrize our ignorance of the exact mechanism of SUSY
breaking by simply setting each of the soft SUSY breaking
parameters in the Lagrangian by hand. In {\Py} this approach can  
be effected by setting \ttt{IMSS(1)=1}, although some simplifications have
already been made to greatly reduce the number of parameters from the initial
more than 100. 

As to specific models, several exist which predict the rich set of measurable
mass and mixing parameters from the assumed soft SUSY breaking scenario with
a much smaller set of free parameters. One example is Supergravity
(\tsc{Sugra}) inspired models, where the number of free parameters is reduced
by imposing universality at some high scale, motivated by the apparent
unification of gauge couplings.  Five parameters fixed at the gauge coupling
unification scale, $\tan\beta, M_0, m_{1/2}, A_0,$ and {\sgnmu}, are then
related to the mass parameters at the scale of Electroweak symmetry breaking
by renormalization group equations (see e.g.\ \cite{Pie97}).

The user who wants to study this and other models in detail can use spectrum
calculation programs (e.g.~\tsc{Isasusy} \cite{Bae93}, \tsc{Softsusy}
\cite{All02}, \tsc{SPheno} \cite{Por03}, or \tsc{Suspect} \cite{Djo02}),
which numerically solve the renormalization group equations (RGE) to
determine the mass and mixing parameters at the weak scale.  These may then
be input to {\Py} via a SUSY Les Houches Accord spectrum file \cite{Ska03}
using \ttt{IMSS(1)=11} and \ttt{IMSS(21)=} the unit number where the spectrum
file has been opened. All of {\Py}'s own internal 
mSUGRA machinery (see below) is then switched off. This means 
that none of the other \ttt{IMSS} switches can be used, except for 
\ttt{IMSS(51:53)} ($R$-parity violation), \ttt{IMSS(10)} (force 
$\tilde{\chi}_2\to\tilde{\chi_1}\gamma$), and 
\ttt{IMSS(11)} (gravitino is the LSP). Note that the 
dependence of the $\b$ and $\t$ quark Yukawa couplings on $\tan\beta$  
and the gluino mass is at present ignored when using \ttt{IMSS(1)=11}.

As an alternative, 
a run-time interface to \tsc{Isasusy} can be accessed by the option 
\ttt{IMSS(1)=12}, in which case the \ttt{SUGRA} routine of 
\tsc{Isasusy} is called by \ttt{PYINIT}. This routine then 
calculates the mSUGRA spectrum of SUSY masses and mixings 
(CP conservation, i.e.\ real--valued parameters, is assumed) and passes the
information run-time rather than in a file. The 
mSUGRA model input parameters should be given in \ttt{RMSS} as for 
\ttt{IMSS(1)=2}, i.e.: \ttt{RMSS(1)}$= M_{1/2}$, \ttt{RMSS(4)}= \sgnmu, 
\ttt{RMSS(5)}$ = \tan\beta$, \ttt{RMSS(8)}$ = M_0$, and 
\ttt{RMSS(16)}$= A_0$. The routine \ttt{PYSUGI} handles the conversion 
between the conventions of {\Py} and \tsc{Isasusy}, so that conventions 
are self-consistent inside \Py. In the call to \ttt{PYSUGI}, 
the \ttt{RMSS} array is filled with the values produced by \tsc{Isasusy} 
as for \ttt{IMSS(1)=1}. In particular, this means that the 
mSUGRA input parameters mentioned above will be overwritten.
Cross sections and decay widths are then calculated by \Py. 
Since {\Py} cannot always be expected to be linked with \tsc{Isajet}, a 
dummy routine and a dummy function have been added to the {\Py} source. 
These are \ttt{SUBROUTINE SUGRA} and \ttt{FUNCTION VISAJE}. 
These must first be given other names and {\Py} recompiled before 
proper linking with \tsc{Isajet} can be achieved. 

A problem is that the size of some \tsc{Isasusy} common blocks has 
been expanded in more recent versions. Thus, starting with version 7.61
of that program, the \ttt{SSPAR} common block has been augmented by 3 
real-valued numbers, and the \ttt{SUGPAS} common block has been enlarged 
by 1 integer and 1 real-valued number. From version 7.67 on, the \ttt{SSPAR} 
common block is expanded by a further three numbers and the \ttt{GSS} 
array in \ttt{SUGMG} by two. Corresponding changes have been implemented 
in the \ttt{PYSUGI} interface routine. {\Py} thus now assumes the 
\ttt{SSPAR}, \ttt{SUGPAS} and \ttt{SUGMG} common blocks to have 
the forms:
\begin{verbatim}
      COMMON/SSPAR/AMGLSS,AMULSS,AMURSS,AMDLSS,AMDRSS,AMSLSS
     $,AMSRSS,AMCLSS,AMCRSS,AMBLSS,AMBRSS,AMB1SS,AMB2SS
     $,AMTLSS,AMTRSS,AMT1SS,AMT2SS,AMELSS,AMERSS,AMMLSS,AMMRSS
     $,AMLLSS,AMLRSS,AML1SS,AML2SS,AMN1SS,AMN2SS,AMN3SS
     $,TWOM1,RV2V1,AMZ1SS,AMZ2SS,AMZ3SS,AMZ4SS,ZMIXSS(4,4)
     $,AMW1SS,AMW2SS
     $,GAMMAL,GAMMAR,AMHL,AMHH,AMHA,AMHC,ALFAH,AAT,THETAT
     $,AAB,THETAB,AAL,THETAL,AMGVSS,MTQ,MBQ,MLQ,FBMA,
     $VUQ,VDQ
      REAL AMGLSS,AMULSS,AMURSS,AMDLSS,AMDRSS,AMSLSS
     $,AMSRSS,AMCLSS,AMCRSS,AMBLSS,AMBRSS,AMB1SS,AMB2SS
     $,AMTLSS,AMTRSS,AMT1SS,AMT2SS,AMELSS,AMERSS,AMMLSS,AMMRSS
     $,AMLLSS,AMLRSS,AML1SS,AML2SS,AMN1SS,AMN2SS,AMN3SS
     $,TWOM1,RV2V1,AMZ1SS,AMZ2SS,AMZ3SS,AMZ4SS,ZMIXSS
     $,AMW1SS,AMW2SS
     $,GAMMAL,GAMMAR,AMHL,AMHH,AMHA,AMHC,ALFAH,AAT,THETAT
     $,AAB,THETAB,AAL,THETAL,AMGVSS,MTQ,MBQ,MLQ,FBMA,VUQ,VDQ
      COMMON /SUGPAS/ XTANB,MSUSY,AMT,MGUT,MU,G2,GP,V,VP,XW,
     $A1MZ,A2MZ,ASMZ,FTAMZ,FBMZ,B,SIN2B,FTMT,G3MT,VEV,HIGFRZ,
     $FNMZ,AMNRMJ,NOGOOD,IAL3UN,ITACHY,MHPNEG,ASM3
      REAL XTANB,MSUSY,AMT,MGUT,MU,G2,GP,V,VP,XW,
     $A1MZ,A2MZ,ASMZ,FTAMZ,FBMZ,B,SIN2B,FTMT,G3MT,VEV,HIGFRZ,
     $FNMZ,AMNRMJ,ASM3
      INTEGER NOGOOD,IAL3UN,ITACHY,MHPNEG
      COMMON /SUGMG/ MSS(32),GSS(31),MGUTSS,GGUTSS,AGUTSS,FTGUT,
     $FBGUT,FTAGUT,FNGUT
      REAL MSS,GSS,MGUTSS,GGUTSS,AGUTSS,FTGUT,FBGUT,FTAGUT,FNGUT
\end{verbatim}
\tsc{Isasusy} users are warned to check that no incompatibilities arise 
between the versions actually used. Unfortunately there is no universal 
solution to this problem: the Fortran standard does not allow you 
dynamically to vary the size of a (named) common block. So if you use an 
earlier \tsc{Isasusy} version, you  have to shrink the size accordingly, 
and for a later you may have to check that the above common blocks
have not been expanded further. 

As a cross check, the option \ttt{IMSS(1)=2}
uses approximate analytical solutions of the renormalization
group equations \cite{Dre95}, which reproduce the output of
\tsc{Isasusy} within $\simeq 10\%$ (based on comparisons
of masses, decay widths, production cross sections, etc.). This option is
intended for debugging only, and does not represent the state--of--the--art.
 
In \tsc{Sugra} and in other models with the SUSY breaking scale of order
$M_{\mrm{GUT}}$, the spin--3/2 superpartner 
of the graviton, the gravitino $\gravitino$ (code 1000039), has
a mass of order $M_\W$ and interacts only gravitationally.
In models of gauge--mediated SUSY breaking \cite{Din96}, however,
the gravitino can play a crucial role in the phenomenology,
and can be the lightest superpartner (LSP). Typically, sfermions decay 
to fermions and gravitinos, and neutralinos, chargino, and gauginos 
decay to gauge or Higgs bosons and gravitinos.
Depending on the gravitino mass, the decay lengths can be substantial
on the scale of colliders.  {\Py} correctly handles finite decay lengths
for all sparticles.

R--parity is a possible symmetry of the SUSY Lagrangian that prevents
problems of rapid proton decay and allows for a viable dark matter
candidate. However, it is also possible to allow a restricted amount of
R--parity violation. At present, there is no theoretical consensus that
R--parity should be conserved, even in string models. 
In the production of superpartners, {\Py} assumes
R--parity conservation (at least on the time and distance
scale of a typical collider experiment), and only lowest order, 
sparticle pair production processes are included. Only those 
processes with $\e^+\e^-, \mu^+\mu^-$, or quark and gluon initial 
states are simulated. Tables~\ref{t:procfive}, \ref{t:procsix} and
\ref{t:procseven} list available SUSY processes. In processes 
210 and 213, $\sell$ refers to both $\se$ and $\smu$.  For ease of 
readability, we have removed the subscript $L$ on $\snu$.
$\tp_i\tm_i, \stau_i\stau_j^*$ and $\stau_i\snu_{\tau}^*$ 
production correctly account for sfermion mixing.  Several processes
are conspicuously absent from the table.  For example, processes
\ttt{255} and \ttt{257} would simulate the associated production 
of right handed squarks with charginos.  Since the right handed squark
only couples to the higgsino component of the chargino, the interaction
strength is proportional to the quark mass, so these processes can
be ignored.    

By default, only R--parity conserving decays are allowed, so that one 
sparticle is stable, either the lightest neutralino, the gravitino, 
or a sneutrino. SUSY decays of the top quark are included, but all 
other SM particle decays are unaltered.

Generally, the decays of the superpartners are calculated using the 
formulae of refs.~\cite{Gun88,Bar86a,Bar86b,Bar95}.  
All decays are spin averaged. Decays involving $\sbo$ and $\st$ 
use the formulae of \cite{Bar95}, so they are valid for large values of
$\tan\beta$.  The one loop decays $\chio_j\to\chio_i\gamma$ and 
$\st\to \c\chio_1$ are also included, but only with approximate formula.
Typically, these decays are only important when other decays are not allowed
because of mixing effects or phase space considerations.

One difference between the SUSY simulation and the other parts of 
the program is that it is not beforehand known which sparticles
may be stable. Normally this would mean either the $\chio^0_1$ 
or the gravitino $\grav$, but in principle also other sparticles could 
be stable. The ones found to be stable have their \ttt{MWID(KC)} and 
\ttt{MDCY(KC,1)} values set zero at initialization. If several
\ttt{PYINIT} calls are made in the same run, with different SUSY
parameters, the ones set zero above are not necessarily set
back to nonzero values, since most original values are not saved 
anywhere. As an exception to this rule, the \ttt{PYMSIN} SUSY 
initialization routine, called by \ttt{PYINIT}, does save and restore
the \ttt{MWID(KC)} and \ttt{MDCY(KC,1)} values of the lightest
SUSY particle. It is therefore possible to combine several \ttt{PYINIT}
calls in a single run, provided that only the lightest SUSY particle is 
stable. If this is not the case, \ttt{MWID(KC)} and \ttt{MDCY(KC,1)} 
values may have to be reset by hand, or else some particles that ought 
to decay will not do that. 

\subsubsection{SUSY examples}

The SUSY routines and common block variables are described in
section~\ref{ss:susycode}. To illustrate the usage of the switches and
parameters, we give five simple examples.

\textit{Example 1:  Light Stop}\\
The first example is an MSSM model with a light neutralino $\chio_1$ 
and a light stop $\st_1$, so that $\t \to \st_1\chio_1$ can occur.  
The input parameters are\\ 
\ttt{IMSS(1)=1}, \ttt{RMSS(1)=70.}, \ttt{RMSS(2)=70.}, 
\ttt{RMSS(3)=225.}, \ttt{RMSS(4)=-40.}, \ttt{RMSS(5)=1.5},\\
\ttt{RMSS(6)=100.}, \ttt{RMSS(7)=125.}, \ttt{RMSS(8)=250.}, 
\ttt{RMSS(9)=250.}, \ttt{RMSS(10)=1500.},\\ 
\ttt{RMSS(11)=1500.}, \ttt{RMSS(12)=-128.}, \ttt{RMSS(13)=100.}, 
\ttt{RMSS(14)=125.}, \ttt{RMSS(15)=800.},\\
\ttt{RMSS(16)=800.}, \ttt{RMSS(17)=0.}, and \ttt{RMSS(19)=400.0.}\\
The top mass is fixed at 175 GeV, \ttt{PMAS(6,1)=175.0}.
The resulting model has $M_{\st_1}=55$ GeV and $M_{\chio_1}=38$ GeV.
\ttt{IMSS(1)=1} turns on the MSSM simulation.
By default, there are no intrinsic relations between the gaugino masses,
so $M_1=70$ GeV, $M_2=70$ GeV, and $M_3=225$ GeV.  The pole mass of
the gluino is slightly higher than the parameter $M_3$, and the
decay $\glu \to\st_1^*\t+\st_1\tbar$ occurs almost 100\% of the time.

\textit{Example 2: SUSY Les Houches Accord spectrum}\\
The second example shows how to input a spectrum file in the SUSY Les
Houches Accord format \cite{Ska03} 
to {\Py}. First, you should set \ttt{IMSS(1)=11} and
open the spectrum file you want to use on some unused Logical Unit Number. 
Then, set \ttt{IMSS(21)} equal to that number, to tell {\Py} where to read
the spectrum file from. This should be done somewhere in your main program 
before calling \ttt{PYINIT}. During the call to \ttt{PYINIT}, {\Py} will read
the spectrum file, perform a number of consistency checks and issue
warning messages if it finds something it does not understand or which seems
inconsistent. E.g.~\ttt{BLOCK GAUGE} will normally be present in the
spectrum file, but since {\Py} currently cannot 
use the information in that block, it will issue a warning that the block
will be ignored. In case a decay table is also desired to be read in, the
Logical Unit Number on which the decay table is opened should be put in
\ttt{IMSS(22)}. To avoid inconsistencies, the spectrum and the
decay table should normally go together, so \ttt{IMSS(22)} should normally be
equal to \ttt{IMSS(21)}. 

\textit{Example 3: Calling \tsc{Isasusy 7.67} at runtime}\\
The third example shows how to use the built-in interface to \tsc{Isasusy}.
First, the {\Py} source code needs to be changed. Rename the function
\ttt{VISAJE} to, for example, \ttt{FDUMMY}, rename the subroutine
\ttt{SUGRA} to e.g.\ \ttt{SDUMMY}, and recompile. In the calling program, 
set \ttt{IMSS(1)=12} and the \ttt{RMSS} input parameters exactly as in
example 4, and compile the executable while linked to both \tsc{Isajet}
and the modified \Py. The resulting mass and mixing spectrum is printed 
in the {\Py} output. 

\textit{Example 4:  Approximate \tsc{SUGRA}}\\
This example shows you how to get a (very) approximate SUGRA model. Note that
this way of obtaining the SUSY spectrum should 
never be used for serious studies.  
The input parameters are\\ 
\ttt{IMSS(1)=2}, \ttt{RMSS(1)=200.}, \ttt{RMSS(4)=1.,} 
\ttt{RMSS(5)=10.}, \ttt{RMSS(8)=800.}, and\\ 
\ttt{RMSS(16)=0.0}.\\
The resulting model has
$M_{\sd_L}=901$ GeV, $M_{\su_R}=890$ GeV, $M_{\st_1}=538$ GeV,
$M_{\se_L}=814$ GeV, $M_{\glu}=560$ GeV, $M_{\chio_1}=80$ GeV,
$M_{\chip_1}=151$ GeV, $M_{\hrm}=110$ GeV, and $M_{A}=883$ GeV.  It 
corresponds to the choice $M_0$=800 GeV, $M_{1/2}=$200 GeV,
$\tan\beta=10$, $A_0=0$, and \sgnmu$>0$.  The output is similar
to an \tsc{Isasusy} run, but there is not exact agreement.

\textit{Example 5: \tsc{Isasusy 7.61} Model}\\
The final example demonstrates how to convert the output of
an \tsc{Isasusy} run directly into the {\Py} format, i.e.~if SUSY Les Houches
Accord output is not available.
This assumes that you already made an \tsc{Isasusy} run, e.g. with the
equivalents of the input parameters above. From the output of this run
you can now extract those physical parameters that need to be handed to
{\Py}, in the above example\\  
\ttt{IMSS(1)=1}, \ttt{IMSS(3)=1}, \ttt{IMSS(8)=0}, \ttt{IMSS(9)=1},
\ttt{RMSS(1)=79.61}, \ttt{RMSS(2)=155.51}, \\
\ttt{RMSS(3)=533.1}, \ttt{RMSS(4)=241.30}, \ttt{RMSS(5)=10.}, 
\ttt{RMSS(6)=808.0}, \ttt{RMSS(7)=802.8}, \\ 
\ttt{RMSS(8)=878.4}, \ttt{RMSS(9)=877.1}, \ttt{RMSS(10)=743.81}, 
\ttt{RMSS(11)=871.26}, \\ 
\ttt{RMSS(12)=569.87}, \ttt{RMSS(13)=803.20}, \ttt{RMSS(14)=794.71}, 
\ttt{RMSS(15)=-554.96}, \\ 
\ttt{RMSS(16)=-383.23}, \ttt{RMSS(17)=-126.11}, \ttt{RMSS(19)=829.94} 
and \ttt{RMSS(22)=878.5}.

\subsubsection{R--Parity Violation}
\label{sss:rparityviol} 

R--parity, defined as R=$(-1)^{2S+3B+L}$, is a discrete multiplicative
symmetry where $S$ is the particle spin, $B$ is the baryon number, and $L$ is
the lepton number.  All SM particles have R=1, while all superpartners have
R=$-1$, so a single SUSY particle cannot decay into just SM particles if
R--parity is conserved.  In this case, the lightest superpartner (LSP) is
absolutely stable.  Astrophysical considerations imply that a stable LSP
should be electrically neutral. Viable candidates are the lightest
neutralino, the lightest sneutrino, or alternatively the gravitino.  Since
the LSP can carry away energy without interacting in a detector, the apparent
violation of momentum conservation is an important part of SUSY
phenomenology.  Also, when R--parity is conserved, superpartners must be
produced in pairs from a SM initial state.  The breaking of the R--parity
symmetry would result in lepton and/or baryon number violating processes.
While there are strong experimental constraints on some classes of R--parity
violating interactions, others are hardly constrained at all.

One simple extension of the MSSM is to break the multiplicative
R--parity symmetry.
Presently, neither experiment nor any theoretical argument 
demand R--parity conservation, so it is natural to consider the
most general case of R--parity breaking.
It is convenient to introduce a function of superfields
called the superpotential, from which the Feynman rules for
R--parity violating processes can be derived.
The R--parity violating (RPV) terms 
which can contribute to the superpotential are:
\begin{equation}
W_{RPV} =  \lambda_{ijk} L^i L^j \bar{E}^k +
           \lambda^{'}_{ijk} L^i Q^j \bar{D}^k +
           \lambda^{''}_{ijk} \bar{U}^i \bar{D}^j \bar{D}^k + 
           \epsilon_i L_iH_2  
\label{eq:superpot}
\end{equation}
where $i,j,k$ are generation indices (1,2,3), $L^i_1 \equiv \nu^i_L$,
$L^i_2=\ell^i_L$ and $Q^i_1=u^i_{L}$, $Q^i_2=d^i_{L}$ are lepton and quark
components of ${\bf SU(2)_L}$ doublet superfields, and $E^i=e^i_{R}$,
$D^i=d^i_{R}$ and $U^i=u^i_R$ are lepton, down and up-- quark ${\bf SU(2)_L}$
singlet superfields, respectively.  The unwritten ${\bf SU(2)_L}$ and ${\bf
  SU(3)_C}$ 
indices imply that the first term is antisymmetric under $i \leftrightarrow
j$, and the third term is antisymmetric under $j \leftrightarrow
k$. Therefore, $i \neq j$ in $L^i L^j \bar{E}^k$ and $j \neq k$ in $\bar{U}^i
\bar{D}^j \bar{D}^k$.  The coefficients $\lambda_{ijk}$, $\lambda^{'}_{ijk}$,
$\lambda^{''}_{ijk}$, and $\epsilon_i$ are Yukawa couplings, and there is no
{\it a priori} generic prediction for their values.  In principle, $W_{RPV}$
contains 48 extra parameters over the $R$--parity--conserving MSSM case. In
{\Py} the effects of the last term in eq.~(\ref{eq:superpot}) are not
included.

Expanding eq.~(\ref{eq:superpot})
as a function of the superfield components, the interaction
Lagrangian derived from the first term is
\begin{equation}
{\cal{L}}_{LLE} = \lambda_{ijk} \left\{ \tilde{\nu}_L^i e_L^j \bar{e}^k_R + 
                \tilde{e}_L^i \nu_L^j \bar{e}^k_R + 
                (\tilde{e}_R^k)^* \nu_L^i e^j_L + h.c. \right\} 
\label{eq:RPVLLE}
\end{equation}
and from the second term,
\begin{eqnarray}
{\cal{L}}_{LQD} = \lambda^{'}_{ijk} \left\{
 \tilde{\nu}_L^i d_L^j \bar{d}^k_R -
                \tilde{e}_L^i u_L^j \bar{d}^k_R +
                \tilde{d}_L^j \nu_L^i \bar{d}^k_R -
                \tilde{u}_L^j e_L^i \bar{d}^k_R + \right. \nonumber \\
                \left. (\tilde{d}_R^k)^* \nu_L^i d^j_L - 
                (\tilde{d}_R^k)^* e_L^i u^j_L + h.c. \right\}
\end{eqnarray}
Both of these sets of interactions violate lepton number. 
The $\bar U\bar D\bar D$ term, instead, violates 
baryon number.
In principle, all types of R--parity violating terms may co--exist,
but this can lead to a proton 
with a lifetime shorter than the present experimental limits. 
The simplest way to avoid this
is to allow only operators which  conserve baryon--number but  violate
lepton--number or vice versa. 

There are several effects on the SUSY phenomenology due to these
new couplings: (1) lepton or baryon number violating processes
are allowed, including the
production of single sparticles (instead of pair production),
(2) the LSP is no longer stable, but
can decay to SM particles within a collider detector,
and 
(3) because it is unstable, 
the LSP need not be the neutralino or sneutrino, but
can be charged and/or colored.

In the current version of {\Py}, decays of supersymmetric particles
to SM particles via two different types of lepton number
violating couplings and one type of baryon number violating couplings 
can be invoked (Details about the $L$-violation implementation and
tests can be found in \cite{Ska01}). 

Complete matrix elements (including $L-R$ mixing for all sfermion 
generations) for all two-body sfermion and three-body neutralino,
chargino, and gluino decays are included
(as given in \cite{Dre00}). The final state fermions are treated as 
massive in the phase space integrations and in the matrix elements
for $\b$, $\t$, and $\tau$.

The existence of $R$--odd couplings also allows for single sparticle
production, i.e.\ there is no requirement that SUSY particles should be
produced in pairs. Single sparticle production cross sections are not yet
included in the program, and it may require some rethinking of the parton
shower to do so. For low-mass sparticles, the
associated error is estimated to be negligible, as long as the $R$--violating
couplings are smaller than the gauge couplings. For higher mass sparticles, 
the reduction of the 
phase space for pair production becomes an important factor, and single
sparticle production could dominate even for very small values of the
$R$-violating couplings. The total SUSY 
production cross sections, as calculated by {\Py} in its current form are
thus underestimated, possibly quite severely for heavy-mass sparticles.

Three possibilities exist for the initializations of the couplings,
representing a fair but not exhaustive range of models. The
first, selected by setting \ttt{IMSS(51)=1} for LLE, \ttt{IMSS(52)=1} for
LQD, and/or \ttt{IMSS(53)=1} for UDD type couplings, sets all the couplings, 
independent of generation, to a
common value of $10^{-\mbox{\scriptsize\ttt{RMSS(51)}}}$,
$10^{-\mbox{\scriptsize\ttt{RMSS(52)}}}$, and/or
$10^{-\mbox{\scriptsize\ttt{RMSS(53)}}}$, depending on which couplings are
activated.  

Taking now LLE couplings as an example, setting 
\ttt{IMSS(51)=2} causes the LLE couplings 
to be initialized (in \ttt{PYINIT}) to so-called `natural' 
generation-hierarchical values, as proposed in \cite{Hin93}. These values, 
inspired by the structure of the Yukawa couplings in the SM, are defined by:
\begin{equation}
\begin{array}{rcl}
  |\lambda_{ijk}|^2 & = & (\mbox{\ttt{RMSS(51)}})^
  2\hat{m}_{e_i}\hat{m}_{e_j}\hat{m}_{e_k}\\ 
  |\lambda'_{ijk}|^2 & = &
  (\mbox{\ttt{RMSS(52)}})^ 2 \hat{m}_{e_i} \hat{m}_{q_j}
  \hat{m}_{d_k}\\
 |\lambda''_{ijk}|^2 & = & (\mbox{\ttt{RMSS(53)}})^
  2\hat{m}_{q_i}\hat{m}_{q_j}\hat{m}_{q_k}\end{array} \hspace*{1cm} ; 
  \hspace*{0.7cm} \hat{m}\equiv
  \frac{m}{v} = \frac{m}{126\mrm{GeV}}
\label{eq:rvnatval}
\end{equation}
where $m_{q_i}$ is the arithmetic mean of $m_{u_i}$ and $m_{d_i}$. 

The third option available is to set \ttt{IMSS(51)=3}, \ttt{IMSS(52)=3},
and/or \ttt{IMSS(53)=3}, 
in which case all the relevant couplings are zero by default (but the
corresponding lepton or baryon number violating processes are turned on) 
and the user is expected to enter the non-zero coupling values by hand. 
(Where antisymmetry is required, half of the entries are automatically
derived from the other half, see \ttt{IMSS(51)=3} and \ttt{IMSS(53)=3}.)
\ttt{RVLAM($i$,$j$,$k$)} contains the $\lambda_{ijk}$,
\ttt{RVLAMP($i$,$j$,$k$)} contains the $\lambda'_{ijk}$ couplings, and
\ttt{RVLAMB($i$,$j$,$k$)} contains the $\lambda''_{ijk}$ couplings.

\subsection{Polarization}
\label{ss:polarization}

In most processes, incoming beams are assumed unpolarized. However,
especially for $\e^+\e^-$ linear collider studies, polarized beams 
would provide further important information on many new physics 
phenomena, and at times help to suppress backgrounds. Therefore a 
few process cross sections are now available
also for polarized incoming beams. The average polarization of the 
two beams is then set by \ttt{PARJ(131)} and \ttt{PARJ(132)}, 
respectively. In some cases, noted below, \ttt{MSTP(50)} need also 
be switched on to access the formulae for polarized beams. 

Process 25, $\W^+ \W^-$ pair production, allows polarized incoming lepton 
beam particles. The polarization effects are included both in the production 
matrix elements and in the angular distribution of the final four fermions. 
Note that the matrix element used \cite{Mah98} is for on-shell $\W$ 
production, with a suppression factor added for finite width effects. 
This polarized cross section expression, evaluated at vanishing polarization, 
disagrees with the standard unpolarized one, which presumably is the more 
accurate of the two. The difference can be quite significant below threshold 
and at very high energies. This can be traced to the simplified description
of off-shell $\W$'s in the polarized formulae. Good agreement is obtained 
either by switching off the $\W$ width with \ttt{MSTP(42)=0} or by 
restricting the $\W$ mass ranges (with \ttt{CKIN(41) - CKIN(44)}) to be 
close to on-shell. It is therefore necessary to set \ttt{MSTP(50)=1}
to switch from the default standard unpolarized formulae to the polarized 
ones. 

Also many SUSY production processes now include the effects from 
polarization of the incoming fermion beams. This applies for scalar pair 
production, with the exception of sneutrino pair production and 
$\hrm^0 \A^0$ and $\H^0 \A^0$ production, this omission being an oversight 
at the time of this release, but easily remedied in the future.

The effect of polarized photons is included in the process
$\gamma \gamma \to \F_k \Fbar_k$, process 85. Here the array values PARJ(131) 
and PARJ(132) are used to define the average longitudinal polarization of 
the two photons.

\subsection{Main Processes by Machine}

In the previous section we have already commented on which processes
have limited validity, or have different meanings (according to
conventional terminology) in different contexts. Let us just repeat 
a few of the main points to be remembered for different machines.

\subsubsection{$\ee$ collisions}

The main annihilation process is number 1, $\ee \to \Z^0$,
where in fact the full $\gamma^*/\Z^0$ interference structure is
included. This process can be used, with some confidence, for
c.m.\ energies from about 4 GeV upwards, i.e.\ at DORIS/CESR,
PETRA/PEP, TRISTAN, LEP, and any future linear colliders.
(To get below 10 GeV, you have to change \ttt{PARP(2)}, however.) 
This is the default process obtained when \ttt{MSEL=1}, i.e.\ 
when you do not change anything yourself. 

Process 141 contains a $\Z'^0$, including 
full interference with the standard $\gammaZ$. With the value 
\ttt{MSTP(44)=4} in fact one is back at the standard $\gammaZ$
structure, i.e.\ the $\Z'^0$ piece has been switched off. Even so,
this process may be useful, since it can simulate e.g.
$\ee \to \hrm^0 \A^0$. Since the $\hrm^0$ may in its turn decay to
$\Z^0 \Z^0$, a decay channel of the ordinary $\Z^0$ to 
$\hrm^0 \A^0$, although physically correct, would be technically
confusing. In particular, it would be messy to set the original 
$\Z^0$ to decay one way and the subsequent ones another. So, in 
this sense, the $\Z'^0$ could be used as a copy of the ordinary
$\Z^0$, but with a distinguishable label.

The process $\ee \to \Upsilon$ does not exist as a separate process
in {\Py}, but can be simulated by using \ttt{PYONIA}, see section
\ref{ss:oniadecays}.

At LEP 2 and even higher energy machines, the simple $s$-channel
process 1 loses out to other processes, such as 
$\ee \to \Z^0 \Z^0$ and $\ee \to \W^+ \W^-$, i.e.\ processes 
22 and 25. The former process in fact includes the structure
$\ee \to (\gammaZ)(\gammaZ)$, which means that the cross section 
is singular if either of the two $\gammaZ$ masses is allowed to 
vanish. A mass cut therefore needs to be introduced, and is 
actually also used in other processes, such as $\ee \to \W^+ \W^-$.
  
For practical
applications, both with respect to cross sections and to event
shapes, it is imperative to include initial-state radiation effects. 
Therefore \ttt{MSTP(11)=1} is the default, wherein exponentiated
electron-inside-electron distributions are used to give the
momentum of the actually interacting electron. By radiative 
corrections to process 1, such processes as $\ee \to \gamma \Z^0$
are therefore automatically generated. If process 19 were to be
used at the same time, this would mean that radiation were to be
double-counted. In the alternative \ttt{MSTP(11)=0}, electrons are 
assumed to deposit their full energy in 
the hard process, i.e.\ initial-state QED radiation is not included.
This option is very useful, since it often corresponds to the
`ideal' events that one wants to correct back to. 

Resolved electrons also means that one may have interactions
between photons. This opens up the whole field of $\gamma\gamma$
processes, which is described in section \ref{ss:photoanddisclass}.
In particular, with \ttt{'gamma/e+','gamma/e-'} as beam and target
particles in a \ttt{PYINIT} call, a flux of photons of different
virtualities is convoluted with a description of direct and resolved
photon interaction processes, including both low-$\pT$ and 
high-$\pT$ processes. This machinery is directed to the description
of the QCD processes, and does e.g.\ not address the production of
gauge bosons or other such particles by the interactions of resolved
photons. For the latter kind of applications, a simpler description
of partons inside photons inside electrons may be obtained with the 
\ttt{MSTP(12)=1} options and $\e^{\pm}$ as beam and target particles. 

The thrust of the {\Py} programs is towards processes that involve
hadron production, one way or another. Because of generalizations
from other areas, also a few completely
non-hadronic processes are available. These include Bhabha
scattering, $\ee \to \ee$ in process 10, and photon pair production,
$\ee \to \gamma \gamma$ in process 18. However, note that the 
precision that could be expected in a {\Py} simulation of those
processes is certainly far less than that of dedicated programs.
For one thing, electroweak loop effects are not included.
For another, nowhere is the electron mass taken into account,
which means that explicit cut-offs at some minimum $\pT$ are always
necessary. 

\subsubsection{Lepton--hadron collisions}

The main option for photoproduction and Deeply Inelastic Scattering
(DIS) physics is provided by the {\galep} option as beam 
or target in a \ttt{PYINIT} call, see section 
\ref{ss:photoanddisclass}. The $Q^2$ range to be covered, and other 
kinematics constraints, can be set by \ttt{CKIN} values. By default, 
when the whole  $Q^2$ range is populated, obviously photoproduction 
dominates.

The older DIS process 10, $\ell \q \to \ell' \q'$, includes
$\gamma^0/\Z^0/\W^{\pm}$ exchange, with full interference, as
described in section \ref{sss:DISclass}. The $\Z^0/\W^{\pm}$
contributions are not implemented in the {\galep} 
machinery. Therefore process 10 is still the main option for 
physics at very high $Q^2$, but has been superseded for lower $Q^2$.
Radiation off the incoming lepton leg is included by \ttt{MSTP(11)=1} 
and off the outgoing one by \ttt{MSTJ(41)=2} (both are default). Note 
that both QED and QCD radiation (off the $\e$ and the $\q$ legs, 
respectively) are allowed to modify the $x$ and $Q^2$ values of the 
process, while the conventional approach in the literature is to allow 
only the former. Therefore an option (on by default) has been added to 
preserve these values by a post-facto rescaling, \ttt{MSTP(23)=1}.  
Further comments on HERA applications are found in \cite{Sjo92b}.

\subsubsection{Hadron--hadron collisions}

The default is to include QCD jet production by $2 \to 2$ processes,
see section \ref{sss:QCDjetclass}. Since the differential 
cross section is divergent for $\pT \to 0$, a lower cut-off has to
be introduced. Normally that cut-off is given by the user-set 
$\pTmin$ value in \ttt{CKIN(3)}. If \ttt{CKIN(3)} is chosen
smaller than a given value of the order of 2 GeV (see \ttt{PARP(81)}
and \ttt{PARP(82)}), then low-$\pT$ events are also switched on. 
The jet cross section is regularized at low $\pT$, so as to 
obtain a smooth joining between the high-$\pT$ and the low-$\pT$
descriptions, see further section \ref{ss:multint}. 
As \ttt{CKIN(3)} is varied, the jump from one 
scenario to another is abrupt, in terms of cross section: in a 
high-energy hadron collider, the cross section for jets down to 
a $\pTmin$ scale of a few GeV can well reach values much 
larger than the total inelastic, non-diffractive cross section.
Clearly this is nonsense; therefore either $\pTmin$ should 
be picked so large that the jet cross section be only a fraction
of the total one, or else one should select $\pTmin = 0$ and 
make use of the full description.

If one switches to \ttt{MSEL=2}, also elastic and diffractive 
processes are switched on, see section \ref{sss:minbiasclass}.
However, the simulation of these processes is fairly primitive,
and should not be used for dedicated studies,
but only to estimate how much they may contaminate the class of
non-diffractive minimum bias events.

Most processes can be simulated in hadron colliders, since the 
bulk of {\Py} processes can be initiated by quarks or gluons.
However, there are limits. Currently we include no photon
or lepton parton distributions, which means that a process like
$\gamma \q \to \gamma \q$ is not accessible. Further, the
possibility of having $\Z^0$ and $\W^{\pm}$ interacting in 
processes such as 71--77 has been hardwired process by process,
and does not mean that there is a generic treatment of $\Z^0$ and 
$\W^{\pm}$ distributions.

The emphasis in the hadron--hadron process description is on high
energy hadron colliders. The program can be used also at 
fixed-target energies, but the multiple interaction model for 
underlying events then breaks down and should not be used. The 
limit of applicability is somewhere at around 100 GeV. Only with 
the simpler model obtained for \ttt{MSTP(82)=1} can one go 
arbitrarily low.
 
\clearpage
 
\section{The Process Generation Program Elements}
 
In the previous two sections, the physics processes and the 
event-generation schemes of {\Py} have been presented. Here, finally,
the event-generation routines and the common block variables are
described. However, routines and variables related to initial- and
final-state showers, beam remnants and underlying events, and
fragmentation and decay are relegated to subsequent sections on
these topics. 
 
In the presentation in this section, information less important for
an efficient use of {\Py} has been put closer to the end.
We therefore begin with the main event generation routines, and
follow this by the main common block variables.
 
It is useful to distinguish three phases in a normal run with {\Py}.
In the first phase, the initialization, the general character of the
run is determined. At a minimum, this requires the specification of the
incoming hadrons and the energies involved. At the discretion of the
user, it is also possible to select specific final states, and to make
a number of decisions about details in the subsequent generation.
This step is finished by a \ttt{PYINIT} call, at which time several
variables are initialized in accordance with the values set. The
second phase consists of the main loop over the number of events,
with each new event being generated by a \ttt{PYEVNT} call. This event
may then be analysed, using information stored in some common blocks,
and the statistics accumulated. In the final phase, results are
presented. This may often be done without the invocation of any {\Py}
routines. From \ttt{PYSTAT}, however, it is possible to obtain a
useful list of cross sections for the different subprocesses.

\subsection{The Main Subroutines}
\label{ss:PYTmainroutines}
 
There are two routines that you must know: \ttt{PYINIT} for
initialization and \ttt{PYEVNT} for the subsequent generation of
each new event. In addition, the cross section and other kinds of
information available with \ttt{PYSTAT} are frequently useful. The
other two routines described here, \ttt{PYFRAM} and \ttt{PYKCUT}, 
are of more specialized interest.
 
\drawbox{CALL PYINIT(FRAME,BEAM,TARGET,WIN)}\label{p:PYINIT}
\begin{entry}
\itemc{Purpose:} to initialize the generation procedure. Normally it 
is foreseen that this call will be followed by many \ttt{PYEVNT} ones,
to generate a sample of the event kind specified by the \ttt{PYINIT}
call. (For problems with cross section estimates in runs of very few
events per \ttt{PYINIT} call, see the description for \ttt{PYSTAT(1)} 
below in this subsection.)
\iteme{FRAME :} a character variable used to specify the frame of the
experiment. Upper-case and lower-case letters may be freely mixed.
\begin{subentry}
\iteme{= 'CMS' :} colliding beam experiment in c.m.\ frame, with beam
momentum in $+z$ direction and target momentum in $-z$ direction.
\iteme{= 'FIXT' :} fixed-target experiment, with beam particle
momentum pointing in $+z$ direction.
\iteme{= '3MOM' :} full freedom to specify frame by giving beam
momentum in \ttt{P(1,1)}, \ttt{P(1,2)} and \ttt{P(1,3)} and target
momentum in \ttt{P(2,1)}, \ttt{P(2,2)} and \ttt{P(2,3)} in
common block \ttt{PYJETS}. Particles are assumed on the mass shell,
and energies are calculated accordingly.
\iteme{= '4MOM' :} as \ttt{'3MOM'}, except also energies should be
specified, in \ttt{P(1,4)} and \ttt{P(2,4)}, respectively. The
particles need not be on the mass shell; effective masses are 
calculated from energy and momentum. (But note that numerical
precision may suffer; if you know the masses the option \ttt{'5MOM'}
below is preferable.)
\iteme{= '5MOM' :} as \ttt{'3MOM'}, except also energies and masses
should be specified, i.e the full momentum information in
\ttt{P(1,1) - P(1,5)} and \ttt{P(2,1) - P(2,5)} should be given for
beam and target, respectively. Particles need not be on the mass
shell. Space-like virtualities should be stored as $-\sqrt{-m^2}$.
Especially useful for physics with virtual photons. (The virtuality 
could be varied from one event to the next, but then it is convenient 
to initialize for the lowest virtuality likely to be encountered.)
Four-momentum and mass information must match. 
\iteme{= 'USER' :} a run primarily intended to involve external,
user-defined processes, see subsection \ref{ss:PYnewproc}. Information
on incoming beam particles and energies is read from the \ttt{HEPRUP}
common block. In this option, the \ttt{BEAM}, \ttt{TARGET} and
\ttt{WIN} arguments are dummy.
\iteme{= 'NONE' :} there will be no initialization of any
processes, but only of resonance widths and a few other
process-independent variables. Subsequent to such a call,
\ttt{PYEVNT} cannot be used to generate events, so this option
is mainly intended for those who will want to construct their own
events afterwards, but still want to have access to some of the
{\Py} facilities. In this option, the \ttt{BEAM}, \ttt{TARGET} and
\ttt{WIN} arguments are dummy.
\end{subentry}
 
\iteme{BEAM, TARGET :} character variables to specify beam and
target particles. Upper-case and lower-case letters may be freely
mixed. An antiparticle can be denoted by `bar' at the end of the name
(`$\sim$' is a valid alternative for reasons of backwards compatibility). 
It is also possible to leave out the underscore (`\_') directly after `nu' 
in neutrino names, and the charge for proton and neutron. The arguments 
are dummy when the \ttt{FRAME} argument above is either \ttt{'USER'} or
\ttt{'NONE'}.
\begin{subentry}
\iteme{= 'e-' :} electron.
\iteme{= 'e+' :} positron.
\iteme{= 'nu\_e' :} $\nu_{\e}$.
\iteme{= 'nu\_ebar' :} $\br{\nu}_{\e}$.
\iteme{= 'mu-' :} $\mu^-$.
\iteme{= 'mu+' :} $\mu^+$.
\iteme{= 'nu\_mu' :} $\nu_{\mu}$.
\iteme{= 'nu\_mubar' :} $\br{\nu}_{\mu}$.
\iteme{= 'tau-' :} $\tau^-$.
\iteme{= 'tau+' :} $\tau^+$.
\iteme{= 'nu\_tau' :} $\nu_{\tau}$.
\iteme{= 'nu\_taubar' :} $\br{\nu}_{\tau}$.
\iteme{= 'gamma' :} photon (real, i.e.\ on the mass shell).
\iteme{= 'gamma/e-' :} photon generated by the virtual-photon 
flux in an electron beam; \ttt{WIN} below refers to electron,
while photon energy and virtuality varies between events according 
to what is allowed by \ttt{CKIN(61) - CKIN(78)}.
\iteme{= 'gamma/e+' :} as above for a positron beam.
\iteme{= 'gamma/mu-' :} as above for a $\mu^-$ beam.
\iteme{= 'gamma/mu+' :} as above for a $\mu^+$ beam.
\iteme{= 'gamma/tau-' :} as above for a $\tau^-$ beam.
\iteme{= 'gamma/tau+' :} as above for a $\tau^+$ beam.
\iteme{= 'pi0' :} $\pi^0$.
\iteme{= 'pi+' :} $\pi^+$.
\iteme{= 'pi-' :} $\pi^-$.
\iteme{= 'n0' :} neutron.
\iteme{= 'nbar0' :} antineutron.
\iteme{= 'p+' :} proton.
\iteme{= 'pbar-' :} antiproton.
\iteme{= 'K+' :} $\K^+$ meson; since parton distributions for strange
hadrons are not available, very simple and untrustworthy recipes are 
used for this and subsequent hadrons, see subsection 
\ref{ss:structfun}.
\iteme{= 'K-' :} $\K^-$ meson.
\iteme{= 'KS0' :} $\K_{\mrm{S}}^0$ meson.
\iteme{= 'KL0' :} $\K_{\mrm{L}}^0$ meson.
\iteme{= 'Lambda0' :} $\Lambda$ baryon.
\iteme{= 'Sigma-' :} $\Sigma^-$ baryon.
\iteme{= 'Sigma0' :} $\Sigma^0$ baryon.
\iteme{= 'Sigma+' :} $\Sigma^+$ baryon.
\iteme{= 'Xi-' :} $\Xi^-$ baryon.
\iteme{= 'Xi0' :} $\Xi^0$ baryon.
\iteme{= 'Omega-' :} $\Omega^-$ baryon.
\iteme{= 'pomeron' :} the pomeron $\pomeron$; since pomeron parton
distribution functions have not been defined this option can not
be used currently.
\iteme{= 'reggeon' :} the reggeon $\reggeon$, with comments as for 
the pomeron above.
\end{subentry}
 
\iteme{WIN :} related to energy of system, exact meaning depends on
\ttt{FRAME}.
\begin{subentry}
\iteme{FRAME='CMS' :} total energy of system (in GeV).
\iteme{FRAME='FIXT' :} momentum of beam particle (in GeV/$c$).
\iteme{FRAME='3MOM', '4MOM', '5MOM' :} dummy (information is taken from 
the \ttt{P} vectors, see above).
\iteme{FRAME='USER' :} dummy (information is taken from the \ttt{HEPRUP} 
common block, see above).
\iteme{FRAME='NONE' :} dummy (no information required).
\end{subentry}
\end{entry}
 
\drawbox{CALL PYEVNT}\label{p:PYEVNT}
\begin{entry}
\itemc{Purpose:} to generate one event of the type specified by the
\ttt{PYINIT} call. (This is the main routine, which calls a number
of other routines for specific tasks.)
\end{entry}
 
\drawbox{CALL PYSTAT(MSTAT)}\label{p:PYSTAT}
\begin{entry}
\itemc{Purpose:} to print out cross-sections statistics, decay
widths, branching ratios, status codes and parameter values.
\ttt{PYSTAT} may be called at any time, after the \ttt{PYINIT} call,
e.g.\ at the end of the run, or not at all.
 
\iteme{MSTAT :} specification of desired information.
\begin{subentry}
\iteme{= 1 :} prints a table of how many events of the different
kinds that have been generated and the corresponding
cross sections. All numbers already include the effects of cuts
required by you in \ttt{PYKCUT}.\\ 
At the bottom of the listing is also given the total number of warnings 
and errors in the current run. (These numbers are reset at each 
\ttt{PYINIT} call.) By default only the ten first warnings and errors 
are written explicitly; here one may easily see whether many further 
occured but were not written in the output. The final number is the 
fraction of events that have failed the fragmentation cuts, i.e.\ where, 
for one reason or another, the program has had problems fragmenting the 
system and has asked for a new hard subprocess.\\ 
Note that no errors are given on the cross sections. In most cases a 
cross section is obtained by Monte Carlo integration during the course 
of the run. (Exceptions include e.g. total and elastic hadron--hadron 
cross sections, which are parameterized and thus known from the very 
onset.) A rule of thumb would then be that the statistical error of a 
given subprocess scales like $\delta \sigma / \sigma \approx 1/\sqrt{n}$, 
where $n$ is the number of events generated of this kind. In principle, 
the numerator of this relation could be decreased by making use of the 
full information accumulated during the run, i.e. also on the cross 
section in those phase space points that are eventually rejected. This
is actually the way the cross section itself is calculated. However, once 
you introduce further cuts so that only some fraction of the generated 
events survive to the final analysis, you would be back to the simple 
$1/\sqrt{n}$ scaling rule for that number of surviving events. Statistical
errors are therefore usually better evaluated within the context of a 
specific analysis. Furthermore, systematic errors often dominate over the 
statistical ones.\\ 
Also note that runs with very few events, in addition to having large 
errors, tend to have a bias towards overestimating the cross sections. 
In a typical case, the average cross section obtained with many runs of 
only one event each may be twice that of the correct answer of a single run 
with many events. The reason is a `quit while you are ahead' phenomenon, 
that an upwards fluctuation in the differential cross section in an early 
try gives an acceptable event and thus terminates the run, while a downwards 
one leads to rejection and a continuation of the run. 
\iteme{= 2 :} prints a table of the resonances defined in the
program, with their particle codes ({\KF}), and all allowed decay
channels. (If the number of generations in \ttt{MSTP(1)} is 3,
however, channels involving fourth-generation particles are not
displayed.) For each decay channel is shown the sequential channel
number (IDC) of the {\Py} decay tables, the decay products (usually
two but sometimes three), the partial decay width,
branching ratio and effective branching ratio (in the event
some channels have been excluded by you).
\iteme{= 3 :} prints a table with the allowed hard interaction
flavours \ttt{KFIN(I,J)} for beam and target particles.
\iteme{= 4 :} prints a table of the kinematical cuts \ttt{CKIN(I)}
set by you in the current run.
\iteme{= 5 :} prints a table with all the values of the status codes
\ttt{MSTP(I)} and the parameters \ttt{PARP(I)} used in the current
run.
\iteme{= 6 :} prints a table of all subprocesses implemented in the
program.
\iteme{= 7 :} prints two tables related to $R$--violating supersymmetry,
where lepton and/or baryon number is not conserved. The first is a 
collection of semi-inclusive branching ratios where the entries have 
a form like \ttt{\~{}chi\_10 --> nu + q + q}, where a sum has 
been performed over all lepton and quark flavours. In the rightmost 
column of the table, the number of modes that went into the sum is given.
The purpose of this table is to give a quick overview of the branching
fractions, since there are currently more than 1500 individual
$R$--violating processes
included in the generator. Note that only the pure $1\to 3$ parts of the
3-body modes are included in this sum. If a process can also proceed via
two successive $1\to 2$ branchings (i.e.\ the intermediate resonance is on
shell) the product of these branchings should be added to the number given
in this table. A small list at the bottom of the table shows 
the total number of $R$--violating processes in the generator, 
the number with non-zero branching ratios in the current run, and the
number with branching ratios larger than $10^{-3}$.
The second table which is printed by this call merely 
lists the $R$--violating $\lambda$, $\lambda'$, and $\lambda''$ couplings. 
\end{subentry}
\end{entry}
 
\drawbox{CALL PYFRAM(IFRAME)}\label{p:PYFRAM}
\begin{entry}
\itemc{Purpose:} to transform an event listing between different 
reference frames, if so desired. The use of this routine assumes 
you do not do any boosts yourself.
\iteme{IFRAME :} specification of frame the event is to be
boosted to.
\begin{subentry}
\iteme{= 1 :} frame specified by you in the \ttt{PYINIT} call.
\iteme{= 2 :} c.m.\ frame of incoming particles.
\iteme{= 3 :} hadronic c.m.\ frame of lepton--hadron interaction 
events. Mainly intended for Deeply Inelastic Scattering, but can also
be used in photoproduction. Is not guaranteed to work with the
{\galep} options, however, and so of limited use. Note that both the 
lepton and any photons radiated off the lepton remain in the event 
listing, and have to be removed separately if you only want to study 
the hadronic subsystem.
\end{subentry}
\end{entry}
 
\drawbox{CALL PYKCUT(MCUT)}\label{p:PYKCUT}
\begin{entry}
\itemc{Purpose:} to enable you to reject a given set of
kinematic variables at an early stage of the generation procedure
(before evaluation of cross sections), so as not to spend unnecessary
time on the generation of events that are not wanted.
The routine will not be called unless you require is by
setting \ttt{MSTP(141)=1}, and never if `minimum-bias'-type events
(including elastic and diffractive scattering) are to be generated
as well. Furthermore it is never called for user-defined external 
processes. A dummy routine \ttt{PYKCUT} is included in the program 
file, so as to avoid unresolved external references when the routine 
is not used.
\iteme{MCUT :} flag to signal effect of user-defined cuts.
\begin{subentry}
\iteme{= 0 :} event is to be retained and generated in full.
\iteme{= 1 :} event is to be rejected and a new one generated.
\end{subentry}
\itemc{Remark :} at the time of selection, several variables in the
\ttt{MINT} and \ttt{VINT} arrays in the \ttt{PYINT1} common block
contain information that can be used to make the decision. The routine
provided in the program file explicitly reads the variables that
have been defined at the time \ttt{PYKCUT} is called, and also
calculates some derived quantities. The information available
includes subprocess type {\ISUB}, $E_{\mrm{cm}}$, $\hat{s}$, $\hat{t}$,
$\hat{u}$, $\hat{p}_{\perp}$, $x_1$, $x_2$, $x_{\mrm{F}}$, $\tau$, $y$,
$\tau'$, $\cos\hat{\theta}$, and a few more. Some of these may not
be relevant for the process under study, and are then set to zero.
\end{entry}
 
\subsection{Switches for Event Type and Kinematics Selection}
\label{ss:PYswitchkin}
 
By default, if {\Py} is run for a hadron collider,
only QCD $2 \to 2$ processes are generated,
composed of hard interactions above $\pTmin=$\ttt{PARP(81)},
with low-$\pT$ processes added on so as to give the full
(parameterized) inelastic, non-diffractive cross section.
In an $\ee$ collider, $\gammaZ$ production is the default, and in
an $\ep$ one it is Deeply Inelastic Scattering. With the help of the
common block \ttt{PYSUBS}, it is possible to select the generation
of another process, or combination of processes. It is also allowed
to restrict the generation to specific incoming partons/particles
at the hard interaction. This often automatically also restricts
final-state flavours but, in processes such as resonance production
or QCD/QED production of new flavours, switches in the {\Py}
program may be used to this end; see section \ref{ss:parapartdat}.
 
The \ttt{CKIN} array may be used to impose specific kinematics cuts.
You should here be warned that, if kinematical variables are
too strongly restricted, the generation time per event may become
very long. In extreme cases, where the cuts effectively close the
full phase space, the event generation may run into an infinite
loop. The generation of $2 \to 1$ resonance production is performed
in terms of the $\hat{s}$ and $y$ variables, and so the ranges
\ttt{CKIN(1) - CKIN(2)} and \ttt{CKIN(7) - CKIN(8)} may be
arbitrarily restricted without a significant loss of speed.
For $2 \to 2$ processes, $\cos\hat{\theta}$ is added as a third
generation variable, and so additionally the range
\ttt{CKIN(27) - CKIN(28)} may be restricted without any loss of
efficiency..

Effects from initial- and final-state radiation
are not included, since they are not known at the time the 
kinematics at the hard interaction is selected. The sharp 
kinematical cut-offs that can be imposed on the generation
process are therefore smeared, both by QCD radiation and by
fragmentation. A few examples of such effects follow.
\begin{Itemize}
\item Initial-state radiation implies that each of the two incoming
partons has a non-vanishing $\pT$ when they interact. The hard
scattering subsystem thus receives a net transverse boost,
and is rotated with respect to the beam directions.
In a $2 \to 2$ process, what typically happens is that one of the
scattered partons receives an increased $\pT$, while the $\pT$
of the other parton is reduced. 
\item Since the initial-state radiation
machinery assigns space-like virtualities to the incoming partons,
the definitions of $x$ in terms of energy fractions and in terms of
momentum fractions no longer coincide, and so the interacting
subsystem may receive a net longitudinal boost compared with 
na\"{\i}ve expectations, as part of the parton-shower machinery. 
\item Initial-state radiation gives rise to 
additional jets, which in extreme cases may be mistaken for either
of the jets of the hard interaction.
\item Final-state radiation gives rise to additional jets, which 
smears the meaning of the basic $2 \to 2$ scattering. The assignment
of soft jets is not unique. The energy of a jet becomes dependent
on the way it is identified, e.g.\ what jet cone size is used. 
\item The beam remnant description assigns primordial $k_{\perp}$
values, which also gives a net $\pT$ shift of the hard-interaction 
subsystem; except at low energies this effect is overshadowed by
initial-state radiation, however. Beam remnants may also add 
further activity under the `perturbative' event.
\item Fragmentation will further broaden jet profiles, and make
jet assignments and energy determinations even more uncertain.   
\end{Itemize}
In a study of events within a given window of
experimentally defined variables, it is up to you to leave
such liberal margins that no events are missed. In other words, cuts
have to be chosen such that a negligible fraction of events migrate
from outside the simulated region to inside the interesting region.
Often this may lead to low efficiency in terms of what fraction of
the generated events are actually of interest to you. 
See also section \ref{ss:PYTstarted}.
 
In addition to the variables found in \ttt{PYSUBS}, also those in the
\ttt{PYPARS} common block may be used to select exactly what one wants
to have simulated. These possibilities will be described in the
following section.
 
The notation used above and in the following is that `$\hat{~}$'
denotes internal variables in the hard scattering subsystem,
while `$^*$' is for variables in the c.m.\ frame of the
event as a whole. 
 
\drawbox{COMMON/PYSUBS/MSEL,MSELPD,MSUB(500),KFIN(2,-40:40),CKIN(200)}%
\label{p:PYSUBS}
\begin{entry}
\itemc{Purpose:} to allow you to run the program with any desired
subset of processes, or restrict flavours or kinematics. If the
default values, denoted below by (D=\ldots), are not satisfactory,
they must be changed before the \ttt{PYINIT} call.
\boxsep
 
\iteme{MSEL :}\label{p:MSEL} (D=1) a switch to select between full 
user control and some preprogrammed alternatives.
\begin{subentry}
\iteme{= 0 :} desired subprocesses have to be switched on in
\ttt{MSUB}, i.e.\ full user control.
\iteme{= 1 :} depending on incoming particles, different
alternatives are used. \\
Lepton--lepton: $\Z$ or $\W$ production ({\ISUB} = 1 or 2).  \\
Lepton--hadron: Deeply Inelastic Scattering ({\ISUB} = 10; this option 
is now out of date for most applications, superseded by the 
{\galep} machinery). \\
Hadron--hadron: QCD high-$\pT$ processes ({\ISUB} = 11, 12, 13, 28,
53, 68); additionally low-$\pT$ production if
\ttt{CKIN(3)}$<$\ttt{PARP(81)} or \ttt{PARP(82)}, depending on
\ttt{MSTP(82)} ({\ISUB} = 95). If low-$\pT$ is switched on, the other
\ttt{CKIN} cuts are not used. \\
A resolved photon counts as hadron. When the photon is not resolved, 
the following cases are possible. \\
Photon--lepton: Compton scattering ({\ISUB} = 34). \\
Photon--hadron: photon-parton scattering ({\ISUB} = 33, 34, 54). \\
Photon--photon: fermion pair production ({\ISUB} = 58).\\
When photons are given by the {\galep} argument in the
\ttt{PYINIT} call, the outcome depends on the \ttt{MSTP(14)} value.
Default is a mixture of many kinds of processes, as described
in section \ref{ss:photoanddisclass}.
\iteme{= 2 :} as \ttt{MSEL = 1} for lepton--lepton, lepton--hadron
and unresolved photons. For hadron--hadron (including resolved 
photons) all QCD processes, including low-$\pT$, single and
double diffractive and elastic scattering, are included ({\ISUB} =
11, 12, 13, 28, 53, 68, 91, 92, 93, 94, 95). The \ttt{CKIN} cuts are
here not used.\\ 
For photons given with the {\galep} argument in the 
\ttt{PYINIT} call, the above processes are replaced by other ones 
that also include the photon virtuality in the cross sections. The
principle remains to include both high- and low-$\pT$ processes,
however. 
\iteme{= 4 :} charm ($\c\cbar$) production with massive matrix
elements ({\ISUB} = 81, 82, 84, 85).
\iteme{= 5 :} bottom ($\b\bbar$) production with massive matrix
elements ({\ISUB} = 81, 82, 84, 85).
\iteme{= 6 :} top ($\t\tbar$) production with massive matrix
elements ({\ISUB} = 81, 82, 84, 85).
\iteme{= 7 :} fourth generation $\b'$ ($\b'\bbar'$) production with
massive matrix elements ({\ISUB} = 81, 82, 84, 85).
\iteme{= 8 :} fourth generation $\t'$ ($\t'\tbar'$) production with 
massive matrix elements ({\ISUB} = 81, 82, 84, 85).
\iteme{= 10 :} prompt photons ({\ISUB} = 14, 18, 29).
\iteme{= 11 :} $\Z^0$ production ({\ISUB} = 1).
\iteme{= 12 :} $\W^{\pm}$ production ({\ISUB} = 2).
\iteme{= 13 :} $\Z^0$ + jet production ({\ISUB} = 15, 30).
\iteme{= 14 :} $\W^{\pm}$ + jet production ({\ISUB} = 16, 31).
\iteme{= 15 :} pair production of different combinations of $\gamma$,
$\Z^0$ and $\W^{\pm}$ (except $\gamma\gamma$; see \ttt{MSEL} = 10)
({\ISUB} = 19, 20, 22, 23, 25).
\iteme{= 16 :} $\hrm^0$ production ({\ISUB} = 3, 102, 103, 123, 124).
\iteme{= 17 :} $\hrm^0 \Z^0$ or $\hrm^0 \W^{\pm}$ ({\ISUB} = 24, 26).
\iteme{= 18 :} $\hrm^0$ production, combination relevant for $\ee$
annihilation ({\ISUB} = 24, 103, 123, 124).
\iteme{= 19 :} $\hrm^0$, $\H^0$ and $\A^0$ production, excepting pair
production ({\ISUB} = 24, 103, 123, 124, 153, 158, 171, 173, 174, 176,
178, 179).
\iteme{= 21 :} $\Z'^0$ production ({\ISUB} = 141).
\iteme{= 22 :} $\W'^{\pm}$ production ({\ISUB} = 142).
\iteme{= 23 :} $\H^{\pm}$ production ({\ISUB} = 143).
\iteme{= 24 :} $\R^0$ production ({\ISUB} = 144).
\iteme{= 25 :} $\L_{\Q}$ (leptoquark) production ({\ISUB} = 145, 162,
163, 164).
\iteme{= 35 :} single bottom production by $\W$ exchange ({\ISUB} = 83).
\iteme{= 36 :} single top production by $\W$ exchange ({\ISUB} = 83).
\iteme{= 37 :} single $\b'$ production by $\W$ exchange ({\ISUB} = 83).
\iteme{= 38 :} single $\t'$ production by $\W$ exchange ({\ISUB} = 83).
\iteme{= 39 :} all MSSM processes except Higgs production.
\iteme{= 40 :} squark and gluino production ({\ISUB} = 243, 244, 258, 259, 
271--280).
\iteme{= 41 :} stop pair production ({\ISUB} = 261--265).
\iteme{= 42 :} slepton pair production ({\ISUB} = 201--214).
\iteme{= 43 :} squark or gluino with chargino or neutralino, 
({\ISUB} = 237--242, 246--256).
\iteme{= 44 :} chargino--neutralino pair production ({\ISUB} = 216--236).
\iteme{= 45 :} sbottom production ({\ISUB} = 281--296).
\iteme{= 50 :} pair production of technipions and gauge bosons by
$\pi^{0,\pm}_{\mrm{tc}}/\omega^0_{\mrm{tc}}$ exchange
({\ISUB} = 361--377).
\iteme{= 51 :}standard QCD $2 \to 2$ processes 381--386, with 
possibility to introduce compositeness/technicolor modifications,
see \ttt{ITCM(5)}.
\end{subentry}
\boxsep
 
\iteme{MSUB :}\label{p:MSUB} (D=500*0) array to be set when 
\ttt{MSEL=0} (for \ttt{MSEL}$\geq 1$ relevant entries are set in 
\ttt{PYINIT}) to choose which subset of subprocesses to include 
in the generation. The ordering follows the {\ISUB} code given in 
section \ref{ss:ISUBcode} (with comments as given there).
\begin{subentry}
\iteme{MSUB(ISUB) = 0 :} the subprocess is excluded.
\iteme{MSUB(ISUB) = 1 :} the subprocess is included.
\itemc{Note:} when \ttt{MSEL=0}, the \ttt{MSUB} values set by 
you are never changed by {\Py}. If you want to combine several
different `subruns', each with its own \ttt{PYINIT} call, into one
single run, it is up to you to remember not only to switch on
the new processes before each new \ttt{PYINIT} call, but also to
switch off the old ones that are no longer desired.
\end{subentry}
\boxsep
 
\iteme{KFIN(I,J) :}\label{p:KFIN} provides an option to selectively 
switch on and
off contributions to the cross sections from the different incoming
partons/particles at the hard interaction. In combination with the
{\Py} resonance decay switches, this also allows you to set
restrictions on flavours appearing in the final state.
\begin{subentry}
\iteme{I :} is 1 for beam side of event and 2 for target side.
\iteme{J :} enumerates flavours according to the {\KF} code; see section
\ref{ss:codes}.
\iteme{KFIN(I,J) = 0 :} the parton/particle is forbidden.
\iteme{KFIN(I,J) = 1 :} the parton/particle is allowed.
\itemc{Note:} By default, the following are switched on: $\d$, $\u$, 
$\s$, $\c$, $\b$, $\e^-$, $\nu_{\e}$, $\mu^-$, $\nu_{\mu}$, $\tau^-$,
$\nu_{\tau}$, $\g$, $\gamma$, $\Z^0$, $\W^+$ and their antiparticles.
In particular, top is off, and has to be switched on explicitly if
needed. 

\end{subentry}
\boxsep
 
\iteme{CKIN :}\label{p:CKIN} kinematics cuts that can be set by you 
before the \ttt{PYINIT} call, and that affect the region of phase space
within which events are generated. Some cuts are `hardwired' while
most are `softwired'. The hardwired ones are directly related to the
kinematical variables used in the event selection procedure,
and therefore have negligible effects on program efficiency.
The most important of these are \ttt{CKIN(1) - CKIN(8)},
\ttt{CKIN(27) - CKIN(28)}, and \ttt{CKIN(31) - CKIN(32)}.
The softwired ones are most of the remaining ones, that cannot
be fully taken into account
in the kinematical variable selection, so that generation in
constrained regions of phase space may be slow. In extreme
cases the phase space may be so small that the maximization
procedure fails to find any allowed points at all (although some
small region might still exist somewhere), and therefore switches
off some subprocesses, or aborts altogether.
 
\iteme{CKIN(1), CKIN(2) :} (D=2.,-1. GeV) range of allowed
$\hat{m} = \sqrt{\hat{s}}$ values. If \ttt{CKIN(2)}$ < 0.$, the upper
limit is inactive.
 
\iteme{CKIN(3), CKIN(4) :} (D=0.,-1. GeV) range of allowed
$\hat{p}_{\perp}$ values for hard $2 \to 2$ processes, with
transverse momentum $\hat{p}_{\perp}$ defined in the rest frame of
the hard interaction. If \ttt{CKIN(4)}$ < 0.$, the upper limit is
inactive. For processes that are singular in the limit
$\hat{p}_{\perp} \to 0$
(see \ttt{CKIN(6)}), \ttt{CKIN(5)} provides an additional constraint.
The \ttt{CKIN(3)} and \ttt{CKIN(4)} limits can also be used in
$2 \to 1 \to 2$ processes. Here, however, the product
masses are not known and hence are assumed to be vanishing in the event
selection. The actual $\pT$ range for massive products is thus
shifted downwards with respect to the nominal one.
\begin{subentry}
\itemc{Note 1:} For processes that are singular in the limit
$\hat{p}_{\perp} \to 0$, a careful choice of \ttt{CKIN(3)} value is 
not only a matter of technical convenience, but a requirement for 
obtaining sensible results. One example is the hadroproduction of
a $\W^{\pm}$ or $\Z^0$ gauge boson together with a jet, discussed
in subsection \ref{sss:WZclass}. Here the point is that this is a
first-order process (in $\alphas$), correcting the zeroth-order 
process of a $\W^{\pm}$ or $\Z^0$ without any jet. A full first-order 
description would also have to include virtual corrections in the 
low-$\hat{p}_{\perp}$ region.\\ 
Generalizing also to other processes, the simple-minded higher-order 
description breaks down when \ttt{CKIN(3)} is selected so small that 
the higher-order process cross section corresponds to a non-negligible 
fraction of the lower-order one. This number will vary depending on 
the process considered and the c.m. energy used, but could easily be 
tens of GeV rather than the default 1 GeV provided as technical 
cut-off in \ttt{CKIN(5)}. Processes singular in $\hat{p}_{\perp} \to 0$
should therefore only be used to describe the high-$\pT$ behaviour,
while the lowest-order process complemented with parton showers
should give the inclusive distribution and in particular the one
at small $\pT$ values.\\ 
Technically the case of QCD production of two jets is slightly more 
complicated, and involves eikonalization to multiple parton--parton
scattering, subsection \ref{ss:multint}, but again the conclusion
is that the processes have to be handled with care at small $\pT$
values. 
\itemc{Note 2:} There are a few situations in which \ttt{CKIN(3)}
may be overwritten; especially when different subprocess classes
are mixed in $\gamma\p$ or $\gamma\gamma$ collisions, see subsection
\ref{ss:photoanddisclass}.
 
\end{subentry} 

\iteme{CKIN(5) :} (D=1. GeV) lower cut-off on $\hat{p}_{\perp}$ values,
in addition to the \ttt{CKIN(3)} cut above, for processes that are
singular in the limit $\hat{p}_{\perp} \to 0$ (see \ttt{CKIN(6)}).
 
\iteme{CKIN(6) :} (D=1. GeV) hard $2 \to 2$ processes, which do not
proceed only via an intermediate resonance (i.e.\ are $2 \to 1 \to 2$
processes), are classified as singular in the limit
$\hat{p}_{\perp} \to 0$ if either or both of the two final-state
products has a mass $m < $\ttt{CKIN(6)}.
 
\iteme{CKIN(7), CKIN(8) :} (D=-10.,10.) range of allowed scattering
subsystem rapidities $y = y^*$ in the c.m.\ frame of the event,
where $y = (1/2)  \ln(x_1/x_2)$. (Following the notation of this
section, the variable should be given as $y^*$, but because of its
frequent use, it was called $y$ in section \ref{ss:kinemtwo}.)
 
\iteme{CKIN(9), CKIN(10) :} (D=-40.,40.) range of allowed (true)
rapidities for the product with largest rapidity in a $2 \to 2$ or a
$2 \to 1 \to 2$ process, defined in the c.m.\ frame of the event,
i.e.\ $\max(y^*_3, y^*_4)$. Note that rapidities are counted with sign,
i.e.\ if $y^*_3 = 1$ and $y^*_4 = -2$ then $\max(y^*_3, y^*_4) = 1$.

\iteme{CKIN(11), CKIN(12) :} (D=-40.,40.) range of allowed (true)
rapidities for the product with smallest rapidity in a $2 \to 2$ or a
$2 \to 1 \to 2$ process, defined in the c.m.\ frame of the event,
i.e.\ $\min(y^*_3, y^*_4)$. Consistency thus requires
\ttt{CKIN(11)}$\leq$\ttt{CKIN(9)} and
\ttt{CKIN(12)}$\leq$\ttt{CKIN(10)}.
 
\iteme{CKIN(13), CKIN(14) :} (D=-40.,40.) range of allowed
pseudorapidities for the product with largest pseudorapidity
in a $2 \to 2$ or a $2 \to 1 \to 2$ process, defined in the c.m.\
frame of the event, i.e.\ $\max(\eta^*_3, \eta^*_4)$. Note that 
pseudorapidities are counted with sign, i.e.\ if $\eta^*_3 = 1$ and 
$\eta^*_4 = -2$ then $\max(\eta^*_3, \eta^*_4) = 1$.
 
\iteme{CKIN(15), CKIN(16) :} (D=-40.,40.) range of allowed
pseudorapidities for the product with smallest pseudorapidity
in a $2 \to 2$ or a $2 \to 1 \to 2$ process, defined in the c.m.\
frame of the event, i.e.\ $\min(\eta^*_3, \eta^*_4)$. Consistency
thus requires \ttt{CKIN(15)}$\leq$\ttt{CKIN(13)} and
\ttt{CKIN(16)}$\leq$\ttt{CKIN(14)}.
 
\iteme{CKIN(17), CKIN(18) :} (D=-1.,1.) range of allowed
$\cos\theta^*$ values for the product with largest $\cos\theta^*$
value in a $2 \to 2$ or a $2 \to 1 \to 2$ process, defined in the
c.m.\ frame of the event, i.e.\ $\max(\cos\theta^*_3,\cos\theta^*_4)$.
 
\iteme{CKIN(19), CKIN(20) :} (D=-1.,1.) range of allowed
$\cos\theta^*$ values for the product with smallest $\cos\theta^*$
value in a $2 \to 2$ or a $2 \to 1 \to 2$ process, defined in the
c.m.\ frame of the event, i.e.\ $\min(\cos\theta^*_3,\cos\theta^*_4)$.
Consistency thus requires \ttt{CKIN(19)}$\leq$\ttt{CKIN(17)} and
\ttt{CKIN(20)}$\leq$\ttt{CKIN(18)}.
 
\iteme{CKIN(21), CKIN(22) :} (D=0.,1.) range of allowed $x_1$ values
for the parton on side 1 that enters the hard interaction.
 
\iteme{CKIN(23), CKIN(24) :} (D=0.,1.) range of allowed $x_2$ values
for the parton on side 2 that enters the hard interaction.
 
\iteme{CKIN(25), CKIN(26) :} (D=-1.,1.) range of allowed Feynman-$x$
values, where $x_{\mrm{F}} = x_1 - x_2$.
 
\iteme{CKIN(27), CKIN(28) :} (D=-1.,1.) range of allowed
$\cos\hat{\theta}$ values in a hard $2 \to 2$ scattering, where
$\hat{\theta}$ is the scattering angle in the rest frame of the
hard interaction.
 
\iteme{CKIN(31), CKIN(32) :} (D=2.,-1. GeV) range of allowed
$\hat{m}' = \sqrt{\hat{s}'}$ values, where $\hat{m}'$ is the mass of
the complete three- or four-body final state in $2 \to 3$ or
$2 \to 4$ processes (while $\hat{m}$, constrained in \ttt{CKIN(1)}
and \ttt{CKIN(2)}, here corresponds to the one- or two-body central
system). If \ttt{CKIN(32)}$ < 0.$, the upper limit is inactive.
 
\iteme{CKIN(35), CKIN(36) :} (D=0.,-1. GeV$^2$) range of allowed
$|\hat{t}| = - \hat{t}$ values in $2 \to 2$ processes. Note that
for Deeply Inelastic Scattering this is nothing but the $Q^2$ scale,
in the limit that initial- and final-state radiation is neglected.
If \ttt{CKIN(36)}$ < 0.$, the upper limit is inactive.
 
\iteme{CKIN(37), CKIN(38) :} (D=0.,-1. GeV$^2$) range of allowed
$|\hat{u}| = - \hat{u}$ values in $2 \to 2$ processes. If
\ttt{CKIN(38)}$ < 0.$, the upper limit is inactive.

\iteme{CKIN(39), CKIN(40) :} (D=4., -1. GeV$^2$) the $W^2$ range 
allowed in DIS processes, i.e.\ subprocess number 10. If 
\ttt{CKIN(40)}$ < 0.$, the upper limit is inactive. Here $W^2$ is 
defined in terms of $W^2 = Q^2 (1-x)/x$. This formula is not quite 
correct, in that \textit{(i)} it neglects the target mass (for a 
proton), and \textit{(ii)} it neglects initial-state photon radiation 
off the incoming electron. It should be good enough for loose cuts, 
however. These cuts are not checked if process 10 is called for two 
lepton beams.
 
\iteme{CKIN(41) - CKIN(44) :} (D=12.,-1.,12.,-1. GeV) range of allowed
mass values of the two (or one) resonances produced in a `true'
$2 \to 2$ process, i.e.\ one not (only) proceeding through a single
$s$-channel resonance ($2 \to 1 \to 2$). (These are the ones listed
as $2 \to 2$ in the tables in section \ref{ss:ISUBcode}.)
Only particles with a width above \ttt{PARP(41)} are considered as
bona fide resonances and tested against the \ttt{CKIN}
limits; particles with a smaller width are put on the mass shell
without applying any cuts. The exact interpretation of the \ttt{CKIN}
variables depends on the flavours of the two produced resonances. \\
For two resonances like $\Z^0 \W^+$ (produced from
$\f_i \fbar_j \to \Z^0\W^+$), which are not identical and which are not
each other's antiparticles, one has \\
\ttt{CKIN(41)}$ < m_1 < $\ttt{CKIN(42)}, and \\
\ttt{CKIN(43)}$ < m_2 < $\ttt{CKIN(44)}, \\
where $m_1$ and $m_2$ are the actually generated masses of the two
resonances, and 1 and 2 are defined by the order in which they are
given in the production process specification.  \\
For two resonances like $\Z^0 \Z^0$, which are identical, or
$\W^+ \W^-$, which are each other's antiparticles, one instead
has \\
\ttt{CKIN(41)}$ < \min(m_1,m_2) < $\ttt{CKIN(42)}, and \\
\ttt{CKIN(43)}$ < \max(m_1,m_2) < $\ttt{CKIN(44)}. \\
In addition, whatever limits are set on \ttt{CKIN(1)} and, in
particular, on \ttt{CKIN(2)} obviously affect the masses actually
selected.
\begin{subentry}
\itemc{Note 1:} If \ttt{MSTP(42)=0}, so that no mass smearing is
allowed, the \ttt{CKIN} values have no effect (the same as for
particles with too narrow a width).
\itemc{Note 2:} If \ttt{CKIN(42)}$ < $\ttt{CKIN(41)} it means that 
the \ttt{CKIN(42)} limit is inactive; correspondingly, if
\ttt{CKIN(44)}$ < $\ttt{CKIN(43)} then \ttt{CKIN(44)} is inactive.
\itemc{Note 3:} If limits are active and the resonances are
identical, it is up to you to ensure that
\ttt{CKIN(41)}$\leq$\ttt{CKIN(43)} and
\ttt{CKIN(42)}$\leq$\ttt{CKIN(44)}.
\itemc{Note 4:} For identical resonances, it is not possible to
preselect which of the resonances is the lighter one; if, for
instance, one
$\Z^0$ is to decay to leptons and the other to quarks, there is no
mechanism to guarantee that the lepton pair has a mass smaller
than the quark one.
\itemc{Note 5:} The \ttt{CKIN} values are applied to all relevant
$2 \to 2$ processes equally, which may not be what one desires if
several processes are generated simultaneously. Some caution is
therefore urged in the use of the \ttt{CKIN(41) - CKIN(44)} values.
Also in other respects, you are recommended to take proper
care: if a $\Z^0$ is only allowed to decay into $\b\bbar$, for example,
setting its mass range to be 2--8 GeV is obviously not a
good idea.
\end{subentry}
 
\iteme{CKIN(45) - CKIN(48) :} (D=12.,-1.,12.,-1. GeV) range of allowed
mass values of the two (or one) secondary resonances produced in
a $2 \to 1 \to 2$ process (like $\g \g \to \hrm^0 \to \Z^0 \Z^0$) or
even a $2 \to 2 \to 4$ (or 3) process (like
$\q \qbar \to \Z^0 \hrm^0 \to \Z^0 \W^+ \W^-$). Note that these
\ttt{CKIN} values only affect the secondary resonances; the
primary ones are constrained by \ttt{CKIN(1)}, \ttt{CKIN(2)} and
\ttt{CKIN(41) - CKIN(44)} (indirectly, of course, the choice of
primary resonance masses affects the allowed mass range for the
secondary ones). What is considered to be a resonance is defined
by \ttt{PARP(41)}; particles with a width smaller than this are
automatically put on the mass shell. The description closely
parallels the one given for \ttt{CKIN(41) - CKIN(44)}. Thus, for
two resonances that are not identical or each other's
antiparticles, one has \\
\ttt{CKIN(45)}$ < m_1 < $\ttt{CKIN(46)}, and \\
\ttt{CKIN(47)}$ < m_2 < $\ttt{CKIN(48)}, \\
where $m_1$ and $m_2$ are the actually generated masses of the two
resonances, and 1 and 2 are defined by the order in which they are
given in the decay channel specification in the program (see e.g.
output from \ttt{PYSTAT(2)} or \ttt{PYLIST(12)}). For two
resonances that are identical or each other's antiparticles,
one instead has \\
\ttt{CKIN(45)}$ < \min(m_1,m_2) < $\ttt{CKIN(46)}, and \\
\ttt{CKIN(47)}$ < \max(m_1,m_2) < $\ttt{CKIN(48)}.
\begin{subentry}
\itemc{Notes 1 - 5:} as for \ttt{CKIN(41) - CKIN(44)}, with trivial
modifications.
\itemc{Note 6:} Setting limits on secondary resonance masses is
possible in any of the channels of the allowed types (see above).
However, so far only $\hrm^0 \to \Z^0 \Z^0$ and $\hrm^0 \to \W^+ \W^-$
have been fully implemented, such that an arbitrary mass range
below the na\"{\i}ve mass threshold may be picked. For other possible
resonances, any restrictions made on the allowed mass range
are not reflected in the cross section; and further it is not
recommendable to pick mass windows that make a decay on the
mass shell impossible. 
\end{subentry}
  
\iteme{CKIN(49) - CKIN(50) :} allow minimum mass limits to be passed 
from \ttt{PYRESD} to \ttt{PYOFSH}. They are used for tertiary and higher 
resonances, i.e.\ those not controlled by \ttt{CKIN(41)-CKIN(48)}. They 
should not be touched by the user.  

\iteme{CKIN(51) - CKIN(56) :} (D=0.,-1.,0.,-1.,0.,-1. GeV) range of
allowed transverse momenta in a true $2 \to 3$ process. This 
means subprocesses such as 121--124 for $\hrm^0$ production, and
their $\H^0$, $\A^0$ and $\H^{\pm\pm}$ equivalents. 
\ttt{CKIN(51) - CKIN(54)} corresponds
to $\pT$ ranges for scattered partons, in order of appearance,
i.e.\ \ttt{CKIN(51) - CKIN(52)} for the parton scattered off the beam
and \ttt{CKIN(53) - CKIN(54)} for the one scattered off the target.
\ttt{CKIN(55)} and \ttt{CKIN(56)} here sets $\pT$ limits for the
third product, the $\hrm^0$, i.e.\ the \ttt{CKIN(3)} and \ttt{CKIN(4)}
values have no effect for this process. Since the $\pT$ of the Higgs
is not one of the primary variables selected, any constraints here
may mean reduced Monte Carlo efficiency, while for these processes
\ttt{CKIN(51) - CKIN(54)} are `hardwired' and therefore do not cost 
anything. As usual, a negative value implies that the upper limit is
inactive.

\iteme{CKIN(61) - CKIN(78) :} allows to restrict the range of kinematics 
for the photons generated off the lepton beams with the 
{\galep} option of \ttt{PYINIT}. In each quartet of numbers, 
the first two corresponds to the range allowed on incoming side 1 (beam) 
and the last two to side 2 (target). The cuts are only applicable for a 
lepton beam. Note that the $x$ and $Q^2$ ($P^2$) variables are the basis
for the generation, and so can be restricted with no loss of
efficiency. For leptoproduction (i.e.\ lepton on hadron) the $W$ is 
uniquely given by the one $x$ value of the problem, so here also $W$ cuts 
are fully efficient. The other cuts may imply a slowdown of the program, 
but not as much as if equivalent cuts only are introduced after events 
are fully  generated. See \cite{Fri00} for details.

\iteme{CKIN(61) - CKIN(64) :} (D=0.0001,0.99,0.0001,0.99) allowed range 
for the energy fractions $x$ that the photon take of the respective 
incoming lepton energy. These fractions are defined in the
c.m.\ frame of the collision, and differ from energy fractions 
as defined in another frame. (Watch out at HERA!) In order to 
avoid some technical problems, absolute lower and upper limits 
are set internally at 0.0001 and 0.9999.

\iteme{CKIN(65) - CKIN(68) :} (D=0.,-1.,0.,-1. GeV$^2$) allowed range for 
the spacelike virtuality of the photon, conventionally called either 
$Q^2$ or $P^2$, depending on process. A negative number means that the
upper limit is inactive, i.e.\ purely given by kinematics. A nonzero
lower limit is implicitly given by kinematics constraints. 

\iteme{CKIN(69) - CKIN(72) :} (D=0.,-1.,0.,-1.) allowed range of the
scattering angle $\theta$ of the lepton, defined in the c.m.\ frame
of the event. (Watch out at HERA!) A negative number means that 
the upper limit is inactive, i.e.\ equal to $\pi$.

\iteme{CKIN(73) - CKIN(76) :} (D=0.0001,0.99,0.0001,0.99) allowed range 
for the light-cone fraction $y$ that the photon take of the respective 
incoming lepton energy. The light-cone is defined by the 
four-momentum of the lepton or hadron on the other side of the
event (and thus deviates from true light-cone fraction by mass
effects that normally are negligible). The $y$ value is related to 
the $x$ and $Q^2$ ($P^2$) values by $y = x + Q^2/s$ if mass terms are
neglected.  
 
\iteme{CKIN(77), CKIN(78) :} (D=2.,-1. GeV) allowed range for $W$, 
i.e.\ either the photon--hadron or photon--photon invariant mass. A 
negative number means that the upper limit is inactive. 
   
\end{entry}
 
\subsection{The General Switches and Parameters}
\label{ss:PYswitchpar}
 
The \ttt{PYPARS} common block contains the status code and parameters
that regulate the performance of the program. All of them are
provided with sensible default values, so that a novice user can
neglect them, and only gradually explore the full range of
possibilities. Some of the switches and parameters in \ttt{PYPARS}
will be described later, in the shower and beam remnants sections.

\drawbox{COMMON/PYPARS/MSTP(200),PARP(200),MSTI(200),PARI(200)}%
\label{p:PYPARS1}
\begin{entry}
 
\itemc{Purpose:} to give access to status code and parameters that
regulate the performance of the program. If the default values,
denoted below by (D=\ldots), are not satisfactory, they must in
general be changed before the \ttt{PYINIT} call. Exceptions, i.e.
variables that can be changed for each new event, are denoted by
(C).
 
\iteme{MSTP(1) :}\label{p:MSTP} (D=3) maximum number of generations. 
Automatically set $\leq 4$.
 
\iteme{MSTP(2) :} (D=1) calculation of $\alphas$ at hard interaction,
in the routine \ttt{PYALPS}.
\begin{subentry}
\iteme{= 0 :} $\alphas$ is fixed at value \ttt{PARU(111)}.
\iteme{= 1 :} first-order running $\alphas$.
\iteme{= 2 :} second-order running $\alphas$.
\end{subentry}
 
\iteme{MSTP(3) :} (D=2) selection of $\Lambda$ value in $\alphas$
for \ttt{MSTP(2)}$\geq 1$.
\begin{subentry}
\iteme{= 1 :} $\Lambda$ is given by \ttt{PARP(1)} for hard
interactions, by \ttt{PARP(61)} for space-like showers,
by \ttt{PARP(72)} for time-like showers not from a resonance decay,
and by \ttt{PARJ(81)} for time-like ones from a resonance decay
(including e.g.\ $\gamma/\Z^0 \to \q\qbar$ decays, i.e.\ conventional
$\ee$ physics). This $\Lambda$ is assumed
to be valid for 5 flavours; for the hard interaction the number of
flavours assumed can be changed by \ttt{MSTU(112)}.
\iteme{= 2 :} $\Lambda$ value is chosen according to the 
parton-distribution-function para\-meteri\-zations. The choice is
always based on the proton parton-distribution set selected, i.e.
is unaffected by pion and photon parton-distribution selection.
All the $\Lambda$ values
are assumed to refer to 4 flavours, and \ttt{MSTU(112)} is set
accordingly. This $\Lambda$ value is used both for the hard
scattering and the initial- and final-state radiation. The
ambiguity in the choice of the $Q^2$ argument still remains (see
\ttt{MSTP(32)}, \ttt{MSTP(64)} and \ttt{MSTJ(44)}). This $\Lambda$
value is used also for \ttt{MSTP(57)=0}, but the sensible choice
here would be to use \ttt{MSTP(2)=0} and have no initial- or
final-state radiation. This option does \textit{not} change the
\ttt{PARJ(81)} value of timelike parton showers in resonance decays,
so that LEP experience on this specific parameter is not overwritten 
unwittingly. Therefore \ttt{PARJ(81)} can be updated completely 
independently.
\iteme{= 3 :} as \ttt{=2}, except that here also \ttt{PARJ(81)} is
overwritten in accordance with the $\Lambda$ value of the proton
parton-distribution-function set. 
\end{subentry}
 
\iteme{MSTP(4) :} (D=0) treatment of the Higgs sector,
predominantly the neutral one.
\begin{subentry}
\iteme{= 0 :} the $\hrm^0$ is given the Standard Model Higgs
couplings, while $\H^0$ and $\A^0$ couplings should be set by
you in \ttt{PARU(171) - PARU(175)} and
\ttt{PARU(181) - PARU(185)}, respectively.
\iteme{= 1 :} you should set couplings for all three Higgs bosons,
for the $\hrm^0$ in \ttt{PARU(161) - PARU(165)}, and for the
$\H^0$ and $\A^0$ as above.
\iteme{= 2 :} the mass of $\hrm^0$ in \ttt{PMAS(25,1)} and the
$\tan\beta$ value in \ttt{PARU(141)} are used to derive
$\H^0$, $\A^0$ and $\H^{\pm}$ masses, and $\hrm^0$, $\H^0$,
$\A^0$ and $\H^{\pm}$ couplings, using the relations of the Minimal
Supersymmetric extension of the Standard Model at Born level
\cite{Gun90}. Existing masses and couplings are overwritten by the
derived values. See section \ref{sss:extneutHclass} for discussion 
on parameter constraints.
\iteme{= 3:} as \ttt{=2}, but using relations at the one-loop level.
This option is not yet implemented as such. However, if you initialize
the SUSY machinery with \ttt{IMSS(1)=1}, then the SUSY 
parameters will be used to calculate also Higgs masses and couplings. 
These are stored in the appropriate slots, and the value of \ttt{MSTP(4)} 
is overwritten to 1. 
\end{subentry}
 
\iteme{MSTP(7) :} (D=0) choice of heavy flavour in subprocesses
81--85. Does not apply for \ttt{MSEL=4-8}, where the \ttt{MSEL} value
always takes precedence. 
\begin{subentry}
\iteme{= 0 :} for processes 81--84 (85) the `heaviest' flavour 
allowed for gluon (photon) splitting into a quark--antiquark 
(fermion--antifermion) pair, as set in the \ttt{MDME} array. 
Note that `heavy' is defined as the one with largest {\KF} code, 
so that leptons take precedence if they are allowed. 
\iteme{= 1 - 8 :} pick this particular quark flavour; e.g., 
\ttt{MSTP(7)=6} means that top will be produced.
\iteme{= 11 - 18 :} pick this particular lepton flavour. Note that
neutrinos are not possible, i.e.\ only 11, 13, 15 and 17 are 
meaningful alternatives. Lepton pair production can only occur in
process 85, so if any of the other processes have been switched on
they are generated with the same flavour as would be obtained in 
the option \ttt{MSTP(7)=0}.
\end{subentry}

\iteme{MSTP(8) :} (D=0) choice of electroweak parameters to use 
in the decay widths of resonances ($\W$, $\Z$, $\hrm$, \ldots) and 
cross sections (production of $\W$'s, $\Z$'s, $\hrm$'s, \ldots).
\begin{subentry} 
\iteme{= 0 :} everything is expressed in terms of a running 
$\alphaem(Q^2)$ and a fixed $\ssintw$, i.e.\ $G_{\F}$ is nowhere 
used.
\iteme{= 1 :} a replacement is made according to 
$\alphaem(Q^2) \to \sqrt{2} G_{\F} m_{\W}^2 \ssintw / \pi$
in all widths and cross sections. If $G_{\F}$ and $m_{\Z}$ are 
considered as given, this means that $\ssintw$ and $m_{\W}$ are the 
only free electroweak parameter.
\iteme{= 2 :} a replacement is made as for \ttt{=1}, but additionally 
$\ssintw$ is constrained by the relation 
$\ssintw = 1 - m_{\W}^2/m_{\Z}^2$.
This means that $m_{\W}$ remains as a free parameter, but that the
$\ssintw$ value in \ttt{PARU(102)} is never used, \textit{except} in
the vector couplings in the combination $v = a - 4 \ssintw e$.
This latter degree of freedom enters e.g.\ for forward-backward
asymmetries in $\Z^0$ decays.
\itemc{Note:} This option does not affect the emission of real photons 
in the initial and final state, where $\alphaem$ is always used. However, 
it does affect also purely electromagnetic hard processes, such as 
$\q \qbar \to \gamma \gamma$.
\end{subentry}

\iteme{MSTP(9) :} (D=0) inclusion of top (and fourth generation) as
allowed remnant flavour $\q'$ in processes that involve $\q \to \q' + \W$
branchings as part of the overall process, and where the matrix elements
have been calculated under the assumption that $\q'$ is massless.
\begin{subentry} 
\iteme{= 0 :} no.
\iteme{= 1 :} yes, but it is possible, as before, to switch off individual
channels by the setting of \ttt{MDME} switches. Mass effects 
are taken into account, in a crude fashion, by rejecting events 
where kinematics becomes inconsistent when the $\q'$ mass is included.
\end{subentry}  
 
\iteme{MSTP(11) :} (D=1) use of electron parton distribution in
$\ee$ and $\ep$ interactions.
\begin{subentry}
\iteme{= 0 :} no, i.e.\ electron carries the whole beam energy.
\iteme{= 1 :} yes, i.e.\ electron carries only a fraction of beam
energy in agreement with next-to-leading electron parton-distribution
function, thereby including the effects of initial-state
bremsstrahlung.
\end{subentry}
 
\iteme{MSTP(12) :} (D=0) use of $\e^-$ (`sea', i.e.\ from
$\e \to \gamma \to \e$), $\e^+$, quark and gluon distribution
functions inside an electron.
\begin{subentry}
\iteme{= 0 :} off.
\iteme{= 1 :} on, provided that \ttt{MSTP(11)}$\geq 1$.
Quark and gluon distributions
are obtained by numerical convolution of the photon content
inside an electron (as given by the bremsstrahlung spectrum of
\ttt{MSTP(11)=1}) with the quark and gluon content inside a
photon. The required numerical precision is set by \ttt{PARP(14)}.
Since the need for numerical integration makes this option
somewhat more time-consuming than ordinary parton-distribution
evaluation, one should only use it when studying processes
where it is needed.
\itemc{Note:} for all traditional photoproduction/DIS physics this
option is superseded by the {\galep} option for
\ttt{PYINIT} calls, but can still be of use for some less standard
processes.
\end{subentry}
 
\iteme{MSTP(13) :} (D=1) choice of $Q^2$ range over which electrons
are assumed to radiate photons; affects normalization of $\e^-$
(sea), $\e^+$, $\gamma$, quark and gluon distributions inside an
electron for \ttt{MSTP(12)=1}.
\begin{subentry}
\iteme{= 1 :} range set by $Q^2$ argument of 
parton-distribution-function call, i.e.\ by $Q^2$ scale of the hard 
interaction. Therefore 
parton distributions are proportional to $\ln(Q^2/m_e^2)$. 
\iteme{= 2 :} range set by the user-determined $Q_{\mmax}^2$, given in
\ttt{PARP(13)}. Parton distributions are assumed to be  proportional to
$\ln((Q_{\mmax}^2/m_e^2)(1-x)/x^2)$. This is normally most appropriate
for photoproduction, where the electron is supposed to go
undetected, i.e.\ scatter less than $Q_{\mmax}^2$.
\itemc{Note:} the choice of effective range is especially touchy for
the quark and gluon distributions. An (almost) on-the-mass-shell 
photon has a VMD piece that dies away for a virtual photon. A simple
convolution of distribution functions does not take this into account
properly. Therefore the contribution from $Q$ values above the 
$\rho$ mass should be suppressed. A choice of 
$Q_{\mmax} \approx 1$~GeV is then appropriate for a 
photoproduction limit description of physics. See also note for
\ttt{MSTP(12)}.
\end{subentry}
 
\iteme{MSTP(14) :} (D=30) structure of incoming photon beam or
target. Historically, numbers up to 10 were set up for real photons,
and subsequent ones have been added also to allow for virtual photon
beams. The reason is that the existing options specify e.g.\ 
direct$\times$VMD, summing over the possibilities of which photon is 
direct and which VMD. This is allowed when the situation is symmetric, 
i.e.\ for two incoming real photons, but not if one is virtual. Some of  
the new options agree with previous ones, but are included to allow a 
more consistent pattern. Further options above 25 have been added also 
to include DIS processes.
\begin{subentry}
\iteme{= 0 :} a photon is assumed to be point-like (a  direct photon), 
i.e.\ can only interact in processes which explicitly contain the 
incoming photon, such as $\f_i \gamma \to \f_i \g$ for $\gamma\p$ 
interactions. In $\gamma\gamma$ interactions both photons are
direct, i.e the main process is $\gamma \gamma \to \f_i \fbar_i$.
\iteme{= 1 :} a photon is assumed to be resolved, i.e.\ can only interact
through its constituent quarks and gluons, giving either high-$\pT$
parton--parton scatterings or low-$\pT$ events. Hard processes are 
calculated with the use of the full photon parton distributions. 
In $\gamma\gamma$ interactions both photons are resolved.
\iteme{= 2 :} a photon is assumed resolved, but only the VMD piece is 
included in the parton distributions, which therefore mainly 
are scaled-down versions of the $\rho^0 / \pi^0$ ones. Both high-$\pT$
parton--parton scatterings and low-$\pT$ events are allowed. In 
$\gamma\gamma$ interactions both photons are VMD-like.
\iteme{= 3 :} a photon is assumed resolved, but only the anomalous 
piece of the photon parton distributions is included. (This event 
class is called either anomalous or GVMD; we will use both 
interchangeably, though the former is more relevant for high-$\pT$
phenomena and the latter for low-$\pT$ ones.)
 In $\gamma\gamma$ 
interactions both photons are anomalous.
\iteme{= 4 :} in $\gamma\gamma$ interactions one photon is direct 
and the other resolved. A typical process is thus 
$\f_i \gamma \to \f_i \g$. Hard processes are calculated with the use 
of the full photon parton distributions for the resolved photon. 
Both possibilities of which photon is direct are included, in event 
topologies and in cross sections. This option cannot be used in 
configurations with only one incoming photon.
\iteme{= 5 :} in $\gamma\gamma$ interactions one photon is direct 
and the other VMD-like. Both possibilities of which photon is direct 
are included, in event topologies and in cross sections. This option 
cannot be used in configurations with only one incoming photon.
\iteme{= 6 :} in $\gamma\gamma$ interactions one photon is direct 
and the other anomalous. Both possibilities of which photon is direct 
are included, in event topologies and in cross sections. This option 
cannot be used in configurations with only one incoming photon.
\iteme{= 7 :} in $\gamma\gamma$ interactions one photon is VMD-like 
and the other anomalous. Only high-$\pT$ parton--parton scatterings 
are allowed. Both possibilities of which photon is VMD-like are 
included, in event topologies and in cross sections. This option 
cannot be used in configurations with only one incoming photon.
\iteme{= 10 :} the VMD, direct and anomalous/GVMD components of the 
photon are automatically mixed. For $\gamma\p$ interactions, this 
means an automatic mixture of the three classes 0, 2 and 3 above 
\cite{Sch93,Sch93a}, for $\gamma\gamma$ ones a mixture of the 
six classes 0, 2, 3, 5, 6 and 7 above \cite{Sch94a}. Various 
restrictions exist for this option, as discussed in section 
\ref{sss:photoprodclass}.  
\iteme{= 11 :} direct$\times$direct (see note 5); 
intended for virtual photons.
\iteme{= 12 :} direct$\times$VMD (i.e.\ first photon direct, second VMD); 
intended for virtual photons. 
\iteme{= 13 :} direct$\times$anomalous; intended for virtual photons.
\iteme{= 14 :} VMD$\times$direct; intended for virtual photons.  
\iteme{= 15 :} VMD$\times$VMD; intended for virtual photons.   
\iteme{= 16 :} VMD$\times$anomalous; intended for virtual photons.  
\iteme{= 17 :} anomalous$\times$direct; intended for virtual photons.  
\iteme{= 18 :} anomalous$\times$VDM; intended for virtual photons.  
\iteme{= 19 :} anomalous$\times$anomalous; intended for virtual photons.
\iteme{= 20 :} a mixture of the nine above components, 11--19, in the same 
spirit as \ttt{=10} provides a mixture for real gammas (or
a virtual gamma on a hadron). For gamma-hadron, this 
option coincides with \ttt{=10}.
\iteme{= 21 :} direct$\times$direct (see note 5).
\iteme{= 22 :} direct$\times$resolved. 
\iteme{= 23 :} resolved$\times$direct.
\iteme{= 24 :} resolved$\times$resolved.     
\iteme{= 25 :} a mixture of the four above components, offering a 
simpler alternative to \ttt{=20} in cases where the parton
distributions of the photon have not been split into VMD 
and anomalous components. For $\gamma$-hadron, only two
components need be mixed.
\iteme{= 26 :} DIS$\times$VMD/$\p$.
\iteme{= 27 :} DIS$\times$anomalous.
\iteme{= 28 :} VMD/$\p$$\times$DIS.
\iteme{= 29 :} anomalous$\times$DIS.
\iteme{= 30 :} a mixture of all the 4 (for $\gamma^*\p$) or 13 (for
$\gamma^*\gamma^*$) components that are available, i.e. (the relevant
ones of) 11--19 and 26--29 above; is as \ttt{=20} with the DIS 
processes mixed in.    
\itemc{Note 1:} The \ttt{MSTP(14)} options apply for a photon defined 
by a \ttt{'gamma'} or {\galep} beam in the \ttt{PYINIT} call, 
but not to those photons implicitly obtained in a \ttt{'lepton'} beam 
with the \ttt{MSTP(12)=1} option. This latter approach to resolved 
photons is more primitive and is no longer recommended for QCD processes. 
\itemc{Note 2:} for real photons our best understanding of how to mix 
event classes is provided by the option 10 above, which also can be 
obtained by combining three (for $\gamma\p$) or six (for $\gamma\gamma$) 
separate runs. In a simpler alternative the VMD and anomalous classes are 
joined into a single resolved class. Then $\gamma\p$ physics only requires 
two separate runs, with 0 and 1, and $\gamma\gamma$ physics requires 
three, with 0, 1 and 4.
\itemc{Note 3:} most of the new options from 11 onwards are not needed and 
therefore not defined for $\e\p$ collisions. The recommended 'best' value 
thus is \ttt{MSTP(14)=30}, which also is the new default value.
\itemc{Note 4:} as a consequence of the appearance of new event classes, 
the \ttt{MINT(122)} and \ttt{MSTI(9)} codes are not the same for 
$\gamma^*\gamma^*$ events as for $\gamma\p$, $\gamma^*\p$ or 
$\gamma\gamma$ ones. Instead the code is $3(i_1 - 1) + i_2$, where 
$i$ is 1 for direct, 2 for VMD and 3 for anomalous/GVMD and indices 
refer to the two incoming photons. For $\gamma^*\p$ code 4 is DIS,
and for $\gamma^* \gamma^*$ codes 10--13 corresponds to the \ttt{MSTP(14)}
codes 26--29. As before, \ttt{MINT(122)} and \ttt{MSTI(9)} are only 
defined when several processes are to be mixed, not when generating one       
at a time. Also the \ttt{MINT(123)} code is modified (not shown here).
\itemc{  Note 5:} The direct$\times$direct event class excludes lepton 
pair production when run with the default \ttt{MSEL=1} option (or 
\ttt{MSEL=2)}, in order not to confuse users. You can obtain lepton 
pairs as well, e.g.\ by running with \ttt{MSEL=0} and switching on the 
desired processes by hand.  
\itemc{  Note 6:} For all non-QCD processes, a photon is assumed unresolved 
when \ttt{MSTP(14)=} 10, 20 or 25. In principle, both the resolved and 
direct possibilities ought to be explored, but this mixing is not currently 
implemented, so picking direct at least will explore one of the two 
main alternatives rather than none. Resolved processes can be accessed 
by the more primitive machinery of having a lepton beam and 
\ttt{MSTP(12)=1}.  
\end{subentry}
 
\iteme{MSTP(15) :} (D=0) possibility to modify the nature of the 
anomalous photon component (as used with the appropriate \ttt{MSTP(14)}
options), in particular with respect to the scale choices and
cut-offs of hard processes. These options are mainly intended for 
comparative studies and should not normally be touched. Some of the
issues are discussed in \cite{Sch93a}, while others have only been used 
for internal studies and are undocumented.
\begin{subentry}
\iteme{= 0 :} none, i.e.\ the same treatment as for the VMD component.
\iteme{= 1 :} evaluate the anomalous parton distributions at a scale
$Q^2/$\ttt{PARP(17)}$^2$.
\iteme{= 2 :} as \ttt{=1}, but instead of \ttt{PARP(17)} use either
\ttt{PARP(81)/PARP(15)} or \ttt{PARP(82)/PARP(15)}, depending on 
\ttt{MSTP(82)} value.
\iteme{= 3 :} evaluate the anomalous parton distribution functions 
of the photon as $f^{\gamma,\mrm{anom}}(x, Q^2, p_0^2) - 
f^{\gamma,\mrm{anom}}(x, Q^2, r^2 Q^2)$ with $r = $\ttt{PARP(17)}.
\iteme{= 4 :} as \ttt{=3}, but instead of \ttt{PARP(17)} use either
\ttt{PARP(81)/PARP(15)} or \ttt{PARP(82)/PARP(15)}, depending on 
\ttt{MSTP(82)} value.
\iteme{= 5 :} use larger $\pTmin$ for the anomalous component than 
for the VMD one, but otherwise no difference.
\end{subentry}

\iteme{MSTP(16) :} (D=1) choice of definition of the fractional momentum 
taken by a photon radiated off a lepton. Enters in the flux
factor for the photon rate, and thereby in cross sections.
\begin{subentry} 
\iteme{= 0 :} $x$, i.e.\ energy fraction in the rest frame of the event.
\iteme{= 1 :} $y$, i.e.\ light-cone fraction.
\end{subentry}

\iteme{MSTP(17) :} (D=4) possibility of a extra factor for processes
involving resolved virtual photons, to approximately take into account 
the effects of longitudinal  photons. Given on the form\\
$R = 1 + \mbox{\ttt{PARP(165)}} \,  r(Q^2,\mu^2) \, 
f_L(y,Q^2)/f_T(y,Q^2)$.\\
Here the 1 represents the basic transverse contribution,
\ttt{PARP(165)} is some arbitrary overall factor, and $f_L/f_T$ 
the (known) ratio of longitudinal to transverse photon 
flux factors. The arbitrary function $r$ depends on the photon  
virtuality $Q^2$ and the hard scale $\mu^2$ of the process.
See \cite{Fri00} for a discussion of the options.
\begin{subentry} 
\iteme{= 0 :} No contribution, i.e.\ $r=0$.
\iteme{= 1 :} $r = 4  \mu^2  Q^2 / (\mu^2 + Q^2)^2$.
\iteme{= 2 :} $r = 4  Q^2 / (\mu^2 + Q^2)$.
\iteme{= 3 :} $r = 4  Q^2 / (m_{\rho}^2 + Q^2)$. 
\iteme{= 4 :} $r = 4  m_V^2  Q^2 / (m_V^2 + Q^2)^2$, where $m_V$ is 
the vector meson mass for VMD and $2\kT$ for GVMD states. Since there 
is no $\mu$ dependence here (as well as for \ttt{=3} and \ttt{=5}) 
also minimum-bias cross sections are affected, where $\mu$ would be 
vanishing. Currently the actual vector meson mass in the VMD case is
replaced by $m_{\rho}$, for simplicity.
\iteme{= 5 :} $r = 4  Q^2 / (m_V^2 + Q^2)$, with $m_V$ and comments 
as above.
\iteme{Note:} For a photon given by the {\galep} option in 
the \ttt{PYINIT} call, the $y$ spectrum is dynamically generated and 
$y$ is thus known from event to event. For a photon beam in the 
\ttt{PYINIT} call, $y$ is unknown from the onset, and has to be provided 
by you if any longitudinal factor is to be included. So long
as these values, in \ttt{PARP(167)} and \ttt{PARP(168)}, are at their
default values, 0, it is assumed they have not been set and thus the 
\ttt{MSTP(17)} and \ttt{PARP(165)} values are inactive.
\end{subentry}

\iteme{MSTP(18) :} (D=3) choice of $\pTmin$ for direct processes.
\begin{subentry}
\iteme{= 1 :} same as for VMD and GVMD states, i.e.\ the $\pTmin(W^2)$
scale. Primarily intended for real photons.
\iteme{= 2 :} $\pTmin$ is chosen to be \ttt{PARP(15)}, i.e.\ the original
old behaviour proposed in \cite{Sch93,Sch93a}. In that case, also parton 
distributions, jet cross sections and $\alphas$ values were dampened for 
small $\pT$, so it may not be easy to obtain full backwards compatibility
with this option.
\iteme{= 3 :} as \ttt{=1}, but if the $Q$ scale of the virtual photon is 
above the VMD/GVMD $\pTmin(W^2)$, $\pTmin$ is chosen equal to $Q$.
This is part of the strategy to mix in DIS processes at $\pT$ below $Q$, 
e.g.\ in \ttt{MSTP(14)=30}.
\end{subentry}

\iteme{MSTP(19) :} (D=4) choice of partonic cross section in the DIS 
process 99.
\begin{subentry}
\iteme{= 0 :} QPM answer $4 \pi^2 \alphaem/Q^2 \, 
\sum_{\q} e_{\q}^2 (x q(x,Q^2) + x \br{q}(x,Q^2))$   
(with parton distributions frozen below the lowest $Q$
allowed in the parameterization). Note that this answer
is divergent for $Q^2 \to 0$ and thus violates gauge invariance.
\iteme{= 1 :} QPM answer is modified by a factor $Q^2/(Q^2 + m_{\rho}^2)$
to provide a finite cross section in the $Q^2 \to 0$ limit.
A minimal regularization recipe.
\iteme{= 2 :} QPM answer is modified by a factor 
$Q^4/(Q^2 + m_{\rho}^2)^2$ to provide a vanishing cross section in 
the $Q^2 \to 0$ limit. Appropriate if one assumes that the normal 
photoproduction description gives the total cross section for $Q^2 = 0$,
without any DIS contribution.
\iteme{= 3 :} as \ttt{= 2}, but additionally suppression by a 
parameterized factor $f(W^2,Q^2)$ (different for $\gamma^*\p$ and 
$\gamma^*\gamma^*$) that avoids double-counting the direct-process 
region where $\pT > Q$. Shower evolution for DIS events is then also
restricted to be at scales below $Q$, whereas evolution all
the way up to $W$ is allowed in the other options above.
\iteme{= 4 :} as \ttt{= 3}, but additionally include factor 
$1/(1-x)$ for conversion from $F_2$ to $\sigma$. This is formally 
required, but is only relevant for small $W^2$ and therefore often 
neglected.
\end{subentry}

\iteme{MSTP(20) :} (D=3) suppression of resolved (VMD or GVMD) cross 
sections, introduced to compensate for an overlap with DIS processes in 
the region of intermediate $Q^2$ and rather small $W^2$.
\begin{subentry}
\iteme{= 0 :} no; used as is.
\iteme{> 1 :} yes, by a factor 
$(W^2/(W^2 + Q_1^2 + Q_2^2))^{\mbox{\ttt{MSTP(20)}}}$
(where $Q_i^2 = 0$ for an incoming hadron).
\iteme{Note:} the suppression factor is joined with the dipole 
suppression stored in \ttt{VINT(317)} and \ttt{VINT(318)}.
\end{subentry}

\iteme{MSTP(21) :} (D=1) nature of fermion--fermion scatterings
simulated in process 10 by $t$-channel exchange.
\begin{subentry}
\iteme{= 0 :} all off (!).
\iteme{= 1 :} full mixture of $\gammaZ$ neutral current and
$\W^{\pm}$ charged current.
\iteme{= 2 :} $\gamma$ neutral current only.
\iteme{= 3 :} $\Z^0$ neutral current only.
\iteme{= 4 :} $\gammaZ$ neutral current only.
\iteme{= 5 :} $\W^{\pm}$ charged current only.
\end{subentry}
 
\iteme{MSTP(22) :} (D=0) special override of normal $Q^2$ definition
used for maximum of parton-shower evolution, intended for Deeply 
Inelastic Scattering in lepton--hadron events, see section
\ref{ss:showrout}.
 
\iteme{MSTP(23) :} (D=1) for Deeply Inelastic Scattering processes
(10 and 83), this option allows the $x$ and $Q^2$ of the original
hard scattering to be retained by the final electron when showers 
are considered (with warnings as below; partly obsolete).
\begin{subentry}
\iteme{= 0 :} no correction procedure, i.e.\ $x$ and $Q^2$ of the
scattered electron differ from the originally generated $x$ and
$Q^2$.
\iteme{= 1 :} post facto correction, i.e.\ the change of electron
momentum, by initial and final QCD radiation, primordial $k_{\perp}$
and beam remnant treatment, is corrected for by a shuffling of
momentum between the electron and hadron side in the final state.
Only process 10 is corrected, while process 83 is not.
\iteme{= 2 :} as \ttt{=1}, except that both process 10 and 83 are
treated. This option is dangerous, especially for top, since it
may well be impossible to `correct' in process 83: the standard DIS 
kinematics definitions are based on the assumption of massless quarks.
Therefore infinite loops are not excluded.
\itemc{Note 1:} the correction procedure will fail for a fraction of
the events, which are thus rejected (and new ones generated in their
place). The correction option is not unambiguous, and should
not be taken too seriously. For very small $Q^2$ values, the $x$
is not exactly preserved even after this procedure.
\itemc{Note 2:} This switch does not affect the recommended DIS 
description obtained with a {\galep} beam/target in 
\ttt{PYINIT}, where $x$ and $Q^2$ are always conserved.
\end{subentry}

\iteme{MSTP(25) :} (D=0) angular decay correlations in Higgs decays 
to $\W^+ \W^-$ or $\Z^0 \Z^0$ to four fermions \cite{Skj93}. 
\begin{subentry}
\iteme{= 0 :} assuming the Higgs decay is pure scalar for 
$\hrm^0$ and $\H^0$ and pure pseudoscalar for $\A^0$.
\iteme{= 1 :} assuming the Higgs decay is always pure scalar (CP-even).
\iteme{= 2 :} assuming the Higgs decay is always pure pseudoscalar
(CP-odd).
\iteme{= 3 :} assuming the Higgs decay is a mixture of the two
(CP-even and CP-odd), including the CP-violating interference term.
The parameter $\eta$, \ttt{PARP(25)} sets the strength of the 
CP-odd admixture, with the interference term being proportional 
to $\eta$ and the CP-odd one to $\eta^2$.
\itemc{Note :} since the decay of an $\A^0$ to $\W^+ \W^-$ or 
$\Z^0 \Z^0$ is vanishing at the Born level, and no loop diagrams 
are included, currently this switch is only relevant for $\hrm^0$ and 
$\H^0$. It is mainly intended to allow `straw man' studies of 
the quantum numbers of a Higgs state, decoupled from the issue of 
branching ratios. 
\end{subentry}
 
\iteme{MSTP(31) :} (D=1) parameterization of total, elastic and
diffractive cross sections.
\begin{subentry}
\iteme{= 0 :} everything is to be set by you yourself in 
the \ttt{PYINT7} common block. For photoproduction, additionally
you need to set \ttt{VINT(281)}. Normally you would set these
values once and for all before the \ttt{PYINIT} call, but if
you run with variable energies (see \ttt{MSTP(171)}) you can
also set it before each new \ttt{PYEVNT} call. 
\iteme{= 1 :} Donnachie--Landshoff for total cross section 
\cite{Don92}, and Schuler--Sj\"ostrand for elastic and diffractive
cross sections \cite{Sch94,Sch93a}.
\end{subentry}
 
\iteme{MSTP(32) :} (D=8) $Q^2$ definition in hard scattering for
$2 \to 2$ processes. For resonance production $Q^2$ is always chosen
to be $\hat{s} = m_R^2$, where $m_R$ is the mass of the resonance.
For gauge boson scattering processes $VV \to VV$ the $\W$ or $\Z^0$
squared mass is used as scale in parton distributions. See 
\ttt{PARP(34)} for a possibility to modify the choice below by 
a multiplicative factor.\\
The newer options 6--10 are specifically intended for processes with 
incoming virtual photons. These are ordered from a `minimal'
dependence on the virtualities to a `maximal' one, based on
reasonable kinematics considerations. The old default value
\ttt{MSTP(32)=2} forms the starting point, with no dependence at 
all, and the new default is some intermediate choice. 
Notation is that $P_1^2$ and $P_2^2$ are the virtualities of the 
two incoming particles, $\pT$ the transverse momentum of the 
scattering process, and $m_3$ and $m_4$ the masses of the two 
outgoing partons. For a direct photon, $P^2$ is the photon
virtuality and $x=1$. For a resolved photon, $P^2$ still refers
to the photon, rather than the unknown virtuality of the 
reacting parton in the photon, and $x$ is the momentum fraction
taken by this parton.
\begin{subentry}
\iteme{= 1 :} $Q^2 = 2 \hat{s} \hat{t} \hat{u} / (\hat{s}^2 +
\hat{t}^2 + \hat{u}^2)$.
\iteme{= 2 :} $Q^2 = (m_{\perp 3}^2 + m_{\perp 4}^2)/2 =
\pT^2 + (m_3^2 + m_4^2)/2$.
\iteme{= 3 :} $Q^2 = \min(-\hat{t}, -\hat{u})$.
\iteme{= 4 :} $Q^2 = \hat{s}$.
\iteme{= 5 :} $Q^2 = -\hat{t}$.
\iteme{= 6 :} $Q^2 = (1 + x_1 P_1^2/\hat{s} + x_2 P_2^2/\hat{s}) 
(\pT^2 + m_3^2/2 + m_4^2/2)$.
\iteme{= 7 :} $Q^2 = (1 + P_1^2/\hat{s} + P_2^2/\hat{s})
(\pT^2 + m_3^2/2 + m_4^2/2)$.
\iteme{= 8 :} $Q^2 = \pT^2 + (P_1^2 + P_2^2 + m_3^2 + m_4^2)/2$.  
\iteme{= 9 :} $Q^2 = \pT^2 + P_1^2 + P_2^2 +m_3^2 + m_4^2$. 
\iteme{= 10 :} $Q^2 = s$ (the full energy-squared of the process).
\itemc{Note:} options 6 and 7 are motivated by assuming that one
wants a scale that interpolates between $\hat{t}$ for small 
$\hat{t}$ and $\hat{u}$ for small $\hat{u}$, such as 
$Q^2 = - \hat{t}\hat{u}/(\hat{t}+\hat{u})$. When kinematics for 
the $2 \to 2$ process is constructed as if an incoming
photon is massless when it is not, it gives rise to a
mismatch factor $1 + P^2/\hat{s}$ (neglecting the other 
masses) in this $Q^2$ definition, which is then what is
used in option 7 (with the neglect of some small 
cross-terms when both photons are virtual). When a
virtual photon is resolved, the virtuality of the
incoming parton can be anything from $xP^2$ and upwards.
So option 6 uses the smallest kinematically possible 
value, while 7 is more representative of the typical 
scale. Option 8 and 9 are more handwaving extensions of
the default option, with 9 specially constructed to 
ensure that the $Q^2$ scale is always bigger than $P^2$.  
\end{subentry}
 
\iteme{MSTP(33) :} (D=0) inclusion of $K$ factors in hard
cross sections for parton--parton interactions (i.e.\ for
incoming quarks and gluons).
\begin{subentry}
\iteme{= 0 :} none, i.e.\ $K = 1$.
\iteme{= 1 :} a common $K$ factor is used, as stored in
\ttt{PARP(31)}.
\iteme{= 2 :} separate factors are used for ordinary
(\ttt{PARP(31)}) and colour annihilation graphs (\ttt{PARP(32)}).
\iteme{= 3 :} A $K$ factor is introduced by a shift in the
$\alphas$ $Q^2$ argument,
$\alphas = \alphas($\ttt{PARP(33)}$Q^2)$.
\end{subentry}
 
\iteme{MSTP(34) :} (D=1) use of interference term in matrix
elements for QCD processes, see section \ref{sss:QCDjetclass}.
\begin{subentry}
\iteme{= 0 :} excluded (i.e.\ string-inspired matrix elements).
\iteme{= 1 :} included (i.e.\ conventional QCD matrix elements).
\itemc{Note:} for the option \ttt{MSTP(34)=1}, i.e.\ interference
terms included, these terms are divided between the different
possible colour configurations according to the pole structure
of the (string-inspired) matrix elements for the different
colour configurations.
\end{subentry}
 
\iteme{MSTP(35) :} (D=0) threshold behaviour for heavy-flavour
production, i.e.\ {\ISUB} = 81, 82, 84, 85, and also for $\Z$ and $\Z'$
decays. The non-standard options are mainly intended for top, but
can be used, with less theoretical reliability, also for charm and
bottom (for $\Z$ and $\Z'$ only top and heavier flavours
are affected). The threshold factors are given in 
eqs.~(\ref{pp:threshenh}) and (\ref{pp:threshsup}).
\begin{subentry}
\iteme{= 0 :} na\"{\i}ve lowest-order matrix-element behaviour.
\iteme{= 1 :} enhancement or suppression close to threshold,
according to the colour structure of the process. The
$\alphas$ value appearing in the threshold factor (which is not
the same as the $\alphas$ of the lowest-order $2 \to 2$ process)
is taken to be fixed at the value given in \ttt{PARP(35)}. The
threshold factor used in an event is stored in \ttt{PARI(81)}.
\iteme{= 2 :} as \ttt{=1}, but the $\alphas$ value appearing in
the threshold factor is taken to be running, with argument
$Q^2 = m_{\Q} \sqrt{ (\hat{m} - 2m_{\Q})^2 + \Gamma_{\Q}^2}$.
Here $m_{\Q}$ is the nominal heavy-quark mass, $\Gamma_{\Q}$ is
the width of the heavy-quark-mass distribution, and $\hat{m}$ is
the invariant mass of the heavy-quark pair. The $\Gamma_{\Q}$ value
has to be stored by you in \ttt{PARP(36)}. The regularization
of $\alphas$ at low $Q^2$ is given by \ttt{MSTP(36)}.
\end{subentry}
 
\iteme{MSTP(36) :} (D=2) regularization of $\alphas$ in the
limit $Q^2 \to 0$ for the threshold factor obtainable in the
\ttt{MSTP(35)=2} option; see \ttt{MSTU(115)} for a list of
the possibilities.
 
\iteme{MSTP(37) :} (D=1) inclusion of running quark masses in
Higgs production ($\q \qbar \to \hrm^0$) and decay
($\hrm^0 \to \q \qbar$) couplings, obtained by calls to the 
\ttt{PYMRUN} function. Also included for charged Higgs
and technipion production and decay.
\begin{subentry}
\iteme{= 0 :} not included, i.e.\ fixed quark masses are used
according to the values in the \ttt{PMAS} array.
\iteme{= 1 :} included, with running starting from the
value given in the \ttt{PMAS} array, at a $Q_0$ 
scale of \ttt{PARP(37)} times the quark mass itself,
up to a $Q$ scale given by the Higgs mass.
This option only works when $\alphas$ is allowed to run (so one can 
define a $\Lambda$ value). Therefore it is only applied if additionally
\ttt{MSTP(2)}$\geq 1$.
\end{subentry}
 
\iteme{MSTP(38) :} (D=5) handling of quark loop masses in the box
graphs $\g \g \to \gamma \gamma$ and $\g \g \to \g \gamma$, and in 
the Higgs production loop graphs $\q \qbar \to \g \hrm^0$, 
$\q \g \to \q \hrm^0$ and $\g \g \to \g \hrm^0$, and their 
equivalents with $\H^0$ or $\A^0$ instead of $\hrm^0$. 
\begin{subentry}
\iteme{= 0 :} for $\g \g \to \gamma \gamma$ and $\g \g \to \g \gamma$
the program will for each flavour automatically choose the massless 
approximation for light quarks and the full massive formulae for heavy 
quarks, with a dividing line between light and heavy quarks that depends 
on the actual $\hat{s}$ scale. For Higgs production, all quark loop
contributions are included with the proper masses. This option is then
correct only in the Standard Model Higgs scenario, and should not be 
used e.g. in the MSSM. 
\iteme{$\geq$1 :} for $\g \g \to \gamma \gamma$ and $\g \g \to \g \gamma$
the program will use the massless approximation throughout, assuming the 
presence of \ttt{MSTP(38)} effectively massless quark species (however, 
at most 8). Normally one would use \ttt{=5} for the inclusion of all quarks 
up to bottom, and \ttt{=6} to include top as well. For Higgs production,
the approximate expressions derived in the $m_{\t} \to \infty$ limit are
used, rescaled to match the correct $\g \g \to \hrm^0/\H^0/\A^0$ cross
sections. This procedure should work, approximately, also for non-standard
Higgs particles.
\itemc{Warning:} for \ttt{=0}, numerical instabilities may arise
in $\g \g \to \gamma \gamma$ and $\g \g \to \g \gamma$ for scattering at 
small angles. You are therefore recommended in this case to set 
\ttt{CKIN(27)} and \ttt{CKIN(28)} so as to exclude the range of scattering 
angles that is not of interest anyway. Numerical problems may also occur 
for Higgs production with \ttt{=0}, and additionally the lengthy expressions 
make the code error-prone.

\end{subentry}

\iteme{MSTP(39) :} (D=2) choice of $Q^2$ scale for parton distributions 
and initial state parton showers in processes $\g\g \to \Q \Qbar \hrm$ 
or $\q \qbar \to \Q \Qbar \hrm$.
\begin{subentry}
\iteme{= 1 :} $m_{\Q}^2$.
\iteme{= 2 :} $\max(m_{\perp\Q}^2,m_{\perp\Qbar}^2 ) = 
m_{\Q}^2 + \max(p_{\perp\Q}^2 , p_{\perp\Qbar}^2)$.
\iteme{= 3 :} $m_{\hrm}^2$, where $m_{\hrm}$ is the actual Higgs mass of 
the event, fluctuating from one event to the next.
\iteme{= 4 :} $\hat{s} = (p_{\hrm} + p_{\Q} + p_{\Qbar})^2$.
\iteme{= 5 :} $m_{\hrm}^2$, where $m_{\hrm}$ is the nominal, fixed 
Higgs mass.
\end{subentry}

\iteme{MSTP(40) :} (D=0) option for Coulomb correction in process 25,
$\W^+\W^-$ pair production, see \cite{Kho96}. The value of the Coulomb 
correction factor for each event is stored in \ttt{VINT(95)}.
\begin{subentry}
\iteme{= 0 :} `no Coulomb'. Is the often-used reference point.
\iteme{= 1 :} `unstable Coulomb', gives the correct first-order
expression valid in the non-relativistic limit. Is the reasonable 
option to use as a `best bet' description of LEP 2 physics.
\iteme{= 2 :} `second-order Coulomb' gives the correct second-order
expression valid in the non-relativistic limit. In principle
this is even better than \ttt{=1}, but the differences are negligible
and computer time does go up because of the need for a 
numerical integration in the weight factor.
\iteme{= 3 :} `dampened Coulomb', where the unstable Coulomb 
expression has been modified by a $(1-\beta)^2$ factor in front of the 
arctan term. This is intended as an alternative to \ttt{=1} within the 
band of our uncertainty in the relativistic limit. 
\iteme{= 4 :} `stable Coulomb', i.e.\ effects are calculated as if
the $\W$'s were stable. Is incorrect, and mainly intended for comparison 
purposes.
\itemc{Note :} Unfortunately the $\W$'s at LEP 2 are not in the 
non-relativistic limit, so the separation of Coulomb effects from other 
radiative corrections is not gauge invariant. The options above should 
therefore be viewed as indicative only, not as the ultimate answer.
\end{subentry}
 
\iteme{MSTP(41) :} (D=2) master switch for all resonance decays
($\Z^0$, $\W^{\pm}$, $\t$, $\hrm^0$, $\Z'^0$, $\W'^{\pm}$, $\H^0$,
$\A^0$, $\H^{\pm}$, $\L_{\Q}$, $\R^0$, $\d^*$, $\u^*$, \ldots).
\begin{subentry}
\iteme{= 0 :} all off.
\iteme{= 1 :} all on.
\iteme{= 2 :} on or off depending on their individual \ttt{MDCY} values.
\itemc{Note:} also for \ttt{MSTP(41)=1} it is possible to switch
off the decays of specific resonances by using the \ttt{MDCY(KC,1)}
switches in {\Py}. However, since the \ttt{MDCY} values are
overwritten in the \ttt{PYINIT} call when \ttt{MSTP(41)=1} (or 0), 
individual resonances should then be switched off after the \ttt{PYINIT} 
call.
\itemc{Warning:} for top, leptoquark and other colour-carrying resonances
it is dangerous to switch off decays if one later on intends to let them 
decay (with \ttt{PYEXEC}); see section \ref{sss:LQclass}.
\end{subentry}
 
\iteme{MSTP(42) :} (D=1) mass treatment in $2 \to 2$ processes, where
the final-state resonances have finite width (see \ttt{PARP(41)}).
(Does not apply for the production of a single $s$-channel resonance,
where the mass appears explicitly in the cross section of the
process, and thus is always selected with width.)
\begin{subentry}
\iteme{= 0 :} particles are put on the mass shell.
\iteme{= 1 :} mass generated according to a Breit--Wigner.
\end{subentry}
 
\iteme{MSTP(43) :} (D=3) treatment of $\Z^0/\gamma^*$ interference
in matrix elements. So far implemented in subprocesses 1, 15, 19, 22,
30 and 35; in other processes what is called a $\Z^0$ is really a 
$\Z^0$ only, without the $\gamma^*$ piece.
\begin{subentry}
\iteme{= 1 :} only $\gamma^*$ included.
\iteme{= 2 :} only $\Z^0$ included.
\iteme{= 3 :} complete $\Z^0/\gamma^*$ structure (with
interference) included.
\end{subentry}
 
\iteme{MSTP(44) :} (D=7) treatment of $\Z'^0/\Z^0/\gamma^*$
interference in matrix elements.
\begin{subentry}
\iteme{= 1 :} only $\gamma^*$ included.
\iteme{= 2 :} only $\Z^0$ included.
\iteme{= 3 :} only $\Z'^0$ included.
\iteme{= 4 :} only $\Z^0/\gamma^*$ (with interference) included.
\iteme{= 5 :} only $\Z'^0/\gamma^*$ (with interference) included.
\iteme{= 6 :} only $\Z'^0/\Z^0$ (with interference) included.
\iteme{= 7 :} complete $\Z'^0/\Z^0/\gamma^*$ structure
(with interference) included.
\end{subentry}
 
\iteme{MSTP(45) :} (D=3) treatment of $\W\W \to \W\W$ structure
({\ISUB} = 77).
\begin{subentry}
\iteme{= 1 :} only $\W^+\W^+ \to \W^+\W^+$ and
$\W^-\W^- \to \W^-\W^-$ included.
\iteme{= 2 :} only $\W^+\W^- \to \W^+\W^-$ included.
\iteme{= 3 :} all charge combinations $\W\W \to \W\W$ included.
\end{subentry}
 
\iteme{MSTP(46) :} (D=1) treatment of $VV \to V'V'$ structures
({\ISUB} = 71--77), where $V$ represents a longitudinal gauge boson.
\begin{subentry}
\iteme{= 0 :} only $s$-channel Higgs exchange included (where
existing). With this option, subprocesses 71--72 and 76--77
will essentially be equivalent to subprocesses 5 and 8,
respectively, with the proper decay channels (i.e.\ only $\Z^0\Z^0$
or $\W^+\W^-$) set via \ttt{MDME}.
The description obtained for subprocesses 5 and 8 in this case
is more sophisticated, however; see section \ref{sss:heavySMHclass}.
\iteme{= 1 :} all graphs contributing to $VV \to V'V'$
processes are included.
\iteme{= 2 :} only graphs not involving Higgs exchange
(either in $s$, $t$ or $u$ channel) are included; this option
then gives the na\"{\i}ve behaviour one would expect if no Higgs 
exists, including unphysical unitarity violations at high energies.
\iteme{= 3 :} the strongly interacting Higgs-like model of
Dobado, Herrero and Terron \cite{Dob91} with Pad\'e unitarization.
Note that to use this option it is necessary to set the Higgs mass
to a large number like 20 TeV (i.e.\ \ttt{PMAS(25,1)=20000}). The
parameter $\nu$ is stored in \ttt{PARP(44)}, but should not have
to be changed.
\iteme{= 4 :} as \ttt{=3}, but with K-matrix unitarization \cite{Dob91}.
\iteme{= 5 :} the strongly interacting QCD-like model of
Dobado, Herrero and Terron \cite{Dob91} with Pad\'e unitarization.
The parameter $\nu$ is stored in \ttt{PARP(44)}, but should not
have to be changed. The effective techni-$\rho$ mass in this model
is stored in \ttt{PARP(45)}; by default it is 2054 GeV, which is
the expected value for three technicolors, based on scaling up
the ordinary $\rho$ mass appropriately.
\iteme{= 6 :} as \ttt{=5}, but with K-matrix unitarization \cite{Dob91}.
While \ttt{PARP(45)} still is a parameter of the model, this type
of unitarization does not give rise to a resonance at a mass of
\ttt{PARP(45)}.
\end{subentry}
 
\iteme{MSTP(47) :} (D=1) (C) angular orientation of decay products
of resonances ($\Z^0$, $\W^{\pm}$, $\t$, $\hrm^0$, $\Z'^0$, $\W'^{\pm}$,
etc.), either when produced singly or in pairs (also
from an $\hrm^0$ decay), or in combination with a single quark,
gluon or photon.
\begin{subentry}
\iteme{= 0 :} independent decay of each resonance, isotropic in c.m.\
frame of the resonance.
\iteme{= 1 :} correlated decay angular distributions according to
proper matrix elements, to the extent these are implemented.
\end{subentry}

\iteme{MSTP(48) :} (D=0) (C) switch for the treatment of 
$\gamma^*/\Z^0$ decay for process 1 in $\ee$ events.
\begin{subentry}
\iteme{= 0 :} normal machinery.
\iteme{= 1 :} if the decay of the $\Z^0$ is to either of the five 
lighter quarks, $\d$, $\u$, $\s$, $\c$ or $\b$, the special treatment 
of $\Z^0$ decay is accessed, including matrix element options,
according to section~\ref{ss:eematrix}.\\
This option is based on the machinery of the \ttt{PYEEVT} and associated 
routines when it comes to the description of QCD multijet structure 
and the angular orientation of jets, but relies on the normal 
\ttt{PYEVNT} machinery for everything else: cross section calculation, 
initial state photon radiation, flavour composition of decays 
(i.e.\ information on channels allowed), etc.\\ 
The initial state has to be $\ee$; forward-backward asymmetries would    
not come out right for quark-annihilation production of the
$\gamma^*/\Z^0$ and therefore the machinery defaults to the standard 
one in such cases.\\
You can set the behaviour for the \ttt{MSTP(48)} option using the normal
matrix element related switches. This especially means \ttt{MSTJ(101)} for
the selection of first- or second-order matrix elements (\ttt{=1} and 
\ttt{=2}, respectively). Further selectivity is obtained with the switches
and parameters \ttt{MSTJ(102)} (for the angular orientation part only), 
\ttt{MSTJ(103)} (except the production threshold factor part), 
\ttt{MSTJ(106)}, \ttt{MSTJ(108) - MSTJ(111)}, \ttt{PARJ(121)}, 
\ttt{PARJ(122)}, and \ttt{PARJ(125) - PARJ(129)}.
Information can be read from \ttt{MSTJ(120)}, \ttt{MSTJ(121)}, 
\ttt{PARJ(150)}, \ttt{PARJ(152) - PARJ(156)}, \ttt{PARJ(168)}, 
\ttt{PARJ(169)}, \ttt{PARJ(171)}.\\
The other $\ee$ switches and parameters should not be touched. In most 
cases they are simply not accessed, since the related part is handled 
by the \ttt{PYEVNT} machinery instead. In other cases they could give
incorrect or misleading results. Beam polarization as set by
\ttt{PARJ(131) - PARJ(134)}, for instance, is only included for the 
angular orientation, but is missing for the cross section information.
\ttt{PARJ(123)} and \ttt{PARJ(124)} for the $\Z^0$ mass and width are 
set in the \ttt{PYINIT} call based on the input mass and calculated 
widths.\\
The cross section calculation is unaffected by the matrix element 
machinery. Thus also for negative \ttt{MSTJ(101)} values, where only 
specific jet multiplicities are generated, the \ttt{PYSTAT} cross 
section is the full one.  
\end{subentry} 

\iteme{MSTP(49) :} (D=1) assumed variation of the Higgs width to massive 
gauge boson pairs, i.e. $\W^+\W^-$, $\Z^0\Z^0$ and $\W^{\pm}\Z^0$, as a 
function of the actual mass $\hat{m} = \sqrt{\hat{s}}$ and the nominal 
mass $m_{\hrm}$. The switch applies both to $\hrm^0$, $\H^0$, $\A^0$ and 
$\H^{\pm}$ decays.  
\begin{subentry}
\iteme{= 0 :} the width is proportional to $\hat{m}^3$; thus the high-mass
tail of the Breit-Wigner is enhanced.
\iteme{= 1 :} the width is proportional to $m_{\hrm}^2  \hat{m}$. For 
a fixed Higgs mass $m_{\hrm}$ this means a width variation across the 
Breit-Wigner more in accord with other resonances (such as the $\Z^0$).  
This alternative gives more emphasis to the low-mass tail, where the 
parton distributions are peaked (for hadron colliders). This option is 
favoured by resummation studies \cite{Sey95a}.
\itemc{Note 1:} the partial width of a Higgs to a fermion pair is always 
taken to be proportional to the actual Higgs mass $\hat{m}$, irrespectively 
of \ttt{MSTP(49)}. Also the width to a gluon or photon pair (via loops)
is unaffected.
\itemc{Note 2:} this switch does not affect processes 71--77, where a 
fixed Higgs width is used in order to control cancellation of divergences.
\end{subentry}

\iteme{MSTP(50) :} (D=0) Switch to allow or not longitudinally polarized 
incoming beams, with the two polarizations stored in \ttt{PARJ(131)} and 
\ttt{PARJ(132)}, respectively. Most cross section expressions with 
polarization reduce to the unpolarized behaviour for the default
\ttt{PARJ(131)=PARJ(132)=0}, and then this switch is not implemented. 
Currently it is only used in process 25, $\f \fbar \to \W^+ \W^-$, 
for reasons explained in subsection \ref{ss:polarization}.
\begin{subentry}
\iteme{= 0 :} no polarization effects, no matter what \ttt{PARJ(131)} 
and \ttt{PARJ(132)} values are set.
\iteme{= 1 :} include polarization information in the cross section of 
the process and for angular correlations.  
\end{subentry}

\iteme{MSTP(51) :} (D=7) choice of proton parton-distribution set; 
see also \ttt{MSTP(52)}.
\begin{subentry}
\iteme{= 1 :} CTEQ 3L (leading order).
\iteme{= 2 :} CTEQ 3M ($\br{\mrm{MS}}$).
\iteme{= 3 :} CTEQ 3D (DIS).
\iteme{= 4 :} GRV 94L (leading order).
\iteme{= 5 :} GRV 94M ($\br{\mrm{MS}}$).
\iteme{= 6 :} GRV 94D (DIS).
\iteme{= 7 :} CTEQ 5L (leading order).
\iteme{= 8 :} CTEQ 5M1 ($\br{\mrm{MS}}$; slightly update version of 
CTEQ 5M).
\iteme{= 11 :} GRV 92L (leading order). 
\iteme{= 12 :} EHLQ set 1 (leading order; 1986 updated version).
\iteme{= 13 :} EHLQ set 2 (leading order; 1986 updated version).
\iteme{= 14 :} Duke--Owens set 1 (leading order).
\iteme{= 15 :} Duke--Owens set 2 (leading order).
\iteme{= 16 :} simple ansatz with all parton distributions of the
form $c/x$, with $c$ some constant; intended for internal debug use 
only.
\itemc{Note 1:} distributions 11--15 are obsolete and should not be 
used for current physics studies. They are only implemented to have 
some sets in common between \Py~5 and 6, for cross-checks. 
\itemc{Note 2:} since all parameterizations have some region of
applicability, the parton distributions are assumed frozen below
the lowest $Q^2$ covered by the parameterizations. In some cases, 
evolution is also frozen above the maximum $Q^2$.

\end{subentry}
 
\iteme{MSTP(52) :} (D=1) choice of proton 
parton-distribution-function library.
\begin{subentry}
\iteme{= 1 :} the internal {\Py} one, with parton distributions
according to the \ttt{MSTP(51)} above.
\iteme{= 2 :} the \tsc{Pdflib} one \cite{Plo93}, with the
\tsc{Pdflib} (version 4) \ttt{NGROUP} and \ttt{NSET} numbers to be 
given as \ttt{MSTP(51) = 1000}$\times$\ttt{NGROUP + NSET}.
\itemc{Note:} to make use of option 2, it is necessary to link 
\tsc{Pdflib}. Additionally, on most computers, the three dummy 
routines \ttt{PDFSET}, \ttt{STRUCTM} and (for virtual photons) 
\ttt{STRUCTP} at the end of the {\Py} file should be
removed or commented out.
\itemc{Warning:} For external parton distribution libraries,
{\Py} does not check whether \ttt{MSTP(51)} corresponds to a
valid code, or if special $x$ and $Q^2$ restrictions exist
for a given set, such that crazy values could be returned.
This puts an extra responsibility on you.
\end{subentry}
 
\iteme{MSTP(53) :} (D=3) choice of pion parton-distribution set;
see also \ttt{MSTP(54)}.
\begin{subentry}
\iteme{= 1 :} Owens set 1.
\iteme{= 2 :} Owens set 2.
\iteme{= 3 :} GRV LO (updated version).
\end{subentry}
 
\iteme{MSTP(54) :} (D=1) choice of pion parton-distribution-function
library.
\begin{subentry}
\iteme{= 1 :} the internal {\Py} one, with parton distributions
according to the \ttt{MSTP(53)} above.
\iteme{= 2 :} the \tsc{Pdflib} one \cite{Plo93}, with the
\tsc{Pdflib} (version 4) \ttt{NGROUP} and \ttt{NSET} numbers to be 
given as \ttt{MSTP(53) = 1000}$\times$\ttt{NGROUP + NSET}.
\itemc{Note:} to make use of option 2, it is necessary to link 
\tsc{Pdflib}. Additionally, on most computers, the three dummy routines
\ttt{PDFSET}, \ttt{STRUCTM} and \ttt{STRUCTP} at the end of the {\Py} 
file should be removed or commented out.
\itemc{Warning:} For external parton distribution libraries,
{\Py} does not check whether \ttt{MSTP(53)} corresponds to a valid
code, or if special $x$ and $Q^2$ restrictions exist for a given
set, such that crazy values could be returned. This puts an extra
responsibility on you.
\end{subentry}

\iteme{MSTP(55)} : (D=5) choice of the parton-distribution 
set of the photon; see also \ttt{MSTP(56)} and \ttt{MSTP(60)}.
\begin{subentry}
\iteme{= 1 :} Drees--Grassie.
\iteme{= 5 :} SaS 1D (in DIS scheme, with $Q_0=0.6$~GeV).
\iteme{= 6 :} SaS 1M (in {\MSbar} scheme, with $Q_0=0.6$~GeV).
\iteme{= 7 :} SaS 2D (in DIS scheme, with $Q_0=2$~GeV).
\iteme{= 8 :} SaS 2M (in {\MSbar} scheme, with $Q_0=2$~GeV).
\iteme{= 9 :} SaS 1D (in DIS scheme, with $Q_0=0.6$~GeV).
\iteme{= 10 :} SaS 1M (in {\MSbar} scheme, with $Q_0=0.6$~GeV).
\iteme{= 11 :} SaS 2D (in DIS scheme, with $Q_0=2$~GeV).
\iteme{= 12 :} SaS 2M (in {\MSbar} scheme, with $Q_0=2$~GeV).
\itemc{Note 1:} sets 5--8 use the parton distributions of the respective
set, and nothing else. These are appropriate for most applications, e.g.\ 
jet production in $\gamma\p$ and $\gamma\gamma$ collisions. Sets 9--12 
instead are appropriate for $\gamma^*\gamma$ processes, i.e.\ DIS 
scattering on a photon, as measured in $F_2^{\gamma}$. Here the anomalous 
contribution for $\c$ and $\b$ quarks are handled by the Bethe-Heitler 
formulae, and the direct term is artificially lumped with the anomalous
one, so that the event simulation more closely agrees with what will be 
experimentally observed in these processes. The agreement with the 
$F_2^{\gamma}$ parameterization is still not perfect, e.g.\ in the treatment 
of heavy flavours close to threshold. 
\itemc{Note 2:} Sets 5--12 contain both VMD pieces and anomalous pieces,
separately parameterized. Therefore the respective piece is automatically 
called, whatever \ttt{MSTP(14)} value is used to select only a part of the
allowed photon interactions. For other sets (set 1 above or \tsc{Pdflib} 
sets), usually there is no corresponding subdivision. Then an option like 
\ttt{MSTP(14)=2} (VMD part of photon only) is based on a rescaling of the 
pion distributions, while \ttt{MSTP(14)=3} gives the SaS anomalous 
parameterization.     
\itemc{Note 3:} Formally speaking, the $k_0$ (or $p_0$) cut-off in 
\ttt{PARP(15)} need not be set in any relation to the $Q_0$ cut-off 
scales used by the various parameterizations. Indeed, due to the 
familiar scale choice ambiguity problem, there could well be some offset 
between the two. However, unless you know what you are doing, it is 
recommended that you let the two agree, i.e.\ set
\ttt{PARP(15)=0.6} for the SaS 1 sets and \ttt{=2.} for the SaS 2 sets.
\end{subentry}
 
\iteme{MSTP(56) :} (D=1) choice of photon parton-distribution-function
library. 
\begin{subentry}
\iteme{= 1 :} the internal {\Py} one, with parton distributions
according to the \ttt{MSTP(55)} above.
\iteme{= 2 :} the \tsc{Pdflib} one \cite{Plo93}, with the
\tsc{Pdflib} (version 4) \ttt{NGROUP} and \ttt{NSET} numbers to be 
given as \ttt{MSTP(55) = 1000}$\times$\ttt{NGROUP + NSET}.
When the VMD and anomalous parts of the photon are split,
like for \ttt{MSTP(14)=10}, it is necessary to specify pion set to be
used for the VMD component, in \ttt{MSTP(53)} and \ttt{MSTP(54)},
while \ttt{MSTP(55)} here is irrelevant.
\iteme{= 3 :} when the parton distributions of the anomalous photon
are requested, the homogeneous solution is provided, evolved from a 
starting value \ttt{PARP(15)} to the requested $Q$ scale. The homogeneous 
solution is normalized so that the net momentum is unity,
i.e.\ any factors of $\alphaem/2\pi$ and charge have been left out. 
The flavour of the original $\q$ is given in \ttt{MSTP(55)} (1, 2, 3, 4 
or 5 for $\d$, $\u$, $\s$, $\c$ or $\b$); the value 0 gives a mixture 
according to squared charge, with the exception that $\c$ and $\b$ 
are only allowed above the respective mass threshold ($Q > m_{\q}$).
The four-flavour $\Lambda$ value is assumed given in \ttt{PARP(1)};
it is automatically recalculated for 3 or 5 flavours at 
thresholds. This option is not intended for standard event 
generation, but is useful for some theoretical studies.
\itemc{Note:} to make use of option 2, it is necessary to link 
\tsc{Pdflib}. Additionally, on most computers, the three dummy routines
\ttt{PDFSET}, \ttt{STRUCTM} and \ttt{STRUCTP} at the end of the {\Py} 
file should be removed or commented out.
\itemc{Warning:} For external parton-distribution libraries, {\Py}
does not check whether \ttt{MSTP(55)} corresponds to a valid code,
or if special $x$ and $Q^2$ restrictions exist for a given set,
such that crazy values could be returned. This puts an extra
responsibility on you.
\end{subentry}
 
\iteme{MSTP(57) :} (D=1) choice of $Q^2$ dependence in 
parton-distribution functions. 
\begin{subentry}
\iteme{= 0 :} parton distributions are evaluated at nominal lower
cut-off value $Q_0^2$, i.e.\ are made $Q^2$-independent.
\iteme{= 1 :} the parameterized $Q^2$ dependence is used.
\iteme{= 2 :} the parameterized parton-distribution behaviour is kept
at large $Q^2$ and $x$, but modified at small $Q^2$ and/or $x$,
so that parton distributions vanish in the limit $Q^2 \to 0$ and
have a theoretically motivated small-$x$ shape \cite{Sch93a}.
This option is only valid for the $\p$ and $\n$. It is obsolete within
the current {\galep} framework.
\iteme{= 3 :} as \ttt{=2}, except that also the $\pi^{\pm}$ is modified 
in a corresponding manner. A VMD photon is not mapped to a pion, but is 
treated with the same photon parton distributions as for other 
\ttt{MSTP(57)} values, but with properly modified behaviour for small 
$x$ or $Q^2$. This option is obsolete within the current {\galep} 
framework.
\end{subentry}
 
\iteme{MSTP(58) :} (D=min(5, 2$\times$\ttt{MSTP(1)})) maximum number of
quark flavours used in parton distributions, and thus also for 
initial-state space-like showers. If some distributions (notably $\t$) 
are absent in the parameterization selected in \ttt{MSTP(51)}, these 
are obviously automatically excluded.
 
\iteme{MSTP(59) :} (D=1) choice of electron-inside-electron parton 
distribution.
\begin{subentry}
\iteme{= 1 :} the recommended standard for LEP 1, next-to-leading
exponentiated, see \cite{Kle89}, p. 34.
\iteme{= 2 :} the recommended `$\beta$' scheme for LEP 2, also
next-to-leading exponentiated, see \cite{Bee96}, p. 130.
\end{subentry}

\iteme{MSTP(60) :} (D=7) extension of the SaS real-photon distributions to
off-shell photons, especially for the anomalous component. See \cite{Sch96}
for an explanation of the options. The starting point is the expression in 
eq.~(\ref{eq:decompvirtga}), which requires a numerical integration of the
anomalous component, however, and therefore is not convenient. Approximately, 
the dipole damping factor can be removed and compensated by a suitably
shifted lower integration limit, whereafter the integral simplifies.
Different `goodness' criteria for the choice of the shifted lower
limit is represented by the options 2--7 below.  
\begin{subentry}
\iteme{= 1 :} dipole dampening by integration; very time-consuming.
\iteme{= 2 :} $P_0^2 = \max( Q_0^2, P^2 )$.
\iteme{= 3 :} ${P'}_0^2 = Q_0^2 + P^2$.
\iteme{= 4 :} $P_{\mrm{eff}}$ that preserves momentum sum. 
\iteme{= 5 :} $P_{\mrm{int}}$ that preserves momentum and average 
evolution range.
\iteme{= 6 :} $P_{\mrm{eff}}$, matched to $P_0$ in $P^2 \to Q^2$ limit.
\iteme{= 7 :} $P_{\mrm{int}}$, matched to $P_0$ in $P^2 \to Q^2$ limit.
\end{subentry}
 
\iteme{MSTP(61) :} (D=1) (C) master switch for initial-state QCD and
QED radiation.
\begin{subentry}
\iteme{= 0 :} off.
\iteme{= 1 :} on.
\end{subentry}
 
\iteme{MSTP(62) - MSTP(69) :} (C) further switches for initial-state 
radiation, see section \ref{ss:showrout}.
 
\iteme{MSTP(71) :} (D=1) (C) master switch for final-state QCD and
QED radiation.
\begin{subentry}
\iteme{= 0 :} off.
\iteme{= 1 :} on.
\itemc{Note:} additional switches (e.g.\ for conventional/coherent
showers) are available in \ttt{MSTJ(38) - MSTJ(50)} and
\ttt{PARJ(80) - PARJ(90)}, see section \ref{ss:showrout}.
\end{subentry}
 
\iteme{MSTP(81) :} (D=1) master switch for multiple interactions.
\begin{subentry}
\iteme{= 0 :} off.
\iteme{= 1 :} on.
\end{subentry}

\iteme{MSTP(82) - MSTP(86) :} further switches for multiple
interactions, see section \ref{ss:multintpar}.

\iteme{MSTP(91) - MSTP(95) :} switches for beam remnant treatment,
see section \ref{ss:multintpar}.
 
\iteme{MSTP(101) :} (D=3) (C) structure of diffractive system.
\begin{subentry}
\iteme{= 1 :} forward moving diquark + interacting quark.
\iteme{= 2 :} forward moving diquark + quark joined via interacting
gluon (`hairpin' configuration).
\iteme{= 3 :} a mixture of the two options above, with a fraction
\ttt{PARP(101)} of the former type.
\end{subentry}

\iteme{MSTP(102) :} (D=1) (C) decay of a $\rho^0$ meson produced by
`elastic' scattering of an incoming $\gamma$, as in
$\gamma \p \to \rho^0 \p$, or the same with the hadron diffractively 
excited.
\begin{subentry}
\iteme{= 0 :} the $\rho^0$ is allowed to decay isotropically, like 
any other $\rho^0$.
\iteme{= 1 :} the decay $\rho^0 \to \pi^+ \pi^-$ is done with an
angular distribution proportional to $\sin^2 \theta$ in its rest frame,
where the $z$ axis is given by the direction of motion of the 
$\rho^0$. The $\rho^0$ decay is then done as part of the hard process, 
i.e.\ also when \ttt{MSTP(111)=0}.
\end{subentry}
  
\iteme{MSTP(110) :} (D=0) switch to allow some or all resonance widths
to be modified by the factor \ttt{PARP(110)}. This is not intended for
serious physics studies. The main application is rather to generate
events with an artificially narrow resonance width in order to study
the detector-related smearing effects on the mass resolution. 
\begin{subentry}
\iteme{> 0 :} rescale the particular resonance with \ttt{KF = MSTP(110)}.
If the resonance has an antiparticle, this one is affected as well.  
\iteme{= -1 :} rescale all resonances, except $\t$, $\tbar$, $\Z^0$ and 
$\W^{\pm}$. 
\iteme{= -2 :} rescale all resonances.
\itemc{Warning:} Only resonances with a width evaluated by \ttt{PYWIDT} 
are affected, and preferentially then those with \ttt{MWID} value 1 or 3.
For other resonances the appearance of effects or not depends on how the 
cross sections have been implemented. So it is important to check that 
indeed the mass distribution is affected as expected. Also beware that,
if a sequential decay chain is involved, the scaling may become more 
complicated. Furthermore, depending on implementational details, a cross 
section may or may not scale with \ttt{PARP(110)} (quite apart from 
differences related to the convolution with parton distributions etc.). 
All in all, it is then an option to be used only with open eyes, and for
very specific applications.
\end{subentry}

\iteme{MSTP(111) :} (D=1) (C) master switch for fragmentation
and decay, as obtained with a \ttt{PYEXEC} call.
\begin{subentry}
\iteme{= 0 :} off.
\iteme{= 1 :} on.
\iteme{= -1 :} only choose kinematical variables for hard scattering,
i.e.\ no jets are defined. This is useful, for instance, to calculate
cross sections (by Monte Carlo integration) without wanting
to simulate events; information obtained with \ttt{PYSTAT(1)}
will be correct.
\end{subentry}
 
\iteme{MSTP(112) :} (D=1) (C) cuts on partonic events; only affects
an exceedingly tiny fraction of events. Normally this concerns what 
happens in the \ttt{PYPREP} routine, if a colour singlet subsystem
has a very small invariant mass and attempts to collapse it to a single
particle fail, see section \ref{sss:smallmasssystem}.
\begin{subentry}
\iteme{= 0 :} no cuts (can be used only with independent
fragmentation, at least in principle).
\iteme{= 1 :} string cuts (as normally required for fragmentation).
\end{subentry}
 
\iteme{MSTP(113) :} (D=1) (C) recalculation of energies of partons
from their momenta and masses, to be done immediately before
and after fragmentation, to partly compensate for some numerical 
problems appearing at high energies.
\begin{subentry}
\iteme{= 0 :} not performed.
\iteme{= 1 :} performed.
\end{subentry}
 
\iteme{MSTP(115) :} (D=0) (C) choice of colour rearrangement scenario
for process 25, $\W^+\W^-$ pair production, when both $\W$'s decay
hadronically. (Also works for process 22, $\Z^0\Z^0$ production,
except when the $\Z$'s are allowed to fluctuate to very small masses.) 
See section \ref{sss:reconnect} for details.
\begin{subentry}
\iteme{= 0 :} no reconnection.
\iteme{= 1 :} scenario I, reconnection inspired by a type I superconductor, 
with the reconnection probability related to the overlap volume in 
space and time between the $\W^+$ and $\W^-$ strings. Related parameters 
are found in \ttt{PARP(115) - PARP(119)}, with \ttt{PARP(117)} of special 
interest.
\iteme{= 2 :} scenario II, reconnection inspired by a type II 
superconductor, with reconnection possible when two string 
cores cross. Related parameter in \ttt{PARP(115)}.
\iteme{= 3 :} scenario II', as model II but with the additional 
requirement that a reconnection will only occur if the 
total string length is reduced by it. 
\iteme{= 5 :} the GH scenario, where the reconnection can occur that
reduces the total string length ($\lambda$ measure) most. 
\ttt{PARP(120)} gives the fraction of such event where a 
reconnection is actually made; since almost all events
could allow a reconnection that would reduce the string
length, \ttt{PARP(120)} is almost the same as the reconnection
probability.
\iteme{= 11 :} the intermediate scenario, where a reconnection is
made at the `origin' of events, based on the subdivision
of all radiation of a $\q\qbar$ system as coming either from 
the $\q$ or the $\qbar$. \ttt{PARP(120)} gives the assumed probability
that a reconnection will occur. A somewhat simpleminded
model, but not quite unrealistic.  
\iteme{= 12 :} the instantaneous scenario, where a reconnection is 
allowed to occur before the parton showers, and showering
is performed inside the reconnected systems with maximum
virtuality set by the mass of the reconnected systems.
\ttt{PARP(120)} gives the assumed probability that a reconnection 
will occur. Is completely unrealistic, but useful as an
extreme example with very large effects.  
\end{subentry}
 
\iteme{MSTP(121) :} (D=0) calculation of kinematics selection
coefficients and differential cross section maxima for
included (by you or default) subprocesses.
\begin{subentry}
\iteme{= 0 :} not known; to be calculated at initialization.
\iteme{= 1 :} not known; to be calculated at initialization;
however, the maximum value then obtained is to be multiplied by
\ttt{PARP(121)} (this may be useful if a violation factor has
been observed in a previous run of the same kind).
\iteme{= 2 :} known; kinematics selection coefficients stored
by you in \ttt{COEF(ISUB,J)} (\ttt{J} = 1--20) in common block
\ttt{PYINT2} and maximum of the corresponding differential
cross section times Jacobians in \ttt{XSEC(ISUB,1)} in
common block \ttt{PYINT5}. This is to be done for each included
subprocess {\ISUB} before initialization, with the sum of all
\ttt{XSEC(ISUB,1)} values, except for {\ISUB} = 95, stored in
\ttt{XSEC(0,1)}.
\end{subentry}
 
\iteme{MSTP(122) :} (D=1) initialization and differential
cross section maximization print-out. Also, less importantly, level 
of information on where in phase space a cross section maximum has 
been violated during the run.
\begin{subentry}
\iteme{= 0 :} none.
\iteme{= 1 :} short message at initialization; only when an error
(i.e.\ not a warning) is generated during the run. 
\iteme{= 2 :} detailed message, including full maximization., at
initialization; always during run.
\end{subentry}
 
\iteme{MSTP(123) :} (D=2) reaction to violation of maximum
differential cross section or to occurence of negative differential
cross sections (except when allowed for external processes, i.e. 
when \ttt{IDWTUP < 0}).
\begin{subentry}
\iteme{= 0 :} stop generation, print message.
\iteme{= 1 :} continue generation, print message for each
subsequently larger violation.
\iteme{= 2 :} as \ttt{=1}, but also increase value of maximum.
\end{subentry}
 
\iteme{MSTP(124) :} (D=1) (C) frame for presentation of event.
\begin{subentry}
\iteme{= 1 :} as specified in \ttt{PYINIT}.
\iteme{= 2 :} c.m.\ frame of incoming particles.
\iteme{= 3 :} hadronic c.m.\ frame for DIS events, with warnings
as given for \ttt{PYFRAM}.
\end{subentry}
 
\iteme{MSTP(125) :} (D=1) (C) documentation of partonic process,
see section \ref{sss:PYrecord} for details.
\begin{subentry}
\iteme{= 0 :} only list ultimate string/particle configuration.
\iteme{= 1 :} additionally list short summary of the hard process.
\iteme{= 2 :} list complete documentation of intermediate steps of
parton-shower evolution.
\end{subentry}
 
\iteme{MSTP(126) :} (D=100) number of lines at the beginning of event
record that are reserved for event-history information; see section
\ref{sss:PYrecord}. This value should never be reduced, but may be
increased at a later date if more complicated processes are included.
 
\iteme{MSTP(127) :} (D=0) possibility to continue run even if none 
of the requested processes have non-vanishing cross sections.
\begin{subentry}
\iteme{= 0 :} no, the run will be stopped in the \ttt{PYINIT} call.
\iteme{= 1 :} yes, the \ttt{PYINIT} execution will finish normally, 
but with the flag \ttt{MSTI(53)=1} set to signal the problem. If 
nevertheless \ttt{PYEVNT} is called after this, the run will be 
stopped, since no events can be generated. If instead a new 
\ttt{PYINIT} call is made, with changed conditions (e.g. modified
supersymmetry parameters in a SUSY run), it may now become possible 
to initialize normally and generate events. 
\end{subentry}

\iteme{MSTP(128) :} (D=0) storing of copy of resonance decay
products in the documentation section of the event record, and
mother pointer (\ttt{K(I,3)}) relation of the actual resonance
decay products (stored in the main section of the event record)
to the documentation copy.
\begin{subentry}
\iteme{= 0 :} products are stored also in the documentation section,
and each product stored in the main section points back
to the corresponding entry in the documentation section.
\iteme{= 1 :} products are stored also in the documentation section,
but the products stored in the main section point back to
the decaying resonance copy in the main section.
\iteme{= 2 :} products are not stored in the documentation section;
the products stored in the main section point back to the
decaying resonance copy in the main section.
\end{subentry}
 
\iteme{MSTP(129) :} (D=10) for the maximization of $2 \to 3$ processes
(\ttt{ISET(ISUB)=5}) each phase-space point in $\tau$, $y$ and $\tau'$
is tested \ttt{MSTP(129)} times in the other dimensions (at randomly
selected points) to determine the effective maximum in the
($\tau$, $y$, $\tau'$) point.
 
\iteme{MSTP(131) :} (D=0) master switch for pile-up events, i.e.\ several
independent hadron--hadron interactions generated in the same
bunch--bunch crossing, with the events following one after the
other in the event record. See subsection \ref{ss:pileup} for details.
\begin{subentry}
\iteme{= 0 :} off, i.e.\ only one event is generated at a time.
\iteme{= 1 :} on, i.e.\ several events are allowed in the same event
record. Information on the processes generated may be found in
\ttt{MSTI(41) - MSTI(50)}.
\end{subentry}

\iteme{MSTP(132) - MSTP(134) :} further switches for pile-up events,
see section \ref{ss:multintpar}.
 
\iteme{MSTP(141) :} (D=0) calling of \ttt{PYKCUT} in the 
event-generation chain, for inclusion of user-specified cuts.
\begin{subentry}
\iteme{= 0 :} not called.
\iteme{= 1 :} called.
\end{subentry}
 
\iteme{MSTP(142) :} (D=0) calling of \ttt{PYEVWT} in the 
event-generation chain, either to give weighted events or to modify
standard cross sections. See \ttt{PYEVWT} description in section
\ref{ss:PYTmainroutines} for further details.
\begin{subentry}
\iteme{= 0 :} not called.
\iteme{= 1 :} called; the distribution of events among subprocesses
and in kinematics variables is modified by the factor \ttt{WTXS},
set by you in the \ttt{PYEVWT} call, but events come with a
compensating weight \ttt{PARI(10)=1./WTXS}, such that total
cross sections are unchanged.
\iteme{= 2 :} called; the cross section itself is modified by the
factor \ttt{WTXS}, set by you in the \ttt{PYEVWT} call.
\end{subentry}

\iteme{MSTP(151) :} (D=0) introduce smeared position of primary vertex
of events.
\begin{subentry}
\iteme{= 0 :} no, i.e.\ the primary vertex of each event is at the 
origin.
\iteme{= 1 :} yes, with Gaussian distributions separately in $x$, $y$,
$z$ and $t$. The respective widths of the Gaussians have to be given 
in \ttt{PARP(151) - PARP(154)}. Also pile-up events obtain separate
primary vertices. No provisions are made for more complicated 
beam-spot shapes, e.g.\ with a spread in $z$ that varies as a 
function of $t$. Note that a large beam spot combined with some of the
\ttt{MSTJ(22)} options may lead to many particles not being allowed to
decay at all.
\end{subentry}

\iteme{MSTP(171) :} (D=0) possibility of variable energies from one 
event to the next. For further details see section \ref{ss:PYvaren}.
\begin{subentry}
\iteme{= 0 :} no; i.e.\ the energy is fixed at the initialization call.
\iteme{= 1 :} yes; i.e.\ a new energy has to be given for each new 
event.
\itemc{Warning:} Variable energies cannot be used in conjunction with
the internal generation of a virtual photon flux obtained by a 
\ttt{PYINIT} call with {\galep} argument. The reason is that 
a variable-energy machinery is now used internally for the $\gamma$-hadron 
or $\gamma\gamma$ subsystem, with some information saved at 
initialization for the full energy.     
\end{subentry}

\iteme{MSTP(172) :} (D=2) options for generation of events with 
variable energies, applicable when \ttt{MSTP(171)=1}. 
\begin{subentry}
\iteme{= 1 :} an event is generated at the requested energy, i.e.\ 
internally a loop is performed over possible event configurations 
until one is accepted. If the requested c.m.\ energy of an event 
is below \ttt{PARP(2)} the run is aborted. Cross-section information 
can not be trusted with this option, since it depends on how you 
decided to pick the requested energies.
\iteme{= 2 :} only one event configuration is tried. If that is 
accepted, the event is generated in full. If not, no event is
generated, and the status code \ttt{MSTI(61)=1} is returned.
You are then expected to give a new energy, looping until an
acceptable event is found. No event is generated if the 
requested c.m.\ energy is below \ttt{PARP(2)}, instead
\ttt{MSTI(61)=1} is set to signal the failure. In principle,
cross sections should come out correctly with this option. 
\end{subentry}

\iteme{MSTP(173) :} (D=0) possibility for you to give in an event 
weight to compensate for a biased choice of beam spectrum.
\begin{subentry}
\iteme{= 0 :} no, i.e.\ event weight is unity.
\iteme{= 1 :} yes; weight to be given for each event in 
\ttt{PARP(173)}, with maximum weight given at initialization
in \ttt{PARP(174)}.
\end{subentry}
 
\iteme{MSTP(181) :} (R) {\Py} version number.
 
\iteme{MSTP(182) :} (R) {\Py} subversion number.
 
\iteme{MSTP(183) :} (R) last year of change for {\Py}.
 
\iteme{MSTP(184) :} (R) last month of change for {\Py}.
 
\iteme{MSTP(185) :} (R) last day of change for {\Py}.

\boxsep
 
\iteme{PARP(1) :}\label{p:PARP} (D=0.25 GeV) nominal 
$\Lambda_{\mrm{QCD}}$ used in running $\alphas$ for hard 
scattering (see \ttt{MSTP(3)}).
 
\iteme{PARP(2) :} (D=10. GeV) lowest c.m.\ energy for the
event as a whole that the program will accept to simulate.
 
\iteme{PARP(13) :} (D=1. GeV$^2$) $Q_{\mmax}^2$ scale, to be set by
you for defining maximum scale allowed for photoproduction when
using the option \ttt{MSTP(13)=2}.
 
\iteme{PARP(14) :} (D=0.01) in the numerical integration of quark
and gluon parton distributions inside an electron, the successive
halvings of evaluation-point spacing is interrupted when two values
agree in relative size, $|$new$-$old$|$/(new$+$old), to better than
\ttt{PARP(14)}. There are hardwired lower and upper limits of 2 and
8 halvings, respectively.

\iteme{PARP(15) :} (D=0.5 GeV) lower cut-off $p_0$ used to define
minimum transverse momentum in branchings $\gamma \to \q\qbar$ in
the anomalous event class of $\gamma\p$ interactions, i.e.\ sets the
dividing line between the VMD and GVMD event classes. 

\iteme{PARP(16) :} (D=1.) the anomalous parton-distribution functions 
of the photon are taken to have the charm and bottom flavour thresholds 
at virtuality \ttt{PARP(16)}$\times m_{\q}^2$.

\iteme{PARP(17) :} (D=1.) rescaling factor used for the $Q$ argument 
of the anomalous parton distributions of the photon, see 
\ttt{MSTP(15)}.

\iteme{PARP(18) :} (D=0.4 GeV) scale $k_{\rho}$, such that the cross 
sections of a GVMD state of scale $\kT$ is suppressed by a factor 
$k_{\rho}^2/\kT^2$ relative to those of a VMD state. Should be of
order $m_{\rho}/2$, with some finetuning to fit data.
 
\iteme{PARP(25) :} (D=0.) parameter $\eta$ describing the admixture 
of CP-odd Higgs decays for \ttt{MSTP(25)=3}.
 
\iteme{PARP(31) :} (D=1.5) common $K$ factor multiplying the
differential cross section for hard parton--parton processes
when \ttt{MSTP(33)=1} or \ttt{2}, with the exception of colour
annihilation graphs in the latter case.
 
\iteme{PARP(32) :} (D=2.0) special $K$ factor multiplying the
differential cross section in hard colour annihilation graphs,
including resonance production, when \ttt{MSTP(33)=2}.
 
\iteme{PARP(33) :} (D=0.075) this factor is used to multiply the
ordinary $Q^2$ scale in $\alphas$ at the hard interaction for
\ttt{MSTP(33)=3}. The effective $K$ factor thus obtained is in
accordance with the results in \cite{Ell86}, modulo the danger of
double counting because of parton-shower corrections to jet rates.

\iteme{PARP(34) :} (D=1.) the $Q^2$ scale defined by \ttt{MSTP(32)} is 
multiplied by \ttt{PARP(34)} when it is used as argument for parton
distributions and $\alphas$ at the hard interaction. It does not affect 
$\alphas$ when \ttt{MSTP(33)=3}, nor does it change the $Q^2$ argument 
of parton showers.
 
\iteme{PARP(35) :} (D=0.20) fix $\alphas$ value that is used in 
the heavy-flavour threshold factor when \ttt{MSTP(35)=1}.
 
\iteme{PARP(36) :} (D=0. GeV) the width $\Gamma_{\Q}$ for the heavy
flavour studied in processes {\ISUB} = 81 or 82; to be used for the
threshold factor when \ttt{MSTP(35)=2}.
 
\iteme{PARP(37) :} (D=1.) for \ttt{MSTP(37)=1} this regulates the
point at which the reference on-shell quark mass in Higgs and 
technicolor couplings is assumed defined in \ttt{PYMRUN} calls; 
specifically the running quark mass is assumed to coincide with the 
fix one at an energy scale \ttt{PARP(37)} times the fix quark mass,
i.e.\ $m_{\mrm{running}}($\ttt{PARP(37)}$\times m_{\mrm{fix}}) = 
m_{\mrm{fix}}$. See discussion at eq.~\ref{eq:runqmass} on ambiguity
of \ttt{PARP(37)} choice.
 
\iteme{PARP(38) :} (D=0.70 GeV$^3$) the squared wave function at the
origin, $|R(0)|^2$, of the $\Jpsi$ wave function. Used for processes
86 and 106--108. See ref. \cite{Glo88}.
 
\iteme{PARP(39) :} (D=0.006 GeV$^3$) the squared derivative of the
wave function at the origin, $|R'(0)|^2/m^2$, of the $\chi_{\c}$
wave functions. Used for processes 87--89 and 104--105. See ref.
\cite{Glo88}.
 
\iteme{PARP(41) :} (D=0.020 GeV) in the process of generating mass
for resonances, and optionally to force that mass to be in a given
range, only resonances with a total width in excess of \ttt{PARP(41)}
are generated according to a Breit--Wigner shape (if allowed by
\ttt{MSTP(42)}), while narrower resonances are put on the mass
shell.
 
\iteme{PARP(42) :} (D=2. GeV) minimum mass of resonances assumed
to be allowed when evaluating total width of $\hrm^0$ to $\Z^0 \Z^0$ or
$\W^+ \W^-$ for cases when the $\hrm^0$ is so light that (at least)
one $\Z/\W$ is forced to be off the mass shell. Also generally used
as safety check on minimum mass of resonance. Note that some 
\ttt{CKIN} values may provide additional constraints. 
 
\iteme{PARP(43) :} (D=0.10) precision parameter used in numerical
integration of width for a channel with at least one daughter off
the mass shell.
 
\iteme{PARP(44) :} (D=1000.) the $\nu$ parameter of the strongly
interacting $\Z/\W$ model of Dobado, Herrero and Terron \cite{Dob91}.
 
\iteme{PARP(45) :} (D=2054. GeV) the effective techni-$\rho$ mass
parameter of the strongly interacting model of Dobado, Herrero and
Terron \cite{Dob91}; see \ttt{MSTP(46)=5}. On physical grounds it
should not be chosen smaller than about 1 TeV or larger than
about the default value.

\iteme{PARP(46) :} (D=123. GeV) the $F_{\pi}$ decay constant that
appears inversely quadratically in all techni-$\eta$ partial decay
widths \cite{Eic84,App92}.

\iteme{PARP(47) :} (D=246. GeV) vacuum expectation value $v$ used 
in the DHT scenario \cite{Dob91} to define the width of the 
techni-$\rho$; this width is inversely proportional $v^2$.

\iteme{PARP(48) :} (D=50.) the Breit-Wigner factor in the cross section 
is set to vanish for masses that deviate from the nominal one by
more than \ttt{PARP(48)} times the nominal resonance width (i.e.\ the
width evaluated at the nominal mass). Is used in most processes 
with a single $s$-channel resonance, but there are some exceptions,
notably $\gamma^*/\Z^0$ and $\W^{\pm}$. The reason for this option 
is that the conventional Breit-Wigner description is at times not 
really valid far away from the resonance position, e.g.\ because of 
interference with other graphs that should then be included. The wings 
of the Breit-Wigner can therefore be removed.

\iteme{PARP(50) :} (D=0.054) dimensionless coupling, which enters 
quadratically in all partial widths of the excited graviton $\G^*$ 
resonance, is 
$\kappa  m_{\G^*} = \sqrt{2} x_1  k /\br{M}_{\mrm{Pl}}$,
where $x_1 \approx 3.83$ is the first zero of the $J_1$ Bessel function
and $\br{M}_{\mrm{Pl}}$ is the modified Planck mass scale
\cite{Ran99,Bij01}.

\iteme{PARP(61) - PARP(65) :} (C) parameters for initial-state 
radiation, see section \ref{ss:showrout}.
 
\iteme{PARP(71) - PARP(72) :} (C) parameter for final-state 
radiation, see section \ref{ss:showrout}.

\iteme{PARP(78) - PARP(90) :} parameters for multiple interactions,
see section \ref{ss:multintpar}.

\iteme{PARP(91) - PARP(100) :} parameters for beam remnant
treatment, see section \ref{ss:multintpar}.

\iteme{PARP(101) :} (D=0.50) fraction of diffractive systems in which
a quark is assumed kicked out by the pomeron rather than a gluon;
applicable for option \ttt{MSTP(101)=3}.
 
\iteme{PARP(102) :} (D=0.28 GeV) the mass spectrum of diffractive 
states (in single and double diffractive scattering) is assumed to
start \ttt{PARP(102)} above the mass of the particle that is 
diffractively excited. In this connection, an incoming $\gamma$
is taken to have the selected VMD meson mass, i.e.\ $m_{\rho}$,
$m_{\omega}$, $m_{\phi}$ or $m_{\Jpsi}$.

\iteme{PARP(103) :} (D=1.0 GeV) if the mass of a diffractive state 
is less than \ttt{PARP(103)} above the mass of the particle that is
diffractively excited, the state is forced to decay isotropically
into a two-body channel. In this connection, an incoming $\gamma$
is taken to have the selected VMD meson mass, i.e.\ $m_{\rho}$,
$m_{\omega}$, $m_{\phi}$ or $m_{\Jpsi}$. If the mass is higher than 
this threshold, the standard string fragmentation machinery is used. 
The forced two-body decay is always carried out, also when 
\ttt{MSTP(111)=0}. 

\iteme{PARP(104) :} (D=0.8 GeV) minimum energy above threshold for 
which hadron--hadron total, elastic and diffractive cross sections 
are defined. Below this energy, an alternative description in terms
of specific few-body channels would have been required, and this
is not modelled in {\Py}.
 
\iteme{PARP(110) :} (D=1.) a rescaling factor for resonance widths,
applied when \ttt{MSTP(110} is switched on.
 
\iteme{PARP(111) :} (D=2. GeV) used to define the minimum invariant
mass of the remnant hadronic system (i.e.\ when interacting partons
have been taken away), together with original hadron masses and
extra parton masses. For a hadron or resolved photon beam, this also 
implies a further constraint that the $x$ of an interacting parton 
be below $1 - 2 \times \mbox{\ttt{PARP(111)}}/E_{\mrm{cm}}$.

\iteme{PARP(115) :} (D=1.5 fm) (C) the average fragmentation time of a 
string, giving the exponential suppression that a reconnection 
cannot occur if strings decayed before crossing. Is implicitly 
fixed by the string constant and the fragmentation function 
parameters, and so a significant change is not recommended.

\iteme{PARP(116) :} (D=0.5 fm) (C) width of the type I string, giving the
radius of the Gaussian distribution in $x$ and $y$ separately.

\iteme{PARP(117) :} (D=0.6) (C) $k_{\mrm{I}}$, the main free parameter in 
the reconnection probability for scenario I; the probability is 
given by \ttt{PARP(117)} times the overlap volume, up to saturation 
effects.

\iteme{PARP(118), PARP(119) :} (D=2.5,2.0) (C) $f_r$ and $f_t$, 
respectively, used in the Monte Carlo sampling of the phase space volume 
in scenario I. There is no real reason to change these numbers. 

\iteme{PARP(120) :} (D=1.0) (D) (C) fraction of events in the GH, 
intermediate and instantaneous scenarios where a reconnection is allowed 
to occur. For the GH one a further suppression of the reconnection
rate occurs from the requirement of reduced string length in a
reconnection.    
 
\iteme{PARP(121) :} (D=1.) the maxima obtained at initial
maximization are multiplied by this factor if \ttt{MSTP(121)=1};
typically \ttt{PARP(121)} would be given as the product of the
violation factors observed (i.e.\ the ratio of final maximum value
to initial maximum value) for the given process(es).
 
\iteme{PARP(122) :} (D=0.4) fraction of total probability that is
shared democratically between the \ttt{COEF} coefficients open for
the given variable, with the remaining fraction distributed according
to the optimization results of \ttt{PYMAXI}.

\iteme{PARP(131) :} parameter for pile-up events, see section
\ref{ss:multintpar}. 

\iteme{PARP(151) - PARP(154) :} (D=4*0.) (C) regulate the assumed
beam-spot size. For \ttt{MSTP(151)=1} the $x$, $y$, $z$ and $t$ 
coordinates of the primary vertex of each event are selected 
according to four independent Gaussians. The widths of these 
Gaussians are given by the four parameters, where the first three 
are in units of mm and the fourth in mm/$c$. 

\iteme{PARP(161) - PARP(164) :} (D=2.20, 23.6, 18.4, 11.5) couplings 
$f_V^2/4\pi$ of the photon to the $\rho^0$, $\omega$, $\phi$ and
$\Jpsi$ vector mesons.

\iteme{PARP(165) :} (D=0.5) a simple multiplicative factor applied to 
the cross section for the transverse resolved photons to take into 
account the effects of longitudinal resolved photons, see 
\ttt{MSTP(17)}. No preferred value, but typically one could use 
\ttt{PARP(165)=1} as main contrast to the no-effect \ttt{=0}, with the 
default arbitrarily chosen in the middle. 

\iteme{PARP(167), PARP(168) :} (D=2*0) the longitudinal energy fraction 
$y$ of an incoming photon, side 1 or 2, used in the $R$ expression 
given for \ttt{MSTP(17)} to evaluate $f_L(y,Q^2)/f_T(y,Q^2)$. Need not 
be supplied when a photon spectrum is generated inside a lepton beam, 
but only when a photon is directly given as argument in the \ttt{PYINIT} 
call.

\iteme{PARP(171) :} to be set, event-by-event, when variable 
energies are allowed, i.e.\ when \ttt{MSTP(171)=1}. If \ttt{PYINIT} is 
called with \ttt{FRAME='CMS'} (\ttt{='FIXT'}), \ttt{PARP(171)} 
multiplies the c.m.\ energy (beam energy) used at initialization. 
For the options \ttt{'3MOM'}, \ttt{'4MOM'} and \ttt{'5MOM'}, 
\ttt{PARP(171)} is dummy, since there the momenta are set in the
\ttt{P} array. It is also dummy for the \ttt{'USER'} option,
where the choice of variable energies is beyond the control of {\Py}.

\iteme{PARP(173) :} event weight to be given by you when 
\ttt{MSTP(173)=1}.  

\iteme{PARP(174) :} (D=1.) maximum event weight that will be 
encountered in \ttt{PARP(173)} during the course of a run with 
\ttt{MSTP(173)=1}; to be used to optimize the efficiency of the 
event generation. It is always allowed to use a larger bound than 
the true one, but with a corresponding loss in efficiency.

\iteme{PARP(181) - PARP(189) :} (D = 0.1, 0.01, 0.01, 0.01, 0.1, 0.01,
0.01, 0.01, 0.3) Yukawa couplings of leptons to $\H^{++}$, assumed 
same for $\H_L^{++}$ and $\H_R^{++}$. Is a symmetric $3 \times 3$ 
array, where \ttt{PARP(177+3*i+j)} gives the coupling to a lepton pair 
with generation indices $i$ and $j$. Thus the default matrix is 
dominated by the diagonal elements and especially by the $\tau\tau$ one.

\iteme{PARP(190) :} (D=0.64) $g_L = e/\sin\theta_W$.

\iteme{PARP(191) :} (D=0.64) $g_R$, assumed same as $g_L$.

\iteme{PARP(192) :} (D=5 GeV) $v_L$ vacuum expectation value of the 
left-triplet. The corresponding $v_R$ is assumed given by
$v_R = \sqrt{2} M_{\W_R} / g_R$ and is not stored explicitly. 
  
\end{entry}

\subsection{Further Couplings}
\label{ss:coupcons}

In this section we collect information on the two routines for 
running $\alphas$ and $\alphaem$, and on other couplings
of standard and non-standard particles found in the \ttt{PYDAT1} and
\ttt{PYTCSM} common blocks. Although originally begun for applications 
within the traditional particle sector, this section of  \ttt{PYDAT1}
has rapidly expanded towards the non-standard aspects, and is thus more 
of interest for applications to specific processes. It could therefore 
equally well have been put somewhere else in this manual. Several other 
couplings indeed appear in the \ttt{PARP} array in the \ttt{PYPARS} 
common block, see section \ref{ss:PYswitchpar}, and the choice between 
the two has largely been dictated by availability of space. The 
improved simulation of the TechniColor Strawman Model, described
in \cite{Lan02,Lan02a}, and the resulting proliferation of model 
parameters, has led to the introduction of the new \ttt{PYTCSM}
common block.
 
\drawbox{ALEM = PYALEM(Q2)}\label{p:PYALEM}
\begin{entry}
\itemc{Purpose:} to calculate the running electromagnetic coupling
constant $\alphaem$. Expressions used are described in
ref. \cite{Kle89}. See \ttt{MSTU(101)}, \ttt{PARU(101)},
\ttt{PARU(103)} and \ttt{PARU(104)}.
\iteme{Q2 :} the momentum transfer scale $Q^2$ at which to evaluate
$\alphaem$.
\end{entry}

\drawbox{ALPS = PYALPS(Q2)}\label{p:PYALPS}
\begin{entry}
\itemc{Purpose:} to calculate the running strong coupling constant 
$\alphas$, e.g.\ in matrix elements and resonance decay widths. 
(The function is not used in parton showers, however, where 
formulae rather are written in terms of the relevant $\Lambda$
values.) The first- and second-order expressions are given by 
eqs.~(\ref{ee:aS3j}) and (\ref{ee:aS4j}). See 
\ttt{MSTU(111) - MSTU(118)} and \ttt{PARU(111) - PARU(118)} for options.
\iteme{Q2 :} the momentum transfer scale $Q^2$ at which to evaluate
$\alphas$.
\end{entry}

\drawbox{PM = PYMRUN(KF,Q2)}\label{p:PYMRUN}
\begin{entry}
\itemc{Purpose:} to give running masses of $\d$, $\u$, $\s$, $\c$, $\b$ 
and $\t$ quarks according to eq.~\ref{eq:runqmass}. For all other 
particles, the \ttt{PYMASS} function is called by \ttt{PYMRUN} to give 
the normal mass. Such running masses appear e.g.\ in couplings of 
fermions to Higgs and technipion states. 
\iteme{KF :} flavour code.
\iteme{Q2 :} the momentum transfer scale $Q^2$ at which to evaluate
$\alphas$.
\itemc{Note:} The nominal values, valid at a reference scale \\
$Q^2_{\mrm{ref}} = \max((\mtt{PARP(37)} m_{\mrm{nominal}})^2 , 4\Lambda^2)$,\\
are stored in \ttt{PARF(91)-PARF(96)}. 
\end{entry}
 
\drawbox{COMMON/PYDAT1/MSTU(200),PARU(200),MSTJ(200),PARJ(200)}
\begin{entry}
\itemc{Purpose:} to give access to a number of status codes and
parameters which regulate the performance of the program as a whole.
Here only those related to couplings are described; the main
description is found in section \ref{ss:JETswitch}.

\boxsep
 
\iteme{MSTU(101) :}\label{p:MSTU101} (D=1) procedure for 
$\alphaem$ evaluation in the \ttt{PYALEM} function.
\begin{subentry}
\iteme{= 0 :} $\alphaem$ is taken fixed at the value 
\ttt{PARU(101)}.
\iteme{= 1 :} $\alphaem$ is running with the $Q^2$ scale, 
taking into account corrections from fermion loops ($\e$, $\mu$, 
$\tau$, $\d$, $\u$, $\s$, $\c$, $\b$).
\iteme{= 2 :} $\alphaem$ is fixed, but with separate values at low 
and high $Q^2$. For $Q^2$ below (above) \ttt{PARU(104)} the value 
\ttt{PARU(101)} (\ttt{PARU(103)}) is used. The former value is
then intended for real photon emission, the latter for 
electroweak physics, e.g.\ of the $\W/\Z$ gauge bosons.
\end{subentry}
 
\iteme{MSTU(111) :} (I, D=1) order of $\alphas$ evaluation in the
\ttt{PYALPS} function. Is overwritten in \ttt{PYEEVT}, \ttt{PYONIA} or
\ttt{PYINIT} calls with the value desired for the process under study.
\begin{subentry}
\iteme{= 0 :} $\alphas$ is fixed at the value \ttt{PARU(111)}.
As extra safety, $\Lambda=$\ttt{PARU(117)} is set in \ttt{PYALPS} so 
that the first-order running $\alphas$ agrees with the desired fixed 
$\alphas$ for the $Q^2$ value used.
\iteme{= 1 :} first-order running $\alphas$ is used.
\iteme{= 2 :} second-order running $\alphas$ is used.
\end{subentry}
 
\iteme{MSTU(112) :} (D=5) the nominal number of flavours assumed in
the $\alphas$ expression, with respect to which $\Lambda$ is defined.
 
\iteme{MSTU(113) :} (D=3) minimum number of flavours that may be
assumed in $\alphas$ expression, see \ttt{MSTU(112)}.
 
\iteme{MSTU(114) :} (D=5) maximum number of flavours that may be
assumed in $\alphas$ expression, see \ttt{MSTU(112)}.
 
\iteme{MSTU(115) :} (D=0) treatment of $\alphas$ singularity for
$Q^2 \to 0$ in \ttt{PYALPS} calls. (Relevant e.g. for QCD $2 \to 2$ 
matrix elements in the $\pT \to 0$ limit, but not for showers, where
\ttt{PYALPS} is not called.)
\begin{subentry}
\iteme{= 0 :} allow it to diverge like $1/\ln(Q^2/\Lambda^2)$.
\iteme{= 1 :} soften the divergence to $1/\ln(1 + Q^2/\Lambda^2)$.
\iteme{= 2 :} freeze $Q^2$ evolution below \ttt{PARU(114)}, i.e.
the effective argument is $\max(Q^2, $\ttt{PARU(114)}$)$.
\end{subentry}
 
\iteme{MSTU(118) :} (I) number of flavours $n_f$ found and used in
latest \ttt{PYALPS} call.

\boxsep

\iteme{PARU(101) :}\label{p:PARU101} (D=0.00729735=1/137.04) 
$\alphaem$, the electromagnetic fine structure constant at 
vanishing momentum transfer.
 
\iteme{PARU(102) :} (D=0.232) $\ssintw$, the weak mixing angle of the
standard electroweak model.

\iteme{PARU(103) :} (D=0.007764=1/128.8) typical $\alphaem$ in
electroweak processes; used for $Q^2 >$\ttt{PARU(104)} in the
option \ttt{MSTU(101)=2} of \ttt{PYALEM}. Although it can technically 
be used also at rather small $Q^2$, this $\alphaem$ value is mainly 
intended for high $Q^2$, primarily $\Z^0$ and $\W^{\pm}$ physics.

\iteme{PARU(104) :} (D=1 GeV$^2$) dividing line between `low' and
`high' $Q^2$ values in the option \ttt{MSTU(101)=2} of \ttt{PYALEM}.

\iteme{PARU(105) :} (D=1.16639E-5 GeV$^{-2}$) $G_{\F}$, the Fermi 
constant of weak interactions.

\iteme{PARU(108) :} (I) the $\alphaem$ value obtained in the 
latest call to the \ttt{PYALEM} function.
 
\iteme{PARU(111) :} (D=0.20) fix $\alphas$ value assumed in
\ttt{PYALPS} when \ttt{MSTU(111)=0} (and also in parton showers
when $\alphas$ is assumed fix there).
 
\iteme{PARU(112) :} (I, D=0.25 GeV) $\Lambda$ used in running $\alphas$
expression in \ttt{PYALPS}. Like \ttt{MSTU(111)}, this value is
overwritten by the calling physics routines, and is therefore purely
nominal.
 
\iteme{PARU(113) :} (D=1.) the flavour thresholds, for the effective
number of flavours $n_f$ to use in the $\alphas$ expression, are
assumed to sit at $Q^2 = $\ttt{PARU(113)}$\times m_{\q}^2$, where
$m_{\q}$ is the quark mass. May be overwritten from the calling
physics routine.
 
\iteme{PARU(114) :} (D=4 GeV$^2$) $Q^2$ value below which the
$\alphas$ value is assumed constant for \ttt{MSTU(115)=2}.
 
\iteme{PARU(115) :} (D=10.) maximum $\alphas$ value that \ttt{PYALPS}
will ever return; is used as a last resort to avoid singularities.
 
\iteme{PARU(117) :} (I) $\Lambda$ value (associated with
\ttt{MSTU(118)} effective flavours) obtained in latest \ttt{PYALPS}
call.
 
\iteme{PARU(118) :} (I) $\alphas$ value obtained in latest
\ttt{PYALPS} call.
 
\iteme{PARU(121) - PARU(130) :} couplings of a new $\Z'^0$; for
fermion default values are given by the Standard Model $\Z^0$ values,
assuming $\ssintw = 0.23$. Since a generation dependence is now
allowed for the $\Z'^0$ couplings to fermions, the variables
\ttt{PARU(121) - PARU(128)} only refer to the first generation,
with the second generation in \ttt{PARJ(180) - PARJ(187)} and
the third in  \ttt{PARJ(188) - PARJ(195)} following exactly the
same pattern. Note that e.g.\ the $\Z'^0$ width contains
squared couplings, and thus depends quadratically on the values below.
\begin{subentry}
\iteme{PARU(121), PARU(122) :} (D=-0.693,-1.) vector and axial
couplings of down type quarks to $\Z'^0$.
\iteme{PARU(123), PARU(124) :} (D=0.387,1.) vector and axial
couplings of up type quarks to $\Z'^0$.
\iteme{PARU(125), PARU(126) :} (D=-0.08,-1.) vector and axial
couplings of leptons to $\Z'^0$.
\iteme{PARU(127), PARU(128) :} (D=1.,1.) vector and axial
couplings of neutrinos to $\Z'^0$.
\iteme{PARU(129) :} (D=1.) the coupling $Z'^0 \to \W^+ \W^-$ is
taken to be \ttt{PARU(129)}$\times$(the Standard Model
$\Z^0 \to \W^+ \W^-$ coupling)$\times (m_{\W}/m_{\Z'})^2$.
This gives a $\Z'^0 \to \W^+ \W^-$ partial width that
increases proportionately to the $\Z'^0$ mass.
\iteme{PARU(130) :} (D=0.) in the decay chain
$\Z'^0 \to \W^+ \W^- \to 4$ fermions, the angular distribution in
the $\W$ decays is supposed to be a mixture, with fraction
\ttt{1-PARU(130)} corresponding to the same angular distribution
between the four final fermions as in $\Z^0 \to \W^+ \W^-$ (mixture
of transverse and longitudinal $\W$'s), and fraction \ttt{PARU(130)}
corresponding to $\hrm^0 \to \W^+ \W^-$ the same way (longitudinal
$\W$'s).
\end{subentry}
 
\iteme{PARU(131) - PARU(136) :} couplings of a new $\W'^{\pm}$;
for fermions default values are given by the Standard Model
$\W^{\pm}$ values (i.e.\ $V-A$). Note that e.g.\ the $\W'^{\pm}$
width contains squared couplings, and
thus depends quadratically on the values below.
\begin{subentry}
\iteme{PARU(131), PARU(132) :} (D=1.,-1.) vector and axial couplings
of a quark--antiquark pair to $\W'^{\pm}$; is further multiplied by the
ordinary CKM factors.
\iteme{PARU(133), PARU(134) :} (D=1.,-1.) vector and axial couplings
of a lepton-neutrino pair to $\W'^{\pm}$.
\iteme{PARU(135) :} (D=1.) the coupling $\W'^{\pm} \to \Z^0 \W^{\pm}$
is taken to be \ttt{PARU(135)}$\times$(the Standard Model
$\W^{\pm} \to \Z^0 \W^{\pm}$ coupling)$\times (m_{\W}/m_{W'})^2$.
This gives a $\W'^{\pm} \to \Z^0 \W^{\pm}$ partial width that
increases proportionately to the $\W'$ mass.
\iteme{PARU(136) :} (D=0.) in the decay chain
$\W'^{\pm} \to \Z^0 \W^{\pm} \to 4$ fermions,
the angular distribution in the $\W/\Z$ decays is supposed to be a
mixture, with fraction \ttt{1-PARU(136)} corresponding to the same
angular distribution between the four final fermions as in
$\W^{\pm} \to \Z^0 \W^{\pm}$ (mixture of transverse and longitudinal
$\W/\Z$'s), and fraction \ttt{PARU(136)} corresponding to
$\H^{\pm} \to \Z^0 \W^{\pm}$ the same way (longitudinal $\W/\Z$'s).
\end{subentry}
 
\iteme{PARU(141) :} (D=5.) $\tan\beta$ parameter of a two Higgs
doublet scenario, i.e.\ the ratio of vacuum expectation values.
This affects mass relations and couplings in the Higgs sector.
If the Supersymmetry simulation is switched on, \ttt{IMSS(1)}
nonvanishing, \ttt{PARU(141)} will be overwritten by \ttt{RMSS(5)}
at initialization, so it is the latter variable that should be set. 
 
\iteme{PARU(142) :} (D=1.) the $\Z^0 \to \H^+ \H^-$ coupling is
taken to be \ttt{PARU(142)}$\times$(the MSSM $\Z^0 \to \H^+ \H^-$
coupling).
 
\iteme{PARU(143) :} (D=1.) the $\Z'^0 \to \H^+ \H^-$ coupling is
taken to be \ttt{PARU(143)}$\times$(the MSSM $\Z^0 \to \H^+ \H^-$
coupling).
 
\iteme{PARU(145) :} (D=1.) quadratically multiplicative factor in the
$\Z'^0 \to \Z^0 \hrm^0$ partial width in left--right-symmetric models,
expected to be unity (see \cite{Coc91}).
 
\iteme{PARU(146) :} (D=1.) $\sin(2\alpha)$ parameter, enters
quadratically as multiplicative factor in the
$\W'^{\pm} \to \W^{\pm} \hrm^0$ partial width in
left--right-symmetric models (see \cite{Coc91}).
 
\iteme{PARU(151) :} (D=1.) multiplicative factor in the
$\L_{\Q} \to \q \ell$ squared Yukawa coupling, and thereby in the
$\L_{\Q}$ partial width and the $\q \ell \to \L_{\Q}$ and other
cross sections. Specifically,
$\lambda^2/(4\pi) = $\ttt{PARU(151)}$\times \alphaem$, i.e.\ it
corresponds to the $k$ factor of \cite{Hew88}.
 
\iteme{PARU(161) - PARU(168) :} (D=5*1.,3*0.) multiplicative factors
that can be used to modify the default couplings of the $\hrm^0$
particle in {\Py}. Note that the factors enter quadratically in the
partial widths. The default values correspond to the couplings given
in the minimal one-Higgs-doublet Standard Model, and are therefore 
not realistic in a two-Higgs-doublet scenario. The default values 
should be changed appropriately by you. Also the last two default
values should be changed; for these the expressions of the
minimal supersymmetric Standard Model (MSSM) are given to show
parameter normalization. Alternatively, the SUSY machinery can 
generate all the couplings for \ttt{IMSS(1)}, see \ttt{MSTP(4)}.
\begin{subentry}
\iteme{PARU(161) :} $\hrm^0$ coupling to down type quarks.
\iteme{PARU(162) :} $\hrm^0$ coupling to up type quarks.
\iteme{PARU(163) :} $\hrm^0$ coupling to leptons.
\iteme{PARU(164) :} $\hrm^0$ coupling to $\Z^0$.
\iteme{PARU(165) :} $\hrm^0$ coupling to $\W^{\pm}$.
\iteme{PARU(168) :} $\hrm^0$ coupling to $\H^{\pm}$ in
$\gamma\gamma \to \hrm^0$ loops, in MSSM
$\sin(\beta-\alpha)+\cos(2\beta)\sin(\beta+\alpha) /
(2\scostw)$.
\end{subentry}
 
\iteme{PARU(171) - PARU(178) :} (D=7*1.,0.) multiplicative factors
that can be used to modify the default couplings of the $\H^0$
particle in {\Py}. Note that the factors enter quadratically in
partial widths. The default values for \ttt{PARU(171) - PARU(175)}
correspond to the couplings given to $\hrm^0$ in the minimal
one-Higgs-doublet Standard Model, and are therefore not realistic
in a two-Higgs-doublet scenario. The default values should
be changed appropriately by you. Also the last two default
values should be changed; for these the expressions of the
minimal supersymmetric Standard Model (MSSM) are given to show
parameter normalization. Alternatively, the SUSY machinery can generate
all the couplings for \ttt{IMSS(1)}, see \ttt{MSTP(4)}.
\begin{subentry}
\iteme{PARU(171) :} $\H^0$ coupling to down type quarks.
\iteme{PARU(172) :} $\H^0$ coupling to up type quarks.
\iteme{PARU(173) :} $\H^0$ coupling to leptons.
\iteme{PARU(174) :} $\H^0$ coupling to $\Z^0$.
\iteme{PARU(175) :} $\H^0$ coupling to $W^{\pm}$.
\iteme{PARU(176) :} $\H^0$ coupling to $\hrm^0 \hrm^0$, in MSSM
$\cos(2\alpha) \cos(\beta+\alpha) - 2 \sin(2\alpha)
\sin(\beta+\alpha)$.
\iteme{PARU(177) :} $\H^0$ coupling to $\A^0 \A^0$, in MSSM
$\cos(2\beta) \cos(\beta+\alpha)$.
\iteme{PARU(178) :} $\H^0$ coupling to $\H^{\pm}$ in
$\gamma \gamma \to \H^0$ loops, in MSSM
$\cos(\beta-\alpha) - \cos(2\beta)\cos(\beta+\alpha) /
(2\scostw)$.
\end{subentry}
 
\iteme{PARU(181) - PARU(190) :} (D=3*1.,2*0.,2*1.,3*0.)
multiplicative factors that can be used to modify the default
couplings of the $\A^0$ particle in {\Py}. Note that the factors 
enter quadratically in partial widths. The default values for 
\ttt{PARU(181) - PARU(183)} correspond
to the couplings given to $\hrm^0$ in the minimal one-Higgs-doublet
Standard Model, and are therefore not realistic in a
two-Higgs-doublet scenario. The default values should
be changed appropriately by you. \ttt{PARU(184)} and \ttt{PARU(185)}
should be vanishing at the tree level, in the absence of CP 
violating phases in the Higgs sector, and are so set;
normalization of these couplings agrees with what is used for
$\hrm^0$ and $\H^0$. Also the other default values should be changed; for
these the expressions of the Minimal Supersymmetric Standard
Model (MSSM) are given to show parameter normalization. Alternatively, 
the SUSY machinery can generate all the couplings for \ttt{IMSS(1)}, 
see \ttt{MSTP(4)}.
\begin{subentry}
\iteme{PARU(181) :} $\A^0$ coupling to down type quarks.
\iteme{PARU(182) :} $\A^0$ coupling to up type quarks.
\iteme{PARU(183) :} $\A^0$ coupling to leptons.
\iteme{PARU(184) :} $\A^0$ coupling to $\Z^0$.
\iteme{PARU(185) :} $\A^0$ coupling to $\W^{\pm}$.
\iteme{PARU(186) :} $\A^0$ coupling to $\Z^0 \hrm^0$ (or
$\Z^*$ to $\A^0 \hrm^0$), in MSSM $\cos(\beta-\alpha)$.
\iteme{PARU(187) :} $\A^0$ coupling to $\Z^0 \H^0$ (or $\Z^*$ to
$\A^0 \H^0$), in MSSM $\sin(\beta-\alpha)$.
\iteme{PARU(188) :} As \ttt{PARU(186)}, but coupling to $\Z'^0$
rather than $\Z^0$.
\iteme{PARU(189) :} As \ttt{PARU(187)}, but coupling to $\Z'^0$
rather than $\Z^0$.
\iteme{PARU(190) :} $\A^0$ coupling to $\H^{\pm}$ in
$\gamma \gamma \to \A^0$ loops, 0 in MSSM.
\end{subentry}
 
\iteme{PARU(191) - PARU(195) :} (D=4*0.,1.) multiplicative factors
that can be used to modify the couplings of the $\H^{\pm}$ particle
in {\Py}. Currently only \ttt{PARU(195)} is in use. See above for
related comments.
\begin{subentry}
\iteme{PARU(195) :} $\H^{\pm}$ coupling to $\W^{\pm} \hrm^0$ (or
$\W^{* \pm}$ to $\H^{\pm} \hrm^0$), in MSSM $\cos(\beta-\alpha)$.
\end{subentry}

\iteme{PARU(197):} (D=0.) $\H^0$ coupling to $\W^{\pm} \H^{\mp}$
within a two-Higgs-doublet model.

\iteme{PARU(198):} (D=0.) $\A^0$ coupling to $\W^{\pm} \H^{\mp}$
within a two-Higgs-doublet model.

\boxsep

\iteme{PARJ(180) - PARJ(187) :}\label{p:PARJ180} couplings of the 
second generation fermions to the $Z'^0$, following the same pattern
and with the same default values as the first one in 
\ttt{PARU(121) - PARU(128)}.

\iteme{PARJ(188) - PARJ(195) :}couplings of the 
third generation fermions to the $Z'^0$, following the same pattern
and with the same default values as the first one in 
\ttt{PARU(121) - PARU(128)}.

\end{entry}
 
\drawbox{COMMON/PYTCSM/ITCM(0:99),RTCM(0:99)}\label{p:PYTCSM}
\begin{entry}
\itemc{Purpose:} to give access to a number of switches and parameters 
which regulate the simulation of the TechniColor Strawman Model
\cite{Lan02,Lan02a}, plus a few further parameters related to
the simulation of compositeness, mainly in earlier incarnations of 
TechniColor.

\boxsep
 
\iteme{ITCM(1) :}\label{p:ITCM} (D=4) $N_{TC}$, number of technicolors; 
fixes the relative values of $g_{\mrm{em}}$ and $g_{\mrm{etc}}$.

\iteme{ITCM(2) :} (D=0) Topcolor model.
\begin{subentry}
\iteme{= 0 :} Standard Topcolor. Third generation quark couplings to 
the coloron are proportional to $\cot\theta_3$, see \ttt{RTCM(21)} 
below; first two generations are proportional to $-\tan\theta_3$.
\iteme{= 1 :} Flavor Universal Topcolor. All quarks couple with 
strength proportional to $\cot\theta_3$.
\end{subentry}

\iteme{ITCM(5) :} (D=0) presence of anomalous couplings in standard model 
processes, see subsection \ref{sss:ancoupclass} for further details.
\begin{subentry}
\iteme{= 0 :} absent.
\iteme{= 1 :} left--left isoscalar model, with only $\u$ and $\d$ 
quarks composite (at the probed scale).
\iteme{= 2 :} left--left isoscalar model, with all quarks composite.
\iteme{= 3 :} helicity-non-conserving model, with only $\u$ and $\d$ 
quarks composite (at the probed scale).
\iteme{= 4 :} helicity-non-conserving model, with all quarks composite.
\iteme{= 5 :} coloured technihadrons, affecting the standard QCD 
$2 \to 2$ cross sections by the exchange of Coloron or Colored Technirho, 
see subsection \ref{sss:technicolorclass}. 
\end{subentry}

\boxsep

\iteme{RTCM(1) :}\label{p:RTCM} (D=82 GeV) $F_T$, the Technicolor decay 
constant.

\iteme{RTCM(2) :} (D=4/3) $Q_U$, charge of up-type technifermion; 
the down-type technifermion has a charge $Q_D=Q_U-1$.   

\iteme{RTCM(3) :} (D=1/3) $\sin\chi$, where $\chi$ is the 
mixing angle between isotriplet technipion interaction and mass 
eigenstates.

\iteme{RTCM(4) :} (D=$1/\sqrt{6}$) $\sin\chi'$, where $\chi'$ is the 
mixing angle between the isosinglet ${\pi'}^0_{\mrm{tc}}$ interaction 
and mass eigenstates. 

\iteme{RTCM(5) :} (D=1) Clebsch for technipi decays to charm. Appears 
squared in decay rates.

\iteme{RTCM(6) :} (D=1) Clebsch for technipi decays to bottom. Appears 
squared in decay rates.

\iteme{RTCM(7) :} (D=0.0182) Clebsch for technipi decays to top,
estimated to be $m_{\b}/m_{\t}$. Appears squared in decay rates.

\iteme{RTCM(8) :} (D=1) Clebsch for technipi decays to $\tau$.
Appears squared in decay rates.

\iteme{RTCM(9) :} (D=0) squared Clebsch for isotriplet technipi decays 
to gluons.

\iteme{RTCM(10) :} (D=4/3) squared Clebsch for isosinglet technipi 
decays to gluons.

\iteme{RTCM(11) :} (D=0.05) technirho--techniomega mixing parameters.  
Allows for isospin violating decays of the techniomega.

\iteme{RTCM(12) :} (D=200 GeV) vector technimeson decay parameter.  
Affects the decay rates of vector technimesons into technipi plus 
transverse gauge boson.

\iteme{RTCM(13) :} (D=200 GeV) axial mass parameter for 
technivector decays to transverse gauge bosons and technipions.

\iteme{RTCM(21) :} (D=$\sqrt{0.08}$) tangent of Topcolor mixing angle,
in the scenario with coloured technihadrons described in subsection 
\ref{sss:technicolorclass} and switched on with \ttt{ITCM(5)=5}. 
For \ttt{ITCM(2)=0}, the coupling of the $\mathrm{V}_8$ to light 
quarks is suppressed by \ttt{RTCM(21)}$^2$ whereas the coupling to 
heavy ($\b$ and $\t$) quarks is enhanced by 1/\ttt{RTCM(21)}$^2$. 
For \ttt{ITCM(21)=1}, the coupling to quarks 
is universal, and given by 1/\ttt{RTCM(21)}$^2$.  

\iteme{RTCM(22) :} (D=$1/\sqrt{2}$) sine of isosinglet technipi mixing 
with Topcolor currents.

\iteme{RTCM(23) :} (D=0) squared Clebsch for color-octet technipi decays 
to charm.

\iteme{RTCM(24) :} (D=0) squared Clebsch for color-octet technipi decays 
to bottom.

\iteme{RTCM(25) :} (D=0) squared Clebsch for color-octet technipi decays 
to top.

\iteme{RTCM(26) :} (D=5/3) squared Clebsch for color-octet technipi 
decays to gluons.

\iteme{RTCM(27) :} (D=250 GeV) color-octet technirho decay parameter for 
decays to technipi plus gluon.

\iteme{RTCM(28) :} (D=250 GeV)  hard mixing parameter between 
color-octet technirhos.

\iteme{RTCM(29) :} (D=$1/\sqrt{2}$) magnitude of $(1,1)$ element of the 
\tbf{U(2)} matrices that diagonalize U-type technifermion condensates.

\iteme{RTCM(30) :} (D=0 Radians) phase for the element described above,
\ttt{RTCM(29)}.

\iteme{RTCM(31) :} (D=$1/\sqrt{2}$) Magnitude of $(1,1)$ element of the 
\tbf{U(2)} matrices that diagonalize D-type technifermion condensates.

\iteme{RTCM(32) :} (D=0 Radians) phase for the element described above,
\ttt{RTCM(31)}.

\iteme{RTCM(33) :} (D=1) if 
$\Gamma_{V_8}(\hat{s}) > \mtt{RTCM(33)}\sqrt{\hat{s}}$, then
$\Gamma_{V_8}(\hat{s})$ is redefined to be $\mtt{RTCM(33)}\sqrt{\hat{s}}$.
It thus prevents the coloron from becoming wider than its mass. 

\iteme{RTCM(41) :}  compositeness scale $\Lambda$, used in processes 
involving excited fermions, and for standard model processes when 
\ttt{ITCM(5)} is between 1 and 4.

\iteme{RTCM(42) :}  (D=1.) sign of the interference term between the
standard cross section and the compositeness term ($\eta$ parameter); 
should be $\pm 1$; used for standard model processes when \ttt{ITCM(5)} 
is between 1 and 4.

\iteme{RTCM(43) - RTCM(45) :} (D=3*1.) strength of the \tbf{SU(2)}, 
\tbf{U(1)} and \tbf{SU(3)} couplings, respectively, in an excited fermion 
scenario; cf. $f$, $f'$ and $f_s$ of \cite{Bau90}.

\iteme{RTCM(46) :}  (D=0.) anomalous magnetic moment of the $\W^{\pm}$ 
in process 20; $\eta = \kappa - 1$, where $\eta = 0$ ($\kappa = 1$) 
is the Standard Model value.
  
\end{entry}

\subsection{Supersymmetry Common Blocks and Routines}
\label{ss:susycode}

The parameters available to the SUSY user are stored in the
common block \ttt{PYMSSM}. In general, options are set by the \ttt{IMSS}
array, while real valued parameters are set by \ttt{RMSS}.  The entries
\ttt{IMSS(0)} and \ttt{RMSS(0)} are not used, but are available
for compatibility with the C programming language. Note also that most
options are only used by {\Py}'s internal SUSY machinery and are ineffective
when external spectrum calculations are used, see section \ref{sss:models}.

\drawbox{COMMON/PYMSSM/IMSS(0:99),RMSS(0:99)}\label{p:PYMSSM}
\begin{entry}
\itemc{Purpose:} to give access to parameters that allow the
simulation of the MSSM.

\iteme{IMSS(1) :}\label{p:IMSS} (D=0) level of MSSM simulation.
\begin{subentry}
\iteme{= 0 :} No MSSM simulation.
\iteme{= 1 :} A general MSSM simulation.  The parameters of the model
are set by the array \ttt{RMSS}.
\iteme{= 2 :} An approximate SUGRA simulation using the analytic
formulae of \cite{Dre95} to reduce the number of free parameters.
In this case, only five input parameters are used.
\ttt{RMSS(1)} is the common gaugino mass
$m_{1/2}$, \ttt{RMSS(8)} is the common scalar mass $m_0$, 
\ttt{RMSS(4)} fixes the sign of the higgsino mass $\mu$, 
\ttt{RMSS(16)} is the common trilinear coupling $A$, and
\ttt{RMSS(5)} is $\tan\beta=v_2/v_1$.
\iteme{= 11 :} Read spectrum from a SUSY Les Houches Accord conformant
file. The Logical Unit Number on which the file is opened should be put in
\ttt{IMSS(21)}. If a decay table should also be read in, the corresponding
Unit Number (normally the same as the spectrum file) should be put in
\ttt{IMSS(22)}. Cross sections are
still calculated by {\Py}, as are decays for those sparticles and higgs
bosons for which a decay table is not found on the file.
\iteme{= 12 :} Invoke a runtime interface to \tsc{Isasusy} \cite{Bae93}
for determining SUSY mass spectrum and mixing parameters. This 
provides a more precise solution of the renormalization group
equations than is offered by the option \ttt{= 2} above. The interface
automatically asks the \ttt{SUGRA} routine (part of \tsc{Isasusy}) to 
solve the RGE's for the weak scale mass spectrum and mixing parameters. 
The mSUGRA input parameters should be given in \ttt{RMSS} as usual, i.e.: 
\ttt{RMSS(1) = }$m_{1/2}$, \ttt{RMSS(4) = }\sgnmu,
\ttt{RMSS(5) = }$\tan\beta$, \ttt{RMSS(8) = }$m_0$, and
\ttt{RMSS(16)= }$A$. As before, we are using the conventions of 
\cite{Hab85, Gun86a} everywhere. Cross sections and decay widths are 
still calculated by {\Py}, using the output provided by \tsc{Isasusy}. 
Note that since {\Py} cannot always be expected to be linked with the 
\tsc{Isajet} library, a new dummy routine and a new dummy function have 
been added. These are \ttt{SUGRA} and \ttt{VISAJE}, located towards the 
very bottom of the {\Py} source code. These routines must be removed and 
{\Py} recompiled before a proper linking with \tsc{Isajet} can be achieved.
Furthermore, the common block sizes and variable positions accessed in
the \ttt{SUGRA} routine have to match those of the\tsc{Isajet} version
used, see subsection \ref{sss:models}.  
\end{subentry}

\iteme{IMSS(2) :} (D=0) treatment of {\bf U(1)}, {\bf SU(2)}, and 
{\bf SU(3)} gaugino mass parameters.
\begin{subentry}
\iteme{= 0 :} The gaugino parameters $M_1, M_2$ and $M_3$ are 
set by \ttt{RMSS(1), RMSS(2),} and \ttt{RMSS(3)}, i.e.\ there is 
no forced relation between them.
\iteme{= 1 :} The gaugino parameters are fixed by the relation 
$(3/5) \, M_1/\alpha_1 =M_2/\alpha_2=M_3/\alpha_3=X$ and the 
parameter \ttt{RMSS(1)}.  If \ttt{IMSS(1)=2}, then
\ttt{RMSS(1)} is treated as the common gaugino mass $m_{1/2}$ 
and \ttt{RMSS(20)} is the GUT scale coupling constant
$\alpha_{GUT}$, so that $X=m_{1/2}/\alpha_{GUT}$.
\iteme{= 2 :} $M_1$ is set by \ttt{RMSS(1)}, $M_2$ by \ttt{RMSS(2)} 
and $M_3 = M_2\alpha_3/\alpha_2$. In such a scenario, 
the {\bf U(1)} gaugino mass behaves anomalously. 
\end{subentry}

\iteme{IMSS(3) :} (D=0) treatment of the gluino mass parameter.
\begin{subentry}
\iteme{= 0 :}  The gluino mass parameter $M_3$ is used to calculate 
the gluino pole mass with the formulae of \cite{Kol96}.  The effects of 
squark loops can significantly shift the mass.  
\iteme{= 1 :}  $M_3$ is the gluino pole mass.   The effects of squark
loops are assumed to have been included in this value. 
\end{subentry}

\iteme{IMSS(4) :} (D=1) treatment of the Higgs sector.
\begin{subentry}
\iteme{= 0 :}  The Higgs sector is determined by 
the approximate formulae of \cite{Car95}  and the pseudoscalar mass 
$M_{\A}$ set by \ttt{RMSS(19)}.
\iteme{= 1 :} The Higgs sector is determined by the exact
formulae of \cite{Car95} and the pseudoscalar mass $M_{\A}$ set by 
\ttt{RMSS(19)}.  The pole mass for $M_{\A}$ is not the same as the input
parameter. 
\iteme{= 2 :} The Higgs sector is fixed by the mixing angle $\alpha$ 
set by \ttt{RMSS(18)} and the mass values \ttt{PMAS(I,1)}, where 
\ttt{I=25,35,36,} and \ttt{37}. 
\end{subentry}

\iteme{IMSS(5) :} (D=0) allows you to set the $\st$, $\sbo$ and 
$\stau$ masses and mixing by hand. 
\begin{subentry}
\iteme{= 0 :} no, the program calculates itself.
\iteme{= 1 :} yes, calculate from given input. The parameters 
\ttt{RMSS(26) - RMSS(28)} specify the mixing angle (in radians) 
for the sbottom, stop, and stau. The parameters \ttt{RMSS(10) - RMSS(14)} 
specify the two stop masses, the one sbottom mass (the other being fixed 
by the other parameters) and the two stau masses.  Note that the masses 
\ttt{RMSS(10), RMSS(11)} and \ttt{RMSS(13)} correspond to the left-left 
entries of the diagonalized matrices, while \ttt{RMSS(12)} and 
\ttt{RMSS(14)} correspond to the right-right entries.  Note that these 
entries need not be ordered in mass.
\end{subentry}

\iteme{IMSS(7) :} (D=0) treatment of the scalar masses in an
extension of SUGRA models.  The presence of additional {\bf U(1)}
symmetries at high energy scales can modify the boundary
conditions for the scalar masses at the unification scale.
\begin{subentry}
\iteme{= 0 :}   No additional $D$--terms are included.  In
SUGRA models, all scalars have the mass $m_0$ at the 
unification scale.
\iteme{= 1 :}   \ttt{RMSS(23) - RMSS(25)} are the values of $D_X, 
D_Y$ and $D_S$ at the unification scale in the model of \cite{Mar94}.
The boundary conditions for the scalar masses are shifted based
on their quantum numbers under the additional {\bf U(1)} symmetries. 
\end{subentry}

\iteme{IMSS(8) :} (D=0) treatment of the $\stau$ mass eigenstates.
\begin{subentry}
\iteme{= 0 :}   The $\stau$ mass eigenstates are calculated 
using the parameters\\ 
\ttt{RMSS(13,14,17)}. 
\iteme{= 1 :}   The $\stau$ mass eigenstates are identical to 
the interaction eigenstates, so they are treated
identically to $\se$ and $\smu$ .
\end{subentry}

\iteme{IMSS(9) :} (D=0) treatment of the right handed squark mass
eigenstates for the first two generations.
\begin{subentry}
\iteme{= 0 :}   The $\sq_R$ masses are fixed by \ttt{RMSS(9)}.  
$\sd_R$ and $\su_R$ are identical except for Electroweak $D$--term 
contributions. 
\iteme{= 1 :}   The masses of $\sd_R$ and $\su_R$ 
are fixed by \ttt{RMSS(9)} and \ttt{RMSS(22)} respectively. 
\end{subentry}

\iteme{IMSS(10) :} (D=0) allowed decays for $\chio_2$.
\begin{subentry}
\iteme{= 0 :}   The second lightest neutralino $\chio_2$ decays 
with a branching ratio calculated from the MSSM parameters.  
\iteme{= 1 :}  $\chio_2$ is forced to decay only to $\chio_1 \gamma$, 
regardless of the actual branching ratio. 
This can be used for detailed studies of this particular final state. 
\end{subentry}

\iteme{IMSS(11) :} (D=0) choice of the lightest superpartner (LSP).
\begin{subentry}
\iteme{= 0 :} $\chio_1$ is the LSP.  
\iteme{= 1 :} $\chio_1$ is the next to lightest superparter (NLSP) 
and the gravitino is the LSP.  The $\chio_1$ 
decay length is calculated from the gravitino 
mass set by \ttt{RMSS(21)} and the $\chio_1$ mass and mixing. 
\end{subentry}

\iteme{IMSS(21) :} (D=0) Logical Unit Number for SUSY Les Houches Accord
spectrum read-in. Only used if \ttt{IMSS(1)=11}. 

\iteme{IMSS(22) :} (D=0) Read-in of SUSY Les Houches Accord decay table.
\begin{subentry}
\iteme{= 0 :} No decays are read in.  The internal {\Py} machinery is
        used to calculate decay rates.
\iteme{> 0 :} Read decays from LHA3 file on unit number
        \ttt{IMSS(22)}. During initialization, decay tables in the file
        will replace the values calculated by {\Py}. Particles for which
        the file \emph{does not} contain a decay table will thus still 
        have their decays calculated by {\Py}.
        In normal usage one would expect \ttt{IMSS(22)} to be equal to 
        \ttt{IMSS(21)}, to ensure that the spectrum and decays are consistent
        with each other, but this is not a strict requirement.
\end{subentry}

\iteme{IMSS(23) :} (D=0) writing of MSSM spectrum data.
\begin{subentry}
\iteme{= 0 :} Don't write out spectrum.
\iteme{> 0 :} Write out spectrum in LHA3 format (calculated by {\Py} or
        otherwise) to file on unit number \ttt{IMSS(23)}.
\end{subentry}

\iteme{IMSS(24) :} (D=0) writing of MSSM particle decay table.
\begin{subentry}
\iteme{= 0 :} Don't write out decay table.
\iteme{> 0 :} Write out decay table in LHA3 format to file on unit number
        \ttt{IMSS(24)}. Not implemented in the code yet.
        In normal usage one would expect \ttt{IMSS(24)} to be equal to
        \ttt{IMSS(23)}, to ensure that the spectrum and decays are consistent
        with each other, but this is not a strict requirement.
\end{subentry}

\iteme{IMSS(51) :} (D=0) Lepton number violation on/off 
(LLE type couplings).
\begin{subentry}
\iteme{= 0 :} All LLE couplings off. LLE decay channels off. 
\iteme{= 1 :} All LLE couplings set to common value given by 
$10^\ttt{\scriptsize-RMSS(51)}$.
\iteme{= 2 :} LLE couplings set to generation-hierarchical 
`natural' values with common normalization \ttt{RMSS(51)} 
(see section \ref{sss:rparityviol}).
\iteme{= 3 :} All LLE couplings set to zero, but LLE decay channels not
switched off. Non-zero couplings should be entered individually into the
array \ttt{RVLAM(I,J,K)}. Because of the antisymmetry in I and J, only
entries with I$<$J need be entered.  
\end{subentry}

\iteme{IMSS(52) :} (D=0) Lepton number violation on/off 
(LQD type couplings).
\begin{subentry}
\iteme{= 0 :} All LQD couplings off. LQD decay channels off. 
\iteme{= 1 :} All LQD couplings set to common value given by
$10^\ttt{\scriptsize-RMSS(52)}$.
\iteme{= 2 :} LQD couplings set to generation-hierarchical 
`natural' values with common normalization \ttt{RMSS(52)} 
(see section \ref{sss:rparityviol}).
\iteme{= 3 :} All LQD couplings set to zero, but LQD decay channels not
switched off. Non-zero couplings should be entered individually into the
array \ttt{RVLAMP(I,J,K)}.
\end{subentry}

\iteme{IMSS(53) :} (D=0) Baryon number violation on/off 
\begin{subentry}
\iteme{= 0 :} All UDD couplings off. UDD decay channels off. 
\iteme{= 1 :} All UDD couplings set to common value given by
$10^\ttt{\scriptsize-RMSS(53)}$.
\iteme{= 2 :} UDD couplings set to generation-hierarchical 
`natural' values with common normalization \ttt{RMSS(53)} 
(see section \ref{sss:rparityviol}).
\iteme{= 3 :} All UDD couplings set to zero, but UDD decay channels not
switched off. Non-zero couplings should be entered individually into the
array \ttt{RVLAMB(I,J,K)}. Because of the antisymmetry in J and K, only
entries with J$<$K need be entered.  
\end{subentry}

\boxsep

\iteme{RMSS(1) :}\label{p:RMSS} (D=80. GeV) If \ttt{IMSS(1)=1} $M_1$, then 
{\bf U(1)} gaugino mass. If \ttt{IMSS(1)=2}, then the common gaugino mass 
$m_{1/2}$.

\iteme{RMSS(2) :} (D=160. GeV) $M_2$, the {\bf SU(2)} gaugino mass.  

\iteme{RMSS(3) :} (D=500. GeV) $M_3$, the {\bf SU(3)} (gluino) 
mass parameter.

\iteme{RMSS(4) :} (D=800. GeV) $\mu$, the higgsino mass parameter.  
If \ttt{IMSS(1)=2}, only the sign of $\mu$ is used.

\iteme{RMSS(5) :} (D=2.) $\tan\beta$, the ratio of Higgs expectation
values.

\iteme{RMSS(6) :} (D=250. GeV) Left slepton mass $M_{\sell_L}$.  
The sneutrino mass is fixed by a sum rule.

\iteme{RMSS(7) :} (D=200. GeV) Right slepton mass $M_{\sell_R}$. 

\iteme{RMSS(8) :} (D=800. GeV) Left squark mass $M_{\sq_L}$. If 
\ttt{IMSS(1)=2}, the common scalar mass $m_0$. 

\iteme{RMSS(9) :} (D=700. GeV) Right squark mass $M_{\sq_R}$. $M_{\sd_R}$ 
when \ttt{IMSS(9)=1}. 

\iteme{RMSS(10) :} (D=800. GeV) Left squark mass for the third
generation $M_{\sq_L}$. When \ttt{IMSS(5)=1}, it is instead the
$\st_2$ mass, and $M_{\sq_L}$ is a derived quantity.

\iteme{RMSS(11) :} (D=700. GeV) Right sbottom mass $M_{\sbo_R}$. When 
\ttt{IMSS(5)=1}, it is instead the $\sbo_1$ mass.

\iteme{RMSS(12) :} (D=500. GeV) Right stop mass $M_{\st_R}$  If
negative, then it is assumed that $M_{\st_R}^2 < 0$. When 
\ttt{IMSS(5)=1}, it is instead the $\st_1$ mass. 

\iteme{RMSS(13) :} (D=250. GeV) Left stau mass $M_{\stau_L}$.  

\iteme{RMSS(14) :} (D=200. GeV) Right stau mass $M_{\stau_R}$. 
 
\iteme{RMSS(15) :} (D=800. GeV) Bottom trilinear coupling $A_{\b}$. When 
\ttt{IMSS(5)=1}, it is a derived quantity.

\iteme{RMSS(16) :} (D=400. GeV) Top trilinear coupling $A_{\t}$.  If 
\ttt{IMSS(1)=2}, the common trilinear coupling $A$. When 
\ttt{IMSS(5)=1}, it is a derived quantity.
 
\iteme{RMSS(17) :} (D=0.) Tau trilinear coupling $A_{\tau}$. When 
\ttt{IMSS(5)=1}, it is a derived quantity.
   
\iteme{RMSS(18) :} (D=0.1) Higgs mixing angle $\alpha$. This is only 
used when all of the Higgs parameters are set by you, i.e  
\ttt{IMSS(4)=2}.  

\iteme{RMSS(19) :} (D=850. GeV) Pseudoscalar Higgs mass parameter $M_{\A}$. 
 
\iteme{RMSS(20) :} (D=0.041) GUT scale coupling constant 
$\alpha_{\mrm{GUT}}$.

\iteme{RMSS(21) :} (D=1.0 eV) The gravitino mass. Note nonconventional
choice of units for this particular mass.
  
\iteme{RMSS(22) :} (D=800. GeV) $\su_R$ mass when \ttt{IMSS(9)=1}. 

\iteme{RMSS(23) :} (D=10$^4$ GeV$^2$) $D_X$ contribution to scalar 
masses when \ttt{IMSS(7)=1}. 

\iteme{RMSS(24) :} (D=10$^4$ GeV$^2$) $D_Y$ contribution to scalar 
masses when \ttt{IMSS(7)=1}. 

\iteme{RMSS(25) :} (D=10$^4$ GeV$^2$) $D_S$ contribution to scalar 
masses when \ttt{IMSS(7)=1}. 

\iteme{RMSS(26) :} (D=0.0 radians) when \ttt{IMSS(5)=1} it is the 
sbottom mixing angle.

\iteme{RMSS(27) :} (D=0.0 radians) when \ttt{IMSS(5)=1} it is the 
stop mixing angle.

\iteme{RMSS(28) :} (D=0.0 radians) when \ttt{IMSS(5)=1} it is the
stau mixing angle. 

\iteme{RMSS(29) :} (D=$2.4 \times 10^{18}$ GeV) The Planck mass,
used for calculating decays to light gravitinos.

\iteme{RMSS(30) - RMSS(33) :} (D=0.0,0.0,0.0,0.0) complex phases for the
mass parameters in \ttt{RMSS(1) - RMSS(4)}, where the latter represent 
the moduli of the mass parameters for the case of nonvanishing phases.

\iteme{RMSS(40), RMSS(41) :} used for temporary storage of the corrections
$\Delta m_{\t}$ and $\Delta m_{\b}$, respectively, in the calculation of
Higgs properties.

\iteme{RMSS(51) :} (D=0.0) when \ttt{IMSS(51)=1} it is the negative 
logarithm of the common value for all lepton number violating 
$\lambda$ couplings (LLE). When \ttt{IMSS(51)=2} it is the constant of
proportionality for generation-hierarchical $\lambda$ couplings. See 
section \ref{sss:rparityviol}.

\iteme{RMSS(52) :} (D=0.0) when \ttt{IMSS(52)=1} it is the negative 
logarithm of the common value for all lepton number violating 
$\lambda'$ couplings (LQD). When \ttt{IMSS(52)=2} it is the constant of
proportionality for generation-hierarchical $\lambda'$ couplings. See 
section \ref{sss:rparityviol}.

\iteme{RMSS(53) :} (D=0.0) when \ttt{IMSS(53)=1} it is the negative 
logarithm of the common value for all baryon number violating 
$\lambda''$ couplings (UDD). When \ttt{IMSS(53)=2} it is the constant of
proportionality for generation-hierarchical $\lambda''$ couplings. See 
section \ref{sss:rparityviol}.

\end{entry}

\drawboxtwo{~COMMON/PYSSMT/ZMIX(4,4),UMIX(2,2),VMIX(2,2),SMZ(4),SMW(2),}%
{\&SFMIX(16,4),ZMIXI(4,4),UMIXI(2,2),VMIXI(2,2)}\label{p:PYSSMT}
\begin{entry}

\itemc{Purpose:} to provide information on the neutralino, chargino,
and sfermion mixing parameters.  The variables should not be changed
by you.

\iteme{ZMIX(4,4) :}\label{p:ZMIX} the real part of the neutralino mixing 
matrix in the Bino--neutral Wino--Up higgsino--Down higgsino basis.

\iteme{UMIX(2,2) :}\label{p:UMIX} the real part of the chargino mixing 
matrix in the charged Wino--charged higgsino basis.

\iteme{VMIX(2,2) :}\label{p:VMIX} the real part of the charged conjugate 
chargino mixing matrix in the wino--charged higgsino basis.

\iteme{SMZ(4) :}\label{p:SMZ} the signed masses of the neutralinos.

\iteme{SMW(2) :}\label{p:SMW} the signed masses of the charginos.

\iteme{SFMIX(16,4) :}\label{p:SFMIX} the sfermion mixing matrices 
\tbf{T} in the L--R basis, identified by the corresponding fermion, i.e.\ 
\ttt{SFMIX(6,I)} is the stop mixing matrix. The four entries for each 
sfermion are $\mrm{T}_{11}, \mrm{T}_{12}, \mrm{T}_{21},$ and 
$\mrm{T}_{22}$.

\iteme{ZMIXI(4,4) :}\label{p:ZMIXI} the imaginary part of the neutralino 
mixing matrix in the Bino--neutral Wino--Up higgsino--Down higgsino basis.

\iteme{UMIXI(2,2) :}\label{p:UMIXI} the imaginary part of the chargino 
mixing matrix in the charged Wino--charged higgsino basis.

\iteme{VMIXI(2,2) :}\label{p:VMIXI} the imaginary part of the charged 
conjugate chargino mixing matrix in the wino--charged higgsino basis.
\end{entry}

\drawbox{~COMMON/PYMSRV/RVLAM(3,3,3), RVLAMP(3,3,3), RVLAMB(3,3,3)}%
\label{p:PYMSRV}
\begin{entry}

\itemc{Purpose:} to provide information on lepton and baryon 
number violating couplings.

\iteme{RVLAM(3,3,3) :}\label{p:RVLAM} the lepton number violating 
$\lambda_{ijk}$ couplings. See \ttt{IMSS(51)}, \ttt{RMSS(51)}.

\iteme{RVLAMP(3,3,3) :}\label{p:RVLAMP} the lepton number violating 
$\lambda'_{ijk}$ couplings. See \ttt{IMSS(52)}, \ttt{RMSS(52)}.

\iteme{RVLAMB(3,3,3) :}\label{p:RVLAMB} the baryon number violating 
$\lambda''_{ijk}$ couplings. Currently not used.
\end{entry}
\boxsep

The following subroutines and functions need not be accessed by the
user, but are described for completeness.

\begin{entry}
\iteme{SUBROUTINE PYAPPS :} uses approximate analytic formulae to 
determine the full set of MSSM parameters from SUGRA inputs.

\iteme{SUBROUTINE PYGLUI :} calculates gluino decay modes.

\iteme{SUBROUTINE PYGQQB :} calculates three body decays of gluinos
into neutralinos or charginos and third generation fermions.  These
routines are valid for large values of $\tan\beta$.

\iteme{SUBROUTINE PYCJDC :} calculates the chargino decay modes.

\iteme{SUBROUTINE PYHEXT :} calculates the non--Standard Model decay
modes of the Higgs bosons.

\iteme{SUBROUTINE PYHGGM :} determines the Higgs boson mass spectrum
using several inputs.

\iteme{SUBROUTINE PYINOM :} finds the mass eigenstates and mixing
matrices for the charginos and neutralinos.

\iteme{SUBROUTINE PYMSIN :} initializes the MSSM simulation.

\iteme{SUBROUTINE PYLHA3 :} to read in or write out SUSY Les Houches Accord
spectra and decay tables. Can also be used stand-alone, before the call to
\ttt{PYINIT}, to read in SUSY Les Houches Accord decay tables for specific
particles. See section \ref{ss:parapartdat} for how to do this. 

\iteme{SUBROUTINE PYNJDC :} calculates neutralino decay modes.

\iteme{SUBROUTINE PYPOLE :} computes the Higgs boson masses using
a renormalization group improved leading--log approximation and
two-loop leading--log corrections.

\iteme{SUBROUTINE PYSFDC :} calculates sfermion decay modes.

\iteme{SUBROUTINE PYSUBH :} computes the Higgs boson masses using
only renormalization group improved formulae.

\iteme{SUBROUTINE PYTBDY :} samples the phase space for three body
decays of neutralinos, charginos, and the gluino.

\iteme{SUBROUTINE PYTHRG :} computes the masses and mixing matrices of
the third generation sfermions.

\iteme{SUBROUTINE PYRVSF :} $R$--violating sfermion decay widths.
\iteme{SUBROUTINE PYRVNE :} $R$--violating neutralino decay widths.
\iteme{SUBROUTINE PYRVCH :} $R$--violating chargino decay widths.
\iteme{SUBROUTINE PYRVGW :} calculates $R$--violating 3-body widths using
\ttt{PYRVI1, PYRVI2, PYRVI3, PYRVG1, PYRVG2, PYRVG3, PYRVG4, PYRVR}, and 
\ttt{PYRVS}. 
\iteme{FUNCTION PYRVSB :} calculates $R$--violating 2-body widths. 

\iteme{SUBROUTINE SUGRA :} dummy routine, to avoid linking problems when
\tsc{Isajet} is not linked; see \ttt{IMSS(1) = 12}.
\iteme{FUNCTION VISAJE :} dummy routine, to avoid linking problems when
\tsc{Isajet} is not linked; see \ttt{IMSS(1) = 12}.

\end{entry}
 
\subsection{General Event Information}
 
When an event is generated with \ttt{PYEVNT}, some information on
it is stored in the \ttt{MSTI} and \ttt{PARI} arrays of the
\ttt{PYPARS} common block (often copied directly from the internal
\ttt{MINT} and \ttt{VINT} variables). Further information is stored
in the complete event record; see section \ref{ss:evrec}.
 
Part of the information is only relevant for some subprocesses; by
default everything irrelevant is set to 0. Kindly note that, like the
\ttt{CKIN} constraints described in section \ref{ss:PYswitchkin},
kinematical variables normally (i.e.\ where it is not explicitly stated
otherwise) refer to the na\"{\i}ve hard scattering, before initial- and
final-state radiation effects have been included.
 
\drawbox{COMMON/PYPARS/MSTP(200),PARP(200),MSTI(200),PARI(200)}%
\label{p:PYPARS2}
\begin{entry}
\itemc{Purpose:} to provide information on latest event generated or,
in a few cases, on statistics accumulated during the run.
 
\iteme{MSTI(1) :}\label{p:MSTI} specifies the general type of 
subprocess that has occurred, according to the {\ISUB} code given 
in section \ref{ss:ISUBcode}.
 
\iteme{MSTI(2) :} whenever \ttt{MSTI(1)} (together with \ttt{MSTI(15)}
and \ttt{MSTI(16)}) are not enough to specify the type of process
uniquely, \ttt{MSTI(2)} provides an ordering of the different
possibilities. This is particularly relevant for the different
colour-flow topologies possible in QCD $2 \to 2$ processes. With
$i = $\ttt{MSTI(15)}, $j = $\ttt{MSTI(16)} and $k = $\ttt{MSTI(2)},
the QCD possibilities are, in the classification scheme of
\cite{Ben84} (cf. section \ref{sss:QCDjetclass}):
\begin{subentry}
\itemn{ISUB = 11,} $i = j$, $\q_i \q_i \to \q_i \q_i$; \\
$k = 1$ : colour configuration $A$. \\
$k = 2$ : colour configuration $B$.
\itemn{ISUB = 11,} $i \neq j$, $\q_i \q_j \to \q_i \q_j$; \\
$k = 1$ : only possibility.
\itemn{ISUB = 12,} $\q_i \qbar_i \to \q_l \qbar_l$; \\
$k = 1$ : only possibility.
\itemn{ISUB = 13,} $\q_i \qbar_i \to \g \g$; \\
$k = 1$ : colour configuration $A$. \\
$k = 2$ : colour configuration $B$.
\itemn{ISUB = 28,} $\q_i \g \to \q_i \g$; \\
$k = 1$ : colour configuration $A$. \\
$k = 2$ : colour configuration $B$.
\itemn{ISUB = 53,} $\g \g \to \q_l \qbar_l$; \\
$k = 1$ : colour configuration $A$. \\
$k = 2$ : colour configuration $B$.
\itemn{ISUB = 68,} $\g \g \to \g \g$; \\
$k = 1$ : colour configuration $A$. \\
$k = 2$ : colour configuration $B$. \\
$k = 3$ : colour configuration $C$.
\itemn{ISUB = 83,} $\f \q \to \f' \Q$ (by $t$-channel $\W$ exchange;
does not distinguish colour flows but result of user selection); \\
$k = 1$ : heavy flavour $\Q$ is produced on side 1. \\
$k = 2$ : heavy flavour $\Q$ is produced on side 2.
\end{subentry}
 
\iteme{MSTI(3) :} the number of partons produced in the hard 
interactions, i.e.\ the number $n$ of the $2 \to n$ matrix elements 
used; it is sometimes 3 or 4 when a basic $2 \to 1$ or $2 \to 2$ 
process has been folded with two $1 \to 2$ initial branchings (like
$\q_i \q_j \to \q_k \q_l \hrm^0$).
 
\iteme{MSTI(4) :} number of documentation lines at the beginning of
the common block \ttt{PYJETS} that are given with \ttt{K(I,1)=21};
0 for \ttt{MSTP(125)=0}.
 
\iteme{MSTI(5) :} number of events generated to date in current
run. In runs with the variable-energy option, \ttt{MSTP(171)=1}
and \ttt{MSTP(172)=2}, only those events that survive (i.e.\ that
do not have \ttt{MSTI(61)=1}) are counted in this number. That
is, \ttt{MSTI(5)} may be less than the total number of \ttt{PYEVNT}
calls.
 
\iteme{MSTI(6) :} current frame of event, cf. \ttt{MSTP(124)}.
 
\iteme{MSTI(7), MSTI(8) :} line number for documentation of
outgoing partons/particles from hard scattering for $2 \to 2$ or
$2 \to 1 \to 2$ processes (else = 0).

\iteme{MSTI(9) :} event class used in current event for $\gamma\p$ or 
$\gamma\gamma$ events. The code depends on which process is being studied.
\begin{subentry}
\iteme{= 0 :} for other processes than the ones listed above.
\itemn{For $\gamma\p$ or $\gast\p$ events,} generated with the 
\ttt{MSTP(14)=10} or \ttt{MSTP(14)=30} options:
\iteme{= 1 :} VMD.
\iteme{= 2 :} direct.
\iteme{= 3 :} anomalous.
\iteme{= 4 :} DIS (only for $\gamma^*\p$, i.e. \ttt{MSTP(14)=30}).
\itemn{For real $\gamma\gamma$ events,} i.e.\ \ttt{MSTP(14)=10}: 
\iteme{= 1 :} VMD$\times$VMD.
\iteme{= 2 :} VMD$\times$direct.
\iteme{= 3 :} VMD$\times$anomalous .
\iteme{= 4 :} direct$\times$direct.
\iteme{= 5 :} direct$\times$anomalous.
\iteme{= 6 :} anomalous$\times$anomalous.
\itemn{For virtual $\gast\gast$ events,} i.e.\ \ttt{MSTP(14)=30}, 
where the two incoming photons are not equivalent and the order 
therefore matters: 
\iteme{= 1 :} direct$\times$direct.
\iteme{= 2 :} direct$\times$VMD.
\iteme{= 3 :} direct$\times$anomalous.
\iteme{= 4 :} VMD$\times$direct.
\iteme{= 5 :} VMD$\times$VMD.
\iteme{= 6 :} VMD$\times$anomalous.
\iteme{= 7 :} anomalous$\times$direct.
\iteme{= 8 :} anomalous$\times$VMD.
\iteme{= 9 :} anomalous$\times$anomalous.
\iteme{= 10 :} DIS$\times$VMD.
\iteme{= 11 :} DIS$\times$anomalous.
\iteme{= 12 :} VMD$\times$DIS.
\iteme{= 13 :} anomalous$\times$DIS.
\end{subentry}

\iteme{MSTI(10) :} is 1 if cross section maximum was violated
in current event, and 0 if not.
 
\iteme{MSTI(11) :} {\KF} flavour code for beam (side 1) particle.
 
\iteme{MSTI(12) :} {\KF} flavour code for target (side 2) particle.
 
\iteme{MSTI(13), MSTI(14) :} {\KF} flavour codes for side 1 and side 2
initial-state shower initiators.
 
\iteme{MSTI(15), MSTI(16) :} {\KF} flavour codes for side 1 and side 2
incoming partons to the hard interaction.
 
\iteme{MSTI(17), MSTI(18) :} flag to signal if particle on side
1 or side 2 has been scattered diffractively; 0 if no, 1 if yes.
 
\iteme{MSTI(21) - MSTI(24) :} {\KF} flavour codes for outgoing partons
from the hard interaction. The number of positions actually used is
process-dependent, see \ttt{MSTI(3)}; trailing positions not used
are set = 0. For events with many outgoing partons, e.g.\ in external
processes, also \ttt{MSTI(25)} and \ttt{MSTI(26)} could be used.
 
\iteme{MSTI(25), MSTI(26) :} {\KF} flavour codes of the products in the
decay of a single $s$-channel resonance formed in the hard interaction.
Are thus only used when \ttt{MSTI(3)=1} and the resonance is allowed
to decay.
 
\iteme{MSTI(31) :} number of hard or semi-hard scatterings that occurred
in the current event in the multiple-interaction scenario; is = 0 for a
low-$\pT$ event.
 
\iteme{MSTI(32) :} information on whether a reconnection occurred in the
current event; is 0 normally but 1 in case of reconnection.
 
\iteme{MSTI(41) :} the number of pile-up events generated in the latest
\ttt{PYEVNT} call (including the first, `hard' event).
 
\iteme{MSTI(42) - MSTI(50) :} {\ISUB} codes for the events 2--10
generated in the pile-up-events scenario. The first event {\ISUB} code is
stored in \ttt{MSTI(1)}. If \ttt{MSTI(41)} is less than 10, only as
many positions are filled as there are pile-up events. If MSTI(41) is
above 10, some {\ISUB} codes will not appear anywhere.

\iteme{MSTI(51) :} normally 0 but set to 1 if a \ttt{UPEVNT} call
did not return an event, such that \ttt{PYEVNT} could not generate
an event. For further details, see section \ref{ss:PYnewproc}.

\iteme{MSTI(52) :} counter for the number of times the current event 
configuration failed in the generation machinery. For accepted events
this is always 0, but the counter can be used inside \ttt{UPEVNT}
to check on anomalous occurrences. For further 
details, see section \ref{ss:PYnewproc}.

\iteme{MSTI(53) :} normally 0, but 1 if no processes with non-vanishing
cross sections were found in a \ttt{PYINIT} call, for the case that
\ttt{MSTP(127)=1}. 

\iteme{MSTI(61) :} status flag set when events are generated. It is 
only of interest for runs with variable energies, \ttt{MSTP(171)=1}, 
with the option \ttt{MSTP(172)=2}.
\begin{subentry}
\iteme{= 0 :} an event has been generated.
\iteme{= 1 :} no event was generated, either because the c.m.\ energy 
was too low or because the Monte Carlo phase space point selection 
machinery rejected the trial point. A new energy is to be picked by 
you.
\end{subentry}

\iteme{MSTI(71), MSTI(72) :} {\KF} code for incoming lepton beam or target 
particles, when a flux of virtual photons are generated internally for
{\galep} beams, while \ttt{MSTI(11)} and \ttt{MSTI(12)} is 
then the photon code.

\boxsep
 
\iteme{PARI(1) :}\label{p:PARI} total integrated cross section for 
the processes
under study, in mb. This number is obtained as a by-product of the
selection of hard-process kinematics, and is thus known with better
accuracy when more events have been generated. The value stored here
is based on all events until the latest one generated.
 
\iteme{PARI(2) :} for unweighted events, \ttt{MSTP(142)=0} or \ttt{=2}, 
it is the ratio \ttt{PARI(1)/MSTI(5)}, i.e.\ the ratio of total 
integrated cross section and number of events generated. Histograms 
should then be filled with unit event weight and, at the end of the run, 
multiplied by \ttt{PARI(2)} and divided by the bin width to convert
results to mb/(dimension of the horizontal axis).
For weighted events, \ttt{MSTP(142)=1}, \ttt{MSTI(5)} is replaced by the 
sum of \ttt{PARI(10)} values. Histograms should then be filled with
event weight \ttt{PARI(10)} and, as before, be multiplied by \ttt{PARI(2)} 
and divided by the bin width at the end of the run.
In runs with the variable-energy option, \ttt{MSTP(171)=1}
and \ttt{MSTP(172)=2}, only those events that survive (i.e.\ that
do not have \ttt{MSTI(61)=1}) are counted.
 
\iteme{PARI(7) :} an event weight, normally 1 and thus uninteresting, 
but for external processes with \ttt{IDWTUP=-1, -2} or \ttt{-3} it can 
be $-1$ for events with negative cross section, with \ttt{IDWTUP=4} it 
can be an arbitrary non-negative weight of dimension mb, and with 
\ttt{IDWTUP=-4} it can be an arbitrary weight of dimension mb.
(The difference being that in most cases a rejection step is involved 
to bring the accepted events to a common weight normalization, up to
a sign, while no rejection need be involved in the last two cases.)
 
\iteme{PARI(9) :} is weight \ttt{WTXS} returned from \ttt{PYEVWT}
call when \ttt{MSTP(142)}$\geq 1$, otherwise is 1.
 
\iteme{PARI(10) :} is compensating weight \ttt{1./WTXS} that should
be associated to events when \ttt{MSTP(142)=1}, else is 1.
 
\iteme{PARI(11) :} $E_{\mrm{cm}}$, i.e.\ total c.m.\ energy
(except when using the {\galep} machinery, see \ttt{PARI(101)}.
 
\iteme{PARI(12) :} $s$, i.e.\ squared total c.m.\ energy
(except when using the {\galep} machinery, see \ttt{PARI(102)}.
 
\iteme{PARI(13) :} $\hat{m} = \sqrt{\hat{s}}$, i.e.\ mass of the 
hard-scattering subsystem.
 
\iteme{PARI(14) :} $\hat{s}$ of the hard subprocess
($2 \to 2$ or $2 \to 1$).
 
\iteme{PARI(15) :} $\hat{t}$ of the hard subprocess
($2 \to 2$ or $2 \to 1 \to 2$).
 
\iteme{PARI(16) :} $\hat{u}$ of the hard subprocess
($2 \to 2$ or $2 \to 1 \to 2$).
 
\iteme{PARI(17) :} $\hat{p}_{\perp}$ of the hard subprocess
($2 \to 2$ or $2 \to 1 \to 2$),
evaluated in the rest frame of the hard interaction.
 
\iteme{PARI(18) :} $\hat{p}_{\perp}^2$ of the hard subprocess;
see \ttt{PARI(17)}.
 
\iteme{PARI(19) :} $\hat{m}'$, the mass of the complete three- or
four-body final state in $2 \to 3$ or $2 \to 4$ processes (while
$\hat{m}$, given in \ttt{PARI(13)}, here corresponds to the one-
or two-body central system).
Kinematically $\hat{m} \leq \hat{m}' \leq E_{\mrm{cm}}$.
 
\iteme{PARI(20) :} $\hat{s}' = \hat{m}'^2$; see \ttt{PARI(19)}.
 
\iteme{PARI(21) :} $Q$ of the hard-scattering subprocess. The exact 
definition is process-dependent, see \ttt{MSTP(32)}.
 
\iteme{PARI(22) :} $Q^2$ of the hard-scattering subprocess; see 
\ttt{PARI(21)}.
 
\iteme{PARI(23) :} $Q$ of the outer hard-scattering subprocess.
Agrees with \ttt{PARI(21)} for a $2 \to 1$ or $2 \to 2$ process.
For a $2 \to 3$ or $2 \to 4$ $\W/\Z$ fusion process, it is set by
the $\W/\Z$ mass scale, and for subprocesses 121 and 122 by the
heavy-quark mass.
 
\iteme{PARI(24) :} $Q^2$ of the outer hard-scattering subprocess;
see \ttt{PARI(23)}.
 
\iteme{PARI(25) :} $Q$ scale used as maximum virtuality in parton
showers. Is equal to \ttt{PARI(23)}, except for 
Deeply Inelastic Scattering processes when \ttt{MSTP(22)}$\geq 1$.
 
\iteme{PARI(26) :} $Q^2$ scale in parton showers; see \ttt{PARI(25)}.
 
\iteme{PARI(31), PARI(32) :} the momentum fractions $x$ of the
initial-state parton-shower initiators on side 1 and 2, respectively.
 
\iteme{PARI(33), PARI(34) :} the momentum fractions $x$ taken by the
partons at the hard interaction, as used e.g.\ in the 
parton-distribution functions.
 
\iteme{PARI(35) :} Feynman-$x$,
$x_{\mrm{F}} = x_1 - x_2 = $\ttt{PARI(33)}$-$\ttt{PARI(34)}.
 
\iteme{PARI(36) :}
$\tau = \hat{s}/s = x_1 \, x_2 = $\ttt{PARI(33)}$\times$\ttt{PARI(34)}.
 
\iteme{PARI(37) :} $y = (1/2) \ln(x_1/x_2)$, i.e.\ rapidity of the
hard-interaction subsystem in the c.m.\ frame of the event as a whole.
 
\iteme{PARI(38)  :} $\tau' = \hat{s}'/s = $\ttt{PARI(20)/PARI(12)}.
 
\iteme{PARI(39), PARI(40) :} the primordial $k_{\perp}$ values
selected in the two beam remnants.
 
\iteme{PARI(41) :} $\cos\hat{\theta}$, where $\hat{\theta}$ is the
scattering angle of a $2 \to 2$ (or $2 \to 1 \to 2$) interaction,
defined in the rest frame of the hard-scattering subsystem.
 
\iteme{PARI(42) :} $x_{\perp}$, i.e.\ scaled transverse momentum of the
hard-scattering subprocess, 
$x_{\perp} = 2 \hat{p}_{\perp}/E_{\mrm{cm}} = 2$%
\ttt{PARI(17)}$/$\ttt{PARI(11)}.
 
\iteme{PARI(43), PARI(44) :} $x_{L3}$ and $x_{L4}$, i.e.\ longitudinal
momentum fractions of the two scattered partons, in the range
$-1 < x_{\mrm{L}} < 1$, in the c.m.\ frame of the event as a whole.
 
\iteme{PARI(45), PARI(46) :} $x_3$ and $x_4$, i.e.\ scaled energy
fractions of the two scattered partons, in the c.m.\ frame of the
event as a whole.
 
\iteme{PARI(47), PARI(48) :} $y^*_3$ and $y^*_4$, i.e.\ rapidities
of the two scattered partons in the c.m.\ frame of the
event as a whole.
 
\iteme{PARI(49), PARI(50) :} $\eta^*_3$ and $\eta^*_4$, i.e.
pseudorapidities of the two scattered partons in the c.m.\ frame
of the event as a whole.
 
\iteme{PARI(51), PARI(52) :} $\cos\theta^*_3$ and $\cos\theta^*_4$,
i.e.\ cosines of the polar angles of the two scattered partons in
the c.m.\ frame of the event as a whole.
 
\iteme{PARI(53), PARI(54) :} $\theta^*_3$ and $\theta^*_4$, i.e.
polar angles of the two scattered partons, defined in the range
$0 < \theta^* < \pi$, in the c.m.\ frame of the event as a whole.
 
\iteme{PARI(55), PARI(56) :} azimuthal angles $\phi^*_3$ and
$\phi^*_4$ of the two scattered partons, defined in the range
$-\pi < \phi^* < \pi$, in the c.m.\ frame of the event as a whole.
 
\iteme{PARI(61) :} multiple interaction enhancement factor for
current event. A large value corresponds to a central collision
and a small value to a peripheral one.
 
\iteme{PARI(65) :} sum of the transverse momenta of partons
generated at the hardest interaction of the event, excluding
initial- and final-state radiation, i.e.\ $2 \times$\ttt{PARI(17)}.
Only intended for $2 \to 2$ or $2 \to 1 \to 2$ processes,
i.e.\ not implemented for $2 \to 3$ ones. 
 
\iteme{PARI(66) :} sum of the transverse momenta of all partons
generated at the hardest interaction, including initial- and 
final-state radiation, resonance decay products, and primordial
$k_{\perp}$.
 
\iteme{PARI(67) :} scalar sum of transverse momenta of partons generated
at hard interactions, excluding the hardest one (see \ttt{PARI(65)}),
and also excluding all initial- and final-state radiation. Is 
non-vanishing only in the multiple-interaction scenario.
 
\iteme{PARI(68) :} sum of transverse momenta of all partons generated
at hard interactions, excluding the hardest one (see \ttt{PARI(66)}),
but including initial- and final-state radiation associated with those
further interactions. Is non-vanishing only in the multiple-interaction 
scenario. Since showering has not yet been added to those additional 
interactions, it currently coincides with \ttt{PARI(67)}. 
 
\iteme{PARI(69) :} sum of transverse momenta of all partons generated
in hard interactions (\ttt{PARI(66) + PARI(68)}) and, additionally,
of all beam remnant partons.
 
\iteme{PARI(71), PARI(72) :} sum of the momentum fractions $x$ taken
by initial-state parton-shower initiators on side 1 and and side 2,
excluding those of the hardest interaction. Is non-vanishing only in
the multiple-interaction scenario.
 
\iteme{PARI(73), PARI(74) :} sum of the momentum fractions $x$ taken
by the partons at the hard interaction on side 1 and side 2, excluding
those of the hardest interaction. Is non-vanishing only in the
multiple-interaction scenario.
 
\iteme{PARI(75), PARI(76) :} the $x$ value of a photon that branches
into quarks or gluons, i.e.\ $x$ at interface between initial-state QED
and QCD cascades, for the old photoproduction machinery..
 
\iteme{PARI(77), PARI(78) :} the $\chi$ values selected for beam
remnants that are split into two objects, describing how the energy
is shared (see \ttt{MSTP(92)} and \ttt{MSTP(94)}); is vanishing if 
no splitting is needed.
 
\iteme{PARI(81) :} size of the threshold factor (enhancement or
suppression) in the latest event with heavy-flavour production;
see \ttt{MSTP(35)}.
 
\iteme{PARI(91) :} average multiplicity $\br{n}$ of pile-up events,
see \ttt{MSTP(133)}. Only relevant for \ttt{MSTP(133)=} 1 or 2.
 
\iteme{PARI(92) :} average multiplicity $\langle n \rangle$ of pile-up 
events as actually simulated, i.e.\ with multiplicity = 0 events 
removed and the high-end tail truncated. Only relevant for 
\ttt{MSTP(133)=} 1 or 2.
 
\iteme{PARI(93) :} for \ttt{MSTP(133)=1} it is the probability that
a beam crossing will produce a pile-up event at all, i.e.\ that there
will be at least one hadron--hadron interaction; for
\ttt{MSTP(133)=2} the probability that a beam crossing will produce
a pile-up event with one hadron--hadron interaction of the desired rare
type. See subsection \ref{ss:pileup}.

\iteme{PARI(101) :} c.m.\ energy for the full collision, while \ttt{PARI(11)} 
gives the $\gamma$-hadron or $\gamma\gamma$ subsystem energy; used for
virtual photons generated internally with the {\galep} option.

\iteme{PARI(102) :} full squared c.m.\ energy, while \ttt{PARI(12)} gives 
the subsystem squared energy; used for virtual photons generated 
internally with the {\galep} option.

\iteme{PARI(103), PARI(104) :} $x$ values, i.e.\ respective photon energy 
fractions of the incoming lepton in the c.m.\ frame of the event; used for
virtual photons generated internally with the {\galep} option.

\iteme{PARI(105), PARI(106) :} $Q^2$ or $P^2$, virtuality of the 
respective photon (thus the square of \ttt{VINT(3)}, \ttt{VINT(4)}); used for
virtual photons generated internally with the {\galep} option.    

\iteme{PARI(107), PARI(108) :} $y$ values, i.e.\ respective photon light-cone 
energy fraction of the incoming lepton; used for virtual photons generated 
internally with the {\galep} option.

\iteme{PARI(109), PARI(110) :} $\theta$, scattering angle of the respective 
lepton in the c.m.\ frame of the event; used for virtual photons generated 
internally with the {\galep} option.   

\iteme{PARI(111), PARI(112) :} $\phi$, azimuthal angle of the respective 
scattered lepton in the c.m.\ frame of the event; used for virtual photons 
generated internally with the {\galep} option.

\iteme{PARI(113), PARI(114):} the $R$ factor defined at \ttt{MSTP(17)},
giving a cross section enhancement from the contribution of resolved
longitudinal photons.
   
\end{entry}

\subsection{How to Generate Weighted Events}
\label{ss:PYwtevts}

By default {\Py} generates unweighted events, i.e.\ all events
in a run are on an equal footing. This means that corners of phase
space with low cross sections are poorly populated, as it should be.
However, sometimes one is interested in also exploring such corners,
in order to gain a better understanding of physics. A typical example
would be the jet cross section in hadron collisions, which is dropping 
rapidly with increasing jet $\pT$, and where it is interesting to trace 
this drop over several orders of magnitude. Experimentally this may 
be solved by prescaling events rates already at the trigger level,
so that all high-$\pT$ events are saved but only a fraction of the
lower-$\pT$ ones. In this section we outline procedures to generate
events in a similar manner.

Basically two approaches can be used. One is to piece together 
results from different subruns, where each subrun is restricted to 
some specific region of phase space. Within each subrun all events 
then have the same weight, but subruns have to be combined according 
to their relative cross sections. The other approach is to let each 
event come with an associated weight, that can vary smoothly as a 
function of $\pT$. These two alternatives correspond to stepwise or
smoothly varying prescaling factors in the experimental analogue. 
We describe them one after the other.

The phase space can be sliced in many different ways. However, for the 
jet rate and many other processes, the most natural variable would be 
$\pT$ itself. (For production of a lepton pair by $s$-channel
resonances, the invariant mass would be a better choice.) It is not
possible to specify beforehand the jet $\pT$'s an event will contain,
since this is a combination of the $\hat{p}_{\perp}$ of the hard 
scattering process with additional showering activity, with 
hadronization, with underlying event and with the jet clustering 
approach actually used. However, one would expect a strong correlation 
between the $\hat{p}_{\perp}$ scale and the jet $\pT$'s. Therefore 
the full $\hat{p}_{\perp}$ range can be subdivided into a set of 
ranges by using the \ttt{CKIN(3)} and \ttt{CKIN(4)} variables as 
lower and upper limits. This could be done e.g.\ for adjacent 
non-overlapping bins 10--20,20--40,40--70, etc.

Only if one would like to cover also very small $\pT$ is there a
problem with this strategy: since the naive jet cross section is 
divergent for $\hat{p}_{\perp} \to 0$, a unitarization procedure is 
implied by setting \ttt{CKIN(3)=0} (or some other low value). This 
unitarization then disregards the actual \ttt{CKIN(3)} and \ttt{CKIN(4)}
values and generates events over the full phase space. In order not to
double count, then events above the intended upper limit of the first 
bin have to be removed by brute force.

A simple but complete example of a code performing this task (with 
some primitive histogramming) is the following:
\begin{verbatim}
C...All real arithmetic in double precision.
      IMPLICIT DOUBLE PRECISION(A-H, O-Z)
C...Three Pythia functions return integers, so need declaring.
      INTEGER PYK,PYCHGE,PYCOMP
C...EXTERNAL statement links PYDATA on most platforms.
      EXTERNAL PYDATA
C...The event record.
      COMMON/PYJETS/N,NPAD,K(4000,5),P(4000,5),V(4000,5)
C...Selection of hard scattering subprocesses.
      COMMON/PYSUBS/MSEL,MSELPD,MSUB(500),KFIN(2,-40:40),CKIN(200)
C...Parameters. 
      COMMON/PYPARS/MSTP(200),PARP(200),MSTI(200),PARI(200)
C...Bins of pT.
      DIMENSION PTBIN(10)
      DATA PTBIN/0D0,10D0,20D0,40D0,70D0,110D0,170D0,250D0,350D0,1000D0/
 
C...Main parameters of run: c.m. energy and number of events per bin.
      ECM=2000D0
      NEV=1000

C...Histograms.
      CALL PYBOOK(1,'dn_ev/dpThat',100,0D0,500D0)
      CALL PYBOOK(2,'dsigma/dpThat',100,0D0,500D0)
      CALL PYBOOK(3,'log10(dsigma/dpThat)',100,0D0,500D0)
      CALL PYBOOK(4,'dsigma/dpTjet',100,0D0,500D0)
      CALL PYBOOK(5,'log10(dsigma/dpTjet)',100,0D0,500D0)
      CALL PYBOOK(11,'dn_ev/dpThat, dummy',100,0D0,500D0)
      CALL PYBOOK(12,'dn/dpTjet, dummy',100,0D0,500D0)

C...Loop over pT bins and initialize.   
      DO 300 IBIN=1,9
        CKIN(3)=PTBIN(IBIN)
        CKIN(4)=PTBIN(IBIN+1)
        CALL PYINIT('CMS','p','pbar',ECM)
 
C...Loop over events. Remove unwanted ones in first pT bin.
        DO 200 IEV=1,NEV
          CALL PYEVNT
          PTHAT=PARI(17)
          IF(IBIN.EQ.1.AND.PTHAT.GT.PTBIN(IBIN+1)) GOTO 200

C...Store pThat. Cluster jets and store variable number of pTjet.
          CALL PYFILL(1,PTHAT,1D0)
          CALL PYFILL(11,PTHAT,1D0)
          CALL PYCELL(NJET)
          DO 100 IJET=1,NJET
            CALL PYFILL(12,P(N+IJET,5),1D0)
  100     CONTINUE 
 
C...End of event loop.
  200   CONTINUE

C...Normalize cross section to pb/GeV and add up.
        FAC=1D9*PARI(1)/(DBLE(NEV)*5D0)     
        CALL PYOPER(2,'+',11,2,1D0,FAC)           
        CALL PYOPER(4,'+',12,4,1D0,FAC)           

C...End of loop over pT bins.
  300 CONTINUE

C...Take logarithm and plot.
      CALL PYOPER(2,'L',2,3,1D0,0D0)           
      CALL PYOPER(4,'L',4,5,1D0,0D0) 
      CALL PYNULL(11)
      CALL PYNULL(12)           
      CALL PYHIST

      END
\end{verbatim}

The alternative to slicing the phase space is to used weighted events.
This is possible by making use of the \ttt{PYEVWT} routine:
 
\drawbox{CALL PYEVWT(WTXS)}\label{p:PYEVWT}
\begin{entry}
\itemc{Purpose:} to allow you to reweight event cross sections,
by process type and kinematics of the hard scattering. There exists
two separate modes of usage, described in the following. \\
For \ttt{MSTP(142)=1}, it is assumed that the cross section of the
process is correctly given by default in {\Py}, but that one
wishes to generate events biased to a specific region of phase
space. While the \ttt{WTXS} factor therefore multiplies the na\"{\i}ve
cross section in the choice of subprocess type and kinematics,
the produced event comes with a compensating weight
\ttt{PARI(10)=1./WTXS}, which should be used when filling histograms
etc. In the \ttt{PYSTAT(1)} table, the cross sections are unchanged
(up to statistical errors) compared with the standard cross sections,
but the relative composition of events may be changed and need
no longer be in proportion to relative cross sections. A typical
example of this usage is if one wishes to enhance the production
of high-$\pT$ events; then a weight like
\ttt{WTXS}$=(\pT/p_{\perp 0})^2$ (with $p_{\perp 0}$ some fixed
number) might be appropriate. See \ttt{PARI(2)} for a discussion of 
overall normalization issues.\\
For \ttt{MSTP(142)=2}, on the other hand, it is assumed that the true
cross section is really to be modified by the multiplicative factor
\ttt{WTXS}. The generated events therefore come with unit weight, just
as usual. This option is really equivalent to replacing the basic
cross sections coded in {\Py}, but allows more flexibility: no
need to recompile the whole of {\Py}. \\
The routine will not be called unless \ttt{MSTP(142)}$\geq 1$, and
never if `minimum-bias'-type events (including elastic and
diffractive scattering) are to be generated as well. Further,
cross sections for additional multiple interactions or pile-up events
are never affected. A dummy routine \ttt{PYEVWT} is included in the
program file, so as to avoid unresolved external references when the
routine is not used.
\iteme{WTXS:} multiplication factor to ordinary event cross section;
to be set (by you) in \ttt{PYEVWT} call.
\itemc{Remark :} at the time of selection, several variables in the
\ttt{MINT} and \ttt{VINT} arrays in the \ttt{PYINT1} common block
contain information that can be used to make the decision. The routine
provided in the program file explicitly reads the variables that
have been defined at the time \ttt{PYEVWT} is called, and also
calculates some derived quantities. The given list of information
includes subprocess type {\ISUB}, $E_{\mrm{cm}}$, $\hat{s}$, $\hat{t}$,
$\hat{u}$, $\hat{p}_{\perp}$, $x_1$, $x_2$, $x_{\mrm{F}}$, $\tau$, $y$,
$\tau'$, $\cos\hat{\theta}$, and a few more. Some of these may
not be relevant for the process under study, and are then set to
zero.
\itemc{Warning:} the weights only apply to the hard scattering
subprocesses. There is no way to reweight the shape of initial- and
final-state showers, fragmentation, or other aspects of the event.
\end{entry}

\boxsep

There are some limitations to the facility. \ttt{PYEVWT} is called
at an early stage of the generation process, when the hard kinematics
is selected, well before the full event is constructed. It then cannot 
be used for low-$\pT$, elastic or diffractive events, for which no 
hard kinematics has been defined. If such processes are included,
the event weighting is switched off. Therefore it is no longer an 
option to run with \ttt{CKIN(3)=0}.  

Which weight expression to use may take some trial and error.
In the above case, a reasonable ansatz seems to be a weight behaving 
like $\hat{p}_{\perp}^6$, where four powers of $\hat{p}_{\perp}$ are 
motivated by the partonic cross section behaving like 
$1/\hat{p}_{\perp}^4$, and the remaining two by the fall-off of
parton densities. An example for the same task as above one would 
then be:
\begin{verbatim}
C...All real arithmetic in double precision.
      IMPLICIT DOUBLE PRECISION(A-H, O-Z)
C...Three Pythia functions return integers, so need declaring.
      INTEGER PYK,PYCHGE,PYCOMP
C...EXTERNAL statement links PYDATA on most platforms.
      EXTERNAL PYDATA
C...The event record.
      COMMON/PYJETS/N,NPAD,K(4000,5),P(4000,5),V(4000,5)
C...Selection of hard scattering subprocesses.
      COMMON/PYSUBS/MSEL,MSELPD,MSUB(500),KFIN(2,-40:40),CKIN(200)
C...Parameters. 
      COMMON/PYPARS/MSTP(200),PARP(200),MSTI(200),PARI(200)
 
C...Main parameters of run: c.m. energy, pTmin and number of events.
      ECM=2000D0
      CKIN(3)=5D0
      NEV=10000

C...Histograms.
      CALL PYBOOK(1,'dn_ev/dpThat',100,0D0,500D0)
      CALL PYBOOK(2,'dsigma/dpThat',100,0D0,500D0)
      CALL PYBOOK(3,'log10(dsigma/dpThat)',100,0D0,500D0)
      CALL PYBOOK(4,'dsigma/dpTjet',100,0D0,500D0)
      CALL PYBOOK(5,'log10(dsigma/dpTjet)',100,0D0,500D0)

C...Initialize with weighted events.
      MSTP(142)=1
      CALL PYINIT('CMS','p','pbar',ECM)
 
C...Loop over events; read out pThat and event weight.
      DO 200 IEV=1,NEV
        CALL PYEVNT
        PTHAT=PARI(17)
        WT=PARI(10)

C...Store pThat. Cluster jets and store variable number of pTjet.
        CALL PYFILL(1,PTHAT,1D0)
        CALL PYFILL(2,PTHAT,WT)
        CALL PYCELL(NJET)
        DO 100 IJET=1,NJET
          CALL PYFILL(4,P(N+IJET,5),WT)
  100   CONTINUE 
 
C...End of event loop.
  200   CONTINUE

C...Normalize cross section to pb/GeV, take logarithm and plot.
      FAC=1D9*PARI(2)/5D0 
      CALL PYFACT(2,FAC)      
      CALL PYFACT(4,FAC)      
      CALL PYOPER(2,'L',2,3,1D0,0D0)           
      CALL PYOPER(4,'L',4,5,1D0,0D0) 
      CALL PYHIST

      END

C*************************************************************
 
      SUBROUTINE PYEVWT(WTXS)
 
C...Double precision and integer declarations.
      IMPLICIT DOUBLE PRECISION(A-H, O-Z)
      IMPLICIT INTEGER(I-N)
      INTEGER PYK,PYCHGE,PYCOMP
C...Common block.
      COMMON/PYINT1/MINT(400),VINT(400)

C...Read out pThat^2 and set weight.
      PT2=VINT(48)
      WTXS=PT2**3
 
      RETURN
      END
\end{verbatim}
Note that, in \ttt{PYEVWT} one cannot look for $\hat{p}_{\perp}$ in
\ttt{PARI(17)}, since this variable is only set at the end of the
event generation. Instead the internal \ttt{VINT(48)} is used.
The dummy copy of the \ttt{PYEVWT} routine found in the {\Py} code
shows what is available and how to access this.

\subsection{How to Run with Varying Energies}
\label{ss:PYvaren}

It is possible to use {\Py} in a mode where the energy can be
varied from one event to the next, without the need to re-initialize
with a new \ttt{PYINIT} call. This allows a significant speed-up of
execution, although it is not as fast as running at a fixed
energy. It can not be used for everything --- we will come to
the fine print at the end --- but it should be applicable for most 
tasks.

The master switch to access this possibility is in \ttt{MSTP(171)}.
By default it is off, so you must set \ttt{MSTP(171)=1} before 
initialization. There are two submodes of running, with 
\ttt{MSTP(172)} being 1 or 2. In the former mode, {\Py} will generate 
an event at the requested energy. This means that you have to know 
which energy you want beforehand. In the latter mode, {\Py}
will often return without having generated an event --- with flag
\ttt{MSTI(61)=1} to signal that --- and you are then requested to give 
a new energy. The energy spectrum of accepted events will then,
in the end, be your naive input spectrum weighted with the 
cross-section of the processes you study. We will come back to 
this. 

The energy can be varied, whichever frame is given in the \ttt{PYINIT}
call. (Except for \ttt{'USER'}, where such information is fed in via 
the \ttt{HEPEUP} common block and thus beyond the control of {\Py}.) 
When the frame is \ttt{'CMS'}, \ttt{PARP(171)} should be filled
with the fractional energy of each event, i.e.\ 
$E_{\mrm{cm}} =$\ttt{PARP(171)}$\times$\ttt{WIN}, where \ttt{WIN} is 
the nominal c.m.\ energy of the \ttt{PYINIT} call. Here \ttt{PARP(171)} 
should normally be smaller than unity, i.e.\ initialization should
be done at the maximum energy to be encountered. For the \ttt{'FIXT'}
frame, \ttt{PARP(171)} should be filled by the fractional beam energy 
of that one, i.e.\ $E_{\mrm{beam}} = $\ttt{PARP(171)}$\times$\ttt{WIN}. 
For the \ttt{'3MOM'}, \ttt{'4MOM'} and \ttt{'5MOM'} options, the
two four-momenta are given in for each event in the same format as
used for the \ttt{PYINIT} call. Note that there is a minimum c.m.\ 
energy allowed, \ttt{PARP(2)}. If you give in values below this, 
the program will stop for \ttt{MSTP(172)=1}, and will return with
\ttt{MSTI(61)=1} for \ttt{MSTP(172)=1}. 

To illustrate the use of the \ttt{MSTP(172)=2} facility, consider the 
case of beamstrahlung in $\ee$ linear colliders. This is just for 
convenience; what is said here can be translated easily into other 
situations. Assume that the beam spectrum is given by $D(z)$, where 
$z$ is the fraction retained by the original $\e$ after beamstrahlung. 
Therefore $0 \leq z \leq 1$ and the integral of $D(z)$ is unity. 
This is not perfectly general; one could imagine branchings 
$\e^- \to \e^- \gamma \to \e^-\e^+\e^-$, which gives a multiplication 
in the number of beam particles. This could either be expressed in 
terms of a $D(z)$ with integral larger than unity or in terms of an 
increased luminosity. We will assume the latter, and use $D(z)$ 
properly normalized. Given a nominal $s = 4E_{\mrm{beam}}^2$, 
the actual $s'$ after beamstrahlung is given by $s' = z_1 z_2 s$. 
For a process with a cross section $\sigma(s)$ the total 
cross section is then
\begin{equation}
\sigma_{\mrm{tot}} = \int_0^1 \int_0^1 D(z_1) \, D(z_2)   
\sigma(z_1 z_2 s) \, \d z_1  \, \d z_2~.
\end{equation} 
The cross section $\sigma$ may in itself be an integral over a number
of additional phase space variables. If the maximum of the differential
cross section is known, a correct procedure to generate events is 
\begin{Enumerate}
\item pick $z_1$ and $z_2$ according to $D(z_1) \, \d z_1$ and 
$D(z_2) \, \d z_2$, respectively;
\item pick a set of phase space variables of the process, for the 
given $s'$ of the event;
\item evaluate $\sigma(s')$ and compare with $\sigma_{\mmax}$;
\item if event is rejected, then return to step 1 to generate new
variables;
\item else continue the generation to give a complete event.
\end{Enumerate}
You as a user are assumed to take care of step 1, and present the
resulting kinematics with incoming $\e^+$ and $\e^-$ of varying energy. 
Thereafter {\Py} will do steps 2--5, and either return an event or
put \ttt{MSTI(61)=1} to signal failure in step 4.

The maximization procedure does search in phase space to find 
$\sigma_{\mmax}$, but it does not vary the $s'$ energy in this process.
Therefore the maximum search in the \ttt{PYINIT} call should be 
performed where the cross section is largest. For processes with 
increasing cross section as a function of energy this means at the 
largest energy that will ever be encountered, i.e.\ $s' = s$ in the 
case above. This is the `standard' case, but often one encounters 
other behaviours, where more complicated procedures are needed. One 
such case would be the process $\ee \to \Z^{*0} \to \Z^0 \hrm^0$, 
which is known to have a cross section that increases near the 
threshold but is decreasing asymptotically. If one already knows
that the maximum, for a given Higgs mass, appears at 300 GeV, say, 
then the \ttt{PYINIT} call should be made with that energy, even if 
subsequently one will be generating events for a 500 GeV collider. 

In general, it may be necessary to modify the selection of $z_1$ and
$z_2$ and assign a compensating event weight. For instance, consider
a process with a cross section behaving roughly like $1/s$. Then the 
$\sigma_{\mrm{tot}}$ expression above may be rewritten as
\begin{equation}
\sigma_{\mrm{tot}} = \int_0^1 \int_0^1 \frac{D(z_1)}{z_1} \, 
\frac{D(z_2)}{z_2} \, z_1 z_2 \sigma(z_1 z_2 s) \, \d z_1 \, \d z_2 ~. 
\end{equation}
The expression $z_1 z_2 \sigma(s')$ is now essentially flat in $s'$, 
i.e.\ not only can $\sigma_{\mmax}$ be found at a convenient energy
such as the maximum one, but additionally the {\Py} generation 
efficiency (the likelihood of surviving step 4) is greatly 
enhanced. The price to be paid is that $z$ has to be selected
according to $D(z)/z$ rather than according to $D(z)$. Note that
$D(z)/z$ is not normalized to unity. One therefore needs to define
\begin{equation}
{\cal I}_D = \int_0^1 \frac{D(z)}{z} \, \d z ~, 
\end{equation}
and a properly normalized 
\begin{equation}
D'(z) = \frac{1}{{\cal I}_D} \, \frac{D(z)}{z} ~.
\end{equation}
Then  
\begin{equation}
\sigma_{\mrm{tot}} = \int_0^1 \int_0^1 D'(z_1) \, D'(z_2) \, 
{\cal I}_D^2 \, z_1 z_2 \sigma(z_1 z_2 s) \, \d z_1 \, \d z_2 ~.
\end{equation}
Therefore the proper event weight is ${\cal I}_D^2 \, z_1 z_2$.
This weight should be stored by you, for each event, in \ttt{PARP(173)}.
The maximum weight that will be encountered should be stored in
\ttt{PARP(174)} before the \ttt{PYINIT} call, and not changed
afterwards. It is not necessary to know the precise maximum;
any value larger than the true maximum will do, but the inefficiency
will be larger the cruder the approximation.
Additionally you must put \ttt{MSTP(173)=1} for the program to 
make use of weights at all. Often $D(z)$ is not known analytically;
therefore ${\cal I}_D$ is also not known beforehand, but may have to 
be evaluated (by you) during the course of the run. Then you should
just use the weight $z_1 z_2$ in \ttt{PARP(173)} and do the overall 
normalization yourself in the end. Since \ttt{PARP(174)=1} by
default, in this case you need not set this variable specially. 
Only the cross sections are affected by the procedure selected for
overall normalization, the events themselves still are properly 
distributed in $s'$ and internal phase space.  

Above it has been assumed tacitly that $D(z) \to 0$ for $z \to 0$. 
If not, $D(z)/z$ is divergent, and it is not possible to define a 
properly normalized $D'(z) = D(z)/z$. If the cross section is truly 
diverging like $1/s$, then a $D(z)$ which is nonvanishing for 
$z \to 0$ does imply an infinite total cross section, whichever way 
things are considered. In cases like that, it is necessary to impose
a lower cut on $z$, based on some physics or detector consideration. 
Some such cut is anyway needed to keep away from the minimum c.m.\ 
energy required for {\Py} events, see above.

The most difficult cases are those with a very narrow and high
peak, such as the $\Z^0$. One could initialize at the energy of 
maximum cross section and use $D(z)$ as is, but efficiency might 
turn out to be very low. One might then be tempted to do more 
complicated transforms of the kind illustrated above. As a rule 
it is then convenient to work in the variables $\tau_z = z_1 z_2$ 
and $y_z = (1/2) \ln (z_1/z_2)$, cf. section \ref{ss:kinemtwo}.

Clearly, the better the behaviour of the cross section can be 
modelled in the choice of $z_1$ and $z_2$, the better the overall
event generation efficiency. Even under the best of circumstances,
the efficiency will still be lower than for runs with fix energy.
There is also a non-negligible time overhead for using variable
energies in the first place, from kinematics reconstruction
and (in part) from the phase space selection. One should therefore
not use variable energies when not needed, and not use a large
range of energies $\sqrt{s'}$ if in the end only a smaller range 
is of experimental interest. 

This facility may be combined with most other aspects of the program.
For instance, it is possible to simulate beamstrahlung as above and
still include bremsstrahlung with \ttt{MSTP(11)=1}. Further, one may
multiply the overall event weight of \ttt{PARP(173)} with a 
kinematics-dependent weight given by \ttt{PYEVWT}, although it is not
recommended (since the chances of making a mistake are also 
multiplied). However, a few things do \textit{not} work.
\begin{Itemize}
\item It is not possible to use pile-up events, i.e.\ you must have 
\ttt{MSTP(131)=0}.
\item The possibility of giving in your own cross-section optimization
coefficients, option \ttt{MSTP(121)=2}, would require more input
than with fixed energies, and this option should therefore not be 
used. You can still use \ttt{MSTP(121)=1}, however.
\item The multiple interactions scenario with \ttt{MSTP(82)}$\geq 2$ 
only works approximately for energies different from the initialization 
one. If the c.m.\ energy spread is smaller than a factor 2, say, the
approximation should be reasonable, but if the spread is larger
one may have to subdivide into subruns of different energy bins.
The initialization should be made at the largest energy to be 
encountered --- whenever multiple interactions are possible (i.e.\ for 
incoming hadrons and resolved photons) this is where the cross sections 
are largest anyway, and so this is no further constraint. There is no 
simple possibility to change \ttt{PARP(82)} during the course of the 
run, i.e.\ an energy-independent $p_{\perp 0}$ must be assumed. The 
default option \ttt{MSTP(82)=1} works fine, i.e.\ does not suffer
from the constraints above. If so desired,
$\pTmin=$\ttt{PARP(81)} can be set differently for each event, 
as a function of c.m.\ energy. Initialization should then be done with 
\ttt{PARP(81)} as low as it is ever supposed to become.
\end{Itemize}
 
\subsection{How to Include External Processes}
\label{ss:PYnewproc}
 
Despite a large repertory of processes in {\Py}, the number of 
interesting missing ones clearly is even larger, and with time this 
discrepancy is likely to increase. There are several reasons why it is 
not practicable to imagine a {\Py} which has `everything'. One is
the amount of time it takes to implement a process for the few
{\Py} authors, compared with the rate of new cross section results
produced by the rather larger matrix-element calculations community.
Another is the length of currently produced matrix-element expressions,
which would make the program very bulky. A third argument is that,
whereas the phase space of $2 \to 1$ and $2 \to 2$ processes can be set
up once and for all according to a reasonably flexible machinery,
processes with more final-state particles are less easy to generate.
To achieve a reasonable efficiency, it is necessary to tailor the
phase-space selection procedure to the dynamics of the given process,
and to the desired experimental cuts.

At times, simple solutions may be found. Some processes may be seen just 
as trivial modifications of already existing ones. For instance, you 
might want to add some extra term, corresponding to contact interactions,
to the matrix elements of a {\Py} $2 \to 2$ process. In that case it is
not necessary to go through the machinery below, but instead you can use
the \ttt{PYEVWT} routine (subsection \ref{ss:PYwtevts}) to introduce an 
additional weight for the event, defined as the ratio of the modified to 
the unmodified differential cross sections. If you use the option 
\ttt{MSTP(142)=2}, this weight is considered as part of the `true' 
cross section of the process, and the generation is changed accordingly.

A {\Py} expert could also consider implementing a new process along the
lines of the existing ones, hardwired in the code. Such a modification 
would have to be ported anytime the {\Py} program is upgraded, however
(unless it is made available to the {\Py} authors and incorporated into
the public distribution). For this and other reasons, we will not
consider this option in detail, but only provide a few generic remarks.
The first step is to pick a process number \ttt{ISUB} among ones not in 
use. The process type needs to be set in \ttt{ISET(ISUB)} and, if the 
final state consists of massive particles, these should be specified in
\ttt{KFPR(ISUB,1)} and \ttt{KFPR(ISUB,2)}. Output is improved if a
process name is set in \ttt{PROC(ISUB)}. The second and main step is to 
code the cross section of the hard scattering subprocess in the 
\ttt{PYSIGH} routine. Usually the best starting point is to use the code 
of an existing similar process as a template for the new code required.
The third step is to program the selection of the final state in 
\ttt{PYSCAT}, normally a simple task, especially if again a similar
process (especially with respect to colour flow) can be used as template. 
In many cases the steps above are enough, in others additional 
modifications are required to \ttt{PYRESD} to handle process-specific 
non-isotropic decays of resonances. Further code may also be required 
e.g. if a process can proceed via an intermediate resonance that can be 
on the mass shell.  
 
The recommended solution, if a desired process is missing, is instead
to include it into {\Py} as an `external' process. In this section we 
will describe how it is possible to specify the parton-level state of some 
hard-scattering process in a common block. (`Parton-level' is not intended
to imply a restriction to quarks and gluons as interacting particles, 
but only that quarks and gluons are given rather than the hadrons they will
produce in the observable final state.) {\Py} will read this common 
block, and add initial- and final-state showers, beam remnants and 
underlying events, fragmentation and decays, to build up an event in as much 
detail as an ordinary {\Py} one. Another common block is to be filled with
information relevant for the run as a whole, where beams and processes 
are specified.  

Such a facility has been available since long, and has been used e.g.\ 
together with the \tsc{CompHEP} package. \tsc{CompHEP} \cite{Puk99} is 
mainly intended for the automatic computation of matrix elements, but also 
allows the sampling of phase space according to these matrix elements 
and thereby the generation of weighted or unweighted events. These 
events can be saved on disk and thereafter read back in to {\Py} for 
subsequent consideration \cite{Bel00}. 

At the Les Houches 2001 workshop it was decided to develop a common 
standard, that could be used by all matrix-elements-based generators to
feed information into any complete event generator \cite{Boo01}. It is 
similar to, but in its details different from, the approach previously 
implemented in {\Py}. Furthermore, it uses the same naming convention:
all names in common blocks end with \ttt{UP}, short for User(-defined) 
Process. This produces some clashes. Therefore the old facility, 
existing up to and including {\Py}~6.1, has been completely removed and 
replaced by the new one. The new code is still under development, and not 
all particulars have yet been implemented. In the description below we 
will emphasize current restrictions to the standard, as well as the 
solutions to aspects not specified by the standard. 

In particular, even with the common block contents defined, it is not
clear where they are to be filled, i.e. how the external supplier of 
parton-level events should synchronize with {\Py}. The solution adopted 
here --- recommended in the standard --- is to introduce two subroutines, 
\ttt{UPINIT} and \ttt{UPEVNT}. The first is called by \ttt{PYINIT} at 
initialization to obtain information about the run itself, and the other 
called by \ttt{PYEVNT} each time a new event configuration is to be fed 
in. We begin by describing these two steps and their related common blocks, 
before proceeding with further details and examples. The description is 
cast in a {\Py}-oriented language, but for the common block contents it 
closely matches the generator-neutral standard in \cite{Boo01}. 
Restrictions to or extensions of the standard should be easily recognized, 
but in case you are vitally dependent on following the standard exactly, 
you should of course check \cite{Boo01}. 

If you want to provide routines based on this standard, free to be used
by a larger community, please inform torbjorn@thep.lu.se. The intention
is to create a list of links to such routines, accessible from the 
standard {\Py} webpage, if there is interest.
 
\subsubsection{Run information}

When \ttt{PYINIT} is called in the main program, with \ttt{'USER'} as first 
argument (which makes the other arguments dummy), it signals that external 
processes are to be implemented. Then \ttt{PYINIT}, as part of its 
initialization tasks, will call the routine \ttt{UPINIT}.
 
\drawbox{CALL UPINIT}\label{p:UPINIT}
\begin{entry}
\itemc{Purpose:} routine to be provided by you when you want to implement 
external processes, wherein the contents of the \ttt{HEPRUP} common block 
are set. This information specifies the character of the run, both beams and 
processes, see further below.
\itemc{Note 1:} alternatively, the \ttt{HEPRUP} common block could be filled
already before \ttt{PYINIT} is called, in which case \ttt{UPINIT} could be
empty. We recommend \ttt{UPINIT} as the logical place to collect the relevant
information, however.
\itemc{Note 2:} a dummy copy of \ttt{UPINIT} is distributed with the 
program, in order to avoid potential problems with unresolved external 
references. This dummy should not be linked when you supply your own 
\ttt{UPINIT} routine.
\end{entry}

\drawboxseven{~INTEGER MAXPUP}%
{~PARAMETER (MAXPUP=100)}%
{~INTEGER IDBMUP,PDFGUP,PDFSUP,IDWTUP,NPRUP,LPRUP}%
{~DOUBLE PRECISION EBMUP,XSECUP,XERRUP,XMAXUP}%
{~COMMON/HEPRUP/IDBMUP(2),EBMUP(2),PDFGUP(2),PDFSUP(2),}%
{\&IDWTUP,NPRUP,XSECUP(MAXPUP),XERRUP(MAXPUP),XMAXUP(MAXPUP),}%
{\&LPRUP(MAXPUP)}\label{p:HEPRUP}
\begin{entry}

\itemc{Purpose:} to contain the initial information necessary for the 
subsequent generation of complete events from externally provided parton 
configurations.  The \ttt{IDBMUP}, \ttt{EBMUP}, \ttt{PDFGUP} and 
\ttt{PDFSUP} variables specify the nature of the two incoming beams. 
\ttt{IDWTUP} is a master switch, selecting the strategy to be used to mix 
different processes. \ttt{NPRUP} gives the number of different external 
processes to mix, and \ttt{XSECUP}, \ttt{XERRUP}, \ttt{XMAXUP} and 
\ttt{LPRUP} information on each of these. The contents in this common block 
must remain unchanged by the user during the course of the run, once set 
in the initialization stage.\\
This common block should be filled in the \ttt{UPINIT} routine or, 
alternatively, before the \ttt{PYINIT} call. During the run, {\Py} may 
update the \ttt{XMAXUP} values as required.

\iteme{MAXPUP :}\label{p:MAXPUP} the maximum number of distinguishable
processes that can be defined. (Each process in itself could consist of 
several subprocesses that have been distinguished in the parton-level
generator, but where this distinction is not carried along.)

\iteme{IDBMUP :}\label{p:IDBMUP} the PDG codes of the two incoming beam 
particles (or, in alternative terminology, the beam and target particles).\\
In {\Py}, this replaces the information normally provided by the \ttt{BEAM}
and \ttt{TARGET} arguments of the \ttt{PYINIT} call. Only particles which 
are acceptable \ttt{BEAM} or \ttt{TARGET} arguments may also be used in 
\ttt{IDBMUP}. The {\galep} options are not available.

\iteme{EBMUP :}\label{p:EBMUP} the energies, in GeV, of the two incoming beam
particles. The first (second) particle is taken to travel in the $+z$ ($-z$) 
direction.\\ 
The standard also allows non-collinear and varying-energy beams to be 
specified, see \ttt{ISTUP = -9} below, but this is not yet implemented 
in {\Py}.

\iteme{PDFGUP, PDFSUP :}\label{p:PDFGUP}\label{p:PDFSUP} the author group 
(\ttt{PDFGUP}) and set (\ttt{PDFSUP}) of the parton distributions of the two
incoming beams, as used in the generation of the parton-level events.
Numbers are based on the \tsc{Pdflib} \cite{Plo93} lists. This enumeration 
may not always be up to date, but it provides the only unique integer labels 
for parton distributions that we have. Where no codes are yet assigned to the
parton distribution sets used, one should do as best as one can, and be 
prepared for more extensive user interventions to interpret the information.
For lepton beams, or when the information is not provided for other
reasons, one should put \ttt{PDFGUP = PDFSUP = -1}.\\ 
By knowing which set has been used, it is possible to reweight cross sections
event by event, to correspond to another set.\\ 
Note that {\Py} does not access the \ttt{PDFGUP} or \ttt{PDFSUP} values 
in its description of internal processes or initial-state showers. If you 
want this to happen, you have to manipulate the \ttt{MSTP(51) - MSTP(56)} 
switches. For instance, to access \tsc{Pdflib} for protons, put 
\ttt{MSTP(51) = 1000*PDFGUP + PDFSUP} and \ttt{MSTP(52) = 2} in \ttt{UPINIT}.
(And remove the dummy \tsc{Pdflib} routines, as described for 
\ttt{MSTP(52)}.)  Also note that \ttt{PDFGUP} and \ttt{PDFSUP} allow an 
independent choice of parton distributions on the two sides of the event, 
whereas {\Py} only allows one single choice for all protons, another for 
all pions and a third for all photons.

\iteme{IDWTUP :}\label{p:IDWTUP} master switch dictating how event weights 
and cross sections should be interpreted. Several different models are 
presented in detail below. There will be tradeoffs between these, e.g. a 
larger flexibility to mix and re-mix several different processes could 
require a larger administrative machinery. Therefore the best strategy 
would vary, depending on the format of the input provided and the output 
desired. In some cases, parton-level configurations have already been 
generated with one specific model in mind, and then there may be no 
choice.\\
\ttt{IDWTUP} significantly affects the interpretation of \ttt{XWGTUP}, 
\ttt{XMAXUP} and \ttt{XSECUP}, as described below, but the basic 
nomenclature is the following. \ttt{XWGTUP} is the event weight for the 
current parton-level event, stored in the 
\ttt{HEPEUP} common block. For each allowed external process \ttt{i}, 
\ttt{XMAXUP(i)} gives the maximum event weight that could be encountered,
while \ttt{XSECUP(i)} is the cross section of the process. Here \ttt{i}
is an integer in the range between 1 and \ttt{NPRUP}; see the \ttt{LPRUP}
description below for comments on alternative process labels.
\begin{subentry}

\iteme{= 1 :} parton-level events come with a weight when input to {\Py}, 
but are then accepted or rejected, so that fully generated events at 
output have a common weight, 
customarily defined as $+1$. The event weight \ttt{XWGTUP} is a non-negative
dimensional quantity, in pb (converted to mb in {\Py}), with a mean value 
converging to the total cross section of the respective process. For each 
process \ttt{i}, the \ttt{XMAXUP(i)} value provides an upper estimate of how 
large \ttt{XWGTUP} numbers can be encountered. There is no need to supply an 
\ttt{XSECUP(i)} value; the cross sections printed with \ttt{PYSTAT(1)} are
based entirely on the averages of the \ttt{XWGTUP} numbers (with a small 
correction for the fraction of events that \ttt{PYEVNT} fails to generate in 
full for some reason).\\ 
The strategy is that \ttt{PYEVNT} selects which process \ttt{i} should be 
generated next, based on the relative size of the \ttt{XMAXUP(i)} values. 
The \ttt{UPEVNT} routine has to fill the \ttt{HEPEUP} common block with a
parton-level event of the requested type, and give its \ttt{XWGTUP} event 
weight. The event is accepted by \ttt{PYEVNT} with probability 
\ttt{XWGTUP/XMAXUP(i)}. In case of rejection, \ttt{PYEVNT} selects a new 
process \ttt{i} and asks for a new event. This ensures that processes 
are mixed in proportion to their average \ttt{XWGTUP} values.\\ 
This model presumes that \ttt{UPEVNT} is able to return a parton-level 
event of the process type requested by \ttt{PYEVNT}. It works well if each 
process is associated with an input stream of its own, either a subroutine 
generating events `on the fly' or a file of already generated events. It 
works less well if parton-level events from different processes already 
are mixed in a 
single file, and therefore cannot easily be returned in the order wanted by 
\ttt{PYEVNT}. In the latter case one should either use another model or else
consider reducing the level of ambition: even if you have mixed several 
different subprocesses on a file, maybe there is no need for {\Py} to 
know this finer classification, in which case we may get back to a 
situation with one `process' per external file. Thus the subdivision into 
processes should be a matter of convenience, not a straight jacket. 
Specifically, the shower and hadronization treatment of a parton-level 
event is independent of the process label assigned to it.
\\ 
If the events of some process are already available unweighted, then a 
correct mixing of this process with others is ensured by putting 
\ttt{XWGTUP = XMAXUP(i)}, where both of these numbers now is the total 
cross section of the process.\\
Each \ttt{XMAXUP(i)} value must be known from the very beginning, e.g. 
from an earlier exploratory run. If a larger value is encountered during 
the course of the run, a warning message will be issued and the 
\ttt{XMAXUP(i)} value (and its copy in \ttt{XSEC(ISUB,1)}) 
increased. Events generated before this time will 
have been incorrectly distributed, both in the process composition and 
in the phase space of the affected process, so that a bad estimate of 
\ttt{XMAXUP(i)} may require a new run with a better starting value.\\
The model described here agrees with the one used for internal {\Py} 
processes, and these can therefore freely be mixed with the external ones. 
Internal processes are switched on with \ttt{MSUB(ISUB) = 1}, as usual, 
either before the \ttt{PYINIT} call or in the \ttt{UPINIT} routine. One 
cannot use \ttt{MSEL} to select a predefined set of processes, for 
technical reasons, wherefore \ttt{MSEL = 0} is 
hardcoded when external processes are included.\\
A reweighting of events is feasible, e.g. by including a 
kinematics-dependent $K$ factor into \ttt{XWGTUP}, so long as 
\ttt{XMAXUP(i)} is also properly modified to take this into account. 
Optionally it is also possible to produce events with non-unit weight, 
making use the \ttt{PYEVWT} facility, see subsection \ref{ss:PYwtevts}. 
This works exactly the same way as for internal {\Py} processes,
except that the event information available inside \ttt{PYEVWT} would 
be different for external processes. You may therefore wish to access 
the \ttt{HEPEUP} common block inside your own copy of \ttt{PYEVWT},
where you calculate the event weight.\\
In summary, this option provides maximal flexibility, but at the price of
potentially requiring the administration of several separate input streams 
of parton-level events.

\iteme{= -1 :} same as \ttt{= 1} above, except that event weights may be 
either positive or negative on input, and therefore can come with an 
output weight of $+1$ or $-1$. This weight is uniquely defined by the 
sign of \ttt{XWGTUP}. It is also stored in \ttt{PARI(7)}. 
The need for negative-weight events arises in some next-to-leading-order 
calculations, but there are inherent dangers, discussed 
in subsection \ref{sss:externcomm} below.\\
In order to allow a correct mixing between processes, a process of 
indeterminate cross section sign has to be split up in two, where one 
always gives a positive or vanishing \ttt{XWGTUP}, and the other always 
gives it negative or vanishing. The \ttt{XMAXUP(i)} value for the latter 
process should give the negative \ttt{XWGTUP} of largest magnitude that 
will be encountered. \ttt{PYEVNT} selects which process \ttt{i} that 
should be generated next, based on the relative size of the 
\ttt{|XMAXUP(i)|} values. A given event is accepted with probability 
\ttt{|XWGTUP|/|XMAXUP(i)|}.  
 
\iteme{= 2 :} parton-level events come with a weight when input to {\Py}, 
but are then accepted or rejected, so that events at output have a common 
weight, customarily defined as $+1$. The non-negative event weight 
\ttt{XWGTUP} and its maximum value \ttt{XMAXUP(i)} may or may not be 
dimensional quantities; it does not matter since only the ratio 
\ttt{XWGTUP/XMAXUP(i)} will be used. Instead \ttt{XSECUP(i)} contains 
the process cross section in pb (converted to mb in {\Py}). It is this 
cross section that appears in the \ttt{PYSTAT(1)} table, only modified 
by the small fraction of events that \ttt{PYEVNT} fails to generate in 
full for some reason.\\ 
The strategy is that \ttt{PYEVNT} selects which process \ttt{i} should be 
generated next, based on the relative size of the \ttt{XSECUP(i)} values.
The \ttt{UPEVNT} routine has to fill the \ttt{HEPEUP} common block with a
parton-level event of the requested type, and give its \ttt{XWGTUP} event 
weight. The event is accepted by \ttt{PYEVNT} with probability 
\ttt{XWGTUP/XMAXUP(i)}. In case of rejection, the process number \ttt{i} 
is retained and  \ttt{PYEVNT} 
asks for a new event of this kind. This ensures that processes are mixed in 
proportion to their \ttt{XSECUP(i)} values.\\  
This model presumes that \ttt{UPEVNT} is able to return a parton-level 
event of the process type requested by \ttt{PYEVNT}, with comments exactly 
as for the \ttt{= 1} option.\\ 
If the events of some process are already available unweighted, then a 
correct mixing of this process with others is ensured by putting 
\ttt{XWGTUP = XMAXUP(i)}.\\
Each \ttt{XMAXUP(i)} and \ttt{XSECUP(i)} value must be known from the very 
beginning, e.g.\ from an earlier integration run. If a larger value is 
encountered during the course of the run, a warning message will be issued 
and the \ttt{XMAXUP(i)} value 
increased. This will not affect the process composition, but events 
generated before this time will have been incorrectly distributed in the 
phase space of the affected process, so that a bad estimate 
of \ttt{XMAXUP(i)} may require a new run with a better starting value.\\
While the generation model is different from the normal internal {\Py} one, 
it is sufficiently close that internal processes can be freely mixed 
with the external ones, exactly as described for the \ttt{= 1} option. In 
such a mix, internal processes are selected according to their equivalents 
of \ttt{XMAXUP(i)} and at rejection a new \ttt{i} is selected, whereas 
external ones are selected according to \ttt{XSECUP(i)} with \ttt{i} 
retained when an event is rejected.\\
A reweighting of individual events is no longer simple, since this would 
change the \ttt{XSECUP(i)} value nontrivially. Thus a new integration run 
with the modified event weights would be necessary to obtain new 
\ttt{XSECUP(i)} and \ttt{XMAXUP(i)} values. An overall rescaling of each 
process separately can be obtained by modifying the \ttt{XSECUP(i)} 
values accordingly, however, e.g.\ by a relevant $K$ factor.\\
In summary, this option is similar to the \ttt{= 1} one. The input of 
\ttt{XSECUP(i)} allows good cross section knowledge also in short test runs, 
but at the price of a reduced flexibility to reweight events. 
 
\iteme{= -2 :} same as \ttt{= 2} above, except that event weights may be 
either positive or negative on input, and therefore can come with an output 
weight of $+1$ or $-1$. This weight is uniquely defined by the sign of 
\ttt{XWGTUP}. It is also stored in \ttt{PARI(7)}. 
The need for negative-weight events arises in some
next-to-leading-order calculations, but there are inherent dangers, 
discussed in subsection \ref{sss:externcomm} below.\\
In order to allow a correct mixing between processes, a process of 
indeterminate cross section sign has to be split up in two, where one 
always gives a positive or vanishing \ttt{XWGTUP}, and the other always 
gives it negative or vanishing. The \ttt{XMAXUP(i)} value for the latter 
process should give the negative \ttt{XWGTUP} of largest magnitude that 
will be encountered, and \ttt{XSECUP(i)} should give the
integrated negative cross section. \ttt{PYEVNT} selects which process 
\ttt{i} that should be generated next, based on the relative size of the 
\ttt{|XSECUP(i)|} values. A given event is accepted with probability 
\ttt{|XWGTUP|/|XMAXUP(i)|}.  

\iteme{= 3 :} parton-level events come with unit weight when input to {\Py}, 
\ttt{XWGTUP = 1}, and are thus always accepted. This makes the 
\ttt{XMAXUP(i)} superfluous, while \ttt{XSECUP(i)} should give the 
cross section of each process.\\
The strategy is that that the next process type \ttt{i} is selected by the 
user inside \ttt{UPEVNT}, at the same time as the \ttt{HEPEUP} common block 
is filled with information about the parton-level event. This event is then 
unconditionally accepted by \ttt{PYEVNT}, except for the small fraction of 
events that \ttt{PYEVNT} fails to generate in full for some reason.\\
This model allows \ttt{UPEVNT} to read events from a file where different
processes already appear mixed. Alternatively, you are free to devise and 
implement your own mixing strategy inside \ttt{UPEVNT}, e.g. to mimic the
ones already outlined for \ttt{PYEVNT} in \ttt{= 1} and \ttt{= 2} above.\\
The \ttt{XSECUP(i)} values should be known from the beginning, in order for
\ttt{PYSTAT(1)} to produce a sensible cross section table. This is the only
place where it matters, however. That is, the processing of events inside
{\Py} is independent of this information.\\   
In this model it is not possible to mix with internal {\Py} processes,
since not enough information is available to perform such a mixing.\\
A reweighting of events is completely in the hands of the \ttt{UPEVNT} 
author. In the case that all events are stored in a single file, and all
are to be handed on to \ttt{PYEVNT}, only a common $K$ factor applied to all 
processes would be possible.\\
In summary, this option puts more power --- and responsibility --- in the
hands of the author of the parton-level generator. It is very convenient for
the processing of unweighted parton-level events stored in a single file. The
price to be paid is a reduced flexibility in the reweighting of events,
or in combining processes at will.

\iteme{= -3 :} same as \ttt{= 3} above, except that event weights may be 
either $+1$ or $-1$. This weight is uniquely defined by the sign of 
\ttt{XWGTUP}. It is also stored in \ttt{PARI(7)}. 
The need for negative-weight events arises in some
next-to-leading-order calculations, but there are inherent dangers, 
discussed in subsection \ref{sss:externcomm} below.\\
Unlike the \ttt{= -1} and \ttt{= -2} options, there is no need to split a 
process in two, each with a definite \ttt{XWGTUP} sign, since \ttt{PYEVNT} 
is not responsible for the mixing of processes. It may well be that the 
parton-level-generator author has enforced such a split, however, to solve a
corresponding mixing problem inside \ttt{UPEVNT}. Information on the relative
cross section in the negative- and positive-weight regions may also be useful
to understand the character and validity of the calculation (large 
cancellations means trouble!).

\iteme{= 4 :} parton-level events come with a weight when input to {\Py}, 
and this weight is to be retained unchanged at output. The event weight 
\ttt{XWGTUP} is a non-negative dimensional quantity, in pb (converted to 
mb in {\Py}, and as such also stored in \ttt{PARI(7)}), with a mean value 
converging to the total cross section of the respective process. When 
histogramming results, one of these event weights would have to be used.\\  
The strategy is exactly the same as \ttt{= 3} above, except that the event 
weight is carried along from \ttt{UPEVNT} to the \ttt{PYEVNT} output. Thus 
again all control is in the hands of the \ttt{UPEVNT} author.\\
A cross section can be calculated from the average of the \ttt{XWGTUP} 
values, as in the \ttt{= 1} option, and is displayed by \ttt{PYSTAT(1)}. 
Here it is of purely informative character, however, and does not influence 
the generation procedure. Neither \ttt{XSECUP(i)} or \ttt{XMAXUP(i)} needs 
to be known or supplied.\\
In this model it is not possible to mix with internal {\Py} processes,
since not enough information is available to perform such a mixing.\\
A reweighting of events is completely in the hands of the \ttt{UPEVNT} 
author, and is always simple, also when events appear sequentially stored 
in a single file.\\
In summary, this option allows maximum flexibility for the 
parton-level-generator author, but potentially at the price of spending 
a significant amount of time processing events of very small weight. Then 
again, in some cases it may be an advantage to have more events in the 
tails of a distribution in order to understand those tails better.

\iteme{= -4 :} same as \ttt{= 4} above, except that event weights in 
\ttt{XWGTUP} may be either positive or negative. In particular, the mean 
value of \ttt{XWGTUP} is converging to the total cross section of the 
respective process. The need for negative-weight events arises in some 
next-to-leading-order calculations, but there are inherent dangers, 
discussed in subsection \ref{sss:externcomm} below.\\
Unlike the \ttt{= -1} and \ttt{= -2} options, there is no need to split 
a process in two, each with a definite \ttt{XWGTUP} sign, since 
\ttt{PYEVNT} does not have
to mix processes. However, as for option \ttt{= -3}, such a split may offer 
advantages in understanding the character and validity of the calculation.  
\end{subentry}

\iteme{NPRUP :}\label{p:NPRUP} the number of different external processes, 
with information stored in the first \ttt{NPRUP} entries of the 
\ttt{XSECUP}, \ttt{XERRUP}, \ttt{XMAXUP} and \ttt{LPRUP} arrays.

\iteme{XSECUP :}\label{p:XSECUP} cross section for each external process, 
in pb. This information is mandatory for \ttt{IDWTUP =}~$\pm 2$, helpful 
for $\pm 3$, and not used for the other options.  

\iteme{XERRUP  :}\label{p:XERRUP} the statistical error on the cross 
section for each external process, in pb.\\
{\Py} will never make use of this information, but if it is available anyway 
it provides a helpful service to the user of parton-level generators.\\
Note that, if a small number $n_{\mrm{acc}}$ of events pass the experimental
selection cuts, the statistical error on this cross section is limited by
$\delta \sigma / \sigma \approx 1/\sqrt{n_{\mrm{acc}}}$, irrespectively of
the quality of the original integration. Furthermore, at least in hadronic
physics, systematic errors from parton distributions and higher orders 
typically are much larger than the statistical errors. 

\iteme{XMAXUP :}\label{p:XMAXUP} the maximum event weight \ttt{XWGTUP} that 
is likely to be encountered for each external process. For 
\ttt{IDWTUP =}~$\pm 1$ it has dimensions pb, while the dimensionality need 
not be specified for $\pm 2$. For the other \ttt{IDWTUP} options it is not 
used.

\iteme{LPRUP :}\label{p:LPRUP} a unique integer identifier of each external 
process, free to be picked by you for your convenience.  
This code is used in the \ttt{IDPRUP} identifier of which process occured.\\
In {\Py}, an external process is thus identified by three different integers.
The first is the {\Py} process number, \ttt{ISUB}. This number is assigned
by \ttt{PYINIT} at the beginning of each run, by scanning the \ttt{ISET} 
array for unused process numbers, and reclaiming such in the order they 
are found. The second is the sequence number \ttt{i}, running from 1 
through \ttt{NPRUP}, used to find information in the cross section arrays.
The third is the \ttt{LPRUP(i)} number, which can be anything that the user
wants to have as a unique identifier, e.g. in a larger database of processes.
For {\Py} to handle conversions, the two \ttt{KFPR} numbers of a given process 
\ttt{ISUB} are overwritten with the second and third numbers above. Thus the 
first external process will land in \ttt{ISUB = 4} (currently), and could 
have \ttt{LPRUP(1) = 13579}. In a \ttt{PYSTAT(1)} call, it would be listed 
as \ttt{User process 13579}.  
 
\end{entry}

\subsubsection{Event information}

Inside the event loop of the main program, \ttt{PYEVNT} will be called to 
generate the next event, as usual. When this is to be an external process, 
the parton-level event configuration and the event weight is found by a call 
from \ttt{PYEVNT} to \ttt{UPEVNT}.
 
\drawbox{CALL UPEVNT}\label{p:UPEVNT}
\begin{entry}
\itemc{Purpose:} routine to be provided by you when you want to implement 
external processes, wherein the contents of the \ttt{HEPEUP} common block 
are set. This information specifies the next parton-level event, and some
additional event information, see further below. How \ttt{UPEVNT} is expected
to solve its task depends on the model selected in \ttt{IDWTUP}, see above.
Specifically, note that the process type \ttt{IDPRUP} has already been
selected for some \ttt{IDWTUP} options (and then cannot be overwritten), 
while it remains to be chosen for others.
\itemc{Note :} a dummy copy of \ttt{UPEVNT} is distributed with the program, 
in order to avoid potential problems with unresolved external references.
This dummy should not be linked when you supply your own \ttt{UPEVNT} 
routine.
\end{entry}

\drawboxseven{~INTEGER MAXNUP}%
{~PARAMETER (MAXNUP=500)}%
{~INTEGER NUP,IDPRUP,IDUP,ISTUP,MOTHUP,ICOLUP}%
{~DOUBLE PRECISION XWGTUP,SCALUP,AQEDUP,AQCDUP,PUP,VTIMUP,SPINUP}%
{~COMMON/HEPEUP/NUP,IDPRUP,XWGTUP,SCALUP,AQEDUP,AQCDUP,IDUP(MAXNUP),}%
{\&ISTUP(MAXNUP),MOTHUP(2,MAXNUP),ICOLUP(2,MAXNUP),PUP(5,MAXNUP),}%
{\&VTIMUP(MAXNUP),SPINUP(MAXNUP)}  
\begin{entry}\label{p:HEPEUP}

\itemc{Purpose :} to contain information on the latest external process
generated in \ttt{UPEVNT}. A part is one-of-a-kind numbers, like the event 
weight, but the bulk of the information is a listing of incoming and
outgoing particles, with history, colour, momentum, lifetime and spin 
information.
 
\iteme{MAXNUP :}\label{p:MAXNUP} the maximum number of particles that can 
be specified by the external process.\\
The maximum of 500 is more than {\Py} is set up to handle. By default, 
\ttt{MSTP(126) = 100}, at most 96 particles could be specified, since
4 additional entries are needed in {\Py} for the two beam particles
and the two initiators of initial-state radiation. If this default is 
not sufficient, \ttt{MSTP(126)} would have to be increased at the 
beginning of the run.

\iteme{NUP :}\label{p:NUP} the number of particle entries in the current 
parton-level event, stored in the \ttt{NUP} first entries of the \ttt{IDUP},
\ttt{ISTUP}, \ttt{MOTHUP}, \ttt{ICOLUP}, \ttt{PUP}, \ttt{VTIMUP} and 
\ttt{SPINUP} arrays. \\
The special value \ttt{NUP = 0} is used to denote the case where 
\ttt{UPEVNT} is unable to provide an event, at least of the type requested 
by \ttt{PYEVNT}, e.g. because all events available in a file have
already been read. For such an event also the error flag 
\ttt{MSTI(51) = 1} instead of the normal \ttt{= 0}.

\iteme{IDPRUP :}\label{p:IDPRUP} the identity of the current process,
as given by the \ttt{LPRUP} codes.\\ 
When \ttt{IDWTUP =} $\pm 1$ or $\pm 2$, \ttt{IDPRUP} is selected by 
\ttt{PYEVNT} and already set when entering \ttt{UPEVNT}. Then 
\ttt{UPEVNT} has to provide an event of the specified process type, 
but cannot change \ttt{IDPRUP}. When \ttt{IDWTUP =} $\pm 3$ or $\pm 4$, 
\ttt{UPEVNT} is free to select the next process, and then should set 
\ttt{IDPRUP} accordingly.

\iteme{XWGTUP :}\label{p:XWGTUP} the event weight. The precise definition
of \ttt{XWGTUP} depends on the value of the \ttt{IDWTUP} master switch.
For \ttt{IDWTUP = 1} or \ttt{= 4} it is a dimensional quantity, in pb, with
a mean value converging to the total cross section of the respective 
process. For \ttt{IDWTUP = 2} the overall normalization is irrelevant.
For \ttt{IDWTUP = 3} only the value \ttt{+1} is allowed. For negative 
\ttt{IDWTUP} also negative weights are allowed, although positive and 
negative weights cannot appear mixed in the same process for 
\ttt{IDWTUP = -1} or \ttt{= -2}.

\iteme{SCALUP :}\label{p:SCALUP} scale $Q$ of the event, as used in the 
calculation of parton distributions (factorization scale). If the scale has 
not been defined, this should be denoted by using the value \ttt{-1}.\\
In {\Py}, this is input to \ttt{PARI(21) - PARI(26)} (and internally
\ttt{VINT(51) - VINT(56)}) When \ttt{SCALUP} is non-positive, the 
invariant mass of the parton-level event is instead used as scale.
Either of these comes to set the maximum virtuality in the initial-state 
parton showers. The same scale is also used for the first final-state 
shower, i.e. the one associated with the hard scattering. As in 
internal events, \ttt{PARP(67)} and \ttt{PARP(71)} offer multiplicative 
factors, whereby the respective initial- or final-state showering 
$Q_{\mrm{max}}^2$ scale can be modified relative to the scale above.
Any subsequent final-state showers are assumed to come from resonance 
decays, where the resonance mass always sets the scale. Since \ttt{SCALUP}
is not directly used inside {\Py} to evaluate parton densities, its role
as regulator of parton-shower activity may be the more important one.

\iteme{AQEDUP :}\label{p:AQEDUP} the QED coupling $\alphaem$ used for this 
event. If $\alphaem$ has not  been defined, this should be denoted by using 
the value \ttt{-1}.\\ 
In {\Py}, this value is stored in \ttt{VINT(57)}. It is not used anywhere,
however.

\iteme{AQCDUP :}\label{p:AQCDUP} the QCD coupling $\alphas$ used for this 
event. If $\alphas$ has not  been defined, this should be denoted by using 
the value \ttt{-1}.\\ 
In {\Py}, this value is stored in \ttt{VINT(58)}. It is not used anywhere,
however.

\iteme{IDUP(i) :}\label{p:IDUP} particle identity code, according to the
PDG convention, for particle \ttt{i}. As an extension to this standard, 
\ttt{IDUP(i) = 0} can be used to designate an intermediate state of 
undefined (and possible non-physical) character, e.g.\ a subsystem with 
a mass to be preserved by parton showers.\\
In the {\Py} event record, this corresponds to the \ttt{KF = K(I,2)} code.
But note that, here and in the following, the positions \ttt{i} in
\ttt{HEPEUP} and \ttt{I} in \ttt{PYJETS}
are likely to be different, since {\Py} normally stores more information in 
the beginning of the event record. Since \ttt{K(I,2) = 0} is forbidden, the 
\ttt{IDUP(i) = 0} code is mapped to \ttt{K(I,2) = 90}.

\iteme{ISTUP(i) :}\label{p:ISTUP} status code of particle \ttt{i}.
\begin{subentry}
\iteme{= -1 :} an incoming particle of the hard-scattering process.\\
In {\Py}, currently it is presumed that the first two particles, 
\ttt{i = 1}  and \ttt{= 2}, are of this character, and none of the others.
If this is not the case, the \ttt{HEPEUP} record will be rearranged to 
put such entries first. If the listing is still not acceptable after this, 
the program execution will stop.
This is a restriction relative to the standard, which allows more
possibilities. It is also presumed that these two particles are given with 
vanishing masses and parallel to the respective incoming beam direction, 
i.e.\ $E = p_z$ for the first and $E = - p_z$ for the second. The assignment 
of spacelike virtualities and nonvanishing $\pT$'s from initial-state 
radiation and primordial $\kT$'s is the prerogative of {\Py}. 
\iteme{= 1 :} an outgoing final-state particle.\\
Such a particle can, of course, be processed further by {\Py}, to add
showers and hadronization, or perform decays of any remaining resonances. 
\iteme{= 2 :} an intermediate resonance, whose mass should be preserved by
parton showers. For instance, in a process such as 
$\e^+\e^- \to \Z^0 \hrm^0 \to \q\qbar \, \b\bbar$, the $\Z^0$ and $\hrm^0$
should both be flagged this way, to denote that the $\q\qbar$ and 
$\b\bbar$ systems should have their individual masses preserved. 
In a more complex example, $\d \ubar \to \W^-\Z^0\g \to %
\ell^-\br{\nu}_{\ell} \, \q \qbar \, \g$, both the $\W^-$ and
$\Z^0$ particles and the $\W^-\Z^0$ pseudoparticle (with \ttt{IDUP(i) = 0})
could be given with status 2.\\ 
Often mass preservation is correlated with colour singlet subsystems, 
but this need not be the case. In 
$\e^+ \e^- \to \t \tbar \to \b \W^+ \, \bbar \W^-$, the $\b$ and 
$\bbar$ would be in a colour singlet state, but \textit{not} with a
preserved mass. Instead the $\t = \b \W^+$ and $\tbar = \bbar \W^-$ 
masses would be preserved, i.e.\ when $\b$ radiates $\b \to \b \g$
the recoil is taken by the $\W^+$. Exact mass preservation also by the 
hadronization stage is only guaranteed for colour singlet subsystems, 
however, at least for string fragmentation, since it is not possible to
define a subset of hadrons that uniquely belong only with a single
coloured particle.\\
The assignment of intermediate states is not always quantum mechanically
well-defined. For instance, 
$\e^+\e^- \to \mu^- \mu^+ \nu_{\mu} \br{\nu}_{\mu}$ can proceed
both through a $\W^+\W^-$ and a $\Z^0\Z^0$ intermediate state, as well
as through other graphs, which can interfere with each other. It is here 
the responsibility of the matrix-element-generator author to pick one
of the alternatives, according to some convenient recipe. One option 
might be to perform two calculations, one complete to select the event
kinematics and calculate the event weight, and a second with all 
interference terms neglected to pick the event history according to the 
relative weight of each graph. Often one particular graph would dominate, 
because a certain pairing of the final-state fermions would give invariant 
masses on or close to some resonance peaks.\\
In {\Py}, the identification of an intermediate resonance is not only a
matter of preserving a mass, but also of improving the modelling of the 
final-state shower evolution, since matrix-element-correction factors
have been calculated for a variety of possible resonance decays and
implemented in the respective parton shower description, see subsection
\ref{sss:PSMEfsmerge}.
\iteme{= 3 :} an intermediate resonance, given for documentation only, 
without any demand that the mass should be preserved in subsequent 
showers.\\
In {\Py}, currently particles defined with this option are not treated 
any differently from the ones with \ttt{= 2}. 
\iteme{= -2 :} an intermediate space-like propagator, defining an $x$ 
and a $Q^2$, in the Deeply Inelastic Scattering terminology, which 
should be preserved.\\
In {\Py}, currently this option is not defined and should not be used.
If it is, the program execution will stop.
\iteme{= -9 :} an incoming beam particle at time $t = -\infty$. Such 
beams are not required in most cases, since the \ttt{HEPRUP} common 
block normally contains the information. The exceptions are 
studies with non-collinear beams and with varying-energy beams
(e.g.\ from beamstrahlung, subsection \ref{sss:estructfun}), where 
\ttt{HEPRUP} does not supply sufficient flexibility. Information given 
with \ttt{= -9} overwrites the one given in \ttt{HEPRUP}.\\
This is an optional part of the standard, since it may be difficult 
to combine with some of the \ttt{IDWTUP} options.\\
Currently it is not recognized by {\Py}. If it is used, the program 
execution will stop.

\end{subentry} 

\iteme{MOTHUP(1,i), MOTHUP(2,i) :}\label{p:MOTHUP} position of the
first and last mother of particle \ttt{i}. Decay products will 
normally have only one mother. Then either \ttt{MOTHUP(2,i) = 0} or 
\ttt{MOTHUP(2,i) = MOTHUP(1,i)}. Particles in the outgoing state of a 
$2 \to n$ process have two mothers. This scheme does not limit the 
number of mothers, so long as these appear consecutively in the listing, 
but in practice there will likely never be more than two mothers per 
particle.\\
As has already been mentioned for \ttt{ISTUP(i) = 2}, the definition of
history is not always unique. Thus, in a case like 
$\e^+\e^- \to \mu^+ \mu^- \gamma$, proceeding via an intermediate 
$\gamma^*/\Z^0$, the squared matrix element contains an interference 
term between initial- and final-state emission of the photon. This 
ambiguity has to be resolved by the matrix-elements-based generator.\\
In {\Py}, only information on the first mother survives into 
\ttt{K(I,3)}. This is adequate for resonance decays, while particles
produced in the primary $2 \to n$ process are given mother code 0, 
as is customary for internal processes. It implies that two particles
are deemed to have the same mothers if the first one agrees; it is 
difficult to conceive of situations where this would not be the case.
Furthermore, it is assumed that the \ttt{MOTHUP(1,i) < i}, i.e.\ that
mothers are stored ahead of their daughters, and that all daughters of
a mother are listed consecutively, i.e.\ without other particles 
interspersed. If this is not the case, the \ttt{HEPEUP} record will 
be rearranged so as to adhere to these principles. \\
{\Py} has a limit of at most 80 particles coming from the same mother,
for the final-state parton shower algorithm to work. In fact, the shower
is optimized for a primary $2 \to 2$ process followed by some sequence
of $1 \to 2$ resonance decays. Then colour coherence with the initial 
state, matrix-element matching to gluon emission in resonance decays,
and other sophisticated features are automatically included. By contrast,
the description of emission in systems with three or more partons is 
less sophisticated. Apart from problems with the algorithm itself,
more information would be needed to do a good job than is provided by 
the standard. Specifically, there is a significant danger of 
double counting or gaps between the radiation already covered by matrix 
elements and the one added by the shower. The omission from \ttt{HEPEUP}
of intermediate resonances known to be there, so that e.g. two 
consecutive $1 \to 2$ decays are bookkept as a single $1 \to 3$
branching, is a simple way to reduce the reliability of your studies!

\iteme{ICOLUP(1,i), ICOLUP(2,i) :}\label{p:ICOLUP} integer tags for the 
colour flow lines passing through the colour and anticolour, respectively, 
of the particle. Any particle with colour (anticolour), such as a quark 
(antiquark) or gluon, will have the first (second) number nonvanishing.\\ 
The tags can be viewed as a numbering of different colours in the 
$N_C \to \infty$ limit of QCD. Any nonzero integer can be used to 
represent a different colour, but the standard recommends to stay 
with positive numbers larger than \ttt{MAXNUP} to avoid confusion 
between colour tags and the position labels \ttt{i} of particles.\\  
The colour and anticolour of a particle is defined with the respect
to the physical time ordering of the process, so as to allow a unique 
definition of colour flow also through intermediate particles. That is,
a quark always has a nonvanishing colour tag \ttt{ICOLUP(1,i)}, whether 
it is in the initial, intermediate or final state. A simple example would 
be $\q\qbar \to \t\tbar \to \b \W^+ \bbar \W^-$, where the same colour
label is to be used for the $\q$, the $\t$ and the $\b$. Correspondingly,
the $\qbar$, $\tbar$ and $\bbar$ share another colour label, now stored 
in the anticolour position \ttt{ICOLUP(2,i)}.\\ 
The colour label in itself does not distinguish between the colour or 
the anticolour of a given kind; that information appears in the usage
either of the \ttt{ICOLUP(1,i)} or of the \ttt{ICOLUP(2,i)} position
for the colour or anticolour, respectively. Thus, in a $\W^+ \to \u \dbar$
decay, the $\u$ and $\dbar$ would share the same colour label, but stored 
in \ttt{ICOLUP(1,i)} for the $\u$ and in \ttt{ICOLUP(2,i)} for the 
$\dbar$.\\
In general, several colour flows are possible in a given subprocess.
This leads to ambiguities, of a character similar to the ones for the
history above, and as is discussed in subsection \ref{sss:QCDjetclass}.
Again it is up to the author of the matrix-elements-based generator to 
find a sensible solution. It is useful to note that all interference terms
between different colour flow topologies vanish in the $N_C \to \infty$ 
limit of QCD. One solution would then be to use a first calculation in
standard QCD to select the momenta and find the weight of the process, 
and a second with $N_C \to \infty$ to pick one specific colour topology 
according to the relative probabilities in this limit.\\
The above colour scheme also allows for baryon number violating processes. 
Such a vertex would appear as `dangling' colour lines, when the 
\ttt{ICOLUP} and \ttt{MOTHUP} information is correlated.
For instance, in $\su \to \dbar \dbar$ the $\su$ inherits an existing
colour label, while the two $\dbar$'s are produced with two different
new labels.\\
Several examples of colour assignments, both with and without baryon
number violation, are given in \cite{Boo01}.\\
In {\Py}, baryon number violation is not yet implemented as part of the
external-process machinery (but exists for internal processes, including
internally handled decays of resonances provided externally). It will 
require substantial extra work to lift this restriction, and this is not 
imminent.

\iteme{PUP(1,i), PUP(2,i), PUP(3,i), PUP(4,i), PUP(5,i) :}\label{p:PUP}
the particle momentum vector $(p_x, p_y, p_z, E, m)$, with units of GeV.
A spacelike virtuality is denoted by a negative sign on the mass.\\
Apart from the index order, this exactly matches the \ttt{P} momentum
conventions in \ttt{PYJETS}.\\ 
{\Py} is forgiving when it comes to using other masses than 
its own, e.g.\ for quarks. Thus the external process can be given 
directly with the $m_{\b}$ used in the calculation, without any worry 
how this matches the {\Py} default. However, remember that the two 
incoming particles with \ttt{ISTUP(i) = -1} have to be massless.

\iteme{VTIMUP(i) :}\label{p:VTIMUP} invariant lifetime $c\tau$ in mm,
i.e. distance from production to decay. Once the primary vertex has
been selected, the subsequent decay vertex positions in space and time
can be constructed step by step, by also making use of the momentum
information. Propagation in vacuum, without any bending e.g.\ by 
magnetic fields, has to be assumed.\\
This exactly corresponds to the \ttt{V(I,5)} component in \ttt{PYJETS}.
Note that it is used in {\Py} to track colour singlet particles through
distances that might be observable in a detector. It is not used to 
trace the motion of coloured partons at fm scales, through the 
hadronization process. Also note that {\Py} will only use this
information for intermediate resonances, not for the 
initial- and final-state particles. For instance, for an undecayed 
$\tau^-$, the lifetime is selected as part of the $\tau^-$ decay 
process, not based on the \ttt{VTIMUP(i)} value. 

\iteme{SPINUP(i) :}\label{p:SPINUP} cosine of the angle between the 
spin vector of a particle and its three-momentum, specified in the 
lab frame, i.e. the frame where the event as a whole is defined. 
This scheme is neither general nor complete, but it is chosen as a 
sensible compromise.\\ 
The main foreseen application is $\tau$'s with a specific helicity. 
Typically a relativistic $\tau^-$ ($\tau^+$) coming from a $\W^-$ 
($\W^+$) decay would have helicity and \ttt{SPINUP(i) =} $-1$ ($+1$). 
This could be changed by the boost from the $\W$ rest frame to the 
lab frame, however. The use of a real number, rather than an integer,
allows for an extension to the non-relativistic case.\\
Particles which are unpolarized or have unknown polarization should be 
given \ttt{SPINUP(i) = 9}.\\
Explicit spin information is not used anywhere in {\Py}. It 
is implicit in many production and decay matrix elements, which often 
contain more correlation information than could be conveyed by the 
simple spin numbers discussed here. Correspondingly, it is to be 
expected that the external generator already performed the decays of 
the $\W$'s, the $\Z$'s and the other resonances, so as to include the 
full spin correlations. If this is not the case, such resonances will 
normally be decayed isotropically. Some correlations could appear in 
decay chains: the {\Py} decay $\t \to \b\W^+$ is isotropic, but the 
subsequent $\W^+ \to \q_1 \qbar_2$ decay contains implicit $\W$ helicity 
information from the $\t$ decay.\\ 
Also $\tau$ decays performed by {\Py} would be isotropic. An interface 
routine \ttt{PYTAUD} (see subsection \ref{ss:JETphysrout}) can be used 
to link to external $\tau$ decay generators, but is based on defining 
the $\tau$ in the rest frame of the decay that produces it, and so is 
not directly applicable here. Eventually, it will be rewritten to 
make use of the \ttt{SPINUP(i)} information. In the meantime, and of 
course also afterwards, a valid option is to perform the $\tau$ decays 
yourself before passing `parton-level' events to {\Py}. 
\end{entry}

One auxiliary routine exists, that formally is part of the 
{\Py} package, but could be used by any generator: 
 \begin{entry}
\iteme{SUBROUTINE PYUPRE :}\label{p:PYUPRE}
called immediately after \ttt{UPEVNT} has been called to provide a 
user-process event. It will rearrange the contents of the  
\ttt{HEPEUP} common block so that afterwards the two incoming 
partons appear in lines 1 and 2, so that all mothers appear ahead 
of their daughters, and so that the daughters of a decay are listed 
consecutively. Such an order can thereby be presumed to exist in 
the subsequent parsing of the event. If the rules already are
obeyed, the routine does not change anything.
\end{entry}

\subsubsection{An example}

To exemplify the above discussion, consider the explicit case of 
$\q\qbar$ or $\g\g \to \t\tbar \to \b\W^+\bbar\W^- \to %
\b\q_1\qbar_2 \, \bbar\q_3\qbar_4$. These two processes are already 
available in {\Py}, but without full spin correlations. One might 
therefore wish to include them from some external generator. A physics 
analysis would then most likely involve angular correlations intended
to set limits on (or reveal traces of) anomalous couplings. However, so 
as to give a simple self-contained example, instead consider the analysis 
of the charged multiplicity distribution. This actually offers a simple 
first cross-check between the internal and external implementations of 
the same process. The main program might then look something like
\begin{verbatim}
      IMPLICIT DOUBLE PRECISION(A-H, O-Z)
      IMPLICIT INTEGER(I-N)

C...User process event common block.
      INTEGER MAXNUP
      PARAMETER (MAXNUP=500) 
      INTEGER NUP,IDPRUP,IDUP,ISTUP,MOTHUP,ICOLUP
      DOUBLE PRECISION XWGTUP,SCALUP,AQEDUP,AQCDUP,PUP,VTIMUP,SPINUP  
      COMMON/HEPEUP/NUP,IDPRUP,XWGTUP,SCALUP,AQEDUP,AQCDUP,IDUP(MAXNUP),
     &ISTUP(MAXNUP),MOTHUP(2,MAXNUP),ICOLUP(2,MAXNUP),PUP(5,MAXNUP),
     &VTIMUP(MAXNUP),SPINUP(MAXNUP)  
      SAVE /HEPEUP/   
 
C...PYTHIA common block.
      COMMON/PYJETS/N,NPAD,K(4000,5),P(4000,5),V(4000,5)
      SAVE /PYJETS/

C...Initialize.
      CALL PYINIT('USER',' ',' ',0D0)

C...Book histogram. Reset event counter.
      CALL PYBOOK(1,'Charged multiplicity',100,-1D0,199D0)
      NACC=0

C...Event loop; check that not at end of run; list first events.
      DO 100 IEV=1,1000
        CALL PYEVNT
        IF(NUP.EQ.0) GOTO 110
        NACC=NACC+1 
        IF(IEV.LE.3) CALL PYLIST(7)
        IF(IEV.LE.3) CALL PYLIST(2) 

C...Analyze event; end event loop.
        CALL PYEDIT(3)
        CALL PYFILL(1,DBLE(N),1D0)  
  100 CONTINUE

C...Statistics and histograms.
  110 CALL PYSTAT(1)
      CALL PYFACT(1,1D0/DBLE(NACC))
      CALL PYHIST

      END
\end{verbatim} 
There \ttt{PYINIT} is called with \ttt{'USER'} as first argument, 
implying that the rest is dummy. The event loop itself looks fairly
familiar, but with two additions. One is that \ttt{NUP} is checked 
after each event, since \ttt{NUP = 0} would signal a premature end 
of the run, with the external generator unable to return more events. 
This would be the case e.g.\ if events have been stored on file, and 
the end of this file is reached. The other is that \ttt{CALL PYLIST(7)}
can be used to list the particle content of the \ttt{HEPEUP} common
block (with some information omitted, on vertices and spin), so that 
one can easily compare this input with the output after {\Py} 
processing, \ttt{CALL PYLIST(2)}. An example of a \ttt{PYLIST(7)}
listing would be
\begin{verbatim}
          Event listing of user process at input (simplified)

 I IST  ID Mothers   Colours    p_x      p_y      p_z       E        m
 1 -1   21   0   0  101  109    0.000    0.000  269.223  269.223    0.000
 2 -1   21   0   0  109  102    0.000    0.000 -225.566  225.566    0.000
 3  2    6   1   2  101    0   72.569  153.924  -10.554  244.347  175.030
 4  2   -6   1   2    0  102  -72.569 -153.924   54.211  250.441  175.565
 5  1    5   3   0  101    0   56.519   33.343   53.910   85.045    4.500
 6  2   24   3   0    0    0   16.050  120.581  -64.464  159.302   80.150
 7  1   -5   4   0    0  102   44.127  -60.882   25.507   79.527    4.500
 8  2  -24   4   0    0    0 -116.696  -93.042   28.705  170.914   78.184
 9  1    2   6   0  103    0   -8.667   11.859   16.063   21.766    0.000
10  1   -1   6   0    0  103   24.717  108.722  -80.527  137.536    0.000
11  1   -2   8   0    0  104  -33.709  -22.471  -26.877   48.617    0.000
12  1    1   8   0  104    0  -82.988  -70.571   55.582  122.297    0.000
\end{verbatim}
Note the reverse listing of \ttt{ID(UP)} and \ttt{IST(UP)} relative to the
\ttt{HEPEUP} order, to have better agreement with the \ttt{PYJETS} one.
(The \ttt{ID} column is wider in real life, to allow for longer codes, 
but has here been reduced to fit the listing onto the page.) 

The corresponding \ttt{PYLIST(2)} listing of course would be considerably 
longer, containing a complete event as it does. Also the particles above 
would there appear boosted by the effects of initial-state radiation and 
primordial $\kT$; copies of them further down in the event record would
also include the effects of final-state radiation. The full story is
available with \ttt{MSTP(125)=2}, while the default listing omits some
of the intermediate steps. 

The \ttt{PYINIT} call will generate a call to the user-supplied
routine \ttt{UPINIT}. It is here that we need to specify the details
of the generation model. Assume, for instance that $\q\qbar$- and
$\g\g$-initiated events have been generated in two separate runs
for Tevatron Run II, with weighted events stored in two separate files.
By the end of each run, cross section and maximum weight information 
has also been obtained, and stored on separate files. Then 
\ttt{UPINIT} could look like     
\begin{verbatim}
      SUBROUTINE UPINIT
 
C...Double precision and integer declarations.
      IMPLICIT DOUBLE PRECISION(A-H, O-Z)
      IMPLICIT INTEGER(I-N)

C...User process initialization common block.
      INTEGER MAXPUP
      PARAMETER (MAXPUP=100)
      INTEGER IDBMUP,PDFGUP,PDFSUP,IDWTUP,NPRUP,LPRUP
      DOUBLE PRECISION EBMUP,XSECUP,XERRUP,XMAXUP
      COMMON/HEPRUP/IDBMUP(2),EBMUP(2),PDFGUP(2),PDFSUP(2),
     &IDWTUP,NPRUP,XSECUP(MAXPUP),XERRUP(MAXPUP),XMAXUP(MAXPUP),
     &LPRUP(MAXPUP)
      SAVE /HEPRUP/

C....Pythia common block - needed for setting PDF's; see below.
      COMMON/PYPARS/MSTP(200),PARP(200),MSTI(200),PARI(200)
      SAVE /PYPARS/      

C...Set incoming beams: Tevatron Run II.
      IDBMUP(1)=2212
      IDBMUP(2)=-2212
      EBMUP(1)=1000D0
      EBMUP(2)=1000D0

C...Set PDF's of incoming beams: CTEQ 5L.
C...Note that Pythia will not look at PDFGUP and PDFSUP.  
      PDFGUP(1)=4
      PDFSUP(1)=46
      PDFGUP(2)=PDFGUP(1)
      PDFSUP(2)=PDFSUP(1)

C...Set use of CTEQ 5L in internal Pythia code.
      MSTP(51)=7 
      
C...If you want Pythia to use PDFLIB, you have to set it by hand.
C...(You also have to ensure that the dummy routines
C...PDFSET, STRUCTM and STRUCTP in Pythia are not linked.)      
C      MSTP(52)=2
C      MSTP(51)=1000*PDFGUP(1)+PDFSUP(1)

C...Decide on weighting strategy: weighted on input, cross section known.
      IDWTUP=2

C...Number of external processes. 
      NPRUP=2

C...Set up q qbar -> t tbar.
      OPEN(21,FILE='qqtt.info',FORM='unformatted',ERR=100)
      READ(21,ERR=100) XSECUP(1),XERRUP(1),XMAXUP(1)
      LPRUP(1)=661
      OPEN(22,FILE='qqtt.events',FORM='unformatted',ERR=100)

C...Set up g g -> t tbar.
      OPEN(23,FILE='ggtt.info',FORM='unformatted',ERR=100)
      READ(23,ERR=100) XSECUP(2),XERRUP(2),XMAXUP(2)
      LPRUP(2)=662
      OPEN(24,FILE='ggtt.events',FORM='unformatted',ERR=100)

      RETURN
C...Stop run if file operations fail.
  100 WRITE(*,*) 'Error! File open or read failed. Program stopped.'
      STOP   

      END
\end{verbatim}
Here unformatted read/write is used to reduce the size of the 
event files, but at the price of a platform dependence. Formatted files 
are preferred if they are to be shipped elsewhere. The rest should be
self-explanatory.

Inside the event loop of the main program, \ttt{PYEVNT} will call
\ttt{UPEVNT} to obtain the next parton-level event. In its simplest 
form, only a single \ttt{READ} statement would be necessary to read 
information on the next event, e.g.\ what is shown in the event
listing earlier in this subsection, with a few additions. Then the 
routine could look like
\begin{verbatim}
      SUBROUTINE UPEVNT
 
C...Double precision and integer declarations.
      IMPLICIT DOUBLE PRECISION(A-H, O-Z)
      IMPLICIT INTEGER(I-N)

C...User process event common block.
      INTEGER MAXNUP
      PARAMETER (MAXNUP=500) 
      INTEGER NUP,IDPRUP,IDUP,ISTUP,MOTHUP,ICOLUP
      DOUBLE PRECISION XWGTUP,SCALUP,AQEDUP,AQCDUP,PUP,VTIMUP,SPINUP  
      COMMON/HEPEUP/NUP,IDPRUP,XWGTUP,SCALUP,AQEDUP,AQCDUP,IDUP(MAXNUP),
     &ISTUP(MAXNUP),MOTHUP(2,MAXNUP),ICOLUP(2,MAXNUP),PUP(5,MAXNUP),
     &VTIMUP(MAXNUP),SPINUP(MAXNUP)  
      SAVE /HEPEUP/   

C...Pick file to read from, based on requested event type.
      IUNIT=22
      IF(IDPRUP.EQ.662) IUNIT=24

C...Read event from this file. (Except that NUP and IDPRUP are known.) 
      NUP=12
      READ(IUNIT,ERR=100,END=100) XWGTUP,SCALUP,AQEDUP,AQCDUP,
     &(IDUP(I),ISTUP(I),MOTHUP(1,I),MOTHUP(2,I),ICOLUP(1,I),
     &ICOLUP(2,I),(PUP(J,I),J=1,5),VTIMUP(I),SPINUP(I),I=1,NUP)

C...Return, with NUP=0 if read failed.
      RETURN
  100 NUP=0
      RETURN
      END    
\end{verbatim}
However, in reality one might wish to save disk space by not storing
redundant information. The \ttt{XWGTUP} and \ttt{SCALUP} numbers are 
vital, while \ttt{AQEDUP} and \ttt{AQCDUP} are purely informational 
and can be omitted. In a $\g\g \to \t\tbar \to \b\W^+\bbar\W^- \to %
\b\q_1\qbar_2 \, \bbar\q_3\qbar_4$ event, only the $\q_1$, $\qbar_2$, 
$\q_3$ and $\qbar_4$ flavours need be given, assuming that the particles 
are always stored in the same order. For a $\q\qbar$ initial state,
the $\q$ flavour should be added to the list. The \ttt{ISTUP}, 
\ttt{MOTHUP} and \ttt{ICOLUP} information is the same in all events
of a given process, except for a twofold ambiguity in the colour flow 
for $\g\g$ initial states. All \ttt{VTIMUP} vanish and
the \ttt{SPINUP} are uninteresting since spins have already been taken
into account by the kinematics of the final fermions. (It would be
different if one of the $\W$'s decayed leptonically to a $\tau$.)  
Only the \ttt{PUP} values of the six final fermions need be given, 
since the other momenta and masses can be reconstructed from this, 
remembering that the two initial partons are massless and along the 
beam pipe. The final fermions are on the mass shell, so their masses
need not be stored event by event, but can be read from a small table. 
The energy of a particle can be 
reconstructed from its momentum and mass. Overall transverse momentum
conservation removes two further numbers. What remains is thus 5 integers 
and 18 real numbers, where the reals could well be stored in single 
precision. Of course, the code needed to unpack information stored
this way would be lengthy but trivial. Even more compact storage 
strategies could be envisaged, e.g. only to save the weight and the 
seed of a dedicated random-number generator, to be used to generate 
the next parton-level event. It is up to you to find
the optimal balance between disk space and coding effort. 

\subsubsection{Further comments}
\label{sss:externcomm}

This section contains additional information on a few different
topics: cross section interpretation, negative-weight events, 
relations with other {\Py} switches and routines, 
and error conditions. 

In several \ttt{IDWTUP} options,
the \ttt{XWGTUP} variable is supposed to give the differential
cross section of the current event, times the phase-space volume
within which events are generated, expressed in picobarns. 
(Converted to millibarns inside {\Py}.) This means that, in the limit 
that many events are generated, the average value of \ttt{XWGTUP} 
gives the total cross section of the simulated process.
Of course, the tricky part is that the differential cross section
usually is strongly peaked in a few regions of the phase space, such
that the average probability to accept an event,
$\langle$\ttt{XWGTUP}$\rangle /$\ttt{XMAXUP(i)} is
small. It may then be necessary to find a suitable set of transformed
phase-space coordinates, for which the correspondingly transformed
differential cross section is better behaved.
 
To avoid confusion, here is a more formal version of the above
paragraph. Call $\d X$ the differential phase space, e.g.\ for a 
$2 \to 2$
process $\d X = \d x_1 \, \d x_2 \, \d \hat{t}$, where $x_1$ and $x_2$ 
are the momentum fractions carried by the two incoming partons and
$\hat{t}$ the Mandelstam variable of the scattering (see subsection
\ref{ss:kinemtwo}). Call $\d \sigma / \d X$ the differential cross section
of the process, e.g.\ for $2 \to 2$: $\d \sigma / \d X = \sum_{ij}
f_i(x_1,Q^2) \, f_j(x_2,Q^2) \, \d \hat{\sigma}_{ij} / \d \hat{t}$,
i.e.\ the product of parton distributions and hard-scattering matrix
elements, summed over all allowed incoming flavours $i$ and $j$.
The physical cross section that one then wants to generate is
$\sigma = \int (\d \sigma / \d X) \, \d X$, where the integral is over 
the allowed phase-space volume. The event generation procedure consists
of selecting an $X$ uniformly in $\d X$ and then evaluating the weight
$\d \sigma / \d X$ at this point. \ttt{XWGTUP} is now simply
\ttt{XWGTUP}$ = \d \sigma / \d X \, \int \d X$, i.e.\ the differential
cross section times the considered volume of phase space. Clearly,
when averaged over many events, \ttt{XWGTUP} will correctly estimate
the desired cross section. If \ttt{XWGTUP} fluctuates too much, one
may try to transform to new variables $X'$,
where events are now picked accordingly to $\d X'$ and
\ttt{XWGTUP}$ = \d \sigma / \d X' \, \int \d X'$.

A warning. It is important that $X$ is indeed uniformly picked within
the allowed phase space, alternatively that any Jacobians are properly 
taken into account. For instance, in the case above, one approach 
would be to pick $x_1$, $x_2$ and $\hat{t}$ uniformly in the ranges
$0 < x_1 < 1$, $0 < x_2 < 1$, and $-s < \hat{t} < 0$, with full phase 
space
volume  $\int \d X = s$. The cross section would only be non-vanishing
inside the physical region given by $-s x_1 x_2 < \hat{t}$ (in the
massless case), i.e.\ Monte Carlo efficiency is likely to be low.
However, if one were to choose $\hat{t}$ values only in the range 
$-\hat{s} < \hat{t} < 0$, small $\hat{s}$ values would be favoured, 
since the density of selected $\hat{t}$ values would be larger there. 
Without the use of a compensating Jacobian $\hat{s}/s$, an incorrect 
answer would be obtained. Alternatively, one could start out with a
phase space like $\d X = \d x_1 \, \d x_2 \, \d (\cos\hat{\theta})$, 
where the limits decouple. Of course, the $\cos\hat{\theta}$ variable 
can be translated back into a $\hat{t}$, which will then always be in 
the desired range $-\hat{s} < \hat{t} < 0$. The transformation itself 
here gives the necessary Jacobian.

At times, it is convenient to split a process into a discrete set of
subprocesses for the parton-level generation, without retaining these
in the \ttt{IDPRUP} classification. For instance, the cross section above 
contains a summation over incoming partons. An alternative would then 
have been to let each subprocess correspond to one unique combination 
of incoming flavours. When an event of process type $i$ is to be
generated, first a specific subprocess $ik$ is selected with probability
$f^{ik}$, where $\sum_k f^{ik} = 1$. For this subprocess an 
\ttt{XWGTUP}$^k$ is generated as above, except that there is no longer 
a summation over incoming flavours. Since only a fraction $f^{ik}$ of all 
events now contain this part of the cross section, a compensating factor 
$1/f^{ik}$ is introduced, i.e.\ \ttt{XWGTUP}$ = $\ttt{XWGTUP}$^k/f^{ik}$. 
Further, one has to define
\ttt{XMAXUP(i)}$ = \mmax_k$ \ttt{XMAXUP}$^{ik}/f^{ik}$ and
\ttt{XSECUP(i)}$ = \sum_k$ \ttt{XSECUP}$^{ik}$. 
The generation efficiency will be maximized for the $f^{ik}$ coefficients
selected proportional to \ttt{XMAXUP}$^{ik}$, but this is no requirement. 

The standard allows external parton-level events to come with negative 
weights, unlike the case for internal {\Py} processes. In order to avoid 
indiscriminate use of this option, some words of caution are in place. 
In next-to-leading-order calculations, events with negative weights 
may occur as part of the virtual corrections. In any physical 
observable quantity, the effect of such events should cancel against
the effect of real events with one more parton in the final state.  
For instance, the next-to-leading order calculation of gluon scattering
contains the real process $\g\g \to \g\g\g$, with a positive divergence
in the soft and collinear regions, while the virtual corrections to
$\g\g \to \g\g$ are negatively divergent. Neglecting the problems of
outgoing gluons collinear with the beams, and those of soft gluons, two 
nearby outgoing gluons in the $\g\g \to \g\g\g$ process can be combined 
into one effective one, such that the divergences can be cancelled. 

If rather widely separated gluons can be combined, the remaining 
negative contributions are not particularly large. Different separation
criteria could be considered; one example would be
$\Delta R = \sqrt{(\Delta \eta)^2 + (\Delta \varphi)^2} \approx 1$.
The recombination of well separated partons is at the price of an 
arbitrariness in the choice of clustering algorithm, when two gluons of 
nonvanishing invariant mass are to be combined into one massless one, 
as required to be able to compare with the kinematics of the massless 
$\g\g \to \g\g$ process when performing the divergence cancellations. 
Alternatively, if a smaller $\Delta R$ cut is used, where the combining 
procedure is less critical, there will be more events with large positive 
and negative weights that are to cancel. 

Without proper care, this cancellation could easily be destroyed by
the subsequent showering description, as follows. The standard for 
external processes does not provide any way to pass information on 
the clustering algorithm used, so the showering routine will have to
make its own choice what region of phase space to populate with radiation.
One choice could be to allow a cone defined by the nearest 
colour-connected parton (see subsection \ref{sss:showermatching} for a 
discussion). There could then arise a significant mismatch in shower 
description between events where two gluons are just below or just 
above the $\Delta R$ cut for being recombined, equivalently between 
$\g\g \to \g\g$ and $\g\g \to \g\g\g$ events. Most of the phase space 
may be open for the former, while only the region below $\Delta R$ may 
be it for the latter. Thus the average `two-parton' events may end up 
containing significantly more jet activity than the corresponding 
`three-parton' ones. The smaller the $\Delta R$ cut, the more severe 
the mismatch, both on an event-by-event basis and in terms of the 
event rates involved.

One solution would be to extend the standard also to specify which 
clustering algorithm has been used in the matrix-element calculation, 
and with what parameter values. Any shower emission that would give 
rise to an extra jet, according to this algorithm, would be vetoed. 
If a small $\Delta R$ cut is used, this is almost equivalent to allowing 
no shower activity at all. (That would still leave us with potential
mismatch problems in the hadronization description. Fortunately the
string fragmentation approach is very powerful in this respect, 
with a smooth transition between two almost parallel gluons and a 
single one with the full energy \cite{Sjo84}.) But we know that the 
unassisted matrix-element description cannot do a good job of the 
internal structure of jets on an event-by-event basis, since 
multiple-gluon emission is the norm in this region. Therefore a 
$\Delta R \sim 1$ will be required, to let the matrix elements 
describe the wide-angle emission and the showers the small-angle one. 
This again suggests a picture with only a small contribution from 
negative-weight events. In summary, the appearance of a large fraction 
of negative-weight events should be a sure warning sign that physics 
is likely to be incorrectly described. 

The above example illustrates that it may, at times, be desirable to
sidestep the standard and provide further information directly in the
{\Py} common blocks. (Currently there is no exact match to the 
clustering task mentioned above, but there are a few simpler ways to 
intervene in the shower evolution.) Then it is useful to note that,
apart from the hard-process generation machinery itself, the external
processes are handled almost exactly as the internal ones. Thus 
essentially all switches and parameter values related to showers,
underlying events and hadronization can be modified at will. This even
applies to alternative listing modes of events and history pointers, 
as set by \ttt{MSTP(128)}. Also some of the information on the hard
scattering is available, such as \ttt{MSTI(3)}, 
\ttt{MSTI(21) - MSTI(26)}, and \ttt{PARI(33) - PARI(38)}. Before using
them, however, it is prudent to check that your variables of interest 
do work as intended for the particular process you study. Several 
differences do remain between internal and external processes, in 
particular related to final-state showers and resonance decays. For
internal processes, the \ttt{PYRESD} routine will perform a shower
(if relevant) directly after each decay. A typical example would be
that a $\t \to \b\W$ decay is immediately followed by a shower,
which could change the momentum of the $\W$ before it decays in its 
turn. For an external process, this decay chain would presumably
already have been carried out. When the equivalent shower to the 
above is performed, it is therefore now necessary also to boost the 
decay products of the $\W$. The special sequence of showers and boosts
for external processes is administrated by the \ttt{PYADSH} routine.
Should the decay chain not have been carried out, e.g if \ttt{HEPEUP}
event record contains an undecayed $\Z^0$, then \ttt{PYRESD} will
be called to let it decay. The decay products will be visible also
in the documentation section, as for internal processes. 
 
You are free to make use of whatever tools you want in your 
\ttt{UPINIT} and \ttt{UPEVNT} routines, and normally there would be 
little or no contact with the rest of {\Py}, except as described above. 
However, several {\Py} tools can be used, if you so wish. One 
attractive possibility is to use \ttt{PYPDFU} for 
parton-distribution-function evaluation. Other possible tools could 
be \ttt{PYR} for random-number generation, \ttt{PYALPS} for $\alphas$ 
evaluation, \ttt{PYALEM} for evaluation of a running $\alphaem$, and 
maybe a few more.

We end with a few comments on anomalous situations. As already 
described, you may put \ttt{NUP=0} inside \ttt{UPEVNT}, e.g.\  
to signal the end of the file from which events are read.
If the program encounters this value at a return from \ttt{UPEVNT}, 
then it will also exit from \ttt{PYEVNT}, without incrementing the 
counters for the number of events generated. It is then up to you to 
have a check on this condition in your main event-generation loop. 
This you do either by looking at \ttt{NUP} or at \ttt{MSTI(51)}; the 
latter is set to 1 if no event was generated.

It may also happen that a parton-level configuration fails elsewhere
in the \ttt{PYEVNT} call. For instance, the beam-remnant treatment
occasionally encounters situations it cannot handle, wherefore the
parton-level event is rejected and a new one generated. This happens also
with ordinary (not user-defined) events, and usually comes about as a
consequence of the initial-state radiation description leaving too
little energy for the remnant. If the same hard scattering were to
be used as input for a new initial-state radiation and beam-remnant
attempt, it could then work fine. There is a possibility to give events
that chance, as follows. \ttt{MSTI(52)} counts the number of times a
hard-scattering configuration has failed to date. If you come in to
\ttt{UPEVNT} with \ttt{MSTI(52)} non-vanishing, this means that the 
latest configuration failed. So long as the contents of the \ttt{HEPEUP} 
common block are not changed, such an event may be given another try.
For instance, a line 
\begin{verbatim}
      IF(MSTI(52).GE.1.AND.MSTI(52).LE.4) RETURN
\end{verbatim}
at the beginning of \ttt{UPEVNT} will give each event up to five 
tries; thereafter a new one would be generated as usual. Note that the
counter for the number of events is updated at each new try. The
fraction of failed configurations is given in the bottom line of
the \ttt{PYSTAT(1)} table.

The above comment only refers to very rare occurrences (less than 
one in a hundred), which are not errors in a strict sense; for instance, 
they do not produce any error messages on output. If you get
warnings and error messages that the program does not understand the 
flavour codes or cannot reconstruct the colour flows, it is due to 
faults of yours, and giving such events more tries is not going to 
help. 
 
\subsection{Interfaces to Other Generators}
\label{ss:PYinterfacegen}

In the previous section an approach to including external processes
in {\Py} was explained. While general enough, it may not always
be the optimal choice. In particular, for $\ee$ annihilation events 
one may envisage some standard cases where simpler approaches could 
be pursued. A few such standard interfaces are described in this 
section. 

In $\ee$ annihilation events, a convenient classification of electroweak
physics is by the number of fermions in the final state. Two fermions
from $\Z^0$ decay is LEP1 physics, four fermions can come e.g.\ from 
$\W^+\W^-$ or $\Z^0\Z^0$ events at LEP2, and at higher energies six 
fermions are produced by three-gauge-boson production or top-antitop. 
Often interference terms are non-negligible, requiring much more complex 
matrix-element expressions than are normally provided in {\Py}. 
Dedicated electroweak generators often exist, however, and the task is 
therefore to interface them to the generic parton showering and 
hadronization machinery available in {\Py}. In the LEP2 workshop 
\cite{Kno96} one possible strategy was outlined to allow reasonably 
standardized interfaces between the electroweak and the QCD generators. 
The \ttt{LU4FRM} routine was provided for the key four-fermion case. This 
routine is now included here, in slightly modified form, together with 
two new siblings for two and six fermions. The former is trivial and 
included mainly for completeness, while the latter is rather more delicate. 

In final states with two or three quark--antiquark pairs, the colour
connection is not unique. For instance, a $\u\d\ubar\dbar$ final
state could either stem from a $\W^+\W^-$ or a $\Z^0\Z^0$ intermediate
state, or even from interference terms between the two. In order to 
shower and fragment the system, it is then necessary to pick one of the
two alternatives, e.g.\ according to the relative matrix element weight
of each alternative, with the interference term dropped. Some different
such strategies are proposed as options below. 

Note that here we discuss purely perturbative ambiguities. One can 
imagine colour reconnection at later stages of the process, e.g.\ if 
the intermediate state indeed is  $\W^+\W^-$, a soft-gluon exchange 
could still result in colour singlets $\u\ubar$ and $\d\dbar$. We are 
then no longer speaking of ambiguities related to the hard process 
itself but rather to the possibility of nonperturbative effects. This 
is an interesting topic in itself, addressed in section 
\ref{sss:reconnect} but not here. 

The fermion-pair routines are not set up to handle QCD four-jet events, 
i.e.\ events of the types $\q\qbar\g\g$ and $\q\qbar\q'\qbar'$ (with 
$\q'\qbar'$ coming from a gluon branching). Such events are generated in 
normal parton showers, but not necessarily at the right rate (a problem 
that may be especially interesting for massive quarks like $\b$). 
Therefore one would like to start a QCD final-state parton shower from 
a given four-parton configuration. Already some time ago, a machinery 
was developed to handle this kind of occurrences \cite{And98a}.
This approach has now been adapted to {\Py}, in a somewhat modified 
form, see section \ref{sss:fourjetmatch}. The main change is that, in the 
original work, the colour flow was 
picked in a separate first step (not discussed in the publication, since 
it is part of the standard 4-parton configuration machinery of \ttt{PYEEVT}), 
which reduces the number of allowed $\q\qbar\g\g$ parton-shower histories. 
In the current implementation, more geared towards completely external 
generators, no colour flow assumptions are made, meaning a few more 
possible shower histories to pick between. Another change is that mass 
effects are better respected by the $z$ definition. The code contains one 
new user routine, \ttt{PY4JET}, two new auxiliary ones, \ttt{PY4JTW} and 
\ttt{PY4JTS}, and significant additions to the \ttt{PYSHOW} showering
routine.

\drawbox{CALL PY2FRM(IRAD,ITAU,ICOM)}\label{p:PY2FRM}
\begin{entry}
\itemc{Purpose:} to allow a parton shower to develop and partons to 
hadronize from a two-fermion starting point. The initial list is 
supposed to be ordered such that the fermion precedes the antifermion. 
In addition, an arbitrary number of photons may be included, e.g.\ from
initial-state radiation; these will not be affected by the operation
and can be put anywhere. The scale for QCD (and QED) final-state radiation 
is automatically set to be the mass of the fermion-antifermion pair. 
(It is thus not suited for Bhabha scattering.)
\iteme{IRAD :} final-state QED radiation.
\begin{subentry}
\iteme{= 0 :} no final-state photon radiation, only QCD showers.
\iteme{= 1 :} photon radiation inside each final fermion pair, also leptons,
in addition to the QCD one for quarks.  
\end{subentry}
\iteme{ITAU :} handling of $\tau$ lepton decay (where {\Py} does not 
include spin effects, although some generators provide the helicity
information that would allow a more sophisticated modelling).
\begin{subentry}
\iteme{= 0 :} $\tau$'s are considered stable (and can therefore be decayed
afterwards).
\iteme{= 1 :} $\tau$'s are allowed to decay.
\end{subentry}
\iteme{ICOM :} place where information about the event (flavours, 
momenta etc.) is stored at input and output.
\begin{subentry}
\iteme{= 0 :} in the \ttt{HEPEVT} common block (meaning that 
information is automatically translated to \ttt{PYJETS} before 
treatment and back afterwards).
\iteme{= 1 :} in the \ttt{PYJETS} common block. All fermions and photons 
can  be given with status code \ttt{K(I,1)=1}, flavour code in 
\ttt{K(I,2)} and five-momentum (momentum, energy, mass) in \ttt{P(I,J)}. 
The \ttt{V} vector and remaining components in the \ttt{K} one are best 
put to zero. Also remember to set the total number of entries \ttt{N}.
\end{subentry}
\end{entry}

\drawbox{CALL PY4FRM(ATOTSQ,A1SQ,A2SQ,ISTRAT,IRAD,ITAU,ICOM)}%
\label{p:PY4FRM}
\begin{entry}
\itemc{Purpose:} to allow a parton shower to develop and partons to 
hadronize from a four-fermion starting point. The initial list of 
fermions is supposed to be ordered in the sequence fermion (1) -- 
antifermion (2) -- fermion (3) -- antifermion (4). The flavour pairs 
should be arranged so that, if possible, the first two could come 
from a $\W^+$ and the second two from a $\W^-$; else each pair should 
have flavours consistent with a $\Z^0$. In addition, an arbitrary number 
of photons may be included, e.g.\ from initial-state radiation; these 
will not be affected by the operation and can be put anywhere. 
Since the colour flow need not be unique, three real and one 
integer numbers are providing further input. Once the colour pairing 
is determined, the scale for final-state QCD (and QED) radiation is 
automatically set to be the mass of the fermion-antifermion pair. 
(This is the relevant choice for normal fermion pair production 
from resonance decay, but is not suited e.g.\ for $\gamma\gamma$ processes 
dominated by small-$t$ propagators.) The pairing is also meaningful 
for QED radiation, in the sense that a four-lepton final state is 
subdivided into two radiating subsystems in the same way. Only if 
the event consists of one lepton pair and one quark pair is the 
information superfluous.
\iteme{ATOTSQ :} total squared amplitude for the event, irrespective of
colour flow.
\iteme{A1SQ :} squared amplitude for the configuration with fermions 
$1 + 2$ and $3 + 4$ as the two colour singlets.
\iteme{A2SQ :} squared amplitude for the configuration with fermions 
$1 + 4$ and $3 + 2$ as the two colour singlets.
\iteme{ISTRAT :} the choice of strategy to select either of the two 
possible colour configurations. Here 0 is supposed to represent a 
reasonable compromise, while 1 and 2 are selected so as to give the 
largest reasonable spread one could imagine.
\begin{subentry}
\iteme{= 0 :} pick configurations according to relative probabilities 
\ttt{A1SQ} : \ttt{A2SQ}.
\iteme{= 1 :} assign the interference contribution to maximize the 
$1 + 2$ and $3 + 4$ pairing of fermions.
\iteme{= 2 :} assign the interference contribution to maximize the 
$1 + 4$ and $3 + 2$ pairing of fermions.
\end{subentry}
\iteme{IRAD :} final-state QED radiation.
\begin{subentry}
\iteme{= 0 :} no final-state photon radiation, only QCD showers.
\iteme{= 1 :} photon radiation inside each final fermion pair, also leptons,
in addition to the QCD one for quarks.  
\end{subentry}
\iteme{ITAU :} handling of $\tau$ lepton decay (where {\Py} does not 
include spin effects, although some generators provide the helicity
information that would allow a more sophisticated modelling).
\begin{subentry}
\iteme{= 0 :} $\tau$'s are considered stable (and can therefore be decayed
afterwards).
\iteme{= 1 :} $\tau$'s are allowed to decay.
\end{subentry}
\iteme{ICOM :} place where information about the event (flavours, 
momenta etc.) is stored at input and output.
\begin{subentry}
\iteme{= 0 :} in the \ttt{HEPEVT} common block (meaning that 
information is automatically translated to \ttt{PYJETS} before 
treatment and back afterwards).
\iteme{= 1 :} in the \ttt{PYJETS} common block. All fermions and photons 
can  be given with status code \ttt{K(I,1)=1}, flavour code in 
\ttt{K(I,2)} and five-momentum (momentum, energy, mass) in \ttt{P(I,J)}. 
The \ttt{V} vector and remaining components in the \ttt{K} one are best 
put to zero. Also remember to set the total number of entries \ttt{N}.
\end{subentry}
\itemc{Comment :} Also colour reconnection phenomena can be studied with the 
\ttt{PY4FRM} routine. \ttt{MSTP(115)} can be used to switch between the 
scenarios, with default being no reconnection. Other reconnection 
parameters also work as normally, including that \ttt{MSTI(32)} can be 
used to find out whether a reconnection occured or not. In order for the 
reconnection machinery to work, the event record is automatically 
complemented with information on the $\W^+ \W^-$ or $\Z^0 \Z^0$
pair that produced the four fermions, based on the rules described
above.\\ 
We remind that the four first parameters of the \ttt{PY4FRM} call are 
supposed to parameterize an ambiguity on the perturbative level of the
process, which has to be resolved before parton showers are performed.
The colour reconnection discussed here is (in most scenarios) occuring
on the nonperturbative level, after the parton showers. 
\end{entry}

\drawbox{CALL PY6FRM(P12,P13,P21,P23,P31,P32,PTOP,IRAD,ITAU,ICOM)}%
\label{p:PY6FRM}
\begin{entry}
\itemc{Purpose:} to allow a parton shower to develop and partons to hadronize
from a six-fermion starting point. The initial list of fermions is 
supposed to be ordered in the sequence fermion (1) -- antifermion (2) -- 
fermion (3) -- antifermion (4) -- fermion (5) -- antifermion (6). The 
flavour pairs should be arranged so that, if possible, the first two 
could come from a $\Z^0$, the middle two from a $\W^+$ and the last two 
from a $\W^-$; else each pair should have flavours consistent with a $\Z^0$.
Specifically, this means that in a $\t\tbar$ event, the $\t$ decay products
would be found in 1 ($\b$) and 3 and 4 (from the $\W^+$ decay) and the 
$\tbar$ ones in 2 ($\bbar$) and 5 and 6 (from the $\W^-$ decay). In 
addition, an 
arbitrary number of photons may be included, e.g.\ from initial-state 
radiation; these will not be affected by the operation and can be put 
anywhere. Since the colour flow need not be unique, further input is
needed to specify this. The number of possible interference 
contributions being much larger than for the four-fermion case, we 
have not tried to implement different strategies. Instead six 
probabilities may be input for the different pairings, that you
e.g.\ could pick as the six possible squared amplitudes, or according 
to some more complicated scheme for how to handle the interference 
terms. The treatment of final-state cascades must be quite different for 
top events and the rest. For a normal three-boson event, each fermion 
pair would form one radiating system, with scale set equal to the 
fermion-antifermion invariant mass. (This is the relevant choice for 
normal fermion pair production from resonance decay, but is not 
suited e.g.\ for $\gamma\gamma$ processes dominated by small-$t$ 
propagators.) 
In the top case, on the other hand, the $\b$ ($\bbar$) would be radiating 
with a recoil taken by the $\W^+$ ($\W^-$) in such a way that the $\t$ 
($\tbar$) mass is preserved, while the $\W$ dipoles would radiate as normal. 
Therefore you need also supply a probability for the event to 
be a top one, again e.g.\ based on some squared amplitude.  
\iteme{P12, P13, P21, P23, P31, P32 :} relative probabilities for the 
six possible pairings of fermions with antifermions. The first (second) 
digit tells which antifermion the first (second) fermion is paired with, 
with the third pairing given by elimination. Thus e.g.\ \ttt{P23} means the 
first fermion is paired with the second antifermion, the second fermion 
with the third antifermion and the third fermion with the first 
antifermion. Pairings are only possible between quarks and leptons 
separately. The sum of probabilities for allowed pairings is 
automatically normalized to unity. 
\iteme{PTOP :} the probability that the configuration is a top one; a 
number between 0 and 1. In this case, it is important that the order 
described above is respected, with the $\b$ and $\bbar$ coming first. 
No colour ambiguity exists if the top interpretation is selected, 
so then the \ttt{P12 - P32} numbers are not used. 
\iteme{IRAD :} final-state QED radiation.
\begin{subentry}
\iteme{= 0 :} no final-state photon radiation, only QCD showers.
\iteme{= 1 :} photon radiation inside each final fermion pair, also leptons,
in addition to the QCD one for quarks.  
\end{subentry}
\iteme{ITAU :} handling of $\tau$ lepton decay (where {\Py} does not 
include spin effects, although some generators provide the helicity
information that would allow a more sophisticated modelling).
\begin{subentry}
\iteme{= 0 :} $\tau$'s are considered stable (and can therefore be decayed
afterwards).
\iteme{= 1 :} $\tau$'s are allowed to decay.
\end{subentry}
\iteme{ICOM :} place where information about the event (flavours, 
momenta etc.) is stored at input and output.
\begin{subentry}
\iteme{= 0 :} in the \ttt{HEPEVT} common block (meaning that 
information is automatically translated to \ttt{PYJETS} before 
treatment and back afterwards).
\iteme{= 1 :} in the \ttt{PYJETS} common block. All fermions and photons 
can  be given with status code \ttt{K(I,1)=1}, flavour code in 
\ttt{K(I,2)} and five-momentum (momentum, energy, mass) in \ttt{P(I,J)}. 
The \ttt{V} vector and remaining components in the \ttt{K} one are best 
put to zero. Also remember to set the total number of entries \ttt{N}.
\end{subentry}
\end{entry}

\drawbox{CALL PY4JET(PMAX,IRAD,ICOM)}\label{p:PY4JET}
\begin{entry}
\itemc{Purpose:} to allow a parton shower to develop and partons to 
hadronize from a $\q\qbar\g\g$ or $\q\qbar\q'\qbar'$ original 
configuration. The partons should be ordered exactly as indicated above, 
with the primary $\q\qbar$ pair first and thereafter the two gluons or 
the secondary $\q'\qbar'$ pair. (Strictly speaking, the definition of 
primary and secondary fermion pair is ambiguous. In practice,
however, differences in topological variables like the pair mass
should make it feasible to have some sensible criterion on an event 
by event basis.) Within each pair, fermion should precede antifermion. 
In addition, an arbitrary number of photons may be included, e.g.\ from
initial-state radiation; these will not be affected by the operation
and can be put anywhere. The program will select a possible 
parton shower history from the given parton configuration, and then 
continue the shower from there on. The history selected is displayed
in lines \ttt{NOLD+1} to \ttt{NOLD+6}, where \ttt{NOLD} is the \ttt{N} 
value before the routine is called. Here the masses and energies of 
intermediate partons are clearly displayed. The lines \ttt{NOLD+7} and 
\ttt{NOLD+8} contain the equivalent on-mass-shell parton pair from which 
the shower is started.  
\iteme{PMAX :} the maximum mass scale (in GeV) from which the shower is 
started in those branches that are not already fixed by the matrix-element 
history. If \ttt{PMAX} is set zero (actually below \ttt{PARJ(82)}, the 
shower cutoff scale), the shower starting scale is instead set to be equal 
to the smallest mass of the virtual partons in the reconstructed
shower history. A fixed \ttt{PMAX} can thus be used to obtain a reasonably
exclusive set of four-jet events (to that \ttt{PMAX} scale), with little 
five-jet contamination, while the \ttt{PMAX=0} option gives a more
inclusive interpretation, with five- or more-jet events possible.
Note that the shower is based on evolution in mass, meaning the cut
is really one of mass, not of $\pT$, and that it may therefore be
advantageous to set up the matrix elements cuts accordingly if one
wishes to mix different event classes. This is not a requirement, 
however.
\iteme{IRAD :} final-state QED radiation.
\begin{subentry}
\iteme{= 0 :} no final-state photon radiation, only QCD showers.
\iteme{= 1 :} photon radiation inside each final fermion pair, also leptons,
in addition to the QCD one for quarks.  
\end{subentry}
\iteme{ICOM :} place where information about the event (flavours, 
momenta etc.) is stored at input and output.
\begin{subentry}
\iteme{= 0 :} in the \ttt{HEPEVT} common block (meaning that 
information is automatically translated to \ttt{PYJETS} before 
treatment and back afterwards).
\iteme{= 1 :} in the \ttt{PYJETS} common block. All fermions and photons 
can  be given with status code \ttt{K(I,1)=1}, flavour code in 
\ttt{K(I,2)} and five-momentum (momentum, energy, mass) in \ttt{P(I,J)}. 
The \ttt{V} vector and remaining components in the \ttt{K} one are best 
put to zero. Also remember to set the total number of entries \ttt{N}.
\end{subentry}
\end{entry}
  
\subsection{Other Routines and Common Blocks}
 
The subroutines and common blocks that you will come in direct
contact with have already been described. A number of other routines
and common blocks exist, and those not described elsewhere are here 
briefly listed for the sake of completeness. The \ttt{PYG***} 
routines are slightly modified versions of the \ttt{SAS***} ones
of the \tsc{SaSgam} library. The common block \ttt{SASCOM} is 
renamed \ttt{PYINT8}. If you want to use the parton distributions 
for standalone purposes, you are encouraged to use the original 
\tsc{SaSgam} routines rather than going the way via the
{\Py} adaptations. 
 
\begin{entry}
 
\iteme{SUBROUTINE PYINRE :}\label{p:PYINRE}
to initialize the widths and effective widths of resonances.

\iteme{SUBROUTINE PYINBM(CHFRAM,CHBEAM,CHTARG,WIN) :}\label{p:PYINBM}
to read in and identify the beam (\ttt{CHBEAM}) and target 
(\ttt{CHTARG}) particles and the frame (\ttt{CHFRAM}) as given in 
the \ttt{PYINIT} call; also to save the original energy (\ttt{WIN}).

\iteme{SUBROUTINE PYINKI(MODKI) :}\label{p:PYINKI}
to set up the event kinematics, either at initialization 
(\ttt{MODKI=0}) or for each separate event, the latter when the 
program is run with varying kinematics (\ttt{MODKI=1}). 

\iteme{SUBROUTINE PYINPR :}\label{p:PYINPR} 
to set up the partonic subprocesses selected with \ttt{MSEL}. 
For $\gamma\p$ and $\gamma\gamma$, also the \ttt{MSTP(14)} value
affects the choice of processes. In particular, options such as 
\ttt{MSTP(14)=10} and \ttt{=30} sets up the several different kinds 
of processes that need to be mixed, with separate cuts for each.

\iteme{SUBROUTINE PYXTOT :}\label{p:PYXTOT}
to give the parameterized total, double diffractive, single
diffractive and elastic cross sections for different energies and
colliding hadrons or photons.
 
\iteme{SUBROUTINE PYMAXI :}\label{p:PYMAXI}
to find optimal coefficients \ttt{COEF} for the selection of
kinematical variables, and to find the related maxima for
the differential cross section times Jacobian factors,
for each of the subprocesses included.
 
\iteme{SUBROUTINE PYPILE(MPILE) :}\label{p:PYPILE}
to determine the number of pile-up events, i.e.\ events
appearing in the same beam--beam crossing.

\iteme{SUBROUTINE PYSAVE(ISAVE,IGA) :}\label{p:PYSAVE}  
saves and restores parameters and cross section values between the 
several $\gamma\p$ and $\gamma\gamma$ components of mixing options
such as \ttt{MSTP(14)=10} and \ttt{=30}. The options for \ttt{ISAVE} 
are (1) a complete save of all parameters specific to a given component,
(2) a partial save of cross-section information, (3) a restoration
of all parameters specific to a given component, (4) as 3 but
preceded by a random selection of component, and (5) a summation of
component cross sections (for \ttt{PYSTAT}). The subprocess code
in \ttt{IGA} is the one described for \ttt{MSTI(9)}; it is input
for options 1, 2 and 3 above, output for 4 and dummy for 5.

\iteme{SUBROUTINE PYGAGA(IGA) :}\label{p:PYGAGA} 
to generate photons according to the virtual photon flux around 
a lepton beam, for the {\galep} option in \ttt{PYINIT}.
\begin{subentry}
\iteme{IGA = 1 :} call at initialization to set up $x$ and $Q^2$ 
limits etc.
\iteme{IGA = 2 :} call at maximization step to give estimate of 
maximal photon flux factor.
\iteme{IGA = 3 :} call at the beginning of the event generation to 
select the kinematics of the photon emission and to give the
flux factor. 
\iteme{IGA = 4 :} call at the end of the event generation to set 
up the full kinematics of the photon emission.
\end{subentry}
  
\iteme{SUBROUTINE PYRAND :}\label{p:PYRAND}
to generate the quantities characterizing a hard scattering
on the parton level, according to the relevant matrix elements.
 
\iteme{SUBROUTINE PYSCAT :}\label{p:PYSCAT}
to find outgoing flavours and to set up the kinematics and
colour flow of the hard scattering.
 
\iteme{SUBROUTINE PYRESD(IRES) :}\label{p:PYRESD}
to allow resonances to decay, including chains of successive decays 
and parton showers. Normally only two-body decays of each resonance,
but a few three-body decays are also implemented.
\begin{subentry}
\iteme{IRES :} The standard call from \ttt{PYEVNT}, for the hard process, 
has \ttt{IRES=0}, and then finds resonances to be treated based on the 
subprocess number {\ISUB}. In case of a nonzero \ttt{IRES} only the 
resonance in position \ttt{IRES} of the event record is considered. This 
is used by \ttt{PYEVNT} and \ttt{PYEXEC} to decay leftover resonances. 
(Example: a $\b \to \W + \t$ branching may give a $\t$ quark as beam 
remnant.)
\end{subentry}
  
\iteme{SUBROUTINE PYMULT(MMUL) :}\label{p:PYMULT}
to generate semi-hard interactions according to the
old multiple interaction formalism.
 
\iteme{SUBROUTINE PYREMN(IPU1,IPU2) :}\label{p:PYREMN}
to add on target remnants and include primordial $k_{\perp}$ according to the
old beam remnant treatment.
 
\iteme{SUBROUTINE PYMIGN(MMUL) :}\label{p:PYMIGN}
to generate multiple interaction kinematics and flavours
according to the new multiple interactions scheme (equivalently to
\ttt{PYMULT} in the old scheme). 

\iteme{SUBROUTINE PYMIHK :}\label{p:PYMIHK}
finds left-behind remnant flavour content and hooks up
the colour flow between the hard scattering and remnants
(part of \ttt{PYREMN} in the old scheme).

\iteme{SUBROUTINE PYCTTR(I,KCS,IEND) :}\label{p:PYCTTR}
auxiliary to \ttt{PYMIHK}. Traces the colour flow in the {\Py} event
      record and translates it into a structure (\ttt{/PYCTAG/}) based on Les
      Houches Accord style colour tags. 

\iteme{SUBROUTINE PYMIHG(JCP1,JCG1,JCP2,JCG2) :}\label{p:PYMIHG}
auxiliary routine to \ttt{PYMIHK} used to test initial state colour
      connections. Keeps track of which colour tags have been collapsed
      with each other and tests whether a given new collapse will result
      in singlet gluons. Uses \ttt{/PYCBLS/} to communicate with
      \ttt{PYMIHK}. 

\iteme{SUBROUTINE PYMIRM :}\label{p:PYMIRM}
picks primordial $\kT$ and shares longitudinal momentum among
      beam remnants (part of \ttt{PYREMN} in the old scheme).
 
\iteme{SUBROUTINE PYDIFF :}\label{p:PYDIFF}
to handle diffractive and elastic scattering events.
 
\iteme{SUBROUTINE PYDISG :}\label{p:PYDISG}
to set up kinematics, beam remnants and showers in the $2 \to 1$ 
DIS process $\gamma^* \f \to \f$. Currently initial-state radiation 
is not yet implemented, while final-state is.
 
\iteme{SUBROUTINE PYDOCU :}\label{p:PYDOCU}
to compute cross sections of processes, based on current
Monte Carlo statistics, and to store event information in the
\ttt{MSTI} and \ttt{PARI} arrays.
 
\iteme{SUBROUTINE PYWIDT(KFLR,SH,WDTP,WDTE) :}\label{p:PYWIDT}
to calculate widths and effective widths of resonances. 
Everything is given in dimensions of GeV.
 
\iteme{SUBROUTINE PYOFSH(MOFSH,KFMO,KFD1,KFD2,PMMO,RET1,RET2) :}%
\label{p:PYOFSH}
to calculate partial widths into channels off the mass shell,
and to select correlated masses of resonance pairs.
 
\iteme{SUBROUTINE PYKLIM(ILIM) :}\label{p:PYKLIM}
to calculate allowed kinematical limits.
 
\iteme{SUBROUTINE PYKMAP(IVAR,MVAR,VVAR) :}\label{p:PYKMAP}
to calculate the value of a kinematical variable when this
is selected according to one of the simple pieces.
 
\iteme{SUBROUTINE PYSIGH(NCHN,SIGS) :}\label{p:PYSIGH}
to give the differential cross section (multiplied by the
relevant Jacobians) for a given subprocess and kinematical
setup. With time, the \ttt{PYSIGH} increased to a size of 
over 7000 lines, which gave some compilers problems. Therefore,
currently, all the phase-space and parton-density generic weights 
remain in \ttt{PYSIGH} itself, whereas the process-specific
matrix elements have been grouped according to: 
\begin{Itemize}
\item PYSGQC: normal QCD processes, plus some similar instead 
with photons;
\item PYSGHF: heavy flavour production, open and closed;
\item PYSGWZ: $\W/\Z$ processes, except as below;
\item PYSGHG: Higgs processes (2 doublets), except Higgs pairs in 
\ttt{PYSGSU}, but including longitudinal $\W\W$ scattering for a 
heavy Higgs;
\item PYSGSU: SUSY processes, including Higgs pair production;
\item PYSGTC: Technicolor processes, including some related 
compositeness ones; 
\item PYSGEX: assorted exotic processes, including new gauge bosons
($\Z'/\W'$), leptoquarks ($\L_{\Q}$), horizontal gauge bosons ($\R^0$), 
a generation of excited fermions ($\d^*/\u^*/\e^{*-}/\nu^*_{\e}$), 
left--right-symmetric scenarios ($\H^{++}/\Z_R/\W_R$), and extra 
dimensions ($\G^*$).
\end{Itemize}

\iteme{SUBROUTINE PYPDFL(KF,X,Q2,XPQ) :}\label{p:PYPDFL}
to give parton distributions for $\p$ and $\n$ in the option
with modified behaviour at small $Q^2$ and $x$, see
\ttt{MSTP(57)}.
 
\iteme{SUBROUTINE PYPDFU(KF,X,Q2,XPQ) :}\label{p:PYPDFU}
to give parton-distribution functions (multiplied by $x$, i.e.\ 
$x f_i(x,Q^2)$) for an arbitrary particle (of those recognized by 
{\Py}). Generic driver routine for the following, specialized ones.
\begin{subentry}
\iteme{KF :} flavour of probed particle, according to {\KF} code.
\iteme{X :} $x$ value at which to evaluate parton distributions.
\iteme{Q2 :} $Q^2$ scale at which to evaluate parton distributions.
\iteme{XPQ :} array of dimensions \ttt{XPQ(-25:25)}, which contains
the evaluated parton distributions $x f_i(x,Q^2)$. Components $i$
ordered according to standard {\KF} code; additionally the gluon is
found in position 0 as well as 21 (for historical reasons).
\itemc{Note:} the above set of calling arguments is enough for
a real photon, but has to be complemented for a virtual one.
This is done by \ttt{VINT(120)}.
\end{subentry}
 
\iteme{SUBROUTINE PYPDEL(KFA,X,Q2,XPEL) :}\label{p:PYPDEL}
to give $\e / \mu / \tau$ parton distributions.

\iteme{SUBROUTINE PYPDGA(X,Q2,XPGA) :}\label{p:PYPDGA}
to give the photon parton distributions for sets other than the SaS 
ones.

\iteme{SUBROUTINE PYGGAM(ISET,X,Q2,P2,IP2,F2GM,XPDFGM) :}\label{p:PYGGAM}
to construct the SaS $F_2$ and parton distributions of the photon
by summing homogeneous (VMD) and inhomogeneous (anomalous) terms.
For $F_2$, $\c$ and $\b$ are included by the Bethe-Heitler formula;
in the `{\MSbar}' scheme additionally a $C^{\gamma}$ term 
is added. \ttt{IP2} sets treatment of virtual photons, with same code
as \ttt{MSTP(60)}. Calls \ttt{PYGVMD}, \ttt{PYGANO}, \ttt{PYGBEH},
and \ttt{PYGDIR}.
 
\iteme{SUBROUTINE PYGVMD(ISET,KF,X,Q2,P2,ALAM,XPGA,VXPGA) :}\label{p:PYGVMD}
~~to evaluate the parton distributions of a VMD photon,
evolved homogeneously from an initial scale $P^2$ to $Q^2$.
 
\iteme{SUBROUTINE PYGANO(KF,X,Q2,P2,ALAM,XPGA,VXPGA) :}\label{p:PYGANO}
to evaluate the parton distributions of the anomalous
photon, inhomogeneously evolved from a scale $P^2$ (where it vanishes)
to $Q^2$.
 
\iteme{SUBROUTINE PYGBEH(KF,X,Q2,P2,PM2,XPBH) :}\label{p:PYGBEH}
to evaluate the Bethe-Heitler cross section for
heavy flavour production.
 
\iteme{SUBROUTINE PYGDIR(X,Q2,P2,AK0,XPGA) :}\label{p:PYGDIR}
to evaluate the direct contribution, i.e.\ the $C^{\gamma}$ term,
as needed in `{\MSbar}' parameterizations.
 
\iteme{SUBROUTINE PYPDPI(X,Q2,XPPI) :}\label{p:PYPDPI}
to give pion parton distributions.
 
\iteme{SUBROUTINE PYPDPR(X,Q2,XPPR) :}\label{p:PYPDPR}
to give proton parton distributions. Calls several auxiliary 
routines: \ttt{PYCTEQ}, \ttt{PYGRVL}, \ttt{PYGRVM}, \ttt{PYGRVD},
\ttt{PYGRVV}, \ttt{PYGRVW}, \ttt{PYGRVS}, \ttt{PYCT5L}, 
\ttt{PYCT5M} and \ttt{PYPDPO}.
 
\iteme{FUNCTION PYHFTH(SH,SQM,FRATT) :}\label{p:PYHFTH}
to give heavy-flavour threshold factor in matrix elements.
 
\iteme{SUBROUTINE PYSPLI(KF,KFLIN,KFLCH,KFLSP) :}\label{p:PYSPLI}
to give hadron remnant or remnants left behind when the reacting
parton is kicked out.
 
\iteme{FUNCTION PYGAMM(X) :}\label{p:PYGAMM}
to give the value of the ordinary $\Gamma (x)$ function
(used in some parton-distribution parameterizations).
 
\iteme{SUBROUTINE PYWAUX(IAUX,EPS,WRE,WIM) :}\label{p:PYWAUX}
to evaluate the two auxiliary functions $W_1$ and $W_2$ appearing
in some cross section expressions in \ttt{PYSIGH}.
 
\iteme{SUBROUTINE PYI3AU(EPS,RAT,Y3RE,Y3IM) :}\label{p:PYI3AU}
to evaluate the auxiliary function $I_3$ appearing
in some cross section expressions in \ttt{PYSIGH}.
 
\iteme{FUNCTION PYSPEN(XREIN,XIMIN,IREIM) :}\label{p:PYSPEN}
to calculate the real and imaginary part of the Spence
function \cite{Hoo79}.
 
\iteme{SUBROUTINE PYQQBH(WTQQBH) :}\label{p:PYQQBH}
to calculate matrix elements for the two processes 
$\g \g \to \Q \Qbar \hrm^0$ and $\q \qbar \to \Q \Qbar \hrm^0$.
 
\iteme{SUBROUTINE PYRECO(IW1,IW2,NSD1,NAFT1)) :}\label{p:PYRECO}
to perform nonperturbative reconnection among strings in 
$\W^+\W^-$ and $\Z^0\Z^0$ events. The physics of this routine 
is described as part of the fragmentation story, section 
\ref{sss:reconnect}, but for technical reasons the code is called
directly in the event generation sequence.
 
\iteme{BLOCK DATA PYDATA :}\label{p:PYDATA}
to give sensible default values to all status codes and
parameters.
 
\end{entry}
 
\drawbox{COMMON/PYINT1/MINT(400),VINT(400)}\label{p:PYINT1}
\begin{entry}
 
\itemc{Purpose:} to collect a host of integer- and real-valued
variables used internally in the program during the initialization
and/or event generation stage. These variables must not be changed
by you.
 
\iteme{MINT(1) :}\label{p:MINT} specifies the general type of 
subprocess that has occurred, according to the {\ISUB} code given in 
section \ref{ss:ISUBcode}.

\iteme{MINT(2) :} whenever \ttt{MINT(1)} (together with \ttt{MINT(15)}
and \ttt{MINT(16)}) are not sufficient to specify the type of process
uniquely, \ttt{MINT(2)} provides an ordering of the different
possibilities, see \ttt{MSTI(2)}. Internally and temporarily, in 
process 53 \ttt{MINT(2)} is increased by 2 or 4 for $\c$ or $\b$, 
respectively. 
 
\iteme{MINT(3) :} number of partons produced in the hard
interactions, i.e.\ the number $n$ of the $2 \to n$ matrix elements
used; is sometimes 3 or 4 when a basic $2 \to 1$ or $2 \to 2$ process
has been convoluted with two $1 \to 2$ initial branchings
(like $\q \q' \to \q'' \q''' \hrm^0$).
 
\iteme{MINT(4) :} number of documentation lines at the beginning of
the common block \ttt{PYJETS} that are given with \ttt{K(I,1)=21};
0 for \ttt{MSTP(125)=0}.
 
\iteme{MINT(5) :} number of events generated to date in current run.
In runs with the variable-energy option, \ttt{MSTP(171)=1}
and \ttt{MSTP(172)=2}, only those events that survive (i.e.\ that
do not have \ttt{MSTI(61)=1}) are counted in this number. That
is, \ttt{MINT(5)} may be less than the total number of \ttt{PYEVNT}
calls.
 
\iteme{MINT(6) :} current frame of event (see \ttt{MSTP(124)} for
possible values).
 
\iteme{MINT(7), MINT(8) :} line number for documentation of outgoing
partons/particles from hard scattering for $2 \to 2$ or
$2 \to 1 \to 2$ processes (else = 0).
\iteme{MINT(10) :} is 1 if cross section maximum was violated in
current event, and 0 if not.
 
\iteme{MINT(11) :} {\KF} flavour code for beam (side 1) particle.
 
\iteme{MINT(12) :} {\KF} flavour code for target (side 2) particle.
 
\iteme{MINT(13), MINT(14) :} {\KF} flavour codes for side 1 and side 2
initial-state shower initiators.
 
\iteme{MINT(15), MINT(16) :} {\KF} flavour codes for side 1 and side 2
incoming partons to the hard interaction. (For use in \ttt{PYWIDT}
calls, occasionally \ttt{MINT(15)=1} signals the presence of an as
yet unspecified quark, but the original value is then restored 
afterwards.) 
 
\iteme{MINT(17), MINT(18) :} flag to signal if particle on side 1 or
side 2 has been scattered diffractively; 0 if no, 1 if yes.
 
\iteme{MINT(19), MINT(20) :} flag to signal initial-state structure
with parton inside photon inside electron on side 1 or side 2;
0 if no, 1 if yes.
 
\iteme{MINT(21) - MINT(24) :} {\KF} flavour codes for outgoing partons
from the hard interaction. The number of positions actually used is
process-dependent, see \ttt{MINT(3)}; trailing positions not used are
set = 0. For events with many outgoing partons, e.g.\ in external
processes, also \ttt{MINT(25)} and \ttt{MINT(26)} could be used.
 
\iteme{MINT(25), MINT(26) :} {\KF} flavour codes of the products in the
decay of a single $s$-channel resonance formed in the hard interaction.
Are thus only used when \ttt{MINT(3)=1} and the resonance is allowed
to decay.
 
\iteme{MINT(30) :}  normally 0, but when 1 or 2 it is a signal for PYPDFU
        that PDF's are to be evaluated taking into account the partons
        already removed from the beam remnant on side 1 or 2 of the event.
 
\iteme{MINT(31) :} number of hard or semi-hard scatterings that occurred
in the current event in the multiple-interaction scenario; is = 0 for a
low-$\pT$ event.
 
\iteme{MINT(32) :} information on whether a nonperturbative colour 
reconnection occurred in the current event; is 0 normally but 1 in case 
of reconnection. 

\iteme{MINT(33) :} Switch to select how to trace colour connections in
\ttt{PYPREP}. 
\begin{subentry}
\iteme{= 0 :} Old method, using \ttt{K(I,4)} and \ttt{K(I,5)} in
\ttt{/PYJETS/}. 
\iteme{= 1 :} New method, using Les Houches Accord style colour tags 
in \ttt{/PYCTAG/}.
\end{subentry}

\iteme{MINT(34) :} counter for the number of final state colour reconnections
       in the new multiple interactions scenario.

\iteme{MINT(41), MINT(42) :} type of incoming beam or target particle;
1 for lepton and 2 for hadron. A photon counts as a lepton if it
is not resolved (direct or DIS) and as a hadron if it is resolved
(VMD or GVMD).
 
\iteme{MINT(43) :} combination of incoming beam and target particles.
A photon counts as a hadron.
\begin{subentry}
\iteme{= 1 :} lepton on lepton.
\iteme{= 2 :} lepton on hadron.
\iteme{= 3 :} hadron on lepton.
\iteme{= 4 :} hadron on hadron.
\end{subentry}
 
\iteme{MINT(44) :} as \ttt{MINT(43)}, but a photon counts as a lepton.
 
\iteme{MINT(45), MINT(46) :} structure of incoming beam and target
particles.
\begin{subentry}
\iteme{= 1 :} no internal structure, i.e.\ a lepton or photon
carrying the full beam energy.
\iteme{= 2 :} defined with parton distributions that are not peaked
at $x = 1$, i.e.\ a hadron or a resolved (VMD or GVMD) photon.
\iteme{= 3 :} defined with parton distributions that are peaked at
$x = 1$, i.e.\ a resolved lepton.
\end{subentry}
 
\iteme{MINT(47) :} combination of incoming beam- and target-particle
parton-distribution function types.
\begin{subentry}
\iteme{= 1 :} no parton distribution either for beam or target.
\iteme{= 2 :} parton distributions for target but not for beam.
\iteme{= 3 :} parton distributions for beam but not for target.
\iteme{= 4 :} parton distributions for both beam and target, but not
both peaked at $x = 1$.
\iteme{= 5 :} parton distributions for both beam and target, with both
peaked at $x = 1$.
\iteme{= 6 :} parton distribution is peaked at $x=1$ for target and no 
distribution at all for beam.
\iteme{= 7 :} parton distribution is peaked at $x=1$ for beam and no 
distribution at all for target.
 \end{subentry}
 
\iteme{MINT(48) :} total number of subprocesses switched on.
 
\iteme{MINT(49) :} number of subprocesses that are switched on, apart
from elastic scattering and single, double and central diffractive.

\iteme{MINT(50) :} combination of incoming particles from a multiple
interactions point of view.
\begin{subentry}
\iteme{= 0 :} the total cross section is not known; therefore no 
multiple interactions are possible.
\iteme{= 1 :} the total cross section is known; therefore multiple
interactions are possible if switched on. Requires beams of hadrons,
VMD photons or GVMD photons.
\end{subentry}
 
\iteme{MINT(51) :} internal flag that event failed cuts.
\begin{subentry}
\iteme{= 0 :} no problem.
\iteme{= 1 :} event failed; new one to be generated.
\iteme{= 2 :} event failed; no new event is to be generated but
instead control is to be given back to the user. Is intended for
user-defined processes, when \ttt{NUP=0}.
\end{subentry}
 
\iteme{MINT(52) :} internal counter for number of lines used
(in \ttt{PYJETS}) before multiple interactions are considered.
 
\iteme{MINT(53) :} internal counter for number of lines used
(in \ttt{PYJETS}) before beam remnants are considered.
 
\iteme{MINT(54) :} internal counter for the number of lines used (in
\ttt{PYJETS}) after beam remnants are considered.
 
\iteme{MINT(55) :} the heaviest new flavour switched on for QCD
processes, specifically the flavour to be generated for 
{\ISUB} = 81, 82, 83 or 84.
 
\iteme{MINT(56) :} the heaviest new flavour switched on for QED
processes, specifically for {\ISUB} = 85. Note that, unlike
\ttt{MINT(55)}, the heaviest flavour may here be a lepton, and
that heavy means the one with largest {\KF} code.

\iteme{MINT(57) :} number of times the beam remnant treatment has 
failed, and the same basic kinematical setup is used to produce 
a new parton shower evolution and beam remnant set. Mainly used
in leptoproduction, for the option when $x$ and $Q^2$ are to be
preserved.   
 
\iteme{MINT(61) :} internal switch for the mode of operation of
resonance width calculations in \ttt{PYWIDT} for $\gammaZ$ or
$\gamma^*/\Z^0/\Z'^0$.
\begin{subentry}
\iteme{= 0 :} without reference to initial-state flavours.
\iteme{= 1 :} with reference to given initial-state flavours.
\iteme{= 2 :} for given final-state flavours.
\end{subentry}
 
\iteme{MINT(62) :} internal switch for use at initialization of
$\hrm^0$ width.
\begin{subentry}
\iteme{= 0 :} use widths into $\Z \Z^*$ or $\W \W^*$ calculated
before.
\iteme{= 1 :} evaluate widths into $\Z \Z^*$ or $\W \W^*$ for
current Higgs mass.
\end{subentry}
 
\iteme{MINT(63) :} internal switch for use at initialization of
the width of a resonance defined with \ttt{MWID(KC)=3}.
\begin{subentry}
\iteme{= 0 :} use existing widths, optionally with simple energy 
rescaling. 
\iteme{= 1 :} evaluate widths at initialization, to be used 
subsequently.
\end{subentry}
 
\iteme{MINT(65) :} internal switch to indicate initialization
without specified reaction.
\begin{subentry}
\iteme{= 0 :} normal initialization.
\iteme{= 1 :} initialization with argument \ttt{'none'} in
\ttt{PYINIT} call.
\end{subentry}
 
\iteme{MINT(71) :} switch to tell whether current process is singular 
for $\pT \to 0$ or not.
\begin{subentry}
\iteme{= 0 :} non-singular process, i.e.\ proceeding via an
$s$-channel resonance or with both products having a mass above
\ttt{CKIN(6)}.
\iteme{= 1 :} singular process.
\end{subentry}

\iteme{MINT(72) :} number of $s$-channel resonances that may
contribute to the cross section.
 
\iteme{MINT(73) :} {\KF} code of first $s$-channel resonance;
0 if there is none.
 
\iteme{MINT(74) :} {\KF} code of second $s$-channel resonance;
0 if there is none.
 
\iteme{MINT(81) :} number of selected pile-up events.
 
\iteme{MINT(82) :} sequence number of currently considered pile-up
event.
 
\iteme{MINT(83) :} number of lines in the event record already
filled by previously considered pile-up events.
 
\iteme{MINT(84) :} \ttt{MINT(83) + MSTP(126)}, i.e.\ number of lines
already filled by previously considered events plus number of lines
to be kept free for event documentation.

\iteme{MINT(91) :} is 1 for a lepton--hadron event and 0
else. Used to determine whether a \ttt{PYFRAM(3)} call is possible.

\iteme{MINT(92) :} is used to denote region in $(x,Q^2)$ plane when
\ttt{MSTP(57)=2}, according to numbering in \cite{Sch93a}. Simply put,
0 means that the modified proton parton distributions were not used,
1 large $x$ and $Q^2$, 2 small $Q^2$ but large $x$,
3 small $x$ but large $Q^2$ and 4 small $x$ and $Q^2$.

\iteme{MINT(93) :} is used to keep track of parton distribution set
used in the latest \ttt{STRUCTM} call to \tsc{Pdflib}. The code for this
set is stored in the form
\ttt{MINT(93) = 1000000}$\times$\ttt{NPTYPE + 1000}$\times$%
\ttt{NGROUP + NSET}.
The stored previous value is compared with the current new value
to decide whether a \ttt{PDFSET} call is needed to switch to another
set.

\iteme{MINT(101), MINT(102) :} is normally 1, but is 4 when a resolved
photon (appearing on side 1 or 2) can be represented by either of the 
four vector mesons $\rho^0$, $\omega$, $\phi$ and $\Jpsi$.

\iteme{MINT(103), MINT(104) :} {\KF} flavour code for the two incoming
particles, i.e.\ the same as \ttt{MINT(11)} and \ttt{MINT(12)}. The 
exception is when a resolved photon is represented by a vector meson 
(a $\rho^0$, $\omega$, $\phi$ or $\Jpsi$). Then the code of the vector 
meson is given.

\iteme{MINT(105) :} is either \ttt{MINT(103)} or \ttt{MINT(104)}, 
depending on which side of the event currently is being studied. 

\iteme{MINT(107), MINT(108) :} if either or both of the two incoming 
particles is a photon, then the respective value gives the nature 
assumed for that photon. The code follows the one used for 
\ttt{MSTP(14)}:
\begin{subentry}
\iteme{= 0 :} direct photon.
\iteme{= 1 :} resolved photon.
\iteme{= 2 :} VMD-like photon.
\iteme{= 3 :} anomalous photon.
\iteme{= 4 :} DIS photon.
\end{subentry}

\iteme{MINT(109) :} is either \ttt{MINT(107)} or \ttt{MINT(108)}, 
depending on which side of the event currently is being studied.
 
\iteme{MINT(111) :} the frame given in \ttt{PYINIT} call, 0--5 for
\ttt{'NONE'}, \ttt{'CMS'}, \ttt{'FIXT'}, \ttt{'3MOM'}, \ttt{'4MOM'} 
and \ttt{'5MOM'}, respectively, and 11 for \ttt{'USER'}. 

\iteme{MINT(121) :} number of separate event classes to initialize 
and mix.
\begin{subentry}
\iteme{= 1 :} the normal value.
\iteme{= 2 - 13 :} for 
$\gamma\p/\gast\p/\gamma\gamma/\gast\gamma/\gast\gast$ 
interaction when \ttt{MSTP(14)} is set to mix different photon 
components.
\end{subentry}

\iteme{MINT(122) :} event class used in current event for $\gamma\p$ or 
$\gamma\gamma$ events generated with one of the \ttt{MSTP(14)} options
mixing several event classes; code as described for \ttt{MSTI(9)}.

\iteme{MINT(123) :} event class used in the current event, with the 
same list of possibilities as for \ttt{MSTP(14)}, except that
options 1, 4 or 10 do not appear. \ttt{= 8} denotes DIS$\times$VMD/$\p$ 
or vice verse, \ttt{= 9} DIS*GVMD or vice versa.
Apart from a different coding, this is exactly the same information as 
is available in \ttt{MINT(122)}. 
 
\iteme{MINT(141), MINT(142) :} for {\galep} beams, {\KF} code 
for incoming lepton beam or target particles, while \ttt{MINT(11)} and 
\ttt{MINT(12)} is then the photon code. A nonzero value is the main 
check whether the photon emission machinery should be called at all.    

\iteme{MINT(143) :} the number of tries before a successful kinematics 
configuration is found in \ttt{PYGAGA}, used for {\galep} 
beams. Used for the cross section updating in \ttt{PYRAND}.

\boxsep
 
\iteme{VINT(1) :}\label{p:VINT} $E_{\mrm{cm}}$, c.m.\ energy.
 
\iteme{VINT(2) :} $s$ ($=E_{\mrm{cm}}^2$) squared mass of 
complete system.
 
\iteme{VINT(3) :} mass of beam particle. Can be negative to denote 
a spacelike particle, e.g.\ a $\gast$.
 
\iteme{VINT(4) :} mass of target particle. Can be negative to denote 
a spacelike particle, e.g.\ a $\gast$.
 
\iteme{VINT(5) :} absolute value of momentum of beam (and target) 
particle in c.m.\ frame.
 
\iteme{VINT(6) - VINT(10) :} $\theta$, $\varphi$ and
\mbox{{\boldmath $\beta$}} for rotation and boost
from c.m.\ frame to user-specified frame.
 
\iteme{VINT(11) :} $\tau_{\mmin}$.
 
\iteme{VINT(12) :} $y_{\mmin}$.
 
\iteme{VINT(13) :} $\cos\hat{\theta}_{\mmin}$ for
$\cos\hat{\theta} \leq 0$.
 
\iteme{VINT(14) :} $\cos\hat{\theta}_{\mmin}$ for
$\cos\hat{\theta} \geq 0$.
 
\iteme{VINT(15) :} $x^2_{\perp \mmin}$.
 
\iteme{VINT(16) :} $\tau'_{\mmin}$.
 
\iteme{VINT(21) :} $\tau$.
 
\iteme{VINT(22) :} $y$.
 
\iteme{VINT(23) :} $\cos\hat{\theta}$.
 
\iteme{VINT(24) :} $\varphi$ (azimuthal angle).
 
\iteme{VINT(25) :} $x_{\perp}^2$.
 
\iteme{VINT(26) :} $\tau'$.
 
\iteme{VINT(31) :} $\tau_{\mmax}$.
 
\iteme{VINT(32) :} $y_{\mmax}$.
 
\iteme{VINT(33) :} $\cos\hat{\theta}_{\mmax}$ for
$\cos\hat{\theta} \leq 0$.
 
\iteme{VINT(34) :} $\cos\hat{\theta}_{\mmax}$ for
$\cos\hat{\theta} \geq 0$.
 
\iteme{VINT(35) :} $x^2_{\perp \mmax}$.
 
\iteme{VINT(36) :} $\tau'_{\mmax}$.
 
\iteme{VINT(41), VINT(42) :} the momentum fractions $x$ taken by the
partons at the hard interaction, as used e.g.\ in the 
parton-distribution functions. For process 99 this agrees with the
Bjorken definition, including target mass corrections, i.e., for a 
proton target, $x = Q^2/(W^2 + Q^2 + m_{\p}^2)$.
 
\iteme{VINT(43) :} $\hat{m} = \sqrt{\hat{s}}$, mass of 
hard-scattering subsystem.
 
\iteme{VINT(44) :} $\hat{s}$ of the hard subprocess ($2 \to 2$ or
$2 \to 1$).
 
\iteme{VINT(45) :} $\hat{t}$ of the hard subprocess ($2 \to 2$ or
$2 \to 1 \to 2$).
 
\iteme{VINT(46) :} $\hat{u}$ of the hard subprocess ($2 \to 2$ or
$2 \to 1 \to 2$).
 
\iteme{VINT(47) :} $\hat{p}_{\perp}$ of the hard subprocess
($2 \to 2$ or $2 \to 1 \to 2$), i.e.\ transverse momentum evaluated
in the rest frame of the scattering.
 
\iteme{VINT(48) :} $\hat{p}_{\perp}^2$ of the hard subprocess;
see \ttt{VINT(47)}.
 
\iteme{VINT(49) :} $\hat{m}'$, the mass of the complete three- or
four-body final state in $2 \to 3$ or $2 \to 4$ processes.
 
\iteme{VINT(50) :} $\hat{s}' = \hat{m}'^2$; see \ttt{VINT(49)}.
 
\iteme{VINT(51) :} $Q$ of the hard subprocess. The exact definition is
process-dependent, see \ttt{MSTP(32)}.
 
\iteme{VINT(52) :} $Q^2$ of the hard subprocess; see \ttt{VINT(51)}.
 
\iteme{VINT(53) :} $Q$ of the outer hard-scattering subprocess,
used as scale for parton distribution function evaluation.
Agrees with \ttt{VINT(51)} for a $2 \to 1$ or $2 \to 2$ process.
For a $2 \to 3$ or $2 \to 4$ $\W/\Z$ fusion process, it is set by
the $\W/\Z$ mass scale, and for
subprocesses 121 and 122 by the heavy-quark mass.
 
\iteme{VINT(54) :} $Q^2$ of the outer hard-scattering subprocess;
see \ttt{VINT(53)}.
 
\iteme{VINT(55) :} $Q$ scale used as maximum virtuality in parton
showers. Is equal to \ttt{VINT(53)}, except for 
DIS processes when \ttt{MSTP(22)}$ > 0$.
 
\iteme{VINT(56) :} $Q^2$ scale in parton showers; see \ttt{VINT(55)}.
 
\iteme{VINT(57) :} $\alphaem$ value of hard process.
 
\iteme{VINT(58) :} $\alphas$ value of hard process.
 
\iteme{VINT(59) :} $\sin\hat{\theta}$ (cf. \ttt{VINT(23)}); used for
improved numerical precision in elastic and diffractive scattering.
 
\iteme{VINT(63), VINT(64) :} nominal $m^2$ values, i.e.\ without 
final-state radiation effects, for the two (or one) partons/particles
leaving the hard interaction. For elastic VMD and GVMD events, this
equals \ttt{VINT(69)}$^2$ or \ttt{VINT(70)}$^2$, and for diffractive
events it is above that. 
 
\iteme{VINT(65) :} $\hat{p}_{\mrm{init}}$, i.e.\ common nominal absolute
momentum of the two partons entering the hard interaction, in their
rest frame.
 
\iteme{VINT(66) :} $\hat{p}_{\mrm{fin}}$, i.e.\ common nominal absolute
momentum of the two partons leaving the hard interaction, in their
rest frame.

\iteme{VINT(67), VINT(68) :} mass of beam and target particle, as 
\ttt{VINT(3)} and \ttt{VINT(4)}, except that an incoming $\gamma$ is
assigned the $\rho^0$, $\omega$ or $\phi$ mass. (This also applies for 
a GVMD photon, where the mass of the VMD state with the equivalent 
flavour content is chosen.) Used for elastic scattering 
$\gamma \p \to \rho^0 \p$ and other similar processes.

\iteme{VINT(69), VINT(70) :} the actual mass of a VMD or GVMD state;
agrees with the above for VMD but is selected as a larger number for 
GVMD, using the approximate association $m = 2 \kT$. Thus the mass
selection for a GVMD state is according to $\d m^2/(m^2+Q^2)^2$ 
between limits $2 k_0 < m < 2 k_1 = 2 \pTmin(W^2)$. Required for 
elastic and diffractive events.
 
\iteme{VINT(71) :} $\pTmin$ of process, i.e.\ \ttt{CKIN(3)} or
\ttt{CKIN(5)}, depending on which is larger, and whether the process
is singular in $\pT \to 0$ or not.
 
\iteme{VINT(73) :} $\tau = m^2/s$ value of first resonance, if any;
see \ttt{MINT(73)}.
 
\iteme{VINT(74) :} $m \Gamma/s$ value of first resonance, if any;
see \ttt{MINT(73)}.
 
\iteme{VINT(75) :} $\tau = m^2/s$ value of second resonance, if any;
see \ttt{MINT(74)}.
 
\iteme{VINT(76) :} $m \Gamma/s$ value of second resonance, if any;
see \ttt{MINT(74)}.
 
\iteme{VINT(80) :} correction factor (evaluated in \ttt{PYOFSH}) for
the cross section of resonances produced in $2 \to 2$ processes, if
only some mass range of the full Breit--Wigner shape is allowed by
user-set mass cuts (\ttt{CKIN(2)}, \ttt{CKIN(45) - CKIN(48)}).
 
\iteme{VINT(95) :} the value of the Coulomb factor in the current event,
see \ttt{MSTP(40)}. For \ttt{MSTP(40)=0} it is $=1$, else it is $>1$.
 
\iteme{VINT(97) :} an event weight, normally 1 and thus uninteresting, 
but for external processes with \ttt{IDWTUP=-1, -2} or \ttt{-3} it can 
be $-1$ for events with negative cross section, with \ttt{IDWTUP=4} it 
can be an arbitrary non-negative weight of dimension mb, and with 
\ttt{IDWTUP=-4} it can be an arbitrary weight of dimension mb.
(The difference being that in most cases a rejection step is involved 
to bring the accepted events to a common weight normalization, up to
a sign, while no rejection need be involved in the last two cases.)

\iteme{VINT(98) :} is sum of \ttt{VINT(100)} values for current run.
 
\iteme{VINT(99) :} is weight \ttt{WTXS} returned from \ttt{PYEVWT} call
when \ttt{MSTP(142)}$\geq 1$, otherwise is 1.
 
\iteme{VINT(100) :} is compensating weight \ttt{1./WTXS} that should be
associated with events when \ttt{MSTP(142)=1}, otherwise is 1.
 
\iteme{VINT(108) :} ratio of maximum differential cross section
observed to maximum differential cross section assumed for the
generation; cf. \ttt{MSTP(123)}.
 
\iteme{VINT(109) :} ratio of minimal (negative!) cross section
observed to maximum differential cross section assumed for the
generation; could only become negative if cross sections are
incorrectly included.
 
\iteme{VINT(111) - VINT(116) :} for \ttt{MINT(61)=1} gives
kinematical factors for the different pieces contributing to
$\gammaZ$ or $\gamma^*/\Z^0/\Z'^0$ production, for \ttt{MINT(61)=2}
gives sum of final-state weights for the same; coefficients are
given in the order pure $\gamma^*$, $\gamma^*$--$\Z^0$ interference,
$\gamma^*$--$\Z'^0$ interference, pure $\Z^0$, $\Z^0$--$\Z'^0$
interference and pure $\Z'^0$.
 
\iteme{VINT(117) :} width of $\Z^0$; needed in
$\gamma^*/\Z^0/\Z'^0$ production.
 
\iteme{VINT(120) :} mass of beam or target particle, i.e.\ coincides 
with \ttt{VINT(3)} or \ttt{VINT(4)}, depending on which side of the 
event is considered. Is used to bring information on the user-defined 
virtuality of a photon beam to the parton distributions of the photon. 
 
\iteme{VINT(131) :} total cross section (in mb) for subprocesses
allowed in the pile-up events scenario according to the
\ttt{MSTP(132)} value.
 
\iteme{VINT(132) :} $\br{n} = $\ttt{VINT(131)}$\times$\ttt{PARP(131)}
of pile-up events, cf. \ttt{PARI(91)}.
 
\iteme{VINT(133) :} $\langle n \rangle = \sum_i i \, {\cal P}_i  /
\sum_i {\cal P}_i$ of pile-up events as actually simulated, i.e.\ 
$1 \leq i \leq 200$ (or smaller), see \ttt{PARI(92)}.
 
\iteme{VINT(134) :} number related to probability to have an event in
a beam--beam crossing; is $\exp(-\br{n}) \sum_i \br{n}^i/i!$ for
\ttt{MSTP(133)=1} and $\exp(-\br{n}) \sum_i \br{n}^i/(i-1)!$
for \ttt{MSTP(133)=2}, cf. \ttt{PARI(93)}.
 
\iteme{VINT(138) :} size of the threshold factor (enhancement or
suppression) in the latest event with heavy-flavour production;
see \ttt{MSTP(35)}.
 
\iteme{VINT(141), VINT(142) :} $x$ values for the parton-shower
initiators of the hardest interaction; used to find what is left
for multiple interactions.
 
\iteme{VINT(143), VINT(144) :} $1 - \sum_i x_i$ for all scatterings;
used for rescaling each new $x$-value in the multiple-interaction
parton-distribution-function evaluation.
 
\iteme{VINT(145) :} estimate of total parton--parton cross section for
multiple interactions; used for \ttt{MSTP(82)}$\geq 2$.
 
\iteme{VINT(146) :} common correction factor $f_c$ in the 
multiple-interaction probability; used for \ttt{MSTP(82)}$\geq 2$ 
(part of $e(b)$, see eq.~(\ref{mi:ebenh})).
 
\iteme{VINT(147) :} average hadronic matter overlap; used for
\ttt{MSTP(82)}$\geq 2$ (needed in evaluation of $e(b)$, see 
eq.~(\ref{mi:ebenh})).
 
\iteme{VINT(148) :} enhancement factor for current event in the
multiple-interaction probability, defined as the actual overlap 
divided by the average one; used for \ttt{MSTP(82)}$\geq 2$ 
(is $e(b)$ of eq.~(\ref{mi:ebenh})).
 
\iteme{VINT(149) :} $x_{\perp}^2$ cut-off or turn-off for multiple
interactions. For \ttt{MSTP(82)}$\leq 1$ it is $4 \pTmin^2/W^2$,
for \ttt{MSTP(82)}$\geq 2$ it is $4 p_{\perp 0}^2/W^2$. For hadronic
collisions, $W^2 = s$, but in photoproduction or $\gamma\gamma$
physics the $W^2$ scale refers to the hadronic subsystem squared
energy. This may vary from event to event, so \ttt{VINT(149)} needs
to be recalculated.
 
\iteme{VINT(150) :} probability to keep the given event in the
multiple-interaction scenario with varying impact parameter,
as given by the exponential factor in eq.~(\ref{mi:hardestvarimp}).
 
\iteme{VINT(151), VINT(152) :} sum of $x$ values for all the
multiple-interaction partons.
 
\iteme{VINT(153) :} current differential cross section value
obtained from \ttt{PYSIGH}; used in multiple interactions only.
 
\iteme{VINT(154) :} current $\pTmin(s)$ or $\pTmin(W^2)$, used
for multiple interactions and also as upper cut-off $k_1$ if the
GVMD $\kT$ spectrum. See comments at \ttt{VINT(149)}.
 
\iteme{VINT(155), VINT(156) :} the $x$ value of a photon that branches
into quarks or gluons, i.e.\ $x$ at interface between initial-state QED
and QCD cascades, in the old photoproduction machinery.
 
\iteme{VINT(157), VINT(158) :} the primordial $k_{\perp}$ values
selected in the two beam remnants.
 
\iteme{VINT(159), VINT(160) :} the $\chi$ values selected for beam
remnants that are split into two objects, describing how the energy
is shared (see \ttt{MSTP(92)} and \ttt{MSTP(94)}); is 0 if no 
splitting is needed.
 
\iteme{VINT(161) - VINT(200) :} sum of Cabibbo--Kobayashi--Maskawa
squared matrix elements that a given flavour is allowed to couple to.
Results are stored in format \ttt{VINT(180+KF)} for quark and lepton
flavours and antiflavours (which need not be the same; see
\ttt{MDME(IDC,2)}). For leptons, these factors are normally unity.
 
\iteme{VINT(201) - VINT(220) :} additional variables needed in 
phase-space selection for $2 \to 3$ processes with \ttt{ISET(ISUB)=5}.
Below indices 1, 2 and 3 refer to scattered partons 1, 2 and 3, except
that the $q$ four-momentum variables are $q_1 + q_2 \to q_1' + q_2' + q_3'$.
All kinematical variables refer to the internal kinematics of
the 3-body final state --- the kinematics of the system as a whole
is described by $\tau'$ and $y$, and the mass distribution of
particle 3 (a resonance) by $\tau$.
\begin{subentry}
\iteme{VINT(201) :} $m_1$.
\iteme{VINT(202) :} $p_{\perp 1}^2$.
\iteme{VINT(203) :} $\varphi_1$.
\iteme{VINT(204) :} $M_1$ (mass of propagator particle).
\iteme{VINT(205) :} weight for the $p_{\perp 1}^2$ choice.
\iteme{VINT(206) :} $m_2$.
\iteme{VINT(207) :} $p_{\perp 2}^2$.
\iteme{VINT(208) :} $\varphi_2$.
\iteme{VINT(209) :} $M_2$ (mass of propagator particle).
\iteme{VINT(210) :} weight for the $p_{\perp 2}^2$ choice.
\iteme{VINT(211) :} $y_3$.
\iteme{VINT(212) :} $y_{3 \mmax}$.
\iteme{VINT(213) :} $\epsilon = \pm 1$; choice between two mirror
solutions $1 \leftrightarrow 2$.
\iteme{VINT(214) :} weight associated to $\epsilon$-choice.
\iteme{VINT(215) :} $t_1 = (q_1 - q_1')^2$.
\iteme{VINT(216) :} $t_2 = (q_2 - q_2')^2$.
\iteme{VINT(217) :} $q_1 q_2'$ four-product.
\iteme{VINT(218) :} $q_2 q_1'$ four-product.
\iteme{VINT(219) :} $q_1' q_2'$ four-product.
\iteme{VINT(220) :} $\sqrt{(m_{\perp 12}^2 - m_{\perp 1}^2 -
m_{\perp 2}^2)^2 - 4 m_{\perp 1}^2 m_{\perp 2}^2}$, where
$m_{\perp 12}$ is the transverse mass of the $q'_1 q'_2$ system.
\end{subentry}
 
\iteme{VINT(221) - VINT(225) :} $\theta$, $\varphi$ and
\mbox{{\boldmath $\beta$}} for rotation and boost
from c.m.\ frame to hadronic c.m.\ frame of a lepton--hadron
event.

\iteme{VINT(231) :} $Q^2_{\mmin}$ scale for current 
parton-distribution function set.

\iteme{VINT(232) :} valence quark distribution of a VMD photon; set in
\ttt{PYPDFU} and used in \ttt{PYPDFL}.

\iteme{VINT(281) :} for resolved photon events, it gives the 
ratio between the total $\gamma X$ cross section and the total 
$\pi^0 X$ cross section, where $X$ represents the target particle.

\iteme{VINT(283), VINT(284) :} virtuality scale at which a
GVMD/anomalous photon on the beam or target side of the event is 
being resolved. More precisely, it gives the $\kT^2$ of the 
$\gamma \to \q\qbar$ vertex. For elastic and diffractive scatterings, 
$m^2/4$ is stored, where $m$ is the mass of the state being diffracted.
For clarity, we point out that elastic and diffractive events are
characterized by the mass of the diffractive states but without
any primordial $\kT$, while jet production involves a primordial $\kT$
but no mass selection. Both are thus not used at the same time,
but for GVMD/anomalous photons, the standard (though approximate) 
identification $\kT^2 = m^2/4$ ensures agreement between the two
applications.

\iteme{VINT(285) :} the \ttt{CKIN(3)} value provided by you
at initialization; subsequently \ttt{CKIN(3)} may be overwritten 
(for \ttt{MSTP(14)=10}) but \ttt{VINT(285)} stays.

\iteme{VINT(289) :} squared c.m.\ energy found in \ttt{PYINIT} call.

\iteme{VINT(290) :} the \ttt{WIN} argument of a \ttt{PYINIT} call.

\iteme{VINT(291) - VINT(300) :} the two five-vectors of the two 
incoming particles, as reconstructed in \ttt{PYINKI}.
These may vary from one event to the next.

\iteme{VINT(301) - VINT(320) :} used when a flux of virtual photons is
being generated by the \ttt{PYGAGA} routine, for {\galep} 
beams. 
\begin{subentry}

\iteme{VINT(301) :} c.m.\ energy for the full collision, while \ttt{VINT(1)} 
gives the $\gamma$-hadron or $\gamma\gamma$ subsystem energy.

\iteme{VINT(302) :} full squared c.m.\ energy, while \ttt{VINT(2)} gives 
the subsystem squared energy.

\iteme{VINT(303), VINT(304) :} mass of the beam or target lepton, while 
\ttt{VINT(3)} or \ttt{VINT(4)} give the mass of a photon  emitted off it.

\iteme{VINT(305), VINT(306) :} $x$ values, i.e.\ respective photon energy 
fractions of the incoming lepton in the c.m.\ frame of the event.

\iteme{VINT(307), VINT(308) :} $Q^2$ or $P^2$, virtuality of the 
respective photon (thus the square of \ttt{VINT(3)},\ttt{ VINT(4)}).    

\iteme{VINT(309), VINT(310) :} $y$ values, i.e.\ respective photon 
light-cone energy fraction of the lepton. 

\iteme{VINT(311), VINT(312) :} $\theta$, scattering angle of the 
respective lepton in the c.m.\ frame of the event.   

\iteme{VINT(313), VINT(314) :} $\phi$, azimuthal angle of the 
respective scattered lepton in the c.m.\ frame of the event. 

\iteme{VINT(315), VINT(316):} the $R$ factor defined at \ttt{MSTP(17)},
giving a cross section enhancement from the contribution of resolved
longitudinal photons.

\iteme{VINT(317) :} dipole suppression factor in \ttt{PYXTOT} for 
current event.

\iteme{VINT(318) :} dipole suppression factor in \ttt{PYXTOT} at 
initialization. 

\iteme{VINT(319) :} photon flux factor in \ttt{PYGAGA} for current event.

\iteme{VINT(320) :} photon flux factor in \ttt{PYGAGA} at initialization.  
\end{subentry}

\end{entry}
 
\drawbox{COMMON/PYINT2/ISET(500),KFPR(500,2),COEF(500,20),ICOL(40,4,2)}%
\label{p:PYINT2}
\begin{entry}
\itemc{Purpose:} to store information necessary for efficient
generation of the different subprocesses, specifically type of
generation scheme and coefficients of the Jacobian. Also to store
allowed colour-flow configurations. These variables must not be
changed by you.
 
\iteme{ISET(ISUB) :}\label{p:ISET} gives the type of 
kinematical-variable selection scheme used for subprocess {\ISUB}.
\begin{subentry}
\iteme{= 0 :} elastic, diffractive and low-$\pT$ processes.
\iteme{= 1 :} $2 \to 1$ processes (irrespective of subsequent decays).
\iteme{= 2 :} $2 \to 2$ processes (i.e.\ the bulk of processes).
\iteme{= 3 :} $2 \to 3$ processes
(like $\q \q' \to \q'' \q''' \hrm^0$).
\iteme{= 4 :} $2 \to 4$ processes
(like $\q \q' \to \q'' \q''' \W^+ \W^-$).
\iteme{= 5 :} `true' $2 \to 3$ processes, one method.
\iteme{= 8 :} $2 \to 1$ process $\gamma^* \f_i \to f_i$ where, unlike
the $2 \to 1$ processes above, $\hat{s}=0$.
\iteme{= 9 :} $2 \to 2$ in multiple interactions ($\pT$ as kinematics
variable).
\iteme{= 11 :} a user-defined process.
\iteme{= -1 :} legitimate process which has not yet been implemented.
\iteme{= -2 :} {\ISUB} is an undefined process code.
\end{subentry}
 
\iteme{KFPR(ISUB,J) :}\label{p:KFPR} give the {\KF} flavour codes for the 
products produced in subprocess {\ISUB}. If there is only one product, the
\ttt{J=2} position is left blank. Also, quarks and leptons assumed
massless in the matrix elements are denoted by 0. The main
application is thus to identify resonances produced
($\Z^0$, $\W^{\pm}$, $\hrm^0$, etc.). For external processes,
\ttt{KFPR} instead stores information on process numbers in the two
external classifications, see subsection \ref{ss:PYnewproc}.  
 
\iteme{COEF(ISUB,J) :}\label{p:COEF} factors used in the Jacobians in 
order to speed
up the selection of kinematical variables. More precisely, the shape
of the cross section is given as the sum of terms with different
behaviour, where the integral over the allowed phase space is
unity for each term. \ttt{COEF} gives the relative strength of these
terms, normalized so that the sum of coefficients for each variable
used is unity. Note that which coefficients are indeed used is
process-dependent.
\begin{subentry}
\iteme{ISUB :} standard subprocess code.
\iteme{J = 1 :} $\tau$ selected according $1/\tau$.
\iteme{J = 2 :} $\tau$ selected according to $1/\tau^2$.
\iteme{J = 3 :} $\tau$ selected according to $1/(\tau(\tau+\tau_R))$,
where $\tau_R = m_R^2/s$ is $\tau$ value of resonance; only used for
resonance production.
\iteme{J = 4 :} $\tau$ selected according to Breit--Wigner of form
$1/((\tau-\tau_R)^2+\gamma_R^2)$, where $\tau_R = m_R^2/s$ is $\tau$
value of resonance and $\gamma_R = m_R \Gamma_R/s$ is its scaled mass
times width; only used for resonance production.
\iteme{J = 5 :} $\tau$ selected according to
$1/(\tau(\tau+\tau_{R'}))$, where $\tau_{R'} = m_{R'}^2/s$ is $\tau$
value of second resonance; only used for simultaneous production of
two resonances.
\iteme{J = 6 :} $\tau$ selected according to second Breit--Wigner of
form $1/((\tau-\tau_{R'})^2+\gamma_{R'}^2)$, where
$\tau_{R'} = m_{R'}^2/s$ is $\tau$ value of second resonance and
$\gamma_{R'} = m_{R'} \Gamma_{R'}/s$ is its scaled
mass times width; is used only for simultaneous production
of two resonances, like $\gamma^*/\Z^0/\Z'^0$.
\iteme{J = 7 :} $\tau$ selected according to $1/(1-\tau)$; only used
when both parton distributions are peaked at $x = 1$.
\iteme{J = 8 :} $y$ selected according to $y - y_{\mmin}$.
\iteme{J = 9 :} $y$ selected according to $y_{\mmax} - y$.
\iteme{J = 10 :} $y$ selected according to $1/\cosh(y)$.
\iteme{J = 11 :} $y$ selected according to $1/(1-\exp(y-y_{\mmax}))$;
only used when beam parton distribution is peaked close to $x = 1$.
\iteme{J = 12 :} $y$ selected according to $1/(1-\exp(y_{\mmin}-y))$;
only used when target parton distribution is peaked close to $x = 1$.
\iteme{J = 13 :} $z = \cos\hat{\theta}$ selected evenly between limits.
\iteme{J = 14 :} $z = \cos\hat{\theta}$ selected according to $1/(a-z)$,
where $a = 1 + 2 m_3^2 m_4^2/\hat{s}^2$, $m_3$ and $m_4$ being the
masses of the two final-state particles.
\iteme{J = 15 :} $z = \cos\hat{\theta}$ selected according to $1/(a+z)$,
with $a$ as above.
\iteme{J = 16 :} $z = \cos\hat{\theta}$ selected according to
$1/(a-z)^2$, with $a$ as above.
\iteme{J = 17 :} $z = \cos\hat{\theta}$ selected according to
$1/(a+z)^2$, with $a$ as above.
\iteme{J = 18 :} $\tau'$ selected according to $1/\tau'$.
\iteme{J = 19 :} $\tau'$ selected according to
$(1 - \tau/\tau')^3/\tau'^2$.
\iteme{J = 20 :} $\tau'$ selected according to $1/(1-\tau')$; only
used when both parton distributions are peaked close to $x = 1$.
\end{subentry}
 
\iteme{ICOL :}\label{p:ICOL} contains information on different 
colour-flow topologies in hard $2 \to 2$ processes.
 
\end{entry}
 
\drawbox{COMMON/PYINT3/XSFX(2,-40:40),ISIG(1000,3),SIGH(1000)}%
\label{p:PYINT3}
\begin{entry}
 
\itemc{Purpose:} to store information on parton distributions,
subprocess cross sections and different final-state relative
weights. These variables must not be changed by you.
 
\iteme{XSFX :}\label{p:XSFX} current values of parton-distribution
functions $xf(x)$ on beam and target side.
 
\iteme{ISIG(ICHN,1) :}\label{p:ISIG} incoming parton/particle on the 
beam side to
the hard interaction for allowed channel number \ttt{ICHN}. The number
of channels filled with relevant information is given by \ttt{NCHN},
one of the arguments returned in a \ttt{PYSIGH} call. Thus only
$1 \leq $\ttt{ICHN}$ \leq $\ttt{NCHN} is filled with relevant
information.
 
\iteme{ISIG(ICHN,2) :} incoming parton/particle on the target side
to the hard interaction for allowed channel number \ttt{ICHN}. See also
comment above.
 
\iteme{ISIG(ICHN,3) :} colour-flow type for allowed channel number
\ttt{ICHN}; see \ttt{MSTI(2)} list. See also above comment. For
`subprocess' 96 uniquely, \ttt{ISIG(ICHN,3)} is also used to
translate information on what is the correct subprocess number
(11, 12, 13, 28, 53 or 68); this is used for reassigning subprocess
96 to either of these.
 
\iteme{SIGH(ICHN) :}\label{p:SIGH} evaluated differential 
cross section for allowed
channel number \ttt{ICHN}, i.e.\ matrix-element value times
parton distributions, for current kinematical setup (in addition,
Jacobian factors are included in the numbers, as used to speed up
generation). See also comment for \ttt{ISIG(ICHN,1)}.
 
\end{entry}
 
\drawbox{COMMON/PYINT4/MWID(500),WIDS(500,5)}\label{p:PYINT4}
\begin{entry}
 
\itemc{Purpose:} to store character of resonance width treatment and
 partial and effective decay widths for the different resonances. 
These variables must not be changed by you.
 
\iteme{MWID(KC) :}\label{p:MWID} gives the character of particle with 
compressed code {\KC}, mainly as used in \ttt{PYWIDT} to calculate widths 
of resonances (not necessarily at the nominal mass).
\begin{subentry}
\iteme{= 0 :} an ordinary particle; not to be treated as resonance.
\iteme{= 1 :} a resonance for which the partial and total widths
(and hence branching ratios) are dynamically calculated 
in \ttt{PYWIDT} calls; i.e.\ special code has to exist for each 
such particle. The effects of allowed/disallowed secondary
decays are included, both in the relative composition
of decays and in the process cross section.
\iteme{= 2 :} The total width is taken to be the one stored in 
\ttt{PMAS(KC,2)} and the relative branching ratios the ones in 
\ttt{BRAT(IDC)} for decay channels \ttt{IDC}. There is then no need 
for any special code in \ttt{PYWIDT} to handle a resonance. During the 
run, the stored \ttt{PMAS(KC,2)} and \ttt{BRAT(IDC)} values are used to 
calculate the total and partial widths of the decay channels. 
Some extra information and assumptions are then used.
Firstly, the stored \ttt{BRAT} values are assumed to be the full 
branching ratios, including all possible channels and
all secondary decays. The actual relative branching fractions 
are modified to take into account that the simulation of some
channels may be switched off (even selectively for a particle
and an antiparticle), as given by \ttt{MDME(IDC,1)}, and that
some secondary channels may not be allowed, as expressed by
the \ttt{WIDS} factors. This also goes into process cross sections.
Secondly, it is assumed that all widths scale like $\sqrt{\hat{s}}/m$, 
the ratio of the actual to the nominal mass. A further nontrivial 
change as a function of the actual mass can be set for each 
channel by the \ttt{MDME(IDC,2)} value, see section 
\ref{ss:parapartdat}.

\iteme{= 3 :} a hybrid version of options 1 and 2 above. At initialization
the \ttt{PYWIDT} code is used to calculate\ttt{ PMAS(KC,2)} and
\ttt{ BRAT(IDC)} at the nominal mass of the resonance. Special code must 
then exist in \ttt{PYWIDT} for the particle. The \ttt{PMAS(KC,2)} and 
\ttt{BRAT(IDC)} values overwrite the default ones. In the subsequent 
generation of events, the simpler scheme of option 2 is used, thus saving
some execution time. 
\itemc{Note:} the $\Z$ and $\Z'$ cannot be used with options 2 and 3, 
since the more complicated interference structure implemented for those
particles is only handled correctly for option 1. 
\end{subentry}
 
\iteme{WIDS(KC,J) :}\label{p:WIDS} gives the dimensionless suppression 
factor to cross sections caused by the closing of some secondary decays, 
as calculated in \ttt{PYWIDT}. It is defined as the ratio of the total 
width of channels switched on to the total width of all possible channels 
(replace width by squared width for a pair of resonances). The on/off 
status of channels is set by the \ttt{MDME} switches; see section 
\ref{ss:parapartdat}. The information in \ttt{WIDS} is used e.g.\ in 
cross-section calculations. Values are built up recursively from the 
lightest particle to the heaviest one at initialization, with the 
exception that $\W$ and $\Z$ are done already from the beginning (since 
these often are forced off the mass shell). \ttt{WIDS} can go wrong in 
case you have perverse situations where the branching ratios vary
rapidly as a function of energy, across the resonance shape.
This then influences process cross sections.  
\begin{subentry}
\iteme{KC :} standard {\KC} code for resonance considered.
\iteme{J = 1 :} suppression when a pair of resonances of type {\KC} are
produced together. When an antiparticle exists, the
particle--antiparticle pair (such as $\W^+ \W^-$) is the relevant
combination, else the particle--particle one (such as $\Z^0 \Z^0$).
\iteme{J = 2 :} suppression for a particle of type {\KF} when produced
on its own, or together with a particle of another type.
\iteme{J = 3 :} suppression for an antiparticle of type {\KF} when produced
on its own, or together with a particle of another type.
\iteme{J = 4 :} suppression when a pair of two identical particles are
produced, for a particle which has a nonidentical antiparticle (e.g.\ 
$\W^+\W^+$).
\iteme{J = 5 :} suppression when a pair of two identical antiparticles are
produced, for a particle which has a nonidentical antiparticle (e.g.\ 
$\W^-\W^-$).
\end{subentry}

\end{entry}
 
\drawbox{COMMON/PYINT5/NGENPD,NGEN(0:500,3),XSEC(0:500,3)}\label{p:PYINT5}
\begin{entry}
 
\itemc{Purpose:} to store information necessary for cross-section
calculation and differential cross-section maximum violation.
These variables must not be changed by you.
 
\iteme{NGEN(ISUB,1) :}\label{p:NGEN} gives the number of times that 
the differential cross section (times Jacobian factors) has been
evaluated for subprocess {\ISUB}, with \ttt{NGEN(0,1)} the sum of
these.
 
\iteme{NGEN(ISUB,2) :} gives the number of times that a kinematical
setup for subprocess {\ISUB} is accepted in the generation procedure,
with \ttt{NGEN(0,2)} the sum of these.
 
\iteme{NGEN(ISUB,3) :} gives the number of times an event of
subprocess type {\ISUB} is generated, with \ttt{NGEN(0,3)} the sum of
these. Usually \ttt{NGEN(ISUB,3) = NGEN(ISUB,2)}, i.e.\ an accepted
kinematical configuration can normally be used to produce an event.
 
\iteme{XSEC(ISUB,1) :}\label{p:XSEC} estimated maximum differential 
cross section (times the Jacobian factors used to speed up the generation 
process) for the different subprocesses in use, with \ttt{XSEC(0,1)} the 
sum of these (except low-$\pT$, i.e.\ {\ISUB} = 95). For 
external processes special rules may apply, see subsection
\ref{ss:PYnewproc}. In particular, negative cross sections and maxima 
may be allowed. In this case, \ttt{XSEC(ISUB,1)} stores the absolute
value of the maximum, since this is the number that allows the 
appropriate mixing of subprocesses.  
 
\iteme{XSEC(ISUB,2) :} gives the sum of differential cross sections
(times Jacobian factors) for the \ttt{NGEN(ISUB,1)} phase-space
points evaluated so far.
 
\iteme{XSEC(ISUB,3) :} gives the estimated integrated cross section
for subprocess {\ISUB}, based on the statistics accumulated so far,
with \ttt{XSEC(0,3)} the estimated total cross section for all
subprocesses included (all in mb). This is exactly the
information obtainable by a \ttt{PYSTAT(1)} call.

\itemc{Warning :} For $\gamma\p$ and $\gamma\gamma$ events, when several
photon components are mixed (see \ttt{MSTP(14)}), a master copy
of these numbers for each component is stored in the \ttt{PYSAVE}
routine. What is then visible after each event is only the numbers
for the last component considered, not the full statistics. A
special \ttt{PYSAVE} call, performed e.g.\ in \ttt{PYSTAT}, is 
required to obtain the sum of all the components.  
 
\end{entry}
 
\drawboxtwo{COMMON/PYINT6/PROC(0:500)}{CHARACTER PROC*28}%
\label{p:PYINT6}
\begin{entry}
 
\itemc{Purpose:} to store character strings for the different
possible subprocesses; used when printing tables.
 
\iteme{PROC(ISUB) :}\label{p:PROC} name for the different 
subprocesses, according to {\ISUB} code. \ttt{PROC(0)} denotes 
all processes.
 
\end{entry}
 
\drawbox{COMMON/PYINT7/SIGT(0:6,0:6,0:5)}\label{p:PYINT7}
\begin{entry}
 
\itemc{Purpose:} to store information on total, elastic and 
diffractive cross sections. These variables should only be set 
by you for the option \ttt{MSTP(31)=0}; else they should not be
touched. All numbers are given in mb. 
 
\iteme{SIGT(I1,I2,J) :}\label{p:SIGT} the cross section, both total
and subdivided by class (elastic, diffractive etc.). For a photon
to be considered as a VMD meson the cross sections are additionally
split into the contributions from the various meson states. 
\begin{subentry}
\iteme{I1, I2 :} allowed states for the incoming particle on side 
1 and 2, respectively.
\begin{subentry}
\iteme{= 0 :} sum of all allowed states. Except for a photon to
be considered as a VMD meson this is the only nonvanishing entry.
\iteme{= 1 :} the contribution from the $\rho^0$ VMD state.
\iteme{= 2 :} the contribution from the $\omega$ VMD state.
\iteme{= 3 :} the contribution from the $\phi$ VMD state.
\iteme{= 4 :} the contribution from the $\Jpsi$ VMD state.
\iteme{= 5, 6 :} reserved for future use.
\end{subentry}
\iteme{J :} the total and partial cross sections. 
\begin{subentry}
\iteme{= 0 :} the total cross section.
\iteme{= 1 :} the elastic cross section.
\iteme{= 2 :} the single diffractive cross section $AB \to XB$.
\iteme{= 3 :} the single diffractive cross section $AB \to AX$.
\iteme{= 4 :} the double diffractive cross section.
\iteme{= 5 :} the inelastic, non-diffractive cross section.
\end{subentry} 
\itemc{Warning:} If you set these values yourself, it is important
that they are internally consistent, since this is not explicitly
checked by the program. Thus the contributions \ttt{J=}1--5 should
add up to the \ttt{J=}0 one and, for VMD photons, the contributions
\ttt{I=}1--4 should add up to the \ttt{I=}0 one.
\end{subentry}

\end{entry}
 
\drawboxtwo{~COMMON/PYINT8/XPVMD(-6:6),XPANL(-6:6),XPANH(-6:6),%
XPBEH(-6:6),}{\&XPDIR(-6:6)}\label{p:PYINT8}
\begin{entry}

\itemc{Purpose:} to store the various components of the photon 
parton distributions when the \ttt{PYGGAM} routine is called.

\iteme{XPVMD(KFL) :}\label{p:XPVMD} gives distributions of the VMD part 
($\rho^0$, $\omega$ and $\phi$). 
     
\iteme{XPANL(KFL) :}\label{p:XPANL} gives distributions of the anomalous 
part of light quarks ($\d$, $\u$ and $\s$).

\iteme{XPANH(KFL) :}\label{p:XPANH} gives distributions of the anomalous 
part of heavy quarks ($\c$ and $\b$).

\iteme{XPBEH(KFL) :}\label{p:XPBEH} gives Bethe-Heitler distributions of 
heavy quarks ($\c$ and $\b$). This provides an alternative to \ttt{XPANH}, 
i.e.\ both should not be used at the same time.

\iteme{XPDIR(KFL) :}\label{p:XPDIR} gives direct correction to the 
production of light quarks ($\d$, $\u$ and $\s$). This term is 
nonvanishing only in the {\MSbar} scheme, and is applicable for 
$F_2^{\gamma}$ rather than for the parton distributions themselves.

\end{entry}
 
\drawbox{~COMMON/PYINT9/VXPVMD(-6:6),VXPANL(-6:6),VXPANH(-6:6),VXPDGM(-6:6)}%
\label{p:PYINT9}
\begin{entry}

\itemc{Purpose:} to give the valence parts of the photon parton distributions
($x$-weighted, as usual) when the \ttt{PYGGAM} routine is called. Companion to 
\ttt{PYINT8}, which gives the total parton distributions.

\iteme{VXPVMD(KFL) :}\label{p:VXPVMD} valence distributions of the VMD part; 
matches \ttt{XPVMD} in \ttt{PYINT8}.
     
\iteme{VXPANL(KFL) :}\label{p:VXPANL} valence distributions of the anomalous 
part of light quarks; matches \ttt{XPANL} in \ttt{PYINT8}.

\iteme{VXPANH(KFL) :}\label{p:VXPANH} valence distributions of the anomalous 
part of heavy quarks; matches \ttt{XPANH} in \ttt{PYINT8}.

\iteme{VXPDGM(KFL) :}\label{p:VXPDGM} gives the sum of valence distributions 
parts; matches \ttt{XPDFGM} in the \ttt{PYGGAM} call.

\itemc{Note 1:} the Bethe-Heitler and direct contributions in \ttt{XPBEH(KFL)} 
and \ttt{XPDIR(KFL)} in \ttt{PYINT8} are pure valence-like, and therefore 
are not duplicated here.
        
\itemc{Note 2:} the sea parts of the distributions can be obtained by 
taking the appropriate differences between the total distributions and the
valence distributions.

\end{entry}

\clearpage
 
\section{Initial- and Final-State Radiation}
\label{s:showinfi}
 
Starting from the hard interaction, initial- and final-state
radiation corrections may be added. This is normally done by
making use of the parton-shower language --- only for the
$\ee \to \q \qbar$ process does {\Py} offer a matrix-element
option (described in section \ref{ss:eematrix}).
The algorithms used to generate initial- and final-state showers
are rather different, and are therefore described separately
below, starting with the conceptually easier final-state one.
Before that, some common elements are introduced.
 
The main references for final-state showers is \cite{Ben87a,Nor01} 
and for initial-state ones \cite{Sjo85,Miu99}.
 
\subsection{Shower Evolution}
 
In the leading-logarithmic picture, a shower may be viewed as a sequence
of $1 \to 2$ branchings $a \to bc$. Here $a$ is called the mother
and $b$ and $c$ the two daughters. Each daughter is free to branch
in its turn, so that a tree-like structure can evolve. We will use
the word `parton' for all the objects $a$, $b$ and $c$ involved
in the branching process, i.e.\ not only for quarks and gluons
but also for leptons and photons. The branchings included in the
program are $\q \to \q \g$, $\g \to \g \g$, $\g \to \q \qbar$,
$\q \to \q \gamma$ and $\ell \to \ell \gamma$. Photon branchings,
i.e.\ $\gamma \to \q \qbar$ and $\gamma \to \ell \br{\ell}$, have not
been included so far, since they are reasonably rare and since no
urgent need for them has been perceived.

A word on terminology may be in place. The algorithms described
here are customarily referred to as leading-log showers. This is 
correct insofar as no explicit corrections from higher orders 
are included, i.e.\ there are no $\mathcal{O}(\alphas^2)$ terms in 
the splitting kernels, neither by new $1 \to 3$ processes nor by
corrections to the $1 \to 2$ ones. However, it is grossly misleading 
if leading-log showers are equated with leading-log analytical 
calculations. In particular, the latter contain no 
constraints from energy--momentum conservation: the radiation off a
quark is described in the approximation that the quark does not
lose any energy when a gluon is radiated, so that the effects of 
multiple emissions factorize. Therefore energy--momentum conservation
is classified as a next-to-leading-log correction. In a Monte Carlo
shower, on the other hand, energy--momentum conservation is explicit
branching by branching. By including coherence phenomena and 
optimized choices of $\alphas$ scales, further information on
higher orders is inserted. While the final product is still not
certified fully to comply with a NLO/NLL standard, it is well above 
the level of an unsophisticated LO/LL analytic calculation.

\subsubsection{The evolution equations}
 
In the shower formulation, the kinematics of each branching is
given in terms of two variables, $Q^2$ and $z$. Slightly different
interpretations may be given to these variables, and indeed this is
one main area where the various programs on the market differ.
$Q^2$ has dimensions of squared mass, and is related to the
mass or transverse momentum scale of the branching. $z$ gives the
sharing of the $a$ energy and momentum between the two daughters,
with parton $b$ taking a fraction $z$ and parton $c$ a fraction
$1-z$. To specify the kinematics, an azimuthal angle
$\varphi$ of the $b$ around the $a$ direction is needed in addition; 
in the simple discussions $\varphi$ is chosen to be isotropically 
distributed, although options for non-isotropic distributions 
currently are the defaults.
 
The probability for a parton to branch is given by the evolution
equations (also called DGLAP or Altarelli--Parisi \cite{Gri72,Alt77}).
It is convenient to introduce
\begin{equation}
t = \ln(Q^2/\Lambda^2) ~~~ \Rightarrow ~~~
\d t = \d \ln(Q^2) = \frac{\d Q^2}{Q^2} ~,
\end{equation}
where $\Lambda$ is the QCD $\Lambda$ scale in $\alphas$. Of course,
this choice is more directed towards the QCD parts of the shower,
but it can be used just as well for the QED ones. In terms of the two
variables $t$ and $z$, the differential probability $\d {\cal P}$ for
parton $a$ to branch is now
\begin{equation}
\d {\cal P}_a = \sum_{b,c} \frac{\alpha_{abc}}{2 \pi} \,
P_{a \to bc}(z) \, \d t \, \d z ~.
\label{sh:Patobc}
\end{equation}
Here the sum is supposed to run over all allowed branchings,
for a quark $\q \to \q \g$ and $\q \to \q \gamma$, and so on. The
$\alpha_{abc}$ factor is $\alphaem$ for QED branchings and
$\alphas$ for QCD ones (to be evaluated at some suitable scale,
see below).
 
The splitting kernels $P_{a \to bc}(z)$ are
\begin{eqnarray}
 P_{\q \to \q\g}(z)&=&C_F \, \frac{1+z^2}{1-z} ~,
                     \nonumber \\
 P_{\g \to \g\g}(z)&=&N_C \, \frac{(1-z(1-z))^2}{z(1-z)} ~,
                     \nonumber \\
 P_{\g \to \q\qbar}(z)&=&T_R \, (z^2 + (1-z)^2) ~,
                     \nonumber \\
 P_{\q \to \q\gamma}(z)&=&e_{\q}^2 \, \frac{1+z^2}{1-z} ~,
                     \nonumber \\
 P_{\ell \to \ell\gamma}(z)&=&e_{\ell}^2 \, \frac{1+z^2}{1-z} ~,
\label{sh:APker}
\end{eqnarray}
with $C_F = 4/3$, $N_C = 3$, $T_R = n_f/2$ (i.e.\ $T_R$ receives
a contribution of $1/2$ for each allowed $\q\qbar$ flavour),
and $e_{\q}^2$ and $e_{\ell}^2$ the squared electric charge
($4/9$ for $\u$-type quarks, $1/9$ for $\d$-type ones, and 1 for
leptons).
 
Persons familiar with analytical calculations may wonder why the
`+ prescriptions' and $\delta(1-z)$ terms of the splitting kernels
in eq.~(\ref{sh:APker}) are missing. These complications fulfil
the task of ensuring flavour and energy conservation in the analytical
equations. The corresponding problem is solved trivially in
Monte Carlo programs, where the shower evolution is traced in detail,
and flavour and four-momentum are conserved at each branching.
The legacy left is the need to introduce a cut-off on the allowed range
of $z$ in splittings, so as to avoid the singular regions corresponding
to excessive production of very soft gluons.
 
Also note that $P_{\g \to \g\g}(z)$ is given here with a factor
$N_C$ in front, while it is sometimes shown with $2 N_C$. The
confusion arises because the final state contains two identical
partons. With the normalization above, $P_{a \to bc}(z)$
is interpreted as the branching probability for the original parton
$a$. On the other hand, one could also write down the probability
that a parton $b$ is produced with a fractional energy $z$. Almost
all the above kernels can be used unchanged also for this purpose,
with the obvious symmetry $P_{a \to bc}(z) = P_{a \to cb}(1-z)$.
For $\g \to \g\g$, however, the total probability to find a gluon
with energy fraction $z$ is the sum of the probability to find either
the first or the second daughter there, and that gives the factor of
2 enhancement.
 
\subsubsection{The Sudakov form factor}
\label{ss:sudakov}
 
The $t$ variable fills the function of a kind of time for the shower
evolution. In final-state showers, $t$ is constrained to be gradually
decreasing away from the hard scattering, in initial-state ones to
be gradually increasing towards the hard scattering.  This does not
mean that an individual parton runs through a range of $t$ values: 
in the end, each parton is associated with a fixed $t$ value, and the
evolution procedure is just a way of picking that value. It is only
the ensemble of partons in many events that evolves continuously with
$t$, cf. the concept of parton distributions.
 
For a given $t$ value we define the integral of the branching
probability over all allowed $z$ values,
\begin{equation}
{\cal I}_{a \to bc}(t) = \int_{z_{-}(t)}^{z_{+}(t)} \d z \,
\frac{\alpha_{abc}}{2 \pi} \, P_{a \to bc}(z) ~.
\end{equation}
The na\"{\i}ve probability that a branching occurs during a small range
of $t$ values, $\delta t$, is given by
$\sum_{b,c} {\cal I}_{a \to bc}(t) \, \delta t$, and thus the
probability for no emission by
$1 - \sum_{b,c} {\cal I}_{a \to bc}(t) \, \delta t$.
 
If the evolution of parton $a$ starts at a `time' $t_0$, the
probability that the parton has not yet branched at a `later time'
$t > t_0$ is given by the product of the probabilities that it did not
branch in any of the small intervals $\delta t$ between $t_0$ and $t$.
In other words, letting $\delta t \to 0$, the no-branching probability
exponentiates:
\begin{equation}
{\cal P}_{\mrm{no-branching}}(t_0,t) =
\exp \left\{ - \int_{t_0}^t \d t' \, \sum_{b,c}
{\cal I}_{a \to bc}(t') \right\} = S_a(t) ~.
\label{sh:Pnobranch}
\end{equation}
Thus the actual probability that a branching of $a$ occurs at $t$
is given by
\begin{equation}
\frac{\d {\cal P}_a}{\d t} =
- \frac{\d {\cal P}_{\mrm{no-branching}}(t_0,t)}{\d t} =
\left( \sum_{b,c} {\cal I}_{a \to bc}(t) \right)
\exp \left\{ - \int_{t_0}^t \d t' \, \sum_{b,c}
{\cal I}_{a \to bc}(t') \right\} ~.
\label{sh:Pbranch}
\end{equation}
 
The first factor is the na\"{\i}ve branching probability, the second
the suppression due to the conservation of total
probability: if a parton has already branched at a `time' $t' < t$,
it can no longer branch at $t$. This is nothing but the exponential
factor that is familiar from radioactive decay. In parton-shower
language the exponential factor
$S_a(t) = {\cal P}_{\mrm{no-branching}}(t_0,t)$ is referred to as
the Sudakov form factor \cite{Sud56}.
 
The ordering in terms of increasing $t$ above
is the appropriate one for initial-state showers. In 
final-state showers the evolution is from an initial $t_{\mmax}$ 
(set by the hard scattering) and towards smaller $t$. In that case 
the integral from $t_0$ to $t$ in eqs.~(\ref{sh:Pnobranch}) 
and (\ref{sh:Pbranch}) is replaced by an integral from $t$ 
to $t_{\mmax}$. Since, by convention,
the Sudakov factor is still defined from the lower cut-off $t_0$,
i.e.\ gives the probability that a parton starting at scale $t$ will
not have branched by the lower cut-off scale $t_0$, the no-branching
factor is actually 
${\cal P}_{\mrm{no-branching}}(t_{\mmax},t) = 
S_a(t_{\mmax})/S_a(t)$.
 
We note that the above structure is exactly of the kind discussed
in section \ref{ss:vetoalg}. The veto algorithm is therefore
extensively used in the Monte Carlo simulation of parton showers.
 
\subsubsection{Matching to the hard scattering}
\label{sss:showermatching}
 
The evolution in $Q^2$ is begun from some maximum scale
$Q_{\mmax}^2$ for final-state parton showers, and is terminated
at (a possibly different) $Q_{\mmax}^2$ for initial-state showers.
In general $Q_{\mmax}^2$ is not known. Indeed, since
the parton-shower language does not guarantee agreement with
higher-order matrix-element results, neither in absolute
shape nor normalization, there is no unique prescription for
a `best' choice. Generically $Q_{\mmax}$ should be of the order
of the hard-scattering scale, i.e.\ the largest virtuality
should be associated with the hard scattering, and initial- and
final-state parton showers should only involve virtualities
smaller than that. This may be viewed just as a matter of sound
book-keeping: in a $2 \to n$ graph, a $2 \to 2$ hard-scattering
subgraph could be chosen several different ways, but if all 
the possibilities were to be generated then the cross section
would be double-counted. Therefore one should define the $2 \to 2$
`hard' piece of a $2 \to n$ graph as the one that involves the 
largest virtuality. 
 
Of course, the issue of double-counting depends a bit on what
processes are actually generated in the program. If one
considers a $\q \qbar \g$ final state in hadron colliders,
this could come either as final-state radiation off a
$\q \qbar$ pair, or by a gluon splitting in a $\q \qbar$ pair,
or many other ways, so that the danger of double-counting is
very real. On the other hand, consider the production of a
low-$\pT$, low-mass Drell--Yan pair of leptons, together with two
quark jets. Such a process in principle could proceed by having
a $\gamma^*$ emitted off a quark leg, with a quark--quark
scattering as hard interaction. However, since this process is
not included in the program, there is no actual danger of
(this particular) double-counting, and so the scale of evolution
could be picked larger than the mass of the Drell--Yan pair, as
we shall see.
 
For most $2 \to 2$ scattering processes in {\Py}, the
$Q^2$ scale of the hard scattering is chosen to be
$Q_{\mrm{hard}}^2 = \pT^2$ (when the final-state particles are
massless, otherwise masses are added). In final-state showers, where
$Q$ is associated with the mass of the branching parton,
transverse momenta generated in the shower are constrained by
$\pT < Q/2$. An ordering that the shower $\pT$ should be smaller 
than the hard-scattering $\pT$ therefore corresponds roughly
to $Q_{\mmax}^2 = 4 Q_{\mrm{hard}}^2$, which is the default 
assumption. The constraints are slightly different for initial-state 
showers, where the spacelike virtuality $Q^2$ attaches better to 
$\pT^2$, and therefore $Q_{\mmax}^2 = Q_{\mrm{hard}}^2$ is a
sensible default. We iterate that these limits, set by \ttt{PARP(71)}
and \ttt{PARP(67)}, respectively, are imagined sensible when there 
is a danger of double counting; if not, large values could well be 
relevant to cover a wider range of topologies, but always with some 
caution. (See also \ttt{MSTP(68)}.)
 
The situation is rather better for the final-state showers in the
decay of any colour-singlet particles, or coloured but reasonably
long-lived ones, such as the $\Z^0$ or the
$\hrm^0$, either as part of a hard $2 \to 1 \to 2$ process, or
anywhere else in the final state. Then we know that $Q_{\mmax}$
has to be put equal to the particle mass. It is also possible
to match the parton-shower evolution to the first-order
matrix-element results. 
 
QCD processes such as $\q\g \to \q\g$ pose a special problem when 
the scattering angle is small. Coherence effects (see below) may
then restrict the emission further than what is just given by 
the $Q_{\mmax}$ scale introduced above. This is most easily viewed
in the rest frame of the $2 \to 2$ hard scattering subprocess.
Some colours flow from the initial to the final state. The radiation
associated with such a colour flow should be restricted to a cone 
with opening angle given by the difference between the original and 
the final colour directions; there is one such cone around the 
incoming parton for initial state radiation and one around the
outgoing parton for final state radiation. Colours that are 
annihilated or created in the process effectively correspond to an
opening angle of 180$^{\circ}$ and therefore the emission is not
constrained for these. For a gluon, which have two colours and 
therefore two different cones, a random choice is made between the
two for the first branching. Further, coherence effects also imply
azimuthal anisotropies of the emission inside the allowed cones.

Finally, we note that there can be some overlap between descriptions
of the same process. Section \ref{sss:WZclass} gives two examples.
One is the correspondence between the description of a single $\W$ or
$\Z$ with additional jet production by showering, or the same picture
obtained by using explicit matrix elements to generate at least one 
jet in association with the $\W/\Z$. The other is the generation of 
$\Z^0\b\bbar$ final states either starting from $\b\bbar \to \Z^0$,
or from $\b\g \to \Z^0\b$ or from $\g\g \to \b\bbar\Z^0$. As a rule
of thumb, to be used with common sense, one would start from as low 
an order as possible for an inclusive description, where the low-$\pT$ 
region is likely to generate most of the cross section, whereas 
higher-order topologies are more relevant for studies of exclusive 
event samples at high $\pT$.
  
\subsection{Final-State Showers}
 
Final-state showers are time-like, i.e.\ all virtualities
$m^2 = E^2 - \mbf{p}^2 \geq 0$. The maximum allowed virtuality
scale $Q^2_{\mmax}$ is set by the hard-scattering process, and
thereafter the virtuality is decreased in each subsequent
branching, down to the cut-off scale $Q_0^2$. This cut-off scale
is used to regulate both soft and collinear divergences in the
emission probabilities.

Many different approaches can be chosen for the parton-shower 
algorithm, e.g.\ in terms of evolution variables, kinematics 
reconstruction and matrix-element corrections. The traditional 
approach in {\Py} is the mass-ordered \ttt{PYSHOW} algorithm.
As an alternative, a $\pT$-ordered \ttt{PYPTFS} algorithm has
recently been introduced. It is not yet fully tried out, and is 
described only briefly at the end of this section.
 
The main points of the \ttt{PYSHOW} showering algorithm are as 
follows.
\begin{Itemize}
\item It is a leading-log algorithm, of the improved, coherent kind,
i.e.\ with angular ordering.
\item It can be used for an arbitrary initial pair of partons
or, in fact, for any number between one and eighty given entities 
(including hadrons and gauge bosons) although only quarks, gluons, 
leptons, squarks and gluinos can initiate a shower.
\item The pair of showering partons may be given in any frame, but
the evolution is carried out in the c.m.\ frame of the showering
partons.
\item Energy and momentum are conserved exactly at each step of the
showering process.
\item If the initial pair comes from the decay of a known resonance, 
an additional rejection technique is used in the gluon emission 
off a parton of the pair, so as to reproduce the lowest-order 
differential 3-jet cross section.
\item In subsequent branchings, angular
ordering (coherence effects) is imposed.
\item Gluon helicity effects, i.e.\ correlations between the
production plane and the decay plane of a gluon, can be included.
\item The first-order
$\alphas$ expression is used, with the $Q^2$ scale given by (an
approximation to) the squared transverse momentum of a branching.
The default $\Lambda_{\mrm{QCD}}$,
which should not be regarded as a proper 
$\Lambda_{\br{\mrm{MS}}}$, is 0.29 GeV.
\item The parton shower is by default
cut off at a mass scale of 1 GeV.
\end{Itemize}
Let us now proceed with a more detailed description.
 
\subsubsection{The choice of evolution variable}
 
In the {\Py} shower algorithm, the evolution variable $Q^2$ is
associated with the squared mass of the branching parton,
$Q^2 = m_a^2$ for a branching $a \to bc$. As a consequence,
$t = \ln(Q^2/\Lambda^2) = \ln(m_a^2/\Lambda^2)$.
This $Q^2$ choice is not unique, and indeed other programs have
other definitions: \tsc{Herwig} uses $Q^2 \approx m^2/(2z(1-z))$
\cite{Mar88} and \tsc{Ariadne}
$Q^2 = \pT^2 \approx z(1-z)m^2$ \cite{Pet88}. Below we will also
modify the $Q^2$ choice to give a better account of mass effects,
e.g.\ for $\b$ quarks.
 
With $Q$ a mass scale, the lower cut-off $Q_0$ is one in mass.
To be more precise, in a QCD shower, the $Q_0$ parameter is used
to derive effective masses
\begin{eqnarray}
m_{\mrm{eff},\g} & = & \frac{1}{2} Q_0 ~, \nonumber \\
m_{\mrm{eff},\q} & = & \sqrt{ m_{\q}^2 + \frac{1}{4} Q_0^2 } ~,
\label{ps:meff}
\end{eqnarray}
where the $m_{\q}$ have been chosen as typical kinematical quark
masses, see section \ref{sss:massdef}. A parton cannot branch unless 
its mass is at least the sum
of the lightest pair of allowed decay products, i.e.\ the minimum
mass scale at which a branching is possible is
\begin{eqnarray}
m_{\mmin,\g} & = & 2 \, m_{\mrm{eff},\g} = Q_0 ~, \nonumber \\
m_{\mmin,\q} & = & m_{\mrm{eff},\q} + m_{\mrm{eff},\g} \geq Q_0 ~.
\label{ps:mmin}
\end{eqnarray}
The above masses are used to constrain the allowed range of
$Q^2$ and $z$ values. However, once it has been decided that a
parton cannot branch any further, that parton is put on the
mass shell, i.e.\ `final-state' gluons are massless.
 
When also photon emission is included, a separate $Q_0$ scale is
introduced for the QED part of the shower, and used to calculate
cut-off masses by analogy with eqs.~(\ref{ps:meff}) and 
(\ref{ps:mmin}) above \cite{Sjo92c}. By default the two $Q_0$ scales 
are chosen equal, and have the value 1 GeV. If anything, one would 
be inclined to allow a cut-off lower for photon emission than for
gluon one. In that case the allowed $z$ range of photon emission
would be larger than that of gluon emission, and at the end of
the shower evolution only photon emission would be allowed.
 
Photon and gluon emission differ fundamentally in that photons appear
as physical particles in the final state, while gluons are confined.
For photon emission off quarks, however, the confinement forces
acting on the quark may provide an effective photon emission
cut-off at larger scales than the bare quark mass. Soft and collinear
photons could also be emitted by the final-state charged hadrons
\cite{Bar94a}; the matching between emission off quarks and off hadrons 
is a delicate issue, and we therefore do not attempt to address the 
soft-photon region.
 
For photon emission off leptons, there is no need to introduce any
collinear emission cut-off beyond what is given by the lepton mass,
but we keep the same cut-off approach as for quarks, although at a 
smaller scale. However, note that, firstly,
the program is not aimed at high-precision studies of lepton
pairs (where interference terms between initial- and final-state
radiation also would have to be included), and, secondly, most 
experimental procedures would include the energy of collinear 
photons into the effective energy of a final-state lepton.
 
\subsubsection{The choice of energy splitting variable}
 
The final-state radiation
machinery is always applied in the c.m.\ frame of the hard scattering,
from which normally emerges a pair of evolving partons.
Occasionally there may be one evolving parton recoiling against a
non-evolving one, as in $\q \qbar \to \g \gamma$, where only the
gluon evolves in the final state, but where the energy of the photon
is modified by the branching activity of the gluon. (With only one
evolving parton and nothing else, it would not be possible to
conserve energy and momentum when the parton is assigned a mass.)
Thus, before the evolution is performed, the parton pair is boosted
to their common c.m.\ frame, and rotated to sit along the $z$ axis.
After the evolution, the full parton shower is rotated and boosted
back to the original frame of the parton pair.
 
The interpretation of the energy and momentum splitting variable
$z$ is not unique, and in fact the program allows the possibility
to switch between four different alternatives \cite{Ben87a},
`local' and `global' $z$ definition combined with `constrained'
or `unconstrained' evolution. In all
four of them, the $z$ variable is interpreted as an energy fraction,
i.e.\ $E_b = z E_a$ and $E_c = (1-z) E_a$. In the `local' choice of
$z$ definition, energy fractions are defined in the rest frame
of the grandmother, i.e.\ the mother of parton $a$. The preferred
choice is the `global' one, in which energies are always evaluated
in the c.m.\ frame of the hard scattering. The two definitions agree
for the branchings of the partons that emerge directly from the hard
scattering, since the hard scattering itself is considered to be the
`mother' of the first generation of partons. For instance, in
$\Z^0 \to \q\qbar$ the $\Z^0$ is considered the mother of the $\q$
and $\qbar$, even though the branching is not handled by the 
parton-showering machinery. The `local' and `global' definitions 
diverge for subsequent branchings, where the `global' tends to allow 
more shower evolution.
 
In a branching $a \to bc$ the kinematically allowed range of
$z = z_a$ values, $z_{-} < z < z_{+}$, is given by
\begin{equation}
z_{\pm} = \frac{1}{2} \left\{ 1 + \frac{m_b^2 - m_c^2}{m_a^2} \pm
\frac{|\mbf{p}_a|}{E_a} \, \frac{\sqrt{ (m_a^2 - m_b^2 - m_c^2)^2
- 4 m_b^2 m_c^2}}{m_a^2} \right\} ~.
\label{sh:zbounds}
\end{equation}
With `constrained' evolution, these bounds are respected in the
evolution. The cut-off masses $m_{\mrm{eff},b}$ and $m_{\mrm{eff},c}$ 
are used to
define the maximum allowed $z$ range, within which $z_a$ is chosen,
together with the $m_a$ value. In the subsequent evolution of $b$ and
$c$, only pairs of $m_b$ and $m_c$ are allowed for which the
already selected $z_a$ fulfils the constraints in 
eq.~(\ref{sh:zbounds}).
 
For `unconstrained' evolution, which is the preferred alternative,
one may start off by assuming the daughters to be massless, so that the
allowed $z$ range is
\begin{equation}
z_{\pm} = \frac{1}{2} \left\{ 1 \pm \frac{|\mbf{p}_a|}{E_a}
\theta(m_a - m_{\mmin,a}) \right\} ~,
\label{sh:zrange}
\end{equation}
where $\theta(x)$ is the step function, $\theta(x) = 1$ for $x > 0$
and $\theta(x) = 0$ for $x < 0$. The decay kinematics into two
massless four-vectors $p_b^{(0)}$ and $p_c^{(0)}$ is now
straightforward. Once $m_b$ and $m_c$ have been found from the
subsequent evolution, subject only to the constraints
$m_b < z_a E_a$, $m_c < (1-z_a) E_a$ and $m_b + m_c < m_a$,
the actual massive four-vectors may be defined as
\begin{equation}
p_{b,c} = p_{b,c}^{(0)} \pm (r_c p_c^{(0)} - r_b p_b^{(0)}) ~,
\label{sh:pshowershift}
\end{equation}
where
\begin{equation}
r_{b,c} = \frac{m_a^2 \pm (m_c^2 -m_b^2) -
\sqrt{ (m_a^2 - m_b^2 - m_c^2)^2 - 4 m_b^2 m_c^2}}{2 m_a^2} ~.
\end{equation}
In other words, the meaning of $z_a$ is somewhat reinterpreted
{\it post facto}. Needless to say, the `unconstrained' option allows
more branchings to take place than the `constrained' one.
In the following discussion we will only refer to the
`global, unconstrained' $z$ choice.
 
\subsubsection{First branchings and matrix-element matching}
 
The final-state evolution is normally started from some initial
parton pair $1 + 2$, at a $Q_{\mmax}^2$ scale determined by
deliberations already discussed. When the evolution of parton 1
is considered, it is assumed that parton 2 is massless,
so that the parton 1 energy and momentum are simple
functions of its mass (and of the c.m.\ energy of the pair, which is
fixed), and hence also the allowed $z_1$ range for splittings is a
function of this mass,
eq.~(\ref{sh:zrange}). Correspondingly, parton 2 is evolved under the
assumption that parton 1 is massless. After both partons have been
assigned masses, their correct energies may be found, which are
smaller than originally assumed. Therefore the allowed $z$ ranges
have shrunk, and it may happen that a branching has been assigned
a $z$ value outside this range. If so, the parton is evolved
downwards in mass from the rejected mass value; if both $z$ values
are rejected, the parton with largest mass is evolved further.
It may also happen that the sum of $m_1$ and $m_2$ is larger
than the c.m.\ energy, in which case the one with the larger mass
is evolved downwards. The checking and evolution steps are
iterated until an acceptable set of $m_1$, $m_2$, $z_1$
and $z_2$ has been found.
 
The procedure is an extension of the veto
algorithm, where an initial overestimation of the allowed $z$
range is compensated by rejection of some branchings. One should
note, however, that the veto algorithm is not strictly
applicable for the coupled evolution in two variables ($m_1$
and $m_2$), and that therefore some arbitrariness is involved.
This is manifest in the choice of which parton will be evolved
further if both $z$ values are unacceptable, or if the mass sum
is too large.
 
For quark and lepton pairs which come from the decay of a
colour-singlet particle, the first branchings are matched to
the explicit first-order matrix elements for gauge boson decays.
 
The matching is based on a mapping of the parton-shower variables
on to the 3-jet phase space. To produce a 3-jet event,
$\gammaZ \to \q(p_1) \qbar(p_2) \g(p_3)$, in the shower language,
one will pass through an intermediate
state, where either the $\q$ or the $\qbar$ is off the mass shell.
If the former is the case then
\begin{eqnarray}
m^2 & = & (p_1 + p_3)^2 = E_{\mrm{cm}}^2 (1 - x_2) ~, \nonumber \\
z   & = & \frac{E_1}{E_1 + E_3} = \frac{x_1}{x_1 + x_3} =
\frac{x_1}{2-x_2} ~,
\label{sh:MEfrPS}
\end{eqnarray}
where $x_i = 2 E_i/E_{\mrm{cm}}$. The $\qbar$ emission case is obtained
with $1 \leftrightarrow 2$. The parton-shower splitting expression
in terms of $m^2$ and $z$, eq.~(\ref{sh:Patobc}), can therefore be
translated into the following differential 3-jet rate:
\begin{eqnarray}
\frac{1}{\sigma} \, \frac{\d \sigma_{\mrm{PS}}}{\d x_1 \, \d x_2} 
& = & \frac{\alphas}{2 \pi} \, C_F \, \frac{1}{(1-x_1)(1-x_2)}
\times
\nonumber \\
& \times &
\left\{ \frac{1-x_1}{x_3} \left( 1 + \left( \frac{x_1}{2-x_2}
\right)^2 \right) + \frac{1-x_2}{x_3} \left( 1 +
\left( \frac{x_2}{2-x_1} \right)^2 \right) \right\} ~,
\label{sh:PSwt}
\end{eqnarray}
where the first term inside the curly bracket comes from emission
off the quark and the second term from emission off the antiquark.
The corresponding expression in matrix-element language is
\begin{equation}
\frac{1}{\sigma} \, \frac{\d \sigma_{\mrm{ME}}}{\d x_1 \, \d x_2} =
\frac{\alphas}{2 \pi} \, C_F \, \frac{1}{(1-x_1)(1-x_2)}
\left\{ x_1^2 +x_2^2 \right\} ~.
\label{sh:MEwt}
\end{equation}
With the kinematics choice of {\Py},
the matrix-element expression is always smaller than the parton-shower
one. It is therefore possible to run the shower as usual,
but to impose an extra weight factor 
$\d \sigma_{\mrm{ME}} / \d \sigma_{\mrm{PS}}$,
which is just the ratio of the expressions in curly brackets.
If a branching is rejected, the evolution is continued from the
rejected $Q^2$ value onwards (the veto algorithm). The weighting
procedure is applied to the first branching of both the $\q$ and the
$\qbar$, in each case with the (nominal) assumption that none of
the other partons branch (neither the sister nor the daughters),
so that the relations of eq.~(\ref{sh:MEfrPS}) are applicable.
 
If a photon is emitted instead of a gluon, the emission rate in
parton showers is given by
\begin{eqnarray}
\frac{1}{\sigma} \, \frac{\d \sigma_{\mrm{PS}}}{\d x_1 \, \d x_2} 
& = & \frac{\alphaem}{2 \pi} \, \frac{1}{(1-x_1)(1-x_2)} 
\times \nonumber \\
& \times &
\left\{ e_{\q}^2 \, \frac{1-x_1}{x_3} \left( 1 + \left(
\frac{x_1}{2-x_2} \right)^2 \right) + e_{\qbar}^2 \, \frac{1-x_2}{x_3}
\left( 1 + \left( \frac{x_2}{2-x_1} \right)^2 \right) \right\} ~,
\end{eqnarray}
and in matrix elements by \cite{Gro81}
\begin{equation}
\frac{1}{\sigma} \, \frac{\d \sigma_{\mrm{ME}}}{\d x_1 \, \d x_2} =
\frac{\alphaem}{2 \pi} \, \frac{1}{(1-x_1)(1-x_2)}
\left\{ \left( e_{\q} \, \frac{1-x_1}{x_3} - e_{\qbar} \,
\frac{1-x_2}{x_3} \right)^2 \left( x_1^2 +x_2^2 \right) \right\} ~.
\end{equation}
As in the gluon emission case, a weighting factor
$\d \sigma_{\mrm{ME}} / \d \sigma_{\mrm{PS}}$ can therefore be applied 
when either the original $\q$ ($\ell$) or the original $\qbar$
($\br{\ell}$) emits a photon. For a
neutral resonance, such as $\Z^0$, where $e_{\qbar} = - e_{\q}$,
the above expressions simplify and one recovers
exactly the same ratio $\d \sigma_{\mrm{ME}} / \d \sigma_{\mrm{PS}}$ 
as for gluon emission.
 
Compared with the standard matrix-element treatment, a few
differences remain. The shower one automatically contains the
Sudakov form factor and an $\alphas$ running as a function
of the $\pT^2$ scale of the branching.
The shower also allows all partons to evolve
further, which means that the na\"{\i}ve kinematics assumed for a
comparison with matrix elements is modified by subsequent
branchings, e.g.\ that the energy of parton 1 is reduced when
parton 2 is assigned a mass. All these effects are formally of
higher order, and so do not affect a first-order comparison.
This does not mean that the corrections need be small, but experimental
results are encouraging: the approach outlined does as
good as explicit second-order matrix elements for the description
of 4-jet production, better in some respects (like overall rate) 
and worse in others (like some angular distributions).
 
\subsubsection{Subsequent branches and angular ordering}
 
The shower evolution is (almost) always done on a pair of partons,
so that energy and momentum can be conserved. In the first
step of the evolution, the two original partons thus undergo
branchings $1 \to 3 + 4$ and $2 \to 5 + 6$. As described
above, the allowed $m_1$, $m_2$, $z_1$ and $z_2$ ranges
are coupled by kinematical constraints. In the second step, 
the pair $3 + 4$ is evolved and,
separately, the pair $5 + 6$. Considering only the former (the
latter is trivially obtained by symmetry), the partons thus have
nominal initial energies $E_3^{(0)} = z_1 E_1$ and
$E_4^{(0)} = (1-z_1) E_1$, and maximum allowed virtualities
$m_{\mmax,3} = \min(m_1,E_3^{(0)})$ and
$m_{\mmax,4} = \min(m_1,E_4^{(0)})$. Initially partons 3 and 4 are
evolved separately, giving masses $m_3$ and $m_4$ and splitting
variables $z_3$ and $z_4$. If $m_3 + m_4 > m_1$,
the parton of 3 and 4 that has the largest ratio of 
$m_i/m_{\mmax,i}$ is
evolved further. Thereafter eq.~(\ref{sh:pshowershift}) is used to
construct corrected energies $E_3$ and $E_4$, and the $z$
values are checked for consistency. If a branching has to be
rejected because the change of parton energy puts $z$ outside the
allowed range, the parton is evolved further.
 
This procedure can then be iterated for the evolution of the two
daughters of parton 3 and for the two of parton 4, etc., until each
parton reaches the cut-off mass $m_{\mmin}$. Then the parton is put on
the mass shell.
 
The model, as described so far, produces so-called conventional
showers, wherein masses are strictly decreasing in the shower
evolution. Emission angles are decreasing only in an average sense,
however, which means that also fairly `late' branchings can give
partons at large angles. Theoretical studies beyond the leading-log
level show that this is not correct \cite{Mue81}, but that destructive
interference effects are large in the region of non-ordered
emission angles. To a good first approximation, these so-called
coherence effects can be taken into account in parton shower
programs by requiring a strict ordering in terms of decreasing
emission angles. (Actually, the fact that the shower described here
is already ordered in mass implies that the additional cut on angle
will be a bit too restrictive. While effects from this should be small
at current energies, some deviations become visible at very high
energies.) 
 
The coherence phenomenon is known already from QED. One
manifestation is the Chudakov effect \cite{Chu55}, discovered
in the study of high-energy cosmic $\gamma$ rays impinging on a
nuclear target. If a $\gamma$ is
converted into a highly collinear $\ee$ pair inside the
emulsion, the $\e^+$ and $\e^-$ in their travel through the
emulsion ionize atoms and thereby produce blackening.
However, near the conversion point the blackening is small:
the $\e^+$ and $\e^-$ then are still close together, so that
an atom traversed by the pair does not resolve the individual
charges of the $\e^+$ and the $\e^-$, but only feels a net
charge close to zero. Only later,
when the $\e^+$ and $\e^-$ are separated by more than a typical
atomic radius, are the two able to ionize independently of
each other.
 
The situation is similar in QCD, but is further extended, since
now also gluons carry colour. For example, in a branching
$\q_0 \to \q\g$ the $\q$ and $\g$ share the newly created pair of
opposite colour--anticolour charges, and therefore the $\q$ and $\g$ 
cannot emit subsequent gluons
incoherently. Again the net effect is to reduce the amount of soft
gluon emission: since a soft gluon (emitted at large angles)
corresponds to a large (transverse) wavelength,
the soft gluon is unable to resolve the separate colour charges
of the $\q$ and the $\g$, and only feels the net charge carried by
the $\q_0$. Such a soft gluon $\g'$ (in the region
$\theta_{\q_0 \g'} > \theta_{\q \g}$)
could therefore be thought of as being
emitted by the $\q_0$ rather than by the $\q$--$\g$ system.
If one considers only emission that should be associated with the
$\q$ or the $\g$, to a good approximation, there is
a complete destructive interference in the
regions of non-decreasing opening angles, while partons
radiate independently of each other inside the regions
of decreasing opening angles ($\theta_{q g'} < \theta_{q g}$ and
$\theta_{g g'} < \theta_{q g}$), once azimuthal angles are averaged
over. The details of the colour interference pattern are
reflected in non-uniform azimuthal emission probabilities.
 
The first branchings of the shower are not affected by the
angular-ordering requirement --- since the evolution is performed
in the c.m.\ frame of the original parton pair, where the original
opening angle is  180$^{\circ}$, any angle would anyway be smaller
than this --- but here instead the matrix-element matching procedure
is used, where applicable. Subsequently, each opening angle is
compared with that of the preceding branching in the shower.
 
For a branching $a \to bc$ the kinematical approximation
\begin{equation}
\theta_a \approx \frac{p_{\perp b}}{E_b} + \frac{p_{\perp c}}{E_c}
\approx \sqrt{z_a (1-z_a)} m_a \left( \frac{1}{z_a E_a} +
\frac{1}{(1-z_a) E_a} \right) = \frac{1}{\sqrt{z_a(1-z_a)}}
\frac{m_a}{E_a}
\end{equation}
is used to derive the opening angle (this is anyway to the same level
of approximation as the one in which angular ordering is derived).
With $\theta_b$ of the $b$ branching calculated similarly, the
requirement $\theta_b < \theta_a$ can be reduced to
\begin{equation}
\frac{z_b (1-z_b)}{m_b^2} > \frac{1-z_a}{z_a m_a^2} ~.
\end{equation}
 
Since photons do not obey angular ordering, the check on angular
ordering is not performed when a photon is emitted.
When a gluon is emitted in the branching after a photon, its emission
angle is restricted by that of the preceding QCD branching in the
shower, i.e.\ the photon emission angle does not enter.
 
\subsubsection{Other final-state shower aspects}
 
The electromagnetic coupling constant for the emission of photons
on the mass shell is
$\alphaem = \alphaem(Q^2 = 0) \approx 1/137$. For 
the strong coupling constant several alternatives are available, the 
default being the first-order expression $\alphas(\pT^2)$, where
$\pT^2$ is defined by the approximate expression
$\pT^2 \approx z(1-z) m^2$. Studies of next-to-leading-order
corrections favour this choice \cite{Ama80}. The other alternatives
are a fixed $\alphas$ and an $\alphas(m^2)$.
 
With the default choice of $\pT^2$ as scale in $\alphas$,
a further cut-off is introduced on the allowed phase space
of gluon emission, not present in the options with fixed
$\alphas$ or with $\alphas(m^2)$, nor in the QED shower.
A minimum requirement, to ensure a well-defined $\alphas$,
is that $\pT / \Lambda > 1.1$, but additionally
{\Py} requires that $\pT > Q_0/2$. This latter
requirement is not a necessity, but it makes sense when
$\pT$ is taken to be the preferred scale of the branching
process, rather than e.g.\ $m$. It reduces the allowed $z$ range,
compared with the purely kinematical constraints.
Since the $\pT$ cut is not
present for photon emission, the relative ratio of photon to gluon
emission off a quark is enhanced at small virtualities compared with
na\"{\i}ve expectations; in actual fact this enhancement is largely
compensated by the running of $\alphas$, which acts in the
opposite direction. The main consequence, however, is that the
gluon energy spectrum is peaked at around $Q_0$ and rapidly
vanishes for energies below that, whilst the photon spectrum
extends all the way to zero energy.
 
Previously it was said that azimuthal angles in branchings are
chosen isotropically. In fact, it is possible to
include some effects of gluon polarization, which correlate the
production and the decay planes of a gluon, such that a
$\g \to \g\g$ branching tends to take place in the production
plane of the gluon, while a decay out of the plane is favoured
for $\g \to \q\qbar$. The formulae are given e.g.\ in ref.
\cite{Web86}, as simple functions of the $z$ value at the
vertex where the gluon is produced and of the $z$ value when
it branches. Also coherence phenomena lead to non-isotropic
azimuthal distributions \cite{Web86}. In either case the $\varphi$ 
azimuthal variable is first chosen isotropically, then the weight 
factor due to polarization times coherence is evaluated, and the 
$\varphi$ value is accepted or rejected. In case of rejection,
a new $\varphi$ is generated, and so on.
 
While the rule is to have an initial pair of partons, there are
a few examples where one or three partons have to be allowed to
shower. If only one parton is given, it is not possible to
conserve both energy and momentum. The choice has been made
to conserve energy and jet direction, but the momentum vector is
scaled down when the radiating parton acquires a mass. The `rest
frame of the system', used e.g.\ in the $z$ definition, is taken to
be whatever frame the jet is given in.
 
In $\Upsilon \to \g\g\g$ decays and other configurations (e.g.\ 
from external processes) with three or more primary parton, one is 
left with the issue how the kinematics from the on-shell matrix 
elements should be reinterpreted for an off-shell multi-parton 
configuration. We have made the arbitrary choice of preserving 
the direction of motion of each parton in the rest frame of the 
system, which means that all three-momenta are scaled down by the 
same amount, and that some particles gain energy at the expense
of others. Mass multiplets outside the allowed phase space are
rejected and the evolution continued.
 
Finally, it should be noted that two toy shower models are
included as options. One is a scalar gluon model, in which the
$\q \to \q\g$ branching kernel is replaced by
$P_{\q \to \q\g}(z) = \frac{2}{3} (1-z)$. The couplings of the
gluon, $\g \to \g\g$ and $\g \to \q\qbar$, have been left as free
parameters, since they depend on the colour structure assumed in the
model. The spectra are flat in $z$ for a spin 0 gluon.
Higher-order couplings of the type $\g \to \g\g\g$ could well
contribute significantly, but are not included.
The second toy model is an Abelian vector one. In this
option $\g \to \g \g$ branchings are absent, and $\g \to \q \qbar$
ones enhanced. More precisely, in the splitting kernels, 
eq.~(\ref{sh:APker}), the Casimir factors are changed as follows:
$C_F = 4/3 \to 1$, $N_C = 3 \to 0$, $T_R = n_f/2 \to 3n_f$.
When using either of these options, one should be aware that also
a number of other components in principle should be changed, from
the running of $\alphas$ to the whole concept of fragmentation.
One should therefore not take them too seriously.
 
\subsubsection{Merging with massive matrix elements}
\label{sss:PSMEfsmerge}

The matching to first-order matrix-elements is well-defined for 
massless quarks, and was originally used unchanged for massive ones. 
A first attempt to include massive matrix elements did not compensate 
for mass effects in the shower kinematics, and therefore came to 
exaggerate the suppression of radiation off heavy quarks 
\cite{Nor01,Bam00}. Now the shower has been modified 
to solve this issue, and also improved and extended to cover 
better a host of different reactions \cite{Nor01}. 

\begin{table}[t]
\captive{The processes that have been calculated, also with one
extra gluon in the final state. Colour is given with 1 for singlet, 
3 for triplet and 8 for octet. See the text for an explanation of 
the $\gamma_5$ column and further comments.
\protect\label{t:massivefinshow}}\\
\vspace{1ex}
\begin{center}
\begin{tabular}{|c|c|c|c|c|@{\protect\rule[-2mm]{0mm}{7mm}}}
\hline
colour & spin & $\gamma_5$ & example & codes \\
\hline
$1 \to 3 + \br{3}$ & --- & --- & (eikonal) & 6 -- 9 \\
$1 \to 3 + \br{3}$ & $1 \to \frac{1}{2} + \frac{1}{2}$ & 
$1,\gamma_5,1\pm\gamma_5$ & $\Z^0 \to \q\qbar$ & 11 -- 14 \\
$3 \to 3 + 1$ & $\frac{1}{2} \to \frac{1}{2} + 1$ & 
$1,\gamma_5,1\pm\gamma_5$ & $\t \to \b\W^+$ & 16 -- 19 \\
$1 \to 3 + \br{3}$ & $0 \to \frac{1}{2} + \frac{1}{2}$ & 
$1,\gamma_5,1\pm\gamma_5$ & $\hrm^0 \to \q\qbar$ & 21 -- 24 \\
$3 \to 3 + 1$ & $\frac{1}{2} \to \frac{1}{2} + 0$ & 
$1,\gamma_5,1\pm\gamma_5$ & $\t \to \b\H^+$ & 26 -- 29 \\
$1 \to 3 + \br{3}$ & $1 \to 0 + 0$ & 
$1$ & $\Z^0 \to \sq\sqbar$ & 31 -- 34 \\
$3 \to 3 + 1$ & $0 \to 0 + 1$ & 
$1$ & $\sq \to \sq'\W^+$ & 36 -- 39 \\
$1 \to 3 + \br{3}$ & $0 \to 0 + 0$ & 
$1$ & $\hrm^0 \to \sq\sqbar$ & 41 -- 44 \\
$3 \to 3 + 1$ & $0 \to 0 + 0$ & 
$1$ & $\sq \to \sq'\H^+$ & 46 -- 49 \\
$1 \to 3 + \br{3}$ & $\frac{1}{2} \to \frac{1}{2} + 0$ & 
$1,\gamma_5,1\pm\gamma_5$ & $\tilde{\chi} \to \q\sqbar$ & 51 -- 54 \\
$3 \to 3 + 1$ & $0 \to \frac{1}{2} + \frac{1}{2}$ & 
$1,\gamma_5,1\pm\gamma_5$ & $\sq \to \q\tilde{\chi}$ & 56 -- 59 \\
$3 \to 3 + 1$ & $\frac{1}{2} \to 0 + \frac{1}{2}$ & 
$1,\gamma_5,1\pm\gamma_5$ & $\t \to \st\tilde{\chi}$ & 61 -- 64 \\
$8 \to 3 + \br{3}$ & $\frac{1}{2} \to \frac{1}{2} + 0$ & 
$1,\gamma_5,1\pm\gamma_5$ & $\sg \to \q\sqbar$ & 66 -- 69 \\
$3 \to 3 + 8$ & $0 \to \frac{1}{2} + \frac{1}{2}$ & 
$1,\gamma_5,1\pm\gamma_5$ & $\sq \to \q\sg$ & 71 -- 74 \\
$3 \to 3 + 8$ & $\frac{1}{2} \to 0 + \frac{1}{2}$ & 
$1,\gamma_5,1\pm\gamma_5$ & $\t \to \st\sg$ & 76 -- 79 \\
$1 \to 8 + 8$ & --- & --- & (eikonal) & 81 -- 84 \\
\hline
\end{tabular}
\end{center}
\end{table}

The starting point is the calculation of processes
$a \to bc$ and $a \to bc\g$, where the ratio
\begin{equation}
W_{\mrm{ME}}(x_1,x_2) =
\frac{1}{\sigma(a \to bc)} \, 
\frac{\d\sigma(a \to bc\g)}{\d x_1 \, \d x_2}
\label{eq:Wmefinshow}
\end{equation} 
gives the process-dependent differential gluon-emission rate. 
Here the phase space variables are $x_1 = 2E_b/m_a$ and  
$x_2 = 2E_c/m_a$, expressed in the rest frame of particle $a$.
Using the standard model and the minimal supersymmetric extension
thereof as templates, a wide selection of colour and spin structures
have been addressed, as shown in Table~\ref{t:massivefinshow}.
When allowed, processes have been calculated for an arbitrary mixture 
of `parities', i.e. without or with a $\gamma_5$ factor, like in the 
vector/axial vector structure of $\gammaZ$. Various combinations of 1 
and $\gamma_5$ may also arise e.g.\ from the wave functions of the 
sfermion partners to the left- and right-handed fermion states. In 
cases where the correct combination is not provided, an equal mixture 
of the two is assumed as a reasonable compromise. All the matrix 
elements are encoded in the new function 
\texttt{PYMAEL(NI,X1,X2,R1,R2,ALPHA)}, where \texttt{NI} 
distinguishes the matrix elements, \texttt{ALPHA} is related to the 
$\gamma_5$ admixture and the mass ratios $r_1 = m_b / m_a$ and 
$r_2 = m_c/m_a$ are free parameters. This routine is 
called by \ttt{PYSHOW}, but might also have an interest on its own.

In order to match to the singularity structure of the massive matrix 
elements, the evolution variable $Q^2$ is changed from $m^2$ to 
$m^2 - m_{\mrm{on-shell}}^2$, i.e.\ $1/Q^2$ is the propagator of a 
massive particle \cite{Nor01}.  For the shower history 
$b \to b\g$ this gives a differential probability
\begin{equation}
W_{\mrm{PS,1}}(x_1,x_2) 
= \frac{\alphas}{2\pi} \, C_F \, \frac{\d Q^2}{Q^2} \, 
\frac{2 \, \d z}{1-z} \, \frac{1}{\d x_1 \, \d x_2}
= \frac{\alphas}{2\pi} \, C_F \,
\frac{2}{x_3 \, (1 + r_2^2 - r_1^2 - x_2)}  ~,
\end{equation} 
where the numerator $1 + z^2$ of the splitting kernel for $\q \to \q \g$ 
has been replaced by a 2 in the shower algorithm. For a process with only 
one radiating parton in the final state, such as $\t \to \b\W^+$, the 
ratio $W_{\mrm{ME}}/W_{\mrm{PS,1}}$ gives the acceptance probability 
for an emission in the shower. The singularity structure exactly agrees 
between ME and PS, giving a well-behaved ratio always below unity. If both 
$b$ and $c$ can radiate, there is a second possible shower history that has 
to be considered. The matrix element is here split in two parts, one 
arbitrarily associated with $b \to b\g$ branchings and the other with 
$c \to c\g$ ones. A convenient choice is 
$W_{\mrm{ME,1}} = W_{\mrm{ME}} (1 + r_1^2 - r_2^2 - x_1)/x_3$ and 
$W_{\mrm{ME,2}} = W_{\mrm{ME}} (1 + r_2^2 - r_1^2 - x_2)/x_3$,
which again gives matching singularity structures in 
$W_{\mrm{ME,}i}/W_{\mrm{PS,}i}$ and thus a
well-behaved Monte Carlo procedure. 

Also subsequent emissions of gluons off the primary particles are 
corrected to $W_{\mrm{ME}}$. To this 
end, a reduced-energy system is constructed, which retains the 
kinematics of the branching under consideration but omits the gluons 
already emitted, so that an effective three-body shower state can be 
mapped to an $(x_1, x_2, r_1, r_2)$ set of variables. For light quarks 
this procedure is almost equivalent with the original one of using the  
simple universal splitting kernels after the first branching. For
heavy quarks it offers an improved modelling of mass effects also in 
the collinear region.

Some further changes have been introduced, a few minor as default and
some more significant ones as non-default options \cite{Nor01}. 
This includes the description of coherence effects and $\alphas$ 
arguments, in general and more specifically for secondary heavy flavour
production by gluon splittings. The problem in the latter area is that
data at LEP1 show a larger rate of secondary charm and bottom production
than predicted in most shower descriptions \cite{Bam00,Man00}, or in
analytical studies \cite{Sey95}. This is based on applying the same kind
of coherence considerations to $\g\to\q\qbar$ branchings as to
$\g\to\g\g$, which is not fully motivated by theory. In the lack of an 
unambiguous answer, it is therefore helpful to allow options that can
explore the range of uncertainty.

Further issues remain to be addressed, e.g.\ radiation off particles
with non-negligible width. In general, however, the new shower should 
allow an improved description of gluon radiation in many different
processes.
 
\subsubsection{Matching to four-parton events}
\label{sss:fourjetmatch}

The shower routine, as described above, is optimized for two objects
forming the showering system, within which energy and momentum should 
be conserved. However, occasionally more than two initial objects are 
given, e.g.\ if one would like to consider the subclass of 
$\e^+\e^- \to \q\qbar\g\g$ events in order to study angular correlations 
as a test of the coupling structure of QCD. Such events are generated in 
the showering of normal $\e^+\e^- \to \q\qbar$ events, but not with high
efficiency within desired cuts, and not with the full angular structure
included in the shower. Therefore four-parton matrix elements may be the 
required starting point but, in order to `dress up' these partons, one 
nevertheless wishes to add shower emission. A possibility to start from  
three partons has existed since long, but only with \cite{And98a} was
an approach for four parton introduced, and with the possibility to 
generalize to more partons, although this latter work has not yet been 
done.

The basic idea is to cast the output of matrix element generators in 
the form of a parton-shower history, that then can be used as input 
for a complete parton shower. In the shower, that normally would be 
allowed to develop at random, some branchings are now fixed to
their matrix-element values while the others are still allowed
to evolve in the normal shower fashion. The preceding history of the 
event is also in these random branchings then reflected e.g.\ in terms 
of kinematical or dynamical (e.g.\ angular ordering) constraints.  

Consider e.g.\ the $\q\qbar\g\g$ case. The 
matrix-element expression contains contributions from five graphs, 
and from interferences between them. The five 
graphs can also be read as five possible parton-shower histories for 
arriving at the same four-parton state, but here without the 
possibility of including interferences. The relative probability for 
each of these possible shower histories can be obtained from the rules 
of shower branchings. For example, the relative probability for the 
history where $\e^+\e^- \to \q(1) \qbar(2)$, followed by 
$\q(1) \to \q(3)\g(4)$ and $\g(4) \to \g(5)\g(6)$,  is given by:
\begin{equation}
{\cal P} = {\cal P}_{1\rightarrow 34} {\cal P}_{4\rightarrow 56}
= \frac{1}{m_{1}^{2}} \frac{4}{3} \frac{1+z^{2}_{34}}{1-z_{34}} 
\cdot \frac{1}{m_{4}^{2}} 3 
\frac{(1-z_{56}(1-z_{56}))^{2}}{z_{56}(1-z_{56})}
\end{equation}
where the probability for each branching contains the mass 
singularity, the colour factor and the momentum splitting
kernel. The masses are given by

\begin{eqnarray}
m_{1}^{2} = p_{1}^{2} & = & (p_{3}+p_{5}+p_{6})^{2}~, \\
m_{4}^{2} = p_{4}^{2} & = & (p_{5}+p_{6})^{2}~, \nonumber
\end{eqnarray} 
and the z values by
\begin{eqnarray} 
z_{bc} = z_{a \to bc} &=& 
\frac{ m^{2}_{a} }{ \lambda }\frac{ E_{b} }{ E_{a} } - 
\frac{m^{2}_{a} - \lambda + m^{2}_{b} - m^{2}_{c}}{2\lambda}
\\
\mathrm{with~~}
\lambda &=& \sqrt{(m^{2}_{a} - m^{2}_{b} - m^{2}_{c})^{2} - 4m^{2}_{b}\,
m^{2}_{c}}~.
\nonumber
\end{eqnarray}
We here assume that the on-shell mass of quarks can be neglected.
The form of the probability then matches the expression used in the
parton-shower algorithm. 

Variants on the above probabilities are imaginable. For instance, in 
the spirit of the matrix-element approach we have assumed a common 
$\alpha_{\mathrm{s}}$ for all graphs, which thus need not be shown, 
whereas the parton-shower language normally assumes $\alpha_{\mathrm{s}}$
to be a function of the transverse momentum of each branching. 
One could also include information on azimuthal anisotropies.

The relative probability ${\cal P}$ for each of the five possible 
parton-shower histories can be used to select one of the possibilities 
at random. (A less appealing alternative would be a `winner takes 
all' strategy, i.e.\ selecting the configuration with the largest 
${\cal P}$.) The selection fixes the values of the $m$, $z$ and 
$\varphi$ at two vertices. The azimuthal angle $\varphi$ is defined 
by the daughter
parton orientation around the mother direction. When the conventional 
parton-shower algorithm is executed, these values are then forced on the
otherwise random evolution. This forcing cannot be exact for the $z$ 
values, since the final partons given by the matrix elements are on the 
mass shell, while the corresponding partons in the parton shower might 
be virtual and branch further. The shift between the 
wanted and the obtained $z$ values are rather small, very seldom
above $10^{-6}$. More significant are the changes of the opening
angle between two daughters: when daughters originally assumed 
massless are given a mass the angle between them tends to be reduced.
This shift has a non-negligible tail even above 0.1 radians. The 
`narrowing' of jets by this mechanism is compensated by the 
broadening caused by the decay of the massive daughters, and thus
overall effects are not so dramatic.

All other branchings of the parton shower are selected at random 
according to the standard evolution scheme. There is an upper
limit on the non-forced masses from internal logic, however.
For instance, for four-parton matrix elements, the singular
regions are typically avoided with a cut $y > 0.01$, where $y$ is
the square of the minimal scaled invariant mass between any pair of 
partons. Larger $y$ values could be used for some purposes, while
smaller ones give so large four-jet rates that the need to
include Sudakov form factors can no longer be neglected. 
The $y > 0.01$ cut roughly corresponds to $m > 9$~GeV at LEP~1
energies, so the hybrid approach must allow branchings at least
below 9~GeV in order to account for the emission missing from the
matrix-element part. Since no 5-parton emission is generated by the 
second-order matrix elements, one could also allow a threshold
higher than 9 GeV in order to account for this potential emission.
However, if any such mass is larger than one of the forced masses, 
the result would be a different history than the chosen one, and one
would risk some double counting issues. So, as an alternative,
one could set the minimum invariant mass between any of the four 
original partons as the maximum scale of the subsequent shower 
evolution.

\subsubsection{A new $\pT$-ordered final-state shower}

The traditional \ttt{PYSHOW} routine described above gives a 
mass-ordered timelike cascade, with angular ordering by veto. 
It offers an alternative to the \tsc{Herwig} angular-ordered shower
\cite{Mar88} and \tsc{Ariadne} $\pT$-ordered dipole emission  
\cite{Gus88,Pet88}. For most properties, comparably 
good descriptions can be obtained of LEP data by all three 
\cite{Kno96}, although \tsc{Ariadne} seems to do slightly better 
than the others.

Recently, the possibility to combine separately generated $n$-parton
configurations from Born-level matrix element expressions has 
attracted attention, see \cite{Cat01}.
This requires the implementation of vetoed parton showers, to forbid
emissions that would lead to double-counting, and trial parton showers,
to generate the appropriate Sudakovs lacking in the matrix elements.
The \ttt{PYSHOW} algorithm is not well suited for this kind of 
applications, since the full evolution process cannot easily be 
factored into a set of evolution steps in well-defined mass ranges 
--- the kinematics is closely tied both to the mother and the daughter 
virtualities of a branching.

As an alternative, the new \ttt{PYPTFS} routine \cite{Sjo03b} offers a 
shower algorithm borrowing several of the dipole ideas, combined with 
many of the old \ttt{PYSHOW} elements. It is a hybrid between the 
traditional parton shower and the dipole emission approaches, in the 
sense that the branching process is associated with the evolution of a 
single parton, like in a shower, but recoil effects occur inside dipoles. 
That is, the daughter partons of a branching are put on-shell. Instead a
recoiling partner is assigned for each branching, and energy and
momentum is ``borrowed'' from this partner to give mass to the parton
about to branch. In this sense, the branching and recoiling partons 
form a dipole. Often the two are colour-connected, i.e.\ the colour
of one matches the anticolour of the other, but this need not be
the case. For instance, in $\t \to \b \W^+$ the $\W^+$ takes the 
recoil when the $\b$ radiates a gluon. Furthermore, the radiation of a 
gluon is split into two dipoles, again normally by colour.

The evolution variable is approximately the $\pT^2$ of a branching,
where $\pT$ is the transverse momentum for each of the two daughters 
with respect to the direction of the mother, in the rest frame of 
the dipole. (The recoiling parton does not obtain any $\pT$ kick in 
this frame; only its longitudinal momentum is affected.) For the simple 
case of massless radiating partons and small virtualities relative to
the kinematically possible ones, and in the limit that recoil 
effects from further emissions can be neglected, it agrees with
the $d_{ij}$ $\pT$-clustering distance defined in the original 
\ttt{LUCLUS} (now \ttt{PYCLUS}) algorithm, see 
section~\ref{sss:clustfindee}.

All emissions are ordered in a single sequence $p_{\perp\mrm{max}} > 
p_{\perp 1} > p_{\perp 2} > \ldots > p_{\perp\mrm{min}}$. That is, all 
initial partons are evolved from the input $p_{\perp\mrm{max}}$ scale, 
and the one with the largest $\pT$ is chosen to undergo the first 
branching. Thereafter, all partons now existing are evolved downwards 
from $p_{\perp 1}$, and a $p_{\perp 2}$ is chosen, and so on, until 
$p_{\perp\mrm{min}}$ is reached. (Technically, the $\pT$ values for 
partons not directly or indirectly affected by a branching need not
be reselected.) As already noted above, the evolution of a gluon is  
split in evolution on two separate sides, with half the branching
kernel each, but with different kinematical constraints since the
two dipoles have different masses. The evolution of a quark is also
split, into one $\pT$ scale for gluon emission and one for photon one,
in general corresponding to different dipoles. 

With the choices above, the evolution factorizes. That is, a set of 
successive calls, where the $p_{\perp\mrm{min}}$ of one call becomes 
the $p_{\perp\mrm{max}}$ of the next, gives the same result (on the 
average) as one single call for the full $\pT$ range. This is the key 
element to allow Sudakovs to be conveniently obtained from trial 
showers, and to veto emissions above some $\pT$ scale, as required 
to combine different $n$-parton configurations.

The formal $\pT$ definition is
\begin{equation} 
p_{\perp\mrm{evol}}^2 = z(1-z)(m^2 - m_0^2) ~, 
\end{equation}
where $p_{\perp\mrm{evol}}$ is the evolution variable, $z$ gives the 
energy sharing in the branching, as selected from the branching 
kernels, $m$ is the off-shell mass of the branching parton and 
$m_0$ its on-shell value. This $p_{\perp\mrm{evol}}$ is also used as
scale for the running $\alphas$.

When a $p_{\perp\mrm{evol}}$ has been selected, this is translated 
to a $m^2 = m_0^2 + p_{\perp\mrm{evol}}^2/(z(1-z))$ (the same formula 
as above, rewritten). From there on, the three-body kinematics of a 
branching is  constructed as in \ttt{PYSHOW} \cite{Nor01}. This includes 
the interpretation of $z$ and the handling of nonzero on-shell masses for 
branching and recoiling partons, which leads to the physical $\pT$ not 
agreeing with the $p_{\perp\mrm{evol}}$ defined here. In this sense, 
$p_{\perp\mrm{evol}}$ becomes a formal variable, while $m$ really is 
a well-defined mass of a parton.

Also the handling of matrix element matching closely follows the 
machinery of \cite{Nor01}, once the $p_{\perp\mrm{evol}}$ has been 
converted to a mass of the branching parton. In general, the ``other'' 
parton used to define the matrix element need not be the same as the
recoiling partner. To illustrate, consider a $\Z^0 \to \q \qbar$ decay. 
In the first branching, say it is the $\q$, obviously the $\qbar$ takes 
the recoil, and the new $\q$, $\g$ and $\qbar$ momenta are used to match 
to the $\q\qbar\g$ matrix element. The next time $\q$ branches, the 
recoil is now taken by the current colour-connected gluon, but the 
matrix element corrections are based on the newly created $\q$ and $\g$ 
momenta together with the $\qbar$ (not the recoiling $\g$!) momentum. 
That way we hope to achieve the most realistic description of mass effects 
in the collinear and soft regions.  

The shower inherits some further elements from \ttt{PYSHOW}, such as 
azimuthal anisotropies in gluon branchings from polarization effects.

The relevant parameters will have to be retuned, since the shower is 
quite different from the mass-ordered one of \ttt{PYSHOW}. In particular,
it appears that the five-flavour $\Lambda_{\mrm{QCD}}$ value in 
\ttt{PARJ(81)} has to be reduced relative to the current default, 
roughly by a factor of two (from 0.29 to 0.14~GeV). 

\subsection{Initial-State Showers}
 
The initial-state shower algorithm in {\Py} is not quite as
sophisticated as the final-state one. This is partly because
initial-state radiation is less well understood theoretically,
partly because the programming task is
more complicated and ambiguous. Still, the program at disposal
is known to do a reasonably good job of describing existing data,
such as $\Z^0$ production properties at hadron colliders
\cite{Sjo85}. It can be used both for QCD showers and for photon
emission off leptons ($\e$, $\mu$ or $\tau$; relative to earlier
versions, the description of incoming $\mu$ and $\tau$ are better
geared to represent the differences in lepton mass, and the 
lepton-inside-lepton parton distributions are properly defined).
 
\subsubsection{The shower structure}
\label{sss:initshowstruc}
 
A fast hadron may be viewed as a cloud of quasi-real partons.
Similarly a fast lepton may be viewed as surrounded by a cloud
of photons and partons; in the program the two situations are on
an equal footing, but here we choose the hadron as example. At
each instant, an individual parton can initiate a virtual cascade,
branching into a number of partons. This cascade can be described
in terms of a tree-like structure, composed of many subsequent
branchings $a \to bc$. Each branching involves some relative
transverse momentum between the two daughters. In a language where
four-momentum is conserved at each vertex, this implies that at
least one of the $b$ and $c$ partons must have a space-like
virtuality, $m^2 < 0$. Since the partons are not on the mass
shell, the cascade only lives a finite time before reassembling, with
those parts of the cascade that are most off the mass shell living the
shortest time.
 
A hard scattering, e.g.\ in deeply inelastic leptoproduction, will
probe the hadron at a given instant. The probe, i.e.\ the virtual
photon in the leptoproduction case, is able to resolve fluctuations
in the hadron up to the $Q^2$ scale of the hard scattering. Thus
probes at different $Q^2$ values will seem to see different parton
compositions in the hadron. The change in parton composition with
$t = \ln(Q^2/\Lambda^2)$ is given by the evolution equations
\begin{equation}
\frac{\d f_b(x,t)}{\d t} = \sum_{a,c} \int \frac{\d x'}{x'} \,
f_a(x',t) \, \frac{\alpha_{abc}}{2 \pi} \,
P_{a \to bc} \left( \frac{x}{x'} \right) ~.
\label{sh:sfevol}
\end{equation}
Here the $f_i(x,t)$ are the parton-distribution functions, expressing
the probability of finding a parton $i$ carrying a fraction $x$
of the total momentum if the hadron is probed at virtuality $Q^2$.
The $P_{a \to bc}(z)$ are given in eq.~(\ref{sh:APker}). As before,
$\alpha_{abc}$ is $\alphas$ for QCD shower and $\alphaem$
for QED ones.
 
Eq.~(\ref{sh:sfevol}) is closely related to eq.~(\ref{sh:Patobc}):
$\d {\cal P}_a$ describes the probability that a given parton $a$ will
branch (into partons $b$ and $c$), $\d f_b$ the influx of partons
$b$ from the branchings of partons $a$. (The expression $\d f_b$ in
principle also should contain a loss term for partons $b$ that
branch; this term is important for parton-distribution evolution,
but does not appear explicitly in what we shall be using eq.
(\ref{sh:sfevol}) for.) The absolute form of
hadron parton distributions cannot be predicted in perturbative
QCD, but rather have to be parameterized at some $Q_0$ scale, with
the $Q^2$ dependence thereafter given by eq.~(\ref{sh:sfevol}).
Available parameterizations are discussed in section \ref{ss:structfun}.
The lepton and photon parton distributions inside a lepton can be
fully predicted, but here for simplicity are treated on equal footing
with hadron parton distributions.
 
If a hard interaction scatters a parton out of the incoming hadron,
the `coherence' \cite{Gri83} of the cascade is broken: the partons can
no longer reassemble completely back to the cascade-initiating parton.
In this semiclassical picture, the partons on the `main chain' of
consecutive branchings that lead directly from the initiating parton
to the scattered parton can no longer reassemble, whereas fluctuations
on the `side branches' to this chain may still disappear. A
convenient description is obtained by assigning a space-like
virtuality to the partons on the main chain, in such a way that
the partons on the side branches may still be on the mass shell.
Since the momentum transfer of the hard process can put the scattered
parton on the mass shell (or even give it a time-like virtuality, so
that it can initiate a final-state shower), one is then guaranteed
that no partons have a space-like virtuality in the final state. (In
real life, confinement effects obviously imply that partons need not
be quite on the mass shell.) If no hard scattering had taken place,
the virtuality of the space-like parton line would still force the
complete cascade to reassemble. Since the virtuality of the cascade
probed is carried by one single parton, it is possible to equate the
space-like virtuality of this parton with the $Q^2$ scale of the
cascade, to be used e.g.\ in the evolution equations. 
Coherence effects \cite{Gri83,Bas83} guarantee that the $Q^2$ values
of the partons along the main chain are strictly ordered, with the
largest $Q^2$ values close to the hard scattering.
 
Further coherence effects have been studied
\cite{Cia87}, with particular implications for the
structure of parton showers at small $x$. None of these additional
complications are implemented in the current algorithm, with the
exception of a few rather primitive options that do not address
the full complexity of the problem.
 
Instead of having a tree-like structure, where all legs are treated
democratically, the cascade is reduced to a single sequence of
branchings $a \to bc$, where the $a$ and $b$ partons are on the
main chain of space-like virtuality, $m_{a,b}^2 < 0$, while the $c$
partons are on the mass shell and do not branch. (Later we will
include the possibility that the $c$ partons may have positive
virtualities, $m_c^2 > 0$, which leads to the appearance of time-like
`final-state' parton showers on the side branches.) This truncation
of the cascade is only possible when it is known which parton
actually partakes in the hard scattering: of all the possible
cascades that exist virtually in the incoming hadron, the hard
scattering will select one.
 
To obtain the correct $Q^2$ evolution of parton distributions,
e.g., it is essential that all branches of the cascade be treated
democratically. In Monte Carlo simulation of space-like showers
this is a major problem. If indeed the evolution of the complete
cascade is to be followed from some small $Q_0^2$ up to the
$Q^2$ scale of the hard scattering, it is not possible at the
same time to handle kinematics exactly, since the virtuality of the
various partons cannot be found until after the hard scattering
has been selected. This kind of `forward evolution' scheme therefore
requires a number of extra tricks to be made to work. Further, in
this approach it is not known e.g.\ what the $\hat{s}$ of the
hard scattering subsystem will be until the evolution has been
carried out, which means that the initial-state evolution
and the hard scattering have to be selected jointly, a not so
trivial task.
 
Instead we use the `backwards evolution' approach \cite{Sjo85},
in which the hard scattering is first selected, and the parton
shower that preceded it is subsequently reconstructed. This
reconstruction is started at the hard interaction, at the
$Q_{\mmax}^2$ scale, and thereafter step by step one moves
`backwards' in  `time', towards smaller $Q^2$, all the way back
to the parton-shower initiator at the cut-off scale $Q_0^2$.
This procedure is possible if evolved parton distributions are
used to select the hard scattering, since the $f_i(x,Q^2)$
contain the inclusive summation of all initial-state parton-shower 
histories that can lead to the appearance of an interacting
parton $i$ at the hard scale. What remains is thus to select an
exclusive history from the set of inclusive ones.
 
\subsubsection{Longitudinal evolution}
 
The evolution equations, eq.~(\ref{sh:sfevol}), express that, during
a small increase $\d t$ there is a probability for parton $a$ with
momentum fraction $x'$ to become resolved into parton $b$ at
$x = z x'$ and another parton $c$ at $x' - x = (1-z) x'$.
Correspondingly, in backwards evolution, during a decrease $\d t$
a parton $b$ may be `unresolved' into parton $a$. The relative
probability $\d {\cal P}_b$ for this to happen is given by
the ratio $\d f_b / f_b$. Using eq.~(\ref{sh:sfevol}) one obtains
\begin{equation}
\d {\cal P}_b = \frac{\d f_b(x,t)}{f_b(x,t)} = |\d t| \, \sum_{a,c}
\int \frac{\d x'}{x'} \, \frac{f_a(x',t)}{f_b(x,t)} \,
\frac{\alpha_{abc}}{2 \pi} \,
P_{a \to bc} \left( \frac{x}{x'} \right) ~.
\end{equation}
Summing up the cumulative effect of many small changes $\d t$, the
probability for no radiation exponentiates. Therefore one may
define a form factor
\begin{eqnarray}
S_b(x,t_{\mmax},t) & = &
\exp \left\{ - \int_t^{t_{\mmax}} \d t' \, \sum_{a,c} \int
\frac{\d x'}{x'} \, \frac{f_a(x',t')}{f_b(x,t')} \,
\frac{\alpha_{abc}(t')}{2\pi} \, P_{a \to bc} \left(
\frac{x}{x'} \right) \right\} \nonumber \\
& = & \exp \left\{ - \int_t^{t_{\mmax}} \d t' \, \sum_{a,c} 
\int \d z \, \frac{\alpha_{abc}(t')}{2\pi} \, P_{a \to bc}(z) \,
\frac{x'f_a(x',t')}{xf_b(x,t')} \right\} ~,
\end{eqnarray}
giving the probability that a parton $b$ remains at $x$ from
$t_{\mmax}$ to a $t < t_{\mmax}$.
 
It may be useful to compare this with the corresponding expression
for forward evolution, i.e.\ with $S_a(t)$ in eq.~(\ref{sh:Pnobranch}).
The most obvious difference is the appearance of parton distributions
in $S_b$. Parton distributions are absent in $S_a$: the probability
for a given parton $a$ to branch, once it exists, is independent of
the density of partons $a$ or $b$. The parton distributions in $S_b$,
on the other hand, express the fact that the probability for a parton
$b$ to come from the branching of a parton $a$ is proportional to
the number of partons $a$ there are in the hadron, and inversely
proportional to the number of partons $b$. Thus the numerator
$f_a$ in the exponential of $S_b$ ensures that the parton composition
of the hadron
is properly reflected. As an example, when a gluon is chosen at the
hard scattering and evolved backwards, this gluon is more likely to
have been emitted by a $\u$ than by a $\d$ if the incoming hadron is
a proton. Similarly, if a heavy flavour is chosen at the hard
scattering, the denominator $f_b$ will vanish at the $Q^2$ threshold
of the heavy-flavour production, which means that the integrand
diverges and $S_b$ itself vanishes, so that no heavy flavour remain
below threshold.
 
Another difference between $S_b$ and $S_a$, already touched upon, is
that the $P_{\g \to \g\g}(z)$ splitting kernel appears with a
normalization $2 N_C$ in $S_b$ but only with $N_C$ in $S_a$, since
two gluons are produced but only one decays in a branching.
 
A knowledge of $S_b$ is enough to reconstruct the parton shower
backwards. At each branching $a \to bc$, three quantities have to
be found: the $t$ value of the branching (which defines the
space-like virtuality $Q_b^2$ of parton $b$), the parton flavour $a$
and the splitting variable $z$. This information may be extracted as
follows:
\begin{Enumerate}
\item If parton $b$ partook in the hard scattering or branched into
other partons at a scale $t_{\mmax}$, the probability that $b$ was
produced in a branching $a \to bc$ at a lower scale $t$ is
\begin{equation}
\frac{\d {\cal P}_b}{\d t} = - \frac{\d S_b(x,t_{\mmax},t)}{\d t} =
\left( \sum_{a,c} \int dz \, \frac{\alpha_{abc}(t')}{2\pi} \,
P_{a \to bc}(z) \, \frac{x'f_a(x',t')}{xf_b(x,t')} \right)
S_b(x,t_{\mmax},t) ~.
\end{equation}
If no branching is found above the cut-off scale $t_0$ the
iteration is stopped and parton $b$ is assumed to be massless.
\item Given the $t$ of a branching, the relative probabilities
for the different allowed branchings $a \to bc$ are given by the
$z$ integrals above, i.e.\ by
\begin{equation}
\int \d z \, \frac{\alpha_{abc}(t)}{2\pi} \, P_{a \to bc}(z) \, 
\frac{x'f_a(x',t)}{xf_b(x,t)} ~.
\label{sh:Pbspace}
\end{equation}
\item Finally, with $t$ and $a$ known, the probability distribution
in the splitting variable $z = x/x' = x_b/x_a$ is given by the
integrand in eq.~(\ref{sh:Pbspace}).
\end{Enumerate}
In addition, the azimuthal angle $\varphi$ of the branching is
selected isotropically, i.e.\ no spin or coherence effects are
included in this distribution.
 
The selection of $t$, $a$ and $z$ is then a standard task of the
kind than can be performed with the help of the veto algorithm.
Specifically, upper and lower bounds for parton distributions are
used to find simple functions that are everywhere larger than the
integrands in eq.~(\ref{sh:Pbspace}). Based on these simple expressions,
the integration over $z$ may be carried out, and $t$, $a$ and $z$
values selected. This set is then accepted with a weight given
by a ratio of the correct integrand in eq.~(\ref{sh:Pbspace}) to
the simple approximation used, both evaluated for the given set.
Since parton distributions, as a rule, are not in a simple
analytical form, it may be tricky to find reasonably good bounds to
parton distributions. It is necessary to make different assumptions
for valence and sea quarks, and be especially attentive close to a
flavour threshold (\cite{Sjo85}). An electron distribution
inside an electron behaves differently from parton distributions
encountered in hadrons, and has to be considered separately.
 
A comment on soft gluon emission. Nominally the range of the $z$
integral in $S_b$ is $x \leq z \leq 1$. The lower limit corresponds
to $x' = x/z = 1$, and parton distributions vanish in this limit,
wherefore no problems are encountered here. At the upper cut-off
$z=1$ the splitting kernels $P_{\q \to \q\g}(z)$ and
$P_{\g \to \g\g}$ diverge. This is the soft gluon singularity:
the energy carried by the emitted gluon is vanishing,
$x_{\g} = x' - x = (1-z) x' = (1-z) x/z \to 0$ for $z \to 1$.
In order to calculate the integral over $z$ in $S_b$, an upper
cut-off $z_{\mmax} = x/(x + x_{\epsilon})$ is introduced, i.e.\ only
branchings with $z \leq z_{\mmax}$ are included in $S_b$. Here
$x_{\epsilon}$ is a small number, typically chosen so that the
gluon energy is above 2 GeV when calculated in the rest frame of 
the hard scattering. That is, the gluon energy
$x_{\g} \sqrt{s}/2 \geq x_{\epsilon} \sqrt{s}/2 = 2~\mrm{GeV}/\gamma$,
where $\gamma$ is the boost factor of the hard scattering. The 
average amount of energy carried away by gluons in the
range $x_{g} < x_{\epsilon}$, over the given range of $t$ values
from $t_a$ to $t_b$, may be estimated \cite{Sjo85}. The finally
selected $z$ value may thus be picked as
$z = z_{\mrm{hard}} \langle z_{\mrm{soft}}(t_a, t_b) \rangle$, 
where $z_{\mrm{hard}}$
is the originally selected $z$ value and $z_{\mrm{soft}}$ is the
correction factor for soft gluon emission.
 
In QED showers, the smallness of $\alphaem$ means that one can
use rather smaller cut-off values without obtaining large amounts of
emission. A fixed small cut-off $x_{\gamma} > 10^{-10}$ is therefore
used to avoid the region of very soft photons. As has been discussed
in section \ref{sss:estructfun}, the electron distribution
inside the electron is cut off at $x_{\e} < 1 - 10^{-10}$, for
numerical reasons, so the two cuts are closely matched.
 
The cut-off scale $Q_0$ may be chosen separately for QCD and QED
showers, just as in final-state radiation. The defaults are
1 GeV and 0.001 GeV, respectively. The former is the typical hadronic
mass scale, below which radiation is not expected resolvable; the
latter is of the order of the electron mass. Photon emission is also
allowed off quarks in hadronic interactions, with the same cut-off 
as for gluon emission, and also in other respects implemented in the
same spirit, rather than according to the pure QED description.
 
Normally QED and QCD showers do not appear mixed. The most notable
exception is resolved photoproduction (in $\ep$) and resolved
$\gamma\gamma$ events
(in $\ee$), i.e.\ shower histories of the type $\e \to \gamma \to \q$.
Here the $Q^2$ scales need not be ordered at the interface, i.e.
the last $\e \to \e\gamma$ branching may well have a larger $Q^2$
than the first $\q \to \q \g$ one, and the branching $\gamma \to \q$
does not even have a strict parton-shower interpretation for the
vector dominance model part of the photon parton distribution. 
This kind of configurations is best described by the 
{\galep} machinery for having a flux of virtual photons
inside the lepton, see section \ref{sss:equivgamma}. In this case,
no initial-state radiation has currently been implemented for the 
electron (or $\mu$ or $\tau$). The one inside the virtual-photon
system is considered with the normal algorithm, but with the lower
cut-off scale modified by the photon virtuality, see \ttt{MSTP(66)}.

An older description still lives on, although no longer as the 
recommended one. There, these issues are currently
not addressed in full. Rather, based on the $x$ selected for the
parton (quark or gluon) at the hard scattering, the $x_{\gamma}$
is selected once and for all in the range $x < x_{\gamma} <1$,
according to the distribution implied by eq.~(\ref{pg:foldqgine}).
The QCD parton shower is then traced backwards from the hard
scattering to the QCD shower initiator at $t_0$. No attempt is
made to perform the full QED shower, but rather the beam remnant
treatment (see section \ref{ss:beamrem}) is used to find the
$\qbar$ (or $\g$) remnant that matches the $\q$ (or $\g$) QCD shower
initiator, with the electron itself considered as a second beam
remnant.
 
\subsubsection{Transverse evolution}
\label{sss:initshowtrans}
 
We have above seen that two parton lines may be defined, stretching
back from the hard scattering to the initial incoming hadron
wavefunctions at small $Q^2$. Specifically, all parton flavours
$i$, virtualities $Q^2$ and energy fractions $x$ may be found.
The exact kinematical interpretation of the $x$ variable is not
unique, however. For partons with small virtualities and transverse
momenta, essentially all definitions agree, but differences may
appear for branchings close to the hard scattering.
 
In first-order QED \cite{Ber85} and in some simple QCD toy models
\cite{Got86}, one may show that the `correct' choice is the
`$\hat{s}$ approach'. Here one requires that $\hat{s} = x_1 x_2 s$,
both at the hard scattering scale and at any lower scale, i.e.
$\hat{s}(Q^2) = x_1(Q^2) \, x_2(Q^2) \, s$, where $x_1$ and $x_2$ are
the $x$ values of the two resolved partons (one from each incoming
beam particle) at the given $Q^2$ scale. In practice this means
that, at a branching with the splitting variable $z$, the total
$\hat{s}$ has to be increased by a factor $1/z$ in the backwards
evolution. It also means that branchings on the two incoming legs
have to be interleaved in a single monotonic sequence of $Q^2$
values of branchings. A problem with this $x$ interpretation is that 
it is not quite equivalent with an {\MSbar} definition of parton 
densities \cite{Col00}, or any other standard definition. In practice, 
effects should not be large from this mismatch.  
 
For a reconstruction of the complete kinematics in this approach,
one should start with the hard scattering, for which $\hat{s}$
has been chosen according to the hard scattering matrix element.
By backwards evolution, the virtualities $Q_1^2 = -m_1^2$ and 
$Q_2^2 = -m_2^2$ of
the two interacting partons are reconstructed. Initially the two
partons are considered in their common c.m.\ frame, coming in along
the $\pm z$ directions. Then the four-momentum vectors have the
non-vanishing components
\begin{eqnarray}
E_{1,2} & = & \frac{ \hat{s} \pm (Q_2^2 - Q_1^2)}{2 \sqrt{\hat{s}}} ~,
\nonumber \\
p_{z1} = - p_{z2} & = & \sqrt{ \frac{ (\hat{s} + Q_1^2 + Q_2^2 )^2
- 4 Q_1^2 Q_2^2 }{ 4 \hat{s} } } ~,
\end{eqnarray}
with $(p_1 + p_2)^2 = \hat{s}$.
 
If, say, $Q_1^2 > Q_2^2$, then the branching $3 \to 1 + 4$, which
produced parton 1, is the one that took place closest to the hard
scattering, and the one to be reconstructed first. With the
four-momentum $p_3$ known, $p_4 = p_3 - p_1$ is automatically
known, so there are four degrees of freedom. One corresponds to
a trivial azimuthal angle around the $z$ axis. The $z$ splitting
variable for the $3 \to 1 + 4$ vertex is found as the same time as
$Q_1^2$, and provides the constraint $(p_3 + p_2)^2 = \hat{s}/z$.
The virtuality $Q_3^2$ is given by backwards evolution of parton 3.
 
One degree of freedom remains to be specified, and this is related
to the possibility that parton 4 initiates a time-like parton shower,
i.e.\ may have a non-zero mass. The maximum allowed squared mass
$m_{\mmax,4}^2$ is found for a collinear branching $3 \to 1 + 4$.
In terms of the combinations
\begin{eqnarray}
s_1 & = & \hat{s} + Q_2^2 + Q_1^2 ~,    \nonumber \\
s_3 & = & \frac{\hat{s}}{z} + Q_2^2 + Q_3^2 ~,  \nonumber \\
r_1 & = & \sqrt{s_1^2 - 4 Q_2^2 Q_1^2} ~, \nonumber \\
r_3 & = & \sqrt{s_3^2 - 4 Q_2^2 Q_3^2} ~,
\end{eqnarray}
one obtains
\begin{equation}
m_{\mmax,4}^2 = \frac{s_1 s_3 - r_1 r_3}{2 Q_2^2} - Q_1^2 - Q_3^2 ~,
\end{equation}
which, for the special case of $Q_2^2 = 0$, reduces to
\begin{equation}
m_{\mmax,4}^2 = \left\{ \frac{Q_1^2}{z} - Q_3^2 \right\}
\left\{ \frac{\hat{s}}{\hat{s} + Q_1^2} -
\frac{\hat{s}}{\hat{s}/z + Q_3^2} \right\} ~.
\label{sh:zrangespace}
\end{equation}
These constraints on $m_4$ are only the kinematical ones, in
addition coherence phenomena could constrain the $m_{\mmax,4}$
values further. Some options of this kind are available; the
default one is to require additionally that $m_4^2 \leq Q_1^2$,
i.e.\ lesser than the space-like virtuality of the sister parton.
 
With the maximum virtuality given, the final-state showering
machinery may be used to give the development of the subsequent
cascade, including the actual mass $m_4^2$, with
$0 \leq m_4^2 \leq m_{\mmax,4}^2$. The evolution is performed in
the c.m.\ frame of the two `resolved' partons, i.e.\ that of
partons 1 and 2 for the
branching $3 \to 1 + 4$, and parton 4 is assumed to have a nominal
energy $E_{\mrm{nom},4} = (1/z - 1) \sqrt{\hat{s}}/2$. (Slight
modifications appear if parton 4 has a non-vanishing mass
$m_{\q}$ or $m_{\ell}$.)
 
Using the relation $m_4^2 = (p_3 - p_1)^2$, the momentum of parton
3 may now be found as
\begin{eqnarray}
E_3 & = & \frac{1}{2 \sqrt{\hat{s}} } \left\{ \frac{\hat{s}}{z}
+ Q_2^2 - Q_1^2 - m_4^2 \right\} ~,  \nonumber \\
p_{z3} & = & \frac{1}{2 p_{z1}} \left\{ s_3 - 2 E_2 E_3 \right\} ~,
\nonumber \\
p_{\perp ,3}^2 & = & \left\{ m_{\mmax,4}^2 - m_4^2 \right\} \,
\frac{ (s_1 s_3 + r_1 r_3)/2 - Q_2^2 (Q_1^2 + Q_3^2 + m_4^2)}{r_1^2} ~.
\end{eqnarray}
 
The requirement that $m_4^2 \geq 0$ (or $\geq m_f^2$ for heavy
flavours) imposes a constraint on allowed $z$ values. This constraint
cannot be included in the choice of $Q_1^2$, where it logically
belongs, since it also depends on $Q_2^2$ and $Q_3^2$, which are
unknown at this point. It is fairly rare (in the order of 10\% of all
events) that a disallowed $z$ value is generated, and when it happens
it is almost always for one of the two branchings closest to the
hard interaction: for $Q_2^2 = 0$ eq.~(\ref{sh:zrangespace}) may be
solved to yield $z \leq \hat{s}/(\hat{s} + Q_1^2 - Q_3^2)$, which is
a more severe cut for $\hat{s}$ small and $Q_1^2$ large. Therefore
an essentially bias-free way of coping is to redo completely any
initial-state cascade for which this problem appears.
 
This completes the reconstruction of the $3 \to 1 + 4$ vertex.
The subsystem made out of partons 3 and 2 may now be boosted to its
rest frame and rotated to bring partons 3 and 2 along the $\pm z$
directions. The partons 1 and 4 now have opposite and compensating
transverse momenta with respect to the event axis. When the next
vertex is considered, either the one that produces parton 3 or the
one that produces parton 2, the 3--2 subsystem will fill the function
the 1--2 system did above, e.g.\ the r\^ole of 
$\hat{s} = \hat{s}_{12}$ in the formulae above is now played by 
$\hat{s}_{32} = \hat{s}_{12}/z$. The internal structure of the 
3--2 system, i.e.\ the branching $3 \to 1 + 4$, appears nowhere 
in the continued description, but has become
`unresolved'. It is only reflected in the successive rotations and
boosts performed to bring back the new endpoints to their common
rest frame. Thereby the hard scattering subsystem 1--2 builds up
a net transverse momentum and also an overall rotation of the
hard scattering subsystem.
 
After a number of steps, the two outermost partons have virtualities
$Q^2 < Q_0^2$ and then the shower is terminated and the endpoints
assigned $Q^2 = 0$. Up to small corrections from primordial
$k_{\perp}$, discussed in section \ref{ss:beamrem}, a final boost
will bring the partons from their c.m.\ frame to the overall c.m.\ 
frame, where the $x$ values of the outermost partons agree also
with the light-cone definition. The combination of several rotations 
and boosts implies that the two colliding partons have a nontrivial
orientation: when boosted back to their rest frame, they will not
be oriented along the $z$ axis. This new orientation is then inherited 
by the final state of the collision, including resonance decay products.
 
\subsubsection{Other initial-state shower aspects}
 
In the formulae above, $Q^2$ has been used as argument for
$\alphas$, and not only as the space-like virtuality of partons.
This is one possibility, but in fact loop calculations tend to
indicate that the proper argument for $\alphas$ is not $Q^2$
but $\pT^2 = (1-z) Q^2$ \cite{Bas83}. The variable $\pT$ does
have the interpretation of transverse momentum, although it
is only exactly so for a branching $a \to bc$
with $a$ and $c$ massless and $Q^2 = - m_b^2$, and with $z$
interpreted as light-cone fraction of energy and momentum.
The use of $\alphas((1-z)Q^2)$ is default in the program.
Indeed, if one wanted to, the complete shower might be
interpreted as an evolution in $\pT^2$ rather than in $Q^2$.

Angular ordering is included in the shower evolution by default. 
However, as already mentioned, the physics is much more complicated 
than for timelike showers, and so this option should only be viewed 
as a first approximation. In the code the quantity ordered is an 
approximation of $\pT/p \approx \sin\theta$. (An alternative would
have been $\pT/p_L \approx \tan\theta$, but this suffers from 
instability problems.)

In flavour excitation processes, a $\c$ (or $\b$) quark enters the 
hard scattering and should be reconstructed by the shower as coming 
from a $\g \to \c \cbar$ (or $\g \to \b \bbar$) branching. Here an 
$x$ value for the incoming $\c$ above $Q_{\c}^2/(Q_{\c}^2 + m_{\c}^2)$,
where $Q_{\c}^2$ is the spacelike virtuality of the $\c$, does not 
allow a kinematical reconstruction of the gluon branching with an 
$x_{\g} < 1$, and is thus outside the allowed phase space. Such
events (with some safety margin) are rejected. Currently they will
appear in \ttt{PYSTAT(1)} listings in the `Fraction of events that fail 
fragmentation cuts', which is partly misleading, but has the correct
consequence of suppressing the physical cross section. Further, the 
$Q^2$ value of the backwards evolution of a $\c$ quark is  by force 
kept above $m_{\c}^2$, so as to ensure that the branching
$\g \to \c \cbar$ is not `forgotten' by evolving $Q^2$ below 
$Q_0^2$. Thereby the possibility of having a $\c$ in the beam remnant 
proper is eliminated \cite{Nor98}. Warning: as a consequence, flavour 
excitation is not at all possible too close to threshold. If the 
\ttt{KFIN} array in \ttt{PYSUBS} is set so as to require a $\c$ 
(or $\b$) on either side, and the phase space is closed for such a 
$\c$ to come from a $\g \to \c \cbar$ branching, the program will 
enter an infinite loop. 

For proton beams, say, any $\c$ or $\b$ quark entering the hard
scattering has to come from a preceding gluon splitting. This is not 
the case for a photon beam, since a photon has a $\c$ and $\b$ valence
quark content. Therefore the above procedure need not be pursued there,
but $\c$ and $\b$ quarks may indeed appear as beam remnants.

As we see, the initial-state showering algorithm leads to a
net boost and rotation of the hard scattering subsystems. The
overall final state is made even more complex by the additional
final-state radiation. In principle, the complexity is very
physical, but it may still have undesirable side effects. One
such, discussed further in section \ref{ss:PYswitchkin}, is that 
it is very difficult to generate events that fulfil specific 
kinematics conditions, since kinematics is smeared and even, 
at times, ambiguous.
 
A special case is encountered in Deeply Inelastic Scattering in
$\ep$ collisions. Here the DIS $x$ and $Q^2$ values are defined
in terms of the scattered electron direction and energy, and
therefore are unambiguous (except for issues of final-state 
photon radiation close to the electron direction).
Neither initial- nor final-state showers preserve the kinematics
of the scattered electron, however, and hence the DIS $x$ and
$Q^2$ are changed. In principle, this is perfectly legitimate,
with the caveat that one then also should use different sets
of parton distributions than ones derived from DIS, since these
are based on the kinematics of the scattered lepton and nothing
else. Alternatively, one might consider showering schemes that
leave $x$ and $Q^2$ unchanged. In \cite{Ben88} detailed
modifications are presented that make a preservation
possible when radiation off the incoming and outgoing electron is
neglected, but these are not included in the current version of
{\Py}. Instead the current {\galep} machinery explicitly 
separates off the $\e \to \e \gamma$ vertex from the continued
fate of the photon. 

The only reason for using the older machinery, such as process 10, 
is that this is still the only place where weak charged and neutral 
current effects can be considered.
What is available there, as an option, is a simple machinery
which preserves $x$ and $Q^2$ from the effects of QCD radiation,
and also from those of primordial $k_{\perp}$ and the beam remnant
treatment, as follows. After the showers have been generated,
the four-momentum of the scattered lepton is changed to the
expected one, based on the nominal $x$ and $Q^2$ values.
The azimuthal angle of the lepton is maintained when the transverse
momentum is adjusted. Photon radiation off the lepton leg is not fully
accounted for, i.e.\ it is assumed that the energy of final-state
photons is added to that of the scattered electron for the definition
of $x$ and $Q^2$ (this is the normal procedure for parton-distribution
definitions). 

The change of three-momentum on the lepton side of the event is 
balanced by the final state partons on the hadron side, excluding 
the beam remnant but including all the partons both from initial-
and final-state showering. The fraction of three-momentum shift taken 
by each parton is proportional to its original light-cone momentum
in the direction of the incoming lepton, i.e.\ to $E \mp p_z$ for a
hadron moving in the $\pm$ direction. This procedure guarantees 
momentum but not energy conservation. For the latter, one additional 
degree of freedom is needed, which is taken to be the longitudinal 
momentum of the initial state shower initiator. As this momentum is 
modified, the change is shared by the final state partons on the 
hadron side, according to the same light-cone fractions as before
(based on the original momenta). Energy conservation 
requires that the total change in final state parton energies plus 
the change in lepton side energy equals the change in initiator 
energy. This condition can be turned into an iterative procedure to 
find the initiator momentum shift.

Sometimes the procedure may break down. For instance, an initiator
with $x > 1$ may be reconstructed. If this should happen, the $x$ and 
$Q^2$ values of the event are preserved, but new initial and final
state showers are generated. After five such failures, the event is 
completely discarded in favour of a new kinematical setup.

Kindly note that the four-momentum of intermediate partons in the
shower history are not being adjusted. In a listing of the complete
event history, energy and momentum need then not be conserved in 
shower branchings. This mismatch could be fixed up, if need be.

The scheme presented above should not be taken too
literally, but is rather intended as a contrast to the more
sophisticated schemes already on the market, if one would like to
understand whether the kind of conservation scheme chosen does affect
the observable physics.
 
\subsubsection{Matrix-element matching}
\label{sss:newinshow}

In {\Py} 6.1, matrix-element matching was introduced for the initial-state
shower description of initial-state
radiation in the production of a single colour-singlet resonance, such
as $\gast/\Z^0/\W^{\pm}$ \cite{Miu99}. The basic idea is to map the 
kinematics between the PS and ME descriptions, and to find a correction 
factor that can be applied to hard emissions in the shower so as to bring
agreement with the matrix-element expression. The {\Py} shower 
kinematics definitions are based on $Q^2$ as the spacelike virtuality of 
the parton produced in a branching and $z$ as the factor by which the 
$\hat{s}$ of the scattering subsystem is reduced by the branching. 
Some simple algebra then shows that the two $\q\qbar' \to \g\W^{\pm}$ 
emission rates disagree by a factor
\begin{equation}
R_{\q\qbar' \to \g\W}(\hat{s},\hat{t}) = 
\frac{(\mrm{d}\hat{\sigma}/\mrm{d}\hat{t})_{\mrm{ME}} }%
     {(\mrm{d}\hat{\sigma}/\mrm{d}\hat{t})_{\mrm{PS}} } = 
\frac{\hat{t}^2+\hat{u}^2+2 m_{\W}^2\hat{s}}{\hat{s}^2+m_{\W}^4} ~,
\label{RqqbargW} 
\end{equation}
which is always between $1/2$ and 1. 
The shower can therefore be improved in two ways, relative to the 
old description. Firstly, the maximum virtuality of emissions is 
raised from $Q^2_{\mrm{max}} \approx m_{\W}^2$ to 
$Q^2_{\mrm{max}} = s$, i.e.\ the shower is allowed to populate the 
full phase space. Secondly, the emission rate for the final (which 
normally also is the hardest) $\q \to \q\g$ emission on each side is 
corrected by the factor $R(\hat{s},\hat{t})$ above, so as to bring 
agreement with the matrix-element rate in the hard-emission region.
In the backwards evolution shower algorithm \cite{Sjo85}, this 
is the first branching considered.

The other possible ${\mathcal{O}}(\alpha_{\mrm{s}})$ graph is 
$\q\g \to \q'\W^{\pm}$, where the corresponding correction factor is
\begin{equation}
R_{\q\g \to \q'\W}(\hat{s},\hat{t}) =
\frac{(\mrm{d}\hat{\sigma}/\mrm{d}\hat{t})_{\mrm{ME}} }%
     {(\mrm{d}\hat{\sigma}/\mrm{d}\hat{t})_{\mrm{PS}} } = 
\frac{\hat{s}^2 + \hat{u}^2 + 2 m_{\W}^2 \hat{t}}{(\hat{s}-m_{\W}^2)^2 
+ m_{\W}^4} ~,
\end{equation}
which lies between 1 and 3. A probable reason for the lower shower 
rate here is that the shower does not explicitly simulate the $s$-channel 
graph $\q\g \to \q^* \to \q'\W$. The $\g \to \q\qbar$ branching 
therefore has to be preweighted by a factor of 3 in the shower, but 
otherwise the method works the same as above. Obviously, the shower 
will mix the two alternative branchings, and the correction factor 
for a final branching is based on the current type.

The reweighting procedure prompts some other changes in the shower. 
In particular, $\hat{u} < 0$ translates into a constraint on the phase
space of allowed branchings, not previously implemented. Here
$\hat{u} = Q^2 - \hat{s}_{\mrm{old}} (1-z)/z = Q^2 - 
\hat{s}_{\mrm{new}} (1-z)$, where the association with the $\hat{u}$ 
variable is relevant if the branching is reinterpreted in terms of a 
$2 \to 2$ scattering. Usually such a requirement comes out of the 
kinematics, and therefore is imposed eventually anyway. The corner of 
emissions that do not respect this requirement is that where the $Q^2$ 
value of the spacelike emitting parton is little changed and the $z$ 
value of the branching is close to unity. (That is, such branchings are 
kinematically allowed, but since the mapping to matrix-element variables 
would assume the first parton to have $Q^2=0$, this mapping gives an 
unphysical $\hat{u}$, and hence no possibility to impose a matrix-element 
correction factor.) The correct behaviour in this region is beyond 
leading-log predictivity. It is mainly important for the hardest emission, 
i.e.\ with largest $Q^2$. The effect of this change is to reduce the total 
amount of emission by a non-negligible amount when no matrix-element 
correction is applied. (This can be confirmed by using the special option 
\ttt{MSTP(68)=-1}.) For matrix-element corrections to be applied, this 
requirement must be used for the hardest branching, and then whether it 
is used or not for the softer ones is less relevant.

Our published comparisons with data on the $p_{\perp\W}$ spectrum 
show quite a good agreement with this improved simulation \cite{Miu99}. 
A worry was that an unexpectedly large primordial $k_{\perp}$, around 
4 GeV, was required to match the data in the low-$p_{\perp\Z}$ region. 
However, at that time we had not realized that the data were not fully 
unsmeared. The required primordial $k_{\perp}$ therefore drops by about 
a factor of two \cite{Bal01}. This number is still uncomfortably large,
but not too dissimilar from what is required in various resummation
descriptions.

The method can also be used for initial-state photon emission, e.g.\ in 
the process $\e^+\e^- \to \gammaZ$. There the old default
$Q^2_{\mrm{max}} = m_{\Z}^2$ allowed no emission at large $\pT$, 
$\pT \gtrsim m_{\Z}$ at LEP2. This is now corrected by the increased
$Q^2_{\mrm{max}} = s$, and using the $R$ of eq.~(\ref{RqqbargW}) with
$m_{\W} \to m_{\Z}$.

The above method does not address the issue of next-to-leading 
order corrections to the total $\W$ cross section. Rather, the implicit
assumption is that such corrections, coming mainly from soft- and 
virtual-gluon effects, largely factorize from the hard-emission effects.
That is, that the $\pT$ shape obtained in our approach will be rather
unaffected by next-to-leading order corrections (when used both for the 
total and the high-$\pT$ cross section). A rescaling by a
common $K$ factor could then be applied by hand at the end of the day.
However, the issue is not clear. Alternative approaches have been 
proposed, where more sophisticated matching procedures are used also
to get the next-to-leading order corrections to the cross section
integrated into the shower formalism \cite{Mre99}.

A matching can also be attempted for other processes than the ones above.
Currently a matrix-element correction factor is also used for 
$\g\to \g\g$ and $\q \to \g \q$ branchings in the $\g\g \to \hrm^0$ 
process, in order to match on to the $\g\g \to \g\hrm^0$ and 
$\q\g \to \q\hrm^0$ matrix elements \cite{Ell88}. The 
loop integrals of Higgs production are quite complex, however, and
therefore only the expressions obtained in the limit of a heavy top 
quark is used as a starting point to define the ratios of 
$\g\g \to \g\hrm^0$ and $\q\g \to \q\hrm^0$ to $\g\g \to \hrm^0$ 
cross sections. (Whereas the $\g\g \to \hrm^0$ cross section by itself
contains the complete expressions.) In this limit, the compact 
correction factors
\begin{equation}
R_{\g\g \to \g\hrm^0}(\hat{s},\hat{t}) = 
\frac{(\mrm{d}\hat{\sigma}/\mrm{d}\hat{t})_{\mrm{ME}} }%
     {(\mrm{d}\hat{\sigma}/\mrm{d}\hat{t})_{\mrm{PS}} } = 
\frac{\hat{s}^4+\hat{t}^4+\hat{u}^4+m_{\hrm}^8}%
{2(\hat{s}^2-m_{\hrm}^2(\hat{s}-m_{\hrm}^2))^2}
\label{Rgghiggs} 
\end{equation}
and
\begin{equation}
R_{\q\g \to \q\hrm^0}(\hat{s},\hat{t}) = 
\frac{(\mrm{d}\hat{\sigma}/\mrm{d}\hat{t})_{\mrm{ME}} }%
     {(\mrm{d}\hat{\sigma}/\mrm{d}\hat{t})_{\mrm{PS}} } = 
\frac{\hat{s}^2+\hat{u}^2}{\hat{s}^2+(\hat{s}-m_{\hrm}^2)^2}
\label{Rqghiggs} 
\end{equation}
can be derived. Even though they are clearly not as reliable as the
above expressions for $\gast/\Z^0/\W^{\pm}$, they should hopefully
represent an improved description relative to having no correction
factor at all. For this reason they are applied not only for the
standard model Higgs, but for all the three Higgs states $\hrm^0$,
$\H^0$ and $\A^0$. The Higgs correction factors are always in the 
comfortable range between $1/2$ and 1. 

Note that a third process, $\q\qbar \to \g\hrm^0$ does not fit into the 
pattern of the other two. The above process cannot be viewed as a 
showering correction to a lowest-order $\q\qbar \to \hrm^0$ one: since
the $\q$ is assumed (essentially) massless there is no pointlike 
coupling. The graph above instead again involved a top loop,
coupled to the initial state by a single s-channel gluon. The 
final-state gluon is necessary to balance colours in the process,
and therefore the cross section is vanishing in the $\pT \to 0$ limit. 
 
\subsection{Routines and Common Block Variables}
\label{ss:showrout}
 
In this section we collect information on how to use the initial-
and final-state showering routines. Of these \ttt{PYSHOW} for 
final-state radiation is the more generally interesting, since it 
can be called to let a user-defined parton configuration shower. 
The same applies for the new \ttt{PYPTFS} routine.
\ttt{PYSSPA}, on the other hand, is so intertwined with the general
structure of a {\Py} event that it is of little use as a 
stand-alone product.
  
\drawbox{CALL PYSHOW(IP1,IP2,QMAX)}\label{p:PYSHOW}
\begin{entry}
\itemc{Purpose:} to generate time-like parton showers, conventional
or coherent. The performance of the program is regulated by the
switches \ttt{MSTJ(38) - MSTJ(50)} and parameters
\ttt{PARJ(80) - PARJ(90)}. In order to keep track of the colour
flow information, the positions \ttt{K(I,4)} and \ttt{K(I,5)} have
to be organized properly for showering partons. Inside the {\Py}
programs, this is done automatically, but for external use
proper care must be taken.
\iteme{IP1 > 0, IP2 = 0 :} generate a time-like parton shower for the
parton in line \ttt{IP1} in common block \ttt{PYJETS}, with maximum
allowed mass \ttt{QMAX}. With only one parton at hand, one cannot
simultaneously conserve both energy and momentum: we here choose to
conserve energy and jet direction, while longitudinal momentum (along
the jet axis) is not conserved.
\iteme{IP1 > 0, IP2 > 0 :} generate time-like parton showers for the
two partons in lines \ttt{IP1} and \ttt{IP2} in the common block
\ttt{PYJETS}, with maximum allowed mass for each parton \ttt{QMAX}.
For shower evolution, the two partons are boosted to their c.m.\ frame.
Energy and momentum is conserved for the pair of partons, although
not for each individually. One of the two partons may be
replaced by a nonradiating particle, such as a photon or a
diquark; the energy and momentum of this particle will then be
modified to conserve the total energy and momentum.
\iteme{IP1 > 0, -80 \mbox{$\leq$} IP2 < 0 :} generate time-like parton 
showers for the \ttt{-IP2} (at most 80) partons in lines \ttt{IP1}, 
\ttt{IP1+1}, \ldots \ttt{IPI-IP2-1} in the common block \ttt{PYJETS}, 
with maximum allowed mass for each parton \ttt{QMAX}. The actions for 
\ttt{IP2=-1} and \ttt{IP2=-2} correspond to what is described above, 
but additionally larger numbers may be used to generate the evolution 
starting from three or more given partons. Then the partons are boosted 
to their c.m.\ frame, the direction of the momentum vector is conserved 
for each parton individually and energy for the system as a whole. It 
should be understood that the uncertainty in this option is larger than
for two-parton systems, and that a number of the sophisticated features 
(such as coherence with the incoming colour flow) are not implemented.
\iteme{IP1 > 0, IP2 = -100 :} generate a four-parton system, where a 
history starting from two partons has already been constructed as 
discussed in subsection \ref{sss:fourjetmatch}. Including intermediate
partons this requires 8 lines. This option is used in \ttt{PY4JET}, 
whereas you would normally not want to use this option directly 
yourself.
\iteme{QMAX :} the maximum allowed mass of a radiating parton, i.e.\
the starting value for the subsequent evolution. (In addition, the
mass of a single parton may not exceed its energy, the mass of a
parton in a system may not exceed the invariant mass of the
system.)
\end{entry}
  
\drawbox{CALL PYPTFS(NPART,IPART,PTMAX,PTMIN,PTGEN)}\label{p:PYPTFS}
\begin{entry}
\itemc{Purpose:} to generate a $\pT$-ordered timelike final-state parton 
shower. The performance of the program is regulated by the switches
\ttt{MSTJ(38)}, \ttt{MSTJ(41)}, \ttt{MSTJ(45)}, \ttt{MSTJ(46)} and 
\ttt{MSTJ(47)}, and parameters \ttt{PARJ(80)} \ttt{PARJ(81)}, 
\ttt{PARJ(82)}, \ttt{PARJ(83)} and \ttt{PARJ(90)}, i.e.\ only a subset 
of the ones available with \ttt{PYSHOW}. In order to keep track of the 
colour flow information, the positions \ttt{K(I,4)} and \ttt{K(I,5)} have
to be organized properly for showering partons. Inside the {\Py}
programs, this is done automatically, but for external use
proper care must be taken.
\iteme{NPART :} the number of partons in the system to be showered. 
Must be at least 2, and can be as large as required (modulo the technical 
limit below). Is updated at return to be the new number of partons, 
after the shower.
\iteme{IPART :} array, of dimension 500 (cf. the Les Houches user process 
accord), in which the positions of the relevant partons are stored. To 
allow the identification of matrix elements, (the showered copies of) the 
original resonance decay products, if any, should be stored in the first 
two slots, \ttt{IPART(1)} and \ttt{IPART(2)}. Is updated at return to be 
the new list of partons, by adding new particles at the end. Thus, for 
$\q \to \q \g$ the quark position is updated and the gluon added, while 
for $\g \to \g \g$ and $\g \to \q \qbar$ the choice of original and new
is arbitrary.
\iteme{PTMAX :} upper scale of shower evolution. An absolute limit is
set by kinematical constraints inside each ``dipole''.
\iteme{PTMIN :} lower scale of shower evolution. For QCD evolution, an 
absolute lower limit is set by \ttt{PARJ(82)}/2 or
$1.1 \times \Lambda_{\mrm{QCD}}^{(3)}$, whichever is larger. For QED 
evolution, an absolute lower limit is set by \ttt{PARJ(83)}/2 or 
\ttt{PARJ(90)}/2. Normally one would therefore set \ttt{PTMIN = 0D0} 
to run the shower to its intended lower cutoff.
\iteme{PTGEN :} returns the hardest $\pT$ generated; if none then 
\ttt{PTGEN = 0}.
\itemc{Note 1:}  The evolution is factorized, so that a set of successive 
calls, where the \ttt{PTMIN} scale and the \ttt{NPART} and \ttt{IPART} 
output of one call becomes the \ttt{PTMAX} scale and the \ttt{NPART} and 
\ttt{IPART} input of the next, gives the same result (on the average) as 
one single call for the full $\pT$ range. In particular, the \ttt{IPART(1)} 
and \ttt{IPART(2)} entries continue to point to (the showered copies of) 
the original decay products of a resonance if they did so to begin with.
\itemc{Note 2:} In order to read a shower listing, note that each branching 
now lists three ``new'' partons. The first two are the daughters of the 
branching, and point back to the branching mother. The third is the 
recoiling parton after it has taken the recoil, and points back to itself 
before the recoil. The total energy and momentum is conserved from the 
mother and original recoil to the three new partons, but not separately 
between the mother and its two daughters.
\itemc{Note 3:} The shower is not (yet) set up to handle final-state 
radiation on the side branches of initial-state cascades, nor radiation in 
baryon-number-violating decays. There are also a few other areas where it 
has not yet been fully developed. 
\itemc{Note 4:} The \ttt{PYPTFS} can also be used as an integrated element 
of a normal {\Py} run, in places where \ttt{PYSHOW} would otherwise be used. 
This is achieved by setting \ttt{MSTJ(41)=11} or \ttt{=12}. Then 
\ttt{PYSHOW} will call \ttt{PYPTFS}, provided that the showering system
consists of at least two partons and that the forced four-parton-shower
option is not used (\ttt{IP2=-100}). \ttt{PTMAX} is then chosen to be half 
of the \ttt{QMAX} scale of the \ttt{PYSHOW} call, and \ttt{PTMIN} is chosen 
to zero (which means the default lower limits will be used). This works 
nicely e.g. in $\e^+\e^-$ annihilation and for resonance decays. Currently 
it is not so convenient for hadronic events: there is not yet a matching to 
avoid double counting between initial- and final-state radiation, and 
sidebranch timelike evolution in spacelike showers is currently handled by 
evolving one parton with \ttt{PYSHOW}, which \ttt{PYPTFS} is not set up for.  
\itemc{Note 5:} For simplicity, all partons are evolved from a common 
\ttt{PTMAX} scale. The formalism easily accommodates separate \ttt{PTMAX} 
scales for each parton, but we have avoided this for now so as not to
complicate the routine unnecessarily for general use.
\end{entry}

\drawbox{FUNCTION PYMAEL(NI,X1,X2,R1,R2,ALPHA)}\label{p:PYMAEL} 
\begin{entry}
\itemc{Purpose:} returns the ratio of the first-order gluon emission
rate normalized to the lowest-order event rate, eq.~(\ref{eq:Wmefinshow}).
An overall factor $C_F \alphas/2\pi$ is omitted, since the running of 
$\alphas$ probably is done better in shower language anyway.
\iteme{NI :} code of the matrix element to be used, see 
Table~\ref{t:massivefinshow}. In each group of four codes in that
table, the first is for the 1 case, the second for the $\gamma_5$ one,
the third for an arbitrary mixture, see \ttt{ALPHA} below, and the last 
for $1 \pm \gamma_5$.
\iteme{X1, X2 :} standard energy fractions of the two daughters.
\iteme{R1, R2 :} mass of the two daughters normalized to the mother mass.
\iteme{ALPHA:} fraction of the no-$\gamma_5$ (i.e.\ vector/scalar/\ldots) 
part of the cross section; a free parameter for the third matrix element 
option of each group in Table~\ref{t:massivefinshow} (13, 18, 23, 28, 
\ldots).  
\end{entry}
 
\drawbox{SUBROUTINE PYADSH(NFIN)}\label{p:PYADSH} 
\begin{entry}
\itemc{Purpose:} to administrate a sequence of final-state showers
for external processes, where the order normally is that all resonances
have decayed before showers are considered, and therefore already
existing daughters have to be boosted when their mothers radiate or
take the recoil from radiation.
\iteme{NFIN :} line in the event record of the last final-state entry 
to consider.  
\end{entry}

\drawbox{SUBROUTINE PYSSPA(IPU1,IPU2)}\label{p:PYSSPA} 
\begin{entry}
\itemc{Purpose:} to generate the space-like showers of the initial-state 
radiation.
\iteme{IPU1, IPU2 :} positions of the two partons entering the hard
scattering, from which the backwards evolution is initiated.
\end{entry}

\drawbox{SUBROUTINE PYMEMX(MECOR,WTFF,WTGF,WTFG,WTGG)}\label{p:PYMEMX} 
\begin{entry}
\itemc{Purpose:} to set the maximum of the ratio of the correct matrix 
element to the one implied by the spacelike parton shower.
\iteme{MECOR :} kind of hard scattering process, 1 for 
$\f + \fbar \to \gamma^*/\Z^0/\W^{\pm}/\ldots$ vector gauge bosons,
2 for $\g + \g \to \hrm^0/\H^0/\A^0$.
\iteme{WTFF, WTGF, WTFG, WTGG :} maximum weights for 
$\f \to \f \, (+ \g/\gamma)$, $\g/\gamma \to \f \, (+ \fbar)$,
$\f \to \g/\gamma \, (+ \f)$ and $\g \to \g \, (+ \g)$, respectively.
\end{entry}
 
\drawbox{SUBROUTINE PYMEWT(MECOR,IFLCB,Q2,Z,PHIBR,WTME)}\label{p:PYMEWT} 
\begin{entry}
\itemc{Purpose:} to calculate the ratio of the correct matrix 
element to the one implied by the spacelike parton shower.
\iteme{MECOR :} kind of hard scattering process, 1 for 
$\f + \fbar \to \gamma^*/\Z^0/\W^{\pm}/\ldots$ vector gauge bosons,
2 for $\g + \g \to \hrm^0/\H^0/\A^0$.
\iteme{IFLCB :} kind of branching, 1 for $\f \to \f \, (+ \g/\gamma)$, 
2 for $\g/\gamma \to \f \, (+ \fbar)$, 3 for $\f \to \g/\gamma \, (+ \f)$ 
and 4 for $\g \to \g \, (+ \g)$.
\iteme{Q2, Z :} $Q^2$ and $z$ values of shower branching under consideration.
\iteme{PHIBR :} $\varphi$ azimuthal angle of the shower branching;
may be overwritten inside routine.
\iteme{WTME :} calculated matrix element correction weight, used in the 
acceptance/rejection of the shower branching under consideration. 
\end{entry}

\drawbox{COMMON/PYDAT1/MSTU(200),PARU(200),MSTJ(200),PARJ(200)}
\begin{entry}
\itemc{Purpose:} to give access to a number of status codes and
parameters which regulate the performance of {\Py}.
Most parameters are described in section \ref{ss:JETswitch};
here only those related to \ttt{PYSHOW} are described.

\boxsep

\iteme{MSTJ(38) :}\label{p:MSTJ38} (D=0) matrix element code \ttt{NI} for 
\ttt{PYMAEL}; as in \ttt{MSTJ(47)}. If nonzero, the \ttt{MSTJ(38)} value 
overrides \ttt{MSTJ(47)}, but is then set \ttt{=0} in the \ttt{PYSHOW} or
\ttt{PYPTFS} call. 
The usefulness of this switch lies in processes where sequential decays 
occur and thus there are several showers, each requiring its matrix element. 
Therefore \ttt{MSTJ(38)} can be set in the calling routine when it is known, 
and when not set one defaults back to the attempted matching procedure of 
\ttt{MSTJ(47)=3} (e.g.).   

\iteme{MSTJ(40) :} (D=0) possibility to suppress the 
branching probability for a branching $\q \to \q\g$ (or 
$\q \to \q\gamma$) of a quark produced in the decay of an unstable 
particle with width $\Gamma$, where this width has to
be specified by you in \ttt{PARJ(89)}. The algorithm used is not 
exact, but still gives some impression of potential effects. This 
switch, valid for \ttt{PYSHOW}, ought to have appeared at the end of the 
current list of shower switches (after \ttt{MSTJ(50)}), but because of lack 
of space it appears immediately before.  
\begin{subentry}
\iteme{= 0 :} no suppression, i.e.\ the standard parton-shower machinery.
\iteme{= 1 :} suppress radiation by a factor 
$\chi(\omega) = \Gamma^2 / (\Gamma^2 + \omega^2)$, where $\omega$ is 
the energy of the gluon (or photon) in the rest frame of the radiating 
dipole. Essentially this means that hard radiation with 
$\omega > \Gamma$ is removed. 
\iteme{= 2 :} suppress radiation by a factor 
$1 - \chi(\omega) = \omega^2 / (\Gamma^2 + \omega^2)$, where $\omega$
is the energy of the gluon (or photon) in the rest frame of the
radiating dipole. Essentially this means that soft radiation with 
$\omega < \Gamma$ is removed. 
\end{subentry}

\iteme{MSTJ(41) :} (D=2) type of branchings allowed in shower.
\begin{subentry}
\iteme{= 0 :} no branchings at all, i.e.\ shower is switched off.
\iteme{= 1 :} QCD type branchings of quarks and gluons.
\iteme{= 2 :} also emission of photons off quarks and leptons; the
photons are assumed on the mass shell.
\iteme{= 3 :} QCD type branchings of quarks and gluons, and also 
emission of photons off quarks, but leptons do not radiate 
(unlike \ttt{=2}).
\iteme{= 10 :} as \ttt{=2}, but enhance photon emission by a factor
\ttt{PARJ(84)}. This option is unphysical, but for moderate values,
\ttt{PARJ(84)}$\leq 10$, it may be used to enhance the prompt photon
signal in $\q\qbar$ events. The normalization of the prompt photon 
rate should then be scaled down by the same factor. The dangers
of an improper use are significant, so do not use this option if you
do not know what you are doing.
\iteme{= 11 :} QCD type branchings of quarks and gluons, like 
\ttt{=1}, but if \ttt{PYSHOW} is called with a parton system that
\ttt{PYPTFS} can handle, the latter routine is called to do the 
shower. If \ttt{PYPTFS} is called directly by the user, this option 
is equivalent to \ttt{=1}.
\iteme{= 12 :} also emission of photons off quarks and leptons, like 
\ttt{=2}, but if \ttt{PYSHOW} is called with a parton system that
\ttt{PYPTFS} can handle, the latter routine is called to do the 
shower. If \ttt{PYPTFS} is called directly by the user, this option 
is equivalent to \ttt{=2}.
\end{subentry}
 
\iteme{MSTJ(42) :} (D=2) branching mode, especially coherence level,
for time-like showers in \ttt{PYSHOW}.
\begin{subentry}
\iteme{= 1 :} conventional branching, i.e.\ without angular ordering.
\iteme{= 2 :} coherent branching, i.e.\ with angular ordering.
\iteme{= 3 :} in a branching $a \to b \g$, where $m_b$ is nonvanishing, 
the decay angle is reduced by a factor 
$(1 + (m_b^2/m_a^2) (1-z)/z)^{-1}$, thereby taking into account 
mass effects in the decay \cite{Nor01}. Therefore more branchings are
acceptable from an angular ordering point of view. 
In the definition of the angle in a $g \to \q \qbar$ 
branchings, the naive massless expression is reduced by a 
factor $\sqrt{1 - 4 m_q^2/m_g^2}$, which can be motivated
by a corresponding actual reduction in the $\pT$ by mass 
effects. The requirement of angular ordering then kills
fewer potential $\g \to \q \qbar$ branchings, i.e.\ the rate of 
such comes up. The $\g \to \g \g$ branchings are not changed 
from \ttt{=2}. This option is fully within the range of
uncertainty that exists.
\iteme{= 4 :} as \ttt{=3} for $a \to b \g$ and $\g \to \g \g$ 
branchings, but no angular ordering requirement conditions at 
all are imposed on $\g \to \q \qbar$ branchings. This is an
unrealistic extreme, and results obtained with it should 
not be overstressed. However, for some studies it is of 
interest. For instance, it not only gives a much higher
rate of charm and bottom production in showers, but also 
affects the kinematical distributions of such pairs.
\iteme{= 5 :} new `intermediate' coherence level \cite{Nor01}, where 
the consecutive gluon emissions off the original pair of branching 
partons is not constrained by angular ordering at all.
The subsequent showering of such a gluon is angular 
ordered, however, starting from its production angle.
At LEP energies, this gives almost no change in the
total parton multiplicity, but this multiplicity now
increases somewhat faster with energy than before, in
better agreement with analytical formulae. (The \ttt{PYSHOW} 
algorithm overconstrains
the shower by ordering emissions in mass and then vetoing   
increasing angles. This is a first simple attempt to redress 
the issue.) Other branchings as in \ttt{=2}.
\iteme{= 6 :} `intermediate' coherence level as \ttt{=5} for primary 
partons, unchanged for $\g \to \g \g$ and reduced angle for 
$\g \to \q \qbar$ and secondary $\q \to \q \g$ as in \ttt{=3}.
\iteme{= 7 :} `intermediate' coherence level as \ttt{=5} for primary 
partons, unchanged for $\g \to \g \g$, reduced angle for secondary 
$\q \to \q \g$ as in \ttt{=3} and no angular ordering for 
$\g \to \q \qbar$ as in \ttt{=4}.
\end{subentry}
 
\iteme{MSTJ(43) :} (D=4) choice of $z$ definition in branchings 
in \ttt{PYSHOW}.
\begin{subentry}
\iteme{= 1 :} energy fraction in grandmother's rest frame (`local,
constrained').
\iteme{= 2 :} energy fraction in grandmother's rest frame assuming
massless daughters, with energy and momentum reshuffled for massive
ones (`local, unconstrained').
\iteme{= 3 :} energy fraction in c.m.\ frame of the showering partons
(`global, constrained').
\iteme{= 4 :} energy fraction in c.m.\ frame of the showering partons
assuming massless daughters, with energy and momentum reshuffled for
massive ones (`global, unconstrained').
\end{subentry}
 
\iteme{MSTJ(44) :} (D=2) choice of $\alphas$ scale for shower 
in \ttt{PYSHOW}.
\begin{subentry}
\iteme{= 0 :} fixed at \ttt{PARU(111)} value.
\iteme{= 1 :} running with $Q^2 = m^2/4$, $m$ mass of decaying
parton, $\Lambda$ as stored in \ttt{PARJ(81)} (natural choice for
conventional showers).
\iteme{= 2 :} running with $Q^2 = z(1-z)m^2$, i.e.\ roughly $\pT^2$
of branching, $\Lambda$ as stored in \ttt{PARJ(81)} (natural choice
for coherent showers).
\iteme{= 3 :} while $\pT^2$ is used as $\alphas$ argument in 
$\q \to \q \g$ and $\g \to \g\g$ branchings, as in \ttt{=2}, instead 
$m^2/4$ is used as argument for $\g \to \q \qbar$ ones. The argument 
is that the soft-gluon resummation results suggesting the $\pT^2$ scale
\cite{Ama80} in the former processes is not valid for the latter one,
so that any multiple of the mass of the branching parton
is a perfectly valid alternative. The $m^2/4$ ones then gives
continuity with $\pT^2$ for $z=1/2$. Furthermore, with this 
choice, it is no longer necessary to have the requirement 
of a minimum $\pT$ in branchings, else required in order to
avoid having $\alphas$ blow up. Therefore, in this option,
that cut has been removed for $\g \to\q\qbar$ branchings. 
Specifically, when combined with \ttt{MSTJ(42)=4}, it is possible 
to reproduce the simple $1 + \cos^2\theta$ angular distribution
of $\g \to \q\qbar$ branchings, which is not possible in any other
approach. (However it may give too high a charm and bottom production 
rate in showers \cite{Nor01}.)   
\iteme{= 4 :} $\pT^2$ as in \ttt{=2}, but scaled down by a factor 
$(1 - m_b^2/m_a^2)^2$ for a branching $a \to b \g$ with 
$b$ massive, in an attempt better to take into account the
mass effect on kinematics.
\iteme{= 5 :} as for \ttt{=4} for $\q \to \q \g$, unchanged for 
$\g \to \g \g$ and as \ttt{=3} for $\g \to \q \qbar$.      
\end{subentry}
 
\iteme{MSTJ(45) :} (D=5) maximum flavour that can be produced in
shower by $\g \to \q \qbar$; also used to determine the maximum
number of active flavours in the $\alphas$ factor in parton showers
(here with a minimum of 3).
 
\iteme{MSTJ(46) :} (D=3) nonhomogeneous azimuthal distributions in
a shower branching.
\begin{subentry}
\iteme{= 0 :} azimuthal angle is chosen uniformly.
\iteme{= 1 :} nonhomogeneous azimuthal angle in gluon decays due to
a kinematics-dependent effective gluon polarization.
Not meaningful for scalar model, i.e.\ then same as \ttt{=0}.
\iteme{= 2 :} nonhomogeneous azimuthal angle in gluon decay due to
interference with nearest neighbour (in colour).
Not meaningful for Abelian model, i.e.\ then same as \ttt{=0}.
\iteme{= 3 :} nonhomogeneous azimuthal angle in gluon decay due to
both polarization (\ttt{=1}) and interference (\ttt{=2}).
Not meaningful for Abelian model, i.e.\ then same as \ttt{=1}.
Not meaningful for scalar model, i.e.\ then same as \ttt{=2}.
\itemc{Note :} \ttt{PYPTFS} only implements nonhomogeneities related
to the gluon spin, and so options 0 and 2 are equivalent, as are
1 and 3.
\end{subentry}
 
\iteme{MSTJ(47) :} (D=3) matrix-element-motivated corrections to the 
gluon shower emission rate in generic processes of the type 
$a \to b c \g$. Also, in the massless fermion approximation, with an 
imagined vector source, to the lowest-order $\q\qbar\gamma$, 
$\ell^+\ell^-\gamma$ or $\ell\nu_{\ell}\gamma$ matrix elements,
i.e.\ more primitive than for QCD radiation.
\begin{subentry}
\iteme{= 0 :} no corrections.
\iteme{= 1 - 5 :} yes; try to match to the most relevant matrix element
and default back to an assumed source (e.g.\ a vector for a $\q\qbar$ pair) 
if the correct mother particle cannot be found.
\iteme{= 6 - :} yes, match to the specific matrix element code
\ttt{NI = MSTJ(47)} of the \ttt{PYMAEL} function; see
Table~\ref{t:massivefinshow}.    
\itemc{Warning :} since a process may contain sequential decays involving
several different kinds of matrix elements, it may be 
dangerous to fix \ttt{MSTJ(47)} to a specialized value $>5$;
see \ttt{MSTJ(38)} above.
\end{subentry}
 
\iteme{MSTJ(48) :} (D=0) possibility to impose maximum angle for the 
first branching in a \ttt{PYSHOW} shower.
\begin{subentry}
\iteme{= 0 :} no explicit maximum angle.
\iteme{= 1 :} maximum angle given by \ttt{PARJ(85)} for single
showering parton, by \ttt{PARJ(85)} and \ttt{PARJ(86)} for pair
of showering partons.
\end{subentry}
 
\iteme{MSTJ(49) :} (D=0) possibility to change the branching
probabilities in \ttt{PYSHOW} according to some alternative toy models 
(note that the $Q^2$ evolution of $\alphas$ may well be different in these
models, but that only the \ttt{MSTJ(44)} options are at your disposal).
\begin{subentry}
\iteme{= 0 :} standard QCD branchings.
\iteme{= 1 :} branchings according to a scalar gluon theory, i.e.\ the
splitting kernels in the evolution equations are,
with a common factor $\alphas/(2\pi)$ omitted,
$P_{\q \to \q\g} = (2/3) (1-z)$, $P_{\g \to \g\g} =$ \ttt{PARJ(87)},
$P_{\g \to \q\qbar} =$ \ttt{PARJ(88)} (for each separate flavour).
The couplings of the gluon have been left as free parameters,
since they depend on the colour structure assumed. Note that,
since a spin 0 object decays isotropically, the gluon splitting
kernels contain no $z$ dependence.
\iteme{= 2 :} branchings according to an Abelian vector gluon theory,
i.e.\ the colour factors are changed (compared with QCD) according to
$C_F = 4/3 \to 1$, $N_C = 3 \to 0$, $T_R = 1/2 \to 3$. Note that an
Abelian model is not expected to contain any coherence effects
between gluons, so that one should normally use \ttt{MSTJ(42)=1} and
\ttt{MSTJ(46)=} 0 or 1. Also, $\alphas$ is expected to increase
with increasing $Q^2$ scale, rather than decrease. No such
$\alphas$ option is available; the one that comes closest
is \ttt{MSTJ(44)=0}, i.e.\ a fix value.
\end{subentry}

\iteme{MSTJ(50) :} (D=3) possibility to introduce colour coherence 
effects in the first branching of a \ttt{PYSHOW} final state shower. 
Only relevant when colour flows through from the initial to the final 
state, i.e.\ mainly for QCD parton--parton scattering processes.
\begin{subentry}
\iteme{= 0 :} none.
\iteme{= 1 :} impose an azimuthal anisotropy. Does not apply when
the intermediate state is a resonance, e.g., in a $\t \to \b \W^+$
decay the radiation off the $b$ quark is not restricted.
\iteme{= 2 :} restrict the polar angle of a branching to be smaller 
than the scattering angle of the relevant colour flow. Does not apply 
when the intermediate state is a resonance.
\iteme{= 3 :} both azimuthal anisotropy and restricted polar angles.
Does not apply when the intermediate state is a resonance.
\iteme{= 4 - 6 :} as \ttt{= 1 - 3}, except that now also decay products
of coloured resonances are restricted in angle.
\itemc{Note:} for subsequent branchings the (polar) angular ordering 
is automatic (\ttt{MSTP(42)=2}) and \ttt{MSTJ(46)=3}).
\end{subentry}

\boxsep 

\iteme{PARJ(80) :}\label{p:PARJ80} (D=0.5) `parity' mixing parameter,
$\alpha$ value for the \ttt{PYMAEL} routine, to be used when 
\ttt{MSTJ(38)} is nonvanishing.

\iteme{PARJ(81) :} (D=0.29 GeV) $\Lambda$ value
in running $\alphas$ for parton showers (see \ttt{MSTJ(44)}). This is
used in all user calls to \ttt{PYSHOW}, in the \ttt{PYEEVT}/\ttt{PYONIA}
$\e^+\e^-$ routines, and in a resonance decay. It is not intended for 
other timelike showers, however, for which \ttt{PARP(72)} is used.
This parameter ought to be reduced by about a factor of two for use
with the \ttt{PYPTFS} routine.

\iteme{PARJ(82) :} (D=1.0 GeV) invariant mass cut-off $m_{\mmin}$ of
parton showers, below which partons are not assumed to radiate.
For $Q^2 = \pT^2$ (\ttt{MSTJ(44)=2}) \ttt{PARJ(82)}/2
additionally gives the minimum $\pT$ of a branching. To avoid
infinite $\alphas$ values, one must have
\ttt{PARJ(82)}$ > 2 \times$\ttt{PARJ(81)} for \ttt{MSTJ(44)}$\geq 1$
(this is automatically checked in the program, with
$2.2 \times$\ttt{PARJ(81)} as the lowest value attainable). When
the \ttt{PYPTFS} routine is called, it is twice the $p_{\perp\mrm{min}}$
cut. 
 
\iteme{PARJ(83) :} (D=1.0 GeV) invariant mass cut-off $m_{\mmin}$ used
for photon emission in parton showers, below which quarks
are not assumed to radiate. The function of \ttt{PARJ(83)} closely
parallels that of \ttt{PARJ(82)} for QCD branchings, but there is a
priori no requirement that the two be equal. The cut-off for photon
emission off leptons is given by \ttt{PARJ(90)}. When the \ttt{PYPTFS} 
routine is called, it is twice the $p_{\perp\mrm{min}}$ cut. 

\iteme{PARJ(84) :} (D=1.) used for option \ttt{MSTJ(41)=10} as a 
multiplicative factor in the prompt photon emission rate in final
state parton showers. Unphysical but useful technical trick, so
beware!
 
\iteme{PARJ(85), PARJ(86) :} (D=10.,10.) maximum opening angles
allowed in the first branching of parton showers; see \ttt{MSTJ(48)}.
 
\iteme{PARJ(87) :} (D=0.) coupling of $\g \to \g\g$ in scalar gluon
shower, see \ttt{MSTJ(49)=1}.
 
\iteme{PARJ(88) :} (D=0.) coupling of $\g \to \q\qbar$ in scalar
gluon shower (per quark species), see \ttt{MSTJ(49)=1}.
 
\iteme{PARJ(89) :} (D=0. GeV) the width of the unstable particle studied 
for the \ttt{MSTJ(40)>0} options; to be set by you (separately 
for each \ttt{PYSHOW} call, if need be). 
 
\iteme{PARJ(90) :} (D=0.0001 GeV) invariant mass cut-off $m_{\mmin}$ 
used for photon emission in parton showers, below which leptons
are not assumed to radiate, cf. \ttt{PARJ(83)} for radiation off quarks.
When the \ttt{PYPTFS} routine is called, it is twice the 
$p_{\perp\mrm{min}}$ cut. By making this separation of cut-off values, 
photon emission off leptons becomes more realistic, covering a larger part 
of the phase space. The emission rate is still not well reproduced for 
lepton--photon invariant masses smaller than roughly twice the lepton 
mass itself. 
\end{entry}

\drawbox{COMMON/PYPARS/MSTP(200),PARP(200),MSTI(200),PARI(200)}
\begin{entry}
 
\itemc{Purpose:} to give access to status code and parameters which
regulate the performance of {\Py}.
Most parameters are described in section \ref{ss:PYswitchpar};
here only those related to \ttt{PYSSPA} and \ttt{PYSHOW} are 
described.
 
\iteme{MSTP(22) :}\label{p:MSTP22} (D=0) special override of normal 
$Q^2$ definition used for maximum of parton-shower evolution. This 
option only affects processes 10 and 83 (Deeply Inelastic Scattering) 
and only in lepton--hadron events.
\begin{subentry}
\iteme{= 0 :} use the scale as given in \ttt{MSTP(32)}.
\iteme{= 1 :} use the DIS $Q^2$ scale, i.e.\ $-\hat{t}$.
\iteme{= 2 :} use the DIS $W^2$ scale, i.e.\ $(-\hat{t})(1-x)/x$.
\iteme{= 3 :} use the DIS $Q \times W$ scale, i.e.\
$(-\hat{t}) \sqrt{(1-x)/x}$.
\iteme{= 4 :} use the scale $Q^2 (1-x) \max(1, \ln(1/x))$, as 
motivated by first order matrix elements \cite{Ing80,Alt78}.
\itemc{Note:} in all of these alternatives, a multiplicative factor is
introduced by \ttt{PARP(67)} and \ttt{PARP(71)}, as usual.
\end{subentry}
 
\iteme{MSTP(61) :}\label{p:MSTP61} (D=2) master switch for 
initial-state QCD and QED radiation.
\begin{subentry}
\iteme{= 0 :} off.
\iteme{= 1 :} on for QCD radiation in hadronic events and QED 
radiation in leptonic ones.
\iteme{= 2 :} on for QCD and QED radiation in hadronic events and 
QED radiation in leptonic ones.
\end{subentry}

\iteme{MSTP(62) :} (D=3) level of coherence imposed on the
space-like parton-shower evolution.
\begin{subentry}
\iteme{= 1 :} none, i.e.\ neither $Q^2$ values nor angles need be
ordered.
\iteme{= 2 :} $Q^2$ values at branches are strictly ordered,
increasing towards the hard interaction.
\iteme{= 3 :} $Q^2$ values and opening angles of emitted
(on-mass-shell or time-like) partons are both strictly ordered,
increasing towards the hard interaction.
\end{subentry}
 
\iteme{MSTP(63) :} (D=2) structure of associated time-like
showers, i.e.\ showers initiated by emission off the incoming
space-like partons.
\begin{subentry}
\iteme{= 0 :} no associated showers are allowed, i.e.\ emitted
partons are put on the mass shell.
\iteme{= 1 :} a shower may evolve, with maximum allowed time-like
virtuality set by the phase space only.
\iteme{= 2 :} a shower may evolve, with maximum allowed time-like
virtuality set by phase space or by \ttt{PARP(71)} times the $Q^2$
value of the space-like parton created in the same vertex, whichever
is the stronger constraint.
\iteme{= 3 :} a shower may evolve, with maximum allowed time-like
virtuality set by phase space, but further constrained to evolve within 
a cone with opening angle (approximately) set by the opening angle of 
the branching where the showering parton was produced.
\end{subentry}
 
\iteme{MSTP(64) :} (D=2) choice of $\alphas$ and $Q^2$ scale
in space-like parton showers.
\begin{subentry}
\iteme{= 0 :} $\alphas$ is taken to be fix at the value
\ttt{PARU(111)}.
\iteme{= 1 :} first-order running $\alphas$ with argument
\ttt{PARP(63)}$Q^2$.
\iteme{= 2 :} first-order running $\alphas$ with argument
\ttt{PARP(64)}$k_{\perp}^2 = $\ttt{PARP(64)}$(1-z)Q^2$.
\end{subentry}
 
\iteme{MSTP(65) :} (D=1) treatment of soft gluon emission in
space-like parton-shower evolution.
\begin{subentry}
\iteme{= 0 :} soft gluons are entirely neglected.
\iteme{= 1 :} soft gluon emission is resummed and included
together with the hard radiation as an effective $z$ shift.
\end{subentry}

\iteme{MSTP(66) :} (D=5) choice of lower cut-off for initial-state
QCD radiation in VMD or anomalous photoproduction events, and matching
to primordial $\kT$.
\begin{subentry}
\iteme{= 0 :} the lower $Q^2$ cutoff is the standard one in 
\ttt{PARP(62)}$^2$.
\iteme{= 1 :} for anomalous photons, the lower $Q^2$ cut-off  is the 
larger of \ttt{PARP(62)}$^2$ and \ttt{VINT(283)} or\ttt{ VINT(284)},
where the latter is the virtuality scale for the 
$\gamma \to \q \qbar$ vertex on the appropriate side of 
the event. The \ttt{VINT} values are selected logarithmically
even between\ttt{ PARP(15)}$^2$ and the $Q^2$ scale of the
parton distributions of the hard process.    
\iteme{= 2 :} extended option of the above, intended for virtual
photons. For VMD photons, the lower $Q^2$ cut-off is the
larger of \ttt{PARP(62)}$^2$ and the $P^2_{\mrm{int}}$ scale of the
SaS parton distributions. For anomalous photons,
the lower cut-off is chosen as for \ttt{=1}, but the
\ttt{VINT(283)} and \ttt{VINT(284)} are here selected logarithmically
even between $P^2_{\mrm{int}}$ and the $Q^2$ scale of the
parton distributions of the hard process.  
\iteme{= 3 :} the $\kT$ of the anomalous/GVMD component is distributed
like $1/\kT^2$ between $k_0$ and $\pTmin(W^2)$. Apart from
the change of the upper limit, this option works just like \ttt{=1}.
\iteme{= 4 :} a stronger damping at large $\kT$, like 
$\d \kT^2/(\kT^2 + Q^2/4)^2$ with $k_0 < \kT < \pTmin(W^2)$. 
Apart from this, it works like \ttt{=1}.
\iteme{= 5 :} a $\kT$ generated as in \ttt{=4} is added vectorially 
with a standard Gaussian $\kT$ generated like for VMD states.      
Ensures that GVMD has typical $\kT$'s above those of VMD,
in spite of the large primordial $\kT$'s implied by hadronic
physics. (Probably attributable to a lack of soft QCD
radiation in parton showers.)
\end{subentry}
 
\iteme{MSTP(67) :} (D=2) possibility to introduce colour coherence 
effects in the first branching of the backwards evolution of an 
initial state shower; mainly of relevance for QCD parton--parton 
scattering processes.
\begin{subentry}
\iteme{= 0 :} none.
\iteme{= 2 :} restrict the polar angle of a branching to be smaller 
than the scattering angle of the relevant colour flow.
\itemc{Note 1:} azimuthal anisotropies have not yet been included.
\itemc{Note 2:} for subsequent branchings, \ttt{MSTP(62)=3} is
used to restrict the (polar) angular range of branchings.
\end{subentry}
 
\iteme{MSTP(68) :} (D=1) choice of maximum virtuality scale and 
matrix-element matching scheme for initial-state radiation.
\begin{subentry}
\iteme{= 0 :} maximum shower virtuality is the same as the $Q^2$ choice 
for the parton distributions, see \ttt{MSTP(32)}. (Except that the
multiplicative extra factor \ttt{PARP(34)} is absent and instead
\ttt{PARP(67)} can be used for this purpose.) No matrix-element 
correction. 
\iteme{= 1 :} as \ttt{=0} for most processes, but new scheme for processes 
1, 2, 141, 142, 144 and 102, i.e.\ single $s$-channel colourless gauge boson
and Higgs production: $\gamma^*/\Z^0$, $\W^{\pm}$, ${\Z'}^0$, ${\W'}^{\pm}$, 
$\R$ and $\hrm^0$. Here the maximum scale of shower evolution is $s$, the 
total squared energy. The nearest branching on either side of the hard 
scattering is corrected by the ratio of the first-order matrix-element 
weight to the parton-shower one, so as to obtain an improved description. 
For gauge boson production, this branching can be of the types 
$\q \to \q + \g$, $\f \to \f + \gamma$, $\g \to \q + \qbar$ or 
$\gamma \to \f + \fbar$, while for Higgs production it is $\g \to \g + \g$.
See section \ref{sss:newinshow} for a detailed description. Note that the 
improvements apply both for incoming hadron and lepton beams.
\iteme{= 2 :} the maximum scale for initial-state shower evolution is 
always selected to be $s$, except for the $2 \to 2$ QCD processes 11, 12, 
13, 28, 53 and 68. The QCD exception is to avoid the double-counting 
issues that could easily arise here. Based on the experience in
\cite{Miu99}, there is reason to assume that this does give an 
improved qualitative description of the high-$\pT$ tail, although
the quantitative agreement is currently beyond our control.
No matrix-element corrections, even for the processes in \ttt{=1}.
\iteme{= -1 :} as \ttt{=0}, except that there is no requirement on 
$\hat{u}$ being negative. 
\end{subentry}
 
\iteme{MSTP(69) :} (D=0) possibility to change $Q^2$ scale for parton 
distributions from the \ttt{MSTP(32)} choice, especially for $\ee$.
\begin{subentry}
\iteme{= 0 :} use \ttt{MSTP(32)} scale.
\iteme{= 1 :} in lepton-lepton collisions, the QED lepton-inside-lepton 
parton distributions are evaluated with $s$, the full squared c.m.\ 
energy, as scale.
\iteme{= 2 :} $s$ is used as parton distribution scale also in other 
processes. 
\end{subentry}

\iteme{MSTP(71) :} (D=1) master switch for final-state QCD and
QED radiation.
\begin{subentry}
\iteme{= 0 :} off.
\iteme{= 1 :} on.
\end{subentry}

\boxsep
 
\iteme{PARP(61) :}\label{p:PARP61} (D=0.25 GeV) $\Lambda$ value 
used in space-like parton shower (see \ttt{MSTP(64)}). This value 
may be overwritten, see \ttt{MSTP(3)}.
 
\iteme{PARP(62) :} (D=1.~GeV) effective cut-off $Q$ or
$k_{\perp}$ value (see \ttt{MSTP(64)}), below which space-like
parton showers are not evolved. Primarily intended for QCD showers 
in incoming hadrons, but also applied to $\q \to \q \gamma$ 
branchings.
 
\iteme{PARP(63) :} (D=0.25) in space-like shower evolution the
virtuality $Q^2$ of a parton is multiplied by \ttt{PARP(63)} for use
as a scale in $\alphas$ and parton distributions when
\ttt{MSTP(64)=1}.
 
\iteme{PARP(64) :} (D=1.) in space-like parton-shower evolution
the squared transverse momentum evolution scale $k_{\perp}^2$ is
multiplied by \ttt{PARP(64)} for use as a scale in $\alphas$ and
parton distributions when \ttt{MSTP(64)=2}.
 
\iteme{PARP(65) :} (D=2.~GeV) effective minimum energy (in c.m.\ 
frame) of time-like or on-shell parton emitted in space-like shower;
see also \ttt{PARP(66)}. For a hard subprocess moving in the rest
frame of the hard process, this number is reduced roughly by a factor
$1/\gamma$ for the boost to the hard scattering rest frame.
 
\iteme{PARP(66) :} (D=0.001) effective lower cut-off on $1-z$ in
space-like showers, in addition to the cut implied by \ttt{PARP(65)}.
 
\iteme{PARP(67) :} (D=1.) the $Q^2$ scale of the hard scattering
(see \ttt{MSTP(32)}) is multiplied by \ttt{PARP(67)} to define the
maximum parton virtuality allowed in space-like showers. This does not
apply to $s$-channel resonances, where the maximum virtuality is set
by $m^2$. The current default is based on arguments from a matching of
scales in heavy-flavour production \cite{Nor98}, and other values 
such as 4 (the previous default) could be imagined from other arguments 
or in other processes. 
 
\iteme{PARP(68) :} (D=0.001~GeV) lower $Q$ cut-off for QED space-like
showers. Comes in addition to a hardcoded cut that the $Q^2$ is
at least $2m_{\e}^2$, $2m_{\mu}^2$ or  $2m_{\tau}^2$, as the case
may be.
 
\iteme{PARP(71) :} (D=4.) the $Q^2$ scale of the hard scattering
(see \ttt{MSTP(32)}) is multiplied by \ttt{PARP(71)} to define the
maximum parton virtuality allowed in time-like showers. This does not
apply to $s$-channel resonances, where the maximum virtuality is set
by $m^2$. Like for \ttt{PARP(67)} this number is uncertain.

\iteme{PARP(72) :} (D=0.25 GeV) $\Lambda$ value used in running 
$\alphas$ for timelike parton showers, except for showers in the
decay of a resonance. (Resonance decay, e.g.\ $\gamma^*/Z^0$ decay,
is instead set by \ttt{PARJ(81)}.) 

\end{entry}

\clearpage
 
\section{Beam Remnants and Underlying Events}
 
Each incoming beam particle may leave behind a beam remnant, which
does not take part in the initial-state radiation or hard scattering
process. If nothing else, the remnants need be reconstructed and
connected to the rest of the event. In hadron--hadron collisions,
the composite nature of the two incoming beam particles implies the
additional possibility that several parton pairs undergo separate
hard or semi-hard scatterings, `multiple interactions'.
This may give a non-negligible
contribution to the `underlying event' structure, and thus to the
total multiplicity. Finally, in high-luminosity colliders, it is
possible to have several collisions between beam particles in one
and the same beam crossing, i.e.\ pile-up events, which further act
to build up the
general particle production activity that is to be observed by
detectors. These three aspects are described in turn, with emphasis
on the middle one, that of multiple interactions within a single
hadron--hadron collision.

Work is ongoing to improve the theoretical framework for multiple
interactions. {\Py} therefore contains two partly separate approaches to this
physics, which also leads to two partly separate descriptions of beam
remnants. We will refer to these as the ``old model'' and the ``new
model''. The old model, based on \cite{Sjo87a}, is available as a backup
solution while the new new one is being developed, and for backward reference
and comparisons. Furthermore, the new model is only being developed for
hadron--hadron collisions. Several of the key multiple interactions aspects
are common between the old and new models, such as the generation of
kinematics of the multiple (semi)hard interactions and the impact-parameter
picture, while other aspects are addressed in a more or less similar spirit,
such as the desire to reduce the string length relative to na\"{\i}ve
expectations. The new model \cite{Sjo03a} attempts a more sophisticated
description of correlations in flavour, colour, longitudinal and (primordial)
transverse momenta between the beam remnants and the shower initiators than
offered by the old one, and also introduces some other
improvements. Therefore the current section starts with a description of the
old model, and thereafter outlines where the new model differs. 
 
\subsection{Beam Remnants --- Old Model}
\label{ss:beamrem}
 
The initial-state radiation algorithm reconstructs one shower
initiator in each beam. (If initial-state radiation is not included,
the initiator is nothing but the incoming parton to the hard
interaction.) Together the two initiators delineate an
interaction subsystem, which contains all the partons that
participate in the initial-state showers, in the hard interaction,
and in the final-state showers. Left behind are two beam remnants
which, to first approximation, just sail through, unaffected by the
hard process. (The issue of additional interactions is covered in
the next section.)
 
A description of the beam remnant structure contains a few components.
First, given the flavour content of a (colour-singlet) beam particle,
and the flavour and colour of the initiator parton, it is possible
to reconstruct the flavour and colour of the
beam remnant. Sometimes the remnant may be
represented by just a single parton or diquark, but often the
remnant has to be subdivided into two separate objects. In the latter
case it is necessary to share the remnant energy and momentum between
the two. Due to Fermi motion inside hadron beams, the initiator parton
may have a `primordial $k_{\perp}$' transverse momentum motion,
which has to be compensated by the beam remnant. If the remnant is
subdivided, there may also be a relative transverse momentum.
In the end, total energy and momentum has to be conserved.
To first approximation, this is ensured within each remnant
separately, but some final global adjustments are necessary to
compensate for the primordial $k_{\perp}$ and any effective beam 
remnant mass.

Consider first a proton (or, with trivial modifications, any other 
baryon or antibaryon).
\begin{Itemize}
\item If the initiator parton is a $\u$ or $\d$ quark, it is
assumed to be a valence quark, and therefore leaves behind a
diquark beam remnant, i.e.\ either a $\u\d$ or a $\u\u$ diquark,
in a colour antitriplet state.
Relative probabilities for different diquark spins are derived within
the context of the non-relativistic {\bf SU(6)} model, i.e.\ flavour
{\bf SU(3)} times spin {\bf SU(2)}. Thus a $\u\d$ is $3/4$ $\u\d_0$
and $1/4$ $\u\d_1$, while a $\u\u$ is always $\u\u_1$.
\item An initiator gluon leaves behind a colour octet $\u\u\d$
state, which is subdivided into a colour triplet quark and a colour
antitriplet diquark. {\bf SU(6)} gives the appropriate subdivision,
$1/2$ of the time into $\u + \u\d_0$, $1/6$ into $\u + \u\d_1$ and
$1/3$ into $\d + \u\u_1$.
\item A sea quark initiator, such as an $\s$, leaves behind a
$\u\u\d\sbar$ four-quark state. The PDG flavour coding scheme and
the fragmentation routines do not foresee such a state, so therefore
it is subdivided into a meson plus a diquark, i.e.\ $1/2$ into
$\u\sbar + \u\d_0$, $1/6$ into $\u\sbar +  \u\d_1$ and $1/3$ into
$\d\sbar + \u\u_1$. Once the flavours of the meson are determined,
the choice of meson multiplet is performed as in the standard
fragmentation description.
\item Finally, an antiquark initiator, such as an $\sbar$, leaves
behind a $\u\u\d\s$ four-quark state, which is subdivided into a
baryon plus a quark. Since, to first approximation, the $\s\sbar$
pair comes from the branching $\g \to \s\sbar$ of a colour octet
gluon, the subdivision $\u\u\d + \s$ is not allowed, since it would
correspond to a colour-singlet $\s\sbar$. Therefore the subdivision
is $1/2$ into $\u\d_0\s + \u$, $1/6$ into $\u\d_1\s + \u$ and $1/3$ 
into $\u\u_1\s + \d$. A baryon is formed among the ones possible for 
the given flavour content and diquark spin, according to the relative
probabilities used in the fragmentation. One could argue for an
additional weighting to count the number of baryon states available
for a given diquark plus quark combination, but this has not been
included.
\end{Itemize}
 
One may note that any $\u$ or $\d$ quark taken out of the proton is
automatically assumed to be a valence quark. Clearly this is
unrealistic,
but not quite as bad as it might seem. In particular, one should
remember that the beam remnant scenario is applied to the 
initial-state shower initiators at a scale of $Q_0 \approx 1$ GeV 
and at an
$x$ value usually much larger than the $x$ at the hard scattering.
The sea quark contribution therefore normally is negligible.
 
For a meson beam remnant, the rules are in the same spirit, but
somewhat easier, since no diquark or baryons need be taken into
account. Thus a valence quark (antiquark) initiator leaves
behind a valence antiquark (quark), a gluon initiator leaves
behind a valence quark plus a valence antiquark, and a sea quark
(antiquark) leaves behind a meson (which contains the partner to
the sea parton) plus a valence antiquark (quark).
 
A resolved photon is similar in spirit to a meson. A VMD photon is
associated with either $\rho^0$, $\omega$, $\phi$ or $\Jpsi$, and so
corresponds to a well-defined valence flavour content. Since the 
$\rho^0$ and $\omega$ are supposed to add coherently, the
$\u\ubar : \d\dbar$ mixing is in the ratio $4:1$. Similarly
a GVMD state is characterized by its $\q\qbar$ classification,
in rates according to $e_{\q}^2$ times a mass suppression for
heavier quarks.

In the older photon physics options, where a quark content inside an 
electron is obtained by a numerical convolution, one does not have to 
make the distinction between valence and sea flavour. Thus any quark
(antiquark) initiator leaves behind the matching antiquark (quark),
and a gluon leaves behind a quark + antiquark pair. The relative
quark flavour composition in the latter case is assumed proportional
to $e_{\q}^2$ among light flavours, i.e.\ $2/3$ into $\u + \ubar$,
$1/6$ into $\d + \dbar$, and $1/6$ into $\s + \sbar$. If one wanted
to, one could also have chosen to represent the remnant by a single
gluon.
 
If no initial-state radiation is assumed, an electron (or, in general,
a lepton or a neutrino) leaves behind no beam remnant. Also when
radiation is included, one would expect to recover a single electron
with the full beam energy when the shower initiator is reconstructed.
This does not have to happen, e.g.\ if the initial-state shower is cut
off at a non-vanishing scale, such that some of the emission at low
$Q^2$ values is not simulated. Further, for purely technical reasons,
the distribution of an electron inside an electron,
$f_{\e}^{\e}(x,Q^2)$, is cut off at $x = 1 - 10^{-10}$. This means that
always, when initial-state radiation is included, a fraction
of at least $10^{-10}$ of the beam energy has to be put into one single
photon along the beam direction, to represent this not simulated
radiation. The physics is here slightly different from the standard
beam remnant concept, but it is handled with the same machinery.
Beam remnants can also appear when the electron is resolved with the
use of parton distributions, but initial-state radiation is switched
off. Conceptually, this is a contradiction, since it is the 
initial-state radiation that builds up the parton distributions, 
but sometimes
the combination is still useful. Finally, since QED radiation has not
yet been included in events with resolved photons inside electrons,
also in this case effective beam remnants have to be assigned by the
program.
 
The beam remnant assignments inside an electron, in either of the cases
above, is as follows.
\begin{Itemize}
\item An $\e^-$ initiator leaves behind a $\gamma$ remnant.
\item A $\gamma$ initiator leaves behind an $\e^-$ remnant.
\item An $\e^+$ initiator leaves behind an $\e^- + \e^-$ remnant.
\item A $\q$ ($\qbar$) initiator leaves behind a
$\qbar + \e^-$ ($\q + \e^-$) remnant.
\item A $\g$ initiator leaves behind a $\g + \e^-$ remnant.
One could argue that, in agreement with the treatment of photon
beams above, the remnant should be $\q + \qbar + \e^-$. The program
currently does not allow for three beam remnant objects, however.
\end{Itemize}

 It is customary to assign a primordial transverse momentum to the 
shower initiator, to take into account the motion of quarks inside 
the original hadron, basically as required by the uncertainty principle.
A number of the order of 
$\langle \kT \rangle \approx m_{\p}/3 \approx 300$~MeV
could therefore be expected. However, in hadronic collisions much 
higher numbers than that are often required to describe data, 
typically of the order of or even above 1 GeV \cite{EMC87,Bal01} if
a Gaussian parameterization is used. (This number is now the default.) 
Thus, an interpretation as a purely nonperturbative motion 
inside a hadron is difficult to maintain. 

Instead a likely culprit is the initial-state shower algorithm. This 
is set up to cover the region of hard emissions, but may miss out on 
some of the softer activity, which inherently borders on 
nonperturbative physics. By default, the shower does not evolve down 
to scales below $Q_0 = 1$~GeV. Any shortfall in shower activity around 
or below this cutoff then has to be compensated by the primordial 
$\kT$ source, which thereby largely loses its original meaning.
One specific reason for such a shortfall is that the current 
initial-state shower algorithm does not include non-order emissions
in $Q^2$, as is predicted to occur especially at small $x$ and $Q^2$
within the BFKL/CCFM framework \cite{Lip76,Cia87}.
 
By the hard scattering and initial-state radiation machinery,
the shower initiator has been assigned some fraction $x$ of the
four-momentum of the beam particle, leaving behind $1-x$ to the
remnant. If the remnant consists of two objects, this energy
and momentum has to be shared, somehow. For an electron in the old
photoproduction machinery, the sharing is given from first principles: 
if, e.g., the initiator is a $\q$, then that $\q$ was produced in the 
sequence of branchings $\e \to \gamma \to \q$, where $x_{\gamma}$ is 
distributed according to the convolution in eq.~(\ref{pg:foldqgine}). 
Therefore the $\qbar$ remnant takes a fraction 
$\chi = (x_{\gamma} -x)/(1-x)$ of the total remnant energy, and the 
$\e$ takes $1 - \chi$.
 
For the other beam remnants, the relative energy-sharing variable
$\chi$ is not known from first principles, but picked according to
some suitable parameterization. Normally several different options are
available, that can be set separately for baryon and meson beams, and
for hadron + quark and quark + diquark (or antiquark) remnants. In one
extreme are shapes in agreement with na\"{\i}ve counting rules, i.e.\
where energy is shared evenly between `valence' partons. For instance,
${\cal P}(\chi) = 2 \, (1-\chi)$ for the energy fraction taken by the
$\q$ in a $\q + \q\q$ remnant. In the other extreme, an uneven 
distribution could be used, like in parton distributions, where the 
quark only takes a small fraction and most is retained by the diquark.
The default for a $\q + \q\q$ remnant is of an intermediate type,
\begin{equation}
{\cal P}(\chi) \propto \frac{(1 - \chi)^3}
{\sqrt[4]{\chi^2 + c_{\mmin}^2}} ~,
\end{equation}
with $c_{\mmin} = 2 \langle m_{\q} \rangle / E_{\mrm{cm}} = (0.6$
GeV)$/ E_{\mrm{cm}}$ providing a lower cut-off. The default when a 
hadron is split off to leave a quark or diquark remnant is to use the 
standard Lund symmetric fragmentation function. In general, the more 
uneven the sharing of the energy, the less the total multiplicity in the
beam remnant fragmentation. If no multiple interactions are allowed,
a rather even sharing is needed to come close to the experimental
multiplicity (and yet one does not quite make it). With an
uneven sharing there is room to generate more of the total multiplicity
by multiple interactions \cite{Sjo87a}.
 
In a photon beam, with a remnant $\q + \qbar$, the $\chi$ variable is
chosen the same way it would have been in a corresponding meson 
remnant. 
 
Before the $\chi$ variable is used to assign remnant momenta, it
is also necessary to consider the issue of primordial $k_{\perp}$.
The initiator partons are thus assigned each a $k_{\perp}$ value,
vanishing for an electron or photon inside an electron, distributed
either according to a Gaussian or an exponential shape for a hadron,
and according to either of these shapes or a power-like shape
for a quark or gluon inside a photon (which may in its turn be inside
an electron). The interaction subsystem is boosted and rotated to bring
it from the frame assumed so far, with each initiator along the $\pm z$
axis, to one where the initiators have the required primordial
$k_{\perp}$ values.
 
The $\pT$ recoil is taken by the remnant. If the remnant is composite,
the recoil is evenly split between the two. In addition, however, the
two beam remnants may be given a relative $\pT$, which is then always 
chosen as for $\q_i \qbar_i$ pairs in the fragmentation description.
 
The $\chi$ variable is interpreted as a sharing of light-cone energy and
momentum, i.e.\ $E + p_z$ for the beam moving in the $+z$ direction and
$E - p_z$ for the other one. When the two transverse masses
$m_{\perp 1}$ and $m_{\perp 2}$ of a composite remnant have been
constructed, the total transverse mass can therefore be found as
\begin{equation}
m_{\perp}^2 = \frac{m_{\perp 1}^2}{\chi} +
\frac{m_{\perp 2}^2}{1 - \chi}   ~,
\end{equation}
if remnant 1 is the one that takes the fraction $\chi$. The choice of
a light-cone interpretation to $\chi$ means the definition is invariant
under longitudinal boosts, and therefore does not depend on the beam
energy itself. A $\chi$ value close to the na\"{\i}ve borders 0 or 1 can
lead to an unreasonably large remnant $m_{\perp}$.
Therefore an additional check is introduced, that the remnant
$m_{\perp}$ be smaller than the na\"{\i}ve c.m.\ frame remnant energy,
$(1-x) E_{\mrm{cm}}/2$. If this is not the case, a new $\chi$ and a 
new relative transverse momentum is selected.
 
Whether there is one remnant parton or two, the transverse mass of
the remnant is not likely to agree with $1-x$ times the mass of the
beam particle, i.e.\ it is not going to be possible to preserve
the energy and momentum
in each remnant separately. One therefore allows a
shuffling of energy and momentum between the beam remnants from
each of the two incoming beams. This may be achieved by performing a
(small) longitudinal boost of each remnant system. Since there are
two boost degrees of freedom, one for each remnant, and two constraints,
one for energy and one for longitudinal momentum, a solution may be
found.
 
Under some circumstances, one beam remnant may be absent or of very
low energy, while the other one is more complicated. One example is
Deeply Inelastic Scattering in $\ep$ collisions, where the electron
leaves no remnant, or maybe only a low-energy photon.
It is clearly then not possible to balance the two beam remnants
against each other. Therefore, if one beam remnant has an energy
below 0.2 of the beam energy, i.e.\ if the initiator parton has
$x > 0.8$, then the two boosts needed to ensure energy and momentum
conservation are instead performed on the other remnant and on the
interaction subsystem. If there is a low-energy remnant at all then,
before that, energy and momentum are assigned to the remnant
constituent(s) so that the appropriate light-cone combination
$E \pm p_z$ is conserved, but not energy or momentum separately.
If both beam remnants have low energy, but both still exist, then
the one with lower $m_{\perp} / E$ is the one that will not be
boosted.
 
\subsection{Multiple Interactions --- Old Model}
\label{ss:multint}
 
In this section we present the model \cite{Sjo87a} used in older 
versions of  {\Py} (and still available with the switch 
\ttt{MSTP(81)=-1}) to describe the possibility that several parton 
pairs undergo hard interactions in a hadron--hadron collision, and 
thereby contribute to the overall event activity, in particular at 
low $\pT$. The same model can also be used to describe the VMD 
$\gamma \p$ events, where the photon interacts like a hadron. 
Many basic features of this model, for instance the 
introduction of a $\pT$ cutoff corresponding to an inverse colour 
screening distance, and the options for a non-trivial transverse 
density structure in the incoming hadrons, carry over to the new 
scenario. It is therefore recommended first to read this section, 
even if the objective should be to learn about the new scenario.

It should from the onset be 
made clear that this is not an easy topic. In fact, in the full event
generation process, probably no other area is as poorly understood
as this one. The whole concept of multiple interactions has been very
controversial, with contradictory experimental conclusions
\cite{AFS87}, but a CDF study \cite{CDF97} some years ago started to 
bring more general acceptance.
 
The multiple interactions scenario presented here \cite{Sjo87a} was
the first detailed model for this kind of physics, and is still one
of the very few available. We will present two related but separate
scenarios, one `simple' model and one somewhat more sophisticated.
In fact, neither of them are all that simple, which may make the
models look unattractive. However, the world of hadron physics
{\it is} complicated, and if we err, it is most likely in being
too unsophisticated. The experience gained with the model(s), in
failures as well as successes, could be used as a guideline in
the evolution of yet more detailed models.
 
Our basic philosophy will be as follows. The total rate of
parton--parton interactions, as a function of the transverse
momentum scale $\pT$, is assumed to be given by perturbative
QCD. This is certainly true for reasonably large $\pT$ values, but
here we shall also extend the perturbative parton--parton
scattering framework into the low-$\pT$ region. A regularization of
the divergence in the cross section for $\pT \to 0$ has to be
introduced, however, which will provide us with the main free
parameter of the model. Since each incoming hadron is a composite
object, consisting of many partons, there should exist the possibility
of several parton pairs interacting when two hadrons collide. It is
not unreasonable to assume that the different pairwise interactions
take place essentially independently of each other, and that therefore
the number of interactions in an event is given by a Poisson
distribution. This is the strategy of the `simple' scenario.
 
Furthermore, hadrons are not only composite but also extended
objects, meaning that collisions range from very central to rather
peripheral ones. Reasonably, the average number of interactions
should be larger in the former than in the latter case. Whereas
the assumption of a Poisson distribution should hold for each
impact parameter separately, the distribution
in number of interactions should be widened by the spread of impact
parameters. The amount of widening depends on the assumed
matter distribution inside the colliding hadrons. In the `complex'
scenario, different matter distributions are therefore introduced.
 
\subsubsection{The basic cross sections}
 
The QCD cross section for hard $2 \to 2$ processes, as a function
of the $\pT^2$ scale, is given by
\begin{equation}
\frac{\d \sigma}{\d \pT^2} = \sum_{i,j,k} \int \d x_1 \int \d x_2 
\int \d \hat{t} \, f_i(x_1,Q^2) \, f_j(x_2,Q^2)
\, \frac{\d \hat{\sigma}_{ij}^k}{\d \hat{t}} \,
\delta \! \left( \pT^2 - \frac{\hat{t}\hat{u}}{\hat{s}} \right) ~,
\label{mi:sigmapt}
\end{equation}
cf. section \ref{ss:kinemtwo}. Implicitly, from now on we are assuming
that the `hardness' of processes is given by the $\pT$ scale of the
scattering. For an application of the formula above to small $\pT$
values, a number of caveats could be made. At low $\pT$, the
integrals receive major contributions from the small-$x$ region,
where parton distributions are poorly understood theoretically
(Regge limit behaviour, dense packing problems etc. \cite{Lev90})
and not yet measured. Different sets of parton distributions can
therefore give numerically rather different results for the
phenomenology of interest. One may also worry about higher-order
corrections to the jet rates, $K$ factors, beyond what is given
by parton-shower corrections --- one simple option we allow here
is to evaluate $\alphas$ of the hard scattering process at an
optimized scale, such as $\alphas(0.075\pT^2)$ \cite{Ell86}.
 
The hard scattering cross section above some given $\pTmin$
is given by
\begin{equation}
\sigma_{\mrm{hard}} (\pTmin) = \int_{\pTmin^2}^{s/4}
\frac{\d \sigma}{\d \pT^2} \, \d \pT^2 ~.
\label{mi:sigmahard}
\end{equation}
Since the differential cross section diverges roughly like
$\d \pT^2 / \pT^4$, $\sigma_{\mrm{hard}}$ is also divergent for
$\pTmin \to 0$. We may compare this with the total
inelastic, non-diffractive cross section $\sigma_{\mrm{nd}}(s)$
--- elastic and diffractive events are not the topic of this
section. At current collider energies
$\sigma_{\mrm{hard}} (\pTmin)$ 
becomes comparable with $\sigma_{\mrm{nd}}$
for $\pTmin \approx$ 1.5--2 GeV. This need not lead to
contradictions: $\sigma_{\mrm{hard}}$ does not give the hadron--hadron
cross section but the parton--parton one. Each of the incoming
hadrons may be viewed as a beam of partons, with the possibility of
having several parton--parton interactions when the hadrons pass
through each other. In this language,
$\sigma_{\mrm{hard}} (\pTmin) / \sigma_{\mrm{nd}}(s)$ 
is simply the
average number of parton--parton scatterings above $\pTmin$
in an event, and this number may well be larger than unity.
 
While the introduction of several interactions per event is the
natural consequence of allowing small $\pTmin$ values
and hence large $\sigma_{\mrm{hard}}$ ones, it is not the solution of
$\sigma_{\mrm{hard}} (\pTmin)$ being divergent for
$\pTmin \to 0$: the average $\hat{s}$ of a scattering
decreases slower with $\pTmin$ than the number of
interactions increases, so na\"{\i}vely the total amount of scattered
partonic energy becomes infinite. One cut-off is therefore
obtained via the need to introduce proper multi-parton correlated
parton distributions inside a hadron. This is not a part of the
standard perturbative QCD formalism and is therefore not built into
eq.~(\ref{mi:sigmahard}). In practice, even correlated 
parton-distribution functions seems to provide too weak a cut, 
i.e.\ one is lead to a
picture with too little of the incoming energy remaining in the
small-angle beam jet region \cite{Sjo87a}.
 
A more credible reason for an effective cut-off is that the incoming
hadrons are colour neutral objects. Therefore, when the $\pT$ of an
exchanged gluon is made small and the transverse wavelength
correspondingly large, the gluon can no longer resolve the individual
colour charges, and the effective coupling is decreased. This
mechanism is not in contradiction to perturbative
QCD calculations, which are always performed assuming scattering
of free partons (rather than partons inside hadrons), but neither does
present knowledge of QCD provide an understanding of how such a
decoupling mechanism would work in detail. In the simple model one
makes use of a sharp cut-off at some scale $\pTmin$, while
a more smooth dampening is assumed for the complex scenario.

One key question is the energy-dependence of $\pTmin$; 
this may be relevant e.g.\ for comparisons of jet rates at different 
Tevatron energies, and even more for any extrapolation to LHC energies. 
The problem actually is more pressing now than at the time of the 
original study \cite{Sjo87a}, since nowadays parton distributions are 
known to be rising more steeply at small $x$ than the flat $xf(x)$ 
behaviour normally assumed for small $Q^2$ before HERA. This 
translates into a more dramatic energy dependence of the 
multiple-interactions rate for a fixed $\pTmin$. 

The larger number of partons should also increase the amount of
screening, however, as confirmed by toy simulations \cite{Dis01}.
As a simple first approximation, $\pTmin$ is assumed
to increase in the same way as the total cross section, i.e.\ with some 
power $\epsilon \approx 0.08$ \cite{Don92} that, via reggeon 
phenomenology, should relate to the behaviour of parton distributions 
at small $x$ and $Q^2$. Thus the default in {\Py} is
\begin{equation}
\pTmin(s) = (1.9~{\mrm{GeV}}) \left(
\frac{s}{1~\mrm{TeV}^2} \right)^{0.08}
\label{eq:ptmin}
\end{equation}
for the simple model, with the same ansatz for $p_{\perp 0}$ in the 
impact-parameter-dependent approach, except that then 1.9~GeV
$\to$ 2.1~GeV. At any energy scale, the simplest criterion to fix
$\pTmin$ is to require the average charged multiplicity to agree with
the experimentally determined one. In general, there is quite a strong
dependence of the multiplicity on $\pTmin$, with a lower $\pTmin$
corresponding to more multiple interactions and therefore a higher 
multiplicity. This is the way the 1.9~GeV and 2.1~GeV numbers are fixed, 
based on a comparison with UA5 data in the energy range 200--900 GeV 
\cite{UA584}. The energy dependence inside this range is also consistent 
with the chosen ansatz. However, clearly, neither the experimental nor 
the theoretical precision is high enough to make too strong statements.
It should also be remembered that the $\pTmin$ values are determined 
within the context of a given calculation of the QCD jet cross section,
and given model parameters within the multiple interactions scenario.
If anything of this is changed, e.g. the parton distributions used,
then $\pTmin$ ought to be retuned accordingly.
 
\subsubsection{The simple model}
 
In an event with several interactions, it is convenient to impose an
ordering. The logical choice is to arrange the scatterings in falling
sequence of $x_{\perp} = 2 \pT / E_{\mrm{cm}}$. The `first' scattering 
is thus the hardest one, with the `subsequent' (`second', `third', etc.)
successively softer. It is important to remember that this terminology
is in no way related to any picture in physical time; we do not know
anything about the latter. In principle, all the scatterings that occur
in an event must be correlated somehow, na\"{\i}vely by momentum 
and flavour conservation for the partons from each incoming hadron, 
less na\"{\i}vely by
various quantum mechanical effects. When averaging over all
configurations of soft partons, however, one should effectively obtain
the standard QCD phenomenology for a hard scattering, e.g.\ in terms of
parton distributions. Correlation effects, known or estimated, can be
introduced in the choice of subsequent scatterings, given that the
`preceding' (harder) ones are already known.
 
With a total cross section of hard interactions
$\sigma_{\mrm{hard}} (\pTmin)$ to be distributed among
$\sigma_{\mrm{nd}}(s)$ (non-diffractive, inelastic) events, the average
number of interactions per event is just the ratio
$\br{n} = \sigma_{\mrm{hard}} (\pTmin) / \sigma_{\mrm{nd}}(s)$. 
As a starting point we will assume that all hadron collisions are
equivalent (no impact parameter dependence), and that the different
parton--parton interactions take place completely independently of
each other. The number of scatterings per event is then distributed
according to a Poisson distribution with mean $\br{n}$. A fit to 
S$\p\pbar$S collider multiplicity data \cite{UA584} gave 
$\pTmin \approx 1.6$ GeV, which corresponds to $\br{n} \approx 1$. 
For Monte Carlo generation of these interactions it is useful to define
\begin{equation}
f(x_{\perp}) = \frac{1}{\sigma_{\mrm{nd}}(s)} \,
\frac{\d \sigma}{\d x_{\perp}}   ~,
\end{equation}
with $\d \sigma / \d x_{\perp}$ obtained by analogy with eq.
(\ref{mi:sigmapt}). Then $f(x_{\perp})$ is simply the probability to
have a parton--parton interaction at $x_{\perp}$, given that the two
hadrons undergo a non-diffractive, inelastic collision.
 
The probability that the hardest interaction, i.e.\ the one with
highest $x_{\perp}$, is at $x_{\perp 1}$, is now given by
\begin{equation}
f(x_{\perp 1}) \exp \left\{ - \int_{x_{\perp 1}}^1
f(x'_{\perp}) \, \d x'_{\perp} \right\}   ~,
\label{mi:hardest}
\end{equation}
i.e.\ the na\"{\i}ve probability to have a scattering at $x_{\perp 1}$
multiplied by the probability that there was no scattering with
$x_{\perp}$ larger than $x_{\perp 1}$. This is the familiar
exponential dampening in radioactive decays, encountered e.g.\ in
parton showers in section \ref{ss:sudakov}. Using the same technique
as in the proof of the veto algorithm, section \ref{ss:vetoalg},
the probability to have an $i$:th scattering at an
$x_{\perp i} < x_{\perp i-1} < \cdots < x_{\perp 1} < 1$ is found
to be
\begin{equation}
f(x_{\perp i}) \, \frac{1}{(i-1)!} \left( \int_{x_{\perp i}}^1
f(x'_{\perp}) \, \d x'_{\perp} \right)^{i-1} \exp \left\{
- \int_{x_{\perp i}}^1 f(x'_{\perp}) \, \d x'_{\perp} \right\}  ~.
\end{equation}
The total probability to have a scattering at a given $x_{\perp}$,
irrespectively of it being the first, the second or whatever,
obviously adds up to give back $f(x_{\perp})$. The multiple
interaction formalism thus retains the correct perturbative
QCD expression for the scattering probability at any given
$x_{\perp}$.
 
With the help of the integral
\begin{equation}
F(x_{\perp}) = \int_{x_{\perp}}^1 f(x'_{\perp}) \, \d x'_{\perp}
= \frac{1}{\sigma_{\mrm{nd}}(s)} \, \int_{s x_{\perp}^2/4}^{s/4}
\frac{\d \sigma}{\d \pT^2} \, \d \pT^2
\end{equation}
(where we assume $F(x_{\perp}) \to \infty$ for $x_{\perp} \to 0$)
and its inverse $F^{-1}$, the iterative procedure to generate
a chain of scatterings
$1 > x_{\perp 1} > x_{\perp 2} > \cdots > x_{\perp i}$
is given by
\begin{equation}
x_{\perp i} = F^{-1}(F(x_{\perp i-1}) - \ln R_i) ~.
\label{mi:iter}
\end{equation}
Here the $R_i$ are random numbers evenly distributed between 0 and 1.
The iterative chain is started with a fictitious $x_{\perp 0} = 1 $
and is terminated when $x_{\perp i}$ is smaller than
$x_{\perp \mmin} = 2 \pTmin / E_{\mrm{cm}}$. Since $F$ and 
$F^{-1}$ are not known analytically, the standard veto algorithm is 
used to generate a much denser set of $x_{\perp}$ values, whereof 
only some are retained in the end. In addition to the $\pT^2$ of an
interaction, it is also necessary to generate the other flavour
and kinematics variables according to the relevant matrix elements.
 
Whereas the ordinary parton distributions should be used for the
hardest scattering, in order to reproduce standard QCD
phenomenology, the parton distributions to be used for subsequent
scatterings must depend on all preceding $x$ values and flavours
chosen. We do not know enough about the hadron wave function to
write down such joint probability distributions. To take
into account the energy `already' used in harder scatterings, a
conservative approach is to evaluate the parton distributions, not at
$x_i$ for the $i$:th scattered parton from hadron, but at
the rescaled value
\begin{equation}
x'_i = \frac{x_i}{1 - \sum_{j=1}^{i-1} x_j} ~.
\label{mi:xresc}
\end{equation}
This is our standard procedure in the program; we have
tried a few alternatives without finding any significantly
different behaviour in the final physics.
 
In a fraction $\exp(-F(x_{\perp \mmin}))$ of the events studied, there
will be no hard scattering above $x_{\perp \mmin}$ when the iterative
procedure in eq.~(\ref{mi:iter}) is applied. It is therefore also
necessary to have a model for what happens in events with no
(semi)hard interactions. The simplest possible way to produce an event
is to have an exchange of a very soft gluon between the two colliding
hadrons. Without (initially) affecting the momentum distribution of
partons, the `hadrons' become colour octet objects rather than colour
singlet ones. If only valence quarks are considered, the colour
octet state of a baryon can be decomposed into a colour triplet quark
and an antitriplet diquark. In a baryon-baryon collision, one would
then obtain a two-string picture, with each string stretched from the
quark of one baryon to the diquark of the other. A baryon-antibaryon
collision would give one string between a quark and an antiquark and
another one between a diquark and an antidiquark.
 
In a hard interaction, the number of possible string drawings are many
more, and the overall situation can become quite complex
when several hard scatterings are present in an event.
Specifically, the string drawing now depends on the relative colour
arrangement, in each hadron individually, of the partons that are about
to scatter. This is a subject about which nothing is known.
To make matters worse, the standard string fragmentation description
would have to be extended, to handle events where two or more valence
quarks have been kicked out of an incoming hadron by separate
interactions. In particular, the position of the baryon number would
be unclear. We therefore here assume that, following the hardest
interaction, all subsequent interactions belong to one of three
classes.
\begin{Itemize}
\item Scatterings of the $\g\g \to \g\g $ type, with
the two gluons in a colour-singlet state, such that a double string is
stretched directly between the two outgoing gluons, decoupled from the
rest of the system.
\item Scatterings $\g\g \to \g\g$, but colour correlations
assumed to be such that each of the gluons is connected to one
of the strings `already' present. Among the different possibilities of
connecting the colours of the gluons, the one which minimizes the total
increase in string length is chosen. This is in contrast to the
previous alternative, which roughly corresponds to a maximization 
(within reason) of the extra string length.
\item Scatterings $\g\g \to \q\qbar$, with the final pair again
in a colour-singlet state, such that a single string is stretched
between the outgoing $\q$ and $\qbar$.
\end{Itemize}
By default, the three possibilities are assumed equally probable.
Note that the total jet rate is maintained at its nominal value, i.e.\
scatterings such as $\q\g \to \q\g$ are included in the cross section,
but are replaced by a mixture of $\g\g$ and $\q\qbar$ events for string
drawing issues. Only the hardest interaction is guaranteed to give
strings coupled to the beam remnants. One should not take this approach
to colour flow too seriously --- clearly
it is a simplification --- but the overall picture does not tend to be
very dependent on the particular choice you make.
 
Since a $\g\g \to \g\g$ or $\q\qbar$ scattering need not remain
of this character if initial- and final-state showers were to be included
(e.g.\ it could turn into a $\q\g$-initiated process), radiation is only
included for the hardest interaction. In practice, this is not a serious
problem: except for the hardest interaction, which can be hard because
of experimental trigger conditions, it is unlikely for a parton
scattering to be so hard that radiation plays a significant r\^ole.
 
In events with multiple interactions, the beam remnant treatment is
slightly modified. First the hard scattering is generated, with its
associated initial- and final-state radiation, and next any additional
multiple interactions. Only thereafter are beam remnants attached to
the initiator partons of the hardest scattering, using the same
machinery as before, except that the energy and momentum already
taken away from the beam remnants also include that of the
subsequent interactions.
 
\subsubsection{A model with varying impact parameters}
 
Up to this point, it has been assumed that the initial state is the
same for all hadron collisions, whereas in fact each collision also
is characterized by a varying impact parameter $b$. Within the
classical framework of the model reviewed here, $b$ is to be thought 
of as a distance of closest approach, not as the Fourier transform 
of the momentum transfer. A small $b$ value corresponds to a large
overlap between the two colliding hadrons, and hence an enhanced
probability for multiple interactions. A large $b$, on the other
hand, corresponds to a grazing collision, with a large probability
that no parton--parton interactions at all take place.
 
In order to quantify the concept of hadronic matter overlap, one may
assume a spherically symmetric distribution of matter inside the
hadron, $\rho(\mbf{x}) \, \d^3 x = \rho(r) \, \d^3 x$.
For simplicity, the same spatial distribution is taken to apply
for all parton species and momenta. Several different matter
distributions have been tried, and are available. We will here
concentrate on the most extreme one, a double Gaussian
\begin{equation}
\rho(r) \propto \frac{1 - \beta}{a_1^3} \, \exp \left\{
- \frac{r^2}{a_1^2} \right\} + \frac{\beta}{a_2^3} \,
\exp \left\{ - \frac{r^2}{a_2^2} \right\} ~.
\label{mi:doubleGauss}
\end{equation}
This corresponds to a distribution with a small core region, of radius
$a_2$ and containing a fraction $\beta$ of the total hadronic matter,
embedded in a larger hadron of radius $a_1$. While it is mathematically
convenient to have the origin of the two Gaussians coinciding, the
physics could well correspond to having three disjoint core regions,
reflecting the presence of three valence quarks, together carrying
the fraction $\beta$ of the proton momentum. One could alternatively
imagine a hard hadronic core surrounded by a pion cloud.
Such details would affect e.g.\ the predictions for the $t$ distribution
in elastic scattering, but are not of any consequence for the
current topics. To be
specific, the values $\beta = 0.5$ and $a_2/a_1 = 0.2$ have been
picked as default values. It should be noted that the overall distance
scale $a_1$ never enters in the subsequent calculations, since the
inelastic, non-diffractive cross section $\sigma_{\mrm{nd}}(s)$ is taken
from literature rather than calculated from the $\rho(r)$.
 
Compared to other shapes, like a simple Gaussian, the double Gaussian
tends to give larger fluctuations, e.g.\ in the multiplicity
distribution of minimum bias events: a collision in which the two
cores overlap tends to have a strongly increased activity, while
ones where they do not are rather less active. One also has a
biasing effect: hard processes are more likely
when the cores overlap, thus hard scatterings are associated with
an enhanced multiple interaction rate. This provides one possible
explanation for the experimental `pedestal effect' \cite{UA187}. 
Recent studies of CDF data \cite{Fie02,Mor02} have confirmed that 
indeed something more peaked than a single Gaussian is required to 
understand the transition from minimum-bias to underlying-event
activity.
 
For a collision with impact parameter $b$, the time-integrated
overlap ${\cal O}(b)$ between the matter distributions of the
colliding hadrons is given by
\begin{eqnarray}
& & {\cal O}(b) \propto \int \d t \int \d^3 x \, \rho(x,y,z) \,
\rho(x+b,y,z+t)      \nonumber \\
& & \propto \frac{(1 - \beta)^2}{2a_1^2} \exp \left\{
- \frac{b^2}{2a_1^2} \right\} +
\frac{2 \beta (1-\beta)}{a_1^2+a_2^2} \exp \left\{
- \frac{b^2}{a_1^2+ a_2^2} \right\} +
\frac{\beta^2}{2a_2^2} \exp \left\{ - \frac{b^2}{2a_2^2} \right\} ~.
\end{eqnarray}
The necessity to use boosted $\rho(\mbf{x})$ distributions
has been circumvented by a suitable scale transformation of the $z$
and $t$ coordinates.
 
The overlap ${\cal O}(b)$ is obviously strongly related to the
eikonal $\Omega(b)$ of optical models. We have kept a separate
notation, since the physics context of the two is slightly
different: $\Omega(b)$ is based on the quantum mechanical
scattering of waves in a potential, and is normally used to
describe the elastic scattering of a hadron-as-a-whole, while
${\cal O}(b)$ comes from a purely classical picture of point-like
partons distributed inside the two colliding hadrons. Furthermore,
the normalization and energy dependence is differently realized
in the two formalisms.

The larger the overlap ${\cal O}(b)$ is, the more likely it is to
have interactions between partons in the two colliding hadrons.
In fact, there should be a linear relationship
\begin{equation}
\langle \tilde{n}(b) \rangle = k {\cal O}(b) ~,
\end{equation}
where $\tilde{n} = 0, 1, 2, \ldots$ counts the number of interactions
when two hadrons pass each other with an impact parameter $b$.
The constant of proportionality, $k$, is related to the
parton--parton cross section and hence increases with c.m.\ energy.
 
For each given impact parameter, the number of interactions is
assumed to be distributed according to a Poisson. If the matter
distribution has a tail to infinity (as the double Gaussian does),
events may be obtained with arbitrarily large $b$ values. In order
to obtain finite total cross sections, it is necessary to assume
that each event contains at least one semi-hard interaction. The
probability that two hadrons, passing each other with an impact
parameter $b$, will actually undergo a collision is then given by
\begin{equation}
{\cal P}_{\mrm{int}}(b) = 1 - \exp ( - \langle \tilde{n}(b) \rangle )
= 1 - \exp (- k {\cal O}(b) ) ~,
\end{equation}
according to Poisson statistics. The average number of
interactions per event at impact parameter $b$ is now
\begin{equation}
\langle n(b) \rangle =
\frac{ \langle \tilde{n}(b) \rangle }{{\cal P}_{\mrm{int}}(b)} =
\frac{ k {\cal O}(b) }{ 1 - \exp (- k {\cal O}(b) ) } ~,
\label{mi:nofb}
\end{equation}
where the denominator comes from the removal of hadron pairs which
pass without colliding, i.e.\ with $\tilde{n} =  0$.
 
The relationship 
$\langle n \rangle = \sigma_{\mrm{hard}}/\sigma_{\mrm{nd}}$
was earlier introduced for the average number of interactions per
non-diffractive, inelastic event. When averaged over all
impact parameters, this relation must still hold true: the
introduction of variable impact parameters may give more interactions
in some events and less in others, but it does not affect either
$\sigma_{\mrm{hard}}$ or $\sigma_{\mrm{nd}}$. 
For the former this is because the
perturbative QCD calculations only depend on the total parton flux,
for the latter by construction. Integrating eq.~(\ref{mi:nofb}) over
$b$, one then obtains
\begin{equation}
\langle n \rangle =
\frac{ \int \langle n(b) \rangle \, {\cal P}_{\mrm{int}}(b) \, \d^2 b}
{ \int {\cal P}_{\mrm{int}}(b) \, \d^2 b} =
\frac{ \int k {\cal O}(b) \, \d^2 b}
{ \int \left( 1 - \exp (- k {\cal O}(b) ) \right) \, \d^2 b} =
\frac{\sigma_{\mrm{hard}}}{\sigma_{\mrm{nd}}}   ~.
\label{mi:kfromsigma}
\end{equation}
For ${\cal O}(b)$, $\sigma_{\mrm{hard}}$ and $\sigma_{\mrm{nd}}$ given, 
with $\sigma_{\mrm{hard}} / \sigma_{\mrm{nd}} > 1$, $k$ can thus always 
be found (numerically) by solving the last equality.
 
The absolute normalization of ${\cal O}(b)$ is not interesting
in itself, but only the relative variation with impact parameter.
It is therefore useful to introduce an `enhancement factor'
$e(b)$, which gauges how the interaction probability for a passage
with impact parameter $b$ compares with the average, i.e.\
\begin{equation}
\langle \tilde{n}(b) \rangle = k{\cal O}(b) =
e(b) \, \langle k{\cal O}(b) \rangle  ~.
\label{mi:ebenh}
\end{equation}
The definition of the average $\langle k{\cal O}(b) \rangle$ is a
bit delicate, since the average number of interactions per event
is pushed up by the requirement that each event contain at least
one interaction. However, an exact meaning can be given \cite{Sjo87a}.
 
With the knowledge of $e(b)$, the $f(x_{\perp})$ function of the
simple model generalizes to
\begin{equation}
f(x_{\perp},b) = e(b) \, f(x_{\perp})     ~.
\end{equation}
The na\"{\i}ve generation procedure is thus to pick a $b$ according 
to the phase space $\d^2 b$, find the relevant $e(b)$ and plug in the
resulting $f(x_{\perp},b)$ in the formalism of the simple model.
If at least one hard interaction is generated, the event is retained,
else a new $b$ is to be found. This algorithm would work fine for
hadronic matter distributions which vanish outside some radius, so
that the $\d^2 b$ phase space which needs to be probed is finite.
Since this is not true for the distributions under study, it is
necessary to do better.
 
By analogy with eq.~(\ref{mi:hardest}), it is possible to ask what
the probability is to find the hardest scattering of an event at
$x_{\perp 1}$. For each impact parameter separately, the probability
to have an interaction at $x_{\perp 1}$ is given by $f(x_{\perp},b)$,
and this should be multiplied by the probability that the event
contains no interactions at a scale $x'_{\perp} > x_{\perp 1}$,
to yield the total probability distribution
\begin{eqnarray}
\frac{\d {\cal P}_{\mrm{hardest}}}{\d^2 b \, \d x_{\perp 1}} & = &
f(x_{\perp 1},b) \, \exp \left\{ - \int_{x_{\perp 1}}^1
f(x'_{\perp},b) \, \d x'_{\perp} \right\}  \nonumber \\
& = & e(b) \, f(x_{\perp 1}) \, \exp \left\{ - e(b)
\int_{x_{\perp 1}}^1 f(x'_{\perp}) \, \d x'_{\perp} \right\} ~.
\label{mi:hardestvarimp}
\end{eqnarray}
If the treatment of the exponential is deferred for a moment,
the distribution in $b$ and $x_{\perp 1}$ appears in factorized form,
so that the two can be chosen independently of each other. In
particular, a high-$\pT$ QCD scattering or any other hard scattering
can be selected with whatever kinematics desired for that process,
and thereafter assigned some suitable `hardness' $x_{\perp 1}$.
With the $b$ chosen according to $e(b) \, \d^2 b$, the neglected
exponential can now be evaluated, and the event retained with a
probability proportional to it. From the $x_{\perp 1}$ scale of the
selected interaction, a sequence of softer $x_{\perp i}$ values may
again be generated as in the simple model, using the known
$f(x_{\perp},b)$. This sequence may be empty, i.e.\ the
event need not contain any further interactions.
 
It is interesting to understand how the algorithm above works.
By selecting $b$ according to $e(b) \, \d^2 b$, i.e.\
${\cal O}(b) \, \d^2 b$, the primary $b$ distribution is
maximally biased towards small impact parameters. If the first
interaction is hard, by choice or by chance, the integral of
the cross section above $x_{\perp 1}$ is small, and the exponential
close to unity. The rejection procedure is therefore very efficient
for all standard hard processes in the program --- one may even
safely drop the weighting with the exponential completely. The large
$e(b)$ value is also likely to lead to the generation of many further,
softer interactions. If, on the other hand, the first interaction is
not hard, the exponential is no longer close to unity, and many
events are rejected. This pulls down the efficiency for `minimum bias'
event generation. Since the exponent is proportional to $e(b)$,
a large $e(b)$ leads to an enhanced probability for rejection,
whereas the chance of acceptance is larger with a small $e(b)$.
Among events where the hardest interaction is soft, the $b$
distribution is therefore biased towards larger values
(smaller $e(b)$), and there is a small probability for yet softer
interactions.
 
To evaluate the exponential factor, the program pretabulates the
integral of $f(x_{\perp})$ at the initialization stage, and further
increases the Monte Carlo statistics of this tabulation as the run
proceeds. The $x_{\perp}$ grid is concentrated towards small
$x_{\perp}$, where the integral is large. For a selected
$x_{\perp 1}$ value, the $f(x_{\perp})$ integral is obtained by
interpolation. After multiplication by the known $e(b)$ factor,
the exponential factor may be found.
 
In this section, nothing has yet been assumed about the form of the
$\d \sigma / \d \pT$ spectrum. Like in the impact parameter independent
case, it is possible to use a sharp cut-off at some given
$\pTmin$ value. However, now each event is required to have
at least one interaction, whereas before events without interactions
were retained and put at $\pT = 0$. It is therefore aesthetically
more appealing to assume a gradual turn-off, so that a (semi)hard
interaction can be rather soft part of the time. The matrix elements
roughly diverge like $\alphas(\pT^2) \, \d \pT^2 / \pT^4$ for
$\pT \to 0$. They could therefore be regularized as follows. Firstly,
to remove the $1/\pT^4$ behaviour, multiply by a factor
$\pT^4 / (\pT^2 + p_{\perp 0}^2)^2$. Secondly, replace the $\pT^2$
argument in $\alphas$ by $\pT^2 + p_{\perp 0}^2$. If one has 
included a $K$ factor by a rescaling of the $\alphas$ argument, 
as mentioned earlier, replace $0.075 \, \pT^2$ by 
$0.075 \, (\pT^2 + p_{\perp 0}^2)$.
 
With these substitutions, a continuous $\pT$ spectrum is obtained,
stretching from $\pT = 0$ to $E_{\mrm{cm}}/2$. For 
$\pT \gg p_{\perp 0}$
the standard perturbative QCD cross section is recovered, while
values $\pT \ll p_{\perp 0}$ are strongly damped. The $p_{\perp 0}$
scale, which now is the main free parameter of the model, in
practice comes out to be of the same order of magnitude as the sharp
cut-off $\pTmin$ did, i.e.\ 1.5--2 GeV, but typically about 10\% higher.
 
Above we have argued that $\pTmin$ and $p_{\perp 0}$ should only have 
a slow energy dependence, and even allowed for the possibility of
fixed values. For the impact parameter independent picture this works 
out fine, with all events being reduced to low-$\pT$ two-string ones 
when the c.m.\ energy is reduced. In the variable impact parameter 
picture, the whole formalism only makes sense
if $\sigma_{\mrm{hard}} > \sigma_{\mrm{nd}}$, see e.g.\ 
eq.~(\ref{mi:kfromsigma}). Since $\sigma_{\mrm{nd}}$ does not vanish 
with decreasing energy, but $\sigma_{\mrm{hard}}$ 
would do that for a fixed $p_{\perp 0}$, this means
that $p_{\perp 0}$ has to be reduced significantly at low energies,
even more than implied by our assumed energy dependence. The more 
`sophisticated' model of this section therefore makes sense at 
collider energies, whereas it is not well suited for applications at 
fixed-target energies. There one should presumably attach to a 
picture of multiple soft Pomeron exchanges.
 
\subsection{Multiple Interactions and Beam Remnants --- New Model}
\label{ss:newmultint}

A new multiple interactions scenario is introduced with {\Py~\textsc{6.3}}
\cite{Sjo03a}. The early versions still represent a tryout phase, where 
extra caution should be exercised. Currently it should only be used for 
the study of single minimum-bias events or events underlying hard 
processes in hadron--hadron collisions. It cannot be used with pile-up, 
for example, nor with SUSY events where baryon number is violated.
 
The basic scheme for the generation of the multiple interactions is kept
from the old models. That is, based on the standard $2\to 2$ QCD matrix
elements convoluted with standard parton densities, a sequence of
$\pT$-ordered interactions is generated: $p_{\perp 1} > p_{\perp2} >
p_{\perp3} > \ldots$ . 
The sequence is stopped at some lower cut-off scale $p_{\perp min}$, or
alternatively the matrix elements are damped smoothly around and below
a scale $p_{\perp 0}$. An impact-parameter-dependent picture of hadron-hadron
interactions can lead to further elements of variability between events.
 
The main limitation of the old approach is that there was no way to
handle complicated beam remnants, e.g.~where two valence quarks had
been kicked out. Therefore the structure of all interactions subsequent
to the first one, i.e.\ the one with largest $\pT$, had to be substantially
simplified. The introduction of junction fragmentation in \cite{Sjo03}
allowed this restriction to be lifted.
 
\subsubsection{Flavour and $x$ Correlations}

In the new approach, the flavour content of the beam remnant is bookkept,
and is used to determine possible flavours in consecutive interactions.
Thus, the standard parton densities are only used to describe the
hardest interaction. Already in the old model, the $x$ scale of parton
densities is rescaled in subsequent interactions, such that the new
$x' = 1$ corresponds to the remaining momentum rather than the original 
beam momentum, eq.~(\ref{mi:xresc}). But now the distributions are not 
only squeezed in this manner, their shapes are also changed:
\begin{Itemize}
\item Whenever a valence quark is kicked out, the number of remaining
  valence quarks of that species is reduced accordingly. Thus, for a
  proton, the valence d distribution is completely removed if the valence
  $\d$ quark has been kicked out, whereas the valence $\u$ distribution is
  halved when one of the two is kicked out. In cases where the valence
  and sea $\u$ and $\d$ quark distributions are not separately provided from
  the PDF libraries, it is assumed that the sea is flavour-antiflavour 
  symmetric, so that one can write e.g.
  \begin{equation}
  u(x,Q^2) = u_{\mrm{val}}(x,Q^2) + u_{\mrm{sea}}(x,Q^2) = 
  u_{\mrm{val}}(x,Q^2) + \overline{u}(x,Q^2).
  \end{equation}
  The parametrized $u$ and $\overline{u}$ distributions are then used to 
  find the relative probability for a kicked-out $\u$ quark to be either
  valence or sea.
\item When a sea quark is kicked out, it must leave behind a corresponding
  antisea parton in the beam remnant, by flavour conservation. We call
  this a companion quark, and bookkeep it separately from the normal sea.
  In the perturbative approximation the sea quark $\qsea$ and its
  companion $\qcmp$ come from a gluon branching
  $\g \to \qsea + \qcmp$, where it is implicitly understood that if \qsea\ is
  a quark, \qcmp\ is its antiquark, and vice versa. 
  This branching often would not be in the perturbative
  regime, but we choose to make a perturbative ansatz, and also to
  neglect subsequent perturbative evolution of the $q_{\c}$ distribution.
  If so, the shape of the companion $q_{\c}$ distribution as a function of
  $x_{\c}$, given a sea parton at $x_{\s}$, becomes
  \begin{equation}
  q_{\c}(x_{\c}; x_{\s}) \propto \frac{g(x_{\c} + x_{\s})}{x_{\c} + x_{\s}} 
  \, P_{\g\to\qsea\qcmp}(z),
  \end{equation}
  where 
\begin{eqnarray}
 z & = & \frac{x_{\c}}{x_{\c} + x_{\s}}, \\
 P_{\g\to \qsea\qcmp}(z) &  = & \frac{1}{2} \left( z^2 + (1-z)^2 \right), \\
 g(x) & \propto & \frac{(1-x)^n}{x}~~~~;~(n=\mrm{MSTP(87)})
\end{eqnarray}
 is chosen.
 (Remember that this is supposed to occur at some low $Q^2$ scale, below
 the parton-shower lower cutoff $Q_0^2$, so
  $g(x)$ above should not be confused with the high-$Q^2$ gluon behaviour.)
\item The normalization of valence and companion distributions is fixed
  by the respective number of quarks, i.e.\ the sum rules
\begin{eqnarray}
\int_0^{x_\mathrm{rem}}\! q_{\mrm{v}}(x) \; \d x & = & n_{\qval},\\
\int_0^{x_\mathrm{rem}}\! q_{\mathrm{c},i}(x;x_{\s,i}) \d x & = & 1 ~~~~
(\mrm{for~each}~i),
\end{eqnarray}
where $x_{\mathrm{rem}}$ is the longitudinal momentum fraction left after the
previous interactions and $n_{\qval}$ is the number of $\q$ valence quarks
remaining. Gluon and sea distributions do 
  not have corresponding requirements. Therefore their normalization
  is adjusted, up or down, so as to obtain overall momentum conservation
  in the parton densities, i.e.\ to fulfil the remaining sum rule:
\begin{equation}
\int_0^{x_\mathrm{rem}}\! \left(\sum_{\q} q(x) + g(x) \right) \d x =
x_{\mrm{rem}} 
\end{equation}
\end{Itemize}

The above parton density strategy is not only used to pick a succession of
hard interactions. Each interaction may now, unlike before, have initial-
and final-state shower activity associated with it. The initial-state
shower is constructed by backwards evolution, using the above parton
densities. Even if the hard scattering does not involve a valence quark,
say, the possibility exists that the shower will reconstruct back to one.

\subsubsection{Colour Topologies}

The second part of the model is to hook up the scattered partons to the
beam remnants, in colours, in transverse momenta and in longitudinal
momenta. The order in which these choices are made is partly intertwined,
and one has to consider several special cases. In this description we
only give an outline, evading many of the details. Especially the
assignment of colours is physically uncertain and technically challenging.
 
Each of the two incoming beam hadrons can be viewed --- post facto --- as
having consisted of a set of ``initiator'' and ``remnant'' partons. An
initiator is the ``original'' parton that starts the initial-state cascade
that leads up to one of the hard interactions. Together the initiators and
remnants make up the original proton, so together they carry the proton net
flavour content, and energy and momentum, and are in a colour singlet
state. Therefore the remnant partons are constrained to carry ``what is
left'' when the initiators are removed. The one exception is that we do not
attempt to conserve longitudinal momentum (and thereby energy) exactly
within each incoming proton, but only for the system as a whole. The
reason here is that we have put all initiators and remnants on mass shell,
whereas a more proper treatment ought to have included (moderate) spacelike
virtualities for them. Such virtualities would have dissipated in the hard
interactions, and so not survived to the final state. The current choice
therefore shortcuts a number of technical details that in the end should
not matter.
 
In an event with $n$ multiple interactions, a corresponding set of $n$
initiator partons is defined for each of the two incoming beams. This also
defines the left-behind flavour content of the remnants. At most there
could be $n+3$ remnant partons for a proton with its 3 valence flavours, but
there could also be none, if all valence quarks have been kicked out and
no unmatched companions have been added. In the latter case a gluon is
inserted to represent the beam remnant and 
carry leftover momentum, but else we do not introduce gluons
in the beam remnant.
 
In the string model, the simplest representation of a baryon is the 
Y-shape topology, where each of the three valence quarks is connected 
via its string piece to the central junction (see section 
\ref{sss:junctiontopo}). This junction does not carry energy
or momentum of its own, but topologically it is the carrier of the
baryon number. Under normal circumstances, the legs in the Y are quite
short. Interactions may deform the topology, however. The simplest example
would be Deeply Inelastic Scattering, where one of the valence quarks is
kicked violently, so that one of the three strings of the Y will stretch
out and fragment into hadrons. In this topology the other two will remain
so close to each other and to the junction that they effectively act as a
single unit, a diquark.
 
In a hadronic interaction, a valence quark is kicked out by a coloured
gluon rather than a colourless photon. This colour exchange implies that
the string from the junction will no longer attach to the scattered quark
but rather to some other quark or gluon, but the other two quarks still
effectively form a diquark. When two or three valence quarks are kicked
out, the junction motion will become more complicated, and must be
considered in full.
 
Multiple valence quark interactions are rare, however. The bulk of
interactions are gluonic. In this case we can imagine that the gluon
originates from an unresolved emission off one of the valence quarks, 
i.e.\ that its anticolour is attached to one of the quarks and its colour
attached to the junction. After the gluon is kicked out, colour lines will
then connect this quark and the junction to partons in the final state of
the interaction. If a second gluon is kicked out, it can, in colour space,
have been located either between the quark and the first gluon, or
between this gluon and the junction, or between the junction and one of the
other two quarks. (We do not include the possibility that this gluon
together with the first one could have been radiated from the system as an
overall colour singlet system, i.e.\ we do not (yet) address diffractive
event topologies in this framework.) If all of these possibilities had equal
probability, the junction would often have two or all three legs reconnected,
and the baryon number could be moved quite dramatically in the event.
Empirically, this does not appear to happen (?), and furthermore it 
could be argued that perturbative and impact-parameter arguments both
allow much of the activity to be correlated in ``hot spot'' regions that
leave much of the rest of the proton unaffected. Therefore a free
suppression parameter is introduced, such that further gluons preferably
connect to a string piece that has already been disturbed. In this way, 
gluons preferentially will be found on one of the three colour lines
to the junction. 

The order in which they appear along that line is more
difficult to make statements about. So far in our description, 
no consideration has been given to the resulting momentum picture,
specified after transverse and longitudinal momenta have been picked
(see below). This in general implies that strings will be stretched
criss-cross in the event, if the initiator gluons are just randomly ordered
along the string piece on which they sit. It is unlikely that such a scenario
catches all the relevant physics. More likely is that, among all the possible
colour topologies for the final state, those that correspond to the
smaller total string length are favoured, all other aspects being the
same. Several options have been introduced that approach this issue
from different directions (see \ttt{MSTP(89)} and \ttt{MSTP(95)}). 
One is to attach the initiator gluons
preferentially in those places that order the hard scattering systems
in rapidity (the default), another to prefer the attachments that will
give rise to the smaller string lengths in the final state. A third
option is to rearrange the colour flow between the final-state partons
themselves, again giving preference to those rearrangements which
minimize the overall string length. So far no preferred scenario has
been identified.

A complication is the following. Two $\g + \g \to \g + \g$ scatterings each on
their own may have a perfectly sensible colour flow. Still, when the two
initial gluons on each side of the event are attached to each other and
to the rest of the remnants, the resulting colour flow may become
unphysical. Specifically the colour flow may ``loop back'' on itself, such
that a single gluon comes to form a separate colour singlet system.
Such configurations are rejected and alternative colour arrangements
are tried.

Another, rare, occurence is that the two junctions of the event can come to
be connected to each other via two strings (graphical representation:
\texttt{-j=j-}, where each dash corresponds to a string piece). Since we have
not (yet) programmed a fragmentation scheme for such events, we simply reject
them and generate a new event instead.
 
So far, interactions with sea quarks have not been mentioned, either a
quark-antiquark pair that both scatter or a single sea scatterer with a
leftover companion quark in the remnant. However, we have already argued
that each such pair can be viewed as coming from the branching of some
initial nonperturbative gluon. This gluon can now be attached to the
beam Y topology just like the gluons considered above, and therefore does
not introduce any new degrees of freedom. When then the colours are traced,
it could well happen that a companion quark together with two remaining
valence quarks form a separate colour singlet system. It is then likely
that this system will be of low mass and collapse to a single baryon.
Such possibilities are optionally allowed (see \ttt{MSTP(88)}), 
and correspondingly a companion antiquark could form a meson together with a
single valence quark. As already mentioned, diquarks can be formed from two
valence quarks.
 
\subsubsection{Primordial $\kT$}
 
Partons are expected to have primordial $\kT$ values of the order of a few
hundred MeV from Fermi motion inside the incoming hadrons. In reality,
one notes a need for a larger input in the context of shower evolution,
either in parton showers or in resummation descriptions. This is not
yet understood, but could e.g. represent a breakdown of the DGLAP evolution
equations at small $Q^2$. Until a better solution has been found, we 
therefore have reason to consider an effective ``primordial $\kT$'', 
at the level of the initiators, larger than the one above. For simplicity, 
a parametrized $Q$-dependent width
\begin{equation}
\sigma(Q) = \mathrm{max}\left(\frac{2.1~\mrm{GeV}}{7~\mrm{GeV} + Q},
  \mathrm{PARJ(21)}\right) 
\end{equation}
is introduced, where $\sigma$ is the width of the two-dimensional Gaussian
distribution of the initiator primordial $\kT$ (so that 
$\left<\kT^2\right> = \sigma^2$), $Q$ is the scale of the hard interaction 
and \ttt{PARJ(21)} is the standard fragmentation $\pT$ width. 
The remnant partons correspond to $Q=0$ and thus
hit the lower limit. Apart from the selection of each individual $\kT$,
there is also the requirement that the total $\kT$ of a beam adds up to zero.
Different strategies can be used here, from sharing the recoil of one
parton uniformly among all other initiator and remnant partons, to only
sharing it among those initiator/remnant partons that have been assigned
as nearest neighbours in colour space.
 
\subsubsection{Beam Remnant Kinematics}
The longitudinal momenta of the initiator partons have been defined by the
$x$ values picked. The remaining longitudinal momentum is shared between
the remnants in accordance with their character. A valence quark receives
an $x$ picked at random according to a small-$Q^2$ valence-like parton 
density, proportional to $(1-x)^a/\sqrt{x}$, where $a = 2$ for a  $\u$ 
quark in a proton and $a = 3.5$ for a $\d$ quark. A sea quark must be the 
companion of one of the initiator quarks, and can have an $x$ picked 
according to the $q_{\c}(x_{\c}; x_{\s})$
distribution introduced above. A diquark would obtain an $x$ given by the
sum of its constituent quarks. (But with the possibility to enhance this
$x$, to reflect the extra gluon cloud that could go with such a bigger
composite object.)  If a baryon or meson is in the remnant, its $x$ is
equated with the $z$ value obtainable from the Lund symmetric fragmentation
function, again with the possibility of enhancing this $x$ as for a diquark. 
A gluon only appears in an otherwise empty remnant, and can
thus be given $x = 1$. Once $x$ values have been picked for each of the
remnants, an overall rescaling is performed such that the remnants
together carry the desired longitudinal momentum.

\subsection{Pile-up Events}
\label{ss:pileup}
 
In high luminosity colliders, there is a non-negligible probability
that one single bunch crossing may produce several separate events,
so-called pile-up events.
This in particular applies to future $\p\p$ colliders like LHC,
but one could also consider e.g.\ $\ee$ colliders with high rates of
$\gamma\gamma$ collisions. The program therefore contains an option,
currently only applicable to hadron--hadron collisions,
wherein several events may be generated and put one after the other
in the event record, to simulate the full amount of particle
production a detector might be facing.
 
The program needs to know the assumed luminosity per bunch--bunch
crossing, expressed in mb$^{-1}$. Multiplied by the cross section
for pile-up processes studied, $\sigma_{\mrm{pile}}$, this gives the 
average number of collisions per beam crossing, $\br{n}$. These pile-up
events are taken to be of the minimum bias type, with diffractive
and elastic events included or not (and a further subdivision of
diffractive events into single and double). This means that
$\sigma_{\mrm{pile}}$ may be either $\sigma_{\mrm{tot}}$,
$\sigma_{\mrm{tot}} - \sigma_{\mrm{el}}$ or
$\sigma_{\mrm{tot}} - \sigma_{\mrm{el}} - \sigma_{\mrm{diffr}}$.
Which option to choose depends on the detector: most detectors
would not be able to observe elastic $\p\p$ scattering, and therefore
it would be superfluous to generate that kind of events.
In addition, we allow for the possibility that one interaction may
be of a rare kind, selected freely by you.
There is no option to generate two `rare' events in the same crossing;
normally the likelihood for that kind of occurrences should be small.
 
If only minimum bias type events are generated, i.e.\ if only one
cross section is involved in the problem, then the number of
events in a crossing is distributed according to a Poisson
with the average number $\br{n}$ as calculated above.
The program actually will simulate only those beam crossings
where at least one event occurs, i.e.\ not consider the fraction
$\exp(-\br{n})$ of zero-event crossings. Therefore the actually
generated average number of pile-up events is
$\langle n \rangle = \br{n}/(1-\exp(-\br{n}))$.
 
Now instead consider the other extreme, where one event is supposed
be rare, with a cross section $\sigma_{\mrm{rare}}$ much smaller 
than $\sigma_{\mrm{pile}}$, i.e.\ 
$f \equiv \sigma_{\mrm{rare}}/\sigma_{\mrm{pile}} \ll 1$.
The probability that a bunch crossing will give $i$ events, whereof
one of the rare kind, now is
\begin{equation}
{\cal P}_i = f \, i \, \exp(-\br{n}) \, \frac{\br{n}^i}{i!} =
f \, \br{n} \exp(-\br{n}) \, \frac{\br{n}^{i-1}}{(i-1)!} ~.
\end{equation}
The na\"{\i}ve Poisson is suppressed by a factor $f$, since one of 
the events is rare rather than of the normal kind, but enhanced by a
factor $i$, since any one of the $i$ events may be the rare one.
As the equality shows, the probability distribution is now a
Poisson in $i-1$:
in a beam crossing which produces one rare event, the multiplicity of
additional pile-up events is distributed according to a Poisson
with average number $\br{n}$. The total average number of events
thus is $\langle n \rangle = \br{n} + 1$.
 
Clearly, for processes with intermediate cross sections,
$\br{n} \, \sigma_{\mrm{rare}}/\sigma_{\mrm{pile}} \simeq 1$, 
also the average number of events will be intermediate, and it is not 
allowed to assume only one event to be of the `rare' type. We do not 
consider that kind of situations.
 
Each pileup event can be assigned a separate collision vertex
within the envelope provided by the colliding beams, see \ttt{MSTP(151)}.
Only simple Gaussian shapes in space and time are implemented internally, 
however. If this is too restrictive, you would have to assign
interaction points yourself, and then shift each event separately
by the required amount in space and time.
 
When the pile-up option is used, one main limitation is that event
records may become very large when several events are put one after
the other, so that the space limit in the \ttt{PYJETS} common block
is reached. It is possible to expand the dimension of the common block,
see \ttt{MSTU(4)} and \ttt{MSTU(5)}, but only up to about 
20\,000 entries, which might not always be enough, e.g.\ for LHC.
Simplifications like switching off $\pi^0$ decay may help keep down 
the size, but also has its limits.
 
For practical reasons, the program will only allow a $\br{n}$ up to
120. The multiplicity distribution is truncated above 200,
or when the probability for a multiplicity has fallen below
$10^{-6}$, whichever occurs sooner. Also low multiplicities with
probabilities below $10^{-6}$ are truncated.

\subsection{Common Block Variables}
\label{ss:multintpar}

Of the routines used to generate beam remnants, multiple interactions 
and pile-up events, none are intended to be called directly by the 
user. The only way to regulate these aspects is therefore via the 
variables in the \ttt{PYPARS} common block. 

\drawbox{COMMON/PYPARS/MSTP(200),PARP(200),MSTI(200),PARI(200)}
\begin{entry}
 
\itemc{Purpose:} to give access to a number of status codes and
parameters which regulate the performance of {\Py}.
Most parameters are described in section \ref{ss:PYswitchpar};
here only those related to beam remnants, multiple interactions
and pile-up events are described. If the default values,
below denoted by (D=\ldots), are not satisfactory, they must in
general be changed before the \ttt{PYINIT} call. Exceptions, i.e.\
variables which can be changed for each new event, are denoted by
(C).

\iteme{MSTP(81) :}\label{p:MSTP81} (D=1) master switch for multiple 
interactions, and also for the associated treatment of beam remnants.
\begin{subentry}
\iteme{= 0 :} off, new beam remnant scenario.
\iteme{= 1 :} on, new scenario for multiple interactions and remnants.
\iteme{=-1 :} on, old scenario for multiple interactions and remnants,
            for backwards compatibility.
\iteme{=-2 :} off, old beam remnant scenario.
\itemc{Warning:} many parameters have to be tuned differently for the
     old and new scenarios, such as \ttt{PARP(81)}--\ttt{PARP(84)}, 
     \ttt{PARP(89)} and \ttt{PARP(90)}, and others are specific 
     to each scenario. In addition, the optimal parameter values depend 
     on the choice of parton densities and so on. Therefore you must 
     pick a consistent set of values, rather than simply changing
     \ttt{MSTP(81)} by itself. 
     An example, for the old multiple interactions scenario, is 
     Rick Field's Tune A \cite{Fie02}:\\ 
     \ttt{MSTP(81)=-1}, \ttt{MSTP(82)=4}, \ttt{PARP(67)=4.0}, 
     \ttt{PARP(82)=2.0},\\ 
     \ttt{PARP(83)=0.5}, \ttt{PARP(84)=0.4}, \ttt{PARP(85)=0.9}, 
     \ttt{PARP(86)=0.95},\\ 
     \ttt{PARP(89)=1800.0} and \ttt{PARP(90)=0.25}, with CTEQ5L (default).
\end{subentry}

\iteme{MSTP(82) :} (D=3) structure of multiple interactions. For QCD
processes, used down to $\pT$ values below $\pTmin$, it also
affects the choice of structure for the one hard/semi-hard interaction.
\begin{subentry}
\iteme{= 0 :} simple two-string model without any hard interactions.
Toy model only!
\iteme{= 1 :} multiple interactions assuming the same probability in
all events, with an abrupt $\pTmin$ cut-off at \ttt{PARP(81)}.
(With a slow energy dependence given by \ttt{PARP(89)} and
\ttt{PARP(90)}.)
\iteme{= 2 :} multiple interactions assuming the same probability in
all events, with a continuous turn-off of the cross section at
$p_{\perp 0} = $\ttt{PARP(82)}.
(With a slow energy dependence given by \ttt{PARP(89)} and
\ttt{PARP(90)}.)
\iteme{= 3 :} multiple interactions assuming a varying impact
parameter and a hadronic matter overlap consistent with a Gaussian
matter distribution, with a continuous turn-off of the cross section
at $p_{\perp 0} = $\ttt{PARP(82)}.
(With a slow energy dependence given by \ttt{PARP(89)} and
\ttt{PARP(90)}.)
\iteme{= 4 :} multiple interactions assuming a varying impact
parameter and a hadronic matter overlap consistent with a double
Gaussian matter distribution given by \ttt{PARP(83)} and
\ttt{PARP(84)}, with a continuous turn-off of the cross section at
$p_{\perp 0} = $\ttt{PARP(82)}.
(With a slow energy dependence given by \ttt{PARP(89)} and
\ttt{PARP(90)}.)
\itemc{Note 1:} For \ttt{MSTP(82)}$\geq 2$ and
\ttt{CKIN(3)}$ > $\ttt{PARP(82)} (modulo the slow energy dependence 
noted above), cross sections
given with \ttt{PYSTAT(1)} may be somewhat too large, since (for
reasons of efficiency) the probability factor that the hard
interaction is indeed the hardest in the event is not
included in the cross sections. It is included in the event
selection, however, so the events generated are correctly
distributed. For \ttt{CKIN(3)} values a couple of times larger than
\ttt{PARP(82)} this ceases to be a problem.
\itemc{Note 2:} The \ttt{PARP(81)} and \ttt{PARP(82)}
values are sensitive to the choice of parton distributions,
$\Lambda_{\mrm{QCD}}$,
etc., in the sense that a change in the latter variables
leads to a net change in the multiple interaction rate, which has
to be compensated by a retuning of \ttt{PARP(81)} or \ttt{PARP(82)}
if one wants to keep the net multiple interaction structure the
same. The default \ttt{PARP(81)} and \ttt{PARP(82)} values are consistent 
with the other default values give, i.e.\ parton distributions of the 
proton etc.
\itemc{Note 3:} The multiple interactions combination of \ttt{MSTP(81)=0} 
and \ttt{MSTP(82)}$\geq 2$ is not so common, and certainly not intended 
to simulate realistic events, but it is an allowed way to obtain events 
with only the hardest interaction of those the events would contain with 
the corresponding \ttt{MSTP(81)=1} scenario. If also \ttt{CKIN(3)} is set 
larger than the respective $p_{\perp 0}$ scale (see \ttt{PARP(82)}), 
however, the program misbehaves. To avoid this, \ttt{MSTP(82)} is now set 
to 1 in such cases, in \ttt{PYINIT}. 
\end{subentry}
 
\iteme{MSTP(83) :} (D=100) number of Monte Carlo generated phase-space 
points per bin (whereof there are 20) in the initialization
(in \ttt{PYMULT}) of multiple interactions for
\ttt{MSTP(82)}$\geq 2$.
 
\iteme{MSTP(84) :} (D=1) switch for initial-state radiation in interactions
        after the first one in the new model.
\begin{subentry}
\iteme{= 0 :} off.
\iteme{= 1 :} on, provided that \ttt{MSTP(61)=1}.
\itemc{Note :} initial-state radiation in the first (hardest) interaction
            is not affected by \ttt{MSTP(84)}, but only by \ttt{MSTP(61)}.
\end{subentry}

\iteme{MSTP(85) :} (D=1) switch for final-state radiation in interactions
        after the first one in the new model.
\begin{subentry}
\iteme{= 0 :} off.
\iteme{= 1 :} on, provided that \ttt{MSTP(71)=1}.
\itemc{Note :} final-state radiation in the first (hardest) interaction
            is not affected by \ttt{MSTP(85)}, but only by \ttt{MSTP(71)}.
\end{subentry}

\iteme{MSTP(86) :} (D=2) requirements on multiple interactions based on 
the hardness scale of the main process.
\begin{subentry}
\iteme{= 1 :} the main collision is harder than all the subsequent 
ones. This is the old behaviour, preserved for reasons of backwards
compatibility, and most of the time quite sensible, but with dangers 
as follows.\\ 
The traditional multiple interactions procedure is to let the main
interaction set the upper $\pT$ scale for subsequent multiple 
interactions. For QCD, this is a matter of avoiding double-counting.
Other processes normally are hard, so the procedure is then also 
sensible. However, for a soft main interaction, further softer
interactions are hardly possible, i.e.\ multiple  interactions are 
more or less killed. Such a behaviour could be motivated by the 
rejected events instead appearing as part of the interactions 
underneath a normal QCD hard interaction, but in practice the latter 
mechanism is not implemented. (And would have been very inefficient to 
work with, had it been.)
For \ttt{MSTP(82)}$\geq 3$ it is even worse, since also the events 
themselves are likely to be rejected in the impact-parameter selection 
stage. Thus the spectrum of main events that survive is biased, with 
the low-$\pT$, soft tail suppressed.  Furthermore, even when events 
are rejected by the impact parameter procedure, this is not reflected
in the cross section for the process, as it should have been. Results
may thus be misleading.
\iteme{= 2 :} when the main process is of the QCD jets type (the same
as those in multiple interactions) subsequent jets are requested to be 
softer, but for other processes no such requirement exists.
\iteme{= 3 :} no requirements at all that multiple interactions have 
to be softer than the main interactions (of dubious use for
QCD processes but intended for cross-checks).
\itemc{Note:} process cross sections are unreliable whenever the 
main process does restrict subsequent interactions, and the
main process can become soft. For QCD jet studies in this 
region it is then better to put \ttt{CKIN(3)=0} and get the 
`correct' total cross section.
\end{subentry}

\iteme{MSTP(87) :} (D=4) when a sea quark (antiquark) is picked from a
        hadron at some $x_{\s}$ value in the new model, there has to be a 
        companion sea antiquark (quark) at some other $x_{\c}$ value. 
        The $x_{\c}$ distribution is assumed given by a convolution 
        of a mother gluon distribution $g(x = x_{\s} + x_{\c})$ with 
        the perturbative $\g \to \q + \qbar$ DGLAP
        splitting kernel. The simple ansatz $g(x) = N (1-x)^n/x$ is used,
        where $N$ is a normalization constant and $n$ = \ttt{MSTP(87)}. 
        \ttt{MSTP(87)} thus
        controls the large-$x$ behaviour of the assumed gluon
        distribution. Only integers \ttt{MSTP(87) $\in \{0, 1, 2, 3, 4\}$} are
        available; values below or above this range are set at the lower or
        upper limit, respectively.

\iteme{MSTP(88) :} (D=1) strategy for the collapse of a quark-quark-junction
        configuration to a diquark, or a quark-quark-junction-quark
        configuration to a baryon, in a beam remnant in the new model
\begin{subentry}
\iteme{=0 :} only allowed when valence quarks only are involved.
\iteme{=1 :} sea quarks can be used for diquark formation, but not for
            baryon formation.
\iteme{=2 :} sea quarks can be used also for baryon formation.
\end{subentry}

\iteme{MSTP(89) :} (D=1) Selection of method for colour connections in the
        initial state of the new model. Note that all options respect the 
        suppression provided by \ttt{PARP(80)}.
\begin{subentry}
\iteme{=0 :} Random.
\iteme{=1 :} The hard scattering systems are ordered in rapidity. The
            initiators on each side are connected so as to minimize the
            rapidity difference between neighbouring systems.
\iteme{=2 :} Each connection is chosen so as to minimize an estimate of
            the total string length resulting from it. (This is the most
            technically complicated, and hence a computationally slow
            approach.)
\end{subentry}

\iteme{MSTP(90) :} (D=0) strategy to compensate the ``primordial $\kT$'' 
        assigned to a parton shower initiator or beam remnant parton 
        in the new model (\ttt{MSTP(81)=1}).
\begin{subentry}
\iteme{=0 :} All other such partons compensate uniformly.
\iteme{=1 :} Compensation spread out across colour chain as $(1/2)^n$,
            where $n$ is number of steps the parton is removed in the chain.
\iteme{=2 :} Nearest colour neighbours only compensate.
\end{subentry}

\iteme{MSTP(91) :} (D=1) (C) primordial $k_{\perp}$ distribution
in hadron. See \ttt{MSTP(93)} for photon. 
Only applicable for \ttt{MSTP(81)=-1,-2}.
\begin{subentry}
\iteme{= 0 :} no primordial $k_{\perp}$.
\iteme{= 1 :} Gaussian, width given in \ttt{PARP(91)}, upper cut-off
in \ttt{PARP(93)}.
\iteme{= 2 :} exponential, width given in \ttt{PARP(92)}, upper
cut-off in \ttt{PARP(93)}.
\end{subentry}
 
\iteme{MSTP(92) :} (D=3) (C) energy partitioning in hadron or 
resolved photon remnant, when this remnant is split into two jets 
in the old model.
(For a splitting into a hadron plus a jet, see \ttt{MSTP(94)}.)
The energy fraction $\chi$ taken by one of the two objects, with
conventions as described for \ttt{PARP(94)} and \ttt{PARP(96)}, 
is chosen according to the different distributions below. Here
$c_{\mmin}  = 0.6~\mrm{GeV}/E_{\mrm{cm}} \approx %
2 \langle m_{\q} \rangle/E_{\mrm{cm}}$.
\begin{subentry}
\iteme{= 1 :} 1 for meson or resolved photon, $2(1-\chi)$ for 
baryon, i.e.\ simple counting rules.
\iteme{= 2 :} $(k+1)(1-\chi)^k$, with $k$ given by
\ttt{PARP(94)} or \ttt{PARP(96)}.
\iteme{= 3 :} proportional to 
$(1-\chi)^k/\sqrt[4]{\chi^2+c_{\mmin}^2}$, with $k$ given by
\ttt{PARP(94)} or \ttt{PARP(96)}.
\iteme{= 4 :} proportional to $(1-\chi)^k/\sqrt{\chi^2+c_{\mmin}^2}$, 
with $k$ given by \ttt{PARP(94)} or \ttt{PARP(96)}.
\iteme{= 5 :} proportional to $(1-\chi)^k/(\chi^2+c_{\mmin}^2)^{b/2}$, 
with $k$ given by \ttt{PARP(94)} or \ttt{PARP(96)}, and 
$b$ by \ttt{PARP(98)}.
\end{subentry}
 
\iteme{MSTP(93) :} (D=1) (C) primordial $k_{\perp}$ distribution
in photon, either it is one of the incoming particles or inside
an electron.
\begin{subentry}
\iteme{= 0 :} no primordial $k_{\perp}$.
\iteme{= 1 :} Gaussian, width given in \ttt{PARP(99)}, upper
cut-off in \ttt{PARP(100)}.
\iteme{= 2 :} exponential, width given in \ttt{PARP(99)}, upper
cut-off in \ttt{PARP(100)}.
\iteme{= 3 :} power-like of the type
$\d k_{\perp}^2/(k_{\perp 0}^2 + k_{\perp}^2)^2$, with
$k_{\perp 0}$ in \ttt{PARP(99)} and upper $k_{\perp}$ cut-off in
\ttt{PARP(100)}.
\iteme{= 4 :} power-like of the type
$\d k_{\perp}^2/(k_{\perp 0}^2 + k_{\perp}^2)$, with
$k_{\perp 0}$ in \ttt{PARP(99)} and upper $k_{\perp}$ cut-off in
\ttt{PARP(100)}.
\iteme{= 5 :} power-like of the type
$\d k_{\perp}^2/(k_{\perp 0}^2 + k_{\perp}^2)$, with
$k_{\perp 0}$ in \ttt{PARP(99)} and upper $k_{\perp}$ cut-off
given by the $\pT$ of the hard process or by
\ttt{PARP(100)}, whichever is smaller.
\itemc{Note:} for options 1 and 2 the \ttt{PARP(100)} value is of
minor importance, once \ttt{PARP(100)}$\gg$\ttt{PARP(99)}. However,
options 3 and 4 correspond to distributions with infinite
$\langle k_{\perp}^2 \rangle$ if the $k_{\perp}$ spectrum is not
cut off, and therefore the \ttt{PARP(100)} value is as important
for the overall distribution as is \ttt{PARP(99)}.
\end{subentry}

\iteme{MSTP(94) :} (D=3) (C) energy partitioning in hadron or 
resolved photon remnant, when this remnant is split into a hadron 
plus a remainder-jet in the old model. The energy fraction $\chi$ 
is taken by one of the two objects, with conventions as described 
below or for \ttt{PARP(95)} and \ttt{PARP(97)}.
\begin{subentry}
\iteme{= 1 :} 1 for meson or resolved photon, $2(1-\chi)$ for 
baryon, i.e.\ simple counting rules.
\iteme{= 2 :} $(k+1)(1-\chi)^k$, with $k$ given by
\ttt{PARP(95)} or \ttt{PARP(97)}.
\iteme{= 3 :} the $\chi$ of the hadron is selected according to the
normal fragmentation function used for the hadron in jet 
fragmentation, see \ttt{MSTJ(11)}. The possibility of a changed 
fragmentation function shape in diquark fragmentation 
(see \ttt{PARJ(45)}) is not included.  
\iteme{= 4 :} as \ttt{=3}, but the shape is changed as allowed in 
diquark fragmentation (see \ttt{PARJ(45)}); this change is here also 
allowed for meson production. (This option is not so natural
for mesons, but has been added to provide the same amount of freedom
as for baryons).  
\end{subentry}

\iteme{MSTP(95) :} (D=1) Selection of method for colour reconnections in the
        final state in the new model. If on, the amount of reconnections is 
        controlled by \ttt{PARP(78)}. 
\begin{subentry}
\iteme{=0 :} Off.
\iteme{=1 :} On.
\iteme{=2 :} On, but only applied to colours present in the initial state.
            This option is similar to choosing MSTP(89)=2, but here the
            colour reconnections happen after primordial $\kT$ assignments.
            Not implemented yet.
\end{subentry}

\iteme{MSTP(131) :}\label{p:MSTP131} (D=0) master switch for pile-up 
events, i.e.\ several
independent hadron--hadron interactions generated in the same
bunch--bunch crossing, with the events following one after the
other in the event record.
\begin{subentry}
\iteme{= 0 :} off, i.e.\ only one event is generated at a time.
\iteme{= 1 :} on, i.e.\ several events are allowed in the same event
record. Information on the processes generated may be found in
\ttt{MSTI(41) - MSTI(50)}.
\end{subentry}
 
\iteme{MSTP(132) :} (D=4) the processes that are switched on for
pile-up events. The first event may be set up completely arbitrarily,
using the switches in the \ttt{PYSUBS} common block, while all the
subsequent events have to be of one of the `inclusive' processes
which dominate the cross section, according to the options below.
It is thus not possible to generate two rare events in the pile-up
option.
\begin{subentry}
\iteme{= 1 :} low-$\pT$ processes ({\ISUB} = 95) only. The
low-$\pT$ model actually used, both in the hard event and in
the pile-up events, is the one set by \ttt{MSTP(81)} etc. This means
that implicitly also high-$\pT$ jets can be generated in the pile-up
events.
\iteme{= 2 :} low-$\pT$ + double diffractive processes
({\ISUB} = 95 and 94).
\iteme{= 3 :} low-$\pT$ + double diffractive + single diffractive
processes ({\ISUB} = 95, 94, 93 and 92).
\iteme{= 4 :} low-$\pT$ + double diffractive + single diffractive
+ elastic processes, together corresponding to the full
hadron--hadron cross section ({\ISUB} = 95, 94, 93, 92 and 91).
\end{subentry}
 
\iteme{MSTP(133) :} (D=0) multiplicity distribution of pile-up events.
\begin{subentry}
\iteme{= 0 :} selected by you, before each \ttt{PYEVNT} call, by
giving the \ttt{MSTP(134)} value.
\iteme{= 1 :} a Poisson multiplicity distribution in the total
number of pile-up events. This is the relevant distribution if the
switches set for the first event in \ttt{PYSUBS} give the same
subprocesses as are implied by \ttt{MSTP(132)}. In that case the
mean number of events per beam crossing is
$\br{n} = \sigma_{\mrm{pile}} \times$\ttt{PARP(131)}, where 
$\sigma_{\mrm{pile}}$ is the sum of the cross section for allowed
processes. Since bunch crossing which do not give any events at all
(probability $\exp(-\br{n})$) are not simulated, the actual average
number per \ttt{PYEVNT} call is
$\langle n \rangle = \br{n}/(1-\exp(-\br{n}))$.
\iteme{= 2 :} a biased distribution, as is relevant when one of the
events to be generated is assumed to belong to an event class
with a cross section much smaller than the total hadronic
cross section. If $\sigma_{\mrm{rare}}$ is the cross section for this
rare process (or the sum of the cross sections of several rare
processes) and $\sigma_{\mrm{pile}}$ the cross section for the 
processes allowed by \ttt{MSTP(132)}, then define
$\br{n} = \sigma_{\mrm{pile}} \times$\ttt{PARP(131)}
and $f = \sigma_{\mrm{rare}}/\sigma_{\mrm{pile}}$. The probability 
that a bunch crossing will give $i$ events is then
${\cal P}_i = f \, i \, \exp(-\br{n}) \, \br{n}^i/i!$,
i.e.\ the na\"{\i}ve Poisson is suppressed by a factor $f$ since
one of the events will be rare rather than frequent, but
enhanced by a factor $i$ since any of the $i$ events may be the
rare one. Only beam crossings which give at least one event
of the required rare type are simulated, and the distribution
above normalized accordingly.
\itemc{Note:} for practical reasons, it is required that
$\br{n} < 120$, i.e.\ that an average beam crossing does not contain
more than 120 pile-up events. The multiplicity distribution is
truncated above 200, or when the probability for a multiplicity
has fallen below $10^{-6}$, whichever occurs sooner. Also low
multiplicities with probabilities below $10^{-6}$ are truncated.
See also \ttt{PARI(91) - PARI(93)}.
\end{subentry}
 
\iteme{MSTP(134) :} (D=1) a user selected multiplicity, i.e.\ total
number of pile-up events, to be generated in the next \ttt{PYEVNT}
call when \ttt{MSTP(133)=0}. May be reset for each new event, but 
must be in the range $1 \leq$\ttt{MSTP(134)}$\leq 200$.
 
\boxsep
 
\iteme{PARP(78) :}\label{p:PARP78} (D=0.025) parameter controlling 
        the amount of colour reconnection in the final state. Try a 
        fraction \ttt{PARP(78)} of the total number of
        possible reconnections. Perform all reconnections which reduce the
        total string length and which are consistent with the choice of
        strategy in \ttt{MSTP(90)}. 
        If at least one reconnection is successfully
        made, loop back and try again, else keep the final state topology
        arrived to at this point. Note that a random colour reconnection is
        tested each time, so that the same reconnection may be tried twice
        and some not at all even when \ttt{PARP(78)=1D0}. Thus, values 
        somewhat larger than \ttt{1D0} are necessary if the ``most extreme 
        case'' scenario
        is desired. Also note that the meaning of this parameter will most
        likely be changed in future updates. At present, it merely
        represents a crude way of turning up and down the amount of colour
        reconnections going on in the final state.

\iteme{PARP(79) :} (D=2.0) Enhancement factor applied when assigning $x$ to
        composite systems (e.g. diquarks) in the beam remnant in the new 
        model. Modulo a global rescaling (=normalization) of all the 
        BR parton $x$ values, the $x$ of a composite system is \ttt{PARP(79)} 
        times the trivial sum over $x$ values of its individual constituents.

\iteme{PARP(80) :} (D=0.1) when colours of partons kicked out from a beam
        remnant are to be attached to the remnant, \ttt{PARP(80)} gives a
        suppression for the probability of attaching those partons to
        the colour line between two partons which themselves both lie
        in the remnant. A smaller value thus corresponds to a smaller
        probability that several string pieces will go back and forth
        between the beam remnant and the hard scattering systems.

\iteme{PARP(81) :} (D=1.9 GeV) effective 
minimum transverse momentum $\pTmin$ for multiple 
interactions with \ttt{MSTP(82)=1}, at the reference energy scale 
\ttt{PARP(89)}, with the degree of energy rescaling given by 
\ttt{PARP(90)}. The optimal value depends on a number of other
assumptions, especially which parton distributions are being used.
The default is intended for CTEQ 5L.
 
\iteme{PARP(82) :} (D=2.0 GeV) regularization scale $p_{\perp 0}$
of the transverse momentum spectrum for multiple interactions with
\ttt{MSTP(82)}$\geq 2$, at the reference energy scale \ttt{PARP(89)}, 
with the degree of energy rescaling given by \ttt{PARP(90)}. (Current 
default based on the \ttt{MSTP(82)=4} option, without any change of 
\ttt{MSTP(2)} or \ttt{MSTP(33)}.) The optimal value depends on a number 
of other assumptions, especially which parton distributions are being 
used. The default is intended for CTEQ 5L.
 
\iteme{PARP(83), PARP(84) :} (D=0.5, 0.2) parameters of an assumed
double Gaussian matter distribution inside the colliding hadrons for
\ttt{MSTP(82)=4}, of the form given in eq.~(\ref{mi:doubleGauss}),
i.e.\ with a core of radius \ttt{PARP(84)} of the main radius and
containing a fraction \ttt{PARP(83)} of the total hadronic matter.
 
\iteme{PARP(85) :} (D=0.33) probability that an additional
interaction in the multiple interaction formalism gives two gluons,
with colour connections to `nearest neighbours' in momentum space.
 
\iteme{PARP(86) :} (D=0.66) probability that an additional
interaction in the multiple interaction formalism gives two gluons,
either as described in \ttt{PARP(85)} or as a closed gluon loop.
Remaining fraction is supposed to consist of quark--antiquark pairs.
 
\iteme{PARP(87), PARP(88) :} (D=0.7, 0.5) in order to account for an
assumed dominance of valence quarks at low transverse momentum scales,
a probability is introduced that a $\g\g$-scattering according to
na\"{\i}ve cross section is replaced by a $\q\qbar$ one; this is used 
only for \ttt{MSTP(82)}$\geq 2$. The probability is parameterized as
${\cal P} = a (1 - (\pT^2/(\pT^2 + b^2)^2)$, where
$a =$\ttt{PARP(87)} and $b =$\ttt{PARP(88)}$\times$\ttt{PARP(82)}
(including the slow energy rescaling of the $p_{\perp 0}$ parameter).

\iteme{PARP(89) :} (D=1800. GeV) reference energy scale, at which 
\ttt{PARP(81)} and \ttt{PARP(82)} give the $\pTmin$ and $p_{\perp 0}$ 
values directly. Has no physical meaning in itself, but is used for 
convenience only. (A form 
$\pTmin = \mathtt{PARP(81)} E_{\mrm{cm}}^{\mathtt{PARP(90)}}$ 
would have been equally possible but then with a less transparent 
meaning of \ttt{PARP(81)}.) For studies of the $\pTmin$ dependence 
at some specific energy it may be convenient to choose \ttt{PARP(89)} 
equal to this energy.

\iteme{PARP(90) :} (D=0.16) power of the energy-rescaling term of the 
$\pTmin$ and $p_{\perp 0}$ parameters, which are assumed proportional to 
$E_{\mrm{cm}}^{\mathtt{PARP(90)}}$. The default value is inspired by 
the rise of the total cross section by the pomeron term, 
$s^{\epsilon} = E_{\mrm{cm}}^{2\epsilon} = E_{\mrm{cm}}^{2\cdot 0.08}$, 
which is not inconsistent with the small-$x$ behaviour.
It is also reasonably consistent with the energy-dependence
implied by a comparison with the UA5 multiplicity distributions
at 200 and 900 GeV \cite{UA584}. \ttt{PARP(90) = 0} is an allowed 
value, i.e.\ it is possible to have energy-independent parameters.
 
\iteme{PARP(91) :} (D=1. GeV/$c$) (C) width of Gaussian primordial
$k_{\perp}$ distribution inside hadron for \ttt{MSTP(91)=1}, i.e.\
$\exp(-k_{\perp}^2/\sigma^2) \, k_{\perp} \, \d k_{\perp}$ with
$\sigma =$\ttt{PARP(91)} and
$\langle k_{\perp}^2 \rangle = $\ttt{PARP(91)}$^2$.
 
\iteme{PARP(92) :} (D=0.40 GeV/$c$) (C) width parameter of 
exponential primordial $k_{\perp}$ distribution inside hadron for
\ttt{MSTP(91)=2}, i.e.\
$\exp(-k_{\perp}/\sigma) \, k_{\perp} \, \d k_{\perp}$ with
$\sigma =$\ttt{PARP(92)} and 
$\langle k_{\perp}^2 \rangle = 6 \times$\ttt{PARP(92)}$^2$.
Thus one should put \ttt{PARP(92)}$\approx$\ttt{PARP(91)}$/\sqrt{6}$
to have continuity with the option above.  
 
\iteme{PARP(93) :} (D=5. GeV/$c$) (C) upper cut-off for primordial
$k_{\perp}$ distribution inside hadron.
 
\iteme{PARP(94) :} (D=1.) (C) for \ttt{MSTP(92)}$\geq 2$ this gives
the value of the parameter $k$ for the case when a meson or
resolved photon remnant is split into two fragments (which is which 
is chosen at random).
 
\iteme{PARP(95) :} (D=0.) (C) for \ttt{MSTP(94)=2} this gives
the value of the parameter $k$ for the case when a meson or
resolved photon remnant is split into a meson and a spectator 
fragment jet, with $\chi$ giving the energy fraction taken by the 
meson.
 
\iteme{PARP(96) :} (D=3.) (C) for \ttt{MSTP(92)}$\geq 2$ this gives
the value of the parameter $k$ for the case when a nucleon remnant
is split into a diquark and a quark fragment, with $\chi$ giving
the energy fraction taken by the quark jet.
 
\iteme{PARP(97) :} (D=1.) (C) for \ttt{MSTP(94)=2} this gives
the value of the parameter $k$ for the case when a nucleon remnant
is split into a baryon and a quark jet or a meson and a diquark jet,
with $\chi$ giving the energy fraction taken by the quark jet or
meson, respectively.
 
\iteme{PARP(98) :} (D=0.75) (C) for \ttt{MSTP(92)=5} this gives
the power of an assumed basic $1/\chi^b$ behaviour in the splitting
distribution, with $b =$\ttt{PARP(98)}.
 
\iteme{PARP(99) :} (D=1. GeV/$c$) (C) width parameter of primordial
$k_{\perp}$ distribution inside photon; exact meaning depends on
\ttt{MSTP(93)} value chosen (cf. \ttt{PARP(91)} and \ttt{PARP(92)}
above).
 
\iteme{PARP(100) :} (D=5. GeV/$c$) (C) upper cut-off for primordial
$k_{\perp}$ distribution inside photon.

\iteme{PARP(131) :}\label{p:PARP131} (D=0.01 mb$^{-1}$) in the 
pile-up events scenario, \ttt{PARP(131)}
gives the assumed luminosity per bunch--bunch crossing, i.e.\
if a subprocess has a cross section $\sigma$, the average number
of events of this type per bunch--bunch crossing is
$\br{n} = \sigma \times$\ttt{PARP(131)}. \ttt{PARP(131)} may be
obtained by dividing the integrated luminosity over a given time
(1 s, say) by the number of bunch--bunch crossings that this
corresponds to. Since the program will not generate more than
200 pile-up events, the initialization procedure will crash if
$\br{n}$ is above 120. 

\end{entry}

\drawboxthree{~COMMON/PYINTM/KFIVAL(2,3),NMI(2),IMI(2,800,2),NVC(2,-6:6),}%
{\&XASSOC(2,-6:6,240),XPSVC(-6:6,-1:240),PVCTOT(2,-1:1),}%
{\&XMI(2,240),Q2MI(240),IMISEP(0:240)}%
\label{p:PYINTM}
\begin{entry}

\itemc{Purpose:} to carry information relevant for multiple
  interactions in the new scheme.

\iteme{KFIVAL(JS,J) :} enumeration of the initial valence quark content on
        side \ttt{JS}. 
        A baryon requires all three slots \ttt{J}, while a meson only
        requires the first two, so that \ttt{KFIVAL(JS,3)=0}. The contents are
        initially zero for a $\gamma$, $\pi^0$, $\K_{\mrm{S}}^0$ and
        $\K_{\mrm{L}}$, and are only set once
        a valence quark is kicked out. (For instance, a $\pi^0$ can be either
        $\u\ubar$ or $\d\dbar$, only a scattering will tell.)

\iteme{NMI(JS) :} total number of scattered and remnant partons on side
        \ttt{JS}, 
        including hypothesized gluons that branches to a sea (anti)quark
        and its companion.

\iteme{IMI(JS,I,1) :} position in the \ttt{/PYJETS/} 
        event record of a scattered
        or remnant parton, or a hypothesized gluon that branches to a
        sea (anti)quark and its companion. Here \ttt{JS} gives the side and
        \ttt{1 $\le$ I $\le$ NMI(JS)} enumerates the partons.

\iteme{IMI(JS,I,2) :} nature of the partons described above.
\begin{subentry}
\iteme{= 0 :} a valence quark or a gluon.
\iteme{> 0 :} parton is part of a sea quark-antiquark pair, and
            \ttt{IMI(JS,I,2)} gives the position in the event record
            of the companion.
\iteme{< 0 :} flag that parton is a sea quark for which no companion
            has yet been defined; useful at intermediate stages but
            should not be found in the final event.
\end{subentry}

\iteme{NVC(JS,IFL) :} total number of unmatched sea quarks on side \ttt{JS}
        of flavour \ttt{IFL}.

\iteme{XASSOC(JS,IFL,I) :} the $x$ value of an unmatched sea quark, on side
        \ttt{JS} 
        of flavour \ttt{IFL} and with $1 \le$ \ttt{I} $\le$
        \ttt{NVC(JS,IFL)}. When a companion
        quark is found then \ttt{XASSOC} is put to zero, while \ttt{NVC}
        is unchanged.

\iteme{XPSVC(IFL,J) :} subdivision of the parton density $xf(x)$ for flavour
        \ttt{IFL} obtained by a \ttt{PYPDFU} call (with \ttt{MINT(30) =} side
        \ttt{JS}). 
\begin{subentry}
\iteme{J =-1 :} sea contribution.
\iteme{J = 0 :} valence contribution.
\iteme{1 $\le$ J $\le$ NVC(JS,IFL) :} 
sea companion contribution from each of the
            above unmatched sea quarks.
\end{subentry}

\iteme{PVCTOT(JS,J) :} momentum fraction, i.e.\ integral of $xf(x)$, inside
        the remaining hadron on side \ttt{JS}.
\begin{subentry}
\iteme{J =-1 :} total carried originally by valence quarks.
\iteme{J = 0 :} total carried by valence quarks that have been removed.
\iteme{J = 1 :} total carried by unmatched sea companion quarks.
\end{subentry}

\iteme{XMI(JS,I) :} $x$ 
        value of incoming parton on side \ttt{JS} for interaction
        number \ttt{I}, \ttt{1} $\le$ \ttt{I} $\le$ \ttt{MINT(31)}. 
        When there is an initial-state
        shower, $x$ refers to the initiator of this shower rather than
        to the incoming parton at the hard interaction. The $x$ values
        are normalized to the full energy of the incoming hadron, i.e.\
        are not rescaled by previous interactions.

\iteme{Q2MI(I) :} the $Q^2$ scale of the hard interaction number \ttt{I},
        \ttt{1} $\le$ \ttt{I} $\le$ \ttt{MINT(31)}. 
        This is the scale stored in \ttt{VINT(54)},
        i.e.\ the $Q^2$ used to evaluate parton densities at the hard
        scattering.

\iteme{IMISEP(I) :} last line in the event record filled by partons belonging
        to the interaction number \ttt{I}, \ttt{1} $\le$ \ttt{I} $\le$ 
        \ttt{MINT(31)}. Thus interaction
        \ttt{I} is stored in lines \ttt{IMISEP(I-1)+1} through
        \ttt{IMISEP(I)}, where 
        \ttt{IMISEP(0)} has been set so that this holds also for \ttt{I=1}.
\end{entry}

\drawbox{~COMMON/PYCBLS/MCO(4000,2),NCC,JCCO(4000,2),JCCN(4000,2),MACCPT}
\label{p:PYCBLS}
\begin{entry}

\itemc{Purpose:} to store temporary colour tag information while
  hooking up initial state, used to communicate between \ttt{PYMIHK} and 
\ttt{PYMIHG}.

\iteme{MCO(I,J) :} Original colour tag \ttt{J} of parton \ttt{I} before
        initial state 
        connection procedure was started.
\iteme{NCC :} Number of colour tags that have so far been collapsed.
\iteme{JCCO(I,J) :} Colour tag collapse array, \ttt{I $\le$ NCC}. 
  The \ttt{I}'th collapse is
        between colour tags \ttt{JCCO(I,1)} and \ttt{JCCO(I,2)}.
\iteme{JCCN(I,J) :} Temporary storage for collapse being tested. If collapse
        is accepted, \ttt{JCCO} is set equal to \ttt{JCCN}.
\end{entry}

\drawbox{~COMMON/PYCTAG/NCT, MCT(4000,2)}%
\label{p:PYCTAG}
\begin{entry}
\itemc{Purpose:} to carry Les Houches Accord style colour tag information
  for partons in the event record.
\iteme{NCT :} Number of colour lines in the event, from \ttt{1} to \ttt{NCT}.
\iteme{MCT(I,1) :} Colour tag of parton \ttt{I} in \ttt{/PYJETS/}.
\iteme{MCT(I,2) :} Anticolour tag of parton \ttt{I} in \ttt{/PYJETS/}.
\end{entry}
 
\clearpage
 
\section{Fragmentation}
 
The main fragmentation option in {\Py}
is the Lund string scheme, but independent fragmentation
options are also available. These latter options should not be
taken too seriously, since we know that independent fragmentation
does not provide a consistent alternative, but occasionally one
may like to compare string fragmentation with something else.
 
The subsequent four sections give further details;
the first one on flavour selection, which is common to the two
approaches, the second on string fragmentation, the third on
independent fragmentation, while the fourth and final contains
information on a few other issues.
 
The Lund fragmentation model is described in \cite{And83}, where
all the basic ideas are presented and earlier papers
\cite{And79,And80,And82,And82a} summarized.
The details given there on how a multiparton jet system is allowed
to fragment are out of date, however, and for this one should turn
to \cite{Sjo84}. Also the `popcorn' baryon production mechanism
is not covered, see \cite{And85}, and \cite{Ede97} for a more
sophisticated version. The most recent comprehensive 
description of the Lund model is found in \cite{And98}. Reviews of 
fragmentation models in general may be found in \cite{Sjo88,Sjo89}.
 
\subsection{Flavour Selection}
\label{ss:flavoursel}
 
In either string or independent fragmentation, an iterative
approach is used to describe the fragmentation process.
Given an initial quark $\q = \q_0$, it is assumed that a new
$\q_1 \qbar_1$
pair may be created, such that a meson $\q_0 \qbar_1$ is
formed, and a $\q_1$ is left behind. This $\q_1$ may at a later
stage pair off with a $\qbar_2$, and so on. What need be
given is thus the relative probabilities to produce the various
possible $\q_i \qbar_i$ pairs, $\u \ubar$, $\d \dbar$,
$\s \sbar$, etc., and the relative probabilities that a given
$\q_{i-1}\qbar_i$ quark pair combination forms a specific meson,
e.g.\ for $\u \dbar$ either $\pi^+$, $\rho^+$ or some higher state.
 
In {\Py}, it is assumed that the two aspects can be factorized,
i.e.\ that it is possible first to select a $\q_i \qbar_i$ pair,
without any reference to allowed physical meson states, and
that, once the $\q_{i-1} \qbar_i$ flavour combination is given,
it can be assigned to a given meson state with total probability
unity. Corrections to this factorized ansatz will come in the 
baryon sector.
 
\subsubsection{Quark flavours and transverse momenta}
 
In order to generate the quark--antiquark pairs $\q_i \qbar_i$ which
lead to string breakups, the Lund model invokes the idea of
quantum mechanical tunnelling, as follows. If the $\q_i$ and
$\qbar_i$ have no (common) mass or transverse momentum, the pair can
classically be created at one point and then be pulled apart
by the field. If the quarks have mass and/or transverse momentum,
however, the $\q_i$ and $\qbar_i$ must classically be produced at a
certain distance so that the field energy between them can be
transformed into the sum of the two transverse masses $m_{\perp}$.
Quantum mechanically, the quarks may be created in one point
(so as to keep the concept of local flavour conservation) and then
tunnel out to the classically allowed region. In terms of a
common transverse mass $m_{\perp}$ of the $\q_i$ and the $\qbar_i$,
the tunnelling probability is given by
\begin{equation}
   \exp \left( -\frac{\pi m_{\perp}^2}{\kappa} \right) =
   \exp \left( -\frac{\pi m^2}{\kappa} \right)
   \exp \left( -\frac{\pi \pT^2}{\kappa} \right) ~.
\label{e:qtunnel}
\end{equation}
 
The factorization of the transverse momentum and the mass terms leads
to a flavour-independent Gaussian spectrum for the $p_x$ and $p_y$
components of $\q_i \qbar_i$ pairs. Since the string is assumed to 
have no transverse excitations, this $\pT$ is locally compensated 
between the quark and the antiquark of the pair. The $\pT$ of a 
meson $\q_{i-1} \qbar_i$ is given by the vector sum of the $\pT$:s 
of the $\q_{i-1}$ and $\qbar_i$ constituents, which implies 
Gaussians in $p_x$ and $p_y$ with a width $\sqrt{2}$ that of the 
quarks themselves. The assumption of a Gaussian shape may be a good 
first approximation, but there remains the possibility of non-Gaussian
tails, that can be important in some situations.
 
In a perturbative QCD framework, a hard scattering is
associated with gluon radiation, and further contributions to what is
na\"{\i}vely called fragmentation $\pT$ comes from unresolved radiation.
This is used as an explanation
why the experimental $\left\langle \pT \right\rangle$ is somewhat
higher than obtained with the formula above.
 
The formula also implies a suppression of heavy quark production
$u : d : s : c \approx$ \mbox{$1 : 1 : 0.3 : 10^{-11}$}. Charm and
heavier quarks
are hence not expected to be produced in the soft fragmentation.
Since the predicted flavour suppressions are in terms of quark masses,
which are notoriously difficult to assign (should it be current algebra,
or constituent, or maybe something in between?), the
suppression of $\s \sbar$ production is left as a free parameter in
the program: $\u \ubar$ : $\d \dbar$ : $\s \sbar$ = 1 : 1 : $\gamma_s$,
where by default $\gamma_s = 0.3$. At least qualitatively, the
experimental value agrees with theoretical prejudice.
There is no production at all of heavier flavours in the fragmentation
process, but only in the hard process or as part of the shower evolution.
 
\subsubsection{Meson production}
\label{sss:mesonprod}
 
Once the flavours $\q_{i-1}$ and $\qbar_i$ have been selected, a choice 
is made between the possible multiplets. The relative composition of
different multiplets is not given from first principles, but must
depend on the details of the fragmentation process. To some
approximation one would expect a negligible fraction of states with
radial excitations or non-vanishing orbital angular momentum. Spin
counting arguments would then suggest a 3:1 mixture between the
lowest lying vector and pseudoscalar multiplets. Wave function
overlap arguments lead to a relative enhancement of the lighter
pseudoscalar states, which is more pronounced the larger the mass
splitting is \cite{And82a}.
 
In the program, six meson multiplets are
included. If the nonrelativistic classification scheme is used, i.e.
mesons are assigned a valence quark spin $S$ and an internal orbital
angular momentum $L$, with the physical spin $s$ denoted $J$,
$\mbf{J} = \mbf{L} + \mbf{S}$, then the multiplets are:
\begin{Itemize}
\item $L = 0$, $S = 0$, $J = 0$: the ordinary pseudoscalar meson
   multiplet;
\item $L = 0$, $S = 1$, $J = 1$: the ordinary vector meson multiplet;
\item $L = 1$, $S = 0$, $J = 1$: an axial vector meson multiplet;
\item $L = 1$, $S = 1$, $J = 0$: the scalar meson multiplet;
\item $L = 1$, $S = 1$, $J = 1$: another axial vector meson multiplet;
    and
\item $L = 1$, $S = 1$, $J = 2$: the tensor meson multiplet.
\end{Itemize}
Each multiplet has the full five-flavour setup of $5 \times 5$
states included in the program. Some simplifications have been made; 
thus there is no mixing included between the two axial vector 
multiplets.
 
In the program, the spin $S$ is first chosen to be either 0 or 1.
This is done according to parameterized relative probabilities,
where the probability for spin 1 by default is taken to be 0.5 for
a meson consisting only of $\u$ and $\d$ quark, 0.6 for one which
contains $\s$ as well, and $0.75$ for quarks with $\c$ or heavier
quark, in accordance with the deliberations above.
 
By default, it is assumed that $L = 0$, such that only pseudoscalar
and vector mesons are produced. For inclusion of $L = 1$ production,
four parameters can be used, one to give the probability that a $S = 0$
state also has $L =1$, the other three for the probability that a
$S = 1$ state has $L = 1$ and $J$ either 0, 1, or 2.
 
For the
flavour-diagonal meson states $\u \ubar$, $\d \dbar$ and $\s \sbar$,
it is also necessary to include mixing into the physical mesons.
This is done according to a parameterization, based on the mixing angles
given in the Review of Particle Properties \cite{PDG88}. In particular,
the default choices correspond to
\begin{eqnarray}
\eta & = & \frac{1}{2} (\u\ubar + \d\dbar) - \frac{1}{\sqrt{2}}
\s\sbar ~;   \nonumber \\
\eta' & = & \frac{1}{2} (\u\ubar + \d\dbar) + \frac{1}{\sqrt{2}}
\s\sbar ~;   \nonumber \\
\omega & = & \frac{1}{\sqrt{2}} (\u\ubar + \d\dbar)    \nonumber \\
\phi & = & \s\sbar ~.
\end{eqnarray}
In the $\pi^0 - \eta - \eta'$ system, no account is therefore
taken of the difference in masses, an approximation which seems to
lead to an overestimate of $\eta'$ rates \cite{ALE92}. Therefore
parameters have been introduced to allow an additional `brute force'
suppression of $\eta$ and $\eta'$ states.
 
\subsubsection{Baryon production}
\label{sss:baryonprod}
 
The mechanism for meson production follows rather naturally from the 
simple picture of a meson as a short piece of string between two 
$\q/\qbar$ endpoints. There is no unique recipe to generalize this 
picture to baryons. The program actually contains three different 
scenarios: diquark, simple popcorn, and advanced popcorn. In the 
diquark model the baryon and antibaryon are always produced as nearest 
neighbours along the string, while mesons may (but need not) be 
produced in between in the popcorn scenarios. The simpler popcorn 
alternative admits at most one intermediate meson, while the advanced 
one allows many. Further differences may be found, but several aspects 
are also common between the three scenarios. Below they are therefore 
described in order of increasing sophistication. Finally the application 
of the models to baryon remnant fragmentation, where a diquark originally 
sits at one endpoint of the string, is discussed.

{\em Diquark picture}
 
Baryon production may, in its simplest form, be obtained by assuming
that any flavour $\q_i$ given above could represent either a quark or
an antidiquark in a colour triplet state. Then the same basic
machinery can be run through as above, supplemented with the probability
to produce various diquark pairs. In principle, there is one
parameter for each diquark, but if tunnelling is still assumed to give
an effective description, mass relations can be used to reduce the
effective number of parameters. There are three main ones
appearing in the program: 
\begin{Itemize}
\item the relative probability to pick a $\qbar\qbar$ diquark
rather than a $\q$; 
\item the extra suppression associated with a diquark
containing a strange quark (over and above the ordinary $\s / \u$
suppression factor $\gamma_s$); and 
\item the suppression of spin 1 diquarks relative to spin 0 ones 
(apart from the factor of 3 enhancement of the former based on 
counting the number of spin states). 
\end{Itemize}
The extra strange diquark suppression factor comes about since what 
appears in the exponent of the tunnelling formula is $m^2$ and not 
$m$, so that the diquark and the strange quark suppressions do not 
factorize.
 
Only two baryon multiplets are included, i.e.\ there are no $L=1$
excited states.  The two multiplets are:
\begin{Itemize}
\item $S = J = 1/2$: the `octet' multiplet of {\bf SU(3)};
\item $S = J = 3/2$: the `decuplet' multiplet of {\bf SU(3)}.
\end{Itemize}
In contrast to the meson case, different flavour combinations have
different numbers of states available: for $\u \u \u$ only
$\Delta^{++}$, whereas $\u \d \s$ may become either $\Lambda$,
$\Sigma^0$ or $\Sigma^{*0}$.
 
An important constraint is that a baryon is a
symmetric state of three quarks, neglecting the colour degree of
freedom. When a diquark and a quark are joined to form a baryon,
the combination is therefore weighted with the probability that
they form a symmetric three-quark state. The program implementation
of this principle is to first select a diquark at random, with
the strangeness and spin 1 suppression factors above included,
but then to accept the selected diquark with a weight proportional to
the number of states available for the quark-diquark combination. This
means that, were it not for the tunnelling suppression factors, all
states in the {\bf SU(6)} (flavour {\bf SU(3)} times spin {\bf SU(2)})
56-multiplet would become equally populated. Of course also heavier
baryons may come from the fragmentation of e.g.\ $\c$ quark jets, but
although the particle classification scheme used in the program is
{\bf SU(10)}, i.e.\ with five flavours, all possible quark-diquark
combinations can be related to {\bf SU(6)} by symmetry arguments.
As in the case for mesons, one could imagine an explicit further
suppression of the heavier spin 3/2 baryons. 

In case of rejection, one again chooses between a diquark or a quark. 
If choosing diquark, a new baryon is selected and tested, etc. (In 
versions earlier than {\Py} 6.106, the algorithm was instead to always 
produce a new diquark if the previous one had been rejected. However, 
the probability that a quark will produce a baryon and a antidiquark 
is then flavour independent, which is not in agreement with the model.) 
Calling the tunnelling factor for diquark $D$ $T_{D}$, the number of 
spin states $\sigma_{D}$ and the {\bf SU(6)} factor for $D$ and a 
quark $\q$ $SU_{D,\q}$, the model prediction for the 
$(\q\rightarrow B + D)/(\q\rightarrow M + \q')$ 
ratio is
\begin{equation} 
S_{\q}=\frac{P(\q\q)}{P(\q)}\sum_{D} T_{D} 
\sigma_{D} SU_{D,\q}   ~.
\end{equation}
(Neglecting this flavour dependence e.g.\ leads to an enhancement of 
the $\Omega^-$ relative to primary proton production with 
approximately a factor $1.2$, using {\Je} 7.4 default values.)
Since the chosen algorithm implies the normalization 
$\sum_{D} T_{D} \sigma_{D} = 1$ and 
$SU_{D,\q} \le 1$, the final diquark production rate is somewhat 
reduced from the $P(\q\q)/P(\q)$ input value.

When a diquark has been fitted into a symmetrical three-particle state, 
it should not suffer any further {\bf SU(6)} suppressions. Thus the 
accompanying antidiquark should `survive' with probability 1. When 
producing a quark to go with a previously produced diquark, this is 
achieved by testing the configuration against the proper {\bf SU(6)} 
factor, but in case of rejection keep the diquark and pick a new quark, 
which then is tested, etc.

There is no obvious corresponding algorithm available when a quark from one 
side and a diquark from the other are joined to form the last hadron of the 
string. In this case the quark is a member of a pair, in which the antiquark 
already has formed a specific hadron. Thus the quark flavour cannot be 
reselected. One could consider the {\bf SU(6)} rejection as a major joining 
failure, and restart the fragmentation of the original string, but then 
the already accepted diquark {\em does} suffer extra {\bf SU(6)} suppression. 
In the program the joining of a quark and a diquark is always accepted.

{\em Simple popcorn}
 
A more general framework for baryon production is the `popcorn' one
\cite{And85}, in which diquarks as such are never produced, but
rather baryons appear from the successive production of several
$\q_i \qbar_i$ pairs. The picture is the following. Assume that the
original $\q$ is red $r$ and the $\qbar$ is $\br{r}$. Normally a
new $\q_1 \qbar_1$ pair produced in the field would also be
$r \br{r}$, so that the $\qbar_1$ is pulled towards the $\q$
end and vice versa, and two separate colour-singlet systems
$\q \qbar_1$ and $\q_1 \qbar$ are formed. Occasionally, the
$\q_1 \qbar_1$ pair may be e.g.\ $g \br{g}$ ($g$ = green), in which
case there is no net colour charge acting on either $\q_1$ or
$\qbar_1$. Therefore, the pair cannot gain energy from the field,
and normally would exist only as a fluctuation. If $\q_1$ moves
towards $\q$ and $\qbar_1$ towards $\qbar$, the net field remaining
between $\q_1$ and $\qbar_1$ is $\br{b} b$ ($b$ = blue;
$g + r = \br{b}$ if only colour triplets are assumed). In this central
field, an additional $\q_2 \qbar_2$ pair can be created, where
$\q_2$ now is pulled towards $\q \q_1$ and $\qbar_2$
towards $\qbar \qbar_1$, with no net colour field between $\q_2$
and $\qbar_2$. If this is all that happens, the baryon $B$ will be
made up out of $\q_1$, $\q_2$ and some $\q_4$ produced between
$\q$ and $\q_1$, and $\br{B}$ of $\qbar_1$, $\qbar_2$ and some
$\qbar_5$, i.e.\ the $B$ and $\br{B}$ will be nearest neighbours in
rank and share two quark pairs. Specifically, $\q_1$ will gain
energy from $\q_2$ in order to end up on mass shell, and the
tunnelling formula for an effective $\q_1 \q_2$ diquark is recovered.
 
Part of the time, several $b \br{b}$ colour pair productions
may take place between the $\q_1$ and $\qbar_1$, however. With two
production vertices $\q_2 \qbar_2$ and $\q_3 \qbar_3$, a central
meson $\qbar_2 \q_3$ may be formed, surrounded by a baryon
$\q_4 \q_1 \q_2$ and an antibaryon $\qbar_3 \qbar_1 \qbar_5$.
We call this a $BM\br{B}$ configuration to distinguish it from the
$\q_4 \q_1 \q_2$ + $\qbar_2 \qbar_1 \qbar_5$ $B\br{B}$ configuration
above. For $BM\br{B}$ the $B$ and $\br{B}$ only share one
quark--antiquark pair, as opposed to two for $B\br{B}$
configurations. The relative probability for a $BM\br{B}$
configuration is given by the uncertainty relation suppression for
having the $\q_1$ and $\qbar_1$ sufficiently far apart that a meson
may be formed in between. The suppression of the $BM\br B$ system is
estimated by
\begin{equation}
|\Delta_F|^2\approx \exp(-2\mu_{\perp} M_{\perp}/\kappa)
\label{e:DeltaF}
\end{equation}
where $\mu_{\perp}$ and $M_{\perp}$ is the transverse mass of $\q_1$ and 
the meson, respectively. Strictly speaking, also configurations
like $BMM\br{B}$, $BMMM\br{B}$, etc. should be possible, but
since the total invariant $M_{\perp}$ grows rapidly with the number of 
mesons, the probability for this is small in the simple model. Further, 
since larger masses corresponds to longer string pieces, the production 
of pseudoscalar mesons is favoured over that of vector ones. If only
$B\br{B}$ and $BM\br{B}$ states are included, and if the probability
for having a vector meson $M$ is not suppressed extra, two partly
compensating errors are made (since a vector meson typically
decays into two or more pseudoscalar ones).
 
In total, the flavour iteration procedure therefore contains the
following possible subprocesses (plus, of course, their charge
conjugates):
\begin{Itemize}
\item $\q_1 \to \q_2 + (\q_1 \qbar_2)$ meson;
\item $\q_1 \to \qbar_2\qbar_3 + (\q_1 \q_2 \q_3)$ baryon;
\item $\q_1 \q_2 \to \qbar_3 + (\q_1 \q_2 \q_3)$ baryon;
\item $\q_1 \q_2 \to \q_1 \q_3 + (\q_2 \qbar_3)$ meson;
\end{Itemize}
with the constraint that the last process cannot be iterated to
obtain several mesons in between the baryon and the antibaryon.
 
When selecting flavours for $\q\q\rightarrow M+\q\q'$, 
the quark coming from the accepted $\q\q$ is kept, and the other member 
of $\q\q'$, as well as the spin of $\q\q'$, is chosen with weights 
taking {\bf SU(6)} symmetry into account. Thus the flavour of $\q\q$ 
is not influenced by {\bf SU(6)} factors for $\q\q'$, but the flavour 
of $M$ is.
 
Unfortunately, the resulting baryon production model has a fair
number of parameters, which would be given by the model only if
quark and diquark masses were known unambiguously.
We have already mentioned the $\s / \u$ ratio and the $\q\q / \q$
one; the latter has to be increased from 0.09 to 0.10 for the
popcorn model, since the total number of possible baryon
production configurations is lower in this case (the particle
produced between the $B$ and $\br{B}$ is constrained to be a
meson). With the improved {\bf SU(6)} treatment introduced in {\Py} 
6.106,  a rejected $\q\rightarrow B +\q\q$ may lead to the splitting 
$\q\rightarrow M+\q'$ instead. This calls for an increase of the 
$\q\q/\q$ input ratio by approximately 10\%. 
For the popcorn model, exactly the same parameters as
already found in the diquark model are needed to describe the
$B\br{B}$ configurations. For $BM\br{B}$ configurations, the square
root of a suppression factor should be applied if the factor is
relevant only for one of the $B$ and $\br{B}$, e.g.\ if the $B$
is formed with a spin 1 `diquark' $\q_1 \q_2$ but the $\br{B}$
with a spin 0 diquark $\qbar_1 \qbar_3$. Additional parameters
include the relative probability for $BM\br{B}$ configurations,
which is assumed to be roughly 0.5 (with the remaining 0.5 being
$B\br{B}$), a suppression factor for having a strange meson $M$
between the $B$ and $\br{B}$ (as opposed to having a lighter
nonstrange one) and a suppression factor for having a $\s \sbar$
pair (rather than a $\u \ubar$ one) shared between the $B$ and
$\br{B}$ of a $BM\br{B}$ configuration. The default parameter
values are based on a combination of experimental observation
and internal model predictions.
 
In the diquark model, a diquark is expected to have exactly the
same transverse momentum distribution as a quark. For $BM\br{B}$
configurations the situation is somewhat more unclear, but we
have checked that various possibilities give very similar results.
The option implemented in the program is to assume no transverse
momentum at all for the $\q_1 \qbar_1$ pair shared by the $B$ and
$\br{B}$, with all other pairs having the standard Gaussian
spectrum with local momentum conservation. This means that the
$B$ and $\br{B}$ $\pT$:s are uncorrelated in a $BM\br{B}$
configuration and (partially) anticorrelated in the $B\br{B}$
configurations, with the same mean transverse momentum for primary
baryons as for primary mesons.

{\em Advanced popcorn}

In \cite{Ede97}, a revised popcorn model is presented, where the 
separate production of the quarks in an effective diquark is taken 
more seriously. The production of a $\q \qbar$ pair which breaks the 
string is in this model determined by eq.~(\ref{e:qtunnel}), also 
when ending up in a diquark. Furthermore, the popcorn model is 
re-implemented in such a way that eq~(\ref{e:DeltaF}) could be used 
explicitly in the Monte Carlo. The two parameters
\begin{equation} 
\beta_{\q} \equiv 2\left\langle \mu_{\perp \q} \right\rangle/\kappa,
~~{\mrm{or}}~~\beta_{\u}~{\mrm{and}}~\Delta\beta \equiv 
\beta_{\s}-\beta_{\u},
 \label{e:betadef} 
\end{equation}
then govern both the diquark and the intermediate meson production. 
In this algorithm, configurations like $BMM\br{B}$ etc. are considered 
in a natural way. The more independent production of the diquark partons 
implies a moderate suppression of spin 1 diquarks. Instead the direct 
suppression of spin 3/2 baryons, in correspondence to the suppression 
of vector mesons relative to pseudo-scalar ones, is assumed to be 
important. Consequently, a suppression of $\Sigma$-states relative to 
$\Lambda^0$ is derived from the spin 3/2 suppression parameter.

Several new routines have been added, and the diquark code has
been extended with information about the curtain quark flavour, i.e.
the $\q\qbar$ pair that is shared between the baryon and antibaryon,
but this is not visible externally. Some parameters are no longer
used, while others have to be given modified values. This is described 
in section \ref{sss:improvedbaryoncode}. 

{\em Baryon remnant fragmentation}
 
Occasionally, the endpoint of a string is not a single parton,
but a diquark or antidiquark, e.g.\ when a quark has been kicked
out of a proton beam particle. One could consider fairly complex
schemes for the resulting fragmentation. One such \cite{And81}
was available in {\Je} version 6 but is no longer found here. 
Instead the same basic scheme is used as for diquark pair
production above. Thus a $\q\q$ diquark endpoint is let to fragment
just as would a $\q\q$ produced in the field behind a matching
$\qbar\qbar$ flavour, i.e.\ either the two quarks of the diquark enter
into the same leading baryon, or else a meson is first produced,
containing one of the quarks, while the other is contained in the
baryon produced in the next step.

Similarly, the revised algorithm for baron production can be applied 
to endpoint diquarks, though this must be made with some care 
\cite{Ede97}. The suppression factor for popcorn mesons is derived from 
the assumption of colour fluctuations violating energy conservation and 
thus being suppressed by the Heisenberg uncertainty principle. When 
splitting an original diquark into two more independent quarks, the same 
kind of energy shift does not obviously emerge. One could still expect 
large separations of the diquark constituents to be suppressed, but the 
shape of this suppression is beyond the scope of the model. For simplicity, 
the same kind of exponential suppression as in the "true popcorn" case is 
implemented in the program. However, there is little reason for the 
strength of the suppression to be exactly the same in the different 
situations. Thus the leading rank meson production in a 
diquark jet is governed by a new $\beta$ parameter, which is independent 
of the popcorn parameters $\beta_{\mrm{u}}$ and $\Delta \beta$ in 
eq.~(\ref{e:betadef}). Furthermore, in the process (original diquark 
$\rightarrow$  baryon+$\qbar$) the spin 3/2 suppression should not 
apply at full strength. This suppression factor stems from the 
normalization of the overlapping $\q$ and $\qbar$ wavefunctions in a 
newly produced $\q\qbar$ pair, but in the process considered here, two 
out of three valence quarks already exist as an initial condition of 
the string.

\subsection{String Fragmentation}
 
An iterative procedure can also be used for other aspects of the
fragmentation. This is possible because, in the string picture,
the various points where the string break by the production of
$\q \qbar$ pairs are causally disconnected. Whereas the space--time
picture in the c.m.\ frame is such that slow particles
(in the middle of the system) are formed first, this ordering is
Lorentz frame dependent and hence irrelevant. One may therefore
make the convenient choice of starting an iteration process at
the ends of the string and proceeding towards the middle.
 
The string fragmentation scheme is rather complicated for a generic
multiparton state. In order to simplify the discussion, we will
therefore start with the simple $\q \qbar$ process, and only later
survey the complications that appear when additional gluons
are present. (This distinction is made for pedagogical reasons,
in the program there is only one general-purpose algorithm).
 
\subsubsection{Fragmentation functions}
 
Assume a $\q \qbar$ jet system, in its c.m.\ frame, with the quark moving
out in the $+z$ direction and the antiquark in the $-z$ one. We have
discussed how it is possible to start the flavour iteration from the
$\q$ end, i.e.\ pick a $\q_1 \qbar_1$ pair, form a hadron $\q \qbar_1$,
etc. It has also been noted that the tunnelling mechanism
is assumed to give a transverse momentum $\pT$ for each new
$\q_i\qbar_i$ pair created, with the $\pT$ locally compensated between
the $\q_i$ and the $\qbar_i$ member of the pair, and with a Gaussian
distribution in $p_x$ and $p_y$ separately. In the program, this is
regulated by one parameter, which gives the root-mean-square $\pT$
of a quark. Hadron transverse momenta are obtained as the sum of
$\pT$:s of the constituent $\q_i$ and $\qbar_{i+1}$, where a diquark
is considered just as a single quark.
 
What remains to be determined is the energy and longitudinal
momentum of the hadron. In fact, only one variable can be selected
independently, since the momentum of the hadron is constrained by
the already determined hadron transverse mass $m_{\perp}$,
\begin{equation}
(E+p_z)(E-p_z) = E^2 - p_z^2 = m_{\perp}^2 = m^2 + p_x^2 + p_y^2 ~.
\label{fr:massconstr}
\end{equation}
In an iteration from the quark end, one is led (by the desire for
longitudinal boost invariance and other considerations)
to select the $z$ variable as the fraction of
$E+p_z$ taken by the hadron, out of the available $E+p_z$.
As hadrons are split off, the $E+p_z$ (and $E-p_z$) left for
subsequent steps is reduced accordingly:
\begin{eqnarray}
(E+p_z)_{\mrm{new}} & = & (1-z) (E+p_z)_{\mrm{old}} ~, \nonumber \\
(E-p_z)_{\mrm{new}} & = & (E-p_z)_{\mrm{old}} -
\frac{m_{\perp}^2}{z (E+p_z)_{\mrm{old}}} ~.
\end{eqnarray}

The fragmentation function $f(z)$, which expresses the probability
that a given $z$ is picked, could in principle be arbitrary --- indeed,
several such choices can be used inside the program, see below.
 
If one, in addition, requires that the fragmentation
process as a whole should look the same,
irrespectively of whether the iterative procedure is performed
from the $\q$ end or the $\qbar$ one, `left--right symmetry',
the choice is essentially unique \cite{And83a}: the `Lund symmetric
fragmentation function',
\begin{equation}
f(z) \propto \frac{1}{z} z^{a_{\alpha}} \left( \frac{1-z}{z}
   \right)^{a_{\beta}} \exp \left( - \frac{bm_{\perp}^2}{z} 
   \right) ~.
\label{fr:LSFFlong}
\end{equation}
There is one separate parameter $a$ for each flavour,
with the index $\alpha$ corresponding to the `old' flavour in the
iteration process, and $\beta$ to the `new' flavour.
It is customary to put all $a_{\alpha,\beta}$ the same, and thus
arrive at the simplified expression
\begin{equation}
 f(z) \propto z^{-1} (1-z)^a \exp (-bm_{\perp}^2/z) ~.
 \label{fr:LSFF}
\end{equation}
In the program, only two separate $a$ values can be given, that
for quark pair production and that for diquark one. In addition, 
there is the $b$ parameter, which is universal.
 
The explicit mass dependence in $f(z)$ implies a harder 
fragmentation function for heavier hadrons. The asymptotic 
behaviour of the mean $z$ value for heavy hadrons is
\begin{equation}
\langle z \rangle \approx 1 - \frac{1+a}{bm_{\perp}^2} ~.
\end{equation}
Unfortunately it seems this predicts a somewhat harder spectrum
for $\B$ mesons than observed in data. However,
Bowler \cite{Bow81} has shown, within the framework of the
Artru--Mennessier model \cite{Art74}, that a massive endpoint quark
with mass $m_Q$ leads to a modification of the symmetric
fragmentation function, due to
the fact that the string area swept out is reduced for massive endpoint
quarks, compared with massless ditto. The Artru--Mennessier model in
principle only applies for clusters with a continuous mass spectrum,
and does not allow an $a$ term (i.e.\ $a \equiv 0$); however, it has
been shown \cite{Mor89} that, for a discrete mass spectrum, one may
still retain an effective $a$ term. In the program an approximate
form with an $a$ term has therefore been used:
\begin{equation}
f(z) \propto \frac{1}{z^{1 + r_Q b m_Q^2}}
   z^{a_{\alpha}} \left( \frac{1-z}{z}
   \right)^{a_{\beta}} \exp \left( - \frac{b m_{\perp}^2}{z} \right) ~.
\label{fr:LSFFBowler}
\end{equation}
In principle the prediction is that $r_Q \equiv 1$, but so as to
be able to extrapolate smoothly between this form and the original 
Lund symmetric one, it is possible to pick $r_Q$ separately for 
$\c$ and $\b$ hadrons.
 
For future reference we note that the derivation of $f(z)$ as a
by-product also gives the probability distribution in proper
time $\tau$ of $\q_i \qbar_i$ breakup vertices. In terms of
$\Gamma = (\kappa \tau)^2$, this distribution is
\begin{equation}
{\cal P}(\Gamma) \, \d \Gamma \propto \Gamma^a \, \exp(-b \Gamma) \,
\d \Gamma ~,
\end{equation}
with the same $a$ and $b$ as above. The exponential decay allows
an interpretation in terms of an area law for the colour flux
\cite{And98}.
 
Many different other fragmentation functions have been proposed,
and a few are available as options in the program.
\begin{Itemize}
\item The Field-Feynman parameterization \cite{Fie78},
\begin{equation}
  f(z) = 1 - a + 3a(1-z)^2 ~,
\end{equation}
with default value $a = 0.77$, is intended to be used only for
ordinary hadrons made out of $\u$, $\d$ and $\s$ quarks.
\item Since there are indications that the shape above is too
strongly peaked at $z = 0$, instead a shape like
\begin{equation}
  f(z) = (1+c) (1-z)^c
\end{equation}
may be used.
\item Charm and bottom data clearly indicate the need for a
harder fragmentation function for heavy flavours.
The best known of these is the Peterson/SLAC~formula \cite{Pet83}
\begin{equation}
  f(z) \propto \frac{1}{ z \left( 1 - 
      \frac{\displaystyle 1}{\displaystyle z} -
      \frac{\displaystyle \epsilon_Q}{\displaystyle 1-z} 
      \right)^2 } ~,
  \label{fr:PetHF}
\end{equation}
where $\epsilon_Q$ is a free parameter, expected to scale between
flavours like $\epsilon_Q \propto 1/m_Q^2$.
\item As a crude alternative, that is also peaked at $z=1$, one may
use
\begin{equation}
  f(z) = (1+c) z^c ~.
\end{equation}
\item In \cite{Ede97}, it is argued that the quarks responsible for the 
colour fluctuations in stepwise diquark production cannot move along the 
light-cones. Instead there is an area of possible starting points for the 
colour fluctuation, which is essentially given by the proper time of the 
vertex squared. By summing over all possible starting points, one obtains 
the total weight for the colour fluctuation. The result is a relative 
suppression of diquark vertices at early times, which is found to be of 
the form $1-\exp(-\rho \Gamma)$, where $\Gamma\equiv\kappa^2\tau^2$ and 
$\rho \approx 0.7{\mrm{GeV}}^{-2}$. This result, and especially the value 
of $\rho$, is independent of the fragmentation function, $f(z)$, used to 
reach a specific $\Gamma$-value. However, if using a $f(z)$ which implies 
a small average value $\langle\Gamma\rangle$, the program implementation is 
such that a large fraction of the $\q\rightarrow B + \q\q$ attempts 
will be rejected. This dilutes the interpretation of the input 
$P(\q\q)/P(\q)$ parameter, which needs to be significantly enhanced to 
compensate for the rejections.\\ 
A property of the Lund Symmetric Fragmentation Function is that the first 
vertices produced near the string ends have a lower $\langle\Gamma\rangle$ 
than central  vertices. Thus an effect of the low-$\Gamma$ suppression 
is a relative reduction of the leading baryons. The effect is smaller if 
the baryon is very heavy, as the large mass implies that the first vertex 
almost reaches the central region. Thus the leading baryon suppression 
effect is reduced for $\c$- and $\b$ jets.
\end{Itemize}
 
\subsubsection{Joining the jets}
 
The $f(z)$ formula above is only valid, for the breakup of a jet
system into a hadron plus a remainder-system, when the remainder
mass is large. If the fragmentation algorithm were to be used all the
way from the $\q$ end to the $\qbar$ one, the mass of the last hadron to
be formed at the $\qbar$ end would be completely constrained by energy
and momentum conservation, and could not be on its mass shell. In theory
it is known how to take such effects into account \cite{Ede00}, but the 
resulting formulae are wholly unsuitable for Monte Carlo implementation.
 
The practical solution to this problem is to carry out the
fragmentation both from the $\q$ and the $\qbar$ end, such that for
each new step in the fragmentation process, a random
choice is made as to from what side the step is to be taken.
If the step is on the $\q$ side, then $z$ is interpreted as fraction
of the remaining $E+p_z$ of the system, while $z$ is interpreted as
$E-p_z$ fraction for a step from the $\qbar$ end. At some point, when
the remaining mass of the system has dropped below a given value,
it is decided that the next breakup will produce two final hadrons,
rather than a hadron and a remainder-system.
Since the momenta of two hadrons are to be selected, rather than
that of one only, there are enough degrees of freedom to have both
total energy and total momentum completely conserved.
 
The mass at which the normal fragmentation
process is stopped and the final two hadrons formed is not actually
a free parameter of the model: it is given by the requirement that
the string everywhere looks the same, i.e.\ that the rapidity spacing
of the final two hadrons, internally and with respect to surrounding
hadrons, is the same as elsewhere in the fragmentation process.
The stopping mass, for a given setup of fragmentation parameters,
has therefore been determined in separate runs. If the fragmentation
parameters are changed, some retuning should be done but, in practice,
reasonable changes can be made without any special arrangements.
 
Consider a fragmentation process which has already split off a number
of hadrons from the $\q$ and $\qbar$ sides, leaving behind a
a $\q_i \qbar_j$ remainder system. When this system breaks by the
production of a $\q_n \qbar_n$ pair, it is decided to make this pair
the final one, and produce the last two hadrons $\q_i\qbar_n$ and
$\q_n\qbar_j$, if
\begin{equation}
( (E+p_z)(E-p_z) )_{\mrm{remaining}} = W_{\mrm{rem}}^2 < 
W_{\mmin}^2 ~.
\end{equation}
The $W_{\mmin}$ is calculated according to
\begin{equation}
W_{\mmin} = ( W_{\mmin 0} + m_{\q i} + m_{\q j} + k \, m_{\q n} )
\, (1 \pm \delta) ~.
\end{equation}
Here $W_{\mmin 0}$ is the main free parameter, typically around 
1 GeV, determined to give a flat
rapidity plateau (separately for each particle species), while the
default $k = 2$ corresponds to the mass of the final pair being taken
fully into account. Smaller values may also be considered, depending
on what criteria are used to define the `best' joining of the $\q$
and the $\qbar$ chain. The factor $1 \pm \delta$, by default evenly
distributed between 0.8 and 1.2, signifies a smearing of the 
$W_{\mmin}$ value, to avoid an abrupt and unphysical cut-off in
the invariant mass distribution of the final two hadrons. Still,
this distribution will be somewhat different from that of any two
adjacent hadrons elsewhere. Due to the cut there will be no tail up to
very high masses; there are also fewer events close to the lower limit,
where the two hadrons are formed at rest with respect to each other.
 
This procedure does not work all that well for heavy flavours, since it
does not fully take into account the harder fragmentation function
encountered. Therefore, in addition to the check above, one further
test is performed for charm and heavier flavours, as follows. If the
check above allows more particle production,
a heavy hadron $\q_i \qbar_n$ is formed, leaving a
remainder $\q_n \qbar_j$.  The range of allowed $z$ values, i.e.\ the
fraction of remaining $E+p_z$ that may be taken by the
$\q_i \qbar_n$ hadron, is constrained away from 0 and 1 by the
$\q_i \qbar_n$ mass and minimal mass of the $\q_n \qbar_j$ system.
The limits of the physical $z$ range is obtained when the
$\q_n \qbar_j$ system only consists of one single particle, which then
has a well-determined
transverse mass $m_{\perp}^{(0)}$. From the $z$ value obtained with the
infinite-energy fragmentation function formulae, a rescaled $z'$ value
between these limits is given by
\begin{equation}
z' = \frac{1}{2} \left\{ 1 + \frac{m_{\perp i n}^2}{W_{\mrm{rem}}^2} - 
\frac{m_{\perp n j}^{(0)2}}{W_{\mrm{rem}}^2}  + 
\sqrt{ \left( 1 - \frac{m_{\perp i n}^2}{W_{\mrm{rem}}^2} - 
\frac{m_{\perp n j}^{(0)2}}{W_{\mrm{rem}}^2} \right)^2 
- 4 \frac{m_{\perp i n}^2}{W_{\mrm{rem}}^2}
\frac{m_{\perp n j}^{(0)2}}{W_{\mrm{rem}}^2} }
\, (2 z -1) \right\} ~.
\end{equation}
{}From the $z'$ value, the actual transverse mass
$m_{\perp n j} \geq m_{\perp n j}^{(0)}$ of the $\q_n \qbar_j$
system may be calculated. For more than one particle
to be produced out of this system, the requirement
\begin{equation}
m_{\perp n j}^2 = (1-z') \, \left( W_{\mrm{rem}}^2 -
\frac{m_{\perp i n}^2}{z'} \right) > (m_{q j} + W_{\mmin 0})^2 
+ \pT^2
\end{equation}
has to be fulfilled. If not, the $\q_n \qbar_j$ system is assumed to
collapse to one single particle.
 
The consequence of the procedure above is that, the more the infinite
energy fragmentation function $f(z)$ is peaked close to $z=1$, the more
likely it is that only two particles are produced. The procedure 
above has been constructed so that
the two-particle fraction can be calculated directly from the shape of
$f(z)$ and the (approximate) mass spectrum, but it is not unique.
For the symmetric Lund fragmentation function, a number of
alternatives tried all give essentially the same result, whereas other
fragmentation functions may be more sensitive to details.
 
Assume now that two final hadrons have been picked. If the
transverse mass of the remainder-system is smaller than the sum of
transverse masses of the final two hadrons, the whole fragmentation
chain is rejected, and started over from the $\q$ and $\qbar$
endpoints. This does not introduce any significant bias, since the
decision to reject a fragmentation chain only depends on what happens
in the very last step, specifically that the next-to-last step took
away too much energy, and not on what happened in the steps before
that.
 
If, on the other hand,  the remainder-mass is large enough, there
are two kinematically allowed solutions for the final two hadrons:
the two mirror images in the rest frame of the remainder-system. Also 
the choice between these two solutions is given by the consistency
requirements, and can be derived from studies of infinite energy jets.
The probability for the reverse ordering, i.e.\ where the rapidity and 
the flavour orderings disagree, is given by the area law as
\begin{equation}
{\cal P}_{\mrm{reverse}} = \frac{1}{1 + e^{b\Delta}} 
~~~\mrm{where}~~~\Delta = \Gamma_2 - \Gamma_1 = 
\sqrt{ (W_{\mrm{rem}}^2 - m_{\perp i n}^2 + m_{\perp n j}^2)^2 -
4 m_{\perp i n}^2 m_{\perp n j}^2} ~.
\label{fr:revord}
\end{equation}
For the Lund symmetric fragmentation function, $b$ is the familiar
parameter, whereas for other functions the $b$ value becomes an
effective number to be fitted to the behaviour when not in the
joining region.
 
When baryon production is included, some particular problems arise
(also see section \ref{sss:baryonprod}). First consider $B\br{B}$ 
situations. In the na\"{\i}ve iterative scheme, away from the middle 
of the event, one already has a quark and is to chose a matching 
diquark flavour or the other way around. In either case the choice 
of the new flavour can be done taking into account the number of 
{\bf SU(6)} states available for the quark-diquark combination. 
For a case where the final $\q_n \qbar_n$ breakup is an
antidiquark-diquark one, the weights for forming $\q_i \qbar_n$ and
$\q_n \qbar_i$ enter at the same time, however. We do not know how to
handle this problem; what is done is to use weights as usual for the
$\q_i \qbar_n$ baryon to select $\q_n$, but then consider
$\q_n \qbar_i$ as given (or the other way around with equal
probability). If $\q_n \qbar_i$ turns out to be
an antidiquark-diquark combination, the whole fragmentation chain is
rejected, since we do not know how to form corresponding hadrons.
A similar problem arises, and is solved in the same spirit, for a
$BM\br{B}$ configuration in which the $B$ (or $\br{B}$) was chosen
as third-last particle. When only two particles remain to be
generated, it is obviously too late to consider having a $BM\br{B}$
configuration. This is as it should, however, as can be found by
looking at all possible ways a hadron of given rank can be a baryon.
 
While some practical compromises have to be accepted in the
joining procedure, the fact that the joining takes place in
different parts of the string in different events means that,
in the end, essentially no visible effects remain.
 
\subsubsection{String motion and infrared stability}
 
We have now discussed the SF scheme for the fragmentation of a simple
$\q \qbar$ jet system. In order to understand how these results
generalize
to arbitrary jet systems, it is first necessary to understand the
string motion for the case when no fragmentation takes place. In the
following we will assume that quarks as well as gluons are massless,
but all arguments can be generalized to massive quarks without too
much problem.
 
For a $\q \qbar$ event viewed in the c.m.\ frame, with total energy $W$,
the partons start moving out back-to-back, carrying half the energy
each. As they move apart, energy and momentum is lost to the string.
When the partons are a distance $W / \kappa$ apart, all the energy
is stored in the string. The partons now turn around and come
together again with the original momentum vectors reversed. This
corresponds to half a period of the full string motion; the second
half the process is repeated, mirror-imaged. For further
generalizations to multiparton systems, a convenient description of
the energy and momentum flow is given in terms of `genes'
\cite{Art83}, infinitesimal packets of the four-momentum given up
by the partons to the string. Genes with $p_z = E$, emitted from the
$\q$ end in the initial stages of the string motion above,
will move in the $\qbar$ direction with the speed of light, whereas
genes with $p_z = -E$ given up by the $\qbar$ will move in the $\q$
direction. Thus, in this simple case, the direction of motion for a
gene is just opposite to that of a free particle with the same
four-momentum. This is due to the string tension. If the system is
not viewed in the c.m.\ frame, the rules are that any parton gives up
genes with four-momentum proportional to its own four-momentum, but
the direction of motion of any gene is given by the momentum direction
of the genes it meets, i.e.\ that were emitted by the parton at the
other end of that particular string piece. When the $\q$ has lost
all its energy, the $\qbar$ genes, which before could not catch up
with $\q$, start impinging on it, and the $\q$ is pulled back,
accreting $\qbar$ genes in the process. When the $\q$ and $\qbar$
meet in the origin again, they have completely traded genes with
respect to the initial situation.
 
A 3-jet $\q \qbar \g$ event initially corresponds to having a
string piece stretched between $\q$ and $\g$ and another between
$\g$ and $\qbar$. Gluon four-momentum genes are thus flowing towards
the $\q$ and $\qbar$. Correspondingly, $\q$ and $\qbar$ genes are
flowing towards the $\g$. When the gluon has lost all its energy,
the $\g$ genes continue moving apart, and instead a third
string region is formed in the `middle' of the total string,
consisting of overlapping $\q$ and $\qbar$ genes. The two `corners'
on the string, separating the three string regions, are not of the
gluon-kink type: they do not carry any momentum.
 
If this third region would only appear at a time later than the typical
time scale for fragmentation, it could not affect the sharing of energy
between different particles. This is true in the limit of high energy,
well separated partons. For a small gluon energy, on the other hand, the
third string region appears early, and the overall drawing of the string
becomes fairly 2-jet-like, since the third string region consists of
$\q$ and $\qbar$ genes and therefore behaves exactly as a sting pulled
out directly between the $\q$ and $\qbar$.
In the limit of vanishing gluon energy,
the two initial string regions collapse to naught, and the ordinary
2-jet event is recovered \cite{Sjo84}. Also for a collinear gluon, i.e.
$\theta_{\q \g}$ (or $\theta_{\qbar \g}$) small, the stretching
becomes 2-jet-like. In particular, the $\q$ string endpoint first
moves out a distance $\mbf{p}_{\q} / \kappa$ losing genes to the
string, and then a further distance $\mbf{p}_{\g} / \kappa$, a first
half accreting genes from the $\g$ and the second half re-emitting
them. (This latter half actually includes yet another
string piece; a corresponding piece appears at the $\qbar$ end, such
that half a period of the system involves five different string
regions.) The end result is, approximately, that a string is drawn out
as if there had only been a single parton with energy
$|\mbf{p}_{\q} + \mbf{p}_{\g}|$, such that the simple 2-jet event
again is recovered in the limit $\theta_{\q \g} \to 0$. These
properties of the string motion are the reason why
the string fragmentation scheme is `infrared safe' with respect to
soft or collinear gluon emission.
 
The discussions for the 3-jet case can be generalized to the motion
of a string with $\q$ and $\qbar$ endpoints and an arbitrary number of
intermediate gluons. For $n$ partons, whereof $n-2$
gluons, the original string contains $n-1$ pieces. Anytime one of the
original gluons has lost its energy, a new string region is formed,
delineated by a pair of `corners'.
As the extra `corners' meet each other, old string regions vanish
and new are created, so that half a period of the string contains
$2n^2 - 6n + 5$ different string regions. Each of these regions can
be understood simply as built up from the overlap of (opposite-moving)
genes from two of the original partons, according to well specified
rules.
 
\subsubsection{Fragmentation of multiparton systems}
 
The full machinery needed for a multiparton system is very
complicated, and is described in detail in \cite{Sjo84}. The
following outline is far from complete, and is complicated
nonetheless. The main message to be conveyed is that a
Lorentz covariant algorithm exists for handling an arbitrary parton
configuration, but that the necessary machinery is more complex
than in either cluster or independent fragmentation.
 
Assume $n$ partons, with ordering along the string, and related
four-momenta, given by
$\q(p_1) \g(p_2) \g(p_3) \cdots \g(p_{n-1}) \qbar(p_n)$.
The initial string then contains $n-1$ separate pieces.
The string piece between the quark and its neighbouring gluon is, in
four-momentum space, spanned by one side with four-momentum
$p_+^{(1)} = p_1$ and another with $p_-^{(1)} = p_2/2$. The factor of
1/2 in the second expression
comes from the fact that the gluon shares its energy between two
string pieces. The indices `$+$' and `$-$' denotes direction towards
the $\q$ and $\qbar$ end, respectively. The next string piece,
counted from the quark end, is spanned by $p_+^{(2)} = p_2/2$ and
$p_-^{(2)} = p_3/2$, and so on, with the last one being
$p_+^{(n-1)} = p_{n-1}/2$ and $p_-^{(n-1)} = p_n$.
 
For the algorithm to work, it is important
that all $p_{\pm}^{(i)}$ be light-cone-like, i.e.\ $p_{\pm}^{(i)2} = 0$.
Since gluons are massless, it is only the two endpoint quarks which
can cause problems. The procedure here is to create new $p_{\pm}$
vectors for each of the two endpoint regions, defined to be linear
combinations of the old $p_{\pm}$ ones for the same region, with
coefficients determined so that the
new vectors are light-cone-like. De facto, this corresponds to
replacing a massive quark at the end of a string piece with a massless
quark at the end of a somewhat longer string piece. With the exception
of the added fictitious piece, which anyway ends up entirely within
the heavy hadron produced from the heavy quark, the string motion
remains unchanged by this.
 
In the continued string motion, when new string regions appear as
time goes by, the first such string regions that appear can be
represented as being spanned by one $p_+^{(j)}$ and another
$p_-^{(k)}$ four-vector, with $j$ and $k$ not necessarily
adjacent. For instance, in the $\q \g \qbar$
case, the `third' string region is spanned by $p_+^{(1)}$ and
$p_-^{(3)}$. Later on in the string evolution history,
it is also possible to have regions made up of two $p_+$ or two
$p_-$ momenta. These appear when an endpoint quark has
lost all its original momentum, has accreted the momentum of an
gluon, and is now re-emitting this momentum. In practice, these
regions may be neglected. Therefore only pieces made up by a
$(p_+^{(j)},p_-^{(k)})$ pair of momenta are considered in the
program.
 
The allowed string regions may be ordered in an abstract parameter
plane, where the $(j,k)$ indices of the four-momentum pairs define
the position of each region along the two (parameter plane)
coordinate axes. In this plane the fragmentation procedure can be
described as a sequence of steps, starting at the quark end,
where $(j,k) = (1,1)$, and ending at the antiquark one,
$(j,k) = (n-1,n-1)$. Each step is taken from an `old' 
$\q_{i-1} \qbar_{i-1}$ pair production vertex, to the 
production vertex of a `new' $\q_i \qbar_i$
pair, and the string piece between these two string
breaks represent a hadron. Some steps may be taken within one and
the same region, while others may have one vertex in one region
and the other vertex in another region. Consistency requirements,
like energy-momentum conservation, dictates that vertex $j$ and
$k$ region values be ordered in a monotonic sequence, and that
the vertex positions are monotonically ordered inside each region.
The four-momentum of each hadron can be read off, for $p_+$ ($p_-$)
momenta, by projecting the separation between the old and the new
vertex on to the $j$ ($k$) axis. If the four-momentum fraction of
$p_{\pm}^{(i)}$ taken by a hadron is denoted $x_{\pm}^{(i)}$,
then the total hadron four-momentum is given by
\begin{equation}
p = \sum_{j=j_1}^{j_2} x_+^{(j)} p_+^{(j)} +
      \sum_{k=k_1}^{k_2} x_-^{(k)} p_-^{(k)} +
      p_{x1} \hat{e}_x^{(j_1 k_1)} + p_{y1} \hat{e}_y^{(j_1 k_1)} +
      p_{x2} \hat{e}_x^{(j_2 k_2)} + p_{y2} \hat{e}_y^{(j_2 k_2)} ~,
\label{fr:fourmom}
\end{equation}
for a step from region $(j_1,k_1)$ to region $(j_2,k_2)$.
By necessity, $x_+^{(j)}$ is unity for a $j_1 < j < j_2$,
and correspondingly for $x_-^{(k)}$.
 
The $(p_x,p_y)$ pairs are the transverse momenta produced at
the two string breaks, and the $(\hat{e}_x,\hat{e}_y)$ pairs 
four-vectors transverse to the string directions in the regions 
of the respective string breaks:
\begin{eqnarray}
 & & \hat{e}_x^{(jk)2} = \hat{e}_y^{(jk)2} = -1 ~,        \nonumber \\
 & & \hat{e}_x^{(jk)} \hat{e}_y^{(jk)} = \hat{e}_{x,y}^{(jk)}
  p_+^{(j)} = \hat{e}_{x,y}^{(jk)} p_-^{(k)} = 0 ~.
\end{eqnarray}
 
The fact that the hadron should be on mass shell, $p^2 = m^2$, puts
one constraint on where a new breakup may be, given that the old
one is already known, just as eq.~(\ref{fr:massconstr}) did in the
simple 2-jet case. The remaining degree of freedom is, as before, 
to be given by
the fragmentation function $f(z)$. The interpretation of the $z$
is only well-defined for a step entirely constrained to one of the
initial string regions, however, which is not enough. In the
2-jet case, the $z$ values can be related to the proper times
of string breaks, as follows. The variable
$\Gamma = (\kappa \tau)^2$, with $\kappa$ the string tension and
$\tau$ the proper time between the production vertex of the
partons and the breakup point, obeys an iterative relation of the
kind
\begin{eqnarray}
\Gamma_0 & = & 0 ~,              \nonumber   \\
\Gamma_i & = & (1-z_i) \left( \Gamma_{i-1} + \frac{m_{\perp i}^2}{z_i}
   \right) ~.
\end{eqnarray}
Here $\Gamma_0$ represents the value at the $\q$ and $\qbar$ endpoints,
and $\Gamma_{i-1}$ and $\Gamma_i$ the values
at the old and new breakup vertices needed to produce
a hadron with transverse mass $m_{\perp i}$, and with the $z_i$ of the
step chosen according to $f(z_i)$. The proper time can be
defined in an unambiguous way, also over boundaries between the
different string regions, so for multijet events the $z$ variable may
be interpreted just as
an auxiliary variable needed to determine the next $\Gamma$ value.
(In the Lund symmetric fragmentation function derivation, the
$\Gamma$ variable actually does appear naturally, so the choice
is not as arbitrary as it may seem here.)
The mass and $\Gamma$ constraints together are sufficient to determine
where the next string breakup is to be chosen, given the preceding
one in the iteration scheme. Actually, several ambiguities remain,
but are of no importance for the overall picture.
 
The algorithm for finding the next breakup then works something
like follows. Pick a hadron, $\pT$, and $z$, and calculate the next
$\Gamma$. If the old breakup is in the region $(j,k)$, and if the
new breakup is also assumed to be in the same region, then the
$m^2$ and $\Gamma$ constraints can be reformulated in terms of
the fractions $x_+^{(j)}$ and $x_-^{(k)}$ the hadron must take of the
total four-vectors $p_+^{(j)}$ and $p_-^{(k)}$:
\begin{eqnarray}
 m^2 &=&
 c_1 + c_2 x_+^{(j)} + c_3 x_-^{(k)} + c_4 x_+^{(j)} x_-^{(k)} ~,
       \nonumber  \\
 \Gamma &=&
 d_1 + d_2 x_+^{(j)} + d_3 x_-^{(k)} + d_4 x_+^{(j)} x_-^{(k)} ~.
\label{fr:mGa}
\end{eqnarray}
Here the coefficients
$c_n$ are fairly simple expressions, obtainable by squaring
eq.~(\ref{fr:fourmom}), while $d_n$ are slightly more complicated 
in that they depend on the position of the old string break,
but both the $c_n$ and the $d_n$ are explicitly calculable. What
remains is an equation system with two unknowns, $x_+^{(j)}$
and $x_-^{(k)}$.
The absence of any quadratic terms is due to
the fact that all $p_{\pm}^{(i)2} = 0$, i.e.\ to the choice of a
formulation based on light-cone-like longitudinal vectors.
Of the two possible solutions to the equation system (elimination of
one variable gives a second degree equation in the other), one is
unphysical and can be discarded outright. The other solution is checked
for
whether the $x_{\pm}$ values are actually inside the physically allowed
region, i.e.\ whether the $x_{\pm}$ values of the current step, plus
whatever has already been used up in previous steps, are less than
unity. If yes, a solution has been found. If no, it is because the
breakup could not take place inside the region studied, i.e.\ because
the equation system was solved for the wrong region. One therefore
has to change either index $j$ or index $k$ above by one step, i.e.
go to the next nearest string region. In this new region, a new equation
system of the type in eq.~(\ref{fr:mGa}) may be written down, with new
coefficients. A new solution is found and tested, and so on until a
physically
acceptable solution is found. The hadron four-momentum is now
given by an expression of the type (\ref{fr:fourmom}). The breakup
found forms the starting point for the new step in the fragmentation
chain, and so on. The final joining in the middle is done as in the
2-jet case, with minor extensions.
 
\subsubsection{Junction topologies}
\label{sss:junctiontopo}

When several valence quarks are kicked out of an incoming proton, or
when baryon number is violated, another kind of string
topology can be produced. In its simplest form, it can be illustrated by
the decay $\chio_10 \to \u\d\d$ which can take place in baryon number
violating supersymmetric theories. If we assume that the `colour triplet'
string kind encountered above is the only basic building block,
we are led to a Y-shaped string topology, with a quark at each end and
a ``junction''
where the strings meet. As the quarks move out, also this
junction would move, so as to minimize the total string energy. It is at
rest in a frame where the opening angle between any pairs of quarks is
120$^{\circ}$, so that the forces acting on the junction cancel.

In such a frame, each of the three strings would fragment pretty much as
ordinary strings, e.g. in a back-to-back $\q\qbar$ pair of jets, at least
as far as reasonably high-momentum particles are concerned. Thus an iterative
procedure can be used, whereby the leading $\q$ is combined with a
newly produced $\qbar_1$, to form a meson and leave behind a remainder-jet
$\q_1$. (As above, this has nothing to do with the ordering in physical
time, where the fragmentation process again starts in the middle and spreads
outwards.) Eventually, when little energy is left, the three remainders
$\q_i \q_j \q_k$ form a single baryon, which thus has a reasonably small
momentum in the rest frame of the junction. We see that the junction
thereby implicitly comes to be the carrier of the net baryon number of the
system. Further baryon production can well occur at higher momenta in
each of the three jets, but then always in pairs of a baryon and an
antibaryon. While the fragmentation principles as such are clear, the 
technical details of the joining of the jets become more complicated
than in the $\q\qbar$  
case. Some approximations can be made that allow a reasonably compact and
efficient algorithm, which  gives sensible results \cite{Sjo03}.
Specifically, two of the strings, preferably the ones with lowest energy,
can be fragmented until their remaining energy is below some cut-off value.
In fact, one of the two is required to have rather little energy left, while
the other could have somewhat more. At this point, the two remainder flavours
are combined into one effective diquark, which is assigned all the remaining
energy and momentum. The final string piece, between this diquark and the 
third quark, can now be considered as described for simple $\q\qbar$ strings 
above.

Among the additional complications are that the diquark formed from the
leftovers may have a larger momentum than energy and thereby nominally may
be spacelike. If only by a little, it normally would not matter, but in
extreme cases the whole final string may come to have a negative squared
mass. Such configurations are rejected on grounds of being unphysical and
the whole fragmentation procedure is restarted from the beginning.

As above, the fragmentation procedure can be formulated in a
Lorentz-frame-independent manner, given the four-vector that describes the
motion of the junction. Therefore, while the fragmentation picture is
simpler to visualize in the rest frame of the junction, one may prefer
to work in the rest frame of the system or in the lab frame, as the case
may be.

Each of the strings considered above normally would not go straight from
the junction to an endpoint quark, rather it would
wind its way via a number of
intermediate gluons, in the neutralino case generated by bremsstrahlung in
the decay. It is straightforward to use the same technology as for
other multiparton systems to extend the description above to such cases.
The one complication is that the motion of the junction may become more
complicated, especially when the emission of reasonably soft gluons is
considered. This can be approximated by a typical mean motion during the
hadronization era \cite{Sjo03}.

The most general string topology foreseen is one with two junctions, i.e.
a $>$\hspace{-1mm}$-$\hspace{-1mm}$<$ topology. Here one junction would
be associated with a baryon number and the other with an antibaryon one.
There would be two quark ends, two antiquark ones, and five string pieces
(including the one between the two junctions) that each could contain an
arbitrary number of intermediate gluons.
 
\subsection{Independent Fragmentation}
 
The independent fragmentation (IF) approach dates back to the early
seventies \cite{Krz72}, and gained widespread popularity with
the Field-Feynman paper \cite{Fie78}. Subsequently, IF was the basis
for two programs widely used in the early PETRA/PEP days, the
Hoyer et al.~\cite{Hoy79} and the Ali et al.~\cite{Ali80} programs.
{\Py} has as (non-default) options a wide selection of independent
fragmentation algorithms.
 
\subsubsection{Fragmentation of a single jet}
 
In the IF approach, it is assumed that the fragmentation of any
system of partons can be described as an incoherent sum of
independent fragmentation procedures for each parton separately.
The process is to be carried out in the overall c.m.\ frame of the
jet system, with each jet fragmentation axis given by the direction
of motion of the corresponding parton in that frame.
 
Exactly as in string fragmentation, an iterative ansatz can be used
to describe the successive production of one hadron after the next.
Assume that a quark is kicked out by some hard interaction, carrying a
well-defined amount of energy and momentum. This quark jet $\q$ is
split into a hadron $\q \qbar_1$ and a remainder-jet $\q_1$, essentially
collinear with each other. New quark and hadron flavours are picked
as already described. The sharing of energy and momentum is given
by some probability distribution $f(z)$, where $z$ is the fraction
taken by the hadron, leaving $1-z$ for the remainder-jet. The
remainder-jet is assumed to be just a scaled-down version of the
original jet, in an average sense. The process of splitting off a
hadron can therefore be iterated, to yield a sequence of hadrons.
In particular, the function $f(z)$ is assumed to be the
same at each step, i.e.\ independent of remaining energy. If $z$ is
interpreted as the fraction of the jet $E+p_{\mrm{L}}$, i.e.\ energy plus
longitudinal momentum with respect to the jet axis, this leads to a
flat central rapidity plateau $dn/dy$ for a large initial energy.
 
Fragmentation functions can be chosen among those listed above for
string fragmentation, but also here the default is the Lund symmetric
fragmentation function.
 
The normal $z$ interpretation means that
a choice of a $z$ value close to $0$ corresponds to a particle
moving backwards, i.e.\ with $p_{\mrm{L}} < 0$.
It makes sense to allow only the production of particles with 
$p_{\mrm{L}} > 0$, but to explicitly constrain $z$ accordingly
would destroy longitudinal invariance. The most
straightforward way out is to allow all $z$ values but discard
hadrons with $p_{\mrm{L}} < 0$. Flavour, transverse momentum and 
$E + p_{\mrm{L}}$
carried by these hadrons are `lost' for the forward jet. The average
energy of the final jet comes out roughly right this way, with a spread
of 1--2 GeV around the mean. The jet longitudinal
momentum is decreased, however, since the jet acquires an effective
mass during the fragmentation procedure. For a 2-jet event this is
as it should be, at least on average, because also the momentum of the
compensating opposite-side parton is decreased.
 
Flavour is conserved locally in each $\q_i \qbar_i$ splitting, 
but not in the jet as a whole. First of all, there is going to be a 
last meson $\q_{n-1}\qbar_n$ generated in the jet, and that will leave 
behind an unpaired quark flavour $\q_n$. Independent fragmentation does 
not specify the fate of this quark. Secondly, also a meson in the
middle of the flavour chain may be selected with such a small $z$
value that it obtains $p_{\mrm{L}} < 0$ and is rejected. Thus a
$\u$ quark jet of charge $2/3$ need not only gives jets of charge 0 
or 1, but also of $-1$ or $+2$, or even higher. Like with the jet 
longitudinal momentum above, one could imagine this compensated by 
other jets in the event. 
 
It is also assumed that transverse momentum is locally conserved, i.e.
the net $\pT$ of the $\q_i \qbar_i$ pair as a whole is assumed to be
vanishing. The $\pT$ of the $\q_i$ is taken to be a Gaussian in the two
transverse degrees of freedom separately, with the transverse momentum
of a hadron obtained by the sum of constituent quark transverse momenta.
The total $\pT$ of a jet can fluctuate for the same two reasons as
discussed above for flavour. Furthermore, in some scenarios one may wish 
to have the same $\pT$ distribution for the first-rank hadron 
$\q \qbar_1$ as for subsequent ones, in which case also the original 
$\q$ should be assigned an unpaired $\pT$ according to a Gaussian. 
 
Within the IF framework, there is no unique recipe for how gluon jet
fragmentation should be handled. One possibility is to treat it exactly
like a quark jet, with the initial quark flavour chosen
at random among $\u$, $\ubar$, $\d$, $\dbar$, $\s$ and
$\sbar$, including the ordinary $\s$ quark suppression factor.
Since the gluon is supposed to fragment more softly
than a quark jet, the fragmentation function may be chosen
independently. Another common option is to split the $\g$ jet into
a pair of parallel $\q$ and $\qbar$ ones, sharing the energy,
e.g.\ as in a perturbative branching $\g \to \q \qbar$, i.e.
$f(z) \propto z^2 + (1-z)^2$.
The fragmentation function could still be chosen
independently, if so desired. Further, in either case the fragmentation
$\pT$ could be chosen to have a different mean.
 
\subsubsection{Fragmentation of a jet system}
 
In a system of many jets, each jet is fragmented independently. Since
each jet by itself does not conserve the flavour, energy and momentum,
as we have seen, then neither does a system of jets. At the end of the 
generation, special algorithms are therefore used to patch this up. The 
choice of approach has major consequences, e.g.\ for event shapes and 
$\alphas$ determinations \cite{Sjo84a}.
 
Little attention is usually given to flavour conservation, and we only
offer one scheme. When the fragmentation of all jets has been performed,
independently of each other, the net initial flavour composition, i.e.
number of $\u$ quarks minus number of $\ubar$ quarks etc., is compared
with the net final flavour composition. In case of an imbalance,
the flavours of the hadron with lowest three-momentum are removed, and
the imbalance is re-evaluated. If the remaining imbalance could be
compensated by a suitable choice of new flavours for this hadron,
flavours are so chosen, a new mass is found and the new energy can be
evaluated, keeping the three-momentum of the original hadron. If
the removal of flavours from the hadron with lowest momentum is not
enough, flavours are removed from the one with next-lowest momentum,
and so on until enough freedom is obtained, whereafter the necessary
flavours are recombined at random to form the new hadrons. Occasionally
one extra $\q_i \qbar_i$ pair must be created, which is then done
according to the customary probabilities.
 
Several different schemes for energy and momentum conservation have
been devised. One \cite{Hoy79} is to conserve transverse momentum
locally within each jet, so that the final momentum vector of a jet 
is always parallel with that of the corresponding parton. Then 
longitudinal momenta
may be rescaled separately for particles within each jet, such that
the ratio of rescaled jet momentum to initial parton momentum is the
same in all jets. Since the initial partons had net vanishing
three-momentum, so do now the hadrons. The rescaling factors may
be chosen such that also energy comes out right. Another common
approach \cite{Ali80} is to boost the event to the frame
where the total hadronic momentum is vanishing. After that, energy
conservation can be obtained by rescaling all particle three-momenta
by a common factor.
 
The number of possible schemes is infinite.
Two further options are available in the program. One is to
shift all particle three-momenta by a common amount to give
net vanishing momentum, and then rescale as before. Another is
to shift all particle three-momenta, for each particle by an
amount proportional to the longitudinal mass with respect to the
imbalance direction, and with overall magnitude selected to give
momentum conservation, and then rescale as before.
In addition, there is a choice of whether to treat separate
colour singlets (like $\q \qbar'$ and $\q' \qbar$ in a
$\q \qbar \q' \qbar'$ event) separately or as one single big system.
 
A serious conceptual weakness of the IF framework is the issue of
Lorentz invariance. The outcome of the fragmentation procedure
depends on the coordinate frame chosen, a problem circumvented by
requiring fragmentation always to be carried out in the c.m.\ frame.
This is a consistent procedure for 2-jet events, but only a
technical trick for multijets.
 
It should be noted, however, that a Lorentz covariant generalization 
of the independent fragmentation model exists, in which separate 
`gluon-type' and `quark-type' strings are used,
the Montvay scheme \cite{Mon79}.
The `quark string' is characterized  by the ordinary string constant
$\kappa$, whereas a `gluon string' is taken to have a string constant
$\kappa_{\g}$. If $\kappa_{\g} > 2 \kappa$ it is always energetically
favourable to split a gluon string into two quark ones, and the
ordinary Lund string model is recovered. Otherwise, for a 3-jet
$\q \qbar \g$ event the three different string pieces are joined at
a junction. The motion of this junction is given by the vector sum of
string tensions acting on it. In particular, it is always possible
to boost an event to a frame where this junction is at rest. In this
frame, much of the standard na\"{\i}ve IF picture holds for the
fragmentation of the three jets; additionally, a correct treatment
would automatically give flavour, momentum and energy conservation.
Unfortunately, the simplicity is lost when studying events with
several gluon jets. In general, each event will contain a number of
different junctions, resulting in a polypod shape with a number of
quark and gluons strings sticking out from a skeleton of gluon
strings. With the shift of emphasis from three-parton to
multi-parton configurations, the simple option existing in {\Je}~6.3 
therefore is no longer included.
 
A second conceptual weakness of IF is the issue of collinear
divergences. In a parton-shower picture, where a quark or gluon is
expected to branch into several reasonably collimated partons,
the independent fragmentation of one single parton or of a bunch of
collinear ones gives quite different outcomes, e.g.\ with a much
larger hadron multiplicity in the latter case. It is conceivable that
a different set of fragmentation functions could be constructed in the
shower case in order to circumvent this problem
(local parton--hadron duality \cite{Dok89} would correspond to having
$f(z) = \delta(z-1)$).
 
\subsection{Other Fragmentation Aspects}
 
Here some further aspects are considered.
 
\subsubsection{Small mass systems}
\label{sss:smallmasssystem}
 
A hadronic event is conventionally subdivided into sets of partons
that form separate colour singlets. These sets are represented by strings,
that e.g.\ stretch from a quark end via a number of intermediate gluons
to an antiquark end. Three string mass regions may be distinguished for 
the hadronization.
\begin{Enumerate}
\item {\em Normal string fragmentation}. In the ideal situation, each 
string has a large invariant mass. Then the standard iterative 
fragmentation scheme above works well. In practice, this approach can be 
used for all strings above some cut-off mass of a few GeV. 
\item {\em Cluster decay}.
If a string is produced with a small invariant mass, maybe only 
two-body final states are kinematically accessible. The traditional 
iterative Lund scheme is then not applicable. We call such a low-mass 
string a cluster, and consider it separately from above. The modelling 
is still intended to give a smooth match on to the standard string 
scheme in the high-cluster-mass limit \cite{Nor98}.
\item {\em Cluster collapse}.
This is the extreme case of the above situation, where the string 
mass is so small that the cluster cannot decay into two hadrons.
It is then assumed to collapse directly into a single hadron, which
inherits the flavour content of the string endpoints. The original 
continuum of string/cluster masses is replaced by a discrete set
of hadron masses. Energy and momentum then cannot be conserved
inside the cluster, but must be exchanged with the rest of the event
\cite{Nor98}.  
\end{Enumerate}

String systems below a threshold mass are handled by the cluster 
machinery. In it, an attempt is first made to produce two hadrons, 
by having the string break in the middle by the production of a new 
$\q\qbar$ pair, with flavours and hadron spins selected according to 
the normal string rules. If the sum of the hadron masses 
is larger than the cluster mass, repeated attempts can be made to find
allowed hadrons; the default is two tries. If an allowed set is found,
the angular distribution of the decay products in the cluster rest
framed is picked isotropically near the threshold, but then gradually
more elongated along the string direction, to provide a smooth match
to the string description at larger masses. This also includes a 
forward--backward asymmetry, so that each hadron is preferentially in 
the same hemisphere as the respective original quark it inherits.

If the attempts to find two hadrons fail, one single hadron is formed 
from the given flavour content. The basic strategy thereafter is to 
exchange some minimal amount of energy and momentum between the
collapsing cluster and other string pieces in the neighbourhood.
The momentum transfer can be in either direction, depending on 
whether the hadron is lighter or heavier than the cluster it comes 
from. When lighter, the excess momentum is split off and put as an
extra `gluon' on the nearest string piece, where `nearest' is 
defined by a space--time history-based distance measure. When the 
hadron is heavier, momentum is instead borrowed from the endpoints 
of the nearest string piece.

The free parameters of the model can be tuned to data, especially
to the significant asymmetries observed between the production of
$\D$ and $\Dbar$ mesons in $\pi^- \p$ collisions, with hadrons
that share some of the $\pi^-$ flavour content very much favoured at 
large $x_F$ in the $\pi^-$ fragmentation region \cite{Ada93}. 
These spectra and asymmetries are closely related to the cluster 
collapse mechanism, and also to other effects of the colour topology
of the event (`beam drag') \cite{Nor98}. The most direct parameters 
are the choice of compensation scheme (\ttt{MSTJ(16)}), the number of  
attempts to find a kinematically valid two-body decay (\ttt{MSTJ(16)})
and the border between cluster and string descriptions
(\ttt{PARJ(32)}). Also many other parameters enter the description, 
however, such as the effective charm mass (\ttt{PMAS(4,1)}), the
quark constituent masses (\ttt{PARF(101) - PARF(105)}), the beam remnant 
structure (\ttt{MSTP(91) - MSTP(94)} and \ttt{PARP(91) - PARP(100)})
and the standard string fragmentation parameters. 

The cluster collapse is supposed to be a part of multiparticle 
production. It is not intended for exclusive production channels,
and may there give quite misleading results. For instance, a 
$\c\cbar$ quark pair produced in a $\gamma\gamma$ collision could well 
be collapsed to a single $\Jpsi$ if the invariant mass is small enough, 
even though the process $\gamma\gamma \to \Jpsi$ in theory is 
forbidden by spin--parity--charge considerations. Furthermore, 
properties such as strong isospin are not considered in the
string fragmentation picture (only its third component, i.e.\ flavour
conservation), neither when one nor when many particles are produced.
For multiparticle states this should matter little, since the
isospin then will be duly randomized, but properly it would forbid the 
production of several one- or two-body states that currently are 
generated.  

\subsubsection{Interconnection Effects}
\label{sss:reconnect}

The widths of the $\W$, $\Z$ and $\t$ are all of the order of 
2 GeV. A standard model Higgs with a mass above 200 GeV, as well 
as many supersymmetric and other beyond the standard model particles
would also have widths in the multi-GeV range. Not far from
threshold, the typical decay times 
$\tau = 1/\Gamma  \approx 0.1 \, {\mrm{fm}} \ll  
\tau_{\mrm{had}} \approx 1 \, \mrm{fm}$.
Thus hadronic decay systems overlap, between a resonance and the
underlying event, or between pairs of resonances, so that the final 
state may not contain independent resonance decays.

So far, studies have mainly been performed in the context of
$\W$ pair production at LEP2. Pragmatically, one may here distinguish 
three main eras for such interconnection:
\begin{Enumerate}
\item Perturbative: this is suppressed for gluon energies 
$\omega > \Gamma$ by propagator/timescale effects; thus only
soft gluons may contribute appreciably.
\item Nonperturbative in the hadroformation process:
normally modelled by a colour rearrangement between the partons 
produced in the two resonance decays and in the subsequent parton
showers.
\item Nonperturbative in the purely hadronic phase: best exemplified 
by Bose--Einstein effects; see next section.
\end{Enumerate}
The above topics are deeply related to the unsolved problems of 
strong interactions: confinement dynamics, $1/N^2_{\mrm{C}}$ 
effects, quantum mechanical interferences, etc. Thus they offer 
an opportunity to study the dynamics of unstable particles,
and new ways to probe confinement dynamics in space and 
time \cite{Gus88a,Sjo94}, {\em but} they also risk 
to limit or even spoil precision measurements.

The reconnection scenarios outlined in \cite{Sjo94a} are now
available, plus also an option along the lines suggested in
\cite{Gus94}. Currently they can only be invoked in process 25,
$\e^+\e^- \to \W^+\W^- \to \q_1\qbar_2\q_3\qbar_4$, which is the most
interesting one for the foreseeable future. (Process 22,
$\e^+\e^- \to \gammaZ \; \gammaZ \to \q_1\qbar_1\q_2\qbar_2$ 
can also be used, but the travel distance is calculated based only 
on the $\Z^0$ propagator part. Thus, the description in scenarios 
I, II and II$'$ below would not be sensible e.g.\ for a light-mass 
$\gamma^*\gamma^*$ pair.) If normally the
event is considered as consisting of two separate colour singlets,
$\q_1\qbar_2$ from the $\W^+$ and $\q_3\qbar_4$ from the $\W^-$,
a colour rearrangement can give two new colour singlets 
$\q_1\qbar_4$ and $\q_3\qbar_2$. It therefore leads to a different
hadronic final state, although differences usually turn out to
be subtle and difficult to isolate \cite{Nor97}. When also
gluon emission is considered, the number of potential reconnection
topologies increases. 

Since hadronization is not understood from first principles, it is
important to remember that we deal with model building rather than 
exact calculations. We will use the standard Lund string fragmentation 
model \cite{And83} as a starting point, but have to extend it 
considerably. The string is here to be viewed as a Lorentz covariant 
representation of a linear confinement field.

The string description is entirely 
probabilistic, i.e.\ any negative-sign interference effects
are absent. This means that the original colour singlets 
$\q_1\qbar_2$ and $\q_3\qbar_4$ may transmute to new singlets
$\q_1\qbar_4$ and $\q_3\qbar_2$, but that any effects e.g.\ of
$\q_1\q_3$ or $\qbar_2\qbar_4$ 
dipoles are absent. In this respect, the nonperturbative discussion is 
more limited in outlook than a corresponding perturbative one. However, 
note that dipoles such as $\q_1\q_3$ do not 
correspond to colour singlets, and can therefore not survive in the 
long-distance limit of the theory, i.e.\ they have to disappear 
in the hadronization phase. 

The imagined time sequence is the following. The $\W^+$ and 
$\W^-$ fly apart from their common production vertex and decay 
at some distance. Around each of these decay vertices, a 
perturbative parton shower evolves from an original $\q\qbar$ 
pair. The typical distance that a virtual parton (of mass
$m \sim 10$~GeV, say, so that it can produce separate jets in the 
hadronic final state) travels before branching is comparable with 
the average $\W^+\W^-$ separation, but shorter than the 
fragmentation time. Each $\W$ can therefore effectively be viewed 
as instantaneously decaying into a string spanned between the 
partons, from a quark end via a number of 
intermediate gluons to the antiquark end. The strings expand, 
both transversely and longitudinally, at a speed limited by that 
of light. They eventually fragment into hadrons and disappear. 
Before that time, however, the string from the $\W^+$ and the one 
from the $\W^-$ may overlap. If so, there is some probability for 
a colour reconnection to occur in the overlap region. The 
fragmentation process is then modified. 
 
The Lund string model does not constrain the nature of the string
fully. At one extreme, the string may be viewed as an elongated bag,
i.e.\ as a flux tube without any pronounced internal structure.
At the other extreme, the string contains a very thin core, a vortex 
line, which carries all the topological information, while the energy is 
distributed over a larger surrounding region. The latter alternative
is the chromoelectric analogue to the magnetic flux lines in a type II 
superconductor, whereas the former one is more akin to the structure
of a type I superconductor. We use them as starting points for 
two contrasting approaches, with nomenclature inspired by the
superconductor analogy.

In scenario~I, the reconnection probability is proportional to the 
space--time volume over which the $\W^+$ and $\W^-$ 
strings overlap, with saturation at unit probability. 
This probability is calculated as 
follows. In the rest frame of a string piece expanding along the 
$\pm z$ direction, the colour field strength is assumed 
to be given by
\begin{equation}
\Omega(\mbox{\bf x},t) = 
\exp \left\{ - (x^2 + y^2)/2r_{\mrm{had}}^2 \right\}
\; \theta(t - |\mbox{\bf x}|) \;
\exp \left\{ - (t^2 - z^2)/\tau_{\mrm{frag}}^2 \right\} ~.
\end{equation}
The first factor gives a Gaussian fall-off in the transverse 
directions, with a string width $r_{\mrm{had}} \approx 0.5$~fm
of typical hadronic dimensions. The time retardation factor 
$\theta(t - |\mbox{\bf x}|)$ ensures that information on the decay of
the $\W$ spreads outwards with the speed of light. The last factor 
gives the probability that the string has not yet fragmented at 
a given proper time along the string axis, with 
$\tau_{\mrm{frag}} \approx 1.5$~fm. For a string piece e.g.\ 
from the $\W^+$ decay, this field strength has to be appropriately
rotated, boosted and displaced to the $\W^+$ decay vertex. 
In addition, since the $\W^+$ string can be made up of many pieces, 
the string field strength $\Omega_{\mrm{max}}^+(\mbox{\bf x},t)$ 
is defined as the maximum of all the contributing $\Omega^+$'s
in the given point. The probability for a reconnection to occur 
is now given by   
\begin{equation}
{\cal P}_{\mrm{recon}} =  1 - \exp \left( - k_{\mrm{I}} 
\int \mrm{d}^3\mbox{\bf x} \, \mrm{d} t \; 
\Omega_{\mrm{max}}^+(\mbox{\bf x},t) \, 
\Omega_{\mrm{max}}^-(\mbox{\bf x},t) \right) ~,
\end{equation} 
where $k_{\mrm{I}}$ is a free parameter. If a reconnection 
occurs, the space--time point for this reconnection is selected
according to the differential probability 
$\Omega_{\mrm{max}}^+(\mbox{\bf x},t) \, 
\Omega_{\mrm{max}}^-(\mbox{\bf x},t)$. 
This defines the string pieces involved 
and the new colour singlets.

In scenario II it is assumed that reconnections can only 
take place when the core regions of two string pieces cross 
each other. This means that the transverse extent of 
strings can be neglected, which leads to considerable 
simplifications compared with the previous scenario.
The position of a string piece at time $t$ is described
by a one-parameter set $\mbox{\bf x}(t,\alpha)$, where 
$0 \leq \alpha \leq 1$ is used to denote the position along the 
string. To find whether two string pieces $i$ and $j$ from the
$\W^+$ and $\W^-$ decays cross, it is sufficient to solve the
equation system $\mbox{\bf x}_i^+(t , \alpha^+) = 
\mbox{\bf x}_j^-(t , \alpha^-)$ and to check that this
(unique) solution is in the physically allowed
domain. Further, it is required
that neither string piece has had time to fragment, which gives
two extra suppression factors of the form
$\exp \{ - \tau^2/\tau_{\mrm{frag}}^2 \}$,
with $\tau$ the proper lifetime of each string piece at the
point of crossing, i.e.\ as in scenario~I. If there are several
string crossings, only the one that occurs first is retained.
The II$'$ scenario is a variant of scenario II, with the 
requirement that a  reconnection is allowed only if
it leads to a reduction of the string length. 

Other models include a simplified implementation of the `GH' model 
\cite{Gus94}, where the reconnection is selected solely 
based on the criterion of a reduced string length. 
The `instantaneous' and `intermediate' scenarios are two toy models. 
In the former (which is equivalent to that in \cite{Gus88a}) the two 
reconnected systems $\q_1\qbar_4$ and $\q_3\qbar_2$ are immediately 
formed and then subsequently shower and fragment  independently of 
each other. In the latter, a  reconnection occurs between the shower 
and fragmentation stages. One has to bear in mind that the last two 
`optimistic' (from the connectometry point of view) toy approaches are 
oversimplified  extremes  and are not supposed to correspond to the
true nature. These scenarios may be useful for reference purposes, 
but are essentially already excluded by data.

While interconnection effects are primarily viewed as hadronization
physics, their implementation is tightly coupled to the event generation 
of a few specific processes, and not to the generic hadronization 
machinery. Therefore the relevant main switch \ttt{MSTP(115)} and 
parameters \ttt{PARP(115) - PARP(120)} are described in subsection
\ref{ss:PYswitchpar}.
 
\subsubsection{Bose--Einstein effects}
 
A crude option for the simulation of Bose--Einstein effects is
included since long, but is turned off by default. In view of its
shortcomings, alternative descriptions have been introduced that
try to overcome at least some of them \cite{Lon95}.

The detailed BE physics is
not that well understood, see e.g.\ \cite{Lor89}. What is
offered is an algorithm, more than just a parameterization (since very
specific assumptions and choices have been made), and yet less than a
true model (since the underlying physics picture is rather fuzzy).
In this scheme, the
fragmentation is allowed to proceed as usual, and so is the decay of
short-lived particles like $\rho$. Then pairs of identical particles,
$\pi^+$ say, are considered one by one. The $Q_{ij}$ value of a pair
$i$ and $j$ is evaluated,
\begin{equation}
 Q_{ij} = \sqrt{ (p_i + p_j)^2 - 4m^2} ~,
\end{equation}
where $m$ is the common particle mass. A shifted (smaller) $Q'_{ij}$
is then to be found such that the (infinite statistics) ratio
$f_2(Q)$ of shifted to unshifted $Q$ distributions is given by the
requested parameterization. The shape may be chosen either
exponential or Gaussian,
\begin{equation}
  f_2(Q) = 1 + \lambda  \exp \left( - (Q/d)^r \right),
  ~~~~r = 1~\mrm{or}~2 ~.
\end{equation}
(In fact, the distribution has to dip slightly below unity at $Q$
values outside the Bose enhancement region, from conservation of
total multiplicity.) If the inclusive distribution of $Q_{ij}$
values is assumed given
just by phase space, at least at small relative momentum then,
with $\d^3 p / E \propto Q^2 \, \d Q / \sqrt{Q^2 + 4m^2}$, 
then $Q'_{ij}$ is found as the solution to the equation
\begin{equation}
\int_0^{Q_{ij}} \frac{Q^2 \, \d Q}{\sqrt{Q^2 + 4 m^2}} =
\int_0^{Q'_{ij}} f_2(Q) \, \frac{Q^2 \, \d Q}{\sqrt{Q^2 + 4 m^2}} ~.
\label{eq:Qshift}
\end{equation}
The change of $Q_{ij}$ can be translated into an effective shift of
the three-momenta of the two particles, if one uses
as extra constraint that the total three-momentum of each pair be
conserved in the c.m.\ frame of the event. Only after all pairwise
momentum shifts have been evaluated, with respect to the original
momenta, are these momenta actually shifted, for each particle by the
(three-momentum) sum of evaluated shifts. The total energy of the 
event is slightly reduced in the process, which is compensated by 
an overall rescaling of all c.m.\ frame momentum vectors. It can be 
discussed which are the particles to involve in this rescaling. 
Currently the only exceptions to using everything are leptons and 
neutrinos coming from resonance decays (such as $\W$'s) and photons 
radiated by leptons (also in initial state radiation). Finally, the 
decay chain is resumed with more long-lived particles like $\pi^0$.
 
Two comments can be made.
The Bose--Einstein effect is here interpreted almost as a
classical force acting on the `final state', rather than
as a quantum mechanical phenomenon on the production amplitude. This
is not a credo, but just an ansatz to make things manageable.
Also, since only pairwise interactions
are considered, the effects associated with three or more nearby
particles tend to get overestimated. (More exact, but also more
time-consuming methods may be found in
\cite{Zaj87}.) Thus the input $\lambda$
may have to be chosen smaller than what one wants to get out.
(On the other hand, many of the pairs of an event contains at least
one particle produced in some secondary vertex, like a $\D$ decay.
This reduces the fraction of pairs which may contribute to the
Bose--Einstein effects, and thus reduces the potential signal.)
This option should therefore be used with caution, and only as a
first approximation to what Bose--Einstein effects can mean.

Probably the largest weakness of the above approach is the issue how
to conserve the total four-momentum. It preserves three-momentum locally, 
but at the expense of not conserving energy. The subsequent rescaling of 
all momenta by a common factor (in the rest frame of the event) to restore
energy conservation is purely {\it ad hoc}. For studies of a single
$\Z^0$ decay, it can plausibly be argued that such a rescaling does
minimal harm. The same need not hold for a pair of resonances.
Indeed, studies \cite{Lon95} show that this global rescaling scheme, 
which we will denote $\BE_0$, introduces an artificial negative shift 
in the reconstructed $\W$ mass, making it difficult (although doable) 
to study the true BE effects in this case. This is one reason to consider 
alternatives.

The global rescaling is also running counter to our philosophy
that BE effects should be local. To be more specific,
we assume that the energy density of the string is a fixed quantity.
To the extent that a pair of particles have their four-momenta slightly 
shifted, the string should act as a `commuting vessel', providing the
difference to other particles produced in the same local region of 
the string. What this means in reality is still not completely  
specified, so further assumptions are necessary. In the
following we discuss four possible algorithms, whereof the last two
are based strictly on the local conservation aspect above, while the 
first two are attempting a slightly different twist to the locality
concept. All are based on
calculating an additional shift $\delta\mbf{r}_k^l$ for some pairs of
particles, where particles $k$ and $l$ need not be identical bosons.
In the end each particle momentum will then be shifted to $\mbf{p}_i' =
\mbf{p}_i + \sum_{j \neq i} \delta \mbf{p}_i^j + \alpha\sum_{k \neq i}
\delta \mbf{r}_i^k$, with the parameter $\alpha$ adjusted separately for 
each event so that the total energy is conserved. 

In the first approach we emulate the criticism of the global
event weight methods with weights always above unity, as being 
intrinsically unstable. It appears more plausible that weights
fluctuate above and below unity. For instance, the simple pair 
symmetrization weight is $1 + \cos (\Delta x \cdot \Delta p)$,
with the $1 + \lambda \exp(-Q^2R^2)$ form only obtained after 
integration over a Gaussian source. Non-Gaussian sources give
oscillatory behaviours

If weights above unity correspond to a shift of pairs towards
smaller relative $Q$ values, the below-unity weights instead give
a shift towards larger $Q$. One therefore is lead to a picture 
where very nearby identical particles are shifted closer, 
those somewhat further are shifted apart, those even further yet
again shifted closer, and so on. Probably the oscillations
dampen out rather quickly, as indicated both by data and by 
the global model studies. We therefore simplify by simulating
only the first peak and dip. Furthermore, to include the desired
damping and to make contact with our normal generation algorithm
(for simplicity), we retain the Gaussian form, but the standard 
$f_2(Q) = 1 + \lambda \exp(-Q^2R^2)$ is multiplied by a further
factor $1 + \alpha \lambda \exp(-Q^2R^2/9)$. The factor $1/9$ in the 
exponential, i.e.\ a factor 3 difference in the $Q$ variable,
is consistent with data and also with what one might expect from 
a dampened $\cos$ form, but should be viewed more as a simple
ansatz than having any deep meaning. 

In the  algorithm, which we denote $\BE_3$, $\delta\mbf{r}_i^j$ is then
non-zero only for pairs of identical bosons, and is calculated in 
the same way as $\delta\mbf{p}_i^j$, with the additional factor $1/9$ in 
the exponential. As explained above, the $\delta\mbf{r}_i^j$ shifts
are then scaled by a common factor $\alpha$ that ensures total energy
conservation. It turns out that the average $\alpha$ needed is 
$\approx -0.2$. The negative sign is exactly what we want to
ensure that $\delta\mbf{r}_i^j$ corresponds to shifting a pair apart, 
while the order of $\alpha$ is consistent with the expected increase 
in the number of affected pairs when a smaller effective radius $R/3$ 
is used. One shortcoming of the method, as implemented here, is
that the input $f_2(0)$ is not quite 2 for $\lambda = 1$ but rather 
$(1 + \lambda) (1 + \alpha \lambda) \approx 1.6$. This could be
solved by starting off with an input $\lambda$ somewhat above unity.

The second algorithm, denoted $\BE_{32}$, is a modification of  
the $\BE_3$ form intended to give $f_2(0) = 1 + \lambda$. The ansatz
is
\begin{equation}
 f_2(Q) = \left\{ 1 + \lambda \exp(-Q^2R^2) \right\}
 \left\{ 1 + \alpha \lambda \exp(-Q^2R^2/9) 
 \left( 1 - \exp(-Q^2R^2/4) \right) \right\} ~,
\end{equation}
applied to identical pairs. The combination
$\exp(-Q^2R^2/9) \left( 1 - \exp(-Q^2R^2/4) \right)$ can be
viewed as a Gaussian, smeared-out representation of the first
dip of the $\cos$ function. As a technical trick, the 
$\delta\mbf{r}_i^j$ are found as in the $\BE_3$ algorithm and 
thereafter scaled down by the $1 - \exp(-Q^2R^2/4)$ factor.
(This procedure does not quite reproduce the formalism of
eq.~(\ref{eq:Qshift}), but comes sufficiently close for our 
purpose, given that the ansatz form in itself is somewhat
arbitrary.)  One should note that, even with the above improvement
relative to the $\BE_3$ scheme, the observable two-particle correlation
is lower at small $Q$ than in the $\BE_0$ algorithm, so some 
further tuning of $\lambda$ could be required. In this scheme, 
$\langle \alpha \rangle \approx -0.25$.

In the other two schemes, the original form of $f_2(Q)$ is retained,
and the energy is instead conserved by picking another pair of particles 
that are shifted apart appropriately. That is, for each pair of identical
particles $i$ and $j$, a pair of non-identical particles, $k$ and $l$,
neither identical to $i$ or $j$, is found in the neighbourhood of $i$
and $j$. For each shift $\delta\mbf{p}_i^j$, a corresponding
$\delta\mbf{r}_k^l$ is found so that the total energy and momentum in the
$i,j,k,l$ system is conserved. However, the actual momentum shift of 
a particle is formed as the vector sum of many contributions, so the 
above pair compensation mechanism is not perfect. The mismatch is 
reflected in a non-unit value $\alpha$ used to rescale the 
$\delta\mbf{r}_k^l$ terms. 

The $k,l$ pair should be the particles `closest' to the pair affected 
by the BE shift, in the spirit of local energy conservation. One option 
would here have been to `look behind the scenes' and use information
on the order of production along the string. However, once decays of
short-lived particles are included, such an approach would still
need arbitrary further rules. We therefore stay with the simplifying 
principle of only using the produced particles.

Looking at $\W^+\W^-$ (or $\Z^0\Z^0$) events and a pair $i,j$ with 
both particles from the same $\W$, it is not obvious whether the pair 
$k,l$ should also be selected only from this $\W$ or if all possible 
pairs should be considered.  Below we have chosen the latter as default 
behaviour, but the former alternative is also studied below.

One obvious measure of closeness is small invariant mass. A first 
choice would then be to pick the combination that minimizes the
invariant mass $m_{ijkl}$ of all four particles. However, such a 
procedure does not reproduce the input $f_2(Q)$ shape very well: both 
the peak height and peak width are significantly reduced, compared with 
what happens in the $\BE_0$ algorithm. The main reason is that either
of $k$ or $l$ may have particles identical to itself in its local
neighbourhood. The momentum compensation shift of $k$ is at random, 
more or less, and therefore tends to smear the BE signal that could
be introduced relative to $k$'s identical partner. Note that, 
if $k$ and its partner are very close in $Q$ to start with, the 
relative change $\delta Q$ required to produce a significant BE effect 
is very small, approximately $\delta Q \propto Q$. The momentum 
compensation shift on $k$ can therefore easily become larger than the
BE shift proper. 

It is therefore necessary to disfavour momentum compensation shifts
that break up close identical pairs. One alternative would have been
to share the momentum conservation shifts suitably inside such pairs.
We have taken a simpler course, by introducing a suppression factor
$1 - \exp(-Q_k^2 R^2)$ for particle $k$, where $Q_k$ is the $Q$ 
value between $k$ and its nearest identical partner. The form is fixed 
such that a $Q_k = 0$ is forbidden and then the rise matches the
shape of the BE distribution itself. Specifically, in the third 
algorithm, $\BE_m$, the pair $k,l$ is chosen so that the measure
\begin{equation}
  W_{ijkl} = \frac{ (1 - \exp(-Q_k^2 R^2))(1 - \exp(-Q_l^2 R^2))}%
{m^2_{ijkl}}
\end{equation}
is maximized. The average $\alpha$ value required to rescale for the 
effect of multiple shifts is 0.73,  i.e.\ somewhat below unity.

The $\BE_{\lambda}$ algorithm is inspired by the so-called $\lambda$ 
measure \cite{And89} (not the be confused with the $\lambda$ parameter of
$f_2(Q)$). It corresponds to a string length in the Lund string 
fragmentation framework. It can be shown that partons in a string are 
colour-connected in a way that tends to minimize this measure. The same is
true for the ordering of the produced hadrons, although with large 
fluctuations. As above, having identical particles nearby to $k,l$ gives
undesirable side effects. Therefore the selection is made so that
\begin{equation}
  W_{ijkl} = \frac{ (1 - \exp(-Q_k^2 R^2))(1 - \exp(-Q_l^2 R^2))}%
{\min_{\mrm{(12~permutations)}} (m_{ij}m_{jk}m_{kl},%
m_{ij}m_{jl}m_{lk},\ldots)}
\end{equation}
is maximized. The denominator is intended to correspond to 
$\exp(\lambda)$. For cases where particles $i$ and $j$ comes from the
same string, this would favour compensating the energy using particles
that are close by and in the same string. 
We find $\langle\alpha\rangle\approx 0.73$, as above. 

The main switches and parameters affecting the Bose--Einstein algorithm 
are \ttt{MSTJ(51) - MSTJ(57)} and \ttt{PARJ(91) - PARJ(96)}.
 
\clearpage
 
\section{Particles and Their Decays}
 
Particles are the building blocks from which events are constructed.
We here use the word `particle' in its broadest sense, i.e.\ including
partons, resonances, hadrons, and so on, subgroups we will describe
in the following. Each particle is characterized by some quantities,
such as charge and mass. In addition, many of the particles are
unstable and subsequently decay. This section contains a survey of
the particle content of the programs, and the particle properties
assumed. In particular, the decay treatment is discussed.  Some
particle and decay properties form part already of the hard
subprocess description, and are therefore described in sections
\ref{s:JETSETproc}, \ref{s:PYTprocgen} and \ref{s:pytproc}.
 
\subsection{The Particle Content}
\label{ss:decpartcont}
 
In order to describe both current and potential future physics,
a number of different particles are needed. A list of some
particles, along with their codes, is given in section \ref{ss:codes}.
Here we therefore emphasize the generality rather than the details.
 
Four full generations of quarks and leptons are included in the
program, although indications from LEP strongly suggest that
only three exist in Nature. The PDG naming convention for the fourth
generation is to repeat the third one but with a prime: $\b'$, $\t'$
$\tau'$ and $\nu'_{\tau}$. Quarks may appear either singly
or in pairs; the latter are called diquarks and are characterized
by their flavour content and their spin. A diquark is always assumed
to be in a colour antitriplet state.
 
The colour neutral hadrons may be build up from the five lighter coloured 
quarks (and diquarks). Six full meson multiplets are included 
and two baryon ones, see section \ref{ss:flavoursel}. In addition, 
$\K_{\mrm{S}}^0$ and $\K_{\mrm{L}}^0$ are considered as separate 
particles coming from the `decay' of $\K^0$ and $\br{\K}^0$ (or, 
occasionally, produced directly).
 
Other particles from the Standard Model include the gluon $\g$, the
photon $\gamma$, the intermediate gauge bosons $\Z^0$ and $\W^{\pm}$,
and the standard Higgs $\hrm^0$. Non-standard particles include
additional gauge bosons, $\Z'^0$ and $\W'^{\pm}$, additional
Higgs bosons $\H^0$, $\A^0$ and $\H^{\pm}$, a leptoquark $\L_{\Q}$
a horizontal gauge boson $\R^0$, technicolor and supersymmetric 
particles, and more.
 
{}From the point of view of usage inside the programs, particles
may be subdivided into three classes, partly overlapping.
\begin{Enumerate}
\item A parton is generically any object which may be found in the
wave function of the incoming beams, and may participate in initial-
or final-state showers. This includes what is normally meant by
partons, i.e.\ quarks and gluons, but here also leptons
and photons. In a few cases other particles may be
classified as partons in this sense.
\item A resonance is an unstable particle produced as part of the
hard process, and where the decay treatment normally is also part
of the hard process. Resonance partial widths are perturbatively
calculable, and therefore it is possible to dynamically recalculate
branching ratios as a function of the mass assigned to a resonance.
Resonances includes particles like the $\Z^0$ and other massive
gauge bosons and Higgs particles, in fact everything with a mass
above the $\b$ quark and additionally also a lighter $\gamma^*$.
\item Hadrons and their decay products, i.e.\ mesons and baryons 
produced either in the fragmentation process, in secondary decays 
or as part of the beam remnant treatment, but not directly as part 
of the hard process (except in a few special cases). Hadrons may be 
stable or unstable. Branching ratios are not assumed perturbatively 
calculable, and can therefore be set freely. Also leptons and photons 
produced in decays belong to this class. In practice, this includes 
everything up to and including $\b$ quarks in mass (except a light 
$\gamma^*$, see above).
\end{Enumerate}
 
Usually the subdivision above is easy to understand and gives you
the control you would expect. Thus the restriction on the allowed
decay modes of a resonance will directly affect the cross section
of a process, while this is not the case for an ordinary hadron,
since in the latter case there is no precise theory knowledge on
the set of decay modes and branching ratios. 
 
\subsection{Masses, Widths and Lifetimes}
 
\subsubsection{Masses}
\label{sss:massdef}
 
Quark masses are not particularly well defined. In the program it is
necessary to make use of three kinds of masses, kinematical, running 
current algebra ones and constituent ones. The first ones are relevant 
for the kinematics in hard processes, e.g.\ in $\g\g \to \c\cbar$,
and are partly fixed by such considerations \cite{Nor98}.
The second define couplings to Higgs particles, and also other
mass-related couplings in models for physics beyond the standard model. 
Both these kinds directly affect cross sections in processes. Constituent 
masses, finally, are used to derive the masses of hadrons, for some
not yet found ones, and e.g.\ to gauge the remainder mass below which
the final two hadrons are to be produced in string fragmentation.
  
The first set of values are the ones stored in the standard mass array 
\ttt{PMAS}. The starting values of the running masses are stored in 
\ttt{PARF(91) - PARF(96)}, with the running calculated in the \ttt{PYMRUN} 
function. Constituent masses are also stored in the \ttt{PARF} array, above
position 101. We maintain this distinction for the five first 
flavours, and partly for top, using the following values by default: \\
\begin{tabular}{cccc@{\protect\rule{0mm}{\tablinsep}}}
quark & kinematical & current algebra mass & constituent mass \\
d & 0.33 GeV & 0.0099 GeV & 0.325 GeV \\
u & 0.33 GeV & 0.0056 GeV & 0.325 GeV \\
s & 0.5 GeV  & 0.199 GeV  & 0.5 GeV \\
c & 1.5 GeV  & 1.35 GeV   & 1.6 GeV \\
b & 4.8 GeV  & 4.5 GeV    & 5.0 GeV \\
t & 175 GeV  & 165 GeV    & ---     \\
\end{tabular} \\
For top no constituent mass is defined, since it does not form hadrons. 
For hypothetical fourth generation quarks only one set of mass values is 
used, namely the one in \ttt{PMAS}. Constituent masses for diquarks are 
defined as the sum of the respective quark masses. The gluon is always 
assumed massless.
 
Particle masses, when known, are taken from ref. \cite{PDG96}.
Hypothesized particles, such as fourth generation fermions and
Higgs bosons, are assigned some not unreasonable set of default
values, in the sense of where you want to search for them in the
not too distant future. Here it is understood that you will go in
and change the default values according to your own opinions at
the beginning of a run.
 
The total number of hadrons in the program is very large, whereof
some are not yet discovered (like charm and bottom baryons). There 
the masses are built up, when needed, from the constituent masses.
For this purpose one uses formulae of the type
\cite{DeR75}
\begin{equation}
m = m_0 + \sum_i m_i + k \, m_{\d}^2 \sum_{i<j} \frac{\langle
\mbox{\boldmath $\sigma$}_i \cdot
\mbox{\boldmath $\sigma$}_j \rangle}{m_i \, m_j} ~,
\end{equation}
i.e.\ one constant term, a sum over constituent masses
and a spin-spin interaction term for each quark pair in the hadron.
The constants $m_0$ and $k$ are fitted from known masses,
treating mesons and baryons separately. For mesons with orbital
angular momentum $L=1$ the spin-spin coupling is assumed vanishing,
and only $m_0$ is fitted.
One may also define `constituent diquarks masses' using the formula
above, with a $k$ value $2/3$ that of baryons. The default values
are: \\
\begin{tabular}{ccc@{\protect\rule{0mm}{\tablinsep}}}
multiplet & $m_0$ & $k$ \\
pseudoscalars and vectors & 0. & 0.16 GeV \\
axial vectors ($S=0$) & 0.50 GeV & 0. \\
scalars & 0.45 GeV & 0. \\
axial vectors ($S=1$) & 0.55 GeV & 0. \\
tensors & 0.60 GeV & 0. \\
baryons  & 0.11 GeV & 0.048 GeV \\
diquarks & 0.077 GeV & 0.048 GeV.\\
\end{tabular} \\
Unlike earlier versions of the program, the actual hadron values are
hardcoded, i.e.\ are unaffected by any change of the charm or bottom
quark masses. 
 
\subsubsection{Widths}
 
A width is calculated perturbatively for those resonances which
appear in the {\Py} hard process generation machinery. The width is
used to select masses in hard processes according to a relativistic
Breit--Wigner shape. In many processes the width is allowed to be
$\hat{s}$-dependent, see section \ref{ss:kinemreson}.
 
Other particle masses, as discussed so far, have been fixed at their
nominal value. We now have to consider the mass broadening for 
short-lived particles
such as $\rho$, $\K^*$ or $\Delta$. Compared to the $\Z^0$, it is
much more difficult to describe the $\rho$ resonance shape, since
nonperturbative and threshold effects act to distort the na\"{\i}ve
shape. Thus the $\rho$ mass is limited from below by its decay
$\rho \to \pi\pi$, but also from above, e.g.\ in the decay
$\phi \to \rho \pi$. Normally thus the allowed mass range is set by 
the most constraining decay chains. Some rare decay modes, specifically 
$\rho^0 \to \eta \gamma$ and $a_2 \to \eta' \pi$, are not allowed to 
have full impact, however. Instead one accepts an imperfect rendering
of the branching ratio, as some low-mass $\rho^0/a_2$ decays of the
above kind are rejected in favour of other decay channels.
In some decay chains, several mass choices are
coupled, like in $\a_2 \to \rho \pi$, where also the $\a_2$ has a
non-negligible width. Finally, there are some extreme cases, like
the $\f_0$, which has a nominal mass below the $\K\K$ threshold, but
a tail extending beyond that threshold, and therefore a
non-negligible branching ratio to the $\K\K$ channel.
 
In view of examples like these, no attempt is made to provide a
full description. Instead a simplified description is used, which
should be enough to give the general smearing of events due to
mass broadening, but maybe not sufficient for detailed studies of
a specific resonance. By default, hadrons are therefore given a
mass distribution according to a non-relativistic Breit--Wigner
\begin{equation}
{\cal P}(m) \, \d m \propto \frac{1}{(m - m_0)^2 + \Gamma^2/4}
\, \d m  ~.
\label{dec:BWlin}
\end{equation}
Leptons and resonances not taken care of by the hard
process machinery are distributed according to a relativistic
Breit--Wigner
\begin{equation}
{\cal P}(m^2) \, \d m^2 \propto \frac{1}{(m^2 - m_0^2)^2 +
m_0^2 \Gamma^2} \, \d m^2 ~.
\label{dec:BWtwo}
\end{equation}
Here $m_0$ and $\Gamma$ are the nominal mass and width of the
particle. The Breit--Wigner shape is truncated symmetrically,
$|m - m_0| < \delta$, with $\delta$ arbitrarily chosen for each
particle so that no problems are encountered in the decay chains.
It is possible to switch off the mass broadening, or to use either
a non-relativistic or a relativistic Breit--Wigners everywhere.
 
The $\f_0$ problem has been `solved' by shifting the $\f_0$ mass to
be slightly above the $\K\K$ threshold and have vanishing width.
Then kinematics in decays $\f_0 \to \K\K$ is reasonably well
modelled. The $\f_0$ mass is too large in the
$\f_0 \to \pi\pi$ channel, but this does not really matter, since
one anyway is far above threshold here.
 
\subsubsection{Lifetimes}
 
Clearly the lifetime and the width of a particle are inversely
related. For practical applications, however, any particle with
a non-negligible width decays too close to its production vertex
for the lifetime to be of any interest. In the program, the two
aspects are therefore considered separately. Particles with a
non-vanishing nominal proper lifetime
$\tau_0 = \langle \tau \rangle$ are
assigned an actual lifetime according to
\begin{equation}
{\cal P}(\tau) \, \d \tau \propto \exp(- \tau / \tau_0 ) \, \d \tau ~,
\end{equation}
i.e.\ a simple exponential decay is assumed. Since the program
uses dimensions where the speed of light $c \equiv 1$, and
space dimensions are in mm, then actually the unit of $c \tau_0$ is
mm and of $\tau_0$ itself mm$/c \approx 3.33\times10^{-12}$ s.
 
If a particle is produced at a vertex $v = (\mbf{x}, t)$ with a
momentum $p = (\mbf{p}, E)$ and a lifetime $\tau$, the
decay vertex position is assumed to be
\begin{equation}
v' = v + \tau \, \frac{p}{m} ~,
\label{dec:newvertex}
\end{equation}
where $m$ is the mass of the particle. With the primary
interaction (normally) in the origin, it is therefore possible to
construct all secondary vertices in parallel with the ordinary
decay treatment.
 
The formula above does not take into account any detector effects,
such as a magnetic field. It is therefore possible to stop the
decay chains at some suitable point, and leave any subsequent
decay treatment to the detector simulation program. One may
select that particles are only allowed to decay if they have a
nominal lifetime $\tau_0$ shorter than some given value or,
alternatively, if their decay vertices $\mbf{x}'$ are inside some
spherical or cylindrical volume around the origin.
 
\subsection{Decays}
\label{ss:partdecays}
 
Several different kinds of decay treatment are used in the program,
depending on the nature of the decay. Not discussed here are the
decays of resonances which are handled as part of the hard process.
 
\subsubsection{Strong and electromagnetic decays}
 
The decays of hadrons containing the `ordinary' $\u$, $\d$ and $\s$
quarks into two or three particles are known, and branching ratios
may be found in \cite{PDG96}. (At least for the lowest-lying states;
the four $L = 1$ meson multiplets are considerably less well known.)
We normally assume that the momentum
distributions are given by phase space. There are a few exceptions,
where the phase space is weighted by a matrix-element expression,
as follows.
 
In $\omega$ and $\phi$ decays to $\pi^+ \pi^- \pi^0$, a matrix element
of the form
\begin{equation}
|{\cal M}|^2 \propto | \mbf{p}_{\pi^+} \times \mbf{p}_{\pi^-} |^2
\label{dec:omegphi}
\end{equation}
is used, with the $\mbf{p}_{\pi}$ the pion momenta in the rest frame
of the decay. (Actually, what is coded is the somewhat more lengthy
Lorentz invariant form of the expression above.)
 
Consider the decay chain $P_0 \to P_1 + V \to P_1 + P_2 + P_3$,
with $P$ representing pseudoscalar mesons and $V$ a vector one. Here
the decay angular distribution of the $V$ in its rest frame is
\begin{equation}
|{\cal M}|^2 \propto \cos^2 \theta_{02} ~,
\label{dec:psvpsps}
\end{equation}
where $\theta_{02}$ is the angle between $P_0$ and $P_2$.
The classical example is $\D \to \K^* \pi \to \K \pi \pi$.
If the $P_1$ is replaced by a $\gamma$, the angular distribution in
the $V$ decay is instead $\propto \sin^2 \theta_{02}$. 
 
In Dalitz decays, $\pi^0$ or $\eta \to \e^+\e^- \gamma$, the mass $m^*$
of the $\ee$ pair is selected according to
\begin{equation}
{\cal P}(m^{*2}) \, \d m^{*2} \propto \frac{\d m^{*2}}{m^{*2}} \,
\left( 1 + \frac{2m_{\e}^2}{m^{*2}} \right) \,
\sqrt{ 1 - \frac{4m_{\e}^2}{m^{*2}} } \,
\left( 1 - \frac{m^{*2}}{m_{\pi,\eta}^2} \right)^3 \,
\frac{1}{ (m_{\rho}^2 - m^{*2})^2 + m_{\rho}^2 \Gamma_{\rho}^2 } ~.
\label{dec:Dalitz}
\end{equation}
The last factor, the VMD-inspired $\rho^0$ propagator, is negligible
for $\pi^0$ decay. Once the $m^*$ has been selected, the angular
distribution of the $\ee$ pair is given by
\begin{equation}
|{\cal M}|^2 \propto (m^{*2} - 2 m_{\e}^2) \left\{
(p_{\gamma} p_{\e^+})^2 + (p_{\gamma} p_{\e^-})^2 \right\} +
4m_{\e}^2 \left\{ (p_{\gamma} p_{\e^+}) (p_{\gamma} p_{\e^-}) +
(p_{\gamma} p_{\e^+})^2 + (p_{\gamma} p_{\e^-})^2 \right\} ~.
\end{equation}
 
Also a number of simple decays involving resonances of heavier
hadrons, e.g.\ $\Sigma_{\c}^0 \to \Lambda_{\c}^+ \pi^-$ or
$\B^{*-} \to \B^- \gamma$ are treated in the same way as the
other two-particle decays.
 
\subsubsection{Weak decays of the $\tau$ lepton}
 
For the $\tau$ lepton, an explicit list of decay channels has been
put together, which includes channels with up to five final-state
particles, some of which may be unstable and subsequently decay to
produce even larger total multiplicities. 
 
The leptonic decays $\tau^- \to \nu_{\tau} \ell^- \br{\nu}_{\ell}$,
where $\ell$ is $\e$ or $\mu$, are distributed according to the
standard $V-A$ matrix element
\begin{equation}
|{\cal M}|^2 = (p_{\tau} p_{\br{\nu}_{\ell}})
(p_{\ell} p_{\nu_{\tau}}) ~.
\end{equation}
(The corresponding matrix element is also used in $\mu$ decays, but
normally the $\mu$ is assumed stable.)
 
In $\tau$ decays to hadrons, the hadrons and the $\nu_{\tau}$ are
distributed according to phase space times the factor
$x_{\nu} \, (3 - x_{\nu})$, where $x_{\nu} = 2E_{\nu}/m_{\tau}$
in the rest frame of the $\tau$. The latter factor is the
$\nu_{\tau}$ spectrum predicted by the parton level $V-A$ matrix
element, and therefore represents an attempt to take into account
that the $\nu_{\tau}$ should take a larger momentum fraction than
given by phase space alone.
 
The probably largest shortcoming of the $\tau$ decay treatment is
that no polarization effects are included, i.e.\ the $\tau$ is
always assumed to decay isotropically. Usually this is not correct,
since a $\tau$ is produced polarized in $\Z^0$ and $\W^{\pm}$ decays.
The \ttt{PYTAUD} routine provides a generic interface to an external
$\tau$ decay library, such as \tsc{Tauola} \cite{Jad91}, where such 
effects could be handled (see also \ttt{MSTJ(28)}). 

\subsubsection{Weak decays of charm hadrons}
 
The charm hadrons have a mass in an intermediate range, where the
effects of the na\"{\i}ve $V-A$ weak decay matrix element is partly but
not fully reflected in the kinematics of final-state particles.
Therefore different decay strategies are combined. We start with
hadronic decays, and subsequently consider semileptonic ones.
 
For the four `main' charm hadrons, $\D^+$, $\D^0$, $\D_{\s}^+$ and
$\Lambda_{\c}^+$, a number of branching ratios are already known.
The known branching ratios have been combined with reasonable
guesses, to construct more or less complete tables of all channels.
For hadronic decays of $\D^0$ and $\D^+$, where rather much is known, 
all channels have an explicitly listed particle content. 
However, only for the two-body decays and some three-body decays is 
resonance production 
properly taken into account. It means that the experimentally measured 
branching ratio for a $\K \pi \pi$ decay channel, say, is represented 
by contributions from a direct $\K \pi \pi$ channel as well as from
indirect ones, such as $\K^* \pi$ and $\K \rho$. For a channel like
$\K \pi \pi \pi \pi$, on the other hand, not all possible combinations
of resonances (many of which would have to be off mass shell to have
kinematics work out) are included. This is more or less in agreement 
with the philosophy adopted in the PDG tables \cite{PDG92}. For 
$\D_{\s}^+$ and $\Lambda_{\c}^+$ 
knowledge is rather incomplete, and only two-body decay channels are
listed. Final states with three or more hadron are only listed in
terms of a flavour content. 
 
The way the program works, it is important to include all the
allowed decay channels up to a given multiplicity. Channels with
multiplicity higher than this may then be generated according to
a simple flavour combination scheme. For instance, in a $\D_{\s}^+$
decay, the normal quark content is $\s\sbar\u\dbar$, where one
$\sbar$ is the spectator quark and the others come from the weak
decay of the $\c$ quark. The spectator quark may also be annihilated,
like in $\D_{\s}^+ \to \u\dbar$. The flavour content to make up
one or two hadrons is therefore present from the onset.
If one decides to generate more hadrons,
this means new flavour-antiflavour pairs have to be generated
and combined with the existing flavours. This is done using the
same flavour approach as in fragmentation, section \ref{ss:flavoursel}.
 
In more detail, the following scheme is used.
\begin{Enumerate}
\item The multiplicity is first selected. The $\D_{\s}^+$ and 
$\Lambda_{\c}^+$ multiplicity is selected according to a distribution 
described further below. The program can also be asked to 
generate decays of a predetermined multiplicity.
\item One of the non-spectator flavours is selected at random.
This flavour is allowed to `fragment' into a hadron plus a new
remaining flavour, using exactly the same flavour generation
algorithm as in the standard jet fragmentation, section
\ref{ss:flavoursel}.
\item Step 2 is iterated until only one or two hadrons remain to
be generated, depending on whether the original number of flavours
is two or four. In each step one `unpaired' flavour is replaced by
another one as a hadron is `peeled off', so the number of unpaired
flavours is preserved.
\item If there are two flavours, these are combined to form the last
hadron. If there are four, then one of the two possible pairings
into two final hadrons is selected at random. To find the hadron
species, the same flavour rules are used as when final flavours are 
combined in the joining of two jets.
\item If the sum of decay product masses is larger than the mass of
the decaying particle, the flavour selection is rejected and the
process is started over at step 1. Normally a new multiplicity is
picked, but for $\D^0$ and $\D^+$ the old multiplicity is retained.
\item Once an acceptable set of hadrons has been found, these are
distributed according to phase space.
\end{Enumerate}
The picture then is one of a number of partons moving apart,
fragmenting almost like jets, but with momenta so low that phase-space 
considerations are enough to give the average behaviour of
the momentum distribution. Like in jet fragmentation, endpoint
flavours are not likely to recombine with each other. Instead
new flavour pairs are created in between them. One should also note
that, while vector and pseudoscalar mesons are produced at their
ordinary relative rates, events with many vectors are likely to
fail in step 5. Effectively, there is therefore a shift towards
lighter particles, especially at large multiplicities.
 
When a multiplicity is to be picked, this is done according to a
Gaussian distribution, centered at $c + n_{\q}/4$ and with a
width $\sqrt{c}$, with the final number rounded off to the nearest
integer. The value for the number of quarks $n_{\q}$ is 2 or 4,
as described above, and
\begin{equation}
c = c_1 \, \ln \left( \frac{m - \sum m_{\q}}{c_2} \right) ~,
\label{dec:multsel}
\end{equation}
where $m$ is the hadron mass and $c_1$ and $c_2$ have been tuned
to give a reasonable description of multiplicities. There is always
some lower limit for the allowed multiplicity; if a number
smaller than this is picked the choice is repeated. Since two-body
decays are explicitly enumerated for $\D_{\s}^+$ and
$\Lambda_{\c}^+$, there the minimum multiplicity  is three.
 
Semileptonic branching ratios are explicitly given in the program
for all the four particles discussed here, i.e.\ it is never
necessary to generate the flavour content using the fragmentation
description. This does not mean that all branching ratios are known;
a fair amount of guesswork is involved for
the channels with higher multiplicities, based
on a knowledge of the inclusive semileptonic branching ratio and
the exclusive branching ratios for low multiplicities.
 
In semileptonic decays it is not appropriate to distribute the
lepton and neutrino momenta according to phase space. Instead the
simple $V-A$
matrix element is used, in the limit that decay product masses may
be neglected and that quark momenta can be replaced by hadron
momenta. Specifically, in the decay $H \to \ell^+ \nu_{\ell} h$,
where $H$ is a charm hadron and $h$ and ordinary hadron, the matrix
element
\begin{equation}
|{\cal M}|^2 = (p_H p_{\ell}) (p_{\nu} p_h)
\end{equation}
is used to distribute the products. It is not clear how to
generalize this formula when several hadrons are present in the final
state. In the program, the same matrix element is used as above,
with $p_h$ replaced by the total four-momentum of all the hadrons.
This tends to favour a low invariant mass for the hadronic system
compared with na\"{\i}ve phase space.

There are a few charm hadrons, such as $\Xi_c$ and $\Omega_c$, which
decay weakly but are so rare that little is known about them. For
these a simplified generic charm decay treatment is used. For
hadronic decays only the quark content is given, and then a
multiplicity and a flavour composition is picked at random, as
already described. Semileptonic decays are assumed to produce only
one hadron, so that $V-A$ matrix element can be simply applied.
 
\subsubsection{Weak decays of bottom hadrons}
 
Some exclusive branching ratios now are known for $\B$
decays. In this version, the $\B^0$, $\B^+$, $\B_{\s}^0$ and  
$\Lambda_{\b}^0$ therefore appear in a similar vein to the one
outlined above for $\D_{\s}^+$ and $\Lambda_{\c}^+$ above. That 
is, all leptonic channels and all hadronic two-body decay channels 
are explicitly listed, while hadronic channels with three or more 
particles are only given in terms of a quark content. The $\B_{\c}$
is exceptional, in that either the bottom or the charm quark may
decay first, and in that annihilation graphs may be non-negligible.
Leptonic and semileptonic channels are here given in full, while 
hadronic channels are only listed in terms of a quark content,
with a relative composition as given in \cite{Lus91}. No separate
branching ratios are set for any of the other weakly decaying
bottom hadrons, but instead a pure `spectator quark' model is assumed, 
where the decay of the $\b$ quark is the same in all hadrons and the 
only difference in final flavour content comes from the spectator quark. 
Compared to the charm decays, the weak decay matrix elements are given
somewhat larger importance in the hadronic decay channels.
 
In semileptonic decays $\b \to \c \ell^- \br{\nu}_{\ell}$ the $\c$
quark is combined with the spectator antiquark or diquark to form
one single hadron. This hadron may be either a pseudoscalar, a vector
or a higher resonance (tensor etc.). The relative fraction of the
higher resonances has been picked to be about 30\%, in order to give
a leptonic spectrum in reasonable experiment with data. (This only 
applies to the main particles  $\B^0$, $\B^+$, $\B_{\s}^0$ and  
$\Lambda_{\b}^0$; for the rest the choice is according to the standard 
composition in the fragmentation.) The overall process is therefore
$H \to h \ell^- \br{\nu}_{\ell}$, where $H$ is a bottom antimeson
or a bottom baryon (remember that $\br{\B}$ is the one that contains
a $\b$ quark), and the matrix element used to distribute momenta is
\begin{equation}
|{\cal M}|^2 = (p_H p_{\nu}) (p_{\ell} p_h)  ~.
\end{equation}
Again decay product masses have been neglected in the matrix element,
but in the branching ratios the $\tau^- \br{\nu}_{\tau}$ channel has
been reduced in rate, compared with $\e^- \br{\nu}_{\e}$ and
$\mu^- \br{\nu}_{\mu}$ ones, according to the expected mass effects.
No CKM-suppressed decays $\b \to \u \ell^- \br{\nu}_{\ell}$ are
currently included.
 
In most multibody hadronic decays, e.g.\ 
$\b \to \c \d \ubar$, the $\c$ quark is
again combined with the spectator flavour to form one single hadron,
and thereafter the hadron and the two quark momenta are distributed
according to the same matrix element as above, with
$\ell^- \leftrightarrow \d$ and
$\br{\nu}_{\ell} \leftrightarrow \ubar$.
The invariant mass of the two quarks is calculated next. If this mass
is so low that two hadrons cannot be formed from the system, the
two quarks are combined into one single hadron. Else the same kind of
approach as in hadronic charm decays is adopted, wherein a
multiplicity  is selected, a number of hadrons are formed and
thereafter momenta are distributed according to phase space. The
difference is that here the charm decay product is distributed according
to the $V-A$ matrix element, and only the rest of the system is
assumed isotropic in its rest frame, while in charm decays all hadrons
are distributed isotropically.
 
Note that the $\c$ quark and the spectator are assumed to form
one colour singlet and the $\d \ubar$ another, separate one.
It is thus assumed that the original colour assignments of the
basic hard process are better retained than in charm decays.
However, sometimes this will not be true, and with about 20\%
probability the colour assignment is flipped around so that
$\c \ubar$ forms one singlet. (In the program, this is achieved by
changing the order in which decay products are given.) In particular,
the decay $\b \to \c \s \cbar$ is allowed to give a $\c\cbar$
colour-singlet state part of the time, and this state may collapse
to a single $\Jpsi$. Two-body decays of this type are explicitly 
listed for $\B^0$, $\B^+$, $\B_{\s}^0$ and $\Lambda_{\b}^0$;
while other $\Jpsi$ production channels appear from the flavour
content specification.
 
The $\B^0$--$\br{\B}^0$ and $\B_{\s}^0$--$\br{\B}_{\s}^0$ systems
mix before decay. This is optionally included. With a probability
\begin{equation}
{\cal P}_{\mrm{flip}} = \sin^2 \left( \frac{x \, \tau}
{ 2\, \langle \tau \rangle} \right)
\end{equation}
a $\B$ is therefore allowed to decay like a $\br{\B}$, and vice versa.
The mixing parameters are by default $x_{\d} = 0.7$ in the
$\B^0$--$\br{\B}^0$ system and $x_{\s} = 10$ in the
$\B_{\s}^0$--$\br{\B}_{\s}^0$ one.

In the past, the generic $\B$ meson and baryon decay properties were 
stored for `particle' 85, now obsolete but not yet removed. This 
particle contains a description of the free $\b$ 
quark decay, with an instruction to find the spectator flavour 
according to the particle code of the actual decaying hadron. 
 
\subsubsection{Other decays}
 
For onia spin 1 resonances, decay channels into a pair of leptons
are explicitly given. Hadronic decays of the $\Jpsi$ are simulated
using the flavour generation model introduced for charm. For
$\Upsilon$ a fraction of the hadronic decays is into $\q\qbar$
pairs, while the rest is into $\g\g\g$ or $\g\g\gamma$, using the
matrix elements of eq.~(\ref{ee:Upsilondec}). The $\eta_c$ and
$\eta_b$ are both allowed to decay into a $\g\g$ pair, which then
subsequently fragments. In $\Upsilon$ and $\eta_b$ decays the partons
are allowed to shower before fragmentation, but energies are too low
for showering to have any impact.

Default branching ratios are given for resonances like the $\Z^0$,
the $\W^{\pm}$, the $\t$ or the $\hrm^0$. When {\Py} is initialized, these
numbers are replaced by branching ratios evaluated from the given
masses. For $\Z^0$ and $\W^{\pm}$ the branching ratios depend only
marginally on the masses assumed, while effects are large e.g.\ for
the $\hrm^0$. In fact, branching ratios may vary over the
Breit--Wigner resonance shape, something which is also taken into
account in the \ttt{PYRESD} description. Therefore the simpler resonance 
treatment of \ttt{PYDECY} is normally not so useful, and should be 
avoided. When it is used, a channel is selected according to the given 
fixed branching ratios. If the decay is into a $\q\qbar$ pair, the 
quarks are allowed to shower and subsequently the parton system is 
fragmented.
 
\clearpage
 
\section{The Fragmentation and Decay Program Elements}

In this section we collect information on most of the routines and 
common block variables found in the fragmentation and decay descriptions
of {\Py}, plus a number of related low-level tasks.
 
\subsection{Definition of Initial Configuration or Variables}
 
With the use of the conventions described for the event record, it
is possible to specify any initial jet/particle configuration. This
task is simplified for a number of often occurring situations by
the existence of the filling routines below. It should be noted that 
many users do not come in direct contact with these routines, since 
that is taken care of by higher-level routines for specific 
processes, particularly \ttt{PYEVNT} and \ttt{PYEEVT}.

Several calls to the routines can be combined in the specification. 
In case one call is enough, the complete
fragmentation/decay chain may be simulated at the same time. At each
call, the value of \ttt{N} is updated to the last line used for
information in the call, so if several calls are used, they should be
made with increasing \ttt{IP} number, or else \ttt{N} should be
redefined by hand afterwards.  

The routine \ttt{PYJOIN} is very useful
to define the colour flow in more complicated parton configurations;
thereby one can bypass the not so trivial rules for how to set the
\ttt{K(I,4)} and \ttt{K(I,5)} colour-flow information. 

The routine \ttt{PYGIVE} contains a facility to set
various common block variables in a controlled and documented fashion.
 
\drawbox{CALL PY1ENT(IP,KF,PE,THE,PHI)}\label{p:PY1ENT}
\begin{entry}
\itemc{Purpose:} to add one entry to the event record, i.e.\ either
a parton or a particle.
\iteme{IP :} normally line number for the parton/particle. There are two
exceptions. \\
If \ttt{IP=0}, line number 1 is used and \ttt{PYEXEC} is called. \\
If \ttt{IP<0}, line \ttt{-IP} is used, with status code 
\ttt{K(-IP,2)=2} rather than 1; thus a parton system may be built up by 
filling all but the last parton of the system with \ttt{IP<0}.
\iteme{KF :} parton/particle flavour code.
\iteme{PE :} parton/particle energy. If \ttt{PE} is smaller than the mass,
the parton/particle is taken to be at rest.
\iteme{THE, PHI :} polar and azimuthal angle for the momentum vector
of the parton/particle.
\end{entry}
 
\drawbox{CALL PY2ENT(IP,KF1,KF2,PECM)}\label{p:PY2ENT}
\begin{entry}
\itemc{Purpose:} to add two entries to the event record, i.e.\ either a
2-parton system or two separate particles.
\iteme{IP :} normally line number for the first parton/particle, with
the second in line \ttt{IP+1}. There are two exceptions. \\
If \ttt{IP=0}, lines 1 and 2 are used and \ttt{PYEXEC} is called. \\
If \ttt{IP<0}, lines \ttt{-IP} and \ttt{-IP+1} are used, with status
code \ttt{K(I,1)=3}, i.e.\ with special colour connection information,
so that a parton shower can be generated by a \ttt{PYSHOW} call,
followed by a \ttt{PYEXEC} call, if so desired (only relevant for partons).
\iteme{KF1, KF2 :} flavour codes for the two partons/particles.
\iteme{PECM :} ($=E_{\mrm{cm}}$) the total energy of the system.
\itemc{Remark:} the system is given in the c.m.\ frame, with the
first parton/particle going out in the $+z$ direction.
\end{entry}
 
\drawbox{CALL PY3ENT(IP,KF1,KF2,KF3,PECM,X1,X3)}\label{p:PY3ENT}
\begin{entry}
\itemc{Purpose:} to add three entries to the event record, i.e.\ either
a 3-parton system or three separate particles.
\iteme{IP :} normally line number for the first parton/particle, with
the other two in \ttt{IP+1} and \ttt{IP+2}. There are two exceptions. \\
If \ttt{IP=0}, lines 1, 2 and 3 are used and \ttt{PYEXEC} is
called. \\
If \ttt{IP<0}, lines \ttt{-IP} through \ttt{-IP+2} are used, with
status code \ttt{K(I,1)=3}, i.e.\ with special colour connection
information, so that a parton shower can be generated by a \ttt{PYSHOW}
call, followed by a \ttt{PYEXEC} call, if so desired (only relevant
for partons).
\iteme{KF1, KF2, KF3:} flavour codes for the three partons/particles.
\iteme{PECM :} ($E_{\mrm{cm}}$) the total energy of the system.
\iteme{X1, X3 :} $x_i = 2E_i/E_{\mrm{cm}}$, i.e.\ twice the energy 
fraction taken by the $i$'th parton. Thus $x_2 = 2 - x_1 - x_3$, and 
need not be given. Note that not all combinations of $x_i$ are 
inside the physically allowed region.
\itemc{Remarks :} the system is given in the c.m.\ frame, in the
$xz$-plane, with the first parton going out in the $+z$ direction and the
third one having $p_x > 0$. A system must be given in the order of colour 
flow, so $\q\g\qbar$ and $\qbar\g\q$ are allowed but $\q\qbar\g$ not.
Thus $x1$ and $x_3$ come to correspond to what is normally called
$x_1$ and $x_2$, i.e.\ the scaled $\q$ and $\qbar$ energies. 
\end{entry}
 
\drawbox{CALL PY4ENT(IP,KF1,KF2,KF3,KF4,PECM,X1,X2,X4,X12,X14)}%
\label{p:PY4ENT}
\begin{entry}
\itemc{Purpose:} to add four entries to the event record, i.e.\ either a
4-parton system or four separate particles (or, for $\q\qbar\q'\qbar'$
events, two 2-parton systems).
\iteme{IP :} normally line number for the first parton/particle, with
the other three in lines \ttt{IP+1}, \ttt{IP+2} and \ttt{IP+3}.
There are two exceptions. \\
If \ttt{IP=0}, lines 1, 2, 3 and 4 are used and \ttt{PYEXEC} is
called. \\
If \ttt{IP<0}, lines \ttt{-IP} through \ttt{-IP+3} are used, with
status code \ttt{K(I,1)=3}, i.e.\ with special colour connection
information, so that a parton shower can be generated by a
\ttt{PYSHOW} call, followed by a \ttt{PYEXEC} call, if so desired
(only relevant for partons).
\iteme{KF1,KF2,KF3,KF4 :} flavour codes for the four partons/particles.
\iteme{PECM :} ($=E_{\mrm{cm}}$) the total energy of the system.
\iteme{X1,X2,X4 :} $x_i = 2E_i/E_{\mrm{cm}}$, i.e.\ twice the energy
fraction taken by the $i$'th parton. Thus $x_3 = 2 - x_1 - x_2 - x_4$,
and need not be given.
\iteme{X12,X14 :} $x_{ij} = 2 p_i p_j/E_{\mrm{cm}}^2$, i.e.\ twice the
four-vector product of the momenta for partons $i$ and $j$, properly
normalized. With the masses known, other $x_{ij}$ may be constructed
from the $x_i$ and $x_{ij}$ given. Note that not all combinations of
$x_i$ and $x_{ij}$ are inside the physically allowed region.
\itemc{Remarks:} the system is given in the c.m.\ frame, with the first
parton going out in the $+z$ direction and the fourth parton lying in the
$xz$-plane with $p_x > 0$. The second parton will have $p_y > 0$ and
$p_y < 0$ with equal probability, with the third parton balancing this
$p_y$ (this corresponds to a random choice between the two possible
stereoisomers). A system must be given in the order of colour flow,
e.g.\ $\q\g\g\qbar$ and $\q\qbar' \q'\qbar$.
\end{entry}
 
\drawbox{CALL PYJOIN(NJOIN,IJOIN)}\label{p:PYJOIN}
\begin{entry}
\itemc{Purpose:} to connect a number of previously defined partons
into a string configuration. Initially the partons must be given with
status codes \ttt{K(I,1)=} 1, 2 or 3. Afterwards the partons
all have status code 3, i.e.\ are given with full colour-flow
information. Compared to the normal way of defining a parton
system, the partons need therefore not appear in the same sequence
in the event record as they are assumed to do along the string. It
is also possible to call \ttt{PYSHOW} for all or some of the
entries making up the string formed by \ttt{PYJOIN}.
\iteme{NJOIN:} the number of entries that are to be joined by one
string.
\iteme{IJOIN:} an one-dimensional array, of size at least \ttt{NJOIN}.
The \ttt{NJOIN} first numbers are the positions of the partons that
are to be joined, given in the order the partons are assumed to appear
along the string. If the system consists entirely of gluons,
the string is closed by connecting back the last to the first
entry.
\itemc{Remarks:} only one string (i.e.\ one colour singlet) may be
defined per call, but one is at liberty to use any number of
\ttt{PYJOIN} calls for a given event. The program will check that the
parton configuration specified makes sense, and not take any action
unless it does. Note, however, that an initially sensible parton
configuration may become nonsensical, if only some of the partons
are reconnected, while the others are left unchanged.
\end{entry}
 
\drawbox{CALL PYGIVE(CHIN)}\label{p:PYGIVE}
\begin{entry}
\itemc{Purpose:} to set the value of any variable residing in the
commmonblocks \ttt{PYJETS}, \ttt{PYDAT1}, \ttt{PYDAT2}, \ttt{PYDAT3},
\ttt{PYDAT4}, \ttt{PYDATR}, \ttt{PYSUBS}, \ttt{PYPARS}, \ttt{PYINT1},
\ttt{PYINT2}, \ttt{PYINT3}, \ttt{PYINT4}, \ttt{PYINT5}, \ttt{PYINT6},
\ttt{PYINT7}, \ttt{PYINT8}, \ttt{PYMSSM}, \ttt{PYMSRV}, or \ttt{PYTCSM}. 
This is done in a more controlled fashion than by directly including 
the common blocks in your program, in that array bounds are checked 
and the old and new values for the variable changed are written to 
the output for reference. An example how \ttt{PYGIVE} can be used to 
parse input from a file is given in subsection \ref{ss:PYTstarted}. 
\iteme{CHIN :} character expression of length at most 100 characters,
with requests for variables to be changed, stored in the form \\
\ttt{variable1=value1;variable2=value2;variable3=value3}\ldots~. \\
Note that an arbitrary number of instructions can be stored in
one call if separated by semicolons, and that blanks may be included
anyplace. An exclamation mark is recognized as the beginning of a comment, 
which is not to be processed. Normal parsing is resumed at the next 
semicolon (if any remain). An example would be\\
\ttt{CALL PYGIVE('MSEL=16!Higgs production;PMAS(25,1)=115.!h0 mass')}\\
The variable$_i$ may be any single variable in the {\Py}
common blocks, and the value$_i$ must be of the correct integer, real
or character (without extra quotes) type. Array indices and values
must be given explicitly, i.e.\ cannot be variables in their own
right. The exception is that the first index can be preceded by
a \ttt{C}, signifying that the index should be translated from normal
{\KF} to compressed {\KC} code with a \ttt{PYCOMP} call; this is allowed for
the \ttt{KCHG}, \ttt{PMAS}, \ttt{MDCY}, \ttt{CHAF} and \ttt{MWID} arrays.\\ 
If a value$_i$ is omitted, i.e.\ with the construction \ttt{variable=}, 
the current value is written to the output, but the variable itself is 
not changed.\\ 
The writing of info can be switched off by \ttt{MSTU(13)=0}.
\itemc{Remark :} The checks on array bounds are hardwired into this
routine. Therefore, if you change array dimensions and
\ttt{MSTU(3)}, \ttt{MSTU(6)} and/or \ttt{MSTU(7)}, as allowed by
other considerations, these changes will not be known to \ttt{PYGIVE}.
Normally this should not be a problem, however.
\end{entry}
 
\subsection{The Physics Routines}
\label{ss:JETphysrout}

Once the initial parton/particle configuration has been specified and
default parameter values changed, if so desired, only a \ttt{PYEXEC}
call is necessary to simulate the whole fragmentation and decay chain.
Therefore a normal user will not directly see any of the other routines
in this section. Some of them could be called directly, but the danger
of faulty usage is then non-negligible.

The \ttt{PYTAUD} routine provides an optional interface to an external 
$\tau$ decay library, where polarization effects could be included. 
It is up to you to write the appropriate calls, as explained at 
the end of this section. 
 
\drawbox{CALL PYEXEC}\label{p:PYEXEC}
\begin{entry}
\itemc{Purpose:} to administrate the fragmentation and decay chain.
\ttt{PYEXEC} may be called several times, but only entries which have
not yet been treated (more precisely, which have 
$1 \leq$\ttt{K(I,1)}$\leq 10$)
can be affected by further calls. This may apply if more partons/particles
have been added by you, or if particles previously considered
stable are now allowed to decay. The actions that will be taken during
a \ttt{PYEXEC} call can be tailored extensively via the
\ttt{PYDAT1}--\ttt{PYDAT3} common blocks, in particular by setting the
\ttt{MSTJ} values suitably.
\end{entry}
 
\boxsep
\begin{entry}
\iteme{SUBROUTINE PYPREP(IP) :}\label{p:PYPREP}
to rearrange parton shower end products (marked with \ttt{K(I,1)=3})
sequentially along strings; also to (optionally) allow small parton
systems to collapse into two particles or one only, in the latter
case with energy and momentum to be shuffled elsewhere in the event;
also to perform checks that e.g.\ flavours of colour-singlet systems
make sense.
 
\iteme{SUBROUTINE PYSTRF(IP) :}\label{p:PYSTRF}
to generate the fragmentation of an arbitrary colour-singlet
parton system according to the Lund string fragmentation model. One of 
the absolutely central routines of {\Py}.
 
\iteme{SUBROUTINE PYINDF(IP) :}\label{p:PYINDF}
to handle the fragmentation of a parton system according to
independent fragmentation models, and implement energy, momentum
and flavour conservation, if so desired. Also the fragmentation of
a single parton, not belonging to a parton system, is considered here
(this is of course physical nonsense, but may sometimes be
convenient for specific tasks).
 
\iteme{SUBROUTINE PYDECY(IP) :}\label{p:PYDECY}
to perform a particle decay, according to known
branching ratios or different kinds of models, depending on our level
of knowledge. Various matrix elements are included for specific
processes.
 
\iteme{SUBROUTINE PYKFDI(KFL1,KFL2,KFL3,KF) :}\label{p:PYKFDI}
to generate a new quark or diquark flavour and to
combine it with an existing flavour to give a hadron.
\begin{subentry}
\iteme{KFL1:} incoming flavour.
\iteme{KFL2:} extra incoming flavour, e.g.\ for formation of final
particle, where the flavours are completely specified. Is normally 0.
\iteme{KFL3:} newly created flavour; is 0 if \ttt{KFL2} is non-zero.
\iteme{KF:} produced hadron. Is 0 if something went wrong (e.g.
inconsistent combination of incoming flavours).
\end{subentry}
 
\iteme{SUBROUTINE PYPTDI(KFL,PX,PY) :}\label{p:PYPTDI}
to give transverse momentum, e.g.\ for a $\q\qbar$ pair created in the
colour field, according to independent Gaussian distributions in
$p_x$ and $p_y$.
 
\iteme{SUBROUTINE PYZDIS(KFL1,KFL3,PR,Z) :}\label{p:PYZDIS}
to generate the longitudinal scaling variable $z$ in jet
fragmentation, either according to the Lund symmetric fragmentation
function, or according to a choice of other shapes.
 
\iteme{SUBROUTINE PYBOEI :}\label{p:PYBOEI}
to include Bose--Einstein effects according to a simple
para\-meteri\-zation. By default, this routine is not called. 
If called from \ttt{PYEXEC}, this is done after the decay of 
short-lived resonances, but before the decay of long-lived ones.
This means the routine should never be called directly by you,
nor would effects be correctly simulated if decays are switched off.
See \ttt{MSTJ(51) - MSTJ(57)} for switches of the routine.
 
\iteme{FUNCTION PYMASS(KF) :}\label{p:PYMASS}
to give the mass for a parton/particle.
 
\iteme{SUBROUTINE PYNAME(KF,CHAU) :}\label{p:PYNAME}
to give the parton/particle name (as a string of type
\ttt{CHARACTER CHAU*16}). The name is read out from the 
\ttt{CHAF} array.
 
\iteme{FUNCTION PYCHGE(KF) :}\label{p:PYCHGE}
to give three times the charge for a parton/particle.
The value is read out from the \ttt{KCHG(KC,1)} array.
 
\iteme{FUNCTION PYCOMP(KF) :}\label{p:PYCOMP}
to give the compressed parton/particle code {\KC} for a given
{\KF} code, as required to find entry into mass and decay data tables.
Also checks whether the given {\KF} code is actually an allowed one
(i.e.\ known by the program), and returns 0 if not. Note that {\KF}
may be positive or negative, while the resulting {\KC} code is never
negative.\\
Internally \ttt{PYCOMP} uses a binary search in a table, with {\KF} codes 
arranged in increasing order, based on the \ttt{KCHG(KC,4)} array. This 
table is constructed the first time \ttt{PYCOMP} is called, at which time 
\ttt{MSTU(20)} is set to 1. In case of a user change of the
\ttt{ KCHG(KC,4)} array one should reset \ttt{MSTU(20)=0} to force a 
re-initialization at the next \ttt{PYCOMP} call (this is automatically 
done in \ttt{PYUPDA} calls). To speed up execution, the latest ({\KF},KC) 
pair is kept in memory and checked before the standard binary search.
 
\iteme{SUBROUTINE PYERRM(MERR,MESSAG) :}\label{p:PYERRM}
to keep track of the number of errors and warnings encountered,
write out information on them, and abort the program in case of too
many errors.
 
\iteme{FUNCTION PYANGL(X,Y) :}\label{p:PYANGL}
to calculate the angle from the $x$ and $y$ coordinates.

\iteme{SUBROUTINE PYLOGO :}\label{p:PYLOGO}
to write a title page for the {\Py} programs. Called by \ttt{PYLIST(0)}.
 
\iteme{SUBROUTINE PYTIME(IDATI) :}\label{p:PYTIME}
to give the date and time, for use in \ttt{PYLOGO} and elsewhere. 
Since Fortran 77 does not contain a standard way of obtaining this 
information, the routine is dummy, to be replaced by you.
Some commented-out examples are given, e.g.\ for Fortran 90 or the
GNU Linux libU77. The 
output is given in an integer array \ttt{ITIME(6)}, with components 
year, month, day, hour, minute and second. If there should be no such 
information available on a system, it is acceptable to put all the 
numbers above to 0.
 
\end{entry}
 
\drawbox{CALL PYTAUD(ITAU,IORIG,KFORIG,NDECAY)}\label{p:PYTAUD}
\begin{entry}
\itemc{Purpose:} to  act as an interface between the standard decay 
routine \ttt{PYDECY} and a user-supplied $\tau$ lepton decay library
such as \tsc{Tauola} \cite{Jad91}. 
The latter library would normally know how to handle polarized $\tau$'s, 
given the $\tau$ helicity as input, so one task of the interface 
routine is to construct the $\tau$ polarization/helicity from the 
information available. Input to the routine (from \ttt{PYDECY}) is 
provided in the first three arguments, while the last argument and 
some event record information have to be set before return. 
To use this facility you have to set the switch \ttt{MSTJ(28)},
include your own interface routine \ttt{PYTAUD} and see to it that 
the dummy routine \ttt{PYTAUD} is not linked. The dummy 
routine is there only to avoid unresolved external references when 
no user-supplied interface is linked.
\iteme{ITAU :} line number in the event record where the $\tau$ is 
stored. The four-momentum of this $\tau$ has first been boosted back 
to the rest frame of the decaying mother and thereafter rotated to move 
out along the $+z$ axis. It would have been possible to also perform
a final boost to the rest frame of the $\tau$ itself, but this has been
avoided so as not to suppress the kinematics aspect of 
close-to-threshold production (e.g.\ in $\B$ decays) vs.\ 
high-energy production (e.g.\ in real $\W$ decays). The choice of frame 
should help the calculation of the helicity configuration. After the 
\ttt{PYTAUD} call the $\tau$ and its decay products will automatically
be rotated and boosted back. However, seemingly, the event record does 
not conserve momentum at this intermediate stage.  
\iteme{IORIG :} line number where the mother particle to the $\tau$ 
is stored. Is 0 if the mother is not stored. This does not have 
to mean the mother is unknown. For instance, in semileptonic $\B$ decays 
the mother is a $\W^{\pm}$ with known four-momentum
$p_{\W} = p_{\tau} + p_{\nu_{\tau}}$, but there is no $\W$ line in the 
event record. When several copies of the mother is stored (e.g.\ one in
the documentation section of the event record and one in the main
section), \ttt{IORIG} points to the last. If a branchings like 
$\tau \to \tau\gamma$ occurs, the `grandmother' is given, i.e.\ the
mother of the direct $\tau$ before branching.
\iteme{KFORIG :} flavour code for the mother particle. Is 0 if the 
mother is unknown. The mother would typically be a resonance such as 
$\gammaZ$ (23), $\W^{\pm}$ ($\pm 24$), $\hrm^0$ (25), or $\H^{\pm}$ 
($\pm 37$). Often the helicity choice would be clear just by the 
knowledge of this mother species, e.g., $\W^{\pm}$ vs.\ $\H^{\pm}$. 
However, sometimes further complications may exist. For instance, 
the {\KF} code 23 represents a mixture of $\gamma^*$ and $\Z^0$; a 
knowledge of the mother mass (in \ttt{P(IORIG,5)}) would here be 
required to make the choice of helicities. Further, a $\W^{\pm}$ or 
$\Z^0$ may either be (predominantly) transverse or longitudinal, 
depending on the production process under study.
\iteme{NDECAY :} the number of decay products of the $\tau$; to be 
given by the user routine. You must also store the {\KF} flavour codes 
of those decay products in the positions \ttt{K(I,2)}, 
\ttt{N+1}$\leq$\ttt{I}$\leq$\ttt{N+NDECAY}, of the event record. 
The corresponding five-momentum (momentum, energy and mass) should be 
stored in the associated \ttt{P(I,J)} positions, 1$\leq$\ttt{J}$\leq$5. 
The four-momenta are expected to add up to the four-momentum of the 
$\tau$ in position \ttt{ITAU}. You should not change the \ttt{N} value 
or any of the other \ttt{K} or \ttt{V} values (neither for the 
$\tau$ nor for its decay products) since this is automatically done 
in \ttt{PYDECY}.  
\end{entry}
  
\subsection{The General Switches and Parameters}
\label{ss:JETswitch}
 
The common block \ttt{PYDAT1} contains the main switches and parameters 
for the fragmentation and decay treatment, but also for some other 
aspects. Here one may control in detail what the program is to do, if 
the default mode of operation is not satisfactory.
 
\drawbox{COMMON/PYDAT1/MSTU(200),PARU(200),MSTJ(200),PARJ(200)}%
\label{p:PYDAT1}
\begin{entry}
\itemc{Purpose:} to give access to a number of status codes and
parameters which regulate the performance of the program as a whole.
Here \ttt{MSTU} and \ttt{PARU} are related to utility functions, as
well as a few parameters of the Standard Model, while \ttt{MSTJ} and
\ttt{PARJ} affect the underlying physics assumptions. Some of the
variables in \ttt{PYDAT1} are described elsewhere, and are therefore 
here only reproduced as references to the relevant sections. This 
in particular applies to many coupling constants, which are found in 
section \ref{ss:coupcons}, and switches of the older dedicated 
$\e^+\e^-$ machinery, section \ref{ss:eeroutines}.
 
\boxsep

\iteme{MSTU(1) - MSTU(3) :}\label{p:MSTU} variables used by the event 
study routines, section \ref{ss:eventstudy}.
 
\iteme{MSTU(4) :} (D=4000) number of lines available in the
common block \ttt{PYJETS}. Should always be changed if the
dimensions of the \ttt{K} and \ttt{P} arrays are changed by you,
but should otherwise never be touched. Maximum allowed value is 10000,
unless \ttt{MSTU(5)} is also changed.
 
\iteme{MSTU(5) :} (D=10000) is used in building up the special
colour-flow information stored in \ttt{K(I,4)} and \ttt{K(I,5)}
for \ttt{K(I,3)=} 3, 13 or 14. The generic form for \ttt{j=} 4 or 5
is \\
\ttt{K(I,j)}$ = 2 \times$\ttt{MSTU(5)}$^2 \times$MCFR$ + 
$\ttt{MSTU(5)}$^2 \times$MCTO$ + $\ttt{MSTU(5)}$\times$ICFR$ + $ICTO, \\
with notation as in section \ref{ss:evrec}.
One should always have \ttt{MSTU(5)}$\geq$\ttt{MSTU(4)}. On a 32 bit
machine, values \ttt{MSTU(5)}$> 20000$ may lead to overflow problems,
and should be avoided.
 
\iteme{MSTU(6) :} (D=500) number of {\KC} codes available in the
\ttt{KCHG}, \ttt{PMAS}, \ttt{MDCY}, and \ttt{CHAF} arrays; should be
changed if these dimensions are changed.
 
\iteme{MSTU(7) :} (D=8000) number of decay channels available in the
\ttt{MDME}, \ttt{BRAT} and \ttt{KFDP} arrays; should be changed if
these dimensions are changed.
 
\iteme{MSTU(10) :} (D=2) use of parton/particle masses in filling
routines (\ttt{PY1ENT}, \ttt{PY2ENT}, \ttt{PY3ENT}, \ttt{PY4ENT}).
\begin{subentry}
\iteme{= 0 :} assume the mass to be zero.
\iteme{= 1 :} keep the mass value stored in \ttt{P(I,5)}, whatever it
is. (This may be used e.g.\ to describe kinematics with
off-mass-shell partons).
\iteme{= 2 :} find masses according to mass tables as usual.
\end{subentry}

\iteme{MSTU(11) - MSTU(12) :} variables used by the event 
study routines, section \ref{ss:eventstudy}.
 
\iteme{MSTU(13) :} (D=1) writing of information on variable values
changed by a \ttt{PYGIVE} call.
\begin{subentry}
\iteme{= 0 :} no information is provided.
\iteme{= 1 :} information is written to standard output.
\end{subentry}

\iteme{MSTU(14) :} variable used by the event study routines, 
section \ref{ss:eventstudy}.

\iteme{MSTU(15) :} (D=0) decides how \ttt{PYLIST} shows empty lines, 
which are interspersed among ordinary particles in the event record.
\begin{subentry}
\iteme{= 0 :} do not print lines with \ttt{K(I,1)}$\leq 0$.
\iteme{= 1 :} do not print lines with \ttt{K(I,1)}$< 0$.
\iteme{= 2 :} print all lines.
\end{subentry}
  
\iteme{MSTU(16) :} (D=1) choice of mother pointers for the particles
produced by a fragmenting parton system.
\begin{subentry}
\iteme{= 1 :} all primary particles of a system point to a line with
{\KF} = 92 or 93, for string or independent fragmentation, respectively,
or to a line with {\KF} = 91 if a parton system has so small a mass
that it is forced to decay into one or two particles. The two
(or more) shower initiators of a showering parton system point
to a line with {\KF} = 94. The entries with {\KF} = 91--94 in their
turn point back to the predecessor partons, so that the
{\KF} = 91--94 entries form a part of the event history proper.
\iteme{= 2 :} although the lines with {\KF} = 91--94 are present, and
contain the correct mother and daughter pointers, they are not part
of the event history proper, in that particles produced in string
fragmentation point directly to either of the two endpoint
partons of the string (depending on the side they were generated
from), particles produced in independent fragmentation point
to the respective parton they were generated from, particles in
small mass systems point to either endpoint parton, and
shower initiators point to the original on-mass-shell
counterparts. Also the daughter pointers bypass the {\KF} = 91--94
entries. In independent fragmentation, a parton need not produce
any particles at all, and then have daughter pointers 0.\\
When junctions are present, the related primary baryon points back to 
the junction around which it is produced. More generally, consider a 
system stored as $\q_1\q_2\J\q_3$. In the listing of primary hadrons 
coming from this system, the first ones will have \ttt{K(I,3)} pointers 
back to $\q_1$, either they come from the hadronization of the $\q_1$ 
or $\q_2$ string pieces. (In principle the hadrons could be classified 
further, but this has not been done.) Then comes the junction baryon, 
which may be followed by hadrons again pointing back to $\q_1$. This 
is because the final string piece, between the junction and $\q_3$, 
is fragmented from both ends with a joining to conserve overall energy 
and momentum. (Typically the two shortest strings are put ahead of the 
junction.) More properly these hadrons also could have pointed back to 
the junction, but then the special status of the junction baryon would 
have been lost. The final hadrons point to $\q_3$, and have fragmented 
off this end of the string piece in to the junction.  The above example 
generalizes easily to topologies with two junctions, e.g. 
$\q_1\q_2\J\J\qbar_3\qbar_4$, where now particles in the central 
$\J\J$ string would point either to $\q_1$ or $\qbar_4$ depending on 
production side.
\itemc{Note :} \ttt{MSTU(16)} should not be changed between the
generation of an event and the translation of this event record with
a \ttt{PYHEPC} call, since this may give an erroneous translation of
the event history.
\end{subentry}
 
\iteme{MSTU(17) :} (D=0) storage option for \ttt{MSTU(90)} and
associated information on $z$ values for heavy-flavour production.
\begin{subentry}
\iteme{= 0 :} \ttt{MSTU(90)} is reset to zero at each \ttt{PYEXEC}
call. This is the appropriate course if \ttt{PYEXEC} is only called
once per event, as is normally the case when you do not
yourself call \ttt{PYEXEC}.
\iteme{= 1 :} you have to reset \ttt{MSTU(90)} to zero yourself
before each new event. This is the appropriate course if several
\ttt{PYEXEC} calls may appear for one event, i.e.\ if you
call \ttt{PYEXEC} directly.
\end{subentry}
 
\iteme{MSTU(19) :} (D=0) advisory warning for unphysical flavour
setups in \ttt{PY2ENT}, \ttt{PY3ENT} or \ttt{PY4ENT} calls.
\begin{subentry}
\iteme{= 0 :} yes.
\iteme{= 1 :} no; \ttt{MSTU(19)} is reset to 0 in such a call.
\end{subentry}
 
\iteme{MSTU(20) :} (D=0) flag for the initialization status of the 
\ttt{PYCOMP} routine. A value 0 indicates that tables should be 
(re)initialized, after which it is set 1. In case you change  
the \ttt{KCHG(KC,4)} array you should reset \ttt{MSTU(20)=0} to force 
a re-initialization at the next \ttt{PYCOMP} call.

\iteme{MSTU(21) :} (D=2) check on possible errors during program
execution. Obviously no guarantee is given that all errors will be
caught, but some of the most trivial user-caused errors may be found.
\begin{subentry}
\iteme{= 0 :} errors do not cause any immediate action, rather the
program will try to cope, which may mean e.g.\ that it
runs into an infinite loop.
\iteme{= 1 :} parton/particle configurations are checked for possible
errors. In case of problem, an exit is made from the
misbehaving subprogram, but the generation of the event is
continued from there on. For the first \ttt{MSTU(22)} errors
a message is printed; after that no messages appear.
\iteme{= 2 :} parton/particle configurations are checked for possible
errors. In case of problem, an exit is made from the
misbehaving subprogram, and subsequently from \ttt{PYEXEC}.
You may then choose to correct the error, and continue the
execution by another \ttt{PYEXEC} call. For the first \ttt{MSTU(22)}
errors a message is printed, after that the last event is
printed and execution is stopped.
\end{subentry}
 
\iteme{MSTU(22) :} (D=10) maximum number of errors that are printed.
 
\iteme{MSTU(23) :} (I) count of number of errors experienced to date.
Is not updated for errors in a string system containing junctions.
(Since errors occasionally do happen there, and are difficult to
eliminate altogether.) 
 
\iteme{MSTU(24) :} (R) type of latest error experienced; reason that
event was not generated in full. Is reset at each \ttt{PYEXEC} call.
\begin{subentry}
\iteme{= 0 :} no error experienced.
\iteme{= 1 :} program has reached end of or is writing outside
\ttt{PYJETS} memory.
\iteme{= 2 :} unknown flavour code or unphysical combination of codes;
may also be caused by erroneous string connection information.
\iteme{= 3 :} energy or mass too small or unphysical kinematical
variable setup.
\iteme{= 4 :} program is caught in an infinite loop.
\iteme{= 5 :} momentum, energy or charge was not conserved (even
allowing for machine precision errors, see \ttt{PARU(11)}); is
evaluated only after event has been generated in full, and does not
apply when independent fragmentation without momentum conservation
was used.
\iteme{= 6 :} error call from outside the fragmentation/decay package
(e.g.\ the $\ee$ routines).
\iteme{= 7 :} inconsistent particle data input in \ttt{PYUPDA}
(\ttt{MUPDA = 2,3}) or other \ttt{PYUPDA}-related problem.
\iteme{= 8 :} problems in more peripheral service routines.
\iteme{= 9 :} various other problems.
\end{subentry}
 
\iteme{MSTU(25) :} (D=1) printing of warning messages.
\begin{subentry}
\iteme{= 0 :} no warnings are written.
\iteme{= 1 :} first \ttt{MSTU(26)} warnings are printed, thereafter
no warnings appear.
\end{subentry}
 
\iteme{MSTU(26) :} (D=10) maximum number of warnings that are printed.
 
\iteme{MSTU(27) :} (I) count of number of warnings experienced to
date.
 
\iteme{MSTU(28) :} (R) type of latest warning given, with codes
parallelling those for \ttt{MSTU(24)}, but of a less serious nature.
 
\iteme{MSTU(29) :} (I) denotes the presence (1) or not (0) of a junction 
in the latest system studied. Used to decide whether to update the 
\ttt{MSTU(23)} counter in case of errors.
 
\iteme{MSTU(30) :} (I) count of number of errors experienced to date,
equivalent to \ttt{MSTU(23)} except that it is also updated for errors 
in a string system containing junctions.
 
\iteme{MSTU(31) :} (I) number of \ttt{PYEXEC} calls in present run.
 
\iteme{MSTU(32) - MSTU(33) :} variables used by the event 
study routines, section \ref{ss:eventstudy}.
 
\iteme{MSTU(41) - MSTU(63) :} switches for event-analysis routines,
see section \ref{ss:evanrout}.
 
\iteme{MSTU(70) - MSTU(80) :} variables used by the event 
study routines, section \ref{ss:eventstudy}.
 
\iteme{MSTU(90) :} number of heavy-flavour hadrons (i.e.\ hadrons
containing charm or bottom) produced in the fragmentation stage of the
current event, for which the positions in the event record are stored 
in \ttt{MSTU(91) - MSTU(98)} and the $z$ values in the fragmentation in
\ttt{PARU(91) - PARU(98)}. At most eight values will be stored
(normally this is no problem). No $z$ values can be stored for those
heavy hadrons produced when a string has so small mass that it
collapses to one or two particles, nor for those produced as one of
the final two particles in the fragmentation of a string. If
\ttt{MSTU(17)=1}, \ttt{MSTU(90)} should be reset to zero by you
before each new event, else this is done automatically.
 
\iteme{MSTU(91) - MSTU(98) :} the first \ttt{MSTU(90)} positions will
be filled with the line numbers of the heavy-flavour hadrons produced
in the current event. See \ttt{MSTU(90)} for additional comments. Note
that the information is corrupted by calls to \ttt{PYEDIT} with options
0--5 and 21--23; calls with options 11--15 work, however.
 
\iteme{MSTU(101) - MSTU(118) :} switches related to couplings, see
section \ref{ss:coupcons}.
 
\iteme{MSTU(121) - MSTU(125) :} internally used in the advanced popcorn 
code, see subsection \ref{sss:improvedbaryoncode}.

\iteme{MSTU(131) - MSTU(140) :} internally used in the advanced popcorn 
code, see subsection \ref{sss:improvedbaryoncode}.
 
\iteme{MSTU(161), MSTU(162) :} information used by event-analysis 
routines, see section \ref{ss:evanrout}.

\boxsep
 
\iteme{PARU(1) :}\label{p:PARU} (R) $\pi \approx 3.141592653589793$.
 
\iteme{PARU(2) :} (R) $2\pi \approx 6.283185307179586$.
 
\iteme{PARU(3) :} (D=0.197327) conversion factor for GeV$^{-1} \to$ fm
or fm$^{-1} \to$ GeV.
 
\iteme{PARU(4) :} (D=5.06773) conversion factor for fm $\to$ GeV$^{-1}$
or GeV $\to$ fm$^{-1}$.
 
\iteme{PARU(5) :} (D=0.389380) conversion factor for GeV$^{-2} \to$ mb
or mb$^{-1} \to$ GeV$^2$.
 
\iteme{PARU(6) :} (D=2.56819) conversion factor for mb $\to$ GeV$^{-2}$
or GeV$^2 \to$ mb$^{-1}$.
 
\iteme{PARU(11) :} (D=0.001) relative error, i.e.\ non-conservation of
momentum and energy divided by total energy, that may be attributable
to machine precision problems before a physics error is suspected
(see \ttt{MSTU(24)=5}).
 
\iteme{PARU(12) :} (D=0.09 GeV$^2$) effective cut-off in squared mass,
below which partons may be recombined to simplify (machine precision
limited) kinematics of string fragmentation. (Default chosen to be of 
the order of a light quark mass, or half a typical light meson mass.)
 
\iteme{PARU(13) :} (D=0.01) effective angular cut-off in radians for
recombination of partons, used in conjunction with \ttt{PARU(12)}.
 
\iteme{PARU(21) :} (I) contains the total energy $W$ of all first
generation partons/particles after a \ttt{PYEXEC} call; to be used by the
\ttt{PYP} function for \ttt{I>0}, \ttt{J=} 20--25.

\iteme{PARU(41) - PARU(63) :} parameters for event-analysis routines,
see section \ref{ss:evanrout}.
 
\iteme{PARU(91) - PARU(98) :} the first \ttt{MSTU(90)} positions will
be filled with the fragmentation $z$ values used internally in the
generation of heavy-flavour hadrons --- how these are translated into
the actual energies and momenta of the observed hadrons is a
complicated function of the string configuration. The particle with
$z$ value stored in \ttt{PARU(i)} is to be found in line \ttt{MSTU(i)}
of the event record. See \ttt{MSTU(90)} and \ttt{MSTU(91) - MSTU(98)}
for additional comments.
 
\iteme{PARU(101) - PARU(195) :} various coupling constants and 
parameters related to couplings, see section \ref{ss:coupcons}.
 
\boxsep
 
\iteme{MSTJ(1) :}\label{p:MSTJ} (D=1) choice of fragmentation scheme.
\begin{subentry}
\iteme{= 0 :} no jet fragmentation at all.
\iteme{= 1 :} string fragmentation according to the Lund model.
\iteme{= 2 :} independent fragmentation, according to specification
in \ttt{MSTJ(2)} and \ttt{MSTJ(3)}.
\end{subentry}
 
\iteme{MSTJ(2) :} (D=3) gluon jet fragmentation scheme in independent
fragmentation.
\begin{subentry}
\iteme{= 1 :} a gluon is assumed to fragment like a random $\d$, $\u$
or $\s$ quark or antiquark.
\iteme{= 2 :} as \ttt{=1}, but longitudinal (see \ttt{PARJ(43)},
\ttt{PARJ(44)} and \ttt{PARJ(59)}) and transverse (see \ttt{PARJ(22)})
momentum properties of quark or antiquark substituting for gluon may
be separately specified.
\iteme{= 3 :} a gluon is assumed to fragment like a pair of a
$\d$, $\u$ or $\s$ quark and its antiquark, sharing the gluon energy
according to the Altarelli-Parisi splitting function.
\iteme{= 4 :} as \ttt{=3}, but longitudinal (see \ttt{PARJ(43)},
\ttt{PARJ(44)} and \ttt{PARJ(59)}) and transverse (see \ttt{PARJ(22)})
momentum properties of quark and antiquark substituting for gluon may
be separately specified.
\end{subentry}
 
\iteme{MSTJ(3) :} (D=0) energy, momentum and flavour conservation
options in independent fragmentation. Whenever momentum conservation
is described below, energy and flavour conservation is also
implicitly assumed.
\begin{subentry}
\iteme{= 0 :} no explicit conservation of any kind.
\iteme{= 1 :} particles share momentum imbalance compensation according
to their energy (roughly equivalent to boosting event to c.m.\ 
frame). This is similar to the approach in the Ali et al.
program \cite{Ali80}.
\iteme{= 2 :} particles share momentum imbalance compensation according
to their longitudinal mass with respect to the imbalance
direction.
\iteme{= 3 :} particles share momentum imbalance compensation equally.
\iteme{= 4 :} transverse momenta are compensated separately within
each jet, longitudinal momenta are rescaled so that ratio
of final jet to initial parton momentum is the same for
all the jets of the event. This is similar to the approach in
the Hoyer et al. program \cite{Hoy79}.
\iteme{= 5 :} only flavour is explicitly conserved.
\iteme{= 6 - 10 :} as \ttt{=1 - 5}, except that above several colour
singlet systems that followed immediately after each other
in the event listing (e.g.\ $\q\qbar\q\qbar$) were treated as
one single system, whereas here they are treated as
separate systems.
\iteme{= -1 :} independent fragmentation, where also particles moving
backwards with respect to the jet direction are kept, and
thus the amount of energy and momentum mismatch may be large.
\end{subentry}
 
\iteme{MSTJ(11) :} (D=4) choice of longitudinal fragmentation
function, i.e.\ how large a fraction of the energy available a
newly-created hadron takes.
\begin{subentry}
\iteme{= 1 :} the Lund symmetric fragmentation function, see
\ttt{PARJ(41) - PARJ(45)}.
\iteme{= 2 :} choice of some different forms for each flavour
separately, see \ttt{PARJ(51) - PARJ(59)}.
\iteme{= 3 :} hybrid scheme, where light flavours are treated with
symmetric Lund (\ttt{=1}), but charm and heavier can be separately
chosen, e.g.\ according to the Peterson/SLAC function (\ttt{=2}).
\iteme{= 4 :} the Lund symmetric fragmentation function (\ttt{=1}),
for heavy endpoint quarks modified according to the Bowler
(Artru--Mennessier, Morris) space--time picture of string evolution,
see \ttt{PARJ(46)}.
\iteme{= 5 :} as \ttt{=4}, but with possibility to interpolate
between Bowler and Lund separately for $\c$ and $\b$; see
\ttt{PARJ(46)} and \ttt{PARJ(47)}.
\end{subentry}
 
\iteme{MSTJ(12) :} (D=2) choice of baryon production model.
\begin{subentry}
\iteme{= 0 :} no baryon-antibaryon pair production at all; initial
diquark treated as a unit.
\iteme{= 1 :} diquark-antidiquark pair production allowed; diquark
treated as a unit.
\iteme{= 2 :} diquark-antidiquark pair production allowed, with
possibility for diquark to be split according to the `popcorn'
scheme.
\iteme{= 3 :} as \ttt{=2}, but additionally the production of first
rank baryons may be suppressed by a factor \ttt{PARJ(19)}.
\iteme{= 4 :} as \ttt{=2}, but diquark vertices suffer an extra 
 suppression of the form $1-\exp(\rho\Gamma)$, where 
 $\rho\approx0.7\mrm{GeV}^{-2}$ is stored in \ttt{PARF(192)}.
\iteme{= 5 :} Advanced version of the popcorn model. Independent of 
\ttt{PARJ(3-7)}. Instead depending on \ttt{PARJ(8-10)}. When using 
 this option \ttt{PARJ(1)} needs to enhanced by approx. a factor 2 
(i.e.\ it losses a bit of its normal meaning), 
and \ttt{PARJ(18)} is suggested to be set to 0.19. See section 
\ref{sss:improvedbaryoncode} for further details.
\end{subentry}
 
\iteme{MSTJ(13) :} (D=0) generation of transverse momentum for
endpoint quark(s) of single quark jet or $\q\qbar$ jet system (in
multijet events no endpoint transverse momentum is ever allowed for).
\begin{subentry}
\iteme{= 0 :} no transverse momentum for endpoint quarks.
\iteme{= 1 :} endpoint quarks obtain transverse momenta like ordinary
$\q\qbar$ pairs produced in the field (see \ttt{PARJ(21)}); for
2-jet systems the endpoints obtain balancing transverse momenta.
\end{subentry}
 
\iteme{MSTJ(14) :} (D=1) treatment of a colour-singlet parton system
with a low invariant mass.
\begin{subentry}
\iteme{= 0 :} no precautions are taken, meaning that problems may
occur in \ttt{PYSTRF} (or \ttt{PYINDF}) later on. Warning messages are 
issued when low masses are encountered, however, or when the flavour or 
colour configuration appears to be unphysical.
\iteme{= 1 :} small parton systems are allowed to collapse into two
particles or, failing that, one single particle. Normally
all small systems are treated this way, starting with the
smallest one, but some systems would require more work and
are left untreated; they include diquark-antidiquark pairs
below the two-particle threshold. See further \ttt{MSTJ(16)} and
\ttt{MSTJ(17)}.
\iteme{= -1 :} special option for \ttt{PYPREP} calls, where no
precautions are taken (as for \ttt{=0}), but, in addition, no checks
are made on the presence of small-mass systems or unphysical flavour or 
colour configurations; i.e.\ \ttt{PYPREP} only rearranges colour strings.
\end{subentry}
 
\iteme{MSTJ(15) :} (D=0) production probability for new flavours.
\begin{subentry}
\iteme{= 0 :} according to standard Lund parameterization, as given
by \ttt{PARJ(1) - PARJ(20)}.
\iteme{= 1 :} according to probabilities stored in
\ttt{PARF(201) - PARF(1960)}; note that no default values exist here,
i.e.\ \ttt{PARF} must be set by you. The \ttt{MSTJ(12)} switch
can still be used to set baryon production mode, with the
modification that \ttt{MSTJ(12)=2} here allows an arbitrary number
of mesons to be produced between a baryon and an antibaryon (since
the probability for diquark $\to$ meson $+$ new diquark is assumed
independent of prehistory).
\end{subentry}

\iteme{MSTJ(16) :} (D=2) mode of cluster treatment (where a cluster is
a low-mass string that can fragment to two particles at the most).
\begin{subentry}
\iteme{= 0 :} old scheme. Cluster decays (to two hadrons) are isotropic.
In cluster collapses (to one hadron),  energy-momentum compensation is 
to/from the parton or hadron furthest away  in mass.
\iteme{= 1 :} intermediate scheme. Cluster decays are anisotropic in a 
way that is intended to mimic the Gaussian $\pT$ suppression and 
string `area law' of suppressed rapidity orderings of ordinary
string fragmentation. In cluster collapses, energy-momentum 
compensation is to/from the string piece most closely moving 
in the same direction as the cluster. Excess energy is put
as an extra gluon on this string piece, while a deficit
is taken from both endpoints of this string piece as a common
fraction of their original momentum.
\iteme{= 2 :} new default scheme. Essentially as \ttt{=1} above, except 
that an energy deficit is preferentially taken from the endpoint of 
the string piece that is moving closest in direction to the cluster.    
\end{subentry}

\iteme{MSTJ(17) :} (D=2) number of attempts made to find two hadrons 
that have a combined mass below the cluster mass, and thus allow a 
cluster to decay to two hadrons rather than collapse to one.
Thus the larger \ttt{MSTJ(17)}, the smaller the fraction of collapses.
At least one attempt is always made, and this was the old default
behaviour.   

\iteme{MSTJ(18) :} (D=10) maximum number of times the junction rest 
frame is evaluated with improved knowledge of the proper energies.
When the boost in an iteration corresponds to a $\gamma < 1.01$ the
iteration would be stopped sooner. An iterative solution is required 
since the rest frame of the junction is defined by a vector sum of 
energies (see \ttt{PARP(48)}) that are assumed already known in this 
rest frame. (In practice, this iteration is normally a minor effect, 
of more conceptual than practical impact.)

\iteme{MSTJ(19) :} (D=0) in a string system containing two junctions
(or, more properly, a junction and an antijunction), there is a 
possibility for these two to disappear, by an `annihilation' that 
gives two separate strings \cite{Sjo03}. That is, a configuration 
like $\q_1 \q_2 \J \J  \qbar_3 \qbar_4$ can collapse to 
$\q_1 \qbar_3$ plus $\q_2 \qbar_4$.
\begin{subentry}
\iteme{= 0 :} the selection between the two alternatives is made 
dynamically, so as to pick the string configuration with the 
smallest length.   
\iteme{= 1 :} the two-junction topology always remains.
\iteme{= 2 :} the two-junction topology always collapses to two 
separate strings.
\itemc{Note:} the above also applies, suitably generalized, when 
parton-shower activity is included in the event. If the shower 
in between the two junctions comes to contain a $\g \to \q\qbar$
branching, however, the system inevitable is split into two 
separate junction systems $\q_1 \q_2 \J \q$ plus 
$\qbar \J  \qbar_3 \qbar_4$ 
\end{subentry}
 
\iteme{MSTJ(21) :} (D=2) form of particle decays.
\begin{subentry}
\iteme{= 0 :} all particle decays are inhibited.
\iteme{= 1 :} a particle declared unstable in the \ttt{MDCY} vector,
and with decay channels defined, may decay within the region given
by \ttt{MSTJ(22)}. A particle may decay into partons, which then fragment
further according to the \ttt{MSTJ(1)} value.
\iteme{= 2 :} as \ttt{=1}, except that a $\q\qbar$ parton system produced 
in a decay (e.g.\ of a $\B$ meson) is always allowed to fragment 
according to string fragmentation, rather than according to the 
\ttt{MSTJ(1)} value (this means that momentum, energy and charge 
are conserved in the decay).
\end{subentry}
 
\iteme{MSTJ(22) :} (D=1) cut-off on decay length for a particle that is
allowed to decay according to \ttt{MSTJ(21)} and the \ttt{MDCY} value.
\begin{subentry}
\iteme{= 1 :} a particle declared unstable is also forced to decay.
\iteme{= 2 :} a particle is decayed only if its average proper
lifetime is smaller than \ttt{PARJ(71)}.
\iteme{= 3 :} a particle is decayed only if the decay vertex is
within a distance \ttt{PARJ(72)} of the origin.
\iteme{= 4 :} a particle is decayed only if the decay vertex is
within a cylindrical volume with radius \ttt{PARJ(73)} in the
$xy$-plane and extent to $\pm$\ttt{PARJ(74)} in the $z$ direction.
\end{subentry}
 
\iteme{MSTJ(23) :} (D=1) possibility of having a shower evolving from
a $\q\qbar$ pair created as decay products. This switch only applies
to decays handled by \ttt{PYDECY} rather than \ttt{PYRESD}, and so
is of less relevance today.
\begin{subentry}
\iteme{= 0 :} never.
\iteme{= 1 :} whenever the decay channel matrix-element code is
\ttt{MDME(IDC,2)=} 4, 32, 33, 44 or 46, the two first decay products
(if they are partons) are allowed to shower, like a colour-singlet
subsystem, with maximum virtuality given by the invariant mass
of the pair.
\end{subentry}
 
\iteme{MSTJ(24) :} (D=2) particle masses.
\begin{subentry}
\iteme{= 0 :} discrete mass values are used.
\iteme{= 1 :} particles registered as having a mass width in the
\ttt{PMAS} vector are given a mass according to a truncated
Breit--Wigner shape, linear in $m$, eq.~(\ref{dec:BWlin}).
\iteme{= 2 :} as \ttt{=1}, but gauge bosons (actually all particles
with $|${\KF}$| \leq 100$) are distributed according to a Breit--Wigner
quadratic in $m$, as obtained from propagators.
\iteme{= 3 :} as \ttt{=1}, but Breit--Wigner shape is always
quadratic in $m$, eq.~(\ref{dec:BWtwo}).
 
\end{subentry}

\iteme{MSTJ(26) :} (D=2) inclusion of $\B$--$\br{\B}$ mixing in
decays.
\begin{subentry}
\iteme{= 0 :} no.
\iteme{= 1 :} yes, with mixing parameters given by \ttt{PARJ(76)}
and \ttt{PARJ(77)}. Mixing decays are not specially marked.
\iteme{= 2 :} yes, as \ttt{=1}, but a $\B$ ($\br{\B}$) that decays
as a $\br{\B}$ ($\B$) is marked as \ttt{K(I,1)=12} rather than the
normal \ttt{K(I,1)=11}.
\end{subentry}

\iteme{MSTJ(28) :} (D=0) call to an external $\tau$ decay library like
\tsc{Tauola}.
For this option to be meaningful, it is up to you to write the 
appropriate interface and include that in the routine \ttt{PYTAUD},
as explained in section \ref{ss:JETphysrout}. 
\begin{subentry}
\iteme{= 0 :} not done, i.e.\ the internal \ttt{PYDECY} treatment is 
used.
\iteme{= 1 :} done whenever the $\tau$ mother particle species can 
be identified, else the internal \ttt{PYDECY} treatment is used. 
Normally the mother particle should always be identified, but it is 
possible for you to remove event history information or to add 
extra $\tau$'s directly to the event record, and then the mother is 
not known.
\iteme{= 2 :} always done. 
\end{subentry}
 
\iteme{MSTJ(38) - MSTJ(50) :} switches for time-like parton showers,
see section \ref{ss:showrout}.
 
\iteme{MSTJ(51) :} (D=0) inclusion of Bose--Einstein effects.
\begin{subentry}
\iteme{= 0 :} no effects included.
\iteme{= 1 :} effects included according to an exponential
parameterization
$f_2(Q) = 1 + $\ttt{PARJ(92)}$\times \exp(-Q/$\ttt{PARJ(93)}$)$,
where $f_2(Q)$ represents the ratio of particle production at
$Q$ with Bose--Einstein effects to that without, and the relative
momentum $Q$ is defined by
$Q^2(p_1,p_2) = -(p_1 - p_2)^2 = (p_1 + p_2)^2 - 4m^2$. Particles
with width broader than \ttt{PARJ(91)} are assumed to have time to
decay before Bose--Einstein effects are to be considered.
\iteme{= 2 :} effects included according to a Gaussian
parameterization
$f_2(Q) = 1 + $\ttt{PARJ(92)}$\times \exp(- (Q/$\ttt{PARJ(93)}$)^2 )$,
with notation and comments as above.
\end{subentry}
 
\iteme{MSTJ(52) :} (D=3) number of particle species for which
Bose--Einstein correlations are to be included, ranged along the
chain $\pi^+$, $\pi^-$, $\pi^0$, $\K^+$, $\K^-$, $\K^0_{\mrm{L}}$,
$\K^0_{\mrm{S}}$, $\eta$ and $\eta'$. Default corresponds to
including all pions ($\pi^+$, $\pi^-$, $\pi^0$), 7 to including all
Kaons as well, and 9 is maximum.

\iteme{MSTJ(53) :} (D=0) In $\e^+\e^- \to \W^+\W^-$, 
$\e^+\e^- \to \Z^0\Z^0$, or if \ttt{PARJ(94)}$> 0$ and there are
several strings in the event, apply BE algorithm 
\begin{subentry}
\iteme{= 0 :} on all pion pairs. 
\iteme{= 1 :} only on pairs were both pions come from the same 
$\W/\Z/$string. 
\iteme{= 2 :} only on pairs were the pions come from different 
$\W/\Z/$strings.
\iteme{= -2 :} when calculating balancing shifts for pions from same
$\W/\Z/$string, only consider pairs from this $\W/\Z/$string. 
\itemc{Note:} if colour reconnections has occurred in an event, the
distinction between pions coming from different $\W/\Z$'s is lost.
\end{subentry}
 
\iteme{MSTJ(54) :} (D=2) Alternative local energy compensation. 
(Notation in brackets refer to the one used in \cite{Lon95}.)
\begin{subentry}
\iteme{= 0 :} global energy compensation ($\BE_0$). 
\iteme{= 1 :} compensate with identical pairs by negative BE 
enhancement with a third of the radius ($\BE_3$).
\iteme{= 2 :} ditto, but with the compensation constrained to vanish
at $Q=0$, by an additional $1-\exp(-Q^2 R^2/4)$ factor ($\BE_{32}$). 
\iteme{= -1 :} compensate with pair giving the smallest invariant mass
($BE_m$). 
\iteme{= -2 :} compensate with pair giving the smallest string length
($BE_{\lambda}$). 
 \end{subentry}

\iteme{MSTJ(55) :} (D=0) Calculation of difference vector. 
\begin{subentry}
\iteme{= 0 :} in the lab frame. 
\iteme{= 1 :} in the c.m.\ of the given pair. 
\end{subentry}
 
\iteme{MSTJ(56) :} (D=0) In $\e^+\e^- \to \W^+\W^-$ or 
$\e^+\e^- \to \Z^0\Z^0$ include distance between $\W/\Z$'s. 
\begin{subentry}
\iteme{= 0 :} radius is the same for all pairs. 
\iteme{= 1 :} radius for pairs from different $\W/\Z$'s is 
$R+\delta R_{\W\W}$ ($R+\delta R_{\Z\Z}$), where $\delta R$ is the
generated distance between the decay vertices. (When considering $\W$ or
$\Z$ pairs with an energy well above threshold, this should give more 
realistic results.)
\end{subentry}
 
\iteme{MSTJ(57) :} (D=1) Penalty for shifting particles with close-by 
identical neighbours in local energy compensation, \ttt{MSTJ(54) < 0}. 
\begin{subentry}
\iteme{= 0 :} no penalty. 
\iteme{= 1 :} penalty. 
\end{subentry}
 
\iteme{MSTJ(91) :} (I) flag when generating gluon jet with options
\ttt{MSTJ(2)=} 2 or 4 (then \ttt{=1}, else \ttt{=0}).
 
\iteme{MSTJ(92) :} (I) flag that a $\q\qbar$ or $\g\g$ pair or a
$\g\g\g$ triplet created in \ttt{PYDECY} should be allowed to shower,
is 0 if no pair or triplet, is the entry number of the first parton
if a pair indeed exists, is the entry number of the first parton,
with a $-$ sign, if a triplet indeed exists.
 
\iteme{MSTJ(93) :} (I) switch for \ttt{PYMASS} action. Is reset to 0
in \ttt{PYMASS} call.
\begin{subentry}
\iteme{= 0 :} ordinary action.
\iteme{= 1 :} light ($\d$, $\u$, $\s$, $\c$, $\b$) quark masses are
taken from \ttt{PARF(101) - PARF(105)} rather than
\ttt{PMAS(1,1) - PMAS(5,1)}. Diquark masses are
given as sum of quark masses, without spin splitting term.
\iteme{= 2 :} as \ttt{=1}. Additionally the constant terms
\ttt{PARF(121)} and \ttt{PARF(122)} are subtracted from quark and
diquark masses, respectively.
\end{subentry}

\iteme{MSTJ(101) - MSTJ(121) :} switches for $\ee$ event generation,
see section \ref{ss:eeroutines}.
 
\boxsep
 
\iteme{PARJ(1) :}\label{p:PARJ} (D=0.10) is 
${\cal P}(\q\q)/{\cal P}(\q)$, the
suppression of diquark-antidiquark pair production in the colour
field, compared with quark--antiquark production.
 
\iteme{PARJ(2) :} (D=0.30) is ${\cal P}(\s)/{\cal P}(u)$, the
suppression of $\s$ quark pair production in the field compared with
$\u$ or $\d$ pair production.
 
\iteme{PARJ(3) :} (D=0.4) is
$({\cal P}(\u\s)/{\cal P}(\u\d))/({\cal P}(\s)/{\cal P}(\d))$,
the extra suppression of strange diquark production  compared with
the normal suppression of strange quarks.
 
\iteme{PARJ(4) :} (D=0.05) is $(1/3){\cal P}(\u\d_1)/{\cal P}(\u\d_0)$,
the suppression of spin 1 diquarks compared with spin 0 ones
(excluding the factor 3 coming from spin counting).
 
\iteme{PARJ(5) :} (D=0.5) parameter determining relative occurrence of
baryon production by $BM\br{B}$ and by $B\br{B}$ configurations in the
simple popcorn baryon production model, roughly
${\cal P}(BM\br{B})/({\cal P}(B\br{B})+{\cal P}(BM\br{B})) =$
\ttt{PARJ(5)}$/(0.5+$\ttt{PARJ(5)}$)$. This and subsequent baryon
parameters are modified in the advanced popcorn scenario, see
subsection \ref{sss:improvedbaryoncode}.
 
\iteme{PARJ(6) :} (D=0.5) extra suppression for having a $\s\sbar$
pair shared by the $B$ and $\br{B}$ of a $BM\br{B}$ situation.
 
\iteme{PARJ(7) :} (D=0.5) extra suppression for having a strange
meson $M$ in a $BM\br{B}$ configuration.

\iteme{PARJ(8) - PARJ(10) :} used in the advanced popcorn scenario, see
subsection \ref{sss:improvedbaryoncode}.
 
\iteme{PARJ(11) - PARJ(17) :} parameters that determine the spin of
mesons when formed in fragmentation or decays.
\begin{subentry}
\iteme{PARJ(11) :} (D=0.5) is the probability that a light meson
(containing $\u$ and $\d$ quarks only) has spin 1 (with
\ttt{1-PARJ(11)} the probability for spin 0).
\iteme{PARJ(12) :} (D=0.6) is the probability that a strange meson
has spin 1.
\iteme{PARJ(13) :} (D=0.75) is the probability that a charm or
heavier meson has spin 1.
\iteme{PARJ(14) :} (D=0.) is the probability that a spin = 0
meson is produced with an orbital angular momentum 1, for a
total spin = 1.
\iteme{PARJ(15) :} (D=0.) is the probability that a spin = 1
meson is produced with an orbital angular momentum 1, for a
total spin = 0.
\iteme{PARJ(16) :} (D=0.) is the probability that a spin = 1
meson is produced with an orbital angular momentum 1, for a
total spin = 1.
\iteme{PARJ(17) :} (D=0.) is the probability that a spin = 1
meson is produced with an orbital angular momentum 1, for a
total spin = 2.
\itemc{Note :} the end result of the numbers above is
that, with \ttt{i} = 11, 12 or 13, depending on flavour content, \\
${\cal P}(S = 0, L = 0, J = 0) = (1 - \mtt{PARJ(i)}) \times 
(1 - \mtt{PARJ(14)})$, \\
${\cal P}(S = 0, L = 1, J = 1) = (1 - \mtt{PARJ(i)}) \times 
\mtt{PARJ(14)}$, \\
${\cal P}(S = 1, L = 0, J = 1) = $ \\
\mbox{~}\hfill$\mtt{PARJ(i)} \times (1 - \mtt{PARJ(15)} - 
\mtt{PARJ(16)} - \mtt{PARJ(17)})$, \\
${\cal P}(S = 1, L = 1, J = 0) = \mtt{PARJ(i)} \times 
\mtt{PARJ(15)}$, \\
${\cal P}(S = 1, L = 1, J = 1) = \mtt{PARJ(i)} \times 
\mtt{PARJ(16)}$, \\
${\cal P}(S = 1, L = 1, J = 2) = \mtt{PARJ(i)} \times 
\mtt{PARJ(17)}$, \\
where $S$ is the quark `true' spin and $J$ is the total spin, usually
called the spin $s$ of the meson.
\end{subentry}
 
\iteme{PARJ(18) :} (D=1.) is an extra suppression factor multiplying
the ordinary {\bf SU(6)} weight for spin $3$/2 baryons, and hence a
means to break {\bf SU(6)} in addition to the dynamic breaking implied
by \ttt{PARJ(2)}, \ttt{PARJ(3)}, \ttt{PARJ(4)}, \ttt{PARJ(6)} and
\ttt{PARJ(7)}.
 
\iteme{PARJ(19) :} (D=1.) extra baryon suppression factor, which
multiplies the ordinary diquark-antidiquark production probability
for the breakup closest to the endpoint of a string, but leaves other
breaks unaffected. Is only used for \ttt{MSTJ(12)=3}.
 
\iteme{PARJ(21) :} (D=0.36 GeV) corresponds to the width $\sigma$ in
the Gaussian $p_x$ and $p_y$ transverse momentum distributions for
primary hadrons. See also \ttt{PARJ(22) - PARJ(24)}.
 
\iteme{PARJ(22) :} (D=1.) relative increase in transverse momentum in
a gluon jet generated with \ttt{MSTJ(2)=} 2 or 4.

\iteme{PARJ(23), PARJ(24) :} (D=0.01, 2.) a fraction \ttt{PARJ(23)}
of the Gaussian transverse momentum distribution is taken to be a
factor \ttt{PARJ(24)} larger than input in \ttt{PARJ(21)}. This
gives a simple parameterization of non-Gaussian tails to the Gaussian
shape assumed above. 

\iteme{PARJ(25) :} (D=1.) extra suppression factor for $\eta$ 
production in fragmentation; if an $\eta$ is rejected a new flavour 
pair is generated and a new hadron formed.

\iteme{PARJ(26) :} (D=0.4) extra suppression factor for $\eta'$ 
production in fragmentation; if an $\eta'$ is rejected a new flavour 
pair is generated and a new hadron formed.
 
\iteme{PARJ(31) :} (D=0.1 GeV) gives the remaining $W_+$ below which
the generation of a single jet is stopped. (It is chosen smaller than a
pion mass, so that no hadrons moving in the forward direction are
missed.)
 
\iteme{PARJ(32) :} (D=1. GeV) is, with quark masses added, used to
define the minimum allowable energy of a colour-singlet parton system.
 
\iteme{PARJ(33) - PARJ(34) :} (D=0.8 GeV, 1.5 GeV) are, together with
quark masses, used to define the remaining energy below which the
fragmentation of a parton system is stopped and two final hadrons
formed. \ttt{PARJ(33)} is normally used, except for \ttt{MSTJ(11)=2},
when \ttt{PARJ(34)} is used.
 
\iteme{PARJ(36) :} (D=2.) represents the dependence on the mass of the
final quark pair for defining the stopping point of the fragmentation.
Is strongly correlated to the choice of \ttt{PARJ(33) - PARJ(35)}.
 
\iteme{PARJ(37) :} (D=0.2) relative width of the smearing of the
stopping point energy.
 
\iteme{PARJ(39) :} (D=0.08 GeV$^{-2}$) refers to the probability
for reverse rapidity ordering of the final two hadrons,
according to eq.~(\ref{fr:revord}), for \ttt{MSTJ(11)=2}
(for other \ttt{MSTJ(11)} values \ttt{PARJ(42)} is used).

\iteme{PARJ(40):} (D=1.) possibility to modify the probability for 
reverse rapidity ordering of the final two hadrons in the fragmentation 
of a string with \ttt{PYSTRF}, or from the only two hadrons of a 
low-mass string considered in \ttt{PYPREP}. Modifies 
eq.~(\ref{fr:revord}) to
${\cal P}_{\mrm{reverse}} = 1/(1 + \exp(\mtt{PARJ(40)}b\Delta))$. 
 
\iteme{PARJ(41), PARJ(42) :} (D=0.3, 0.58 GeV$^{-2}$) give the $a$ and
$b$ parameters of the symmetric Lund fragmentation function for
\ttt{MSTJ(11)=}1, 4 and 5 (and \ttt{MSTJ(11)=3} for ordinary hadrons).
 
\iteme{PARJ(43), PARJ(44) :} (D=0.5, 0.9 GeV$^{-2}$) give the $a$ and
$b$ parameters as above for the special case of a gluon jet generated
with IF and \ttt{MSTJ(2)=} 2 or 4.
 
\iteme{PARJ(45) :} (D=0.5) the amount by which the effective $a$
parameter in the Lund flavour dependent symmetric fragmentation
function is assumed to be larger than the normal $a$ when diquarks
are produced. More specifically, referring to eq.~(\ref{fr:LSFFlong}),
$a_{\alpha} = $\ttt{PARJ(41)} when considering the
fragmentation of a quark and = \ttt{PARJ(41) + PARJ(45)} for
the fragmentation of a diquark, with corresponding expression for
$a_{\beta}$ depending on whether the newly created object is a quark
or diquark (for an independent gluon jet generated with
\ttt{MSTJ(2)=} 2 or 4, replace \ttt{PARJ(41)} by \ttt{PARJ(43)}).
In the popcorn model, a meson created in between the baryon and
antibaryon has $a_{\alpha} = a_{\beta} = $\ttt{PARJ(41) + PARJ(45)}.
 
\iteme{PARJ(46), PARJ(47) :} (D=2*1.) modification of the Lund
symmetric fragmentation for heavy endpoint quarks according to the
recipe by Bowler, available when
\ttt{MSTJ(11)=} 4 or 5 is selected. The shape is given
by eq.~(\ref{fr:LSFFBowler}). If \ttt{MSTJ(11)=4} then
$r_{\Q} = $\ttt{PARJ(46)} for both $\c$ and $\b$, while if
\ttt{MSTJ(11)=5} then $r_{\c} = $\ttt{PARJ(46)} and
$r_{\b} = $\ttt{PARJ(47)}. \ttt{PARJ(46)} and \ttt{PARJ(47)} thus 
provide a possibility to interpolate between the `pure' Bowler 
shape, $r = 1$, and the normal Lund one, $r = 0$. The additional
modifications made in \ttt{PARJ(43) - PARJ(45)}
are automatically taken into account, if necessary.
 
\iteme{PARJ(48) :} (D=1.5 GeV) in defining the junction rest frame, 
the effective pull direction of a chain of partons is defined
by the vector sum of their momenta, multiplied by a factor
$\exp (- \sum E / \mtt{PARJ(48)})$, where the energy sum runs over 
all partons on the string between (but excluding) the given one 
and the junction itself. The energies should be defined in the 
junction rest frame, which requires an iterative approximation,
see \ttt{MSTJ(18)}. 
 
\iteme{PARJ(49) :} (D=1. GeV) retry (up to 10 times) when both strings,
to be joined in a junction to form a new string endpoint, have a 
remaining energy above \ttt{PARJ(49)} (evaluated in the junction 
rest frame) after having been fragmented.
 
\iteme{PARJ(50) :} (D=10. GeV) retry as above when either of the strings
have a remaining energy above a random energy evenly distributed between 
\ttt{PARJ(49)} and \ttt{PARJ(49) + PARJ(50)} (drawn anew for each test).
 
\iteme{PARJ(51) - PARJ(55) :} (D=3*0.77, $-0.05$, $-0.005$) give a 
choice of four possible ways to parameterize the 
fragmentation function for \ttt{MSTJ(11)=2} (and \ttt{MSTJ(11)=3} 
for charm and heavier). The fragmentation of each flavour {\KF} may 
be chosen separately; for a diquark the flavour of the heaviest 
quark is used. With $c = $\ttt{PARJ(50+KF)}, the parameterizations 
are: \\
$0 \leq c \leq 1$ : Field-Feynman, $f(z) = 1 - c + 3c(1-z)^2$; \\
$-1 \leq c < 0$ : Peterson/SLAC, $f(z) = 1/(z(1-1/z-(-c)/(1-z))^2)$; \\
$c > 1$ : power peaked at $z=0$, $f(z) = (1-z)^{c-1}$; \\
$c < -1$ : power peaked at $z=1$, $f(z) = z^{-c-1}$.
 
\iteme{PARJ(59) :} (D=1.) replaces \ttt{PARJ(51) - PARJ(53)} for
gluon jet generated with \ttt{MSTJ(2)=} 2 or 4.
 
\iteme{PARJ(61) - PARJ(63) :} (D=4.5, 0.7, 0.) parameterizes the
energy dependence of the primary multiplicity distribution in
phase-space decays. The former two correspond to $c_1$ and $c_2$ 
of eq.~(\ref{dec:multsel}), while the latter allows a further
additive term in the multiplicity specifically for onium decays.
 
\iteme{PARJ(64) :} (0.003 GeV) minimum kinetic energy in decays
(safety margin for numerical precision errors). When violated,
typically new masses would be selected if particles have a 
Breit-Wigner width, or a new decay channel where that is relevant.
 
\iteme{PARJ(65) :} (D=0.5 GeV) mass which, in addition to the
spectator quark or diquark mass, is not assumed to partake in the
weak decay of a heavy quark in a hadron. This parameter was mainly 
intended for top decay and is currently not in use.
 
\iteme{PARJ(66) :} (D=0.5) relative probability that colour is
rearranged when two singlets are to be formed from decay products.
Only applies for \ttt{MDME(IDC,2)=} 11--30, i.e.\ low-mass 
phase-space decays.
 
\iteme{PARJ(71) :} (D=10 mm) maximum average proper lifetime $c\tau$
for particles allowed to decay in the \ttt{MSTJ(22)=2} option. With
the default value, $\K_{\mrm{S}}^0$, $\Lambda$, $\Sigma^-$, $\Sigma^+$,
$\Xi^-$, $\Xi^0$ and $\Omega^-$ are stable (in addition to those
normally taken to be stable), but charm and bottom do still decay.
 
\iteme{PARJ(72) :} (D=1000 mm) maximum distance from the origin at
which a decay is allowed to take place in the \ttt{MSTJ(22)=3}
option.
 
\iteme{PARJ(73) :} (D=100 mm) maximum cylindrical distance
$\rho = \sqrt{x^2 + y^2}$ from the origin at which a decay is allowed
to take place in the \ttt{MSTJ(22)=4} option.
 
\iteme{PARJ(74) :} (D=1000 mm) maximum z distance from the origin
at which a decay is allowed to take place in the \ttt{MSTJ(22)=4}
option.
 
\iteme{PARJ(76) :} (D=0.7) mixing parameter $x_d = \Delta M/\Gamma$
in $\B^0$--$\br{\B}^0$ system.
 
\iteme{PARJ(77) :} (D=10.) mixing parameter $x_s = \Delta M/\Gamma$
in $\B_s^0$--$\br{\B}_s^0$ system.
 
\iteme{PARJ(80) - PARJ(90) :} parameters for time-like parton showers,
see section \ref{ss:showrout}.
 
\iteme{PARJ(91) :} (D=0.020 GeV) minimum particle width in
\ttt{PMAS(KC,2)}, above which particle decays are assumed to take
place before the stage where Bose--Einstein effects are introduced.
 
\iteme{PARJ(92) :} (D=1.) nominal strength of Bose--Einstein effects
for $Q = 0$, see \ttt{MSTJ(51)}. This parameter, often denoted
$\lambda$, expresses the amount of incoherence in particle
production. Due to the simplified picture used for the
Bose--Einstein effects, in particular for effects from three
nearby identical particles, the actual $\lambda$
of the simulated events may be larger than the input value.
 
\iteme{PARJ(93) :} (D=0.20 GeV) size of the Bose--Einstein effect
region in terms of the $Q$ variable, see \ttt{MSTJ(51)}. The more
conventional measure, in terms of the radius $R$ of the production
volume, is given by
$R = \hbar/$\ttt{PARJ(93)}$ \approx 0.2$
fm$\times$GeV/\ttt{PARJ(93)}$ = $\ttt{PARU(3)}/\ttt{PARJ(93)}.

\iteme{PARJ(94) :} (D=0.0 GeV) Increase radius for pairs from different 
$\W/\Z/$strings.
\begin{subentry}
\iteme{< 0 :} if \ttt{MSTJ(56) = 1}, the radius for pairs from 
different $\W/\Z$'s is increased to 
$R+\delta R_{\W\W}+\mbox{\ttt{PARU(3)}}/\mrm{abs}(\mbox{\ttt{PARJ(94)}})$.
\iteme{> 0 :} the radius for pairs from different strings is 
increased to\\ 
$R+\mbox{\ttt{PARU(3)}}/\mbox{\ttt{PARJ(94)}}$.
\end{subentry}

\iteme{PARJ(95) :} (R) Set to the energy imbalance after the BE algorithm, 
before rescaling of momenta.
 
\iteme{PARJ(96) :} (R) Set to the $\alpha$ needed to retain energy-momentum 
conservation in each event for relevant models. 

\iteme{PARJ(121) - PARJ(171) :} parameters for $\ee$ event generation,
see section \ref{ss:eeroutines}.
 
\iteme{PARJ(180) - PARJ(195) :} various coupling constants and 
parameters related to couplings, see section \ref{ss:coupcons}.
 
\end{entry}

\subsubsection{The advanced popcorn code for baryon production}
\label{sss:improvedbaryoncode}

In section \ref{sss:baryonprod} a new advanced popcorn code for 
baryon production model was presented, based on \cite{Ede97}. It partly 
overwrites and redefines the meaning of some of the parameters above. 
Therefore the full description of these new options are given separately 
in this section, together with a listing of the new routines involved.

\boxsep

In order to use the new options, a few possibilities are open.
\begin{Itemize}
\item Use of the old diquark and popcorn models, \ttt{MSTJ(12) = 1} and 
\ttt{2}, is essentially unchanged. Note, however, that \ttt{PARJ(19)} is 
available for an ad-hoc suppression of first-rank baryon production.
\item Use of the old popcorn model with new {\bf SU(6)} weighting:
\begin{Itemize}
\item Set \ttt{MSTJ(12)=3}.
\item Increase \ttt{PARJ(1)} by approximately a factor 1.2 to retain 
about the same effective baryon production rate as in \ttt{MSTJ(12)=2}.
\item Note: the new {\bf SU(6)} weighting e.g.\ implies that the total 
production rate of charm and bottom baryons is reduced.
\end{Itemize}
\item Use of the old flavour model with new {\bf SU(6)} treatment and 
modified fragmentation function for diquark vertices (which softens 
baryon spectra):
\begin{Itemize}
\item Set \ttt{MSTJ(12)=4}.
\item Increase \ttt{PARJ(1)} by about a factor 1.7 and \ttt{PARJ(5)} by 
about a factor 1.2 to restore the baryon and popcorn rates of the 
\ttt{MSTJ(12)=2} default.
\end{Itemize}
\item Use of the new flavour model (automatically with modified diquark 
fragmentation function.)
\begin{Itemize}
\item Set \ttt{MSTJ(12)=5}.
\item Increase \ttt{PARJ(1)} by approximately a factor 2.
\item Change \ttt{PARJ(18)} from 1 to approx. 0.19.
\item Instead of \ttt{PARJ(3) - PARJ(7)}, tune \ttt{PARJ(8)}, \ttt{PARJ(9)}, 
\ttt{PARJ(10)} and \ttt{PARJ(18)}. (Here \ttt{PARJ(10)} is used 
only in collisions having remnants of baryon beam particles.)
\item Note: the proposed parameter values are based on a global fit to
all baryon production rates. This e.g.\ means that the proton rate 
is lower than in the \ttt{MSTJ(12)=2} option, with current data 
somewhere in between. The \ttt{PARJ(1)} value would have to be about
3 times higher in \ttt{MSTJ(12)=5} than in \ttt{=2} to have the same total
baryon production rate (=proton+neutron), but then other baryon
rates would not match at all. 
\end{Itemize}
\item The new options \ttt{MSTJ(12)=4} and \ttt{=5} (and, to some extent,
\ttt{=3}) soften baryon spectra in such a way that \ttt{PARJ(45)} 
(the change of $a$ for diquarks in the Lund symmetric fragmentation 
function) is available for a retune. It affects i.e.\ baryon-antibaryon 
rapidity correlations and the baryon excess over antibaryons in quark jets.
\end{Itemize}

\boxsep

The changes in and additions to the common blocks are as follows.
\begin{entry}
\iteme{MSTU(121) - MSTU(125) :} Internal flags and counters; only 
MSTU(123) may be touched by you.
\begin{subentry}
\iteme{MSTU(121) :} Popcorn meson counter.
\iteme{MSTU(122) :} Points at the proper diquark production weights, to 
distinguish between ordinary popcorn and rank 0 diquark 
systems. Only needed if \ttt{MSTJ(12)=5}.
\iteme{MSTU(123) :} Initialization flag. If \ttt{MSTU(123)} is 0 in a 
\ttt{PYKFDI} call, \ttt{PYKFIN} is called and \ttt{MSTU(123)} set to 1. 
Would need to be reset by you if flavour parameters are changed in 
the middle of a run. 
\iteme{MSTU(124) :} First parton flavour in decay call, stored to easily 
find random flavour partner in a popcorn system.
\iteme{MSTU(125) :} Maximum number of popcorn mesons allowed in decay flavour 
generation. If a larger popcorn system passes the fake string 
suppressions, the error \ttt{KF=0} is returned and the flavour 
generation for the decay is restarted.
\end{subentry}
\iteme{MSTU(131) - MSTU(140) :} Store of popcorn meson flavour codes in 
decay algorithm. Purely internal.
 
\boxsep

\iteme{MSTJ(12) :} (D=2) Main switch for choice of baryon production model.
Suppression of rank 1 baryons by a parameter \ttt{PARJ(19)} is no longer 
governed by the \ttt{MSTJ(12)} switch, but instead turned on by setting 
\ttt{PARJ(19)<1}. Three new options are available: 
\begin{subentry}
\iteme{= 3 :} as \ttt{=2}, but additionally the production of first
rank baryons may be suppressed by a factor \ttt{PARJ(19)}.
\iteme{= 4 :} as \ttt{=2}, but diquark vertices suffers an extra 
 suppression of the form $1-\exp(\rho\Gamma)$, where 
 $\rho\approx 0.7 \mrm{GeV}^{-2}$ is stored in \ttt{PARF(192)}.
\iteme{= 5 :} Advanced version of the popcorn model. Independent of 
\ttt{PARJ(3-7)}. Instead depending on \ttt{PARJ(8-10)}. When using 
this option \ttt{PARJ(1)} needs to enhanced by approx. a factor 2 
(i.e.\ it losses a bit of its normal meaning), 
and \ttt{PARJ(18)} is suggested to be set to 0.19.
\end{subentry}
 
\boxsep

\iteme{PARJ(8), PARJ(9) :} (D=0.6, 1.2 GeV$^{-1}$) The new popcorn 
parameters $\beta_{\u}$ and $\delta \beta = \beta_{\s} - \beta_{\u}$. 
Used to suppress popcorn mesons of total invariant mass $M_{\perp}$ by 
$\exp(-\beta_q*M_{\perp})$. Larger \ttt{PARJ(9)} leads to a stronger 
suppression of popcorn systems surrounded by an $\s\sbar$ pair, and 
also a little stronger suppression of strangeness in diquarks. 
\iteme{PARJ(10) :} (D=0.6 GeV$^{-1}$) Corresponding parameter for 
suppression of leading rank mesons of transverse mass $M_{\perp}$ in the 
fragmentation of diquark jets, used if \ttt{MSTJ(12)}=5. The 
treatment of original diquarks is flavour independent, i.e.\ \ttt{PARJ(10)} 
is used even if the diquark contains $\s$ or heavier quarks.
 
\boxsep

\iteme{PARF(131) - PARF(190) :} Different diquark and popcorn weights, 
calculated in \ttt{PYKFIN}, which is automatically called from 
\ttt{PYKFDI}.
\begin{subentry}
\iteme{PARF(131) :} Popcorn ratio $BM\br{B}/B\br{B}$ in the old model.
\iteme{PARF(132-134) :} Leading rank meson ratio $MB/B$ in the old 
 model, for original diquark with 0, 1 and 2 $\s$-quarks, respectively.
\iteme{PARF(135-137) :} Colour fluctuation quark ratio, i.e.\ the 
 relative probability that the heavier quark in a diquark fits into the 
 baryon at the opposite side of the popcorn meson. For $\s\q$, original 
 $\s\q$ and original $\c\q$ diquarks, respectively.
\iteme{PARF(138) :} The extra suppression of strange colour fluctuation 
 quarks, due to the requirement of surrounding a popcorn meson. (In the 
 old model, it is simply \ttt{PARJ(6)}.)
\iteme{PARF(139) :} Preliminary suppression of a popcorn meson in the new 
 model. A system of $N$ popcorn mesons is started with weight proportional 
 to \ttt{PARF(139)}$^N$. It is then tested against the correct weight, 
 derived from the mass of the system. For strange colour fluctuation 
 quarks, the weight is \ttt{PARF(138)}*\ttt{PARF(139)}.
\iteme{PARF(140) :} Preliminary suppression of  leading rank mesons in 
 diquark strings, irrespective of flavour. Corresponds to \ttt{PARF(139)}.
\iteme{PARF(141-145) :} Maximal {\bf SU(6)} factors for different types 
 of diquarks.
\iteme{PARF(146) :}  $\Sigma/\Lambda$ suppression if \ttt{MSTJ(12)=5}, 
derived from \ttt{PARJ(18)}.
\iteme{PARF(151-190) :} Production ratios for different diquarks. Stored 
 in four groups, handling $\q\rightarrow B\br{B}$, 
 $\q\rightarrow BM...\br{B}$, $\q\q\rightarrow MB$ and finally 
 $\q\q\rightarrow MB$ in the case of original diquarks. In each group 
 is stored:
\begin{subentry} 
\item{1 :} $\s/\u$ colour fluctuation ratio. 
\item{2,3 :} $\s/\u$ ratio for the vertex quark if the colour fluctuation 
 quark is light or strange, respectively.
\item{4 :} $\q/\q'$ vertex quark ratio if the colour fluctuation quark is 
light and $=\q$. 
\item{5-7 :} (spin 1)/(spin 0) ratio for $\s\u$, $\u\s$ and $\u\d$, where 
the first flavour is the colour fluctuation quark. 
\item{8-10 :} Unused.
\end{subentry}
\end{subentry}
\iteme{PARF(191) :} (D=0.2) Non-constituent mass in GeV of a $\u\d_0$ 
diquark. Used in combination with diquark constituent mass differences 
to derive relative production rates for different diquark flavours in 
the \ttt{MSTJ(12)}=5 option.
\iteme{PARF(192) :} (D=0.5) Parameter for the low-$\Gamma$ suppression 
of diquark vertices in the \ttt{MSTJ(12)}$\ge 4$ options. \ttt{PARF(192)} 
represents $e^{-\rho}$, i.e.\ the suppression 
is of the form 1-\ttt{PARF(192)}$^\Gamma$, $\Gamma$ in GeV$^2$.
\iteme{PARF(193,194) :} (I) Store of some popcorn weights used by the 
present popcorn system.
\iteme{PARF(201-1400) :} (I) Weights for every possible popcorn meson 
construction in the \ttt{MSTJ(12)}=5 option.  Calculated from input 
parameters and meson masses in \ttt{PYKFIN}. When 
$\q_1\q_2 \rightarrow M+\q_1\q_3$, the weights for M and the new 
diquark depends not only on $\q_1$ and $\q_2$. It is also important if 
this is a `true' popcorn system, or a system which started with a diquark 
at the string end, and if M is the final meson of the popcorn system, 
i.e.\ if the $\q_1\q_3$ diquark will go into a baryon or not. With five 
possible flavours for $\q_1$ and $\q_2$ this gives 80 different situations 
when selecting $M$ and $\q_3$. However, quarks heavier than $\s$ only exist 
in the string endpoints, and if more popcorn mesons are to be produced, 
the $\q_1\q_3$ diquark does not influence the weights and the $\q_1$ 
dependence reduces to what $\beta$ factor (\ttt{PARJ(8-10)}) that is used.
Then 40 distinct situations remains, i.e.:\\
\begin{minipage}{10ex}\begin{tabbing}
`true' popcorn~~\=final meson~~\=d,u,s,$>$s~~\= \kill
`true popcorn'\>final meson\>$\q_1$\>$\q_2$ \\
YES\>YES\>d,u,s\>d,u,s\\
\>NO\>$<$s,s\>d,u,s\\
NO\>YES\>d,u,s,$>$s\>d,u,s,c,b\\
\>NO\>1 case\>d,u,s,c,b
\end{tabbing}\end{minipage}\\
This table also shows the order in which the situations are stored. E.g.\ 
situation no. 1 is `YES,YES,d,d', situation no.11 is `YES,NO,$<s$,u'.\\
In every situation $\q_3$ can be $\d$, $\u$ or $\s$. if $\q_3=\q_2$ there 
are in the program three possible flavour mixing states available for the 
meson. This gives five possible meson flavours, and for each one of them 
there are six possible $L,S$ spin states. Thus 30 \ttt{PARF} positions are 
reserved for each  situation, and these are used as follows:\\
For each spin multiplet (in the same order as in \ttt{PARF(1-60)}) five 
positions are reserved. First are stored the weights for the the 
$\q_3\ne\q_2$ mesons, with  $\q_3$ in increasing order. If $\q_2>\s$, this 
occupies three spots, and the final two are unused. If $\q_2\le\s$, the 
final three spots are used for the diagonal states when $\q_3=\q_2$.
\end{entry}

\boxsep
 
In summary, all common block variables are completely internal, except 
\ttt{MSTU(123)}, \ttt{MSTJ(12)}, \ttt{PARJ(8) - PARJ(10)} and 
\ttt{PARF(191), PARF(192)}. Among these, \ttt{PARF(191)} and \ttt{PARF(192)} 
should not need to be changed. \ttt{MSTU(123)} should be 0 when starting, 
and reset to 0 whenever changing a switch or parameter which influences 
flavour weight With \ttt{MSTJ(12)=4}, \ttt{PARJ(5)} may need to increase.
With \ttt{MSTJ(12)=5}, a preliminary tune suggests \ttt{PARJ(8) = 0.6}, 
\ttt{PARJ(9) = 1.2}, \ttt{PARJ(10) = 0.6}, \ttt{PARJ(1) = 0.20} and 
\ttt{PARJ(18)=0.19}.

\boxsep

Three new subroutines are added, but are only needed for internal use.
\begin{entry}

\iteme{SUBROUTINE PYKFIN :}\label{p:PYKFIN}
to calculate a large set of diquark and popcorn weights from input 
parameters. Is called from \ttt{PYKFDI} if \ttt{MSTU(123)}=0. Sets 
\ttt{MSTU(123)} to 1.

\iteme{SUBROUTINE PYNMES(KFDIQ) :}\label{p:PYNMES}
to calculate number of popcorn mesons to be generated in a popcorn 
system, or the number of leading rank mesons when fragmenting a 
diquark string. Stores the number in \ttt{MSTU(121)}.
Always returns 0 if \ttt{MSTJ(12)}$<2$. Returns 0 or 1 if 
\ttt{MSTJ(12)}$<5$.
\begin{subentry}
\iteme{KFDIQ :} Flavour of the diquark in a diquark string. If starting 
a popcorn system inside a string, \ttt{KFDIQ} is 0.
\end{subentry}

\iteme{SUBROUTINE PYDCYK(KFL1,KFL2,KFL3,KF) :}\label{p:PYDCYK}
to generate flavours in the phase space model of hadron decays, and 
in cluster decays. Is essentially the same as a \ttt{PYKFDI} call, but 
also takes into account the effects of string dynamics in flavour 
production in the \ttt{MSTJ(12)}$\ge 4$ options. This is done in order 
to get a reasonable interpretation of the input parameters also for 
hadron decays with these options.
\begin{subentry}
\iteme{KFL1,KFL2,KFL3,KF :} See \ttt{SUBROUTINE PYKFDI}.
\end{subentry}
\end{entry}

\boxsep

Internally the diquark codes have been extended to store the necessary
further popcorn information. As before, an initially existing diquark 
has a code of the type $1000 q_a + 100 q_b + 2s+1$, where $q_a > q_b$. 
Diquarks created in the fragmentation process now have the longer code
$10000 q_c + 1000 q_a + 100 q_b + 2s+1$, i.e.\ one further digit is set.
Here $q_c$ is the curtain quark, i.e.\ the flavour of the quark-antiquark
pair that is shared between the baryon and the antibaryon, either
$q_a$ or $q_b$. The non-curtain quark, the other of $q_a$ and $q_b$, may 
have its antiquark partner in a popcorn meson. In case there are no popcorn 
mesons this information is not needed, but is still set at random to be 
either of $q_a$ and $q_b$. The extended code is used internally in 
\ttt{PYSTRF} and \ttt{PYDECY} and in some routines called by them, but 
is not visible in any event listings.  
 
\subsection{Further Parameters and Particle Data}
\label{ss:parapartdat}
 
The \ttt{PYUPDA} routine is the main tool for updating particle data
tables. However, for decay tables in the SUSY Les Houches Accord
format, the \ttt{PYLHA3} routine should be used instead.  The following
common blocks are maybe of a more peripheral interest, with the exception of
the \ttt{MDCY} array, which allows a selective inhibiting of particle decays,
and setting masses of not yet discovered particles, such as \ttt{PMAS(25,1)},
the (Standard Model) Higgs mass.
 
\drawbox{CALL PYUPDA(MUPDA,LFN)}\label{p:PYUPDA}
\begin{entry}
\itemc{Purpose:} to give you the ability to update particle data, or to keep
several versions of modified particle data for special purposes (e.g.\ bottom
studies).  \iteme{MUPDA :} gives the type of action to be taken.
\begin{subentry}
\iteme{= 1 :} write a table of particle data, that you then can edit at
leisure. For ordinary listing of decay data, \ttt{PYLIST(12)} should be used,
but that listing could not be read back in by the program. \\ 
For each compressed flavour code {\KC} = 1--500, one line is written 
containing the corresponding uncompressed {\KF} flavour code (\ttt{1X,I9}) in
\ttt{KCHG(KC,4)}, the particle and antiparticle names (\ttt{2X,A16,2X,A16})
in \ttt{CHAF}, the electric (\ttt{I3}), colour charge (\ttt{I3}) and
particle/antiparticle distinction (\ttt{I3}) codes in \ttt{KCHG}, the mass
(\ttt{F12.5}), the mass width (\ttt{F12.5}), maximum broadening (\ttt{F12.5})
and average proper lifetime (\ttt{1P,E13.5}) in \ttt{PMAS}, the resonance
width treatment (\ttt{I3}) in \ttt{MWID} and the on/off decay switch
(\ttt{I3}) in \ttt{MDCY(KC,1).} \\ 
After a {\KC} line follows one line for
each possible decay channel, containing the \ttt{MDME} codes (\ttt{10X,2I5}),
the branching ratio (\ttt{F12.6}) in \ttt{BRAT}, and the \ttt{KFDP} codes for
the decay products (\ttt{5I10}), with trailing 0's if the number of decay
products is smaller than 5.  
\iteme{= 2 :} read in particle data, as written
with \ttt{=1} and thereafter edited by you, and use this data subsequently in
the current run. This also means e.g.\ the mapping between the full {\KF} and
compressed {\KC} flavour codes. Reading is done with fixed format, which
means that you have to preserve the format codes described for \ttt{=1}
during the editing. A number of checks will be made to see if input looks
reasonable, with warnings if not.  If some decay channel is said not to
conserve charge, it should be taken seriously. Warnings that decay is
kinematically disallowed need not be as serious, since that particular decay
mode may not be switched on unless the particle mass is increased.  
\iteme{= 3 :} read in particle data, like option 2, but use it as a 
complement to
rather than a replacement of existing data. The input file should therefore
only contain new particles and particles with changed data. New particles are
added to the bottom of the {\KC} and decay channel tables. Changed particles
retain their {\KC} codes and hence the position of particle data, but their
old decay channels are removed, this space is recuperated, and new decay
channels are added at the end.  Thus also the decay channel numbers of
unchanged particles are affected.  
\iteme{= 4 :} write current particle data
as data lines, which can be edited into \ttt{BLOCK DATA PYDATA} for a
permanent replacement of the particle data. This option is intended for the
program author only, not for you.
\end{subentry}
\iteme{LFN :} the file number which the data should be written to or read
from. You must see to it that this file is properly opened for read or write
(since the definition of file names is platform dependent).
\end{entry}
 
\drawbox{CALL PYLHA3(MUPDA,KF,IFAIL)}\label{p:PYLHA3}
\begin{entry}
\itemc{Purpose:} to read and write SUSY spectra and/or decay tables
conforming to the SUSY Les Houches Accord format \cite{Ska03}.  This
normally happens automatically during initialization, when the appropriate
switches have been set, see the examples in section \ref{ss:susyclass}, but
\ttt{PYLHA3} may also be used stand-alone, to update the decay table for any
particle. To do this, the user should call \ttt{PYLHA3} manually before the
call to \ttt{PYINIT}, with \ttt{MUPDA=2} (see below) and \ttt{KF} the flavour
code of the particle whose decay table is to be updated. In case several
decay tables are to be updated, separate calls must be made for each
particle.  \iteme{MUPDA :} gives the type of action to be taken.
\begin{subentry}
\iteme{= 1 :} read in spectrum from Logical Unit Number \ttt{IMSS(21)}. This
call happens automatically during initialization when \ttt{IMSS(1)=11}. It is
not intended for manual use.  
\iteme{= 2 :} read in decay table for particle
code \ttt{KF} from Logical Unit Number \ttt{IMSS(22)}. Is performed
automatically when \ttt{IMSS(1)=11} and \ttt{IMSS(22)$\ne$0}, but only for
sparticles and higgs bosons. This call may also be used stand-alone to update
the decay table of a particle.  
\iteme{= 3 :} write out spectrum on Logical
Unit Number \ttt{IMSS(23)}. This call happens automatically during
initialization when \ttt{IMSS(1)$\ne$0} and \ttt{IMSS(23)$\ne$0}. It is not
intended for manual use.  
\iteme{= 4 :} write out decay table for particle
code \ttt{KF} on Logical Unit Number \ttt{IMSS(24)}.  This option is not
supported yet.
\end{subentry}
\iteme{KF :} the flavour code of the particle for which decay table read-in
or write-out is desired, only relevant for \ttt{MUPDA=2,4}.  
\iteme{IFAIL :} return code specifying whether the desired operation 
succeeded.
\begin{subentry}
\iteme{= 0 :} Operation succeeded.  
\iteme{= 1 :} Operation failed, e.g.\ if no decay table in the SUSY 
Les Houches Accord format matching the desired \ttt{KF} code was found on 
the file.
\end{subentry}
\end{entry}

\drawbox{COMMON/PYDAT2/KCHG(500,4),PMAS(500,4),PARF(2000),VCKM(4,4)}%
\label{p:PYDAT2}
\begin{entry}
\itemc{Purpose:} to give access to a number of flavour treatment
constants or parameters and particle/parton data. Particle data is
stored by compressed code {\KC} rather than by the full {\KF} code.
You are reminded that the way to know the {\KC} value is to use the 
\ttt{PYCOMP} function, i.e.\ \ttt{KC = PYCOMP(KF)}.
 
\boxsep
 
\iteme{KCHG(KC,1) :}\label{p:KCHG} three times particle/parton charge 
for compressed code {\KC}.
 
\iteme{KCHG(KC,2) :} colour information for compressed code {\KC}.
\begin{subentry}
\iteme{= 0 :} colour-singlet particle.
\iteme{= 1 :} quark or antidiquark.
\iteme{= -1 :} antiquark or diquark.
\iteme{= 2 :} gluon.
\end{subentry}
 
\iteme{KCHG(KC,3) :} particle/antiparticle distinction for
compressed code {\KC}.
\begin{subentry}
\iteme{= 0 :} the particle is its own antiparticle.
\iteme{= 1 :} a nonidentical antiparticle exists.
\end{subentry}
 
\iteme{KCHG(KC,4) :} equals the uncompressed particle code {\KF}
(always with a positive sign). This gives the inverse mapping of 
what is provided by the \ttt{PYCOMP} routine.
 
\boxsep
 
\iteme{PMAS(KC,1) :}\label{p:PMAS} particle/parton mass $m$ (in GeV) 
for compressed code {\KC}.
 
\iteme{PMAS(KC,2) :} the total width $\Gamma$ (in GeV) of an assumed
symmetric Breit--Wigner mass shape for compressed particle code {\KC}.
 
\iteme{PMAS(KC,3) :} the maximum deviation (in GeV) from the
\ttt{PMAS(KC,1)} value at which the Breit--Wigner shape above is
truncated. Is used in ordinary particle decays, but not in the 
resonance treatment; cf. the \ttt{CKIN} variables.
 
\iteme{PMAS(KC,4) :} the average lifetime $\tau$ for compressed
particle code {\KC}, with $c \tau$ in mm, i.e.\ $\tau$ in units of
about $3.33 \times 10^{-12}$ s.
 
\boxsep
 
\iteme{PARF(1) - PARF(60) :}\label{p:PARF} give a parameterization of 
the $\d\dbar$--$\u\ubar$--$\s\sbar$ flavour mixing in production of
flavour-diagonal mesons. Numbers are stored in groups of 10, for
the six multiplets pseudoscalar, vector, axial vector ($S=0$),
scalar, axial vector ($S=1$) and tensor, in this order; see
section \ref{sss:mesonprod}. Within each group, the first two 
numbers determine the fate of a $\d\dbar$ flavour state, the 
second two that of a $\u\ubar$ one, the next two that of an 
$\s\sbar$ one, while the last four are unused. Call the numbers 
of a pair $p_1$ and $p_2$. Then the probability to produce the 
state with smallest {\KF} code is $1-p_1$, the probability for the 
middle one is $p_1 - p_2$ and the probability for the one with 
largest code is $p_2$, i.e.\ $p_1$ is the probability to produce 
either of the two `heavier' ones.
 
\iteme{PARF(61) - PARF(80) :} give flavour {\bf SU(6)} weights for
the production of a spin 1/2 or spin 3/2 baryon from a given
diquark--quark combination. Should not be changed.

\iteme{PARF(91) - PARF(96) :} (D  = 0.0099, 0.0056, 0.199, 1.23, 
4.17, 165 GeV) default nominal quark masses, used to give the starting 
value for running masses calculated in \ttt{PYMRUN}.
 
\iteme{PARF(101) - PARF(105) :} contain  $\d$, $\u$, $\s$,
$\c$ and $\b$ constituent masses, in the past used in mass formulae
for undiscovered hadrons, and should not be changed. 
 
\iteme{PARF(111), PARF(112) :} (D=0.0, 0.11 GeV) constant terms in the
mass formulae for heavy mesons and baryons, respectively (with diquark
getting 2/3 of baryon).
 
\iteme{PARF(113), PARF(114) :} (D=0.16, 0.048 GeV) factors which,
together with Clebsch-Gordan coefficients and quark constituent
masses, determine the mass splitting due to spin-spin interactions
for heavy mesons and baryons, respectively. The latter factor is
also used for the splitting between spin 0 and spin 1 diquarks.
 
\iteme{PARF(115) - PARF(118) :} (D=0.50, 0.45, 0.55, 0.60 GeV),
constant mass terms, added to the constituent masses, to get the
mass of heavy mesons with orbital angular momentum $L = 1$. The four
numbers are for pseudovector mesons with quark spin 0, and for scalar,
pseudovector and tensor mesons with quark spin 1, respectively.
 
\iteme{PARF(121), PARF(122) :} (D=0.1, 0.2 GeV) constant terms, which
are subtracted for quark and diquark masses, respectively, in defining
the allowed phase space in particle decays into partons (e.g. 
$\B^0 \to \cbar \d \u \dbar$).
 
\iteme{PARF(201) - PARF(1960) :} (D=1760*0) relative probabilities
for flavour production in the \ttt{MSTJ(15)=1} option; to be
defined by you before any {\Py} calls. (The standard meaning is 
changed in the advanced baryon popcorn code, described in subsection
\ref{sss:improvedbaryoncode}, where many of the \ttt{PARF} numbers
are used for other purposes.)\\
The index in \ttt{PARF} is of the compressed form \\
$120 + 80 \times$KTAB1$ + 25 \times$KTABS$ + $KTAB3. \\
Here KTAB1 is the
old flavour, fixed by preceding fragmentation history, while KTAB3
is the new flavour, to be selected according to the relevant relative
probabilities (except for the very last particle, produced when
joining two jets, where both KTAB1 and KTAB3 are known).
Only the most frequently appearing quarks/diquarks are defined,
according to the code $1 = \d$, $2 = \u$, $3 = \s$, $4 = \c$,
$5 = \b$, $6 = \t$ (obsolete!), $7 = \d\d_1$, $8 = \u\d_0$, 
$9 = \u\d_1$, $10 = \u\u_1$, $11 = \s\d_0$, $12 = \s\d_1$, 
$13 = \s\u_0$, $14 = \s\u_1$, $15 = \s\s_1$, $16 = \c\d_0$, 
$17 = \c\d_1$, $18 = \c\u_0$, $19 = \c\u_1$, $20 = \c\s_0$, 
$21 = \c\s_1$, $22 = \c\c_1$.
These are thus the only possibilities for the new flavour to be
produced; for an occasional old flavour not on this list, the
ordinary relative flavour production probabilities will be used. \\
Given the initial and final flavour, the intermediate hadron that
is produced is almost fixed. (Initial and final diquark here
corresponds to `popcorn' production of mesons intermediate between
a baryon and an antibaryon). The additional index KTABS gives the
spin type of this hadron, with \\
0 = pseudoscalar meson or $\Lambda$-like spin 1/2 baryon, \\
1 = vector meson or $\Sigma$-like spin 1/2 baryon, \\
2 = tensor meson or spin 3/2 baryon. \\
(Some meson multiplets, not frequently produced, are not accessible
by this parameterization.) \\
Note that some combinations of KTAB1, KTAB3 and KTABS do not
correspond to a physical particle (a $\Lambda$-like baryon must
contain three different quark flavours, a $\Sigma$-like one at least
two), and that you must see to it that the corresponding
\ttt{PARF} entries are vanishing. One additional complication exist
when KTAB3 and KTAB1 denote the same flavour content (normally
KTAB3$ = $KTAB1, but for diquarks the spin freedom may give
KTAB3$ = $KTAB1$\pm 1$): then a flavour neutral meson is to be
produced, and here $\d\dbar$, $\u\ubar$ and $\s\sbar$ states mix
(heavier flavour states do not, and these are therefore no problem).
For these cases the ordinary KTAB3 value gives the total probability
to produce either of the mesons possible, while KTAB3$ = $23 gives the
relative probability to produce the lightest meson state ($\pi^0$,
$\rho^0$, $\a_2^0$), KTAB3$ = $24 relative probability for the middle
meson ($\eta$, $\omega$, $\f_2^0$), and KTAB3 = 25 relative
probability for the heaviest one ($\eta'$, $\phi$, $f'^0_2$). Note
that, for simplicity, these relative probabilities are assumed the
same whether initial and final diquark have the same spin or not; the
total probability may well be assumed different, however. \\
As a general comment, the sum of \ttt{PARF} values for a given KTAB1
need not be normalized to unity, but rather the program will find the
sum of relevant weights and normalize to that. The same goes for
the KTAB3$ = $23--25 weights. This makes it straightforward to use
one common setup of \ttt{PARF} values and still switch between
different \ttt{MSTJ(12)} baryon production modes, with the exception 
of the advanced popcorn scenarios.
 
\boxsep
 
\iteme{VCKM(I,J) :}\label{p:VCKM} squared matrix elements of the
Cabibbo-Kobayashi-Maskawa flavour mixing matrix.
\begin{subentry}
\iteme{I :} up type generation index, i.e.\ $1 = \u$, $2 = \c$,
$3 = \t$ and $4 = \t'$.
\iteme{J :} down type generation index, i.e.\ $1 = \d$, $2 = \s$,
$3 = \b$ and $4 = \b'$.
\end{subentry}
 
\end{entry}
 
\drawbox{COMMON/PYDAT3/MDCY(500,3),MDME(8000,2),BRAT(8000),%
KFDP(8000,5)}\label{p:PYDAT3}
\begin{entry}
\itemc{Purpose:} to give access to particle decay data and
parameters. In particular, the \ttt{MDCY(KC,1)} variables may be 
used to switch on or off the decay of a given particle species, 
and the \ttt{MDME(IDC,1)} ones
to switch on or off an individual decay channel of a particle.
For quarks, leptons and gauge bosons, a number of decay channels
are included that are not allowed for on-mass-shell particles, see
\ttt{MDME(IDC,2)=102}. These channels are not directly used to
perform decays, but rather to denote allowed couplings in a more
general sense, and to switch on or off such couplings, as
described elsewhere. Particle data is
stored by compressed code {\KC} rather than by the full {\KF} code.
You are reminded that the way to know the {\KC} value is to use the 
\ttt{PYCOMP} function, i.e.\ \ttt{KC = PYCOMP(KF)}.
 
\boxsep
 
\iteme{MDCY(KC,1) :}\label{p:MDCY} switch to tell whether a particle 
with compressed code {\KC} may be allowed to decay or not. 
\begin{subentry}
\iteme{= 0 :} the particle is not allowed to decay.
\iteme{= 1 :} the particle is allowed to decay (if decay information
is defined below for the particle).
\itemc{Warning:} these values may be overwritten for resonances in a 
\ttt{PYINIT} call, based on the \ttt{MSTP(41)} option you have selected. 
If you want to allow resonance decays in general but switch off the decay 
of one particular resonance, this is therefore better done after the 
\ttt{PYINIT} call. 
\end{subentry} 
 
\iteme{MDCY(KC,2) :} gives the entry point into the decay channel
table for compressed particle code {\KC}. Is 0 if no decay channels 
have been defined.
 
\iteme{MDCY(KC,3) :} gives the total number of decay channels defined
for compressed particle code {\KC}, independently of whether they have
been assigned a non-vanishing branching ratio or not. Thus the
decay channels are found in positions \ttt{MDCY(KC,2)} to
\ttt{MDCY(KC,2)+MDCY(KC,3)-1}.
 
\boxsep
 
\iteme{MDME(IDC,1) :}\label{p:MDME} on/off switch for individual 
decay channel IDC. In addition, a channel may be left selectively 
open; this has some special applications in the event generation
machinery. Effective branching ratios are 
automatically recalculated for the decay channels left open. 
Also process cross sections are affected; see section 
\ref{sss:resdecaycross}. If a particle is allowed to decay by the 
\ttt{MDCY(KC,1)} value, at least one channel must be left open
by you. A list of decay channels with current IDC numbers
may be obtained with \ttt{PYLIST(12)}.
\begin{subentry}
\iteme{= -1 :} this is a non-Standard Model decay mode, which by
default is assumed not to exist. Normally, this option is used for
decays involving fourth generation or $\H^{\pm}$ particles.
\iteme{= 0 :} channel is switched off.
\iteme{= 1 :} channel is switched on.
\iteme{= 2 :} channel is switched on for a particle but off for an
antiparticle. It is also on for a particle which is its own 
antiparticle, i.e.\ here it means the same as \ttt{=1}.
\iteme{= 3 :} channel is switched on for an antiparticle but off for
a particle. It is off for a particle which is its own antiparticle.
\iteme{= 4 :} in the production of a pair of equal or charge
conjugate resonances in {\Py}, say $\hrm^0 \to \W^+ \W^-$, either one
of the resonances is allowed to decay according to this group of
channels, but not both. If the two particles of the pair
are different, the channel is on. For ordinary particles, not 
resonances, this option only means that the channel is switched off.
\iteme{= 5 :} as \ttt{=4}, but an independent group of channels, such
that in a pair of equal or charge conjugate resonances the decay of
either resonance may be specified independently. If the two
particles in the pair are different, the channel is off.
For ordinary particles, not  resonances, this option only means 
that the channel is switched off.
\itemc{Warning:} the two values -1 and 0 may look similar, but in fact
are quite different. In neither case the channel so set is
generated, but in the latter case the channel still contributes
to the total width of a resonance, and thus affects both
simulated line shape and the generated cross section when
{\Py} is run. The value 0 is appropriate to a channel
we assume exists, even if we are not currently simulating it,
while -1 should be used for channels we believe do not exist.
In particular, you are warned unwittingly to set fourth
generation channels 0 (rather than -1), since by now the
support for a fourth generation is small.
\itemc{Remark:} all the options above may be freely mixed. The
difference, for those cases where both make sense, between using
values 2 and 3 and using 4 and 5 is that the latter automatically
include charge conjugate states, e.g.
$\hrm^0 \to \W^+ \W^- \to \e^+ \nu_e \d \ubar$ or
$\dbar \u \e^- \br{\nu}_e$, but the former only one
of them. In calculations of the joint branching ratio, this
makes a factor 2 difference.
\itemc{Example:} to illustrate the above options, consider the case
of a $\W^+ \W^-$ pair. One might then set the following combination
of switches for the $\W$:\\
\begin{tabular}{ccl@{\protect\rule{0mm}{\tablinsep}}}
channel & value & comment \\
$\u\dbar$ & 1 & allowed for $\W^+$ and $\W^-$ in any combination, \\
$\u\sbar$ & 0 & never produced but contributes to $\W$ width, \\
$\c\dbar$ & 2 & allowed for $\W^+$ only, \\
$\c\sbar$ & 3 & allowed for $\W^-$ only, i.e.\ properly 
$\W^- \to \cbar\s$, \\
$\t\bbar$ & 0 & never produced but contributes to $\W$ width \\
  &  & if the channel is kinematically allowed, \\
$\nu_{\e}\e^+$ & 4 & allowed for one of $\W^+$ or $\W^-$, but not 
both, \\
$\nu_{\mu}\mu^+$ & 4 & allowed for one of $\W^+$ or $\W^-$, but not 
both, \\
  &  & and not in combination with $\nu_{\e}\e^+$, \\
$\nu_{\tau}\tau^+$ & 5 & allowed for the other $\W$, but not both, \\
$\nu_{\chi}\chi^-$ & $-1$ & not produced and does not contribute to 
$\W$ width.\\
\end{tabular}\\
A $\W^+\W^-$ final state $\u\dbar + \cbar\s$  is allowed, but not its
charge conjugate $\ubar\d + \c\sbar$, since the latter decay mode is 
not allowed for a $\W^+$. The combination 
$\nu_{\e}\e^+ + \bar{\nu}_{\tau}\tau^-$ is allowed, since the two 
channels belong to different groups, but not 
$\nu_{\e}\e^+ + \bar{\nu}_{\mu}\mu^-$, where both belong to the same.
Both $\u\dbar + \bar{\nu}_{\tau}\tau^-$ and 
$\ubar\d + \nu_{\tau}\tau^+$ are allowed, since there is no clash.
The full rulebook, for this case, is given by 
eq.~(\ref{eq:WWallchancombgen}). A term $r_i^2$ means channel 
$i$ is allowed
for $\W^+$ and $\W^-$ simultaneously, a term $r_i r_j$ that channels
$i$ and $j$ may be combined, and a term $2 r_i r_j$ that channels 
$i$ and $j$ may be combined two ways, i.e.\ that also a charge 
conjugate combination is allowed. 
\end{subentry}
 
\iteme{MDME(IDC,2) :} information on special matrix-element treatment
for decay channel IDC. Is mainly intended for the normal-particle
machinery in \ttt{PYDECY}, so many of the codes are superfluous in the
more sophisticated resonance decay treatment by \ttt{PYRESD}, see 
section \ref{sss:resdecintro}. 
In addition to the outline below, special rules
apply for the order in which decay products should be given, so that
matrix elements and colour flow is properly treated. One such example
is the weak matrix elements, which only will be correct if decay
products are given in the right order. The program does not police 
this, so if you introduce channels of your own and use these codes,
you should be guided by the existing particle data. 
\begin{subentry}
\iteme{= 0 :} no special matrix-element treatment; partons and
particles are copied directly to the event record, with momentum
distributed according to phase space.
\iteme{= 1 :} $\omega$ and $\phi$ decays into three pions, eq.
(\ref{dec:omegphi}).
\iteme{= 2 :} $\pi^0$ or $\eta$ Dalitz decay to $\gamma \e^+ \e^-$,
eq.~(\ref{dec:Dalitz}).
\iteme{= 3 :} used for vector meson decays into two pseudoscalars,
to signal non-isotropic decay angle according to eq.
(\ref{dec:psvpsps}), where relevant.
\iteme{= 4 :} decay of a spin 1 onium resonance to three gluons or
to a photon and two gluons, eq.~(\ref{ee:Upsilondec}). The gluons may
subsequently develop a shower if \ttt{MSTJ(23)=1}.
\iteme{= 11 :} phase-space production of hadrons from the quarks
available.
\iteme{= 12 :} as \ttt{=11}, but for onia resonances, with the option
of modifying the multiplicity distribution separately.
\iteme{= 13 :} as \ttt{=11}, but at least three hadrons to be produced
(useful when the two-body decays are given explicitly).
\iteme{= 14 :} as \ttt{=11}, but at least four hadrons to be produced.
\iteme{= 15 :} as \ttt{=11}, but at least five hadrons to be produced.
\iteme{= 22 - 30 :} phase-space production of hadrons from the quarks
available, with the multiplicity fixed to be \ttt{MDME(IDC,2)-20},
i.e.\ 2--10.
\iteme{= 31 :} two or more quarks and particles are distributed
according to phase space. If three or more products, the last product
is a spectator quark, i.e.\ sitting at rest with respect to the
decaying hadron.
\iteme{= 32 :} a $\q\qbar$ or $\g\g$ pair, distributed according to
phase space (in angle), and allowed to develop a shower if
\ttt{MSTJ(23)=1}.
\iteme{= 33 :} a triplet $\q X \qbar$, where $X$ is either a gluon or
a colour-singlet particle; the final particle ($\qbar$) is assumed to
sit at rest with respect to the decaying hadron, and the
two first particles ($\q$ and $X$) are allowed to develop a shower if
\ttt{MSTJ(23)=1}. Nowadays superfluous.
\iteme{= 41 :} weak decay, where particles are distributed according
to phase space, multiplied by a factor from the expected shape of
the momentum spectrum of the direct product of the weak decay
(the $\nu_{\tau}$ in $\tau$ decay).
\iteme{= 42 :} weak decay matrix element for quarks and leptons.
Products may be given either in terms of quarks or hadrons, or
leptons for some channels. If the spectator system is given in
terms of quarks, it is assumed to collapse into one particle from
the onset. If the virtual $\W$ decays into quarks, these quarks
are converted to particles, according to phase space in the
$\W$ rest frame, as in \ttt{=11}. Is intended for $\tau$, charm and
bottom.
\iteme{= 43 :} as \ttt{=42}, but if the $\W$ decays into quarks, these
will either appear as jets or, for small masses, collapse into a
one- or two-body system. Nowadays superfluous.
\iteme{= 44 :} weak decay matrix element for quarks and leptons, where
the spectator system may collapse into one particle for a small
invariant mass. If the first two decay products are a $\q\qbar'$
pair, they may develop a parton shower if \ttt{MSTJ(23)=1}.
Was intended for top and beyond, but is nowadays superfluous.
\iteme{= 46 :} $\W$ ({\KF} = 89) decay into $\q\qbar'$ or
$\ell \nu_{\ell}$ according to relative probabilities given
by couplings (as stored in the \ttt{BRAT} vector)
times a dynamical phase-space factor given by the current $\W$
mass. In the decay, the correct $V-A$ angular distribution is
generated if the $\W$ origin is known (heavy quark or lepton).
This is therefore the second step of a decay with \ttt{MDME=45}.
A $\q\qbar'$ pair may subsequently develop a shower if 
\ttt{MSTJ(23)=1}. Was intended for top and beyond, but is nowadays 
superfluous.
\iteme{= 48 :} as \ttt{=42}, but require at least three decay
products.
\iteme{= 50 :} (default behaviour, also obtained for any other code value
apart from the ones listed below) do not include any special
threshold factors. That is, a decay channel is left open even 
if the sum of daughter nominal masses is above the mother 
actual mass, which is possible if at least one of the daughters 
can be pushed off the mass shell. Is intended for decay treatment
in \ttt{PYRESD} with \ttt{PYWIDT} calls, and has no special meaning
for ordinary \ttt{PYDECY} calls.
\iteme{= 51 :} a step threshold, i.e.\ a channel is switched off when
the sum of daughter nominal masses is above the mother actual
mass. Is intended for decay treatment in \ttt{PYRESD} with \ttt{PYWIDT} 
calls, and has no special meaning for ordinary \ttt{PYDECY} calls.
\iteme{= 52 :} a $\beta$-factor threshold, i.e.\ 
$\sqrt{ (1-m_1^2/m^2-m_2^2/m^2)^2 - 4 m_1^2 m_2^2/m^4}$,
assuming that the values stored in \ttt{PMAS(KC,2)} and \ttt{BRAT(IDC)}
did not include any threshold effects at all. Is intended for decay 
treatment in \ttt{PYRESD} with \ttt{PYWIDT} calls, and has no special 
meaning for ordinary \ttt{PYDECY} calls.     
\iteme{= 53 :} as \ttt{=52}, but assuming that \ttt{PMAS(KC,2)} and 
\ttt{BRAT(IDC)} did include the threshold effects, so that the weight 
should be the ratio of the $\beta$ value at the actual mass to that at 
the nominal one. Is intended for decay treatment in \ttt{PYRESD} with 
\ttt{PYWIDT} calls, and has no special meaning for ordinary \ttt{PYDECY} 
calls.
\iteme{= 101 :} this is not a proper decay channel, but only to be
considered as a continuation line for the decay product listing
of the immediately preceding channel. Since the \ttt{KFDP} array can
contain five decay products per channel, with this code it is
possible to define channels with up to ten decay products. It is
not allowed to have several continuation lines after each other.
\iteme{= 102 :} this is not a proper decay channel for a decaying
particle on the mass shell (or nearly so), and is therefore assigned
branching ratio 0. For a particle off the mass shell, this decay
mode is allowed, however. By including this channel among the
others, the switches \ttt{MDME(IDC,1)} may be used to allow or
forbid these channels in hard processes, with cross sections
to be calculated separately. As an example, $\gamma \to \u \ubar$
is not possible for a massless photon, but is an allowed
channel in $\ee$ annihilation.
\end{subentry}
 
\boxsep
 
\iteme{BRAT(IDC) :}\label{p:BRAT} give branching ratios for the 
different decay
channels. In principle, the sum of branching ratios for a given
particle should be unity. Since the program anyway has to calculate
the sum of branching ratios left open by the \ttt{MDME(IDC,1)} values
and normalize to that, you need not explicitly ensure this
normalization, however. (Warnings are printed in \ttt{PYUPDA(2)} or
\ttt{PYUPDA(3)} calls if the sum is not unity, but this is entirely 
intended as a help for finding user mistypings.) For decay channels 
with \ttt{MDME(IDC,2)}$> 80$ the \ttt{BRAT} values are dummy.
 
\boxsep
 
\iteme{KFDP(IDC,J) :}\label{p:KFDP} contain the decay products in the 
different channels, with five positions \ttt{J=} 1--5 reserved for each
channel IDC. The decay products are given following the standard {\KF}
code for partons and particles, with 0 for trailing empty positions. Note
that the \ttt{MDME(IDC+1,2)=101} option allows you to double the
maximum number of decay product in a given channel from 5 to 10,
with the five latter products stored \ttt{KFDP(IDC+1,J)}.
 
\end{entry}
 
\drawboxtwo{COMMON/PYDAT4/CHAF(500,2)}{CHARACTER CHAF*16}\label{p:PYDAT4}
\begin{entry}
\itemc{Purpose:} to give access to character type variables.
 
\boxsep
 
\iteme{CHAF(KC,1) :}\label{p:CHAF} particle name according to {\KC} code.
 
\iteme{CHAF(KC,2) :} antiparticle name according to {\KC} code
when an antiparticle exists, else blank.

\end{entry}
 
\subsection{Miscellaneous Comments}
 
The previous sections have dealt with the subroutine options and
variables one at a time. This is certainly important, but for a full
use of the capabilities of the program, it is also necessary
to understand how to make different pieces work together. This is
something that cannot be explained fully in a manual, but must also
be learnt by trial and error. This section contains some examples of
relationships between subroutines, common blocks and parameters.
It also contains comments on issues that did not fit in naturally
anywhere else, but still might be useful to have on record.
 
\subsubsection{Interfacing to detector simulation}
 
Very often, the output of the program is to be fed into a subsequent
detector simulation program. It therefore becomes necessary to set up
an interface between the \ttt{PYJETS} common block and the detector
model. Preferably this should be done via the \ttt{HEPEVT} standard
common block, see section \ref{ss:HEPEVT}, but sometimes this may not
be convenient. If a \ttt{PYEDIT(2)} call is made, the remaining
entries exactly correspond to those an ideal detector could see: all
non-decayed particles, with the exception of neutrinos. The translation
of momenta should be trivial (if need be, a \ttt{PYROBO} call can be
made to rotate the `preferred' $z$ direction to whatever is the
longitudinal direction of the detector), and so should the translation
of particle codes. In particular, if the
detector simulation program also uses the standard Particle Data Group
codes, no conversion at all is needed. The problem then is to select
which particles are allowed to decay, and how decay vertex information
should be used.
 
Several switches regulate which particles are allowed to decay. First,
the master switch \ttt{MSTJ(21)} can be used to switch on/off all
decays (and it also contains a choice of how fragmentation should
be interfaced). Second, a particle must have decay modes defined for
it, i.e.\ the corresponding \ttt{MDCY(KC,2)} and \ttt{MDCY(KC,3)}
entries must be non-zero for compressed code \ttt{KC = PYCOMP(KF)}.
This is true for all colour neutral particles except the neutrinos,
the photon, the proton and the neutron. (This statement is actually
not fully correct, since irrelevant `decay modes' with
\ttt{MDME(IDC,2)=102} exist in some cases.) Third, the
individual switch in \ttt{MDCY(KC,1)} must be on. Of all the particles
with decay modes defined, only $\mu^{\pm}$, $\pi^{\pm}$, $\K^{\pm}$
and $\K_{\mrm{L}}^0$ are by default considered stable.
 
Finally, if \ttt{MSTJ(22)} does not have its default value 1, checks
are also made on the lifetime of a particle before it is allowed to
decay. In the simplest alternative, \ttt{MSTJ(22)=2}, the comparison
is based on the average lifetime, or rather $c \tau$, measured in mm.
Thus if the limit \ttt{PARJ(71)} is (the default) 10 mm, then decays
of $\K_{\mrm{S}}^0$, $\Lambda$, $\Sigma^-$, $\Sigma^+$, $\Xi^-$, 
$\Xi^0$ and $\Omega^-$ are all switched off, but charm and bottom 
still decay. No $c \tau$ values below 1 $\mu$m are defined. With the 
two options \ttt{MSTJ(22)=} 3 or 4, a spherical or cylindrical volume 
is defined around the origin, and all decays taking place inside this
volume are ignored.
 
Whenever a particle is in principle allowed to decay, i.e.
\ttt{MSTJ(21)} and \ttt{MDCY} on, an proper lifetime is selected
once and for all and stored in \ttt{V(I,5)}. The \ttt{K(I,1)} is then
also changed to 4. For \ttt{MSTJ(22)=1}, such a particle will also
decay, but else it could remain in the event record. It is then
possible, at a later stage, to expand the volume inside which decays
are allowed, and do a new \ttt{PYEXEC} call to have particles
fulfilling the new conditions (but not the old) decay. As a further
option, the \ttt{K(I,1)} code may be put to 5, signalling that the
particle will definitely decay in the next \ttt{PYEXEC} call, at the
vertex position given (by you) in the \ttt{V} vector.
 
This then allows the {\Py} decay routines to be used inside a
detector simulation program, as follows. For a particle which did not
decay before entering the detector, its point of decay is still well
defined (in the absence of deflections by electric or magnetic fields),
eq.~(\ref{dec:newvertex}). If it interacts before that point, the 
detector simulation program is left to handle things. If not, the 
\ttt{V} vector is updated according to the formula above, \ttt{K(I,1)} 
is set to 5, and \ttt{PYEXEC} is called, to give a set of decay 
products, that can again be tracked.
 
A further possibility is to force particles to decay into specific
decay channels; this may be particularly interesting for charm or
bottom physics. The choice of channels left open is determined by the
values of the switches \ttt{MDME(IDC,1)} for decay channel IDC
(use \ttt{PYLIST(12)} to obtain the full listing). One or several
channels may be left open;
in the latter case effective branching ratios are automatically
recalculated without the need for your intervention. It is also
possible to differentiate between which channels are left open for
particles and which for antiparticles. Lifetimes are not affected by
the exclusion of some decay channels. Note that, whereas forced decays
can enhance the efficiency for several kinds of studies, it can
also introduce unexpected biases, in particular when events may contain
several particles with forced decays, cf. section 
\ref{sss:resdecaycross}.
 
\subsubsection{Parameter values}
 
A non-trivial question is to know which parameter values to use. The
default values stored in the program are based on comparisons with 
LEP $\ee \to \Z^0$ data at around 91 GeV \cite{LEP90}, using a 
parton-shower picture followed by string fragmentation. Some examples
of more recent parameter sets are found in \cite{Kno96}. If 
fragmentation is indeed a universal phenomenon, as
we would like to think, then the same parameters should also apply at
other energies and in other processes. The former aspect, at least, 
seems to be borne out by comparisons with lower-energy PETRA/PEP
data and higher-energy LEP2 data. Note, however, that the choice of 
parameters is intertwined with 
the choice of perturbative QCD description. If instead matrix elements 
are used, a best fit to 30 GeV data would require the values 
\ttt{PARJ(21)=0.40}, \ttt{PARJ(41)=1.0} and \ttt{PARJ(42)=0.7}. 
With matrix elements one does not expect an energy
independence of the parameters, since the effective minimum invariant
mass cut-off is then energy dependent, i.e.\ so is the amount of soft
gluon emission effects lumped together with the fragmentation
parameters. This is indeed confirmed by the LEP data.
A mismatch in the perturbative QCD treatment could also
lead to small differences between different processes.
 
It is often said that the string fragmentation model contains a wealth
of parameters. This is certainly true, but it must be remembered that
most of these deal with flavour properties, and to a large extent
factorize from the treatment of the general event shape. In a fit to
the latter it is therefore usually enough to consider the parameters of
the perturbative QCD treatment, like $\Lambda$ in $\alphas$ and a
shower cut-off $Q_0$ (or $\alphas$ itself and $y_{\mmin}$, if matrix
elements are used), the $a$ and $b$ parameter of the Lund symmetric
fragmentation function (\ttt{PARJ(41)} and \ttt{PARJ(42)}) and the
width of the transverse momentum distribution
($\sigma = $\ttt{PARJ(21)}). In addition, the $a$ and $b$ parameters
are very strongly correlated by the requirement of having the correct
average multiplicity, such that in a typical $\chi^2$ plot, the
allowed region corresponds to a very narrow but very long valley,
stretched diagonally from small ($a$,$b$) pairs to large ones.
As to the flavour parameters, these are certainly many more, but most
of them are understood qualitatively within one single framework, that
of tunnelling pair production of flavours.
 
Since the use of independent fragmentation has fallen in disrespect, it
should be pointed out that the default parameters here are not
particularly well tuned to the data. This especially applies if one,
in addition to asking for independent fragmentation, also asks for
another setup of fragmentation functions, i.e.\ other than the standard
Lund symmetric one. In particular, note that most fits to the popular
Peterson/SLAC heavy-flavour fragmentation function are based
on the actual observed spectrum. In a Monte Carlo simulation, one must
then start out with something harder, to compensate for the energy lost
by initial-state photon radiation and gluon bremsstrahlung. Since
independent fragmentation is not collinear safe (i.e, the emission of
a collinear gluon changes the properties of the final event), the tuning
is strongly dependent on the perturbative QCD treatment chosen. All the
parameters needed for a tuning of independent fragmentation are
available, however.
 
\subsection{Examples}
 
A 10 GeV $\u$ quark jet going out along the $+z$ axis is generated
with
\begin{verbatim}
      CALL PY1ENT(0,2,10D0,0D0,0D0)
\end{verbatim}
Note that such a single jet is not required to conserve energy,
momentum or flavour. In the generation scheme, particles with
negative $p_z$ are produced as well, but these are automatically
rejected unless \ttt{MSTJ(3)=-1}. While frequently used in former
days, the one-jet generation option is not of much current interest.
 
In e.g.\ a leptoproduction event a typical situation could be a $\u$
quark going out in the $+z$ direction and a $\u\d_0$ target remnant
essentially at rest. (Such a process can be simulated by {\Py}, but
here we illustrate how to do part of it yourself.) The simplest
procedure is probably to treat the process in the c.m.\ frame
and boost it to the lab frame afterwards. Hence, if the c.m.\ energy is
20 GeV and the boost $\beta_z = 0.996$ (corresponding to
$x_B = 0.045$), then
\begin{verbatim}
      CALL PY2ENT(0,2,2101,20D0)
      CALL PYROBO(0,0,0D0,0D0,0D0,0D0,0.996D0)
\end{verbatim}
The jets could of course also be defined and allowed to fragment in the
lab frame with
\begin{verbatim}
      CALL PY1ENT(-1,2,223.15D0,0D0,0D0)
      CALL PY1ENT(2,12,0.6837D0,3.1416D0,0D0)
      CALL PYEXEC
\end{verbatim}
Note here that the target diquark is required to move in the backwards
direction with $E-p_z = m_{\p}(1-x_B)$ to obtain the correct
invariant mass for the system. This is, however, only an artifact of
using a fixed diquark mass to represent a varying target remnant mass,
and is of no importance for the fragmentation. If one wants a
nicer-looking event record, it is possible to use the following
\begin{verbatim}
      CALL PY1ENT(-1,2,223.15D0,0D0,0D0)
      MSTU(10)=1
      P(2,5)=0.938D0*(1D0-0.045D0)
      CALL PY1ENT(2,2101,0D0,0D0,0D0)
      MSTU(10)=2
      CALL PYEXEC
\end{verbatim}
 
A 30 GeV $\u\ubar\g$ event with $E_{\u} = 8$ GeV and
$E_{\ubar} = 14$ GeV is simulated with
\begin{verbatim}
      CALL PY3ENT(0,2,21,-2,30D0,2D0*8D0/30D0,2D0*14D0/30D0)
\end{verbatim}
The event will be given in a standard orientation with the $\u$
quark along the $+z$ axis and the $\ubar$ in the $\pm z, +x$ half plane.
Note that the flavours of the three partons have to be given in the
order they are found along a string, if string fragmentation options
are to work. Also note that, for 3-jet events, and particularly
4-jet ones, not all setups of kinematical variables $x$ lie within
the kinematically allowed regions of phase space.
 
All common block variables can obviously be changed by including the
corresponding common block in the user-written main program.
Alternatively, the routine \ttt{PYGIVE} can be used to feed in
values, with some additional checks on array bounds then performed.
A call
\begin{verbatim}
      CALL PYGIVE('MSTJ(21)=3;PMAS(C663,1)=210.;CHAF(401,1)=funnyino;'//
     &'PMAS(21,4)=')
\end{verbatim}
will thus change the value of \ttt{MSTJ(21)} to 3, the value of
\ttt{PMAS(PYCOMP(663),1) = PMAS(136,1)} to 210., the value of
\ttt{CHAF(401,1)} to 'funnyino', and print the current value of
\ttt{PMAS(21,4)}. Since old and new values of parameters changed are
written to output, this may offer a convenient way of documenting
non-default values used in a given run. On the other hand, if a
variable is changed back and forth frequently, the resulting
voluminous output may be undesirable, and a direct usage
of the common blocks is then to be recommended (the output can also be
switched off, see \ttt{MSTU(13)}).
 
A general rule of thumb is that none of the physics routines
(\ttt{PYSTRF}, \ttt{PYINDF}, \ttt{PYDECY}, etc.) should ever be
called directly, but only via \ttt{PYEXEC}. This routine may be
called repeatedly for one single event. At each call only those
entries that are allowed to fragment or decay, and have not yet
done so, are treated. Thus
\begin{verbatim}
      CALL PY2ENT(1,1,-1,20D0)       ! fill 2 partons without fragmenting
      MSTJ(1)=0                      ! inhibit jet fragmentation
      MSTJ(21)=0                     ! inhibit particle decay
      MDCY(PYCOMP(111),1)=0          ! inhibit pi0 decay
      CALL PYEXEC                    ! will not do anything
      MSTJ(1)=1                      !
      CALL PYEXEC                    ! partons will fragment, no decays
      MSTJ(21)=2                     !
      CALL PYEXEC                    ! particles decay, except pi0
      CALL PYEXEC                    ! nothing new can happen
      MDCY(PYCOMP(111),1)=1          !
      CALL PYEXEC                    ! pi0:s decay
\end{verbatim}
 
A partial exception to the rule above is \ttt{PYSHOW}. Its main
application is for internal use by \ttt{PYEEVT}, \ttt{PYDECY}, and
\ttt{PYEVNT}, but it can also be directly called by you. Note that a
special format for storing colour-flow information in \ttt{K(I,4)}
and \ttt{K(I,5)} must then be used. For simple cases, the \ttt{PY2ENT}
can be made to take care of that automatically, by calling with the
first argument negative.
\begin{verbatim}
      CALL PY2ENT(-1,1,-2,80D0)      ! store d ubar with colour flow
      CALL PYSHOW(1,2,80D0)          ! shower partons
      CALL PYEXEC                    ! subsequent fragmentation/decay
\end{verbatim}
For more complicated configurations, \ttt{PYJOIN} should be used.
 
It is always good practice to list one or a few events during a run to
check that the program is working as intended. With
\begin{verbatim}
      CALL PYLIST(1)
\end{verbatim}
all particles will be listed and in addition total charge, momentum and
energy of stable entries will be given. For string fragmentation these
quantities should be conserved exactly (up to machine precision errors),
and the same goes when running independent fragmentation with one of
the momentum conservation options. \ttt{PYLIST(1)} gives a format that
comfortably fits in an 80 column window, at the price of not giving
the complete story. With \ttt{PYLIST(2)} a more extensive listing is
obtained, and \ttt{PYLIST(3)} also gives vertex information. Further
options are available, like \ttt{PYLIST(12)}, which gives a list of
particle data.
 
An event, as stored in the \ttt{PYJETS} common block, will contain the
original partons and the whole decay chain, i.e.\ also particles which 
subsequently decayed. If parton showers are used, the amount of parton 
information is also considerable: first the on-shell partons before 
showers have been considered, then a \ttt{K(I,1)=22} line with total 
energy of the showering
subsystem, after that the complete shower history tree-like structure,
starting off with the same initial partons (now off-shell), and finally
the end products of the shower rearranged along the string directions.
This detailed record is useful in many connections, but if one only
wants to retain the final particles, superfluous information may be
removed with \ttt{PYEDIT}. Thus e.g.
\begin{verbatim}
      CALL PYEDIT(2)
\end{verbatim}
will leave you with the final charged and neutral particles, except
for neutrinos.
 
The information in \ttt{PYJETS} may be used directly to study an event.
Some useful additional quantities derived from these, such as charge
and rapidity, may easily be found via the \ttt{PYK} and \ttt{PYP}
functions. Thus electric charge \ttt{=PYP(I,6)} (as integer,
three times charge \ttt{=PYK(I,6)}) and true rapidity $y$ with respect
to the $z$ axis \ttt{= PYP(I,17)}.
 
A number of utility (\ttt{MSTU}, \ttt{PARU}) and physics (\ttt{MSTJ},
\ttt{PARJ}) switches and parameters are available in common block
\ttt{PYDAT1}. All of these have sensible default values. Particle data
is stored in common blocks \ttt{PYDAT2}, \ttt{PYDAT3} and \ttt{PYDAT4}.
Note that the data in the arrays \ttt{KCHG}, \ttt{PMAS}, \ttt{MDCY}
\ttt{CHAF} and \ttt{MWID} is not stored by {\KF} code, but by the compressed
code {\KC}. This code is not to be learnt by heart, but instead accessed
via the conversion function \ttt{PYCOMP}, \ttt{KC = PYCOMP(KF)}.
 
In the particle tables, the following particles are considered stable:
the photon, $\e^{\pm}$, $\mu^{\pm}$, $\pi^{\pm}$, $\K^{\pm}$, 
$\K_{\mrm{L}}^0$,
$\p$, $\pbar$, $\n$, $\br{\n}$ and all the neutrinos. It is, however,
always possible to inhibit the decay of any given particle by putting
the corresponding \ttt{MDCY} value zero or negative, e.g.
\ttt{MDCY(PYCOMP(310),1)=0} makes $\K_{\mrm{S}}^0$ and
\ttt{MDCY(PYCOMP(3122),1)=0} $\Lambda$ stable. It is also possible
to select stability based on the average lifetime (see \ttt{MSTJ(22)}),
or based on whether the decay takes place within a given spherical
or cylindrical volume around the origin.
 
The Field-Feynman jet model \cite{Fie78} is available in the program
by changing the following values: \ttt{MSTJ(1)=2} (independent
fragmentation), \ttt{MSTJ(3)=-1} (retain particles with $p_z < 0$;
is not mandatory), \ttt{MSTJ(11)=2} (choice of longitudinal
fragmentation function, with the $a$ parameter stored in
\ttt{PARJ(51) - PARJ(53)}), \ttt{MSTJ(12)=0} (no baryon production),
\ttt{MSTJ(13)=1} (give endpoint quarks $\pT$ as quarks created
in the field), \ttt{MSTJ(24)=0} (no mass broadening of resonances),
\ttt{PARJ(2)=0.5} ($\s/\u$ ratio for the production of new $\q\qbar$
pairs), \ttt{PARJ(11)=PARJ(12)=0.5}  (probability for mesons to
have spin 1) and \ttt{PARJ(21)=0.35} (width of Gaussian transverse
momentum distribution). In addition only $\d$, $\u$ and $\s$ single
quark jets may be generated following the FF recipe. Today the FF
`standard jet' concept is probably dead and buried, so the numbers
above should more be taken as an example of the flexibility of the
program, than as something to apply in practice.
 
A wide range of independent fragmentation options are implemented,
to be accessed with the master switch \ttt{MSTJ(1)=2}. In particular,
with \ttt{MSTJ(2)=1} a gluon jet is assumed to fragment like a
random $\d$, $\dbar$, $\u$, $\ubar$, $\s$ or $\sbar$ jet, while with
\ttt{MSTJ(2)=3} the gluon is split into a $\d\dbar$, $\u\ubar$ or
$\s\sbar$ pair of jets sharing the energy according to the
Altarelli-Parisi splitting function. Whereas energy, momentum and
flavour is not explicitly conserved in independent fragmentation, a
number of options are available in \ttt{MSTJ(3)} to ensure this
`post facto', e.g.\ \ttt{MSTJ(3)=1} will boost the event to ensure
momentum conservation and then (in the c.m.\ frame) rescale momenta by
a common factor to obtain energy conservation, whereas
\ttt{MSTJ(3)=4} rather uses a method of stretching the jets in
longitudinal momentum along the respective jet axis to keep angles
between jets fixed.
 
\clearpage
 
\section{Event Study and Analysis Routines}

After an event has been generated, one may wish to list it, or 
process it further in various ways. The first section below 
describes some simple study routines of this kind, while the 
subsequent ones describe more sophisticated analysis routines. 

To describe the complicated geometries encountered in multihadronic
events, a number of event measures have been introduced. These
measures are intended to provide a global view of the properties
of a given event, wherein the full information content of the event
is condensed into one or a few numbers. A steady stream of such
measures are proposed for different purposes. Many are rather
specialized or never catch on, but a few become standards, and are
useful to have easy access to. {\Py} therefore contains a
number of routines that can be called for any event, and that will
directly access the event record to extract the required information.
 
In the presentation below, measures have been grouped in three kinds.
The first contains simple event shape quantities, such as sphericity
and thrust. The second is jet finding algorithms.
The third is a mixed bag of particle multiplicities and compositions,
factorial moments and energy--energy correlations, put together
in a small statistics package.
 
None of the measures presented here are Lorentz invariant. The
analysis will be performed in whatever frame the event happens to
be given in. It is therefore up to you to decide whether the
frame in which events were generated is the right one, or whether
events beforehand should be boosted, e.g.\ to the c.m.\ frame. You
can also decide which particles you want to have affected by the
analysis.
 
\subsection{Event Study Routines}
\label{ss:eventstudy}
 
After a \ttt{PYEVNT} call, or another similar physics routine call,
the event generated is stored in the
\ttt{PYJETS} common block, and whatever physical variable is desired
may be constructed from this record. An event may be rotated, boosted
or listed, and particle data may be listed or modified. Via the
functions \ttt{PYK} and \ttt{PYP} the values of some frequently
appearing variables may be obtained more easily. As described in
subsequent sections, also more detailed event shape analyses
may be performed simply.
 
\drawbox{CALL PYROBO(IMI,IMA,THE,PHI,BEX,BEY,BEZ)}\label{p:PYROBO}
\begin{entry}
\itemc{Purpose:} to perform rotations and Lorentz boosts (in that
order, if both in the same call) of jet/particle momenta and vertex
position variables.
\iteme{IMI, IMA :} range of entries affected by transformation,
\ttt{IMI}$\leq$\ttt{I}$\leq$\ttt{IMA}. If 0 or below, \ttt{IMI}
defaults to 1 and \ttt{IMA} to \ttt{N}. Lower and upper bounds given
by positive \ttt{MSTU(1)} and \ttt{MSTU(2)} override the \ttt{IMI} 
and \ttt{IMA} values, and also the 1 to \ttt{N} constraint.
\iteme{THE, PHI :} standard polar coordinates $\theta, \varphi$,
giving the rotated direction of a momentum vector initially along
the $+z$ axis.
\iteme{BEX, BEY, BEZ :} gives the direction and size
\mbox{\boldmath $\beta$} of a Lorentz boost,
such that a particle initially at rest will have
$\mbf{p}/E = $\mbox{\boldmath $\beta$} afterwards.
\itemc{Remark:} all entries in the range \ttt{IMI}--\ttt{IMA} (or 
modified as described above) are affected, unless the status code 
of an entry is \ttt{K(I,1)}$\leq 0$.
\end{entry}
 
\drawbox{CALL PYEDIT(MEDIT)}\label{p:PYEDIT}
\begin{entry}
\itemc{Purpose:} to exclude unstable or undetectable jets/particles
from the event record. One may also use \ttt{PYEDIT} to store spare
copies of events (specifically initial parton configuration) that can
be recalled to allow e.g.\ different fragmentation schemes to be run
through with one and the same parton configuration. Finally, an
event which has been analyzed with \ttt{PYSPHE}, \ttt{PYTHRU} or
\ttt{PYCLUS} (see section \ref{ss:evanrout}) may be rotated to align 
the event axis with the $z$ direction.
\iteme{MEDIT :} tells which action is to be taken.
\begin{subentry}
\iteme{= 0 :} empty (\ttt{K(I,1)=0}) and documentation 
(\ttt{K(I,1)>20}) lines
are removed. The jets/particles remaining are compressed in the
beginning of the \ttt{PYJETS} common block and the \ttt{N} value is
updated accordingly. The event history is lost, so that information
stored in \ttt{K(I,3)}, \ttt{K(I,4)} and \ttt{K(I,5)} is no longer
relevant.
\iteme{= 1 :} as \ttt{=0}, but in addition all jets/particles that have
fragmented/decayed (\ttt{K(I,1)>10}) are removed.
\iteme{= 2 :} as \ttt{=1}, but also all neutrinos and unknown particles
(i.e.\ compressed code {\KC}$ = 0$) are removed.
\iteme{= 3 :} as \ttt{=2}, but also all uncharged, colour neutral
particles are removed, leaving only charged, stable particles (and
unfragmented partons, if fragmentation has not been performed).
\iteme{= 5 :} as \ttt{=0}, but also all partons which have branched or
been rearranged in a parton shower and all particles which have
decayed are removed, leaving only the fragmenting parton
configuration and the final-state particles.
\iteme{= 11 :} remove lines with \ttt{K(I,1)<0}. Update event
history information (in \ttt{K(I,3) - K(I,5)}) to refer to remaining
entries.
\iteme{= 12 :} remove lines with \ttt{K(I,1)=0}. Update event
history information (in \ttt{K(I,3) - K(I,5)}) to refer to remaining
entries.
\iteme{= 13 :} remove lines with \ttt{K(I,1)}$ = 11$, 12 or 15, except
for any line with \ttt{K(I,2)=94}. Update event history information
(in \ttt{K(I,3) - K(I,5)}) to refer to remaining entries. In
particular, try to trace origin of daughters, for which the mother
is decayed, back to entries not deleted.
\iteme{= 14 :} remove lines with \ttt{K(I,1)}$ = 13$ or 14, and also
any line with \ttt{K(I,2)=94}. Update event history information
(in \ttt{K(I,3) - K(I,5)}) to refer to remaining entries. In
particular, try to trace origin of rearranged jets back through
the parton-shower history to the shower initiator.
\iteme{= 15 :} remove lines with \ttt{K(I,1)>20}. Update event
history information (in \ttt{K(I,3) - K(I,5)}) to refer to remaining
entries.
\iteme{= 16 :} try to reconstruct missing daughter pointers of 
decayed particles from the mother pointers of decay products.
These missing pointers typically come from the need to use
\ttt{K(I,4)} and \ttt{K(I,5)} also for colour flow information.
\iteme{= 21 :} all partons/particles in current event record are
stored (as a spare copy) in bottom of common block \ttt{PYJETS}
(is e.g.\ done to save original partons before calling \ttt{PYEXEC}).
\iteme{= 22 :} partons/particles stored in bottom of event record with
\ttt{=21} are placed in beginning of record again, overwriting previous
information there (so that e.g.\ a different fragmentation scheme
can be used on the same partons). Since the copy at bottom is
unaffected, repeated calls with \ttt{=22} can be made.
\iteme{= 23 :} primary partons/particles in the beginning of event
record are marked as not fragmented or decayed, and number of entries
\ttt{N} is updated accordingly. Is a simple substitute for \ttt{=21} plus
\ttt{=22} when no fragmentation/decay products precede any of the
original partons/particles.
\iteme{= 31 :} rotate largest axis, determined by \ttt{PYSPHE},
\ttt{PYTHRU} or \ttt{PYCLUS}, to sit along the $z$ direction, and the
second largest axis into the $xz$ plane. For \ttt{PYCLUS} it can be
further specified to $+z$ axis and $xz$ plane with $x > 0$,
respectively. Requires that one of these routines has been called
before.
\iteme{= 32 :} mainly intended for \ttt{PYSPHE} and \ttt{PYTHRU}, this
gives a further alignment of the event, in addition to the one implied
by \ttt{=31}. The `slim' jet, defined as the side ($z > 0$ or $z < 0$)
with the smallest summed $\pT$ over square root of number of particles,
is rotated into the $+z$ hemisphere. In the opposite hemisphere
(now $z < 0$), the side of $x > 0$ and $x < 0$ which has the largest
summed $|p_z|$ is rotated into the $z < 0, x > 0$ quadrant.
Requires that \ttt{PYSPHE} or \ttt{PYTHRU} has been called before.
\end{subentry}
\itemc{Remark:} all entries 1 through \ttt{N} are affected by the
editing. For options 0--5 lower and upper bounds can be explicitly 
given by \ttt{MSTU(1)} and \ttt{MSTU(2)}.
\end{entry}
 
\drawbox{CALL PYLIST(MLIST)}\label{p:PYLIST}
\begin{entry}
\itemc{Purpose:} to list an event, jet or particle data, or current
parameter values.
\iteme{MLIST :} determines what is to be listed.
\begin{subentry}
\iteme{= 0 :} writes a title page with program version number and last 
date of change; is mostly for internal use.
\iteme{= 1 :} gives a simple list of current event record, in an 80
column format suitable for viewing directly in a terminal window.
For each entry, the following information is given: the entry
number \ttt{I}, the parton/particle name (see below), the status code
(\ttt{K(I,1)}), the flavour code {\KF} (\ttt{K(I,2)}), the line number
of the mother (\ttt{K(I,3)}), and the three-momentum, energy and mass
(\ttt{P(I,1) - P(I,5)}). If \ttt{MSTU(3)} is non-zero, lines immediately
after the event record proper are also listed. A final line
contains information on total charge, momentum, energy and
invariant mass. \\
The particle name is given by a call to the routine \ttt{PYNAME}.
For an entry which has decayed/fragmented (\ttt{K(I,1)=} 11--20), 
this particle name is given within parentheses. Similarly, a
documentation line (\ttt{K(I,1)=} 21--30) has the name enclosed in
expression signs (!\ldots!) and an event/jet axis information line
the name within inequality signs ($<$\ldots$>$). If the last
character of the name is a `?', it is a signal that the complete name
has been truncated to fit in, and can therefore not be trusted;
this is very rare. For partons which have been arranged along
strings (\ttt{K(I,1)=} 1, 2, 11 or 12), the end of the parton name
column contains information about the colour string arrangement:
an \ttt{A} for the first entry of a string, an \ttt{I} for all
intermediate ones, and a \ttt{V} for the final one (a poor man's
rendering of a vertical doublesided arrow, $\updownarrow$). \\
It is possible to insert lines just consisting of sequences
of \ttt{======} to separate different sections of the event record,
see \ttt{MSTU(70) - MSTU(80)}.
\iteme{= 2 :} gives a more extensive list of the current event record,
in a 132 column format, suitable for wide terminal windows.
For each entry, the following information is given: the entry
number \ttt{I}, the parton/particle name (with padding as described
for \ttt{=1}), the status code (\ttt{K(I,1)}), the flavour code
{\KF} (\ttt{K(I,2)}), the line number of the mother (\ttt{K(I,3)}), the
decay product/colour-flow pointers (\ttt{K(I,4), K(I,5)}), and the
three-momentum, energy and mass (\ttt{P(I,1) - P(I,5)}). If
\ttt{MSTU(3)} is non-zero, lines immediately after the event record
proper are also listed. A final line contains information on total
charge, momentum, energy and invariant mass. Lines with only
\ttt{======} may be inserted as for \ttt{=1}.
\iteme{= 3 :} gives the same basic listing as \ttt{=2}, but with an
additional line for each entry containing information on production
vertex position and time (\ttt{V(I,1) - V(I,4)}) and, for unstable
particles, proper lifetime (\ttt{V(I,5)}).
\iteme{= 4 :} is a simpler version of the listing obtained with 
\ttt{=2},but excluding the momentum information, and instead 
providing the alternative Les Houches style colour tags used for 
final-state colour reconnection in the new multiple interactions 
scenario. It thus allows colour flow to be studied and debugged in a 
"leaner" format.
\iteme{= 5 :} gives a simple listing of the event record stored in
the \ttt{HEPEVT} common block. This is mainly intended as a tool to 
check how conversion with the \ttt{PYHEPC} routine works. The listing 
does not contain vertex information, and the flavour code is not 
displayed as a name.  
\iteme{= 7 :} gives a simple listing of the parton-level event record 
for an external process, as stored in the \ttt{HEPEUP} common block. 
This is mainly intended as a tool to check how reading from 
\ttt{HEPEUP}  input works. The listing does not contain lifetime or 
spin information, and the flavour code is not displayed as a name. 
It also does not show other \ttt{HEPEUP} numbers, such as the event 
weight.  
\iteme{= 11 :} provides a simple list of all parton/particle codes
defined in the program, with {\KF} code and corresponding particle 
name. The list is grouped by particle kind, and only within each 
group in ascending order.
\iteme{= 12 :} provides a list of all parton/particle and decay data
used in the program. Each parton/particle code is represented by one
line containing {\KF} flavour code, {\KC} compressed code, particle
name, antiparticle name (where appropriate), electrical and
colour charge and presence or not of an antiparticle (stored in 
\ttt{KCHG}), mass, resonance width and maximum broadening, average 
proper lifetime (in \ttt{PMAS}) and
whether the particle is considered stable or not (in \ttt{MDCY}).
Immediately after a particle, each decay channel gets one line,
containing decay channel number (\ttt{IDC} read from \ttt{MDCY}),
on/off switch for the channel, matrix element type (\ttt{MDME}),
branching ratio (\ttt{BRAT}), and decay products (\ttt{KFDP}).
The \ttt{MSTU(1)} and \ttt{MSTU(2)} flags can be used to set the 
range of {\KF} codes for which particles are listed.
\iteme{= 13 :} gives a list of current parameter values for
\ttt{MSTU}, \ttt{PARU}, \ttt{MSTJ} and \ttt{PARJ}, and the first
200 entries of \ttt{PARF}. This is useful to keep check of which
default values were changed in a given run.
\end{subentry}
\itemc{Remark:} for options 1--3 and 12 lower and upper bounds of the
listing can be explicitly given by \ttt{MSTU(1)} and \ttt{MSTU(2)}.
\end{entry}
 
\drawbox{KK = PYK(I,J)}\label{p:PYK}
\begin{entry}
\itemc{Purpose:} to provide various integer-valued event data. Note
that many of the options available (in particular \ttt{I}$ > 0$,
\ttt{J}$\geq 14$) which refer to event history will not work after a
\ttt{PYEDIT} call. Further, the options 14--18 depend on the way the
event history has been set up, so with the explosion of different
allowed formats these options are no longer as safe as they may have
been. For instance, option 16 can only work if \ttt{MSTU(16)=2}. 
\iteme{I=0, J= :} properties referring to the complete event.
\begin{subentry}
\iteme{= 1 :} \ttt{N}, total number of lines in event record.
\iteme{= 2 :} total number of partons/particles remaining after
fragmentation and decay.
\iteme{= 6 :} three times the total charge of remaining (stable)
partons and particles.
\end{subentry}
\iteme{I>0, J= :} properties referring to the entry in line no.
\ttt{I} of the event record.
\begin{subentry}
\iteme{= 1 - 5 :} \ttt{K(I,1) - K(I,5)}, i.e.\ parton/particle status 
KS, flavour code {\KF} and origin/decay product/colour-flow information.
\iteme{= 6 :} three times parton/particle charge.
\iteme{= 7 :} 1 for a remaining entry, 0 for a decayed, fragmented or
documentation entry.
\iteme{= 8 :} {\KF} code (\ttt{K(I,2)}) for a remaining entry, 0 for a
decayed, fragmented or documentation entry.
\iteme{= 9 :} {\KF} code (\ttt{K(I,2)}) for a parton (i.e.\ not colour
neutral entry), 0 for a particle.
\iteme{= 10 :} {\KF} code (\ttt{K(I,2)}) for a particle (i.e.\ colour
neutral entry), 0 for a parton.
\iteme{= 11 :} compressed flavour code {\KC}.
\iteme{= 12 :} colour information code, i.e.\ 0 for colour neutral,
1 for colour triplet, -1 for antitriplet and 2 for octet.
\iteme{= 13 :} flavour of `heaviest' quark or antiquark (i.e.\ with
largest code) in hadron or diquark (including sign for antiquark),
0 else.
\iteme{= 14 :} generation number. Beam particles or virtual exchange
particles are generation 0, original jets/particles generation
1 and then 1 is added for each step in the fragmentation/decay
chain.
\iteme{= 15 :} line number of ancestor, i.e.\ predecessor in first
generation (generation 0 entries are disregarded).
\iteme{= 16 :} rank of a hadron in the jet it belongs to. Rank denotes
the ordering in flavour space, with hadrons containing the original
flavour of the jet having rank 1, increasing by 1 for each step
away in flavour ordering. All decay products inherit the rank
of their parent. Whereas the meaning of a first-rank hadron
in a quark jet is always well-defined, the definition of higher
ranks is only meaningful for independently fragmenting quark
jets. In other cases, rank refers to the ordering in the actual
simulation, which may be of little interest.
\iteme{= 17 :} generation number after a collapse of a parton system into
one particle, with 0 for an entry not coming from a collapse, and
-1 for entry with unknown history. A particle formed in a
collapse is generation 1, and then one is added in each decay
step.
\iteme{= 18 :} number of decay/fragmentation products (only defined
in a collective sense for fragmentation).
\iteme{= 19 :} origin of colour for showering parton, 0 else.
\iteme{= 20 :} origin of anticolour for showering parton, 0 else.
\iteme{= 21 :} position of colour daughter for showering parton,
0 else.
\iteme{= 22 :} position of anticolour daughter for showering parton,
0 else.
\end{subentry}
\end{entry}
 
\drawbox{PP = PYP(I,J)}\label{p:PYP}
\begin{entry}
\itemc{Purpose:} to provide various real-valued event data. Note that
some of the options available (\ttt{I}$ > 0$, \ttt{J}$ = 20$--25),
which are primarily intended for studies of systems in their
respective c.m.\ frame, requires that a \ttt{PYEXEC} call has been
made for the current initial parton/particle configuration, but that
the latest \ttt{PYEXEC} call has not been followed by a \ttt{PYROBO}
one.
\iteme{I=0, J= :} properties referring to the complete event.
\begin{subentry}
\iteme{= 1 - 4 :} sum of $p_x$, $p_y$, $p_z$ and $E$, respectively,
for the stable remaining entries.
\iteme{= 5 :} invariant mass of the stable remaining entries.
\iteme{= 6 :} sum of electric charge of the stable remaining entries.
\end{subentry}
\iteme{I>0, J= :} properties referring to the entry in line no.
\ttt{I} of the event record.
\begin{subentry}
\iteme{= 1 - 5 :} \ttt{P(I,1) - P(I,5)}, i.e.\ normally $p_x$, $p_y$,
$p_z$, $E$ and $m$ for jet/particle.
\iteme{= 6 :} electric charge $e$.
\iteme{= 7 :} squared momentum $|\mbf{p}|^2 = p_x^2 + p_y^2 + p_z^2$.
\iteme{= 8 :} absolute momentum $|\mbf{p}|$.
\iteme{= 9 :} squared transverse momentum $\pT^2 = p_x^2 + p_y^2$.
\iteme{= 10 :} transverse momentum $\pT$.
\iteme{= 11 :} squared transverse mass
$m_{\perp}^2 = m^2 + p_x^2 + p_y^2$.
\iteme{= 12 :} transverse mass $m_{\perp}$.
\iteme{= 13 - 14 :} polar angle $\theta$ in radians (between 0 and
$\pi$) or degrees, respectively.
\iteme{= 15 - 16 :} azimuthal angle $\varphi$ in radians (between
$-\pi$ and $\pi$)  or degrees, respectively.
\iteme{= 17 :} true rapidity $y = (1/2) \, \ln((E+p_z)/(E-p_z))$.
\iteme{= 18 :} rapidity $y_{\pi}$ obtained by assuming that the
particle is a pion when calculating the energy $E$, to be used in the
formula above, from the (assumed known) momentum $\mbf{p}$.
\iteme{= 19 :} pseudorapidity $\eta = (1/2) \, \ln((p+p_z)/(p-p_z))$.
\iteme{= 20 :} momentum fraction $x_p = 2|\mbf{p}|/W$, where $W$
is the total energy of the event, i.e. of the initial jet/particle 
configuration.
\iteme{= 21 :} $x_{\mrm{F}} = 2p_z/W$ (Feynman-$x$ if system is 
studied in the c.m.\ frame).
\iteme{= 22 :} $x_{\perp} = 2\pT/W$.
\iteme{= 23 :} $x_E = 2E/W$.
\iteme{= 24 :} $z_+ = (E+p_z)/W$.
\iteme{= 25 :} $z_- = (E-p_z)/W$.
\end{subentry}
\end{entry}
 
\drawbox{COMMON/PYDAT1/MSTU(200),PARU(200),MSTJ(200),PARJ(200)}%
\begin{entry}
\itemc{Purpose:} to give access to a number of status codes that
regulate the behaviour of the event study routines. The main 
reference for \ttt{PYDAT1} is in section \ref{ss:JETswitch}.
 
\boxsep

\iteme{MSTU(1),MSTU(2) :}\label{p:MSTU1} (D=0,0) can be used to replace 
the ordinary
lower and upper limits (normally 1 and \ttt{N}) for the action of
\ttt{PYROBO}, and most \ttt{PYEDIT} and \ttt{PYLIST} calls. Are reset
to 0 in a \ttt{PYEXEC} call.
 
\iteme{MSTU(3) :} (D=0) number of lines with extra information added
after line \ttt{N}. Is reset to 0 in a \ttt{PYEXEC} call, or in an
\ttt{PYEDIT} call when particles are removed.

\iteme{MSTU(11) :} (D=6) file number to which all program output is
directed. It is your responsibility to see to it that the
corresponding file is also opened for output.
 
\iteme{MSTU(12) :} (D=1) writing of title page (version number and last
date of change for {\Py}) on output file.
\begin{subentry}
\iteme{= 0 :} not done.
\iteme{= 1 :} title page is written at first occasion, at which time
\ttt{MSTU(12)} is set =0.
\end{subentry}

\iteme{MSTU(32) :} (I) number of entries stored with
\ttt{PYEDIT(21)} call.
 
\iteme{MSTU(33) :} (I) if set 1 before a \ttt{PYROBO} call, the
\ttt{V} vectors (in the particle range to be rotated/boosted) are
set 0 before the rotation/boost. \ttt{MSTU(33)} is set back to 0 in
the \ttt{PYROBO} call.
 
\iteme{MSTU(70) :} (D=0) the number of lines consisting only of equal
signs (\ttt{======}) that are inserted in the event listing obtained
with \ttt{PYLIST(1)}, \ttt{PYLIST(2)} or \ttt{PYLIST(3)}, so as to
distinguish different sections of the event record on output. At most
10 such lines can be inserted; see \ttt{MSTU(71) - MSTU(80)}. Is reset
at \ttt{PYEDIT} calls with arguments 0--5.
 
\iteme{MSTU(71) - MSTU(80) :} line numbers below which lines
consisting only of equal signs (\ttt{======}) are inserted in event
listings. Only the first \ttt{MSTU(70)} of the 10 allowed positions
are enabled.

\end{entry}
 
\subsection{Event Shapes}
\label{ss:evshape}
 
In this section we study general event shape variables: sphericity,
thrust, Fox-Wolfram moments, and jet masses. These measures are
implemented in the routines \ttt{PYSPHE}, \ttt{PYTHRU}, \ttt{PYFOWO}
and \ttt{PYJMAS}, respectively.
 
Each event is assumed characterized by the particle four-momentum
vectors $p_i = (\mbf{p}_i, E_i)$, with $i = 1, 2, \cdots , n$ an
index running over the particles of the event.
 
\subsubsection{Sphericity}
 
The sphericity tensor is defined as \cite{Bjo70}
\begin{equation}
S^{\alpha \beta} = \frac{\displaystyle \sum_i p^{\alpha}_{i} \,
   p^{\beta}_{i} } {\displaystyle \sum_i |\mbf{p}_i|^2 }  ~,
\end{equation}
where $\alpha, \beta = 1, 2, 3$ corresponds to the $x$, $y$ and $z$
components.
By standard diagonalization of $S^{\alpha \beta}$ one may find three
eigenvalues $\lambda_1 \geq \lambda_2 \geq \lambda_3$, with
$\lambda_1 + \lambda_2 + \lambda_3 = 1$. The sphericity of the event
is then defined as
\begin{equation}
S = \frac{3}{2} \, (\lambda_2 + \lambda_3) ~,
\end{equation}
so that $0 \leq S \leq 1$. Sphericity is essentially a measure of the
summed $\pT^2$ with respect to the event axis; a 2-jet event
corresponds to $S \approx 0$ and an isotropic event to $S \approx 1$.
 
The aplanarity $A$, with definition $A = \frac{3}{2} \lambda_3$, is
constrained to the range $0 \leq A \leq \frac{1}{2}$. It measures the
transverse momentum component out of the event plane: a planar event
has $A \approx 0$ and an isotropic one $A \approx \frac{1}{2}$.
 
Eigenvectors $\mbf{v}_j$ can be found that correspond to the three
eigenvalues $\lambda_j$ of the sphericity tensor. The $\mbf{v}_1$
one is called the sphericity axis (or event axis, if it is clear
from the context that sphericity has been used), while the sphericity
event plane is spanned by $\mbf{v}_1$ and $\mbf{v}_2$.
 
The sphericity tensor is quadratic in particle momenta. This means
that the sphericity
value is changed if one particle is split up into two collinear ones
which share the original momentum. Thus sphericity is
not an infrared safe quantity in QCD perturbation theory. A useful
generalization of the sphericity tensor is
\begin{equation}
S^{(r) \alpha \beta} = \frac{\displaystyle \sum_i |\mbf{p}_i|^{r-2} \,
    p^{\alpha}_{i} \, p^{\beta}_{i} }{\displaystyle
    \sum_i |\mbf{p}_i|^r }  ~,
\end{equation}
where $r$ is the power of the momentum dependence. While $r = 2$
thus corresponds to sphericity, $r = 1$ corresponds to linear
measures calculable in perturbation theory \cite{Par78}:
\begin{equation}
S^{(1) \alpha \beta} = \frac{\displaystyle \sum_i \frac{\displaystyle
    p^{\alpha}_{i} \, p^{\beta}_{i}}{\displaystyle |\mbf{p}_i|} }
    {\displaystyle \sum_i |\mbf{p}_i| } ~.
\end{equation}
 
Eigenvalues and eigenvectors may be defined exactly as before, and
therefore also equivalents of $S$ and $A$. These have no standard
names; we may call them linearized sphericity $S_{\mrm{lin}}$ and
linearized aplanarity $A_{\mrm{lin}}$. Quantities derived from the
linear matrix that are standard in the literature are instead the 
combinations \cite{Ell81}
\begin{eqnarray}
   C & = & 3 ( \lambda_1 \lambda_2 + \lambda_1 \lambda_3 +
   \lambda_2 \lambda_3 ) ~,    \\
   D & = & 27 \lambda_1 \lambda_2 \lambda_3 ~.
\end{eqnarray}
Each of these is constrained to be in the range between 0 and 1.
Typically, $C$ is used to measure the 3-jet structure and $D$
the 4-jet one, since $C$ is vanishing for a perfect 2-jet event
and $D$ is vanishing for a planar event. The $C$ measure is related
to the second Fox-Wolfram moment (see below), $C = 1 - H_2$.
 
Noninteger $r$ values may also be used, and corresponding generalized
sphericity and aplanarity measures calculated. While perturbative
arguments favour $r = 1$, we know that the fragmentation `noise', e.g.
from transverse momentum fluctuations, is
proportionately larger for low momentum particles, and so $r > 1$
should be better for experimental event axis determinations. The use
of too large an $r$ value, on the other hand, puts all the emphasis
on a few high-momentum particles, and therefore involves a loss of
information. It should then come as no surprise that intermediate
$r$ values, of around 1.5, gives the best performance for event
axis determinations in 2-jet events, where the theoretical
meaning of the event axis is well-defined. The gain in accuracy
compared with the more conventional choices $r=2$ or $r=1$ is
rather modest, however.
 
\subsubsection{Thrust}
 
The quantity thrust $T$ is defined by \cite{Bra64}
\begin{equation}
T = \max_{|\mbf{n}| = 1} \,
\frac{\displaystyle \sum_i |\mbf{n} \cdot \mbf{p}_i|}
{\displaystyle \sum_i |\mbf{p}_i|} ~,
\label{em:thrust}
\end{equation}
and the thrust axis $\mbf{v}_1$ is given by the $\mbf{n}$ vector
for which maximum is attained. The allowed range is
$1/2 \leq T \leq 1$, with a 2-jet event corresponding
to $T \approx 1$ and an isotropic event to $T \approx 1/2$.
 
In passing, we note that this is not the only definition found in
the literature. The definitions agree for events studied in the 
c.m.\ frame and where all particles are detected. However, a 
definition like
\begin{equation}
T = 2 \, \max_{|\mbf{n}| = 1} \,
\frac{\displaystyle \left| \sum_i \theta(\mbf{n} \cdot \mbf{p}_i) \,
\mbf{p}_i \right| }{\displaystyle \sum_i |\mbf{p}_i|} =
2 \, \max_{\theta_i = 0,1} \,
\frac{\displaystyle \left| \sum_i \theta_i \, \mbf{p}_i \right| }
{\displaystyle \sum_i |\mbf{p}_i|}
\label{em:thrusttwo}
\end{equation}
(where $\theta(x)$ is the step function, $\theta(x) = 1 $ if $x > 0$,
else $\theta(x) = 0$) gives different results than the one above
if e.g.\ only charged particles are detected. It would even be possible
to have $T > 1$; to avoid such problems, often an extra fictitious
particle is introduced to balance the total momentum \cite{Bra79}.
 
Eq.~(\ref{em:thrust}) may be rewritten as
\begin{equation}
T = \max_{\epsilon_i = \pm 1} \,
\frac{\displaystyle \left| \sum_i \epsilon_i \, \mbf{p}_i \right| }
{\displaystyle \sum_i |\mbf{p}_i|} ~.
\label{em:thrustthree}
\end{equation}
(This may also be viewed as applying eq.~(\ref{em:thrusttwo}) to an
event with $2 n$ particles, $n$ carrying the momenta $\mbf{p}_i$
and $n$ the momenta $- \mbf{p}_i$, thus automatically balancing
the momentum.) To find the thrust value and axis this way, $2^{n-1}$
different possibilities would have to be tested. The reduction by a
factor of 2 comes from $T$ being unchanged when all
$\epsilon_i \to - \epsilon_i$. Therefore this approach rapidly becomes
prohibitive. Other exact methods exist, which `only' require about
$4n^2$ combinations to be tried.
 
In the implementation in {\Py}, a faster alternative method is
used, in which the thrust axis is iterated from a starting direction
$\mbf{n}^{(0)}$ according to
\begin{equation}
\mbf{n}^{(j+1)} = \frac{\displaystyle \sum_i \epsilon(
\mbf{n}^{(j)} \cdot \mbf{p}_i) \, \mbf{p}_i }
{\displaystyle \left| \sum_i \epsilon(
\mbf{n}^{(j)} \cdot \mbf{p}_i) \, \mbf{p}_i \right| }
\end{equation}
(where $\epsilon(x) = 1$ for $x > 0$ and $\epsilon(x) = -1$ for
$x < 0$). It is easy to show that the related thrust value will
never decrease, $T^{(j+1)} \geq T^{(j)}$. In fact, the method
normally converges in 2--4 iterations. Unfortunately, this
convergence need not be towards the correct thrust axis but is
occasionally only towards a local maximum of the thrust function
\cite{Bra79}. We know of no foolproof way around this complication,
but the danger of an error may be lowered if several different
starting axes $\mbf{n}^{(0)}$ are tried and found to agree. These
$\mbf{n}^{(0)}$ are suitably constructed from the $n'$ (by default
4) particles with the largest momenta in the event, and the
$2^{n' -1}$ starting directions $\sum_i \epsilon_i \, \mbf{p}_i$
constructed from these are tried in falling order of the
corresponding absolute momentum values. When a predetermined number
of the starting axes have given convergence towards the same
(best) thrust axis this one is accepted.
 
In the plane perpendicular to the thrust axis, a major \cite{MAR79}
axis and value may be defined in just the same fashion as thrust, i.e.
\begin{equation}
M_a = \max_{|\mbf{n}| = 1, \, \mbf{n} \cdot \mbf{v}_1 = 0} \,
\frac{\displaystyle \sum_i |\mbf{n} \cdot \mbf{p}_i|}
{\displaystyle \sum_i |\mbf{p}_i| }    ~.
\end{equation}
In a plane more efficient methods can be used to find an axis than in
three dimensions \cite{Wu79}, but for simplicity we use the same
method as above. Finally, a third axis, the minor axis, is defined
perpendicular to the thrust and major ones, and a minor value $M_i$
is calculated just as thrust and major.
The difference between major and minor is called
oblateness, $O = M_a -M_i$. The upper limit on oblateness depends
on the thrust value in a not so simple way. In general
$O \approx 0$ corresponds to an event symmetrical around the thrust
axis and high $O$ to a planar event.
 
As in the case of sphericity, a generalization to arbitrary momentum
dependence may easily be obtained, here by replacing the $\mbf{p}_i$
in the formulae above by $|\mbf{p}_i|^{r-1} \, \mbf{p}_i$. This
possibility is included, although so far it has not found any
experimental use.
 
\subsubsection{Fox-Wolfram moments}
 
The Fox-Wolfram moments $H_l$, $l = 0, 1, 2, \ldots$, are defined by
\cite{Fox79}
\begin{equation}
H_l = \sum_{i,j} \frac{ |\mbf{p}_i| \, |\mbf{p}_j| }
{E_{\mrm{vis}}^2} \, P_l (\cos \theta_{ij}) ~,
\end{equation}
where $\theta_{ij}$ is the opening angle between hadrons $i$ and
$j$ and $E_{\mrm{vis}}$ the total visible energy of the event. Note
that also autocorrelations, $i = j$, are included. The $P_l(x)$ are 
the Legendre polynomials,
\begin{eqnarray}
P_0(x) & = & 1 ~,   \nonumber \\
P_1(x) & = & x  ~,  \nonumber \\
P_2(x) & = & \frac{1}{2} \, (3x^2 -1) ~,    \nonumber \\
P_3(x) & = & \frac{1}{2} \, (5x^3 - 3x) ~,  \nonumber \\
P_4(x) & = & \frac{1}{8} \, (35x^4 - 30x^2 + 3)  ~.
\end{eqnarray}
To the extent that particle masses may be neglected, $H_0 \equiv 1$.
It is customary to normalize the results to $H_0$, i.e.\ to give
$H_{l0} = H_l / H_0$. If momentum is balanced then $H_1 \equiv 0$.
2-jet events tend to give $H_l \approx 1$ for $l$ even and
$\approx 0$ for $l$ odd.
 
\subsubsection{Jet masses}
 
The particles of an event may be divided into two classes.
For each class a squared invariant mass may be calculated, $M_1^2$
and $M_2^2$. If the assignment of particles is adjusted such that
the sum $M_1^2 + M_2^2$ is minimized, the two masses thus obtained
are called heavy and light jet mass, $M_{\mrm{H}}$ and $M_{\mrm{L}}$. 
It has been shown that these quantities are well behaved in perturbation
theory \cite{Cla79}. In $\ee$ annihilation, the heavy jet mass
obtains a contribution from $\q\qbar\g$ 3-jet events, whereas
the light mass is non-vanishing only when 4-jet events also
are included. In the c.m.\ frame of an event one has the limits
$0 \leq M_{\mrm{H}}^2 \leq E_{\mrm{cm}}^2/3$.
 
In general, the subdivision of particles tends to be into two
hemispheres, separated by a plane perpendicular to an event axis.
As with thrust, it is time-consuming to find the exact solution.
Different approximate strategies may therefore be used. In the
program, the sphericity axis is used to perform a fast subdivision
into two hemispheres, and thus into two preliminary jets. Thereafter
one particle at a time is tested to determine whether the sum
$M_1^2 + M_2^2$ would be decreased if that particle were to be
assigned to the other jet. The procedure is stopped when no
further significant change is obtained. Often the original
assignment is retained as it is, i.e.\ the sphericity axis gives
a good separation. This is not a full guarantee, since the program
might get stuck in a local minimum which is not the global one.
 
\subsection{Cluster Finding}
\label{ss:clustfind}
 
Global event measures, like sphericity or thrust, can only be used to
determine the jet axes for back-to-back 2-jet events. To determine
the individual jet axes in events with three or more jets,
or with two (main) jets which are not back-to-back, cluster
algorithms are customarily used. In these, nearby particles are
grouped together into a variable number of clusters. Each cluster
has a well-defined direction, given by a suitably weighted average
of the constituent particle directions.
 
The cluster algorithms traditionally used in $\ee$ and in $\pp$
physics differ in several respects. The former tend to be
spherically symmetric, i.e.\ have no preferred axis in space,
and normally all particles have to be assigned to some jet. The
latter pick the beam axis as preferred direction, and make use
of variables related to this choice, such as rapidity and transverse
momentum; additionally only a fraction of all particles are assigned
to jets.
 
This reflects a difference in the underlying physics: in
$\pp$ collisions, the beam remnants found at low transverse momenta
are not related to any hard processes, and therefore only provide an
unwanted noise to many studies. (Of course, also hard processes may
produce particles at low transverse momenta, but at a rate much
less than that from soft or semi-hard processes.) Further, the
kinematics of hard processes is, to a good approximation, factorized
into the hard subprocess itself, which is boost invariant in rapidity,
and parton-distribution effects, which determine the overall position
of a hard scattering in rapidity. Hence rapidity, azimuthal angle and
transverse momentum is a suitable coordinate frame to describe hard
processes in.
 
In standard $\ee$ annihilation events, on the other hand, the hard
process c.m.\ frame tends to be almost at rest, and the event
axis is just about randomly distributed in space, i.e.\ with no 
preferred r\^ole for the axis defined by the incoming $\e^{\pm}$.
All particle production is initiated by and related to the hard
subprocess. Some of the particles may be less easy to associate to
a specific jet, but there is no compelling reason to remove any
of them from consideration.
 
This does not mean that the separation above is always required.
$2\gamma$ events in $\ee$ may have a structure with `beam jets' and
`hard scattering' jets, for which the $\pp$ type algorithms might
be well suited. Conversely, a heavy particle produced in $\pp$
collisions could profitably be studied, in its own rest frame,
with $\ee$ techniques.
 
In the following, particles are only characterized by their
three-momenta or, alternatively, their energy and direction of
motion. No knowledge is therefore assumed of particle types, or
even of mass and charge. Clearly, the more is known, the more
sophisticated clustering algorithms can be used. The procedure
then also becomes more detector-dependent, and therefore less
suitable for general usage.
 
{\Py} contains two cluster finding routines. \ttt{PYCLUS} is of
the $\ee$ type and \ttt{PYCELL} of the $\pp$ one. Each of them
allows some variations of the basic scheme.
 
\subsubsection{Cluster finding in an $\ee$ type of environment}
\label{sss:clustfindee}
 
The usage of cluster algorithms for $\ee$ applications started in
the late 1970's. A number of different approaches were proposed
\cite{Bab80}, see review in \cite{Mor98}. Of these, we will here 
only discuss those based on binary joining. In this kind of
approach, initially each final-state particle is considered to be a
cluster. Using some distance measure, the two nearest clusters are
found. If their distance is smaller than some cut-off value,
the two clusters are joined into one. In this new configuration, the
two clusters that are now nearest are found and joined, and so on until
all clusters are separated by a distance larger than the cut-off.
The clusters remaining at the end are often also called jets. Note that,
in this approach, each single particle belongs to exactly one
cluster. Also note that the resulting jet picture explicitly depends
on the cut-off value used. Normally the number of clusters is allowed to
vary from event to event, but occasionally it is more useful to have
the cluster algorithm find a predetermined number of jets (like 3).
 
The obvious choice for a distance measure is to use squared invariant
mass, i.e.\ for two clusters $i$ and $j$ to define the
distance to be
\begin{equation}
   m^2_{ij} = (E_i + E_j)^2 - (\mbf{p}_i + \mbf{p}_j)^2 ~.
\end{equation}
(Equivalently, one could have used the invariant mass as measure
rather than its square; this is just a matter of convenience.)
In fact, a number of people (including one of the authors)
tried this measure long ago and gave up on it, since
it turns out to have severe instability problems. The reason is well
understood: in general, particles tend to cluster closer in
invariant mass in the region of small momenta. The clustering process
therefore tends to start in the center of the event, and only
subsequently spread outwards to encompass also the fast particles.
Rather than clustering slow particles around the fast ones (where the
latter na\"{\i}vely should best represent the jet directions), the 
invariant mass measure will tend to cluster fast particles around 
the slow ones.
 
Another instability may be seen by considering the clustering in a
simple 2-jet event. By the time that clustering has reached the
level of three clusters, the `best' the clustering algorithm can
possibly have achieved, in terms of finding three low-mass clusters,
is to have one fast cluster around each jet, plus a third slow cluster
in the middle. In the last step this third cluster would be joined with
one of the fast ones, to produce two final asymmetric clusters:
one cluster would contain all the slow particles, also those that
visually look like belonging to the opposite jet. A simple binary
joining process, with no possibility to reassign particles between
clusters, is therefore not likely to be optimal.
 
The solution adopted \cite{Sjo83} is to reject invariant
mass as distance measure. Instead a jet is defined as a collection
of particles which have a limited transverse momentum with respect to
a common jet axis, and hence also with respect to each other. This
picture is clearly inspired by the standard fragmentation picture,
e.g.\ in string fragmentation. A distance measure $d_{ij}$ between two
particles (or clusters) with momenta $\mbf{p}_i$ and $\mbf{p}_j$ should
thus not depend critically on the longitudinal momenta but only on the
relative transverse momentum. A number of such measures were tried,
and the one eventually selected was
\begin{equation}
d_{ij}^2 = \frac{1}{2} \, (|\mbf{p}_i| \, |\mbf{p}_j| -
\mbf{p}_i \cdot \mbf{p}_j) \, \frac{4 \, |\mbf{p}_i| \, |\mbf{p}_j|}
{(|\mbf{p}_i| + |\mbf{p}_j|)^2} =
\frac{4 \, |\mbf{p}_i|^2 \, |\mbf{p}_j|^2 \, \sin^2(\theta_{ij}/2)}
{(|\mbf{p}_i| + |\mbf{p}_j|)^2}   ~.
\end{equation}
 
For small relative angle $\theta_{ij}$, where
$2 \sin(\theta_{ij}/2) \approx \sin\theta_{ij}$ and
$\cos \theta_{ij} \approx 1$, this measure reduces to
\begin{equation}
d_{ij} \approx \frac{|\mbf{p}_i \times \mbf{p}_j|}
{|\mbf{p}_i + \mbf{p}_j|}  ~,
\end{equation}
where `$\times$' represents the cross product. We therefore see that
$d_{ij}$ in this limit has the simple physical interpretation as the
transverse momentum of either particle with respect to the direction
given by the sum of the two particle momenta. Unlike the approximate
expression, however, $d_{ij}$ does not vanish for two back-to-back
particles, but is here more related to the invariant mass
between them.
 
The basic scheme is of the binary joining type, i.e.\ initially each
particle is assumed to be a cluster by itself. Then the two clusters
with smallest relative distance $d_{ij}$ are found and, if
$d_{ij} < d_{\mrm{join}}$, with $d_{\mrm{join}}$ 
some predetermined distance, the two clusters are joined
to one, i.e.\ their four-momenta are added vectorially to give the
energy and momentum of the new cluster. This is repeated until the 
distance between any two clusters is $> d_{\mrm{join}}$. The number 
and momenta of these final clusters then represent our reconstruction 
of the initial jet configuration, and each particle is assigned to one 
of the clusters.
 
To make this scheme workable, two further ingredients are necessary,
however. Firstly, after two clusters have been joined, some particles
belonging to the new cluster may actually be closer to
another cluster. Hence, after each joining, all particles in the event
are reassigned to the closest of the clusters. For particle $i$, this
means that the distance $d_{ij}$ to all clusters $j$ in the event has
to be evaluated and compared. After all particles
have been considered, and only then, are cluster momenta recalculated
to take into account any reassignments. To save time, the assignment
procedure is not iterated until a stable configuration is reached,
but, since all particles are reassigned at each step, such an
iteration is effectively taking place in parallel with the cluster
joining. Only at the very end, when all $d_{ij} > d_{\mrm{join}}$, is 
the reassignment procedure iterated to convergence --- still with the
possibility to continue the cluster joining if some $d_{ij}$ should
drop below $d_{\mrm{join}}$ due to the reassignment.
 
Occasionally, it may occur that the reassignment step leads to an empty
cluster, i.e.\ one to which no particles are assigned. Since such a
cluster has a distance $d_{ij} = 0$ to any other cluster, it is
automatically removed in the next cluster joining. However, it is
possible to run the program in a mode where a minimum number of jets
is to be reconstructed. If this minimum is reached with one cluster
empty, the particle is found which has largest distance to the cluster
it belongs to. That cluster is then split into two, namely the
large-distance particle and a remainder. Thereafter the reassignment
procedure is continued as before.
 
Secondly, the large multiplicities normally encountered means that,
if each particle initially is to be treated as a separate cluster, the
program will become very slow. Therefore a smaller number of clusters,
for a normal $\ee$ event typically 8--12, is constructed as a starting
point for the iteration above, as follows. The particle with the
highest momentum is found, and thereafter all particles within a
distance $d_{ij} < d_{\mrm{init}}$ from it, where 
$d_{\mrm{init}} \ll d_{\mrm{join}}$.
Together these are allowed to form a single cluster. For the
remaining particles, not assigned to this cluster, the procedure is
iterated, until all particles have been used up. Particles in the
central momentum region, $|\mbf{p}| < 2d_{\mrm{init}}$ are treated
separately; if their vectorial momentum sum is above $2d_{\mrm{init}}$
they are allowed to form one cluster, otherwise they are left
unassigned in the initial configuration. The value of $d_{\mrm{init}}$,
as long as reasonably small, has no physical importance, in that the
same final cluster configuration will be found as if each particle
initially is assumed to be a cluster by itself: the particles
clustered at this step are so nearby anyway that they almost
inevitably must enter the same jet; additionally the reassignment
procedure allows any possible `mistake' to be corrected in later
steps of the iteration.
 
Thus the jet reconstruction depends on one single parameter, 
$d_{\mrm{join}}$, with a clearcut physical meaning of a transverse 
momentum `jet-resolution power'. Neglecting smearing from 
fragmentation, $d_{ij}$ between two clusters of equal energy 
corresponds to half the invariant mass of the two original partons. 
If one only wishes to reconstruct well separated jets, a large 
$d_{\mrm{join}}$ should be chosen, while a small $d_{\mrm{join}}$ 
would allow the separation of close jets, at the
cost of sometimes artificially dividing a single jet into two. In
particular, $\b$ quark jets may here be a nuisance. The value of
$d_{\mrm{join}}$ to use for a fixed jet-resolution power in principle
should be independent of the c.m.\ energy of events, although
fragmentation effects may give a contamination of spurious extra jets
that increases slowly with $E_{\mrm{cm}}$ for fixed $d_{\mrm{join}}$. 
Therefore a $d_{\mrm{join}} = 2.5$ GeV was acceptable at PETRA/PEP, 
while 3--4 GeV may be better for applications at LEP and beyond.
 
This completes the description of the main option of the \ttt{PYCLUS}
routine. Variations are possible. One such is to skip the reassignment
step, i.e.\ to make use only of the simple binary joining procedure,
without any possibility to reassign particles between jets.
(This option is included mainly as a reference, to check how important
reassignment really is.) The other main alternative is
to replace the distance measure used above with the one used in the
JADE algorithm \cite{JAD86}.
 
The JADE cluster algorithm is an attempt to save the invariant mass
measure. The distance measure is defined to be
\begin{equation}
   y_{ij} = \frac{2 E_i E_j (1-\cos \theta_{ij})}{E^2_{\mrm{vis}}} ~.
\end{equation}
Here $E_{\mrm{vis}}$ is the total visible energy of the event. The
usage of $E^2_{\mrm{vis}}$ in the denominator rather than 
$E_{\mrm{cm}}^2$ tends to make the measure less
sensitive to detector acceptance corrections; in addition the
dimensionless nature of $y_{ij}$ makes it well suited for a comparison
of results at different c.m.\ energies. For the subsequent discussions,
this normalization will be irrelevant, however.
 
The $y_{ij}$ measure is very closely related to the squared mass
distance measure: the two coincide (up to the difference in
normalization) if $m_i = m_j = 0$. However, consider a pair of
particles or clusters with non-vanishing individual masses and a fixed
pair mass. Then, the larger the net momentum of the pair, the
smaller the $y_{ij}$ measure. This somewhat tends to favour clustering
of fast particles, and makes the algorithm less unstable than the one
based on true invariant mass.
 
The successes of the JADE algorithm are well known: one obtains a good
agreement between the number of partons generated on the matrix-element
(or parton-shower) level and the number of clusters reconstructed from
the hadrons, such that QCD aspects like the running of $\alphas$ can
be studied with a small dependence on fragmentation effects. Of
course, the insensitivity to fragmentation effects depends on the choice
of fragmentation model. Fragmentation effects are small in the string
model, but not necessarily in independent fragmentation scenarios.
Although independent fragmentation in itself is not credible,
this may be seen as a signal for caution.
 
One should note that the JADE measure still suffers from some of the
diseases of the simple mass measure (without reassignments), namely
that particles which go in opposite directions may well be joined
into the same cluster. Therefore, while the JADE algorithm is a good
way to find the number of jets, it is inferior to the standard
$d_{ij}$ measure for a determination of jet directions and energies
\cite{Bet92}. The $d_{ij}$ measure also gives narrower jets, which
agree better with the visual impression of jet structure.
 
Later, the `Durham algorithm' was introduced \cite{Cat91},
which works as the JADE one but with a distance measure
\begin{equation}
\tilde{y}_{ij} = \frac{2 \min(E_i^2,E_j^2) (1-\cos \theta_{ij})}
{E^2_{cm}} ~.
\end{equation}
Like the $d_{ij}$ measure, this is a transverse momentum, but
$\tilde{y}_{ij}$ has the geometrical interpretation as the
transverse momentum of the softer particle with respect to the
direction of the harder one, while $d_{ij}$ is the transverse
momentum of either particle with respect to the common direction
given by the momentum vector sum. The two definitions agree when
one cluster is much softer than the other, so the soft gluon
exponentiation proven for the Durham measure also holds for the
$d_{ij}$ one.
 
The main difference therefore is that the standard \ttt{PYCLUS}
option allows reassignments, while the Durham algorithm does not.
The latter is therefore more easily calculable on the perturbative
parton level. This point is sometimes overstressed, and one could
give counterexamples why reassignments in fact may bring better
agreement with the underlying perturbative level. In particular,
without reassignments, one will make the recombination that seems
the `best' in the current step, even when that forces you to make
`worse' choices in subsequent steps. With reassignments, it is
possible to correct for mistakes due to the too local sensitivity of
a simple binary joining scheme.
 
\subsubsection{Cluster finding in a $\pp$ type of environment}
 
The \ttt{PYCELL} cluster finding routines is of the kind pioneered
by UA1 \cite{UA183}, and commonly used in $\pp$ physics. It is based
on a choice of pseudorapidity $\eta$, azimuthal angle $\varphi$ and
transverse momentum $\pT$ as the fundamental coordinates.
This choice is discussed in the introduction to cluster finding
above, with the proviso that the theoretically preferred true
rapidity has to be replaced by pseudorapidity, to make contact with
the real-life detector coordinate system.
 
A fix detector grid is assumed, with the pseudorapidity range
$|\eta| < \eta_{\mmax}$ and the full azimuthal range each divided
into a number of equally large bins, giving a rectangular grid.
The particles of an event impinge on this detector grid. For each
cell in ($\eta$,$\varphi$) space, the transverse energy (normally
$\approx \pT$) which enters that cell is summed up to give a total 
cell $E_{\perp}$ flow.
 
Clearly the model remains very primitive in a number of
respects, compared with a real detector. There is no magnetic field
allowed for, i.e.\ also charged particles move in straight tracks.
The dimensions of the detector are not specified; hence the positions
of the primary vertex and any secondary vertices are neglected when
determining which cell a particle belongs to. The rest mass of
particles is not taken into account, i.e.\ what is used is really
$\pT = \sqrt{p_x^2 + p_y^2}$, while in a real detector some
particles would decay or annihilate, and then deposit additional
amounts of energy.
 
To take into account the energy resolution of the detector, it is
possible to smear the $E_{\perp}$ contents, bin by bin. This is done
according to a Gaussian, with a width assumed proportional to the
$\sqrt{E_{\perp}}$ of the bin. The Gaussian is cut off at zero and
at some predetermined multiple of the unsmeared $E_{\perp}$, by default
twice it. Alternatively, the smearing may be performed in $E$
rather than in $E_{\perp}$. To find the $E$, it is assumed
that the full energy of a cell is situated at its center, so
that one can translate back and forth with
$E = E_{\perp} \cosh\eta_{\mrm{center}}$.
 
The cell with largest $E_{\perp}$ is taken as a jet initiator if its
$E_{\perp}$ is above some threshold. A candidate jet is defined to
consist of all cells which are within some given radius $R$ in the
($\eta$,$\varphi$) plane, i.e.\ which have 
$(\eta - \eta_{\mrm{initiator}})^2
 + (\varphi - \varphi_{\mrm{initiator}})^2 < R^2$.
Coordinates are always given with respect to the center of the
cell. If the summed $E_{\perp}$ of the jet is above the
required minimum jet energy, the candidate jet is accepted, and all
its cells are removed from further consideration. If not, the
candidate is rejected. The sequence is now repeated with the
remaining cell of highest $E_{\perp}$, and so on until no single
cell fulfils the jet initiator condition.
 
The number of jets reconstructed can thus vary from none to a maximum
given by purely geometrical considerations, i.e.\ how many circles of
radius $R$ are needed to cover the allowed ($\eta$,$\varphi$) plane.
Normally only a fraction of the particles are assigned to jets.
 
One could consider to iterate the jet assignment process, using the
$E_{\perp}$-weighted center of a jet to draw a new circle of radius
$R$. In the current algorithm there is no such iteration step.
For an ideal jet assignment it would also be necessary to improve
the treatment when two jet circles partially overlap.
 
A final technical note. A natural implementation of a cell finding
algorithm is based on having a two-dimensional array of $E_{\perp}$
values, with dimensions to match the detector grid.
Very often most of the cells would then
be empty, in particular for low-multiplicity events in fine-grained
calorimeters. Our implementation is somewhat atypical, since cells are
only reserved space (contents and position) when they are shown to be
non-empty. This means that all non-empty cells have to be looped over
to find which are within the required distance $R$ of a potential jet
initiator. The algorithm is therefore faster than the ordinary kind
if the average cell occupancy is low, but slower if it is high.
 
\subsection{Event Statistics}
 
All the event-analysis routines above are defined on an
event-by-event basis. Once found, the quantities are about equally
often used to define inclusive distributions as to select specific
classes of events for continued study. For instance, the thrust
routine might be used either to find the inclusive $T$ distribution
or to select events with $T < 0.9$. Other measures, although still
defined for the individual event, only make sense to discuss in
terms of averages over many events. A small set of such measures
is found in \ttt{PYTABU}. This routine
has to be called once after each event to accumulate statistics,
and once in the end to print the final tables. Of course, among the
wealth of possibilities imaginable, the ones collected here are only
a small sample, selected because the authors at some point have found
a use for them.
 
\subsubsection{Multiplicities}
 
Three options are available to collect information on multiplicities
in events. One gives the flavour content of the final state in hard
interaction processes, e.g.\ the relative composition of
$\d\dbar / \u\ubar / \s\sbar / \c\cbar / \b\bbar$ in $\ee$
annihilation events. Additionally it gives the total parton multiplicity
distribution at the end of parton showering. Another gives the
inclusive rate of all the different particles produced in events,
either as intermediate resonances or as final-state particles.
The number is subdivided into particles produced from fragmentation
(primary particles) and those produced in decays (secondary
particles).
 
The third option tabulates the rate of exclusive final states, after
all allowed decays have occurred. Since only events with up to 8 
final-state particles are analyzed, this is clearly not intended for the
study of complete high-energy events. Rather the main application is
for an analysis of the decay modes of a single particle. For instance,
the decay data for $\D$ mesons is given in terms of channels that
also contain unstable particles, such as $\rho$ and $\eta$, which
decay further. Therefore a given final state may receive
contributions from several tabulated decay channels; e.g.
$\K \pi \pi$ from $\K^* \pi$ and $\K \rho$, and so on.
 
\subsubsection{Energy-Energy Correlation}
 
The Energy-Energy Correlation is defined by \cite{Bas78}
\begin{equation}
\mrm{EEC}(\theta) = \sum_{i < j} 
\frac{2 E_i E_j}{E_{\mrm{vis}}^2} \,
\delta(\theta - \theta_{ij})  ~,
\end{equation}
and its Asymmetry by
\begin{equation}
\mrm{EECA}(\theta) = \mrm{EEC}(\pi - \theta) - \mrm{EEC}(\theta) ~.
\end{equation}
Here $\theta_{ij}$ is the opening angle between the two particles
$i$ and $j$, with energies $E_i$ and $E_j$. In principle,
normalization should be to $E_{\mrm{cm}}$, but if not all particles 
are detected it is convenient to normalize to the total visible
energy $E_{\mrm{vis}}$. Taking into account the autocorrelation term
$i = j$, the total $\mrm{EEC}$ in an event then is unity.
The $\delta$ function peak is smeared out by the
finite bin width $\Delta \theta$ in the histogram, i.e.,
it is replaced by a contribution $1 / \Delta \theta$ to the bin
which contains $\theta_{ij}$.
 
The formulae above refer to an individual event, and are to be
averaged over all events to suppress statistical fluctuations, and
obtain smooth functions of $\theta$.
 
\subsubsection{Factorial moments}
 
Factorial moments may be used to search for intermittency in events
\cite{Bia86}. The whole field has been much studied in past years,
and a host of different measures have been proposed. We only implement
one of the original prescriptions.
 
To calculate the factorial moments, the full rapidity (or
pseudorapidity) and
azimuthal ranges are subdivided into bins of successively smaller
size, and the multiplicity distributions in bins is studied. The
program calculates pseudorapidity with respect to the $z$ axis;
if desired, one could first find
an event axis, e.g.\ the sphericity or thrust axis, and subsequently
rotate the event to align this axis with the $z$ direction.
 
The full rapidity range $|y| < y_{\mmax}$ (or pseudorapidity range
$|\eta| < \eta_{\mmax}$) and
azimuthal range $0 < \varphi < 2\pi$ are subdivided into $m_y$ and
$m_{\varphi}$ equally large bins. In fact, the whole analysis is
performed thrice: once with $m_{\varphi}=1$ and the $y$ (or $\eta$)
range gradually
divided into 1, 2, 4, 8, 16, 32, 64, 128, 256 and 512 bins, once
with $m_y = 1$ and the $\varphi$ range subdivided as above, and
finally once with $m_y = m_{\varphi}$ according to the same binary
sequence. Given the multiplicity $n_j$ in bin $j$, the $i$:th
factorial moment is defined by
\begin{equation}
F_i = (m_y m_{\varphi})^{i-1} \, \sum_j
\frac{n_j(n_j-1)\cdots(n_j-i+1)}{n(n-1)\cdots(n-i+1)}  ~.
\end{equation}
Here $n = \sum_j n_j$ is the total multiplicity of the event within
the allowed $y$ (or $\eta$) limits. The calculation is performed for
the second through the fifth moments, i.e.\ $F_2$ through $F_5$.
 
The $F_i$ as given here are defined for the individual event,
and have to be averaged over many events to give a reasonably smooth
behaviour. If particle production is uniform and uncorrelated
according to Poisson statistics, one expects
$\langle F_i \rangle \equiv 1$ for all moments and all bin sizes.
If, on the other hand, particles are locally clustered, factorial
moments should increase when bins are made smaller, down to the
characteristic dimensions of the clustering.
 
\subsection{Routines and Common Block Variables}
\label{ss:evanrout}
 
The six routines \ttt{PYSPHE}, \ttt{PYTHRU}, \ttt{PYCLUS},
\ttt{PYCELL}, \ttt{PYJMAS} and \ttt{PYFOWO} give you the possibility
to find some global event shape properties. The routine \ttt{PYTABU}
performs a statistical analysis of a number of different quantities
like particle content, factorial moments and the energy--energy
correlation.
 
Note that, by default, all remaining partons/particles except
neutrinos are used in the analysis. Neutrinos may be included with
\ttt{MSTU(41)=1}. Also note that axes determined are stored in
\ttt{PYJETS}, but are not proper four-vectors and, as a general rule
(with some exceptions), should therefore not be rotated or boosted.
 
\drawbox{CALL PYSPHE(SPH,APL)}\label{p:PYSPHE}
\begin{entry}
\itemc{Purpose:} to diagonalize the momentum tensor, i.e.\ find the
eigenvalues $\lambda_1 > \lambda_2 > \lambda_3$, with sum unity,
and the corresponding eigenvectors. \\
Momentum power dependence is given by \ttt{PARU(41)}; default
corresponds to sphericity, while \ttt{PARU(41)=1.} gives measures 
linear in momenta. Which particles (or partons) are used in the 
analysis is determined by the \ttt{MSTU(41)} value.
\iteme{SPH :} $\frac{3}{2} (\lambda_2 + \lambda_3)$, i.e.\ sphericity
(for \ttt{PARU(41)=2.}).
\begin{subentry}
\iteme{= -1. :} analysis not performed because event contained less
than two particles (or two exactly back-to-back particles, in
which case the two transverse directions would be undefined).
\end{subentry}
\iteme{APL :} $\frac{3}{2} \lambda_3$, i.e.\ aplanarity (for
\ttt{PARU(41)=2.}).
\begin{subentry}
\iteme{= -1. :} as \ttt{SPH=-1.}
\end{subentry}
\itemc{Remark:} the lines \ttt{N+1} through \ttt{N+3} (\ttt{N-2}
through \ttt{N} for \ttt{MSTU(43)=2}) in \ttt{PYJETS} will, after
a call, contain the following information: \\
\ttt{K(N+i,1) =} 31; \\
\ttt{K(N+i,2) =} 95; \\
\ttt{K(N+i,3) :} $i$, the axis number, $i=1,2,3$; \\
\ttt{K(N+i,4), K(N+i,5) =} 0; \\
\ttt{P(N+i,1) - P(N+i,3) :} the $i$'th eigenvector, $x$, $y$ and $z$
components; \\
\ttt{P(N+i,4) :} $\lambda_i$, the $i$'th eigenvalue; \\
\ttt{P(N+i,5) =} 0; \\
\ttt{V(N+i,1) - V(N+i,5) =} 0. \\
Also, the number of particles used in the analysis is given in
\ttt{MSTU(62)}.
\end{entry}
 
\drawbox{CALL PYTHRU(THR,OBL)}\label{p:PYTHRU}
\begin{entry}
\itemc{Purpose:} to find the thrust, major and minor axes and
corresponding projected momentum quantities, in particular thrust
and oblateness. The performance of the program is affected by
\ttt{MSTU(44)}, \ttt{MSTU(45)}, \ttt{PARU(42)} and \ttt{PARU(48)}.
In particular, \ttt{PARU(42)} gives the momentum dependence, with
the default value \ttt{=1} corresponding to linear dependence.
Which particles (or partons) are used in the analysis
is determined by the \ttt{MSTU(41)} value.
\iteme{THR :} thrust (for \ttt{PARU(42)=1.}).
\begin{subentry}
\iteme{= -1. :} analysis not performed because event contained
less than two particles.
\iteme{= -2. :} remaining space in \ttt{PYJETS} (partly used as
working area) not large enough to allow analysis.
\end{subentry}
\iteme{OBL :} oblateness (for \ttt{PARU(42)=1.}).
\begin{subentry}
\iteme{= -1., -2. :} as for \ttt{THR}.
\end{subentry}
\itemc{Remark:} the lines \ttt{N+1} through \ttt{N+3} (\ttt{N-2}
through \ttt{N} for \ttt{MSTU(43)=2}) in \ttt{PYJETS} will, after
a call, contain the following information: \\
\ttt{K(N+i,1) =} 31; \\
\ttt{K(N+i,2) =} 96; \\
\ttt{K(N+i,3) :} $i$, the axis number, $i=1,2,3$; \\
\ttt{K(N+i,4), K(N+i,5) =} 0; \\
\ttt{P(N+i,1) - P(N+i,3) :} the thrust, major and minor axis,
respectively, for $i = 1, 2$ and 3; \\
\ttt{P(N+i,4) :} corresponding thrust, major and minor value; \\
\ttt{P(N+i,5) =} 0; \\
\ttt{V(N+i,1) - V(N+i,5) =} 0. \\
Also, the number of particles used in the analysis is given in
\ttt{MSTU(62)}.
\end{entry}
 
\drawbox{CALL PYCLUS(NJET)}\label{p:PYCLUS}
\begin{entry}
\itemc{Purpose:} to reconstruct an arbitrary number of jets using a
cluster analysis method based on particle momenta. \\
Three different distance measures are available, see section
\ref{ss:clustfind}. The choice is controlled by \ttt{MSTU(46)}. The 
distance scale $d_{\mrm{join}}$, above which two clusters may not be
joined, is normally given by \ttt{PARU(44)}. In general, 
$d_{\mrm{join}}$
may be varied to describe different `jet-resolution powers';
the default value, 2.5 GeV, is fairly well suited for $\ee$ physics
at 30--40 GeV. With the alternative mass distance measure,
\ttt{PARU(44)} can be used to set the absolute maximum cluster mass,
or \ttt{PARU(45)} to set the scaled one, i.e.\ in 
$y = m^2/E_{\mrm{cm}}^2$, where $E_{\mrm{cm}}$ is the total 
invariant mass of the particles being considered. \\
It is possible to continue the cluster search from the configuration
already found, with a new higher $d_{\mrm{join}}$ scale, by selecting
\ttt{MSTU(48)} properly. In \ttt{MSTU(47)} one can also require a
minimum number of jets to be reconstructed; combined with an
artificially large $d_{\mrm{join}}$ this can be used to reconstruct a
predetermined number of jets. \\
Which particles (or partons) are used in the analysis is determined
by the \ttt{MSTU(41)} value, whereas assumptions about particle masses
is given by \ttt{MSTU(42)}. The parameters \ttt{PARU(43)} and
\ttt{PARU(48)} regulate more technical details (for events at high
energies and large multiplicities, however, the choice of a larger
\ttt{PARU(43)} may be necessary to obtain reasonable reconstruction
times).
\iteme{NJET :} the number of clusters reconstructed.
\begin{subentry}
\iteme{= -1 :} analysis not performed because event contained less than
\ttt{MSTU(47)} (normally 1) particles, or analysis failed to
reconstruct the requested number of jets.
\iteme{= -2 :} remaining space in \ttt{PYJETS} (partly used as working
area) not large enough to allow analysis.
\end{subentry}
\itemc{Remark:} if the analysis does not fail, further information is
found in \ttt{MSTU(61) - MSTU(63)} and \ttt{PARU(61) - PARU(63)}.
In particular, \ttt{PARU(61)} contains the invariant mass for the
system analyzed, i.e.\ the number used in determining the denominator
of $y = m^2/E_{\mrm{cm}}^2$. \ttt{PARU(62)} gives the generalized 
thrust, i.e.\ the sum of (absolute values of) cluster momenta divided 
by the sum of particle momenta (roughly the same as multicity 
\cite{Bra79}). \ttt{PARU(63)} gives the minimum distance $d$ (in $\pT$ 
or $m$) between two clusters in the final cluster configuration, 0 in 
case of only one cluster. \\
Further, the lines \ttt{N+1} through \ttt{N+NJET} (\ttt{N-NJET+1}
through \ttt{N} for \ttt{MSTU(43)=2}) in \ttt{PYJETS}
will, after a call, contain the following information: \\
\ttt{K(N+i,1) =} 31; \\
\ttt{K(N+i,2) =} 97; \\
\ttt{K(N+i,3) :} $i$, the jet number, with the jets arranged in
falling order of absolute momentum; \\
\ttt{K(N+i,4) :} the number of particles assigned to jet $i$; \\
\ttt{K(N+i,5) =} 0; \\
\ttt{P(N+i,1) - P(N+i,5) :} momentum, energy and invariant mass of
jet $i$; \\
\ttt{V(N+i,1) - V(N+i,5) =} 0. \\
Also, for a particle which was used in the analysis,
\ttt{K(I,4)}$ = i$, where \ttt{I} is the particle number and $i$
the number of the jet it has been assigned to. Undecayed particles
not used then have \ttt{K(I,4)=0}. An exception is made for lines
with \ttt{K(I,1)=3} (which anyhow are not normally interesting for
cluster search), where the colour-flow information stored in
\ttt{K(I,4)} is left intact.\\
\ttt{MSTU(3)} is only set equal to the number of jets for positive
\ttt{NJET} and \ttt{MSTU(43)=1}.
\end{entry}
 
\drawbox{CALL PYCELL(NJET)}\label{p:PYCELL}
\begin{entry}
\itemc{Purpose:} to provide a simpler cluster routine more in line
with what is currently used in the study of high-$\pT$ collider
events. \\
A detector is assumed to stretch in pseudorapidity between
\ttt{-PARU(51)} and \ttt{+PARU(51)} and be segmented in
\ttt{MSTU(51)} equally large $\eta$ (pseudorapidity) bins and
\ttt{MSTU(52)} $\varphi$ (azimuthal) bins. Transverse
energy $E_{\perp}$ for undecayed entries are summed up in each bin.
For \ttt{MSTU(53)} non-zero, the energy is smeared by calorimetric
resolution effects, cell by cell. This is done according to a Gaussian
distribution; if \ttt{MSTU(53)=1} the standard deviation for the
$E_{\perp}$ is \ttt{PARU(55)}$\times \sqrt{E_{\perp}}$, if
\ttt{MSTU(53)=2} the standard deviation for the $E$ is
\ttt{PARU(55)}$\times \sqrt{E}$, $E_{\perp}$ and $E$ expressed in GeV.
The Gaussian is cut off at 0 and at a factor \ttt{PARU(56)} times the
correct $E_{\perp}$ or $E$. Cells with an  $E_{\perp}$ below a given 
threshold \ttt{PARU(58)} are removed from further consideration;
by default \ttt{PARU(58)=0.} and thus all cells are kept.
\\
All bins with $E_{\perp} > $\ttt{PARU(52)} are taken to be possible
initiators of jets, and are tried in falling $E_{\perp}$ sequence to
check whether the total $E_{\perp}$ summed over cells no more distant
than \ttt{PARU(54)} in $\sqrt{(\Delta\eta)^2 + (\Delta\varphi)^2}$
exceeds \ttt{PARU(53)}. If so, these cells define one jet, and are
removed from further consideration. Contrary to \ttt{PYCLUS}, not all
particles need be assigned to jets. Which particles (or partons) are
used in the analysis is determined by the \ttt{MSTU(41)} value.
\iteme{NJET :} the number of jets reconstructed (may be 0).
\begin{subentry}
\iteme{= -2 :} remaining space in \ttt{PYJETS} (partly used as
working area) not large enough to allow analysis.
\end{subentry}
\itemc{Remark:} the lines \ttt{N+1} through \ttt{N+NJET}
(\ttt{N-NJET+1} through \ttt{N} for \ttt{MSTU(43)=2}) in
\ttt{PYJETS} will, after a call, contain the following information: \\
\ttt{K(N+i,1) =} 31; \\
\ttt{K(N+i,2) =} 98; \\
\ttt{K(N+i,3) :} $i$, the jet number, with the jets arranged in
falling order in $E_{\perp}$; \\
\ttt{K(N+i,4) :} the number of particles assigned to jet $i$; \\
\ttt{K(N+i,5) =} 0; \\
\ttt{V(N+i,1) - V(N+i,5) =} 0. \\
Further, for \ttt{MSTU(54)=1} \\
\ttt{P(N+i,1), P(N+i,2) =} position in $\eta$ and $\varphi$ of the
center of the jet initiator cell, i.e.\ geometrical center of jet; \\
\ttt{P(N+i,3), P(N+i,4) =} position in $\eta$ and $\varphi$ of the
$E_{\perp}$-weighted center of the jet, i.e.\ the center of gravity
of the jet; \\
\ttt{P(N+i,5) =} sum $E_{\perp}$ of the jet; \\
while for \ttt{MSTU(54)=2} \\
\ttt{P(N+i,1) - P(N+i,5) :} the jet momentum vector, constructed
from the summed $E_{\perp}$ and the $\eta$ and $\varphi$ of the
$E_{\perp}$-weighted center of the jet as \\
$(p_x, p_y, p_z, E, m)=E_{\perp} (\cos\varphi, \sin\varphi,
\sinh\eta, \cosh\eta, 0)$; \\
and for \ttt{MSTU(54)=3} \\
\ttt{P(N+i,1) - P(N+i,5) :} the jet momentum vector, constructed by
adding vectorially the momentum of each cell assigned to the jet,
assuming that all the $E_{\perp}$ was deposited at the center of the
cell, and with the jet mass in \ttt{P(N+i,5)} calculated from the
summed $E$ and $\mbf{p}$ as $m^2 = E^2 - p_x^2 - p_y^2 - p_z^2$. \\
Also, the number of particles used in the analysis is given in
\ttt{MSTU(62)}, and the number of cells hit in \ttt{MSTU(63)}.\\
\ttt{MSTU(3)} is only set equal to the number of jets for positive
\ttt{NJET} and \ttt{MSTU(43)=1}.
\end{entry}
 
\drawbox{CALL PYJMAS(PMH,PML)}\label{p:PYJMAS}
\begin{entry}
\itemc{Purpose:} to reconstruct high and low jet mass of an event.
A simplified algorithm is used, wherein a preliminary division of
the event into two hemispheres is done transversely to the sphericity
axis. Then one particle at a time is reassigned to the other
hemisphere if that reduces the sum of squares of the two jet masses,
$m_{\mrm{H}}^2 + m_{\mrm{L}}^2$. The procedure is stopped when no 
further significant change (see \ttt{PARU(48)}) is obtained. Often, the
original assignment is retained as it is. Which particles (or partons)
are used in the analysis is determined by the \ttt{MSTU(41)} value,
whereas assumptions about particle masses is given by \ttt{MSTU(42)}.
\iteme{PMH :} heavy jet mass (in GeV).
\begin{subentry}
\iteme{= -2. :} remaining space in \ttt{PYJETS} (partly used as
working area) not large enough to allow analysis.
\end{subentry}
\iteme{PML :} light jet mass (in GeV).
\begin{subentry}
\iteme{= -2. :} as for \ttt{PMH=-2.}
\end{subentry}
\itemc{Remark:} After a successful call, \ttt{MSTU(62)} contains the
number of particles used in the analysis, and \ttt{PARU(61)} the
invariant mass of the system analyzed. The latter number is helpful
in constructing scaled jet masses.
\end{entry}
 
\drawbox{CALL PYFOWO(H10,H20,H30,H40)}\label{p:PYFOWO}
\begin{entry}
\itemc{Purpose:} to do an event analysis in terms of the Fox-Wolfram
moments. The moments $H_i$ are normalized to the lowest one, $H_0$.
Which particles (or partons) are used in the analysis is determined
by the \ttt{MSTU(41)} value.
\iteme{H10 :} $H_1/H_0$. Is $=0$ if momentum is balanced.
\iteme{H20 :} $H_2/H_0$.
\iteme{H30 :} $H_3/H_0$.
\iteme{H40 :} $H_4/H_0$.
\itemc{Remark:} the number of particles used in the analysis is given
in \ttt{MSTU(62)}.
\end{entry}
 
\drawbox{CALL PYTABU(MTABU)}\label{p:PYTABU}
\begin{entry}
\itemc{Purpose:} to provide a number of event-analysis options which
can be be used on each new event, with accumulated statistics to be
written out on request. When errors are quoted, these refer to
the uncertainty in the average value for the event sample as a
whole, rather than to the spread of the individual events, i.e.
errors decrease like one over the square root of the number of
events analyzed. For a correct use of \ttt{PYTABU}, it is not
permissible to freely mix generation and analysis of different
classes of events, since only one set of statistics counters exists.
A single run may still contain sequential `subruns', between
which statistics is reset. Whenever an event is analyzed, the
number of particles/partons used is given in \ttt{MSTU(62)}.
\iteme{MTABU :} determines which action is to be taken. Generally, a
last digit equal to 0 indicates that the statistics counters for this
option is to be reset; since the counters are reset (by \ttt{DATA}
statements) at the beginning of a run, this is not used normally. Last
digit 1 leads to an analysis of current event with respect to the
desired properties. Note that the resulting action may depend on how
the event generated has been rotated, boosted or edited before this
call. The statistics accumulated is output in tabular form with
last digit 2, while it is dumped in the \ttt{PYJETS} common block for
last digit 3. The latter option may be useful for interfacing to
graphics output.
\itemc{Warning:} this routine cannot be used on weighted events,
i.e.\ in the statistics calculation all events are assumed to come
with the same weight.
\begin{subentry}
\iteme{= 10 :} statistics on parton multiplicity is reset.
\iteme{= 11 :} the parton content of the current event is analyzed,
classified according to the flavour content of the hard
interaction and the total number of partons. The flavour
content is assumed given in \ttt{MSTU(161)} and \ttt{MSTU(162)};
these are automatically set e.g.\ in \ttt{PYEEVT} and \ttt{PYEVNT}
calls.
\iteme{= 12 :} gives a table on parton multiplicity distribution.
\iteme{= 13 :} stores the parton multiplicity distribution of events
in \ttt{PYJETS}, using the following format: \\
\ttt{N =} total number of different channels found; \\
\ttt{K(I,1) =} 32; \\
\ttt{K(I,2) =} 99; \\
\ttt{K(I,3), K(I,4) =} the two flavours of the flavour content; \\
\ttt{K(I,5) =} total number of events found with flavour content of
\ttt{K(I,3)} and \ttt{K(I,4)}; \\
\ttt{P(I,1) - P(I,5) =} relative probability to find given flavour
content and a total of 1, 2, 3, 4 or 5 partons, respectively; \\
\ttt{V(I,1) - V(I,5) =} relative probability to find given flavour
content and a total of 6--7, 8--10, 11--15, 16--25 or above 25
partons, respectively. \\
In addition, \ttt{MSTU(3)=1} and \\
\ttt{K(N+1,1) =} 32; \\
\ttt{K(N+1,2) =} 99; \\
\ttt{K(N+1,5) =} number of events analyzed.
\iteme{= 20 :} statistics on particle content is reset.
\iteme{= 21 :} the particle/parton content of the current event is
analyzed, also for particles which have subsequently decayed and
partons which have fragmented (unless this has been made impossible
by a preceding \ttt{PYEDIT} call). Particles are subdivided into
primary and secondary ones, the main principle being that primary
particles are those produced in the fragmentation of a string,
while secondary come from decay of other particles. 
\iteme{= 22 :} gives a table of particle content in events.
\iteme{= 23 :} stores particle content in events in \ttt{PYJETS},
using the following format: \\
\ttt{N =} number of different particle species found; \\
\ttt{K(I,1) =} 32; \\
\ttt{K(I,2) =} 99; \\
\ttt{K(I,3) =} particle {\KF} code; \\
\ttt{K(I,5) =} total number of particles and antiparticles of this
species; \\
\ttt{P(I,1) =} average number of primary particles per event; \\
\ttt{P(I,2) =} average number of secondary particles per event; \\
\ttt{P(I,3) =} average number of primary antiparticles per event; \\
\ttt{P(I,4) =} average number of secondary antiparticles per event; \\
\ttt{P(I,5) =} average total number of particles or antiparticles
per event. \\
In addition, \ttt{MSTU(3)=1} and \\
\ttt{K(N+1,1) =} 32; \\
\ttt{K(N+1,2) =} 99; \\
\ttt{K(N+1,5) =} number of events analyzed; \\
\ttt{P(N+1,1) =} average primary multiplicity per event; \\
\ttt{P(N+1,2) =} average final multiplicity per event; \\
\ttt{P(N+1,3) =} average charged multiplicity per event.
\iteme{= 30 :} statistics on factorial moments is reset.
\iteme{= 31 :} analyzes the factorial moments of the multiplicity
distribution in different bins of rapidity and azimuth.
Which particles (or partons) are used in the analysis is
determined by the \ttt{MSTU(41)} value. The selection between usage
of true rapidity, pion rapidity or pseudorapidity is regulated
by \ttt{MSTU(42)}. The $z$ axis is assumed to be event axis; if this
is not desirable find an event axis e.g.\ with \ttt{PYSPHE} or
\ttt{PYTHRU} and use \ttt{PYEDIT(31)}. Maximum (pion-, pseudo-)
rapidity, which sets the limit for the rapidity plateau or the
experimental acceptance, is given by \ttt{PARU(57)}.
\iteme{= 32 :} prints a table of the first four factorial moments
for various bins of pseudorapidity and azimuth. The moments are
properly normalized so that they would be unity (up to
statistical fluctuations) for uniform and uncorrelated particle
production according to Poisson statistics, but increasing
for decreasing bin size in case of `intermittent' behaviour.
The error on the average value is based on the actual
statistical sample (i.e.\ does not use any assumptions on the
distribution to relate errors to the average values of higher
moments). Note that for small bin sizes, where the average
multiplicity is small and the factorial moment therefore only
very rarely is non-vanishing, moment values may fluctuate wildly
and the errors given may be too low.
\iteme{= 33 :} stores the factorial moments in \ttt{PYJETS},
using the format: \\
\ttt{N =} 30, with \ttt{I = }$i=1$--10 corresponding to results for
slicing the rapidity range in $2^{i-1}$ bins, \ttt{I = }$i = 11$--20
to slicing the azimuth in $2^{i-11}$ bins, and \ttt{I = }$i = 21$--30
to slicing both rapidity and azimuth, each in $2^{i-21}$ bins; \\
\ttt{K(I,1) =} 32; \\
\ttt{K(I,2) =} 99; \\
\ttt{K(I,3) =} number of bins in rapidity; \\
\ttt{K(I,4) =} number of bins in azimuth; \\
\ttt{P(I,1) =} rapidity bin size; \\
\ttt{P(I,2) - P(I,5) =} $\langle F_2 \rangle$--$\langle F_5 \rangle$,
i.e.\ mean of second, third, fourth and fifth factorial moment; \\
\ttt{V(I,1) =} azimuthal bin size; \\
\ttt{V(I,2) - V(I,5) =} statistical errors on
$\langle F_2 \rangle$--$\langle F_5 \rangle$. \\
In addition, \ttt{MSTU(3) =} 1 and \\
\ttt{K(31,1) =} 32; \\
\ttt{K(31,2) =} 99; \\
\ttt{K(31,5) =} number of events analyzed.
\iteme{= 40 :} statistics on energy--energy correlation is reset.
\iteme{= 41 :} the energy--energy correlation $\mrm{EEC}$ of the 
current
event is analyzed. Which particles (or partons) are used in the
analysis is determined by the \ttt{MSTU(41)} value. Events are
assumed given in their c.m.\ frame. The weight assigned to a pair
$i$ and $j$ is $2 E_i E_j/E_{\mrm{vis}}^2$, where $E_{\mrm{vis}}$ 
is the sum of energies of
all analyzed particles in the event. Energies are determined from
the momenta of particles, with mass determined according to the
\ttt{MSTU(42)} value. Statistics is accumulated for the relative
angle $\theta_{ij}$, ranging between 0 and 180 degrees, subdivided
into 50 bins.
\iteme{= 42 :} prints a table of the energy--energy correlation 
$\mrm{EEC}$ and its asymmetry $\mrm{EECA}$, with errors. 
The definition of errors is not unique. In our approach each event 
is viewed as one observation, i.e.\ an $\mrm{EEC}$ and 
$\mrm{EECA}$ distribution is obtained by
summing over all particle pairs of an event, and then the
average and spread of this event-distribution is calculated
in the standard fashion. The quoted error is therefore inversely
proportional to the square root of the number of events. It could
have been possible to view each single particle pair as one
observation, which would have given somewhat lower errors, but then
one would also be forced to do a complicated correction procedure
to account for the pairs in an event not being uncorrelated
(two hard jets separated by a given angle typically corresponds
to several pairs at about that angle). Note, however, that
in our approach the squared error on an $\mrm{EECA}$ bin is smaller
than the sum of the squares of the errors on the corresponding 
$\mrm{EEC}$ bins (as it should be). Also note that it is not possible 
to combine the errors of two nearby bins by hand from the
information given, since nearby bins are correlated (again a
trivial consequence of the presence of jets).
\iteme{= 43 :} stores the $\mrm{EEC}$ and $\mrm{EECA}$ in 
\ttt{PYJETS}, using the format: \\
\ttt{N =} 25; \\
\ttt{K(I,1) =} 32; \\
\ttt{K(I,2) =} 99; \\
\ttt{P(I,1) =} $\mrm{EEC}$ for angles between \ttt{I-1} and \ttt{I}, 
in units of $3.6^{\circ}$; \\
\ttt{P(I,2) =} $\mrm{EEC}$ for angles between \ttt{50-I} and 
\ttt{51-I}, in units of $3.6^{\circ}$; \\
\ttt{P(I,3) =} $\mrm{EECA}$ for angles between \ttt{I-1} and 
\ttt{I}, in units of $3.6^{\circ}$; \\
\ttt{P(I,4), P(I,5) :} lower and upper edge of angular range of bin
\ttt{I}, expressed in radians; \\
\ttt{V(I,1) - V(I,3) :} errors on the $\mrm{EEC}$ and $\mrm{EECA}$ 
values stored in \ttt{P(I,1) - P(I,3)} (see \ttt{=42} for comments); \\
\ttt{V(I,4), V(I,5) :} lower and upper edge of angular range of bin
\ttt{I}, expressed in degrees. \\
In addition, \ttt{ MSTU(3)=1} and \\
\ttt{K(26,1) =} 32; \\
\ttt{K(26,2) =} 99; \\
\ttt{K(26,5) =} number of events analyzed.
\iteme{= 50 :} statistics on complete final states is reset.
\iteme{= 51 :} analyzes the particle content of the final state of
the current event record. During the course of the run, statistics
is thus accumulated on how often different final states appear.
Only final states with up to 8 particles are analyzed, and there
is only reserved space for up to 200 different final states.
Most high energy events have multiplicities far above 8, so the
main use for this tool is to study the effective branching
ratios obtained with a given decay model for e.g.\ charm or bottom
hadrons. Then \ttt{PY1ENT} may be used to generate one decaying
particle at a time, with a subsequent analysis by \ttt{PYTABU}.
Depending on at what level this studied is to be carried out,
some particle decays may be switched off, like $\pi^0$.
\iteme{= 52 :} gives a list of the (at most 200) channels with up
to 8 particles in the final state, with their relative branching
ratio. The ordering is according to multiplicity, and within
each multiplicity according to an ascending order of {\KF} codes.
The {\KF} codes of the particles belonging to a given channel are
given in descending order.
\iteme{= 53 :} stores the final states and branching ratios found in
\ttt{PYJETS}, using the format: \\
\ttt{N =} number of different explicit final states found (at most
200); \\
\ttt{K(I,1) =} 32; \\
\ttt{K(I,2) =} 99; \\
\ttt{K(I,5) =} multiplicity of given final state, a number between
1 and 8; \\
\ttt{P(I,1) - P(I,5), V(I,1) - V(I,3) :} the {\KF} codes of the up to 8
particles of the given final state, converted to real
numbers, with trailing zeroes for positions not used; \\
\ttt{V(I,5) :} effective branching ratio for the given final state. \\
In addition, \ttt{MSTU(3)=1} and \\
\ttt{K(N+1,1) =} 32; \\
\ttt{K(N+1,2) =} 99; \\
\ttt{K(N+1,5) =} number of events analyzed; \\
\ttt{V(N+1,5) =} summed branching ratio for finals states not given
above, either because they contained more than 8 particles
or because all 200 channels have been used up.
\end{subentry}
\end{entry}
  
\drawbox{COMMON/PYDAT1/MSTU(200),PARU(200),MSTJ(200),PARJ(200)}
\begin{entry}
\itemc{Purpose:} to give access to a number of status codes and
parameters which regulate the performance of fragmentation and event 
analysis routines. Most parameters are described in section 
\ref{ss:JETswitch}; here only those related to the event-analysis 
routines are described.

\boxsep
 
\iteme{MSTU(41) :}\label{p:MSTU41} (D=2) partons/particles used in 
the event-analysis routines \ttt{PYSPHE}, \ttt{PYTHRU}, \ttt{PYCLUS}, 
\ttt{PYCELL}, \ttt{PYJMAS}, \ttt{PYFOWO} and \ttt{PYTABU} 
(\ttt{PYTABU(11)} excepted).
\begin{subentry}
\iteme{= 1 :} all partons/particles that have not fragmented/decayed.
\iteme{= 2 :} ditto, with the exception of neutrinos and unknown
particles.
\iteme{= 3 :} only charged, stable particles, plus any partons
still not fragmented.
\end{subentry}
 
\iteme{MSTU(42) :} (D=2) assumed particle masses, used in
calculating energies $E^2 = \mbf{p}^2 + m^2$, as subsequently
used in \ttt{PYCLUS}, \ttt{PYJMAS} and \ttt{PYTABU}
(in the latter also for pseudorapidity, pion rapidity or true
rapidity selection).
\begin{subentry}
\iteme{= 0 :} all particles are assumed massless.
\iteme{= 1 :} all particles, except the photon, are assumed to have
the charged pion mass.
\iteme{= 2 :} the true masses are used.
\end{subentry}
 
\iteme{MSTU(43) :} (D=1) storing of event-analysis information (mainly
jet axes), in \ttt{PYSPHE}, \ttt{PYTHRU}, \ttt{PYCLUS} and
\ttt{PYCELL}.
\begin{subentry}
\iteme{= 1 :} stored after the event proper, in positions \ttt{N+1}
through \ttt{N+MSTU(3)}. If several of the routines are used in
succession, all but the latest information is overwritten.
\iteme{= 2 :} stored with the event proper, i.e.\ at the end of the
event listing, with \ttt{N} updated accordingly. If several of the
routines are used in succession, all the axes determined are
available.
\end{subentry}
 
\iteme{MSTU(44) :} (D=4) is the number of the fastest (i.e.\ with
largest momentum) particles used to construct the (at most) 10 most
promising starting configurations for the thrust axis determination.
 
\iteme{MSTU(45) :} (D=2) is the number of different starting
configurations above, which have to converge to the same (best) value
before this is accepted as the correct thrust axis.
 
\iteme{MSTU(46) :} (D=1) distance measure used for the joining of
clusters in \ttt{PYCLUS}.
\begin{subentry}
\iteme{= 1 :} $d_{ij}$, i.e.\ approximately relative transverse 
momentum. Anytime two clusters have been joined, particles are 
reassigned to the cluster they now are closest to. The distance 
cut-off $d_{\mrm{join}}$ is stored in \ttt{PARU(44)}.
\iteme{= 2 :} distance measure as in \ttt{=1}, but particles are
never reassigned to new jets.
\iteme{= 3 :} JADE distance measure $y_{ij}$, but with dimensions
to correspond approximately to total invariant mass. Particles may
never be reassigned between clusters. The distance cut-off 
$m_{\mmin}$ is stored in \ttt{PARU(44)}.
\iteme{= 4 :} as \ttt{=3}, but a scaled JADE distance $y_{ij}$ is 
used instead of $m_{ij}$. The distance cut-off $y_{\mmin}$ is stored 
in \ttt{PARU(45)}.
\iteme{= 5 :} Durham distance measure $\tilde{y}_{ij}$, but with 
dimensions to correspond approximately to transverse momentum. Particles 
may never be reassigned between clusters. The distance cut-off 
$\pTmin$ is stored in \ttt{PARU(44)}.
\iteme{= 6 :} as \ttt{=5}, but a scaled Durham distance $\tilde{y}_{ij}$ 
is used instead of $p_{\perp ij}$. The distance cut-off 
$\tilde{y}_{\mmin}$ is stored in \ttt{PARU(45)}.
\end{subentry}
 
\iteme{MSTU(47) :} (D=1) the minimum number of clusters to be
reconstructed by \ttt{PYCLUS}.
 
\iteme{MSTU(48) :} (D=0) mode of operation of the \ttt{PYCLUS}
routine.
\begin{subentry}
\iteme{= 0 :} the cluster search is started from scratch.
\iteme{= 1 :} the clusters obtained in a previous cluster search on the
same event (with \ttt{MSTU(48)=0}) are to be taken as the starting
point for subsequent cluster joining. For this call to have any effect,
the joining scale in \ttt{PARU(44)} or \ttt{PARU(45)} must have been
changed. If the event record has been modified after the last
\ttt{PYCLUS} call, or if any other cluster search parameter setting
has been changed, the subsequent result is unpredictable.
\end{subentry}
 
\iteme{MSTU(51) :} (D=25) number of pseudorapidity bins that the range
between \ttt{-PARU(51)} and \ttt{+PARU(51)} is divided into to define
cell size for \ttt{PYCELL}.
 
\iteme{MSTU(52) :} (D=24) number of azimuthal bins, used to define the
cell size for \ttt{PYCELL}.
 
\iteme{MSTU(53) :} (D=0) smearing of correct energy, imposed
cell-by-cell in \ttt{PYCELL}, to simulate calorimeter resolution
effects.
\begin{subentry}
\iteme{= 0 :} no smearing.
\iteme{= 1 :} the transverse energy in a cell, $E_{\perp}$, is smeared
according to a Gaussian distribution with standard deviation
\ttt{PARU(55)}$\times \sqrt{E_{\perp}}$, where $E_{\perp}$ is given in
GeV. The Gaussian is cut off so that $0 < E_{\perp \mrm{smeared}} 
< $\ttt{PARU(56)}$\times E_{\perp \mrm{true}}$.
\iteme{= 2 :} as \ttt{=1}, but it is the energy $E$ rather than the
transverse energy $E_{\perp}$ that is smeared.
\end{subentry}
 
\iteme{MSTU(54) :} (D=1) form for presentation of information about
reconstructed clusters in \ttt{PYCELL}, as stored in \ttt{PYJETS}
according to the \ttt{MSTU(43)} value.
\begin{subentry}
\iteme{= 1 :} the \ttt{P} vector in each line contains $\eta$ and
$\varphi$ for the geometric origin of the jet, $\eta$ and $\varphi$
for the weighted center of the jet, and jet $E_{\perp}$, respectively.
\iteme{= 2 :} the \ttt{P} vector in each line contains a massless
four-vector giving the direction of the jet, obtained as \\
$(p_x,p_y,p_z,E,m) = E_{\perp}
(\cos\varphi,\sin\varphi,\sinh\eta,\cosh\eta,0)$, \\
where $\eta$ and $\varphi$ give the weighted center of a jet and
$E_{\perp}$ its transverse energy.
\iteme{= 3 :} the \ttt{P} vector in each line contains a massive
four-vector, obtained by adding the massless four-vectors of all cells
that form part of the jet, and calculating the jet mass from
$m^2 = E^2-p_x^2-p_y^2-p_z^2$. For each cell, the total $E_{\perp}$ is
summed up, and then translated into a massless four-vector
assuming that all the $E_{\perp}$ was deposited in the center of the
cell.
\end{subentry}
 
\iteme{MSTU(61) :} (I) first entry for storage of event-analysis
information in last event analyzed with \ttt{PYSPHE}, \ttt{PYTHRU},
\ttt{PYCLUS} or \ttt{PYCELL}.
 
\iteme{MSTU(62) :} (R) number of particles/partons used in the last
event analysis with \ttt{PYSPHE}, \ttt{PYTHRU}, \ttt{PYCLUS},
\ttt{PYCELL}, \ttt{PYJMAS}, \ttt{PYFOWO} or \ttt{PYTABU}.
 
\iteme{MSTU(63) :} (R) in a \ttt{PYCLUS} call, the number of
preclusters constructed in order to speed up analysis (should be
equal to \ttt{MSTU(62)} if \ttt{PARU(43)=0.}). In a \ttt{PYCELL}
call, the number of cells hit.
 
\iteme{MSTU(161), MSTU(162) :}\label{p:MSTU161} hard flavours involved 
in current event,
as used in an analysis with \ttt{PYTABU(11)}. Either or both may be set
0, to indicate the presence of one or none hard flavours in event.
Is normally set by high-level routines, like \ttt{PYEEVT} or
\ttt{PYEVNT}, but can also be set by you.

\boxsep
 
\iteme{PARU(41) :}\label{p:PARU41} (D=2.) power of momentum-dependence 
in \ttt{PYSPHE}, default corresponds to sphericity, \ttt{=1.} to linear 
event measures.
 
\iteme{PARU(42) :} (D=1.) power of momentum-dependence in \ttt{PYTHRU},
default corresponds to thrust.
 
\iteme{PARU(43) :} (D=0.25 GeV) maximum distance $d_{\mrm{init}}$ 
allowed in
\ttt{PYCLUS} when forming starting clusters used to speed up
reconstruction. The meaning of the parameter is in $\pT$ for
\ttt{MSTU(46)}$\leq 2$ or $\geq 5$ and in $m$ else. If
\ttt{=0.}, no preclustering is obtained.
If chosen too large, more joining may be generated at this stage
than is desirable. The main application is at high energies,
where some speedup is imperative, and the small details are not
so important anyway.
 
\iteme{PARU(44) :} (D=2.5 GeV) maximum distance $d_{\mrm{join}}$, 
below which it is allowed to join two clusters into one in 
\ttt{PYCLUS}. Is used for \ttt{MSTU(46)}$\leq 3$ and \ttt{=5}, i.e.\ 
both for $\pT$ and mass distance measure.
 
\iteme{PARU(45) :} (D=0.05) maximum distance 
$y_{\mrm{join}} = m^2/E_{\mrm{vis}}^2$ or ditto with $m^2 \to \pT^2$,
below which it is allowed to join two clusters into one in
\ttt{PYCLUS} for \ttt{MSTU(46)=4, =6}.
 
\iteme{PARU(48) :} (D=0.0001) convergence criterion for thrust (in
\ttt{PYTHRU}) or generalized thrust (in \ttt{PYCLUS}), or relative
change of $m_{\mrm{H}}^2 + m_{\mrm{L}}^2$ (in \ttt{PYJMAS}), 
i.e.\ when the value changes by less than this amount between two 
iterations the process is stopped.
 
\iteme{PARU(51) :} (D=2.5) defines maximum absolute pseudorapidity
used for detector assumed in \ttt{PYCELL}.
 
\iteme{PARU(52) :} (D=1.5 GeV) gives minimum $E_{\perp}$ for a cell
to be considered as a potential jet initiator by \ttt{PYCELL}.
 
\iteme{PARU(53) :} (D=7.0 GeV) gives minimum summed $E_{\perp}$ for
a collection of cells to be accepted as a jet.
 
\iteme{PARU(54) :} (D=1.) gives the maximum distance in
$R = \sqrt{(\Delta\eta)^2 + (\Delta\varphi)^2}$ from cell initiator
when grouping cells to check whether they qualify as a jet.
 
\iteme{PARU(55) :} (D=0.5) when smearing the transverse energy 
(or energy, see \ttt{MSTU(53)}) in \ttt{PYCELL}, the calorimeter 
cell resolution is taken to be
\ttt{PARU(55)}$\times \sqrt{E_{\perp}}$ (or
\ttt{PARU(55)}$\times \sqrt{E}$) for $E_{\perp}$ (or $E$) in GeV.
 
\iteme{PARU(56) :} (D=2.) maximum factor of upward fluctuation in
transverse energy or energy in a given cell when calorimeter
resolution is included in \ttt{PYCELL} (see \ttt{MSTU(53)}).
 
\iteme{PARU(57) :} (D=3.2) maximum rapidity (or pseudorapidity or
pion rapidity, depending on \ttt{MSTU(42)}) used in the factorial 
moments analysis in \ttt{PYTABU}.

\iteme{PARU(58) :} (D=0. GeV) in a \ttt{PYCELL} call, cells with a 
transverse energy $E_{\perp}$ below \ttt{PARP(58)} are removed from
further consideration. This may be used to represent a threshold
in an actual calorimeter, or may be chosen just to speed up the 
algorithm in a high-multiplicity environment.
 
\iteme{PARU(61) :} (I) invariant mass $W$ of a system analyzed with
\ttt{PYCLUS} or \ttt{PYJMAS}, with energies calculated according to
the \ttt{MSTU(42)} value.
 
\iteme{PARU(62) :} (R) the generalized thrust obtained after a
successful \ttt{PYCLUS} call, i.e.\ ratio of summed cluster momenta
and summed particle momenta.
 
\iteme{PARU(63) :} (R) the minimum distance $d$ between two clusters
in the final cluster configuration after a successful \ttt{PYCLUS}
call; is 0 if only one cluster left.

\end{entry}

\subsection{Histograms}
\label{ss:histogram}
The GBOOK package was written in 1979, at a time when HBOOK \cite{Bru87} 
was not available in Fortran 77. It has been used since as a small and 
simple histogramming program. For this version of {\Py} the program has
been updated to run together with {\Py} in double precision. Only the
one-dimensional histogram part has been retained, and subroutine names
have been changed to fit {\Py} conventions. These modified routines 
are now distributed together with {\Py}. They would not be used for
final graphics, but may be handy for simple checks, and are extensively
used to provide free-standing examples of analysis programs, to be found
on the {\Py} web page.

There is a maximum of 1000 histograms at your disposal, 
numbered in the range 1 to 1000. Before a histogram can be filled, 
space must be reserved (booked) for it, and histogram information 
provided. Histogram contents are stored in a common block of dimension 
20000, in the order they are booked. Each booked histogram requires 
NX+28 numbers, where NX is the number of x bins and the 28 include limits, 
under/overflow and the title. If you run out of space, the program 
can be recompiled with larger dimensions. The histograms can be 
manipulated with a few routines. Histogram output is `line printer' 
style, i.e.\ no graphics. 

\drawbox{CALL PYBOOK(ID,TITLE,NX,XL,XU)}\label{p:PYBOOK}
\begin{entry}
\itemc{Purpose:} to book a one-dimensional histogram.
\iteme{ID :} histogram number, integer between 1 and 1000.
\iteme{TITLE :} histogram title, at most 60 characters.
\iteme{NX :} number of bins in the histogram; integer between 1 and 100.
\iteme{XL, XU :} lower and upper bound, respectively, on the $x$ range
      covered by the histogram.
\end{entry}

\drawbox{CALL PYFILL(ID,X,W)}\label{p:PYFILL}
\begin{entry}
\itemc{Purpose:} to fill a one-dimensional histogram.
\iteme{ID :} histogram number.
\iteme{X :} $x$ coordinate of point.
\iteme{W :} weight to be added in this point.
\end{entry}

\drawbox{CALL PYFACT(ID,F)}\label{p:PYFACT}
\begin{entry}
\itemc{Purpose:} to rescale the contents of a histogram.
\iteme{ID :} histogram number.
\iteme{F :} rescaling factor, i.e.\ a factor that all bin contents (including
      overflow etc.) are multiplied by.
\itemc{Remark:} a typical rescaling factor could be $f$ = 
      1/(bin size * number of events) = 
      \ttt{NX/(XU-XL)} * 1/(number of events).
\end{entry}

\drawbox{CALL PYOPER(ID1,OPER,ID2,ID3,F1,F2)}\label{p:PYOPER}
\begin{entry}
\itemc{Purpose:} this is a general-purpose routine for editing one or several
      histograms, which all are assumed to have the same number of
      bins. Operations are carried out bin by bin, including overflow
      bins etc.
\iteme{OPER:} gives the type of operation to be carried out, a one-character
      string or a \ttt{CHARACTER*1} variable.
\begin{subentry}
\iteme{= '+', '-', '*', '/' :}  add, subtract, multiply or divide the
      contents in \ttt{ID1} and \ttt{ID2} and put the result in \ttt{ID3}. 
      \ttt{F1} and \ttt{F2}, if not 1D0, give factors by which the \ttt{ID1} 
      and \ttt{ID2} bin contents are multiplied before the indicated 
      operation. (Division with vanishing bin content will give 0.)
\iteme{= 'A', 'S', 'L' :} for \ttt{'S'} the square root of the content in 
      \ttt{ID1} is taken (result 0 for negative bin contents) and for 
      \ttt{'L'} the 10-logarithm is taken (a nonpositive bin content is 
      before that replaced by 0.8 times the smallest positive bin content).
      Thereafter, in all three cases, the content is multiplied by \ttt{F1}
      and added with \ttt{F2}, and the result is placed in \ttt{ID3}. Thus 
      \ttt{ID2} is dummy in these cases.
\iteme{= 'M' :} intended for statistical analysis, bin-by-bin mean and
      standard deviation of a variable, assuming that \ttt{ID1} contains
      accumulated weights, \ttt{ID2} accumulated weight*variable and
      \ttt{ID3} accumulated weight*variable-squared. Afterwards \ttt{ID2} will
      contain the mean values (\ttt{=ID2/ID1}) and \ttt{ID3} the standard
      deviations ($=\sqrt{\mathtt{ID3/ID1}-\mathtt{(ID2/ID1)}^2}$). 
      In the end, \ttt{F1} multiplies \ttt{ID1} (for normalization purposes), 
      while \ttt{F2} is dummy.
\end{subentry}
\iteme{ID1, ID2, ID3 :} histogram numbers, used as described above.
\iteme{F1, F2 :} factors or offsets, used as described above.
\end{entry}

\drawbox{CALL PYHIST}\label{p:PYHIST}
\begin{entry}
\itemc{Purpose:} to print all histograms that have been filled, and
      thereafter reset their bin contents to 0.
\end{entry}

\drawbox{CALL PYPLOT(ID)}\label{p:PYPLOT}
\begin{entry}
\itemc{Purpose:} to print out a single histogram.
\iteme{ID :} histogram to be printed.
\end{entry}

\drawbox{CALL PYNULL(ID)}\label{p:PYNULL}
\begin{entry}
\itemc{Purpose:} to reset all bin contents, including overflow etc., to 0.
\iteme{ID :} histogram to be reset.
\end{entry}

\drawbox{CALL PYDUMP(MDUMP,LFN,NHI,IHI)}\label{p:PYDUMP}
\begin{entry}
\itemc{Purpose:} to dump the contents of existing histograms on an external 
      file, from which they could be read in to another program.
\iteme{MDUMP :} the action to be taken.
\begin{subentry}
\iteme{= 1 :} dump histograms, each with the first line giving histogram 
           number and title, the second the number of $x$ bins and lower 
           and upper limit, the third the total number of entries and
           under-, inside- and overflow, and subsequent ones the bin 
           contents grouped five per line. If \ttt{NHI=0} all existing 
           histograms are dumped and \ttt{IHI} is dummy, else the \ttt{NHI} 
           histograms with numbers \ttt{IHI(1)} through \ttt{IHI(NHI)} 
           are dumped.
\iteme{= 2 :} read in histograms dumped with \ttt{MDUMP=1} and book and 
           fill histograms according to this information. (With
           modest modifications this option could instead be used
           to write the info to HBOOK/HPLOT format, or whatever.) 
           \ttt{NHI} and \ttt{IHI} are dummy.
\iteme{= 3 :} dump histogram contents in column style, where the
           first column contains the $x$ values (average of respective
           bin) of the first histogram, and subsequent columns the
           histogram contents. All histograms dumped this way must
           have the same number of $x$ bins, but it is not checked whether
           the $x$ range is also the same. If \ttt{NHI=0} all existing 
           histograms are dumped and \ttt{IHI} is dummy, else the \ttt{NHI} 
           histograms with numbers \ttt{IHI(1)} through \ttt{IHI(NHI}) are 
           dumped. A file written this way can be read e.g.\ by 
           \tsc{Gnuplot} \cite{Gnu99}.
\end{subentry}
\iteme{LFN :} the file number to which the contents should be written. 
      You must see to it that this file is properly opened for write
      (since the definition of file names is platform dependent). 
\iteme{NHI :} number of histograms to be dumped; if 0 then all existing
      histograms are dumped.
\iteme{IHI :} array containing histogram numbers in the first \ttt{NHI} 
      positions for \ttt{NHI} nonzero.        
\end{entry}

\drawbox{COMMON/PYBINS/IHIST(4),INDX(1000),BIN(20000)}\label{p:PYBINS}
\begin{entry}
\itemc{Purpose:} to contain all information on histograms.
\iteme{IHIST(1) :} (D=1000) maximum allowed histogram number, i.e.\ dimension
      of the \ttt{INDX} array. 
\iteme{IHIST(2) :} (D=20000) size of histogram storage, i.e.\ dimension of 
      the \ttt{BIN} array.
\iteme{IHIST(3) :} (D=55) maximum number of lines per page assumed for 
      printing histograms. 18 lines are reserved for title, 
      bin contents and statistics, while the rest can be used for the
      histogram proper.
\iteme{IHIST(4) :} internal counter for space usage in the \ttt{BIN} array.
\iteme{INDX :} gives the initial address in \ttt{BIN} for each histogram.
      If this array is expanded, also \ttt{IHIST(1)} should be changed.
\iteme{BIN :} gives bin contents and some further histogram information for 
      the booked histograms.  If this array is expanded, also \ttt{IHIST(2)} 
      should be changed.
\end{entry}

\clearpage
 
\section{Summary and Outlook}

A complete description of the {\Py} program would have to cover
four aspects:  
\begin{Enumerate}
\item the basic philosophy and principles underlying the programs;
\item the detailed physics scenarios implemented, with all the
necessary compromises and approximations;
\item the structure of the implementation, including program flow,
internal variable names and programming tricks; and  
\item the manual, which describes how to use the programs.
\end{Enumerate}
Of these aspects, the first has been dealt with in reasonable detail. 
The second is unevenly covered: in depth for aspects which are not 
discussed anywhere else, more summarily for areas where separate 
up-to-date papers already exist. The third is not included at all, 
but `left as an exercise' for the reader, to figure out from the code 
itself. The fourth, finally, should be largely covered, although 
many further comments could have been made, in particular about the 
interplay between different parts of the programs. Still, in the 
end, no manual, however complete, can substitute for `hands on' 
experience.

The {\Py} program is continuously being developed. We are aware
of many shortcomings, some of which hopefully will be addressed in 
the future. Mainly this is a matter of including new interesting 
physics scenarios and improving the existing ones, but also some 
cleanup and reorganization would be appropriate.
No timetable is set up for such future changes, however. 
After all, this is not a professionally maintained software product, 
but part of a small physics research project. Very often, developments
of the programs have come about as a direct response to the evolution
of the physics stage, i.e.\ experimental results and studies for 
future accelerators. Hopefully, the program will keep on evolving
in step with the new challenges opening up.

In the longer future, a radically new version of the program is required. 
Given the decisions by the big laboratories and collaborations to 
discontinue the use of Fortran and instead adopt C++, it is 
natural to attempt to move also event generators in that direction. 
User-friendly interfaces will have to hide the considerable underlying 
complexity from the non-expert. The {\Py}~7 project got going in the
beginning of 1998, and is an effort to reformulate the event generation 
process in object oriented language. Even if much of the physics will
be carried over unchanged, none of the existing code will survive.
The structure of the event record and the whole administrative
apparatus is completely different from the current one, in order
to allow a much more general and flexible formulation of the
event generation process. A strategy document \cite{Lon99} 
was followed by a first `proof of concept version' in June 2000 
\cite{Ber01}, containing the generic event generation machinery, some 
processes, and the string fragmentation routines. In the next few years, 
the hope is to produce useful versions, even if still limited in scope. 
Due to the considerable complexity of the undertaking, it will still be 
several years before the C++ version of {\Py} will contain more 
and better physics than the Fortran one. The two
versions therefore will coexist for several years, with the
Fortran one used for physics `production' and the C++
one for exploration of the object-oriented approach that will
be standard at the LHC.

\clearpage

\section*{Subprocess Summary Table}
\addcontentsline{toc}{section}{Subprocess Summary Table}

This index is intended to give a quick reference to the 
different physics processes implemented in the program. 
Further details are to be found elsewhere in the manual, 
especially in section \ref{s:pytproc}. A trailing '+' on a few
SUSY processes indicates inclusion of charge-conjugate modes 
as well.

\begin{tabular}[t]{|rl|@{\protect\rule[-2mm]{0mm}{6mm}}}
\hline
No. & Subprocess  \\ 
\hline
\multicolumn{2}{|l|@{\protect\rule[-2mm]{0mm}{7mm}}}{Hard QCD processes:} \\
 11 & $\f_i \f_j \to \f_i \f_j$ \\
 12 & $\f_i \fbar_i \to \f_k \fbar_k$ \\
 13 & $\f_i \fbar_i \to \g \g$  \\
 28 & $\f_i \g \to \f_i \g$   \\
 53 & $\g \g \to \f_k \fbar_k$  \\
 68 & $\g \g \to \g \g$  \\ 
\hline
\multicolumn{2}{|l|@{\protect\rule[-2mm]{0mm}{7mm}}}{Soft QCD processes:} \\
 91 & elastic scattering  \\
 92 & single diffraction ($XB$)  \\
 93 & single diffraction ($AX$)  \\
 94 & double diffraction  \\
 95 & low-$p_{\perp}$ production   \\ 
\hline
\multicolumn{2}{|l|@{\protect\rule[-2mm]{0mm}{7mm}}}{Open heavy flavour:} \\
\multicolumn{2}{|l|@{\protect\rule[-2mm]{0mm}{6mm}}}{(also fourth 
generation)} \\
 81 & $\f_i \fbar_i \to \Q_k \Qbar_k$  \\
 82 & $\g \g \to \Q_k \Qbar_k$  \\
 83 & $\q_i \f_j \to \Q_k \f_l$  \\
 84 & $\g \gamma \to \Q_k \Qbar_k$  \\
 85 & $\gamma \gamma \to \F_k \Fbar_k$  \\
\hline
\multicolumn{2}{|l|@{\protect\rule[-2mm]{0mm}{7mm}}}{Closed heavy flavour:} \\
 86 & $\g \g \to \J/\psi \g$  \\
 87 & $\g \g \to \chi_{0\c} \g$   \\
 88 & $\g \g \to \chi_{1\c} \g$  \\
 89 & $\g \g \to \chi_{2\c} \g$   \\
104 & $\g \g \to \chi_{0\c}$ \\   
105 & $\g \g \to \chi_{2\c}$ \\ 
106 & $\g \g \to \J/\psi \gamma$  \\
107 & $\g \gamma \to \J/\psi \g$  \\
108 & $\gamma \gamma \to \J/\psi \gamma$  \\
\hline
\end{tabular}
\hfill
\begin{tabular}[t]{|rl|@{\protect\rule[-2mm]{0mm}{6mm}}}
\hline
No. & Subprocess \\ 
\hline
\multicolumn{2}{|l|@{\protect\rule[-2mm]{0mm}{7mm}}}{$\W / \Z$ production:} \\
  1 & $\f_i \fbar_i \to \gammaZ$  \\
  2 & $\f_i \fbar_j \to \W^{\pm}$  \\
 22 & $\f_i \fbar_i \to \Z^0 \Z^0$   \\
 23 & $\f_i \fbar_j \to \Z^0 \W^{\pm}$   \\
 25 & $\f_i \fbar_i \to \W^+ \W^-$  \\
 15 & $\f_i \fbar_i \to \g \Z^0$  \\
 16 & $\f_i \fbar_j \to \g \W^{\pm}$   \\
 30 & $\f_i \g \to \f_i \Z^0$  \\
 31 & $\f_i \g \to \f_k \W^{\pm}$   \\
 19 & $\f_i \fbar_i \to \gamma \Z^0$   \\
 20 & $\f_i \fbar_j \to \gamma \W^{\pm}$   \\
 35 & $\f_i \gamma \to \f_i \Z^0$   \\
 36 & $\f_i \gamma \to \f_k \W^{\pm}$  \\
 69 & $\gamma \gamma \to \W^+ \W^-$  \\
 70 & $\gamma \W^{\pm} \to \Z^0 \W^{\pm}$  \\
\hline 
\multicolumn{2}{|l|@{\protect\rule[-2mm]{0mm}{7mm}}}{Prompt photons:} \\
 14 & $\f_i \fbar_i \to \g \gamma$  \\
 18 & $\f_i \fbar_i \to \gamma \gamma$   \\
 29 & $\f_i \g \to \f_i \gamma$   \\
114 & $\g \g \to \gamma \gamma$  \\
115 & $\g \g \to \g \gamma$   \\
\hline
\multicolumn{2}{|l|@{\protect\rule[-2mm]{0mm}{7mm}}}{Deeply Inel. Scatt.:} \\
 10 & $\f_i \f_j \to \f_k \f_l$  \\
 99 & $\gast \q \to \q$ \\
\hline
\multicolumn{2}{|l|@{\protect\rule[-2mm]{0mm}{7mm}}}{Photon-induced:} \\
 33 & $\f_i \gamma \to \f_i \g$  \\
 34 & $\f_i \gamma \to \f_i \gamma$   \\
 54 & $\g \gamma \to \f_k \fbar_k$  \\
 58 & $\gamma \gamma \to \f_k \fbar_k$   \\
131 & $\f_i \gast_{\mrm{T}} \to \f_i \g$ \\
\hline
\end{tabular}
\hfill
\begin{tabular}[t]{|rl|@{\protect\rule[-2mm]{0mm}{6mm}}}
\hline
No. & Subprocess \\ 
\hline
132 & $\f_i \gast_{\mrm{L}} \to \f_i \g$ \\
133 & $\f_i \gast_{\mrm{T}} \to \f_i \gamma$ \\
134 & $\f_i \gast_{\mrm{L}} \to \f_i \gamma$ \\
135 & $\g \gast_{\mrm{T}} \to \f_i \fbar_i$ \\
136 & $\g \gast_{\mrm{L}} \to \f_i \fbar_i$ \\
137 & $\gast_{\mrm{T}} \gast_{\mrm{T}} \to \f_i \fbar_i$ \\
138 & $\gast_{\mrm{T}} \gast_{\mrm{L}} \to \f_i \fbar_i$ \\
139 & $\gast_{\mrm{L}} \gast_{\mrm{T}} \to \f_i \fbar_i$ \\
140 & $\gast_{\mrm{L}} \gast_{\mrm{L}} \to \f_i \fbar_i$ \\
 80 & $\q_i \gamma \to \q_k \pi^{\pm}$ \\
\hline
\multicolumn{2}{|l|@{\protect\rule[-2mm]{0mm}{7mm}}}{Light SM Higgs:} \\
  3 & $\f_i \fbar_i \to \hrm^0$  \\
 24 & $\f_i \fbar_i \to \Z^0 \hrm^0$  \\
 26 & $\f_i \fbar_j \to \W^{\pm} \hrm^0$  \\
 32 & $\f_i \g \to \f_i \hrm^0$   \\
102 & $\g \g \to \hrm^0$   \\
103 & $\gamma \gamma \to \hrm^0$  \\
110 & $\f_i \fbar_i \to \gamma \hrm^0$  \\
111 & $\f_i \fbar_i \to \g \hrm^0$  \\
112 & $\f_i \g \to \f_i \hrm^0$  \\
113 & $\g \g \to \g \hrm^0$  \\
121 & $\g \g \to \Q_k \Qbar_k \hrm^0$  \\
122 & $\q_i \qbar_i \to \Q_k \Qbar_k \hrm^0$  \\
123 & $\f_i \f_j \to \f_i \f_j \hrm^0$  \\
124 & $\f_i \f_j \to \f_k \f_l \hrm^0$  \\
\hline
\multicolumn{2}{|l|@{\protect\rule[-2mm]{0mm}{7mm}}}{Heavy SM Higgs:} \\
  5 & $\Z^0 \Z^0 \to \hrm^0$  \\
  8 & $\W^+ \W^- \to \hrm^0$  \\
 71 & $\Z^0_{\mrm{L}} \Z^0_{\mrm{L}} \to \Z^0_{\mrm{L}} 
\Z^0_{\mrm{L}}$ \\
 72 & $\Z^0_{\mrm{L}} \Z^0_{\mrm{L}} \to \W^+_{\mrm{L}} 
\W^-_{\mrm{L}}$ \\
 73 & $\Z^0_{\mrm{L}} \W^{\pm}_{\mrm{L}} \to \Z^0_{\mrm{L}} 
\W^{\pm}_{\mrm{L}}$  \\
 76 & $\W^+_{\mrm{L}} \W^-_{\mrm{L}} \to \Z^0_{\mrm{L}} 
\Z^0_{\mrm{L}}$  \\
 77 & $\W^{\pm}_{\mrm{L}} \W^{\pm}_{\mrm{L}} \to 
\W^{\pm}_{\mrm{L}} \W^{\pm}_{\mrm{L}}$  \\
\hline
\end{tabular}

\newpage

\begin{tabular}[t]{|rl|@{\protect\rule[-2mm]{0mm}{6mm}}}
\hline
No. & Subprocess \\ 
\hline
\multicolumn{2}{|l|@{\protect\rule[-2mm]{0mm}{7mm}}}{BSM Neutral Higgs:} \\
151 & $\f_i \fbar_i \to \H^0$ \\
152 & $\g \g \to \H^0$  \\
153 & $\gamma \gamma \to \H^0$  \\
171 & $\f_i \fbar_i \to \Z^0 \H^0$  \\
172 & $\f_i \fbar_j \to \W^{\pm} \H^0$  \\
173 & $\f_i \f_j \to \f_i \f_j \H^0$   \\
174 & $\f_i \f_j \to \f_k \f_l \H^0$  \\
181 & $\g \g \to \Q_k \Qbar_k \H^0$  \\
182 & $\q_i \qbar_i \to \Q_k \Qbar_k \H^0$   \\
183 & $\f_i \fbar_i \to \g \H^0$  \\
184 & $\f_i \g \to \f_i \H^0$  \\
185 & $\g \g \to \g \H^0$  \\
156 & $\f_i \fbar_i \to \A^0$  \\
157 & $\g \g \to \A^0$  \\
158 & $\gamma \gamma \to \A^0$  \\
176 & $\f_i \fbar_i \to \Z^0 \A^0$   \\
177 & $\f_i \fbar_j \to \W^{\pm} \A^0$  \\
178 & $\f_i \f_j \to \f_i \f_j \A^0$  \\
179 & $\f_i \f_j \to \f_k \f_l \A^0$   \\
186 & $\g \g \to \Q_k \Qbar_k \A^0$ \\
187 & $\q_i \qbar_i \to \Q_k \Qbar_k \A^0$  \\
188 & $\f_i \fbar_i \to \g \A^0$  \\
189 & $\f_i \g \to \f_i \A^0$  \\
190 & $\g \g \to \g \A^0$  \\
\hline
\multicolumn{2}{|l|@{\protect\rule[-2mm]{0mm}{7mm}}}{Charged Higgs:} \\
143 & $\f_i \fbar_j \to \H^+$  \\
161 & $\f_i \g \to \f_k \H^+$  \\
\hline
\multicolumn{2}{|l|@{\protect\rule[-2mm]{0mm}{7mm}}}{Higgs pairs:} \\
297 & $\f_i \fbar_j \to \H^{\pm} \hrm^0$ \\ 
298 & $\f_i \fbar_j \to \H^{\pm} \H^0$ \\ 
299 & $\f_i \fbar_i \to \A^0 \hrm^0$ \\ 
300 & $\f_i \fbar_i \to \A^0 \H^0$ \\ 
301 & $\f_i \fbar_i \to \H^+ \H^-$ \\ 
\hline
\end{tabular}
\hfill
\begin{tabular}[t]{|rl|@{\protect\rule[-2mm]{0mm}{6mm}}}
\hline
No. & Subprocess \\ 
\hline
\multicolumn{2}{|l|@{\protect\rule[-2mm]{0mm}{7mm}}}{New gauge bosons:} \\
141 & $\f_i \fbar_i \to \gamma/\Z^0/{\Z'}^0$ \\
142 & $\f_i \fbar_j \to {\W'}^+$ \\
144 & $\f_i \fbar_j \to \R$  \\
\hline
\multicolumn{2}{|l|@{\protect\rule[-2mm]{0mm}{7mm}}}{Technicolor:} \\
149 & $\g \g \to \eta_{\mrm{tc}}$   \\
191 & $\f_i \fbar_i \to \rho_{\mrm{tc}}^0$   \\
192 & $\f_i \fbar_j \to \rho_{\mrm{tc}}^+$   \\
193 & $\f_i \fbar_i \to \omega_{\mrm{tc}}^0$   \\
194 & $\f_i \fbar_i \to \f_k \fbar_k$   \\
195 & $\f_i \fbar_j \to \f_k \fbar_l$   \\
361 & $\f_i \fbar_i \to \W^+_{\mrm{L}} \W^-_{\mrm{L}} $  \\
362 & $\f_i \fbar_i \to \W^{\pm}_{\mrm{L}} \pi^{\mp}_{\mrm{tc}}$   \\
363 & $\f_i \fbar_i \to \pi^+_{\mrm{tc}} \pi^-_{\mrm{tc}}$   \\
364 & $\f_i \fbar_i \to \gamma \pi^0_{\mrm{tc}} $   \\
365 & $\f_i \fbar_i \to \gamma {\pi'}^0_{\mrm{tc}} $   \\
366 & $\f_i \fbar_i \to \Z^0 \pi^0_{\mrm{tc}} $   \\
367 & $\f_i \fbar_i \to \Z^0 {\pi'}^0_{\mrm{tc}} $   \\
368 & $\f_i \fbar_i \to \W^{\pm} \pi^{\mp}_{\mrm{tc}}$ \\
370 & $\f_i \fbar_j \to \W^{\pm}_{\mrm{L}} \Z^0_{\mrm{L}} $  \\
371 & $\f_i \fbar_j \to \W^{\pm}_{\mrm{L}} \pi^0_{\mrm{tc}}$   \\
372 & $\f_i \fbar_j \to \pi^{\pm}_{\mrm{tc}} \Z^0_{\mrm{L}} $ \\
373 & $\f_i \fbar_j \to \pi^{\pm}_{\mrm{tc}} \pi^0_{\mrm{tc}} $   \\
374 & $\f_i \fbar_j \to \gamma \pi^{\pm}_{\mrm{tc}} $   \\
375 & $\f_i \fbar_j \to \Z^0 \pi^{\pm}_{\mrm{tc}} $   \\
376 & $\f_i \fbar_j \to \W^{\pm} \pi^0_{\mrm{tc}} $   \\
377 & $\f_i \fbar_j \to \W^{\pm} {\pi'}^0_{\mrm{tc}}$ \\
381 & $\q_i \q_j \to \q_i \q_j$  \\
382 & $\q_i \qbar_i \to \q_k \qbar_k$ \\
383 & $\q_i \qbar_i \to \g \g$  \\
384 & $\f_i \g \to \f_i \g$  \\
385 & $\g \g \to \q_k \qbar_k$   \\
386 & $\g \g \to \g \g$ \\
387 & $\f_i \fbar_i \to \Q_k \Qbar_k$  \\
388 & $\g \g \to \Q_k \Qbar_k$  \\
\hline
\end{tabular}
\hfill
\begin{tabular}[t]{|rl|@{\protect\rule[-2mm]{0mm}{6mm}}}
\hline
No. & Subprocess \\ 
\hline
\multicolumn{2}{|l|@{\protect\rule[-2mm]{0mm}{7mm}}}{Compositeness:} \\
146 & $\e \gamma \to \e^*$  \\
147 & $\d \g \to \d^*$  \\
148 & $\u \g \to \u^*$  \\
167 & $\q_i \q_j \to \d^* \q_k$  \\
168 & $\q_i \q_j \to \u^* \q_k$  \\
169 & $\q_i \qbar_i \to \e^{\pm} \e^{*\mp}$  \\
165 & $\f_i \fbar_i (\to \gast/\Z^0) \to \f_k \fbar_k$  \\
166 & $\f_i \fbar_j (\to \W^{\pm}) \to \f_k \fbar_l$ \\
\hline
\multicolumn{2}{|l|@{\protect\rule[-2mm]{0mm}{7mm}}}{Leptoquarks:} \\
145 & $\q_i \ell_j \to \L_{\Q}$  \\
162 & $\q \g \to \ell \L_{\Q}$  \\
163 & $\g \g \to \L_{\Q} \br{\L}_{\Q}$  \\
164 & $\q_i \qbar_i \to \L_{\Q} \br{\L}_{\Q}$ \\
\hline
\multicolumn{2}{|l|@{\protect\rule[-2mm]{0mm}{7mm}}}{Left--right symmetry:} \\
341 & $\ell_i \ell_j \to \H_L^{\pm\pm}$ \\
342 & $\ell_i \ell_j \to \H_R^{\pm\pm}$ \\
343 & $\ell_i^{\pm} \gamma \to \H_L^{\pm\pm} \e^{\mp}$ \\
344 & $\ell_i^{\pm} \gamma \to \H_R^{\pm\pm} \e^{\mp}$ \\
345 & $\ell_i^{\pm} \gamma \to \H_L^{\pm\pm} \mu^{\mp}$ \\
346 & $\ell_i^{\pm} \gamma \to \H_R^{\pm\pm} \mu^{\mp}$ \\
347 & $\ell_i^{\pm} \gamma \to \H_L^{\pm\pm} \tau^{\mp}$ \\
348 & $\ell_i^{\pm} \gamma \to \H_R^{\pm\pm} \tau^{\mp}$ \\
349 & $\f_i \fbar_i \to \H_L^{++} \H_L^{--}$ \\ 
350 & $\f_i \fbar_i \to \H_R^{++} \H_R^{--}$ \\ 
351 & $\f_i \f_j \to \f_k \f_l \H_L^{\pm\pm}$  \\
352 & $\f_i \f_j \to \f_k \f_l \H_R^{\pm\pm}$  \\
353 & $\f_i \fbar_i \to \Z_R^0$ \\
354 & $\f_i \fbar_j \to \W_R^{\pm}$ \\
\hline
\multicolumn{2}{|l|@{\protect\rule[-2mm]{0mm}{7mm}}}{Extra Dimensions:} \\
391 & $\f \fbar \to \G^*$ \\
392 & $\g \g \to \G^*$ \\
393 & $\q \qbar \to \g \G^*$ \\
394 & $\q \g \to \q \G^*$ \\
395 & $\g \g \to \g \G^*$ \\
\hline
\end{tabular}

\newpage

\begin{tabular}[t]{|rl|@{\protect\rule[-2mm]{0mm}{6mm}}}
\hline
No. & Subprocess \\ 
\hline
\multicolumn{2}{|l|@{\protect\rule[-2mm]{0mm}{7mm}}}{SUSY:} \\
201 & $\f_i \fbar_i \to \se_L \se_L^*$  \\
202 & $\f_i \fbar_i \to \se_R \se_R^*$  \\
203 & $\f_i \fbar_i \to \se_L \se_R^* +$ \\
204 & $\f_i \fbar_i \to \smu_L \smu_L^*$ \\
205 & $\f_i \fbar_i \to \smu_R \smu_R^*$ \\
206 & $\f_i \fbar_i\to\smu_L \smu_R^* +$ \\
207 & $\f_i \fbar_i\to\stau_1 \stau_1^*$  \\
208 & $\f_i \fbar_i\to\stau_2 \stau_2^*$  \\
209 & $\f_i \fbar_i\to\stau_1 \stau_2^* +$ \\
210 & $\f_i \fbar_j\to \sell_L {\snu}_{\ell}^* +$\\
211 & $\f_i \fbar_j\to \stau_1\tilde{\nu}_{\tau}^* +$ \\
212 & $\f_i \fbar_j\to \stau_2\tilde{\nu}_{\tau}^* +$\\
213 & $\f_i \fbar_i\to \tilde{\nu_{\ell}} \tilde{\nu_{\ell}}^*$  \\
214 & $\f_i \fbar_i\to \tilde{\nu}_{\tau} \tilde{\nu}_{\tau}^*$ \\
216 & $\f_i \fbar_i \to \chio_1 \chio_1$  \\
217 & $\f_i \fbar_i \to \chio_2 \chio_2$  \\
218 & $\f_i \fbar_i \to \chio_3 \chio_3$  \\
219 & $\f_i \fbar_i \to \chio_4 \chio_4$  \\
220 & $\f_i \fbar_i \to \chio_1 \chio_2$  \\
221 & $\f_i \fbar_i \to \chio_1 \chio_3$  \\
222 & $\f_i \fbar_i \to \chio_1 \chio_4$  \\
223 & $\f_i \fbar_i \to \chio_2 \chio_3$  \\
224 & $\f_i \fbar_i \to \chio_2 \chio_4$  \\
225 & $\f_i \fbar_i \to \chio_3 \chio_4$  \\
226 & $\f_i \fbar_i \to \chip_1 \chim_1$  \\
227 & $\f_i \fbar_i \to \chip_2 \chim_2$  \\
228 & $\f_i \fbar_i \to \chip_1 \chim_2$  \\
229 & $\f_i \fbar_j \to \chio_1 \chip_1$  \\
\hline
\end{tabular}
\hfill
\begin{tabular}[t]{|rl|@{\protect\rule[-2mm]{0mm}{6mm}}}
\hline
No. & Subprocess \\ 
\hline
230 & $\f_i \fbar_j \to \chio_2 \chip_1$  \\
231 & $\f_i \fbar_j \to \chio_3 \chip_1$  \\
232 & $\f_i \fbar_j \to \chio_4 \chip_1$  \\
233 & $\f_i \fbar_j \to \chio_1 \chip_2$  \\
234 & $\f_i \fbar_j \to \chio_2 \chip_2$  \\
235 & $\f_i \fbar_j \to \chio_3 \chip_2$  \\
236 & $\f_i \fbar_j \to \chio_4 \chip_2$  \\
237 & $\f_i \fbar_i \to \glu \chio_1$    \\
238 & $\f_i \fbar_i \to \glu \chio_2$    \\
239 & $\f_i \fbar_i \to \glu \chio_3$    \\
240 & $\f_i \fbar_i \to \glu \chio_4$    \\
241 & $\f_i \fbar_j \to \glu \chip_1$    \\
242 & $\f_i \fbar_j \to \glu \chip_2$    \\
243 & $\f_i \fbar_i \to \glu \glu$\\
244 & $\g \g \to \glu \glu$\\
246 & $\f_i \g \to {\sq_i}{}_L \chio_1$\\
247 & $\f_i \g \to {\sq_i}{}_R \chio_1$\\
248 & $\f_i \g \to {\sq_i}{}_L \chio_2$\\
249 & $\f_i \g \to {\sq_i}{}_R \chio_2$\\
250 & $\f_i \g \to {\sq_i}{}_L \chio_3$\\
251 & $\f_i \g \to {\sq_i}{}_R \chio_3$\\
252 & $\f_i \g \to {\sq_i}{}_L \chio_4$\\
253 & $\f_i \g \to {\sq_i}{}_R \chio_4$\\
254 & $\f_i \g \to {\sq_j}{}_L \chip_1$\\
256 & $\f_i \g \to {\sq_j}{}_L \chip_2$\\
258 & $\f_i \g \to {\sq_i}{}_L \glu$ \\
259 & $\f_i \g \to {\sq_i}{}_R \glu$ \\
261 & $\f_i \fbar_i \to \tp_1 \tm_1$\\
262 & $\f_i \fbar_i \to \tp_2 \tm_2$\\
\hline
\end{tabular}
\hfill
\begin{tabular}[t]{|rl|@{\protect\rule[-2mm]{0mm}{6mm}}}
\hline
No. & Subprocess \\ 
\hline
263 & $\f_i \fbar_i \to \tp_1 \tm_2 +$\\
264 & $\g \g \to \tp_1 \tm_1$\\
265 & $\g \g \to \tp_2 \tm_2$\\
271 & $\f_i \f_j \to {\sq_i}{}_L {\sq_j}{}_L$\\
272 & $\f_i \f_j \to {\sq_i}{}_R {\sq_j}{}_R$\\
273 & $\f_i \f_j \to {\sq_i}{}_L {\sq_j}{}_R +$\\
274 & $\f_i \fbar_j \to {\sq_i}{}_L {\sqs_j}{}_L$\\
275 & $\f_i \fbar_j \to {\sq_i}{}_R {\sqs_j}{}_R$\\
276 & $\f_i \fbar_j \to {\sq_i}{}_L {\sqs_j}{}_R +$\\
277 & $\f_i \fbar_i \to {\sq_j}{}_L {\sqs_j}{}_L$\\
278 & $\f_i \fbar_i \to {\sq_j}{}_R {\sqs_j}{}_R$\\
279 & $\g \g \to {\sq_i}{}_L {\sqs_i}{}_L$\\
280 & $\g \g \to {\sq_i}{}_R {\sqs_i}{}_R$\\
281  & $\b \q_i \to \sbo_1 {\sq_i}{}_L$\\
282  & $\b \q_i \to \sbo_2 {\sq_i}{}_R$\\
283  & $\b \q_i \to \sbo_1 {\sq_i}{}_R + \sbo_2 {\sq_i}{}_L$\\
284  & $\b \qbar_i \to \sbo_1 {\sqs_i}{}_L$\\
285  & $\b \qbar_i \to \sbo_2 {\sqs_i}{}_R$\\
286  & $\b \qbar_i \to \sbo_1 {\sqs_i}{}_R + \sbo_2 {\sqs_i}{}_L$\\
287  & $\f_i \fbar_i \to \sbo_1 \sbs_1$\\
288  & $\f_i \fbar_i \to \sbo_2 \sbs_2$\\
289  & $\g \g \to \sbo_1 \sbs_1$\\
290  & $\g \g \to \sbo_2 \sbs_2$\\
291  & $\b \b \to \sbo_1 \sbo_1$\\
292  & $\b \b \to \sbo_2 \sbo_2$\\
293  & $\b \b \to \sbo_1 \sbo_2$\\
294  & $\b \g \to \sbo_1 \glu$\\
295  & $\b \g \to \sbo_2 \glu$\\
296  & $\b \bbar \to \sbo_1 \sbs_2 +$\\ 
\hline
\end{tabular}

\clearpage

\section*{Index of Subprograms and Common Block Variables}
\addcontentsline{toc}{section}%
{Index of Subprograms and Common Block Variables}

This index is not intended to be complete, but gives the page where
the main description begins of a subroutine, function, block data, 
common block, variable or array. For common block variables also the
name of the common block is given. When some components of an array 
are described in a separate place, a special reference (indented with
respect to the main one) is given for these components.

\boxsep

\noindent
\begin{minipage}[t]{\halfpagewid}
\begin{tabular*}{\halfpagewid}[t]{@{}l@{\extracolsep{\fill}}r@{}}
\ttt{AQCDUP} in \ttt{HEPEUP}     & \pageref{p:AQCDUP} \\
\ttt{AQEDUP} in \ttt{HEPEUP}     & \pageref{p:AQEDUP} \\
\ttt{BRAT} in \ttt{PYDAT3}       & \pageref{p:BRAT} \\
\ttt{CHAF} in \ttt{PYDAT4}       & \pageref{p:CHAF} \\
\ttt{CKIN} in \ttt{PYSUBS}       & \pageref{p:CKIN} \\
\ttt{COEF} in \ttt{PYINT2}       & \pageref{p:COEF} \\
\ttt{EBMUP} in \ttt{HEPRUP}      & \pageref{p:EBMUP} \\
\ttt{HEPEUP} common block        & \pageref{p:HEPEUP} \\
\ttt{HEPEVT} common block        & \pageref{p:HEPEVT} \\
\ttt{HEPRUP} common block        & \pageref{p:HEPRUP} \\
\ttt{ICOL} in \ttt{PYINT2}       & \pageref{p:ICOL} \\
\ttt{ICOLUP} in \ttt{HEPEUP}     & \pageref{p:ICOLUP} \\
\ttt{IDBMUP} in \ttt{HEPRUP}     & \pageref{p:IDBMUP} \\
\ttt{IDPRUP} in \ttt{HEPEUP}     & \pageref{p:IDPRUP} \\
\ttt{IDUP} in \ttt{HEPEUP}       & \pageref{p:IDUP} \\
\ttt{IDWTUP} in \ttt{HEPRUP}     & \pageref{p:IDWTUP} \\
\ttt{IMSS} in \ttt{PYMSSM}       & \pageref{p:IMSS} \\
\ttt{ISET} in \ttt{PYINT2}       & \pageref{p:ISET} \\
\ttt{ISIG} in \ttt{PYINT3}       & \pageref{p:ISIG} \\
\ttt{ISTUP} in \ttt{HEPEUP}      & \pageref{p:ISTUP} \\
\ttt{ITCM} in \ttt{PYTCSM}       & \pageref{p:ITCM} \\
\ttt{K} in \ttt{PYJETS}          & \pageref{p:K} \\
\ttt{KCHG} in \ttt{PYDAT2}       & \pageref{p:KCHG} \\
\ttt{KFDP} in \ttt{PYDAT3}       & \pageref{p:KFDP} \\
\ttt{KFIN} in \ttt{PYSUBS}       & \pageref{p:KFIN} \\
\ttt{KFPR} in \ttt{PYINT2}       & \pageref{p:KFPR} \\
\ttt{LPRUP} in \ttt{HEPRUP}      & \pageref{p:LPRUP} \\
\ttt{MAXNUP} in \ttt{HEPEUP}     & \pageref{p:MAXNUP} \\
\ttt{MAXPUP} in \ttt{HEPRUP}     & \pageref{p:MAXPUP} \\
\ttt{MDCY} in \ttt{PYDAT3}       & \pageref{p:MDCY} \\
\ttt{MDME} in \ttt{PYDAT3}       & \pageref{p:MDME} \\
\ttt{MINT} in \ttt{PYINT1}       & \pageref{p:MINT} \\
\ttt{MOTHUP} in \ttt{HEPEUP}     & \pageref{p:MOTHUP} \\
\ttt{MRPY} in \ttt{PYDATR}       & \pageref{p:MRPY} \\
\ttt{MSEL} in \ttt{PYSUBS}       & \pageref{p:MSEL} \\
\ttt{MSTI} in \ttt{PYPARS}       & \pageref{p:MSTI} \\
\ttt{MSTJ} in \ttt{PYDAT1}, main & \pageref{p:MSTJ} \\
~~~\ttt{MSTJ(38) - MSTJ(50)}     & \pageref{p:MSTJ38} \\
~~~\ttt{MSTJ(101) - MSTJ(121)}   & \pageref{p:MSTJ101} \\
\ttt{MSTP} in \ttt{PYPARS}, main & \pageref{p:MSTP} \\
~~~\ttt{MSTP(22)}                & \pageref{p:MSTP22} \\
~~~\ttt{MSTP(61) - MSTP(71)}     & \pageref{p:MSTP61} \\
~~~\ttt{MSTP(81) - MSTP(95)}     & \pageref{p:MSTP81} \\
\end{tabular*}
\end{minipage}%
\hfill%
\begin{minipage}[t]{\halfpagewid}
\begin{tabular*}{\halfpagewid}[t]{@{}l@{\extracolsep{\fill}}r@{}}
~~~\ttt{MSTP(131) - MSTP(134)}   & \pageref{p:MSTP131} \\
\ttt{MSTU} in \ttt{PYDAT1}, main & \pageref{p:MSTU} \\
~~~\ttt{MSTU(1)} and some more   & \pageref{p:MSTU1} \\
~~~\ttt{MSTU(41) - MSTU(63)}     & \pageref{p:MSTU41} \\
~~~\ttt{MSTU(101) - MSTU(118)}   & \pageref{p:MSTU101} \\
~~~\ttt{MSTU(161) - MSTU(162)}   & \pageref{p:MSTU161} \\
\ttt{MSUB} in \ttt{PYSUBS}       & \pageref{p:MSUB} \\
\ttt{MWID} in \ttt{PYINT4}       & \pageref{p:MWID} \\
\ttt{N} in \ttt{PYJETS}          & \pageref{p:N} \\
\ttt{NGEN} in \ttt{PYINT5}       & \pageref{p:NGEN} \\
\ttt{NPRUP} in \ttt{HEPRUP}      & \pageref{p:NPRUP} \\
\ttt{NUP} in \ttt{HEPEUP}        & \pageref{p:NUP} \\
\ttt{P} in \ttt{PYJETS}          & \pageref{p:P} \\
\ttt{PARF} in \ttt{PYDAT2}       & \pageref{p:PARF} \\
\ttt{PARI} in \ttt{PYPARS}       & \pageref{p:PARI} \\
\ttt{PARJ} in \ttt{PYDAT1}, main & \pageref{p:PARJ} \\
~~~\ttt{PARJ(80) - PARJ(90)}     & \pageref{p:PARJ80} \\
~~~\ttt{PARJ(121) - PARJ(171)}   & \pageref{p:PARJ121} \\
~~~\ttt{PARJ(180) - PARJ(195)}   & \pageref{p:PARJ180} \\
\ttt{PARP} in \ttt{PYPARS}, main & \pageref{p:PARP} \\
~~~\ttt{PARP(61) - PARP(72)}     & \pageref{p:PARP61} \\
~~~\ttt{PARP(78) - PARP(100)}    & \pageref{p:PARP78} \\
~~~\ttt{PARP(131)}               & \pageref{p:PARP131} \\
\ttt{PARU} in \ttt{PYDAT1}, main & \pageref{p:PARU} \\
~~~\ttt{PARU(41) - PARU(63)}     & \pageref{p:PARU41} \\
~~~\ttt{PARU(101) - PARU(195)}   & \pageref{p:PARU101} \\
\ttt{PDFGUP} in \ttt{HEPRUP}     & \pageref{p:PDFGUP} \\
\ttt{PDFSUP} in \ttt{HEPRUP}     & \pageref{p:PDFSUP} \\
\ttt{PMAS} in \ttt{PYDAT2}       & \pageref{p:PMAS} \\
\ttt{PROC} in \ttt{PYINT6}       & \pageref{p:PROC} \\
\ttt{PUP} in \ttt{HEPEUP}        & \pageref{p:PUP} \\
\ttt{PY1ENT} subroutine          & \pageref{p:PY1ENT} \\
\ttt{PY2ENT} subroutine          & \pageref{p:PY2ENT} \\
\ttt{PY3ENT} subroutine          & \pageref{p:PY3ENT} \\
\ttt{PY4ENT} subroutine          & \pageref{p:PY4ENT} \\
\ttt{PY2FRM} subroutine          & \pageref{p:PY2FRM} \\
\ttt{PY4FRM} subroutine          & \pageref{p:PY4FRM} \\
\ttt{PY6FRM} subroutine          & \pageref{p:PY6FRM} \\
\ttt{PY4JET} subroutine          & \pageref{p:PY4JET} \\
\ttt{PYADSH} function            & \pageref{p:PYADSH} \\
\ttt{PYALEM} function            & \pageref{p:PYALEM} \\
\ttt{PYALPS} function            & \pageref{p:PYALPS} \\
\ttt{PYANGL} function            & \pageref{p:PYANGL} \\
\end{tabular*}
\end{minipage} 

\newpage

\noindent
\begin{minipage}[t]{\halfpagewid}
\begin{tabular*}{\halfpagewid}[t]{@{}l@{\extracolsep{\fill}}r@{}}
\ttt{PYBINS} common block        & \pageref{p:PYBINS} \\
\ttt{PYBOEI} subroutine          & \pageref{p:PYBOEI} \\
\ttt{PYBOOK} subroutine          & \pageref{p:PYBOOK} \\
\ttt{PYCBLS} common block        & \pageref{p:PYCBLS} \\
\ttt{PYCELL} subroutine          & \pageref{p:PYCELL} \\
\ttt{PYCHGE} function            & \pageref{p:PYCHGE} \\
\ttt{PYCLUS} subroutine          & \pageref{p:PYCLUS} \\
\ttt{PYCOMP} function            & \pageref{p:PYCOMP} \\
\ttt{PYCTAG} common block        & \pageref{p:PYCTAG} \\
\ttt{PYCTTR} subroutine          & \pageref{p:PYCTTR} \\
\ttt{PYDAT1} common block        & \pageref{p:PYDAT1} \\
\ttt{PYDAT2} common block        & \pageref{p:PYDAT2} \\
\ttt{PYDAT3} common block        & \pageref{p:PYDAT3} \\
\ttt{PYDAT4} common block        & \pageref{p:PYDAT4} \\
\ttt{PYDATA} block data          & \pageref{p:PYDATA} \\
\ttt{PYDATR} common block        & \pageref{p:PYDATR} \\
\ttt{PYDCYK} subroutine          & \pageref{p:PYDCYK} \\
\ttt{PYDECY} subroutine          & \pageref{p:PYDECY} \\
\ttt{PYDIFF} subroutine          & \pageref{p:PYDIFF} \\
\ttt{PYDISG} subroutine          & \pageref{p:PYDISG} \\
\ttt{PYDOCU} subroutine          & \pageref{p:PYDOCU} \\
\ttt{PYDUMP} subroutine          & \pageref{p:PYDUMP} \\
\ttt{PYEDIT} subroutine          & \pageref{p:PYEDIT} \\
\ttt{PYEEVT} subroutine          & \pageref{p:PYEEVT} \\
\ttt{PYERRM} subroutine          & \pageref{p:PYERRM} \\
\ttt{PYEVNT} subroutine          & \pageref{p:PYEVNT} \\
\ttt{PYEVWT} subroutine          & \pageref{p:PYEVWT} \\
\ttt{PYEXEC} subroutine          & \pageref{p:PYEXEC} \\
\ttt{PYFACT} subroutine          & \pageref{p:PYFACT} \\
\ttt{PYFILL} subroutine          & \pageref{p:PYFILL} \\
\ttt{PYFOWO} subroutine          & \pageref{p:PYFOWO} \\
\ttt{PYFRAM} subroutine          & \pageref{p:PYFRAM} \\
\ttt{PYGAGA} subroutine          & \pageref{p:PYGAGA} \\
\ttt{PYGAMM} function            & \pageref{p:PYGAMM} \\
\ttt{PYGANO} function            & \pageref{p:PYGANO} \\
\ttt{PYGBEH} function            & \pageref{p:PYGBEH} \\
\ttt{PYGDIR} function            & \pageref{p:PYGDIR} \\
\ttt{PYGGAM} function            & \pageref{p:PYGGAM} \\
\ttt{PYGIVE} subroutine          & \pageref{p:PYGIVE} \\
\ttt{PYGVMD} function            & \pageref{p:PYGVMD} \\
\ttt{PYHEPC} subroutine          & \pageref{p:PYHEPC} \\
\ttt{PYHFTH} function            & \pageref{p:PYHFTH} \\
\ttt{PYHIST} subroutine          & \pageref{p:PYHIST} \\
\ttt{PYI3AU} subroutine          & \pageref{p:PYI3AU} \\
\ttt{PYINBM} subroutine          & \pageref{p:PYINBM} \\
\ttt{PYINDF} subroutine          & \pageref{p:PYINDF} \\
\ttt{PYINIT} subroutine          & \pageref{p:PYINIT} \\
\ttt{PYINKI} subroutine          & \pageref{p:PYINKI} \\
\ttt{PYINPR} subroutine          & \pageref{p:PYINPR} \\
\ttt{PYINRE} subroutine          & \pageref{p:PYINRE} \\
\ttt{PYINT1} common block        & \pageref{p:PYINT1} \\
\end{tabular*}
\end{minipage}%
\hfill%
\begin{minipage}[t]{\halfpagewid}
\begin{tabular*}{\halfpagewid}[t]{@{}l@{\extracolsep{\fill}}r@{}}
\ttt{PYINT2} common block        & \pageref{p:PYINT2} \\
\ttt{PYINT3} common block        & \pageref{p:PYINT3} \\
\ttt{PYINT4} common block        & \pageref{p:PYINT4} \\
\ttt{PYINT5} common block        & \pageref{p:PYINT5} \\
\ttt{PYINT6} common block        & \pageref{p:PYINT6} \\
\ttt{PYINT7} common block        & \pageref{p:PYINT7} \\
\ttt{PYINT8} common block        & \pageref{p:PYINT8} \\
\ttt{PYINT9} common block        & \pageref{p:PYINT9} \\
\ttt{PYINTM} common block        & \pageref{p:PYINTM} \\
\ttt{PYJETS} common block        & \pageref{p:PYJETS} \\
\ttt{PYJMAS} subroutine          & \pageref{p:PYJMAS} \\
\ttt{PYJOIN} subroutine          & \pageref{p:PYJOIN} \\
\ttt{PYK}    function            & \pageref{p:PYK} \\
\ttt{PYKCUT} subroutine          & \pageref{p:PYKCUT} \\
\ttt{PYKFDI} subroutine          & \pageref{p:PYKFDI} \\
\ttt{PYKFIN} subroutine          & \pageref{p:PYKFIN} \\
\ttt{PYKLIM} subroutine          & \pageref{p:PYKLIM} \\
\ttt{PYKMAP} subroutine          & \pageref{p:PYKMAP} \\
\ttt{PYLHA3} subroutine          & \pageref{p:PYLHA3} \\
\ttt{PYLIST} subroutine          & \pageref{p:PYLIST} \\
\ttt{PYLOGO} subroutine          & \pageref{p:PYLOGO} \\
\ttt{PYMAEL} function            & \pageref{p:PYMAEL} \\
\ttt{PYMASS} function            & \pageref{p:PYMASS} \\
\ttt{PYMAXI} subroutine          & \pageref{p:PYMAXI} \\
\ttt{PYMEMX} subroutine          & \pageref{p:PYMEMX} \\
\ttt{PYMEWT} subroutine          & \pageref{p:PYMEWT} \\
\ttt{PYMIGN} subroutine          & \pageref{p:PYMIGN} \\
\ttt{PYMIHG} subroutine          & \pageref{p:PYMIHG} \\
\ttt{PYMIHK} subroutine          & \pageref{p:PYMIHK} \\
\ttt{PYMIRM} subroutine          & \pageref{p:PYMIRM} \\
\ttt{PYMRUN} function            & \pageref{p:PYMRUN} \\
\ttt{PYMSRV} common block        & \pageref{p:PYMSRV} \\
\ttt{PYMSSM} common block        & \pageref{p:PYMSSM} \\
\ttt{PYMULT} subroutine          & \pageref{p:PYMULT} \\
\ttt{PYNAME} subroutine          & \pageref{p:PYNAME} \\
\ttt{PYNMES} subroutine          & \pageref{p:PYNMES} \\
\ttt{PYNULL} subroutine          & \pageref{p:PYNULL} \\
\ttt{PYONIA} subroutine          & \pageref{p:PYONIA} \\
\ttt{PYOPER} subroutine          & \pageref{p:PYOPER} \\
\ttt{PYOFSH} subroutine          & \pageref{p:PYOFSH} \\
\ttt{PYP}    function            & \pageref{p:PYP} \\
\ttt{PYPARS} common block        & \pageref{p:PYPARS1},
                                   \pageref{p:PYPARS2} \\ 
\ttt{PYPDEL} subroutine          & \pageref{p:PYPDEL} \\
\ttt{PYPDFL} subroutine          & \pageref{p:PYPDFL} \\
\ttt{PYPDFU} subroutine          & \pageref{p:PYPDFU} \\
\ttt{PYPDGA} subroutine          & \pageref{p:PYPDGA} \\
\ttt{PYPDPI} subroutine          & \pageref{p:PYPDPI} \\
\ttt{PYPDPR} subroutine          & \pageref{p:PYPDPR} \\
\ttt{PYPILE} subroutine          & \pageref{p:PYPILE} \\
\ttt{PYPLOT} subroutine          & \pageref{p:PYPLOT} \\
\ttt{PYPREP} subroutine          & \pageref{p:PYPREP} \\
\end{tabular*}
\end{minipage} 

\newpage

\noindent
\begin{minipage}[t]{\halfpagewid}
\begin{tabular*}{\halfpagewid}[t]{@{}l@{\extracolsep{\fill}}r@{}}
\ttt{PYPTDI} subroutine          & \pageref{p:PYPTDI} \\
\ttt{PYPTFS} subroutine          & \pageref{p:PYPTFS} \\
\ttt{PYQQBH} subroutine          & \pageref{p:PYQQBH} \\
\ttt{PYR} function               & \pageref{p:PYR} \\
\ttt{PYRGET} subroutine          & \pageref{p:PYRGET} \\
\ttt{PYRSET} subroutine          & \pageref{p:PYRSET} \\
\ttt{PYRADK} subroutine          & \pageref{p:PYRADK} \\
\ttt{PYRAND} subroutine          & \pageref{p:PYRAND} \\
\ttt{PYRECO} subroutine          & \pageref{p:PYRECO} \\
\ttt{PYREMN} subroutine          & \pageref{p:PYREMN} \\
\ttt{PYRESD} subroutine          & \pageref{p:PYRESD} \\
\ttt{PYROBO} subroutine          & \pageref{p:PYROBO} \\
\ttt{PYSAVE} subroutine          & \pageref{p:PYSAVE} \\
\ttt{PYSCAT} subroutine          & \pageref{p:PYSCAT} \\
\ttt{PYSHOW} subroutine          & \pageref{p:PYSHOW} \\
\ttt{PYSIGH} subroutine          & \pageref{p:PYSIGH} \\
\ttt{PYSPEN} function            & \pageref{p:PYSPEN} \\
\ttt{PYSPHE} subroutine          & \pageref{p:PYSPHE} \\
\ttt{PYSPLI} subroutine          & \pageref{p:PYSPLI} \\
\ttt{PYSSMT} common block        & \pageref{p:PYSSMT} \\
\ttt{PYSSPA} subroutine          & \pageref{p:PYSSPA} \\
\ttt{PYSTAT} subroutine          & \pageref{p:PYSTAT} \\
\ttt{PYSTRF} subroutine          & \pageref{p:PYSTRF} \\
\ttt{PYSUBS} common block        & \pageref{p:PYSUBS} \\
\ttt{PYTABU} subroutine          & \pageref{p:PYTABU} \\
\ttt{PYTAUD} subroutine          & \pageref{p:PYTAUD} \\
\ttt{PYTCSM} common block        & \pageref{p:PYTCSM} \\
\ttt{PYTEST} subroutine          & \pageref{p:PYTEST} \\
\ttt{PYTHRU} subroutine          & \pageref{p:PYTHRU} \\
\ttt{PYTIME} subroutine          & \pageref{p:PYTIME} \\
\ttt{PYUPDA} subroutine          & \pageref{p:PYUPDA} \\
\ttt{PYUPRE} subroutine          & \pageref{p:PYUPRE} \\
\ttt{PYWAUX} subroutine          & \pageref{p:PYWAUX} \\
\ttt{PYWIDT} subroutine          & \pageref{p:PYWIDT} \\
\ttt{PYX3JT} subroutine          & \pageref{p:PYX3JT} \\
\ttt{PYX4JT} subroutine          & \pageref{p:PYX4JT} \\
\ttt{PYXDIF} subroutine          & \pageref{p:PYXDIF} \\
\ttt{PYXJET} subroutine          & \pageref{p:PYXJET} \\
\ttt{PYXKFL} subroutine          & \pageref{p:PYXKFL} \\
\ttt{PYXTEE} subroutine          & \pageref{p:PYXTEE} \\
\ttt{PYXTOT} subroutine          & \pageref{p:PYXTOT} \\
\ttt{PYZDIS} subroutine          & \pageref{p:PYZDIS} \\
\end{tabular*}
\end{minipage}%
\hfill%
\begin{minipage}[t]{\halfpagewid}
\begin{tabular*}{\halfpagewid}[t]{@{}l@{\extracolsep{\fill}}r@{}}
\ttt{RMSS} in \ttt{PYMSSM}       & \pageref{p:RMSS} \\
\ttt{RRPY} in \ttt{PYDATR}       & \pageref{p:RRPY} \\
\ttt{RTCM} in \ttt{PYTCSM}       & \pageref{p:RTCM} \\
\ttt{RVLAM} in \ttt{PYMSRV}      & \pageref{p:RVLAM} \\
\ttt{RVLAMB} in \ttt{PYMSRV}     & \pageref{p:RVLAMB} \\
\ttt{RVLAMP} in \ttt{PYMSRV}     & \pageref{p:RVLAMP} \\
\ttt{SCALUP} in \ttt{HEPEUP}     & \pageref{p:SCALUP} \\
\ttt{SFMIX} in \ttt{PYSSMT}      & \pageref{p:SFMIX} \\
\ttt{SIGH} in \ttt{PYINT3}       & \pageref{p:SIGH} \\
\ttt{SIGT} in \ttt{PYINT7}       & \pageref{p:SIGT} \\
\ttt{SMW} in \ttt{PYSSMT}        & \pageref{p:SMW} \\
\ttt{SMZ} in \ttt{PYSSMT}        & \pageref{p:SMZ} \\
\ttt{SPINUP} in \ttt{HEPEUP}     & \pageref{p:SPINUP} \\
\ttt{UPEVNT} subroutine          & \pageref{p:UPEVNT} \\
\ttt{UPINIT} subroutine          & \pageref{p:UPINIT} \\
\ttt{V} in \ttt{PYJETS}          & \pageref{p:V} \\
\ttt{VCKM} in \ttt{PYDAT2}       & \pageref{p:VCKM} \\
\ttt{VINT} in \ttt{PYINT1}       & \pageref{p:VINT} \\
\ttt{VTIMUP} in \ttt{HEPEUP}     & \pageref{p:VTIMUP} \\
\ttt{VXPANH} in \ttt{PYINT9}     & \pageref{p:VXPANH} \\
\ttt{VXPANL} in \ttt{PYINT9}     & \pageref{p:VXPANL} \\
\ttt{VXPDGM} in \ttt{PYINT9}     & \pageref{p:VXPDGM} \\
\ttt{VXPVMD} in \ttt{PYINT9}     & \pageref{p:VXPVMD} \\
\ttt{WIDS} in \ttt{PYINT4}       & \pageref{p:WIDS} \\
\ttt{XERRUP} in \ttt{HEPRUP}     & \pageref{p:XERRUP} \\
\ttt{XMAXUP} in \ttt{HEPRUP}     & \pageref{p:XMAXUP} \\
\ttt{XPANH} in \ttt{PYINT8}      & \pageref{p:XPANH} \\
\ttt{XPANL} in \ttt{PYINT8}      & \pageref{p:XPANL} \\
\ttt{XPBEH} in \ttt{PYINT8}      & \pageref{p:XPBEH} \\
\ttt{XPDIR} in \ttt{PYINT8}      & \pageref{p:XPDIR} \\
\ttt{XPVMD} in \ttt{PYINT8}      & \pageref{p:XPVMD} \\
\ttt{XSEC} in \ttt{PYINT5}       & \pageref{p:XSEC} \\
\ttt{XSECUP} in \ttt{HEPRUP}     & \pageref{p:XSECUP} \\
\ttt{XSFX} in \ttt{PYINT3}       & \pageref{p:XSFX} \\
\ttt{XWGTUP} in \ttt{HEPEUP}     & \pageref{p:XWGTUP} \\
\ttt{UMIX} in \ttt{PYSSMT}       & \pageref{p:UMIX} \\
\ttt{UMIXI} in \ttt{PYSSMT}      & \pageref{p:UMIXI} \\
\ttt{VMIX} in \ttt{PYSSMT}       & \pageref{p:VMIX} \\
\ttt{VMIXI} in \ttt{PYSSMT}      & \pageref{p:VMIXI} \\
\ttt{ZMIX} in \ttt{PYSSMT}       & \pageref{p:ZMIX} \\
\ttt{ZMIXI} in \ttt{PYSSMT}      & \pageref{p:ZMIXI} \\
\end{tabular*}
\end{minipage}
 
\end{document}